# Human noise at the fingertip: Positional (non)control under varying haptic × musical conditions

**Staas de Jong**[1]

[1]Universiteit Leiden, Honours Academy

## ABSTRACT

As technologies and interfaces for the instrumental control of musical sound get ever better at tracking aspects of human position and motion in space, a fundamental problem emerges: Unintended or even counter-intentional control may result when humans themselves become a source of positional noise. A clear case of what is meant by this, is the “stillness movement” of a body part, occurring despite the simultaneous explicit intention for that body part to remain still.

In this paper, we present the results of a randomized, controlled experiment investigating this phenomenon along a vertical axis relative to the human fingertip. The results include characterizations of both the spatial distribution and frequency distribution of the stillness movement observed.

Also included are results indicating a possible role for constant forces and viscosities in reducing stillness movement amplitude, thereby potentially enabling the implementation of more positional control of musical sound within the same available spatial range.

Importantly, the above is summarized in a form that is directly interpretable for anyone designing technologies, interactions, or performances that involve fingertip control of musical sound.

Also, a complete data set of the experimental results is included in the separate Appendices to this paper, again in a format that is directly interpretable.

# 1. Introduction

Often when learning to play an instrument, old or new, the goal will be to obtain some level of intentional control over musical sound. Even where the *ultimate* goal is to (partially) relinquish control, as in the uncontrollability being developed in feedback instruments (see e.g. [4] and [2]), the necessity of mastering a level of intentional control still holds. Often, too, whether in traditional instruments or novel interfaces, such control will be directly related to the position of (parts of) the human body in space.

Here, however, a fundamental problem comes into focus, as technologies and interfaces get ever better at tracking aspects of human position and motion in space: Movement will occur, even during episodes in which a person has no intention for the part of their anatomy in question to move – or, more strongly, even when the explicit intention for that body part is to remain still. Here, we will call such movement *stillness movement*.

For new interfaces for musical expression, stillness movement is potentially problematic: It may lead to *unintended control,* where control changes occur which do not correspond to any user intention (e.g. “I have no intention of moving right now”). Stillness movement may also lead to *counter-intentional control,* where control changes occur which *go against* user intention (e.g. “I want to remain completely still right now”).

In short, from the perspective of intentional control, human stillness movement introduces *noise*.

To properly inform control implementations in which control should be intentional, it is therefore useful to experimentally obtain data sets which after analysis will enable statements that, first of all, quantitatively characterize stillness movement noise.

Additionally, it is useful to extend such experimentation to include musical haptics (see e.g. [3]): This simply because in haptic systems with precise position inputs, this kind of noise will be present; also, however, because haptic devices and algorithms with appropriate force outputs may play a role in dealing with human noise where it is a problem, due to their inherent ability to influence the mechanics of movement and touch. A useful extension of experimentation, therefore, would be to also quantify the effects of different haptic conditions on stillness movement, and on the simultaneously resulting musical control.

Below, we present the execution and results of this kind of experimentation, when focusing on a specific type of anatomical movement involving the fingertip: *vertical fingertip movement,* which is here defined as movement that is approximately orthogonal to both the surface of the human fingerpad itself and to some second surface of known position. This anatomical movement is of course very important in human manipulation and control, and integrated in the critical path of many forms of instrumental control of musical sound, from ancient to recent.

To be sure, however, such experimentation will not occur in a contextual vacuum in more ways than this. Hand and finger tremors have been long studied in the medical literature, and this also in the absence of pathology. For example, based on [5], in healthy individuals, we may reasonably expect such finger movement to be mainly occurring in the 1 - 30 Hz frequency range, and to be largely determined by the mechanical properties of the finger itself. There also already has been considerable research into involuntary micromotion in a musical context, for example in [6], where, during standstill, the effect of listening to different types of musical sound on head motion was tracked and studied. A different example is [7], where involuntary micromotion elsewhere in the human body was used to generate auditory feedback, with the aim of inducing an awareness of the related kinaesthetic phenomena themselves. Below, however, we will focus on fingertip movement in the context of the control of musical sound.

# 2. Experiment

## *2.1 Type, setup, and raw measurements*

The randomized, controlled experiment was executed using a Ghostfinger system for computed fingertip haptics (see [1]). This system provides haptic programming primitives for up/down fingertip movement which are simultaneously visualized as stereoscopic holograms, together with a floating cursor tracking the user's fingertip position (see Figure 1). As is illustrated, the system both tracks vertical fingertip movement (in mm) and applies vertical forces (in N) along a $z$ axis above the device surface, via the fingerpad transducer shown.

The weight of the fingerpad transducer was 10 g, with applied forces added to or subtracted from this weight. Tracking of transducer position was based on pulsed reflective infrared sensing, while forces were applied electromagnetically. Peak-to-peak positional amplitude noise was less than 0.2 mm (signal recordings are in the Appendices). The total haptic roundtrip latency from input to output was 4.0 ms, when measured using strict magnetic field strength thresholds of ≥ 99% of target level. Force output had a resolution of ± 0.003 N, this with noise performance aided by the technology completely avoiding the use of connected mechanical parts moving relative to the target anatomical site, and with an underlying capability to produce accurate wave output up to 1000 Hz. For more information, including on how handling the occurrence of lateral fingertip slippage was incorporated in the basic design of the technology, please see [1] and its references.

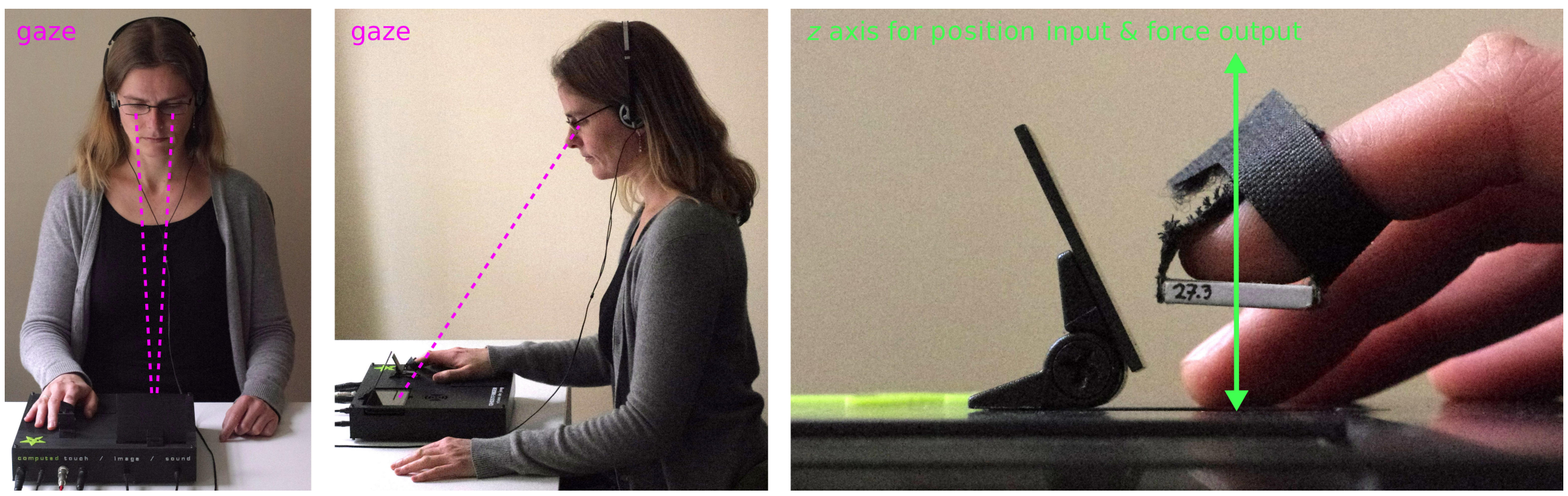

**Figure 1.** Experimental setup: the Ghostfinger system.

Correspondingly, the raw measurements made were fingerpad *z position* (mm) and *z speed* (mm/s) time series, sampled at 4000 Hz. Simultaneously, the system's outputs were used to apply a *target z force* (N) to the fingerpad at a sample rate of 1000 Hz, and produce *headphone audio* at 96 kHz. As regards proper periodic sampling of the *z* (mm) signal, it was experimentally double-checked that any aliased additions (i.e. at frequencies > 2 kHz) added along the various analog and digital stages of the sensing chain stayed below the peak-to-peak threshold used for amplitude noise in general.

## *2.2 Conditions*

A shared property of all experimental conditions, however, was that subjects did *not* receive (computed) visual output during the actual measurement of stillness movement. The idea here was that facilitating an attentional focus on haptic and audio output might help in measuring effects specific to these modalities.

Another property shared across conditions was that fingerpad *z* position always stayed *between* 0.0 and 10.0 mm above device surface. This was the range where the haptic and musical-control conditions that were used existed, and where measurement took place. This also implied that a subject's fingertip was always held suspended, and did not rest on the device surface.

A third shared property was that, as part of the experimental task, subjects moved their fingertip both up and down, doing so before each measurement but with its condition already activated. The idea here was that subjects could perceive the current haptic and musical-control response to movement before measurement of stillness movement took place; and that this measurement was then part of an episode of continuous positional control.

A final shared property across all conditions was that during the measurement itself, subjects explicitly focused on holding their fingertip as still as possible, and this for a duration of 4 seconds, chosen as a reasonable duration for a short real-life control episode.

The properties that were specific to each of the experimental conditions are then described in Table 1.

**Index to the experimental conditions *n,m***

| *n* | short name | description | to-fingerpad force output |
|---|---|---|---|
| 0 | zero force | constant zero force | 0.00 N |
| 1 | positive force | constant upward force | +0.25 N |
| 2 | negative force | constant downward force | -0.25 N |
| 3 | viscosity | constant vertical viscosity | -0.0030 N / (mm / s) |
| 4 | anti-viscosity | constant “anti-viscosity” | +0.0008 N / (mm / s) |
| 5 | positional markers | force level alternating each 0.2 mm | +0.022 N or -0.022 N |

| *m* | short name | description | headphone audio output |
|---|---|---|---|
| 0 | no musical control | bandpass-filtered white noise | width 1 kHz, center 220 Hz |
| 1 | musical control | pitch-controlled sine wave | at 440 Hz + 4.0 semitones / mm |

**Table 1.** The 12 experimental conditions used.

Regarding the haptic component (“*n*”) of the conditions listed, *zero force* was intended as part of the control condition. The specific parametrizations shown for *positive force, negative force, viscosity, anti-viscosity,* and *positional markers* were then intended to help preserve the subjects' ability to exercise continuous positional control while testing these variations.

Here, the amplitude for positional markers was based on a trade-off between easy perceptibility and limited infraction on continuous positional control during their occurrence. Anti-viscosity was limited in amplitude so as to avoid the occurrence of wild positional oscillations having a negative impact on both people and technology.

Regarding the musical-control component (“*m*”) of the conditions in Table 1, *no musical control,* too, was intended as part of the control condition. The parametrization of *no musical control* versus *musical control* was then intended to have this component vary not the presence, but the control of sound.

In addition to this, the parametrization of *musical control* specifically was also intended to aim for both ease and sensitivity during the control of sound: After limited dry runs, control had to be executable also by subjects with little to no musical training; and very small fingertip movement was still to result in audibly perceivable changes during stillness movement.

Further details about the experimental conditions can be found by looking at the code archived as part of the separate Appendices to this paper.

## 2.3 Protocol

Details about the experimental protocol followed also can be found by looking at the code in the Appendices, e.g. regarding the specific dry runs used. Overall, the protocol can be summarized by describing three repeatedly occurring phases:

• *Pre-measurement phase:* The subject rests the palmar side of the right hand on the device surface, while keeping the right index fingerpad above surface. Across the spatial measurement range, the current haptic × musical-control condition is activated. The subject is asked and observed to move their fingerpad up and down, ending up mid-range (i.e. near 5.0 mm at a position of their own choosing); and to close their eyes.

• *Measurement phase:* The subject is now focusing on holding as still as possible both their fingerpad and the output of musical control, if the latter is present. (For all experimental runs, finger posture was observed to be as illustrated in Figure 1.)

• *Post-measurement phase:* The subject is asked and observed to open their eyes; haptic output transitions to 0 N, and audio output to silence.

In the above way, each condition was tested 3 times with each subject, for a total of 36 randomized experimental runs per subject.

### *2.4 Test subjects*

There were 8 test subjects, 3 male, 5 female, 15-19 years old, with 2 individuals being left-handed. The test subjects followed a university course explicitly designed to combine research education with simultaneous participation in actual experimental research. Their age range was considered a suitable one, for being a typical age at which people (first) study the mastery of musical expression. 7 out of 8 subjects reported having had between 1.5 and 10 (average: 6.2) years of lessons on a finger-operated musical instrument. 6 out of 8 reported regularly making music by playing their finger-operated musical instrument (being either an acoustic, electric, or bass guitar, or a cello or violin), and this on both amateur (4×) and semi-professional (2×) levels. No finger length measurements were taken.

In general, all test subjects reported having neither hearing problems nor problems with manual control. However, during the experiment 1 subject reported not being able to concentrate well; 1 reported being not well-rested; 2 reported being a bit nervous; and 2 reported having cold hands (room temperature throughout the experiment was 18 to 21 degrees Celsius).

## 3. Results and analysis

### *3.1 Per-run visualization and analysis*

In Figure 2, a visualization and quantitative analysis is shown of the signal I/O during a single experimental run recording 4 seconds of stillness movement by a fingertip. This format was used to analyze and archive all experimental runs.

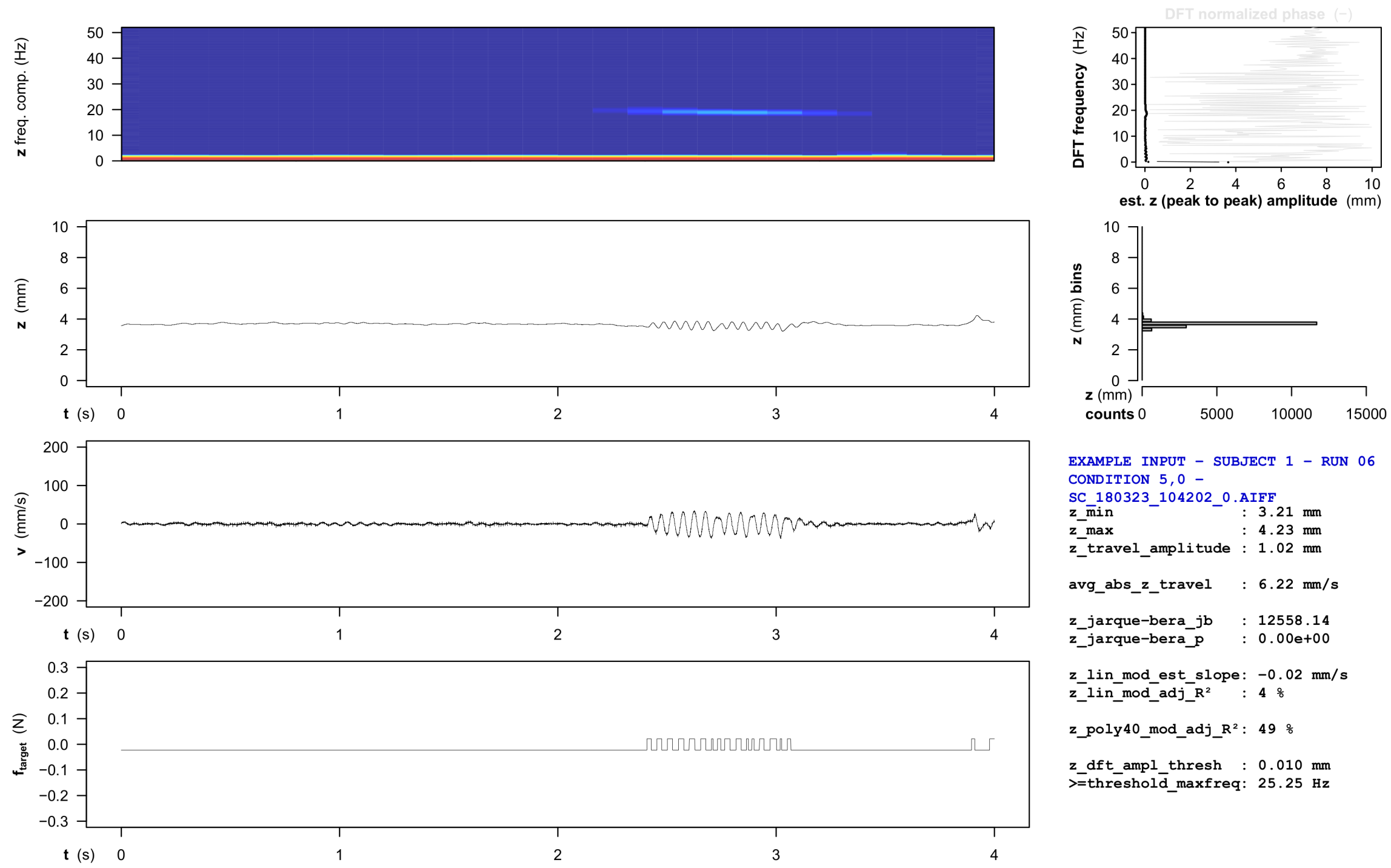

**Figure 2.** Example of per-run visualization and statistical analysis. All 288 experimental runs can be found, in this format, in the Appendices. (Please note that, for purposes of direct interpretation, the DFT results are presented along *linear* axes.)

The starting point in this format is the graph on the left that is second from top: It shows the fingertip *z* (mm) time series. Directly below it, along the same time scale, but of course using different physical amplitude scales, are the simultaneous fingertip speed (mm/s) and target force (N) time series.

Directly to the right of the *z* (mm) time series is a histogram view of its amplitude distribution, using an identical vertical scale, for easy visual comparison, and with submillimeter bins.

Directly above the *z* (mm) time series is a spectrogram view of it, again placed along the same time scale, but intended only to provide visual clues about changes over time in the frequency domain.

To its right, however, is a graph showing a custom Discrete Fourier Transform (DFT) of the recorded segment of stillness movement as a whole. For easy visual comparison, this is again plotted using an identical vertical scale (now, in Hz). To support interpretability, the black dots (with lines inbetween) here show either the *z* signal average (at 0 Hz) or the estimated peak-to-peak amplitude (above 0 Hz) directly in millimeters.

So, for reasons of direct interpretability, the DFT analysis here (and below) intentionally was carefully implemented and tested to produce its results in terms of real-world units, measured along *linear* measurement axes: This enables easy comparison to real-world fingertip movements, e.g. when using the results during the design of fingertip-based musical control.

Finally, directly below the histogram in Figure 2 is an identifier of the experimental run and its source recording. The printed statistics below this identifier as well as the DFT will be further discussed in the next sections.

## 3.2 Measured stillness movement: characterization when not applying a force

When looking at the spatial distributions of fingertip stillness movement recorded under conditions where a zero force was applied (see Appendices), it often would seem clearly incorrect to summarize what is visible in the time series and histograms as movement around a single mean, with variation distributed symmetrically.

This was confirmed by performing a Jarque-Bera statistical test on each of the *z* (mm) time series: The $p$ values computed for the null hypothesis of normality were always $< 0.005$.

As another means, then, to quantitatively summarize observed stillness movement, *z travel amplitude* (mm) was used: i.e. the difference between the highest and lowest *z* positions recorded during a single run. This derived measure has the possible disadvantage of being sensitive to outliers in the underlying time series. An advantage, however, is the direct interpretability when designing fingertip interactions.

Figure 3 gives an overview of *z* travel amplitude as observed across experimental runs: The dot plot shows the individual values, and, introducing more structure, a visually matching histogram shows the distribution of these travel amplitudes. Overall, the observed segments of stillness movement had travel amplitudes between 0.27 and 3.52 mm. As can be seen in the histogram, within this general range, travel amplitude most often was between 0.5 and 1.5 mm.

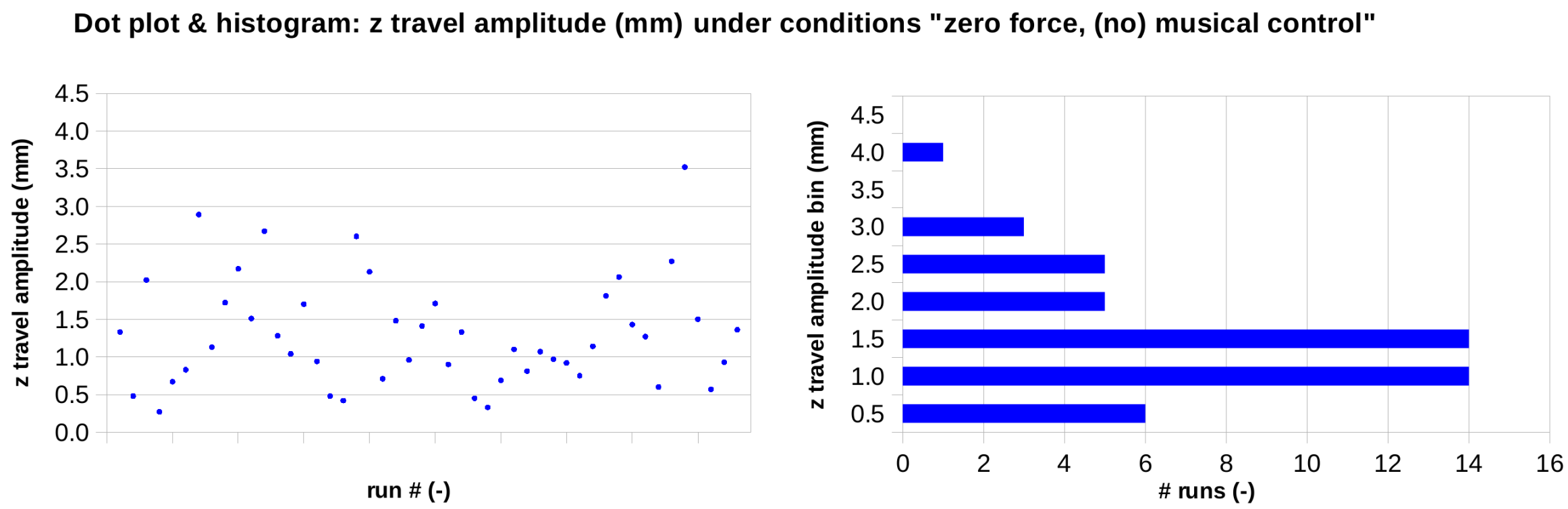

**Figure 3.** Dot plot, and corresponding histogram, showing the travel amplitudes (in mm) of the vertical stillness movement that was recorded during the 48 experimental runs which applied a zero force to the subject's fingertip.

When now moving to the frequency characteristics of observed stillness movement, it is important to first briefly discuss the interpretability of the *z* (mm) DFTs introduced previously.

The DFT used was verified to correctly measure peak-to-peak movement amplitudes at frequencies matching the underlying process of cosine correlation. However, just as for DFT in general, for cosine signal input with periods that did not exactly match test periods, DFT yielded false-positive, non-zero amplitudes at other frequencies; a highest DFT amplitude that could be lower than the wave amplitude in the time series; and addition of frequency-adjacent DFT amplitudes also not accurately yielding the wave amplitude.

Importantly, and more domain-specifically, the spread of the “false positive” frequencies and the corresponding amplitude levels seemed to be non-negligible when analysing the type of fingertip distance time series acquired in this experiment. Therefore, in per-run visualization and interpretation, DFT peak-to-peak amplitude in mm was regarded only as an estimate, not an exact measurement.

Having said this, under the conditions where zero force was applied, the peak-to-peak amplitude of fingertip movement was always measured as less than 0.01 mm at frequencies above 30.0 Hz.

For an example view of what then was happening between 30 Hz and 0 Hz, Figure 4a plots the DFTs of two experimental runs, one with and one without musical control (and with the frequency axis now oriented horizontally).

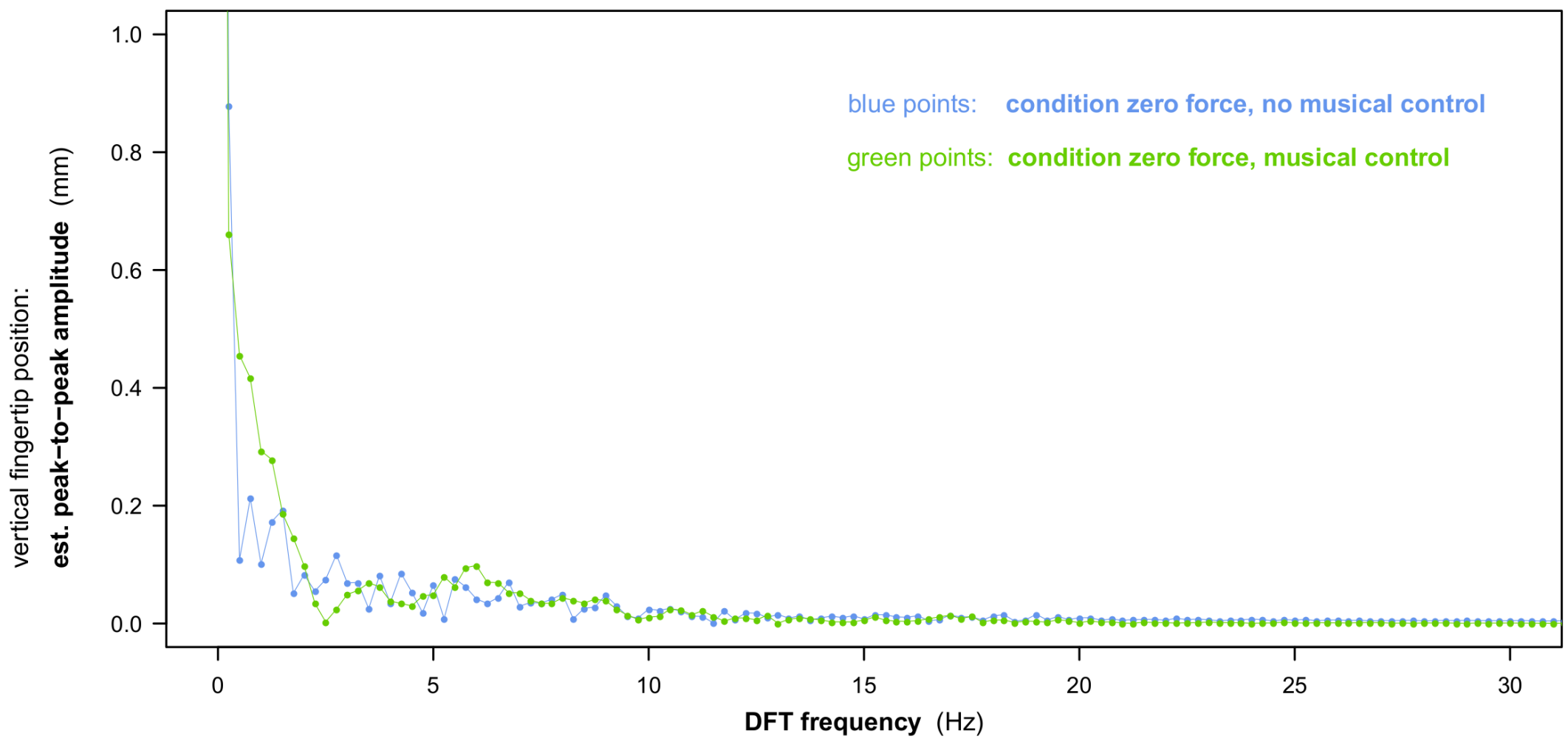

**Figure 4a.** The peak-to-peak amplitudes (in mm), across frequency, of vertical stillness movement, estimated by Discrete Fourier Transform (DFT), during two example zero-force experimental runs, one with and one without musical control. (Please note that for purposes of direct interpretation, the DFT results are presented along *linear* axes.)

As can be seen, for both examples, the estimated peak-to-peak amplitude increases from values below 0.1 mm to values above 0.5 mm when going toward 0 Hz, doing so steeply, but irregularly. When averaging DFTs across *all* zero-force experimental runs, however, a less random picture of fingertip stillness movement emerges: As is shown in Figure 4b, both for the case with and the case without musical control, the incline has become smoother, and can be approximated by a $3/17 \cdot 1/f$ model.

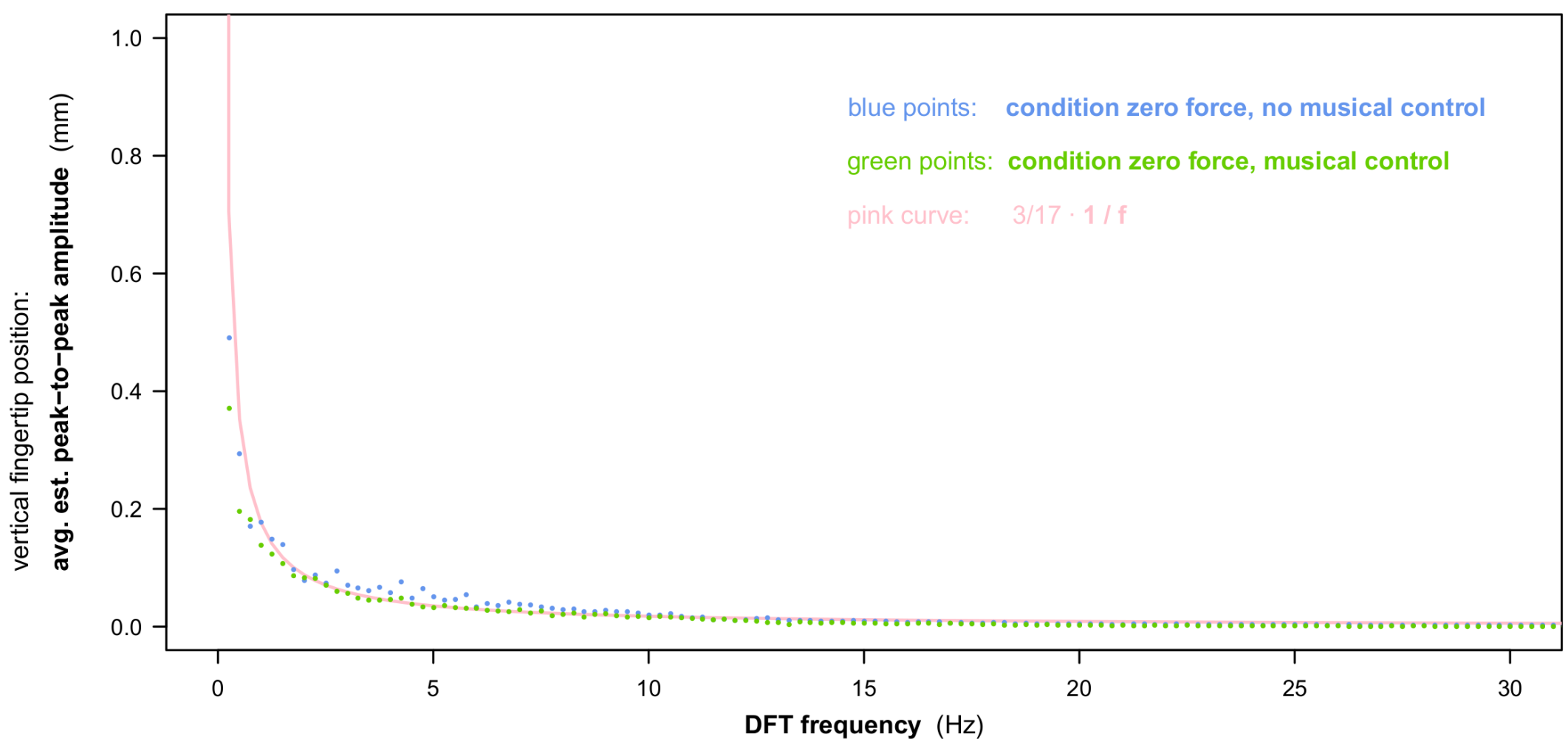

**Figure 4b.** The peak-to-peak amplitudes (in mm), across frequency, of vertical stillness movement, estimated by DFT, then averaged across all zero-force experimental runs: 24 with, and 24 without musical control. Added for comparison is the curve of a 3/17 · 1/*f* model. (Please note that for purposes of direct interpretation, the DFT results are presented along *linear* axes.)

### 3.3 Measured stillness movement: comparison across haptic conditions

The observed stillness movement was also compared across haptic conditions, using the same *z travel amplitude* (mm) measure that was introduced in the previous section. In Figure 5a, the results of this are first summarized for the experimental runs with *no* musical control: For each haptic condition, the observed minimum, maximum, and mean *z* travel amplitude are shown, and also the observed standard deviation.

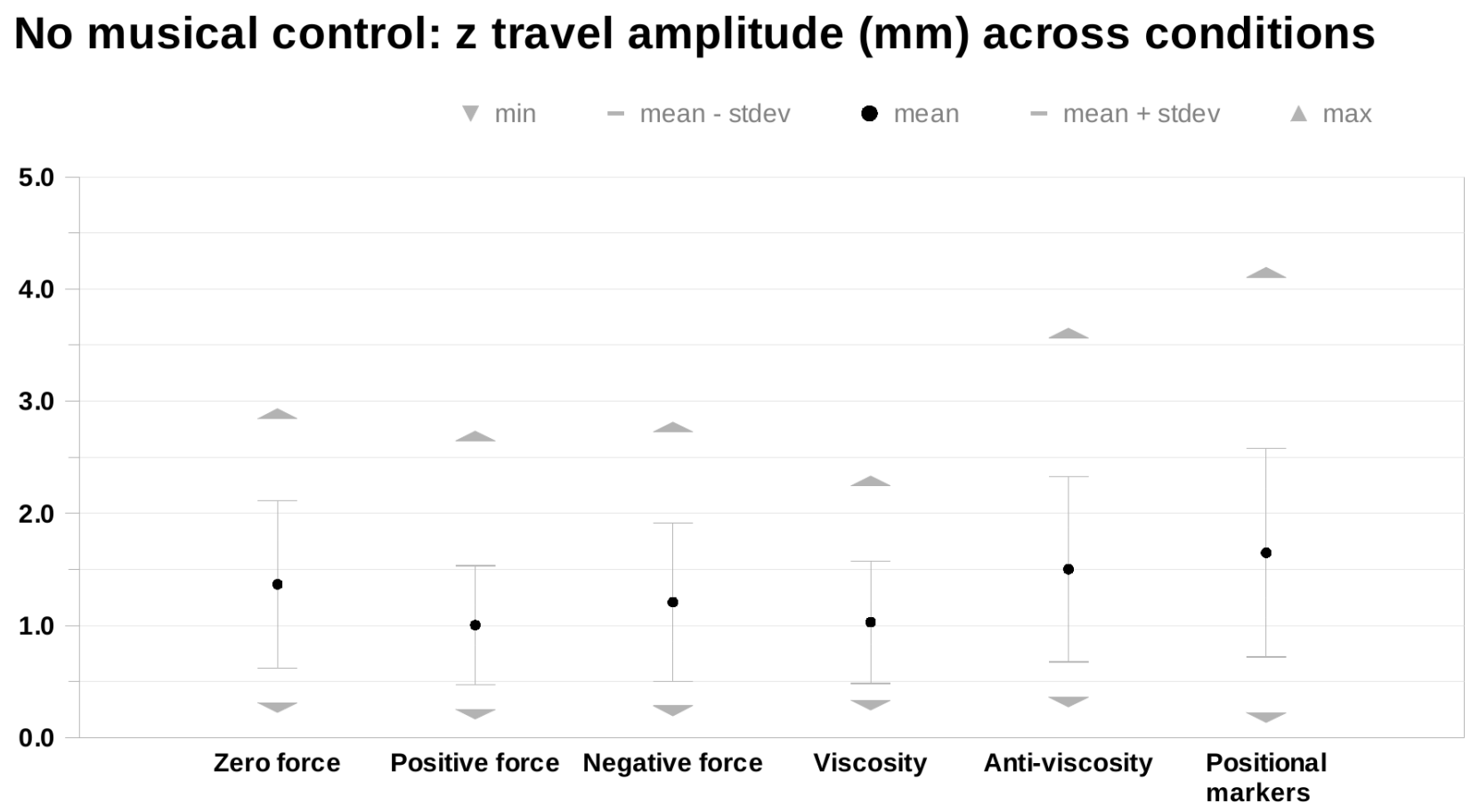

**Figure 5a.** Comparing travel amplitude (in mm) of fingertip vertical stillness movement across the experimental conditions with no musical control. For each condition, the observed extremes, standard deviation and mean are plotted.

As can be seen, under conditions *positive force* and *viscosity,* fingertip stillness movement on average had a smaller travel amplitude than under the *zero force* control condition.

Of course, chance plays a role in such relative outcomes, and the worry would be that on repeated execution the result, by chance, would become different. Consider, however, the results for the experimental runs *with* musical control, shown in Figure 5b.

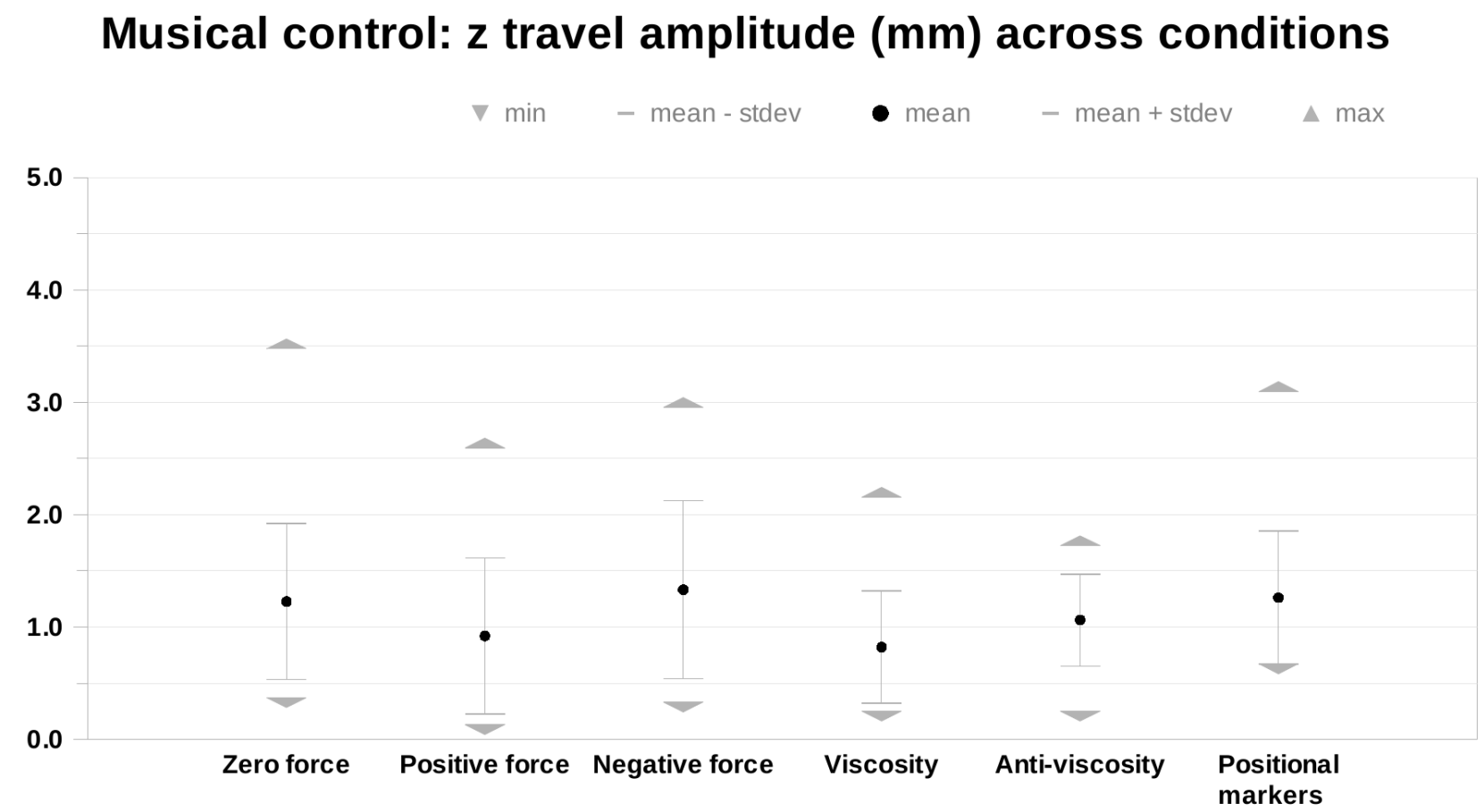

**Figure 5b.** Comparing travel amplitude (in mm) of fingertip vertical stillness movement across the experimental conditions with musical control. For each condition, the observed extremes, standard deviation and mean are plotted.

Here too, *positive force* and *viscosity* on average yielded a smaller travel amplitude than *zero force*.

As part of the a priori fixed experimental design, an analysis of variance (ANOVA) across the experimental conditions was initiated – now of course hopefully also to further investigate the reliability of this specific difference. As its first step, a preliminary test of normality was performed, to ascertain that both group and population measurements were approximately normally distributed. As there were < 50 measurements per group, a Shapiro-Wilk statistical test was used for this. Its input and outcomes can be found in the Appendices.

In summary, however, the within-group measurements did not show a statistically significant likeness to the normal distribution, as for both the experimental conditions with and without musical control, 4 out of 6 yielded a $p < 0.05$. The ANOVA therefore could not be meaningfully executed, only yielding the result that future tests based on the *travel amplitude* measure should be of a different type.

## 4. Discussion

It was hoped that *positional markers* would help in reducing stillness travel amplitude relative to the control condition of not applying a force to the fingertip, by haptically signaling small movements to the user. As can be seen in Figures 5a and 5b, however, this did not happen.

This was also reflected within the results of a subjective questionnaire, filled in before and after the measurement runs by each test subject: 3 out of 8 mentioned *positional markers* as explicitly *unhelpful* in performing the experimental task, writing down “spatial markers [...] made keeping still harder”; or even, “spatial markers made executing the task hardest”; and finally, “did not prefer spatial markers, as these amplified small movements”.

Indeed, when we observe the reflection of $f_{target}$ in $z$ in the experimental run recorded in Figure 2, the sentence quoted last seems an astute observation. In general, parametrizing *positional markers* to produce force step changes of 0.044 N amplitude (at each 0.2 mm positional transition) may have been too large a value to achieve the desired effect.

## 5. Conclusion

Above, we have described the motivation, execution, and results of a randomized controlled experiment in which 8 test subjects repeatedly performed the task of holding their right index fingertip as still as possible, during a series of 4-second episodes, and under varying conditions of fingertip haptics and musical control.

The vertical stillness movement that was experimentally observed while not applying a force to the fingertip can be characterized by the following main points:

- spatially, it did not follow a normal distribution;
- its peak-to-peak amplitude lay between 0.2 and 3.6 mm;
- most often, between 0.5 and 1.5 mm (see Figure 3 for the distribution profile);
- in the frequency region above 30 Hz, peak-to-peak amplitude was always measured as less than 0.01 mm;

- between 0 and 30 Hz, the peak-to-peak amplitude on average could be approximated by a $3/17 \cdot 1/f$ model (see Figure 4b);
- the average peak-to-peak amplitude decreased when a *constant upward force* to the fingertip, or *constant viscosity* was activated – both when musical control was present and when it was absent (see Figures 5a and 5b).

The final point listed above, although without claims of statistical significance, can be seen as an empirical hint that applying constant upward forces and viscosities on average may yield smaller fingertip stillness movement amplitudes, which in turn may enable implementing more positional control within the same available spatial range.

Also more generally, the above points can be used as a directly interpretable reference when designing technologies, interactions, or performances that involve the fingertip control of musical sound.

Finally, in the spirit of open data, the data set consisting of the complete per-run experimental results – archived in the separate Appendices to this paper – is also presented as a research outcome. And, notwithstanding that this data set is directly representative of the (larger-size) raw signal recordings, the author of course can also be contacted for the latter.

## Acknowledgments

I would like to thank Annebeth Simonsz and the Honours Academy for supporting the development and execution of the course (of which this research was an integral part) every step of the way.

My heartfelt thanks also to René Overgauw and Arno van Amersfoort, at the Electronics Department at LION, for their important and continued support.

Sincere thanks also to Anne Ouweneel, Aquilla van den Berg, Bardia Nikookar, Dave Aulia, Dennis Vianen, Julia Wesseling, Kiki Abels, Laurentine Rietvelt, Simone Flipse, Sraman Chatterjee, Stella Berbiers, Stijn Huijgen, Timo Alderliesten, and Wouter Bakker, for their assistance while executing this research.

Last but not least, I would like to thank the anonymous reviewers and meta-reviewer, for their valuable comments and guidance in processing these.

## Appendices

For purposes of reference, reproducibility, and transparency, the Appendices, in PDF format, also contain the complete experimental results. Their visualization has been carefully designed so that one experimental run can still be inspected while fitting on a single (A4-size or larger) electronic screen. This supports the quick and easy comparison, across subjects and conditions, of multiple individual runs.

pdf Appendices.pdf 63 MB

# Appendices

to

# Human noise at the fingertip: Positional (non)control under varying haptic × musical conditions

Staas de Jong
Universiteit Leiden, Honours Academy
apajong@xs4all.nl

**TABLE OF CONTENTS**

**Index to the experimental conditions *n,m***

| ***n*** | **short name** | **description** | **to-fingerpad force output** |
|---|---|---|---|
| 0 | zero force | constant zero force | 0.00 N |
| 1 | positive force | constant upward force | +0.25 N |
| 2 | negative force | constant downward force | -0.25 N |
| 3 | viscosity | constant vertical viscosity | -0.0030 N / (mm / s) |
| 4 | anti-viscosity | constant “anti-viscosity” | +0.0008 N / (mm / s) |
| 5 | positional markers | force level alternating each 0.2 mm | +0.022 N or -0.022 N |

| ***m*** | **short name** | **description** | **headphone audio output** |
|---|---|---|---|
| 0 | no musical control | bandpass-filtered white noise | width 1 kHz, center 220 Hz |
| 1 | musical control | pitch-controlled sine wave | at 440 Hz + 4.0 semitones / mm |

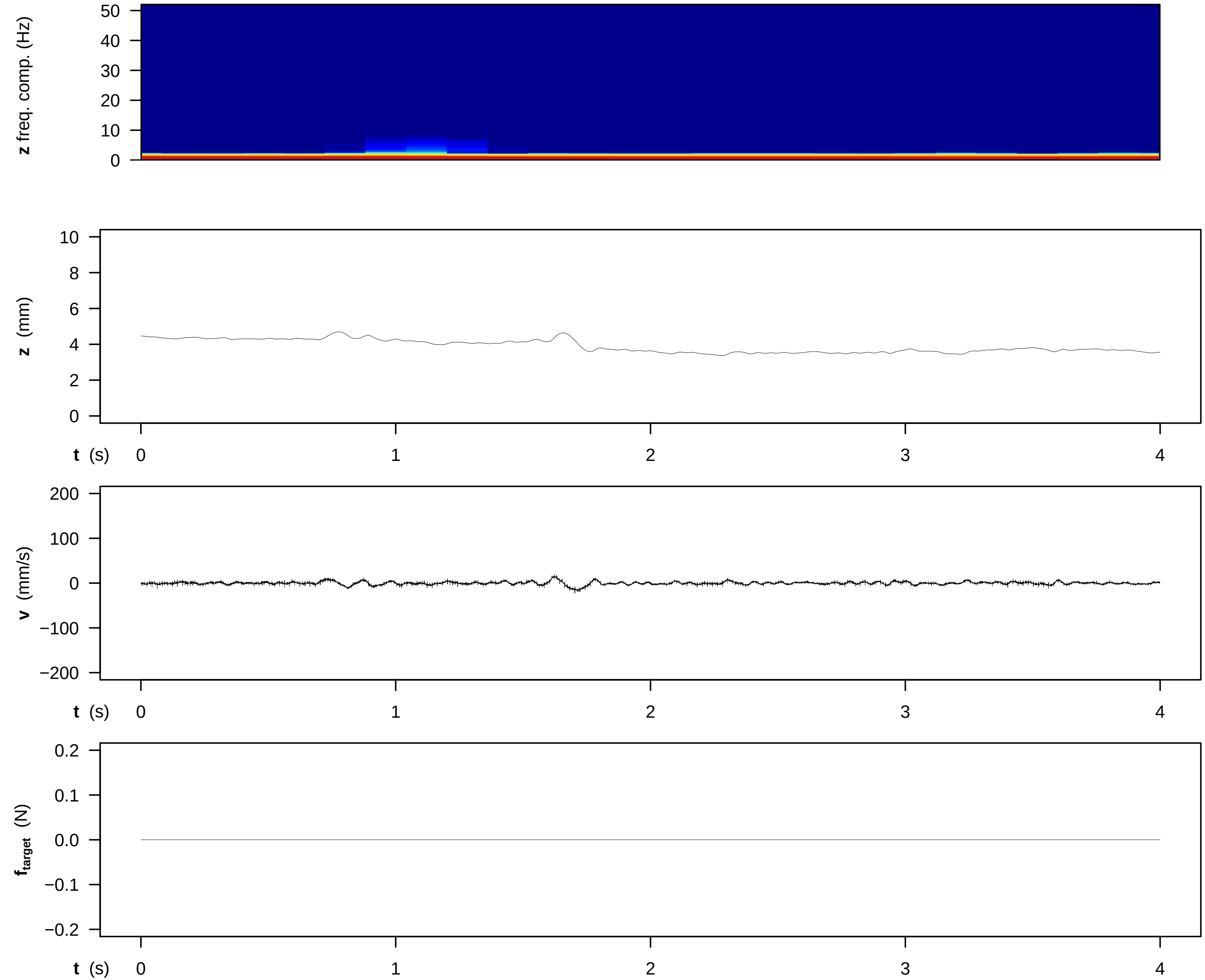

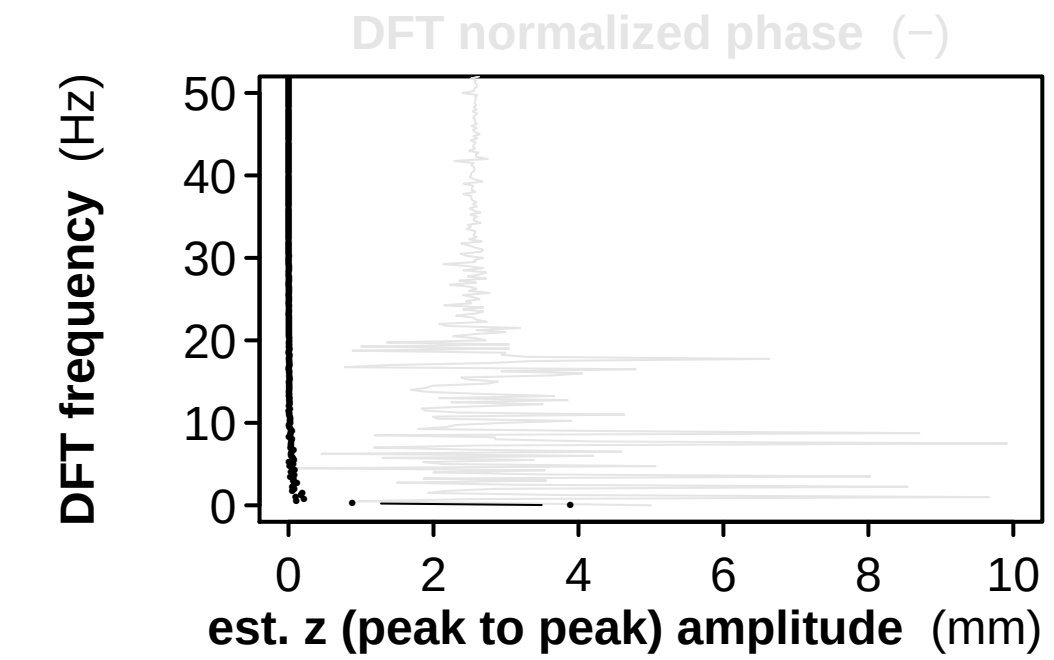

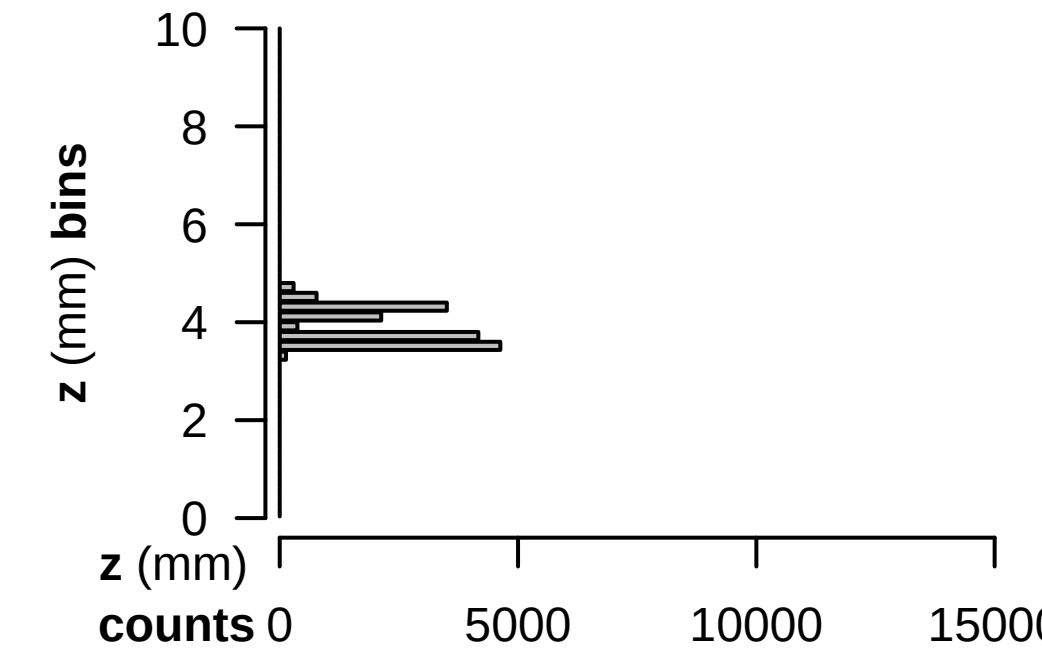

**SUBJECT 1 − RUN 10 − CONDITION 0,0**
**SC_180323_104417_0.AIFF**

```
z_min               : 3.36 mm
z_max               : 4.70 mm
z_travel_amplitude  : 1.33 mm

avg_abs_z_travel    : 4.46 mm/s

z_jarque-bera_jb    : 1606.59
z_jarque-bera_p     : 0.00e+00

z_lin_mod_est_slope: −0.25 mm/s
z_lin_mod_adj_R²    : 65 %

z_poly40_mod_adj_R²: 94 %

z_dft_ampl_thresh   : 0.010 mm
>=threshold_maxfreq: 19.50 Hz
```

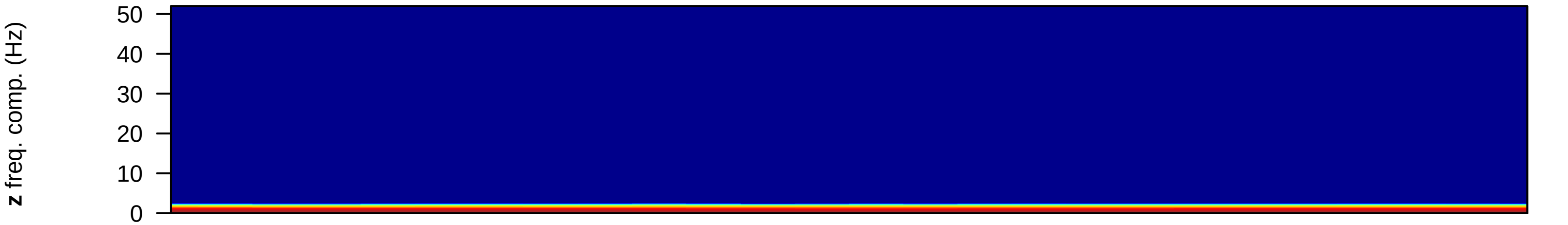

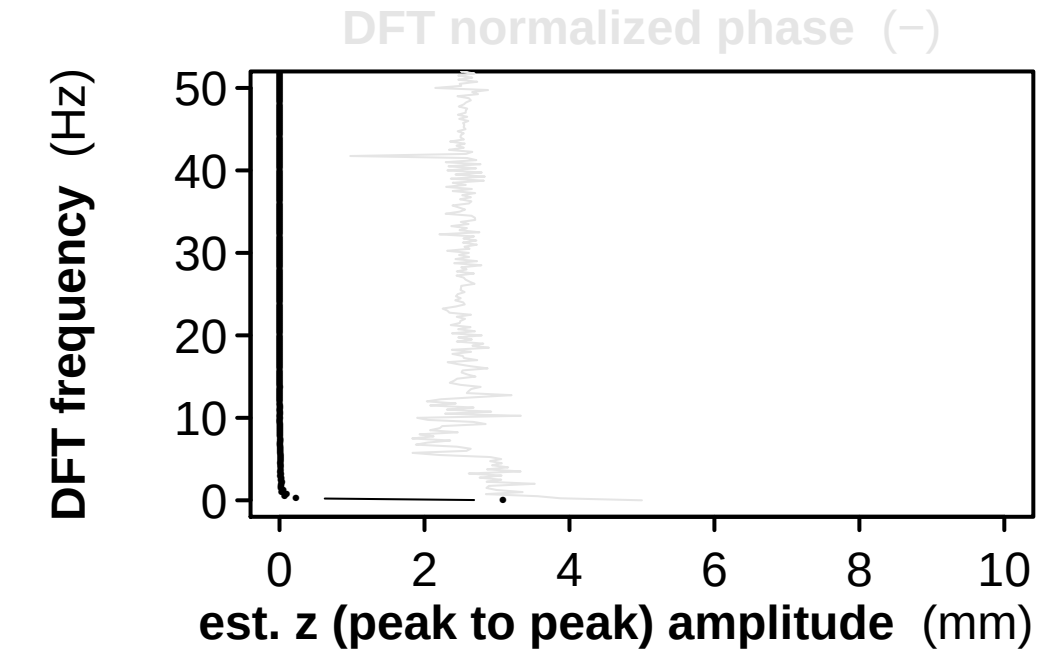

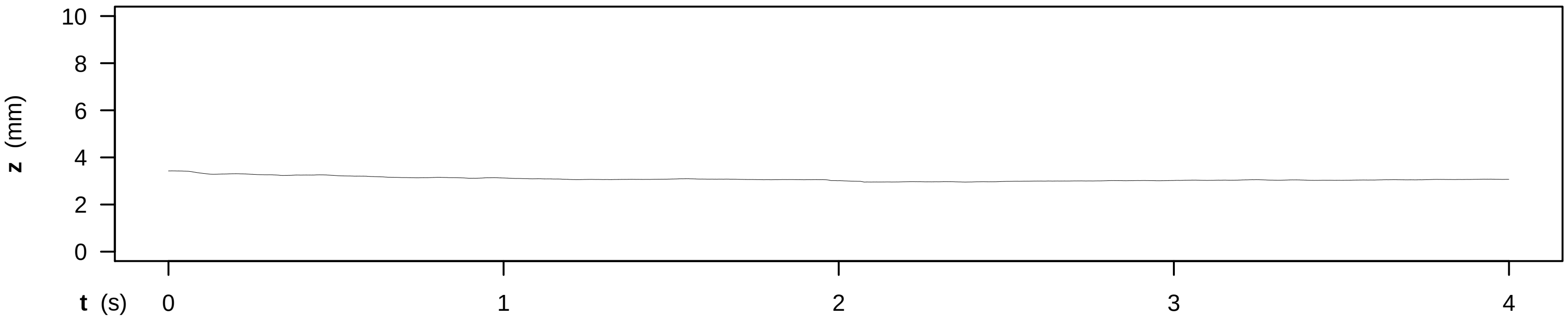

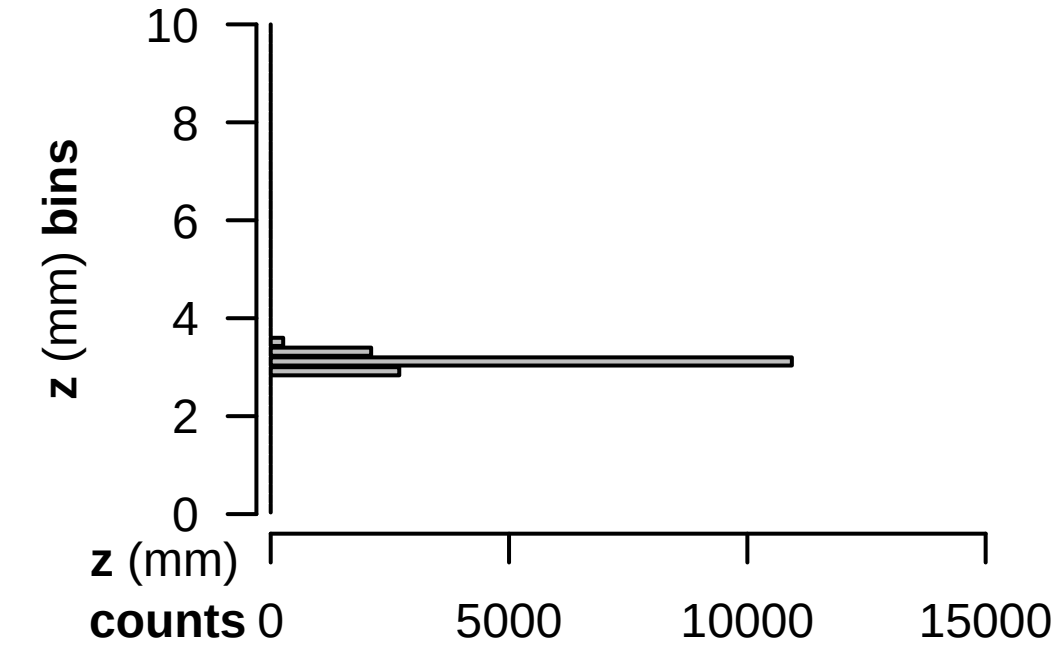

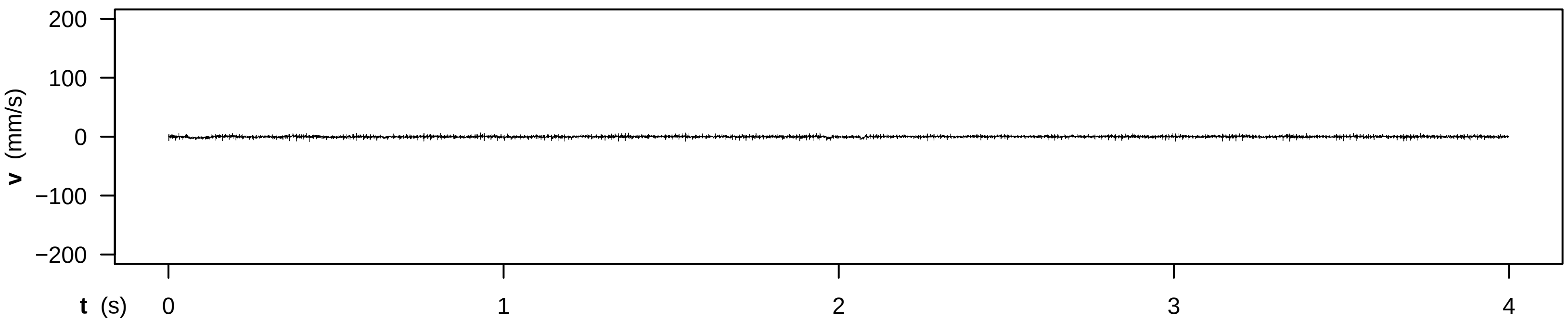

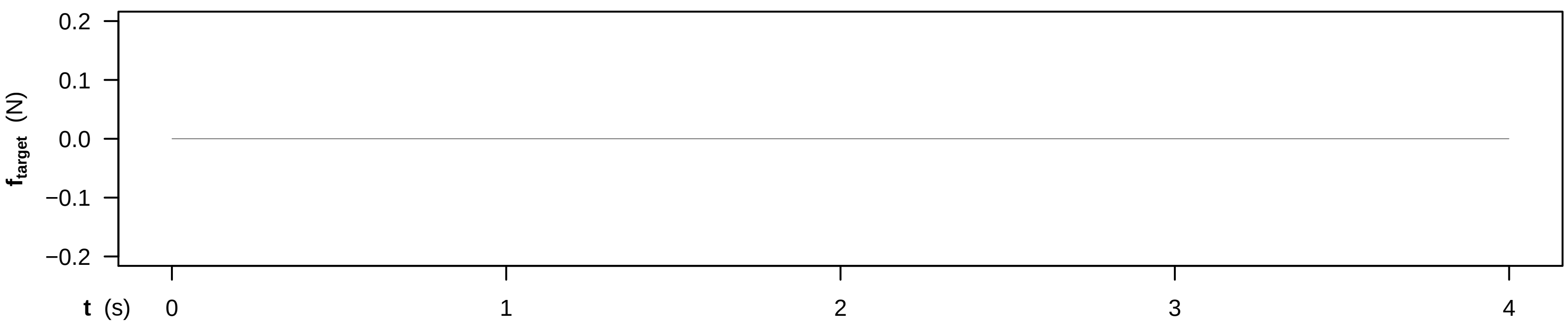

**SUBJECT 1 − RUN 34 − CONDITION 0,0**
**SC_180323_110046_0.AIFF**

```
z_min               : 2.95 mm
z_max               : 3.43 mm
z_travel_amplitude  : 0.48 mm

avg_abs_z_travel    : 1.76 mm/s

z_jarque-bera_jb    : 6527.56
z_jarque-bera_p     : 0.00e+00

z_lin_mod_est_slope: −0.06 mm/s
z_lin_mod_adj_R²    : 52 %

z_poly40_mod_adj_R²: 99 %

z_dft_ampl_thresh   : 0.010 mm
>=threshold_maxfreq: 7.50 Hz
```

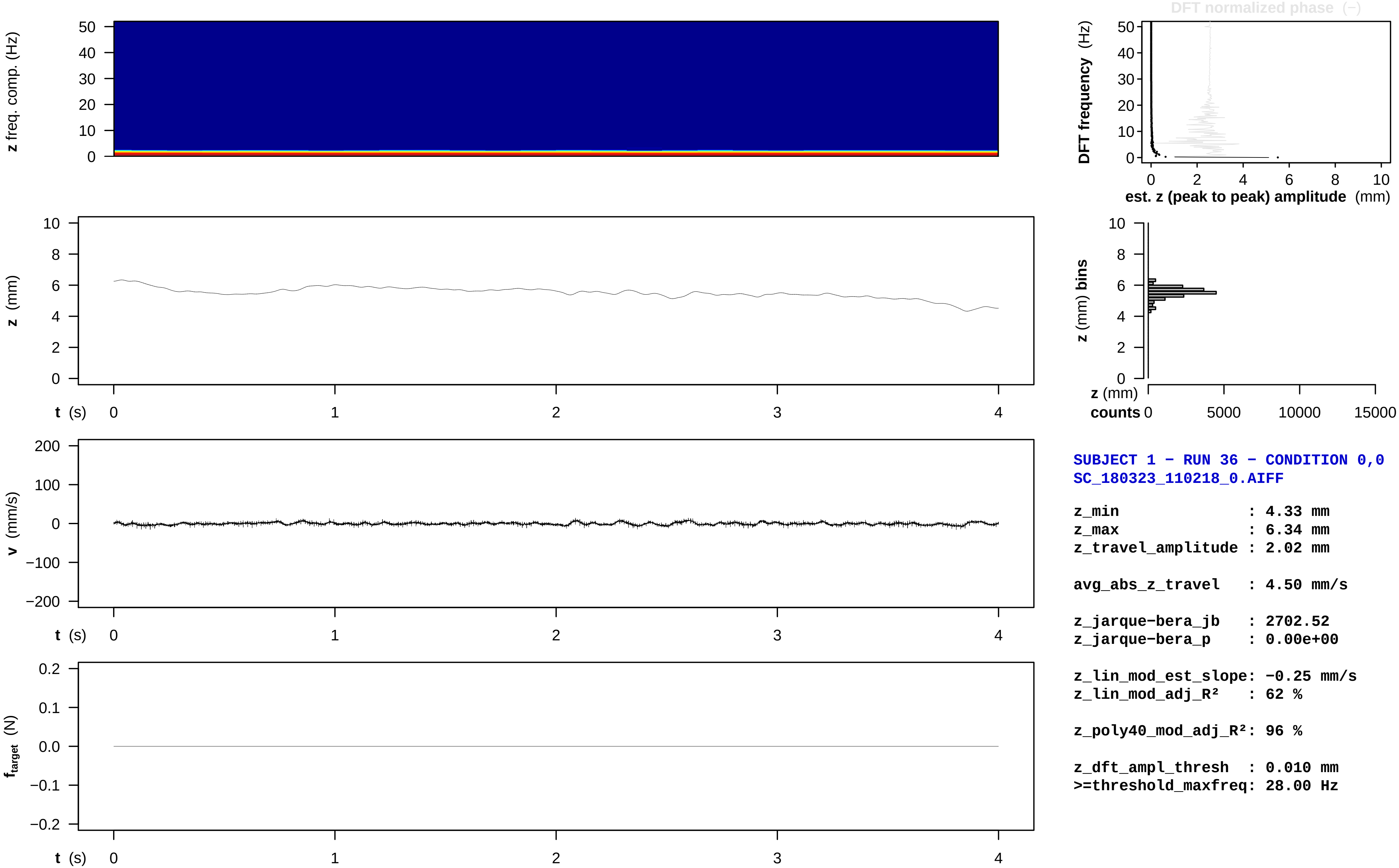

**SUBJECT 1 − RUN 36 − CONDITION 0,0**
**SC_180323_110218_0.AIFF**

```
z_min               : 4.33 mm
z_max               : 6.34 mm
z_travel_amplitude  : 2.02 mm

avg_abs_z_travel    : 4.50 mm/s

z_jarque-bera_jb    : 2702.52
z_jarque-bera_p     : 0.00e+00

z_lin_mod_est_slope: −0.25 mm/s
z_lin_mod_adj_R²    : 62 %

z_poly40_mod_adj_R²: 96 %

z_dft_ampl_thresh   : 0.010 mm
>=threshold_maxfreq: 28.00 Hz
```

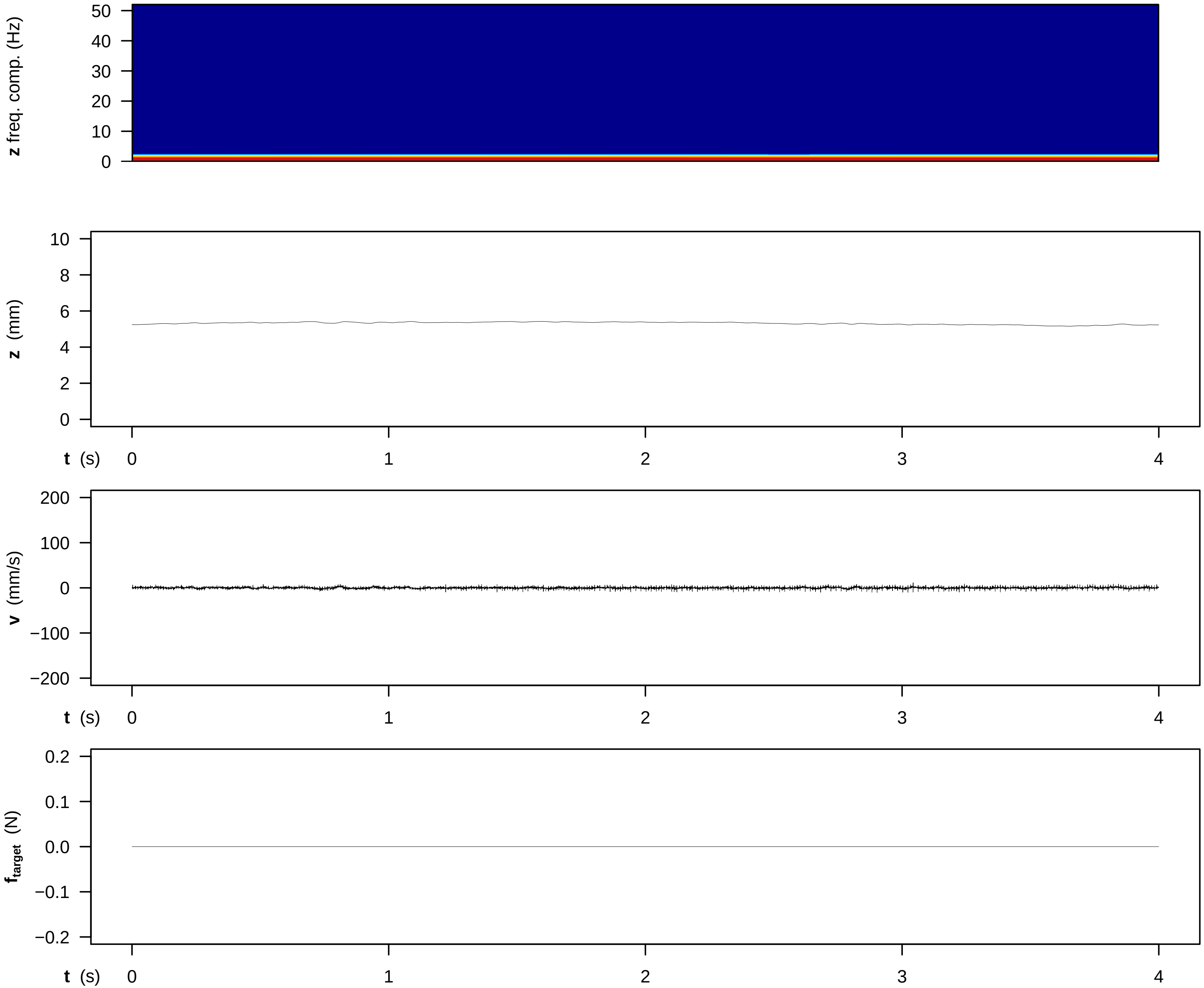

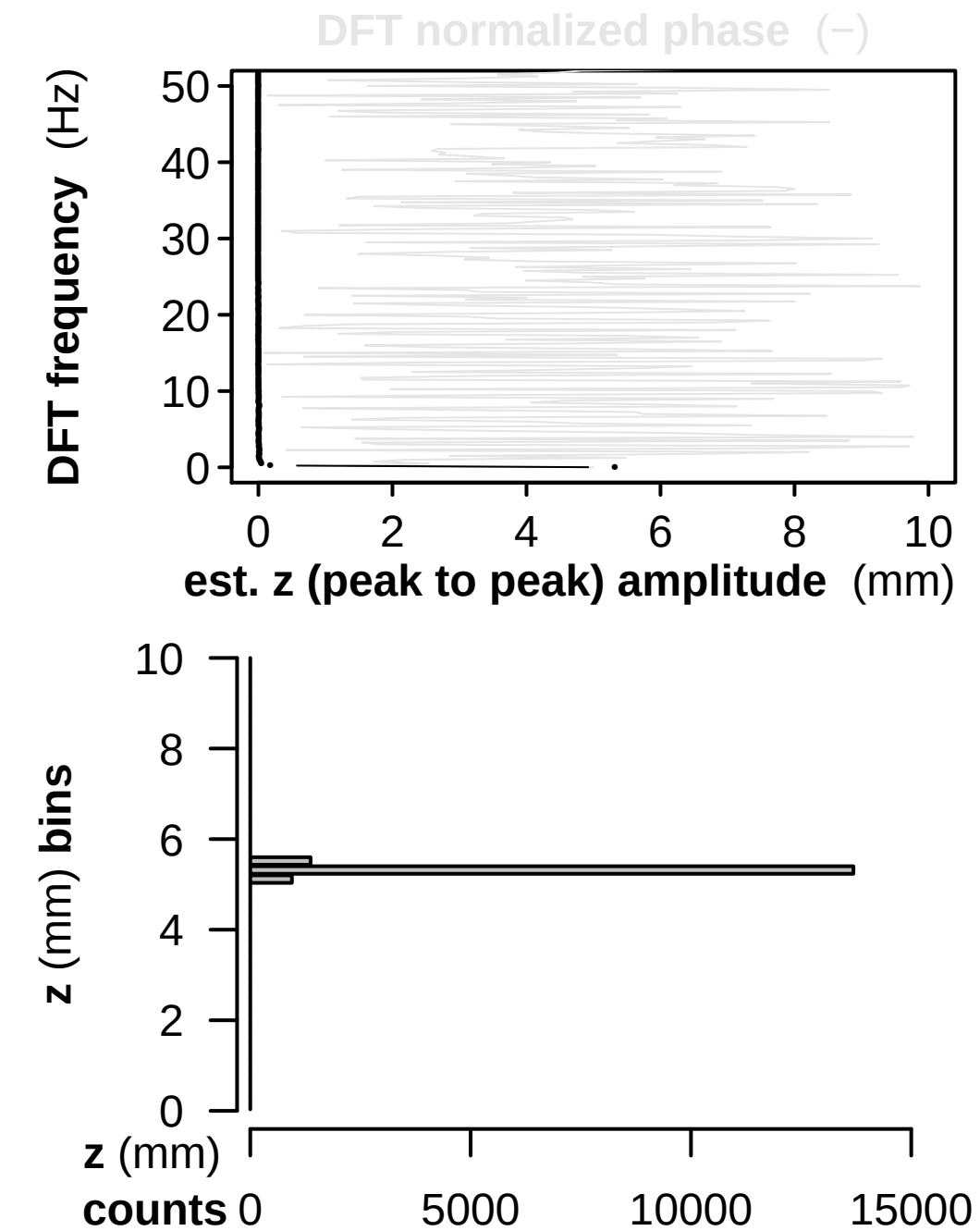

**SUBJECT 2 − RUN 02 − CONDITION 0,0**
**SC_180323_111605_0.AIFF**

**z_min : 5.16 mm**
**z_max : 5.42 mm**
**z_travel_amplitude : 0.27 mm**

**avg_abs_z_travel : 2.19 mm/s**

**z_jarque-bera_jb : 1201.82**
**z_jarque-bera_p : 0.00e+00**

**z_lin_mod_est_slope: −0.04 mm/s**
**z_lin_mod_adj_R² : 44 %**

**z_poly40_mod_adj_R²: 94 %**

**z_dft_ampl_thresh : 0.010 mm**
**>=threshold_maxfreq: 8.25 Hz**

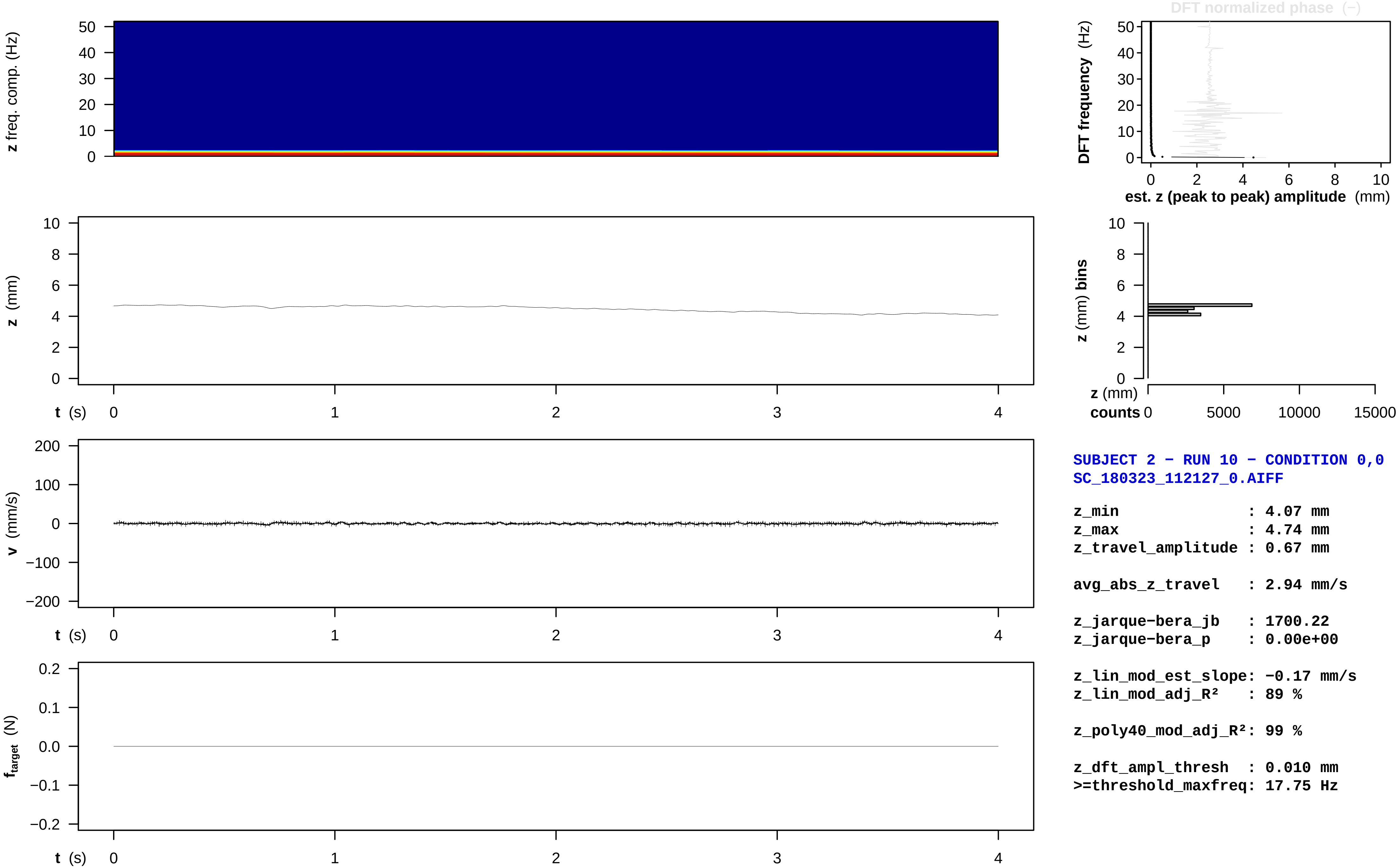

**SUBJECT 2 − RUN 10 − CONDITION 0,0**
**SC_180323_112127_0.AIFF**

```
z_min               : 4.07 mm
z_max               : 4.74 mm
z_travel_amplitude  : 0.67 mm

avg_abs_z_travel    : 2.94 mm/s

z_jarque-bera_jb    : 1700.22
z_jarque-bera_p     : 0.00e+00

z_lin_mod_est_slope: −0.17 mm/s
z_lin_mod_adj_R²    : 89 %

z_poly40_mod_adj_R²: 99 %

z_dft_ampl_thresh   : 0.010 mm
>=threshold_maxfreq: 17.75 Hz
```

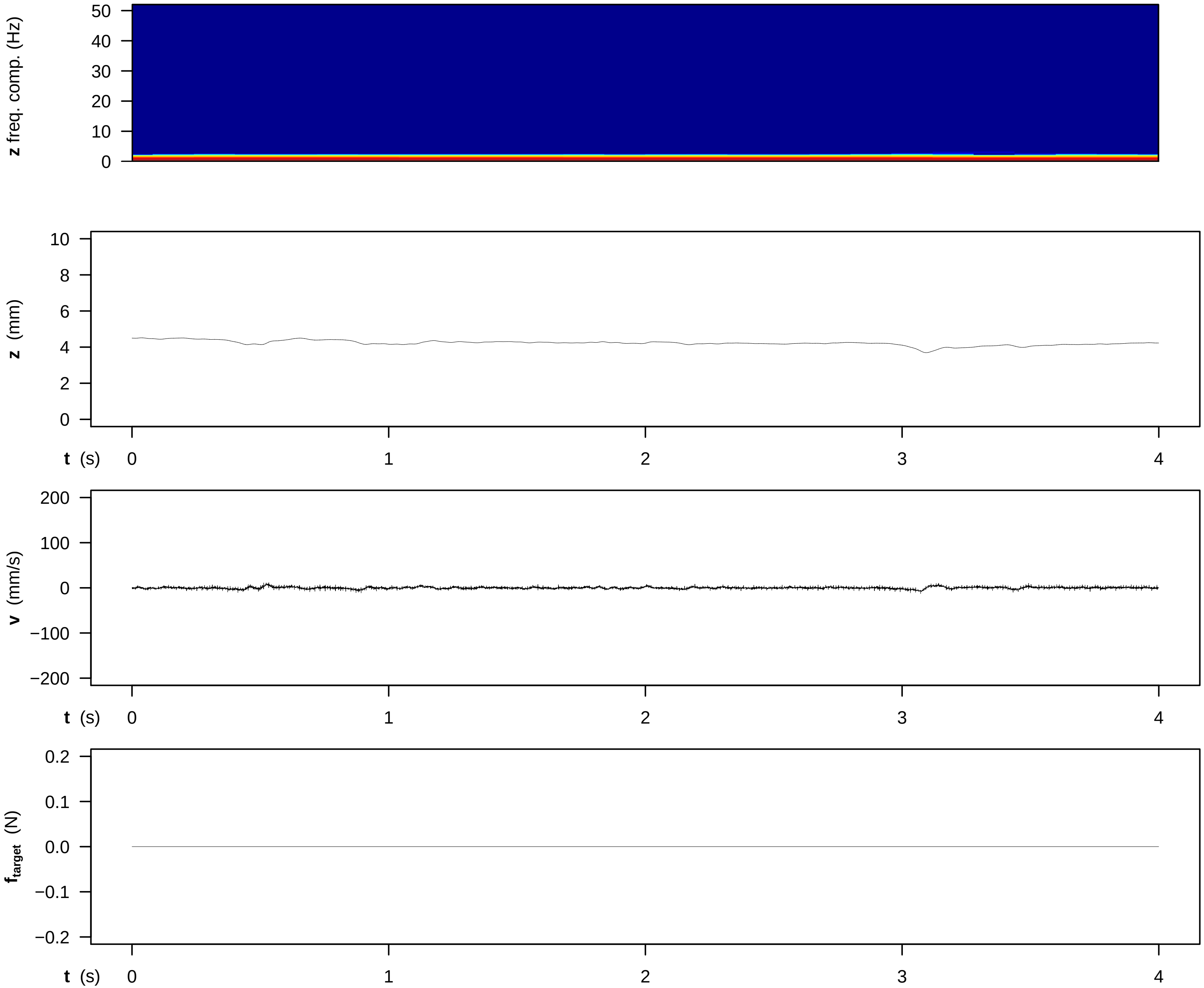

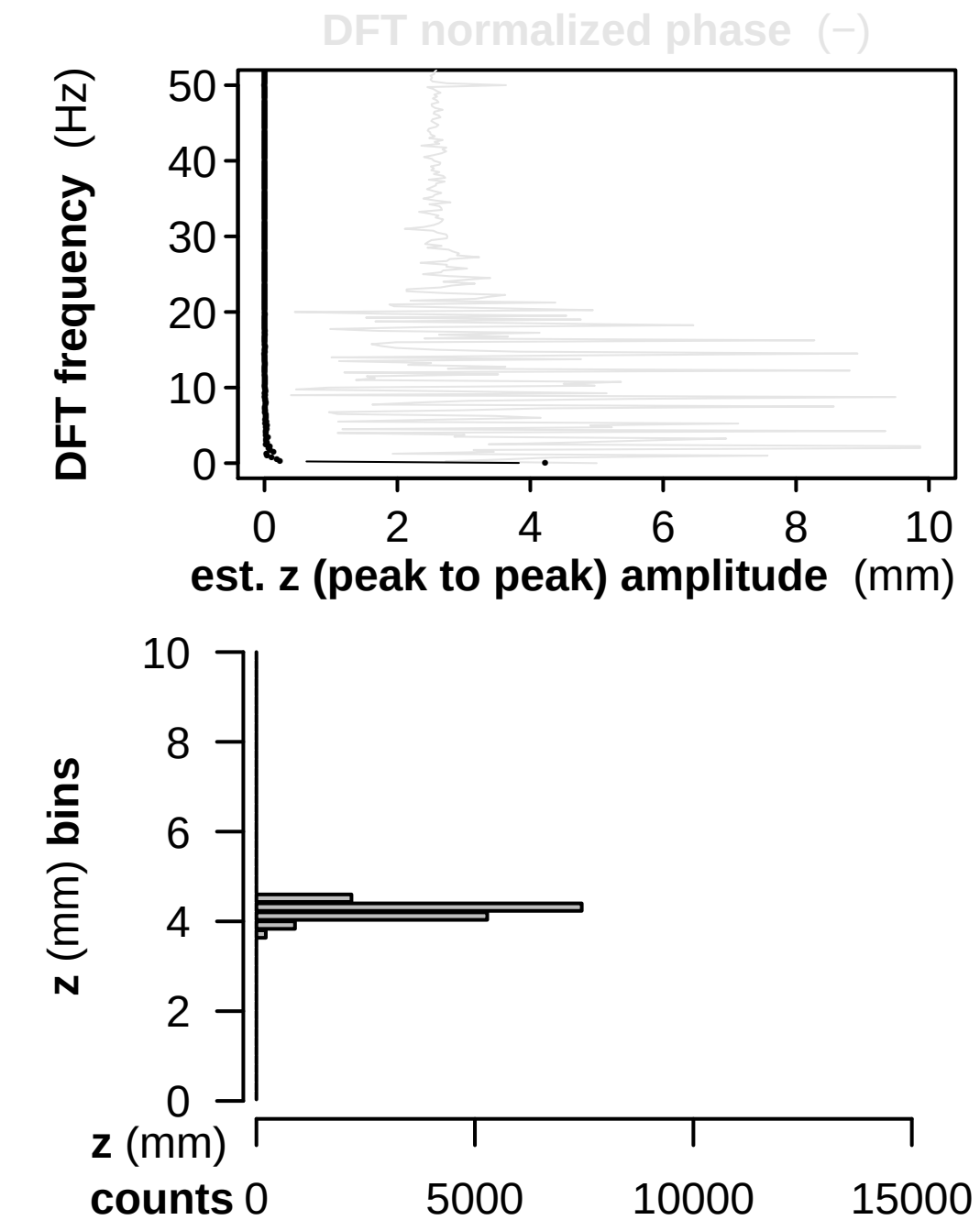

**SUBJECT 2 − RUN 23 − CONDITION 0,0**
**SC_180323_112848_0.AIFF**

```
z_min              : 3.69 mm
z_max              : 4.52 mm
z_travel_amplitude : 0.83 mm

avg_abs_z_travel   : 3.55 mm/s

z_jarque-bera_jb   : 2169.45
z_jarque-bera_p    : 0.00e+00

z_lin_mod_est_slope: -0.09 mm/s
z_lin_mod_adj_R²   : 50 %

z_poly40_mod_adj_R²: 90 %

z_dft_ampl_thresh  : 0.010 mm
>=threshold_maxfreq: 9.75 Hz
```

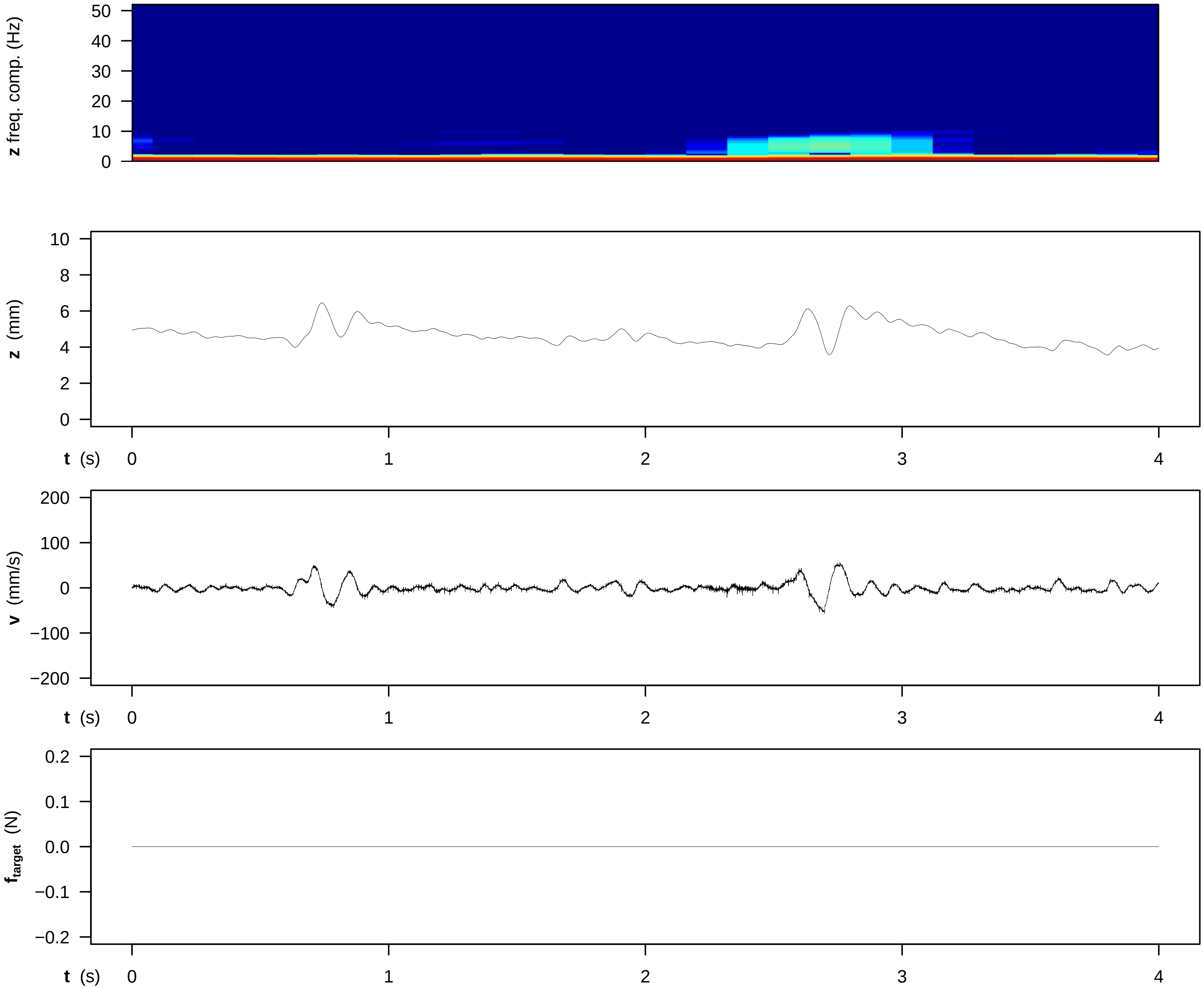

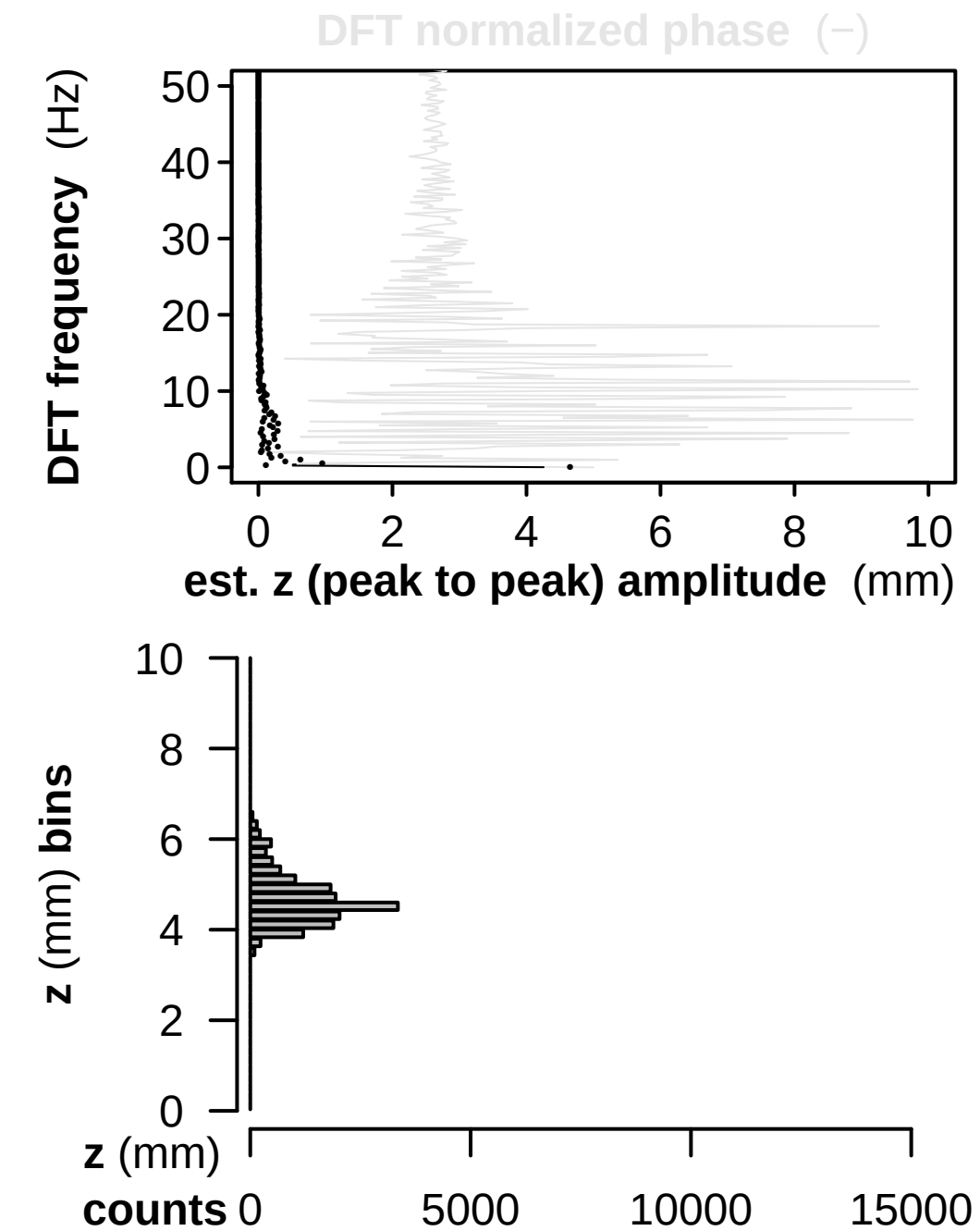

**SUBJECT 3 − RUN 09 − CONDITION 0,0**
**SC_180323_120009_0.AIFF**

```
z_min               : 3.57 mm
z_max               : 6.45 mm
z_travel_amplitude : 2.89 mm

avg_abs_z_travel    : 8.65 mm/s

z_jarque-bera_jb    : 1983.22
z_jarque-bera_p     : 0.00e+00

z_lin_mod_est_slope: −0.13 mm/s
z_lin_mod_adj_R²    : 8 %

z_poly40_mod_adj_R²: 69 %

z_dft_ampl_thresh   : 0.010 mm
>=threshold_maxfreq: 22.25 Hz
```

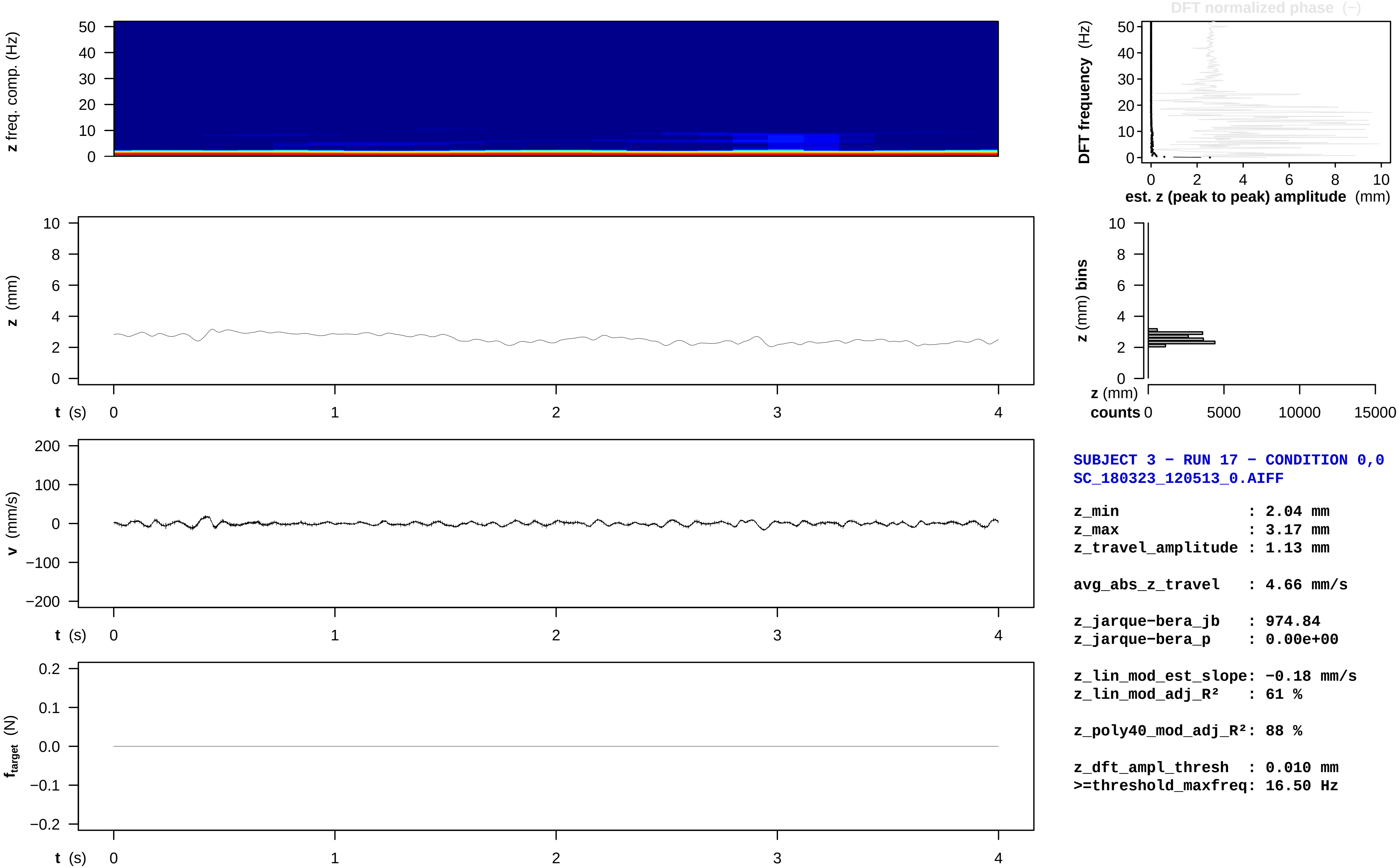

**SUBJECT 3 − RUN 17 − CONDITION 0,0**
**SC_180323_120513_0.AIFF**

```
z_min              : 2.04 mm
z_max              : 3.17 mm
z_travel_amplitude : 1.13 mm

avg_abs_z_travel   : 4.66 mm/s

z_jarque-bera_jb   : 974.84
z_jarque-bera_p    : 0.00e+00

z_lin_mod_est_slope: -0.18 mm/s
z_lin_mod_adj_R²   : 61 %

z_poly40_mod_adj_R²: 88 %

z_dft_ampl_thresh  : 0.010 mm
>=threshold_maxfreq: 16.50 Hz
```

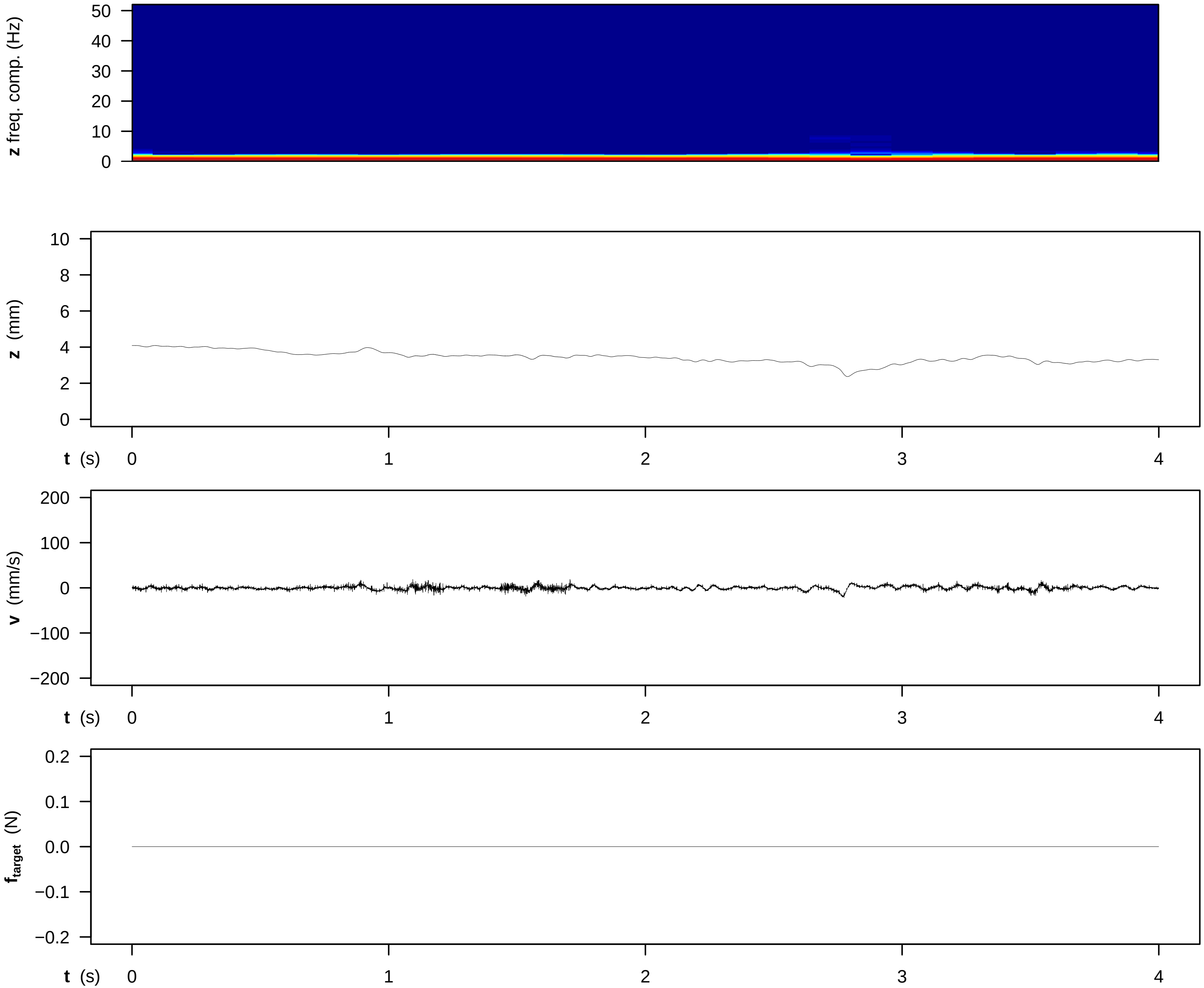

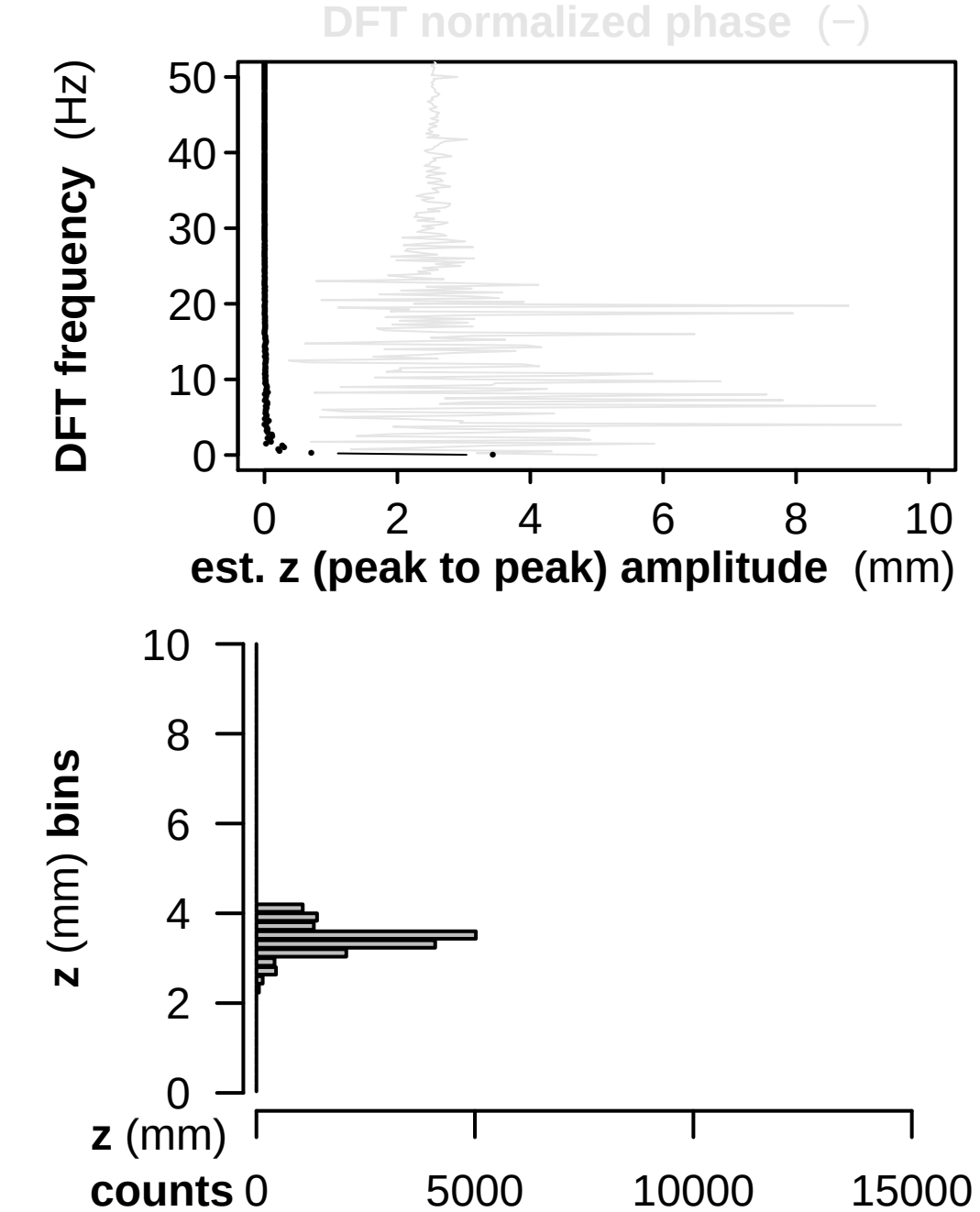

**SUBJECT 3 − RUN 30 − CONDITION 0,0**
**SC_180323_121221_0.AIFF**

```
z_min               : 2.37 mm
z_max               : 4.09 mm
z_travel_amplitude  : 1.72 mm

avg_abs_z_travel    : 3.75 mm/s

z_jarque-bera_jb    : 126.53
z_jarque-bera_p     : 0.00e+00

z_lin_mod_est_slope: −0.22 mm/s
z_lin_mod_adj_R²    : 61 %

z_poly40_mod_adj_R²: 94 %

z_dft_ampl_thresh   : 0.010 mm
>=threshold_maxfreq: 21.75 Hz
```

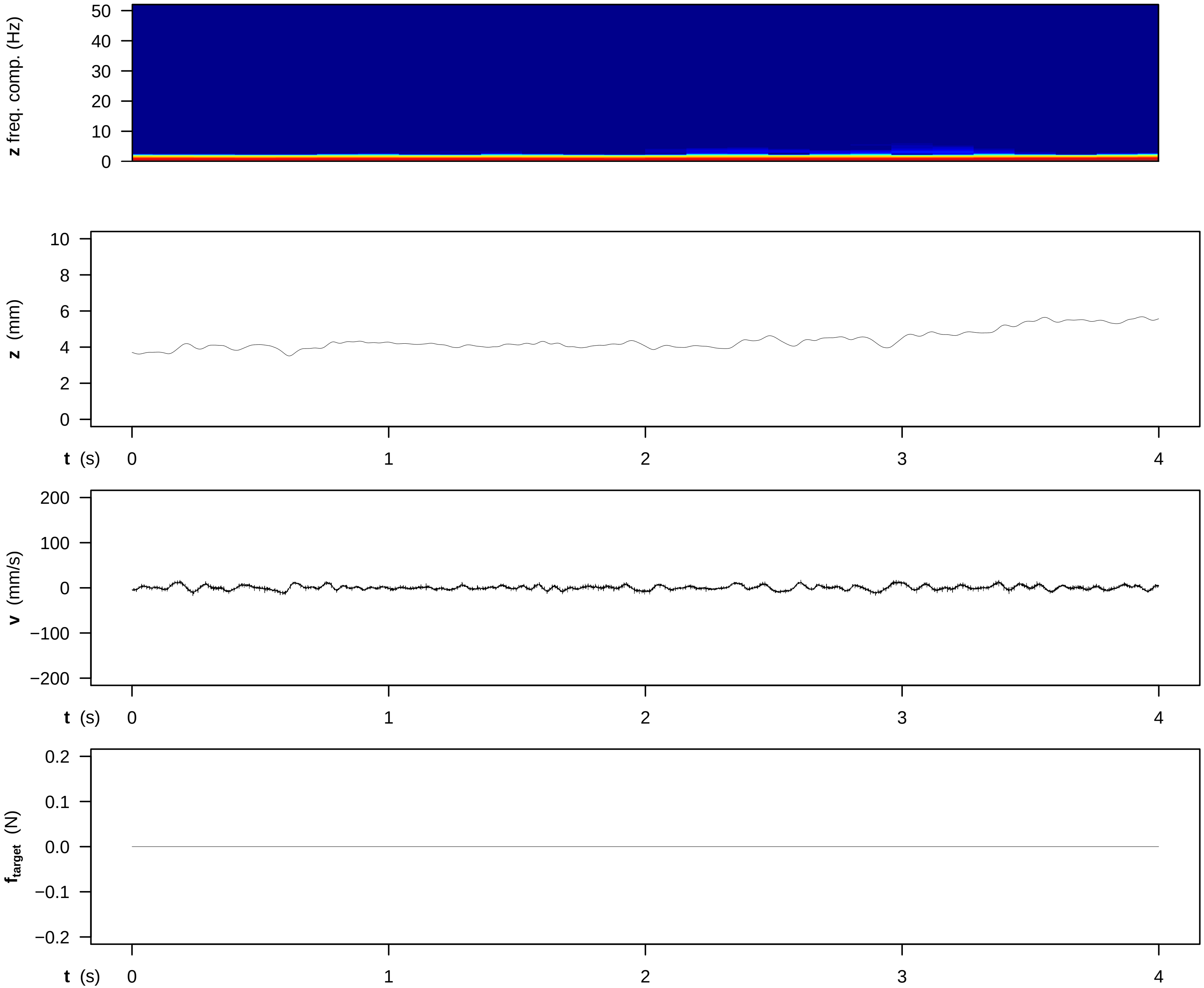

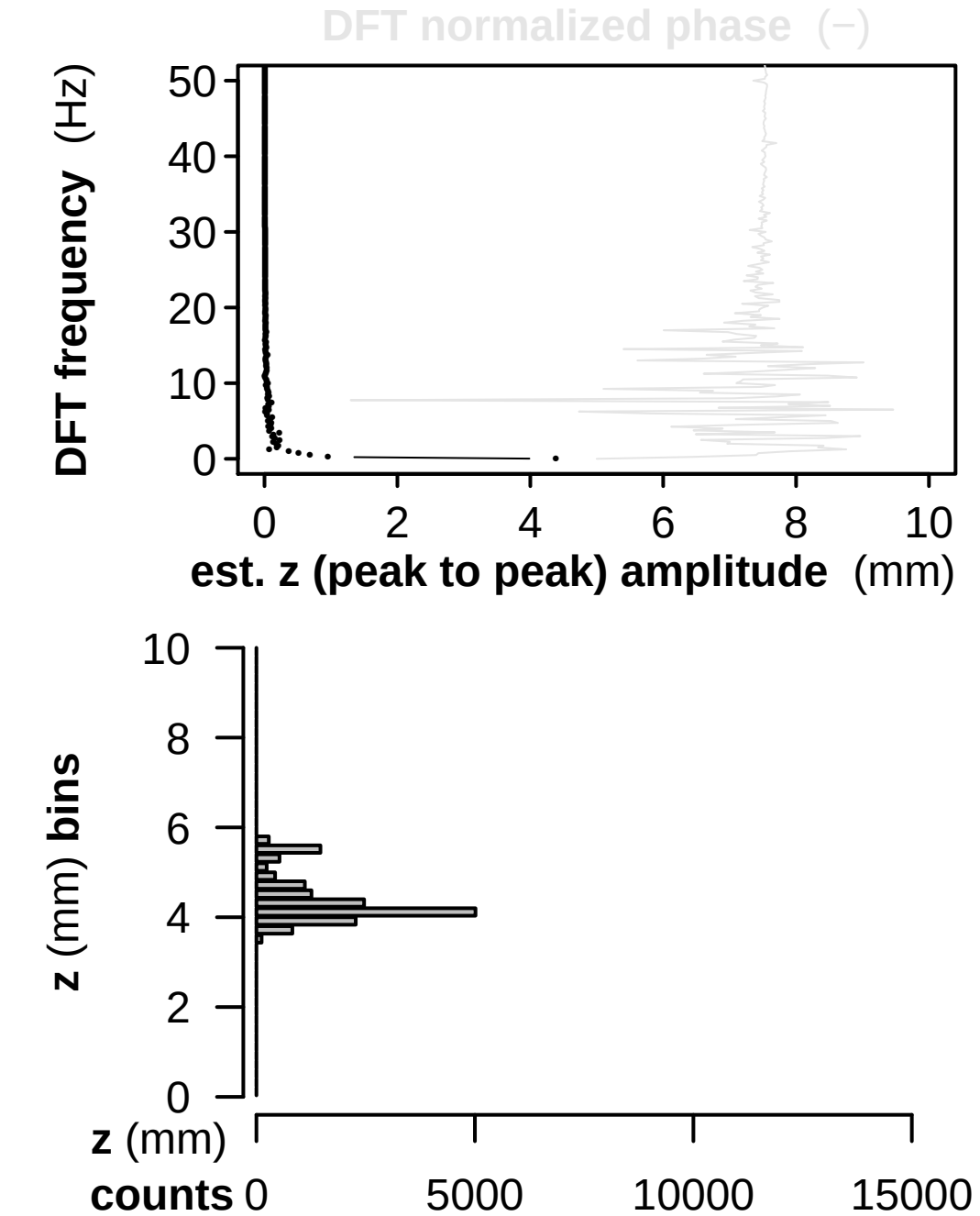

**SUBJECT 4 − RUN 05 − CONDITION 0,0**
**SC_180323_123300_0.AIFF**

```
z_min               : 3.51 mm
z_max               : 5.68 mm
z_travel_amplitude  : 2.17 mm

avg_abs_z_travel    : 4.22 mm/s

z_jarque-bera_jb    : 2993.43
z_jarque-bera_p     : 0.00e+00

z_lin_mod_est_slope: 0.38 mm/s
z_lin_mod_adj_R²    : 68 %

z_poly40_mod_adj_R²: 95 %

z_dft_ampl_thresh   : 0.010 mm
>=threshold_maxfreq: 30.00 Hz
```

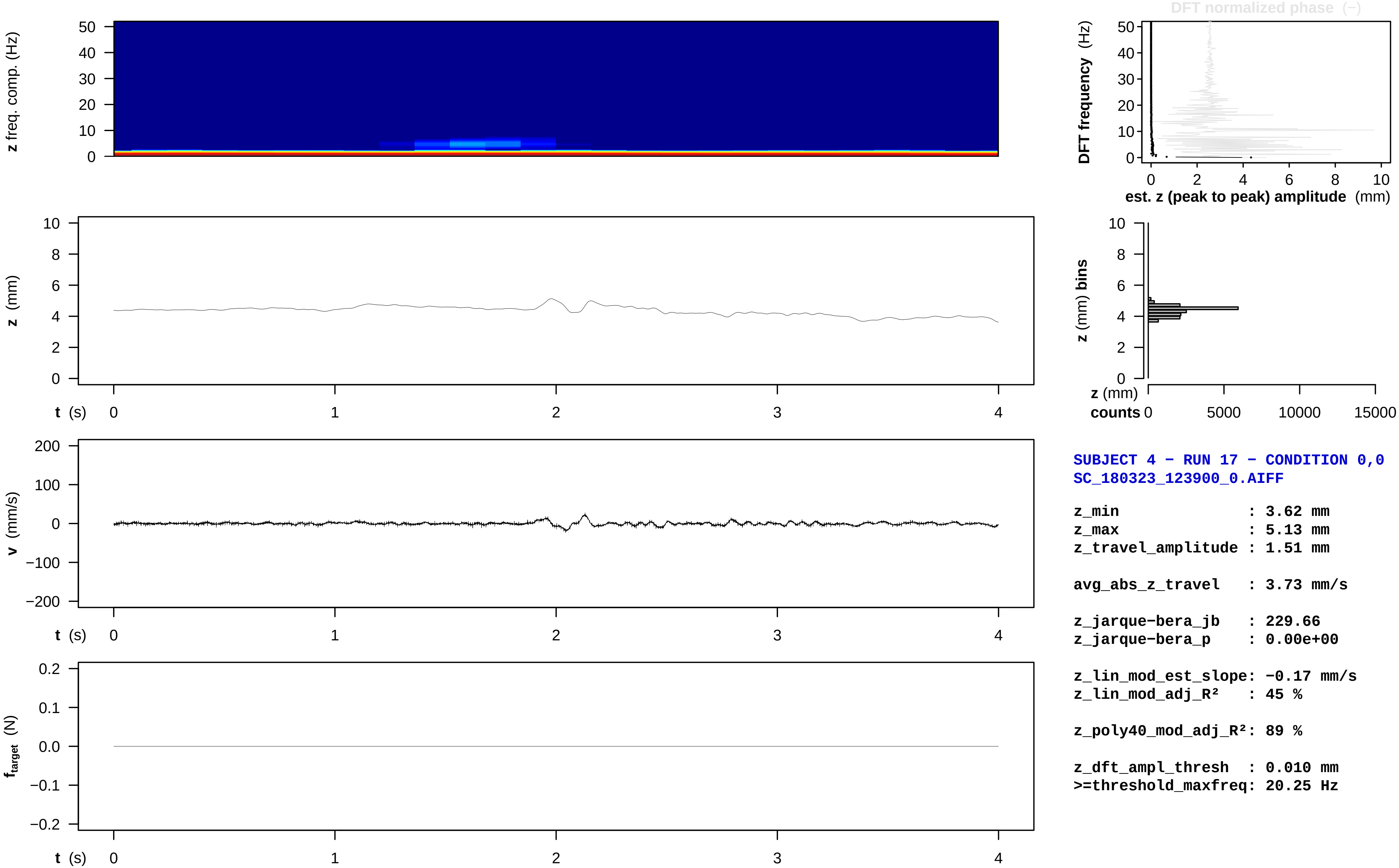

**SUBJECT 4 − RUN 17 − CONDITION 0,0**
**SC_180323_123900_0.AIFF**

```
z_min               : 3.62 mm
z_max               : 5.13 mm
z_travel_amplitude  : 1.51 mm

avg_abs_z_travel    : 3.73 mm/s

z_jarque-bera_jb    : 229.66
z_jarque-bera_p     : 0.00e+00

z_lin_mod_est_slope: −0.17 mm/s
z_lin_mod_adj_R²    : 45 %

z_poly40_mod_adj_R²: 89 %

z_dft_ampl_thresh   : 0.010 mm
>=threshold_maxfreq: 20.25 Hz
```

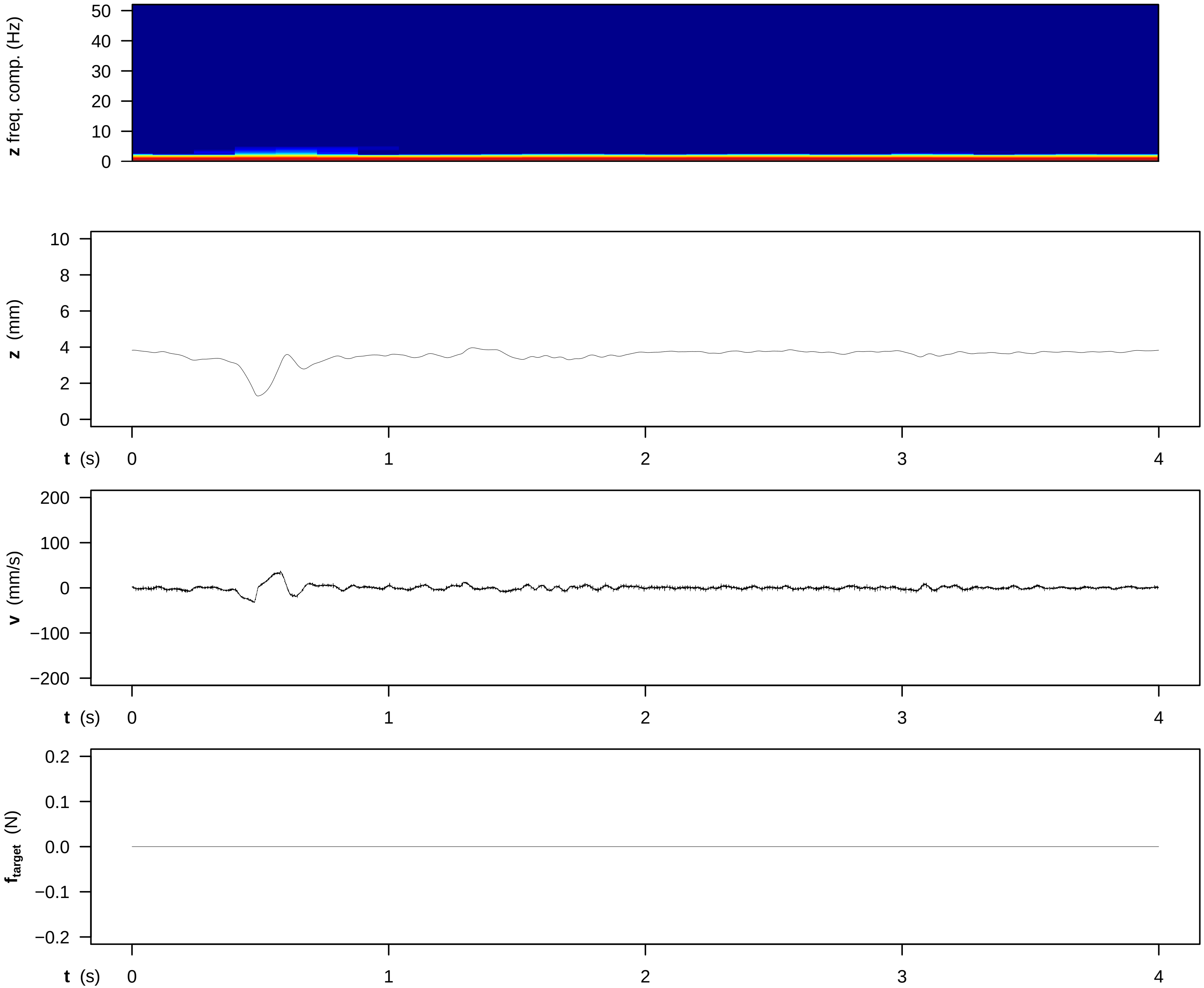

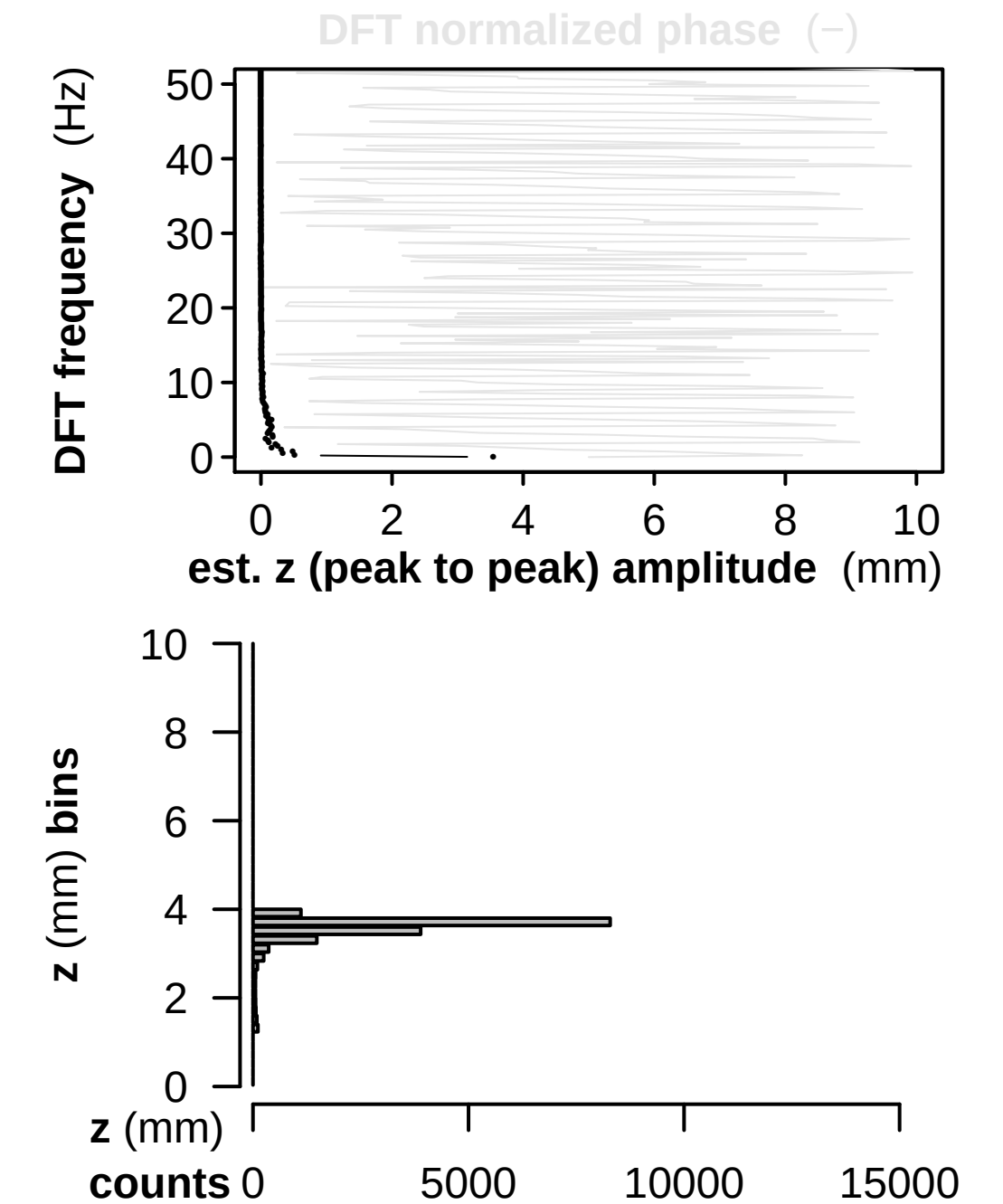

**SUBJECT 4 − RUN 22 − CONDITION 0,0**
**SC_180323_124120_0.AIFF**

```
z_min               : 1.29 mm
z_max               : 3.97 mm
z_travel_amplitude  : 2.67 mm

avg_abs_z_travel    : 4.12 mm/s

z_jarque-bera_jb    : 193367.96
z_jarque-bera_p     : 0.00e+00

z_lin_mod_est_slope: 0.16 mm/s
z_lin_mod_adj_R²    : 22 %

z_poly40_mod_adj_R²: 76 %

z_dft_ampl_thresh   : 0.010 mm
>=threshold_maxfreq: 16.75 Hz
```

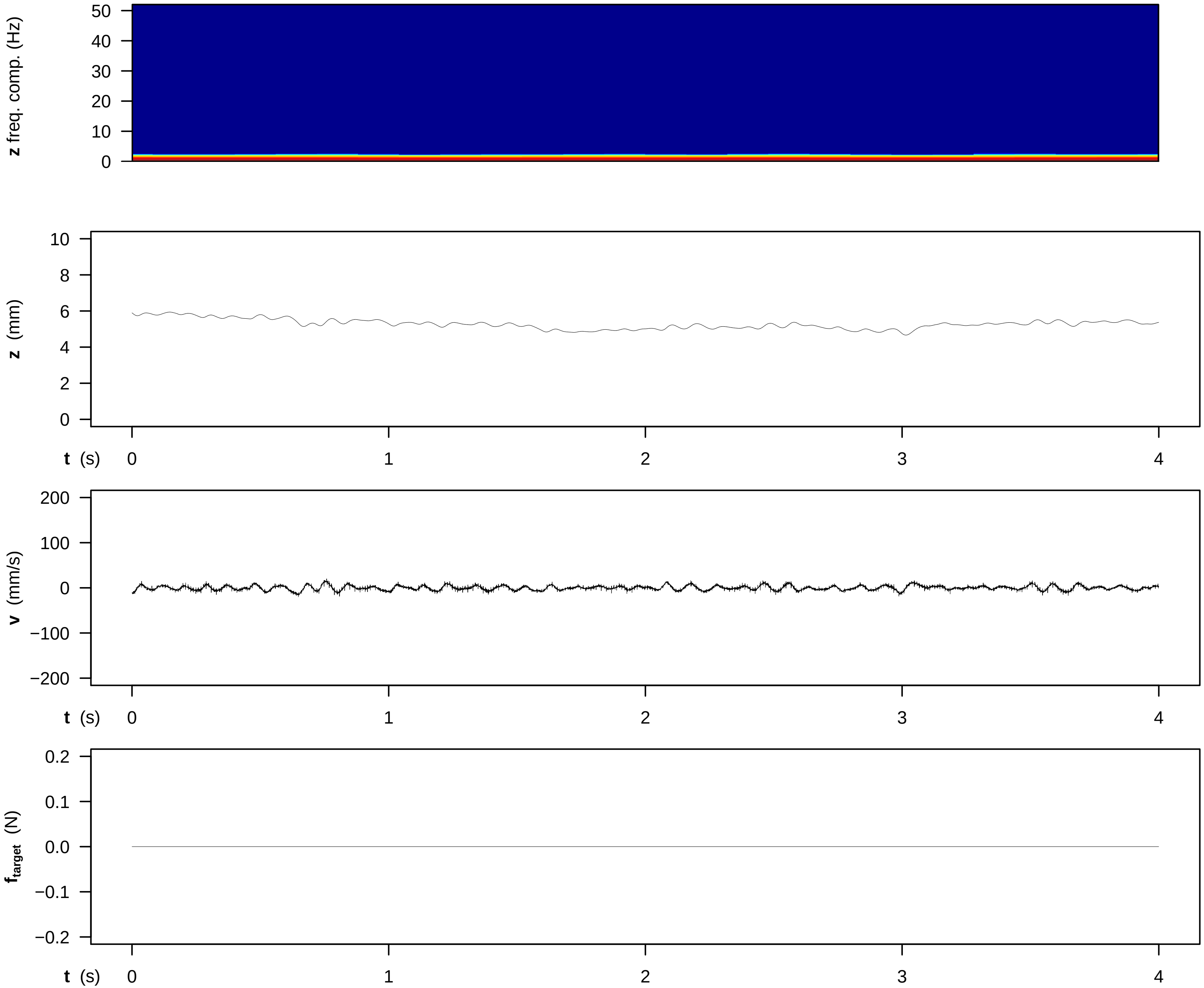

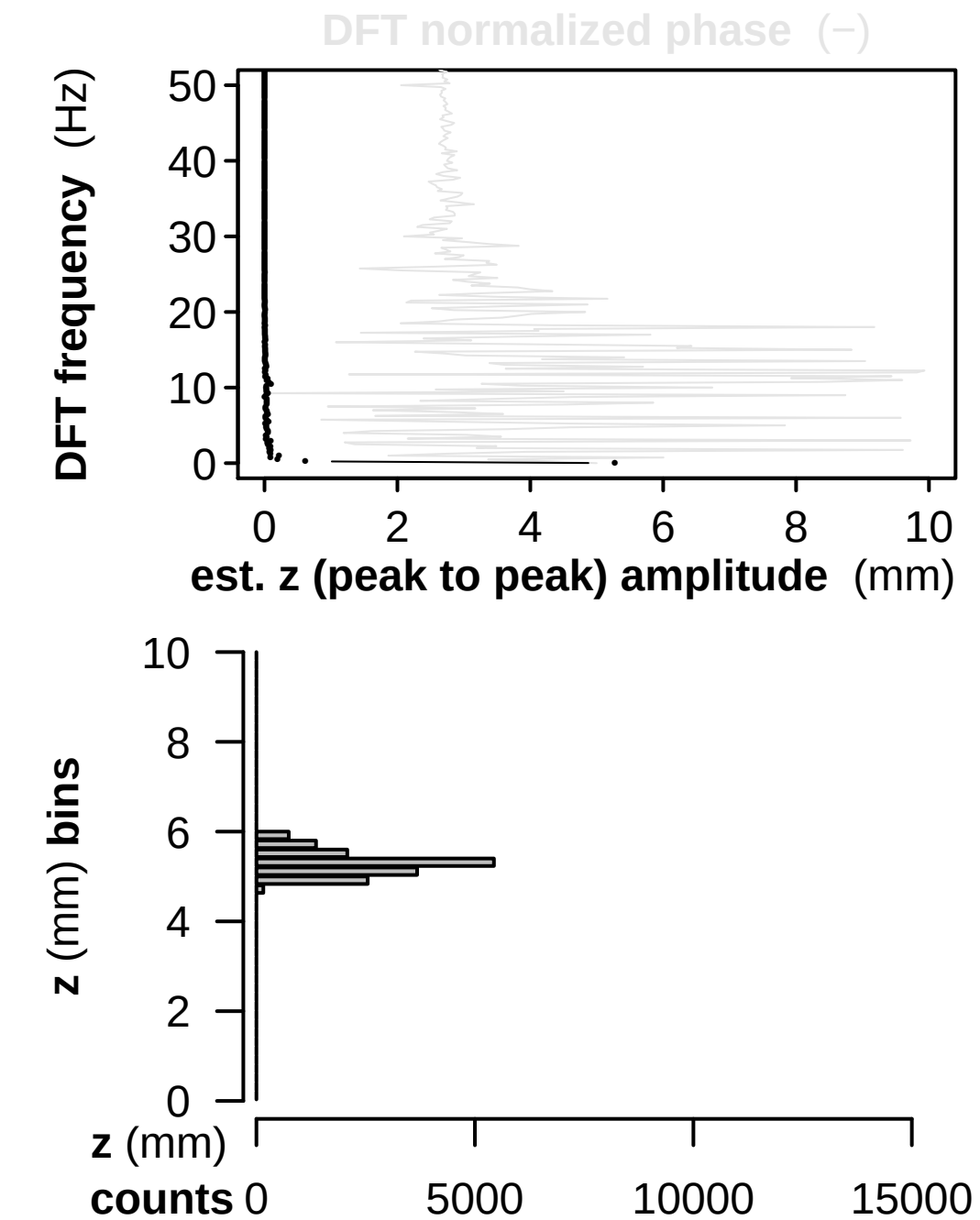

**SUBJECT 5 − RUN 17 − CONDITION 0,0**
**SC_180323_132458_0.AIFF**

```
z_min              : 4.67 mm
z_max              : 5.94 mm
z_travel_amplitude : 1.28 mm

avg_abs_z_travel   : 4.71 mm/s

z_jarque-bera_jb   : 463.01
z_jarque-bera_p    : 0.00e+00

z_lin_mod_est_slope: −0.11 mm/s
z_lin_mod_adj_R²   : 21 %

z_poly40_mod_adj_R²: 89 %

z_dft_ampl_thresh  : 0.010 mm
>=threshold_maxfreq: 20.25 Hz
```

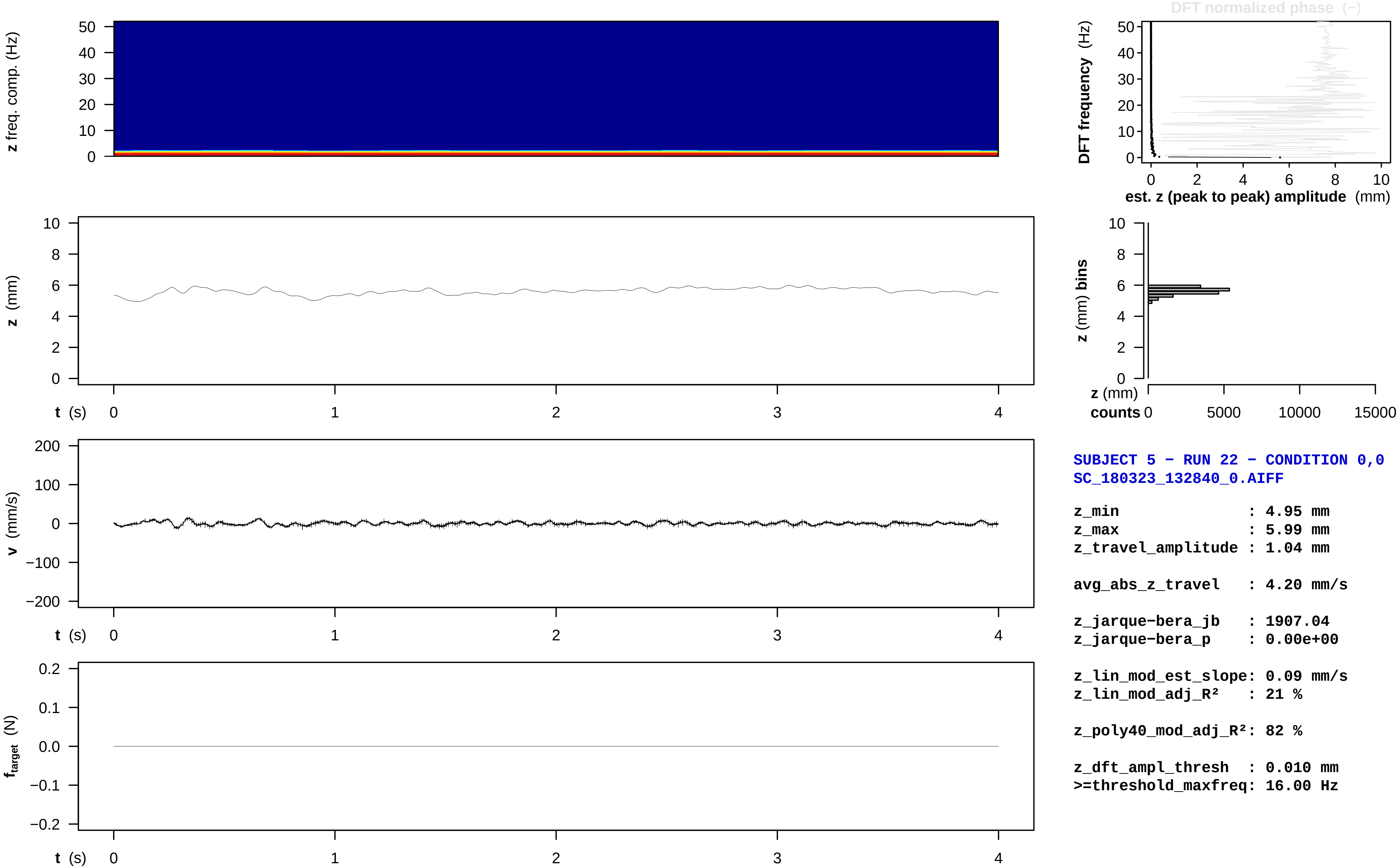

**SUBJECT 5 − RUN 22 − CONDITION 0,0**
**SC_180323_132840_0.AIFF**

```
z_min              : 4.95 mm
z_max              : 5.99 mm
z_travel_amplitude : 1.04 mm

avg_abs_z_travel   : 4.20 mm/s

z_jarque-bera_jb   : 1907.04
z_jarque-bera_p    : 0.00e+00

z_lin_mod_est_slope: 0.09 mm/s
z_lin_mod_adj_R²   : 21 %

z_poly40_mod_adj_R²: 82 %

z_dft_ampl_thresh  : 0.010 mm
>=threshold_maxfreq: 16.00 Hz
```

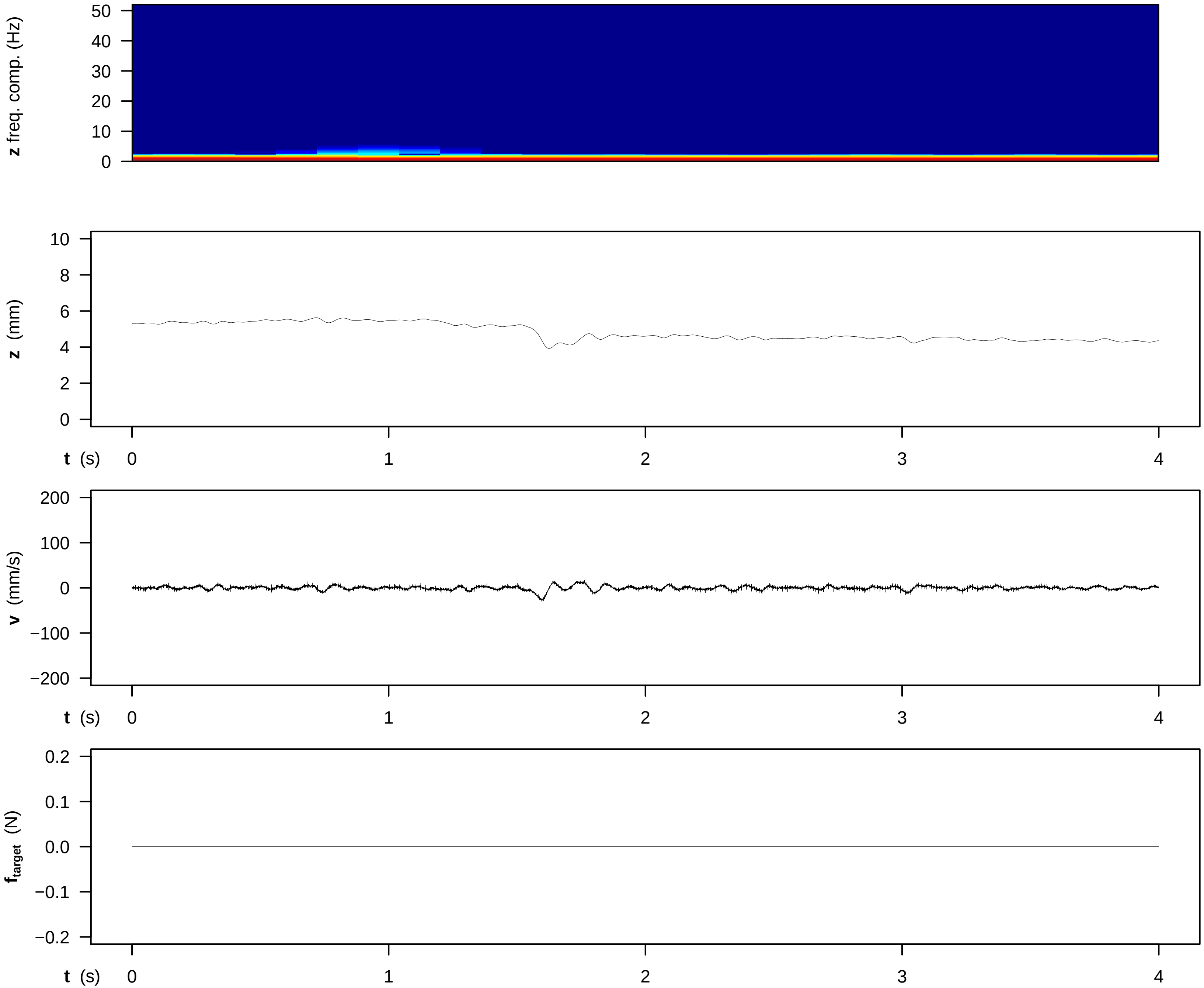

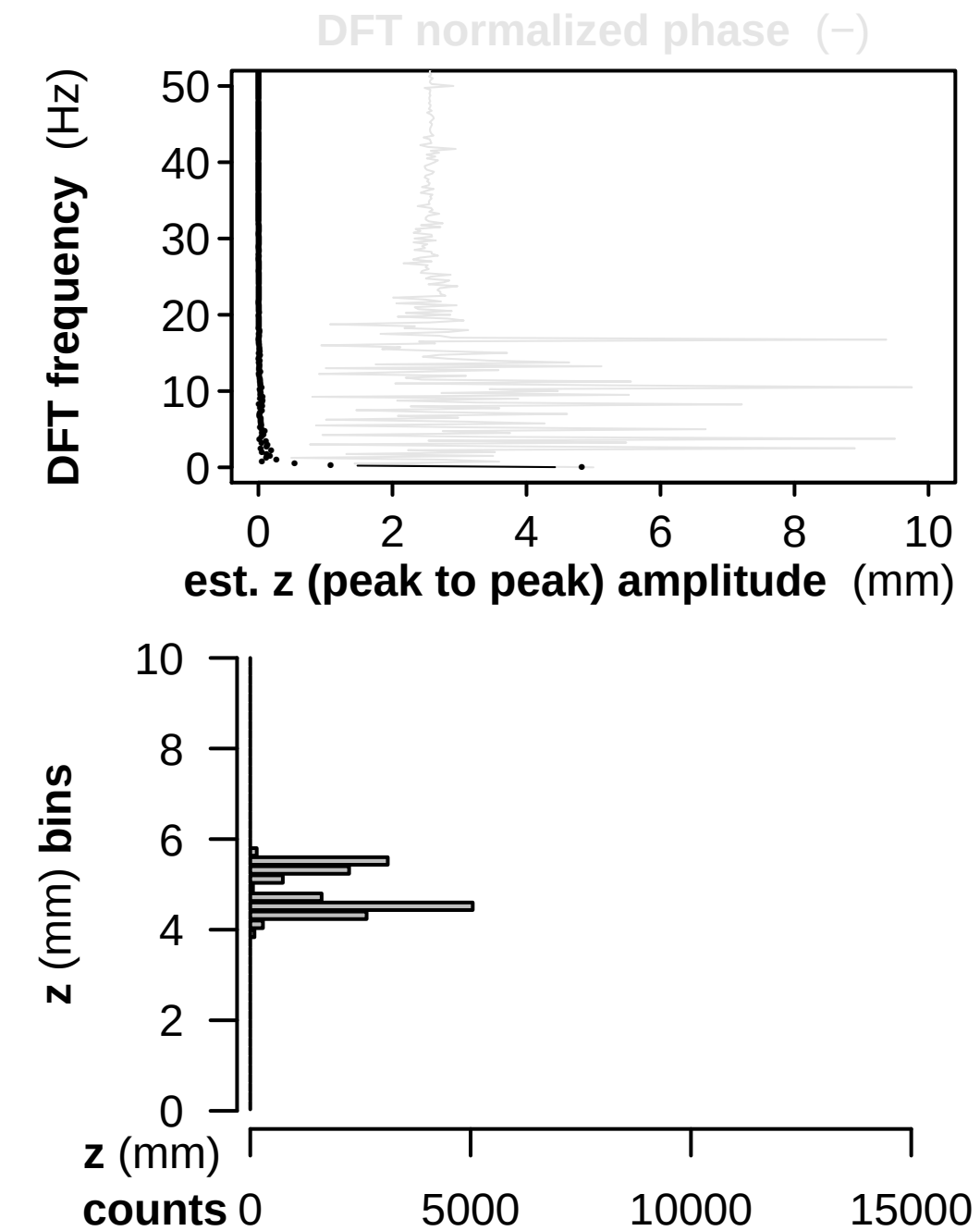

**SUBJECT 5 − RUN 28 − CONDITION 0,0**
**SC_180323_133313_0.AIFF**

```
z_min               : 3.92 mm
z_max               : 5.63 mm
z_travel_amplitude  : 1.70 mm

avg_abs_z_travel    : 5.28 mm/s

z_jarque-bera_jb    : 1847.37
z_jarque-bera_p     : 0.00e+00

z_lin_mod_est_slope: −0.35 mm/s
z_lin_mod_adj_R²    : 74 %

z_poly40_mod_adj_R²: 94 %

z_dft_ampl_thresh   : 0.010 mm
>=threshold_maxfreq: 18.00 Hz
```

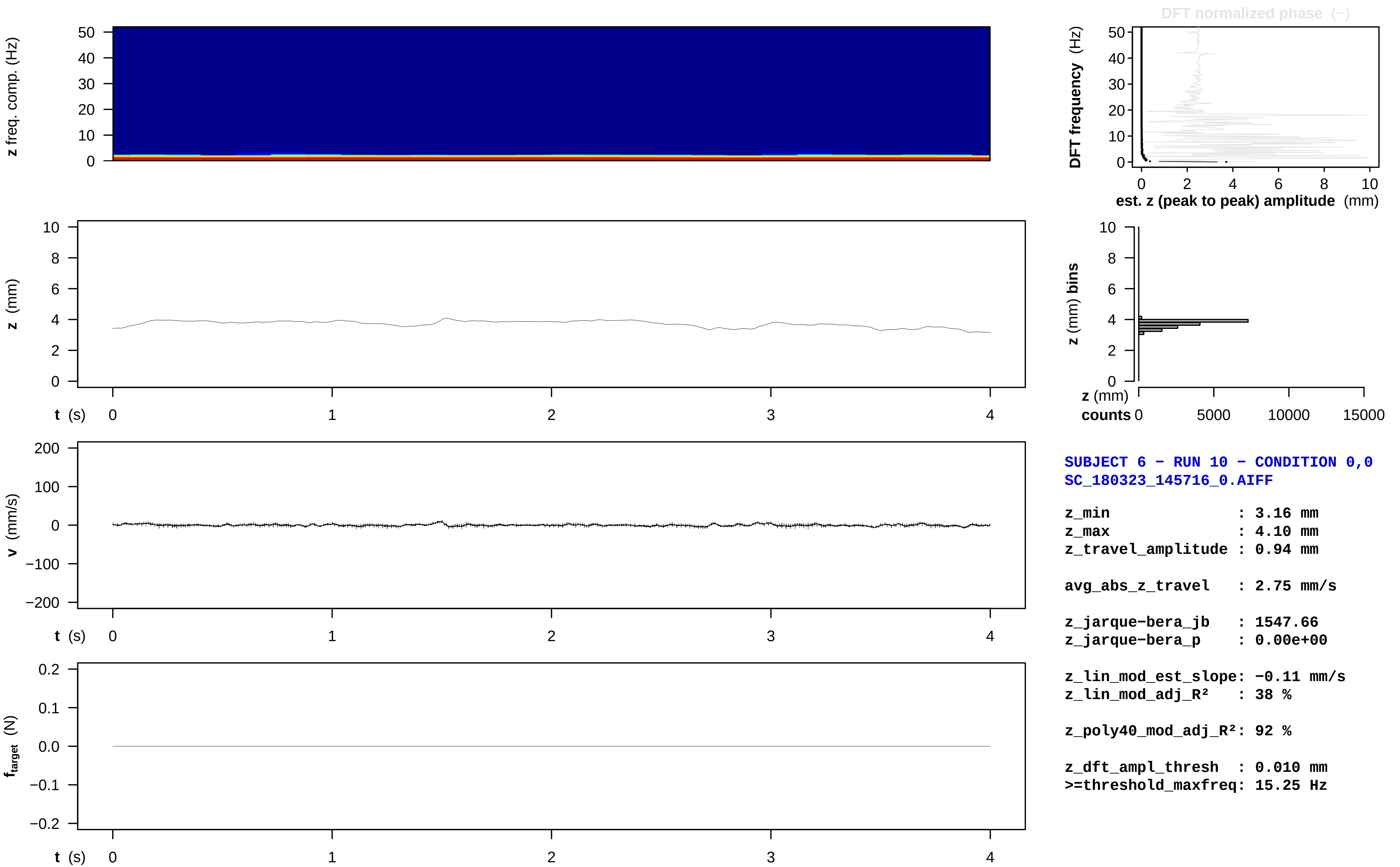

**SUBJECT 6 − RUN 10 − CONDITION 0,0**
**SC_180323_145716_0.AIFF**

```
z_min               : 3.16 mm
z_max               : 4.10 mm
z_travel_amplitude  : 0.94 mm

avg_abs_z_travel    : 2.75 mm/s

z_jarque-bera_jb    : 1547.66
z_jarque-bera_p     : 0.00e+00

z_lin_mod_est_slope: −0.11 mm/s
z_lin_mod_adj_R²    : 38 %

z_poly40_mod_adj_R²: 92 %

z_dft_ampl_thresh   : 0.010 mm
>=threshold_maxfreq: 15.25 Hz
```

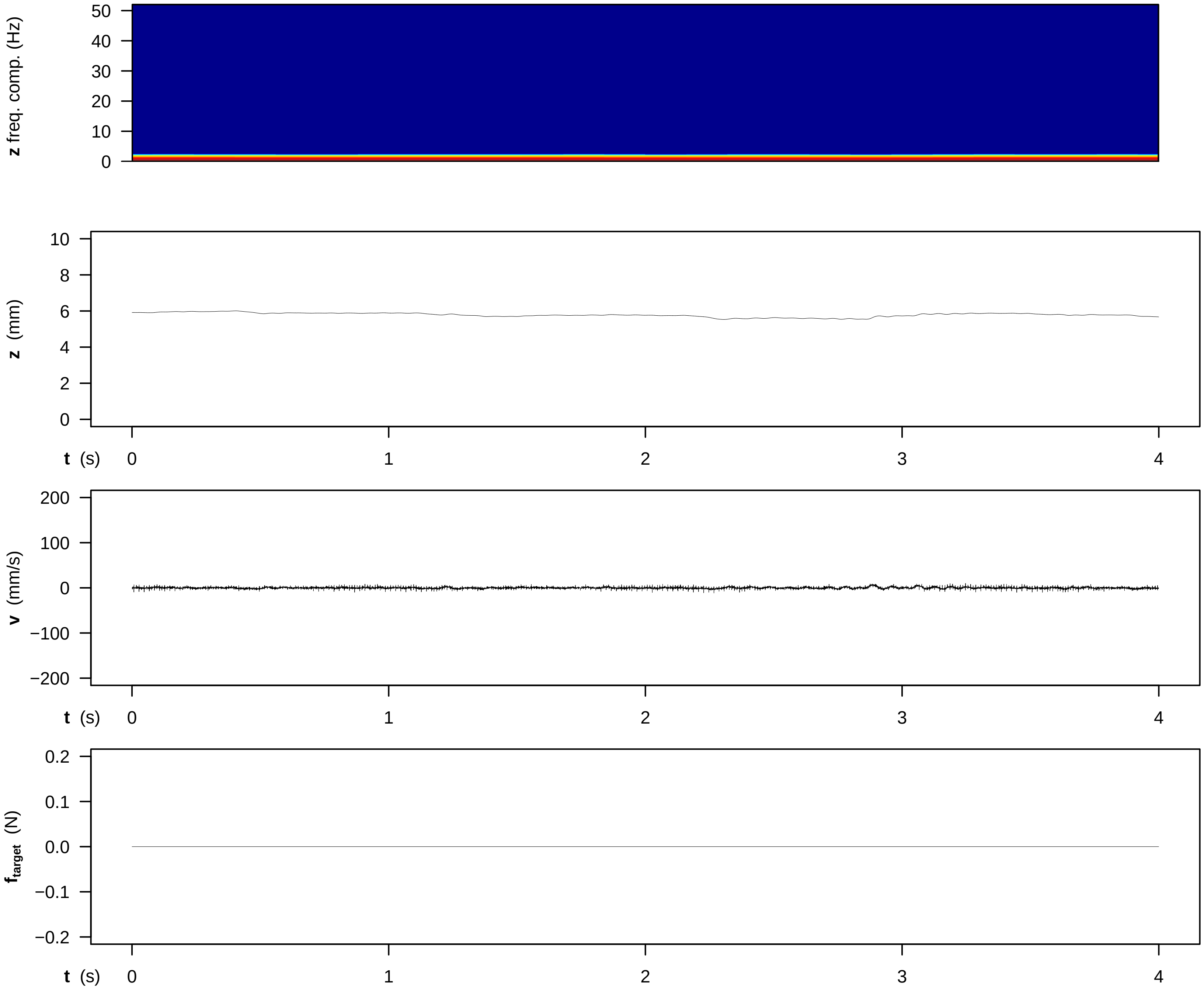

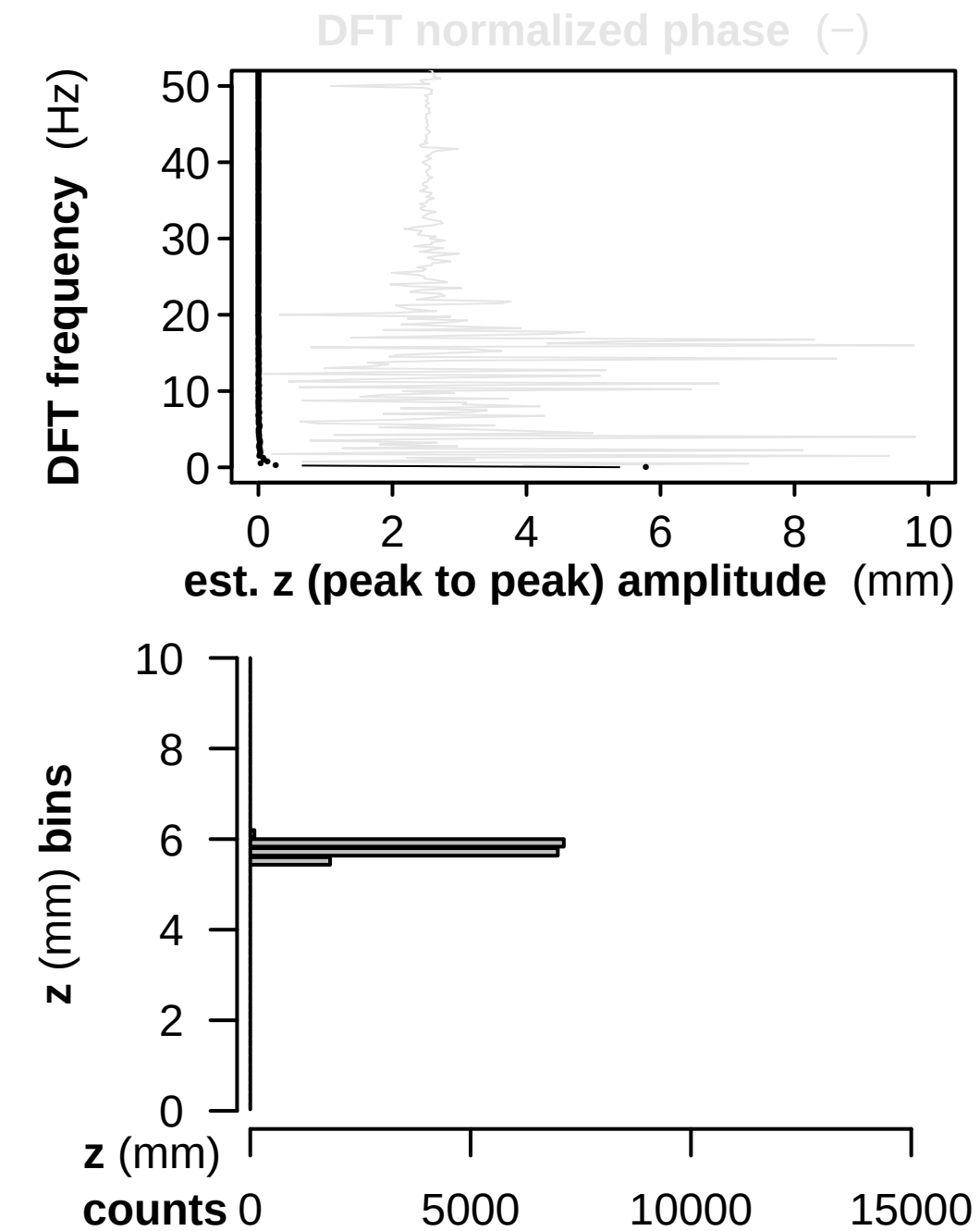

**SUBJECT 6 − RUN 33 − CONDITION 0,0**
**SC_180323_151159_0.AIFF**

```
z_min               : 5.53 mm
z_max               : 6.01 mm
z_travel_amplitude  : 0.48 mm

avg_abs_z_travel    : 2.42 mm/s

z_jarque-bera_jb    : 588.75
z_jarque-bera_p     : 0.00e+00

z_lin_mod_est_slope: −0.05 mm/s
z_lin_mod_adj_R²    : 25 %

z_poly40_mod_adj_R²: 96 %

z_dft_ampl_thresh   : 0.010 mm
>=threshold_maxfreq: 10.75 Hz
```

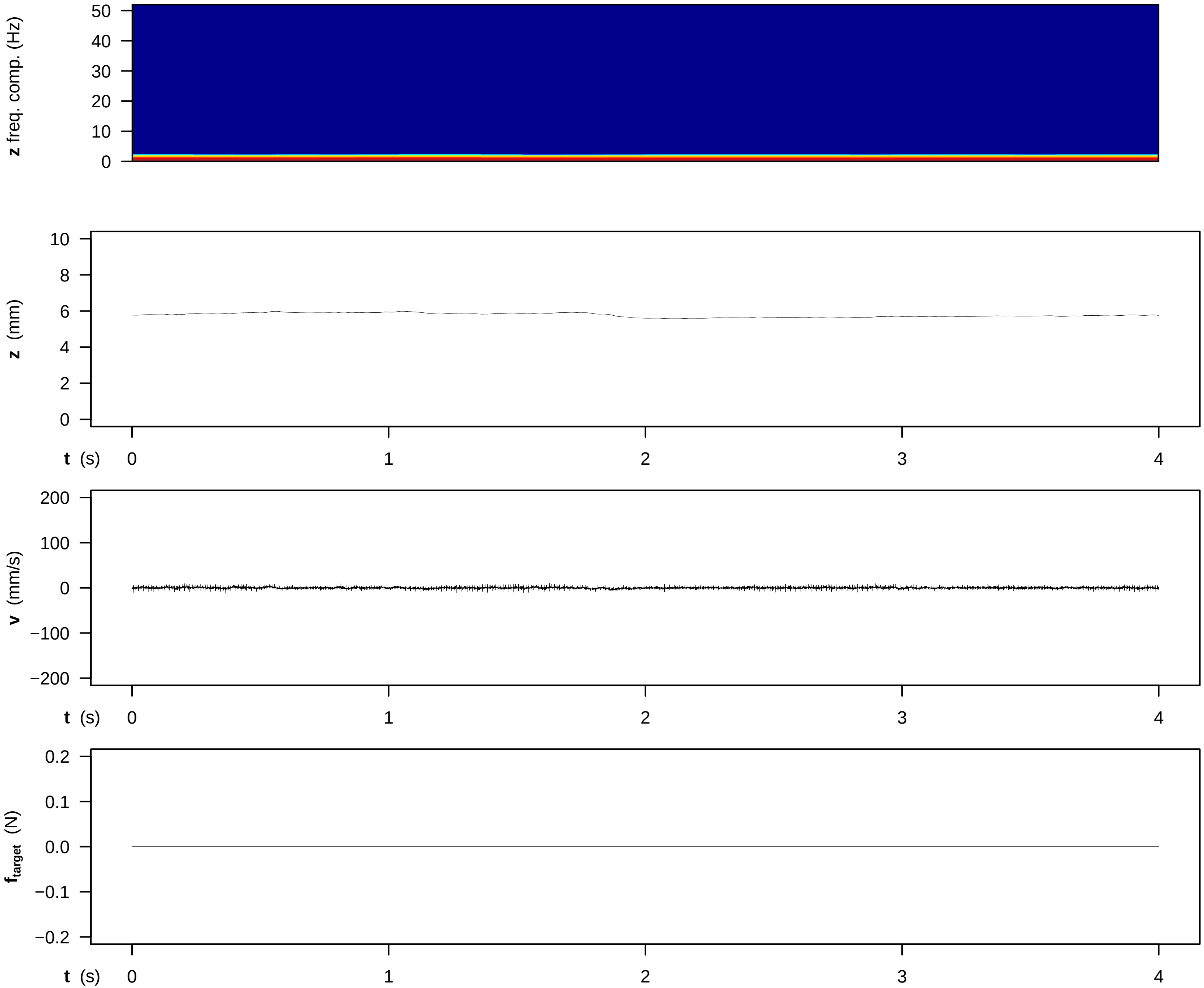

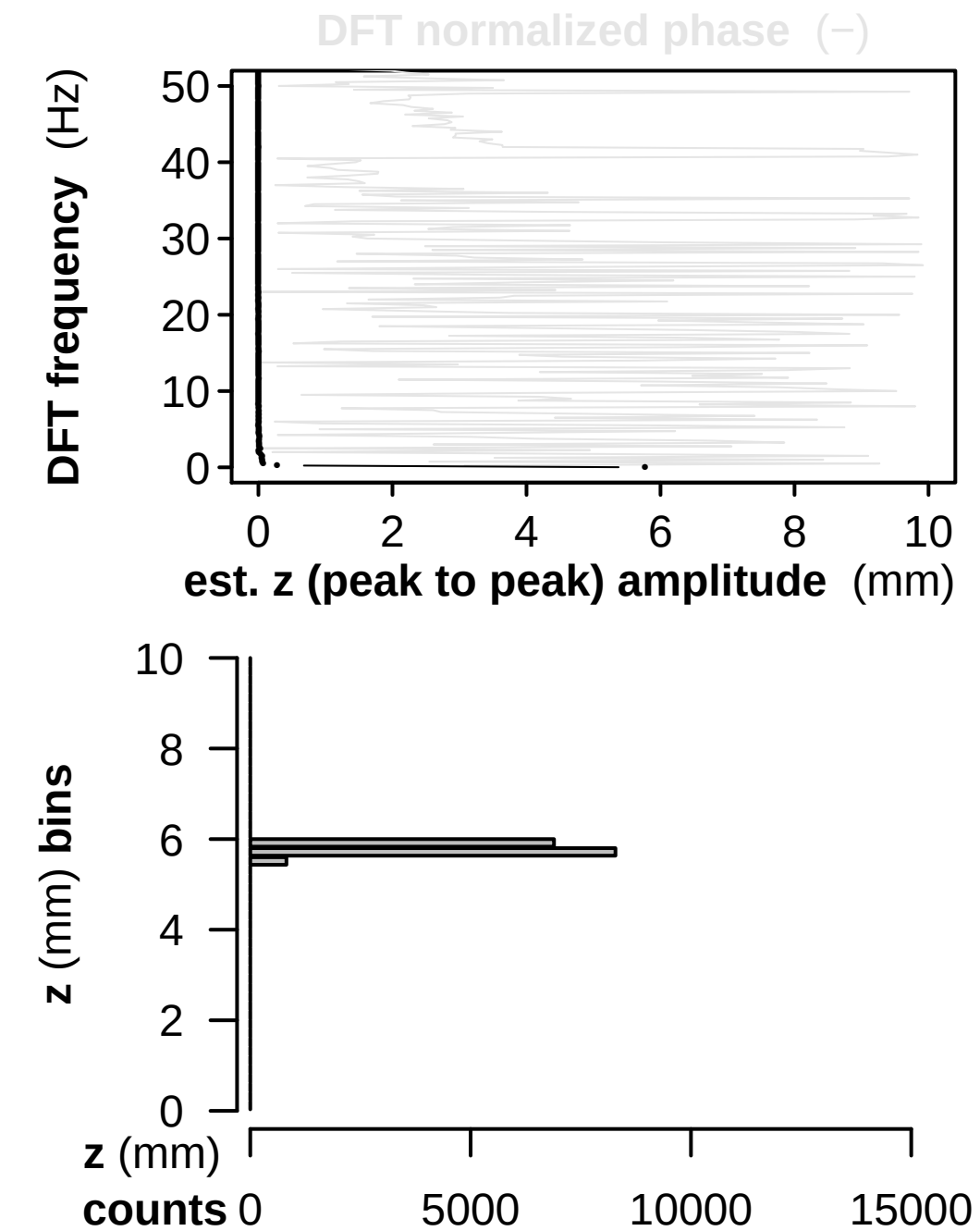

**SUBJECT 6 − RUN 34 − CONDITION 0,0**
**SC_180323_151226_0.AIFF**

```
z_min              : 5.57 mm
z_max              : 5.99 mm
z_travel_amplitude : 0.42 mm

avg_abs_z_travel   : 4.14 mm/s

z_jarque-bera_jb   : 1074.98
z_jarque-bera_p    : 0.00e+00

z_lin_mod_est_slope: −0.06 mm/s
z_lin_mod_adj_R²   : 40 %

z_poly40_mod_adj_R²: 97 %

z_dft_ampl_thresh  : 0.010 mm
>=threshold_maxfreq: 4.25 Hz
```

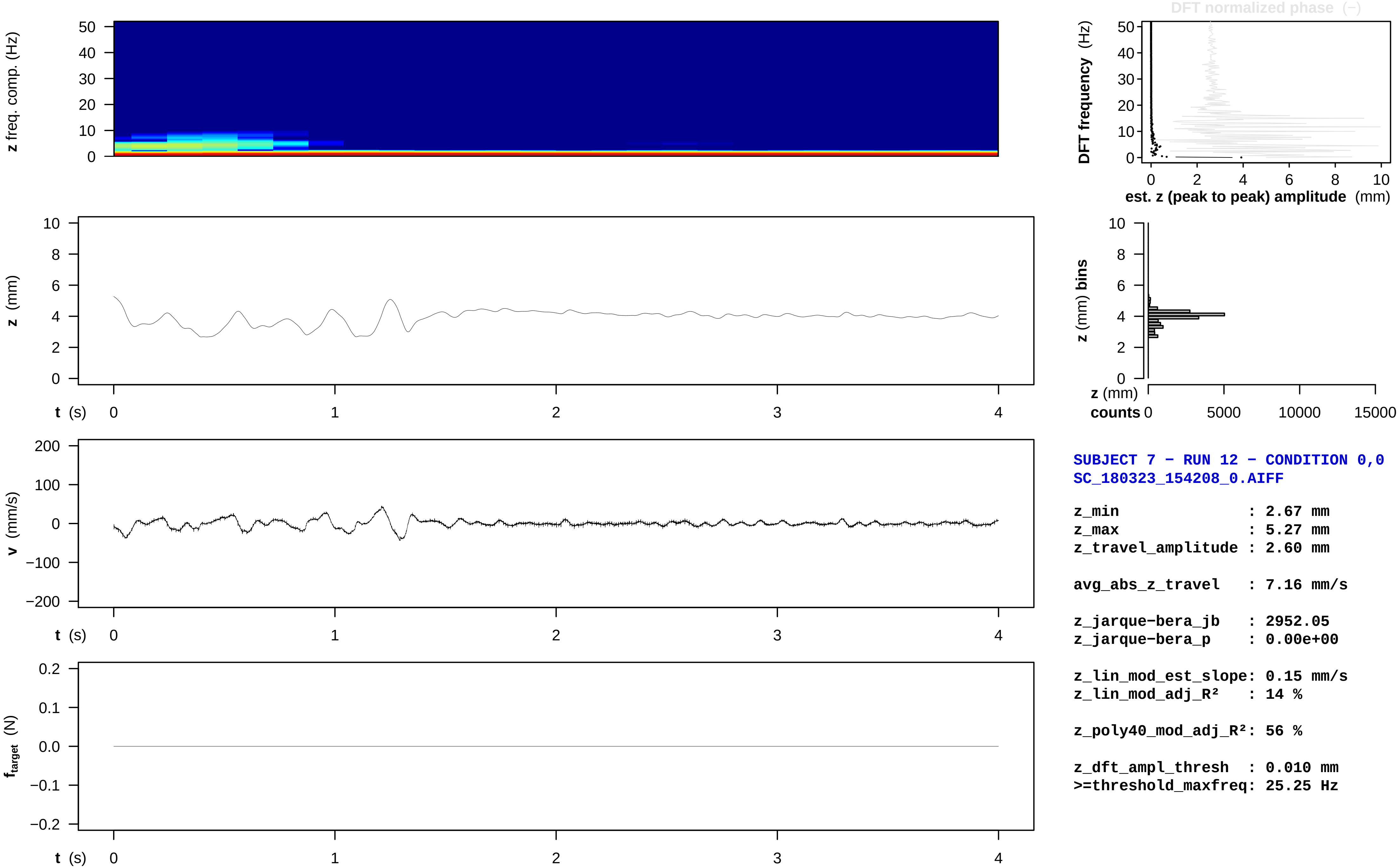

SUBJECT 7 − RUN 12 − CONDITION 0,0
SC_180323_154208_0.AIFF

```
z_min               : 2.67 mm
z_max               : 5.27 mm
z_travel_amplitude : 2.60 mm

avg_abs_z_travel   : 7.16 mm/s

z_jarque-bera_jb   : 2952.05
z_jarque-bera_p    : 0.00e+00

z_lin_mod_est_slope: 0.15 mm/s
z_lin_mod_adj_R²   : 14 %

z_poly40_mod_adj_R²: 56 %

z_dft_ampl_thresh  : 0.010 mm
>=threshold_maxfreq: 25.25 Hz
```

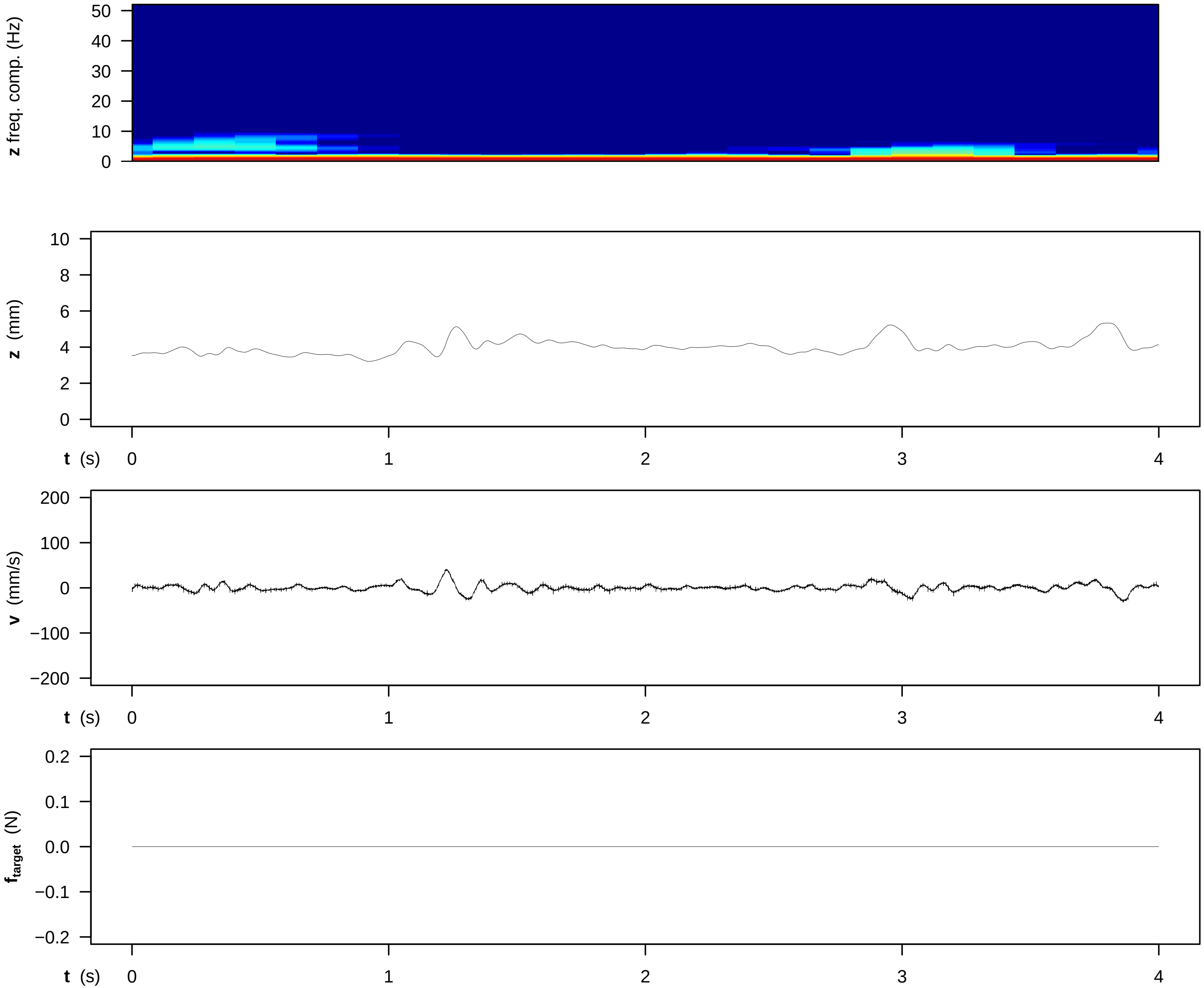

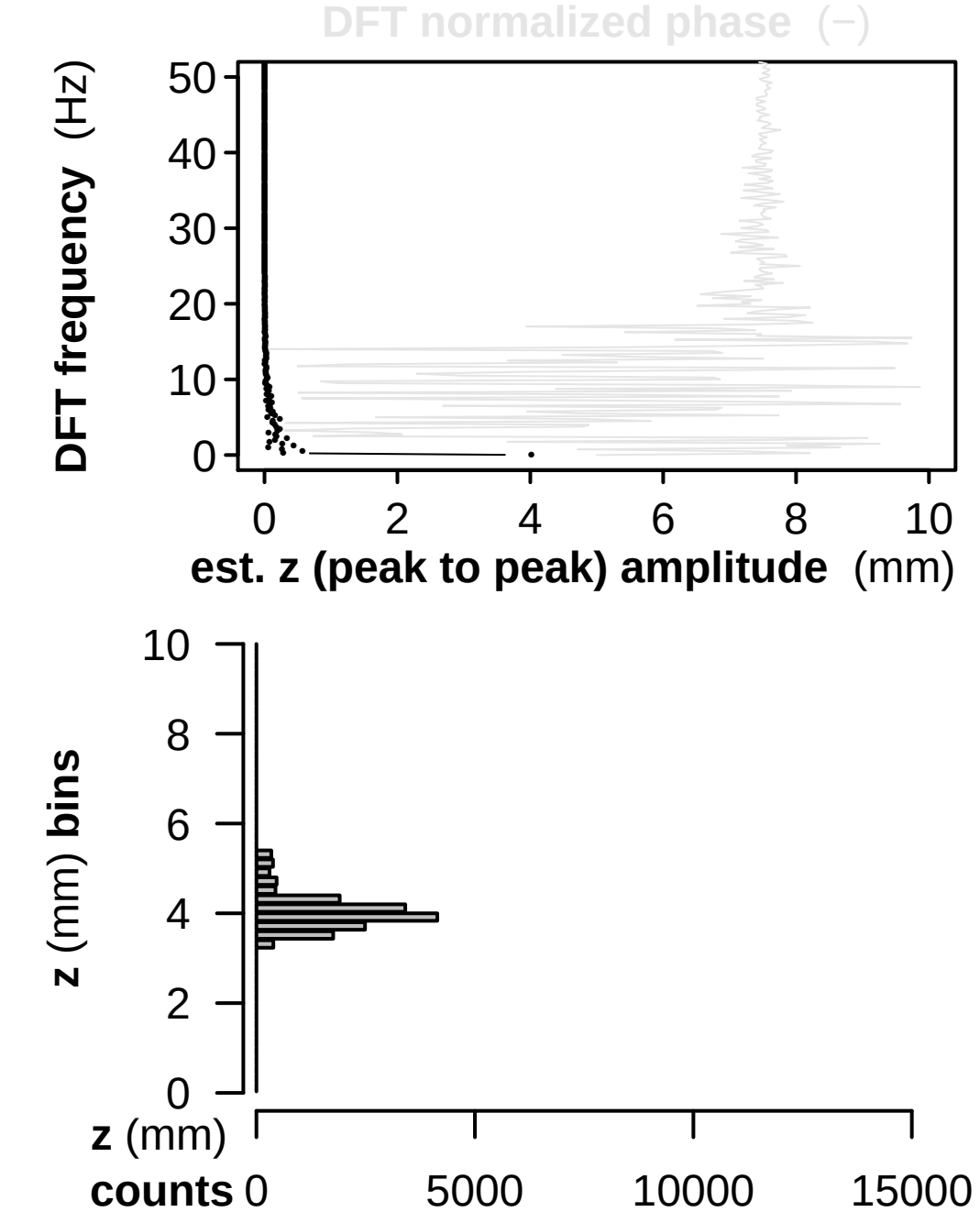

**SUBJECT 7 − RUN 21 − CONDITION 0,0**
**SC_180323_154809_0.AIFF**

```
z_min               : 3.21 mm
z_max               : 5.33 mm
z_travel_amplitude  : 2.13 mm

avg_abs_z_travel    : 6.01 mm/s

z_jarque-bera_jb    : 4744.19
z_jarque-bera_p     : 0.00e+00

z_lin_mod_est_slope: 0.15 mm/s
z_lin_mod_adj_R²    : 19 %

z_poly40_mod_adj_R²: 74 %

z_dft_ampl_thresh   : 0.010 mm
>=threshold_maxfreq: 18.75 Hz
```

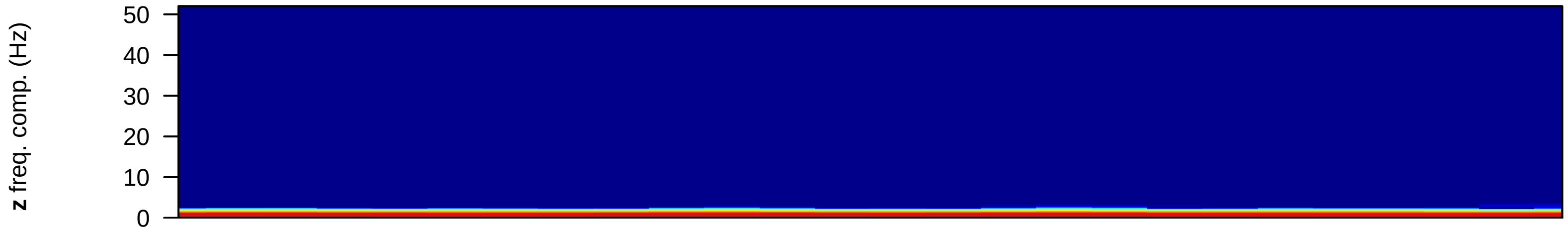

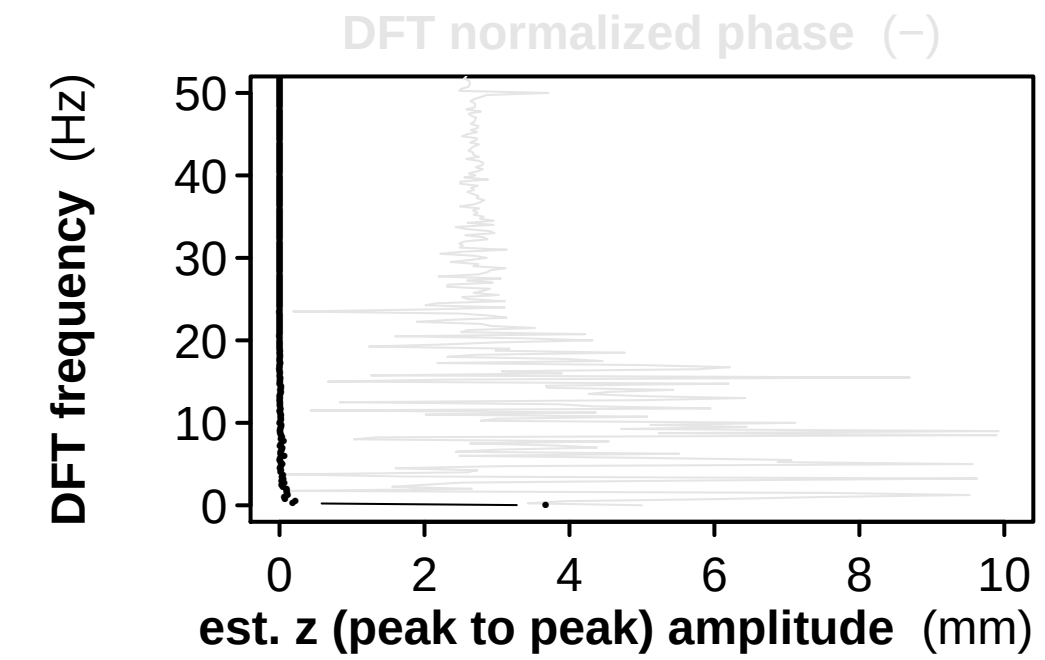

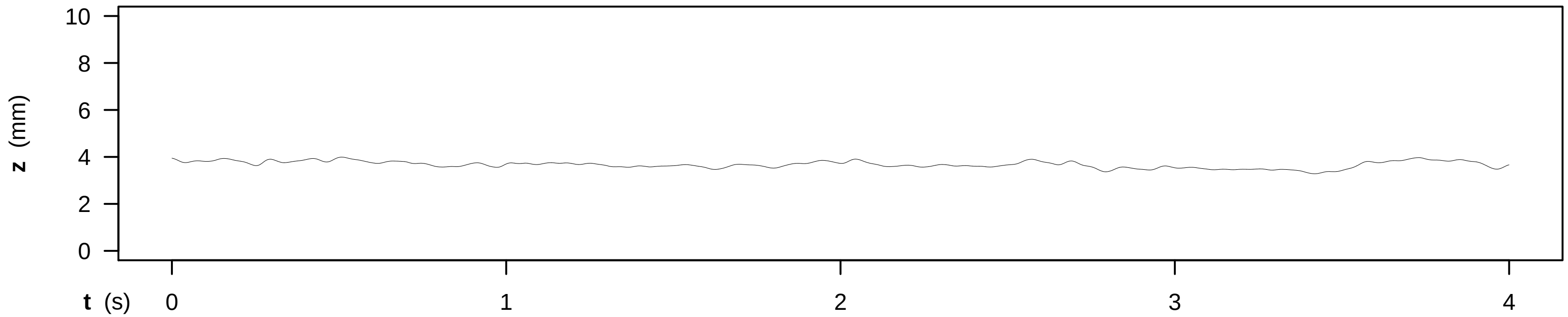

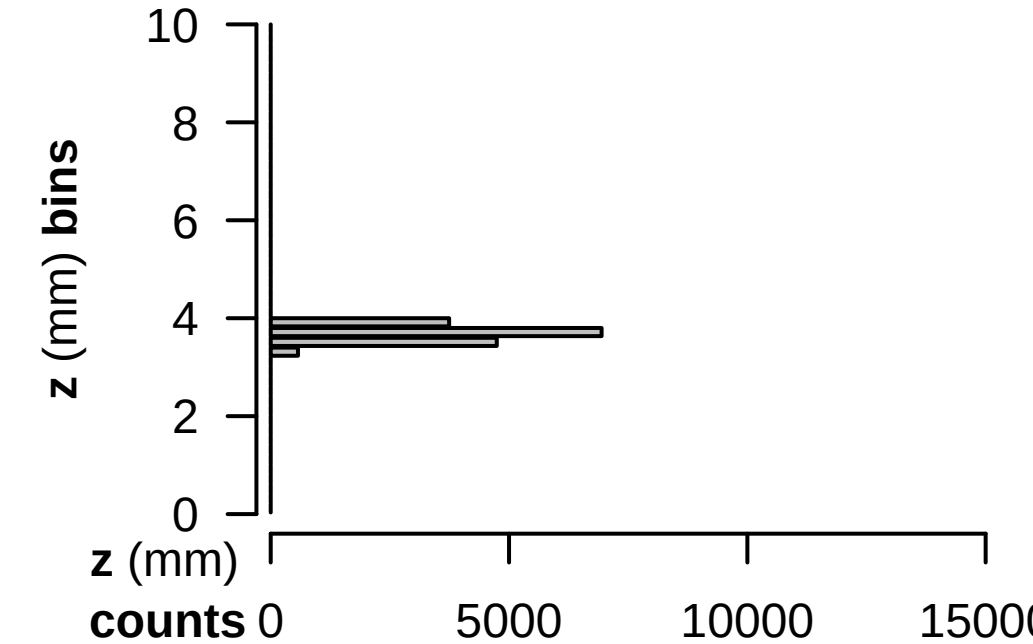

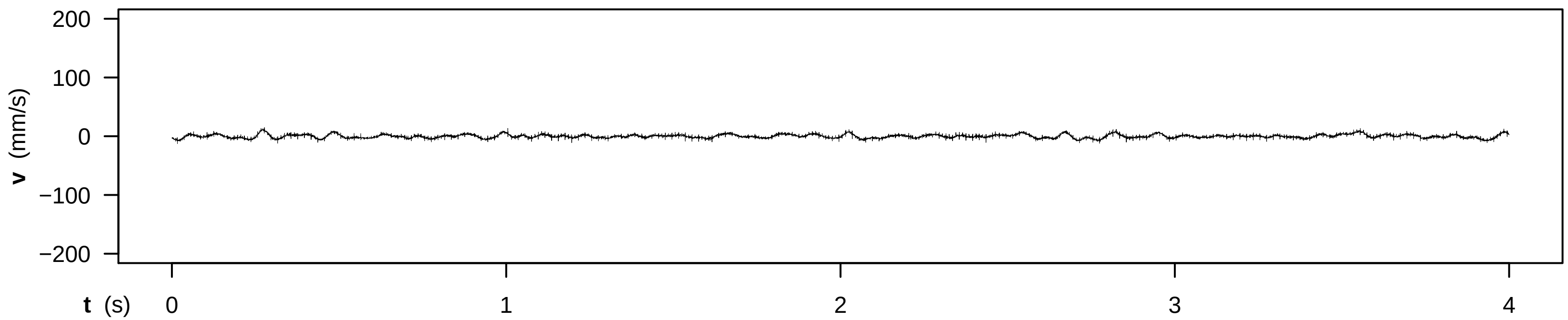

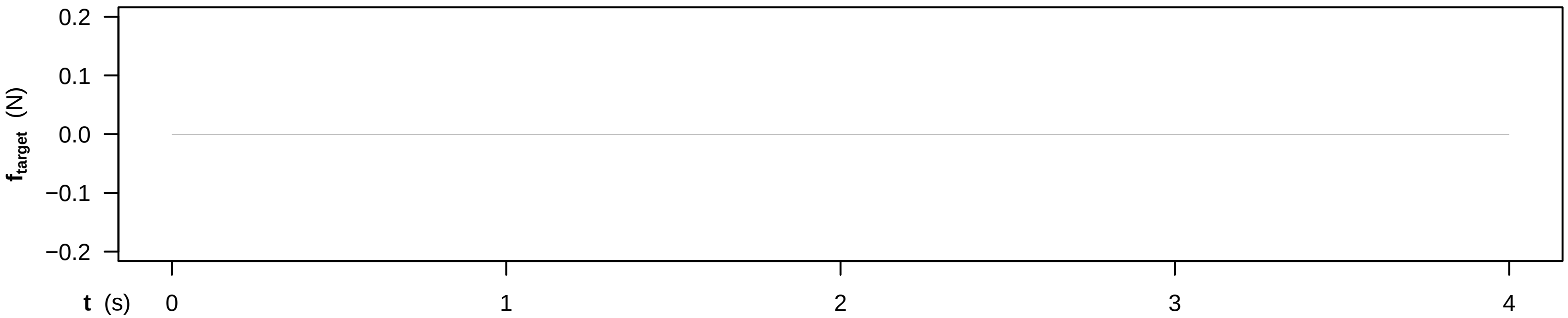

**SUBJECT 7 − RUN 31 − CONDITION 0,0**
**SC_180323_155707_0.AIFF**

```
z_min               : 3.28 mm
z_max               : 3.99 mm
z_travel_amplitude  : 0.71 mm

avg_abs_z_travel    : 3.06 mm/s

z_jarque-bera_jb    : 300.57
z_jarque-bera_p     : 0.00e+00

z_lin_mod_est_slope: −0.05 mm/s
z_lin_mod_adj_R²    : 15 %

z_poly40_mod_adj_R²: 82 %

z_dft_ampl_thresh   : 0.010 mm
>=threshold_maxfreq: 17.25 Hz
```

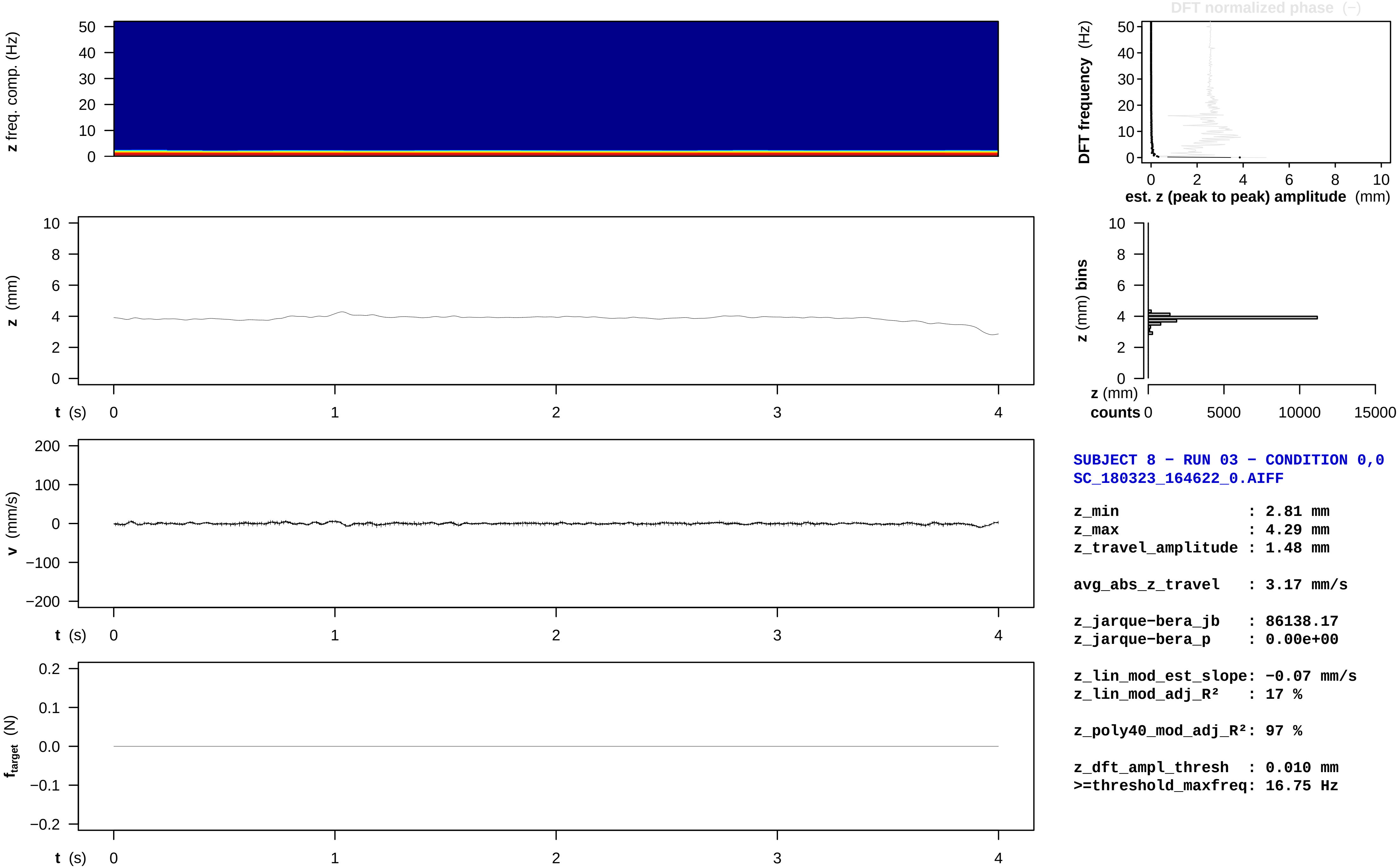

**SUBJECT 8 − RUN 03 − CONDITION 0,0**
**SC_180323_164622_0.AIFF**

```
z_min               : 2.81 mm
z_max               : 4.29 mm
z_travel_amplitude  : 1.48 mm

avg_abs_z_travel    : 3.17 mm/s

z_jarque-bera_jb    : 86138.17
z_jarque-bera_p     : 0.00e+00

z_lin_mod_est_slope: −0.07 mm/s
z_lin_mod_adj_R²    : 17 %

z_poly40_mod_adj_R²: 97 %

z_dft_ampl_thresh   : 0.010 mm
>=threshold_maxfreq: 16.75 Hz
```

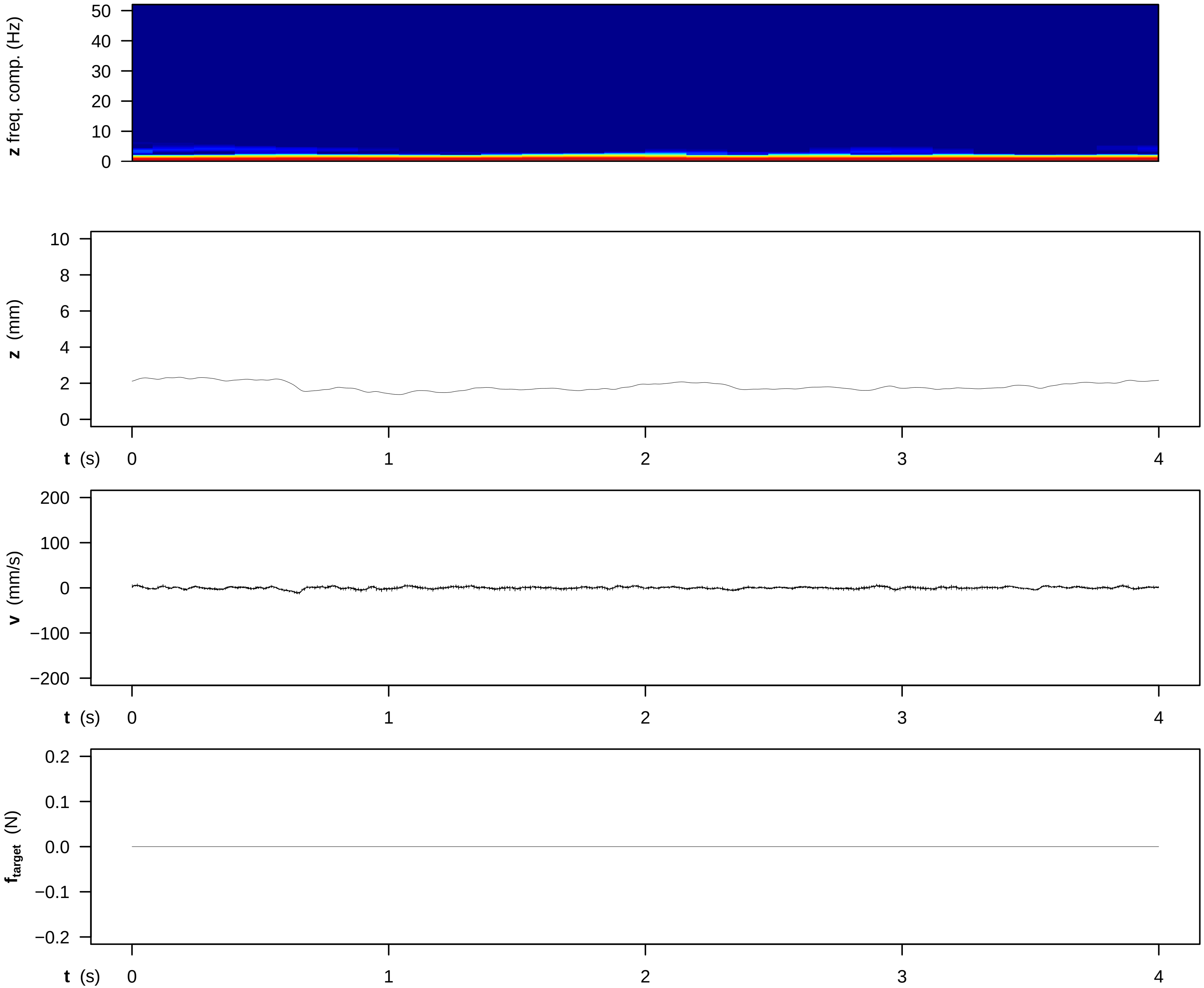

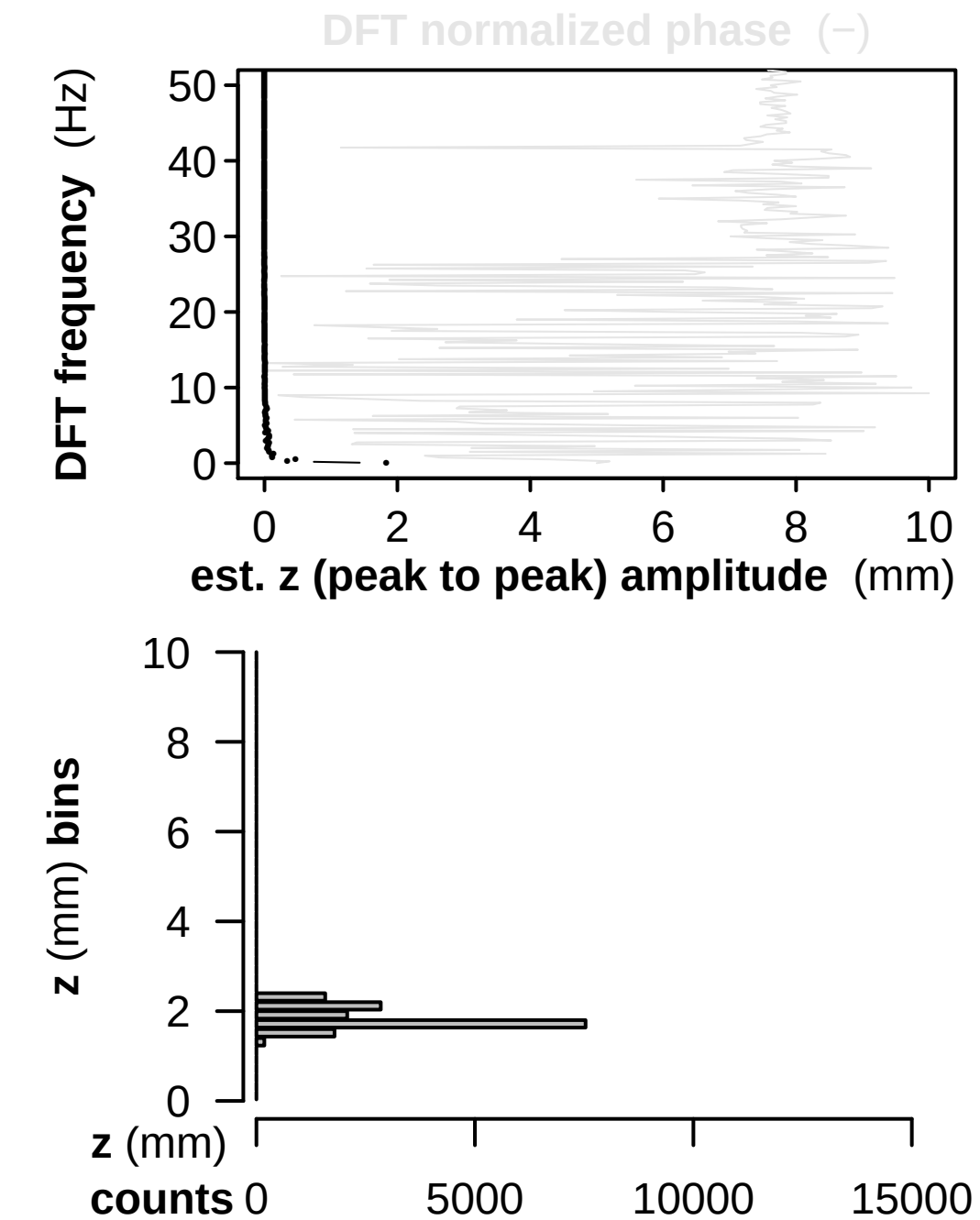

**SUBJECT 8 − RUN 06 − CONDITION 0,0**
**SC_180323_164803_0.AIFF**

```
z_min               : 1.37 mm
z_max               : 2.33 mm
z_travel_amplitude  : 0.96 mm

avg_abs_z_travel    : 2.25 mm/s

z_jarque-bera_jb    : 1068.98
z_jarque-bera_p     : 0.00e+00

z_lin_mod_est_slope: −0.03 mm/s
z_lin_mod_adj_R²    : 2 %

z_poly40_mod_adj_R²: 91 %

z_dft_ampl_thresh   : 0.010 mm
>=threshold_maxfreq: 13.25 Hz
```

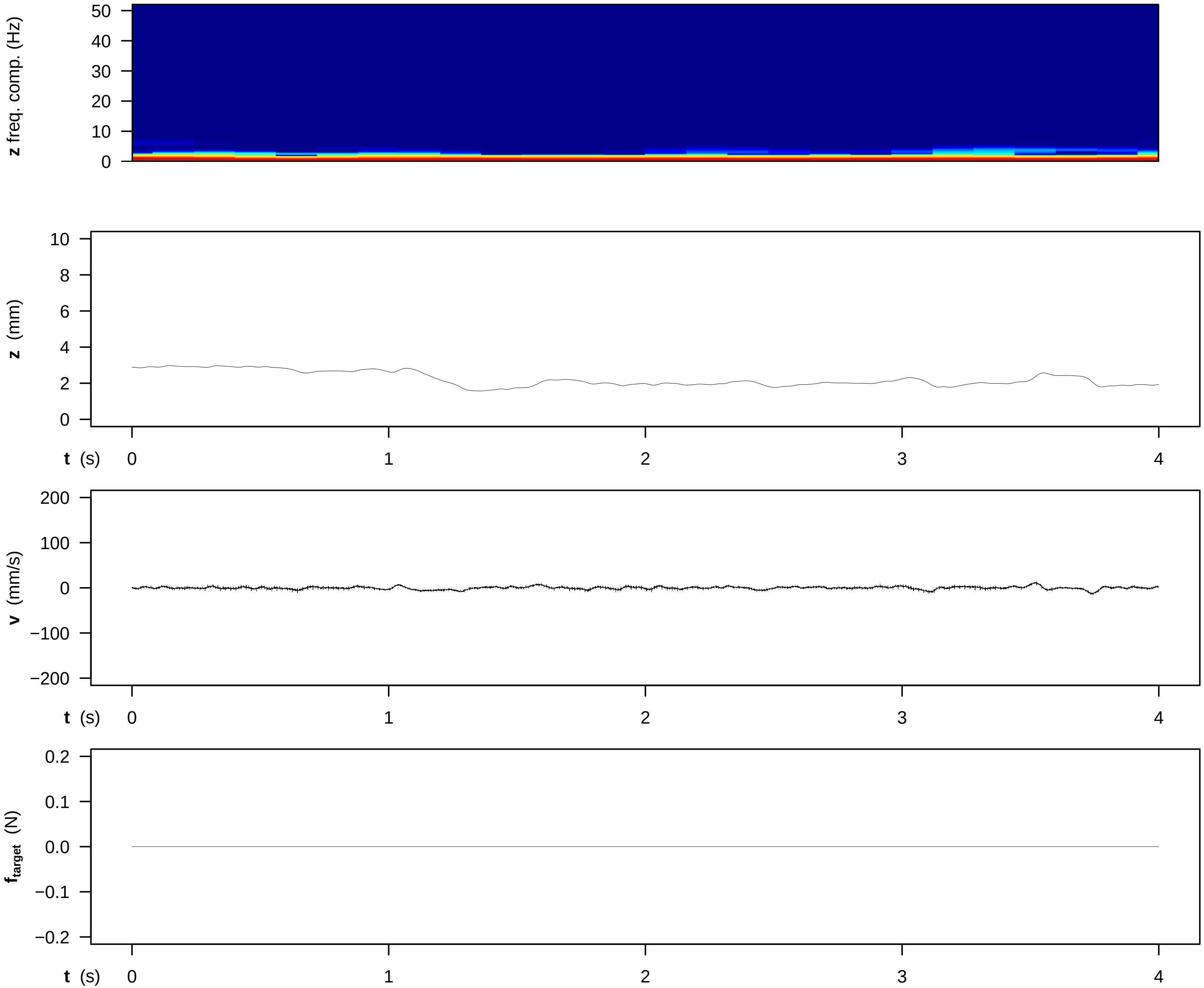

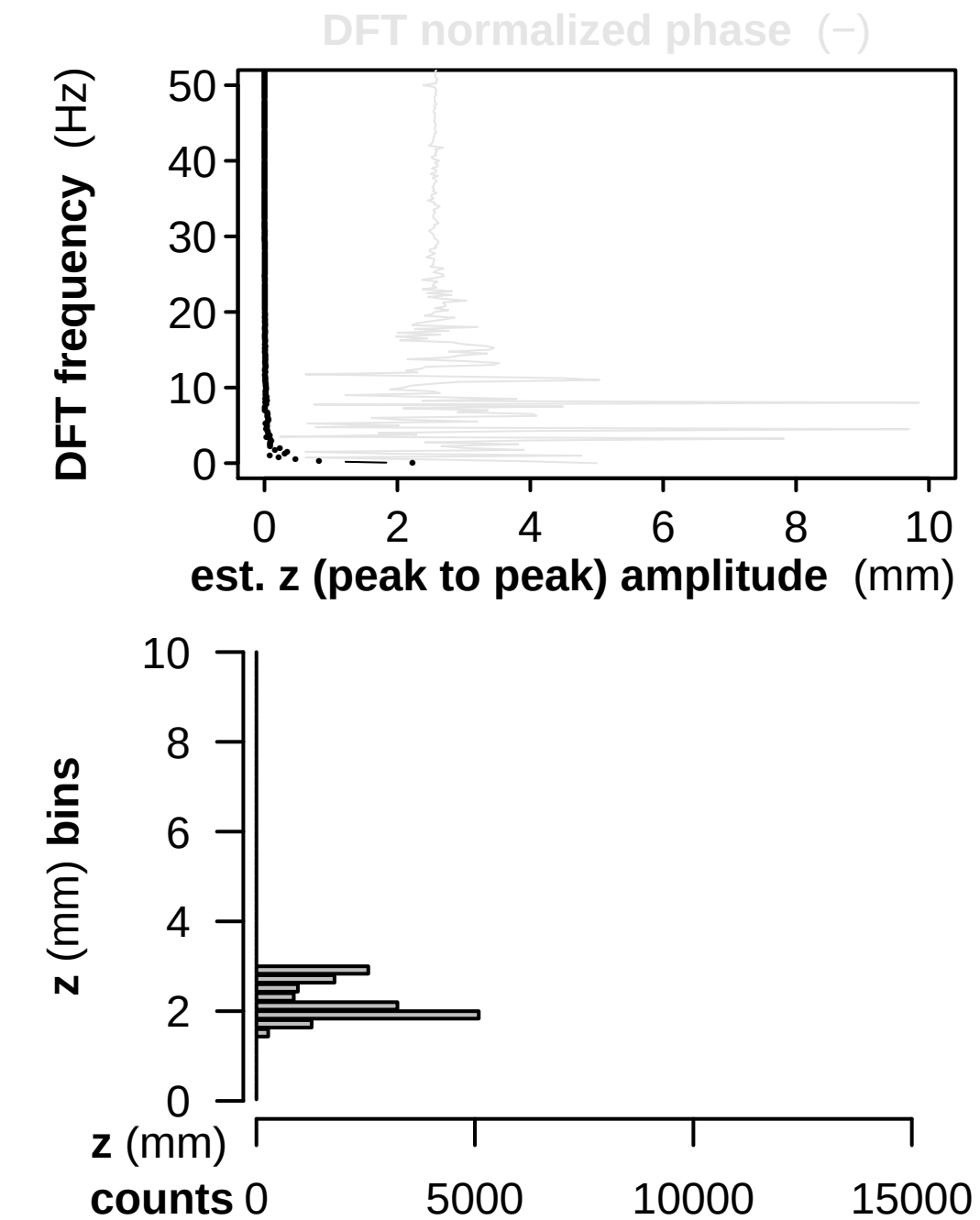

**SUBJECT 8 − RUN 12 − CONDITION 0,0**
**SC_180323_165213_0.AIFF**

```
z_min               : 1.57 mm
z_max               : 2.98 mm
z_travel_amplitude : 1.41 mm

avg_abs_z_travel    : 3.45 mm/s

z_jarque-bera_jb    : 1592.28
z_jarque-bera_p     : 0.00e+00

z_lin_mod_est_slope: -0.23 mm/s
z_lin_mod_adj_R²    : 42 %

z_poly40_mod_adj_R²: 96 %

z_dft_ampl_thresh   : 0.010 mm
>=threshold_maxfreq: 19.75 Hz
```

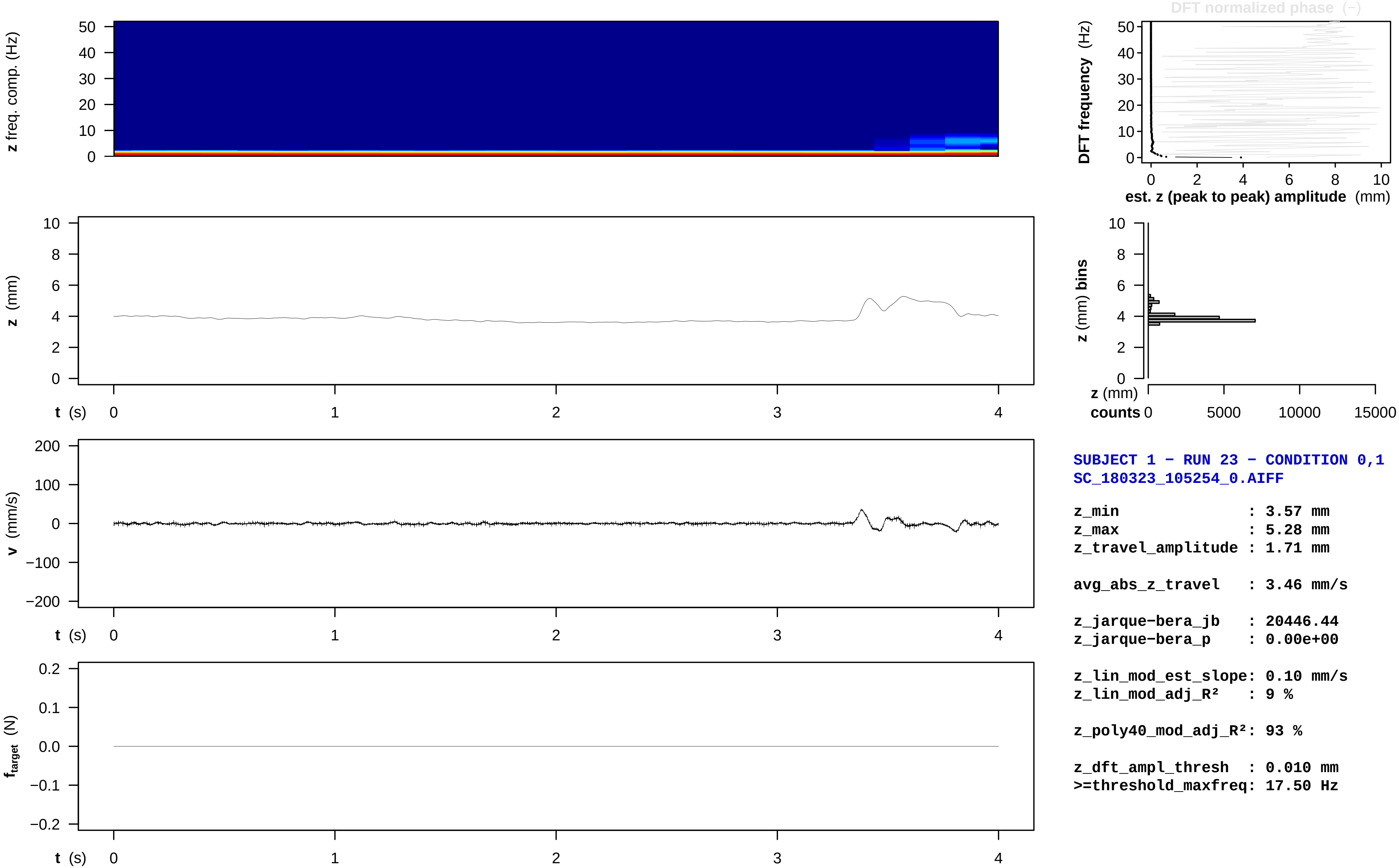

**SUBJECT 1 − RUN 23 − CONDITION 0,1**
**SC_180323_105254_0.AIFF**

```
z_min               : 3.57 mm
z_max               : 5.28 mm
z_travel_amplitude  : 1.71 mm

avg_abs_z_travel    : 3.46 mm/s

z_jarque-bera_jb    : 20446.44
z_jarque-bera_p     : 0.00e+00

z_lin_mod_est_slope: 0.10 mm/s
z_lin_mod_adj_R²    : 9 %

z_poly40_mod_adj_R²: 93 %

z_dft_ampl_thresh   : 0.010 mm
>=threshold_maxfreq: 17.50 Hz
```

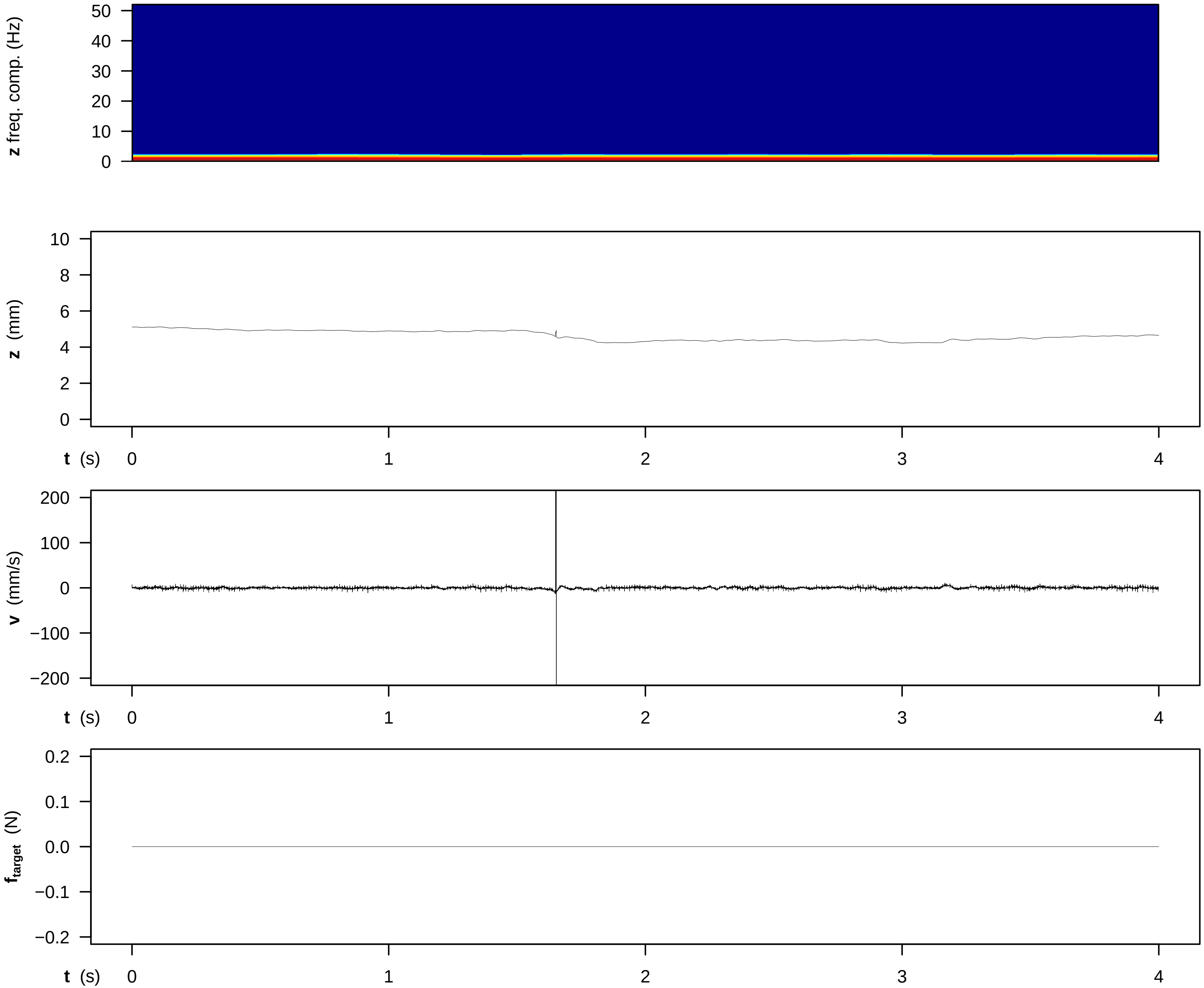

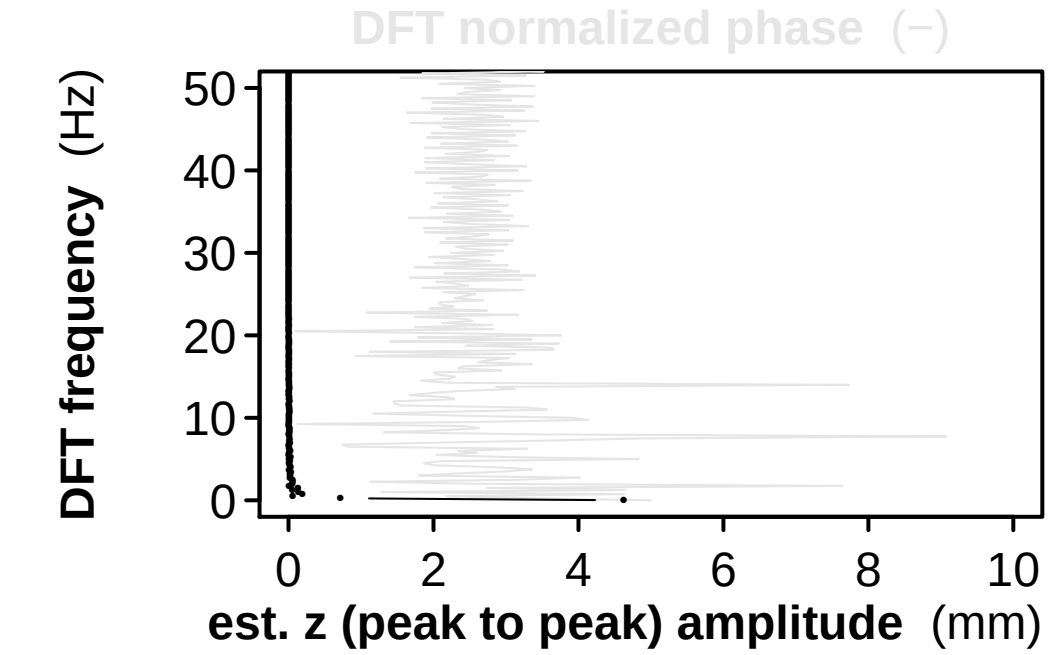

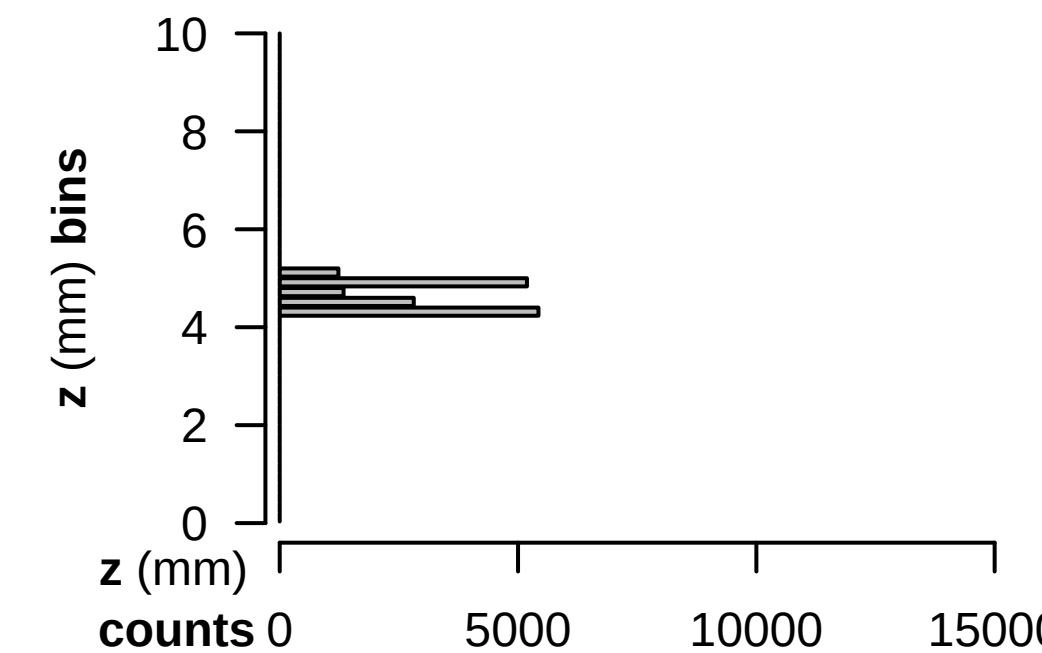

SUBJECT 1 − RUN 26 − CONDITION 0,1
SC_180323_105413_0.AIFF

```
z_min               : 4.22 mm
z_max               : 5.12 mm
z_travel_amplitude  : 0.90 mm

avg_abs_z_travel    : 3.15 mm/s

z_jarque-bera_jb    : 1590.07
z_jarque-bera_p     : 0.00e+00

z_lin_mod_est_slope: −0.18 mm/s
z_lin_mod_adj_R²    : 55 %

z_poly40_mod_adj_R²: 99 %

z_dft_ampl_thresh   : 0.010 mm
>=threshold_maxfreq: 13.75 Hz
```

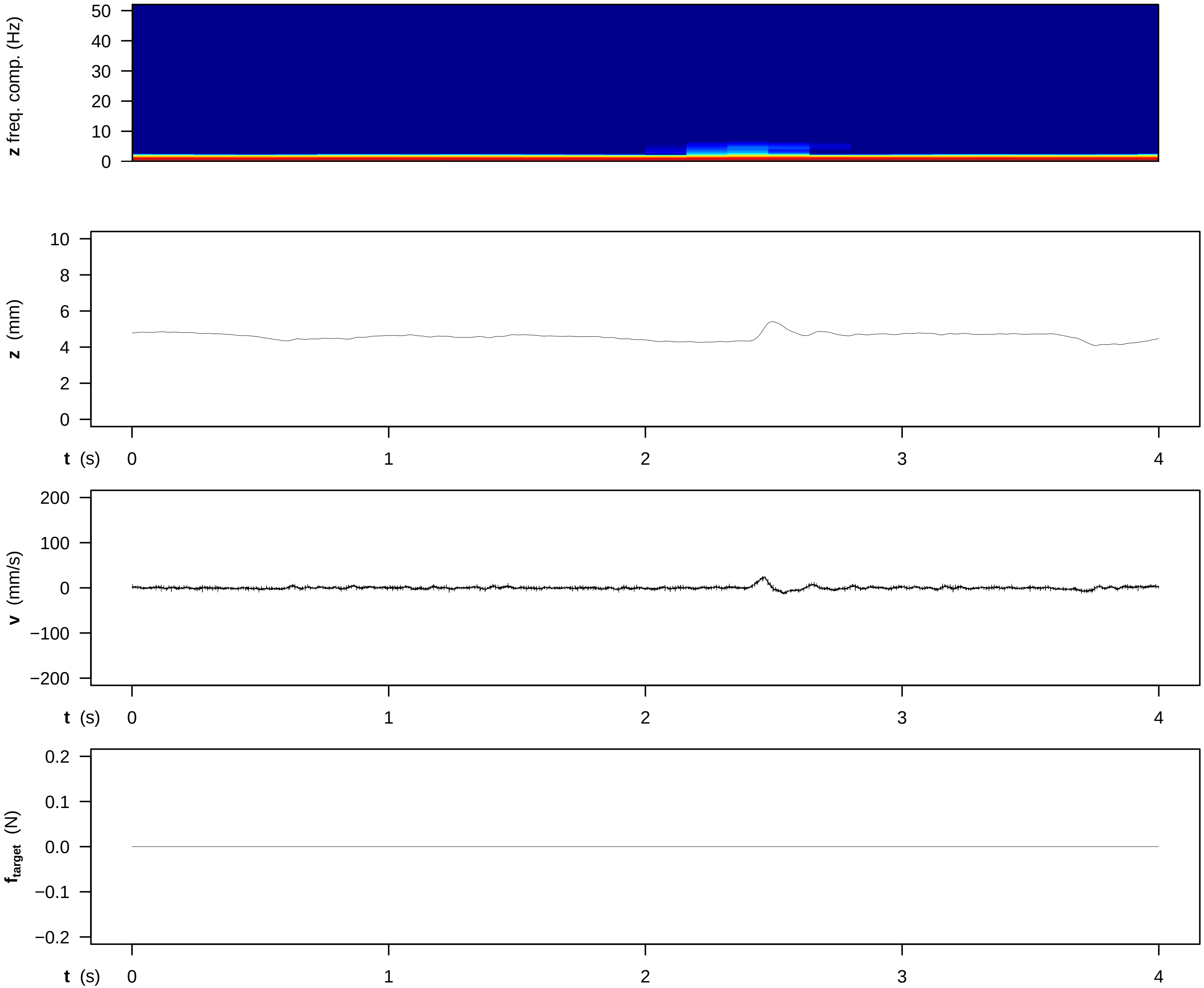

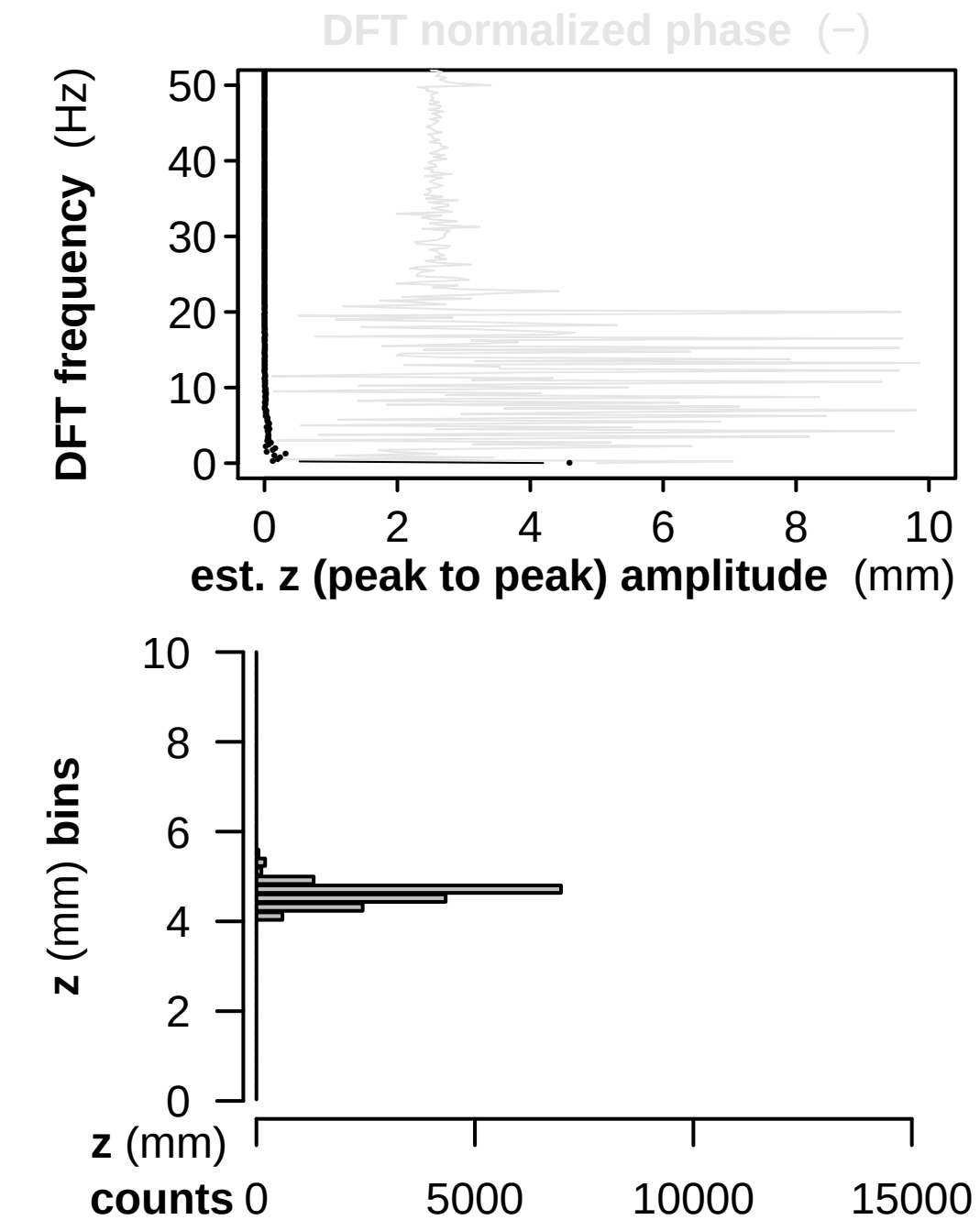

**SUBJECT 1 − RUN 35 − CONDITION 0,1**
**SC_180323_110131_0.AIFF**

```
z_min               : 4.09 mm
z_max               : 5.41 mm
z_travel_amplitude  : 1.33 mm

avg_abs_z_travel    : 3.61 mm/s

z_jarque-bera_jb    : 1859.65
z_jarque-bera_p     : 0.00e+00

z_lin_mod_est_slope: −0.02 mm/s
z_lin_mod_adj_R²    : 2 %

z_poly40_mod_adj_R²: 80 %

z_dft_ampl_thresh   : 0.010 mm
>=threshold_maxfreq: 17.00 Hz
```

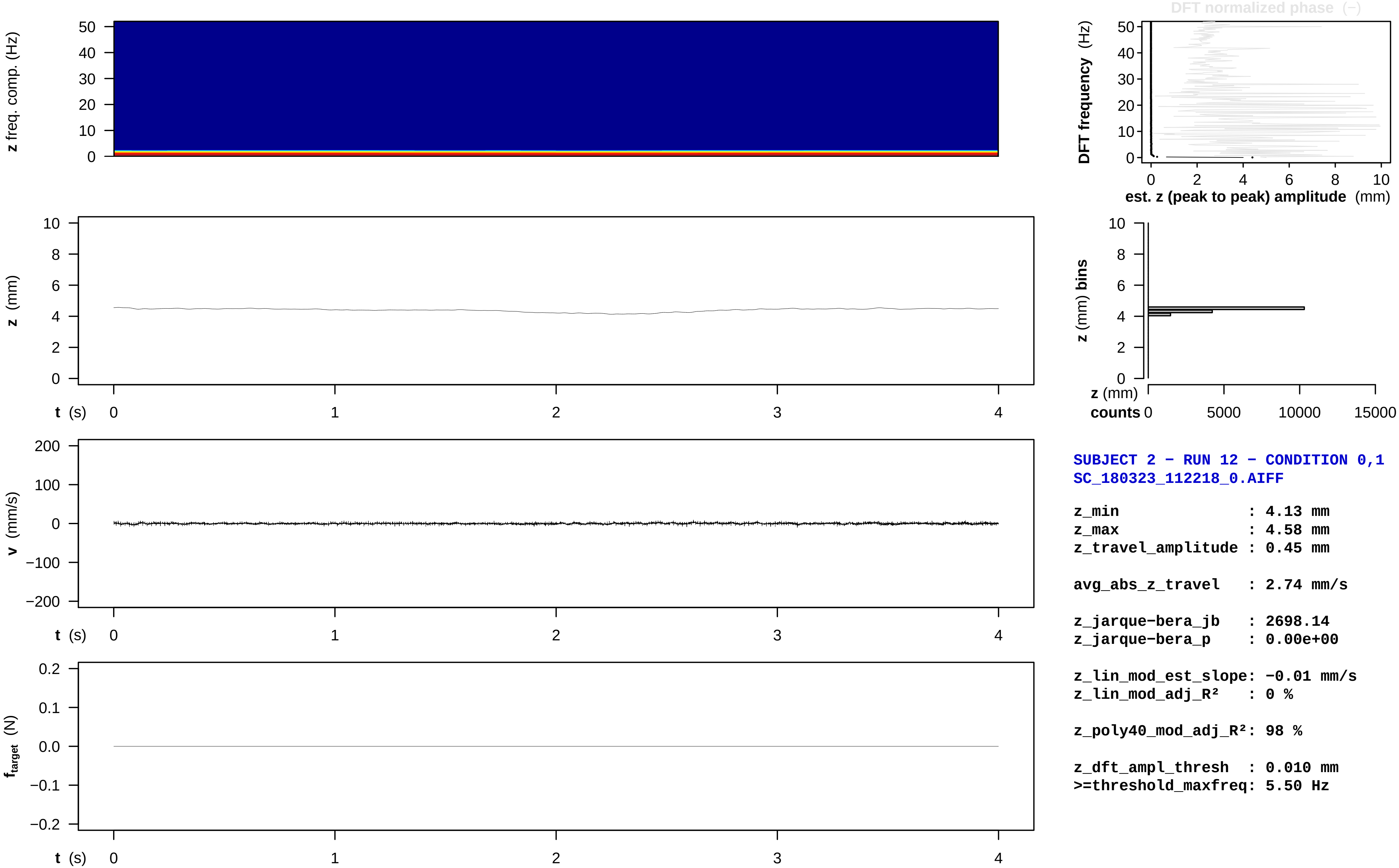

**SUBJECT 2 − RUN 12 − CONDITION 0,1**
**SC_180323_112218_0.AIFF**

```
z_min               : 4.13 mm
z_max               : 4.58 mm
z_travel_amplitude : 0.45 mm

avg_abs_z_travel   : 2.74 mm/s

z_jarque-bera_jb   : 2698.14
z_jarque-bera_p    : 0.00e+00

z_lin_mod_est_slope: -0.01 mm/s
z_lin_mod_adj_R²   : 0 %

z_poly40_mod_adj_R²: 98 %

z_dft_ampl_thresh  : 0.010 mm
>=threshold_maxfreq: 5.50 Hz
```

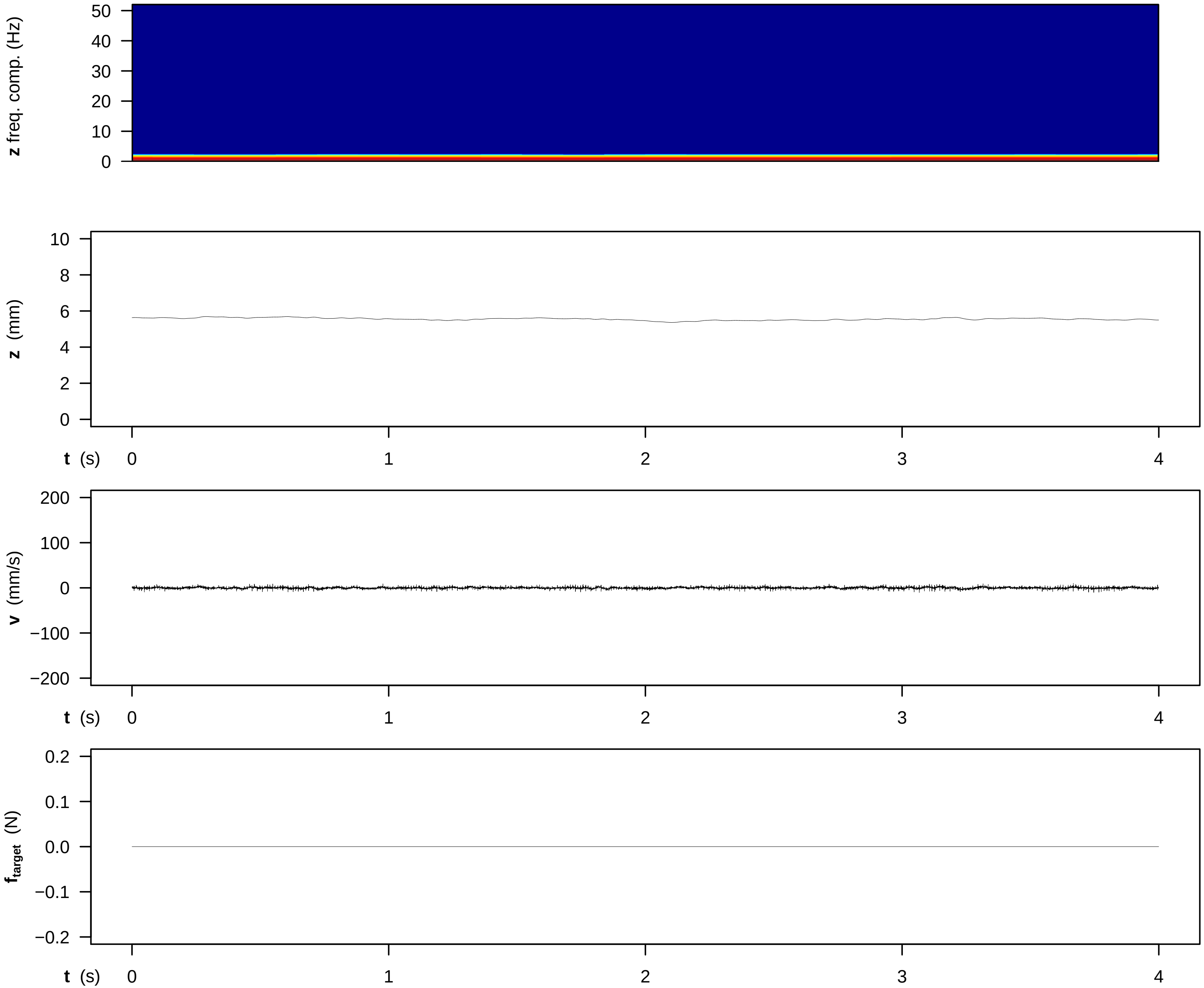

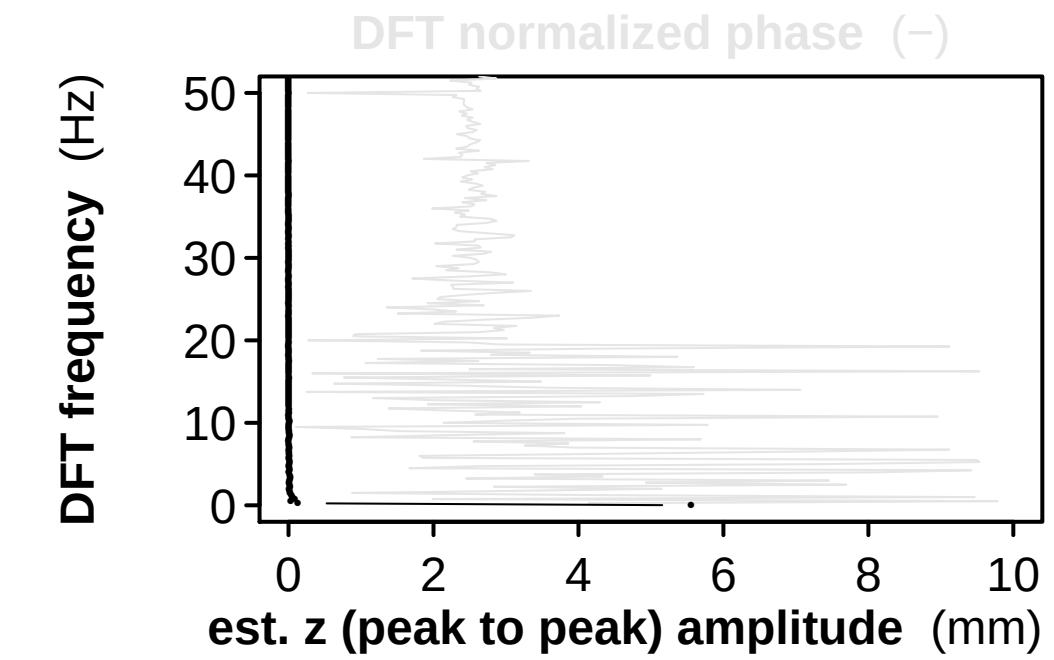

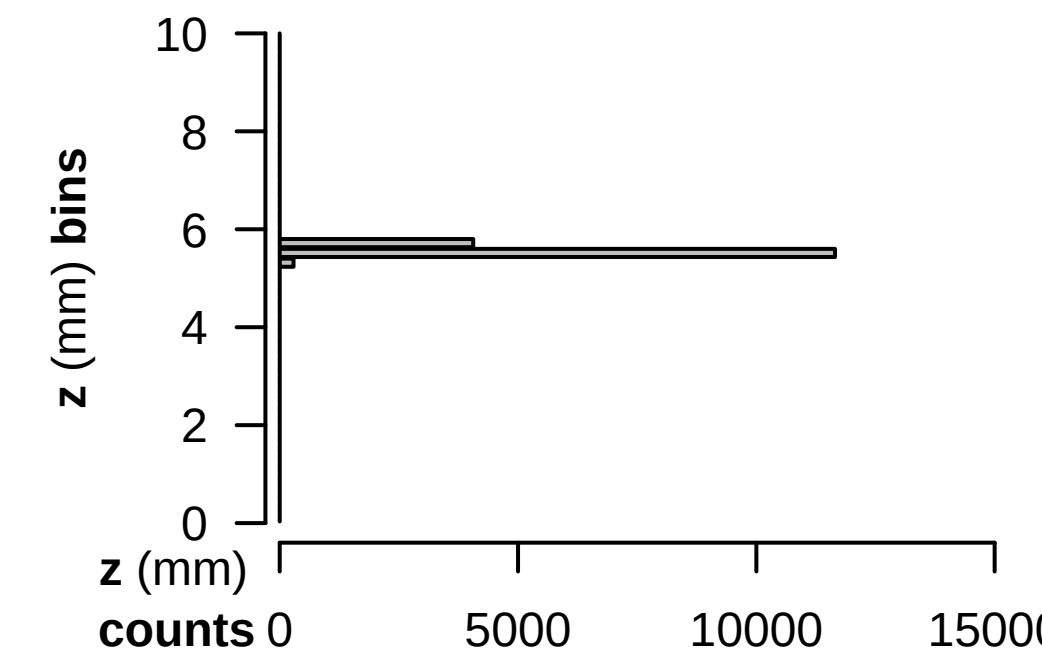

**SUBJECT 2 − RUN 18 − CONDITION 0,1**
**SC_180323_112606_0.AIFF**

```
z_min               : 5.36 mm
z_max               : 5.69 mm
z_travel_amplitude  : 0.33 mm

avg_abs_z_travel    : 2.81 mm/s

z_jarque-bera_jb    : 189.65
z_jarque-bera_p     : 0.00e+00

z_lin_mod_est_slope: −0.02 mm/s
z_lin_mod_adj_R²    : 18 %

z_poly40_mod_adj_R²: 86 %

z_dft_ampl_thresh   : 0.010 mm
>=threshold_maxfreq: 10.25 Hz
```

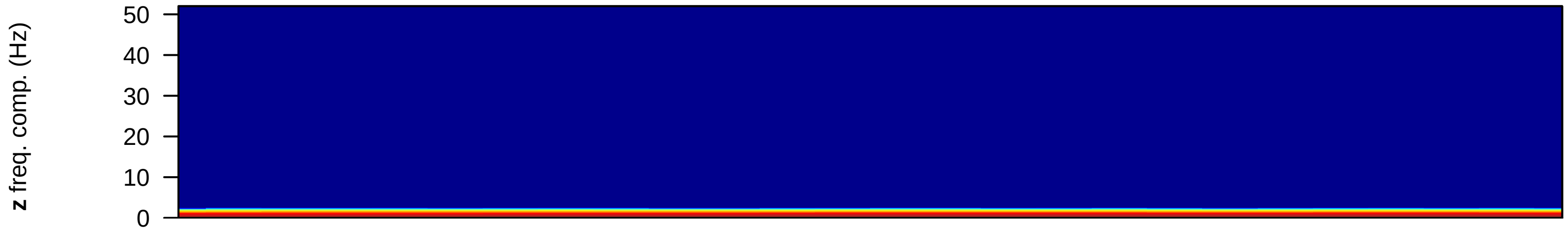

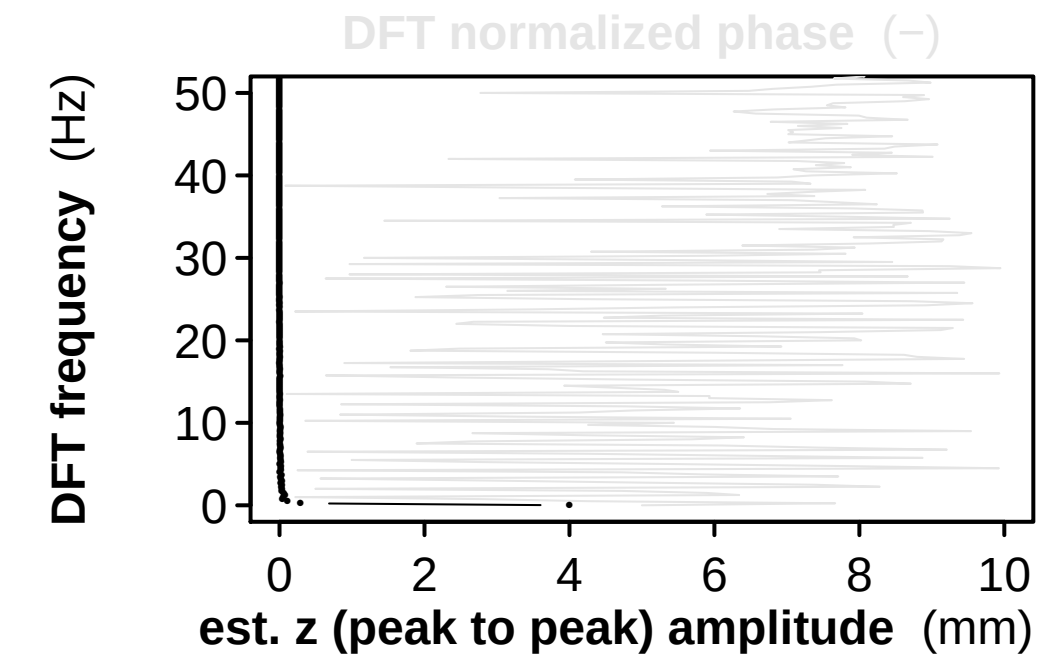

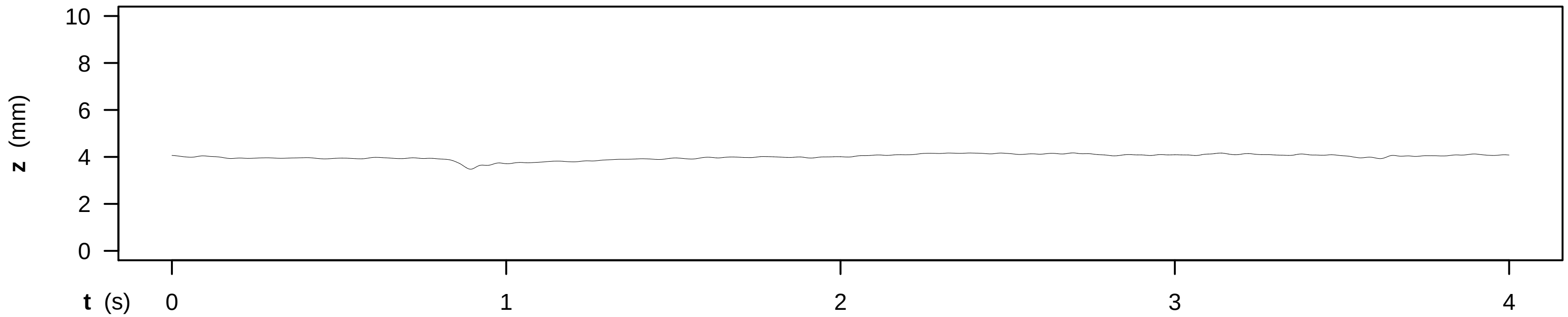

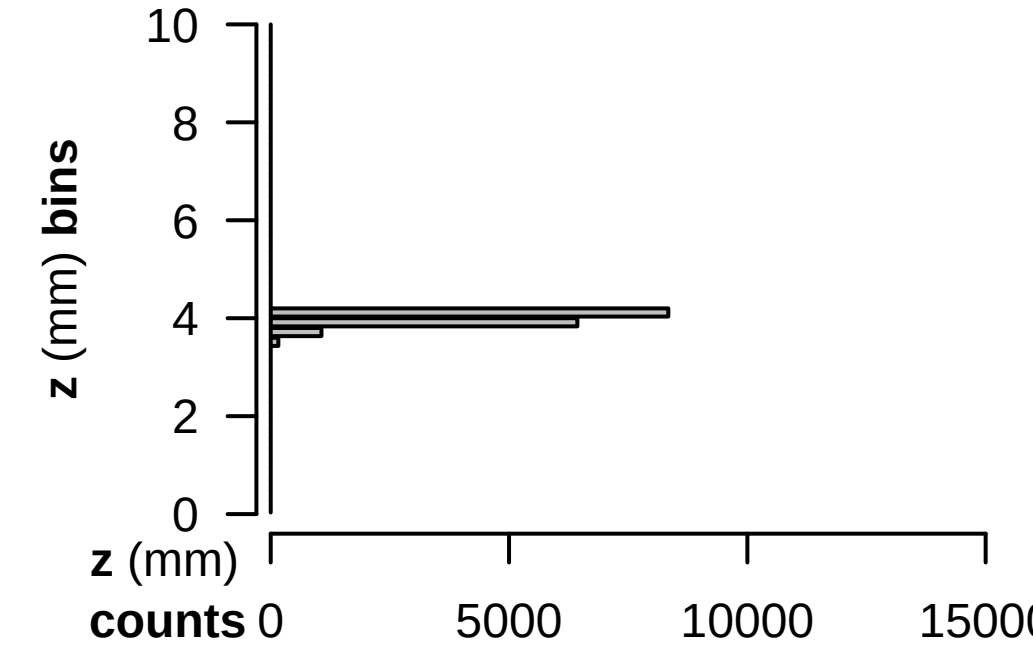

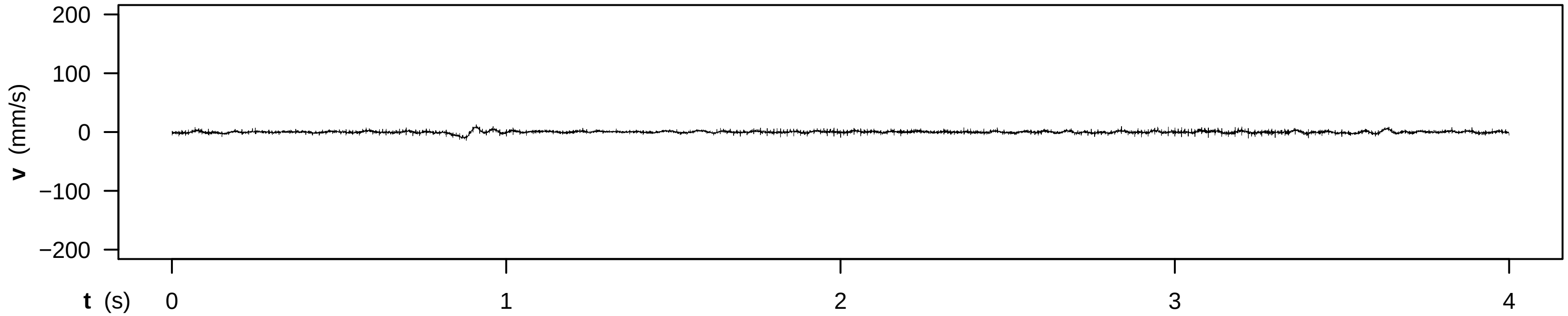

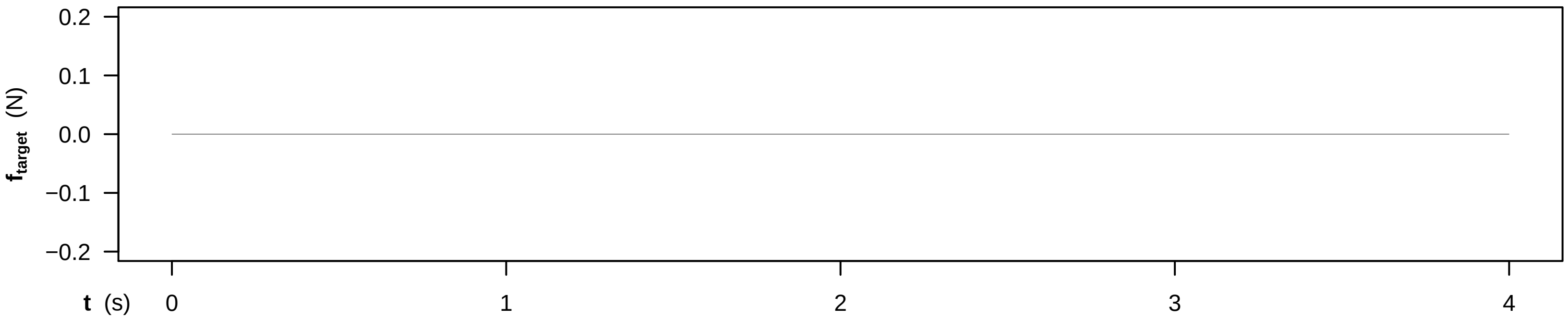

**SUBJECT 2 − RUN 20 − CONDITION 0,1**
**SC_180323_112656_0.AIFF**

```
z_min              : 3.48 mm
z_max              : 4.17 mm
z_travel_amplitude : 0.69 mm

avg_abs_z_travel   : 2.59 mm/s

z_jarque-bera_jb   : 6020.26
z_jarque-bera_p    : 0.00e+00

z_lin_mod_est_slope: 0.06 mm/s
z_lin_mod_adj_R²   : 36 %

z_poly40_mod_adj_R²: 91 %

z_dft_ampl_thresh  : 0.010 mm
>=threshold_maxfreq: 15.75 Hz
```

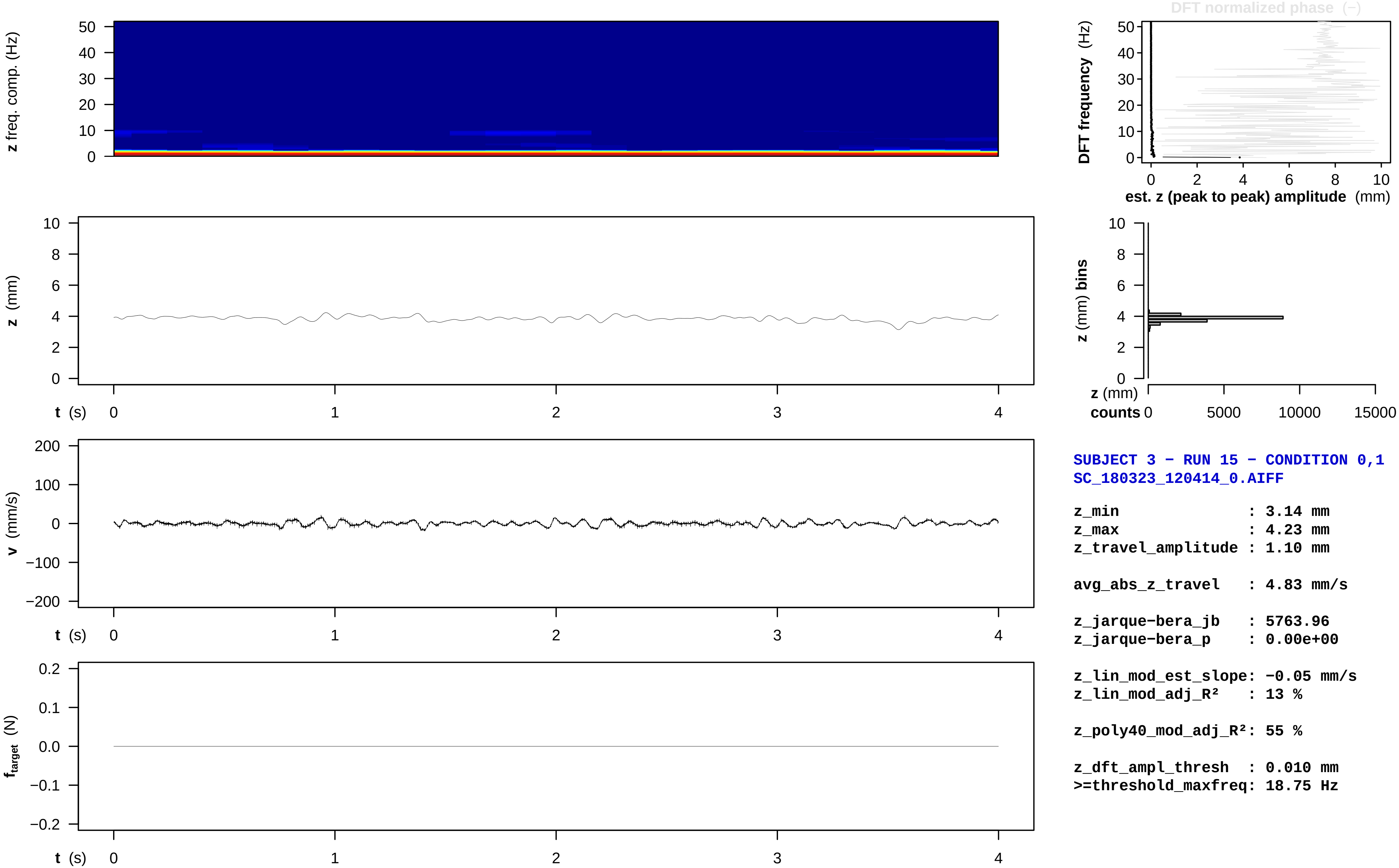

**SUBJECT 3 − RUN 15 − CONDITION 0,1**
**SC_180323_120414_0.AIFF**

```
z_min              : 3.14 mm
z_max              : 4.23 mm
z_travel_amplitude : 1.10 mm

avg_abs_z_travel   : 4.83 mm/s

z_jarque-bera_jb   : 5763.96
z_jarque-bera_p    : 0.00e+00

z_lin_mod_est_slope: -0.05 mm/s
z_lin_mod_adj_R²   : 13 %

z_poly40_mod_adj_R²: 55 %

z_dft_ampl_thresh  : 0.010 mm
>=threshold_maxfreq: 18.75 Hz
```

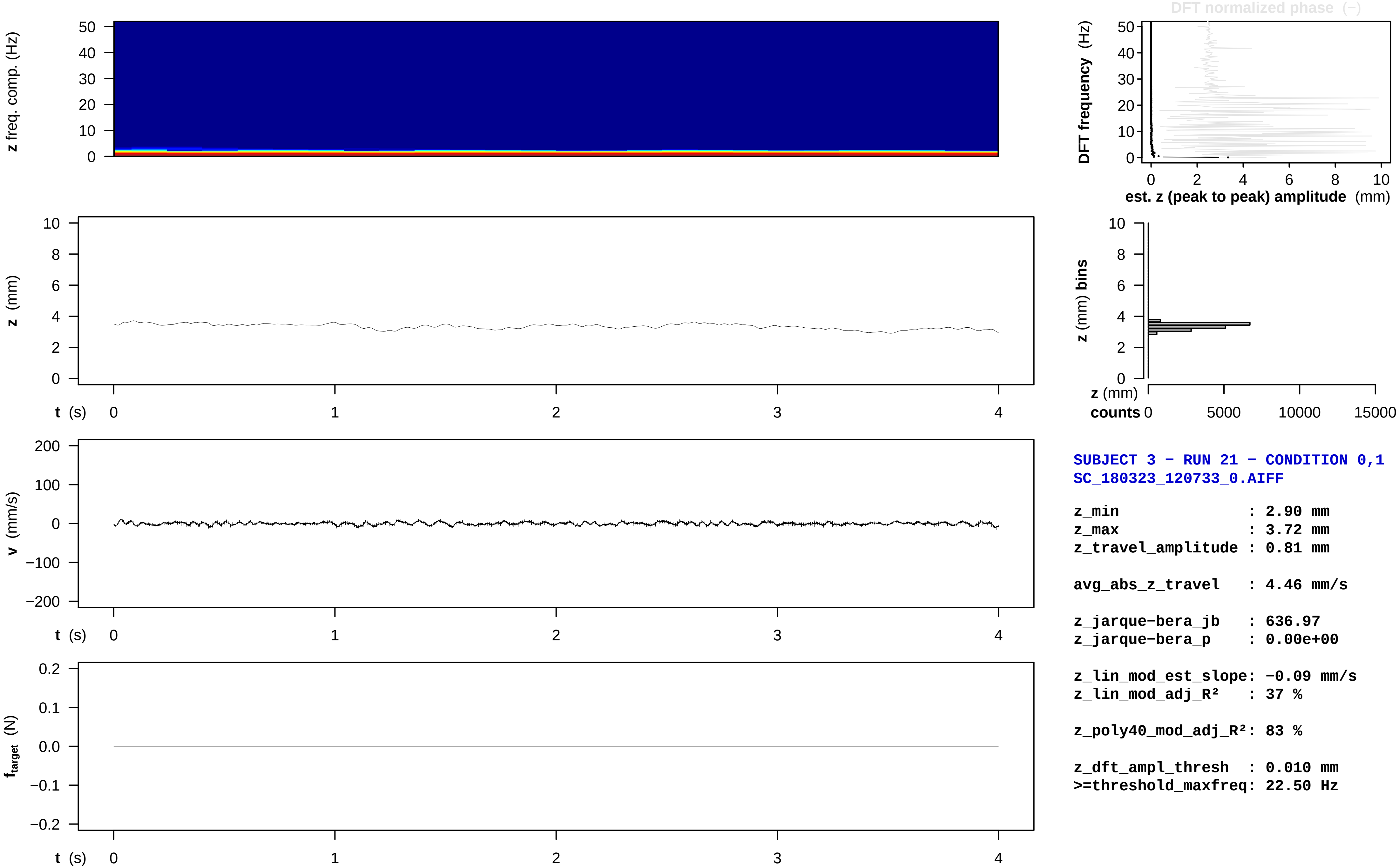

**SUBJECT 3 − RUN 21 − CONDITION 0,1**
**SC_180323_120733_0.AIFF**

```
z_min               : 2.90 mm
z_max               : 3.72 mm
z_travel_amplitude  : 0.81 mm

avg_abs_z_travel    : 4.46 mm/s

z_jarque-bera_jb    : 636.97
z_jarque-bera_p     : 0.00e+00

z_lin_mod_est_slope: -0.09 mm/s
z_lin_mod_adj_R²    : 37 %

z_poly40_mod_adj_R²: 83 %

z_dft_ampl_thresh   : 0.010 mm
>=threshold_maxfreq: 22.50 Hz
```

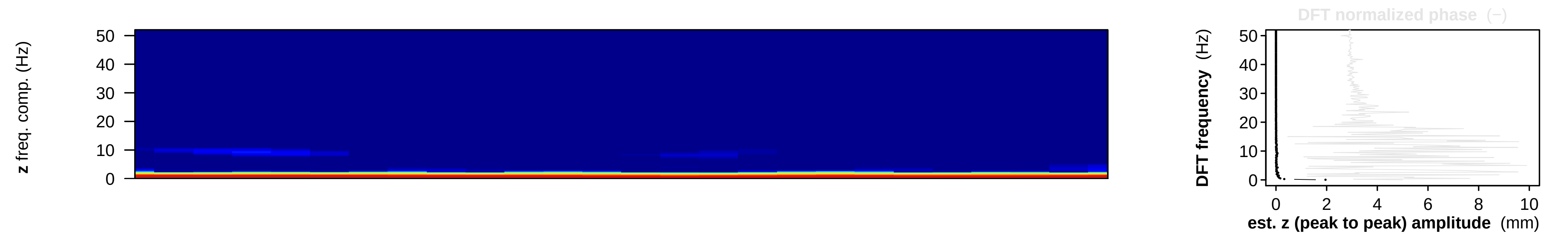

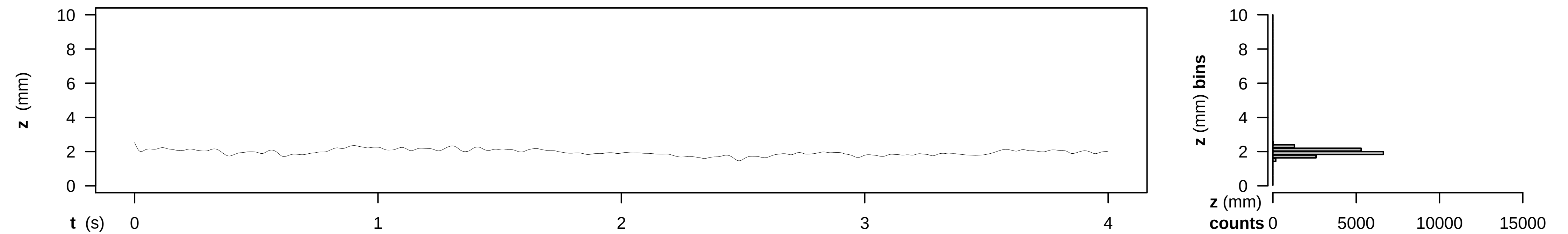

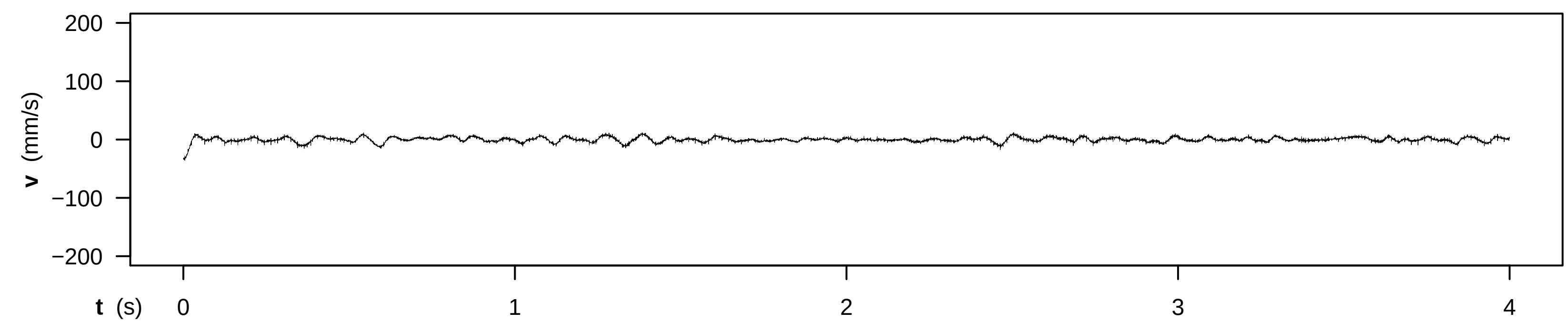

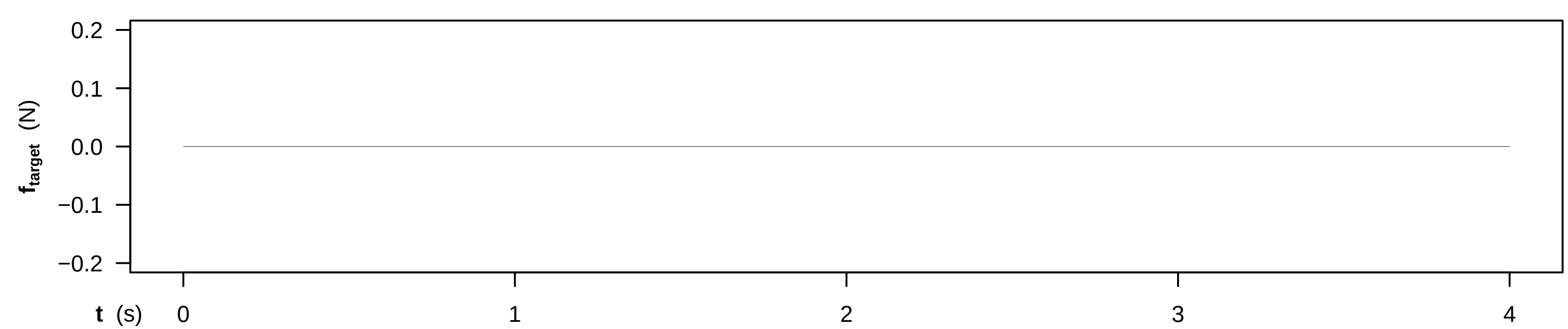

**SUBJECT 3 − RUN 26 − CONDITION 0,1**
**SC_180323_121007_0.AIFF**

```
z_min              : 1.46 mm
z_max              : 2.53 mm
z_travel_amplitude : 1.07 mm

avg_abs_z_travel   : 4.31 mm/s

z_jarque-bera_jb   : 149.77
z_jarque-bera_p    : 0.00e+00

z_lin_mod_est_slope: −0.06 mm/s
z_lin_mod_adj_R²   : 18 %

z_poly40_mod_adj_R²: 84 %

z_dft_ampl_thresh  : 0.010 mm
>=threshold_maxfreq: 18.00 Hz
```

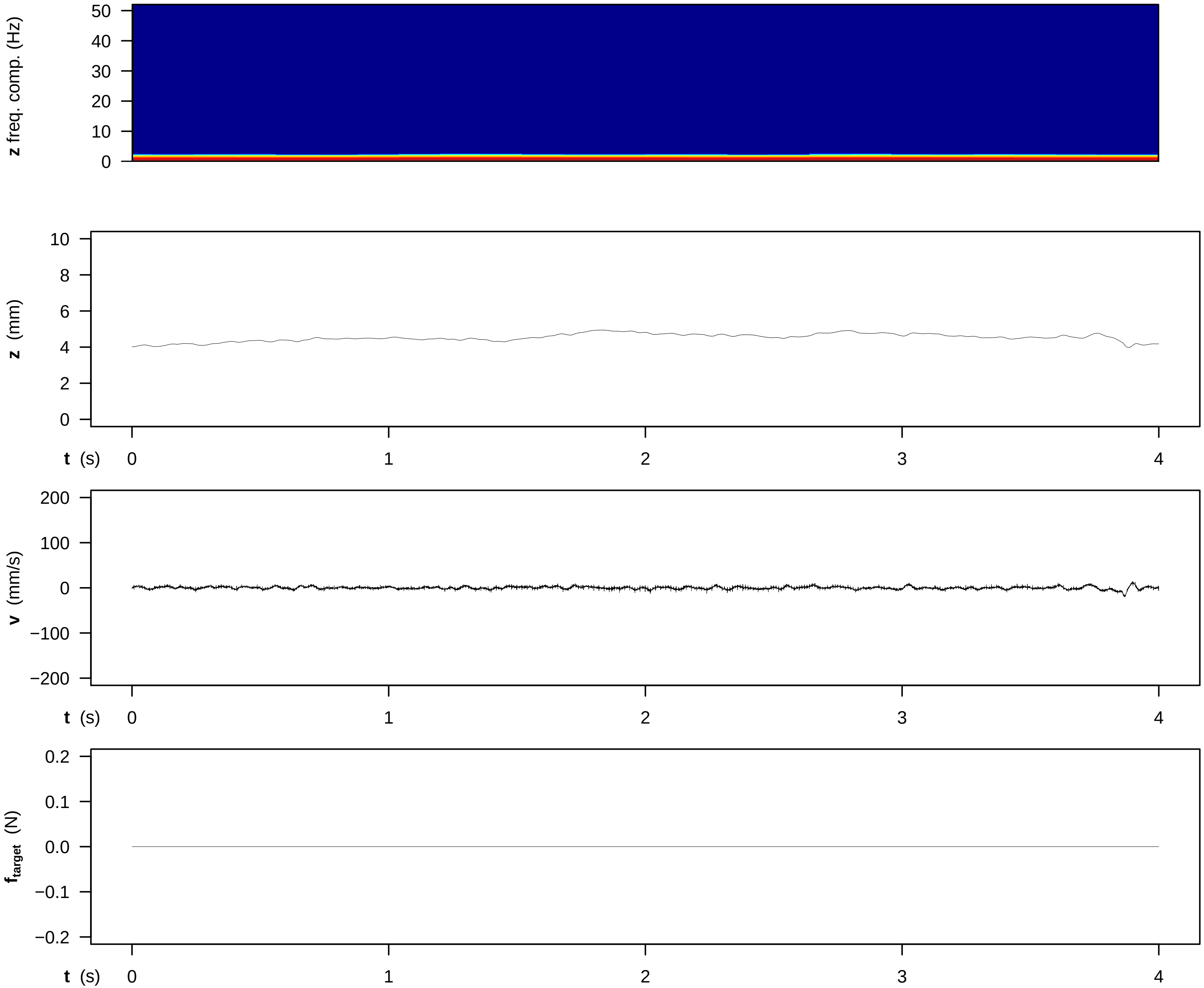

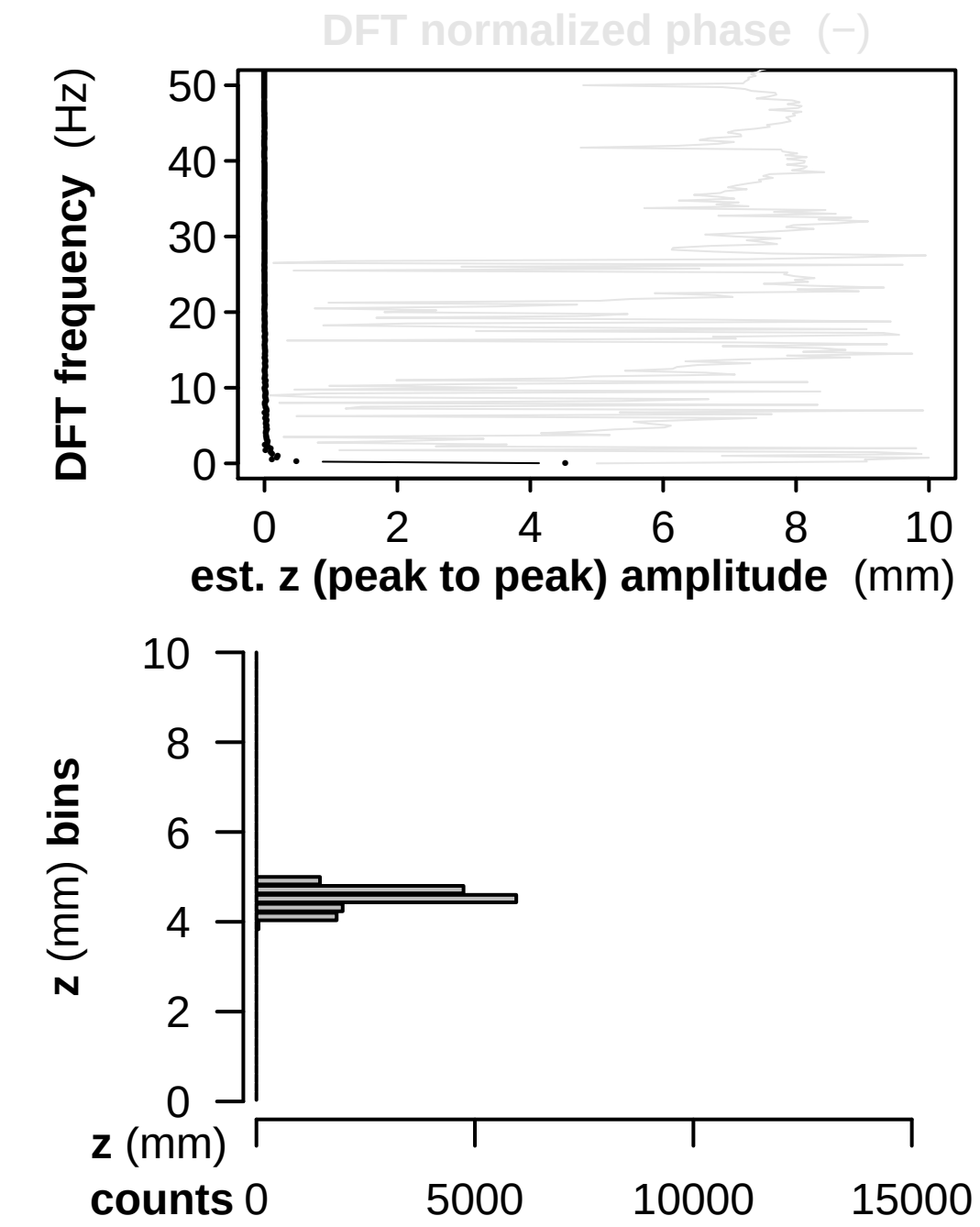

**SUBJECT 4 − RUN 10 − CONDITION 0,1**
**SC_180323_123533_0.AIFF**

```
z_min               : 3.98 mm
z_max               : 4.95 mm
z_travel_amplitude  : 0.97 mm

avg_abs_z_travel    : 3.71 mm/s

z_jarque-bera_jb    : 434.16
z_jarque-bera_p     : 0.00e+00

z_lin_mod_est_slope: 0.08 mm/s
z_lin_mod_adj_R²    : 19 %

z_poly40_mod_adj_R²: 93 %

z_dft_ampl_thresh   : 0.010 mm
>=threshold_maxfreq: 17.25 Hz
```

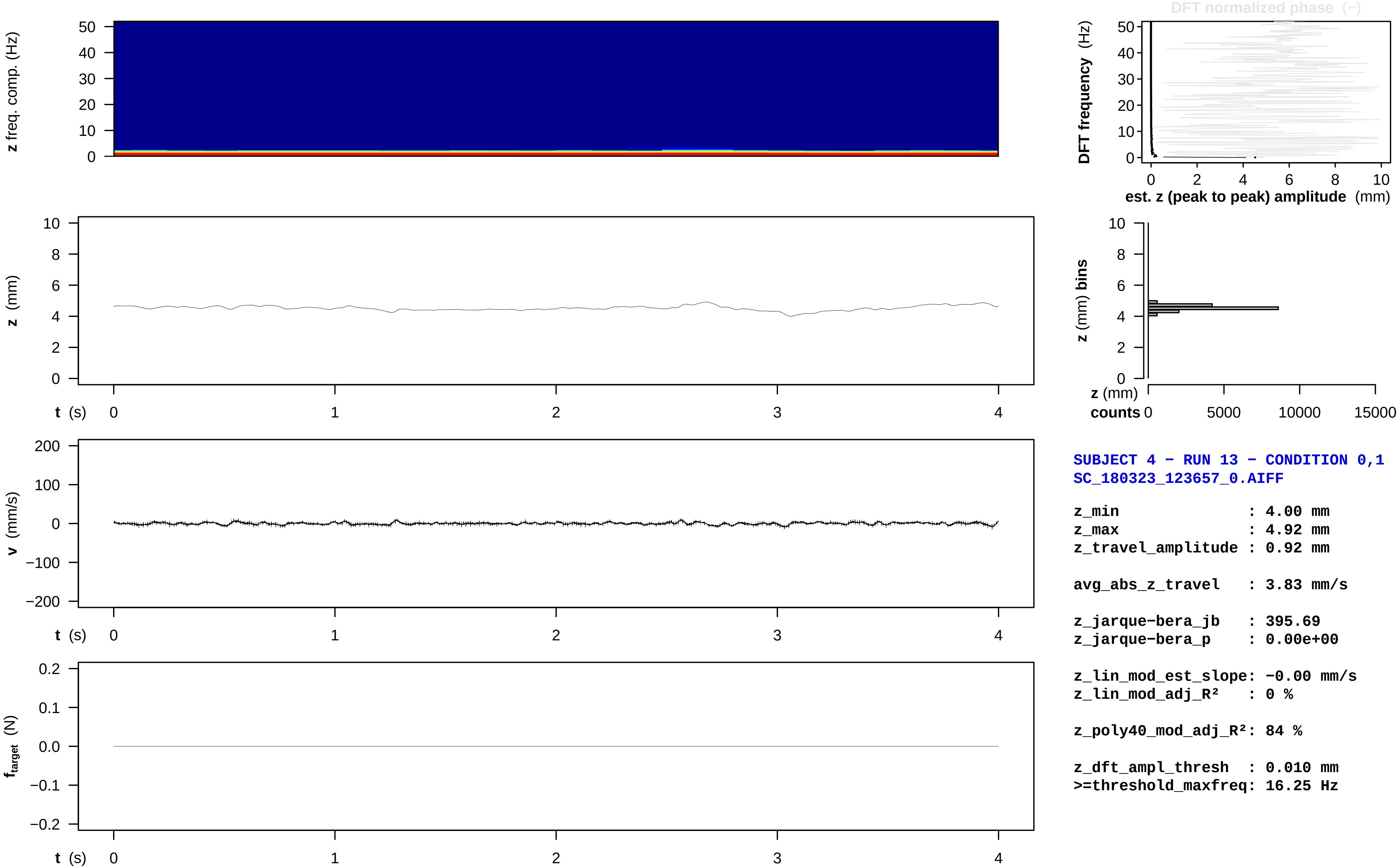

**SUBJECT 4 − RUN 13 − CONDITION 0,1**
**SC_180323_123657_0.AIFF**

```
z_min               : 4.00 mm
z_max               : 4.92 mm
z_travel_amplitude : 0.92 mm

avg_abs_z_travel   : 3.83 mm/s

z_jarque-bera_jb   : 395.69
z_jarque-bera_p    : 0.00e+00

z_lin_mod_est_slope: −0.00 mm/s
z_lin_mod_adj_R²   : 0 %

z_poly40_mod_adj_R²: 84 %

z_dft_ampl_thresh  : 0.010 mm
>=threshold_maxfreq: 16.25 Hz
```

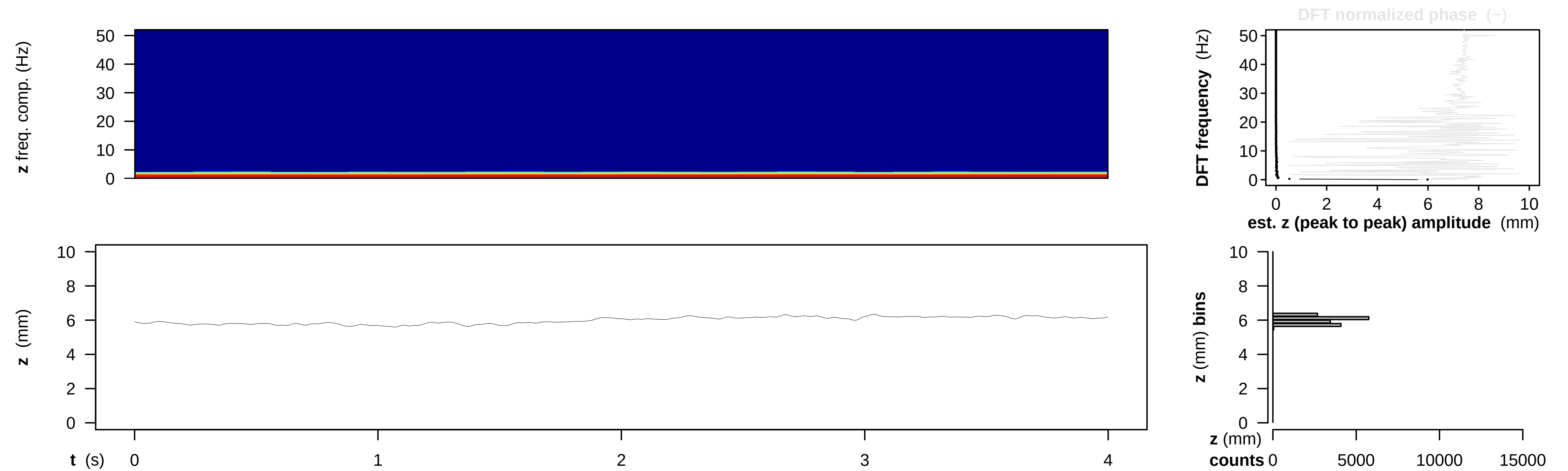

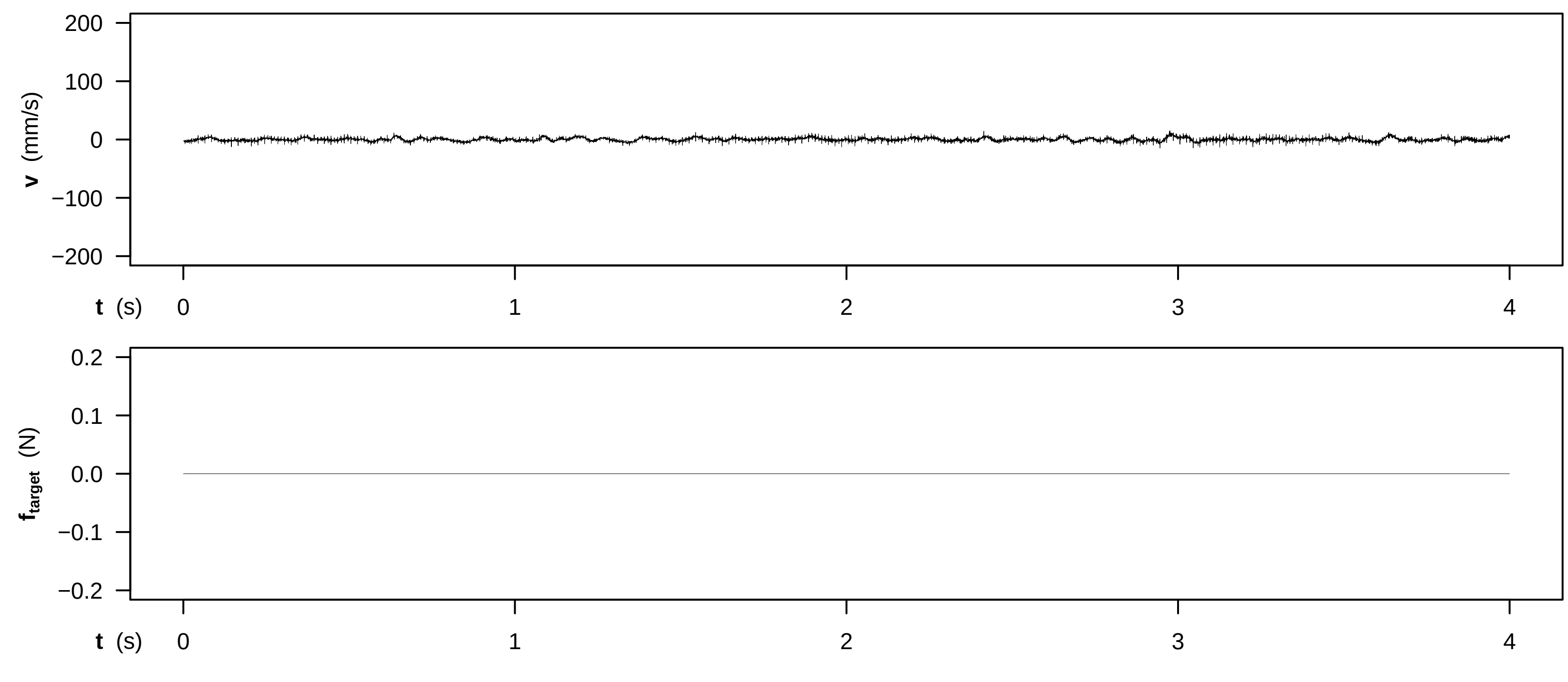

**SUBJECT 4 − RUN 23 − CONDITION 0,1**
**SC_180323_124152_0.AIFF**

```
z_min               : 5.59 mm
z_max               : 6.34 mm
z_travel_amplitude  : 0.75 mm

avg_abs_z_travel    : 3.69 mm/s

z_jarque-bera_jb    : 1529.70
z_jarque-bera_p     : 0.00e+00

z_lin_mod_est_slope: 0.15 mm/s
z_lin_mod_adj_R²    : 73 %

z_poly40_mod_adj_R²: 93 %

z_dft_ampl_thresh   : 0.010 mm
>=threshold_maxfreq: 18.00 Hz
```

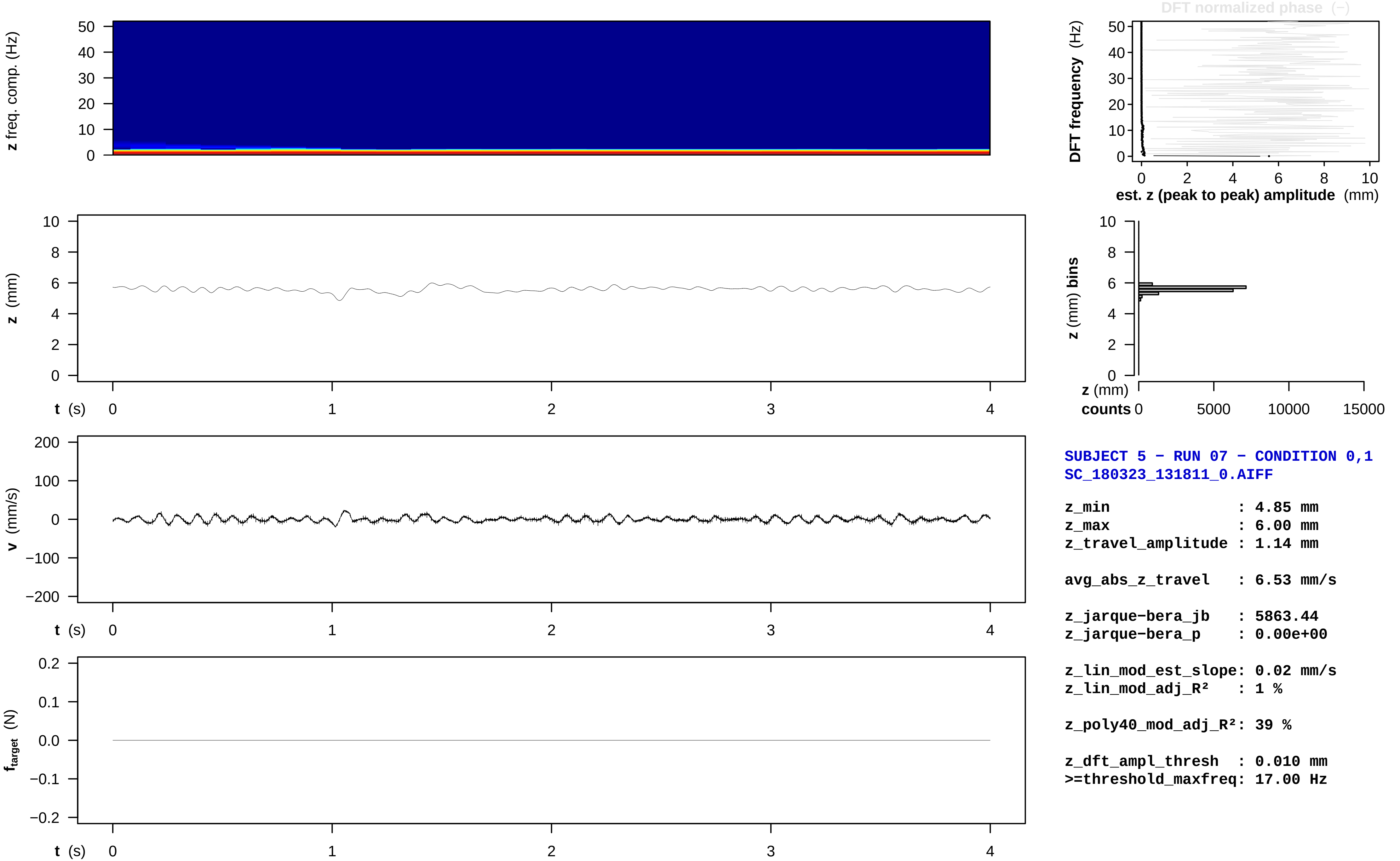

SUBJECT 5 − RUN 07 − CONDITION 0,1
SC_180323_131811_0.AIFF

```
z_min               : 4.85 mm
z_max               : 6.00 mm
z_travel_amplitude  : 1.14 mm

avg_abs_z_travel    : 6.53 mm/s

z_jarque-bera_jb    : 5863.44
z_jarque-bera_p     : 0.00e+00

z_lin_mod_est_slope: 0.02 mm/s
z_lin_mod_adj_R²    : 1 %

z_poly40_mod_adj_R²: 39 %

z_dft_ampl_thresh   : 0.010 mm
>=threshold_maxfreq: 17.00 Hz
```

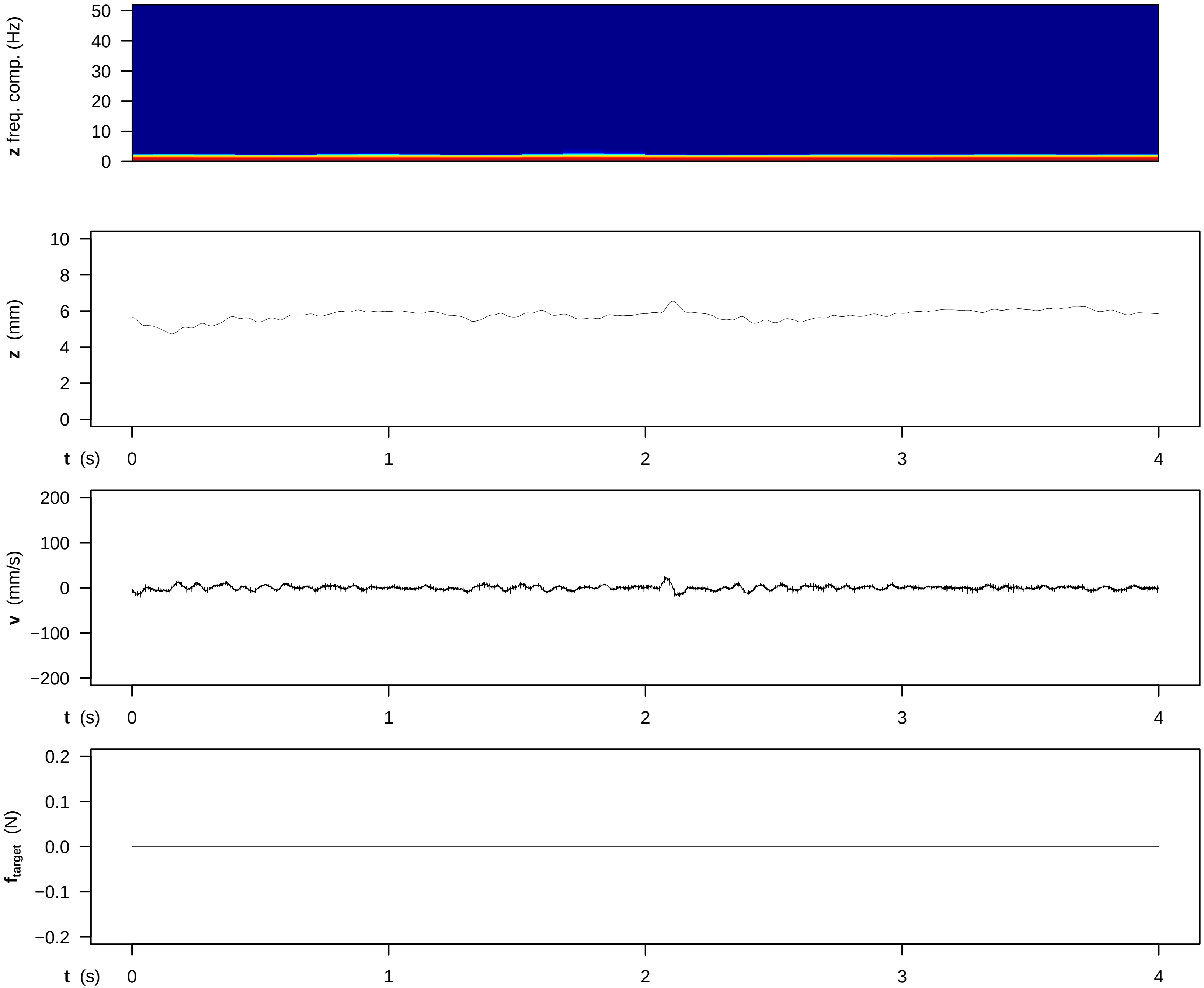

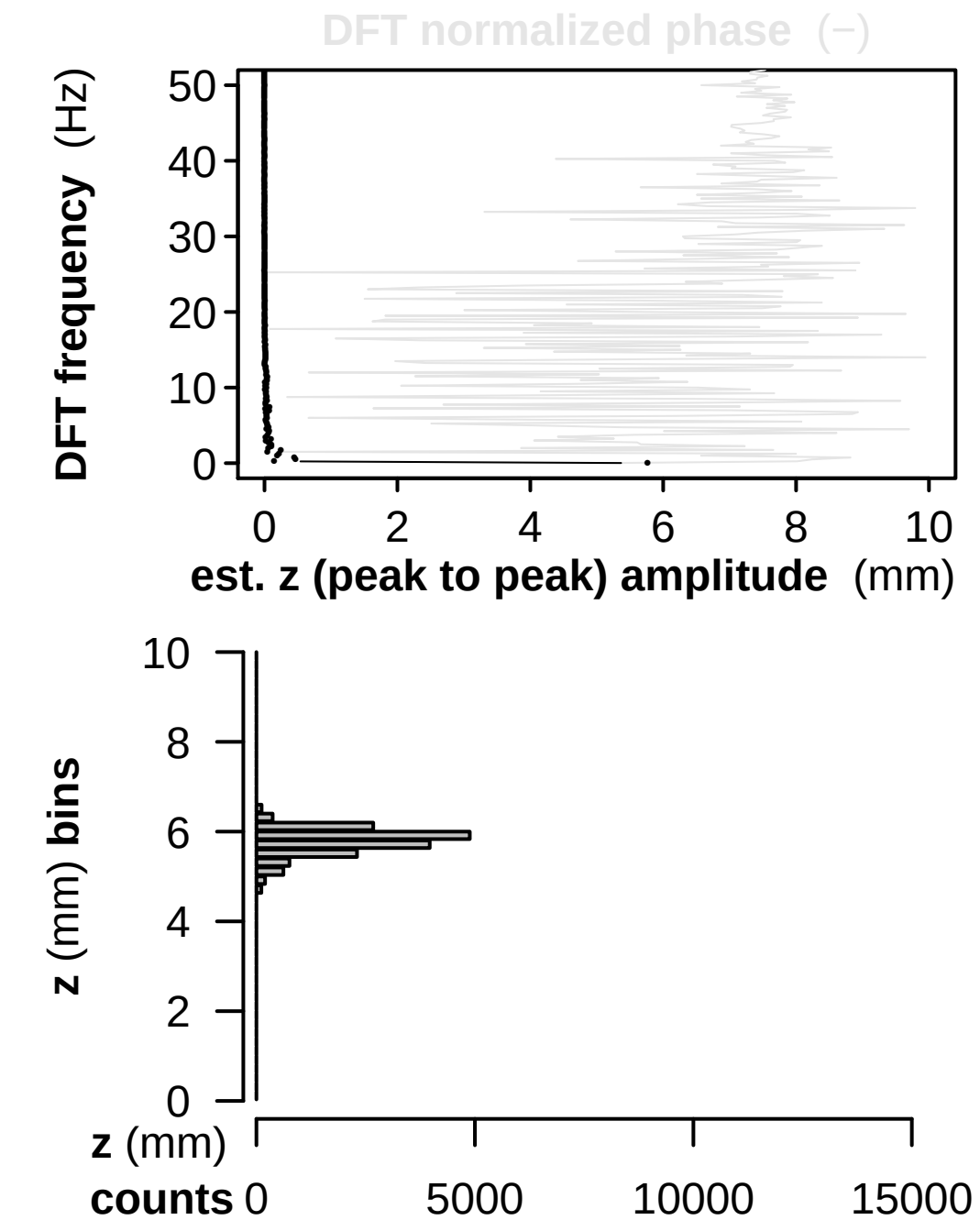

**SUBJECT 5 − RUN 29 − CONDITION 0,1**
**SC_180323_133419_0.AIFF**

```
z_min              : 4.73 mm
z_max              : 6.54 mm
z_travel_amplitude : 1.81 mm

avg_abs_z_travel   : 5.64 mm/s

z_jarque-bera_jb   : 2622.57
z_jarque-bera_p    : 0.00e+00

z_lin_mod_est_slope: 0.14 mm/s
z_lin_mod_adj_R²   : 31 %

z_poly40_mod_adj_R²: 88 %

z_dft_ampl_thresh  : 0.010 mm
>=threshold_maxfreq: 16.25 Hz
```

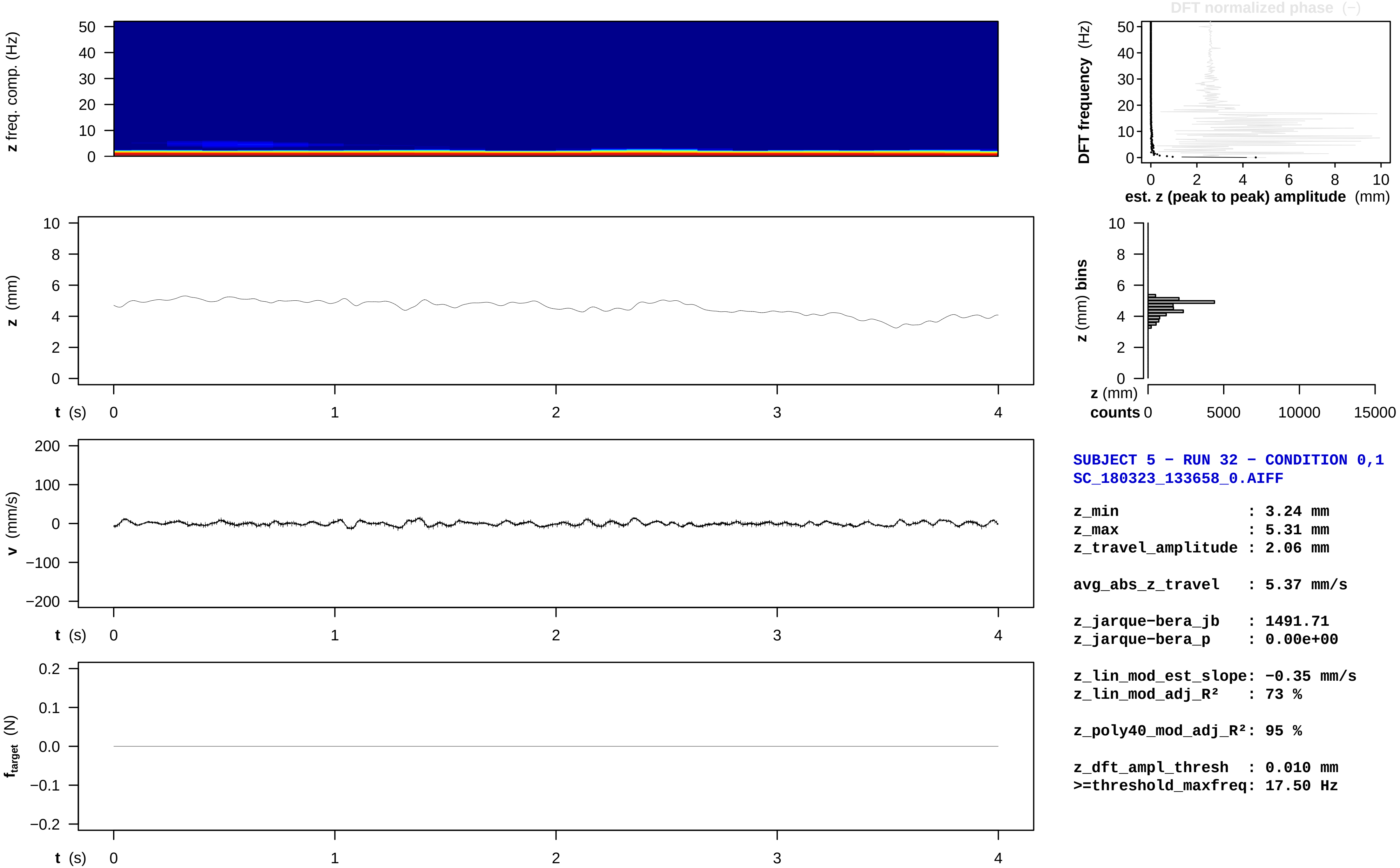

**SUBJECT 5 − RUN 32 − CONDITION 0,1**
**SC_180323_133658_0.AIFF**

```
z_min              : 3.24 mm
z_max              : 5.31 mm
z_travel_amplitude : 2.06 mm

avg_abs_z_travel   : 5.37 mm/s

z_jarque-bera_jb   : 1491.71
z_jarque-bera_p    : 0.00e+00

z_lin_mod_est_slope: -0.35 mm/s
z_lin_mod_adj_R²   : 73 %

z_poly40_mod_adj_R²: 95 %

z_dft_ampl_thresh  : 0.010 mm
>=threshold_maxfreq: 17.50 Hz
```

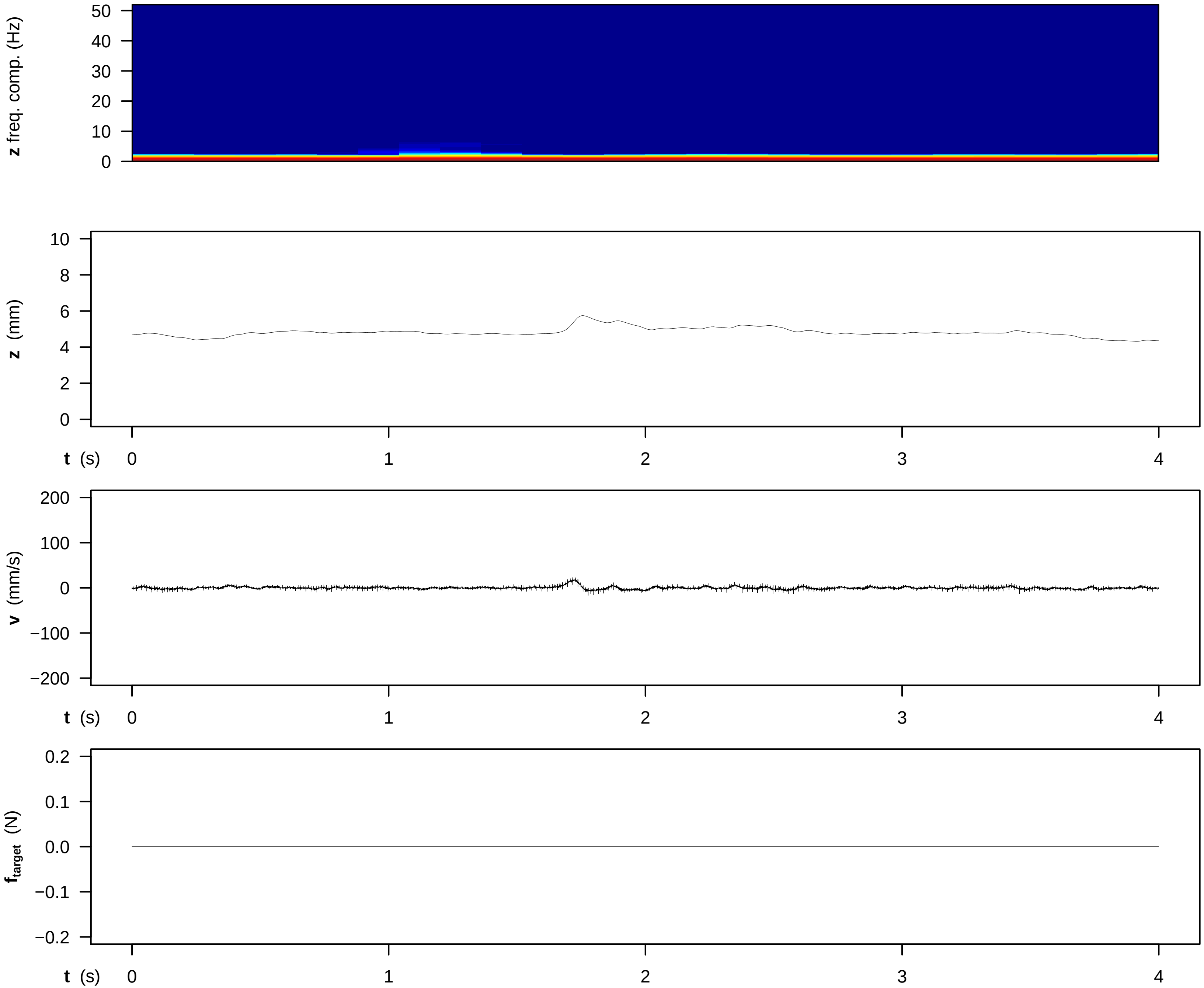

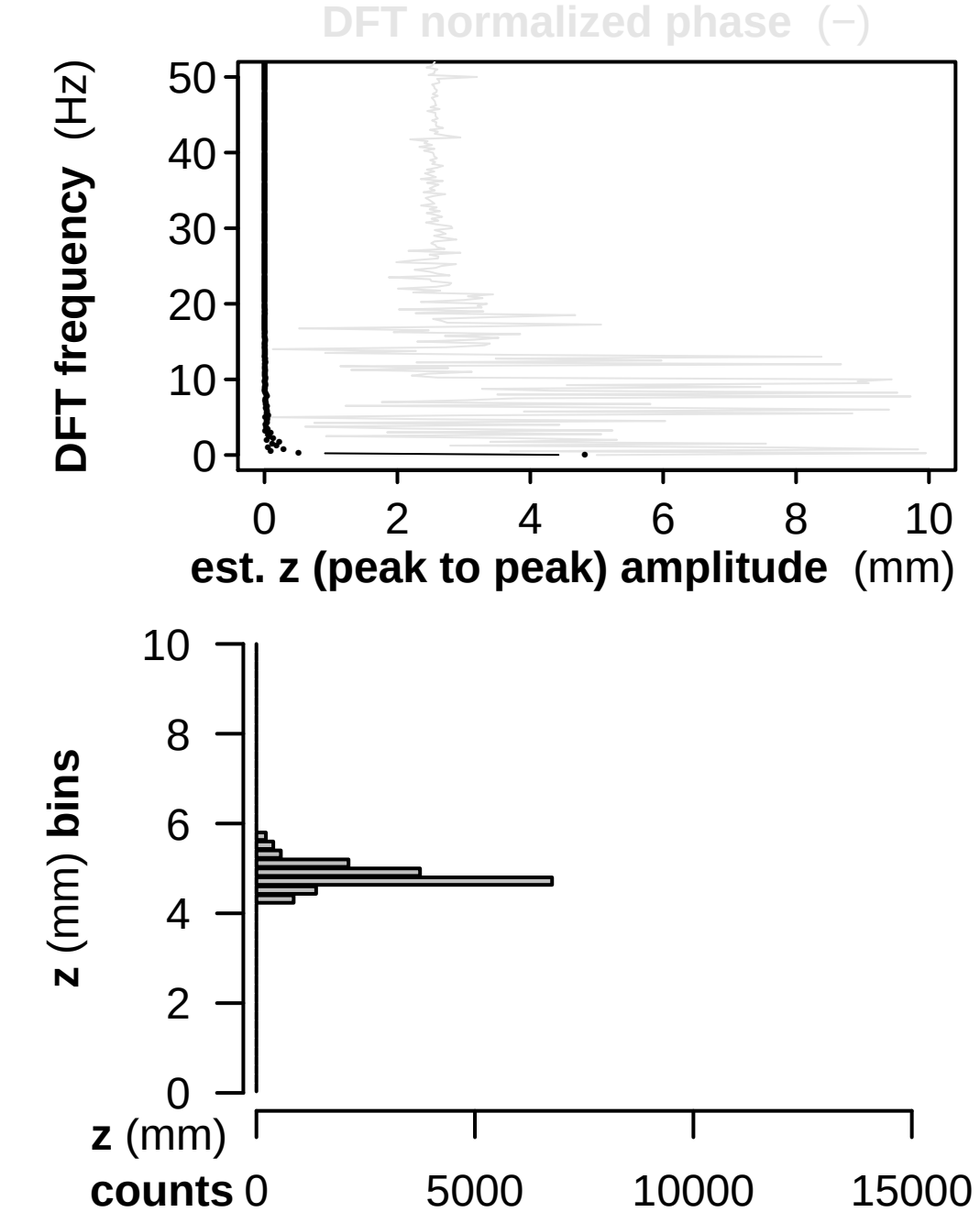

**SUBJECT 6 − RUN 11 − CONDITION 0,1**
**SC_180323_145746_0.AIFF**

```
z_min               : 4.32 mm
z_max               : 5.75 mm
z_travel_amplitude  : 1.43 mm

avg_abs_z_travel    : 3.22 mm/s

z_jarque-bera_jb    : 3527.42
z_jarque-bera_p     : 0.00e+00

z_lin_mod_est_slope: -0.02 mm/s
z_lin_mod_adj_R²    : 1 %

z_poly40_mod_adj_R²: 90 %

z_dft_ampl_thresh   : 0.010 mm
>=threshold_maxfreq: 15.25 Hz
```

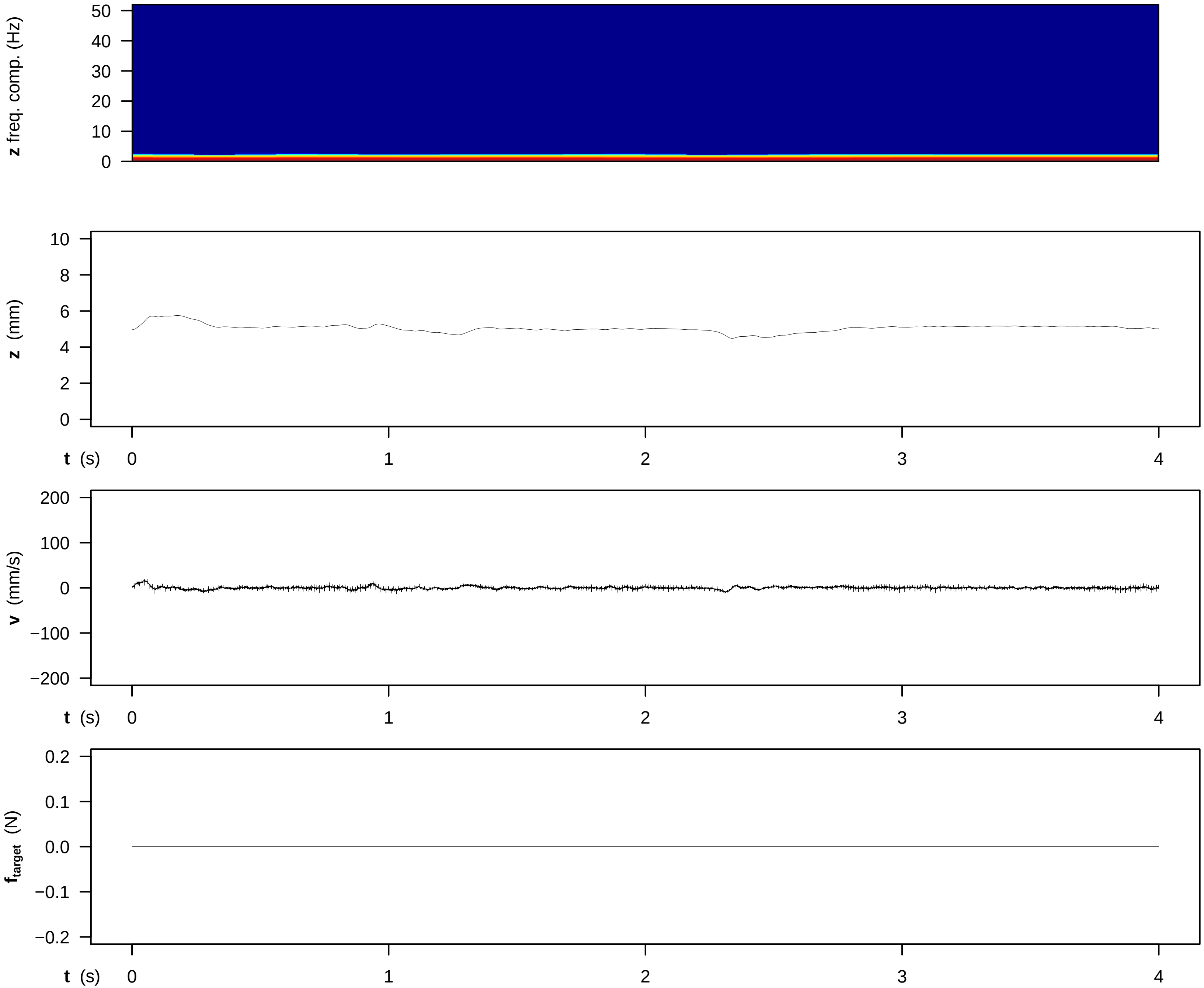

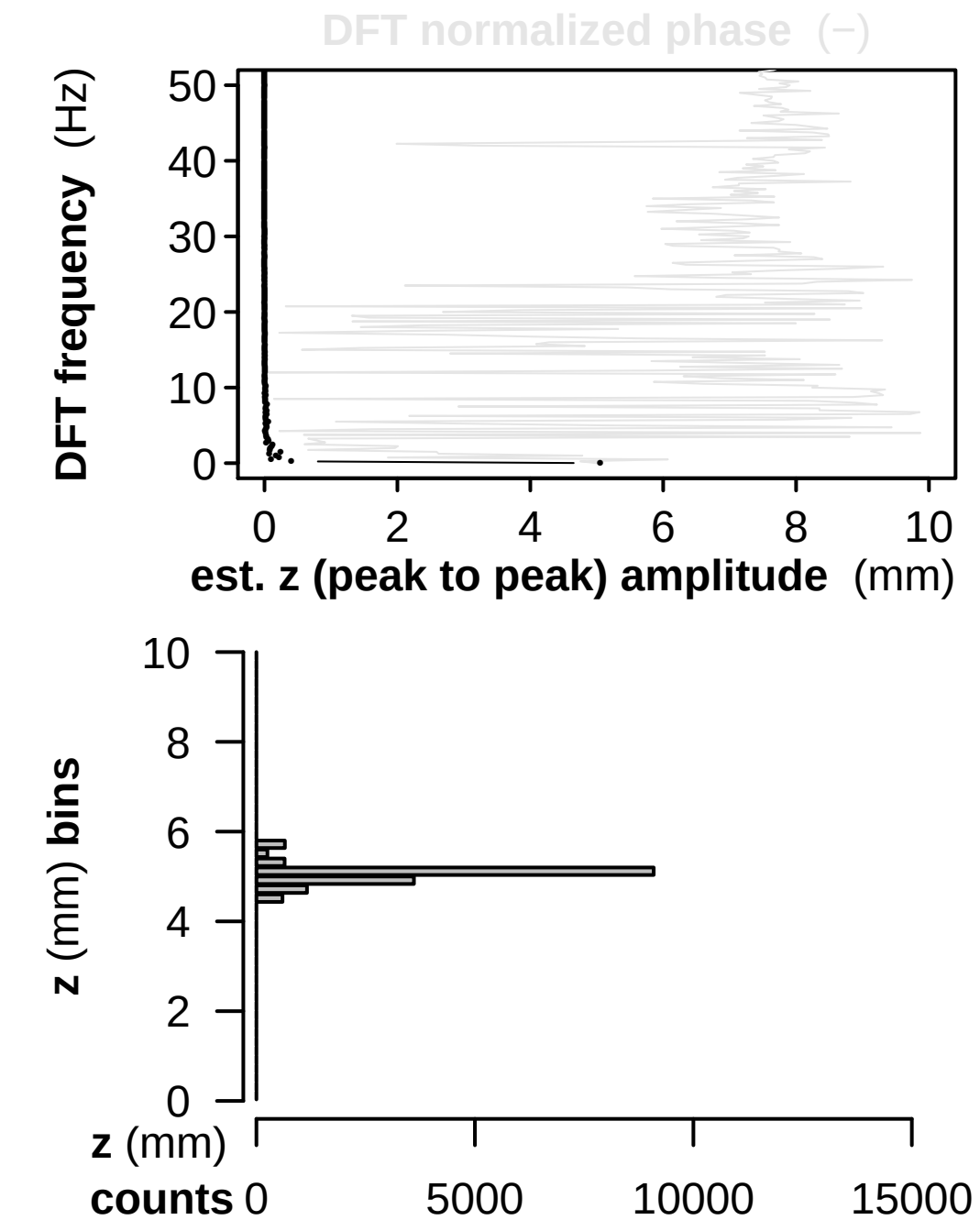

**SUBJECT 6 − RUN 22 − CONDITION 0,1**
**SC_180323_150429_0.AIFF**

**z_min : 4.49 mm**
**z_max : 5.75 mm**
**z_travel_amplitude : 1.27 mm**

**avg_abs_z_travel : 2.97 mm/s**

**z_jarque-bera_jb : 4296.31**
**z_jarque-bera_p : 0.00e+00**

**z_lin_mod_est_slope: −0.05 mm/s**
**z_lin_mod_adj_R² : 6 %**

**z_poly40_mod_adj_R²: 94 %**

**z_dft_ampl_thresh : 0.010 mm**
**>=threshold_maxfreq: 10.25 Hz**

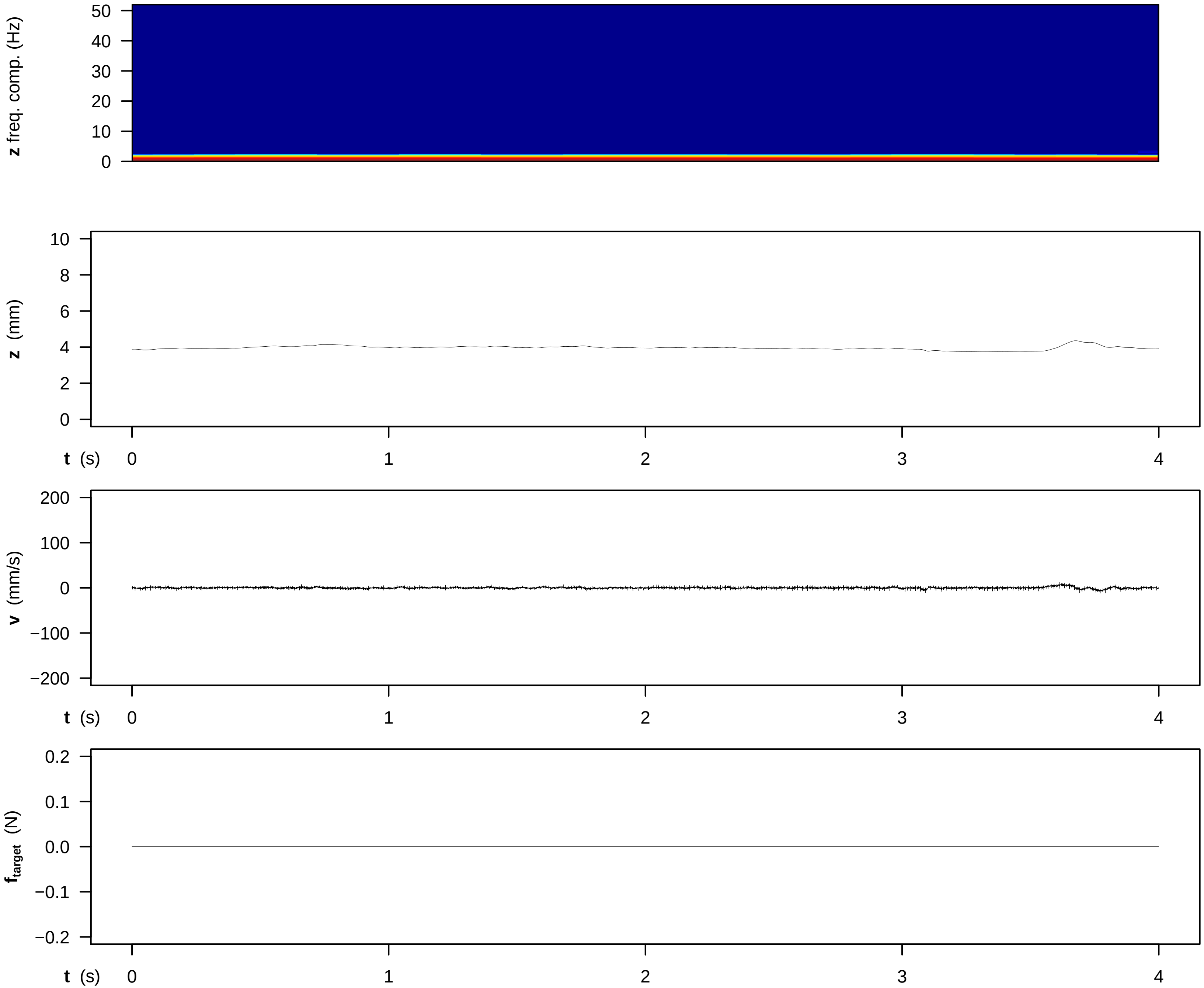

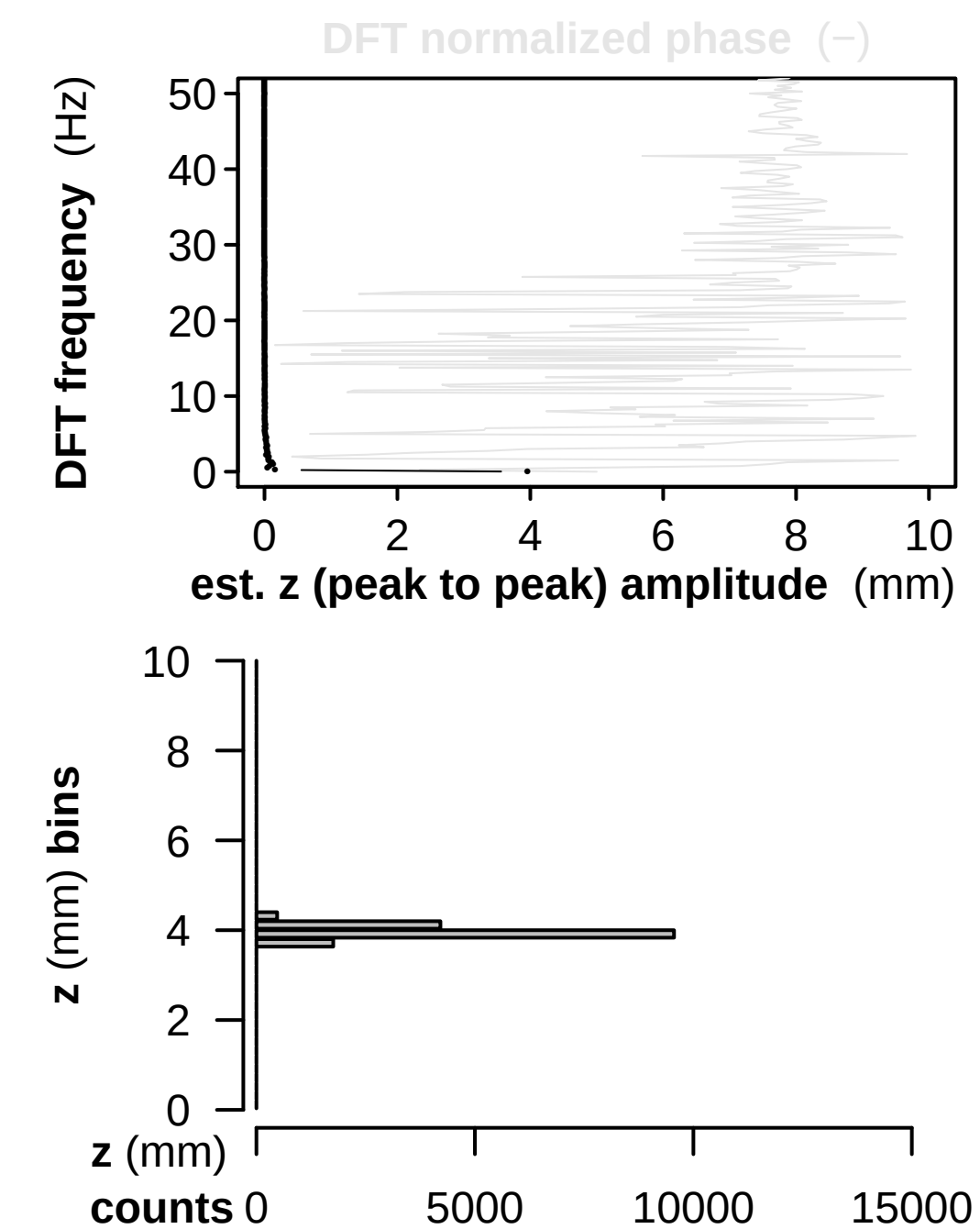

**SUBJECT 6 − RUN 26 − CONDITION 0,1**
**SC_180323_150724_0.AIFF**

```
z_min               : 3.75 mm
z_max               : 4.35 mm
z_travel_amplitude  : 0.60 mm

avg_abs_z_travel    : 2.26 mm/s

z_jarque-bera_jb    : 2508.30
z_jarque-bera_p     : 0.00e+00

z_lin_mod_est_slope: −0.02 mm/s
z_lin_mod_adj_R²    : 6 %

z_poly40_mod_adj_R²: 92 %

z_dft_ampl_thresh   : 0.010 mm
>=threshold_maxfreq: 9.00 Hz
```

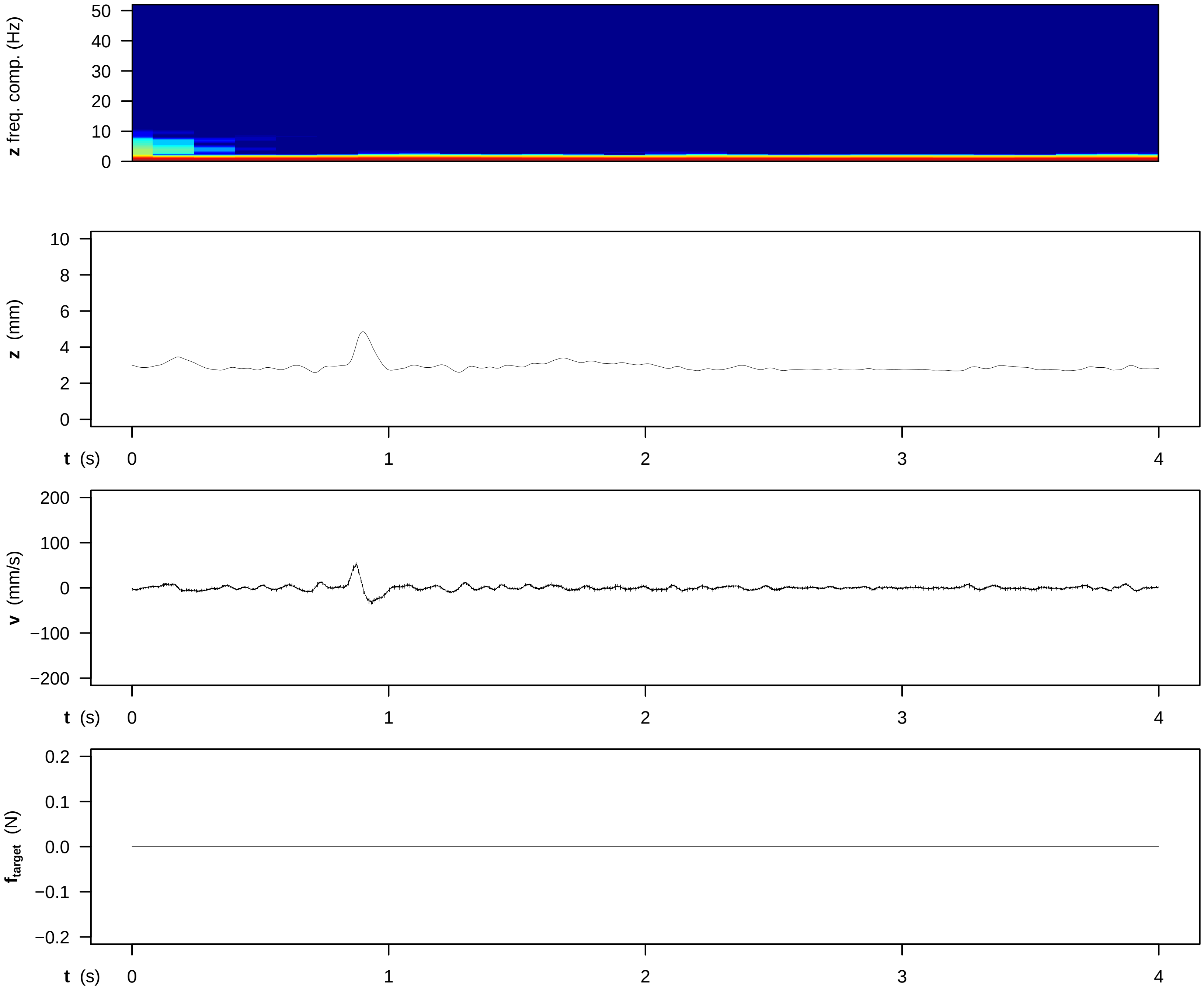

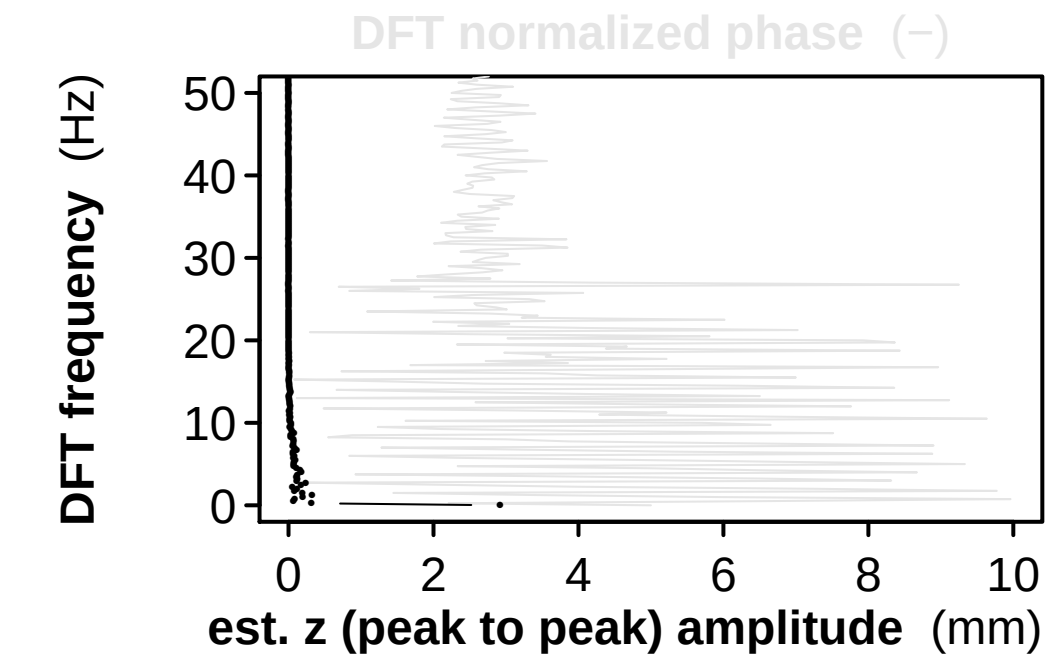

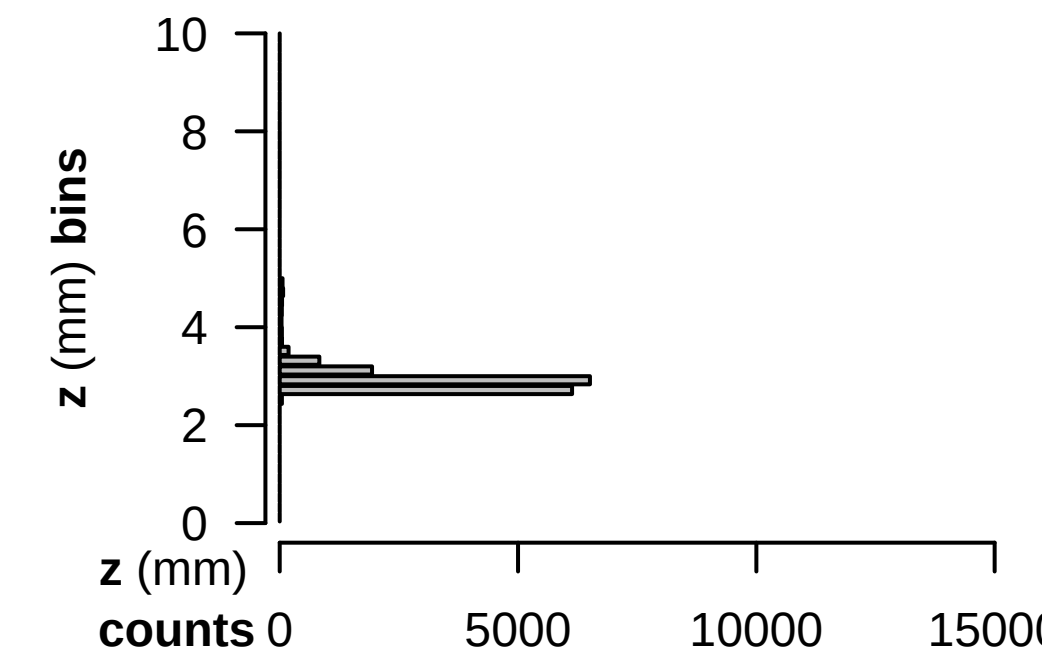

**SUBJECT 7 − RUN 11 − CONDITION 0,1**
**SC_180323_154140_0.AIFF**

```
z_min               : 2.59 mm
z_max               : 4.86 mm
z_travel_amplitude  : 2.27 mm

avg_abs_z_travel    : 4.05 mm/s

z_jarque-bera_jb    : 301749.21
z_jarque-bera_p     : 0.00e+00

z_lin_mod_est_slope: −0.08 mm/s
z_lin_mod_adj_R²    : 11 %

z_poly40_mod_adj_R²: 49 %

z_dft_ampl_thresh   : 0.010 mm
>=threshold_maxfreq: 17.50 Hz
```

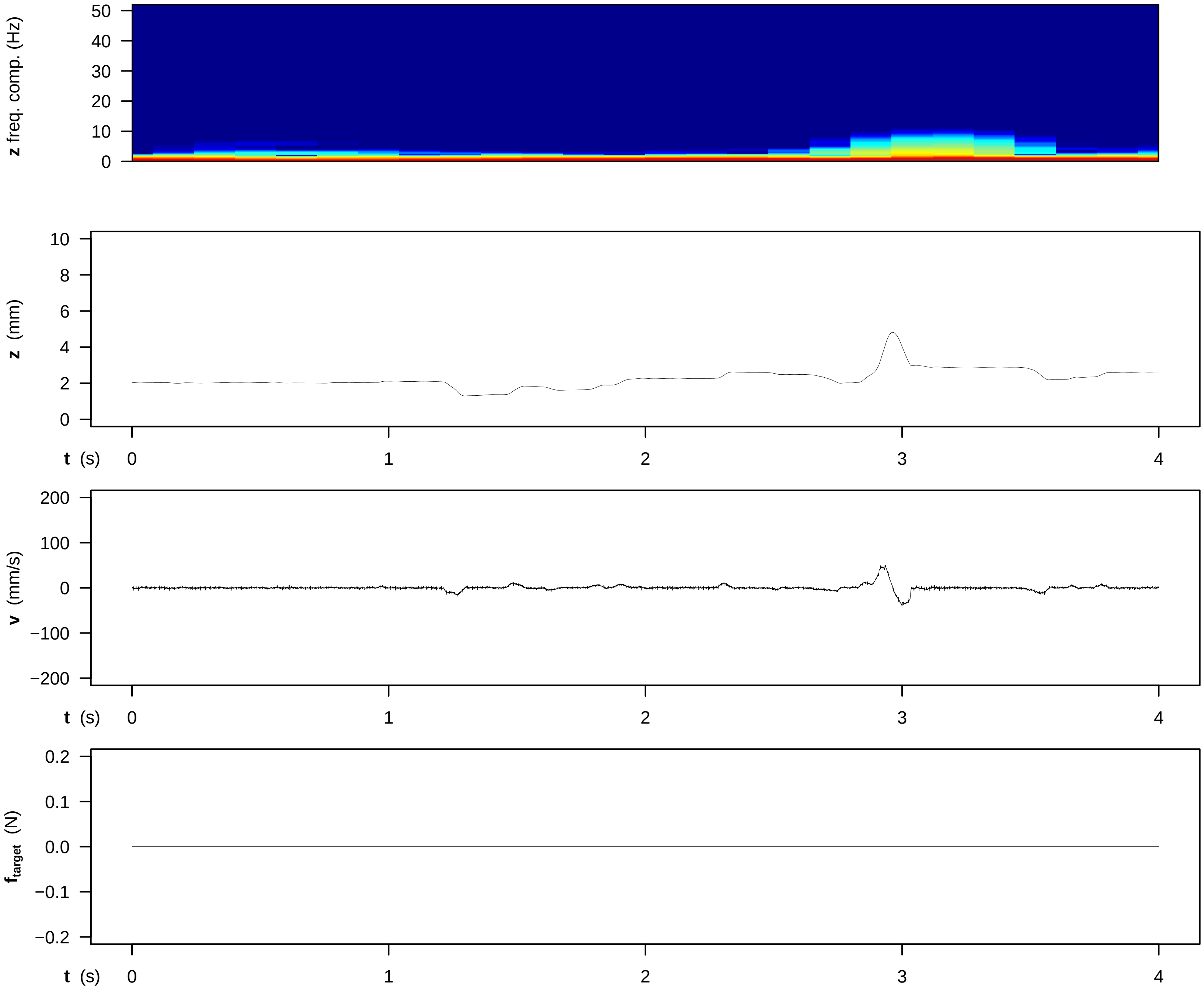

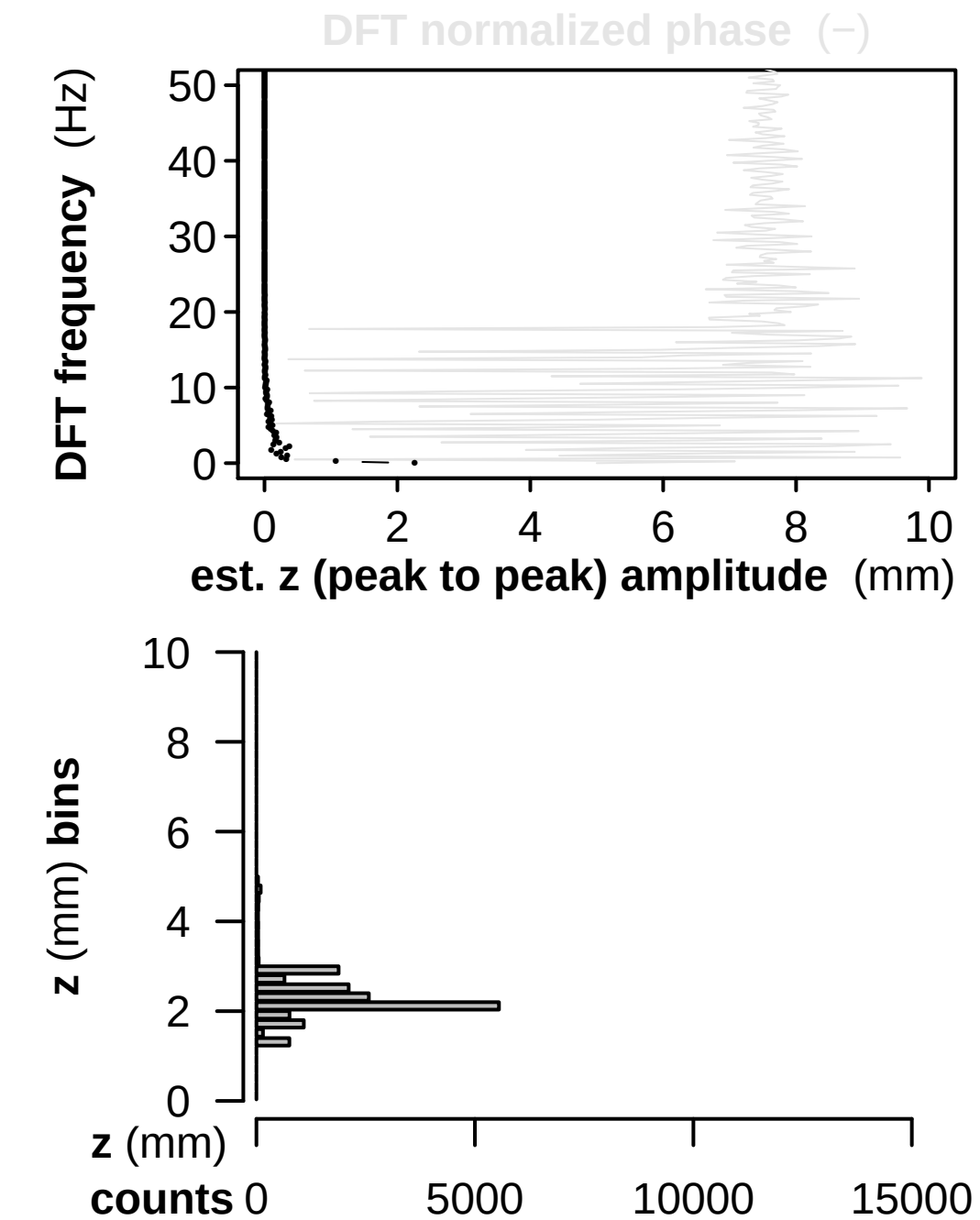

**SUBJECT 7 − RUN 23 − CONDITION 0,1**
**SC_180323_155043_0.AIFF**

```
z_min               : 1.29 mm
z_max               : 4.82 mm
z_travel_amplitude  : 3.52 mm

avg_abs_z_travel    : 4.24 mm/s

z_jarque-bera_jb    : 25628.94
z_jarque-bera_p     : 0.00e+00

z_lin_mod_est_slope: 0.26 mm/s
z_lin_mod_adj_R²    : 34 %

z_poly40_mod_adj_R²: 79 %

z_dft_ampl_thresh   : 0.010 mm
>=threshold_maxfreq: 15.25 Hz
```

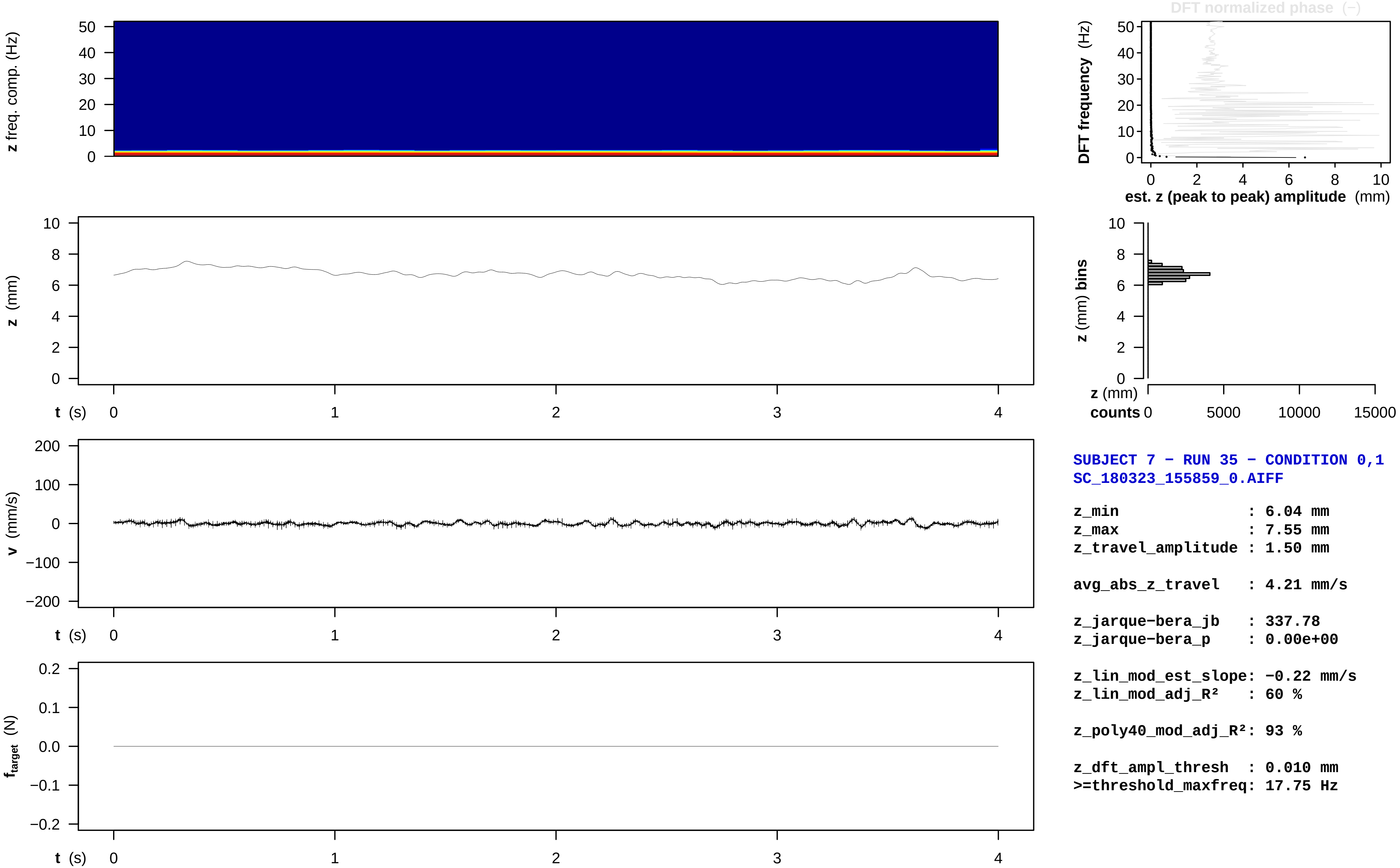

SUBJECT 7 − RUN 35 − CONDITION 0,1
SC_180323_155859_0.AIFF

z_min : 6.04 mm
z_max : 7.55 mm
z_travel_amplitude : 1.50 mm

avg_abs_z_travel : 4.21 mm/s

z_jarque-bera_jb : 337.78
z_jarque-bera_p : 0.00e+00

z_lin_mod_est_slope: −0.22 mm/s
z_lin_mod_adj_R² : 60 %

z_poly40_mod_adj_R²: 93 %

z_dft_ampl_thresh : 0.010 mm
>=threshold_maxfreq: 17.75 Hz

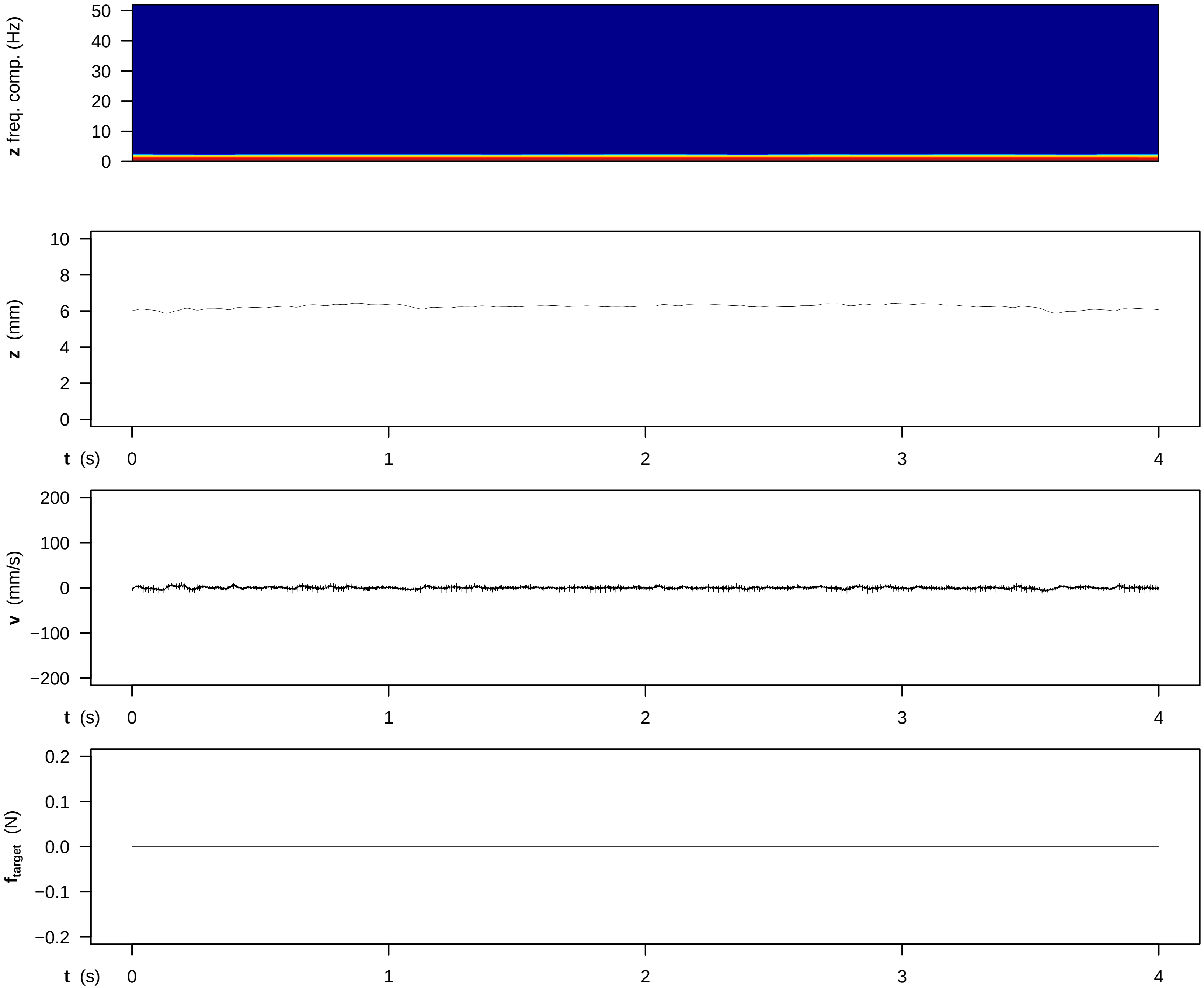

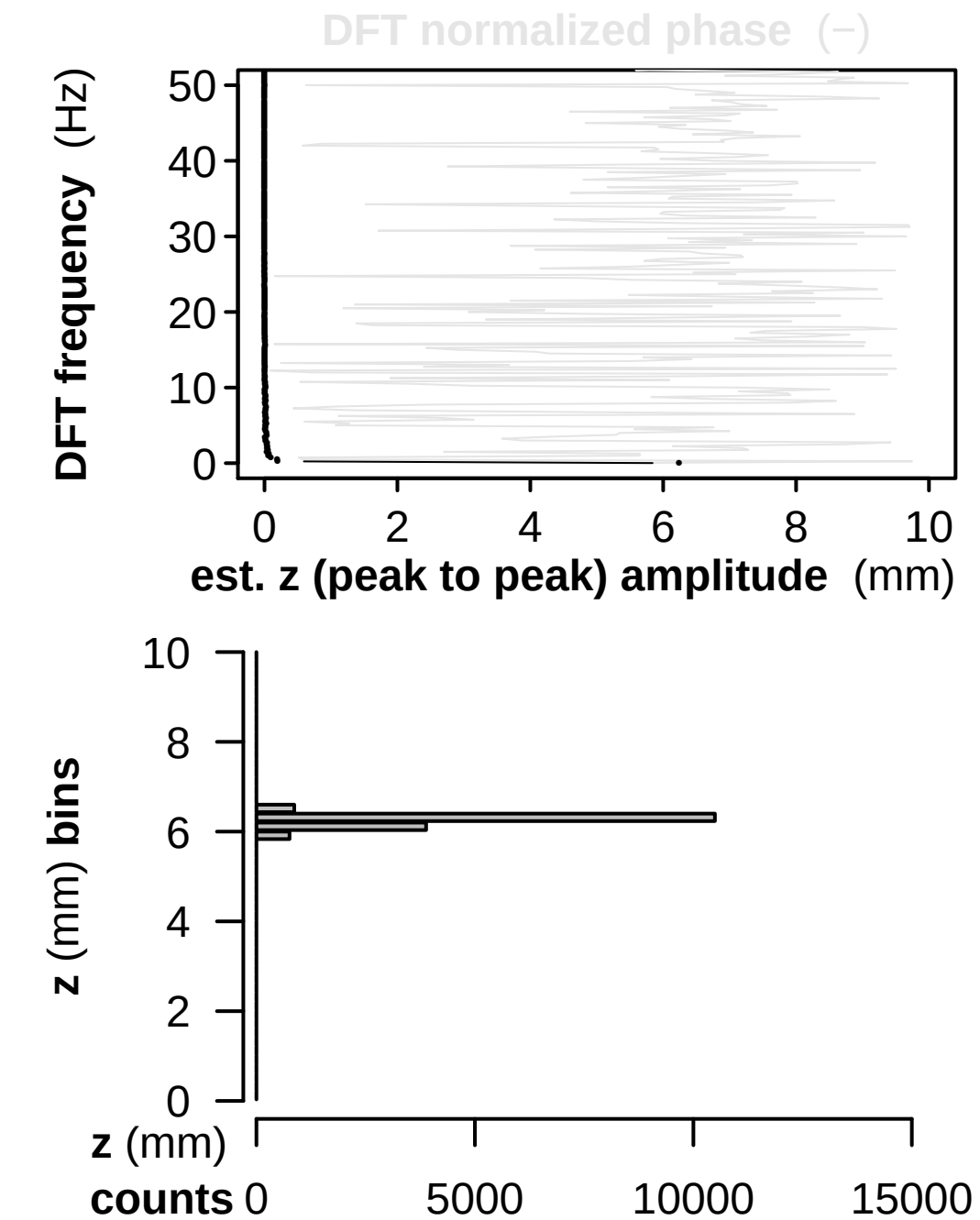

**SUBJECT 8 − RUN 08 − CONDITION 0,1**
**SC_180323_164921_0.AIFF**

```
z_min              : 5.87 mm
z_max              : 6.44 mm
z_travel_amplitude : 0.57 mm

avg_abs_z_travel   : 3.11 mm/s

z_jarque-bera_jb   : 1858.70
z_jarque-bera_p    : 0.00e+00

z_lin_mod_est_slope: 0.00 mm/s
z_lin_mod_adj_R²   : 0 %

z_poly40_mod_adj_R²: 88 %

z_dft_ampl_thresh  : 0.010 mm
>=threshold_maxfreq: 15.75 Hz
```

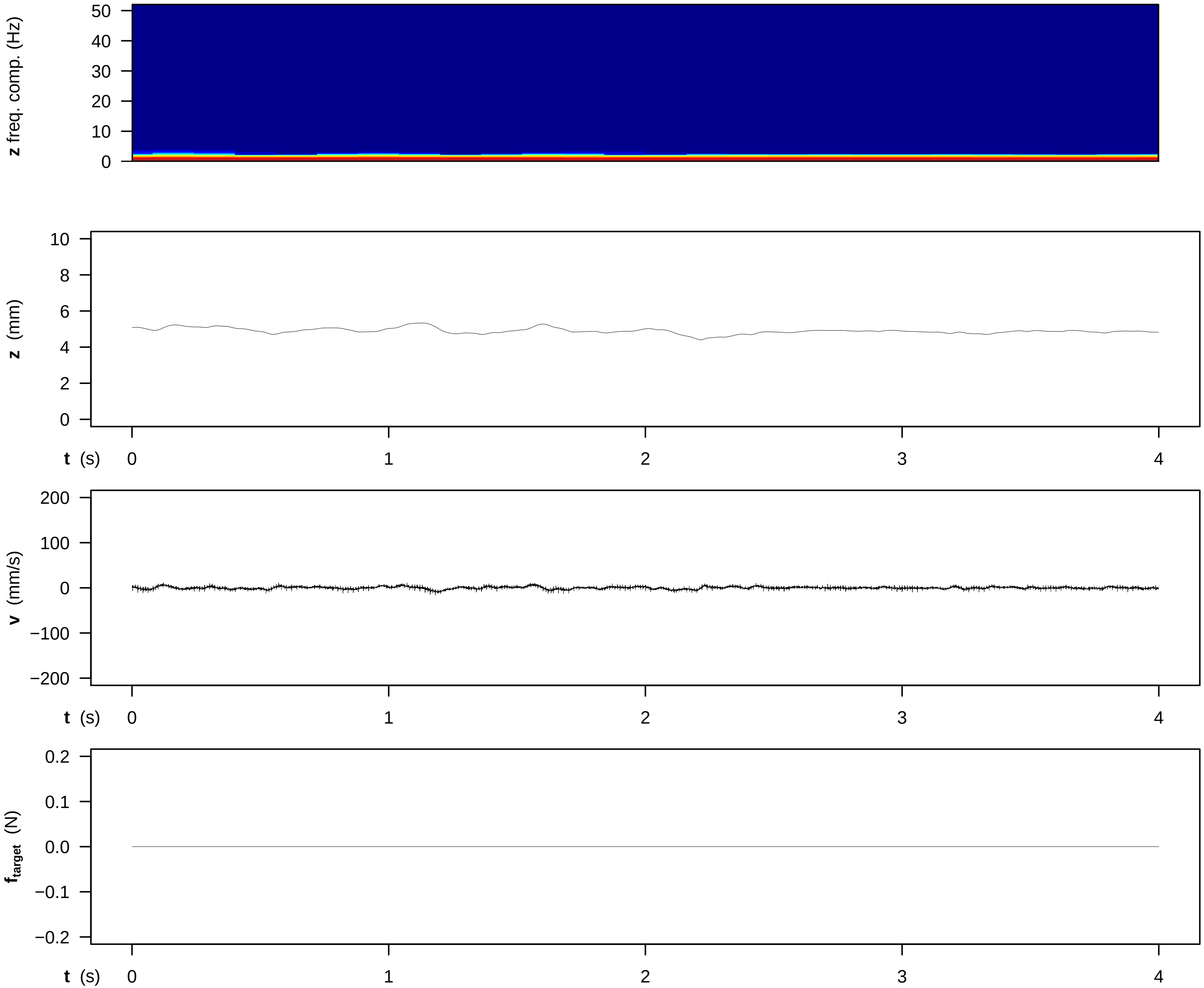

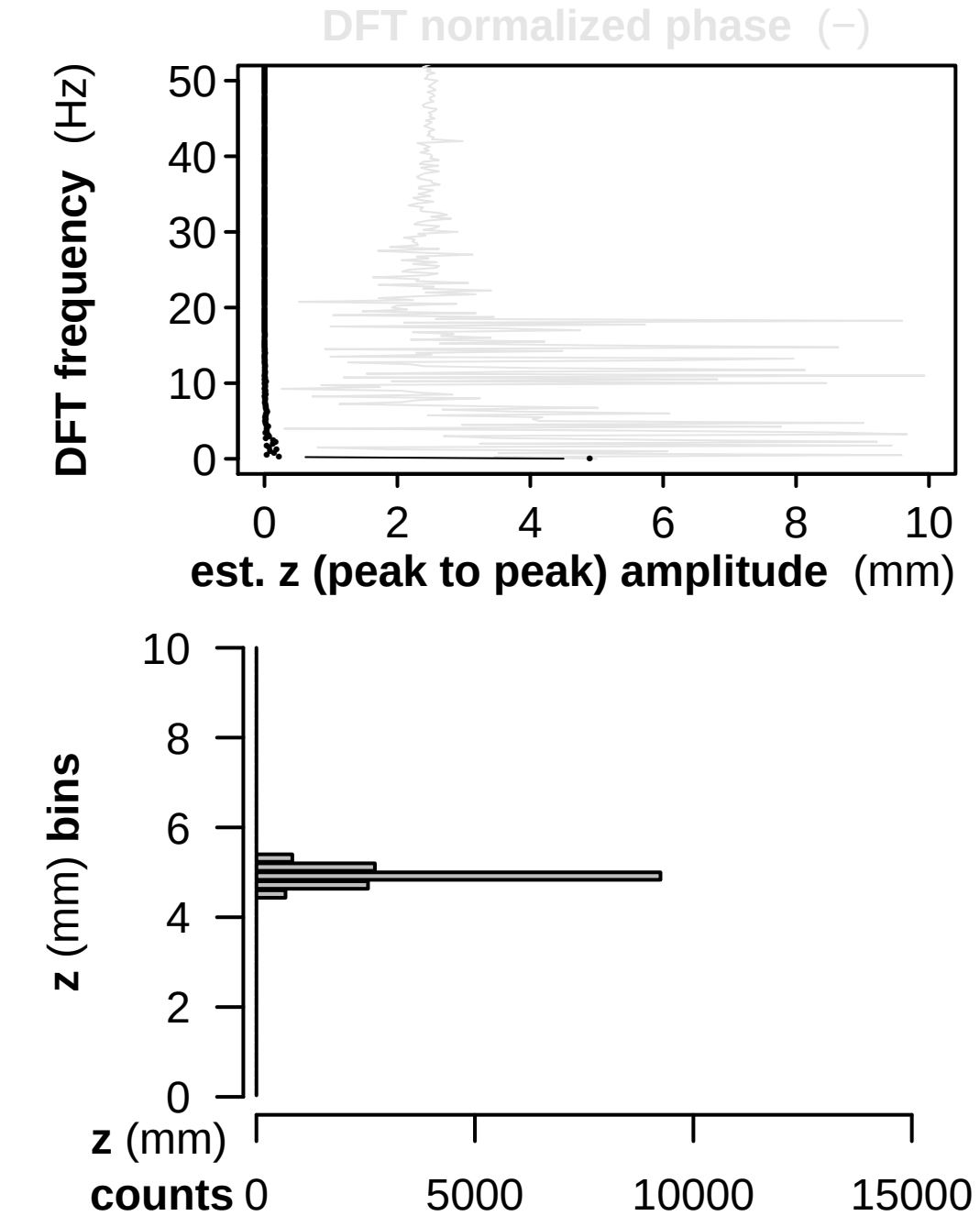

**SUBJECT 8 − RUN 31 − CONDITION 0,1**
**SC_180323_170749_0.AIFF**

```
z_min               : 4.40 mm
z_max               : 5.34 mm
z_travel_amplitude  : 0.93 mm

avg_abs_z_travel    : 3.33 mm/s

z_jarque-bera_jb    : 640.98
z_jarque-bera_p     : 0.00e+00

z_lin_mod_est_slope: -0.06 mm/s
z_lin_mod_adj_R²    : 19 %

z_poly40_mod_adj_R²: 59 %

z_dft_ampl_thresh   : 0.010 mm
>=threshold_maxfreq: 14.00 Hz
```

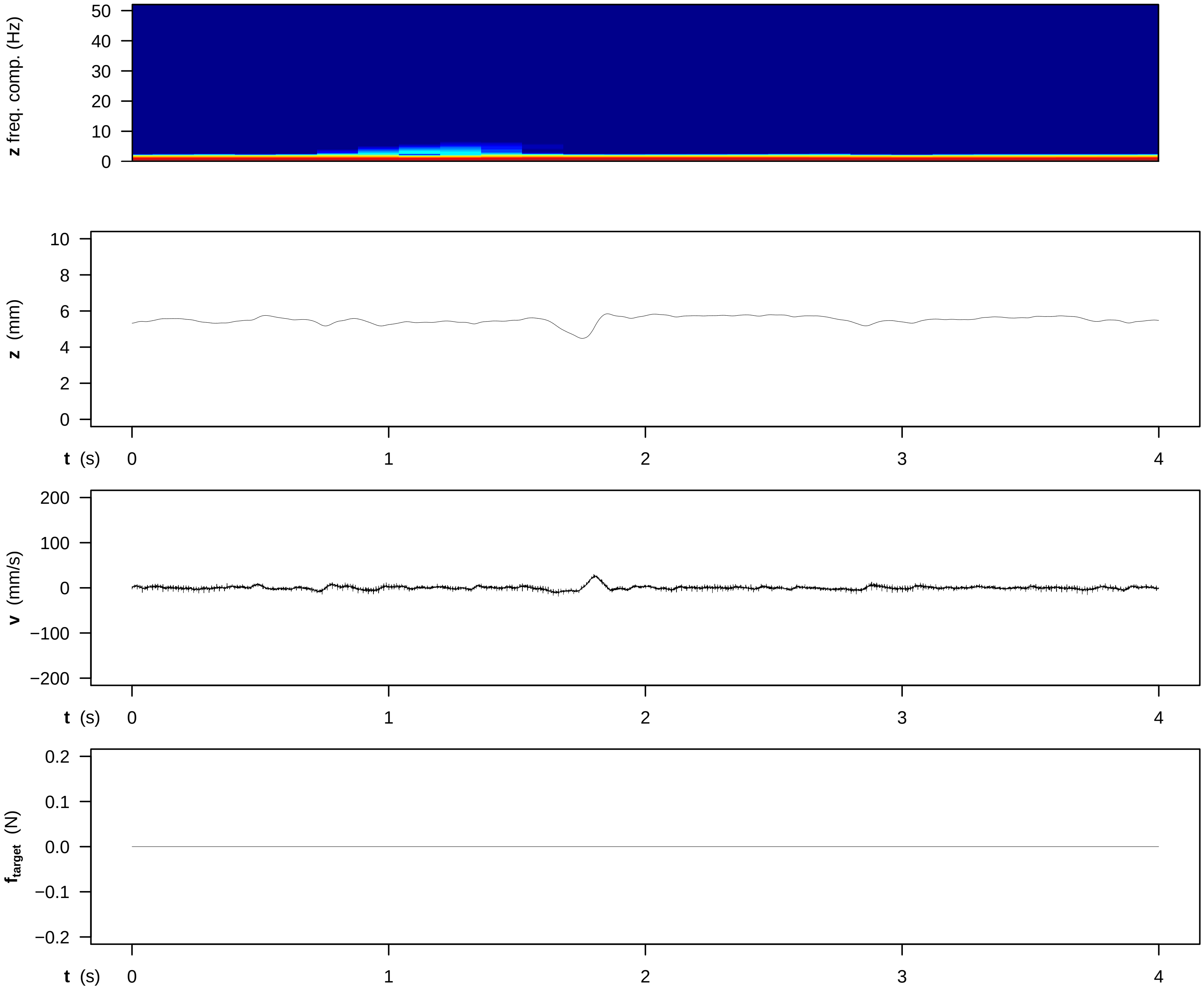

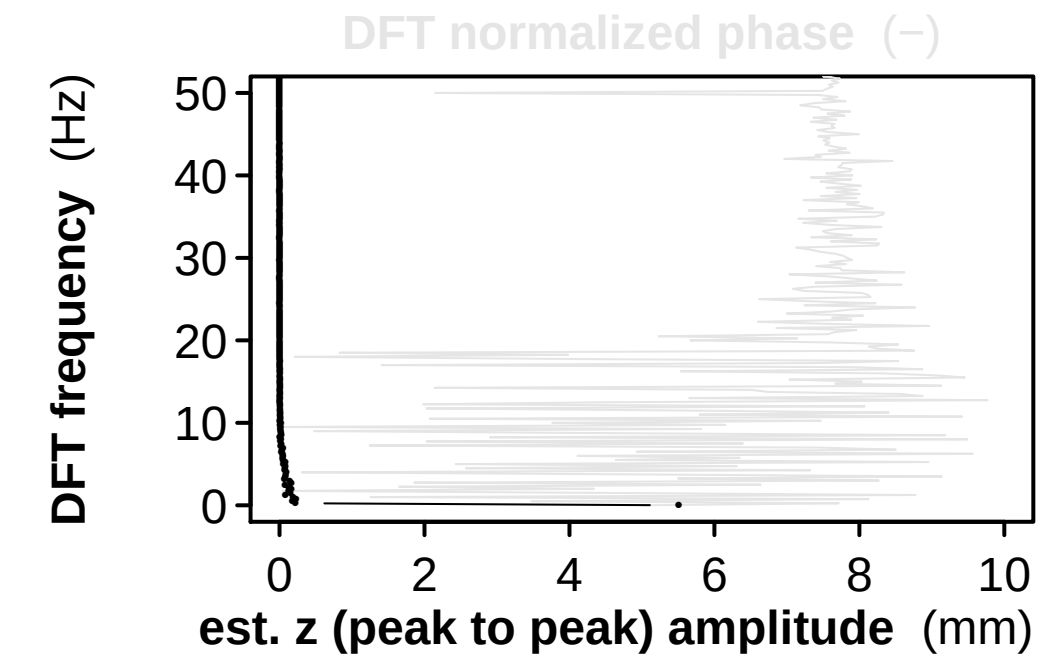

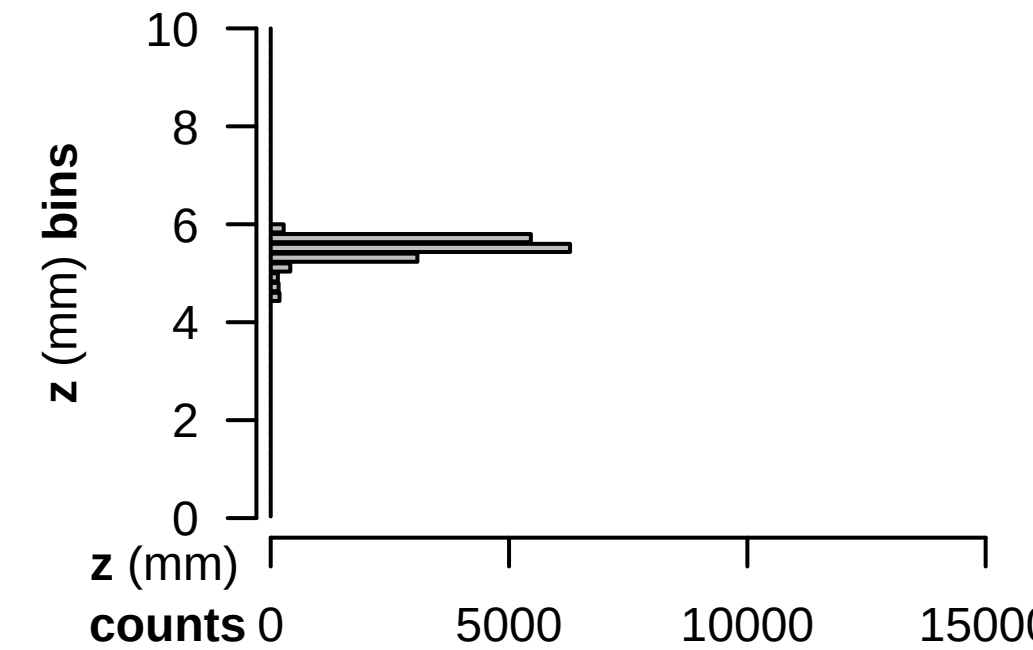

**SUBJECT 8 − RUN 36 − CONDITION 0,1**
**SC_180323_171035_0.AIFF**

```
z_min               : 4.48 mm
z_max               : 5.85 mm
z_travel_amplitude  : 1.36 mm

avg_abs_z_travel    : 3.46 mm/s

z_jarque-bera_jb    : 24025.24
z_jarque-bera_p     : 0.00e+00

z_lin_mod_est_slope: 0.04 mm/s
z_lin_mod_adj_R²    : 5 %

z_poly40_mod_adj_R²: 58 %

z_dft_ampl_thresh   : 0.010 mm
>=threshold_maxfreq: 11.25 Hz
```

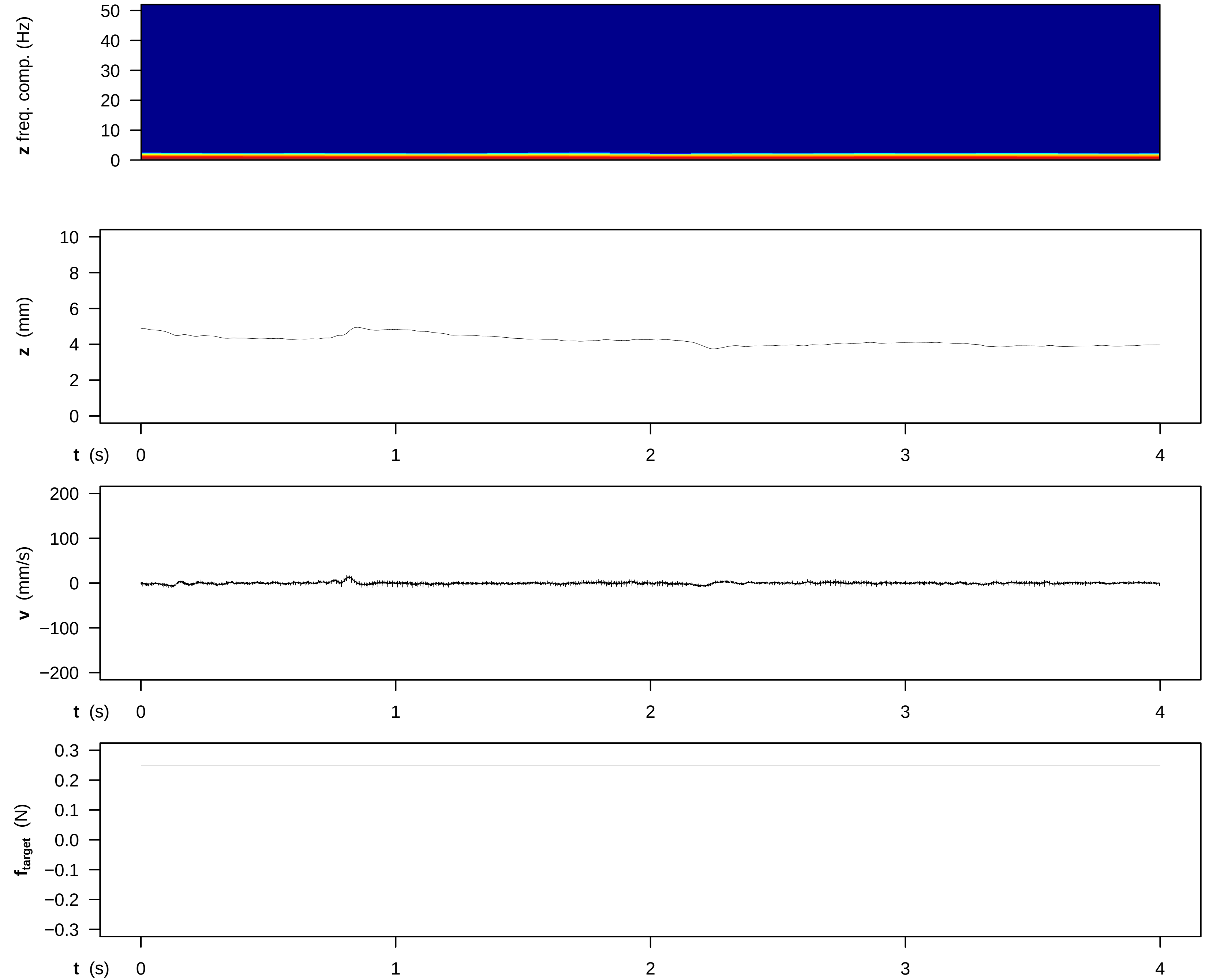

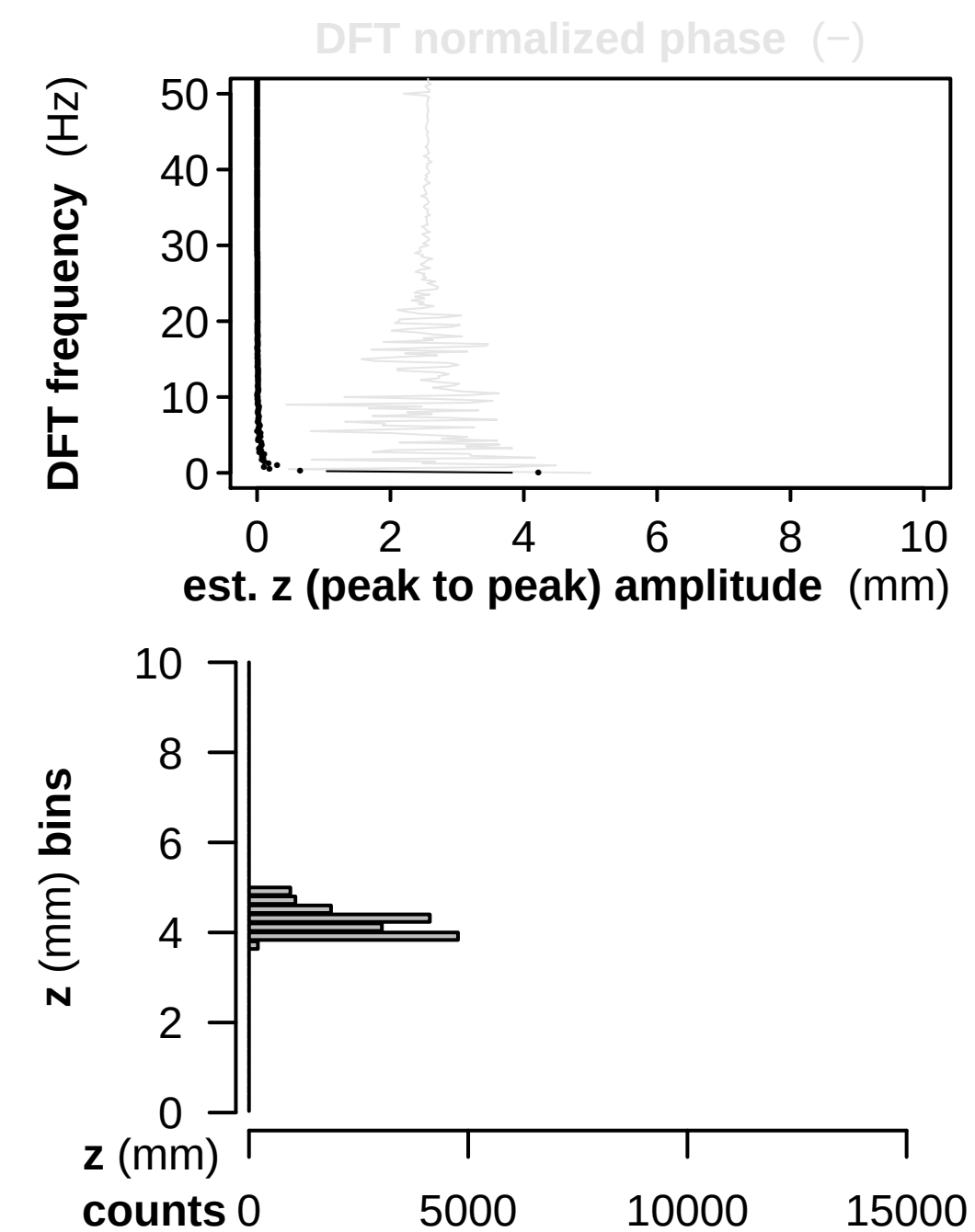

**SUBJECT 1 − RUN 21 − CONDITION 1,0**
**SC_180323_105133_0.AIFF**

```
z_min               : 3.74 mm
z_max               : 4.95 mm
z_travel_amplitude  : 1.21 mm

avg_abs_z_travel    : 3.91 mm/s

z_jarque-bera_jb    : 1272.90
z_jarque-bera_p     : 0.00e+00

z_lin_mod_est_slope: −0.20 mm/s
z_lin_mod_adj_R²    : 66 %

z_poly40_mod_adj_R²: 97 %

z_dft_ampl_thresh   : 0.010 mm
>=threshold_maxfreq: 18.25 Hz
```

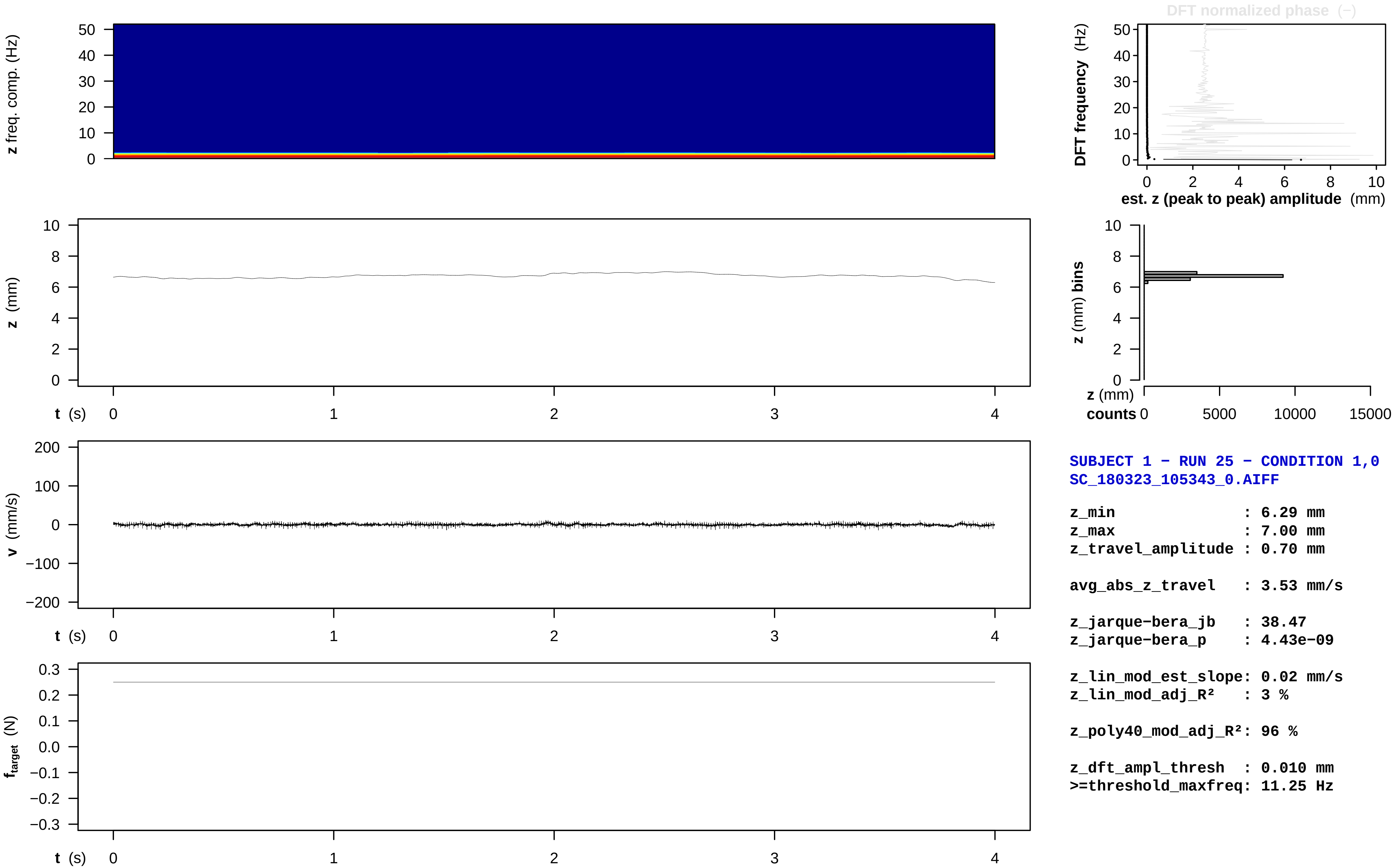

**SUBJECT 1 − RUN 25 − CONDITION 1,0**
**SC_180323_105343_0.AIFF**

```
z_min               : 6.29 mm
z_max               : 7.00 mm
z_travel_amplitude  : 0.70 mm

avg_abs_z_travel    : 3.53 mm/s

z_jarque-bera_jb    : 38.47
z_jarque-bera_p     : 4.43e-09

z_lin_mod_est_slope: 0.02 mm/s
z_lin_mod_adj_R²    : 3 %

z_poly40_mod_adj_R²: 96 %

z_dft_ampl_thresh   : 0.010 mm
>=threshold_maxfreq: 11.25 Hz
```

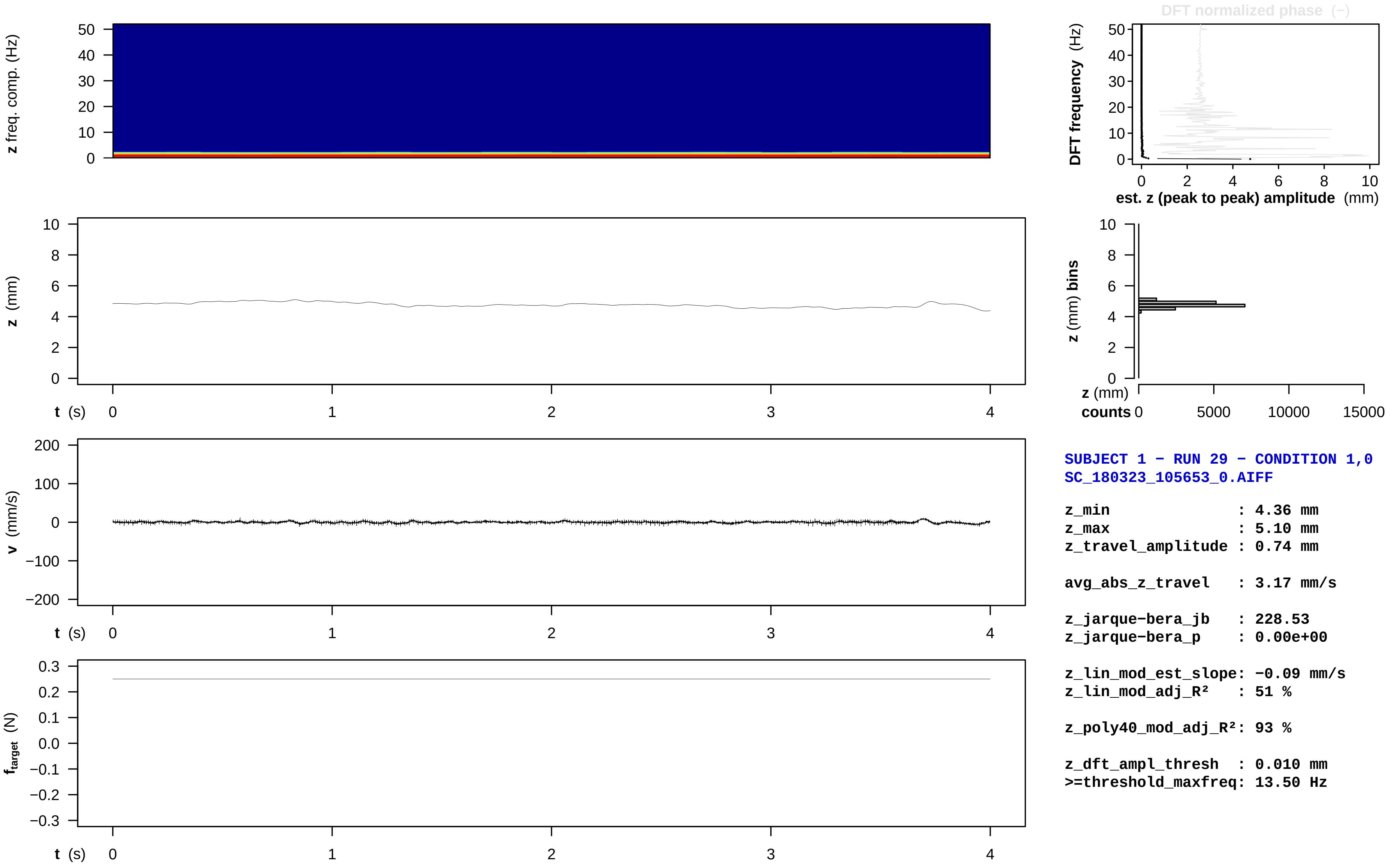

**SUBJECT 1 − RUN 29 − CONDITION 1,0**
**SC_180323_105653_0.AIFF**

```
z_min               : 4.36 mm
z_max               : 5.10 mm
z_travel_amplitude  : 0.74 mm

avg_abs_z_travel    : 3.17 mm/s

z_jarque-bera_jb    : 228.53
z_jarque-bera_p     : 0.00e+00

z_lin_mod_est_slope: −0.09 mm/s
z_lin_mod_adj_R²    : 51 %

z_poly40_mod_adj_R²: 93 %

z_dft_ampl_thresh   : 0.010 mm
>=threshold_maxfreq: 13.50 Hz
```

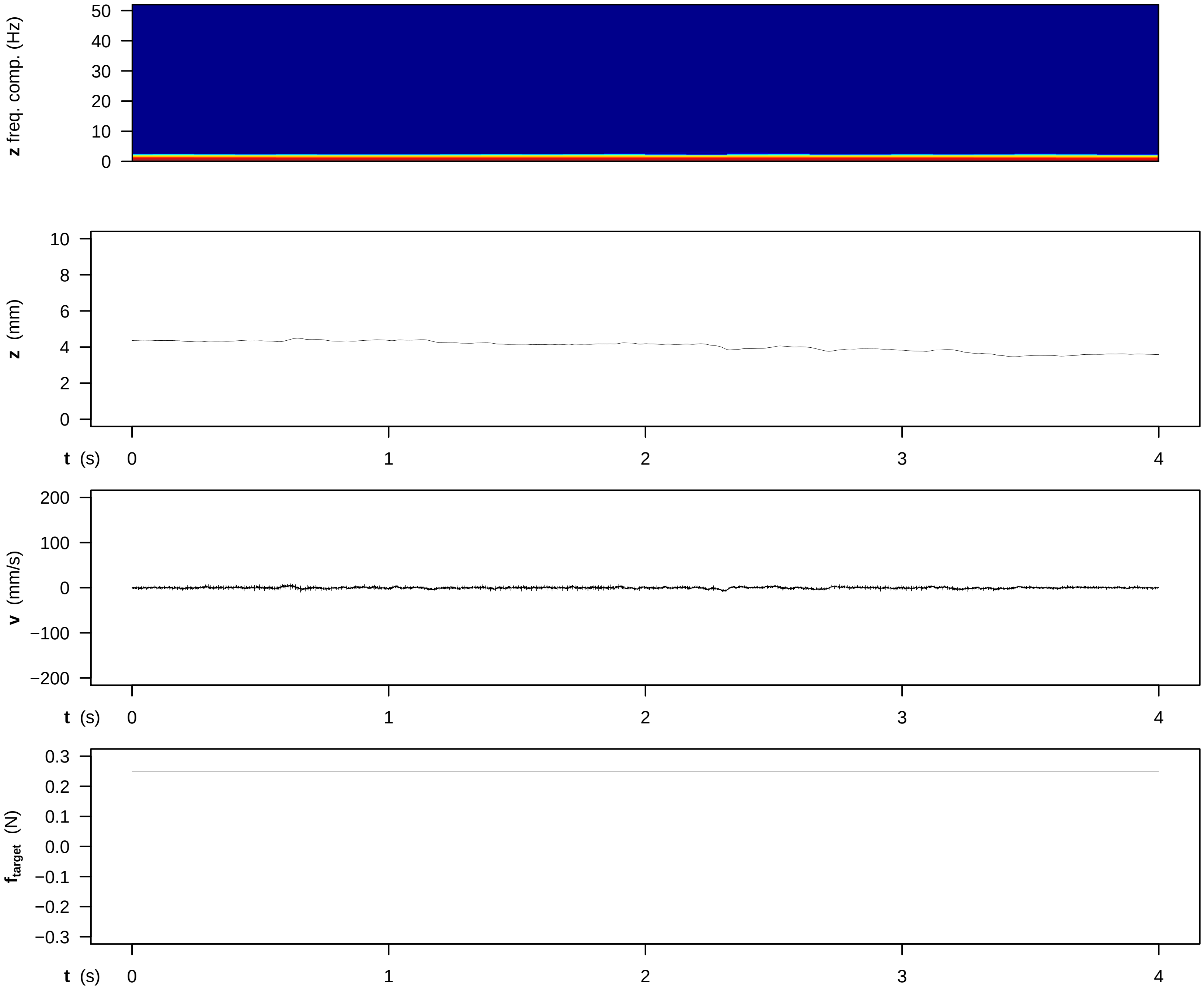

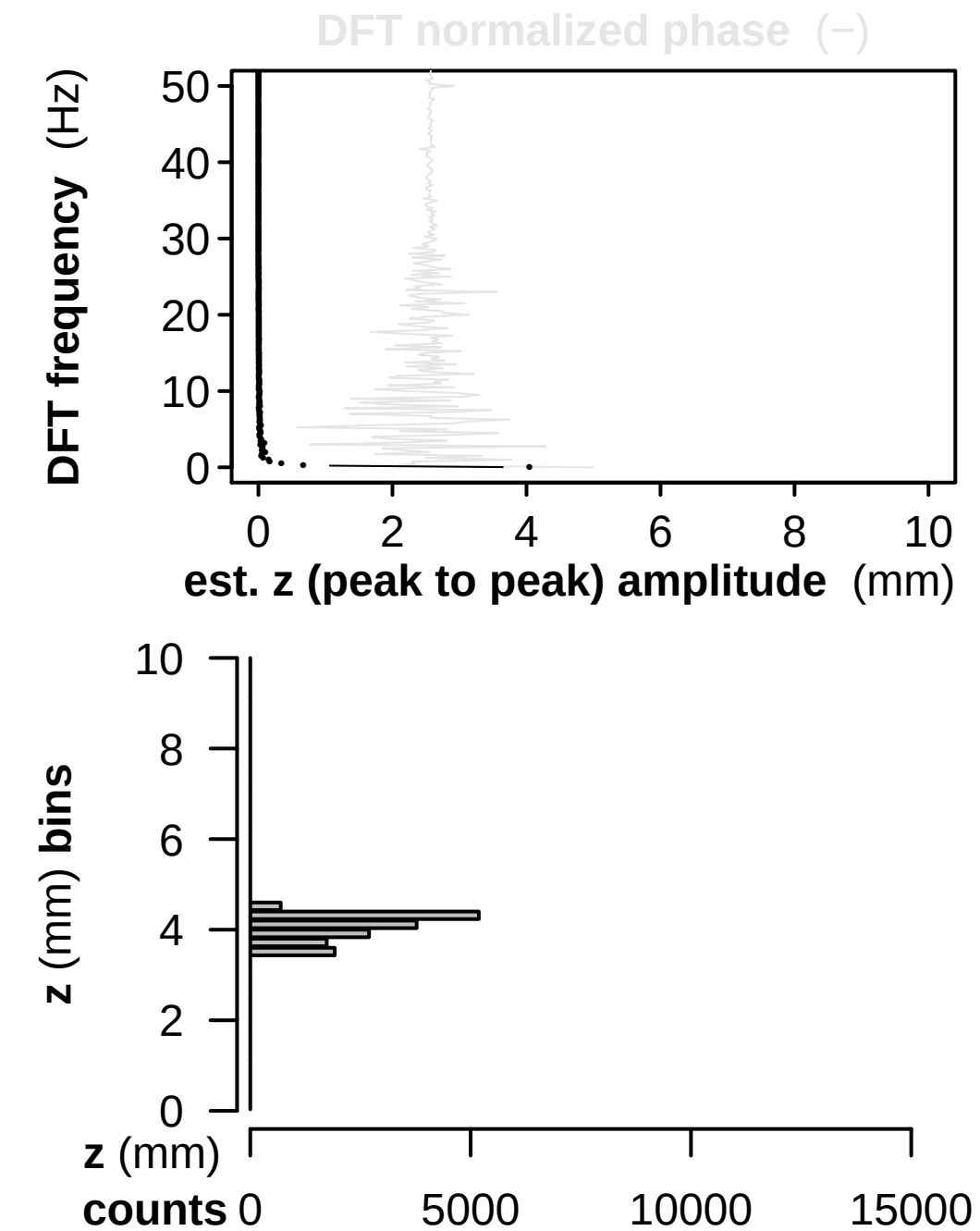

**SUBJECT 2 – RUN 22 – CONDITION 1,0**
**SC_180323_112829_0.AIFF**

```
z_min               : 3.46 mm
z_max               : 4.49 mm
z_travel_amplitude  : 1.04 mm

avg_abs_z_travel    : 2.09 mm/s

z_jarque-bera_jb    : 1412.42
z_jarque-bera_p     : 0.00e+00

z_lin_mod_est_slope: −0.24 mm/s
z_lin_mod_adj_R²    : 89 %

z_poly40_mod_adj_R²: 98 %

z_dft_ampl_thresh   : 0.010 mm
>=threshold_maxfreq: 16.75 Hz
```

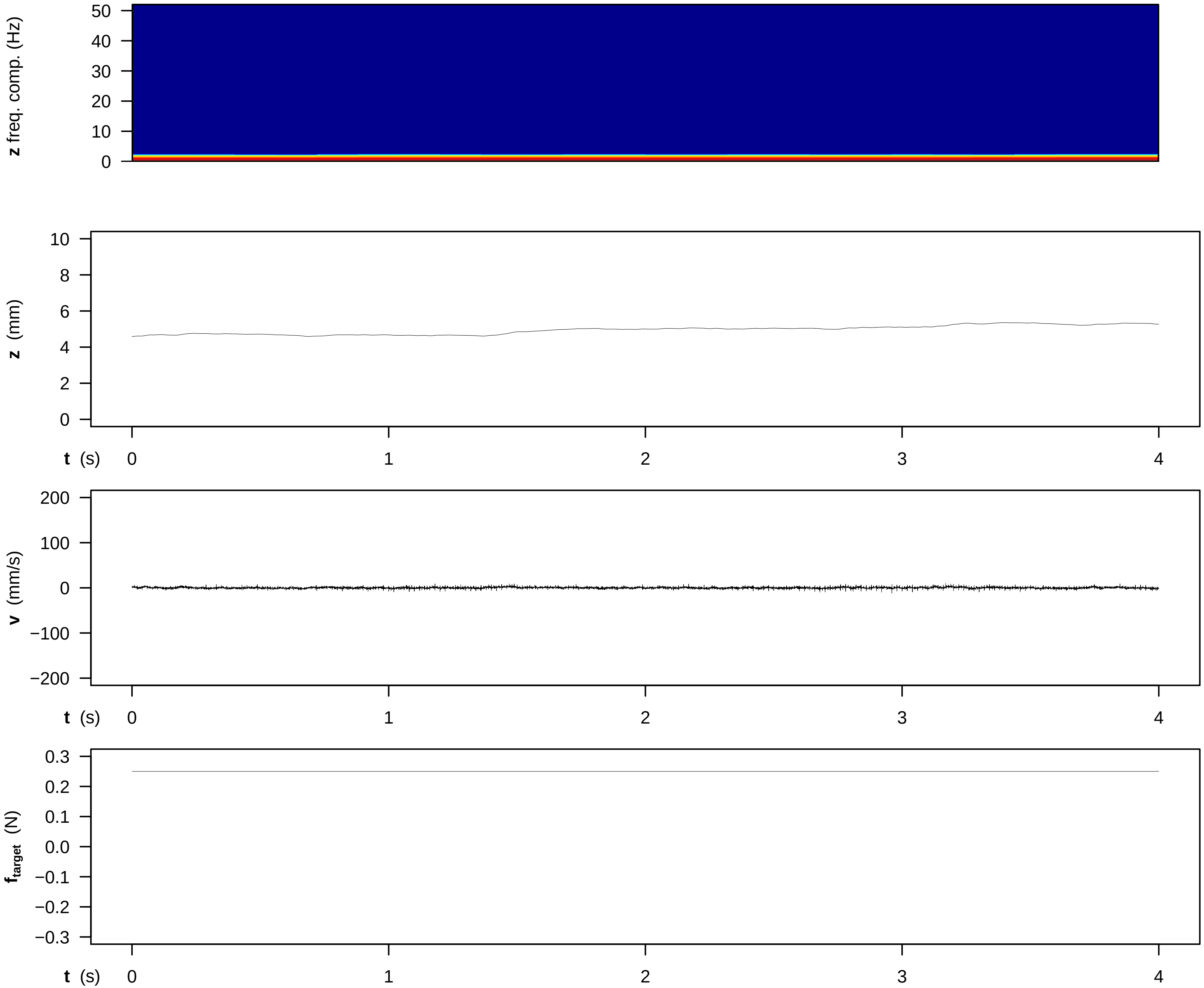

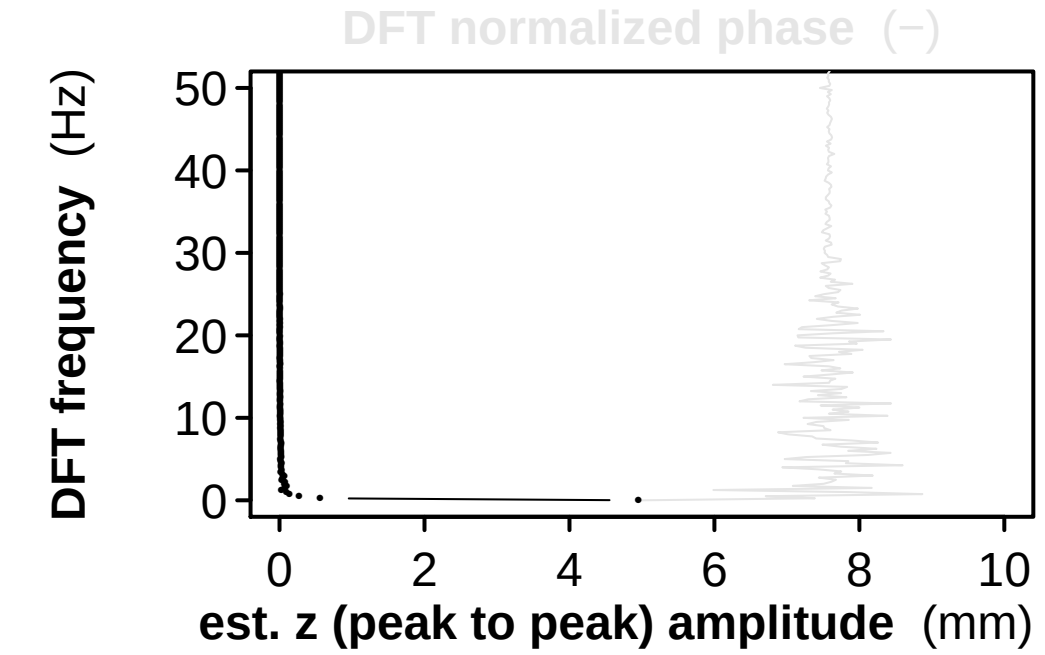

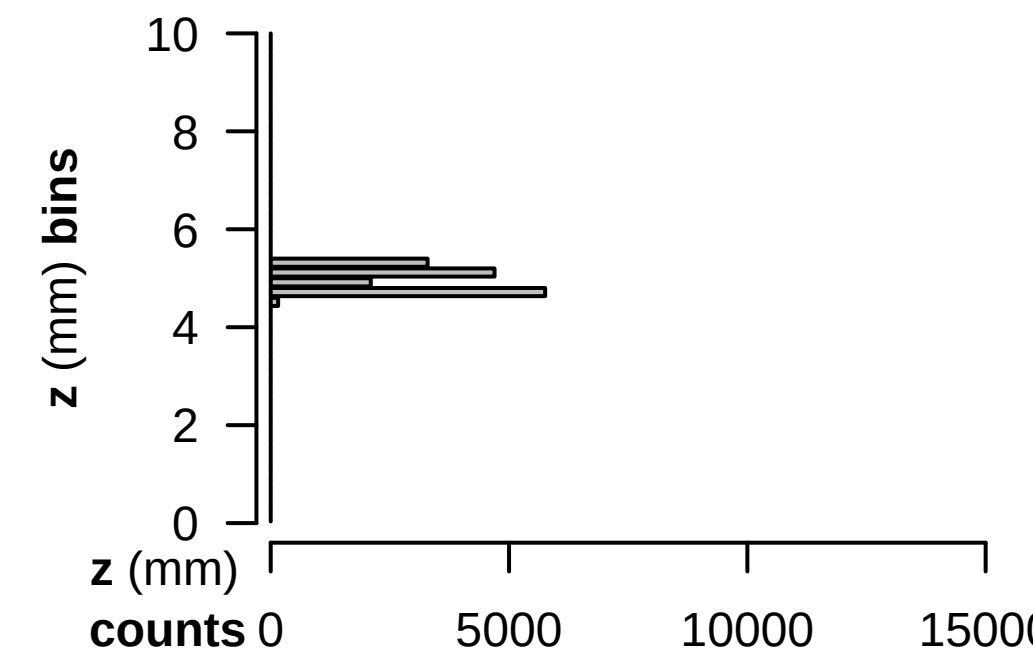

**SUBJECT 2 − RUN 30 − CONDITION 1,0**
**SC_180323_113316_0.AIFF**

```
z_min               : 4.58 mm
z_max               : 5.36 mm
z_travel_amplitude  : 0.78 mm

avg_abs_z_travel    : 2.46 mm/s

z_jarque-bera_jb    : 1148.77
z_jarque-bera_p     : 0.00e+00

z_lin_mod_est_slope: 0.20 mm/s
z_lin_mod_adj_R²    : 88 %

z_poly40_mod_adj_R²: 99 %

z_dft_ampl_thresh   : 0.010 mm
>=threshold_maxfreq: 16.50 Hz
```

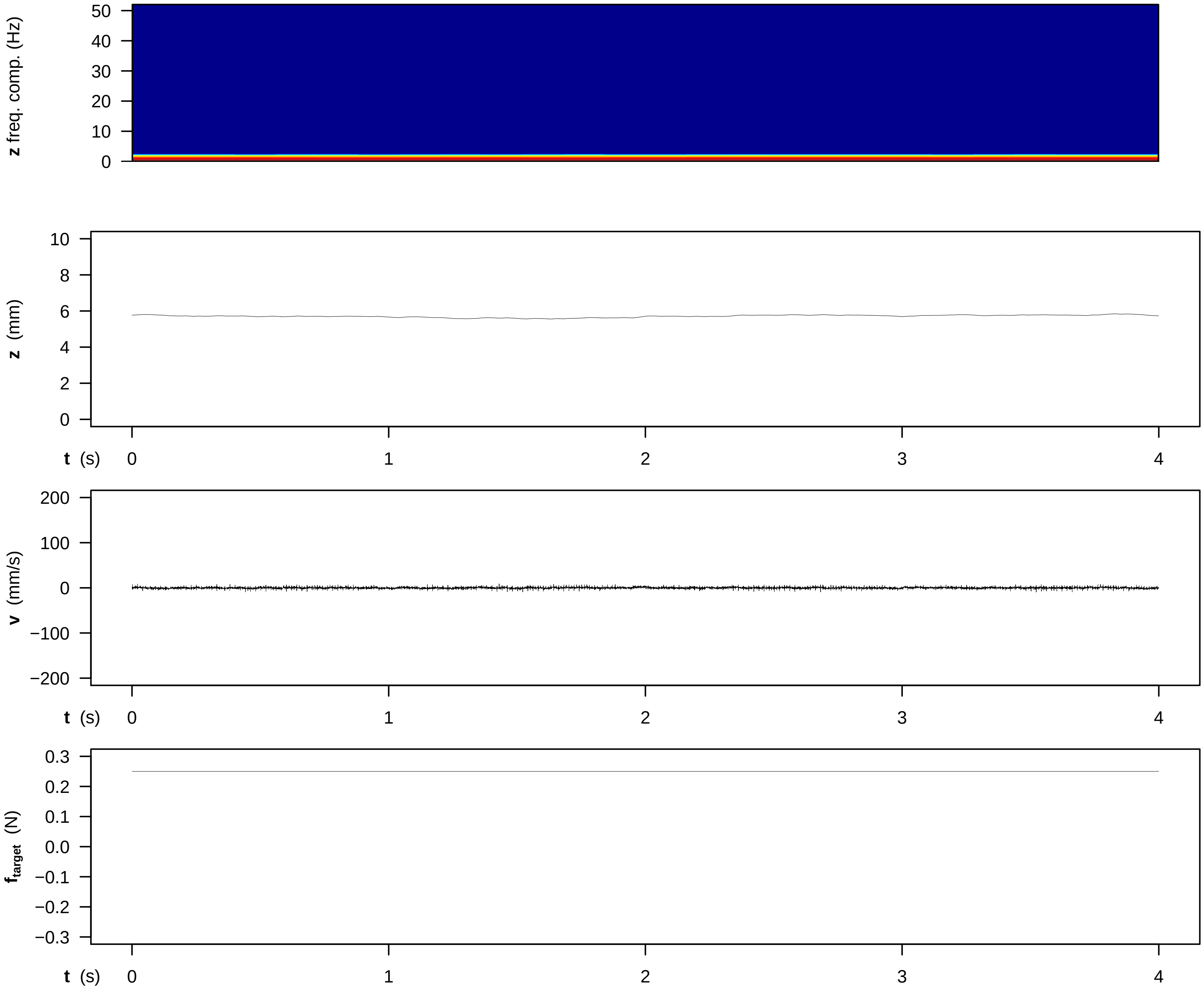

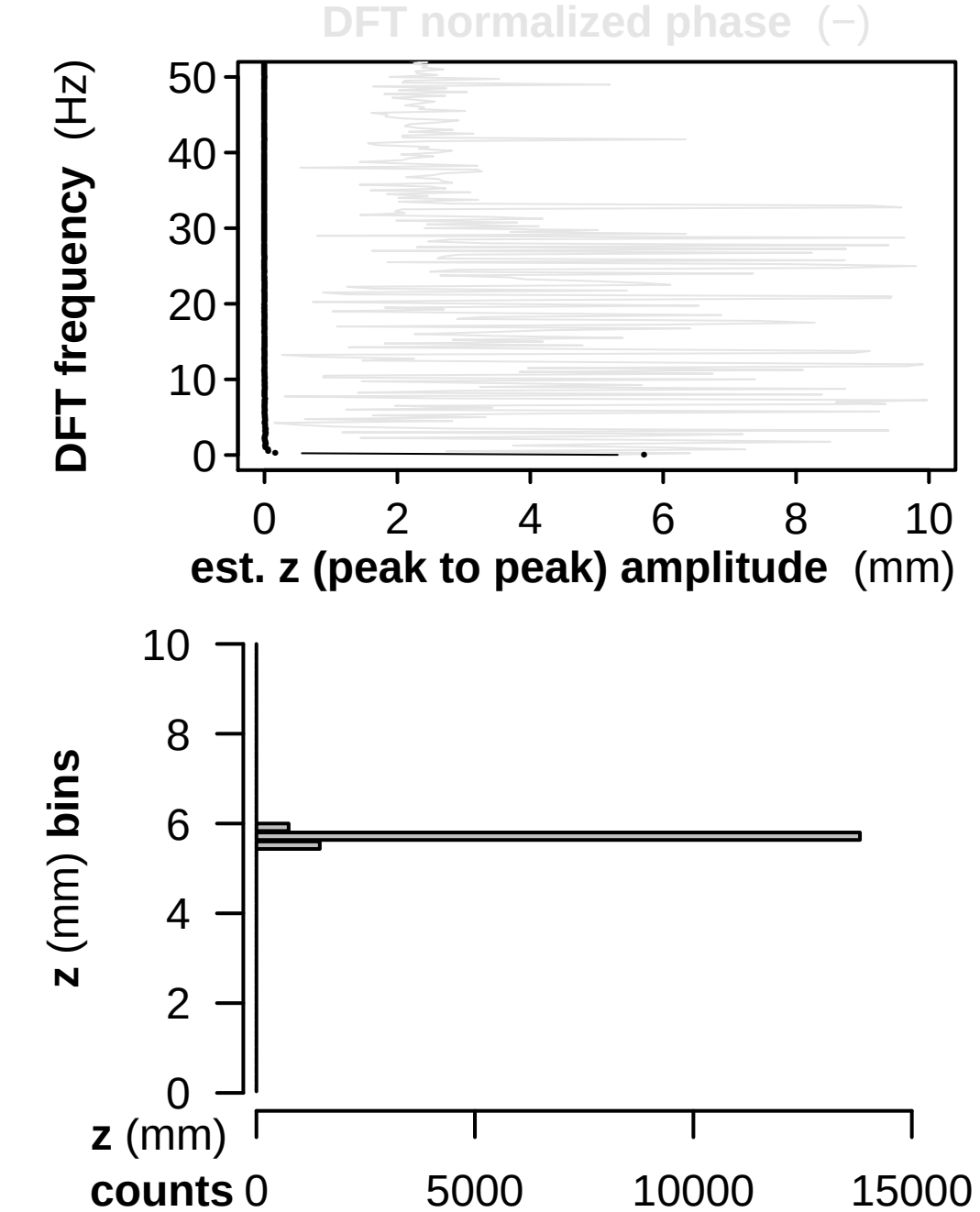

**SUBJECT 2 − RUN 31 − CONDITION 1,0**
**SC_180323_113337_0.AIFF**

```
z_min               : 5.56 mm
z_max               : 5.85 mm
z_travel_amplitude  : 0.29 mm

avg_abs_z_travel    : 2.49 mm/s

z_jarque-bera_jb    : 1052.44
z_jarque-bera_p     : 0.00e+00

z_lin_mod_est_slope: 0.03 mm/s
z_lin_mod_adj_R²    : 23 %

z_poly40_mod_adj_R²: 95 %

z_dft_ampl_thresh   : 0.010 mm
>=threshold_maxfreq: 4.75 Hz
```

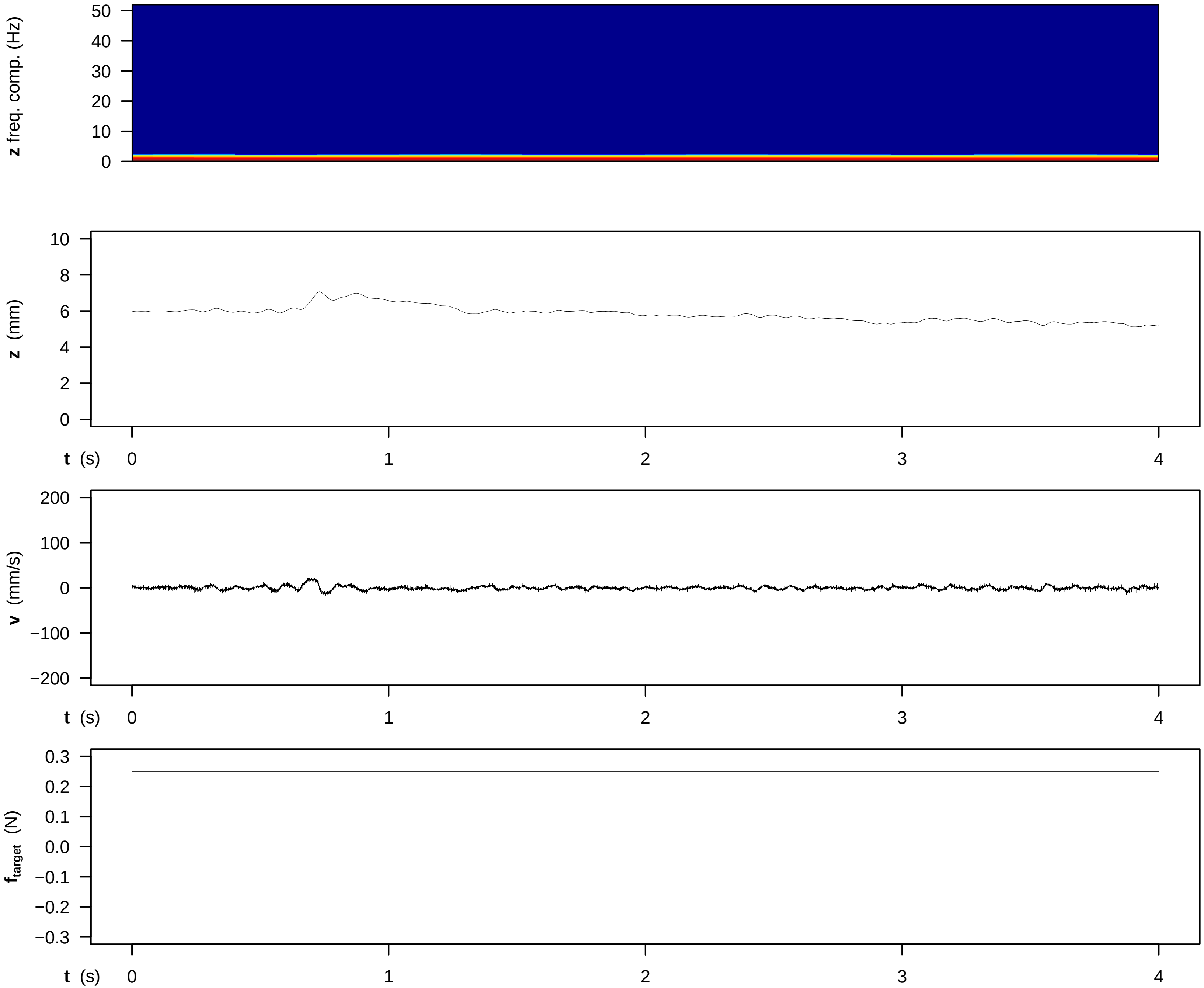

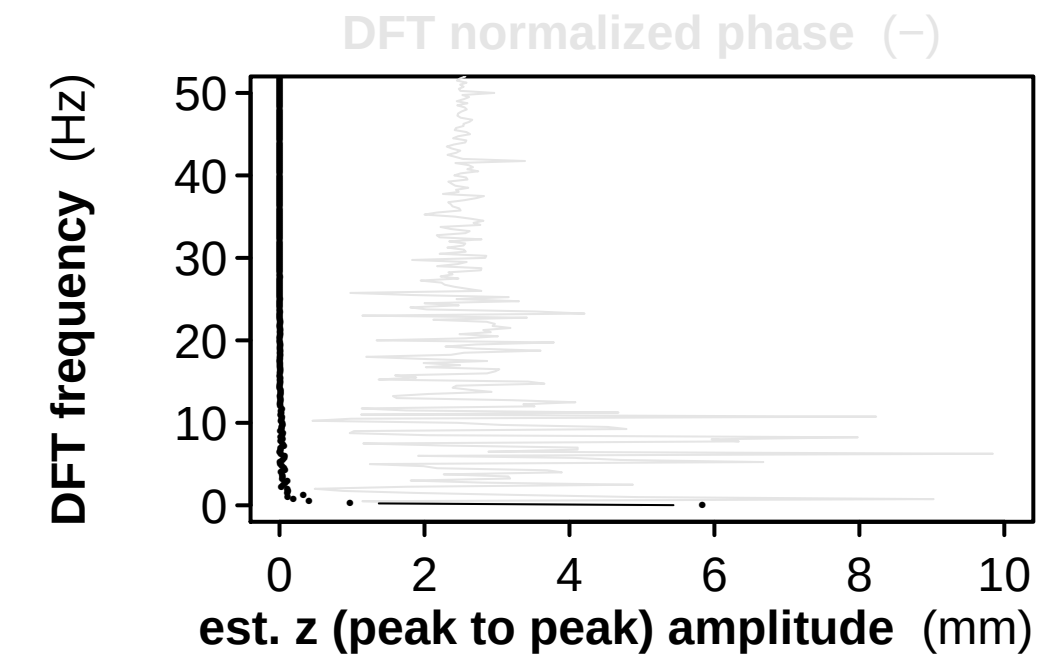

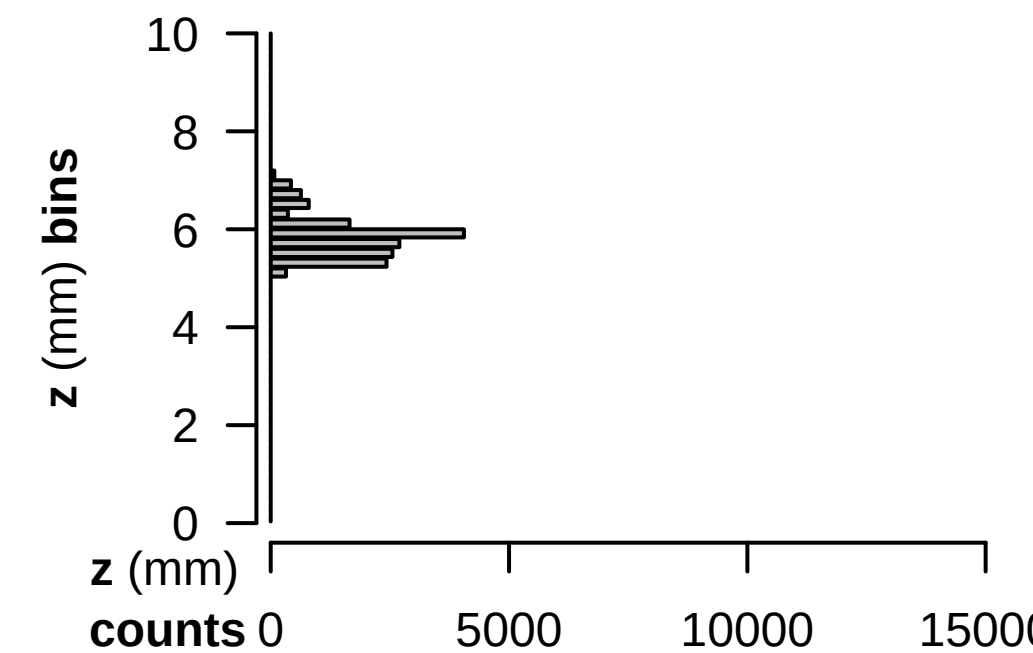

**SUBJECT 3 − RUN 02 − CONDITION 1,0**
**SC_180323_115535_0.AIFF**

```
z_min               : 5.13 mm
z_max               : 7.06 mm
z_travel_amplitude  : 1.93 mm

avg_abs_z_travel    : 4.85 mm/s

z_jarque-bera_jb    : 1491.88
z_jarque-bera_p     : 0.00e+00

z_lin_mod_est_slope: −0.29 mm/s
z_lin_mod_adj_R²    : 64 %

z_poly40_mod_adj_R²: 96 %

z_dft_ampl_thresh   : 0.010 mm
>=threshold_maxfreq: 20.75 Hz
```

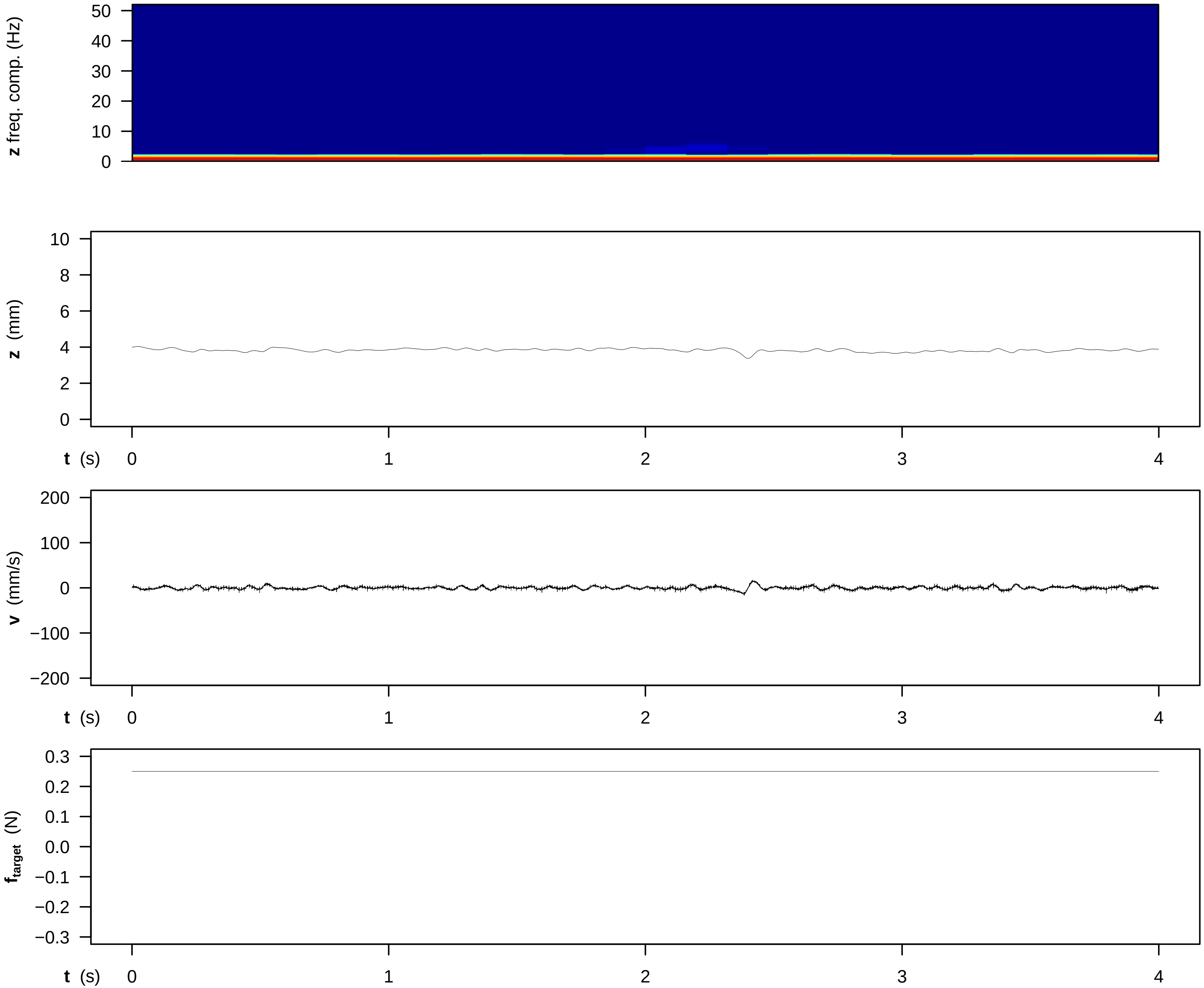

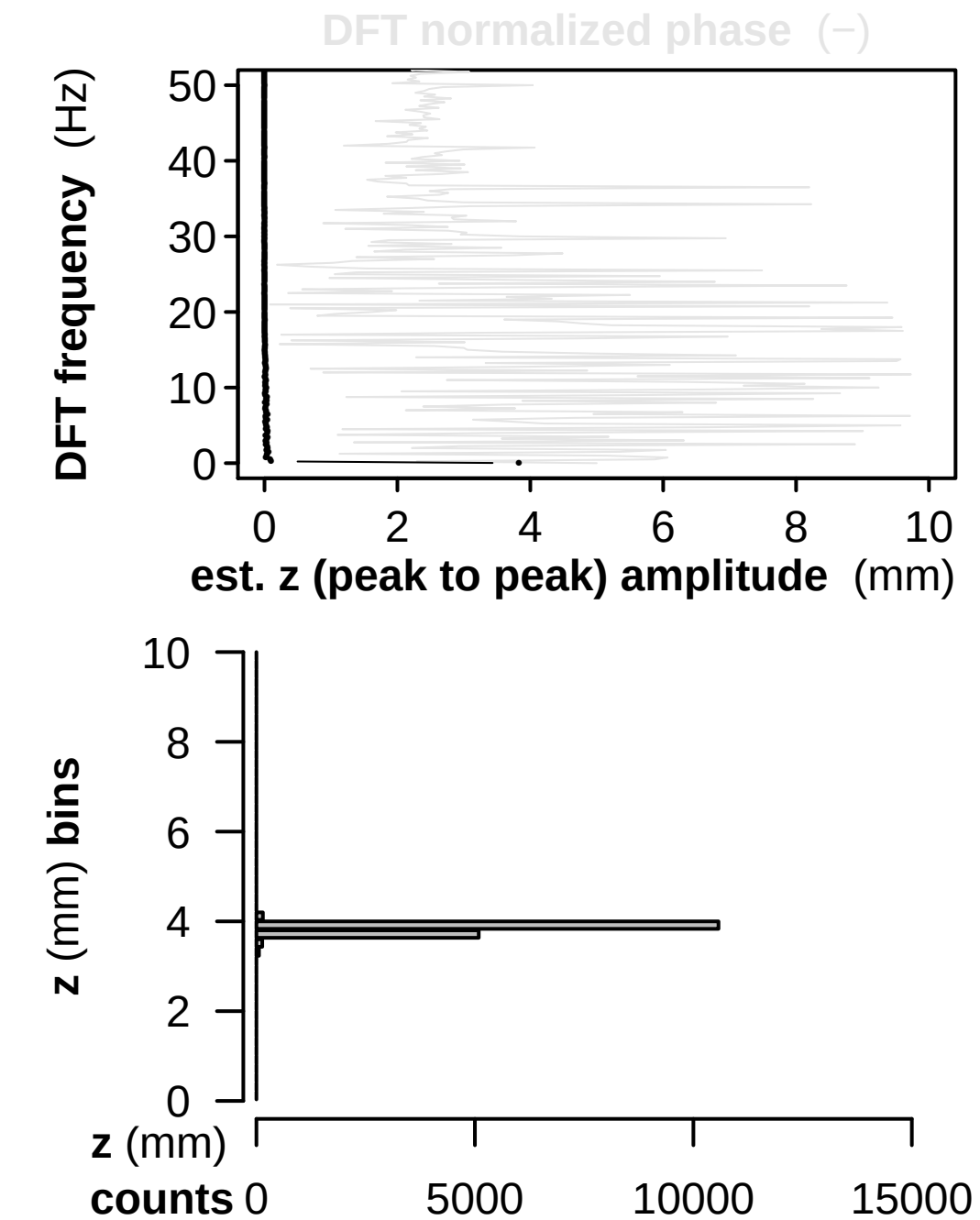

**SUBJECT 3 − RUN 06 − CONDITION 1,0**
**SC_180323_115843_0.AIFF**

```
z_min               : 3.37 mm
z_max               : 4.04 mm
z_travel_amplitude  : 0.67 mm

avg_abs_z_travel    : 3.86 mm/s

z_jarque-bera_jb    : 9308.09
z_jarque-bera_p     : 0.00e+00

z_lin_mod_est_slope: −0.02 mm/s
z_lin_mod_adj_R²    : 8 %

z_poly40_mod_adj_R²: 46 %

z_dft_ampl_thresh   : 0.010 mm
>=threshold_maxfreq: 15.75 Hz
```

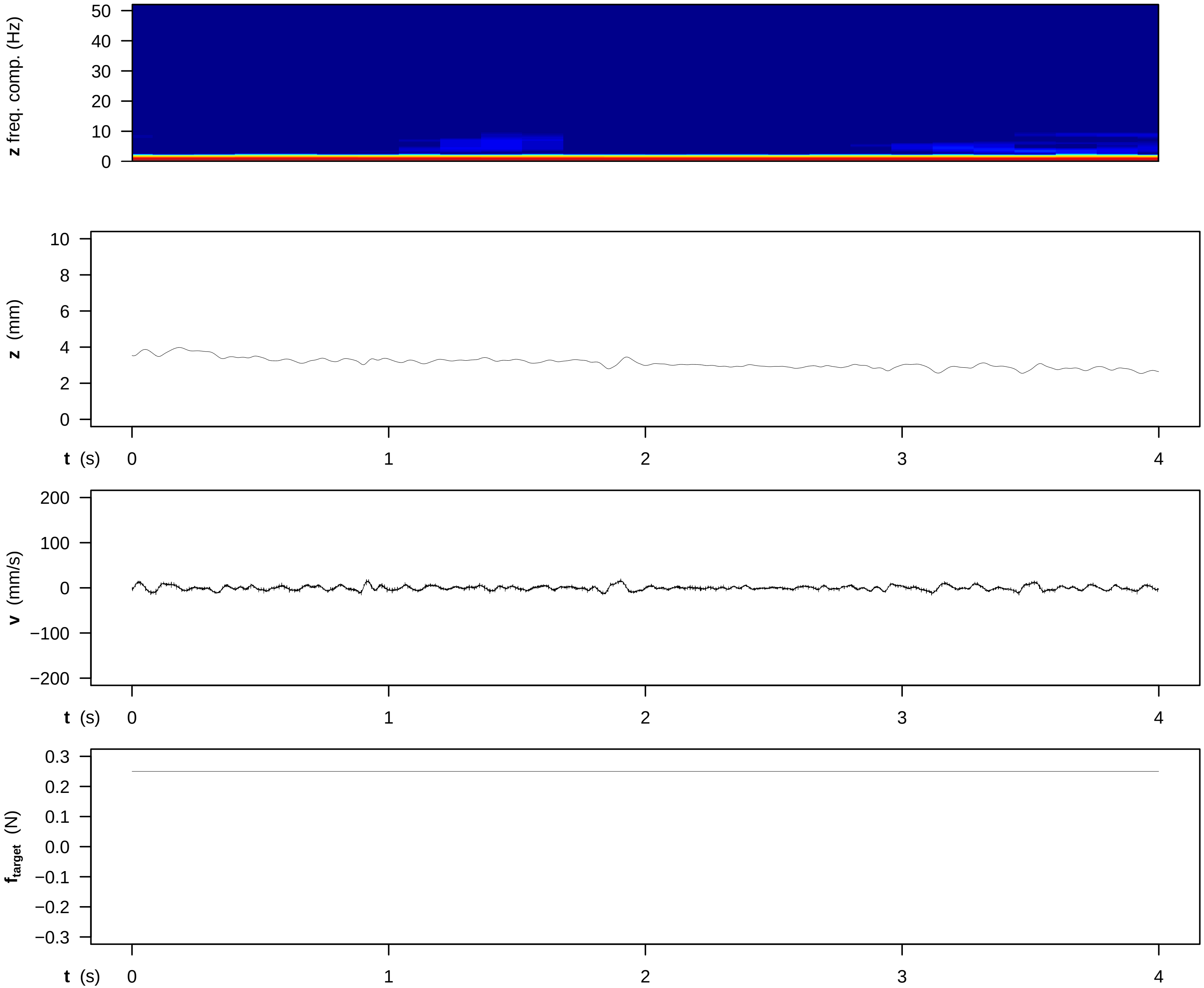

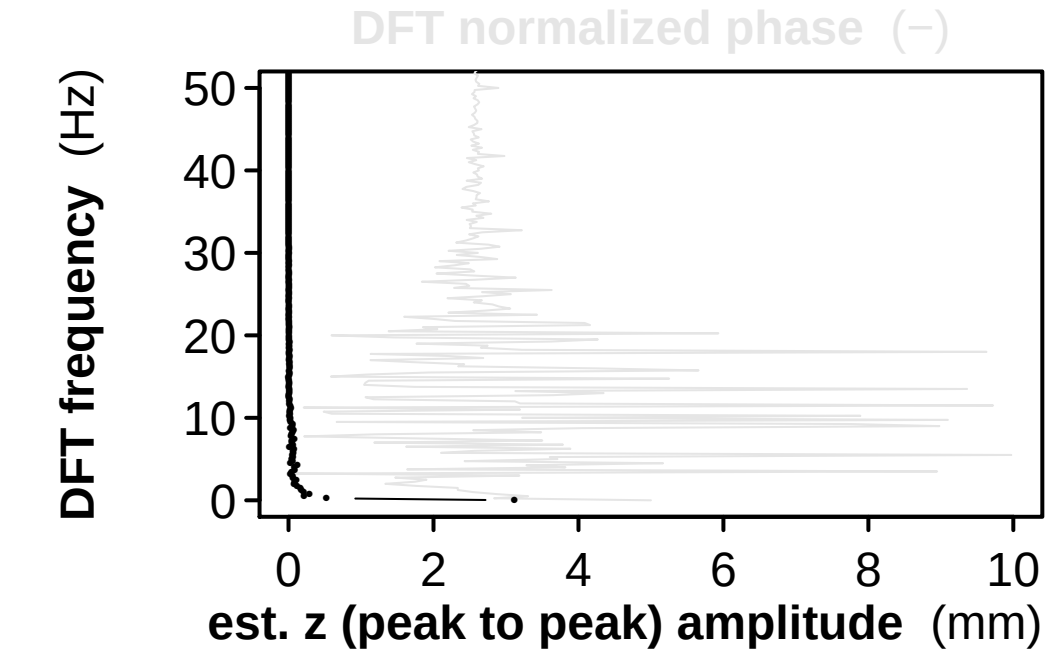

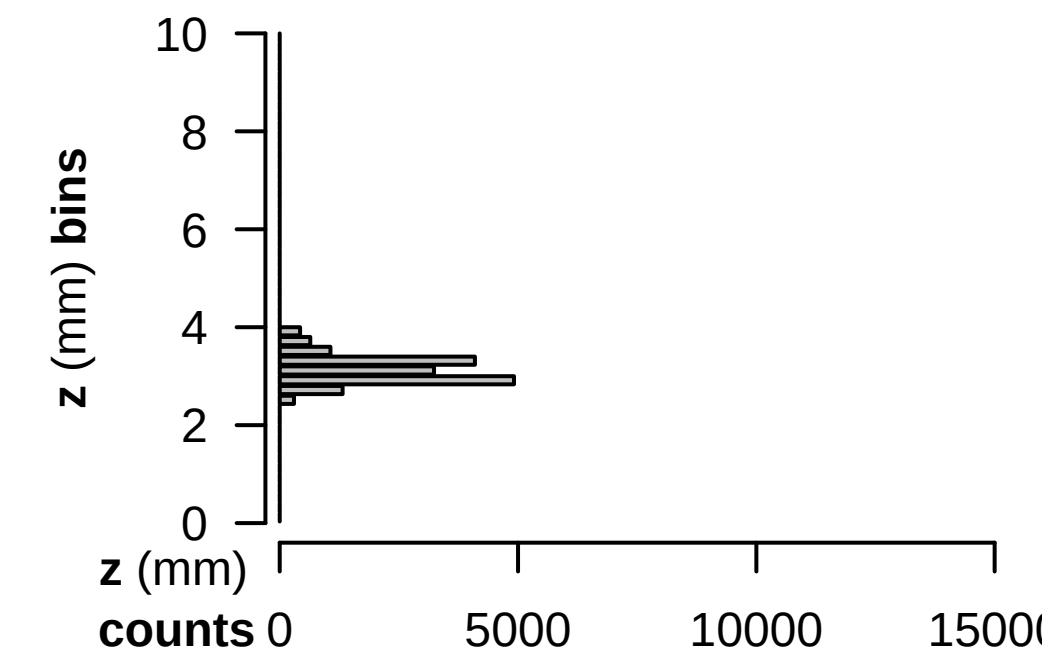

**SUBJECT 3 − RUN 13 − CONDITION 1,0**
**SC_180323_120223_0.AIFF**

```
z_min              : 2.54 mm
z_max              : 3.98 mm
z_travel_amplitude : 1.44 mm

avg_abs_z_travel   : 4.36 mm/s

z_jarque-bera_jb   : 1186.34
z_jarque-bera_p    : 0.00e+00

z_lin_mod_est_slope: −0.22 mm/s
z_lin_mod_adj_R²   : 77 %

z_poly40_mod_adj_R²: 88 %

z_dft_ampl_thresh  : 0.010 mm
>=threshold_maxfreq: 21.00 Hz
```

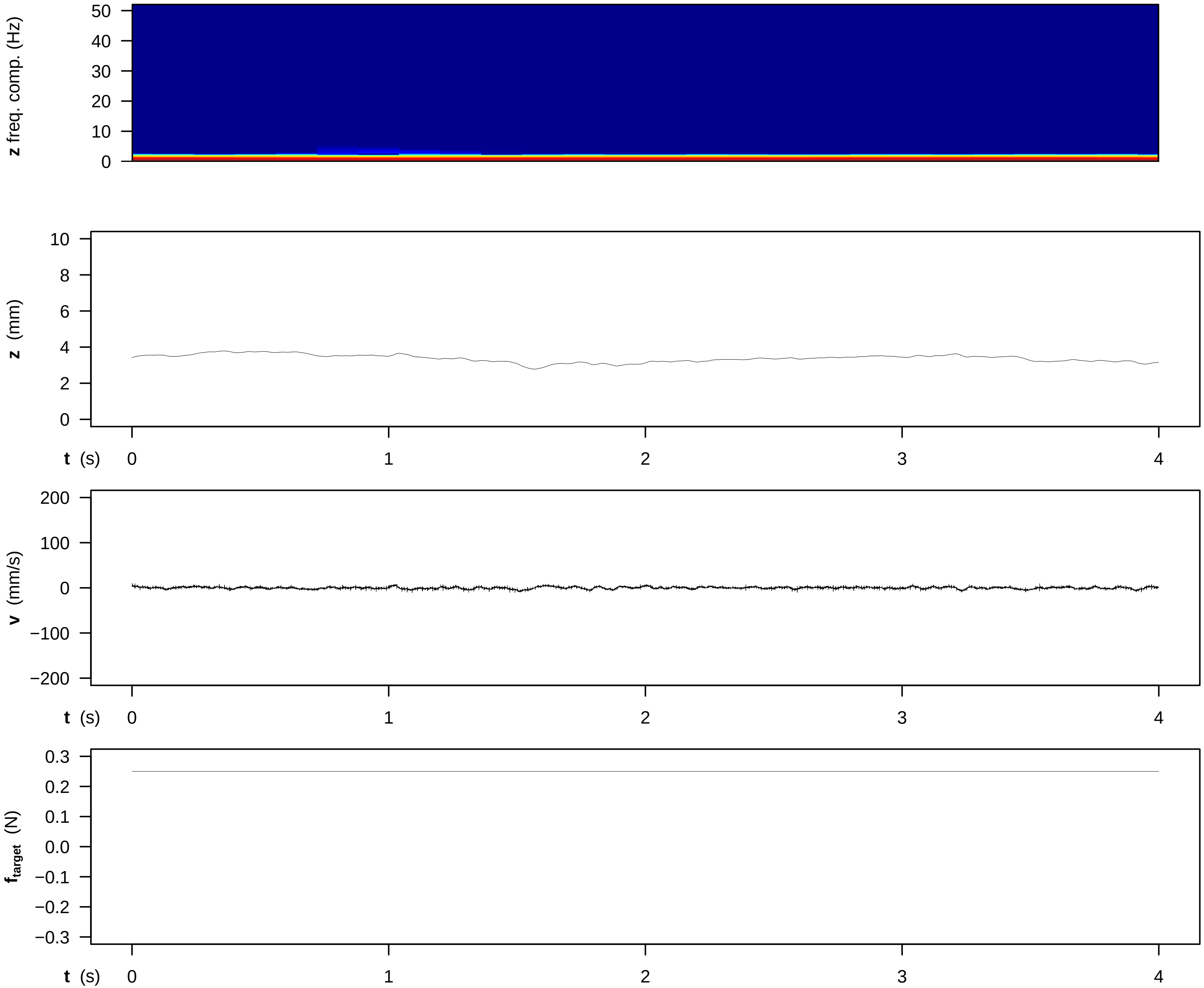

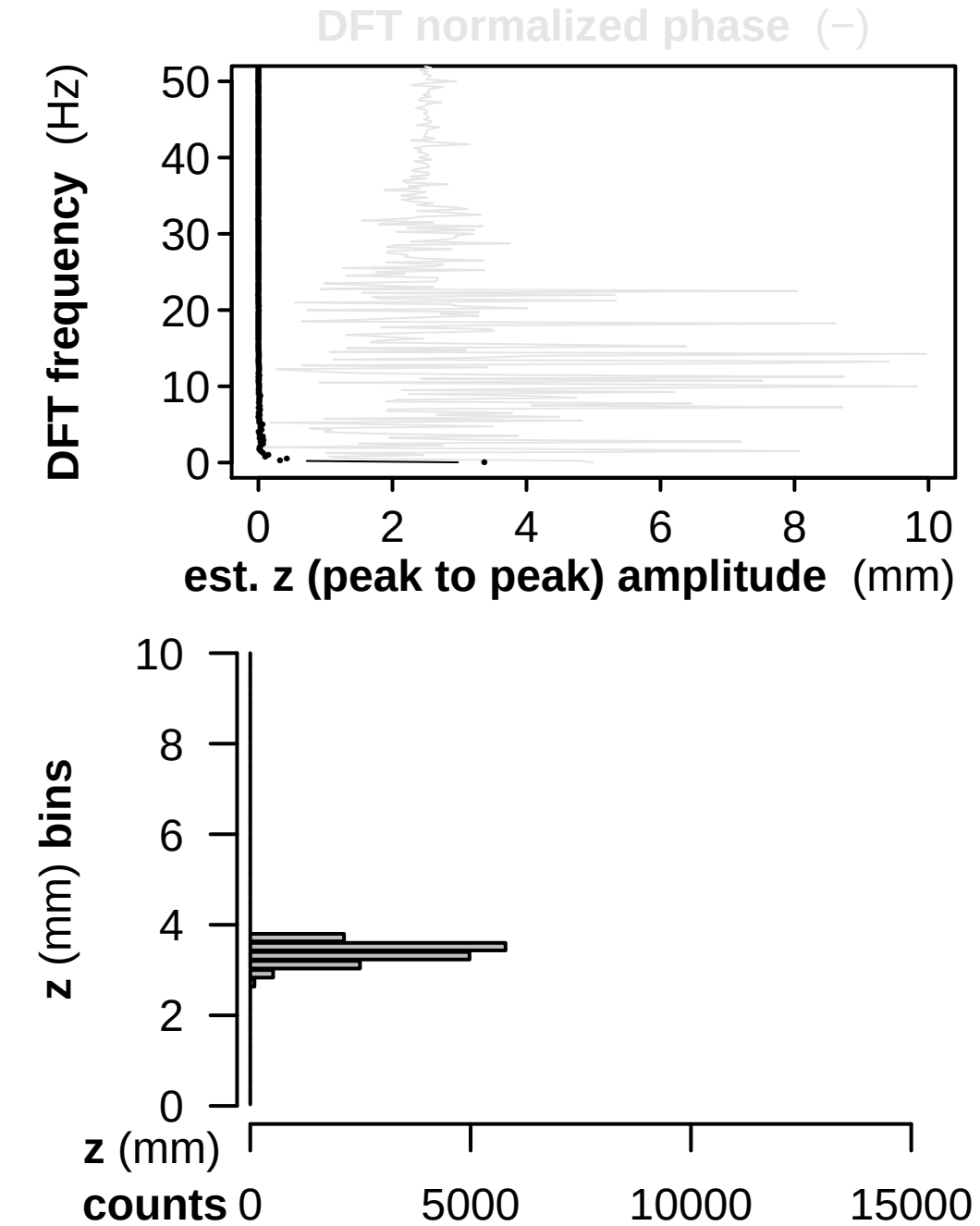

**SUBJECT 4 − RUN 06 − CONDITION 1,0**
**SC_180323_123336_0.AIFF**

```
z_min               : 2.78 mm
z_max               : 3.79 mm
z_travel_amplitude  : 1.01 mm

avg_abs_z_travel    : 3.27 mm/s

z_jarque-bera_jb    : 249.04
z_jarque-bera_p     : 0.00e+00

z_lin_mod_est_slope: −0.07 mm/s
z_lin_mod_adj_R²    : 15 %

z_poly40_mod_adj_R²: 93 %

z_dft_ampl_thresh   : 0.010 mm
>=threshold_maxfreq: 12.50 Hz
```

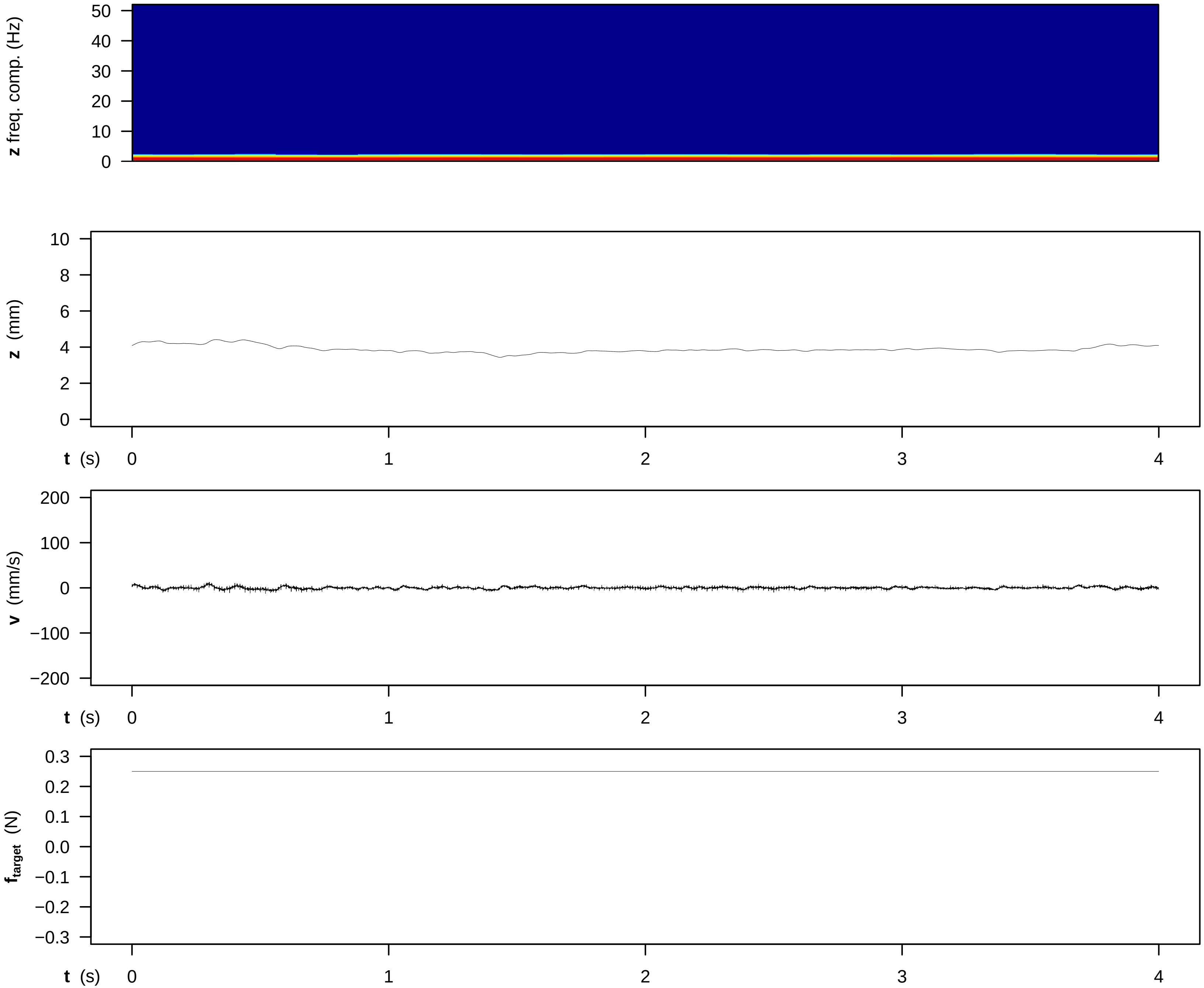

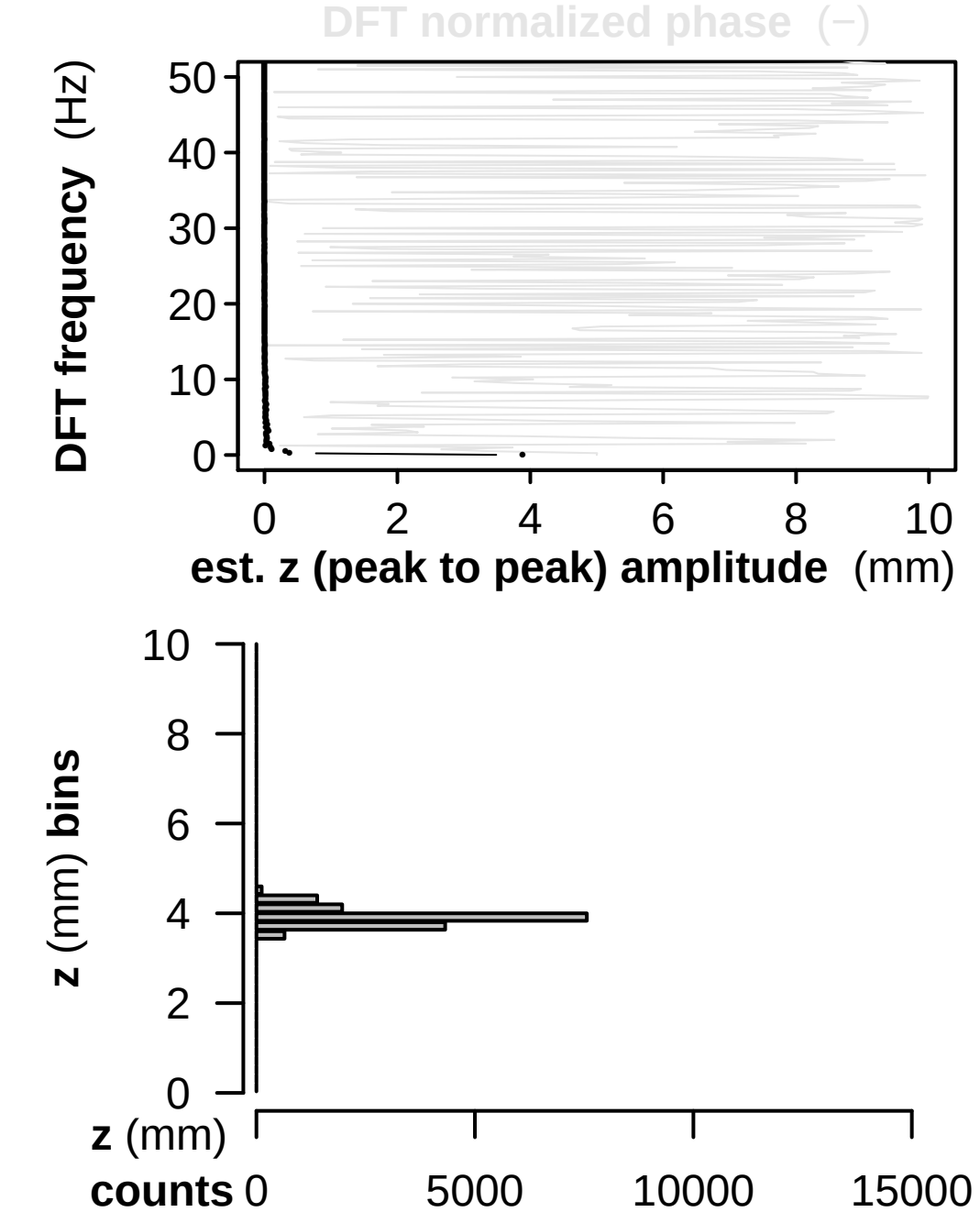

**SUBJECT 4 − RUN 18 − CONDITION 1,0**
**SC_180323_123924_0.AIFF**

```
z_min               : 3.43 mm
z_max               : 4.42 mm
z_travel_amplitude : 0.99 mm

avg_abs_z_travel    : 2.86 mm/s

z_jarque-bera_jb    : 2132.10
z_jarque-bera_p     : 0.00e+00

z_lin_mod_est_slope: −0.04 mm/s
z_lin_mod_adj_R²    : 7 %

z_poly40_mod_adj_R²: 94 %

z_dft_ampl_thresh   : 0.010 mm
>=threshold_maxfreq: 10.50 Hz
```

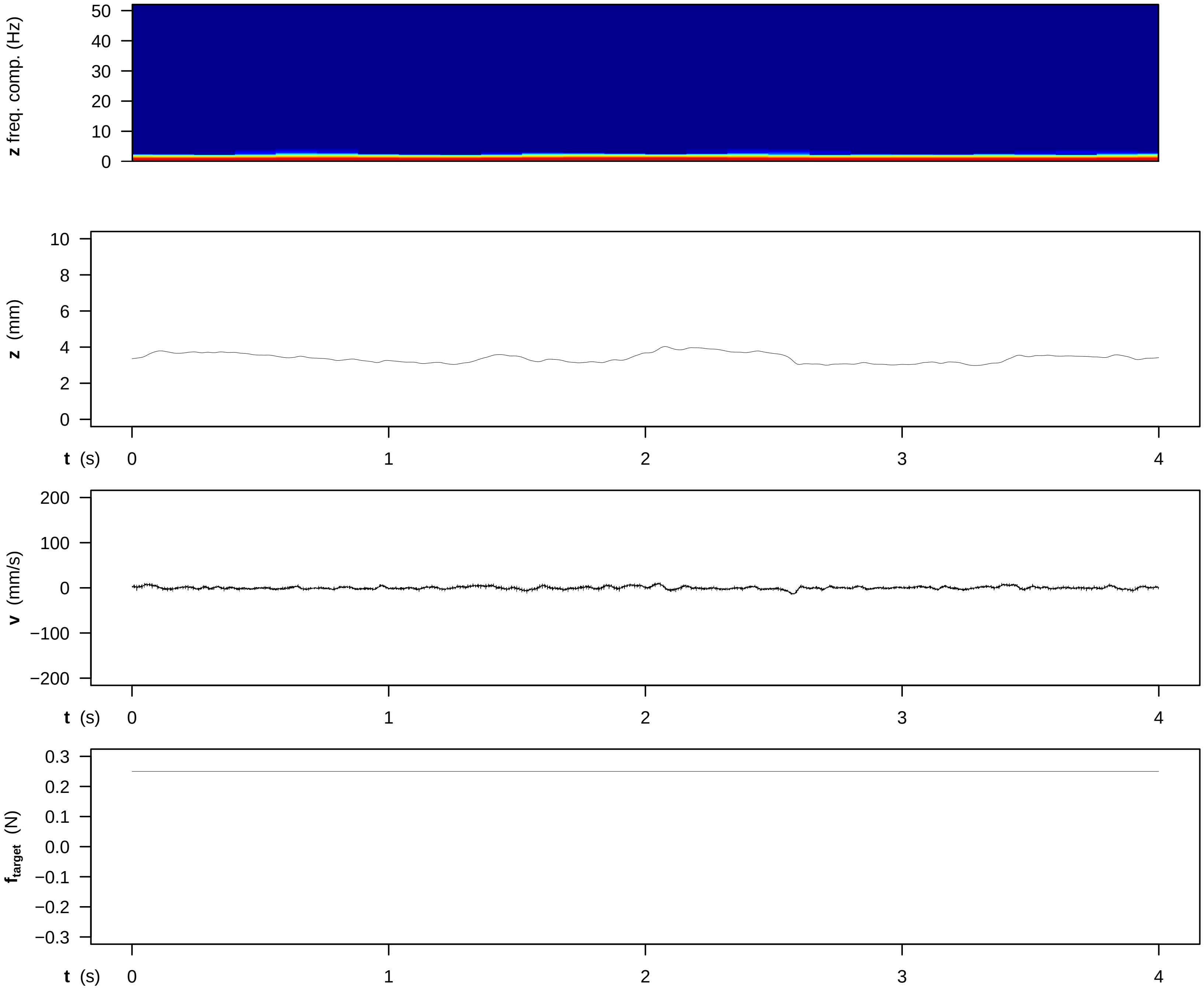

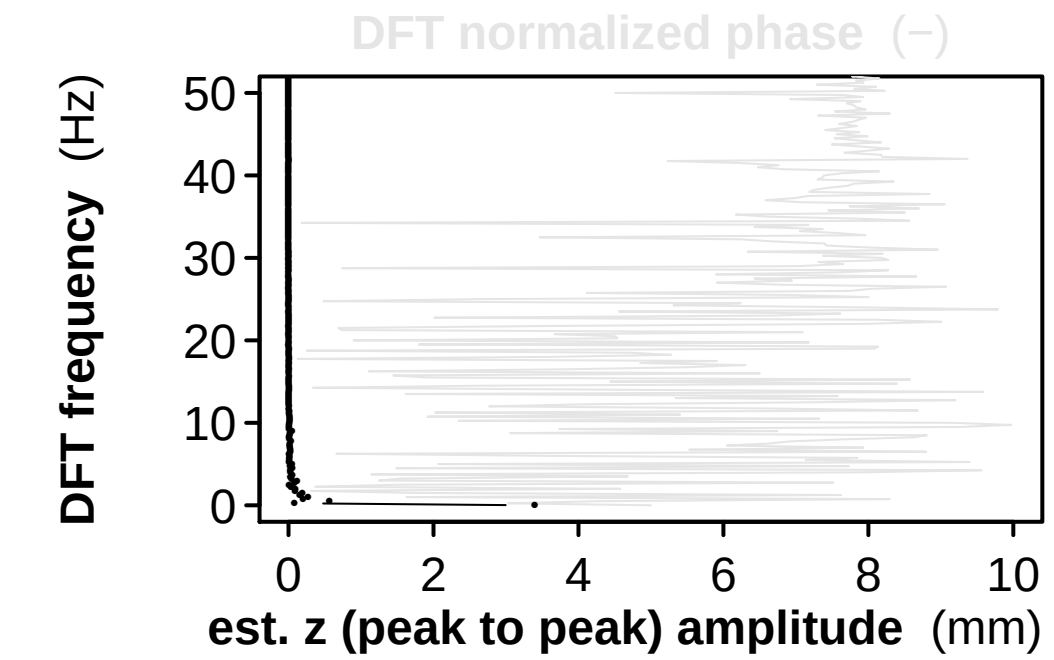

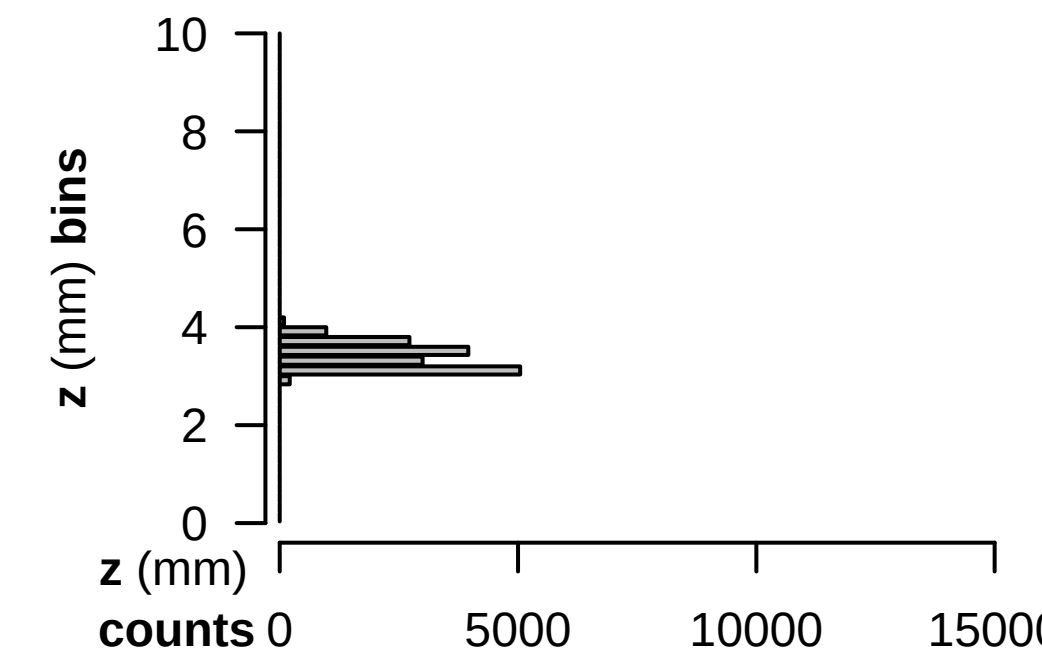

**SUBJECT 4 − RUN 19 − CONDITION 1,0**
**SC_180323_123951_0.AIFF**

```
z_min               : 2.98 mm
z_max               : 4.03 mm
z_travel_amplitude  : 1.05 mm

avg_abs_z_travel    : 3.13 mm/s

z_jarque-bera_jb    : 848.25
z_jarque-bera_p     : 0.00e+00

z_lin_mod_est_slope: −0.05 mm/s
z_lin_mod_adj_R²    : 5 %

z_poly40_mod_adj_R²: 92 %

z_dft_ampl_thresh   : 0.010 mm
>=threshold_maxfreq: 11.50 Hz
```

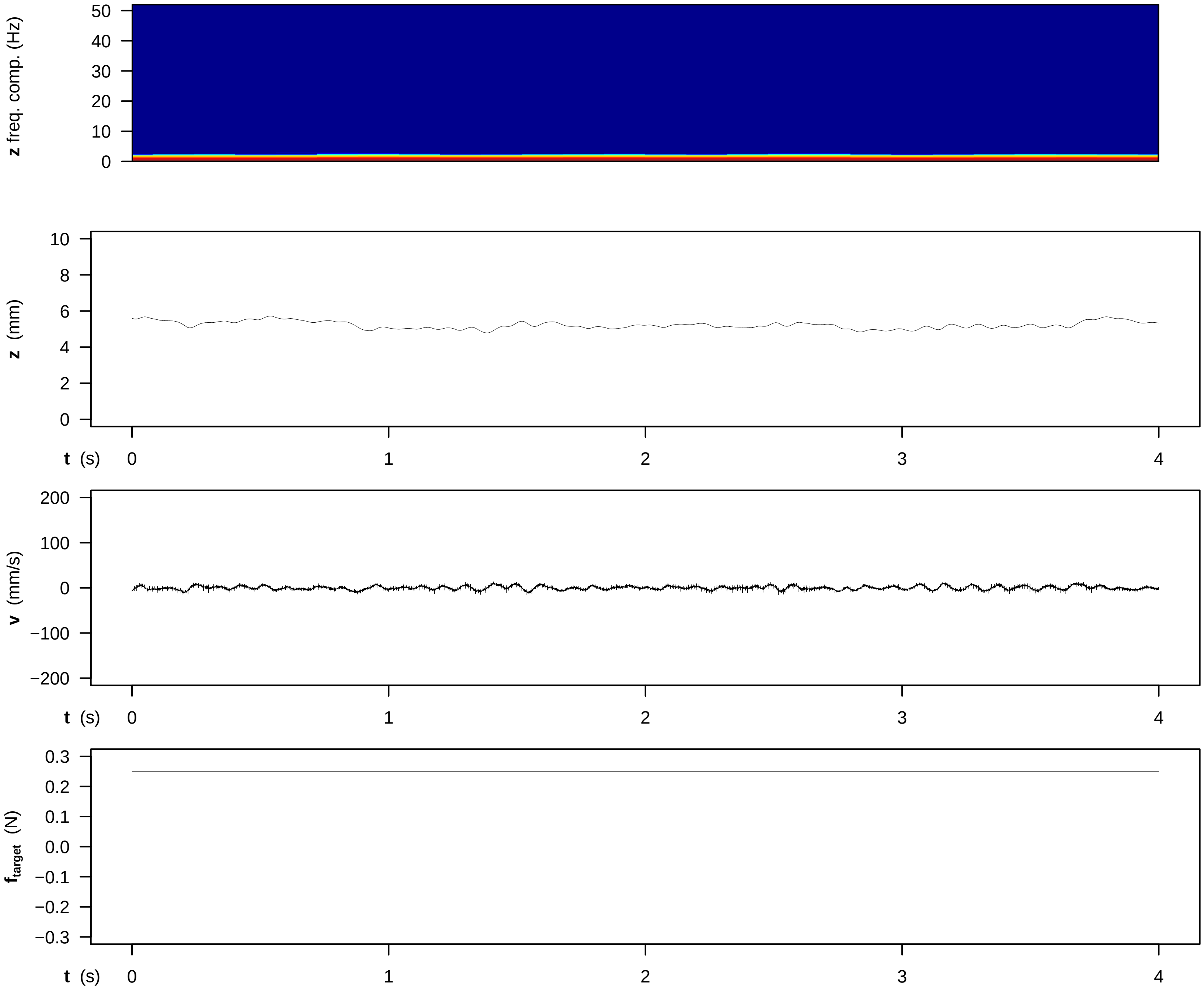

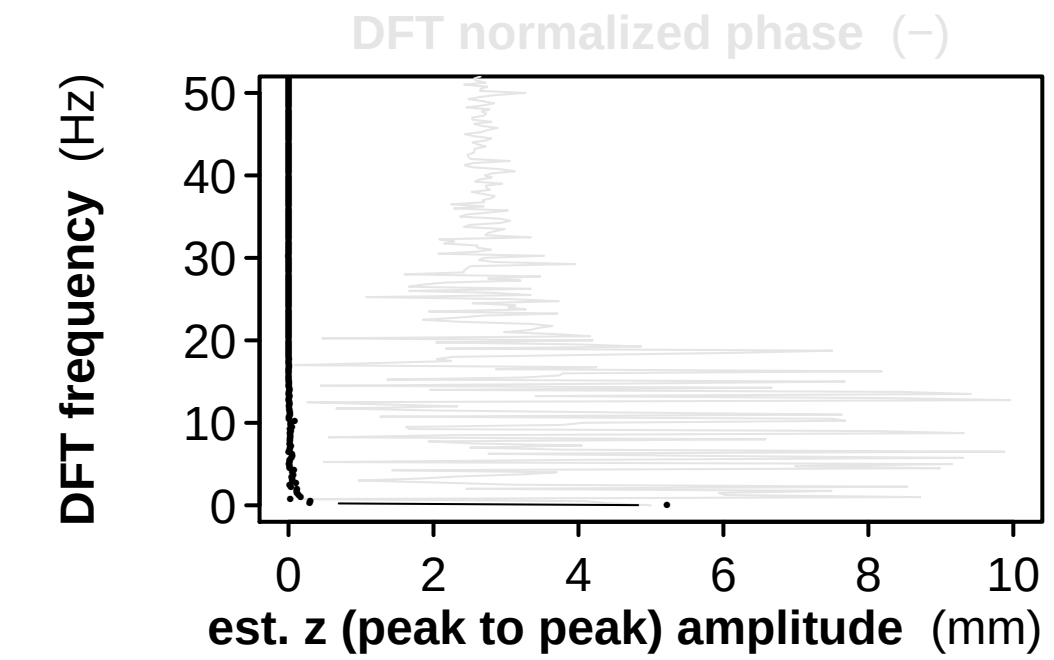

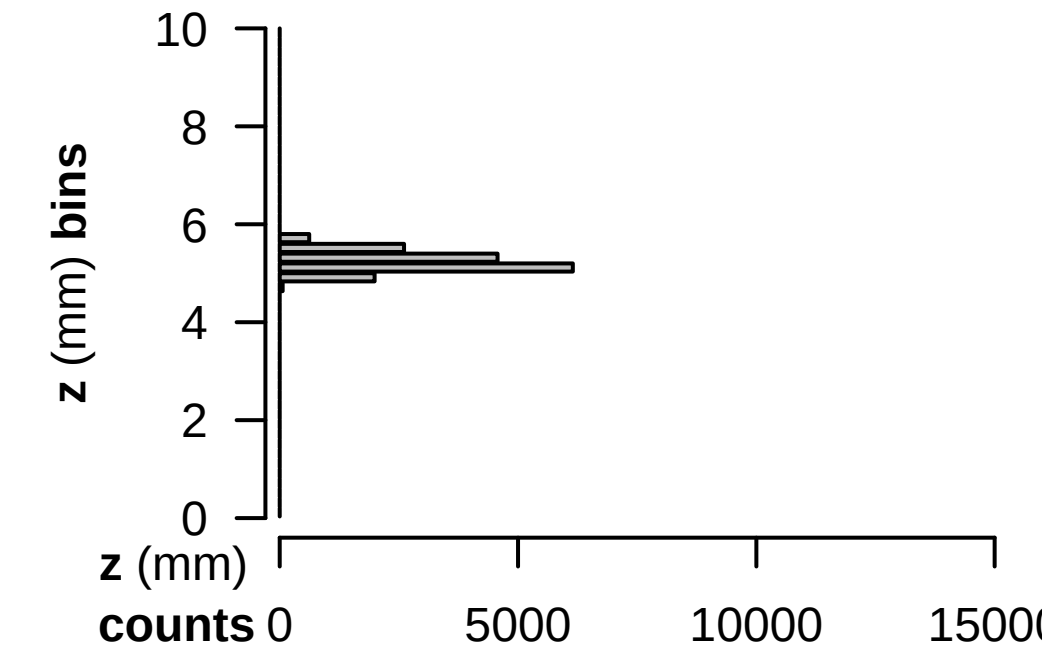

**SUBJECT 5 − RUN 14 − CONDITION 1,0**
**SC_180323_132323_0.AIFF**

```
z_min              : 4.79 mm
z_max              : 5.72 mm
z_travel_amplitude : 0.94 mm

avg_abs_z_travel   : 5.22 mm/s

z_jarque-bera_jb   : 617.10
z_jarque-bera_p    : 0.00e+00

z_lin_mod_est_slope: −0.04 mm/s
z_lin_mod_adj_R²   : 5 %

z_poly40_mod_adj_R²: 82 %

z_dft_ampl_thresh  : 0.010 mm
>=threshold_maxfreq: 14.25 Hz
```

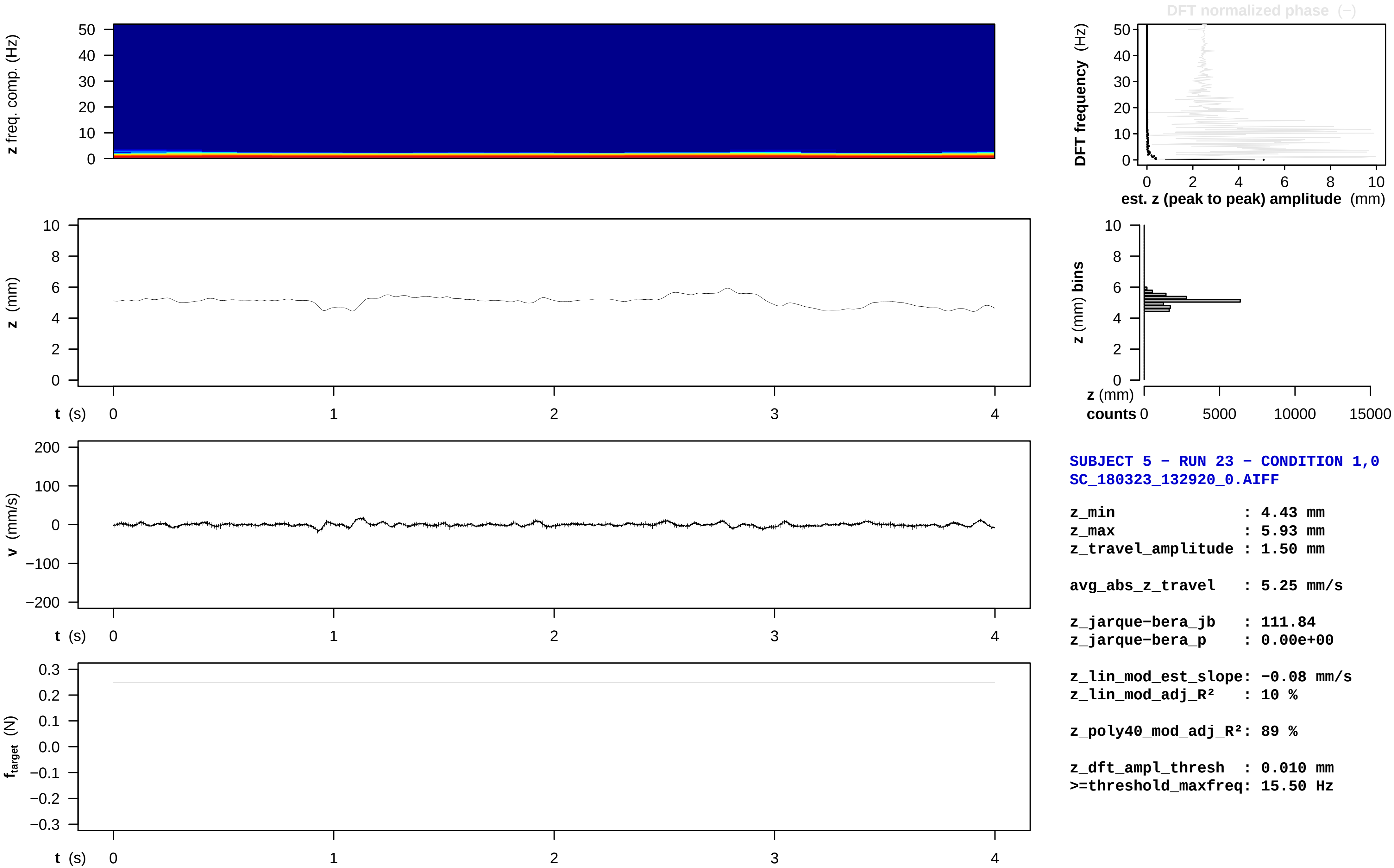

**SUBJECT 5 − RUN 23 − CONDITION 1,0**
**SC_180323_132920_0.AIFF**

```
z_min               : 4.43 mm
z_max               : 5.93 mm
z_travel_amplitude  : 1.50 mm

avg_abs_z_travel    : 5.25 mm/s

z_jarque-bera_jb    : 111.84
z_jarque-bera_p     : 0.00e+00

z_lin_mod_est_slope: −0.08 mm/s
z_lin_mod_adj_R²    : 10 %

z_poly40_mod_adj_R²: 89 %

z_dft_ampl_thresh   : 0.010 mm
>=threshold_maxfreq: 15.50 Hz
```

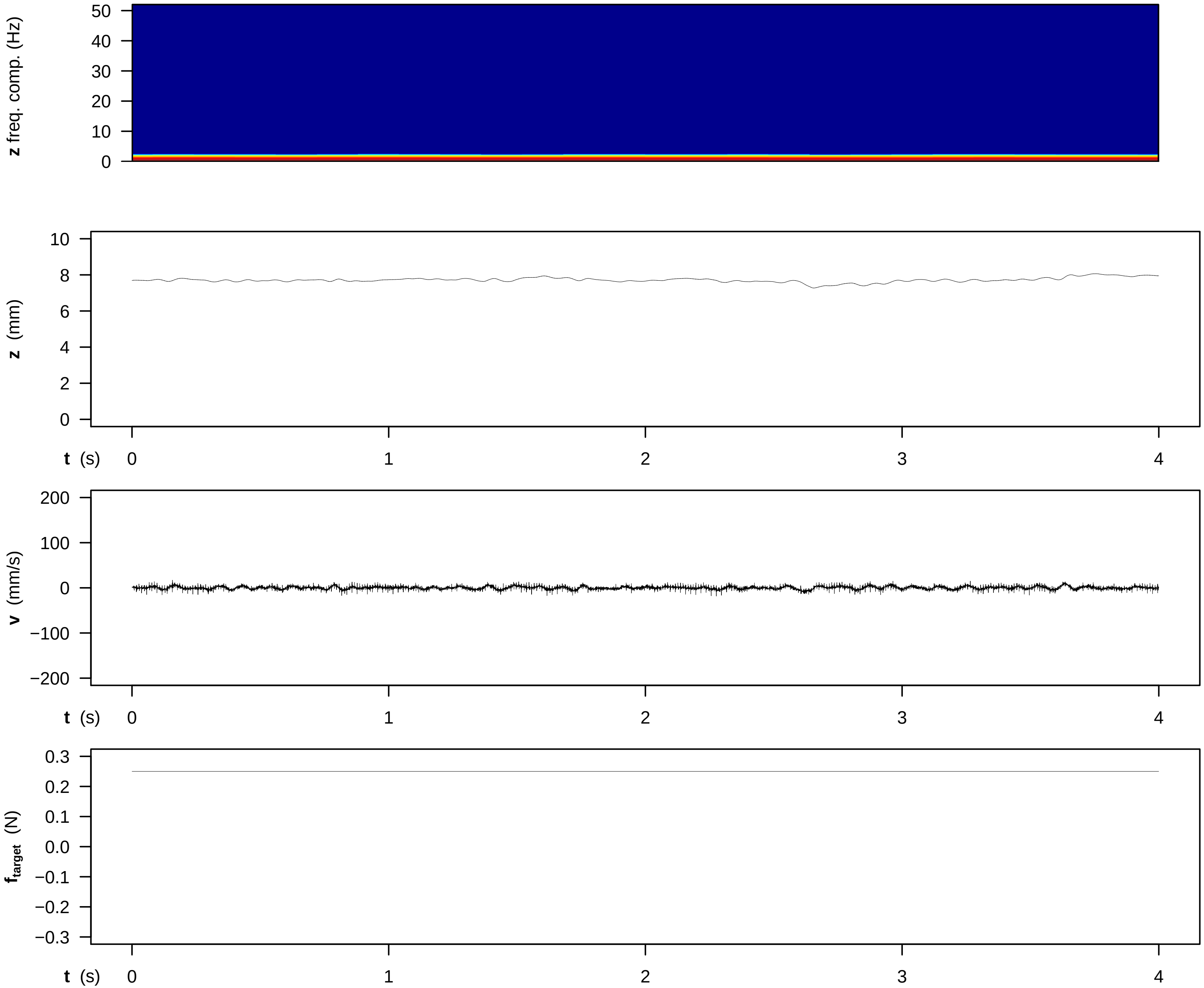

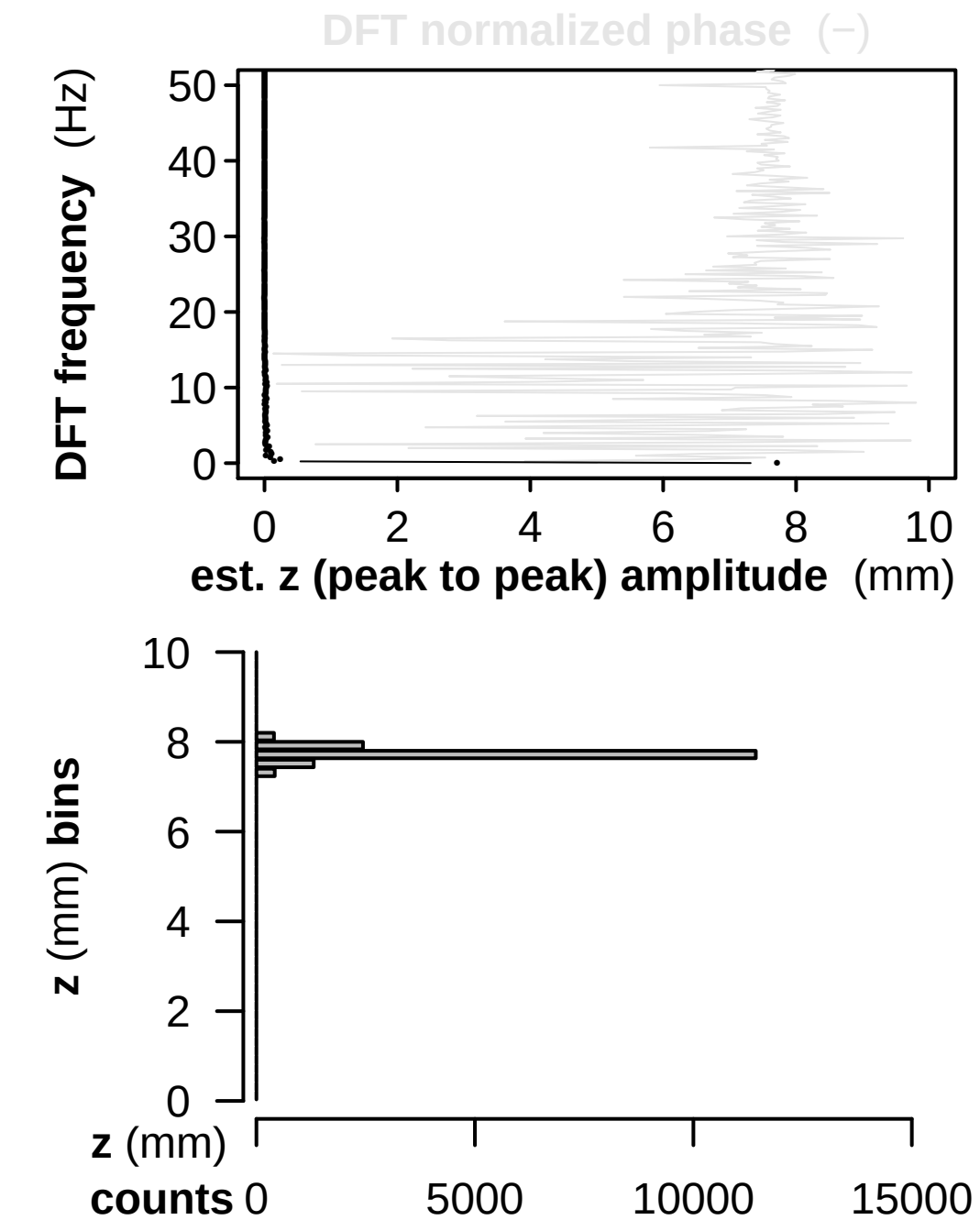

**SUBJECT 5 − RUN 34 − CONDITION 1,0**
**SC_180323_133801_0.AIFF**

```
z_min               : 7.27 mm
z_max               : 8.07 mm
z_travel_amplitude  : 0.80 mm

avg_abs_z_travel    : 4.50 mm/s

z_jarque-bera_jb    : 1210.35
z_jarque-bera_p     : 0.00e+00

z_lin_mod_est_slope: 0.02 mm/s
z_lin_mod_adj_R²    : 3 %

z_poly40_mod_adj_R²: 81 %

z_dft_ampl_thresh   : 0.010 mm
>=threshold_maxfreq: 15.50 Hz
```

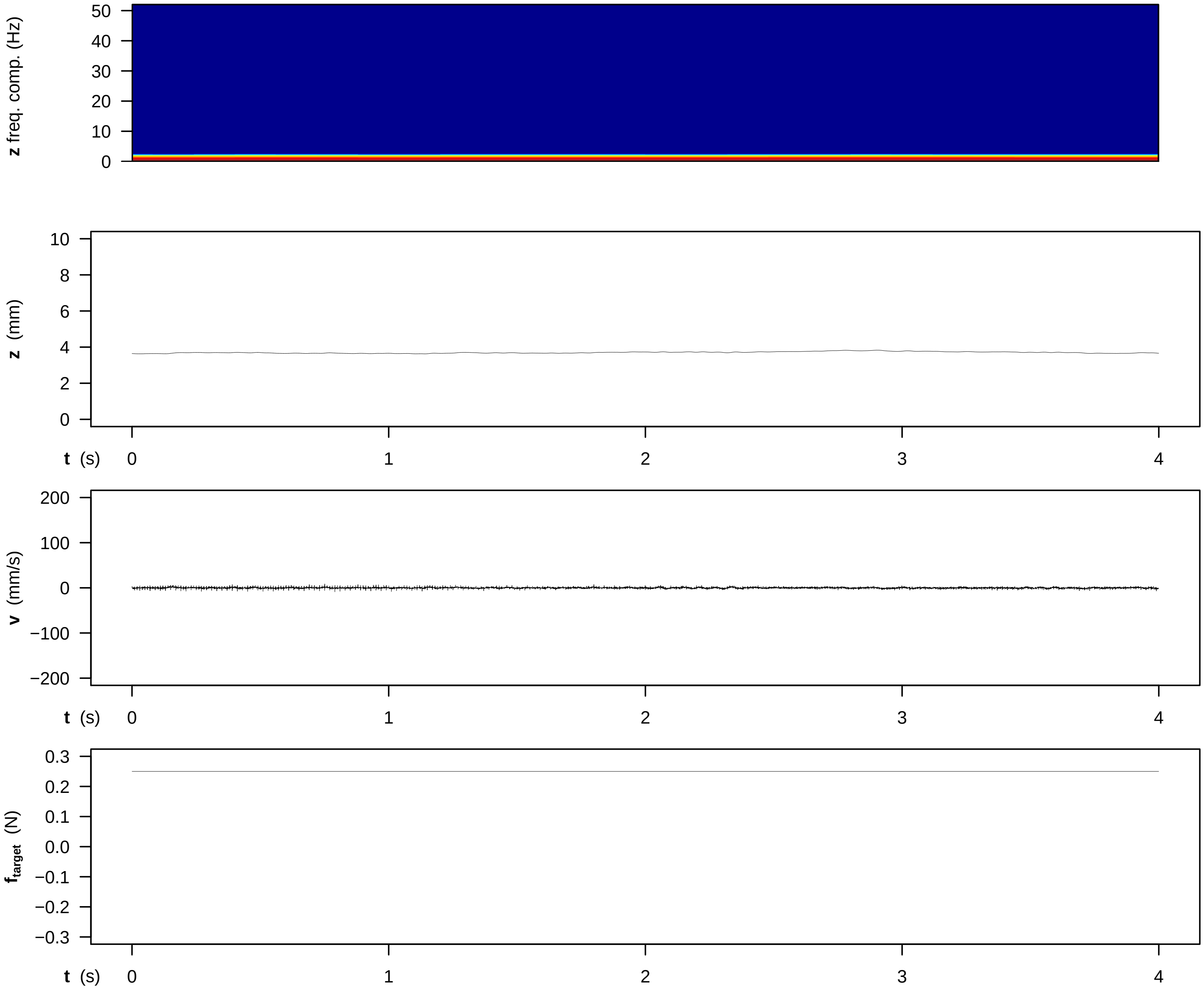

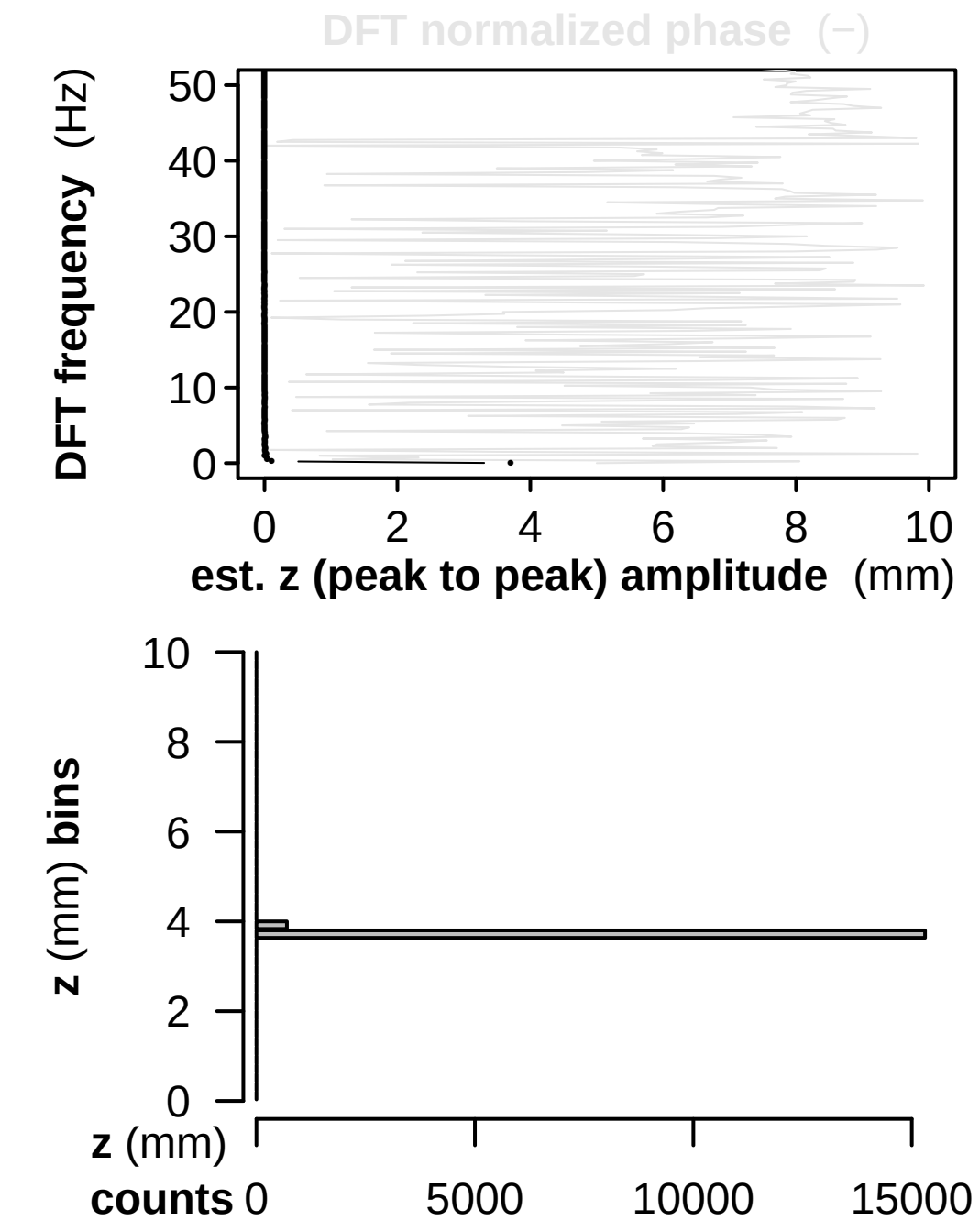

**SUBJECT 6 − RUN 14 − CONDITION 1,0**
**SC_180323_150010_0.AIFF**

```
z_min               : 3.62 mm
z_max               : 3.83 mm
z_travel_amplitude  : 0.21 mm

avg_abs_z_travel    : 1.90 mm/s

z_jarque-bera_jb    : 1034.13
z_jarque-bera_p     : 0.00e+00

z_lin_mod_est_slope: 0.02 mm/s
z_lin_mod_adj_R²    : 26 %

z_poly40_mod_adj_R²: 94 %

z_dft_ampl_thresh   : 0.010 mm
>=threshold_maxfreq: 3.75 Hz
```

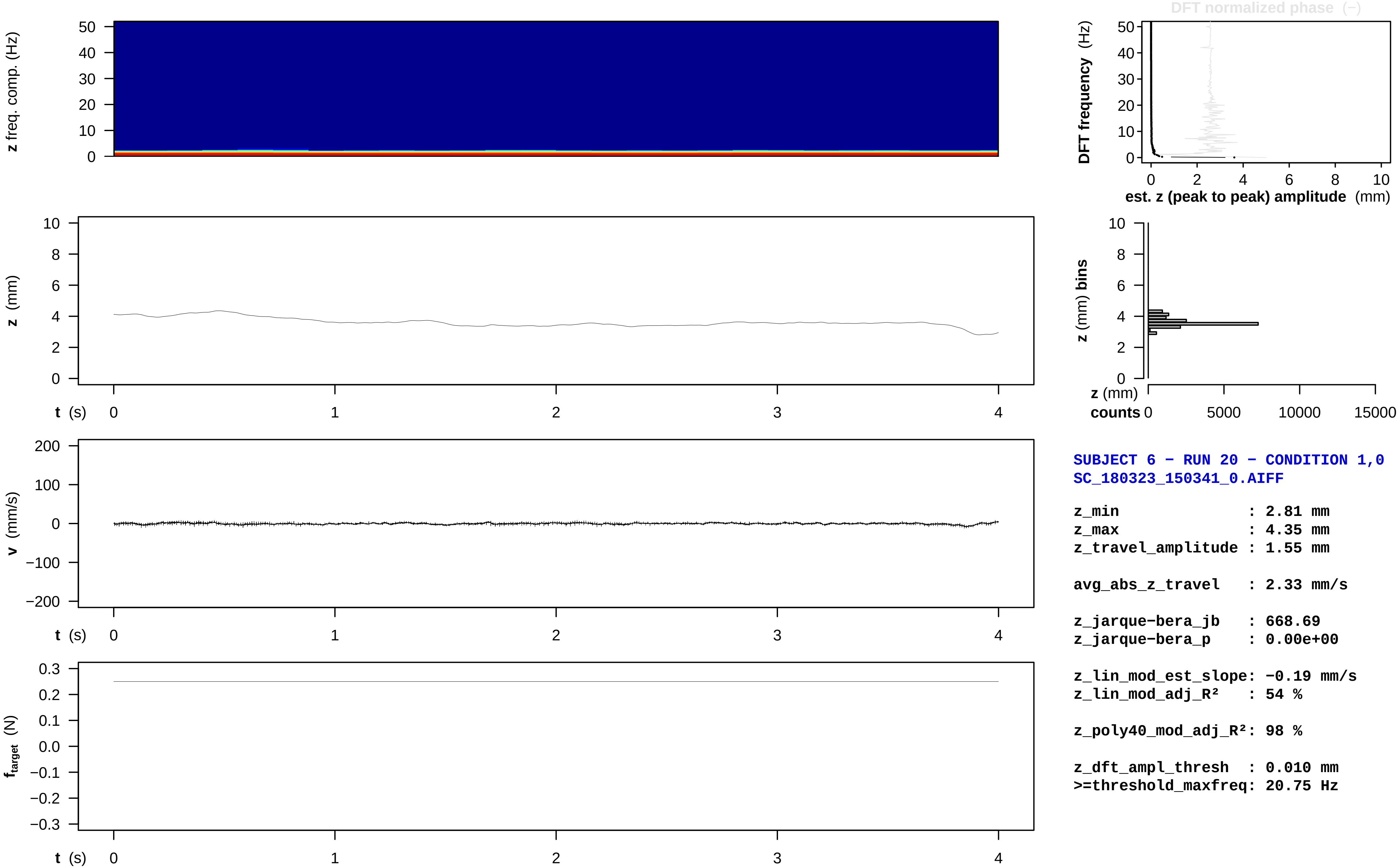

**SUBJECT 6 − RUN 20 − CONDITION 1,0**
**SC_180323_150341_0.AIFF**

```
z_min              : 2.81 mm
z_max              : 4.35 mm
z_travel_amplitude : 1.55 mm

avg_abs_z_travel   : 2.33 mm/s

z_jarque-bera_jb   : 668.69
z_jarque-bera_p    : 0.00e+00

z_lin_mod_est_slope: −0.19 mm/s
z_lin_mod_adj_R²   : 54 %

z_poly40_mod_adj_R²: 98 %

z_dft_ampl_thresh  : 0.010 mm
>=threshold_maxfreq: 20.75 Hz
```

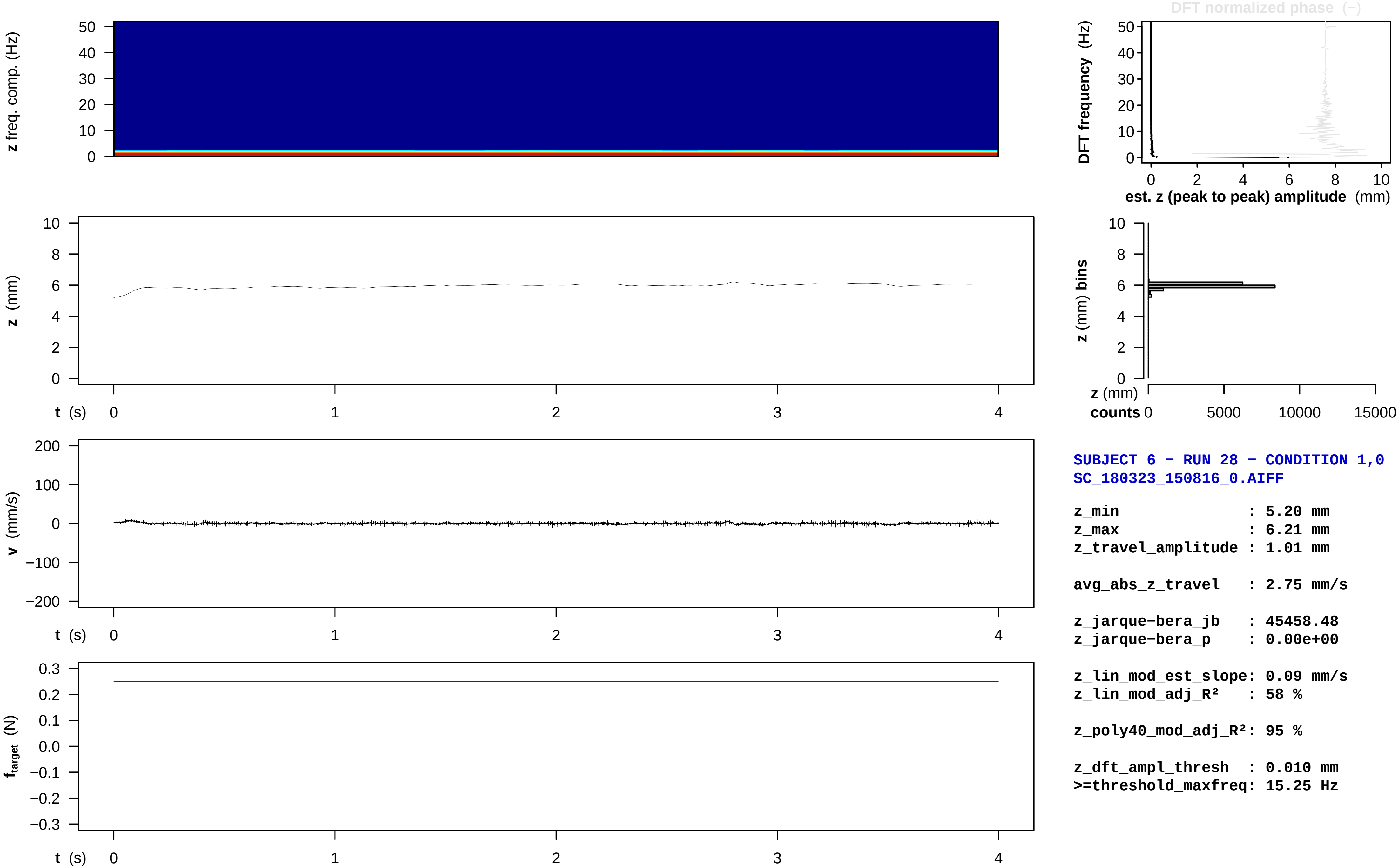

**SUBJECT 6 − RUN 28 − CONDITION 1,0**
**SC_180323_150816_0.AIFF**

```
z_min               : 5.20 mm
z_max               : 6.21 mm
z_travel_amplitude : 1.01 mm

avg_abs_z_travel    : 2.75 mm/s

z_jarque-bera_jb    : 45458.48
z_jarque-bera_p     : 0.00e+00

z_lin_mod_est_slope: 0.09 mm/s
z_lin_mod_adj_R²    : 58 %

z_poly40_mod_adj_R²: 95 %

z_dft_ampl_thresh   : 0.010 mm
>=threshold_maxfreq: 15.25 Hz
```

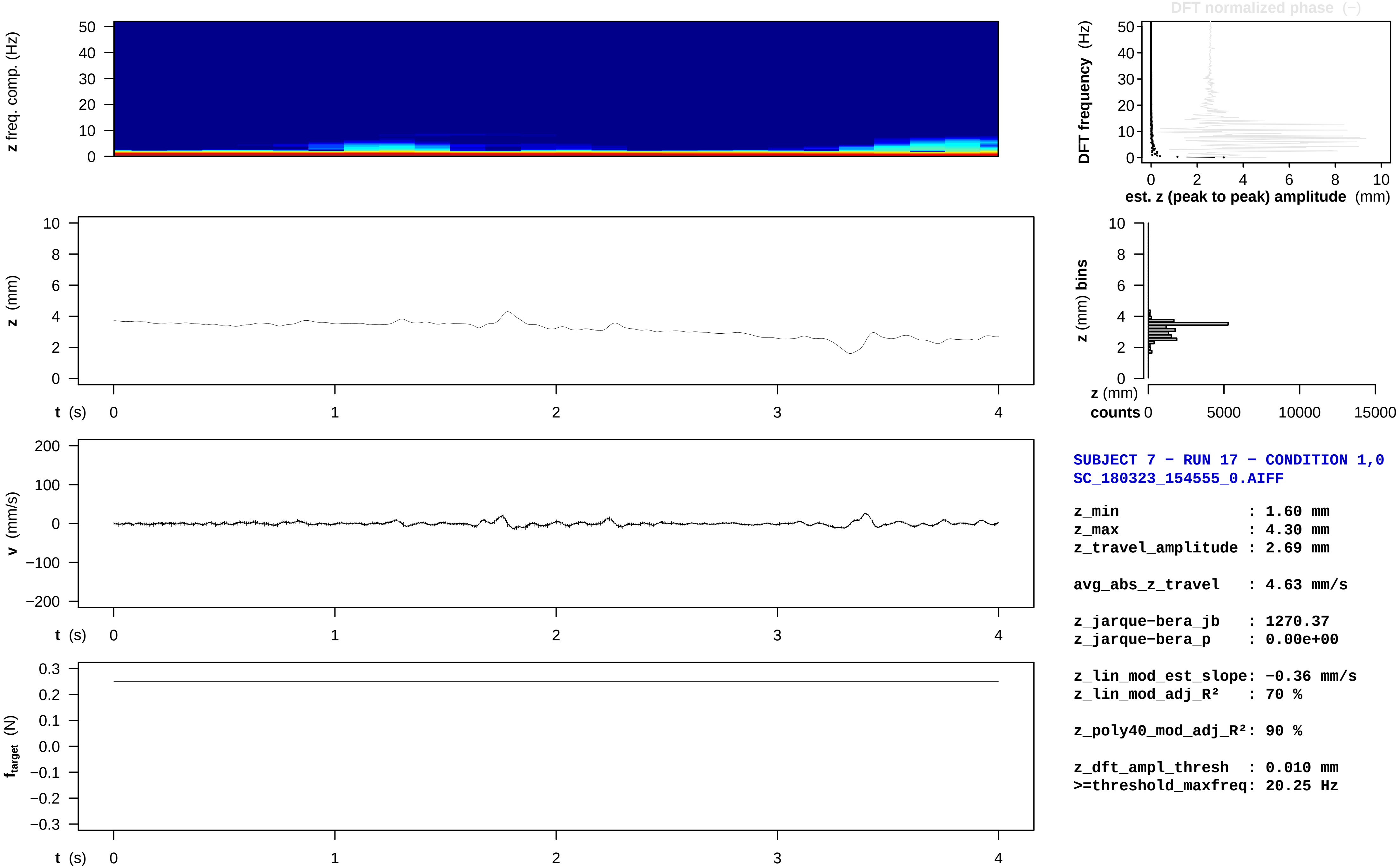

**SUBJECT 7 − RUN 17 − CONDITION 1,0**
**SC_180323_154555_0.AIFF**

```
z_min               : 1.60 mm
z_max               : 4.30 mm
z_travel_amplitude  : 2.69 mm

avg_abs_z_travel    : 4.63 mm/s

z_jarque-bera_jb    : 1270.37
z_jarque-bera_p     : 0.00e+00

z_lin_mod_est_slope: −0.36 mm/s
z_lin_mod_adj_R²    : 70 %

z_poly40_mod_adj_R²: 90 %

z_dft_ampl_thresh   : 0.010 mm
>=threshold_maxfreq: 20.25 Hz
```

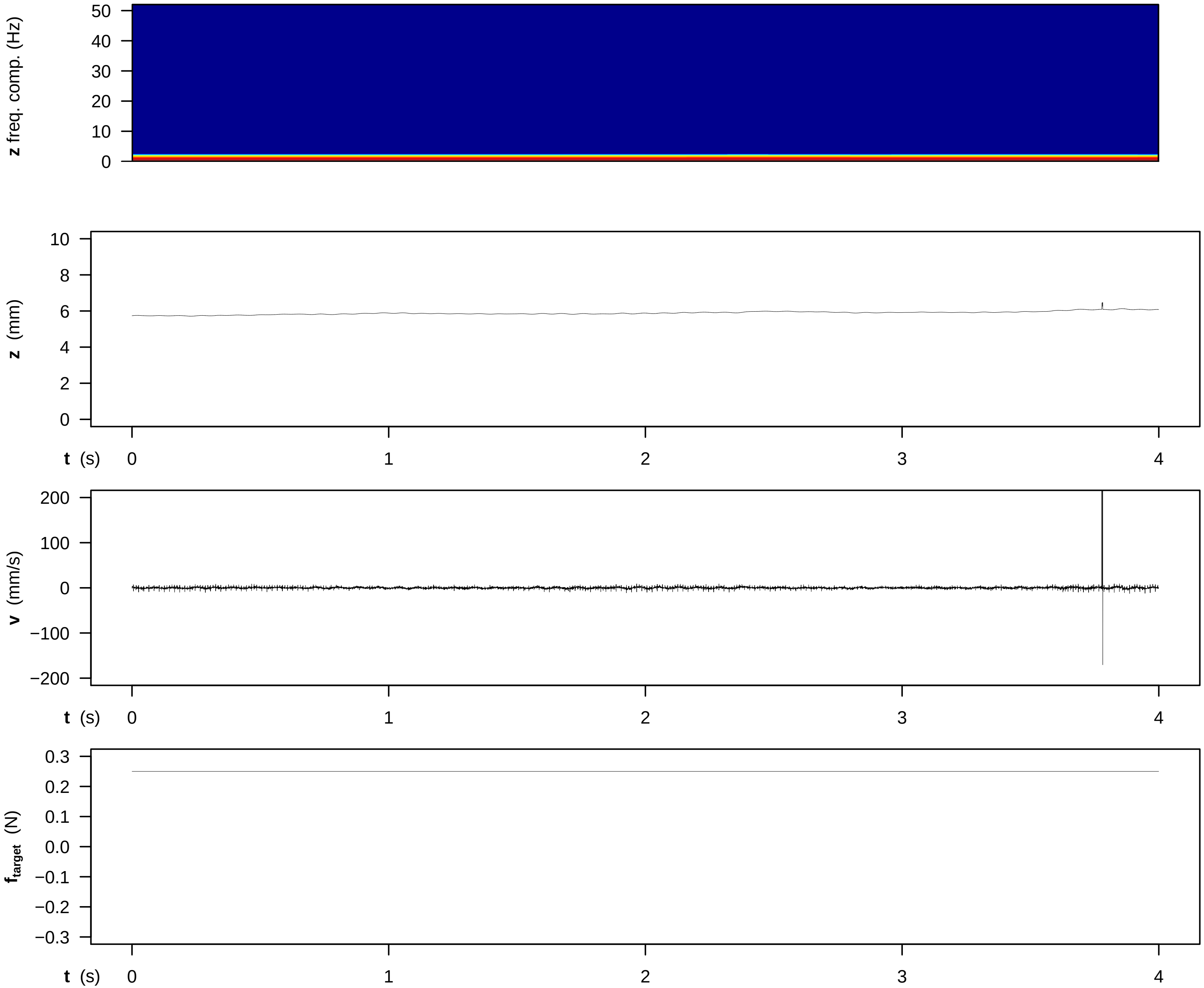

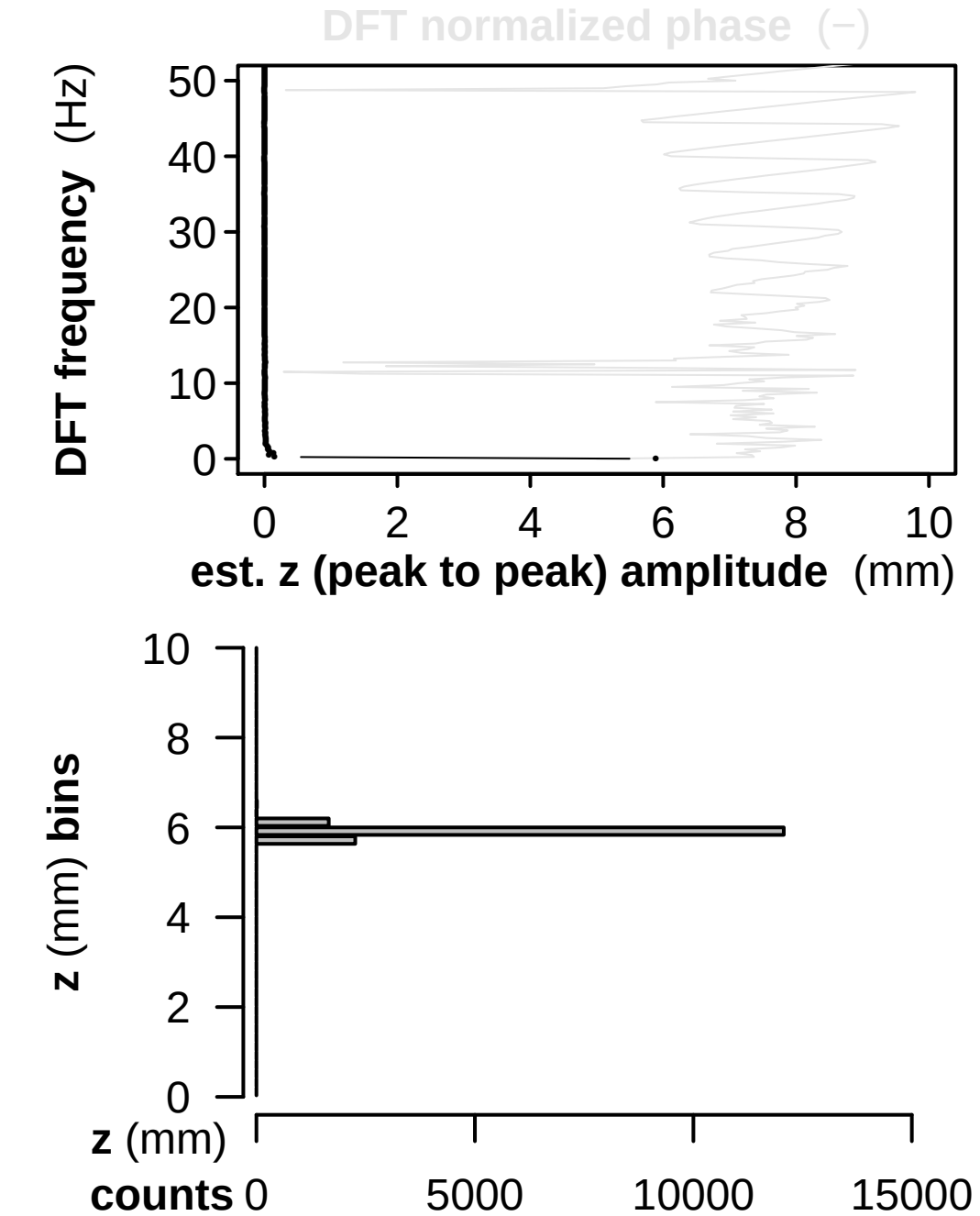

**SUBJECT 7 − RUN 24 − CONDITION 1,0**
**SC_180323_155136_0.AIFF**

```
z_min               : 5.71 mm
z_max               : 6.47 mm
z_travel_amplitude  : 0.76 mm

avg_abs_z_travel    : 2.86 mm/s

z_jarque-bera_jb    : 1401.20
z_jarque-bera_p     : 0.00e+00

z_lin_mod_est_slope: 0.07 mm/s
z_lin_mod_adj_R²    : 81 %

z_poly40_mod_adj_R²: 96 %

z_dft_ampl_thresh   : 0.010 mm
>=threshold_maxfreq: 12.75 Hz
```

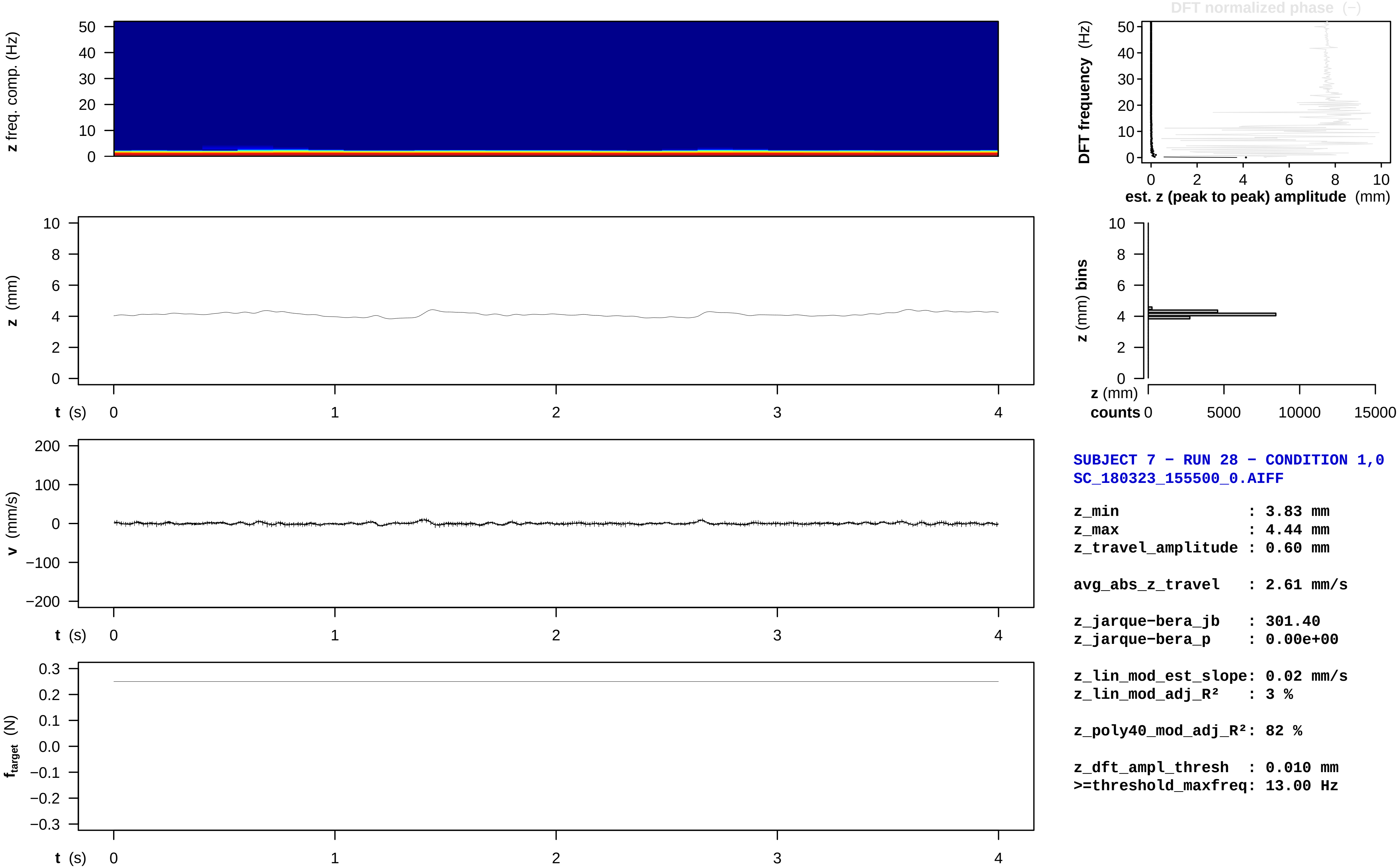

**SUBJECT 7 − RUN 28 − CONDITION 1,0**
**SC_180323_155500_0.AIFF**

```
z_min               : 3.83 mm
z_max               : 4.44 mm
z_travel_amplitude : 0.60 mm

avg_abs_z_travel   : 2.61 mm/s

z_jarque-bera_jb   : 301.40
z_jarque-bera_p    : 0.00e+00

z_lin_mod_est_slope: 0.02 mm/s
z_lin_mod_adj_R²   : 3 %

z_poly40_mod_adj_R²: 82 %

z_dft_ampl_thresh  : 0.010 mm
>=threshold_maxfreq: 13.00 Hz
```

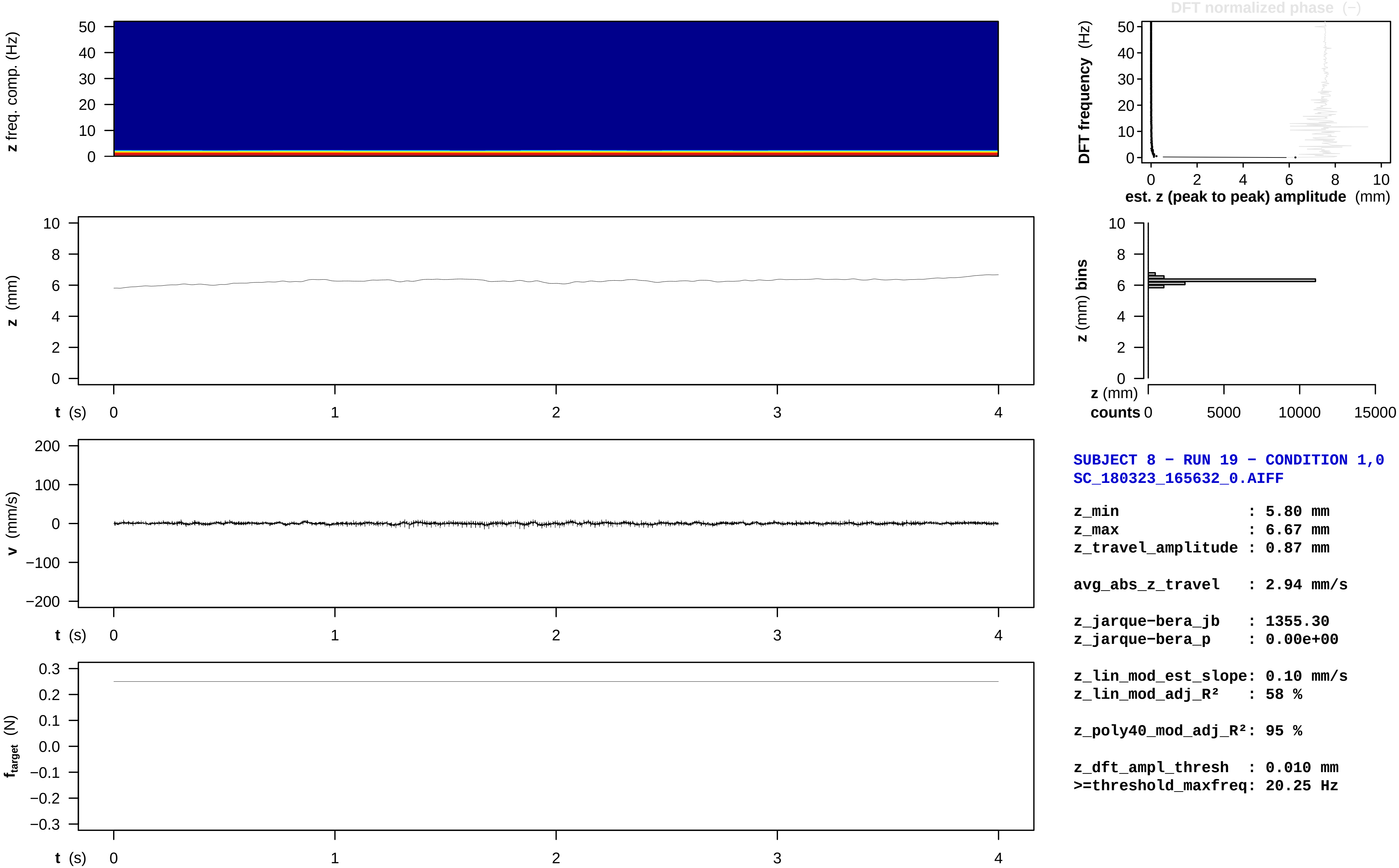

**SUBJECT 8 − RUN 19 − CONDITION 1,0**
**SC_180323_165632_0.AIFF**

```
z_min               : 5.80 mm
z_max               : 6.67 mm
z_travel_amplitude : 0.87 mm

avg_abs_z_travel   : 2.94 mm/s

z_jarque-bera_jb   : 1355.30
z_jarque-bera_p    : 0.00e+00

z_lin_mod_est_slope: 0.10 mm/s
z_lin_mod_adj_R²   : 58 %

z_poly40_mod_adj_R²: 95 %

z_dft_ampl_thresh  : 0.010 mm
>=threshold_maxfreq: 20.25 Hz
```

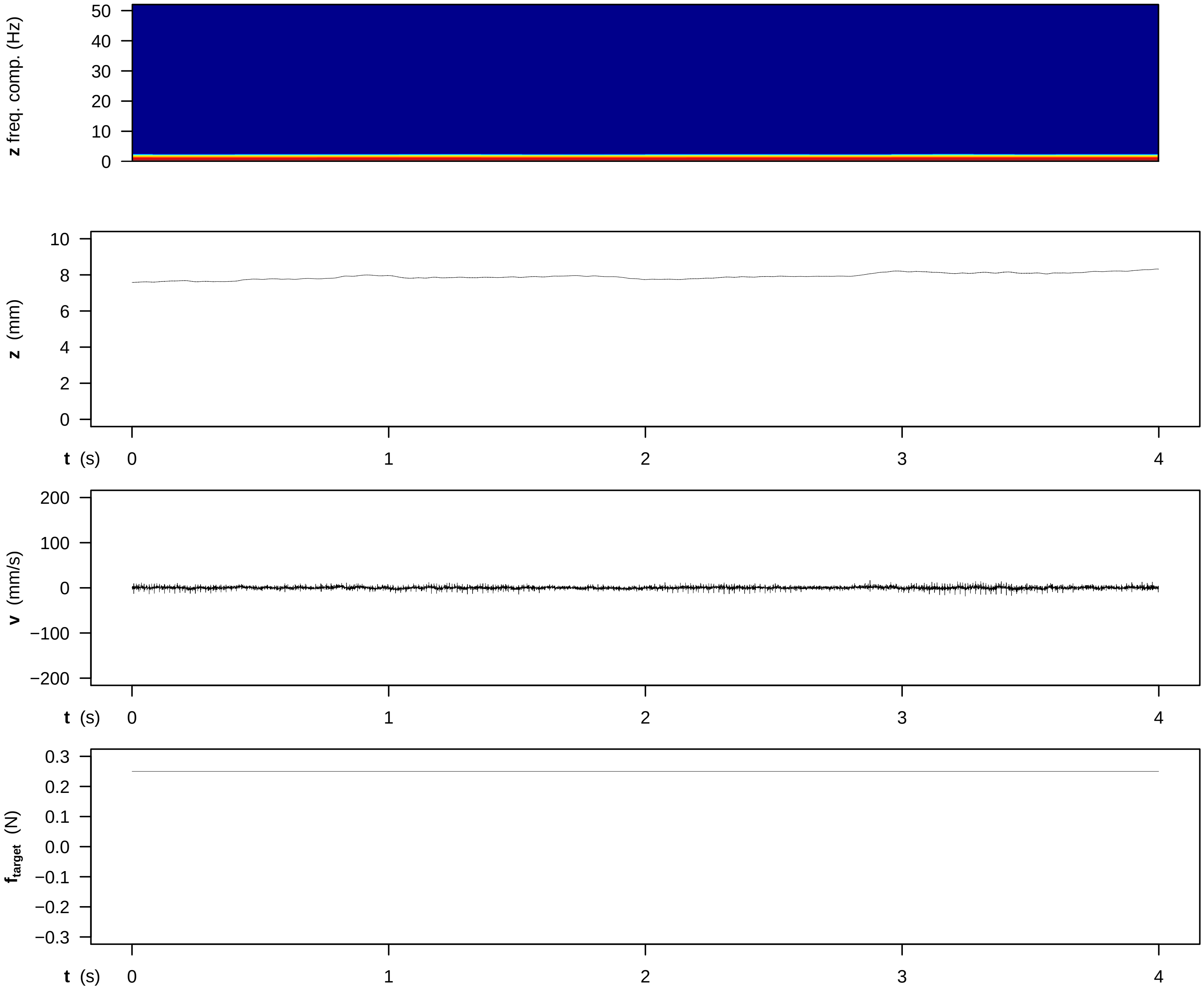

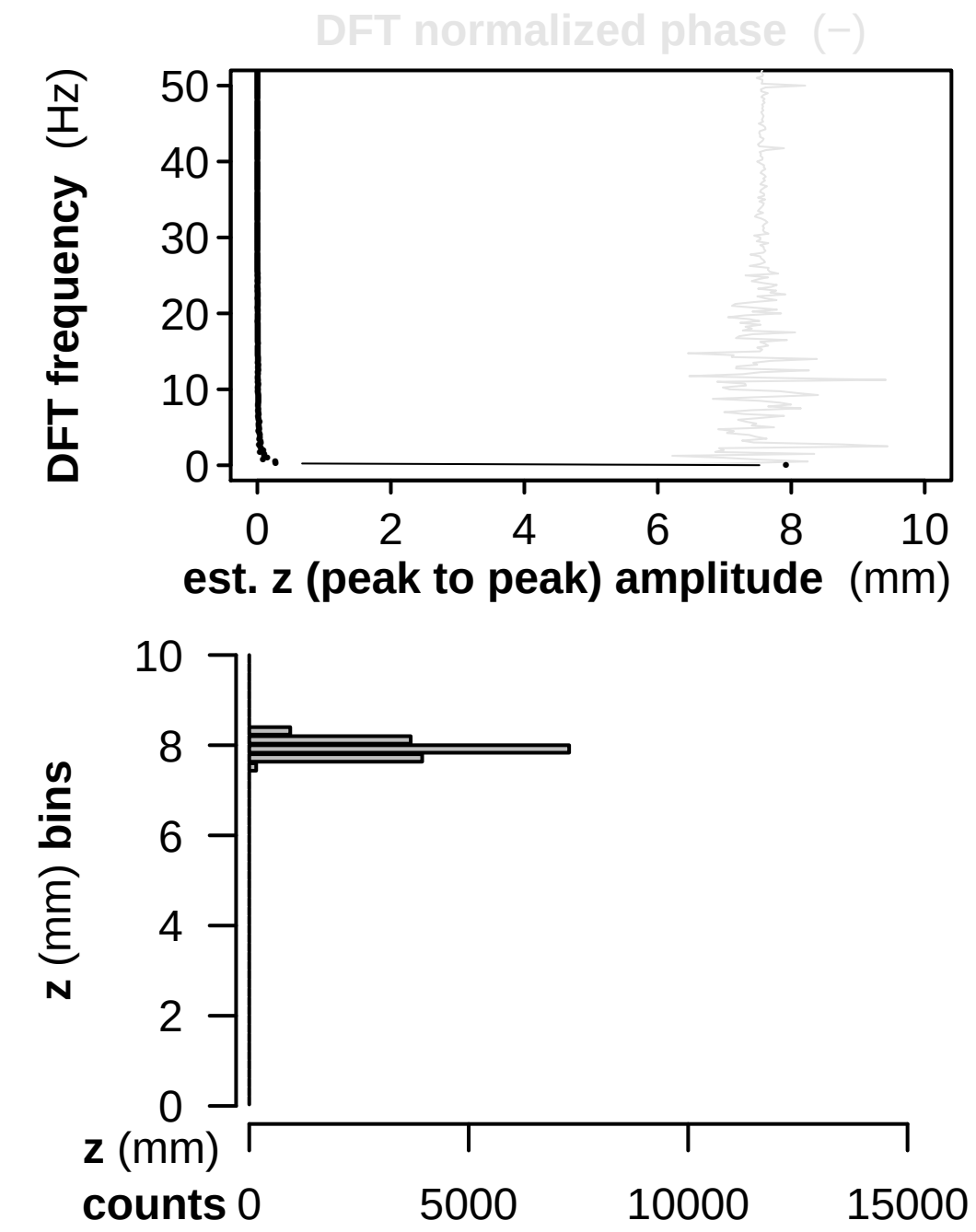

**SUBJECT 8 − RUN 21 − CONDITION 1,0**
**SC_180323_165726_0.AIFF**

```
z_min               : 7.57 mm
z_max               : 8.33 mm
z_travel_amplitude  : 0.75 mm

avg_abs_z_travel    : 6.28 mm/s

z_jarque-bera_jb    : 436.04
z_jarque-bera_p     : 0.00e+00

z_lin_mod_est_slope: 0.13 mm/s
z_lin_mod_adj_R²    : 76 %

z_poly40_mod_adj_R²: 97 %

z_dft_ampl_thresh   : 0.010 mm
>=threshold_maxfreq: 14.25 Hz
```

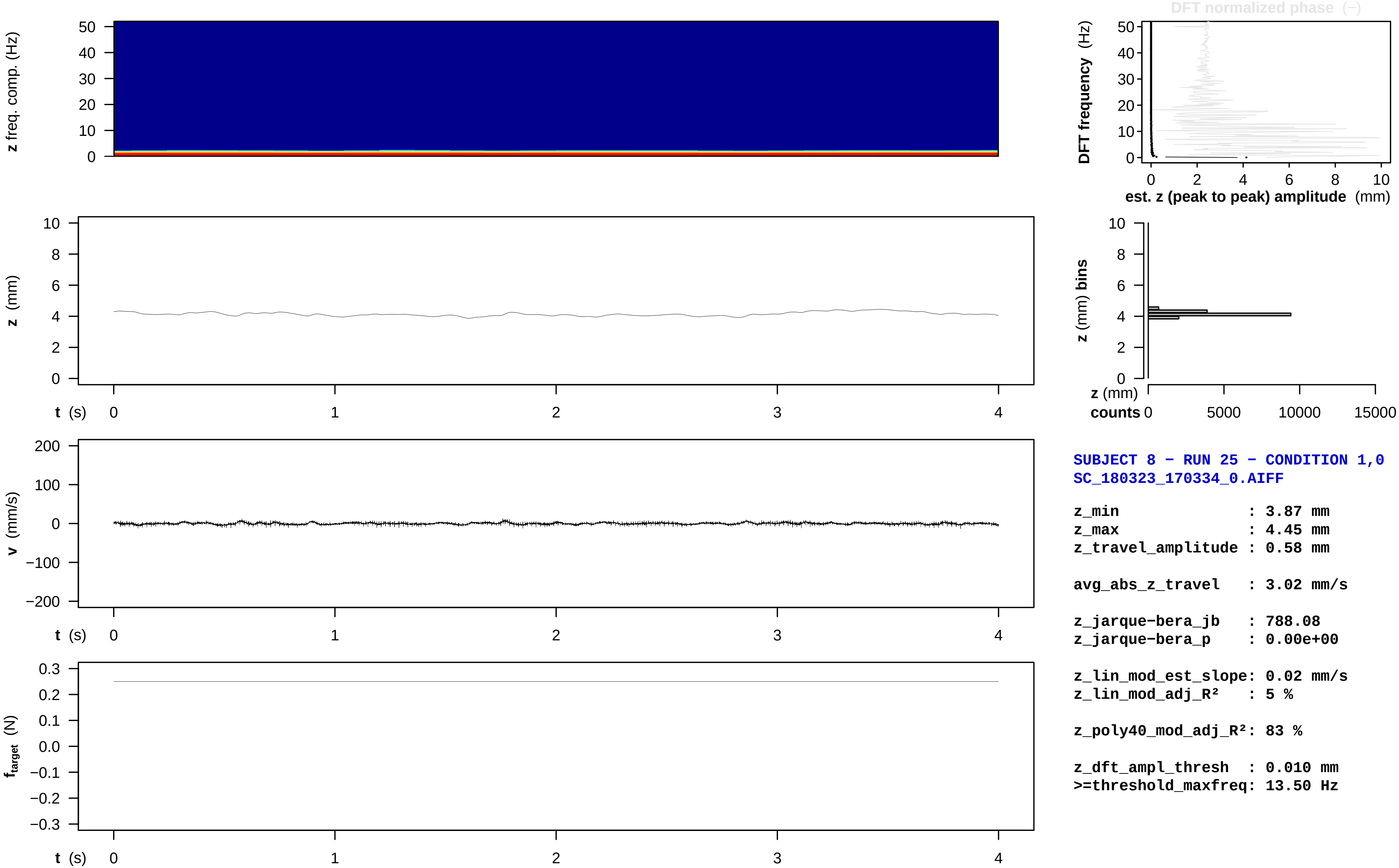

**SUBJECT 8 − RUN 25 − CONDITION 1,0**
**SC_180323_170334_0.AIFF**

```
z_min               : 3.87 mm
z_max               : 4.45 mm
z_travel_amplitude  : 0.58 mm

avg_abs_z_travel    : 3.02 mm/s

z_jarque-bera_jb    : 788.08
z_jarque-bera_p     : 0.00e+00

z_lin_mod_est_slope: 0.02 mm/s
z_lin_mod_adj_R²    : 5 %

z_poly40_mod_adj_R²: 83 %

z_dft_ampl_thresh   : 0.010 mm
>=threshold_maxfreq: 13.50 Hz
```

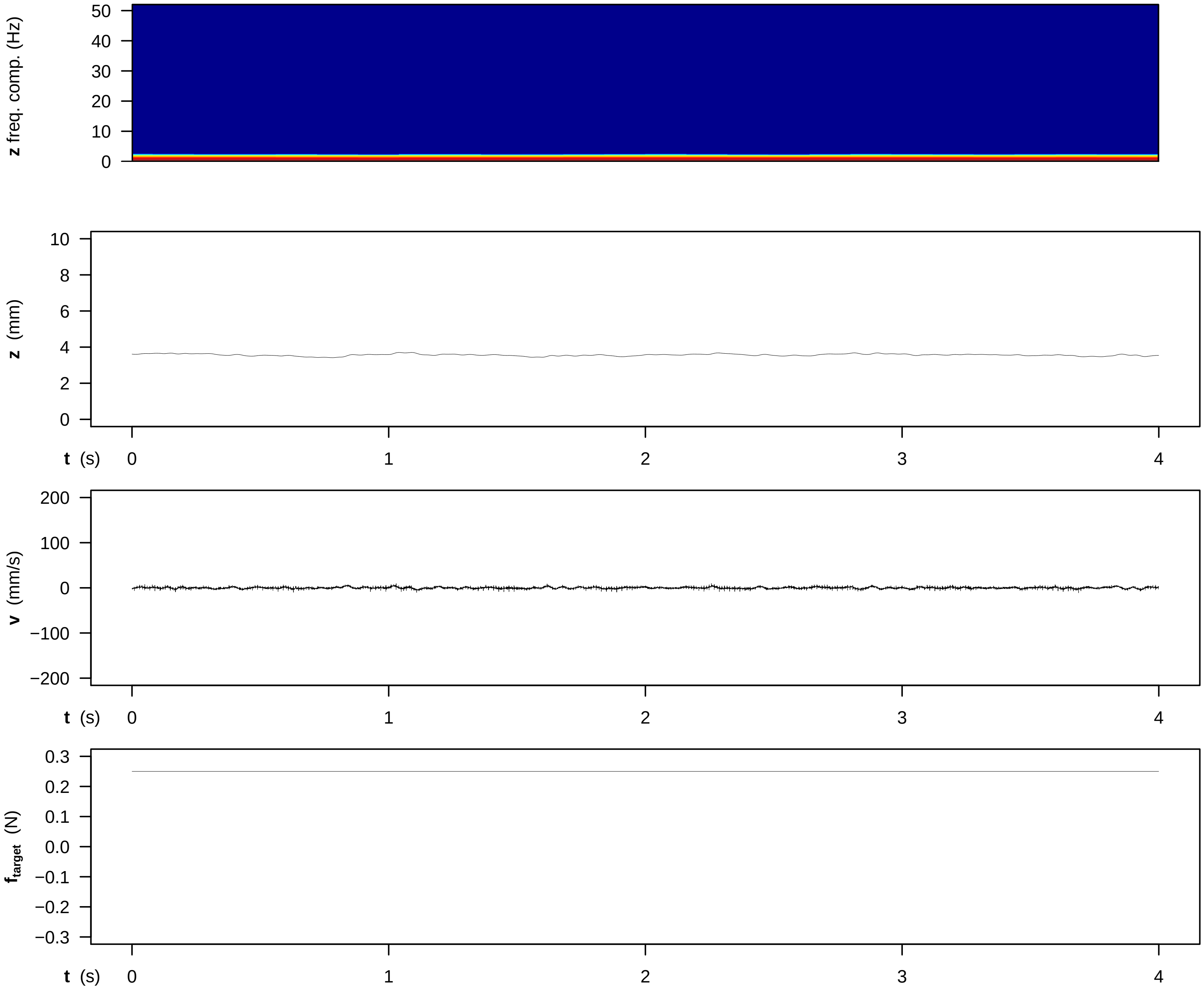

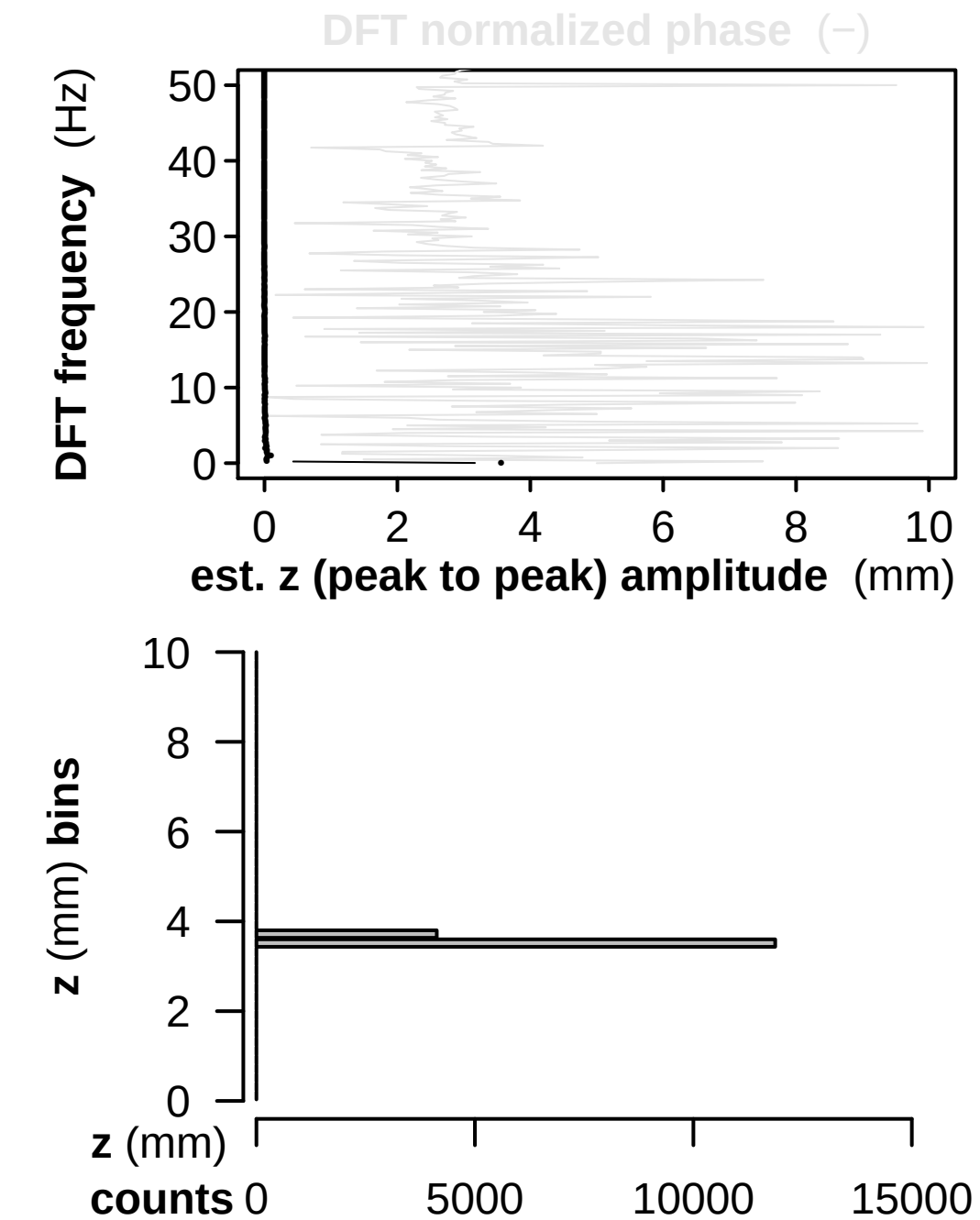

**SUBJECT 1 − RUN 07 − CONDITION 1,1**
**SC_180323_104247_0.AIFF**

```
z_min               : 3.41 mm
z_max               : 3.71 mm
z_travel_amplitude  : 0.29 mm

avg_abs_z_travel    : 2.34 mm/s

z_jarque-bera_jb    : 105.13
z_jarque-bera_p     : 0.00e+00

z_lin_mod_est_slope: −0.00 mm/s
z_lin_mod_adj_R²    : 1 %

z_poly40_mod_adj_R²: 71 %

z_dft_ampl_thresh   : 0.010 mm
>=threshold_maxfreq: 16.75 Hz
```

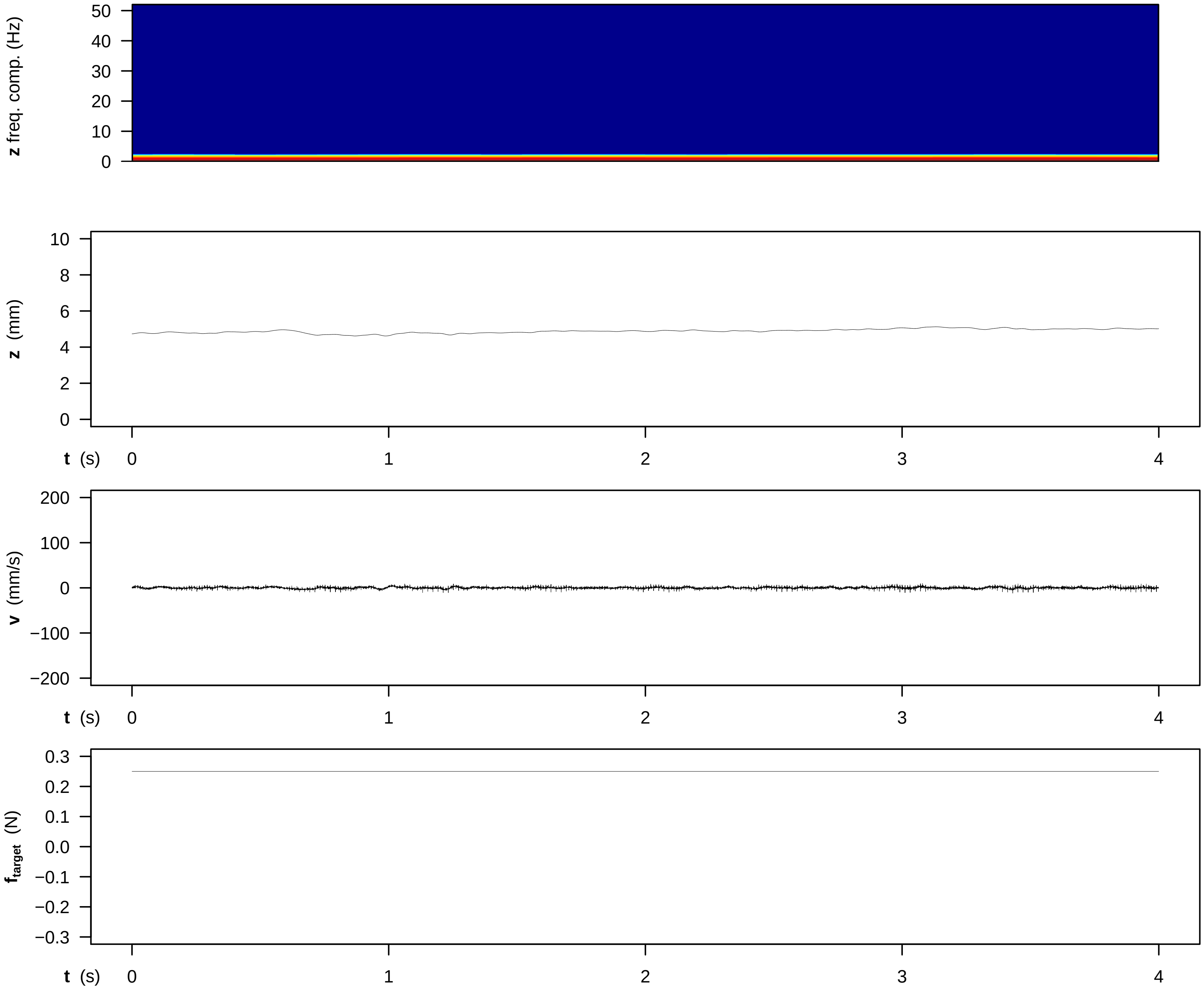

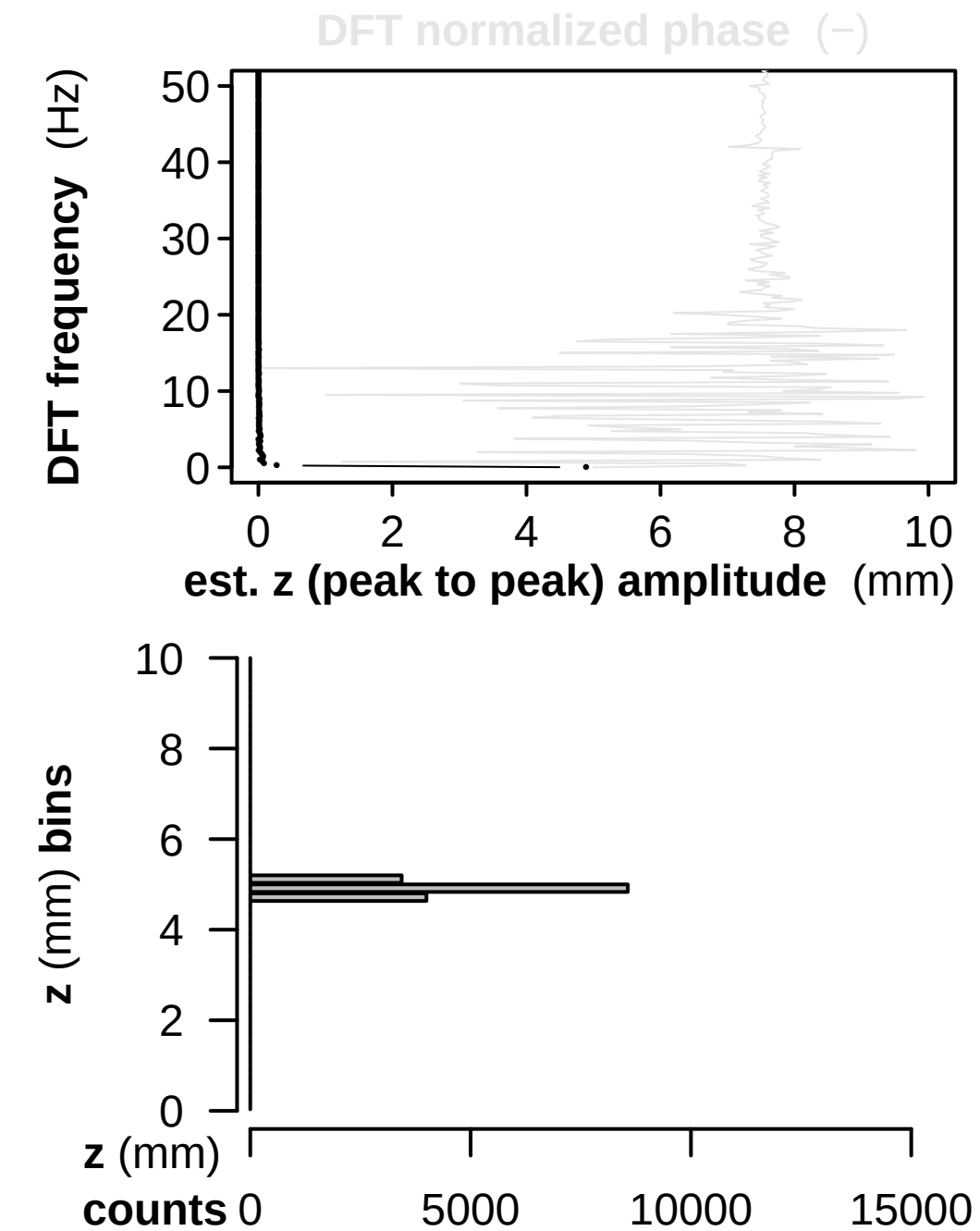

**SUBJECT 1 − RUN 08 − CONDITION 1,1**
**SC_180323_104317_0.AIFF**

```
z_min               : 4.61 mm
z_max               : 5.13 mm
z_travel_amplitude  : 0.51 mm

avg_abs_z_travel    : 2.59 mm/s

z_jarque-bera_jb    : 378.92
z_jarque-bera_p     : 0.00e+00

z_lin_mod_est_slope: 0.08 mm/s
z_lin_mod_adj_R²    : 67 %

z_poly40_mod_adj_R²: 93 %

z_dft_ampl_thresh   : 0.010 mm
>=threshold_maxfreq: 12.25 Hz
```

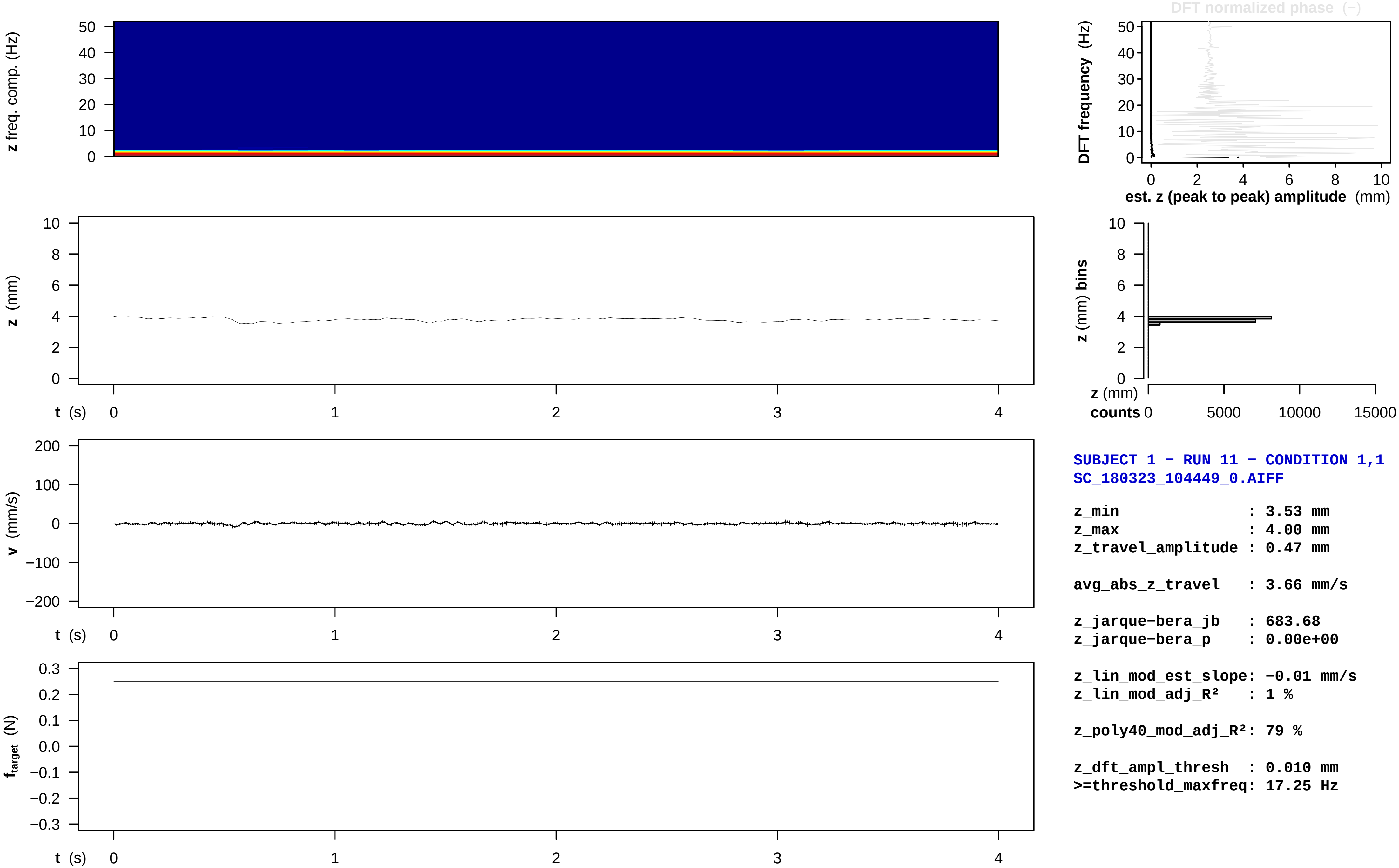

**SUBJECT 1 − RUN 11 − CONDITION 1,1**
**SC_180323_104449_0.AIFF**

```
z_min               : 3.53 mm
z_max               : 4.00 mm
z_travel_amplitude  : 0.47 mm

avg_abs_z_travel    : 3.66 mm/s

z_jarque-bera_jb    : 683.68
z_jarque-bera_p     : 0.00e+00

z_lin_mod_est_slope: −0.01 mm/s
z_lin_mod_adj_R²    : 1 %

z_poly40_mod_adj_R²: 79 %

z_dft_ampl_thresh   : 0.010 mm
>=threshold_maxfreq: 17.25 Hz
```

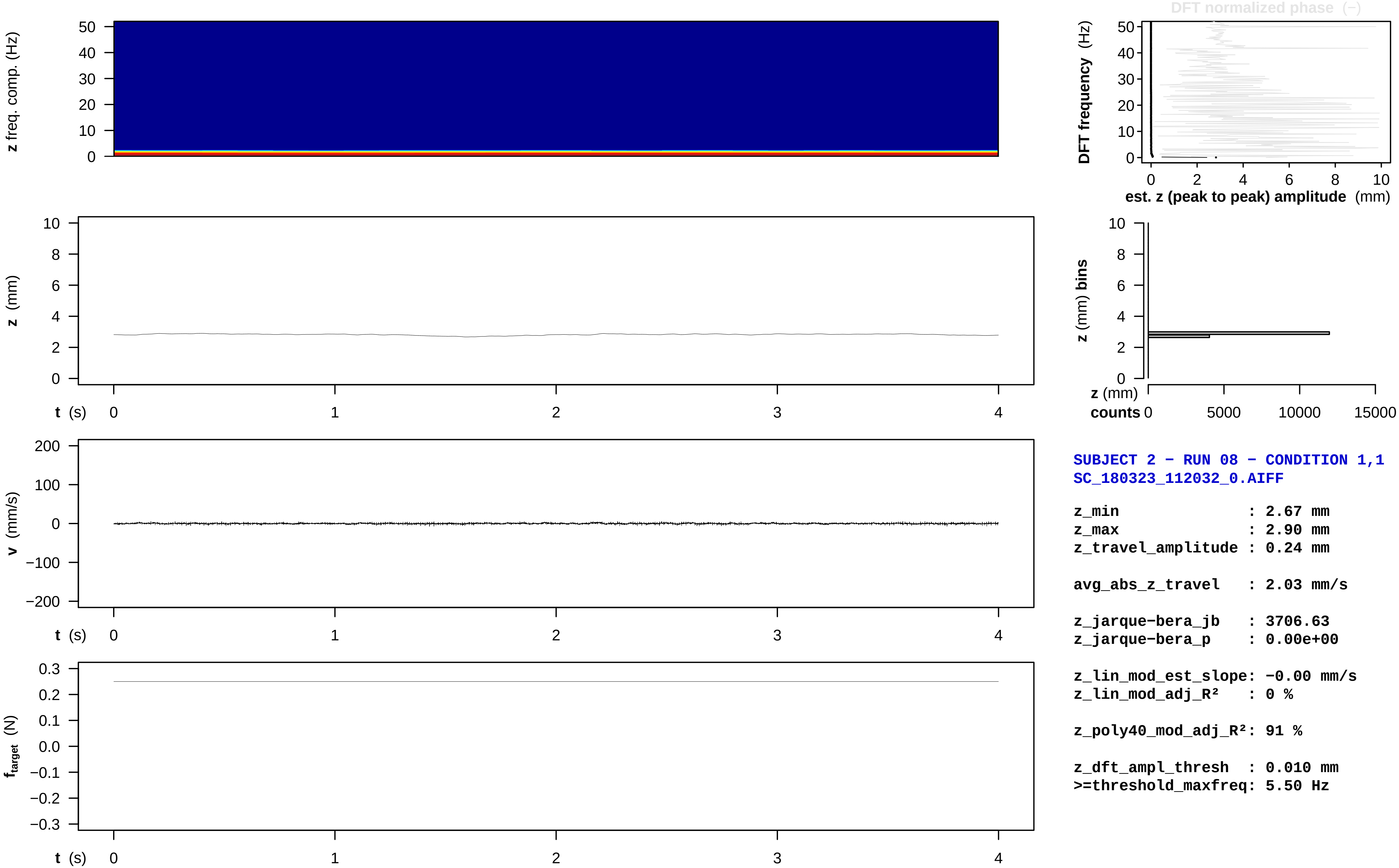

**SUBJECT 2 − RUN 08 − CONDITION 1,1**
**SC_180323_112032_0.AIFF**

| | |
|---|---|
| z_min | : 2.67 mm |
| z_max | : 2.90 mm |
| z_travel_amplitude | : 0.24 mm |
| avg_abs_z_travel | : 2.03 mm/s |
| z_jarque-bera_jb | : 3706.63 |
| z_jarque-bera_p | : 0.00e+00 |
| z_lin_mod_est_slope | : −0.00 mm/s |
| z_lin_mod_adj_R² | : 0 % |
| z_poly40_mod_adj_R² | : 91 % |
| z_dft_ampl_thresh | : 0.010 mm |
| >=threshold_maxfreq | : 5.50 Hz |

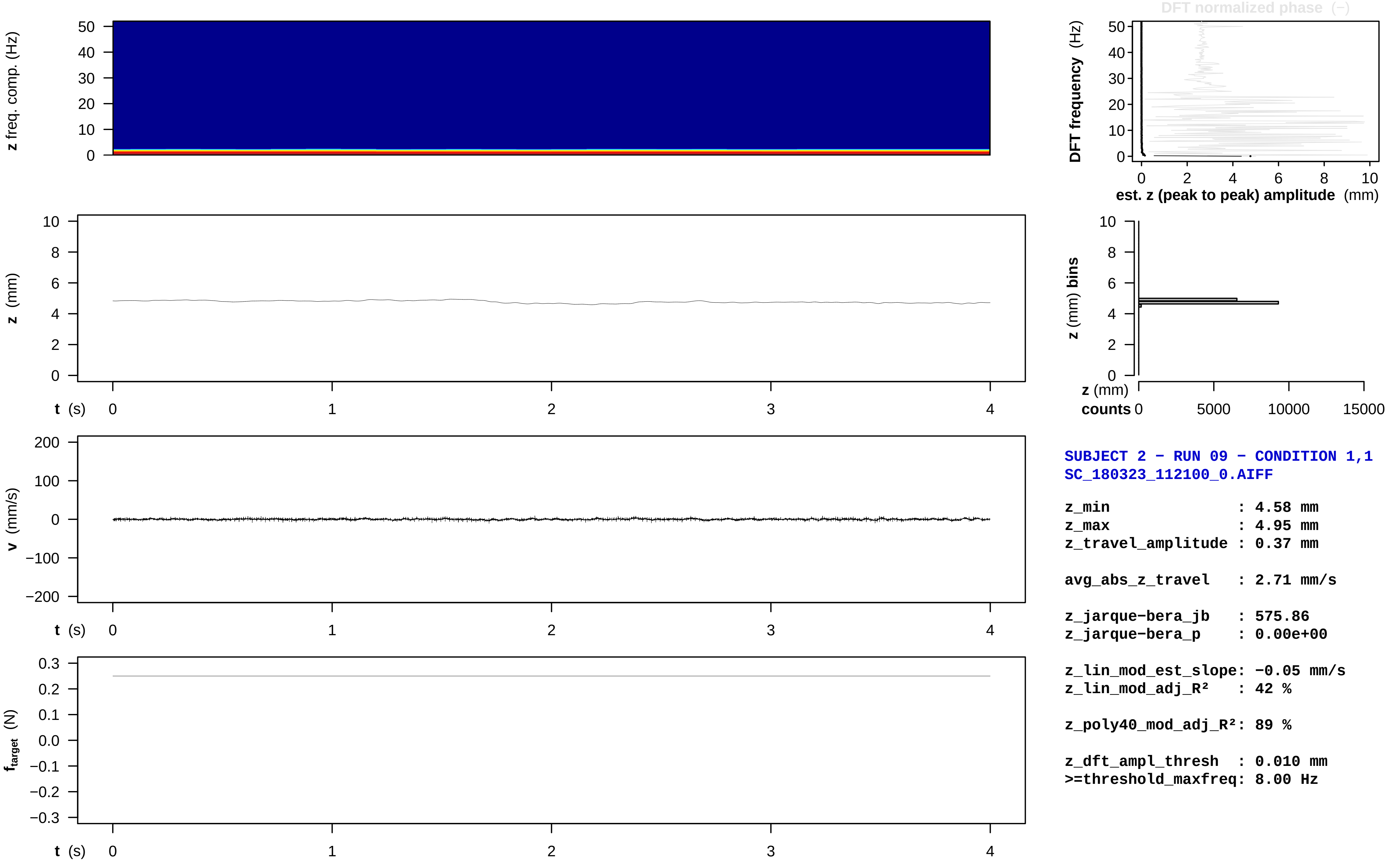
DFT normalized phase (−)
z freq. comp. (Hz)
DFT frequency (Hz)
est. z (peak to peak) amplitude (mm)
z (mm)
z (mm) bins
counts
t (s)
v (mm/s)
f_target (N)
SUBJECT 2 − RUN 09 − CONDITION 1,1
SC_180323_112100_0.AIFF
z_min : 4.58 mm
z_max : 4.95 mm
z_travel_amplitude : 0.37 mm
avg_abs_z_travel : 2.71 mm/s
z_jarque-bera_jb : 575.86
z_jarque-bera_p : 0.00e+00
z_lin_mod_est_slope: −0.05 mm/s
z_lin_mod_adj_R² : 42 %
z_poly40_mod_adj_R²: 89 %
z_dft_ampl_thresh : 0.010 mm
>=threshold_maxfreq: 8.00 Hz

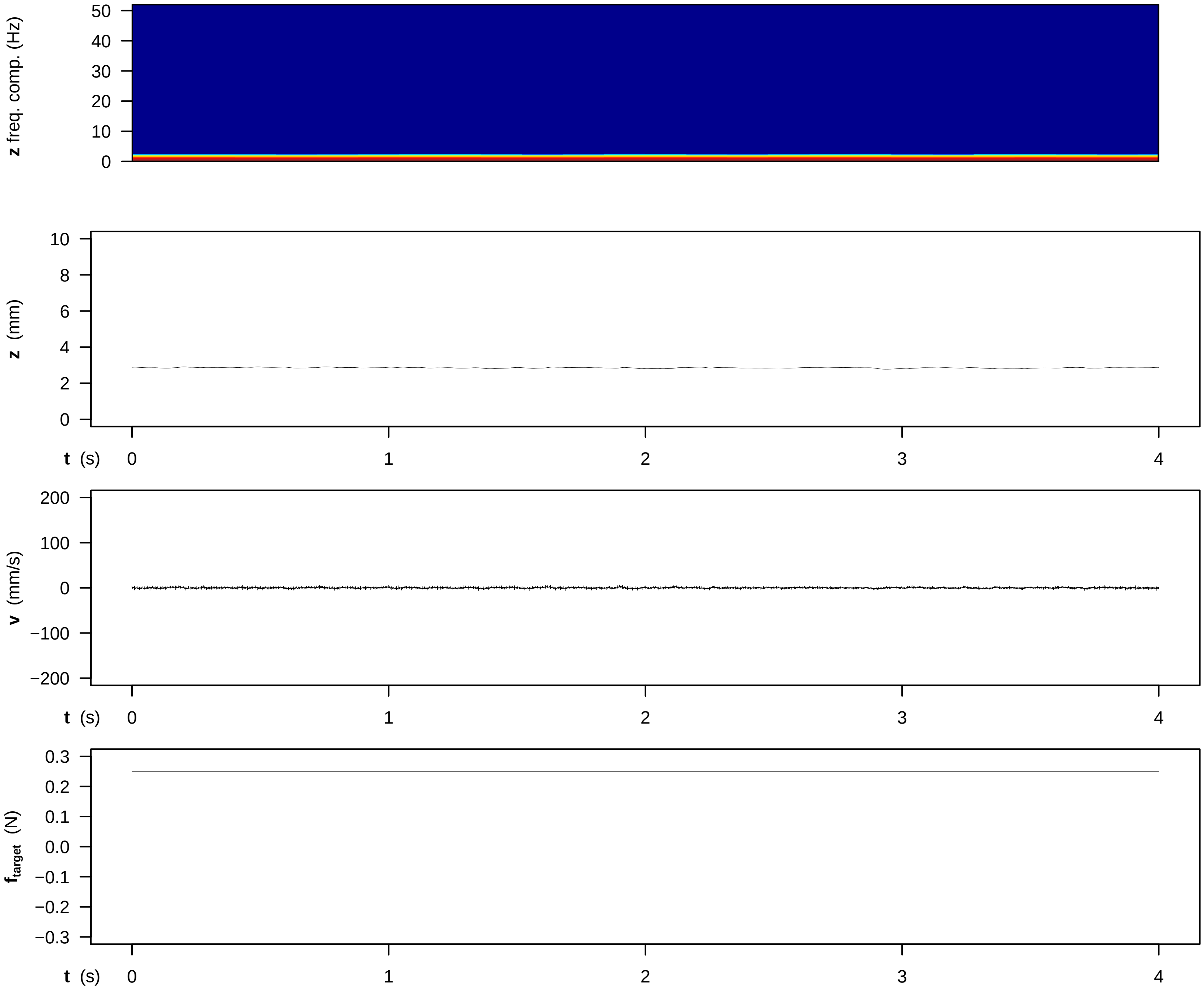

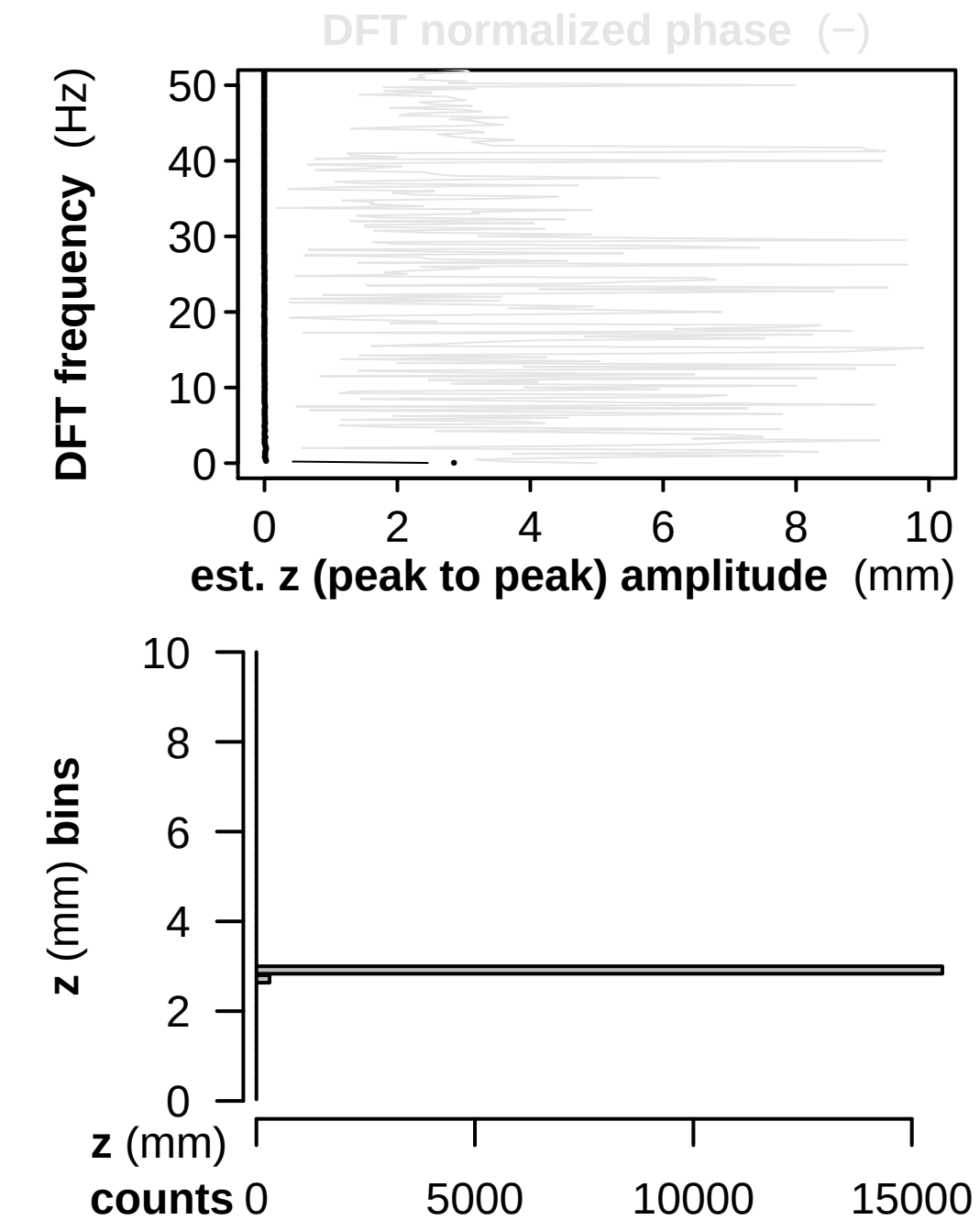

**SUBJECT 2 − RUN 21 − CONDITION 1,1**
**SC_180323_112726_0.AIFF**

```
z_min               : 2.78 mm
z_max               : 2.91 mm
z_travel_amplitude  : 0.13 mm

avg_abs_z_travel    : 1.60 mm/s

z_jarque-bera_jb    : 1059.61
z_jarque-bera_p     : 0.00e+00

z_lin_mod_est_slope: −0.01 mm/s
z_lin_mod_adj_R²    : 9 %

z_poly40_mod_adj_R²: 46 %

z_dft_ampl_thresh   : 0.010 mm
>=threshold_maxfreq: 7.50 Hz
```

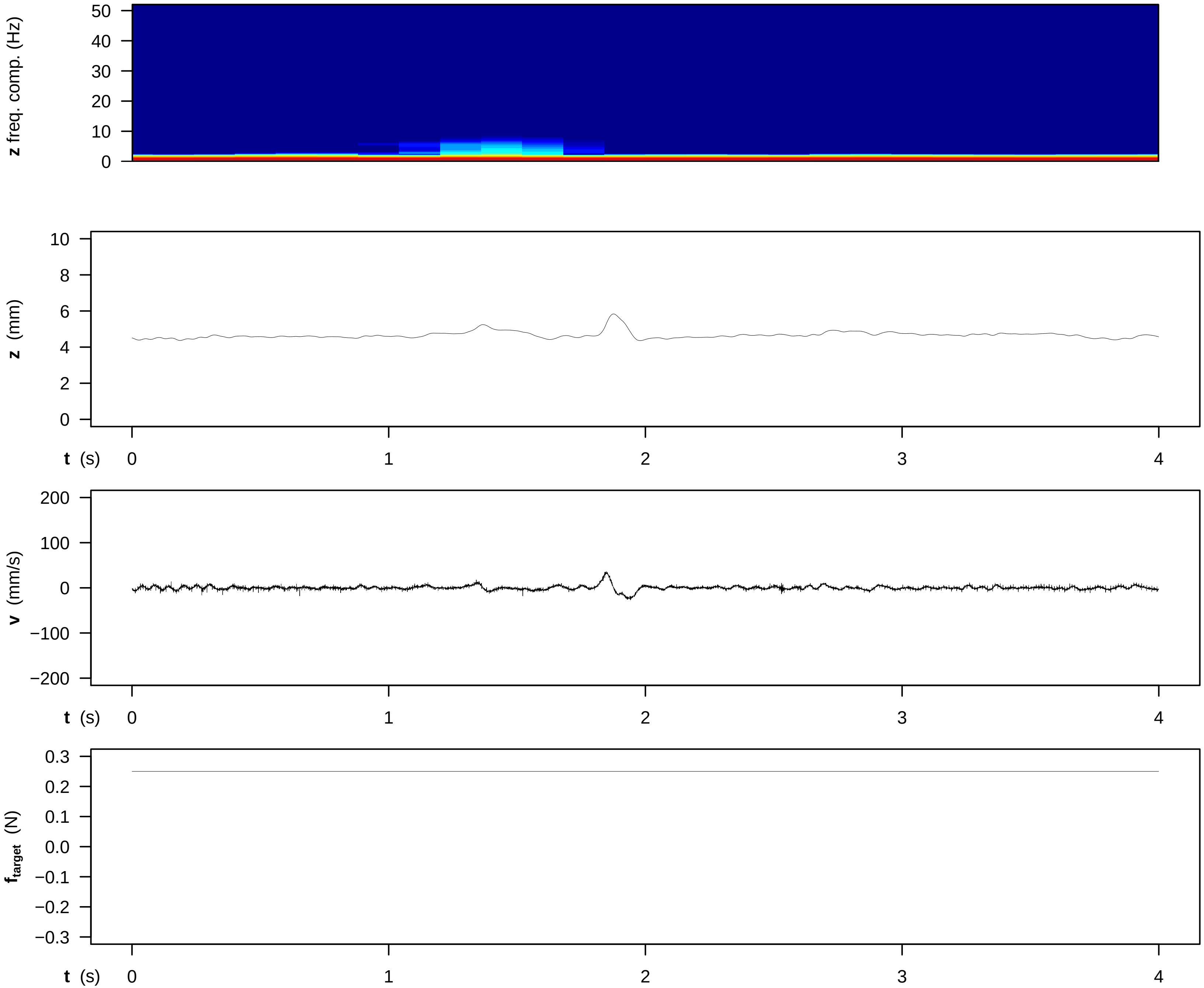

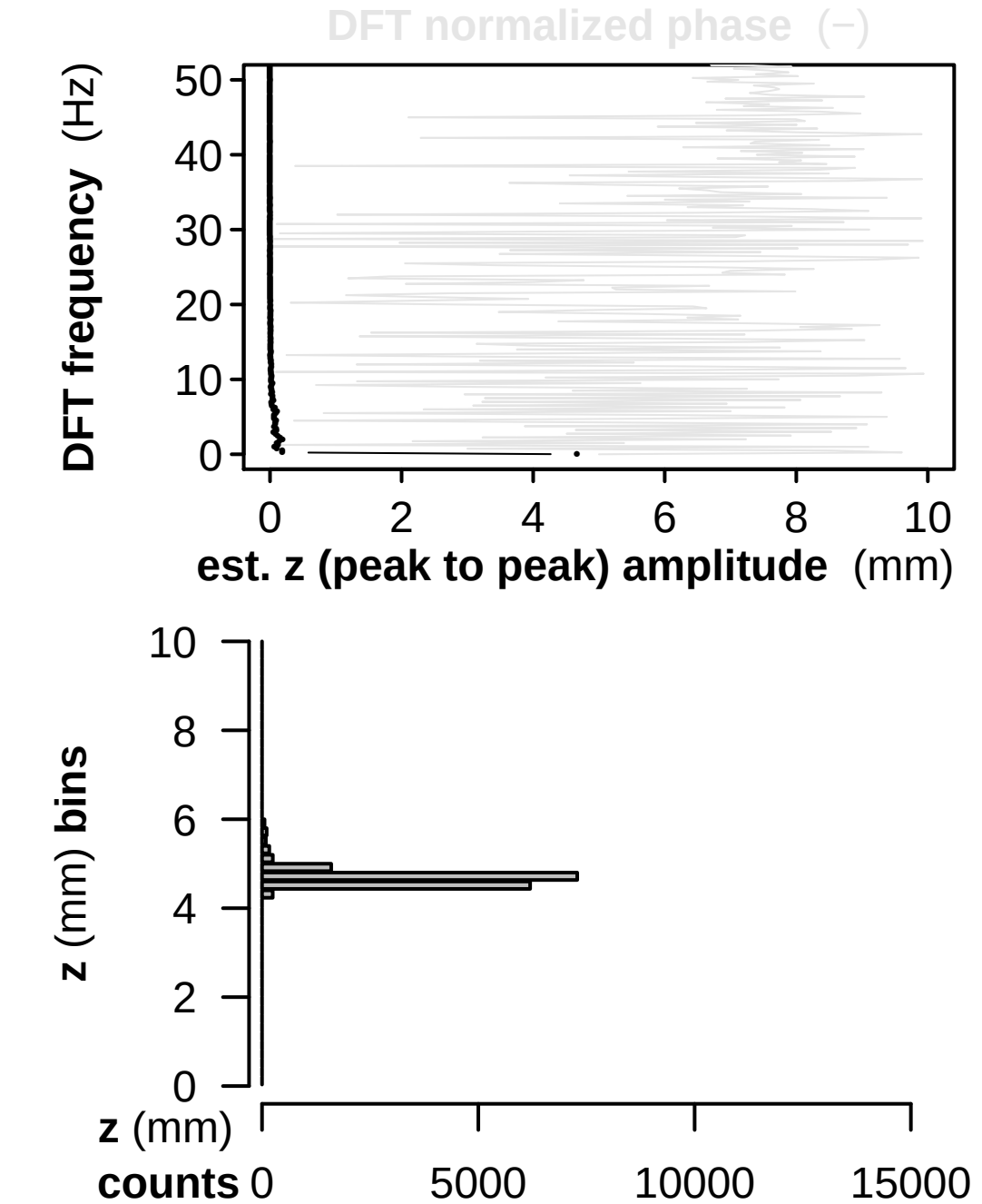

**SUBJECT 3 − RUN 04 − CONDITION 1,1**
**SC_180323_115705_0.AIFF**

```
z_min               : 4.36 mm
z_max               : 5.84 mm
z_travel_amplitude  : 1.48 mm

avg_abs_z_travel    : 3.78 mm/s

z_jarque-bera_jb    : 82080.23
z_jarque-bera_p     : 0.00e+00

z_lin_mod_est_slope: 0.03 mm/s
z_lin_mod_adj_R²    : 2 %

z_poly40_mod_adj_R²: 40 %

z_dft_ampl_thresh   : 0.010 mm
>=threshold_maxfreq: 19.25 Hz
```

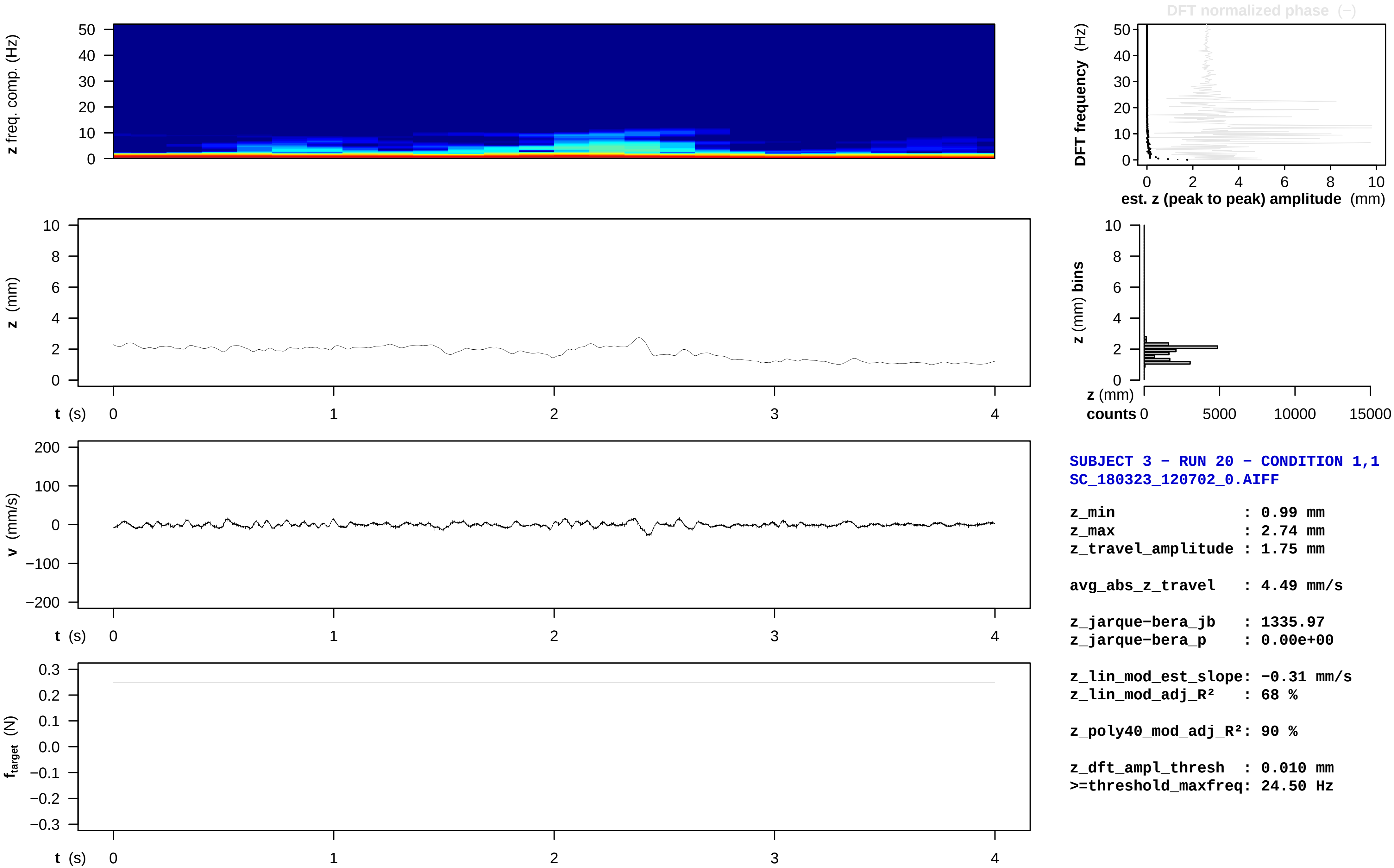

**SUBJECT 3 − RUN 20 − CONDITION 1,1**
**SC_180323_120702_0.AIFF**

```
z_min               : 0.99 mm
z_max               : 2.74 mm
z_travel_amplitude  : 1.75 mm

avg_abs_z_travel    : 4.49 mm/s

z_jarque-bera_jb    : 1335.97
z_jarque-bera_p     : 0.00e+00

z_lin_mod_est_slope: −0.31 mm/s
z_lin_mod_adj_R²    : 68 %

z_poly40_mod_adj_R²: 90 %

z_dft_ampl_thresh   : 0.010 mm
>=threshold_maxfreq: 24.50 Hz
```

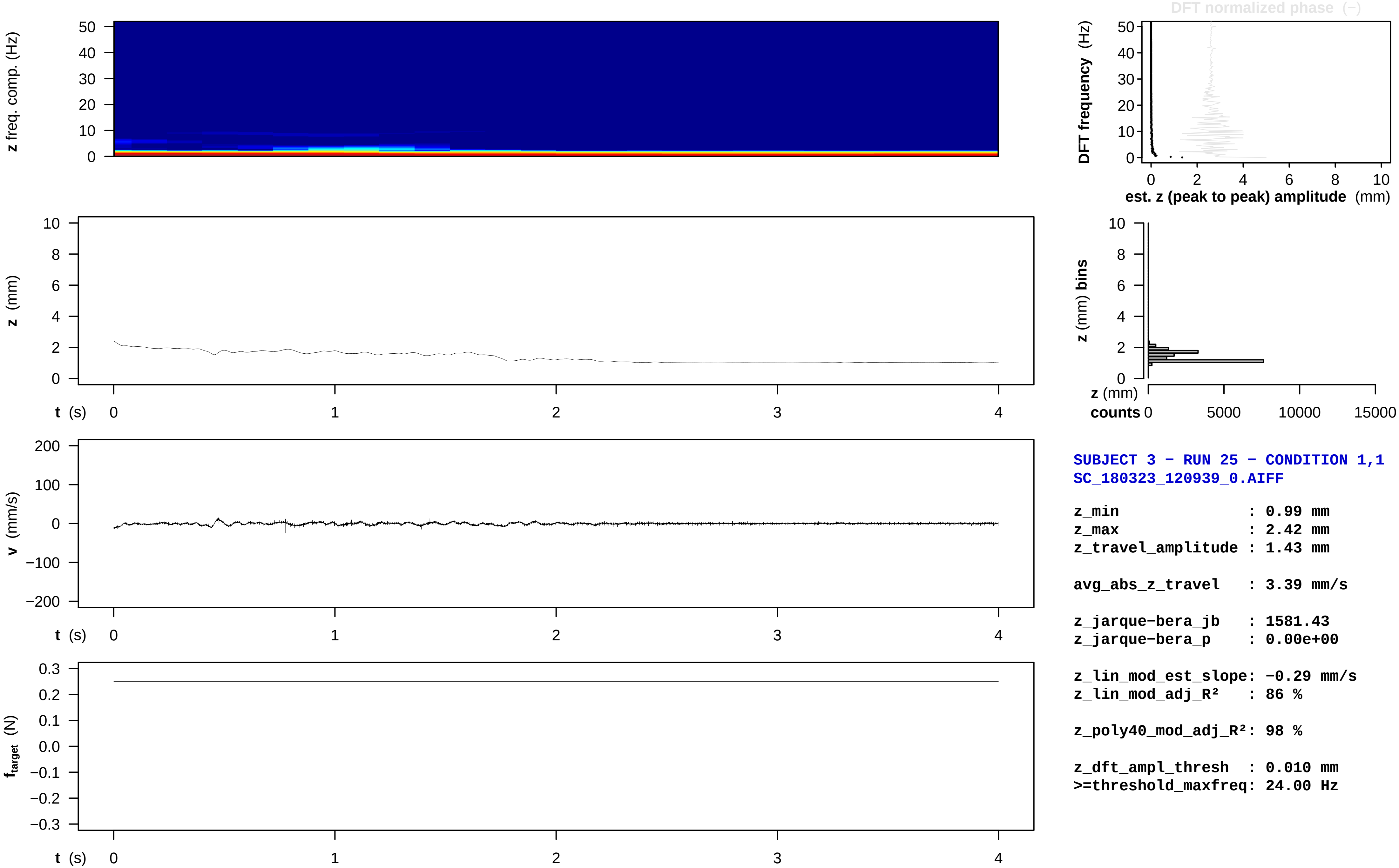

**SUBJECT 3 − RUN 25 − CONDITION 1,1**
**SC_180323_120939_0.AIFF**

```
z_min               : 0.99 mm
z_max               : 2.42 mm
z_travel_amplitude  : 1.43 mm

avg_abs_z_travel    : 3.39 mm/s

z_jarque-bera_jb    : 1581.43
z_jarque-bera_p     : 0.00e+00

z_lin_mod_est_slope: −0.29 mm/s
z_lin_mod_adj_R²    : 86 %

z_poly40_mod_adj_R²: 98 %

z_dft_ampl_thresh   : 0.010 mm
>=threshold_maxfreq: 24.00 Hz
```

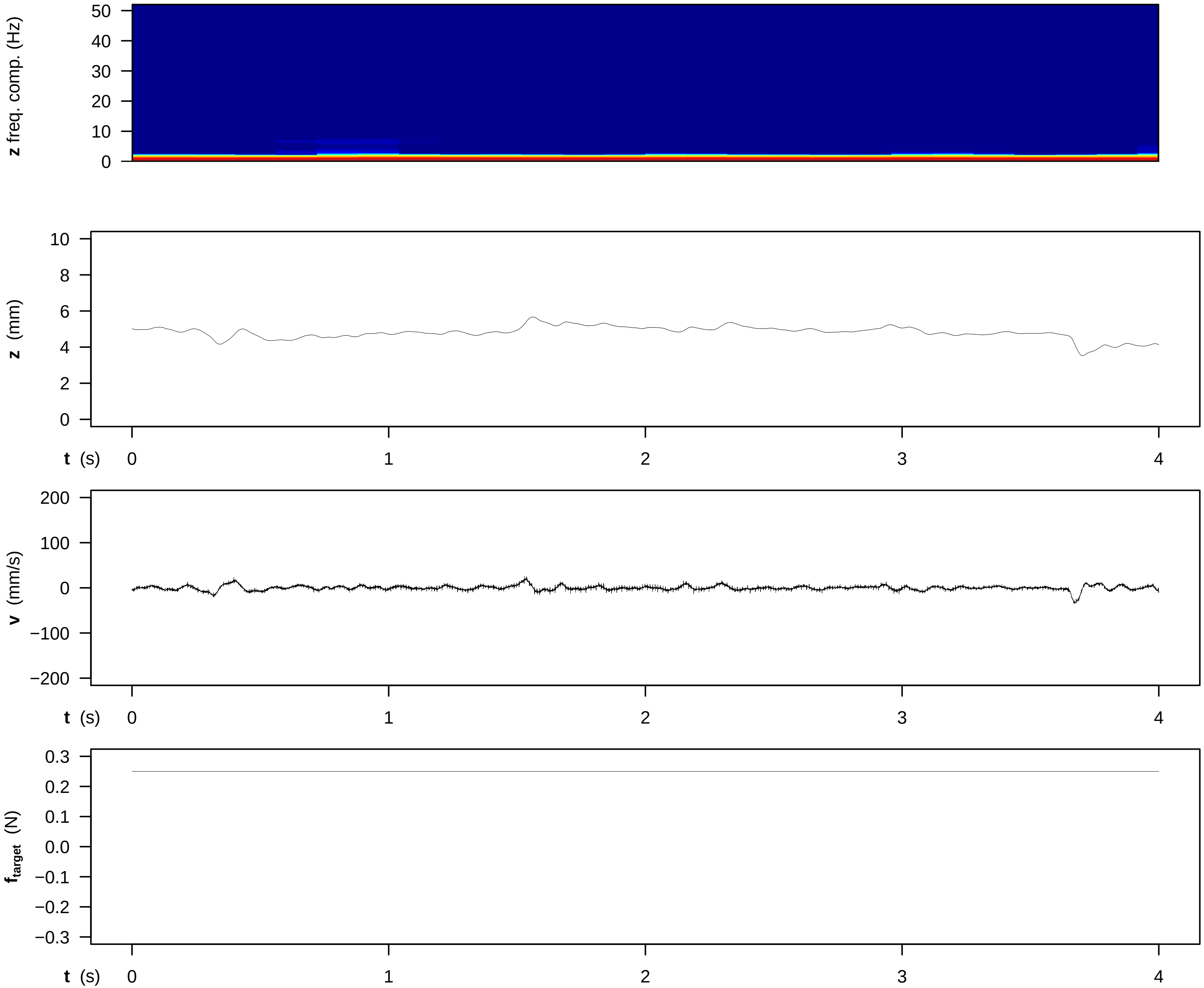

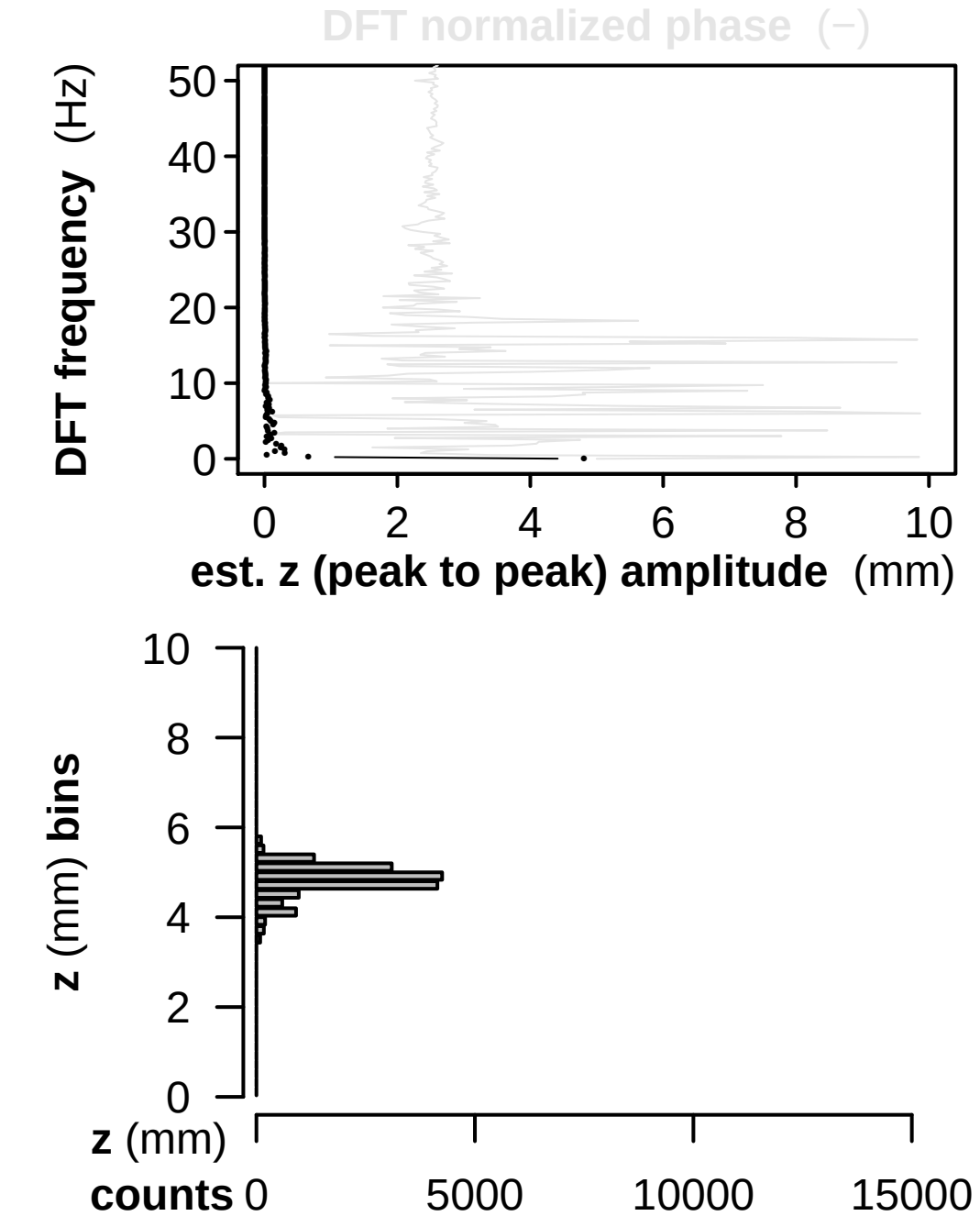

**SUBJECT 4 − RUN 24 − CONDITION 1,1**
**SC_180323_124219_0.AIFF**

```
z_min              : 3.52 mm
z_max              : 5.67 mm
z_travel_amplitude : 2.15 mm

avg_abs_z_travel   : 5.29 mm/s

z_jarque-bera_jb   : 3536.68
z_jarque-bera_p    : 0.00e+00

z_lin_mod_est_slope: −0.07 mm/s
z_lin_mod_adj_R²   : 5 %

z_poly40_mod_adj_R²: 87 %

z_dft_ampl_thresh  : 0.010 mm
>=threshold_maxfreq: 17.50 Hz
```

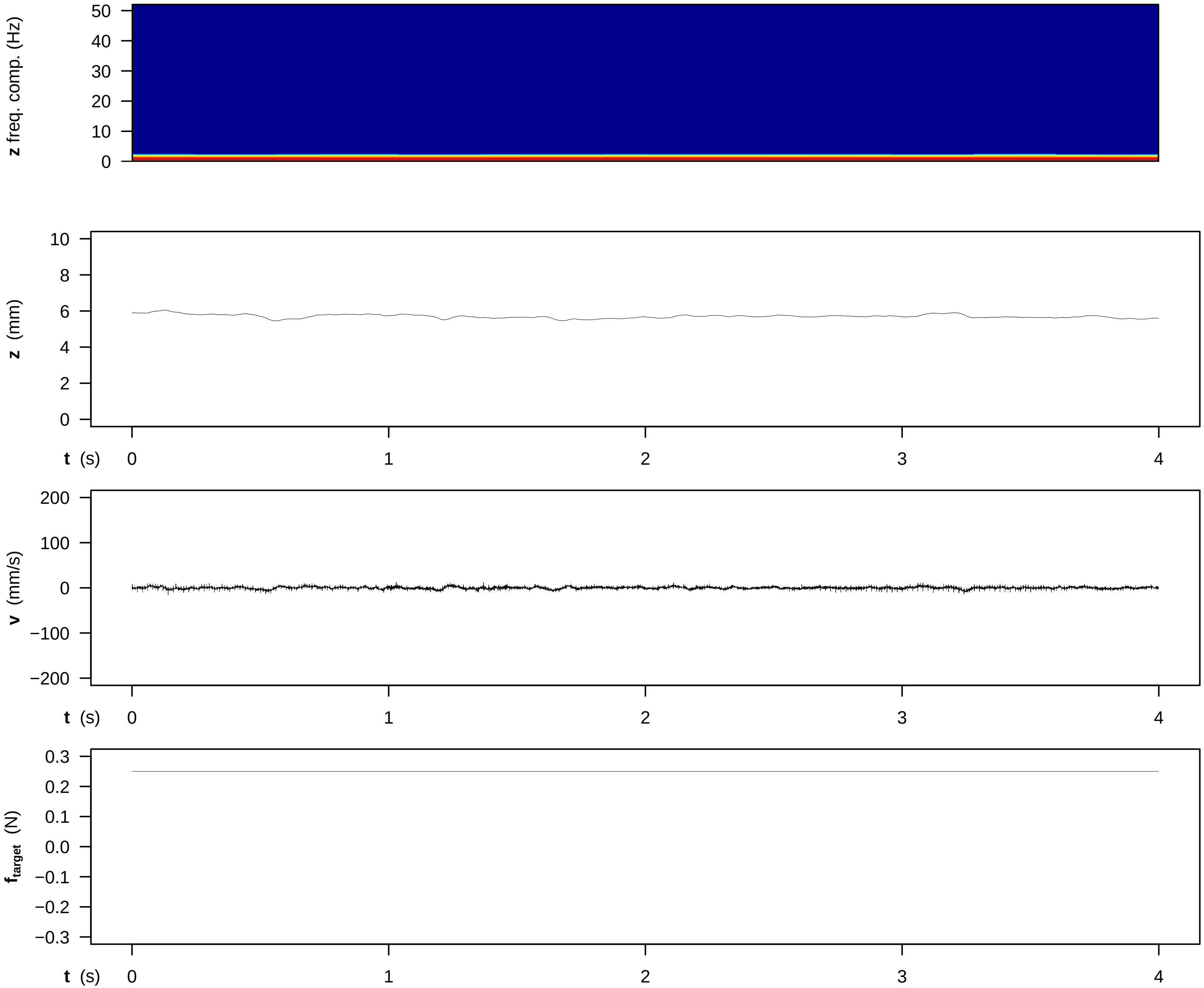

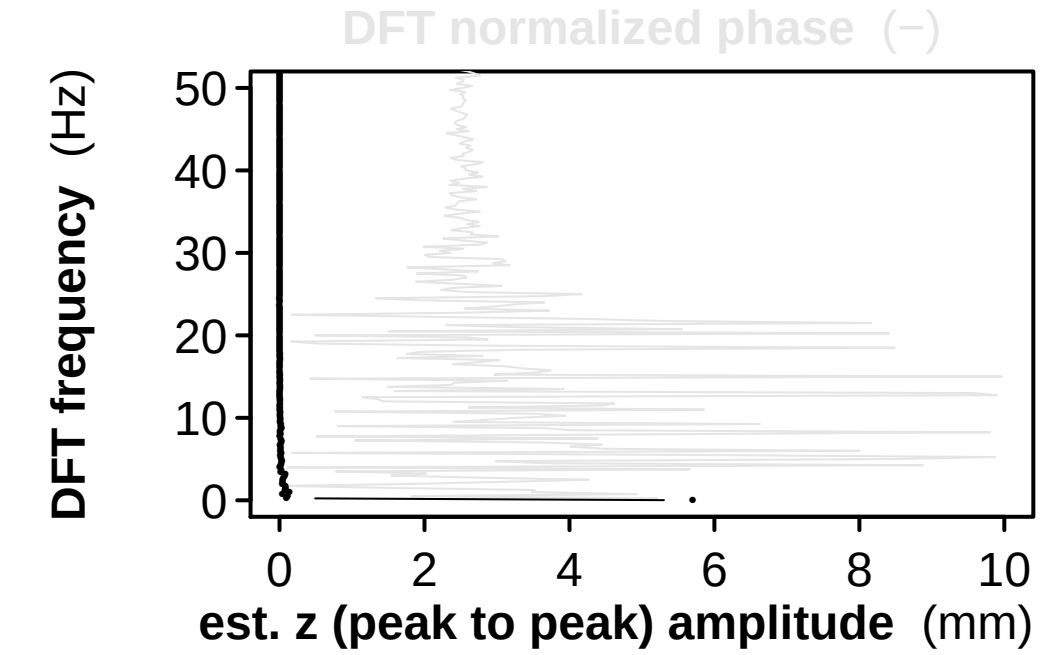

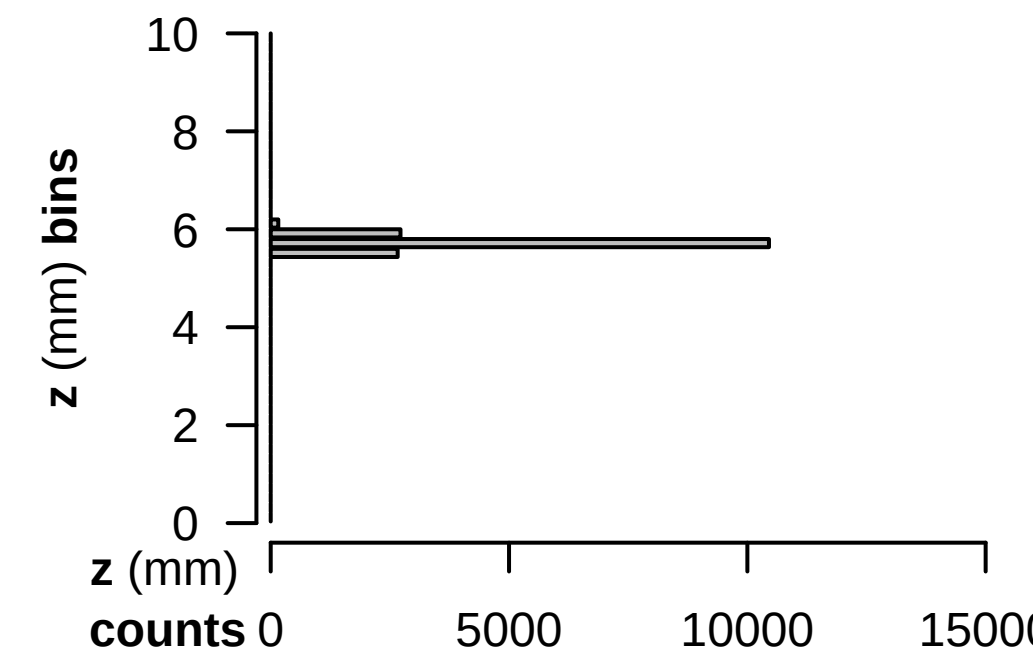

**SUBJECT 4 − RUN 29 − CONDITION 1,1**
**SC_180323_124524_0.AIFF**

```
z_min              : 5.45 mm
z_max              : 6.06 mm
z_travel_amplitude : 0.61 mm

avg_abs_z_travel   : 4.57 mm/s

z_jarque-bera_jb   : 400.94
z_jarque-bera_p    : 0.00e+00

z_lin_mod_est_slope: −0.03 mm/s
z_lin_mod_adj_R²   : 10 %

z_poly40_mod_adj_R²: 79 %

z_dft_ampl_thresh  : 0.010 mm
>=threshold_maxfreq: 10.75 Hz
```

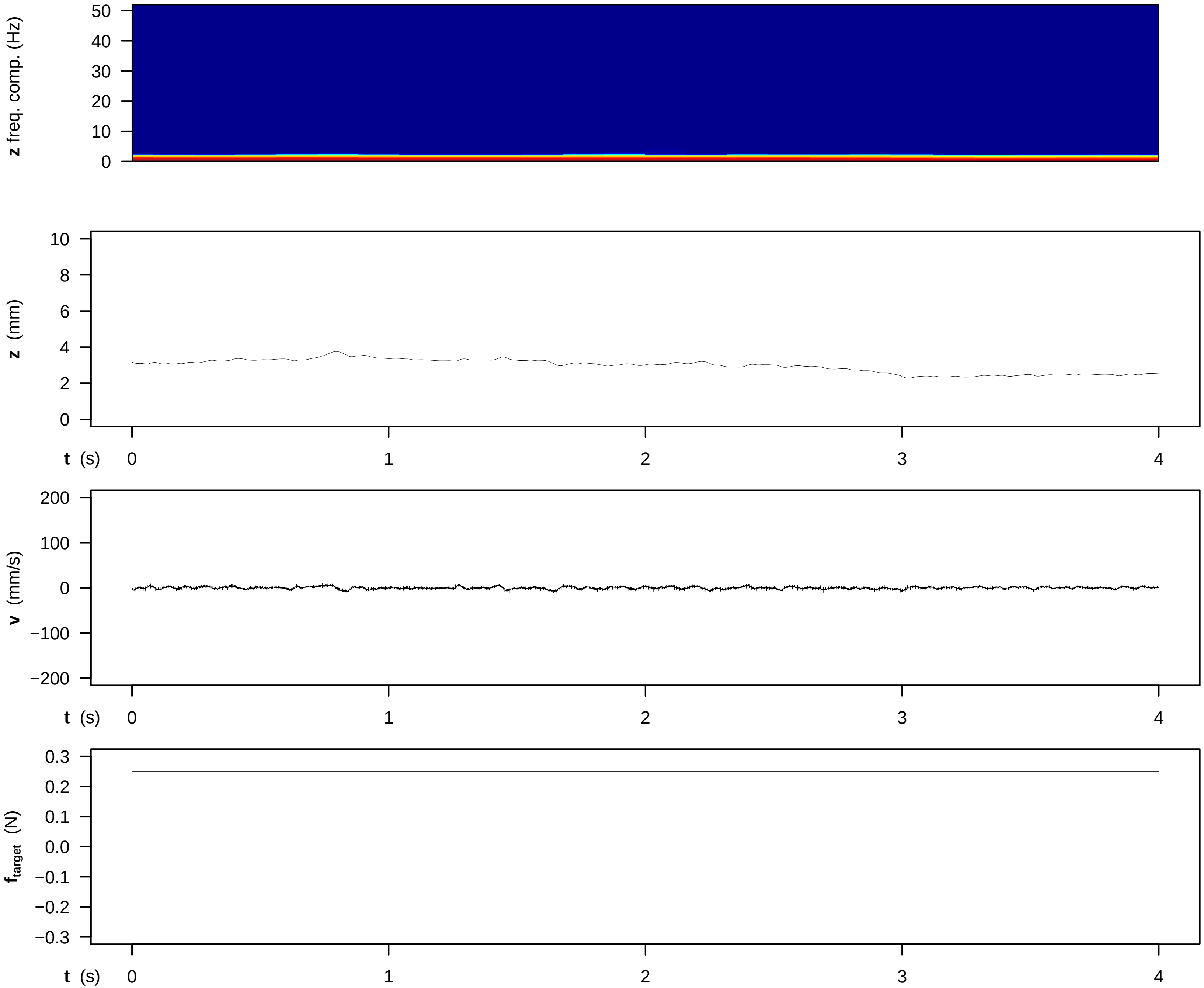

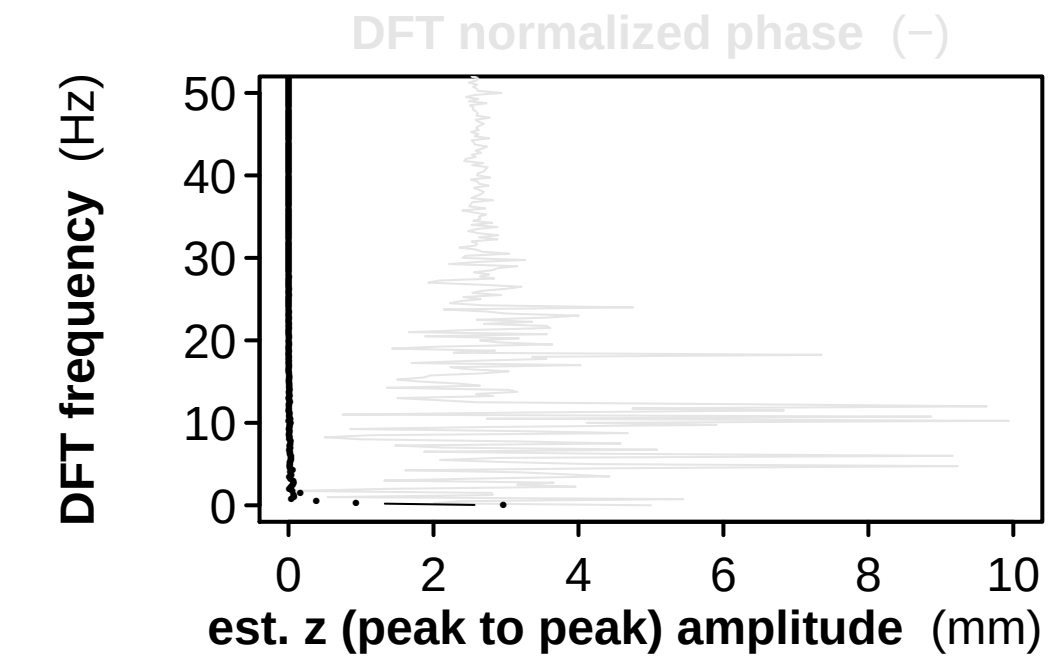

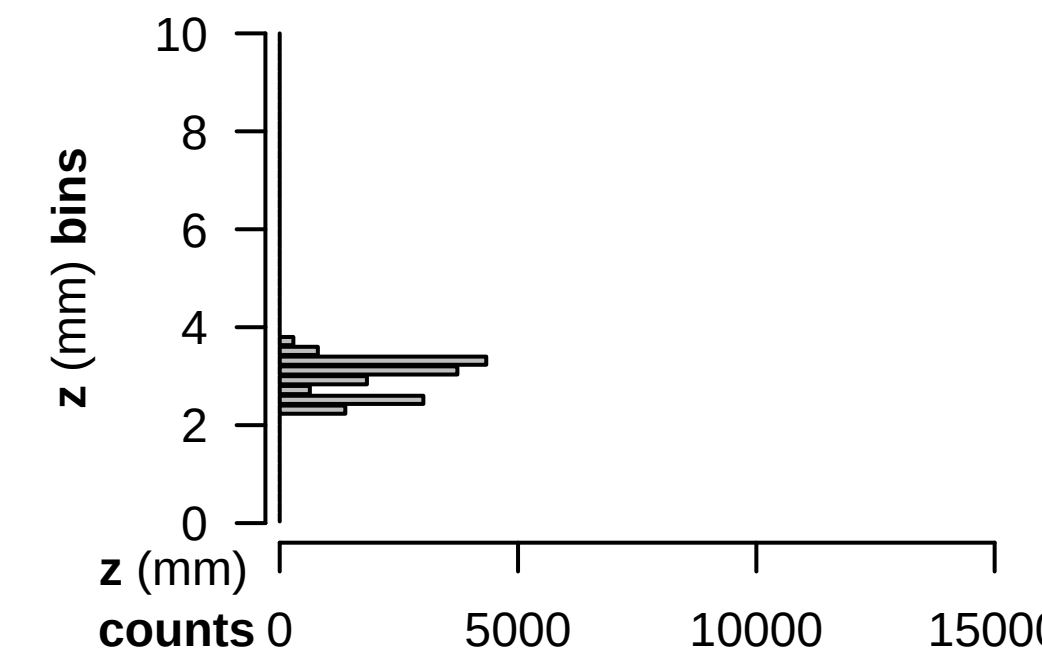

**SUBJECT 4 − RUN 36 − CONDITION 1,1**
**SC_180323_124935_0.AIFF**

**z_min : 2.29 mm**
**z_max : 3.76 mm**
**z_travel_amplitude : 1.47 mm**

**avg_abs_z_travel : 2.64 mm/s**

**z_jarque-bera_jb : 1073.24**
**z_jarque-bera_p : 0.00e+00**

**z_lin_mod_est_slope: −0.28 mm/s**
**z_lin_mod_adj_R² : 76 %**

**z_poly40_mod_adj_R²: 98 %**

**z_dft_ampl_thresh : 0.010 mm**
**>=threshold_maxfreq: 15.75 Hz**

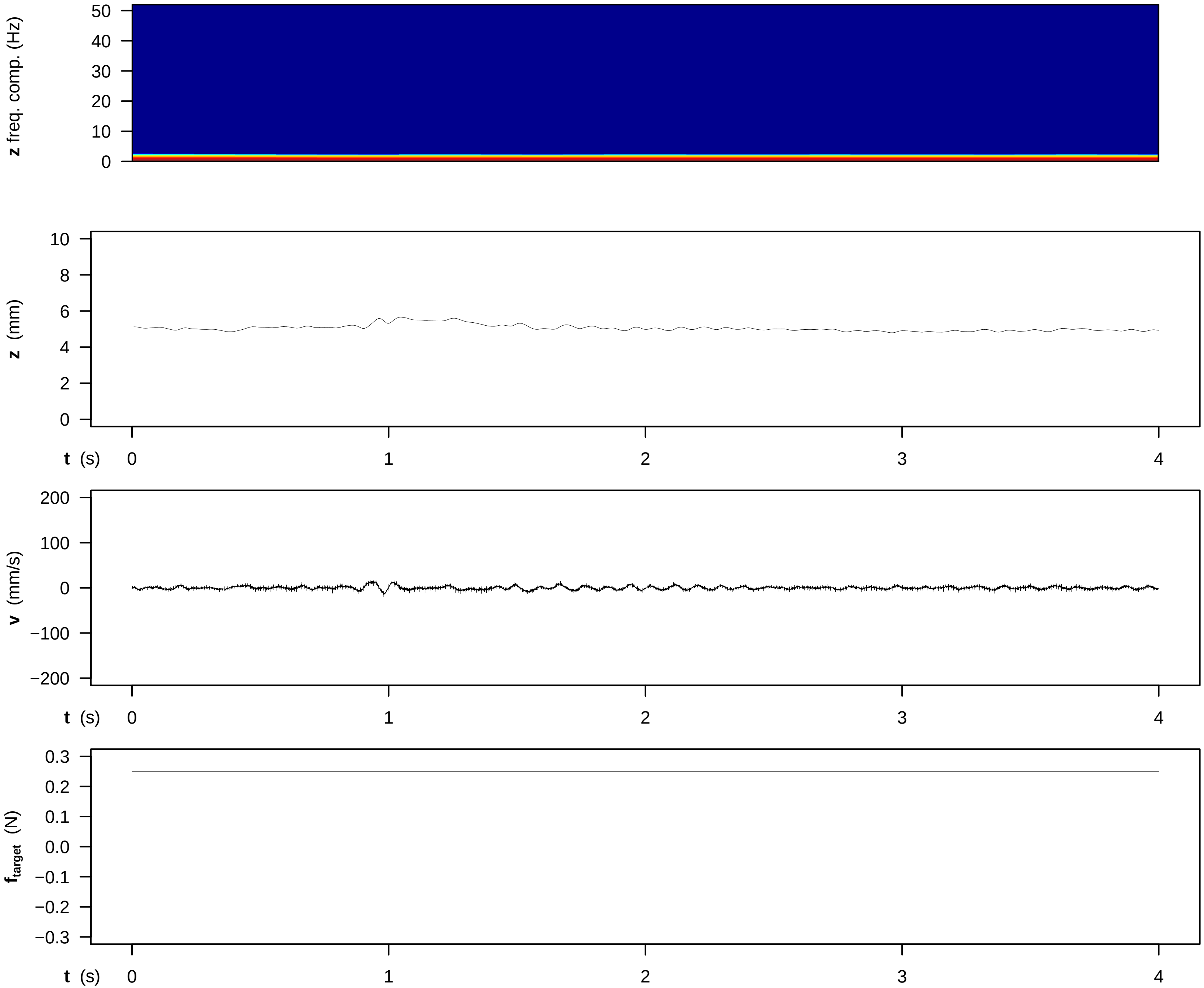

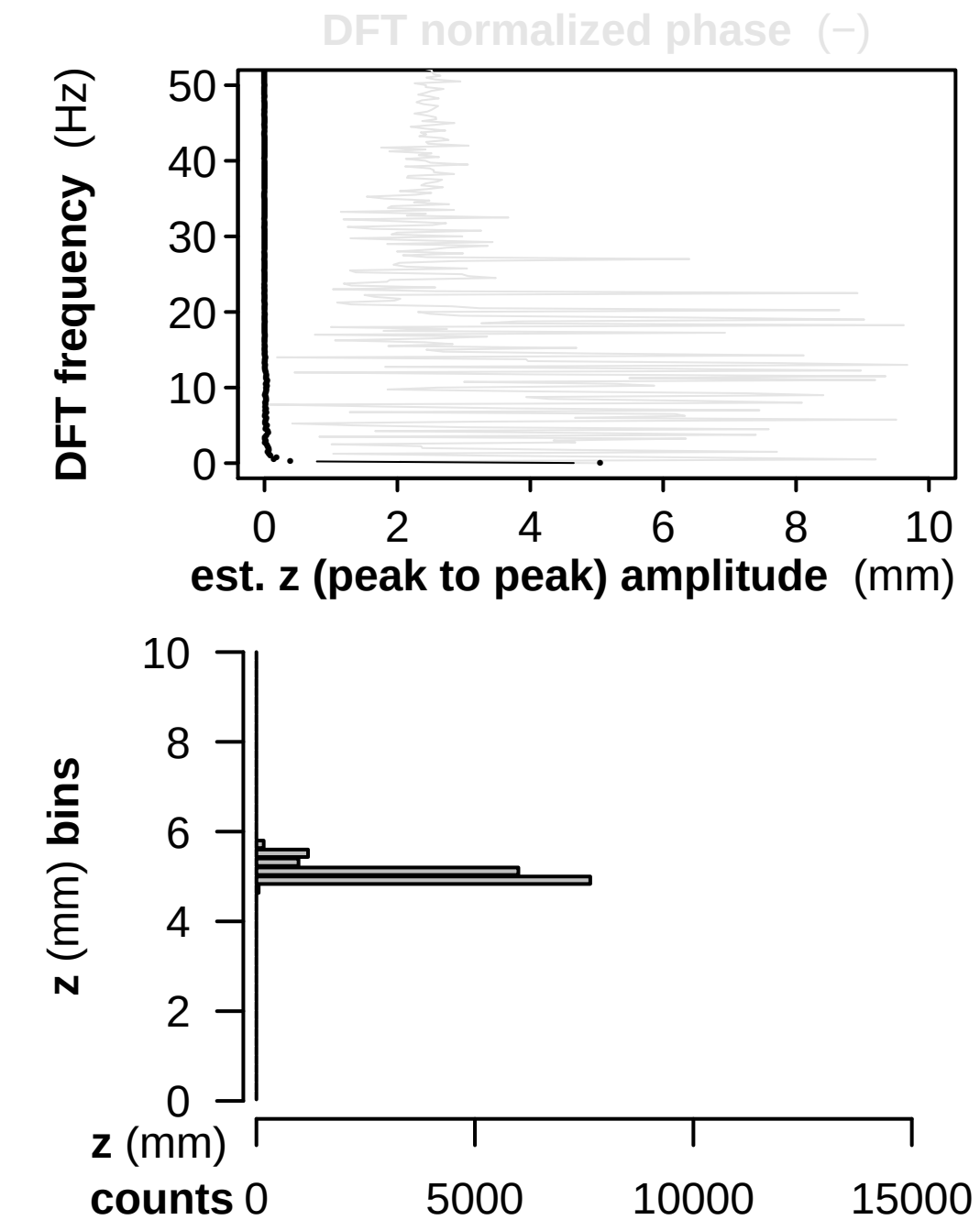

**SUBJECT 5 − RUN 02 − CONDITION 1,1**
**SC_180323_131519_0.AIFF**

```
z_min               : 4.79 mm
z_max               : 5.66 mm
z_travel_amplitude  : 0.87 mm

avg_abs_z_travel    : 3.97 mm/s

z_jarque-bera_jb    : 7625.25
z_jarque-bera_p     : 0.00e+00

z_lin_mod_est_slope: −0.08 mm/s
z_lin_mod_adj_R²    : 26 %

z_poly40_mod_adj_R²: 89 %

z_dft_ampl_thresh   : 0.010 mm
>=threshold_maxfreq: 14.00 Hz
```

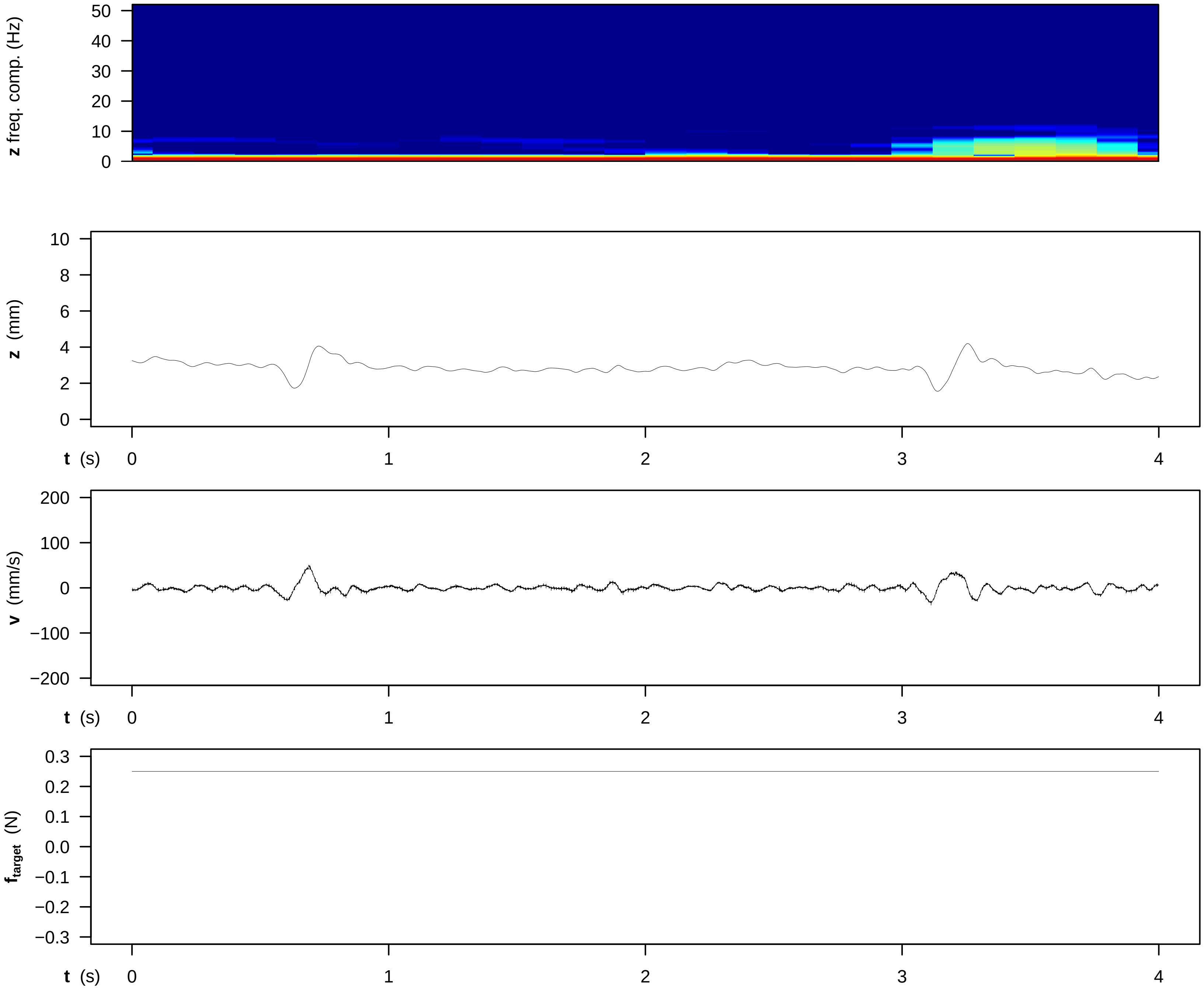

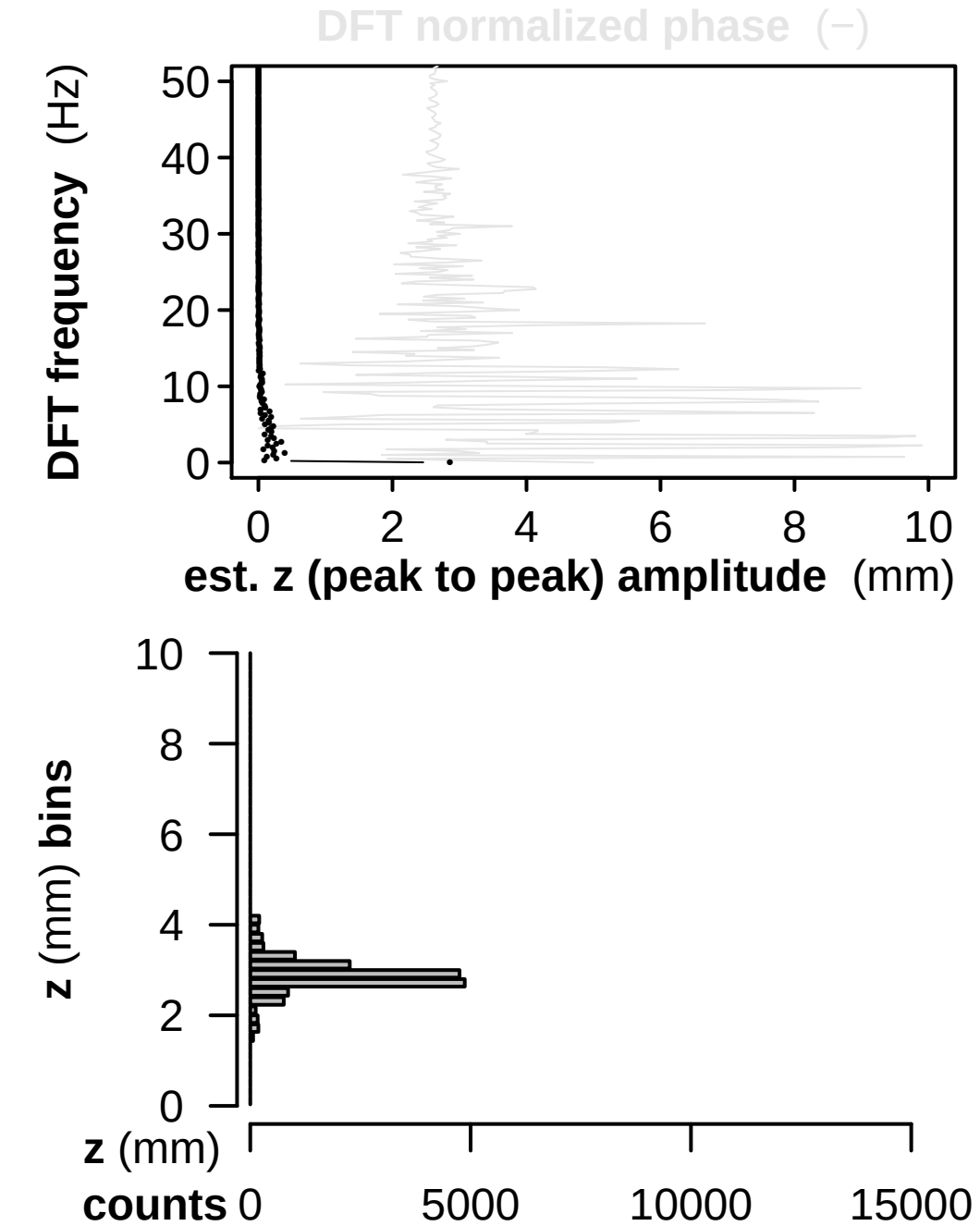

**SUBJECT 5 − RUN 10 − CONDITION 1,1**
**SC_180323_132104_0.AIFF**

```
z_min               : 1.55 mm
z_max               : 4.20 mm
z_travel_amplitude  : 2.64 mm

avg_abs_z_travel    : 6.19 mm/s

z_jarque-bera_jb    : 4943.85
z_jarque-bera_p     : 0.00e+00

z_lin_mod_est_slope: −0.10 mm/s
z_lin_mod_adj_R²    : 10 %

z_poly40_mod_adj_R²: 43 %

z_dft_ampl_thresh   : 0.010 mm
>=threshold_maxfreq: 22.25 Hz
```

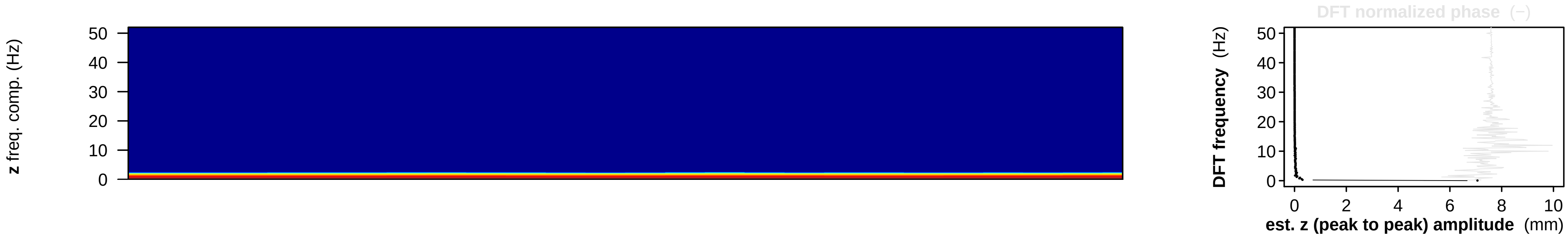

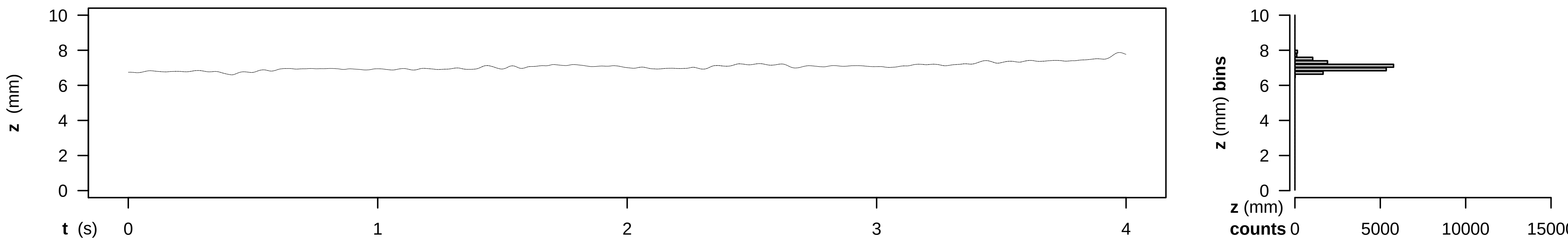

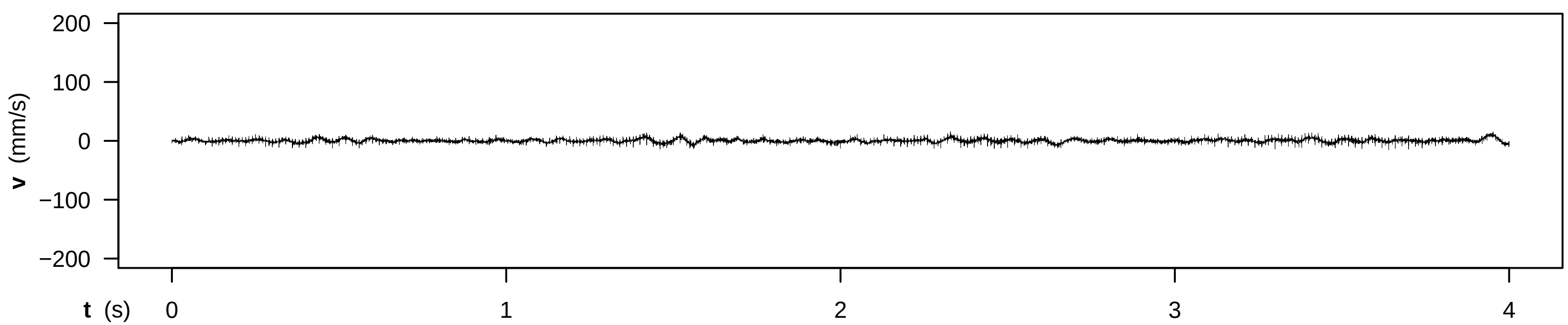

**SUBJECT 5 − RUN 19 − CONDITION 1,1**
**SC_180323_132649_0.AIFF**

```
z_min               : 6.60 mm
z_max               : 7.87 mm
z_travel_amplitude  : 1.27 mm

avg_abs_z_travel    : 5.56 mm/s

z_jarque-bera_jb    : 2670.62
z_jarque-bera_p     : 0.00e+00

z_lin_mod_est_slope: 0.16 mm/s
z_lin_mod_adj_R²    : 77 %

z_poly40_mod_adj_R²: 96 %

z_dft_ampl_thresh   : 0.010 mm
>=threshold_maxfreq: 19.25 Hz
```

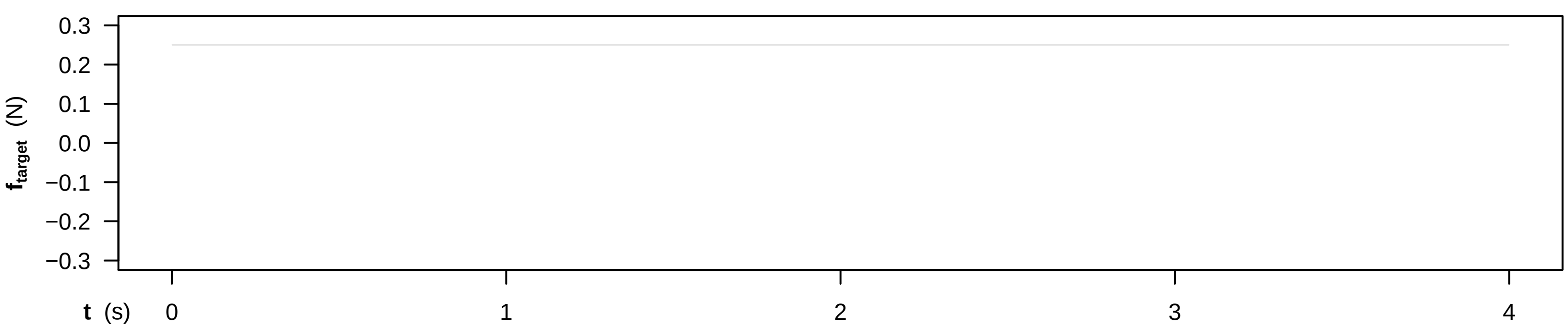

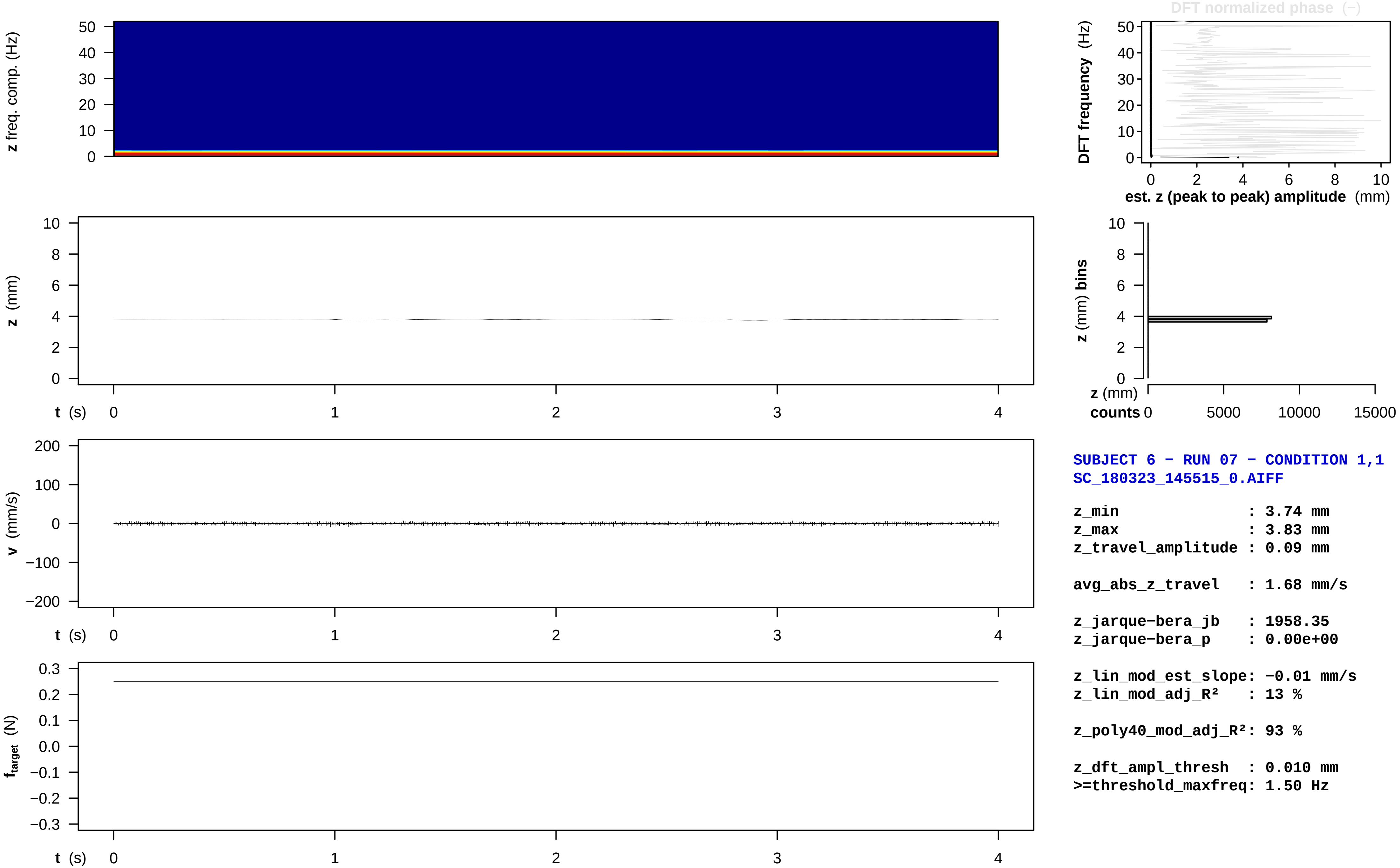

**SUBJECT 6 − RUN 07 − CONDITION 1,1**
**SC_180323_145515_0.AIFF**

```
z_min               : 3.74 mm
z_max               : 3.83 mm
z_travel_amplitude  : 0.09 mm

avg_abs_z_travel    : 1.68 mm/s

z_jarque-bera_jb    : 1958.35
z_jarque-bera_p     : 0.00e+00

z_lin_mod_est_slope: −0.01 mm/s
z_lin_mod_adj_R²    : 13 %

z_poly40_mod_adj_R²: 93 %

z_dft_ampl_thresh   : 0.010 mm
>=threshold_maxfreq: 1.50 Hz
```

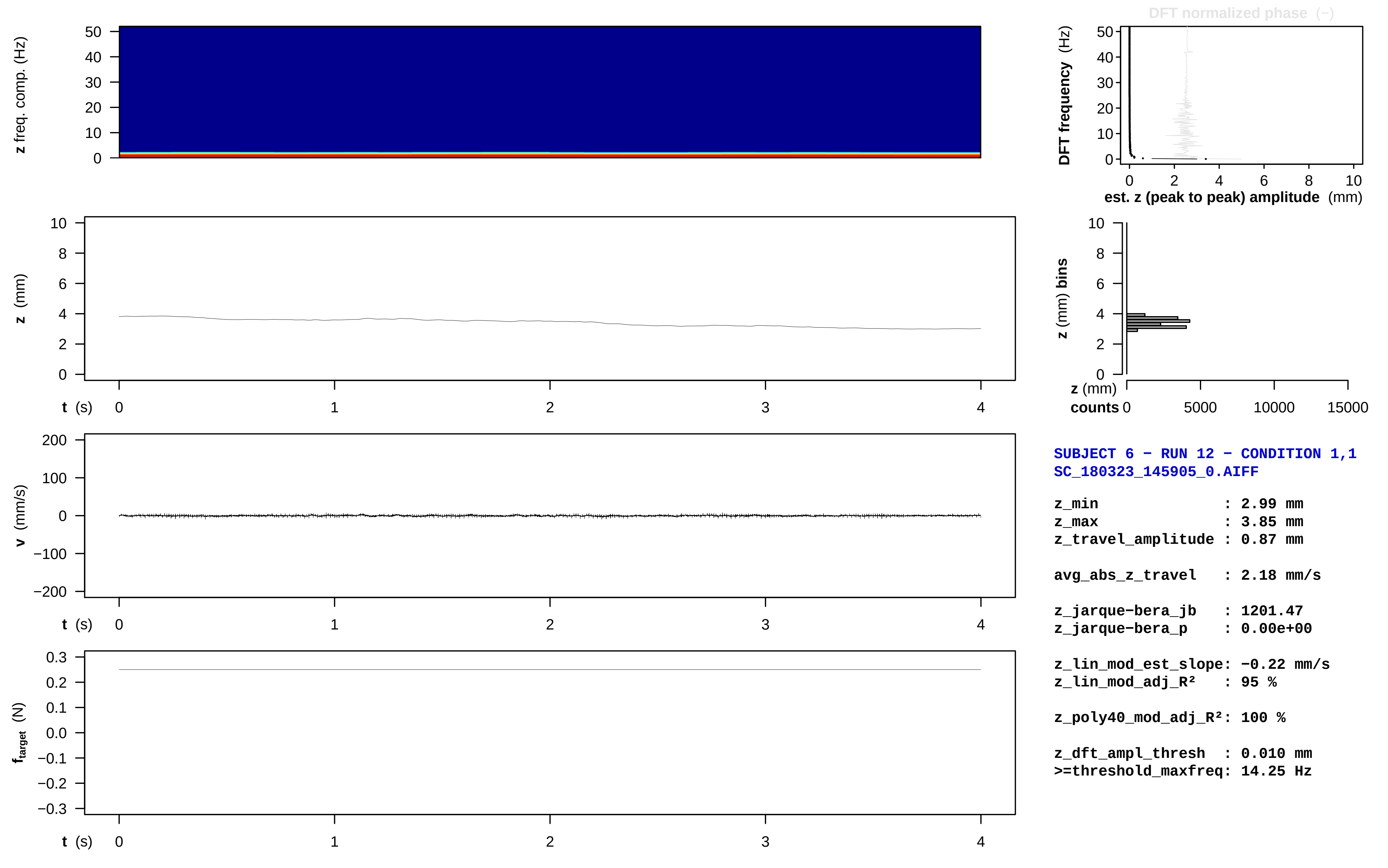

**SUBJECT 6 − RUN 12 − CONDITION 1,1**
**SC_180323_145905_0.AIFF**

```
z_min               : 2.99 mm
z_max               : 3.85 mm
z_travel_amplitude  : 0.87 mm

avg_abs_z_travel    : 2.18 mm/s

z_jarque-bera_jb    : 1201.47
z_jarque-bera_p     : 0.00e+00

z_lin_mod_est_slope: −0.22 mm/s
z_lin_mod_adj_R²    : 95 %

z_poly40_mod_adj_R²: 100 %

z_dft_ampl_thresh   : 0.010 mm
>=threshold_maxfreq: 14.25 Hz
```

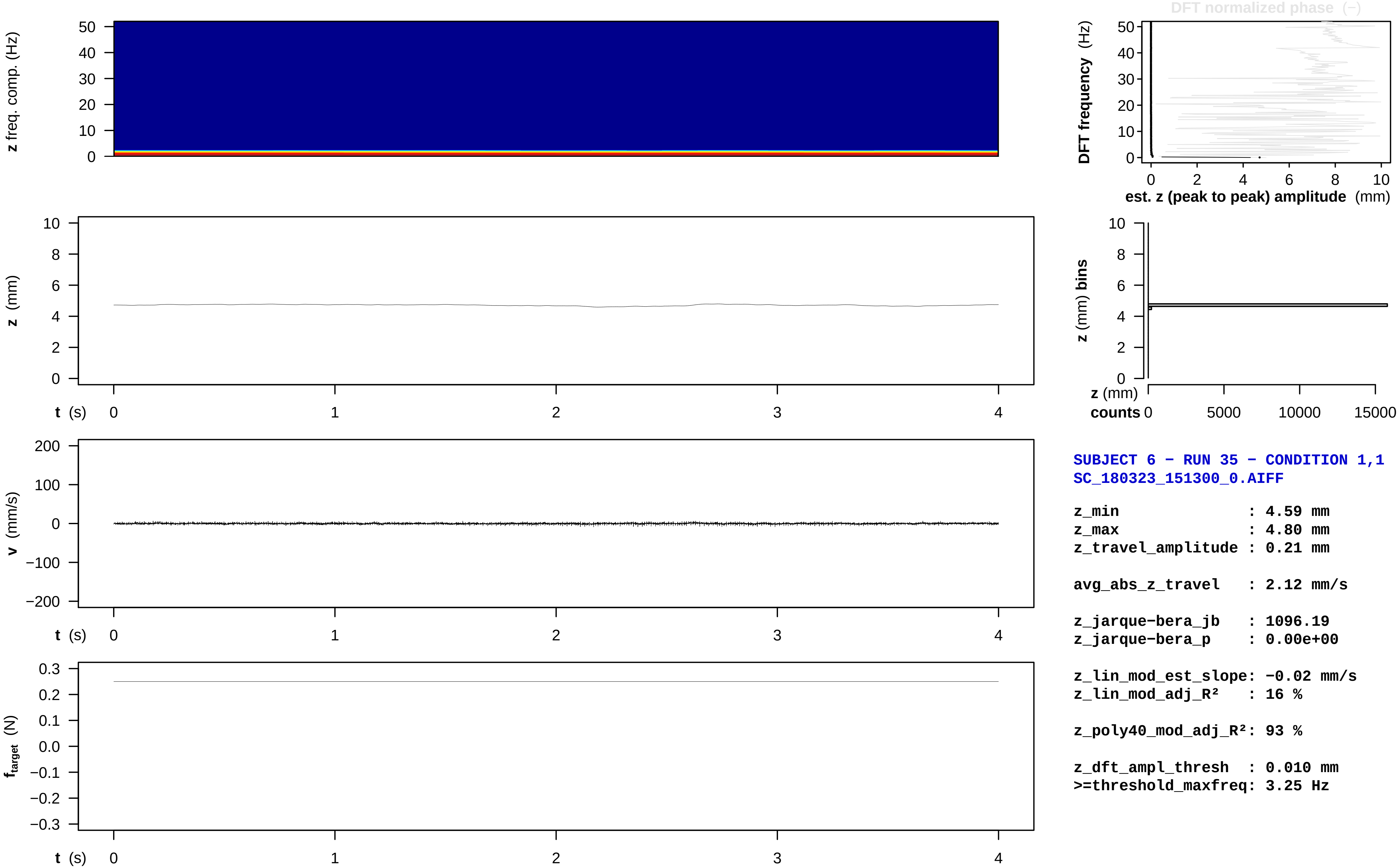

**SUBJECT 6 − RUN 35 − CONDITION 1,1**
**SC_180323_151300_0.AIFF**

```
z_min              : 4.59 mm
z_max              : 4.80 mm
z_travel_amplitude : 0.21 mm

avg_abs_z_travel   : 2.12 mm/s

z_jarque-bera_jb   : 1096.19
z_jarque-bera_p    : 0.00e+00

z_lin_mod_est_slope: −0.02 mm/s
z_lin_mod_adj_R²   : 16 %

z_poly40_mod_adj_R²: 93 %

z_dft_ampl_thresh  : 0.010 mm
>=threshold_maxfreq: 3.25 Hz
```

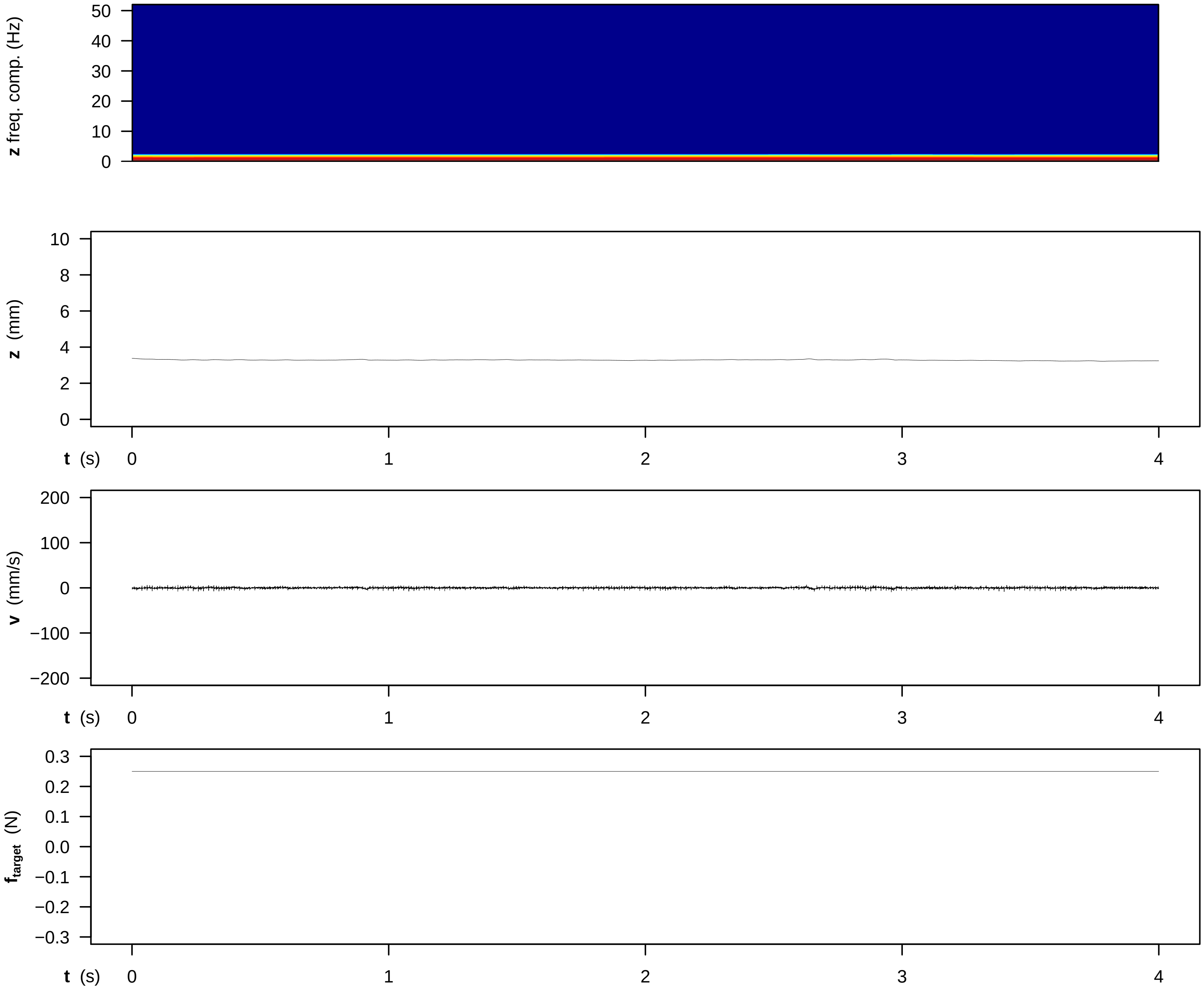

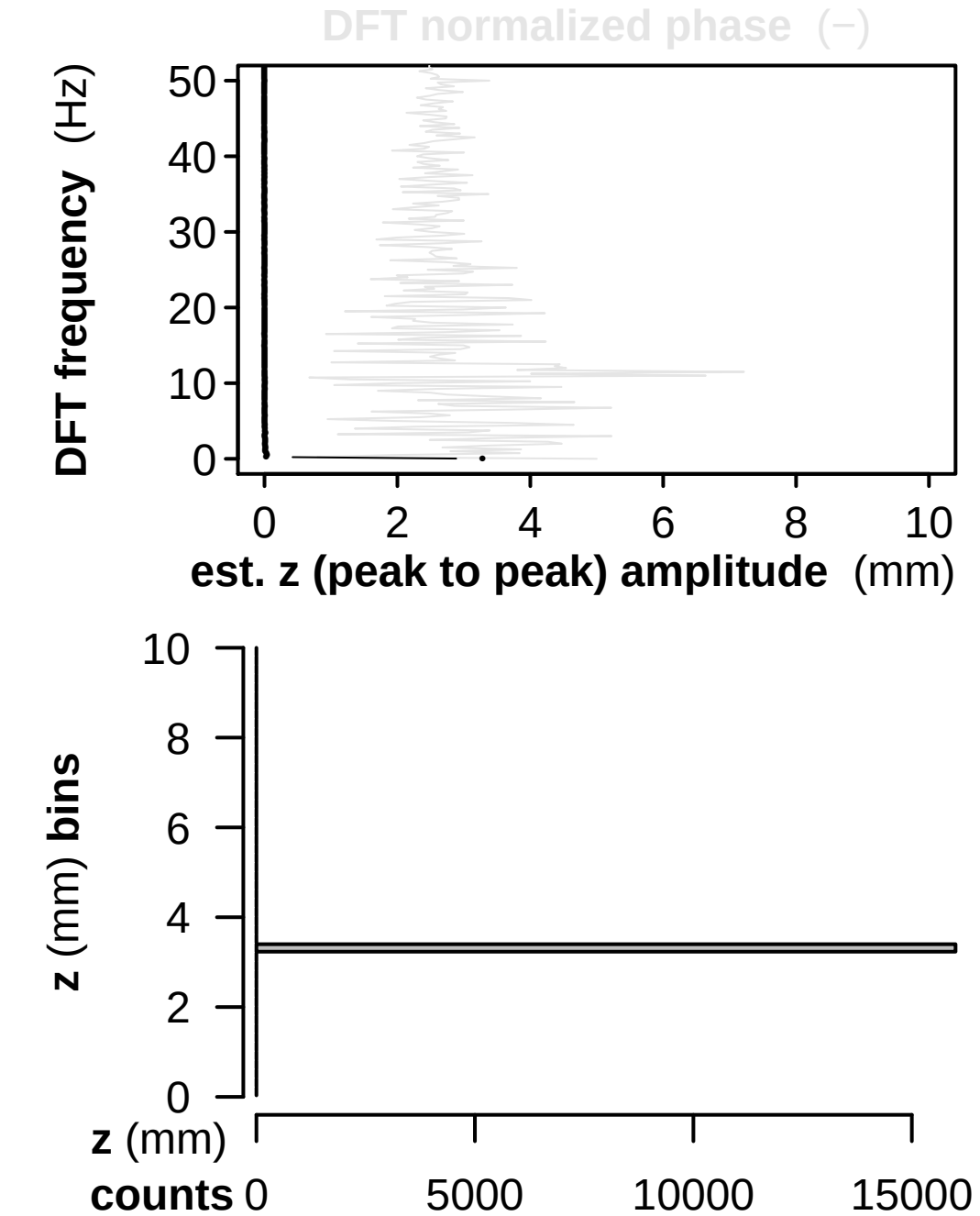

**SUBJECT 7 − RUN 10 − CONDITION 1,1**
**SC_180323_154058_0.AIFF**

```
z_min              : 3.21 mm
z_max              : 3.38 mm
z_travel_amplitude : 0.17 mm

avg_abs_z_travel   : 1.83 mm/s

z_jarque-bera_jb   : 281.45
z_jarque-bera_p    : 0.00e+00

z_lin_mod_est_slope: −0.01 mm/s
z_lin_mod_adj_R²   : 35 %

z_poly40_mod_adj_R²: 86 %

z_dft_ampl_thresh  : 0.010 mm
>=threshold_maxfreq: 3.50 Hz
```

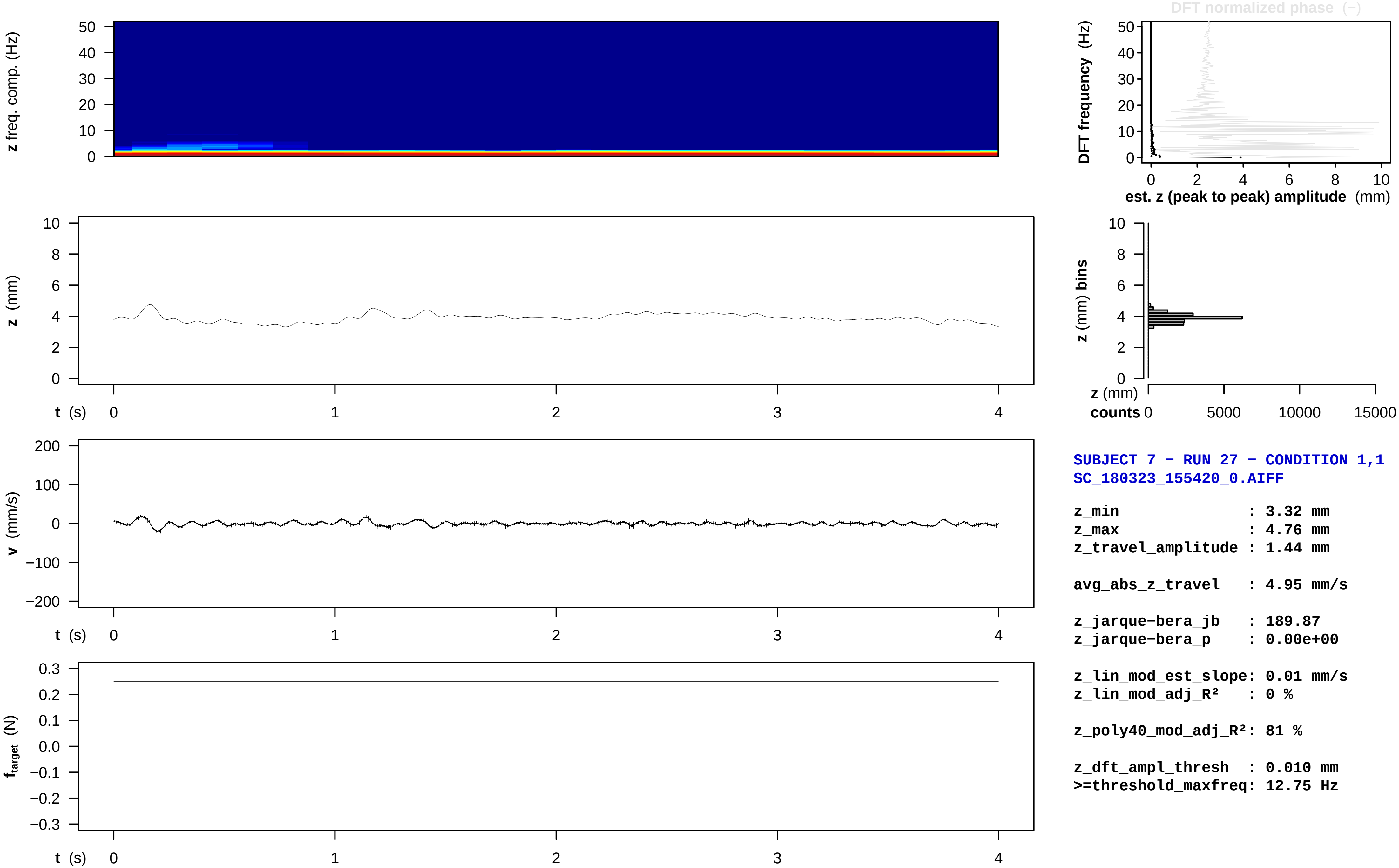

**SUBJECT 7 − RUN 27 − CONDITION 1,1**
**SC_180323_155420_0.AIFF**

```
z_min              : 3.32 mm
z_max              : 4.76 mm
z_travel_amplitude : 1.44 mm

avg_abs_z_travel   : 4.95 mm/s

z_jarque-bera_jb   : 189.87
z_jarque-bera_p    : 0.00e+00

z_lin_mod_est_slope: 0.01 mm/s
z_lin_mod_adj_R²   : 0 %

z_poly40_mod_adj_R²: 81 %

z_dft_ampl_thresh  : 0.010 mm
>=threshold_maxfreq: 12.75 Hz
```

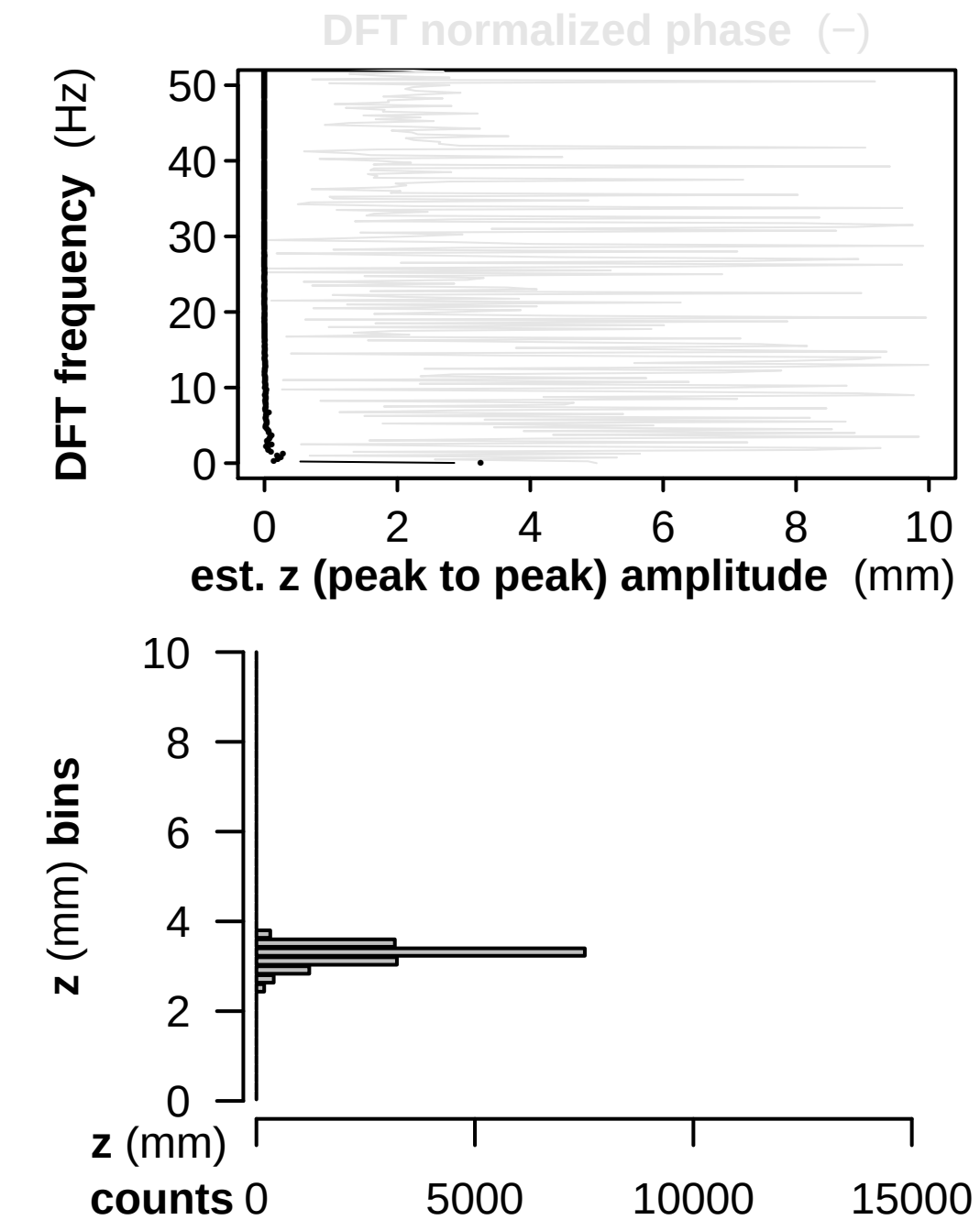

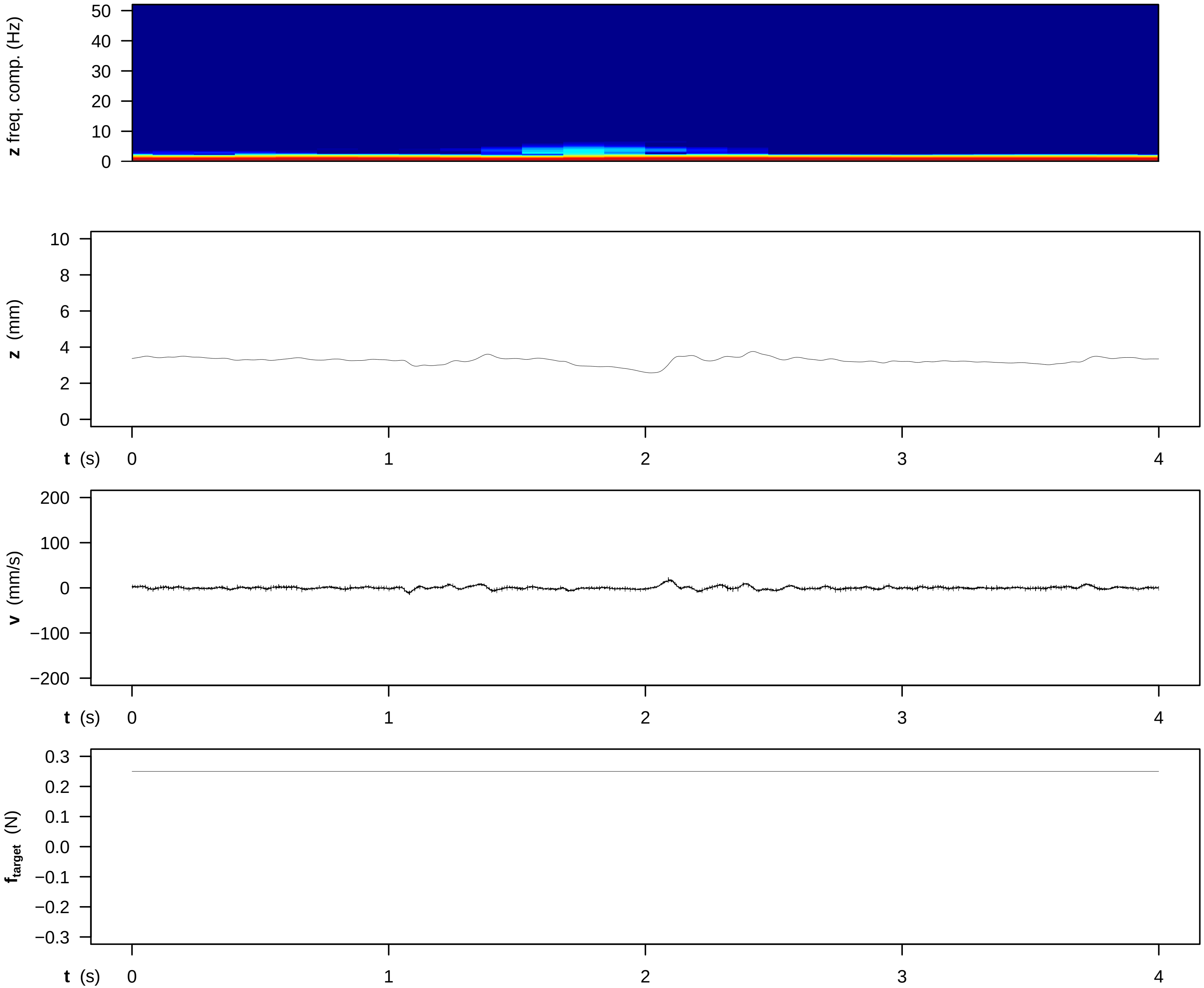

**SUBJECT 7 − RUN 33 − CONDITION 1,1**
**SC_180323_155758_0.AIFF**

```
z_min               : 2.57 mm
z_max               : 3.76 mm
z_travel_amplitude  : 1.19 mm

avg_abs_z_travel    : 2.79 mm/s

z_jarque-bera_jb    : 3994.23
z_jarque-bera_p     : 0.00e+00

z_lin_mod_est_slope: −0.03 mm/s
z_lin_mod_adj_R²    : 2 %

z_poly40_mod_adj_R²: 80 %

z_dft_ampl_thresh   : 0.010 mm
>=threshold_maxfreq: 14.25 Hz
```

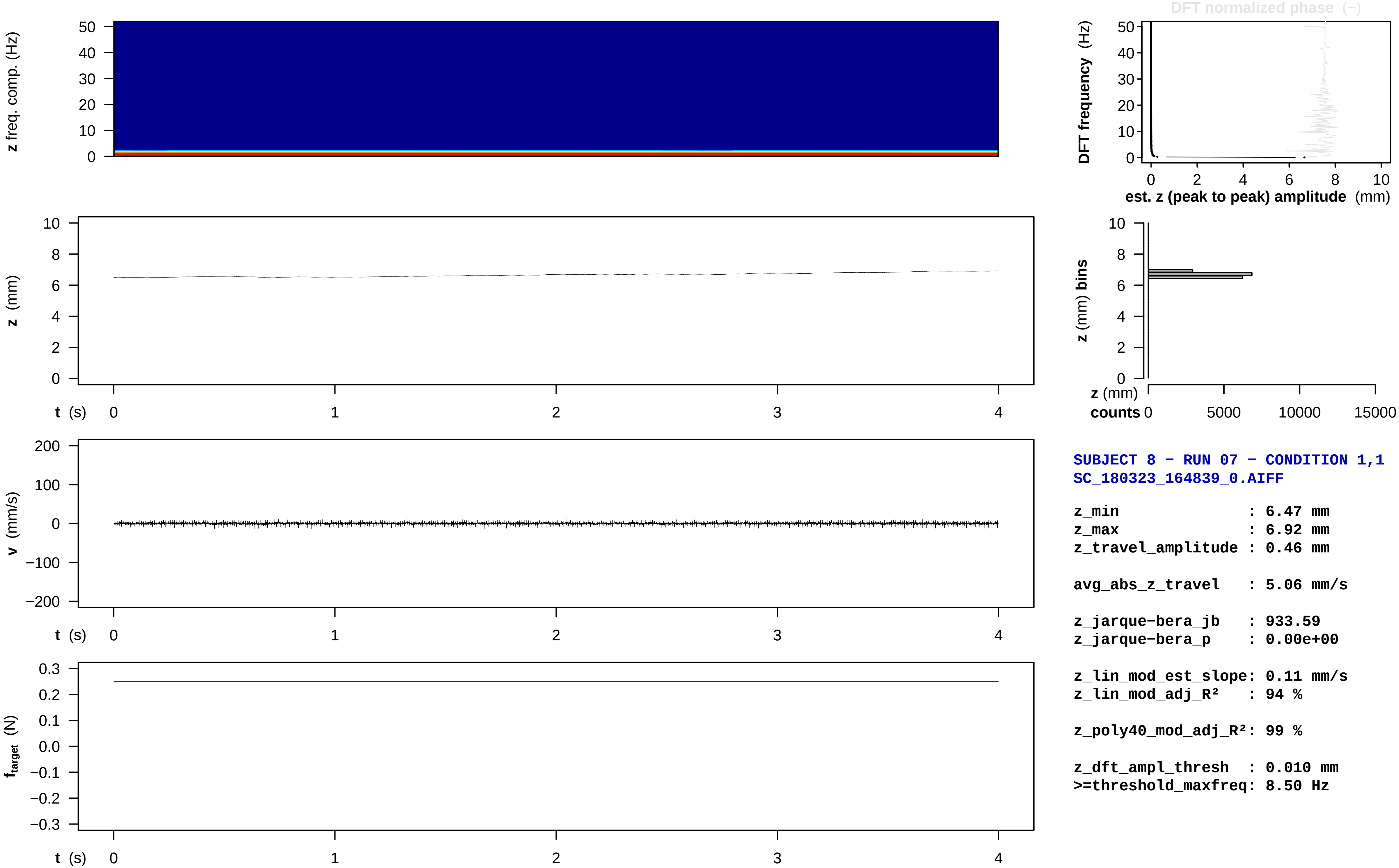

**SUBJECT 8 − RUN 07 − CONDITION 1,1**
**SC_180323_164839_0.AIFF**

```
z_min               : 6.47 mm
z_max               : 6.92 mm
z_travel_amplitude  : 0.46 mm

avg_abs_z_travel    : 5.06 mm/s

z_jarque-bera_jb    : 933.59
z_jarque-bera_p     : 0.00e+00

z_lin_mod_est_slope: 0.11 mm/s
z_lin_mod_adj_R²    : 94 %

z_poly40_mod_adj_R²: 99 %

z_dft_ampl_thresh   : 0.010 mm
>=threshold_maxfreq: 8.50 Hz
```

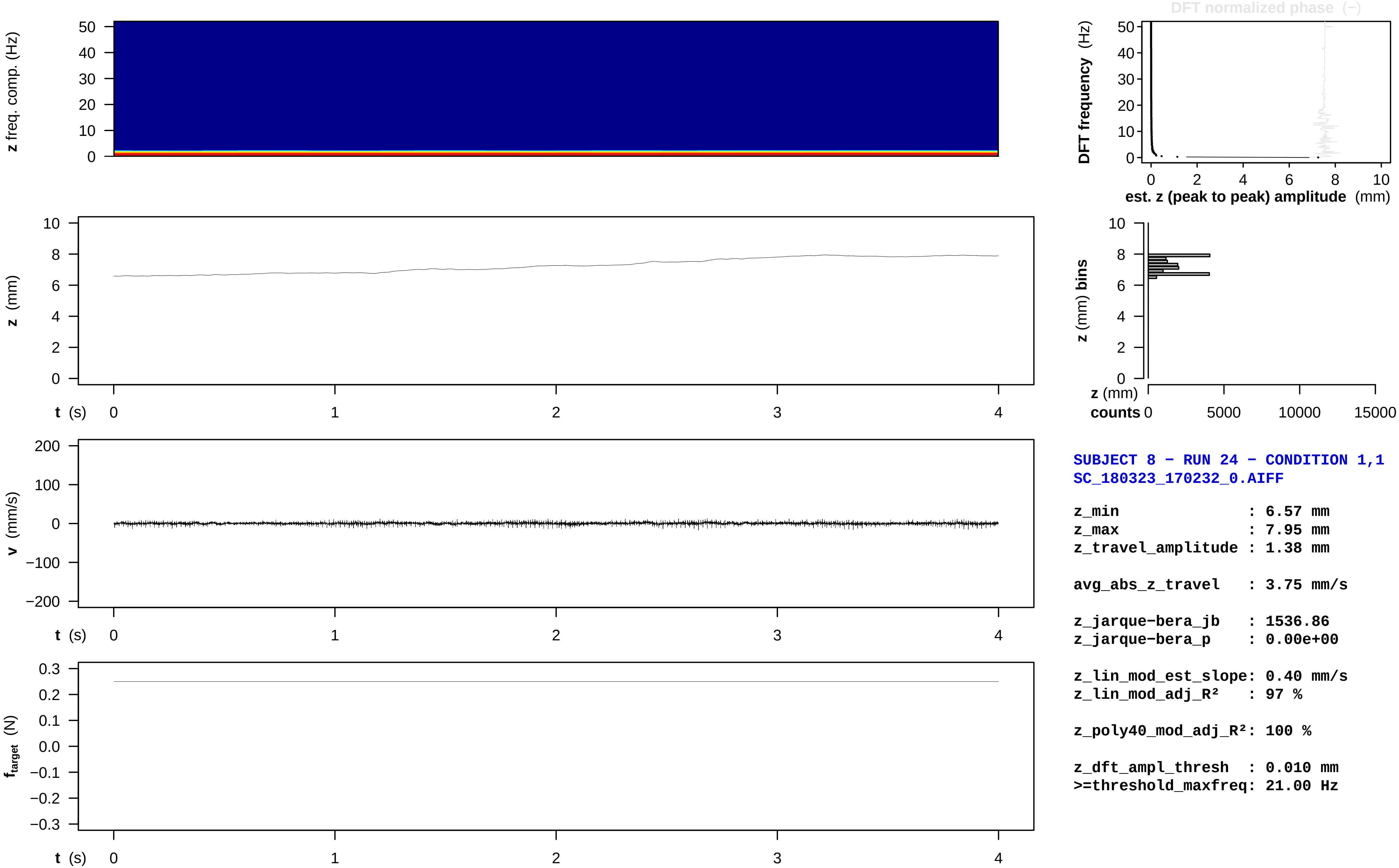

**SUBJECT 8 − RUN 24 − CONDITION 1,1**
**SC_180323_170232_0.AIFF**

```
z_min               : 6.57 mm
z_max               : 7.95 mm
z_travel_amplitude  : 1.38 mm

avg_abs_z_travel    : 3.75 mm/s

z_jarque-bera_jb    : 1536.86
z_jarque-bera_p     : 0.00e+00

z_lin_mod_est_slope: 0.40 mm/s
z_lin_mod_adj_R²    : 97 %

z_poly40_mod_adj_R²: 100 %

z_dft_ampl_thresh   : 0.010 mm
>=threshold_maxfreq: 21.00 Hz
```

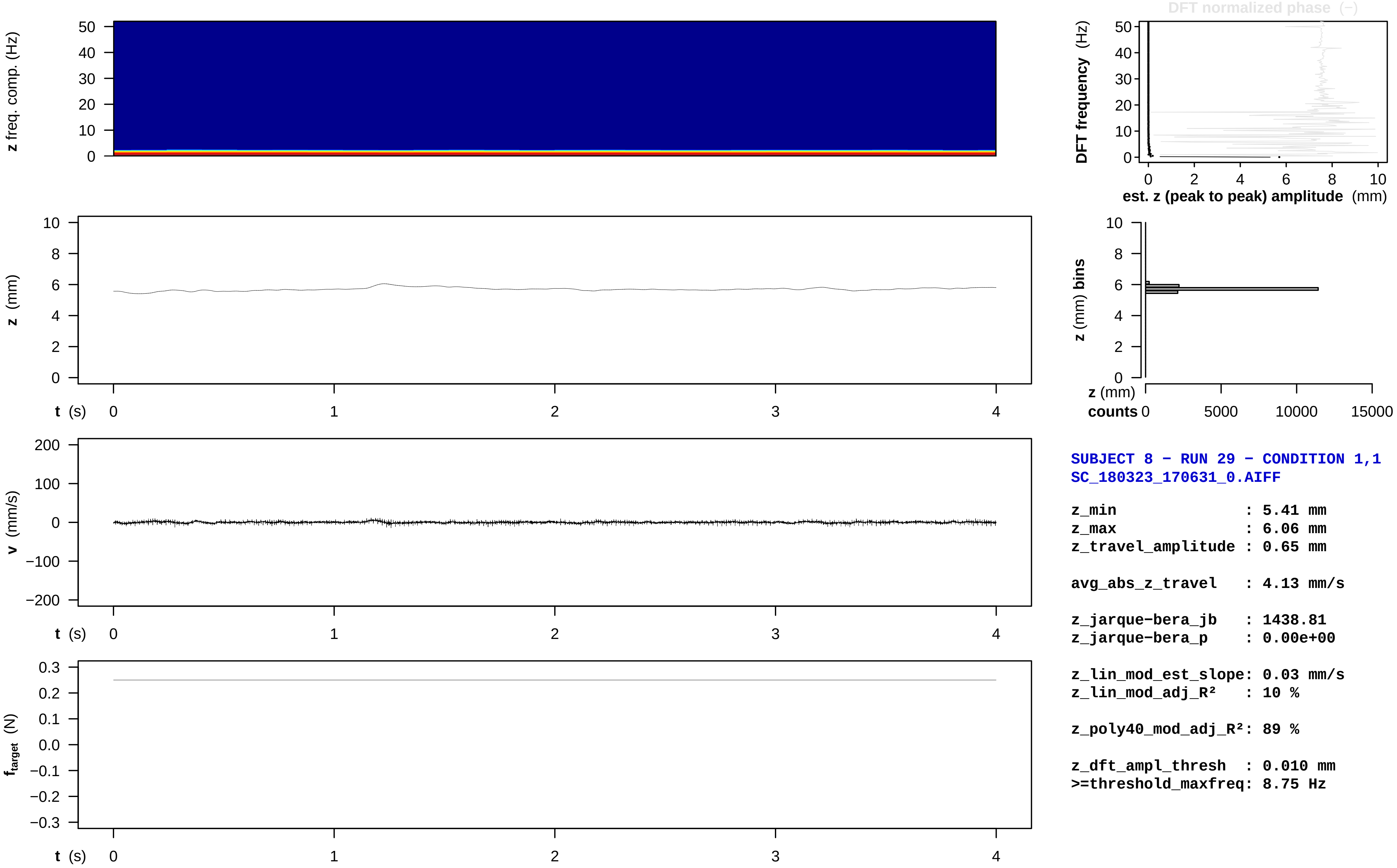

**SUBJECT 8 − RUN 29 − CONDITION 1,1**
**SC_180323_170631_0.AIFF**

```
z_min               : 5.41 mm
z_max               : 6.06 mm
z_travel_amplitude  : 0.65 mm

avg_abs_z_travel    : 4.13 mm/s

z_jarque-bera_jb    : 1438.81
z_jarque-bera_p     : 0.00e+00

z_lin_mod_est_slope: 0.03 mm/s
z_lin_mod_adj_R²    : 10 %

z_poly40_mod_adj_R²: 89 %

z_dft_ampl_thresh   : 0.010 mm
>=threshold_maxfreq: 8.75 Hz
```

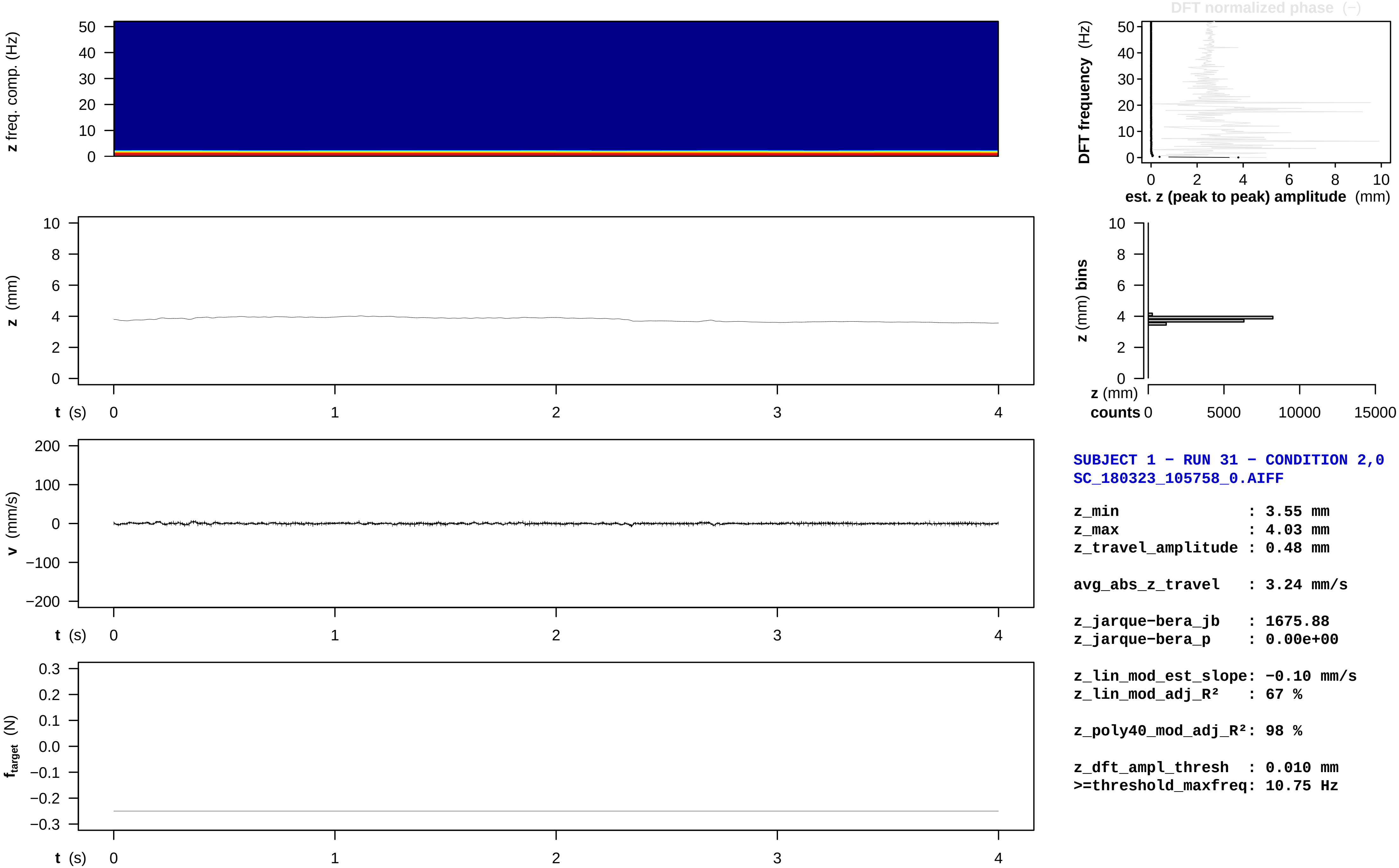

**SUBJECT 1 − RUN 31 − CONDITION 2,0**
**SC_180323_105758_0.AIFF**

```
z_min              : 3.55 mm
z_max              : 4.03 mm
z_travel_amplitude : 0.48 mm

avg_abs_z_travel   : 3.24 mm/s

z_jarque-bera_jb   : 1675.88
z_jarque-bera_p    : 0.00e+00

z_lin_mod_est_slope: −0.10 mm/s
z_lin_mod_adj_R²   : 67 %

z_poly40_mod_adj_R²: 98 %

z_dft_ampl_thresh  : 0.010 mm
>=threshold_maxfreq: 10.75 Hz
```

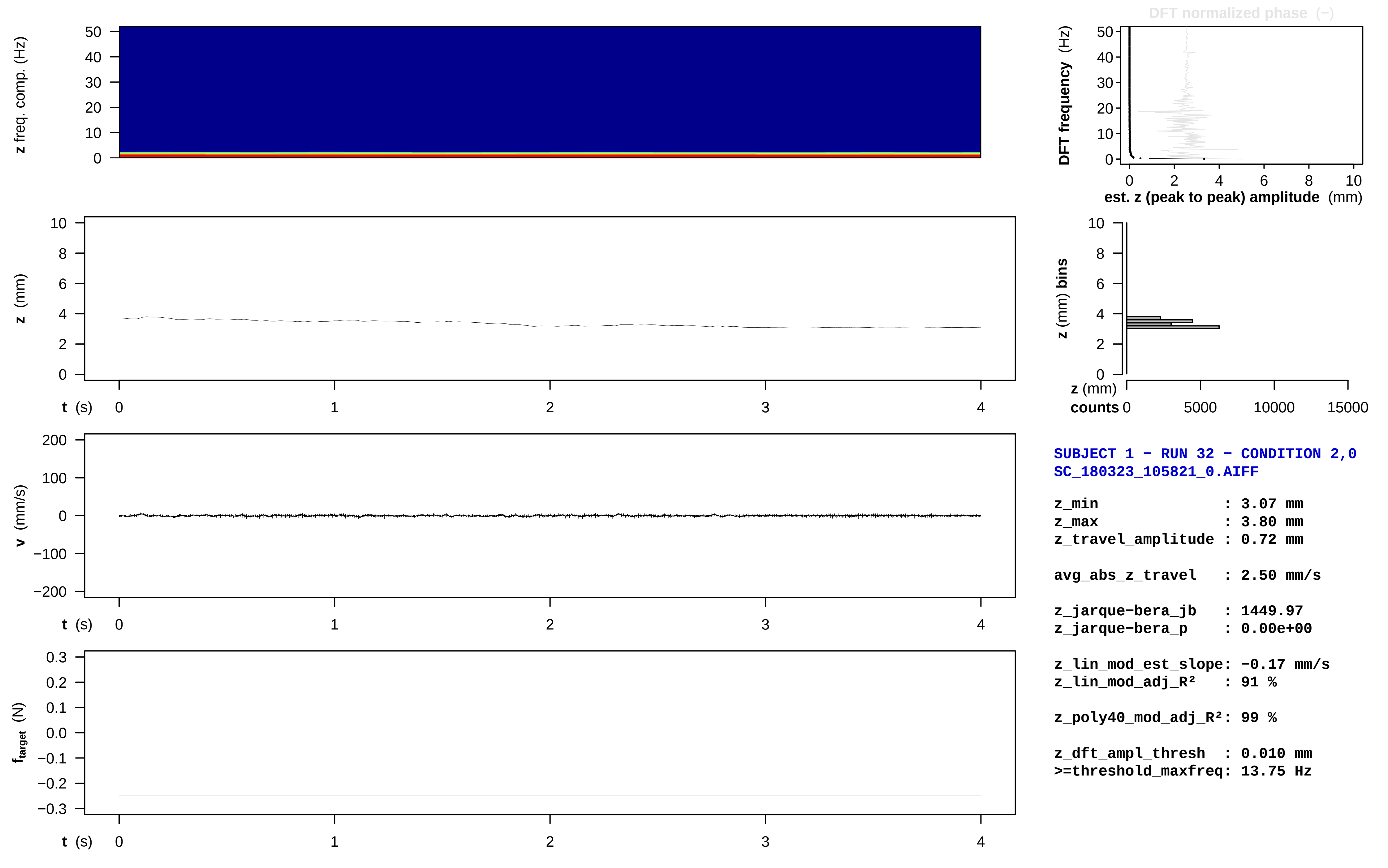

**SUBJECT 1 − RUN 32 − CONDITION 2,0**
**SC_180323_105821_0.AIFF**

```
z_min              : 3.07 mm
z_max              : 3.80 mm
z_travel_amplitude : 0.72 mm

avg_abs_z_travel   : 2.50 mm/s

z_jarque-bera_jb   : 1449.97
z_jarque-bera_p    : 0.00e+00

z_lin_mod_est_slope: −0.17 mm/s
z_lin_mod_adj_R²   : 91 %

z_poly40_mod_adj_R²: 99 %

z_dft_ampl_thresh  : 0.010 mm
>=threshold_maxfreq: 13.75 Hz
```

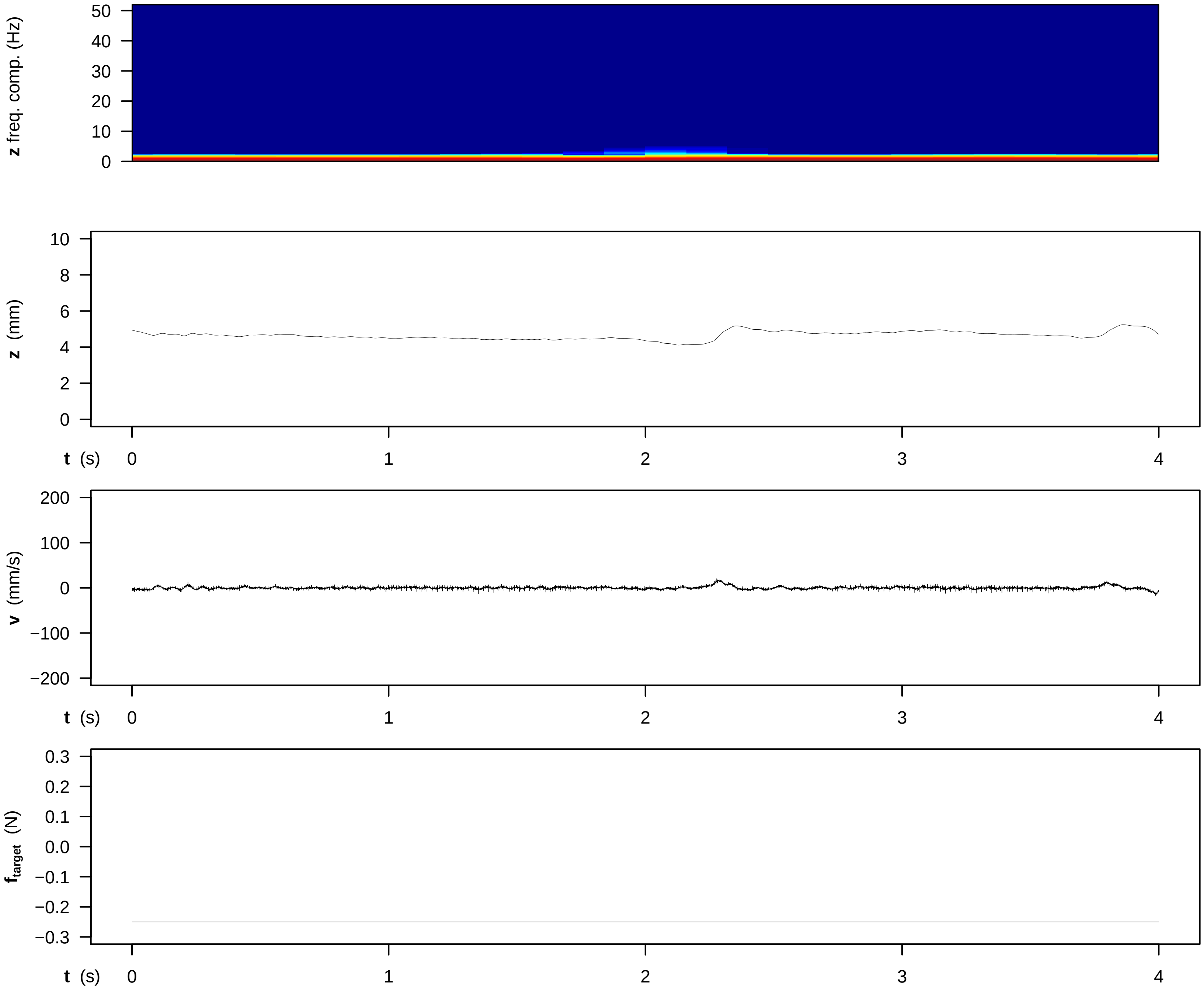

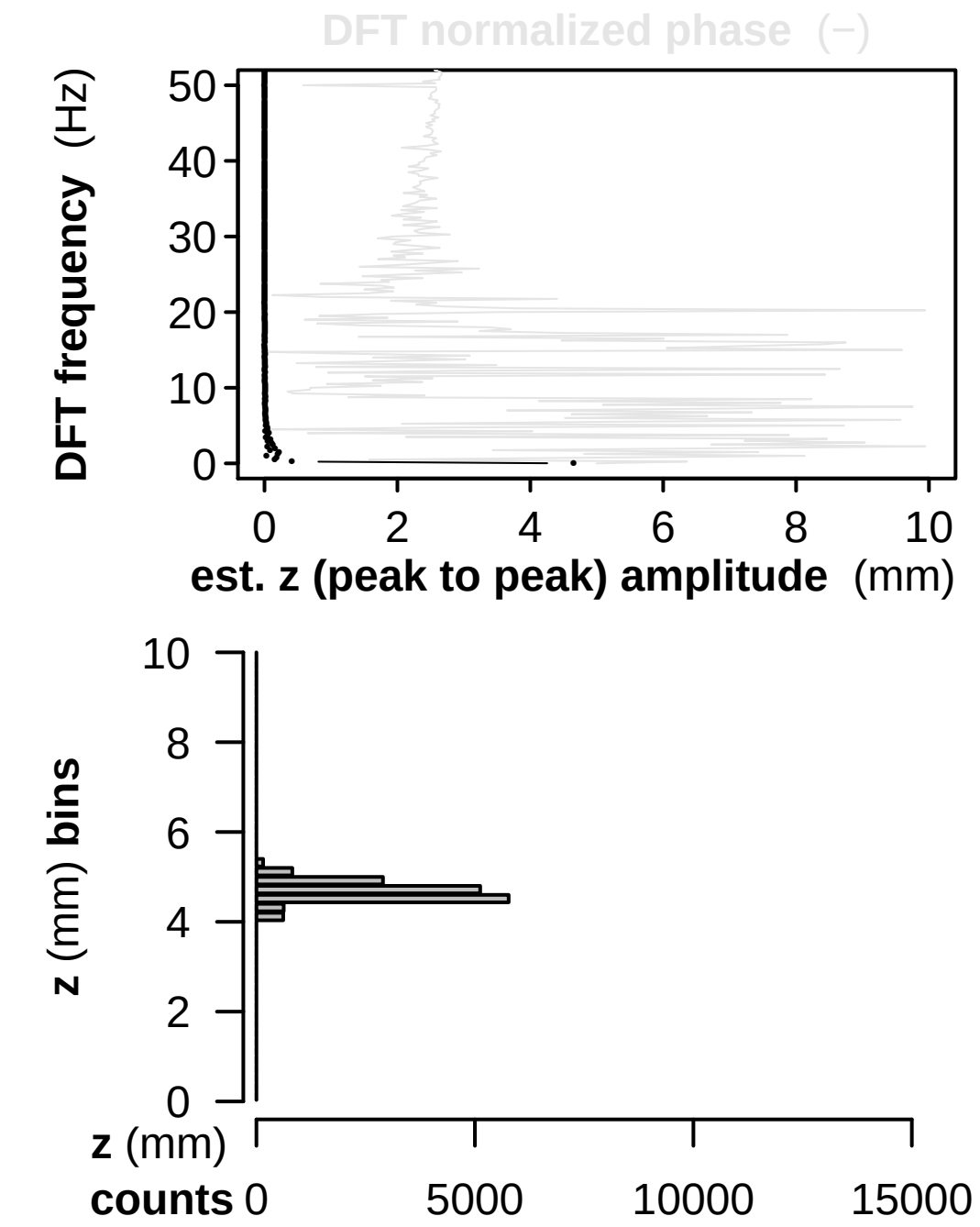

**SUBJECT 1 − RUN 33 − CONDITION 2,0**
**SC_180323_105843_0.AIFF**

```
z_min               : 4.11 mm
z_max               : 5.24 mm
z_travel_amplitude  : 1.13 mm

avg_abs_z_travel    : 2.90 mm/s

z_jarque-bera_jb    : 70.93
z_jarque-bera_p     : 4.44e−16

z_lin_mod_est_slope: 0.07 mm/s
z_lin_mod_adj_R²    : 14 %

z_poly40_mod_adj_R²: 87 %

z_dft_ampl_thresh   : 0.010 mm
>=threshold_maxfreq: 14.50 Hz
```

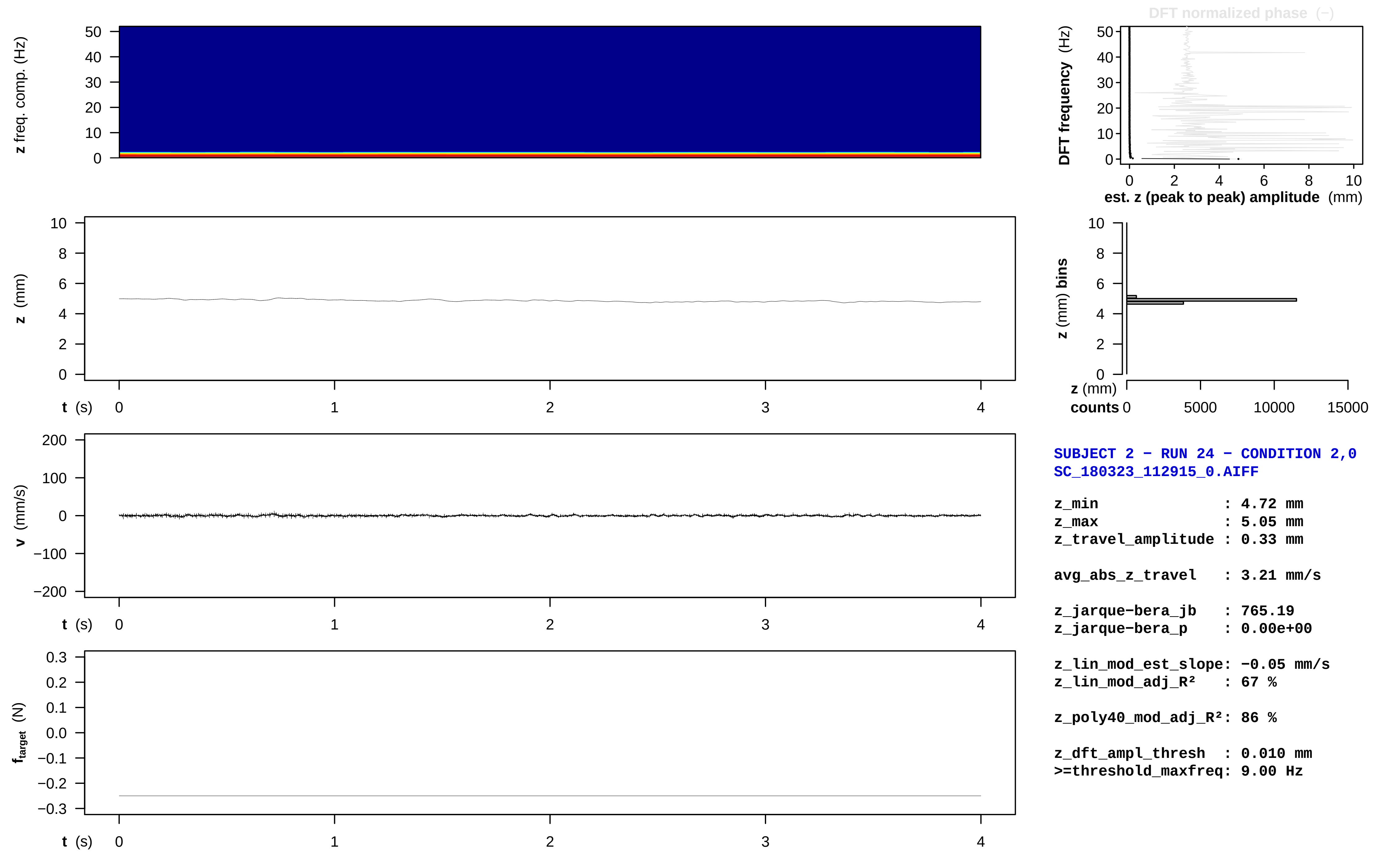
DFT normalized phase (–)
z freq. comp. (Hz)
DFT frequency (Hz)
est. z (peak to peak) amplitude (mm)
z (mm)
z (mm) bins
z (mm)
counts
t (s)
v (mm/s)
f target (N)
SUBJECT 2 – RUN 24 – CONDITION 2,0
SC_180323_112915_0.AIFF
z_min : 4.72 mm
z_max : 5.05 mm
z_travel_amplitude : 0.33 mm
avg_abs_z_travel : 3.21 mm/s
z_jarque-bera_jb : 765.19
z_jarque-bera_p : 0.00e+00
z_lin_mod_est_slope: −0.05 mm/s
z_lin_mod_adj_R² : 67 %
z_poly40_mod_adj_R²: 86 %
z_dft_ampl_thresh : 0.010 mm
>=threshold_maxfreq: 9.00 Hz

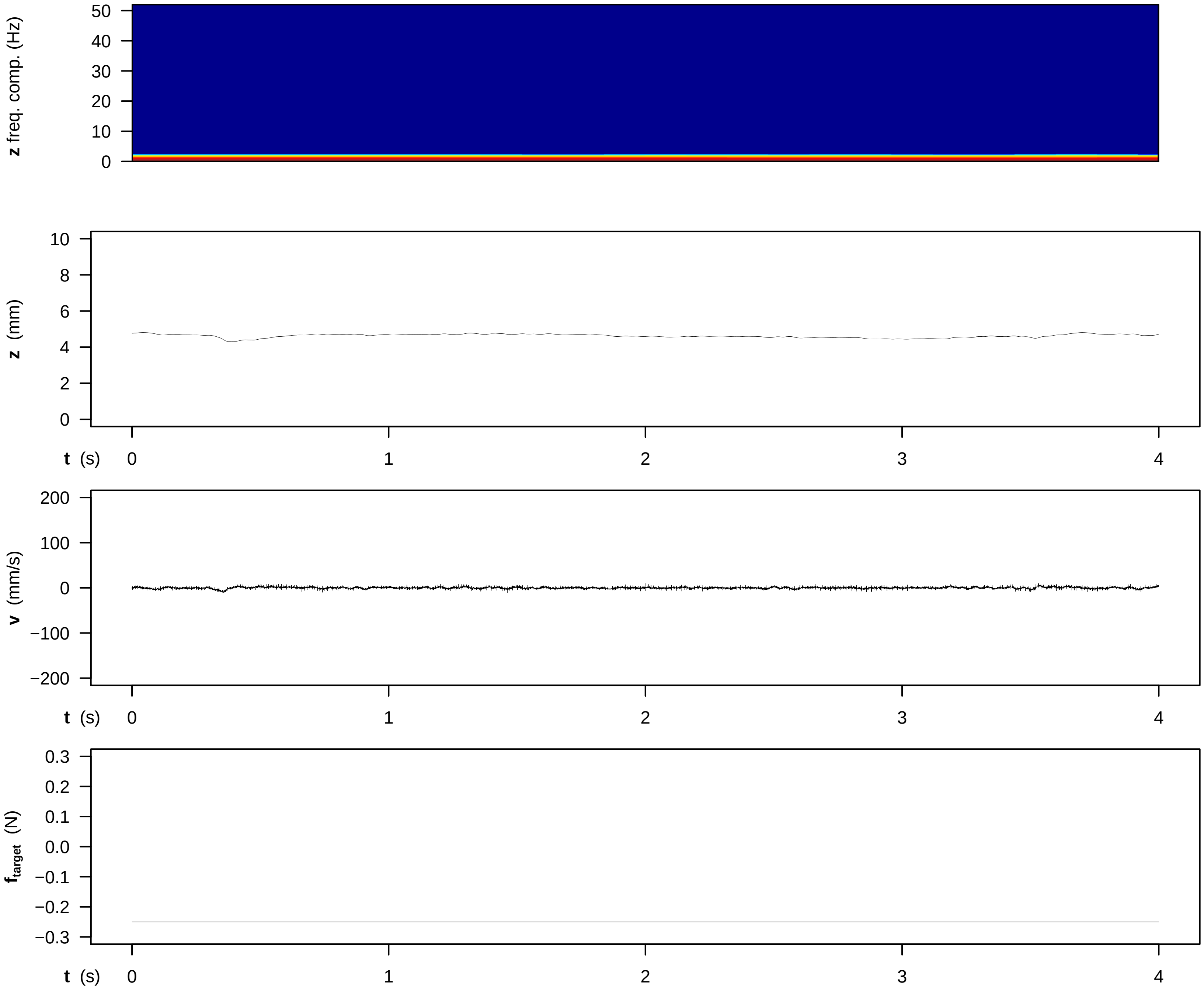

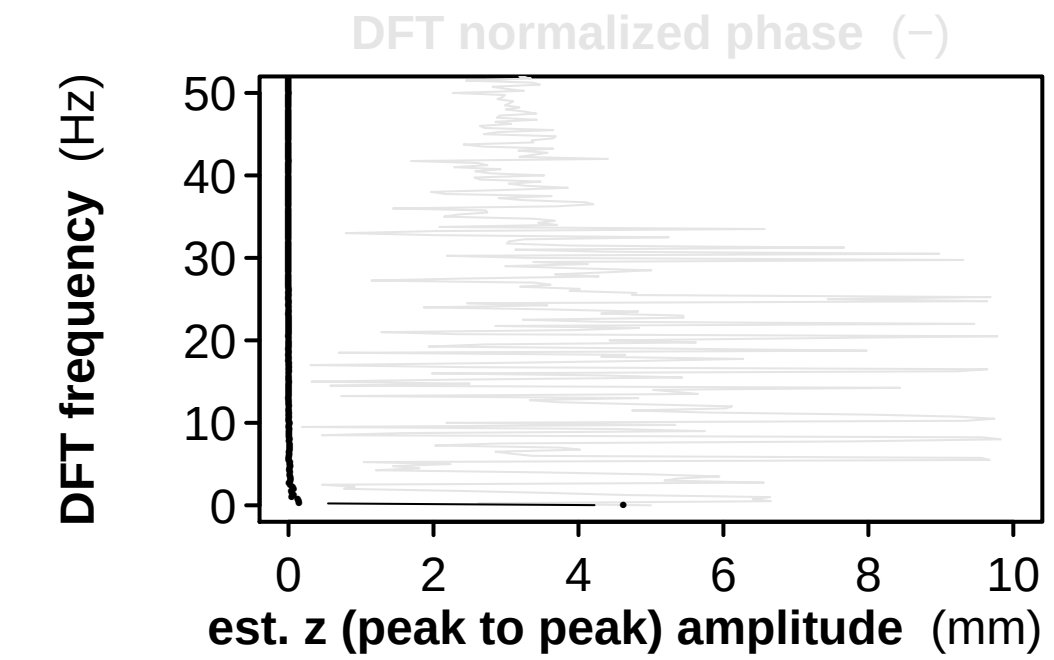

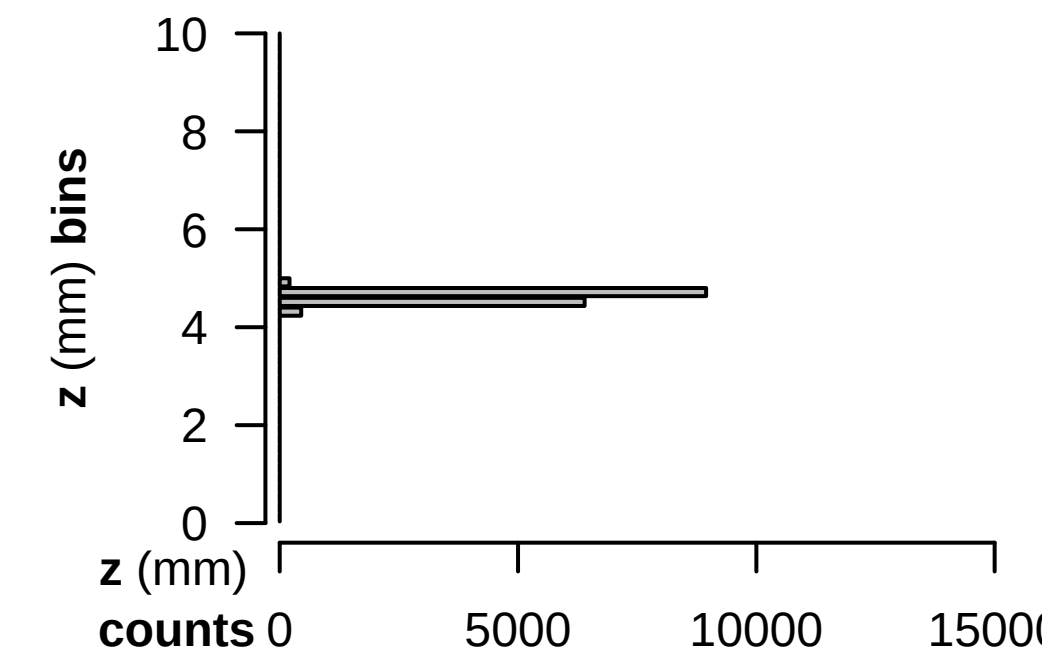

**SUBJECT 2 − RUN 28 − CONDITION 2,0**
**SC_180323_113049_0.AIFF**

```
z_min              : 4.30 mm
z_max              : 4.81 mm
z_travel_amplitude : 0.51 mm

avg_abs_z_travel   : 2.31 mm/s

z_jarque-bera_jb   : 927.39
z_jarque-bera_p    : 0.00e+00

z_lin_mod_est_slope: −0.02 mm/s
z_lin_mod_adj_R²   : 4 %

z_poly40_mod_adj_R²: 92 %

z_dft_ampl_thresh  : 0.010 mm
>=threshold_maxfreq: 10.00 Hz
```

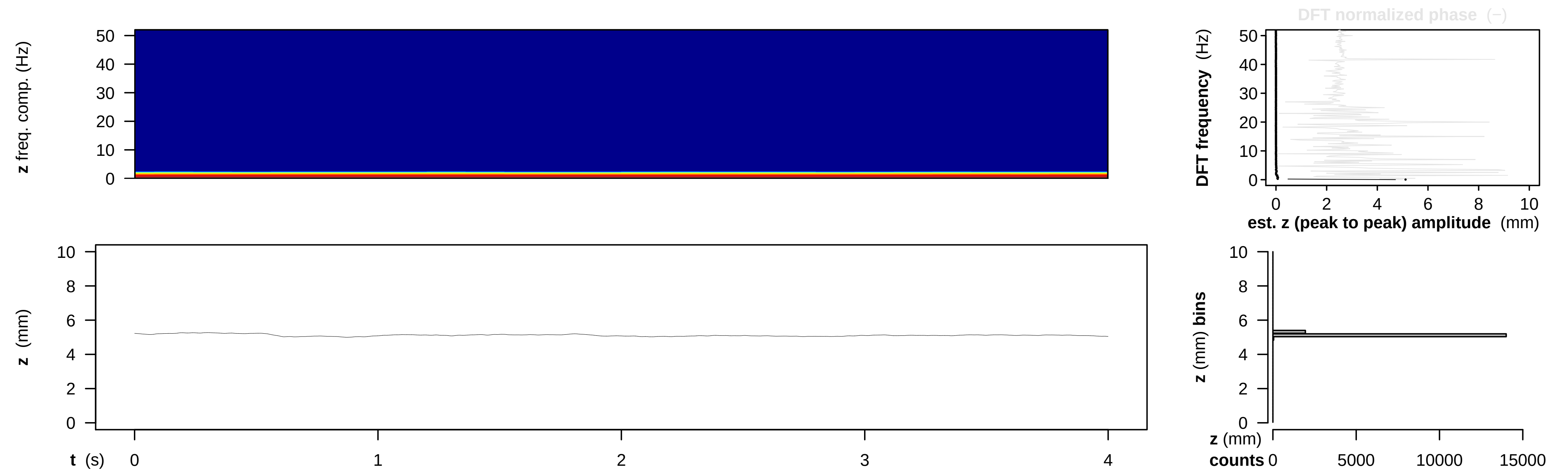

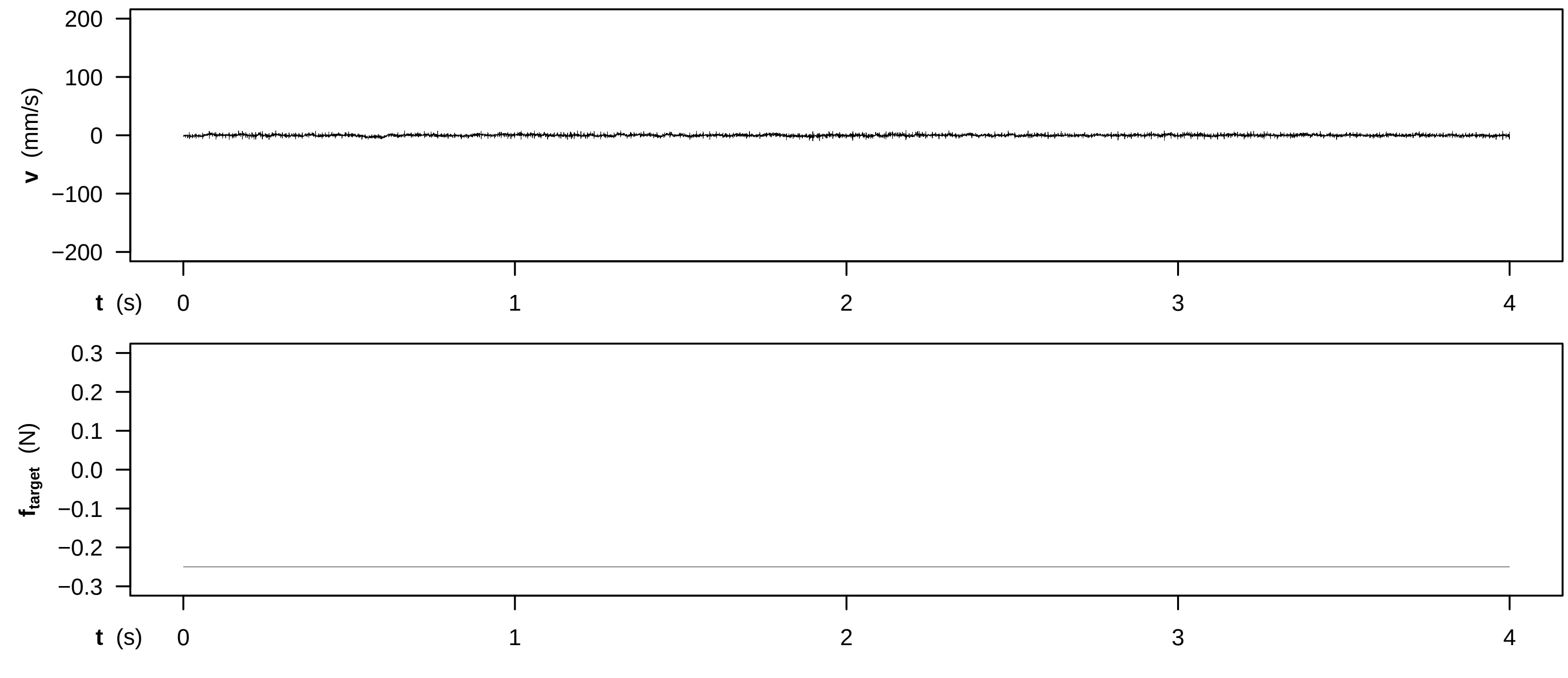

**SUBJECT 2 − RUN 34 − CONDITION 2,0**
**SC_180323_113510_0.AIFF**

```
z_min               : 5.00 mm
z_max               : 5.27 mm
z_travel_amplitude  : 0.28 mm

avg_abs_z_travel    : 2.57 mm/s

z_jarque-bera_jb    : 1228.38
z_jarque-bera_p     : 0.00e+00

z_lin_mod_est_slope: −0.02 mm/s
z_lin_mod_adj_R²    : 15 %

z_poly40_mod_adj_R²: 90 %

z_dft_ampl_thresh   : 0.010 mm
>=threshold_maxfreq: 4.75 Hz
```

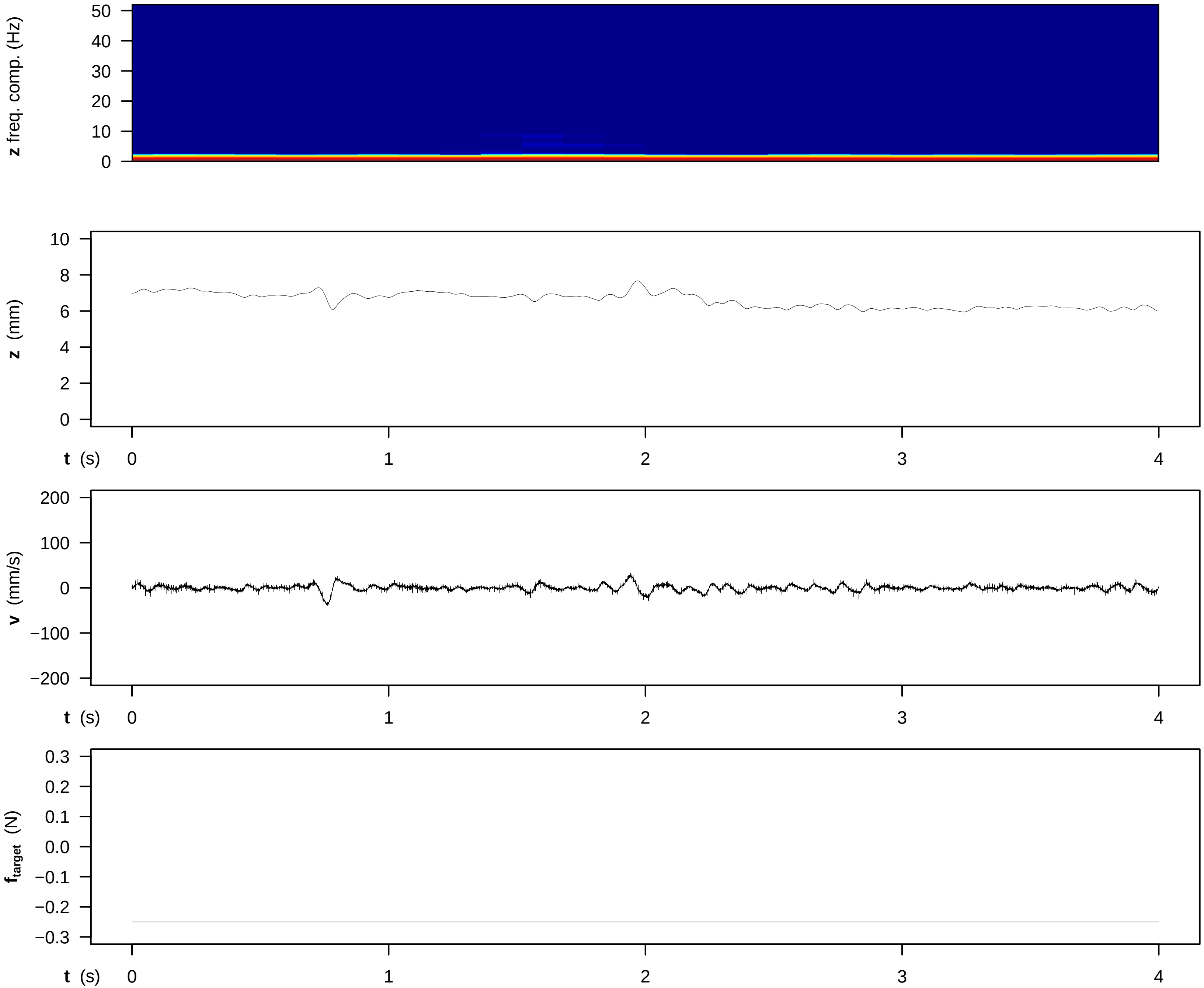

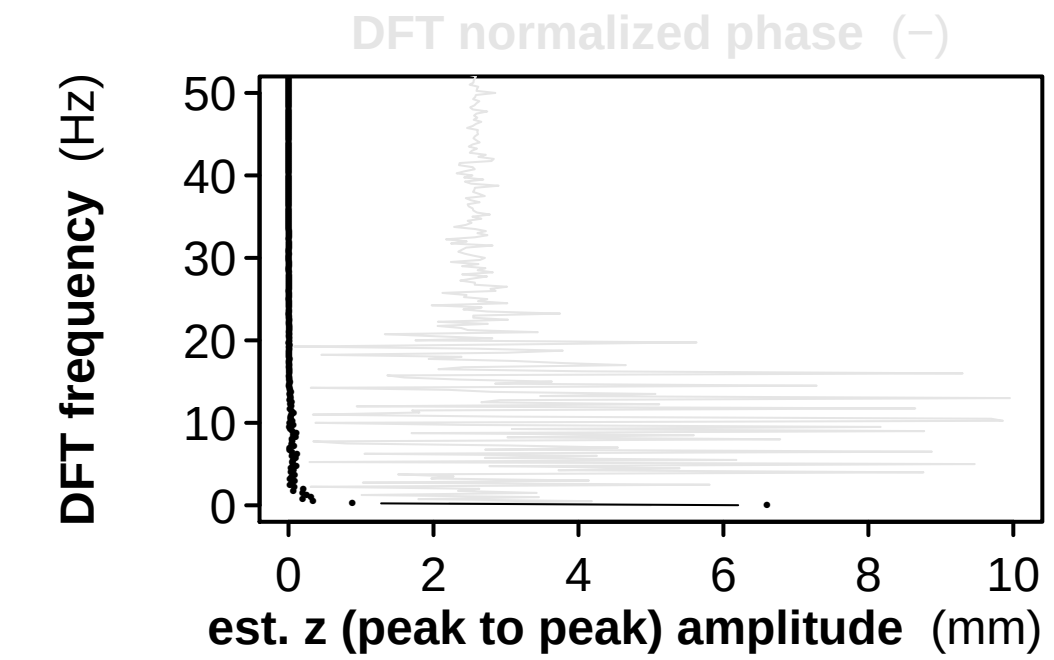

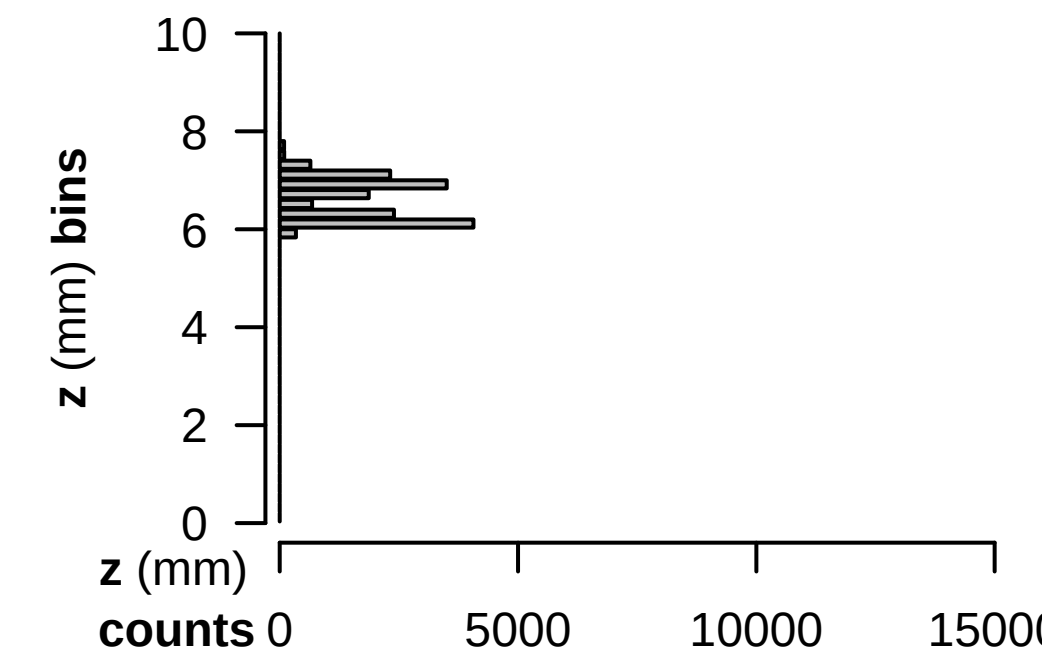

**SUBJECT 3 − RUN 08 − CONDITION 2,0**
**SC_180323_115941_0.AIFF**

```
z_min               : 5.95 mm
z_max               : 7.67 mm
z_travel_amplitude  : 1.73 mm

avg_abs_z_travel    : 6.71 mm/s

z_jarque-bera_jb    : 1131.75
z_jarque-bera_p     : 0.00e+00

z_lin_mod_est_slope: −0.29 mm/s
z_lin_mod_adj_R²    : 67 %

z_poly40_mod_adj_R²: 89 %

z_dft_ampl_thresh   : 0.010 mm
>=threshold_maxfreq: 22.50 Hz
```

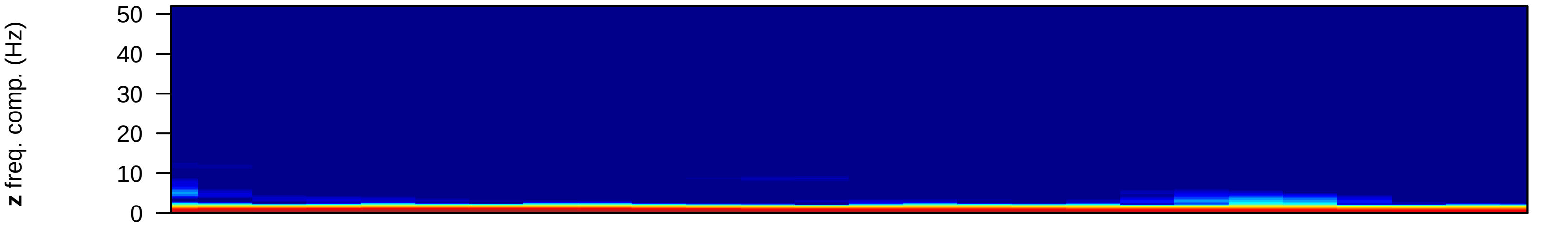

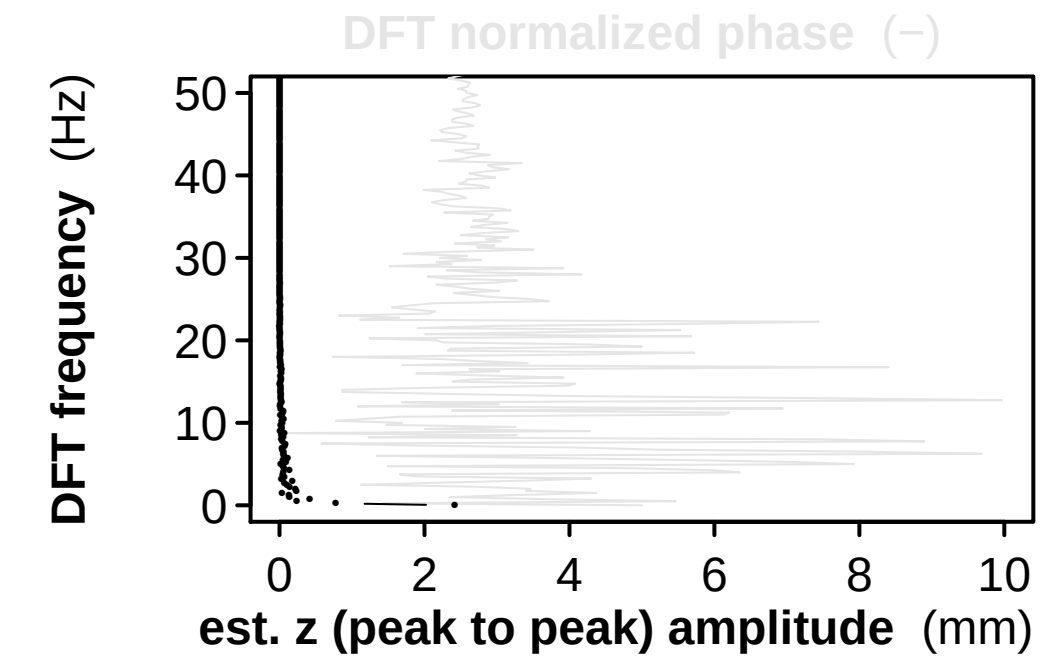

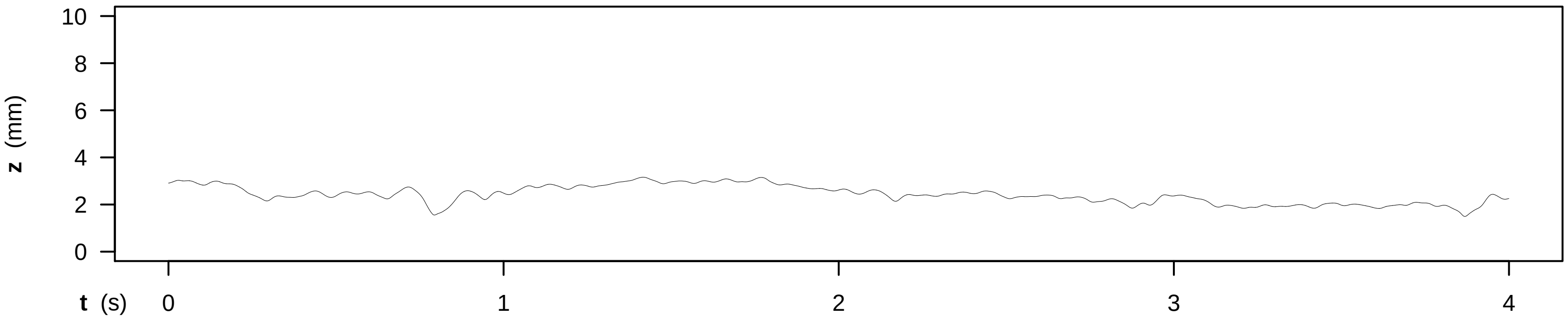

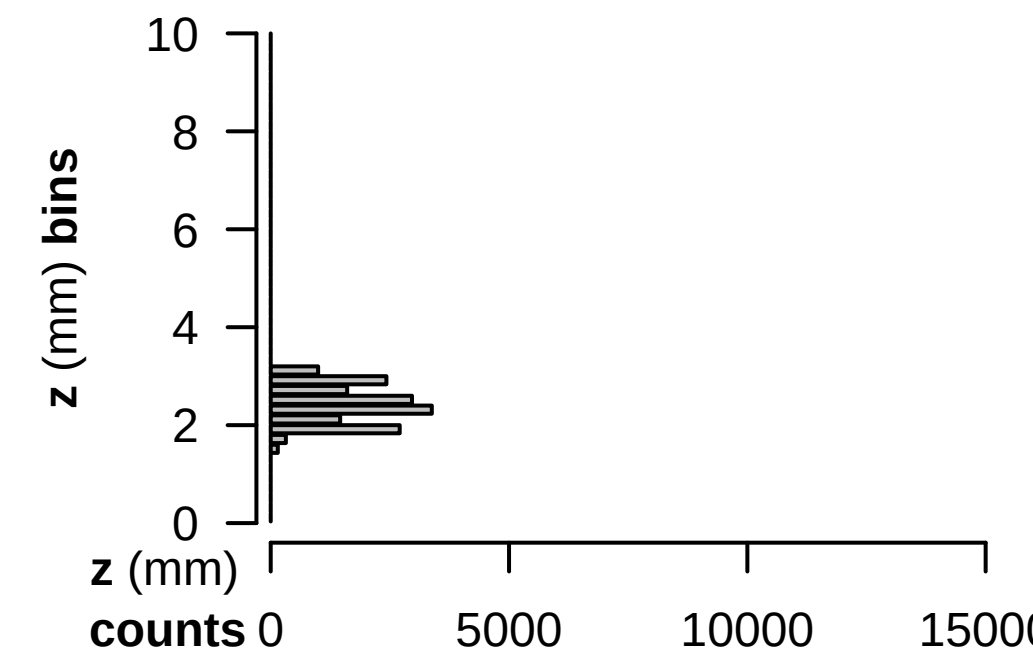

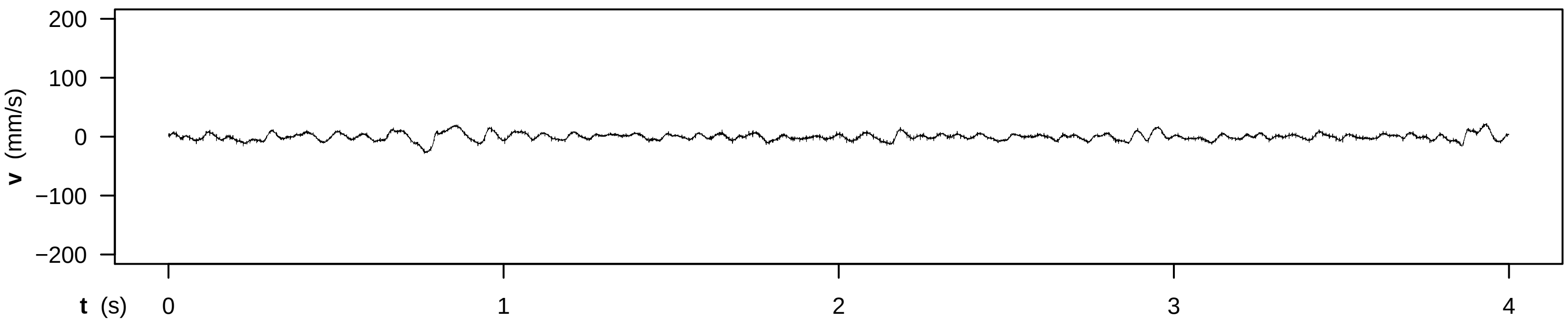

SUBJECT 3 − RUN 12 − CONDITION 2,0
SC_180323_120148_0.AIFF

```
z_min               : 1.49 mm
z_max               : 3.16 mm
z_travel_amplitude  : 1.67 mm

avg_abs_z_travel    : 5.39 mm/s

z_jarque-bera_jb    : 530.79
z_jarque-bera_p     : 0.00e+00

z_lin_mod_est_slope: −0.20 mm/s
z_lin_mod_adj_R²    : 38 %

z_poly40_mod_adj_R²: 89 %

z_dft_ampl_thresh   : 0.010 mm
>=threshold_maxfreq: 24.25 Hz
```

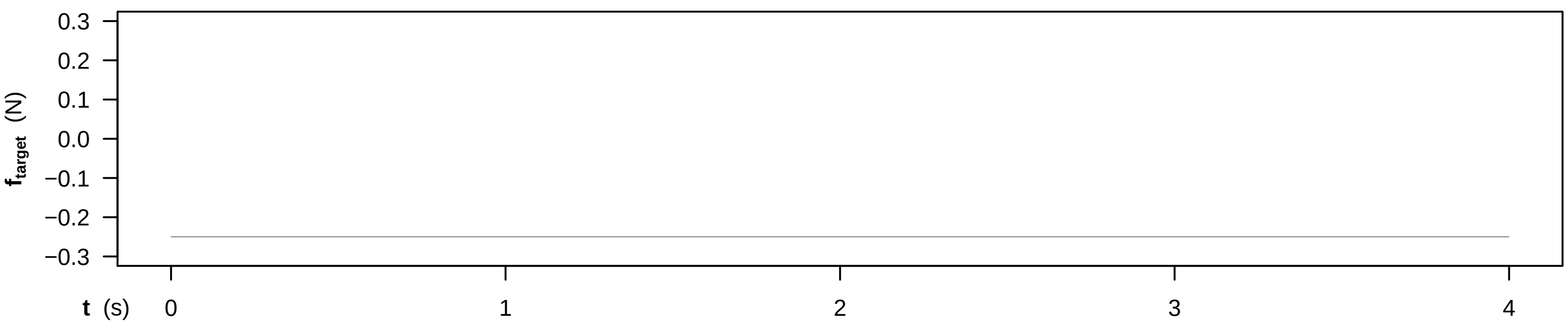

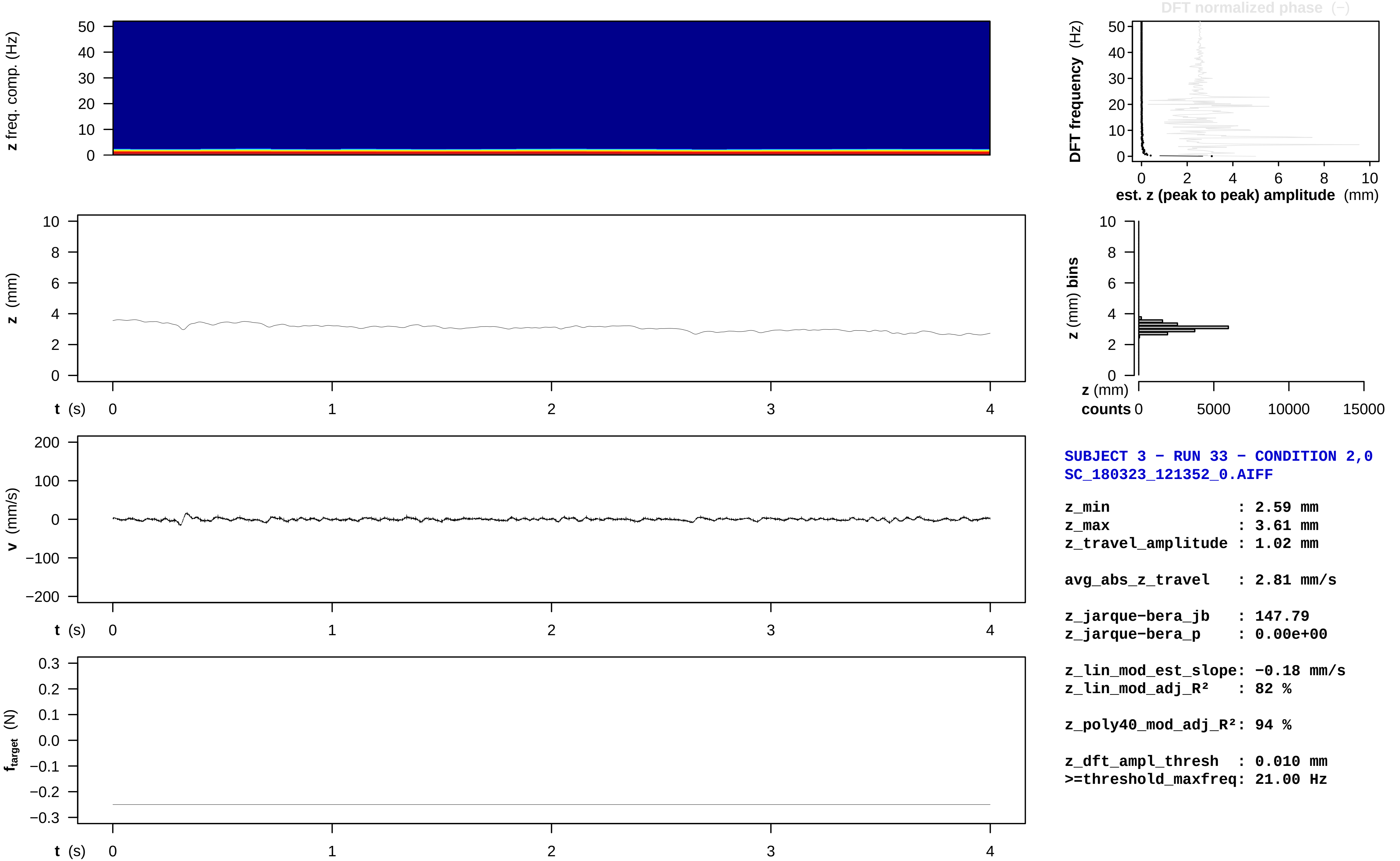

**SUBJECT 3 − RUN 33 − CONDITION 2,0**
**SC_180323_121352_0.AIFF**

```
z_min               : 2.59 mm
z_max               : 3.61 mm
z_travel_amplitude  : 1.02 mm

avg_abs_z_travel    : 2.81 mm/s

z_jarque-bera_jb    : 147.79
z_jarque-bera_p     : 0.00e+00

z_lin_mod_est_slope: −0.18 mm/s
z_lin_mod_adj_R²    : 82 %

z_poly40_mod_adj_R²: 94 %

z_dft_ampl_thresh   : 0.010 mm
>=threshold_maxfreq: 21.00 Hz
```

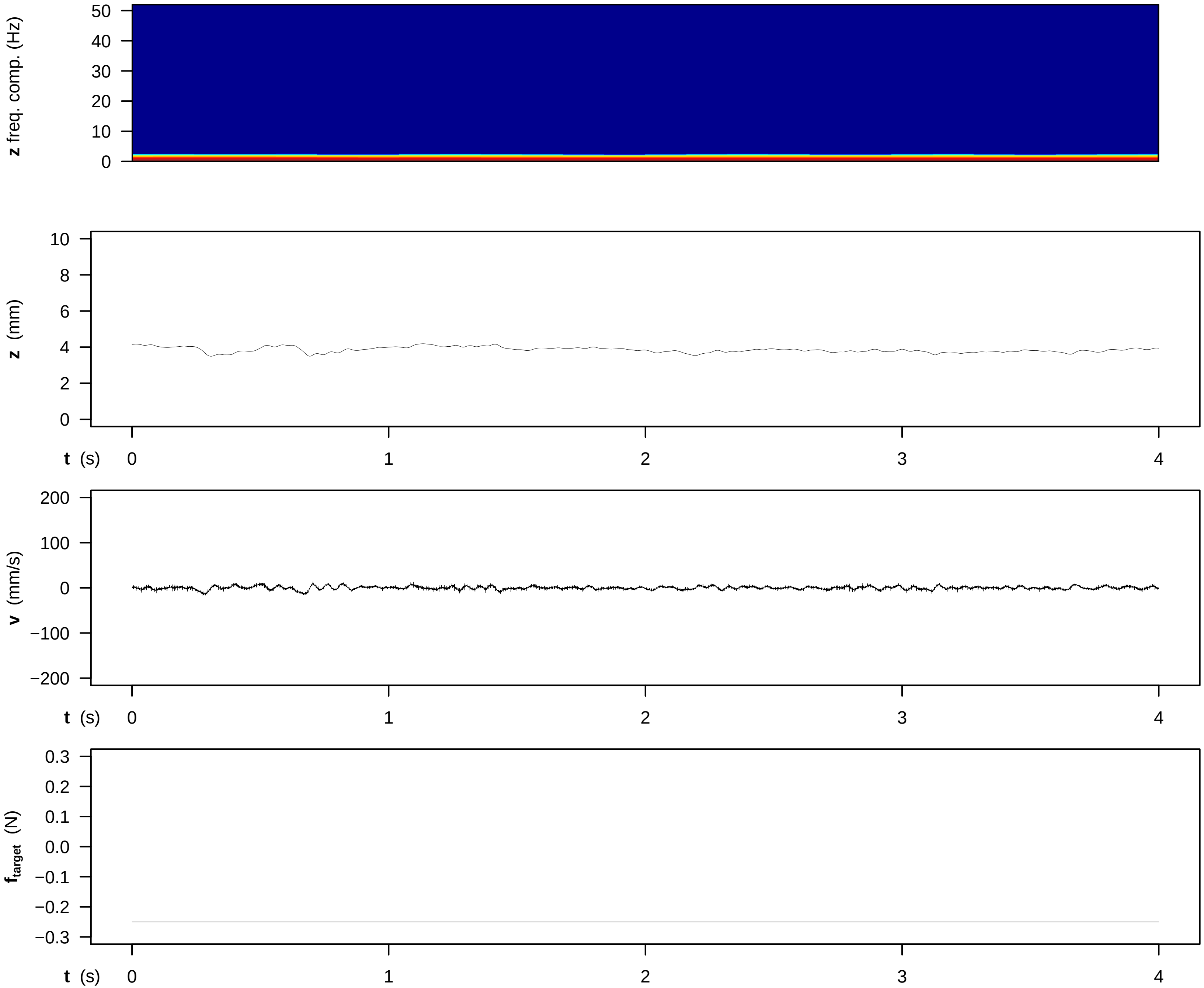

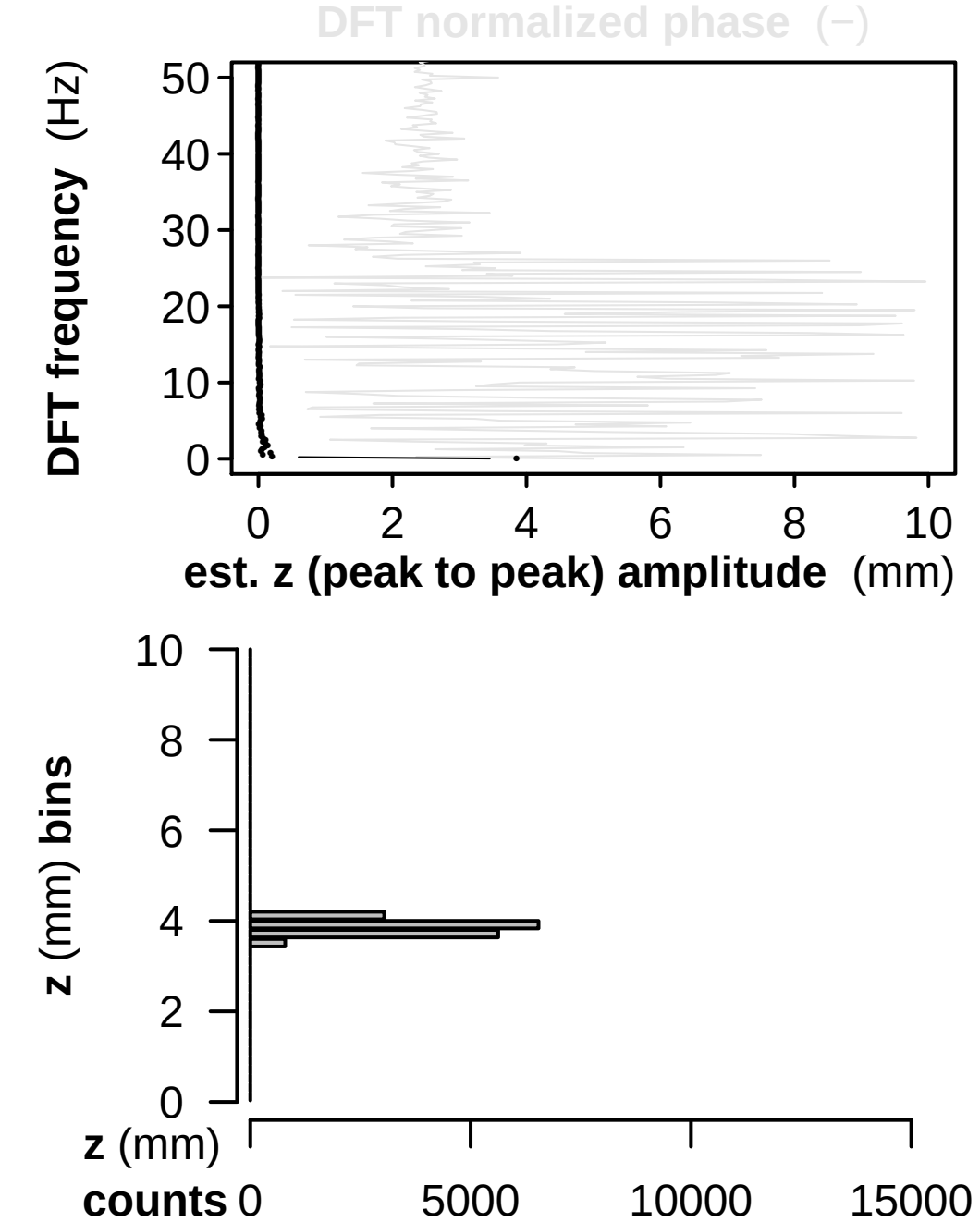

**SUBJECT 4 − RUN 03 − CONDITION 2,0**
**SC_180323_123151_0.AIFF**

```
z_min               : 3.49 mm
z_max               : 4.19 mm
z_travel_amplitude  : 0.70 mm

avg_abs_z_travel    : 3.15 mm/s

z_jarque-bera_jb    : 227.15
z_jarque-bera_p     : 0.00e+00

z_lin_mod_est_slope: −0.05 mm/s
z_lin_mod_adj_R²    : 16 %

z_poly40_mod_adj_R²: 75 %

z_dft_ampl_thresh   : 0.010 mm
>=threshold_maxfreq: 19.50 Hz
```

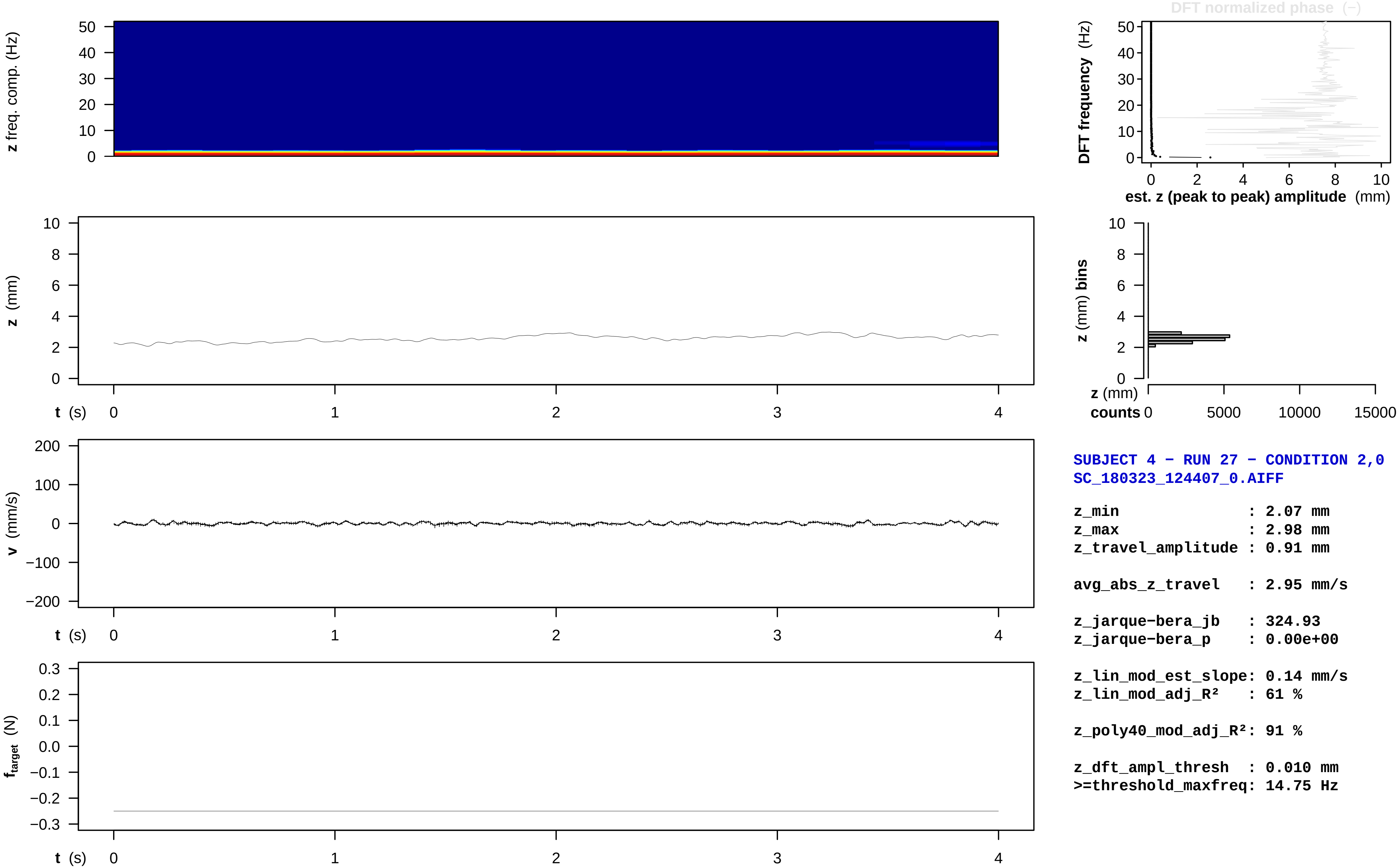

**SUBJECT 4 − RUN 27 − CONDITION 2,0**
**SC_180323_124407_0.AIFF**

```
z_min               : 2.07 mm
z_max               : 2.98 mm
z_travel_amplitude  : 0.91 mm

avg_abs_z_travel    : 2.95 mm/s

z_jarque-bera_jb    : 324.93
z_jarque-bera_p     : 0.00e+00

z_lin_mod_est_slope: 0.14 mm/s
z_lin_mod_adj_R²    : 61 %

z_poly40_mod_adj_R²: 91 %

z_dft_ampl_thresh   : 0.010 mm
>=threshold_maxfreq: 14.75 Hz
```

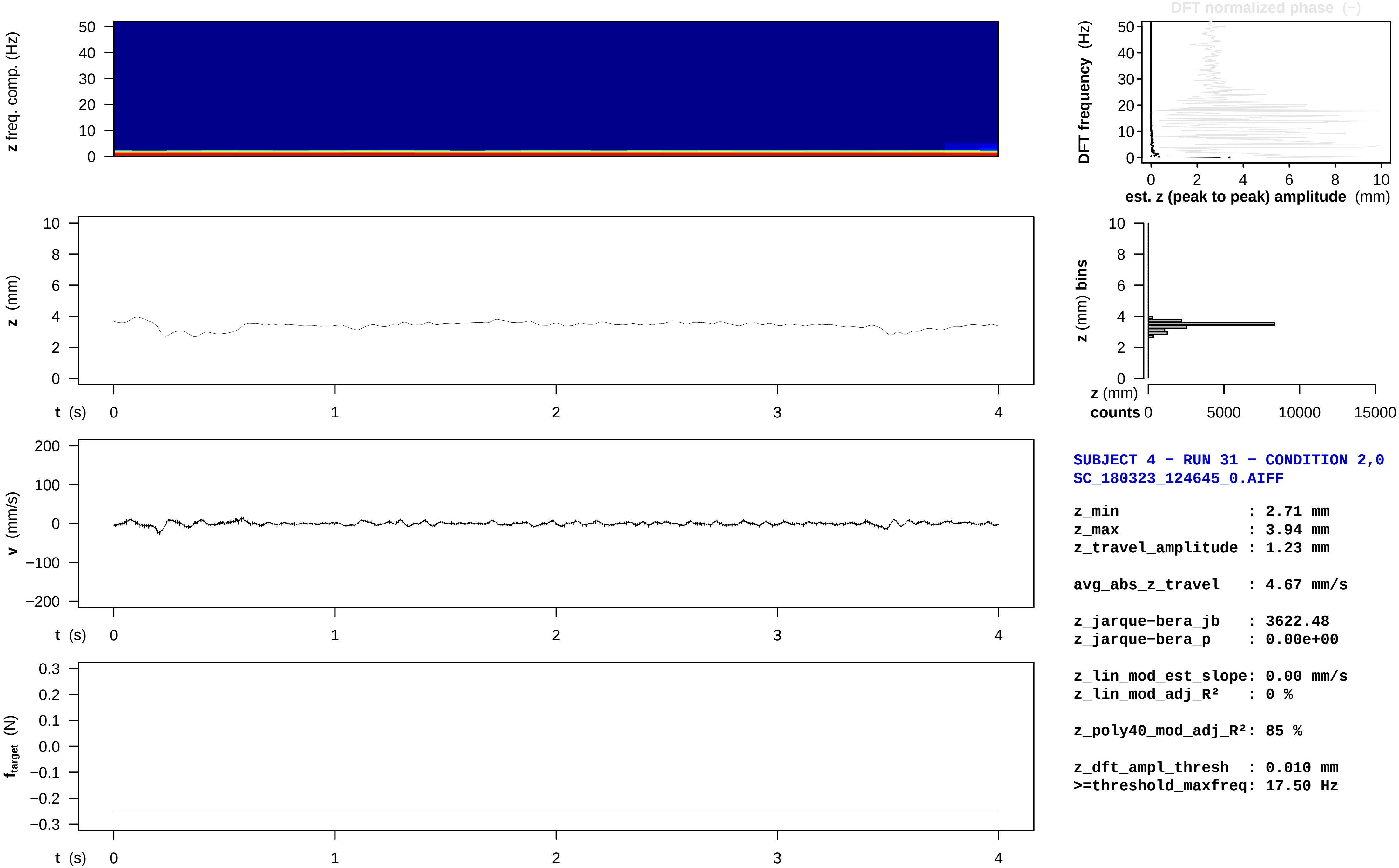

**SUBJECT 4 − RUN 31 − CONDITION 2,0**
**SC_180323_124645_0.AIFF**

```
z_min              : 2.71 mm
z_max              : 3.94 mm
z_travel_amplitude : 1.23 mm

avg_abs_z_travel   : 4.67 mm/s

z_jarque-bera_jb   : 3622.48
z_jarque-bera_p    : 0.00e+00

z_lin_mod_est_slope: 0.00 mm/s
z_lin_mod_adj_R²   : 0 %

z_poly40_mod_adj_R²: 85 %

z_dft_ampl_thresh  : 0.010 mm
>=threshold_maxfreq: 17.50 Hz
```

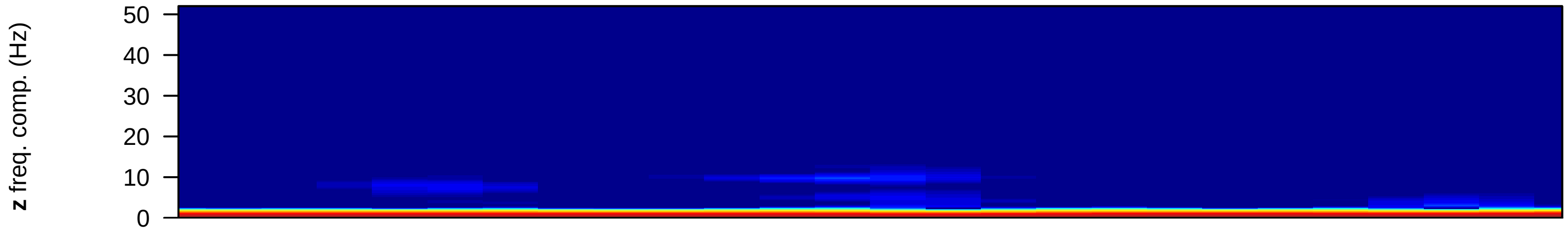

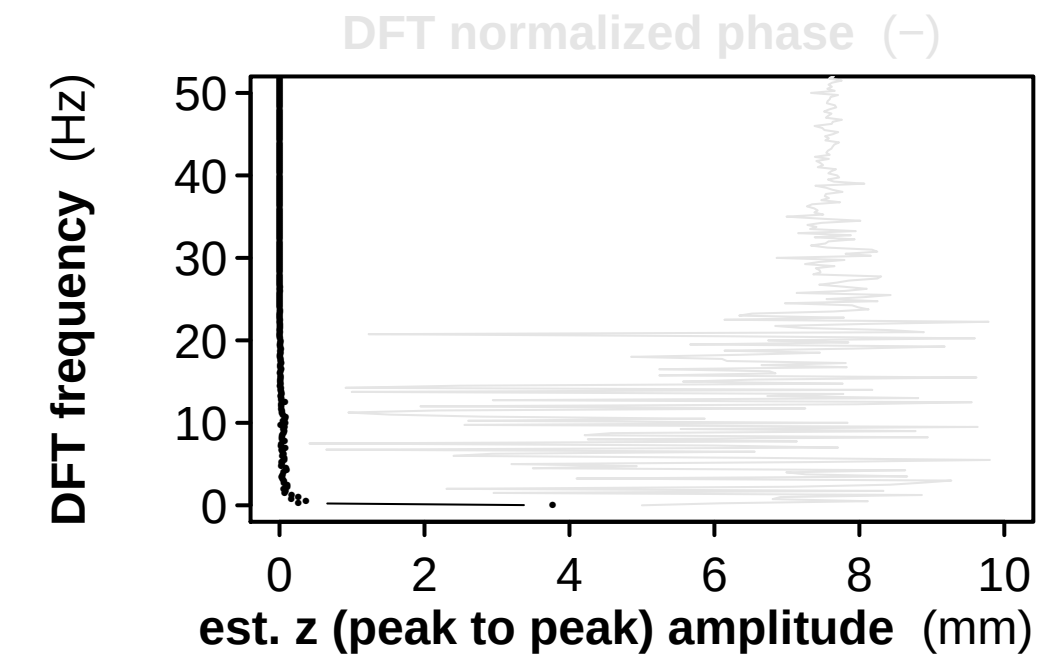

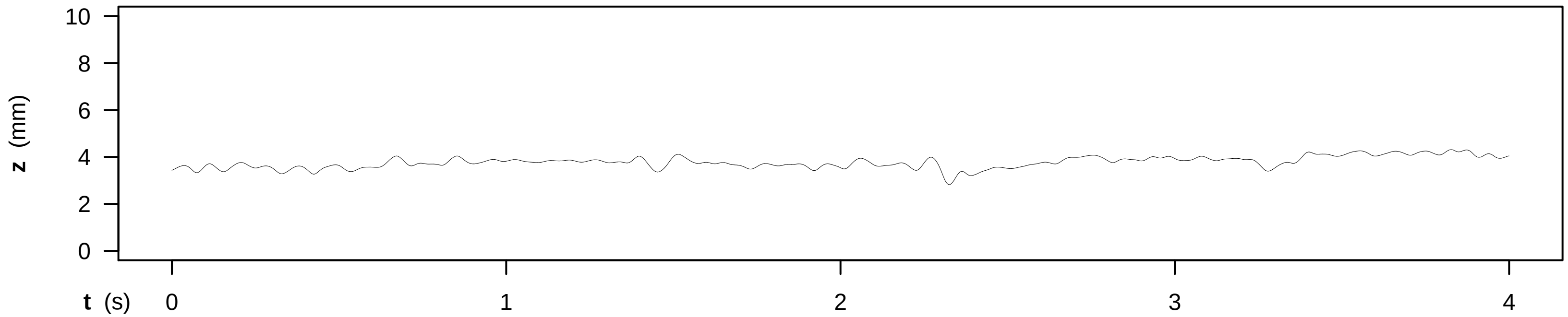

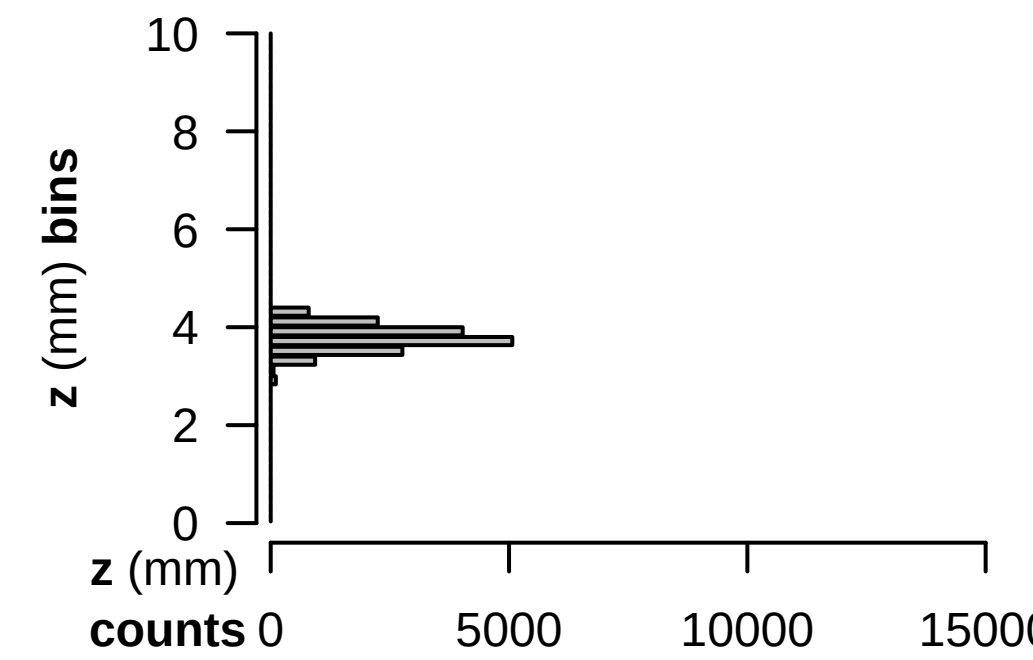

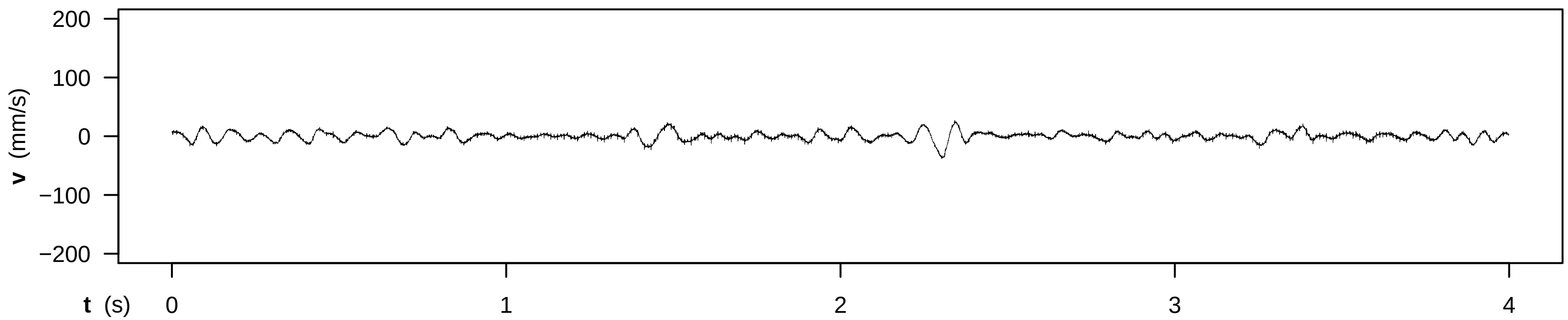

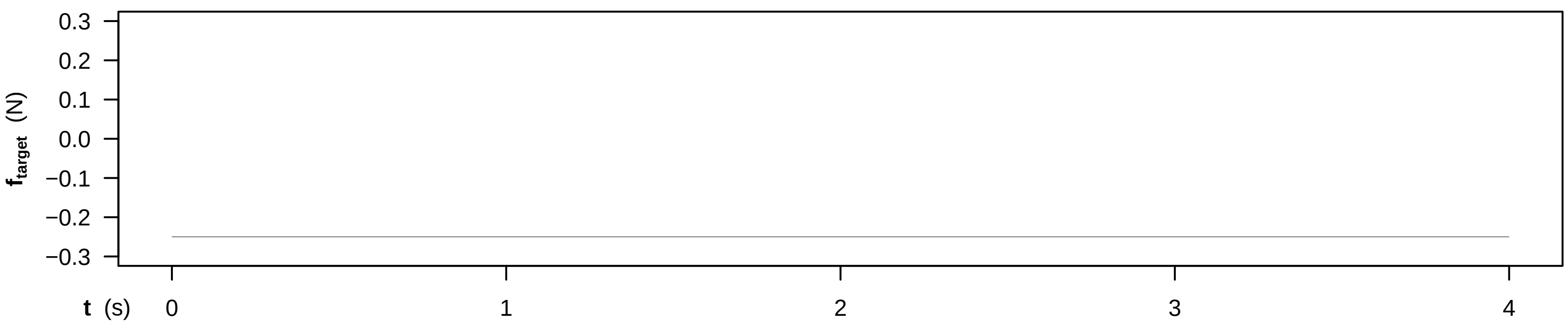

**SUBJECT 5 − RUN 04 − CONDITION 2,0**
**SC_180323_131632_0.AIFF**

```
z_min              : 2.82 mm
z_max              : 4.31 mm
z_travel_amplitude : 1.49 mm

avg_abs_z_travel   : 6.00 mm/s

z_jarque-bera_jb   : 180.00
z_jarque-bera_p    : 0.00e+00

z_lin_mod_est_slope: 0.12 mm/s
z_lin_mod_adj_R²   : 31 %

z_poly40_mod_adj_R²: 72 %

z_dft_ampl_thresh  : 0.010 mm
>=threshold_maxfreq: 20.00 Hz
```

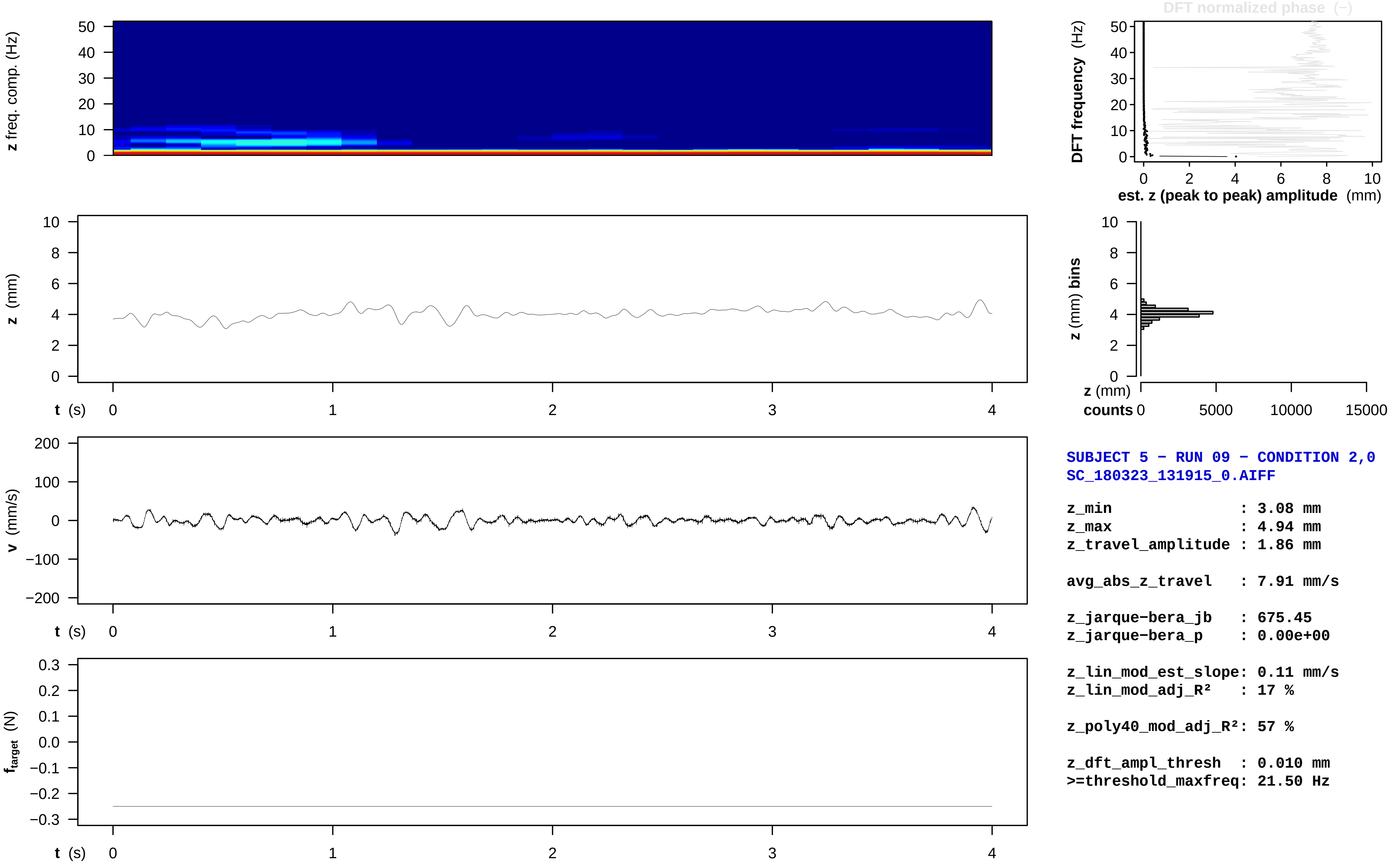

**SUBJECT 5 − RUN 09 − CONDITION 2,0**
**SC_180323_131915_0.AIFF**

```
z_min               : 3.08 mm
z_max               : 4.94 mm
z_travel_amplitude  : 1.86 mm

avg_abs_z_travel    : 7.91 mm/s

z_jarque-bera_jb    : 675.45
z_jarque-bera_p     : 0.00e+00

z_lin_mod_est_slope: 0.11 mm/s
z_lin_mod_adj_R²    : 17 %

z_poly40_mod_adj_R²: 57 %

z_dft_ampl_thresh   : 0.010 mm
>=threshold_maxfreq: 21.50 Hz
```

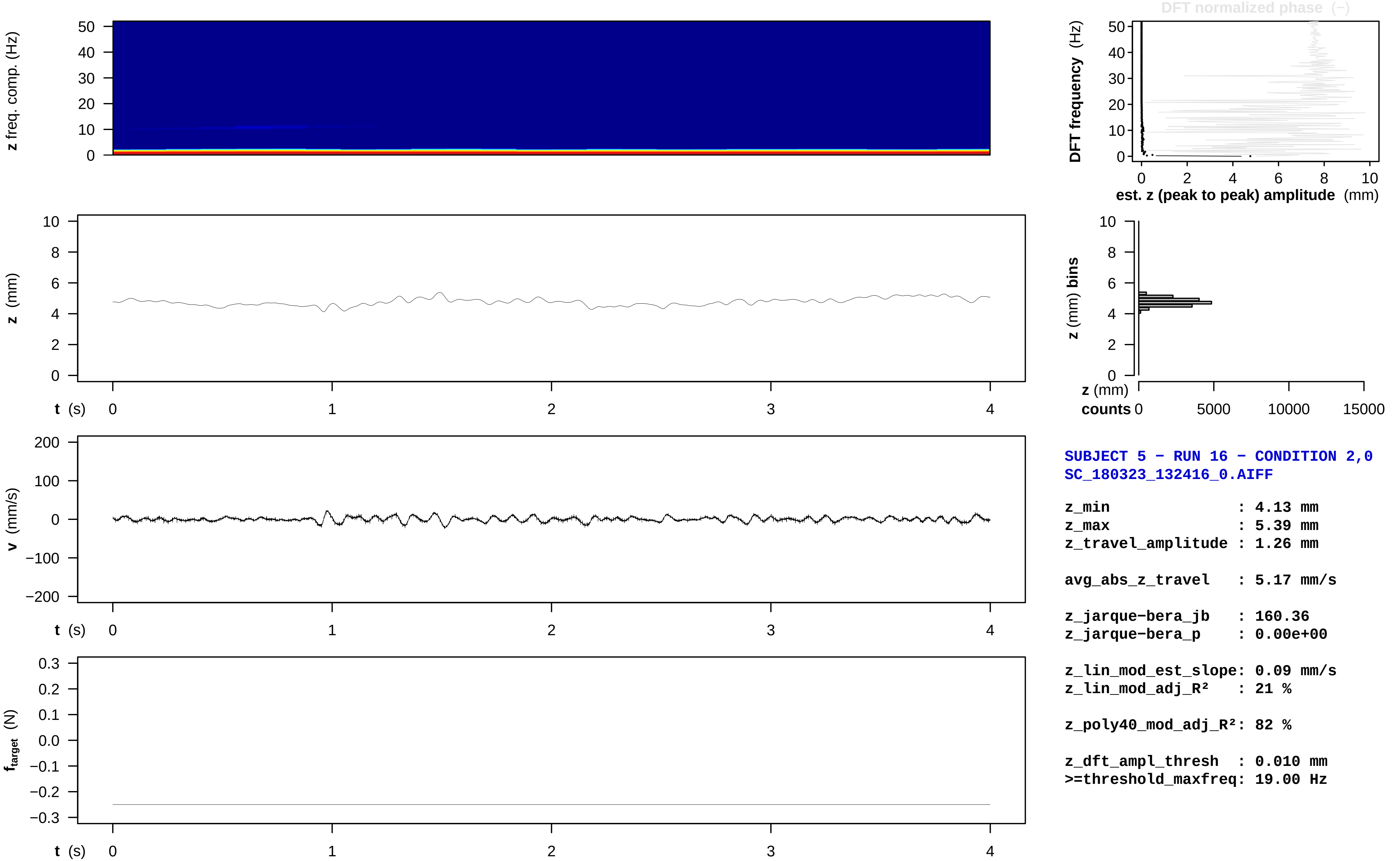

**SUBJECT 5 − RUN 16 − CONDITION 2,0**
**SC_180323_132416_0.AIFF**

```
z_min               : 4.13 mm
z_max               : 5.39 mm
z_travel_amplitude  : 1.26 mm

avg_abs_z_travel    : 5.17 mm/s

z_jarque-bera_jb    : 160.36
z_jarque-bera_p     : 0.00e+00

z_lin_mod_est_slope: 0.09 mm/s
z_lin_mod_adj_R²    : 21 %

z_poly40_mod_adj_R²: 82 %

z_dft_ampl_thresh   : 0.010 mm
>=threshold_maxfreq: 19.00 Hz
```

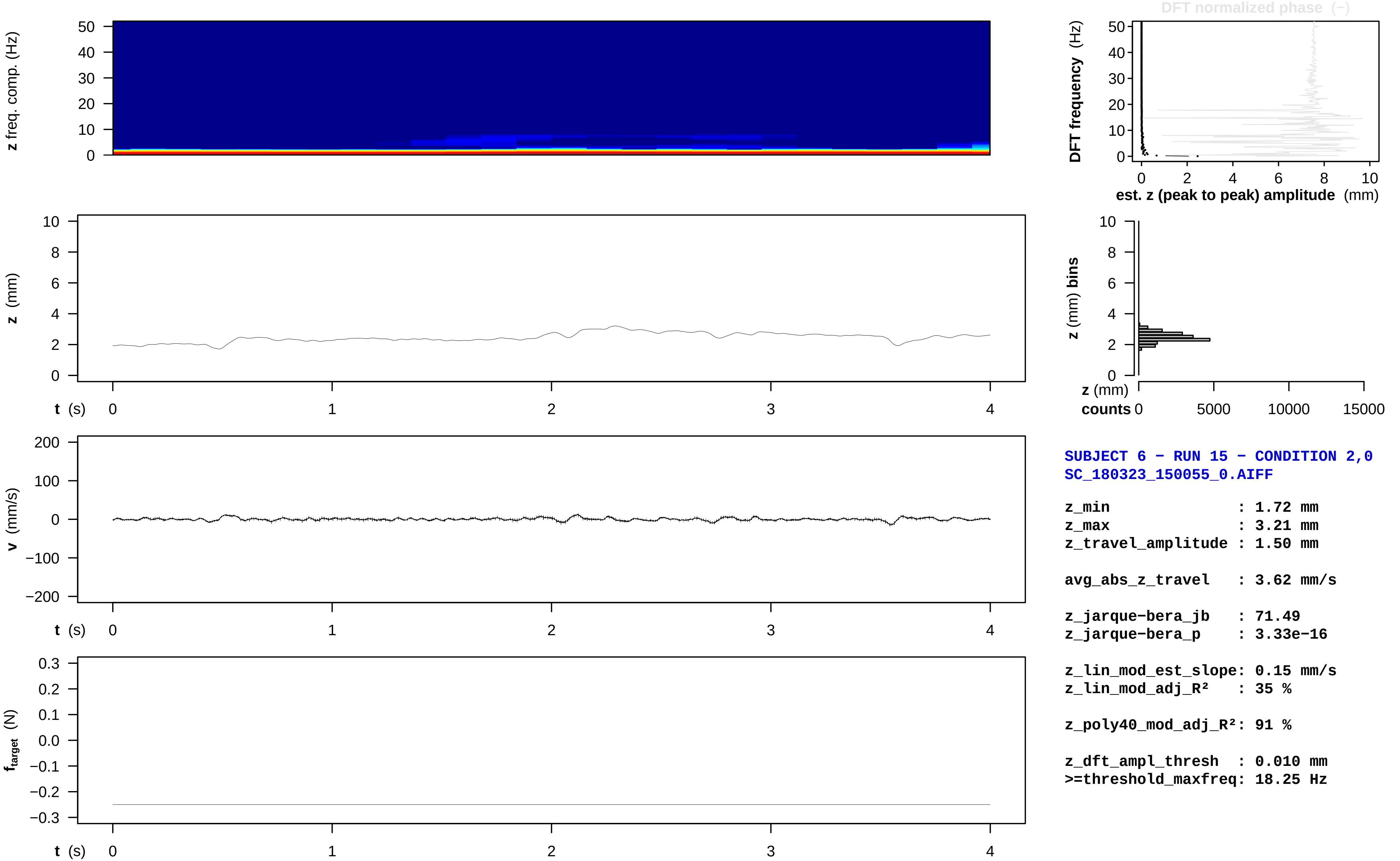

**SUBJECT 6 − RUN 15 − CONDITION 2,0**
**SC_180323_150055_0.AIFF**

```
z_min               : 1.72 mm
z_max               : 3.21 mm
z_travel_amplitude  : 1.50 mm

avg_abs_z_travel    : 3.62 mm/s

z_jarque-bera_jb    : 71.49
z_jarque-bera_p     : 3.33e−16

z_lin_mod_est_slope: 0.15 mm/s
z_lin_mod_adj_R²    : 35 %

z_poly40_mod_adj_R²: 91 %

z_dft_ampl_thresh   : 0.010 mm
>=threshold_maxfreq: 18.25 Hz
```

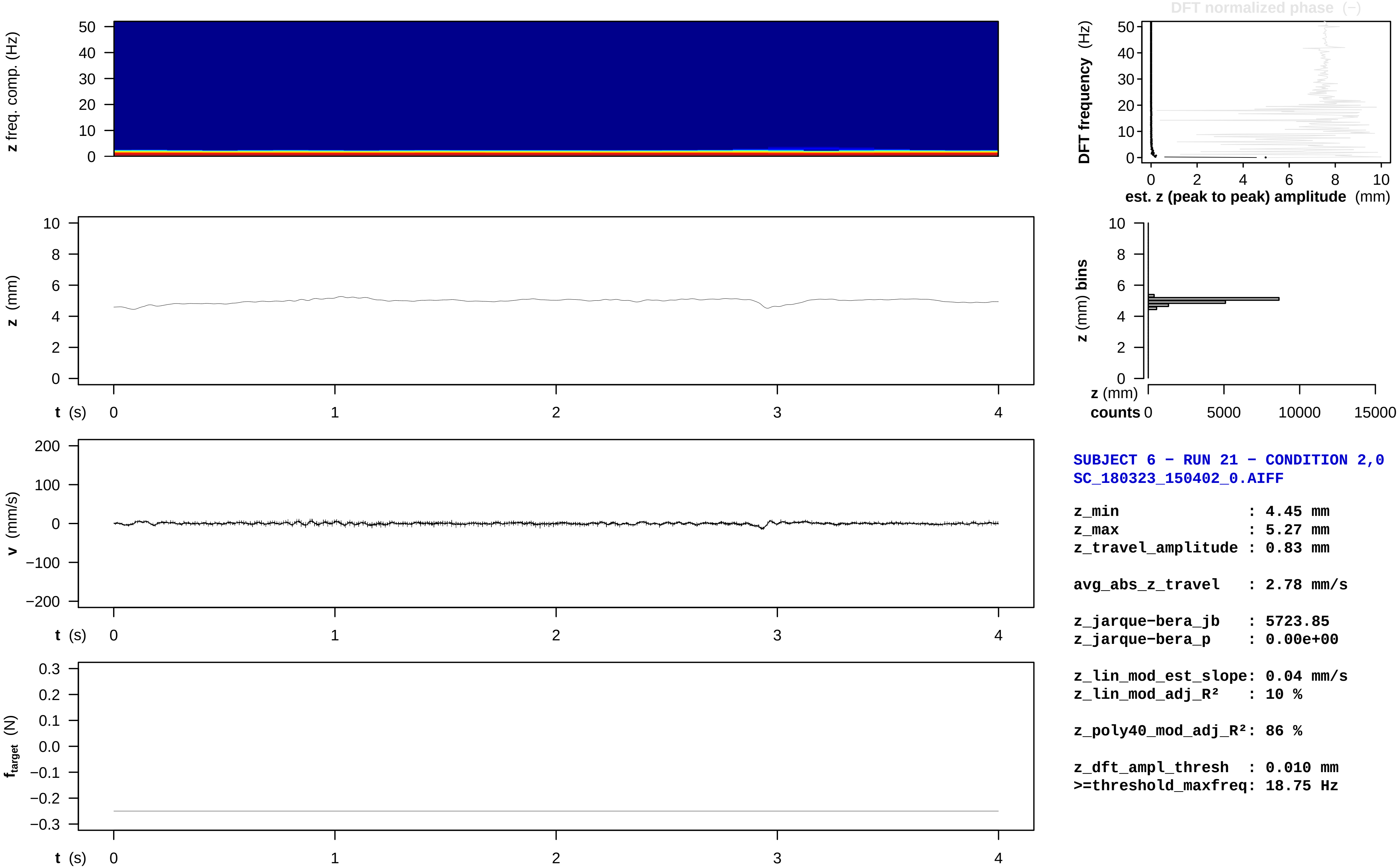

**SUBJECT 6 − RUN 21 − CONDITION 2,0**
**SC_180323_150402_0.AIFF**

```
z_min               : 4.45 mm
z_max               : 5.27 mm
z_travel_amplitude  : 0.83 mm

avg_abs_z_travel    : 2.78 mm/s

z_jarque-bera_jb    : 5723.85
z_jarque-bera_p     : 0.00e+00

z_lin_mod_est_slope: 0.04 mm/s
z_lin_mod_adj_R²    : 10 %

z_poly40_mod_adj_R²: 86 %

z_dft_ampl_thresh   : 0.010 mm
>=threshold_maxfreq: 18.75 Hz
```

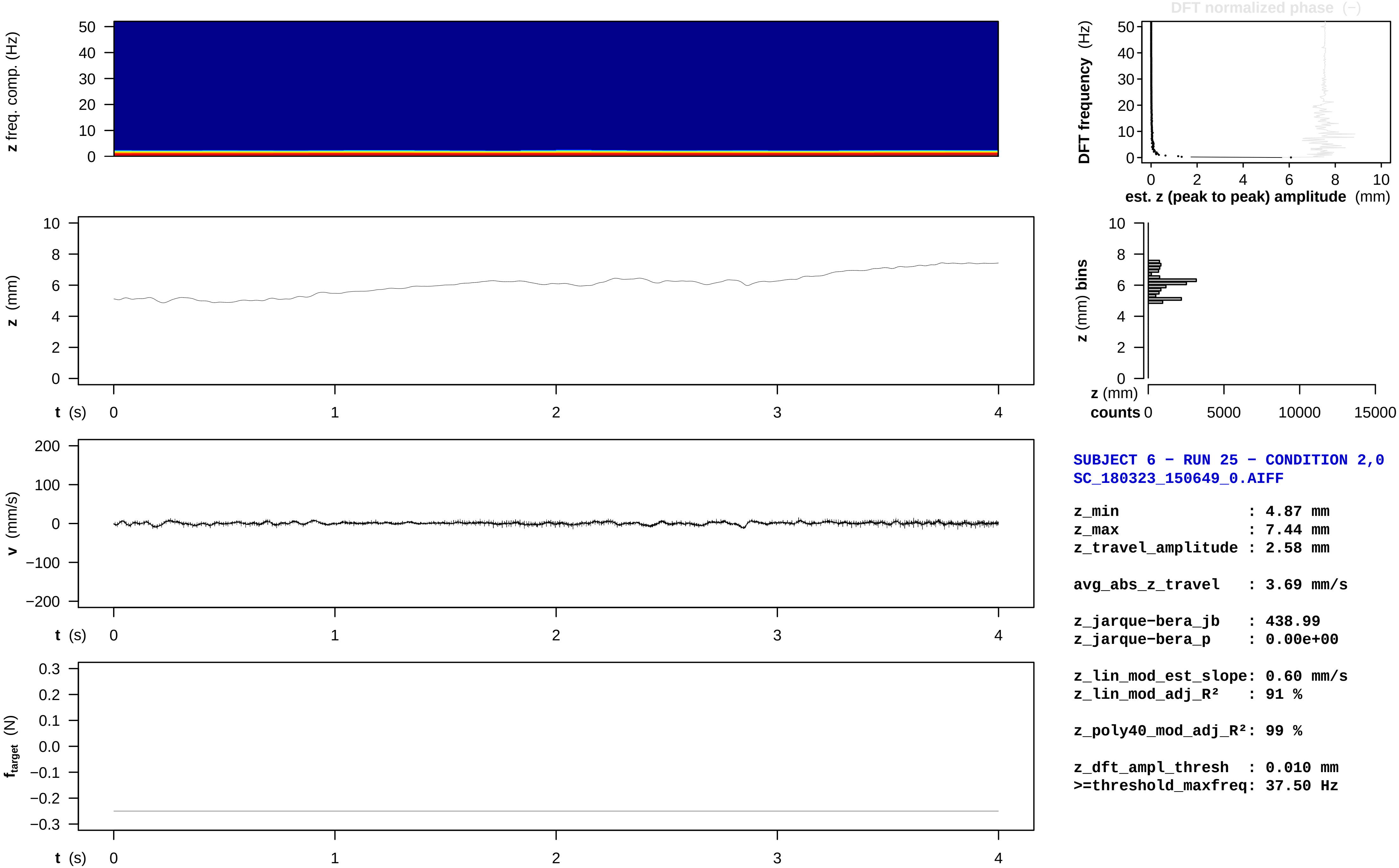

**SUBJECT 6 − RUN 25 − CONDITION 2,0**
**SC_180323_150649_0.AIFF**

```
z_min               : 4.87 mm
z_max               : 7.44 mm
z_travel_amplitude  : 2.58 mm

avg_abs_z_travel    : 3.69 mm/s

z_jarque-bera_jb    : 438.99
z_jarque-bera_p     : 0.00e+00

z_lin_mod_est_slope: 0.60 mm/s
z_lin_mod_adj_R²    : 91 %

z_poly40_mod_adj_R²: 99 %

z_dft_ampl_thresh   : 0.010 mm
>=threshold_maxfreq: 37.50 Hz
```

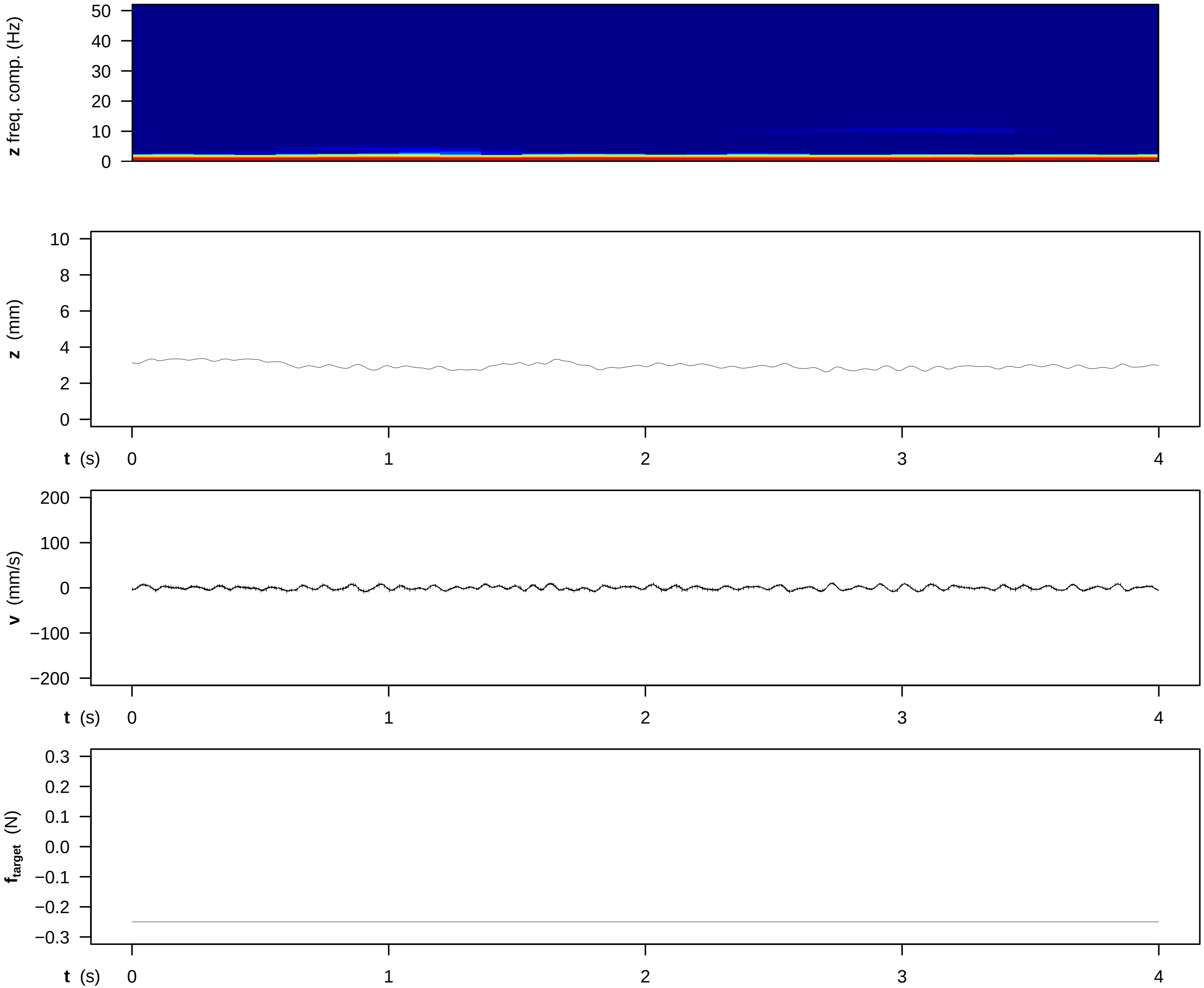

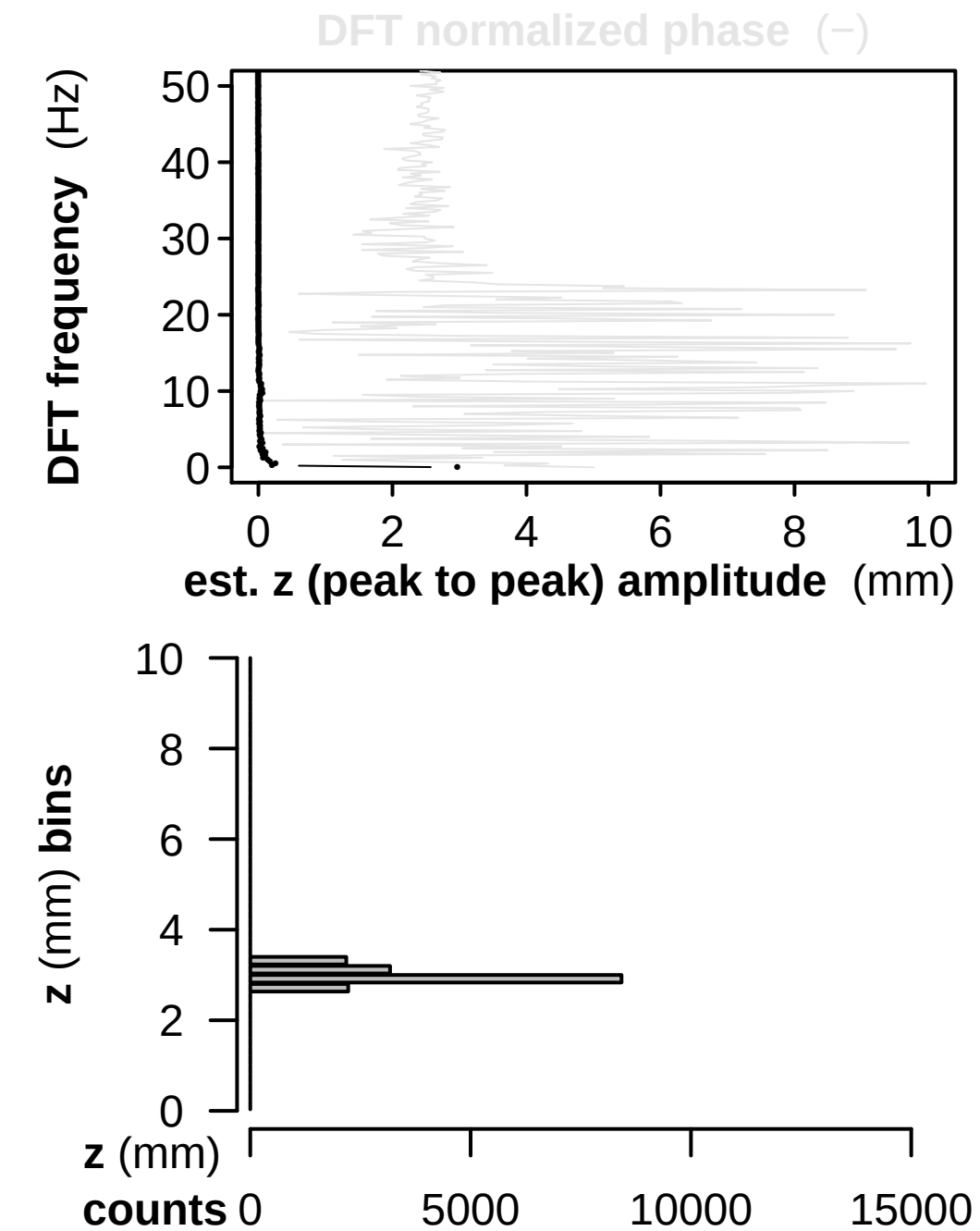

**SUBJECT 7 − RUN 14 − CONDITION 2,0**
**SC_180323_154413_0.AIFF**

```
z_min                : 2.63 mm
z_max                : 3.37 mm
z_travel_amplitude : 0.74 mm

avg_abs_z_travel   : 3.65 mm/s

z_jarque-bera_jb   : 1309.04
z_jarque-bera_p    : 0.00e+00

z_lin_mod_est_slope: −0.08 mm/s
z_lin_mod_adj_R²   : 27 %

z_poly40_mod_adj_R²: 79 %

z_dft_ampl_thresh  : 0.010 mm
>=threshold_maxfreq: 15.75 Hz
```

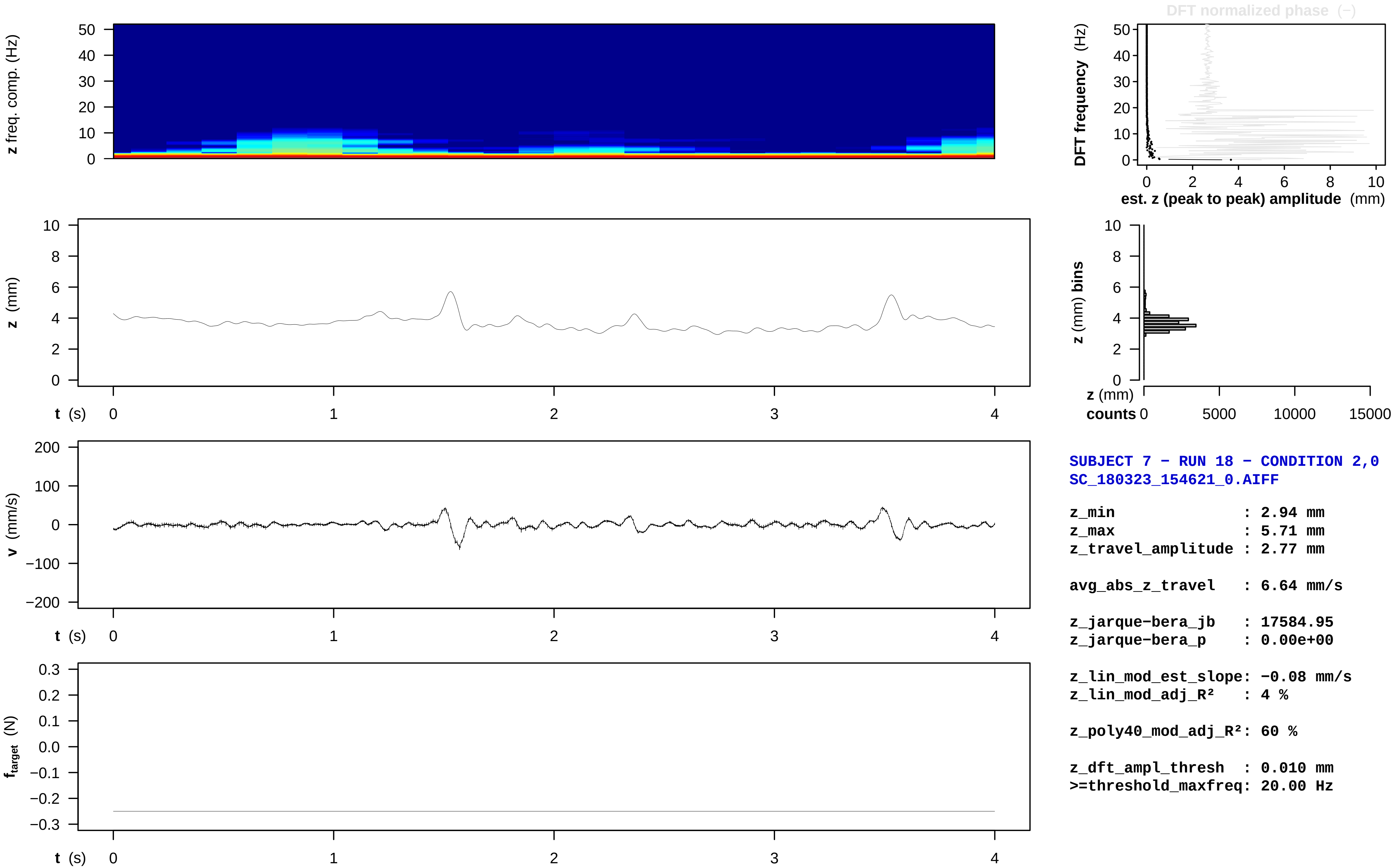

**SUBJECT 7 − RUN 18 − CONDITION 2,0**
**SC_180323_154621_0.AIFF**

```
z_min               : 2.94 mm
z_max               : 5.71 mm
z_travel_amplitude  : 2.77 mm

avg_abs_z_travel    : 6.64 mm/s

z_jarque-bera_jb    : 17584.95
z_jarque-bera_p     : 0.00e+00

z_lin_mod_est_slope: -0.08 mm/s
z_lin_mod_adj_R²    : 4 %

z_poly40_mod_adj_R²: 60 %

z_dft_ampl_thresh   : 0.010 mm
>=threshold_maxfreq: 20.00 Hz
```

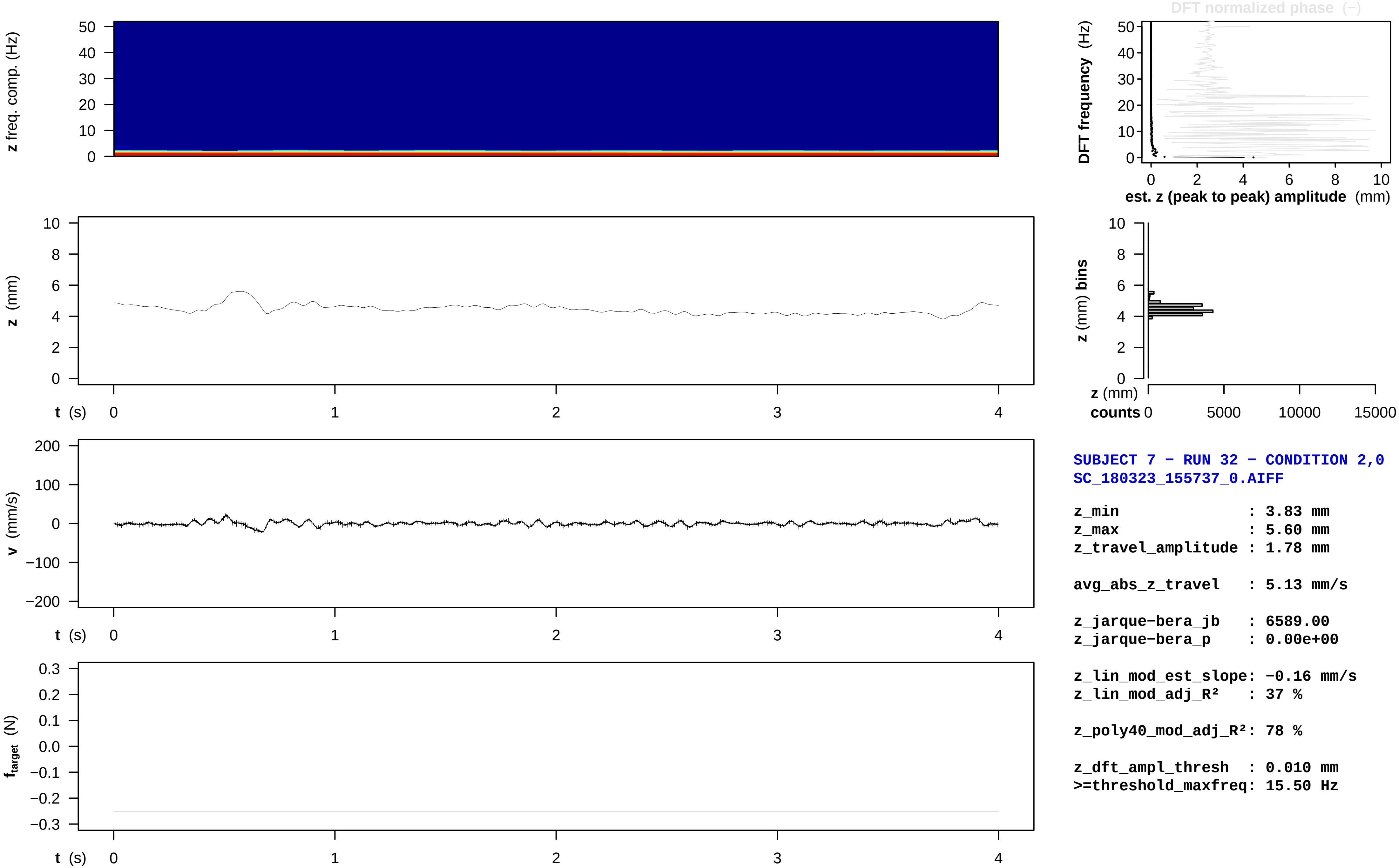

SUBJECT 7 − RUN 32 − CONDITION 2,0
SC_180323_155737_0.AIFF

z_min : 3.83 mm
z_max : 5.60 mm
z_travel_amplitude : 1.78 mm

avg_abs_z_travel : 5.13 mm/s

z_jarque-bera_jb : 6589.00
z_jarque-bera_p : 0.00e+00

z_lin_mod_est_slope: −0.16 mm/s
z_lin_mod_adj_R² : 37 %

z_poly40_mod_adj_R²: 78 %

z_dft_ampl_thresh : 0.010 mm
>=threshold_maxfreq: 15.50 Hz

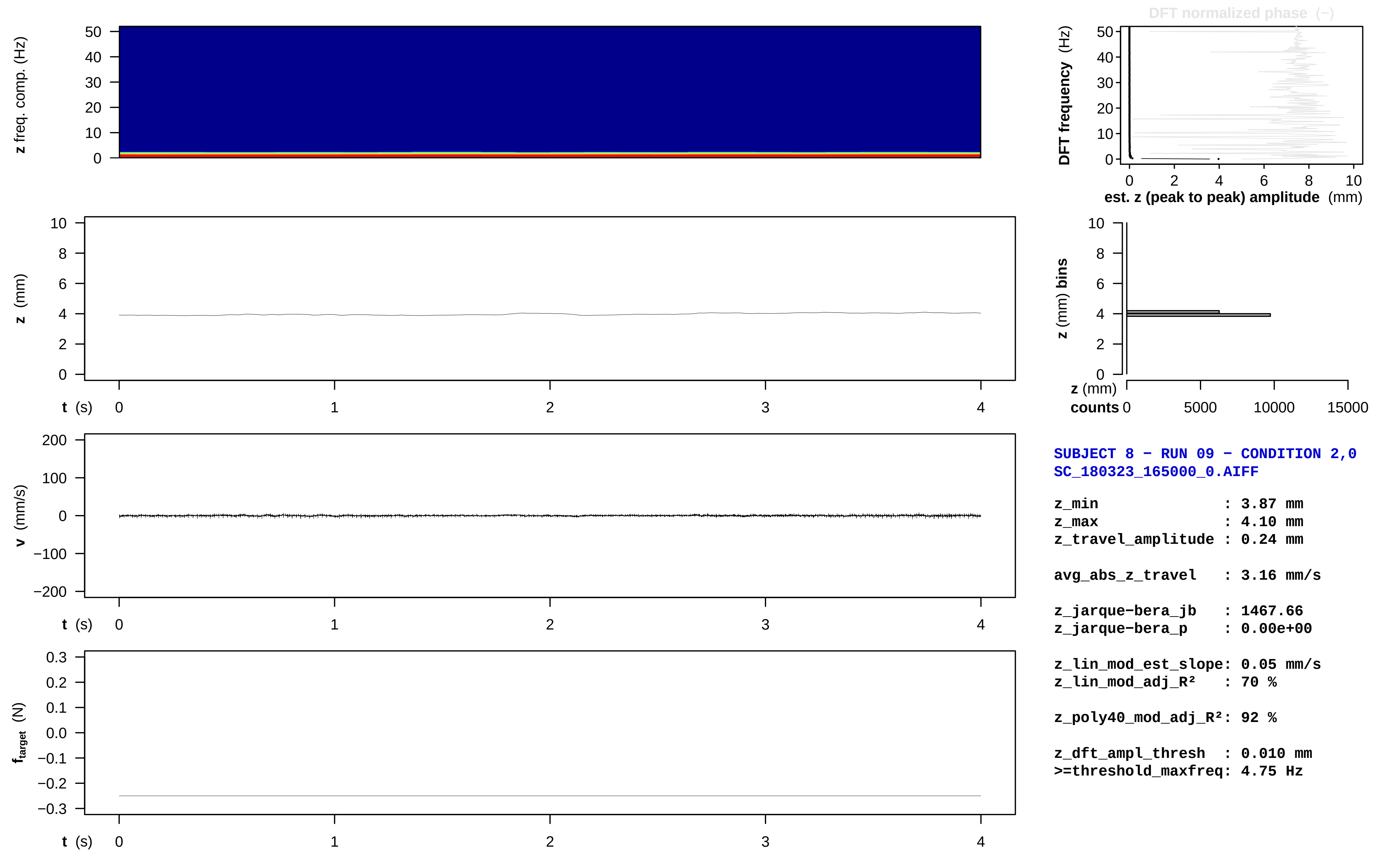

z freq. comp. (Hz)
DFT normalized phase (−)
DFT frequency (Hz)
est. z (peak to peak) amplitude (mm)
z (mm)
t (s)
z (mm) bins
z (mm)
counts
v (mm/s)
f_target (N)
SUBJECT 8 − RUN 09 − CONDITION 2,0
SC_180323_165000_0.AIFF
z_min : 3.87 mm
z_max : 4.10 mm
z_travel_amplitude : 0.24 mm
avg_abs_z_travel : 3.16 mm/s
z_jarque-bera_jb : 1467.66
z_jarque-bera_p : 0.00e+00
z_lin_mod_est_slope: 0.05 mm/s
z_lin_mod_adj_R² : 70 %
z_poly40_mod_adj_R²: 92 %
z_dft_ampl_thresh : 0.010 mm
>=threshold_maxfreq: 4.75 Hz

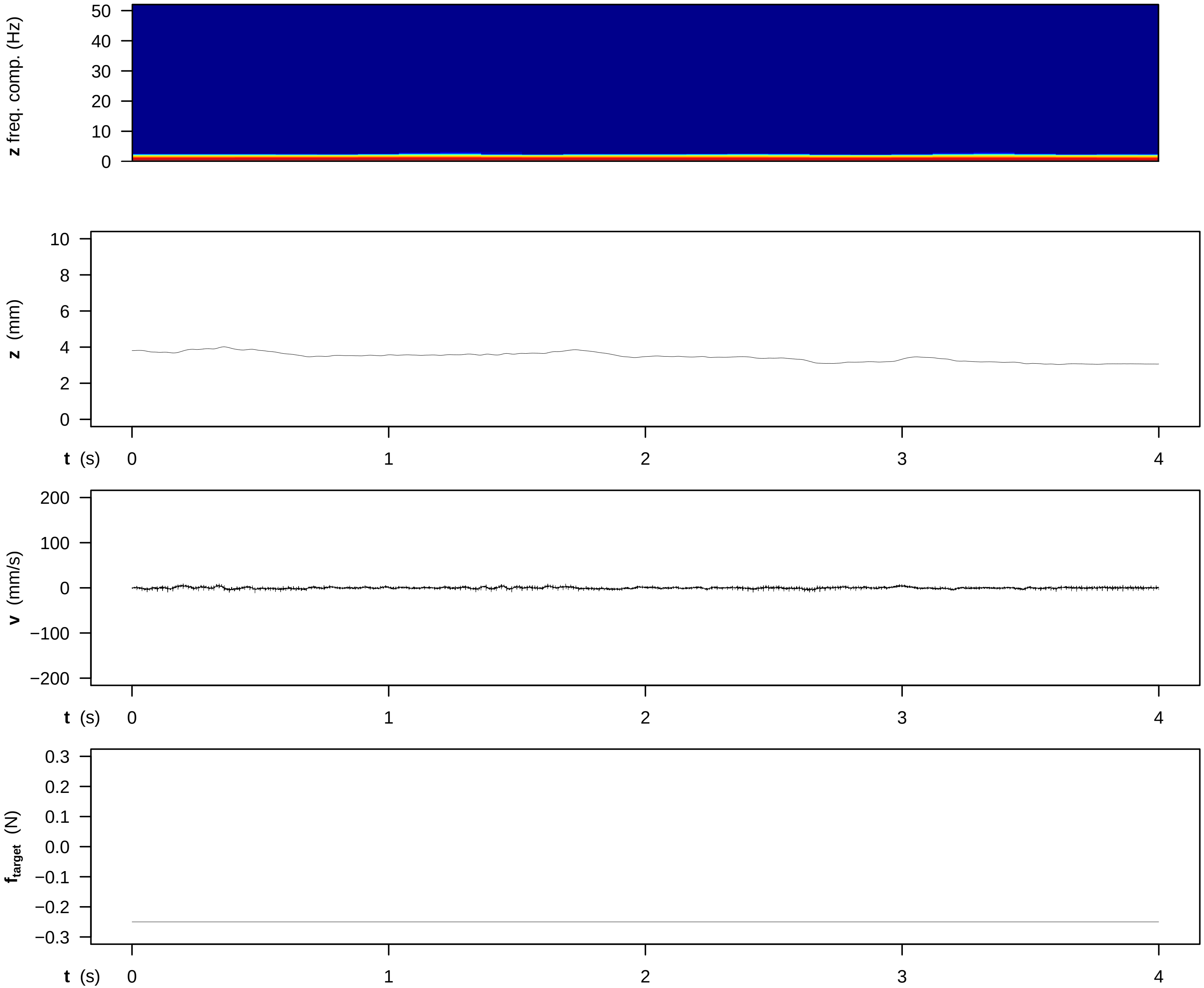

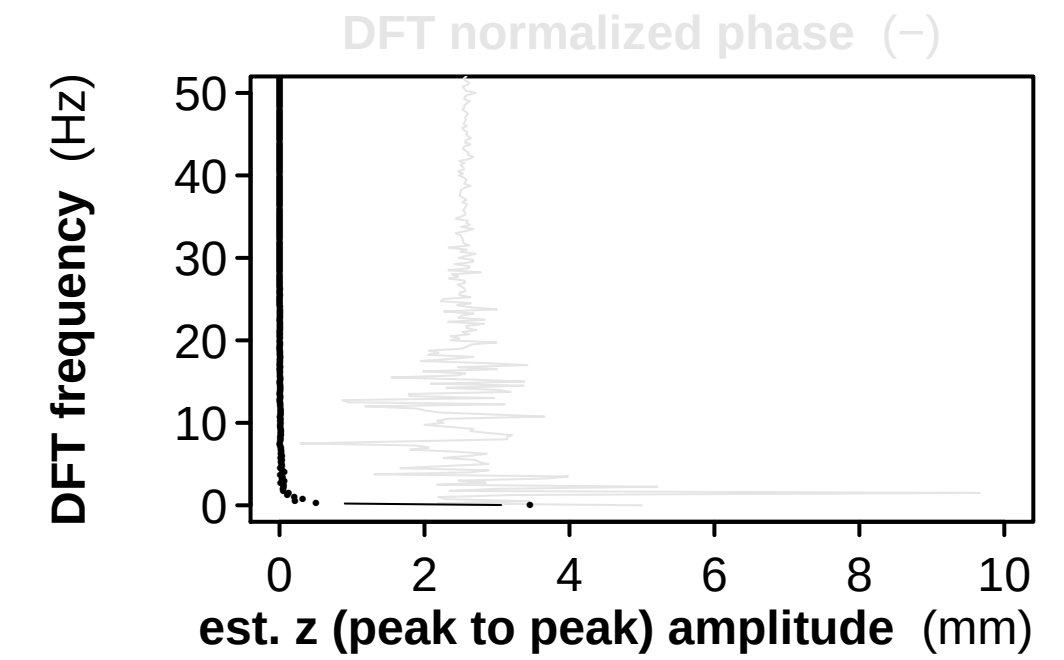

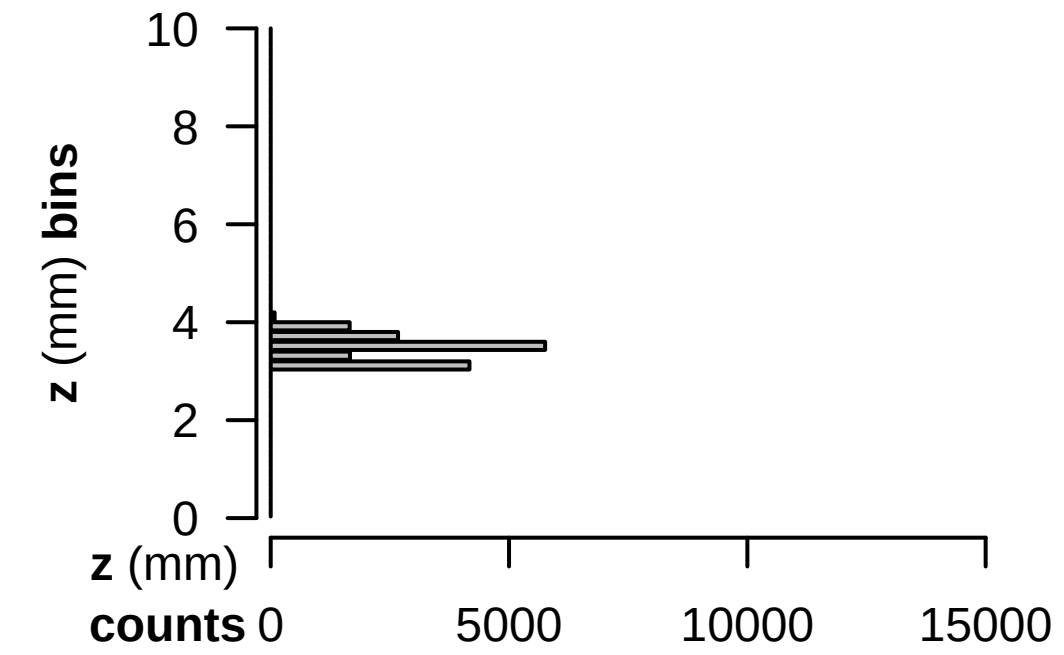

**SUBJECT 8 − RUN 11 − CONDITION 2,0**
**SC_180323_165110_0.AIFF**

```
z_min               : 3.03 mm
z_max               : 4.02 mm
z_travel_amplitude  : 0.98 mm

avg_abs_z_travel    : 2.17 mm/s

z_jarque-bera_jb    : 605.07
z_jarque-bera_p     : 0.00e+00

z_lin_mod_est_slope: -0.20 mm/s
z_lin_mod_adj_R²    : 79 %

z_poly40_mod_adj_R²: 98 %

z_dft_ampl_thresh   : 0.010 mm
>=threshold_maxfreq: 15.50 Hz
```

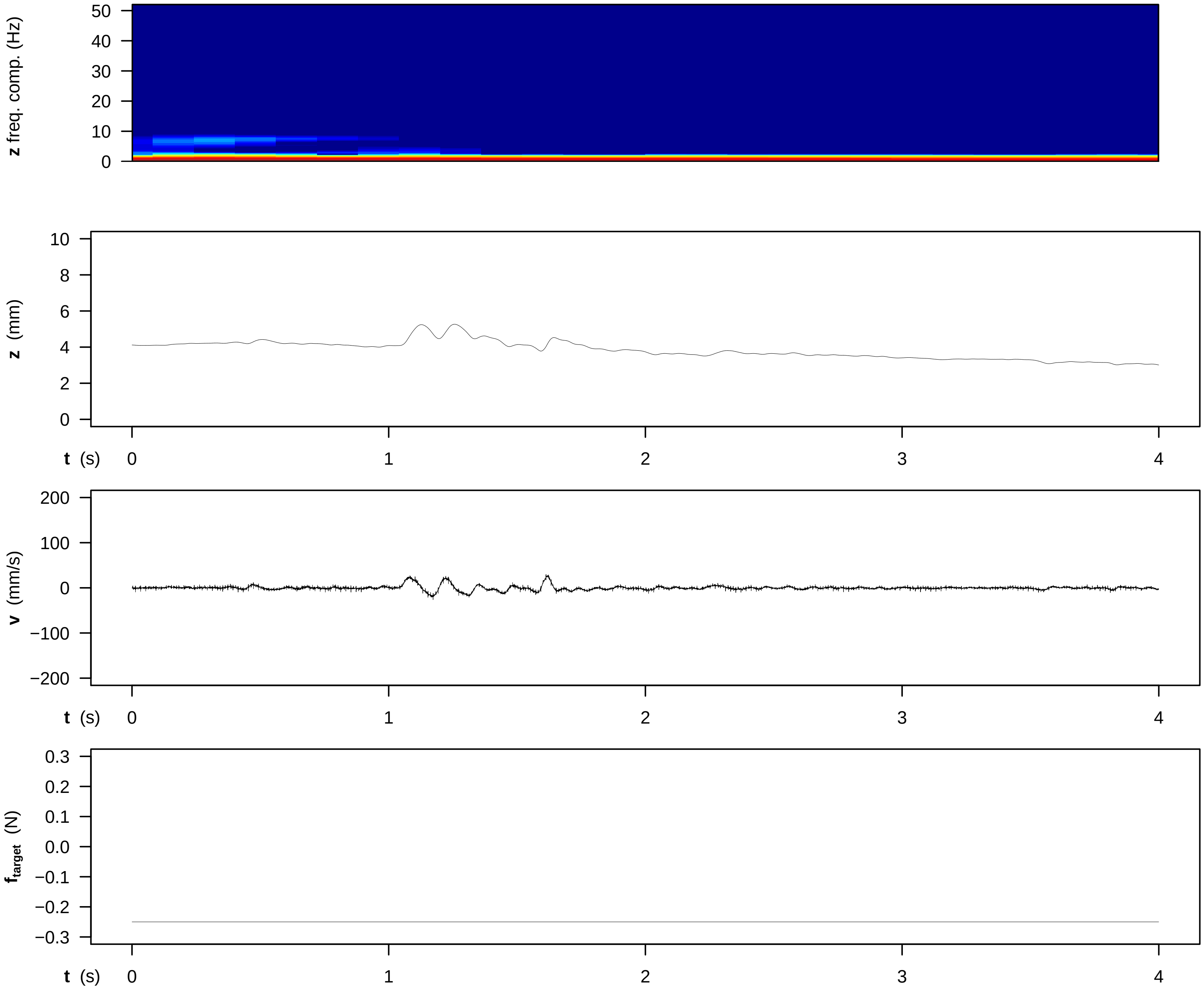

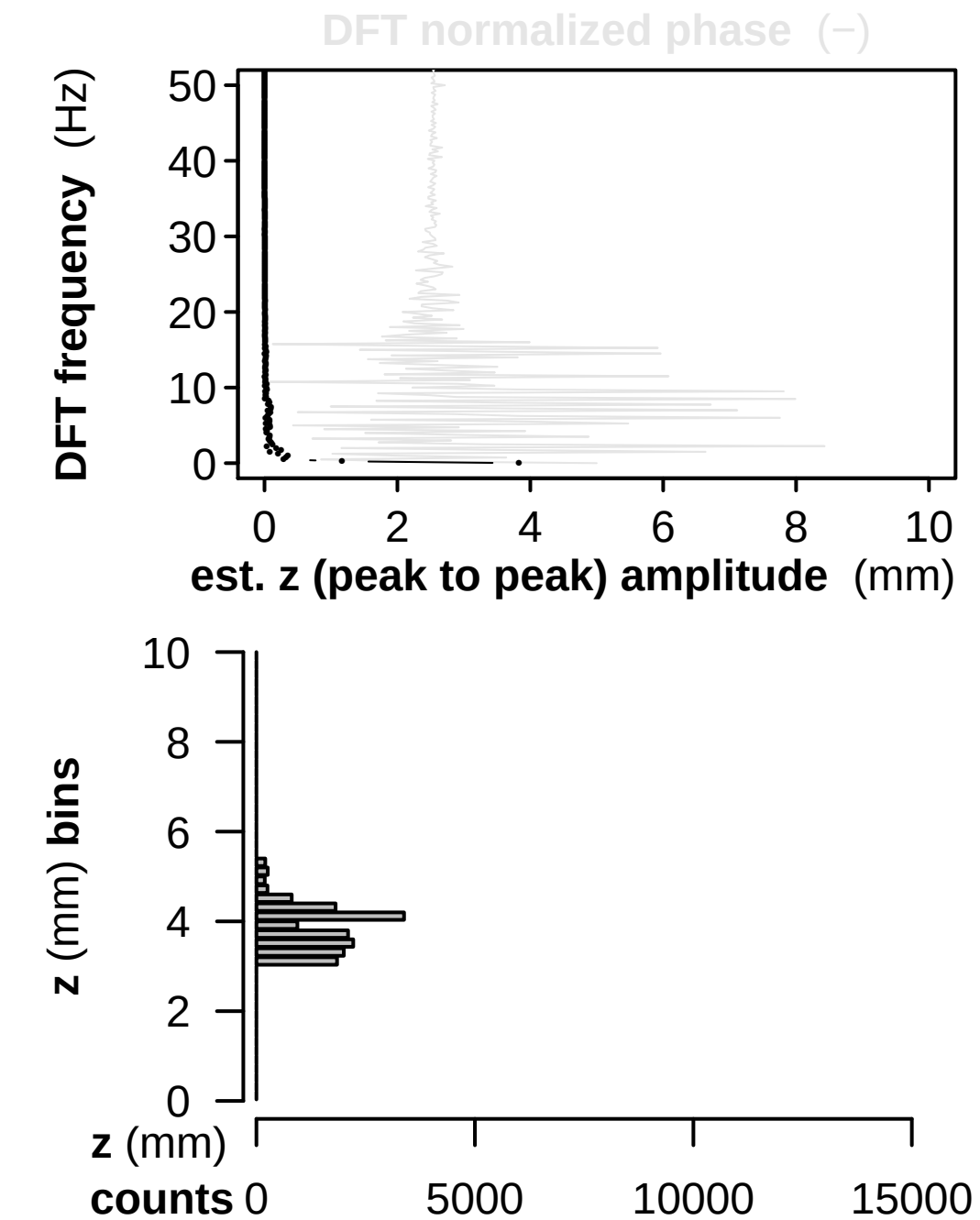

**SUBJECT 8 − RUN 33 − CONDITION 2,0**
**SC_180323_170838_0.AIFF**

```
z_min               : 3.01 mm
z_max               : 5.27 mm
z_travel_amplitude  : 2.26 mm

avg_abs_z_travel    : 3.49 mm/s

z_jarque-bera_jb    : 653.82
z_jarque-bera_p     : 0.00e+00

z_lin_mod_est_slope: −0.36 mm/s
z_lin_mod_adj_R²    : 70 %

z_poly40_mod_adj_R²: 94 %

z_dft_ampl_thresh   : 0.010 mm
>=threshold_maxfreq: 21.50 Hz
```

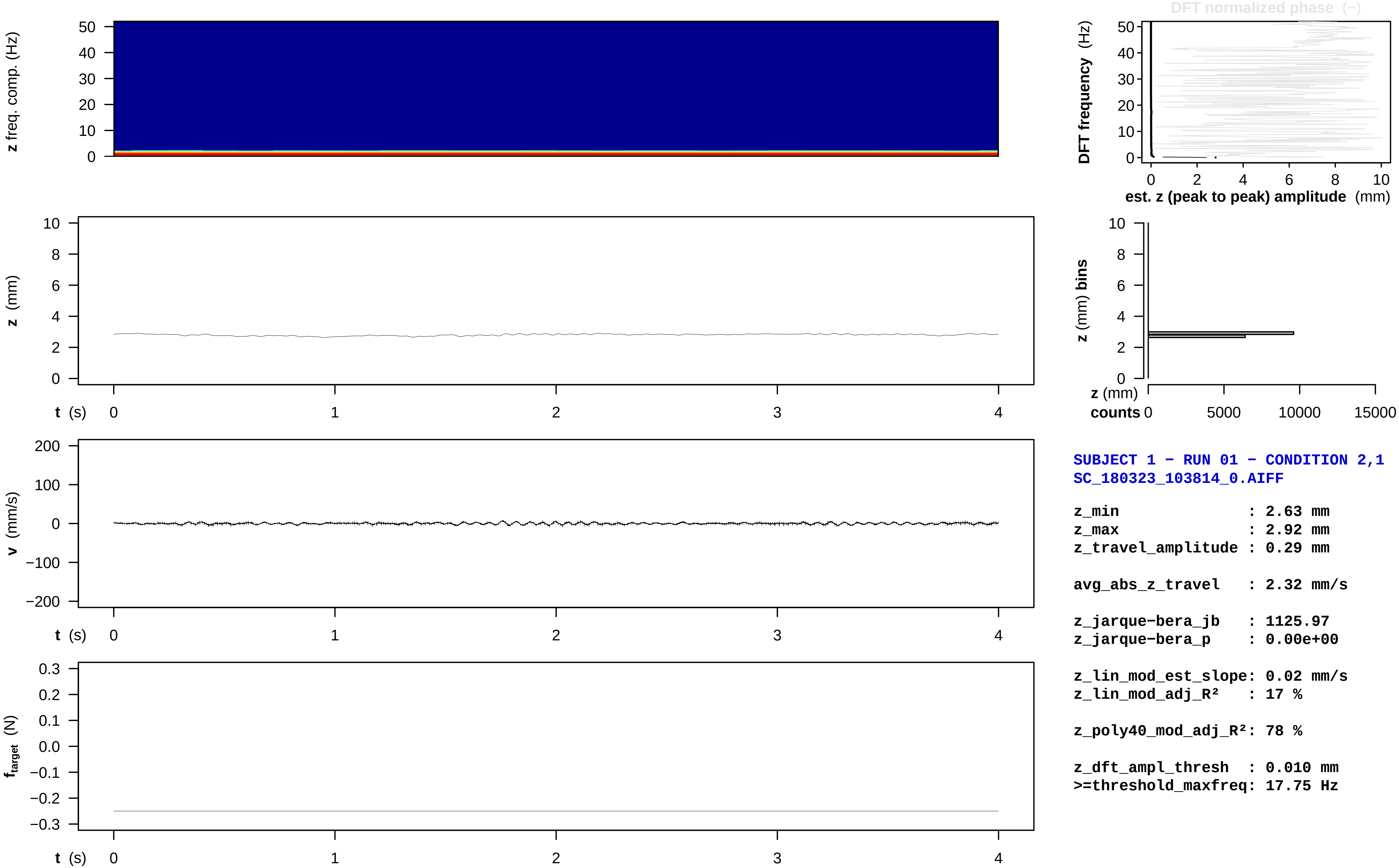

**SUBJECT 1 − RUN 01 − CONDITION 2,1**
**SC_180323_103814_0.AIFF**

```
z_min               : 2.63 mm
z_max               : 2.92 mm
z_travel_amplitude : 0.29 mm

avg_abs_z_travel   : 2.32 mm/s

z_jarque-bera_jb   : 1125.97
z_jarque-bera_p    : 0.00e+00

z_lin_mod_est_slope: 0.02 mm/s
z_lin_mod_adj_R²   : 17 %

z_poly40_mod_adj_R²: 78 %

z_dft_ampl_thresh  : 0.010 mm
>=threshold_maxfreq: 17.75 Hz
```

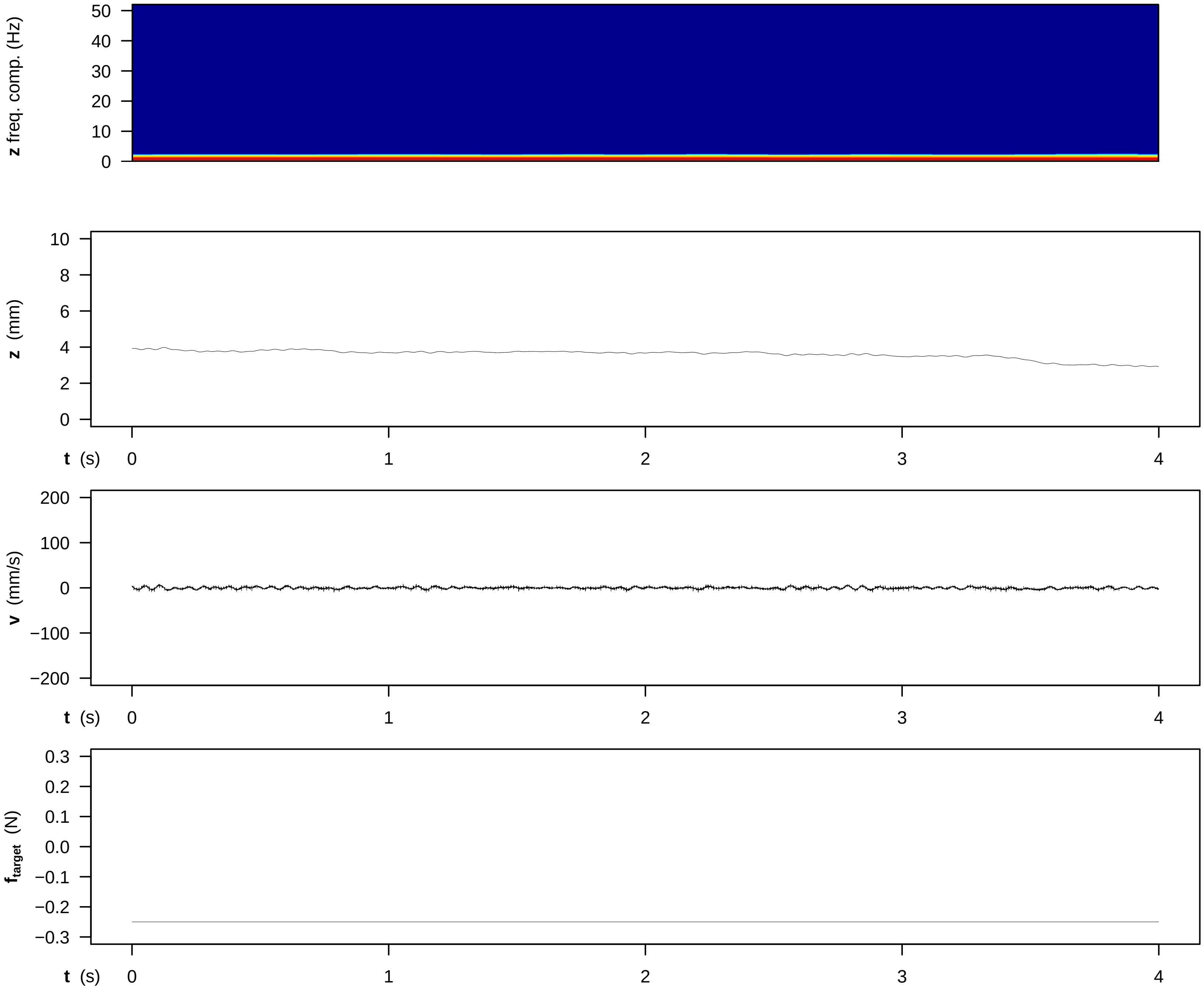

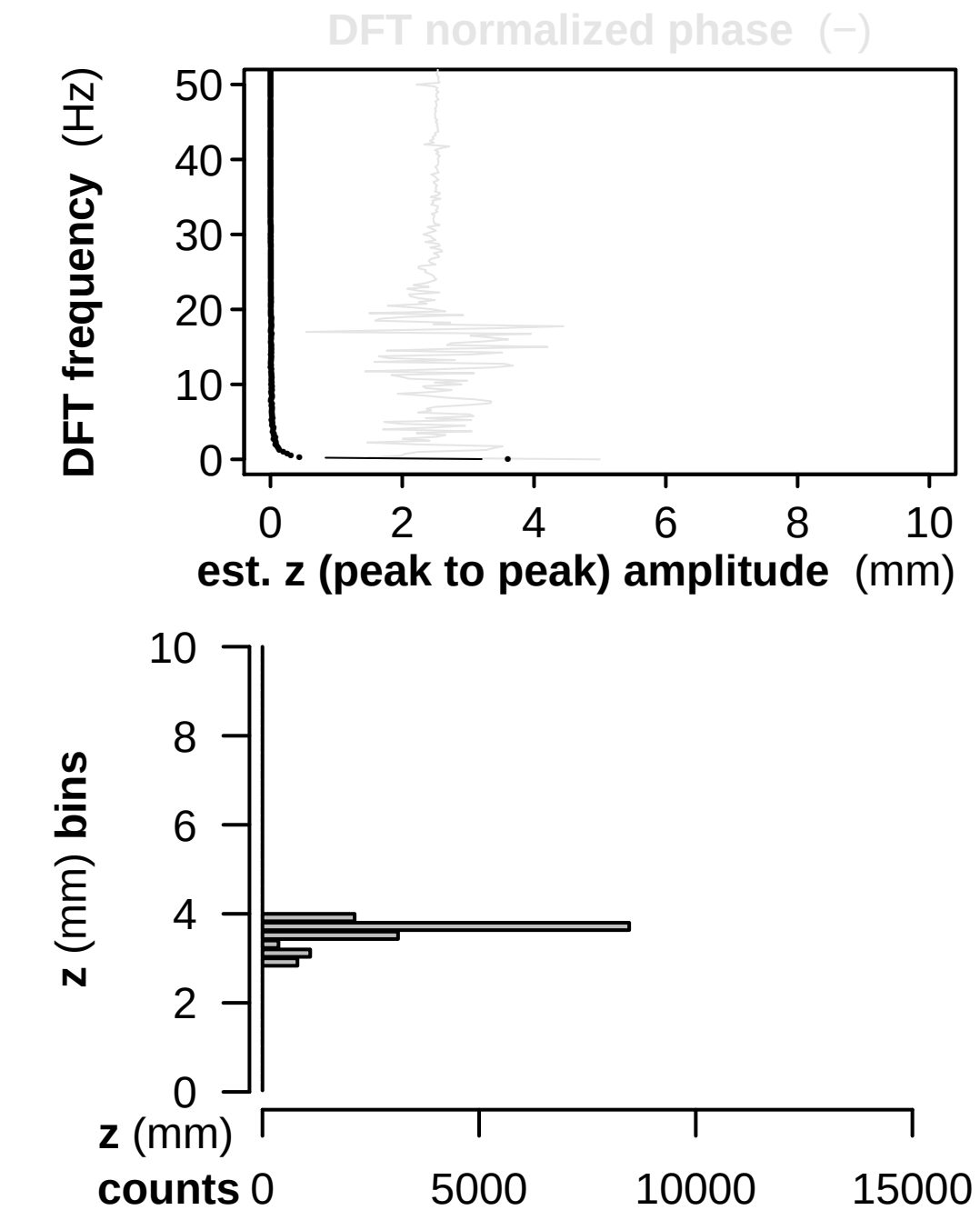

**SUBJECT 1 − RUN 15 − CONDITION 2,1**
**SC_180323_104753_0.AIFF**

```
z_min               : 2.92 mm
z_max               : 3.98 mm
z_travel_amplitude  : 1.06 mm

avg_abs_z_travel    : 3.69 mm/s

z_jarque-bera_jb    : 6394.70
z_jarque-bera_p     : 0.00e+00

z_lin_mod_est_slope: -0.18 mm/s
z_lin_mod_adj_R²    : 72 %

z_poly40_mod_adj_R²: 99 %

z_dft_ampl_thresh   : 0.010 mm
>=threshold_maxfreq: 21.50 Hz
```

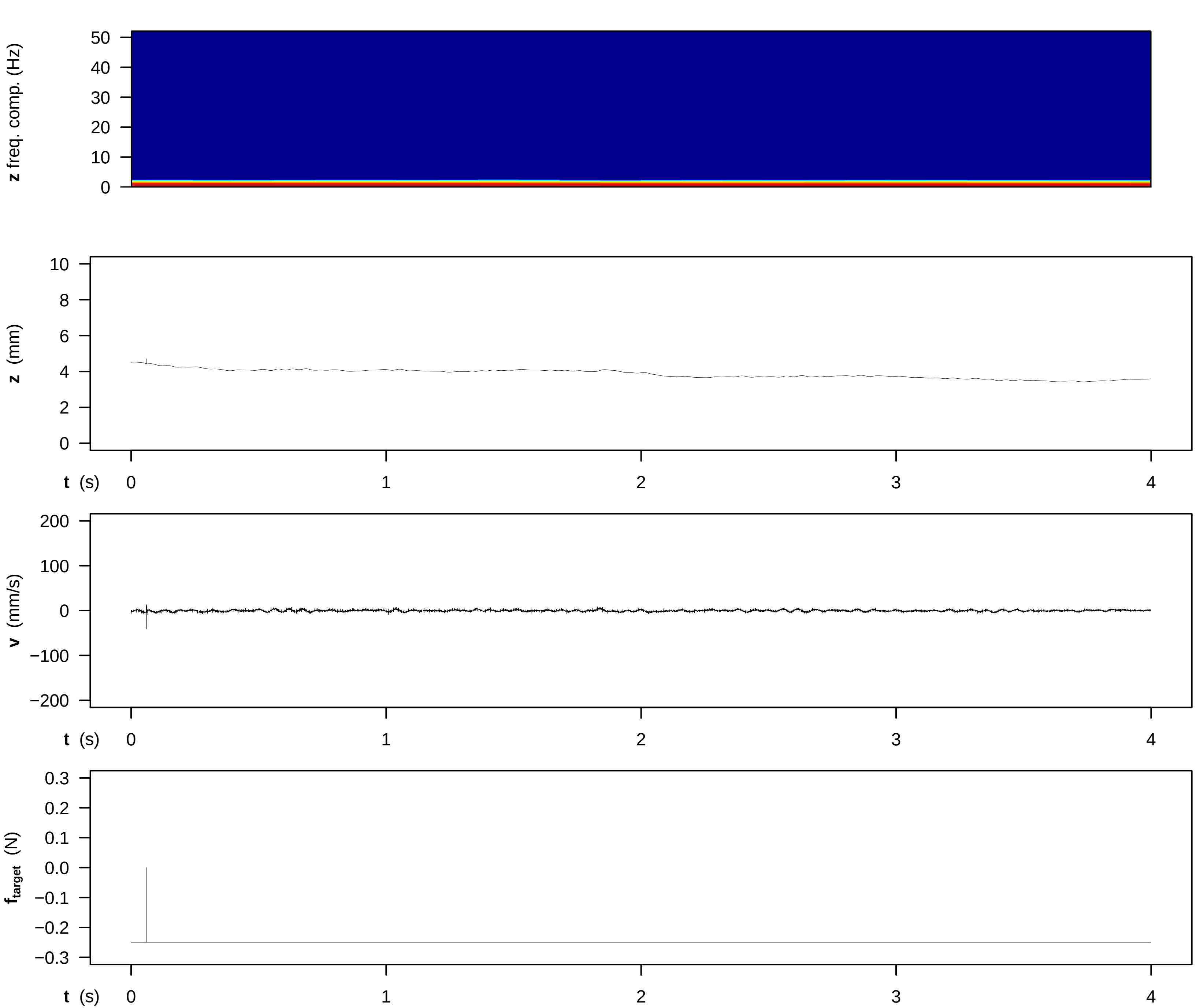

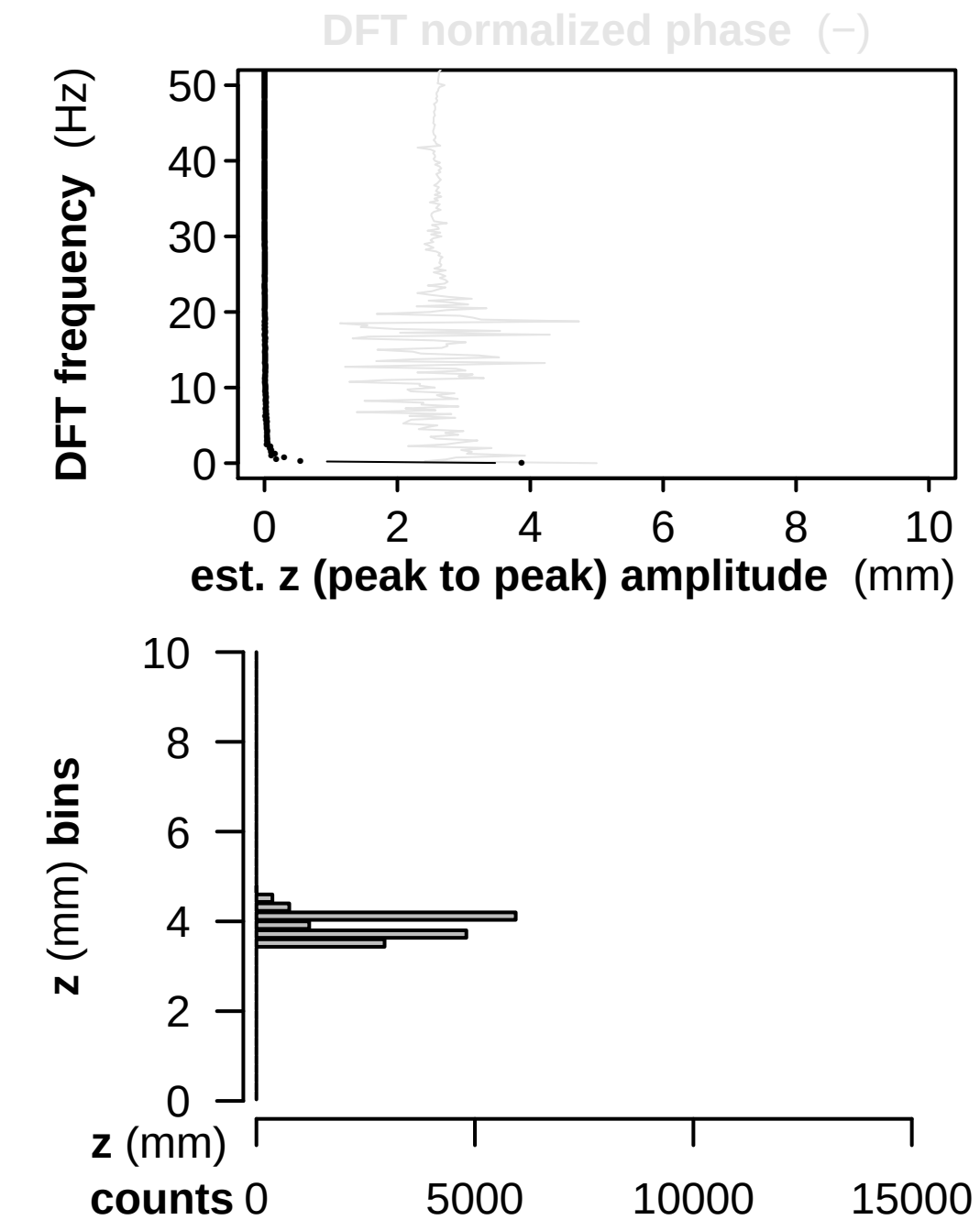

**SUBJECT 1 − RUN 20 − CONDITION 2,1**
**SC_180323_105104_0.AIFF**

```
z_min               : 3.42 mm
z_max               : 4.71 mm
z_travel_amplitude : 1.29 mm

avg_abs_z_travel   : 3.28 mm/s

z_jarque-bera_jb   : 585.01
z_jarque-bera_p    : 0.00e+00

z_lin_mod_est_slope: −0.21 mm/s
z_lin_mod_adj_R²   : 90 %

z_poly40_mod_adj_R²: 99 %

z_dft_ampl_thresh  : 0.010 mm
>=threshold_maxfreq: 19.00 Hz
```

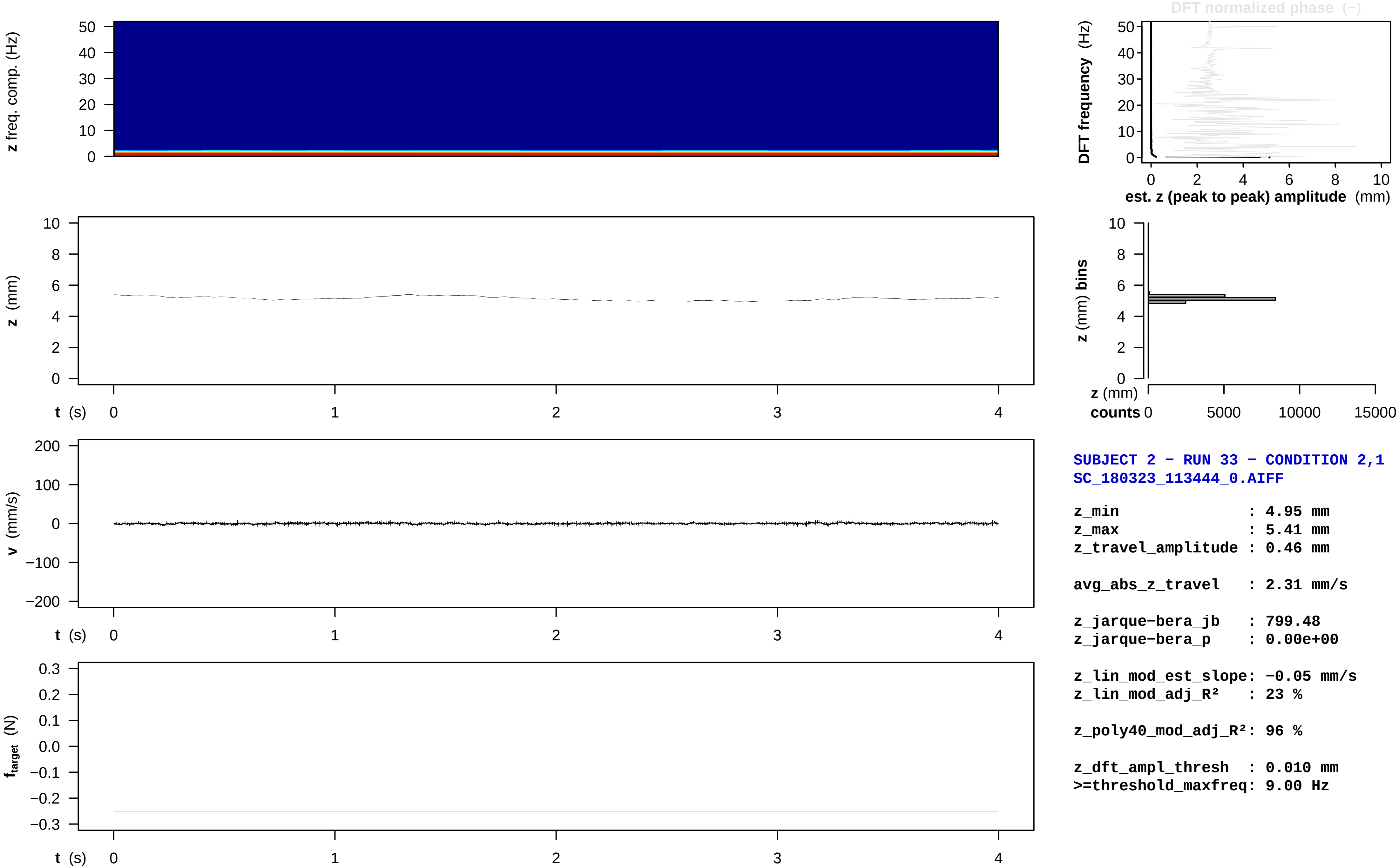

**SUBJECT 2 − RUN 33 − CONDITION 2,1**
**SC_180323_113444_0.AIFF**

```
z_min               : 4.95 mm
z_max               : 5.41 mm
z_travel_amplitude  : 0.46 mm

avg_abs_z_travel    : 2.31 mm/s

z_jarque-bera_jb    : 799.48
z_jarque-bera_p     : 0.00e+00

z_lin_mod_est_slope: −0.05 mm/s
z_lin_mod_adj_R²    : 23 %

z_poly40_mod_adj_R²: 96 %

z_dft_ampl_thresh   : 0.010 mm
>=threshold_maxfreq: 9.00 Hz
```

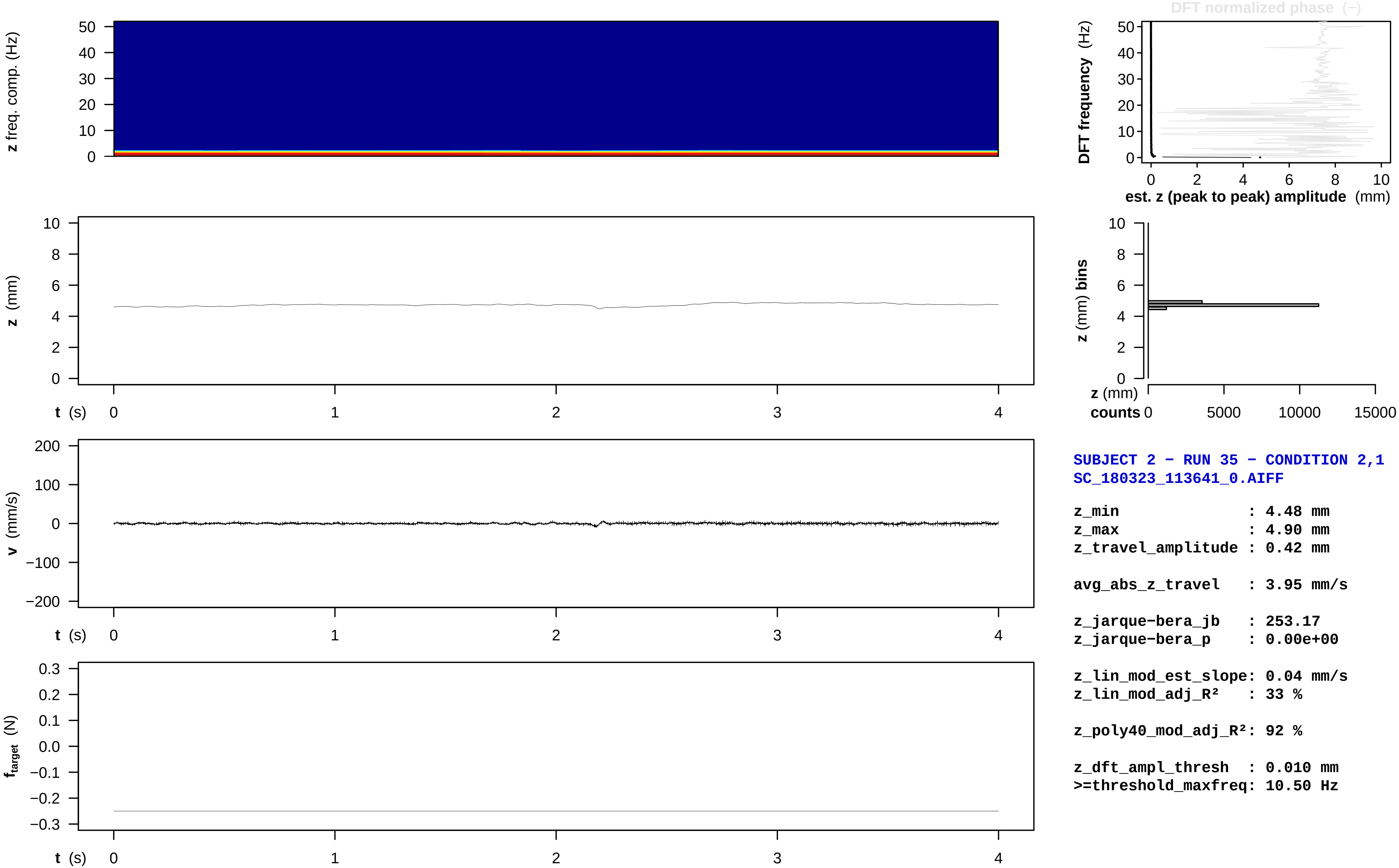

**SUBJECT 2 − RUN 35 − CONDITION 2,1**
**SC_180323_113641_0.AIFF**

```
z_min              : 4.48 mm
z_max              : 4.90 mm
z_travel_amplitude : 0.42 mm

avg_abs_z_travel   : 3.95 mm/s

z_jarque-bera_jb   : 253.17
z_jarque-bera_p    : 0.00e+00

z_lin_mod_est_slope: 0.04 mm/s
z_lin_mod_adj_R²   : 33 %

z_poly40_mod_adj_R²: 92 %

z_dft_ampl_thresh  : 0.010 mm
>=threshold_maxfreq: 10.50 Hz
```

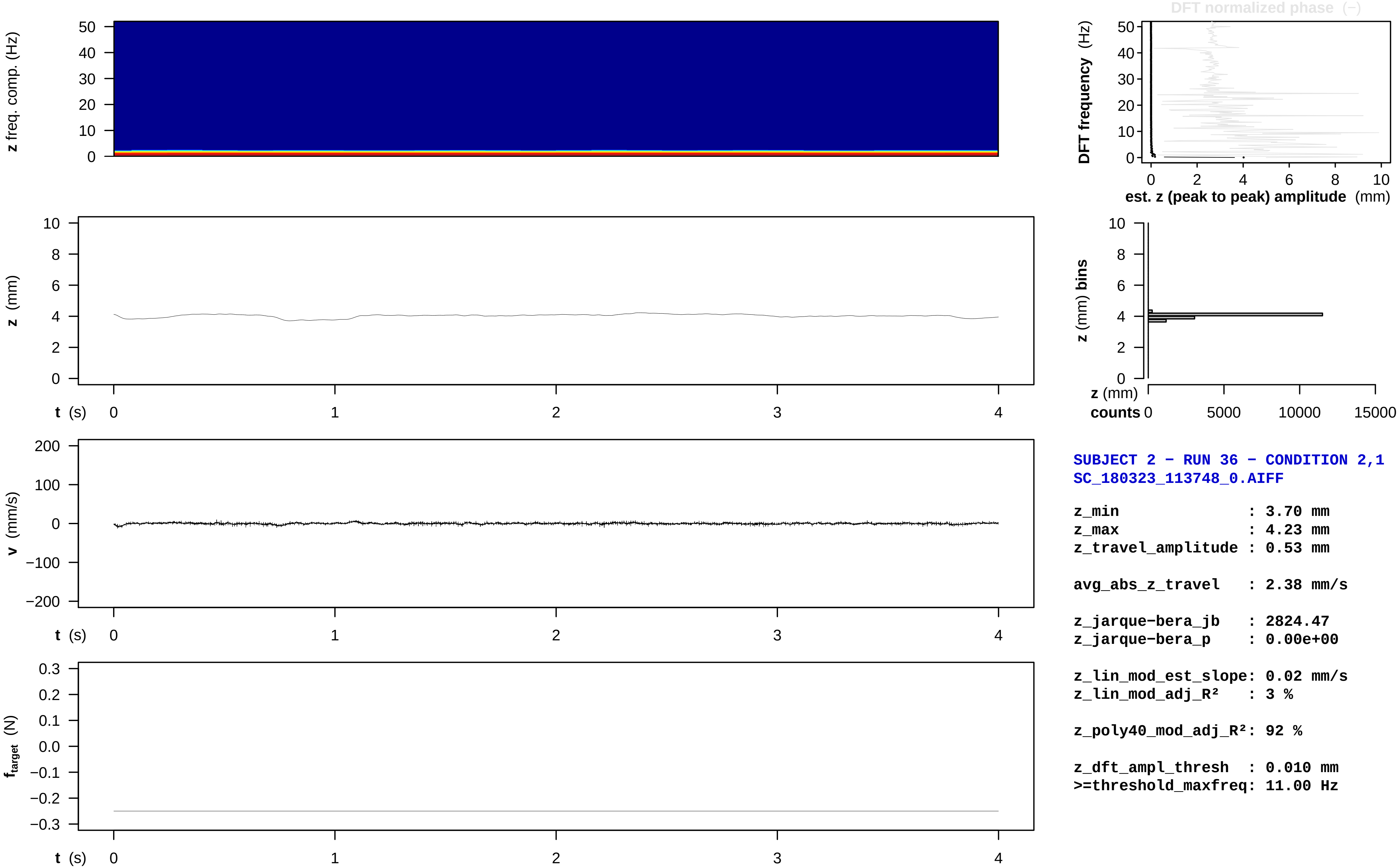

**SUBJECT 2 − RUN 36 − CONDITION 2,1**
**SC_180323_113748_0.AIFF**

```
z_min               : 3.70 mm
z_max               : 4.23 mm
z_travel_amplitude  : 0.53 mm

avg_abs_z_travel    : 2.38 mm/s

z_jarque-bera_jb    : 2824.47
z_jarque-bera_p     : 0.00e+00

z_lin_mod_est_slope : 0.02 mm/s
z_lin_mod_adj_R²    : 3 %

z_poly40_mod_adj_R² : 92 %

z_dft_ampl_thresh   : 0.010 mm
>=threshold_maxfreq : 11.00 Hz
```

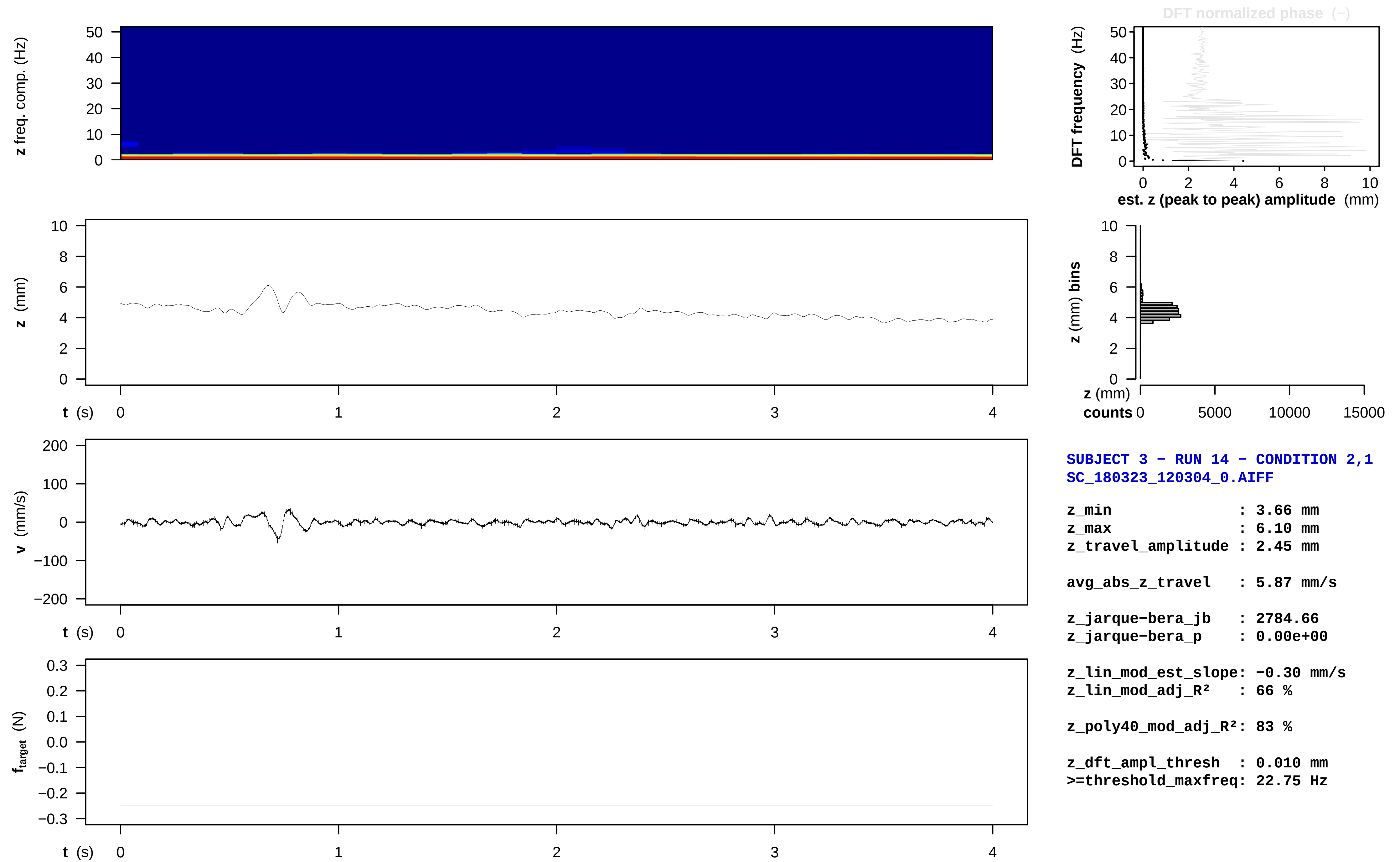

**SUBJECT 3 − RUN 14 − CONDITION 2,1**
**SC_180323_120304_0.AIFF**

```
z_min               : 3.66 mm
z_max               : 6.10 mm
z_travel_amplitude  : 2.45 mm

avg_abs_z_travel    : 5.87 mm/s

z_jarque-bera_jb    : 2784.66
z_jarque-bera_p     : 0.00e+00

z_lin_mod_est_slope: −0.30 mm/s
z_lin_mod_adj_R²    : 66 %

z_poly40_mod_adj_R²: 83 %

z_dft_ampl_thresh   : 0.010 mm
>=threshold_maxfreq: 22.75 Hz
```

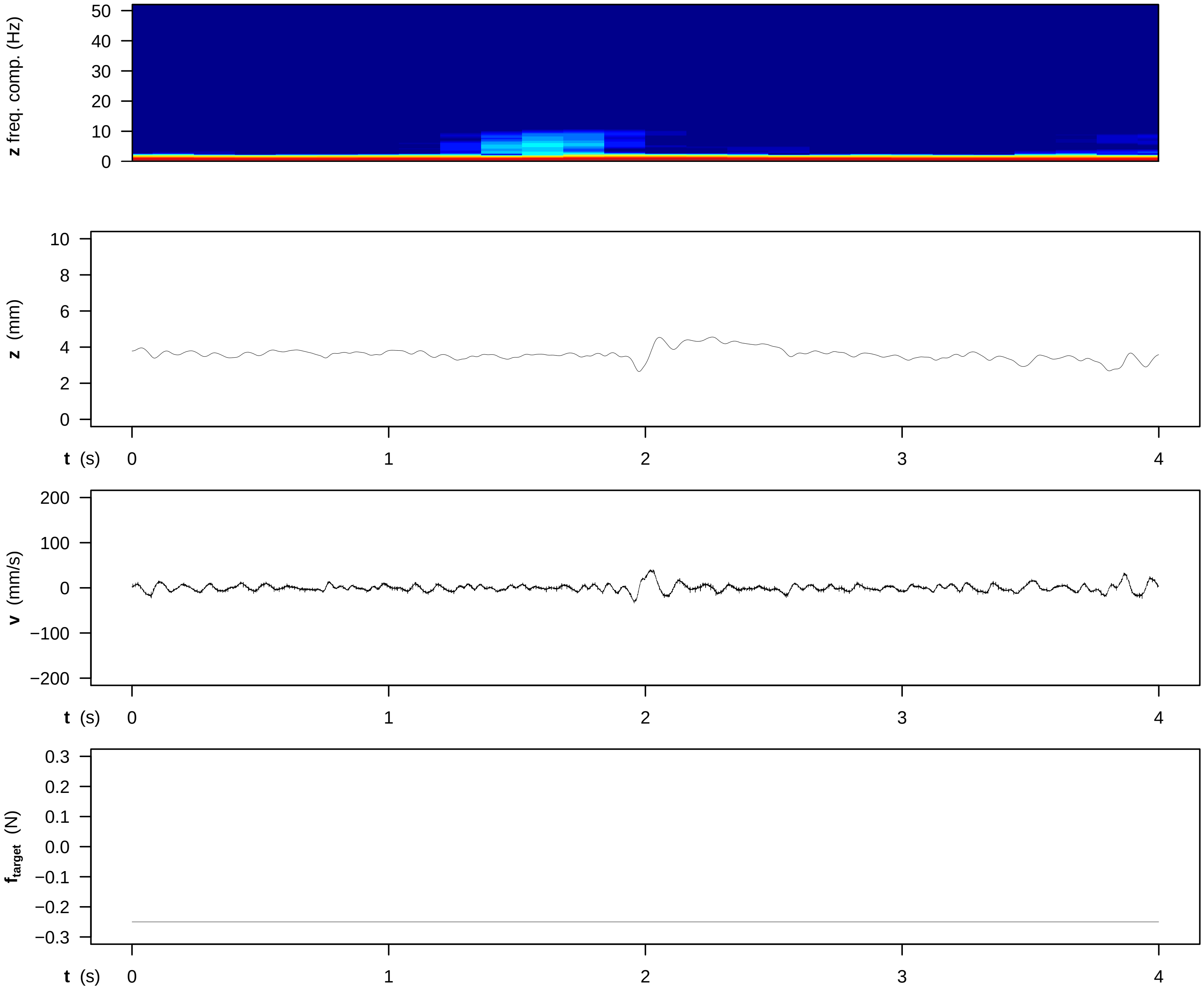

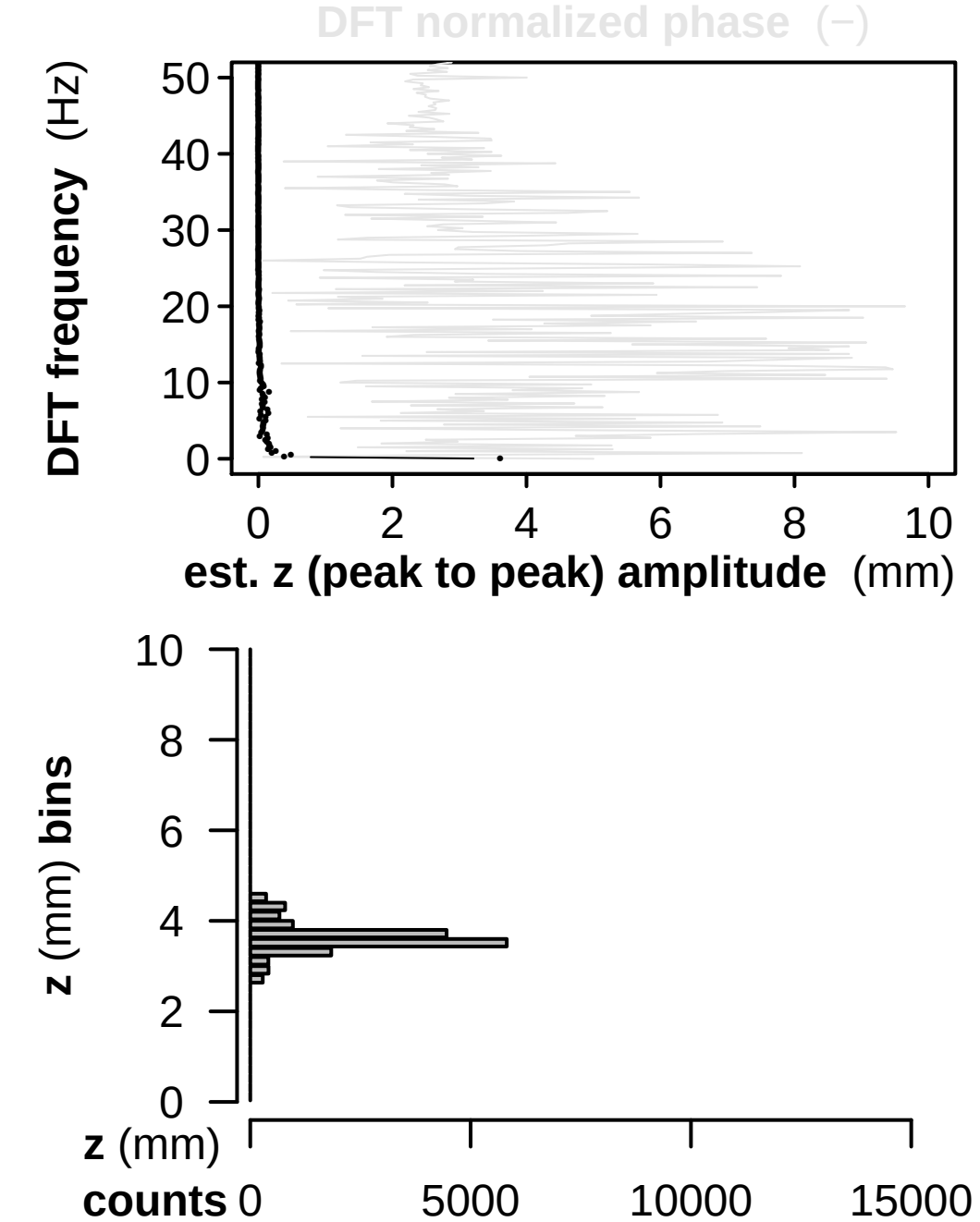

**SUBJECT 3 − RUN 16 − CONDITION 2,1**
**SC_180323_120445_0.AIFF**

```
z_min              : 2.65 mm
z_max              : 4.56 mm
z_travel_amplitude : 1.91 mm

avg_abs_z_travel   : 6.12 mm/s

z_jarque-bera_jb   : 1535.41
z_jarque-bera_p    : 0.00e+00

z_lin_mod_est_slope: −0.08 mm/s
z_lin_mod_adj_R²   : 8 %

z_poly40_mod_adj_R²: 74 %

z_dft_ampl_thresh  : 0.010 mm
>=threshold_maxfreq: 19.50 Hz
```

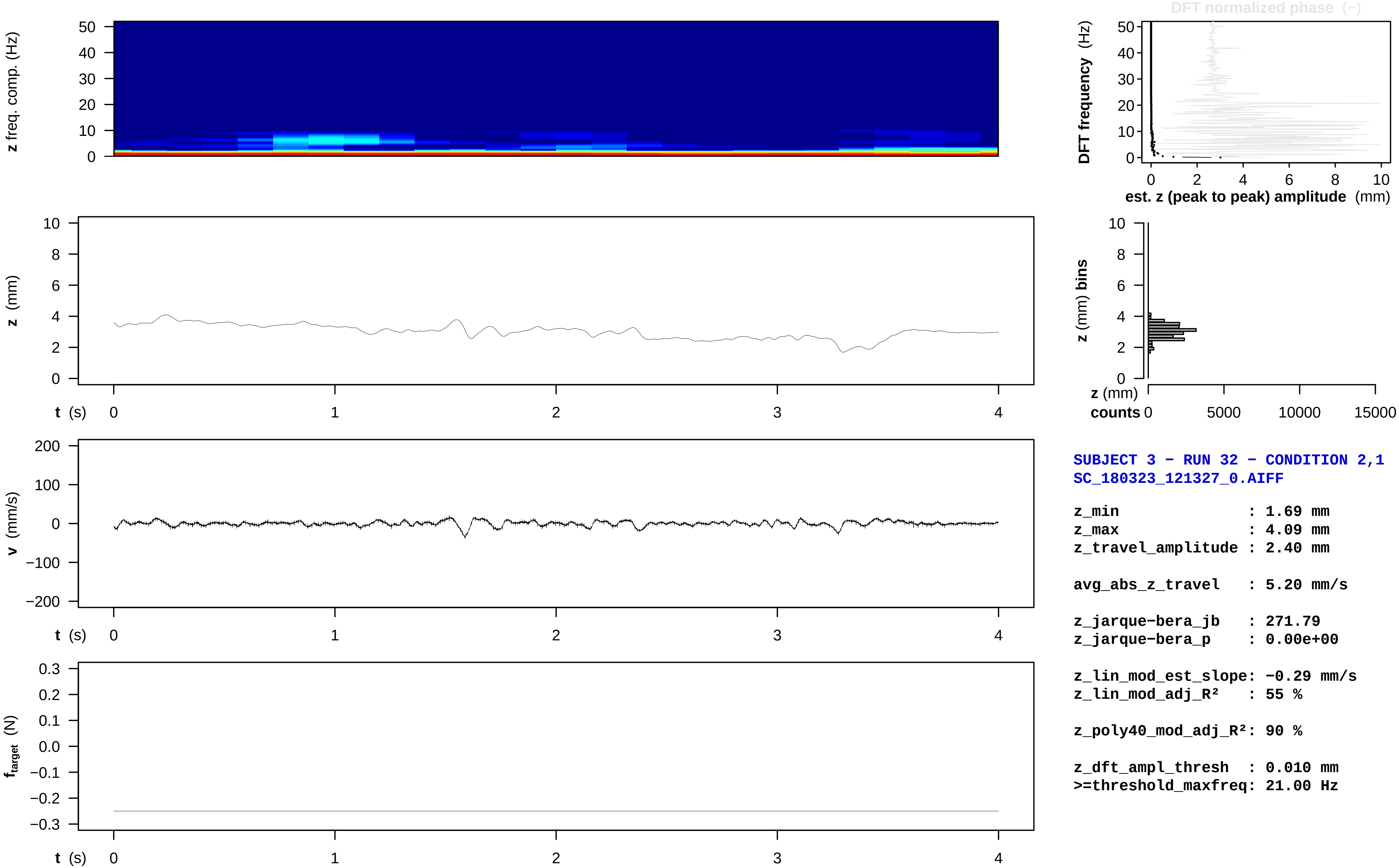

**SUBJECT 3 − RUN 32 − CONDITION 2,1**
**SC_180323_121327_0.AIFF**

```
z_min               : 1.69 mm
z_max               : 4.09 mm
z_travel_amplitude  : 2.40 mm

avg_abs_z_travel    : 5.20 mm/s

z_jarque-bera_jb    : 271.79
z_jarque-bera_p     : 0.00e+00

z_lin_mod_est_slope: −0.29 mm/s
z_lin_mod_adj_R²    : 55 %

z_poly40_mod_adj_R²: 90 %

z_dft_ampl_thresh   : 0.010 mm
>=threshold_maxfreq: 21.00 Hz
```

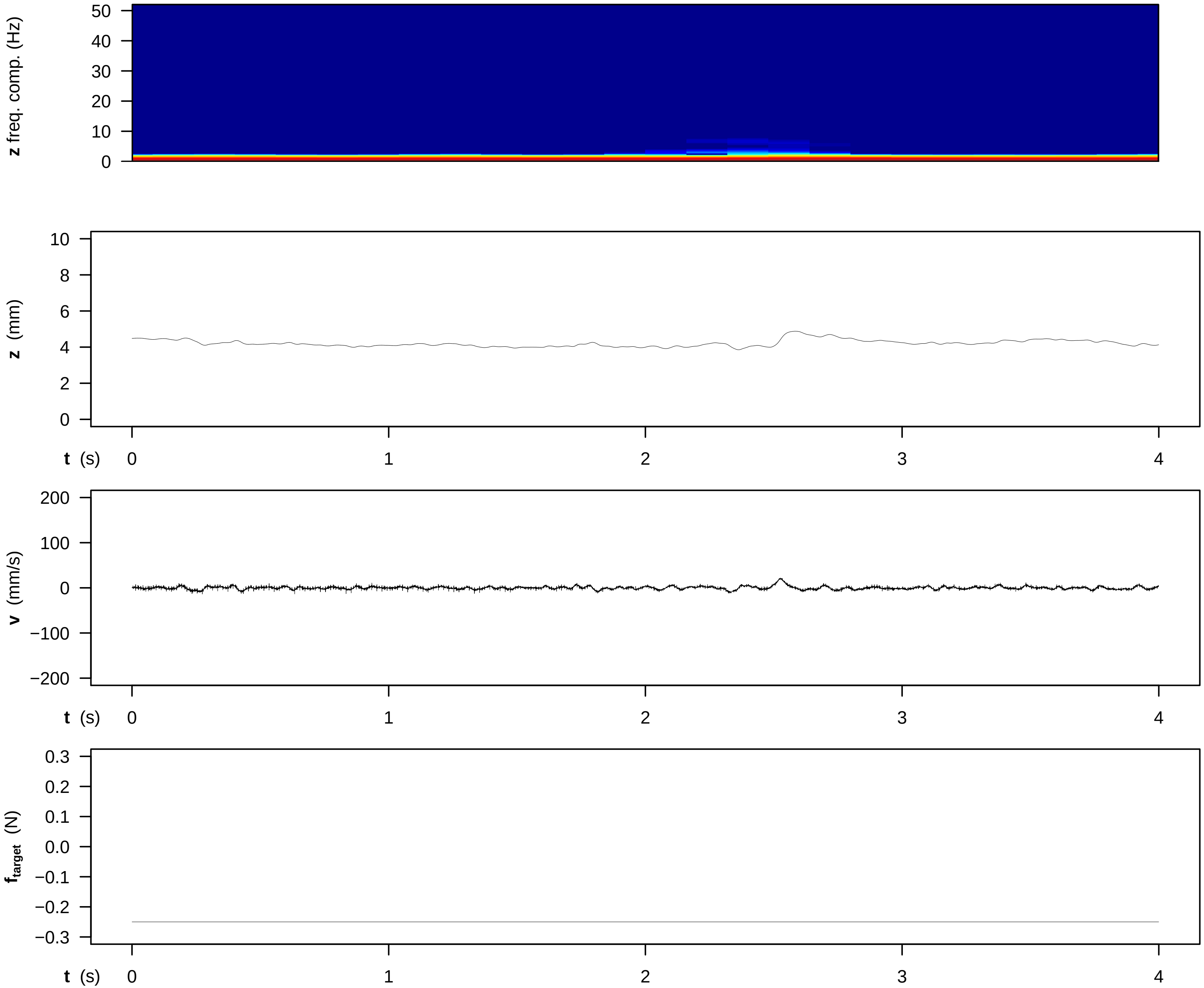

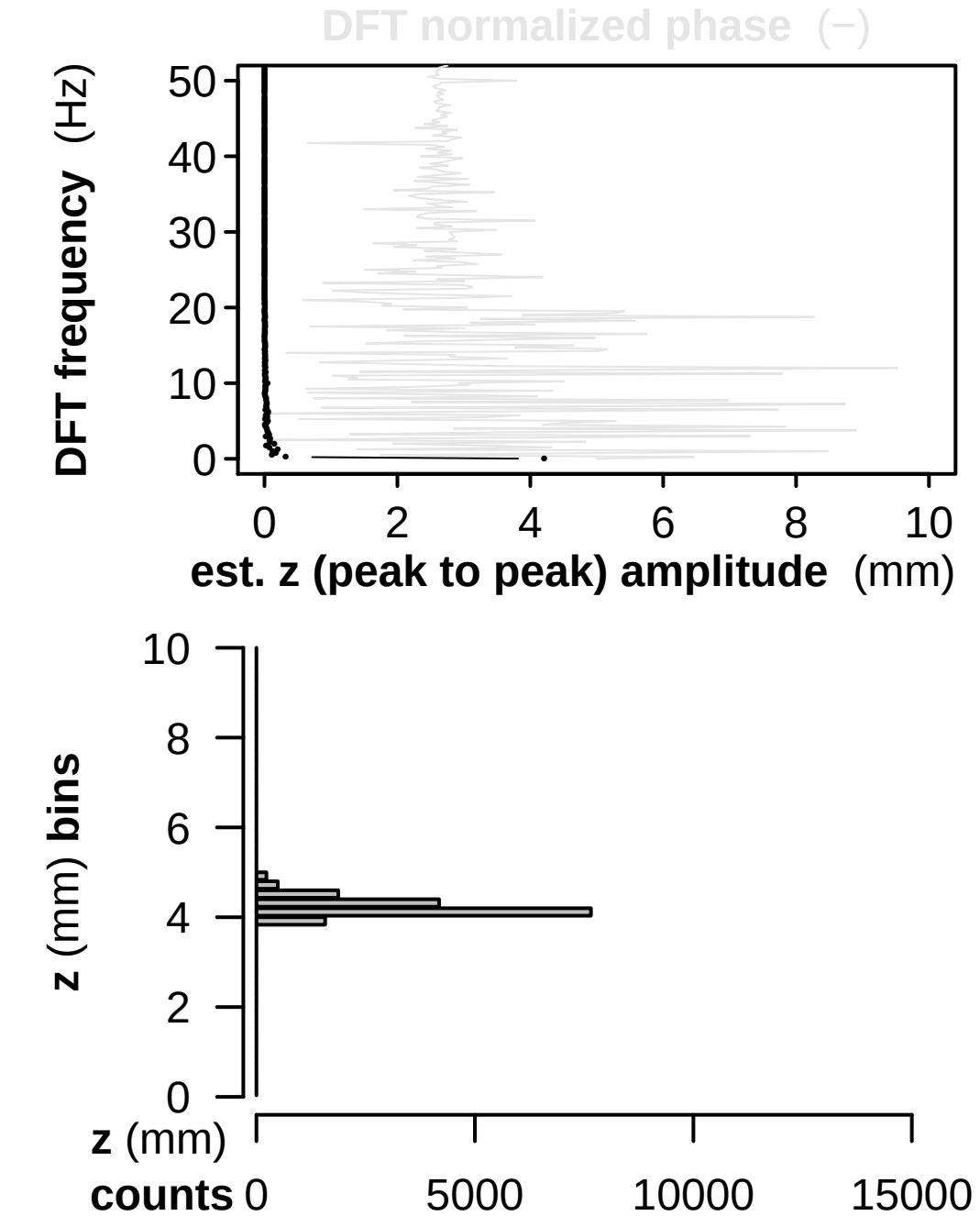

**SUBJECT 4 − RUN 02 − CONDITION 2,1**
**SC_180323_123118_0.AIFF**

```
z_min               : 3.85 mm
z_max               : 4.88 mm
z_travel_amplitude  : 1.03 mm

avg_abs_z_travel    : 3.10 mm/s

z_jarque-bera_jb    : 3430.17
z_jarque-bera_p     : 0.00e+00

z_lin_mod_est_slope: 0.03 mm/s
z_lin_mod_adj_R²    : 4 %

z_poly40_mod_adj_R²: 78 %

z_dft_ampl_thresh   : 0.010 mm
>=threshold_maxfreq: 18.75 Hz
```

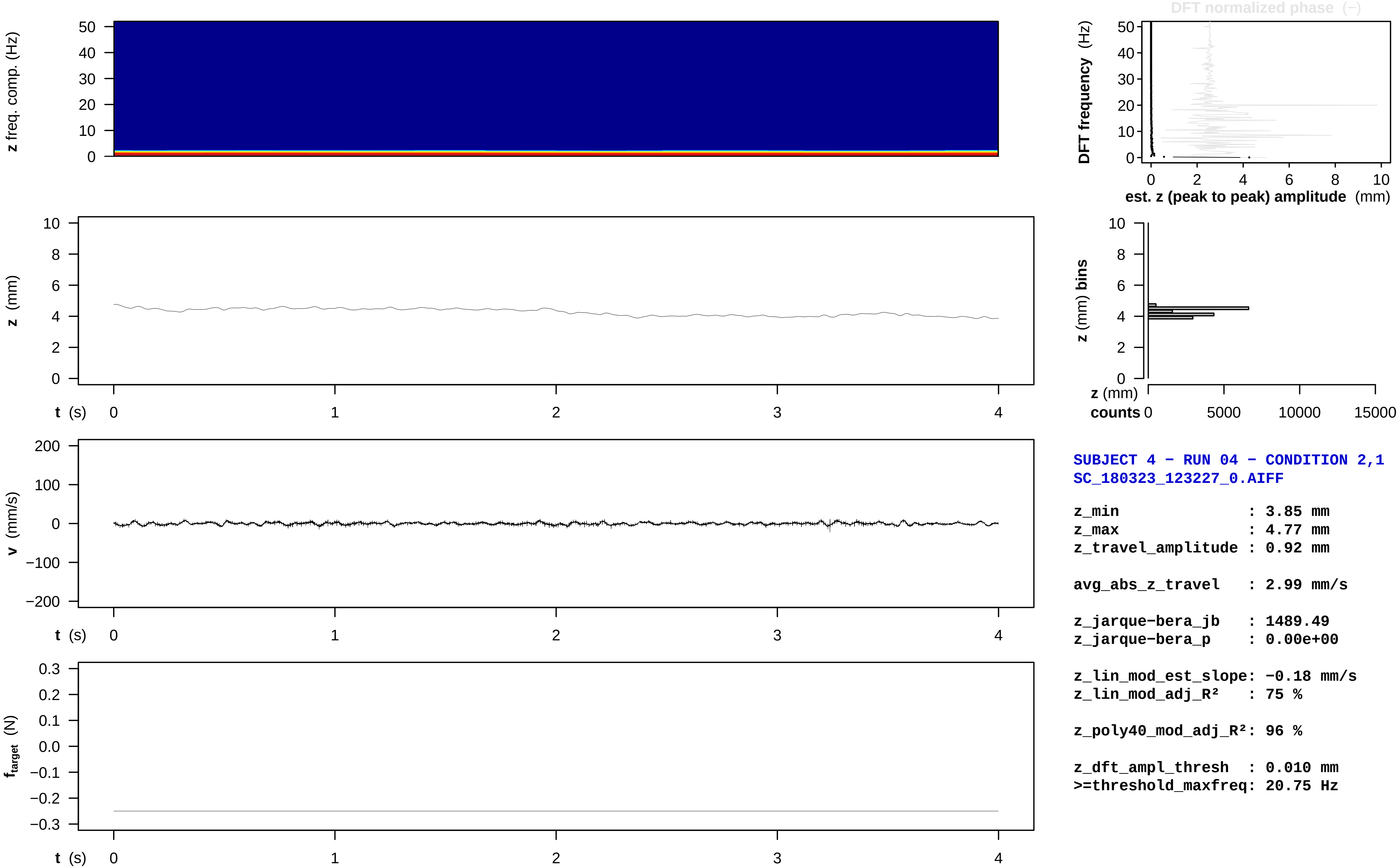

**SUBJECT 4 − RUN 04 − CONDITION 2,1**
**SC_180323_123227_0.AIFF**

```
z_min              : 3.85 mm
z_max              : 4.77 mm
z_travel_amplitude : 0.92 mm

avg_abs_z_travel   : 2.99 mm/s

z_jarque-bera_jb   : 1489.49
z_jarque-bera_p    : 0.00e+00

z_lin_mod_est_slope: -0.18 mm/s
z_lin_mod_adj_R²   : 75 %

z_poly40_mod_adj_R²: 96 %

z_dft_ampl_thresh  : 0.010 mm
>=threshold_maxfreq: 20.75 Hz
```

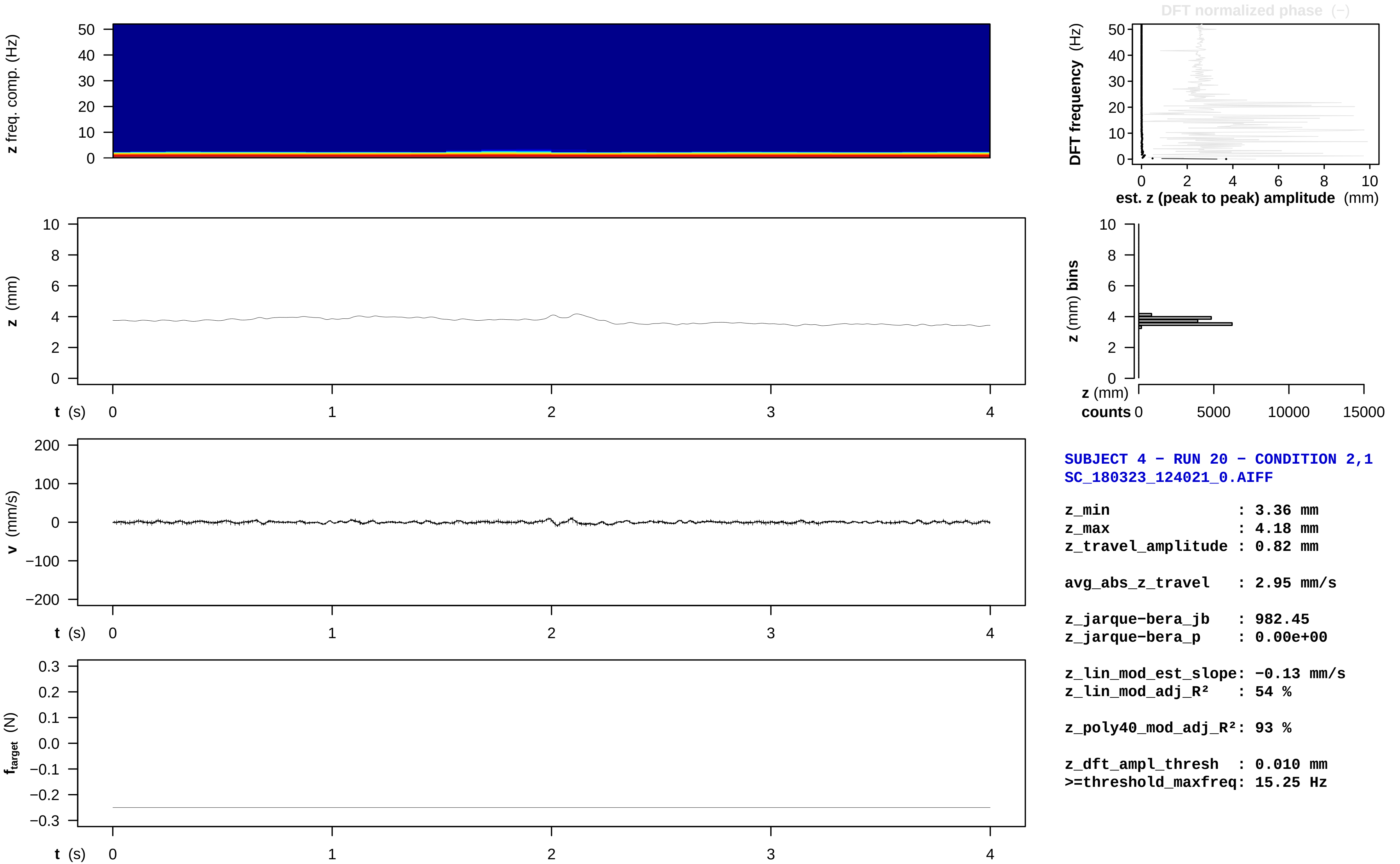

**SUBJECT 4 − RUN 20 − CONDITION 2,1**
**SC_180323_124021_0.AIFF**

```
z_min               : 3.36 mm
z_max               : 4.18 mm
z_travel_amplitude  : 0.82 mm

avg_abs_z_travel    : 2.95 mm/s

z_jarque-bera_jb    : 982.45
z_jarque-bera_p     : 0.00e+00

z_lin_mod_est_slope: −0.13 mm/s
z_lin_mod_adj_R²    : 54 %

z_poly40_mod_adj_R²: 93 %

z_dft_ampl_thresh   : 0.010 mm
>=threshold_maxfreq: 15.25 Hz
```

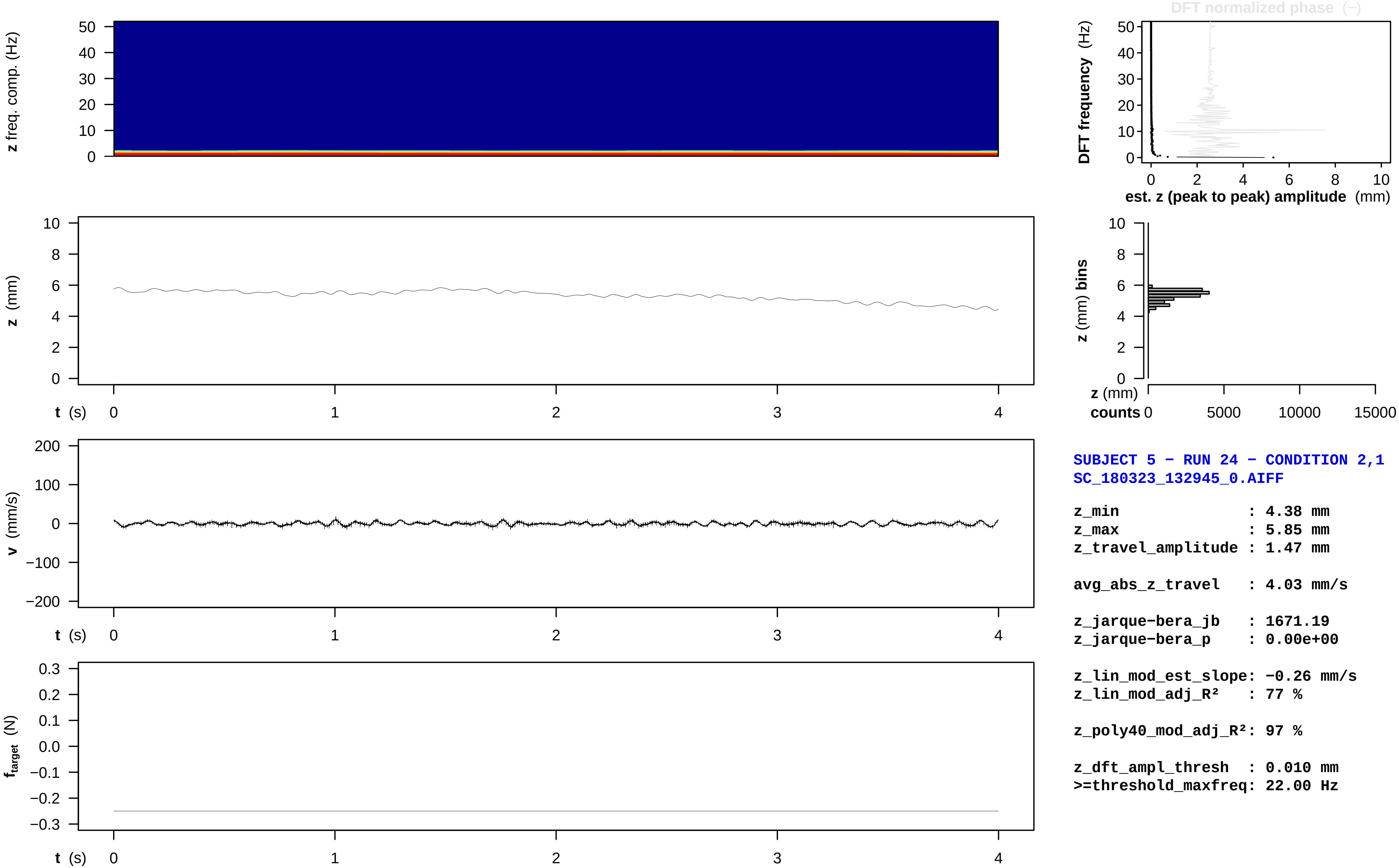

**SUBJECT 5 − RUN 24 − CONDITION 2,1**
**SC_180323_132945_0.AIFF**

```
z_min               : 4.38 mm
z_max               : 5.85 mm
z_travel_amplitude  : 1.47 mm

avg_abs_z_travel    : 4.03 mm/s

z_jarque-bera_jb    : 1671.19
z_jarque-bera_p     : 0.00e+00

z_lin_mod_est_slope: −0.26 mm/s
z_lin_mod_adj_R²    : 77 %

z_poly40_mod_adj_R²: 97 %

z_dft_ampl_thresh   : 0.010 mm
>=threshold_maxfreq: 22.00 Hz
```

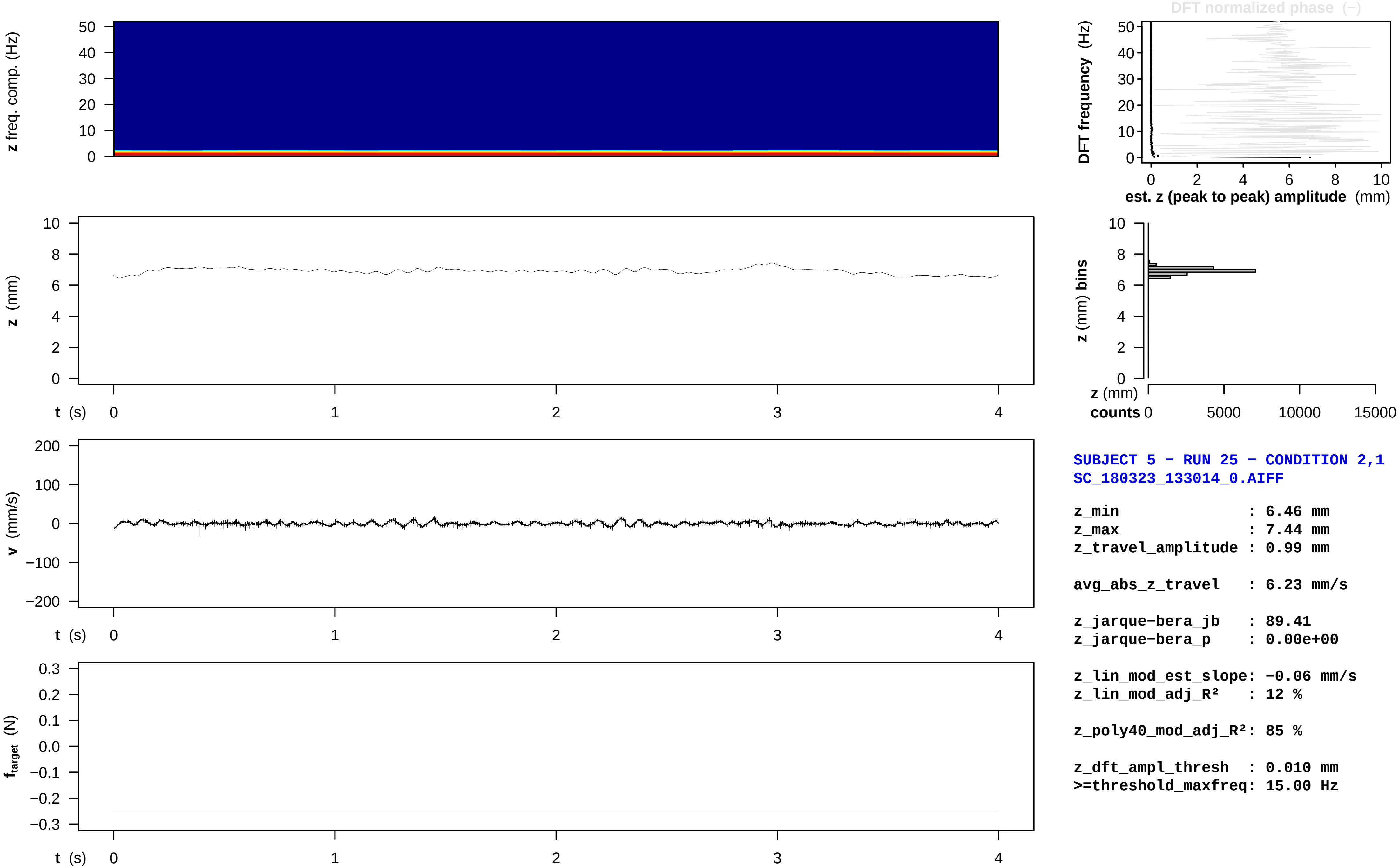

**SUBJECT 5 − RUN 25 − CONDITION 2,1**
**SC_180323_133014_0.AIFF**

```
z_min               : 6.46 mm
z_max               : 7.44 mm
z_travel_amplitude  : 0.99 mm

avg_abs_z_travel    : 6.23 mm/s

z_jarque-bera_jb    : 89.41
z_jarque-bera_p     : 0.00e+00

z_lin_mod_est_slope: −0.06 mm/s
z_lin_mod_adj_R²    : 12 %

z_poly40_mod_adj_R²: 85 %

z_dft_ampl_thresh   : 0.010 mm
>=threshold_maxfreq: 15.00 Hz
```

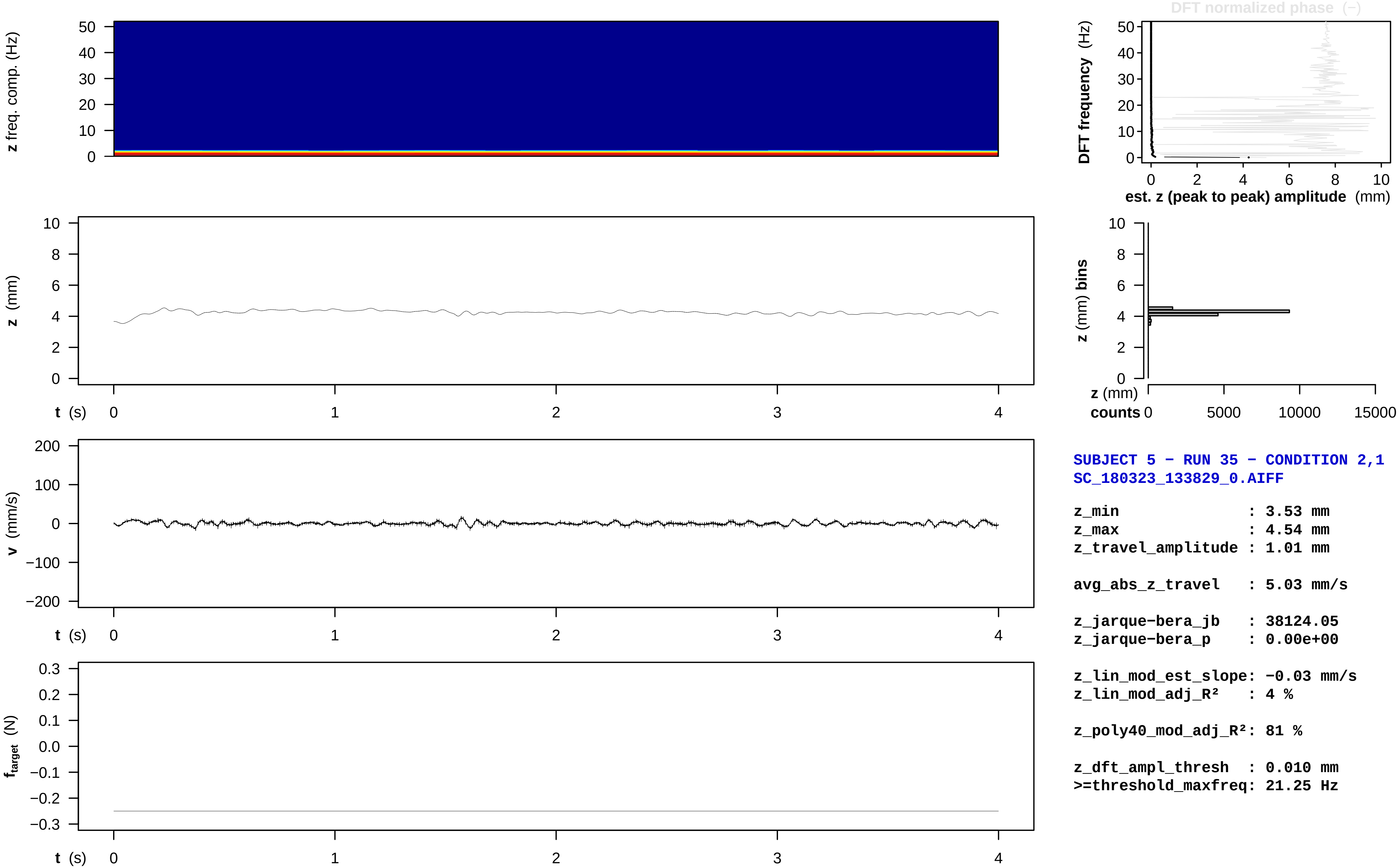

**SUBJECT 5 − RUN 35 − CONDITION 2,1**
**SC_180323_133829_0.AIFF**

```
z_min               : 3.53 mm
z_max               : 4.54 mm
z_travel_amplitude  : 1.01 mm

avg_abs_z_travel    : 5.03 mm/s

z_jarque-bera_jb    : 38124.05
z_jarque-bera_p     : 0.00e+00

z_lin_mod_est_slope: −0.03 mm/s
z_lin_mod_adj_R²    : 4 %

z_poly40_mod_adj_R²: 81 %

z_dft_ampl_thresh   : 0.010 mm
>=threshold_maxfreq: 21.25 Hz
```

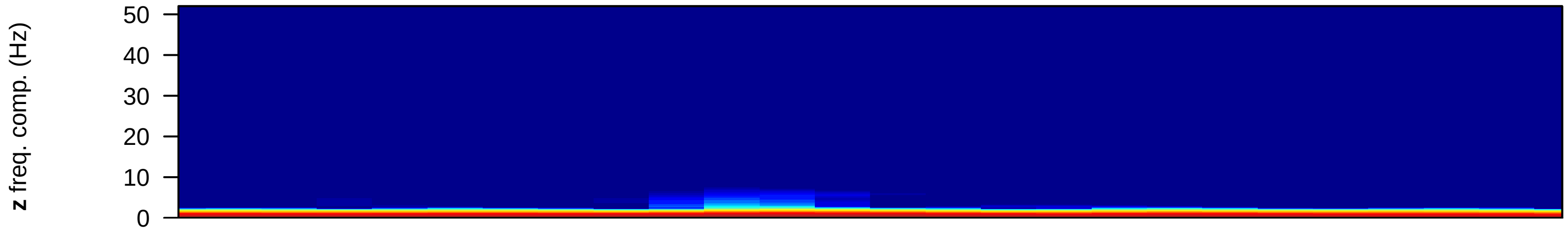

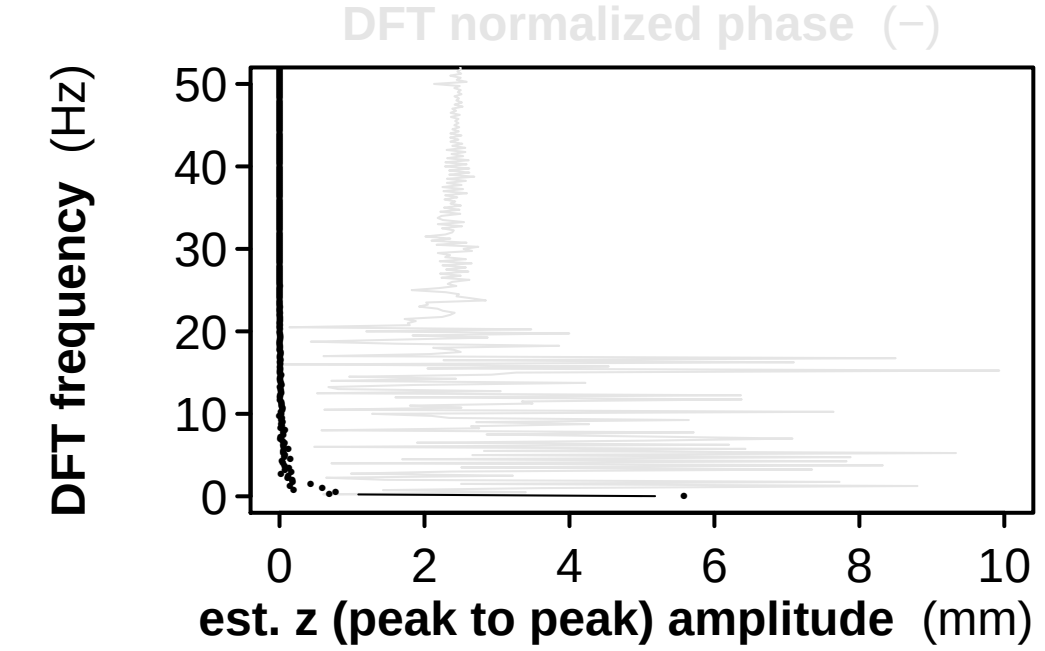

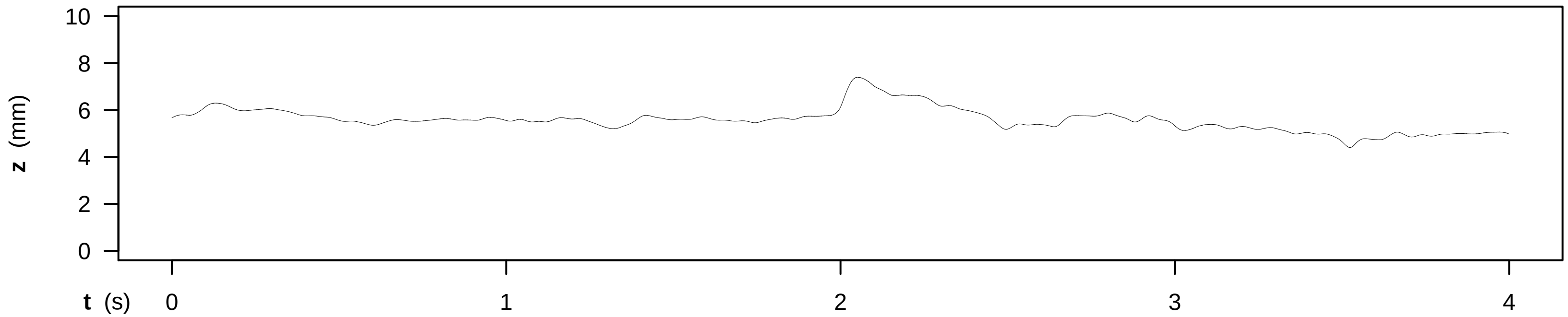

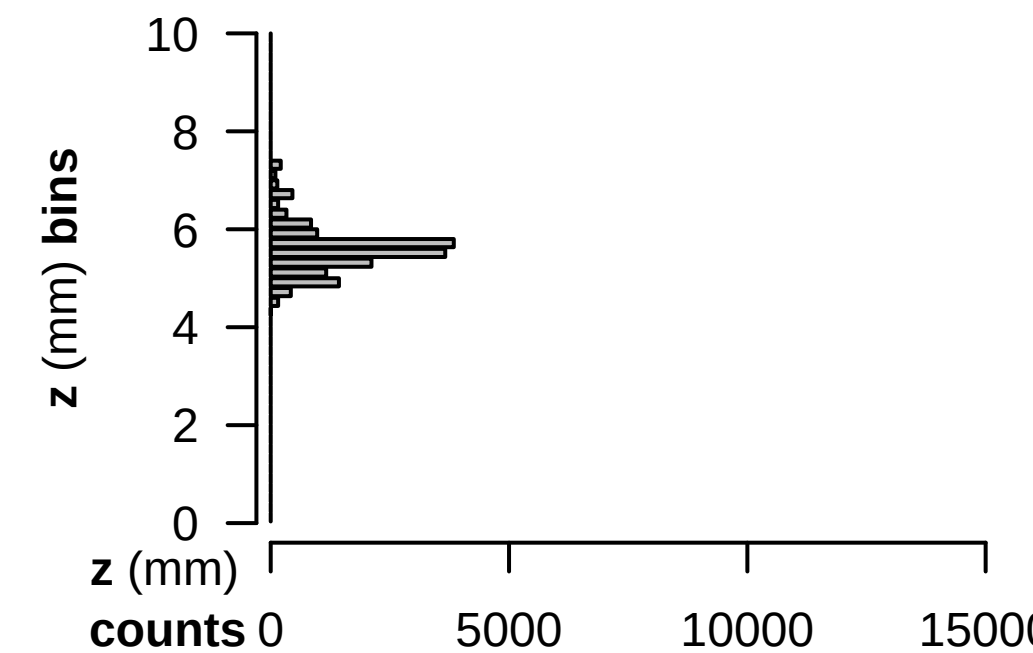

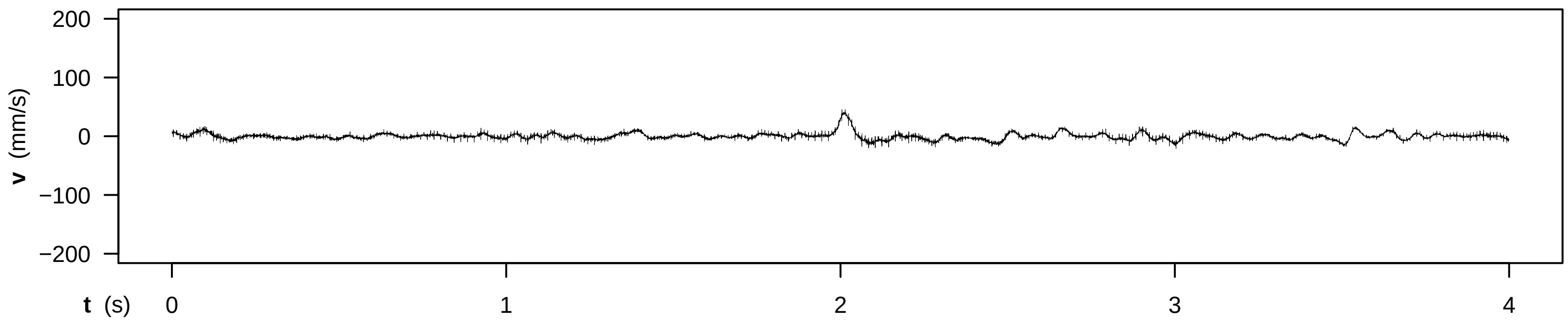

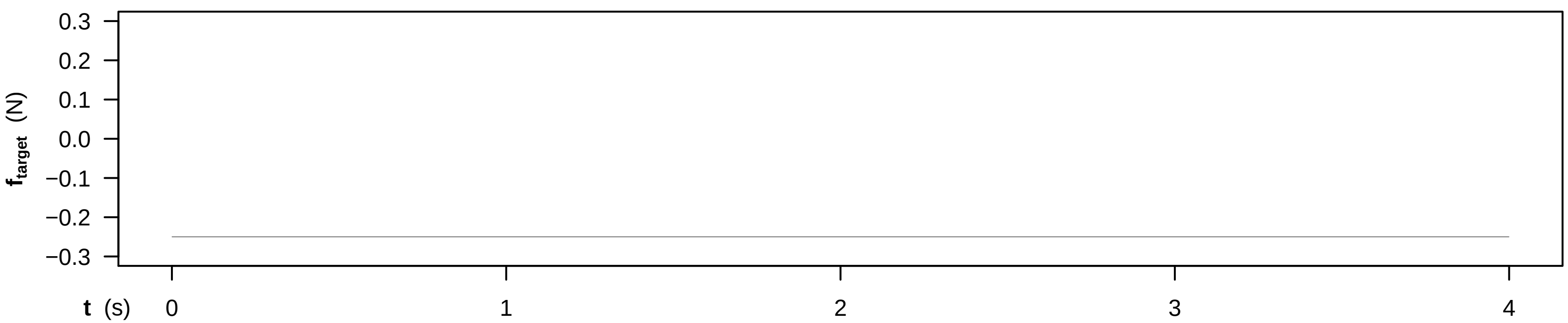

**SUBJECT 6 − RUN 02 − CONDITION 2,1**
**SC_180323_145147_0.AIFF**

```
z_min               : 4.40 mm
z_max               : 7.40 mm
z_travel_amplitude  : 3.00 mm

avg_abs_z_travel    : 5.47 mm/s

z_jarque-bera_jb    : 4741.59
z_jarque-bera_p     : 0.00e+00

z_lin_mod_est_slope: −0.20 mm/s
z_lin_mod_adj_R²    : 22 %

z_poly40_mod_adj_R²: 91 %

z_dft_ampl_thresh   : 0.010 mm
>=threshold_maxfreq: 19.50 Hz
```

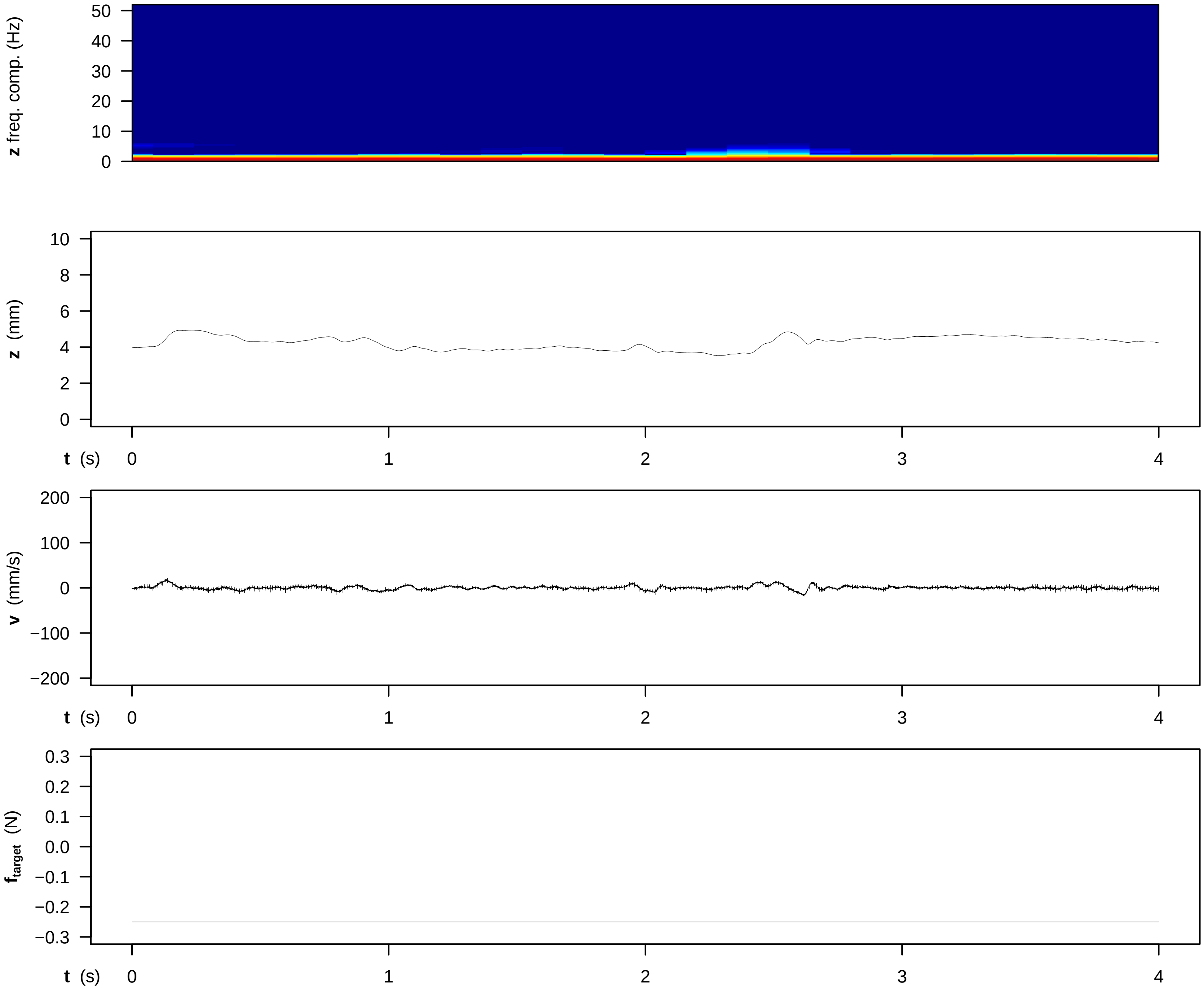

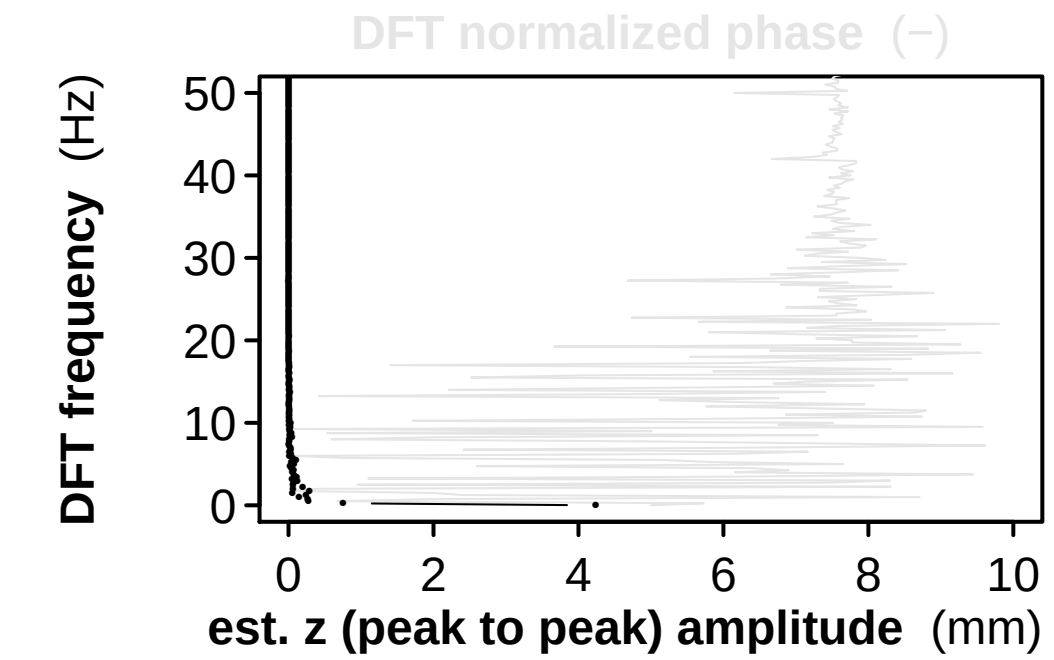

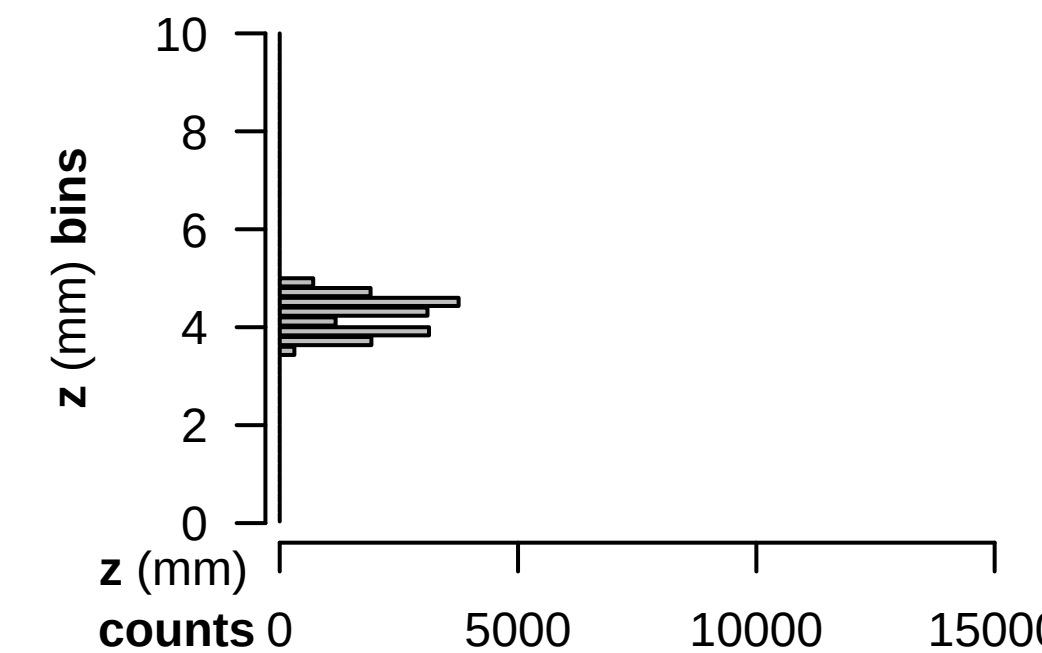

**SUBJECT 6 − RUN 27 − CONDITION 2,1**
**SC_180323_150748_0.AIFF**

z_min : 3.54 mm
z_max : 4.94 mm
z_travel_amplitude : 1.41 mm

avg_abs_z_travel : 4.73 mm/s

z_jarque-bera_jb : 838.48
z_jarque-bera_p : 0.00e+00

z_lin_mod_est_slope: 0.05 mm/s
z_lin_mod_adj_R² : 3 %

z_poly40_mod_adj_R²: 90 %

z_dft_ampl_thresh : 0.010 mm
>=threshold_maxfreq: 16.00 Hz

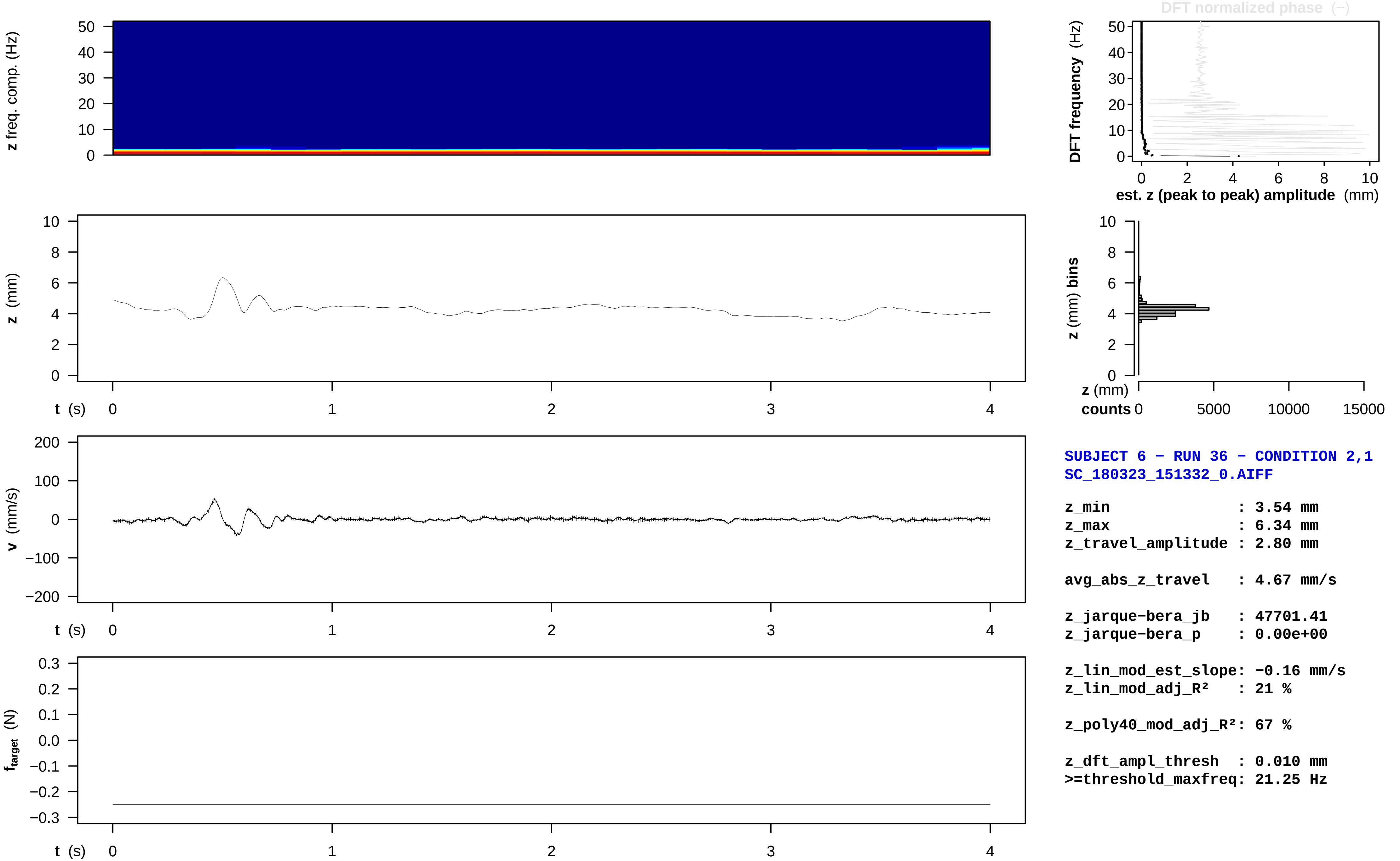

**SUBJECT 6 − RUN 36 − CONDITION 2,1**
**SC_180323_151332_0.AIFF**

```
z_min               : 3.54 mm
z_max               : 6.34 mm
z_travel_amplitude  : 2.80 mm

avg_abs_z_travel    : 4.67 mm/s

z_jarque-bera_jb    : 47701.41
z_jarque-bera_p     : 0.00e+00

z_lin_mod_est_slope: −0.16 mm/s
z_lin_mod_adj_R²    : 21 %

z_poly40_mod_adj_R²: 67 %

z_dft_ampl_thresh   : 0.010 mm
>=threshold_maxfreq: 21.25 Hz
```

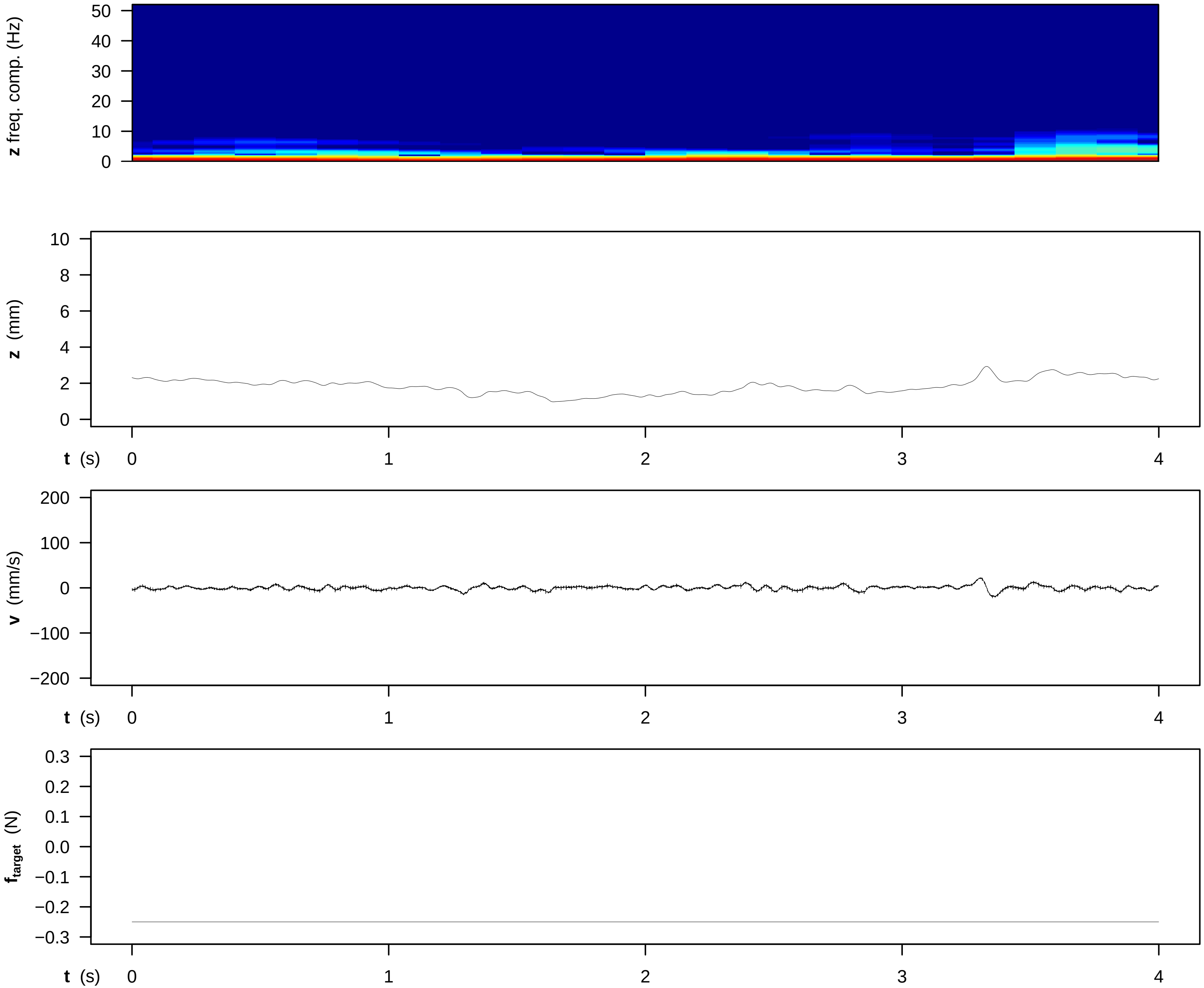

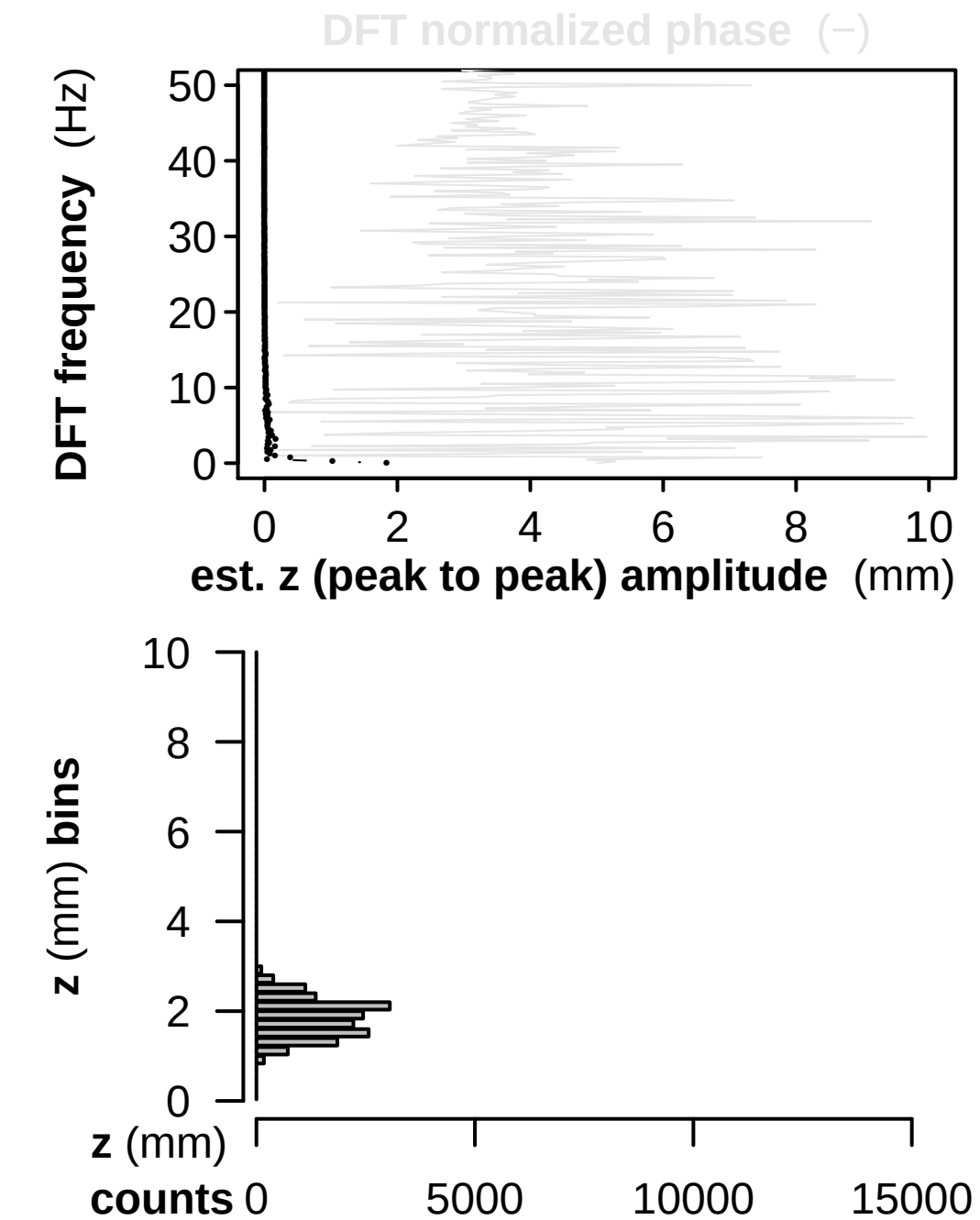

**SUBJECT 7 − RUN 01 − CONDITION 2,1**
**SC_180323_153423_0.AIFF**

```
z_min              : 0.97 mm
z_max              : 2.93 mm
z_travel_amplitude : 1.96 mm

avg_abs_z_travel   : 3.75 mm/s

z_jarque-bera_jb   : 287.01
z_jarque-bera_p    : 0.00e+00

z_lin_mod_est_slope: 0.07 mm/s
z_lin_mod_adj_R²   : 4 %

z_poly40_mod_adj_R²: 90 %

z_dft_ampl_thresh  : 0.010 mm
>=threshold_maxfreq: 16.50 Hz
```

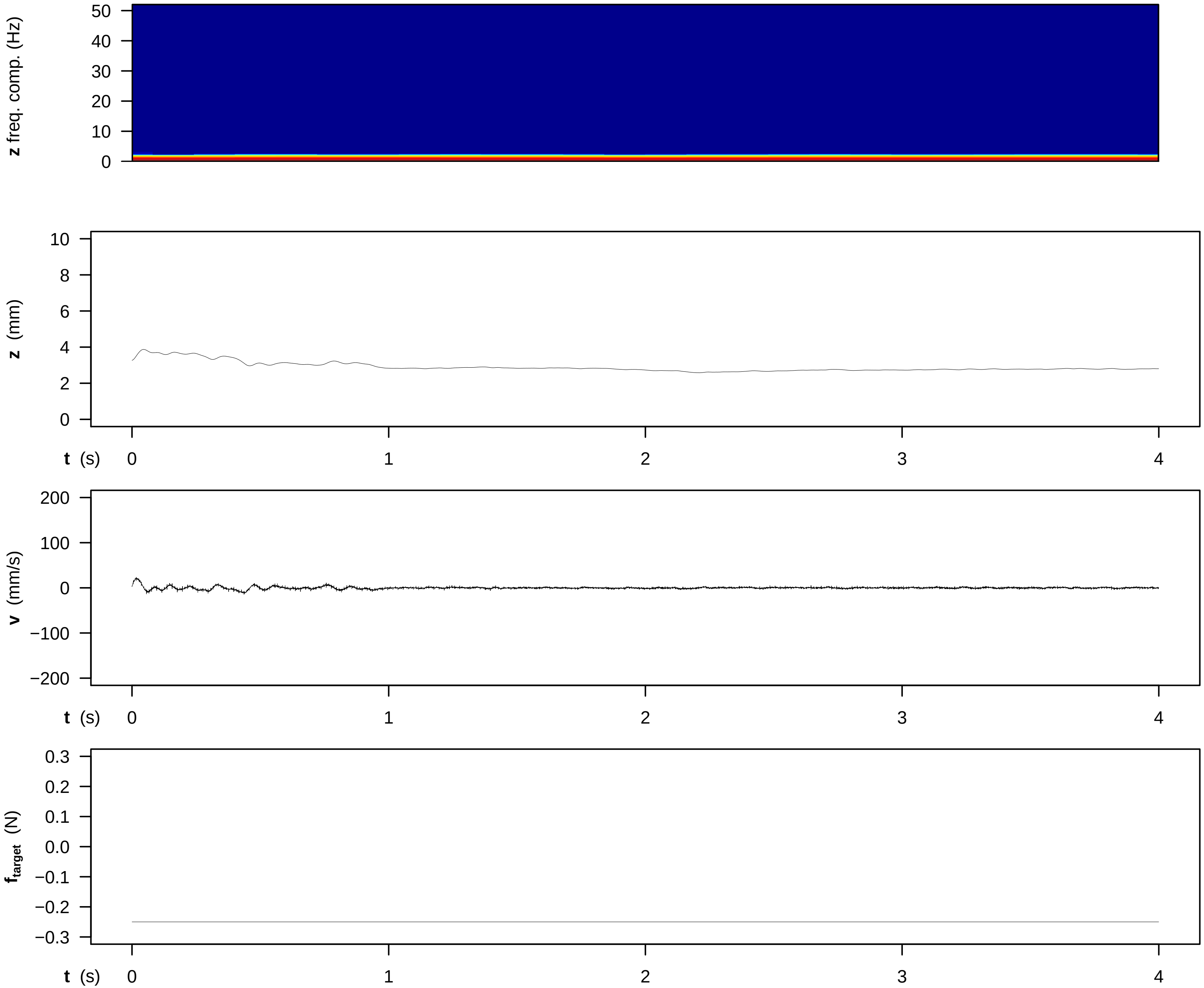

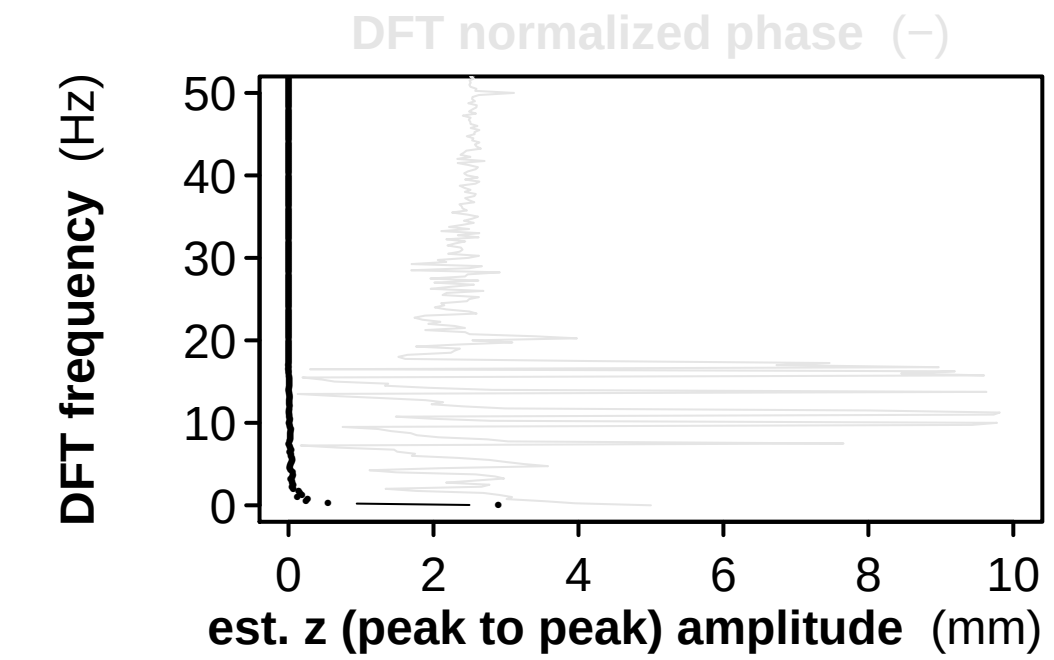

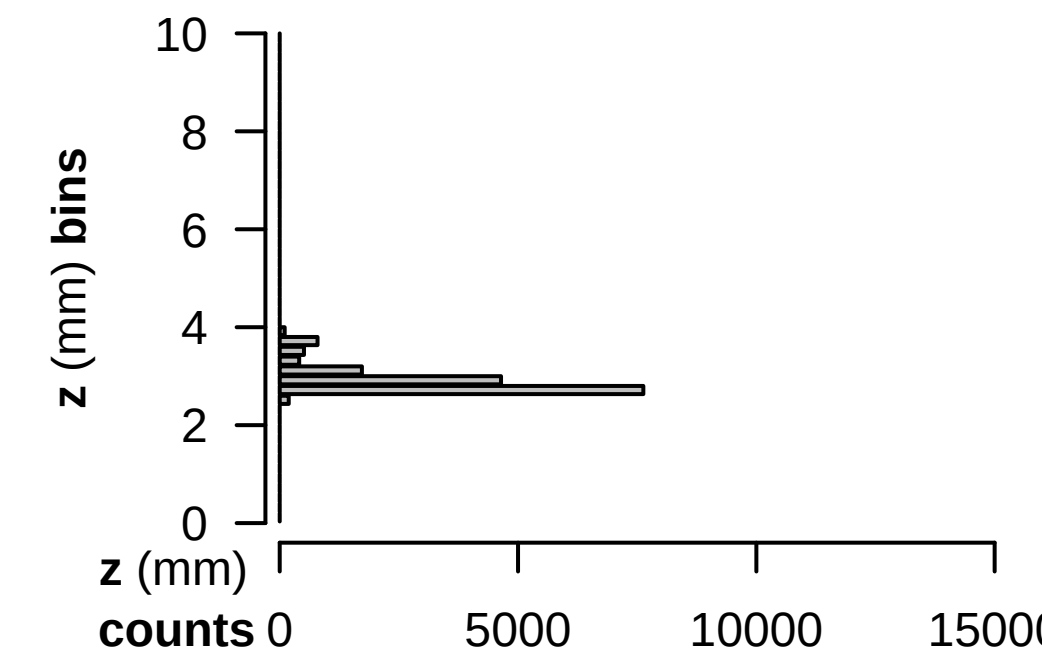

**SUBJECT 7 − RUN 20 − CONDITION 2,1**
**SC_180323_154722_0.AIFF**

```
z_min               : 2.58 mm
z_max               : 3.88 mm
z_travel_amplitude  : 1.30 mm

avg_abs_z_travel    : 2.18 mm/s

z_jarque-bera_jb    : 14297.21
z_jarque-bera_p     : 0.00e+00

z_lin_mod_est_slope: −0.16 mm/s
z_lin_mod_adj_R²    : 50 %

z_poly40_mod_adj_R²: 98 %

z_dft_ampl_thresh   : 0.010 mm
>=threshold_maxfreq: 15.00 Hz
```

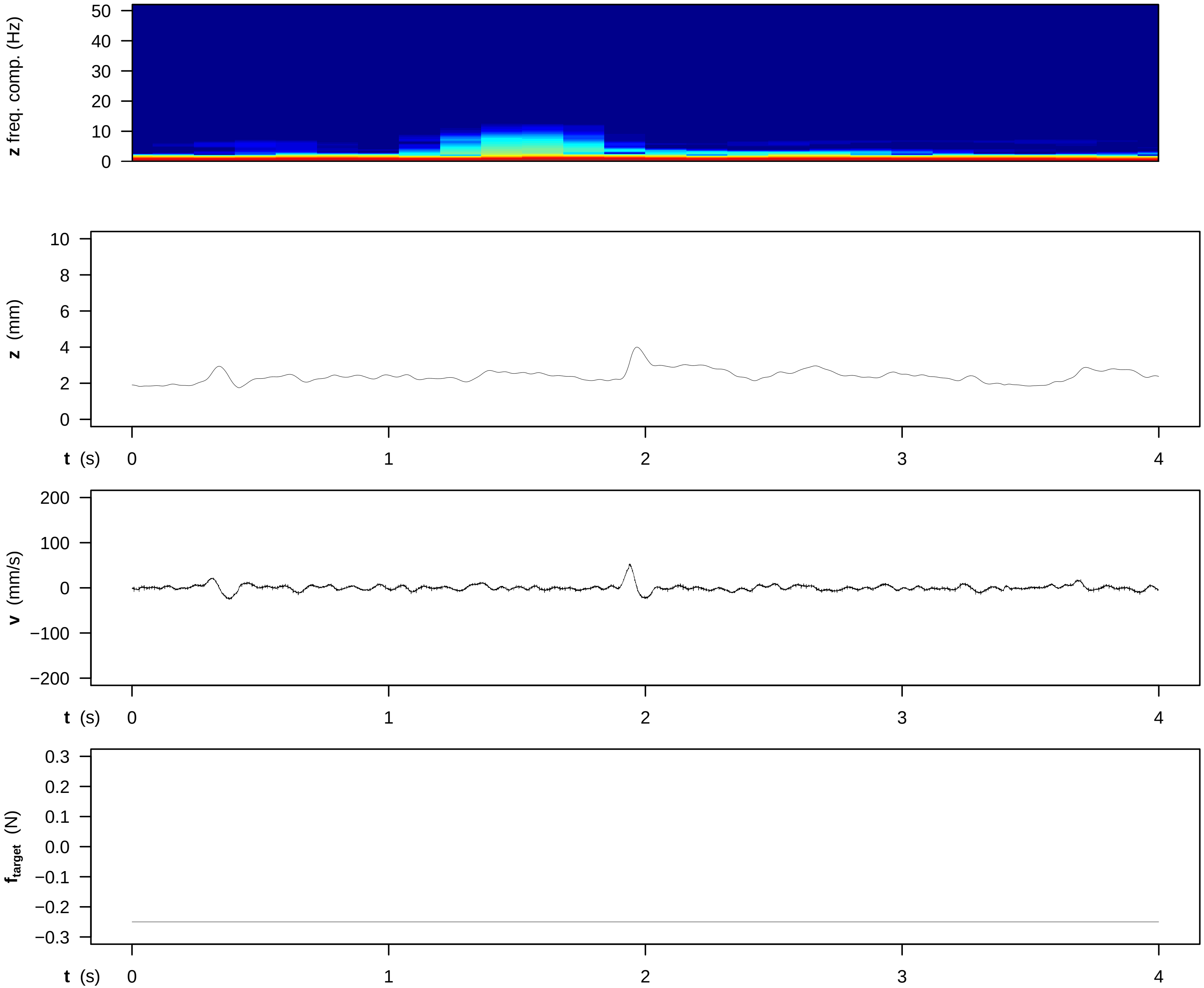

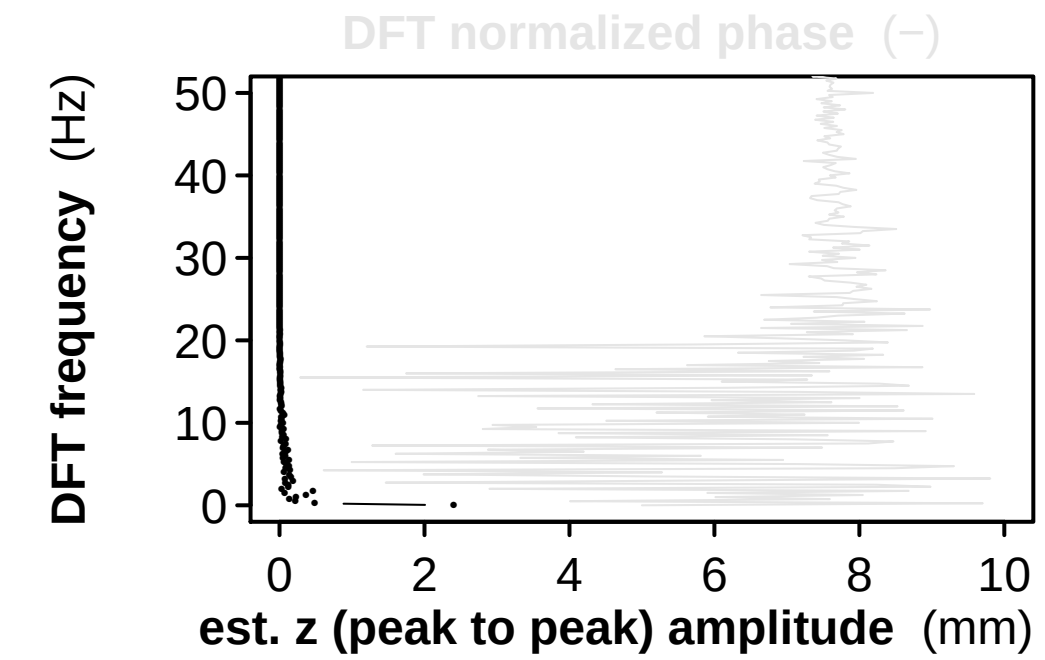

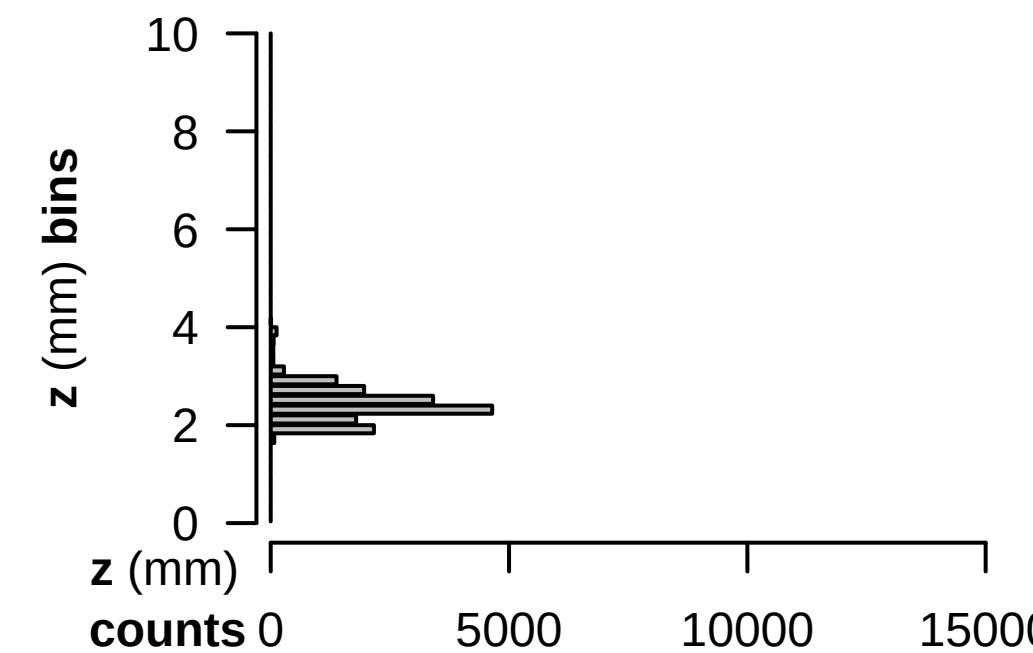

**SUBJECT 7 − RUN 22 − CONDITION 2,1**
**SC_180323_154956_0.AIFF**

```
z_min               : 1.75 mm
z_max               : 4.00 mm
z_travel_amplitude  : 2.25 mm

avg_abs_z_travel    : 4.79 mm/s

z_jarque-bera_jb    : 5951.06
z_jarque-bera_p     : 0.00e+00

z_lin_mod_est_slope: 0.06 mm/s
z_lin_mod_adj_R²    : 4 %

z_poly40_mod_adj_R²: 69 %

z_dft_ampl_thresh   : 0.010 mm
>=threshold_maxfreq: 17.75 Hz
```

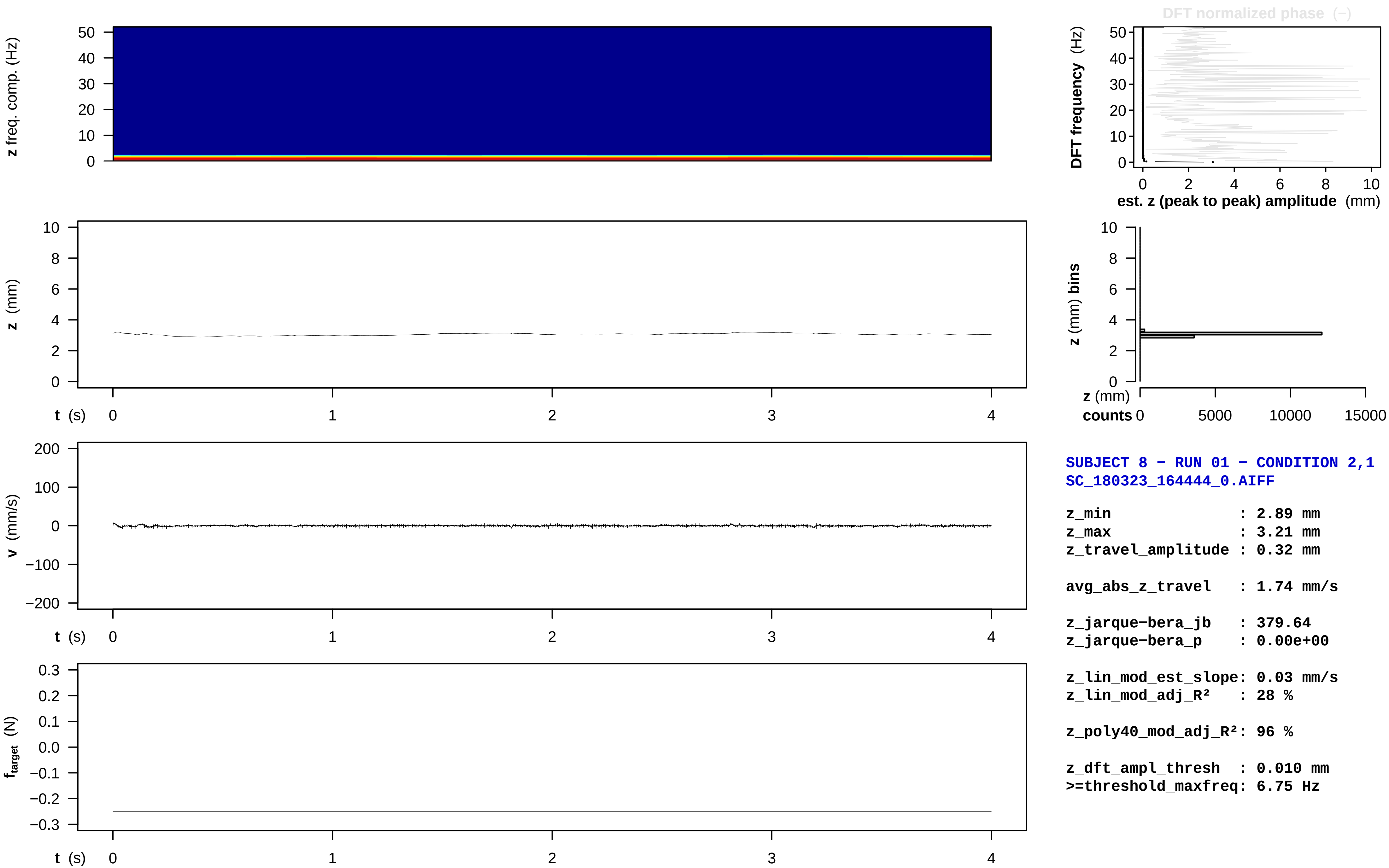

**SUBJECT 8 − RUN 01 − CONDITION 2,1**
**SC_180323_164444_0.AIFF**

```
z_min               : 2.89 mm
z_max               : 3.21 mm
z_travel_amplitude  : 0.32 mm

avg_abs_z_travel    : 1.74 mm/s

z_jarque-bera_jb    : 379.64
z_jarque-bera_p     : 0.00e+00

z_lin_mod_est_slope: 0.03 mm/s
z_lin_mod_adj_R²    : 28 %

z_poly40_mod_adj_R²: 96 %

z_dft_ampl_thresh   : 0.010 mm
>=threshold_maxfreq: 6.75 Hz
```

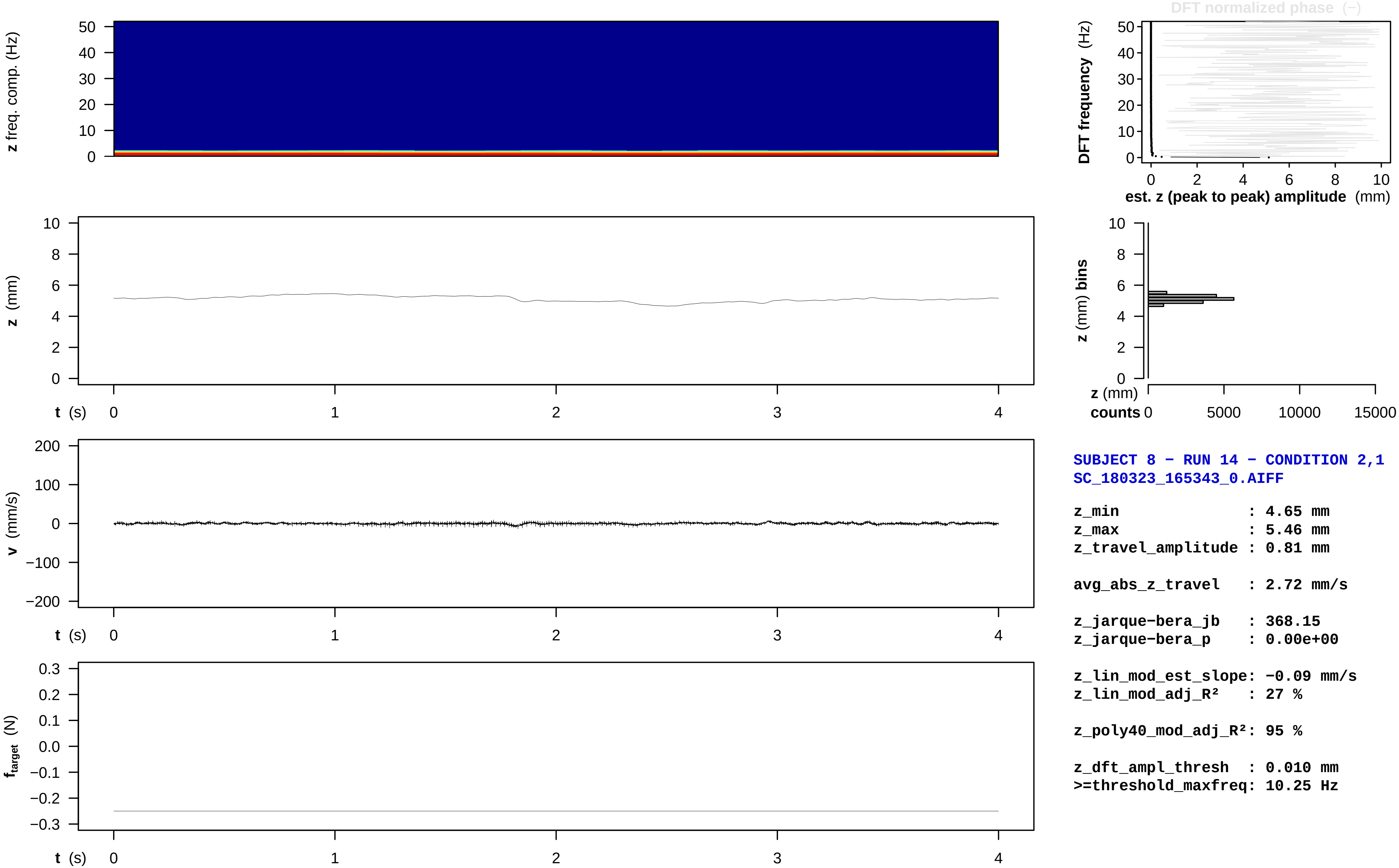

**SUBJECT 8 − RUN 14 − CONDITION 2,1**
**SC_180323_165343_0.AIFF**

z_min : 4.65 mm
z_max : 5.46 mm
z_travel_amplitude : 0.81 mm

avg_abs_z_travel : 2.72 mm/s

z_jarque-bera_jb : 368.15
z_jarque-bera_p : 0.00e+00

z_lin_mod_est_slope: −0.09 mm/s
z_lin_mod_adj_R² : 27 %

z_poly40_mod_adj_R²: 95 %

z_dft_ampl_thresh : 0.010 mm
>=threshold_maxfreq: 10.25 Hz

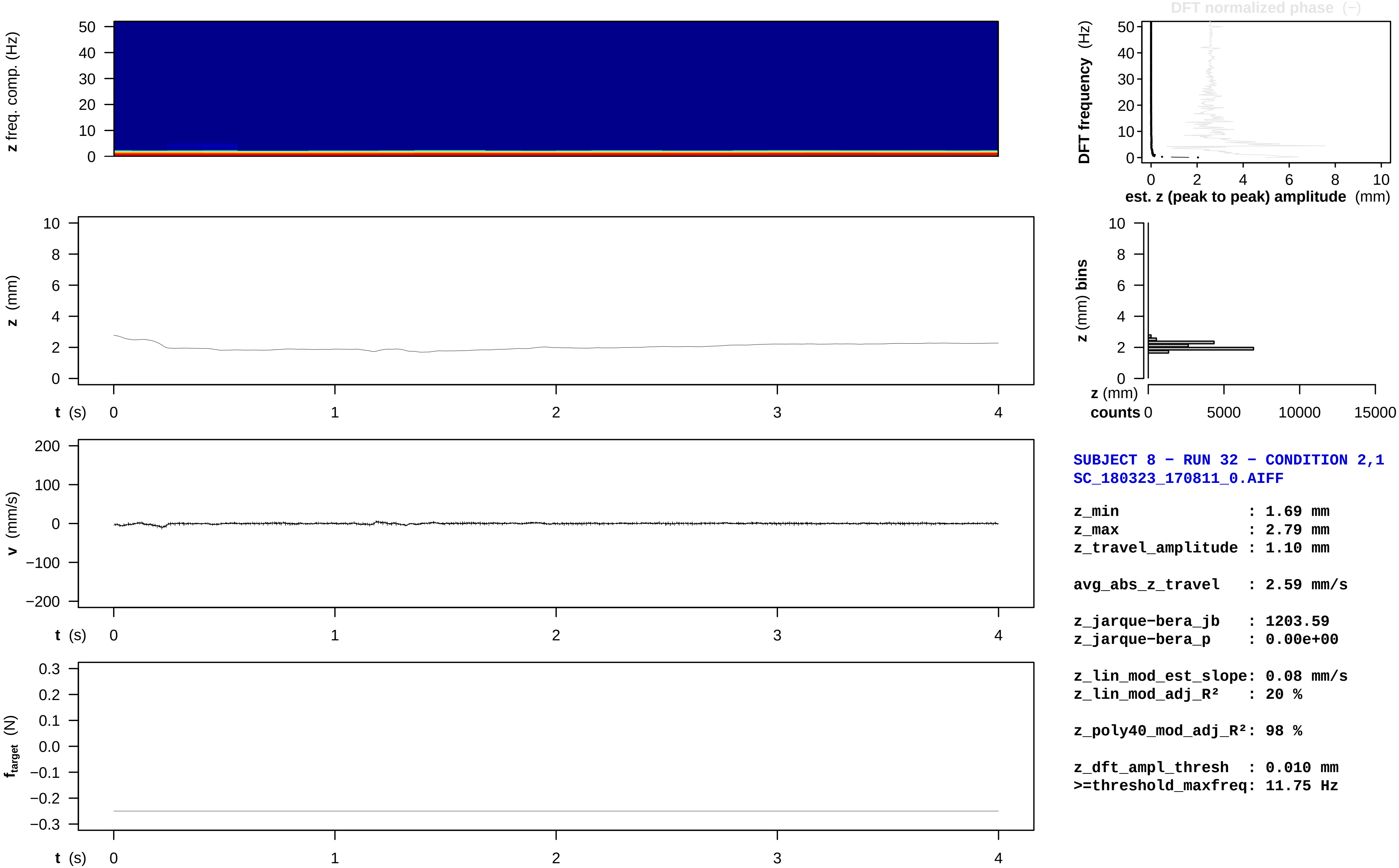

**SUBJECT 8 − RUN 32 − CONDITION 2,1**
**SC_180323_170811_0.AIFF**

```
z_min               : 1.69 mm
z_max               : 2.79 mm
z_travel_amplitude  : 1.10 mm

avg_abs_z_travel    : 2.59 mm/s

z_jarque-bera_jb    : 1203.59
z_jarque-bera_p     : 0.00e+00

z_lin_mod_est_slope: 0.08 mm/s
z_lin_mod_adj_R²    : 20 %

z_poly40_mod_adj_R²: 98 %

z_dft_ampl_thresh   : 0.010 mm
>=threshold_maxfreq: 11.75 Hz
```

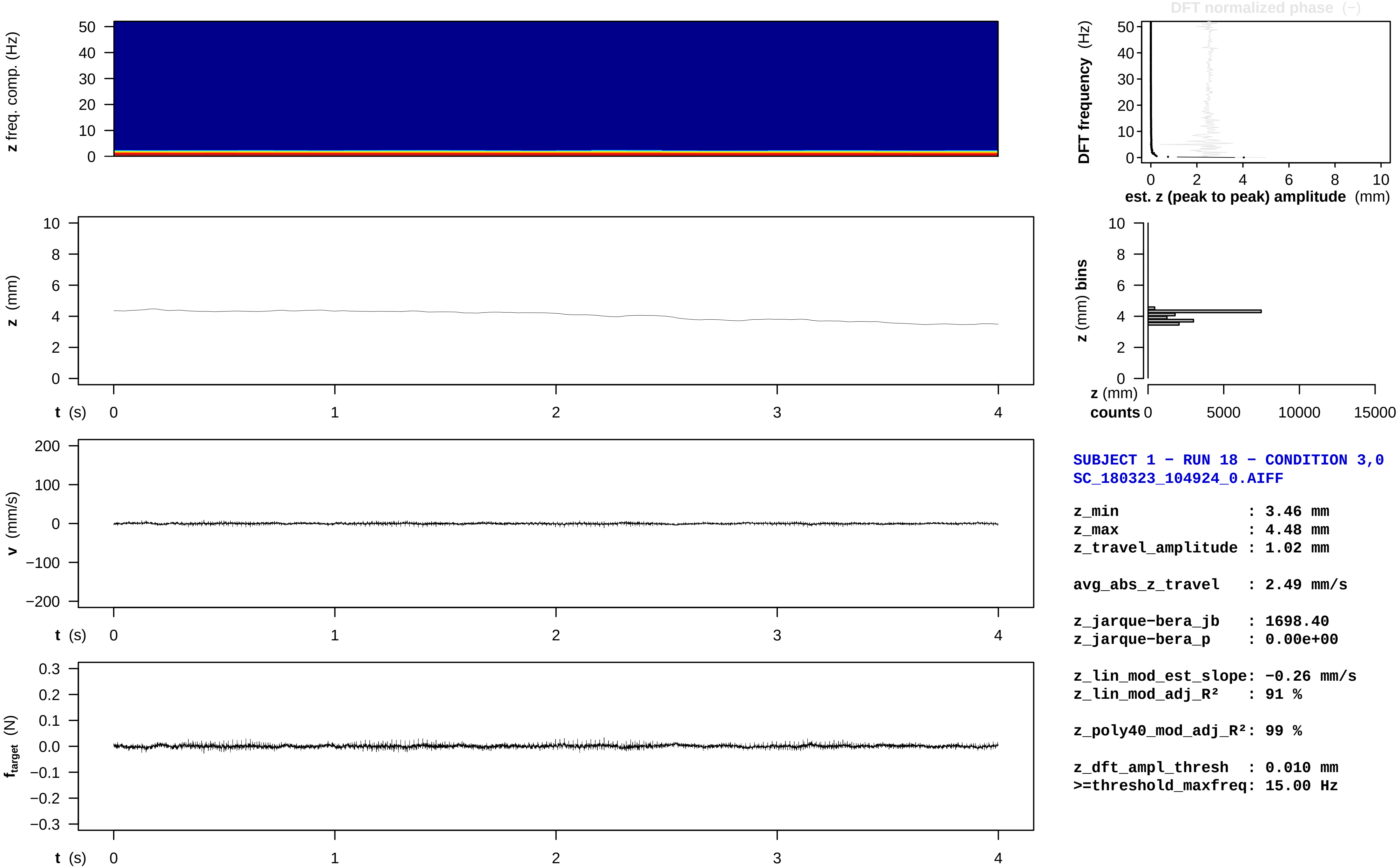

**SUBJECT 1 − RUN 18 − CONDITION 3,0**
**SC_180323_104924_0.AIFF**

```
z_min              : 3.46 mm
z_max              : 4.48 mm
z_travel_amplitude : 1.02 mm

avg_abs_z_travel   : 2.49 mm/s

z_jarque-bera_jb   : 1698.40
z_jarque-bera_p    : 0.00e+00

z_lin_mod_est_slope: -0.26 mm/s
z_lin_mod_adj_R²   : 91 %

z_poly40_mod_adj_R²: 99 %

z_dft_ampl_thresh  : 0.010 mm
>=threshold_maxfreq: 15.00 Hz
```

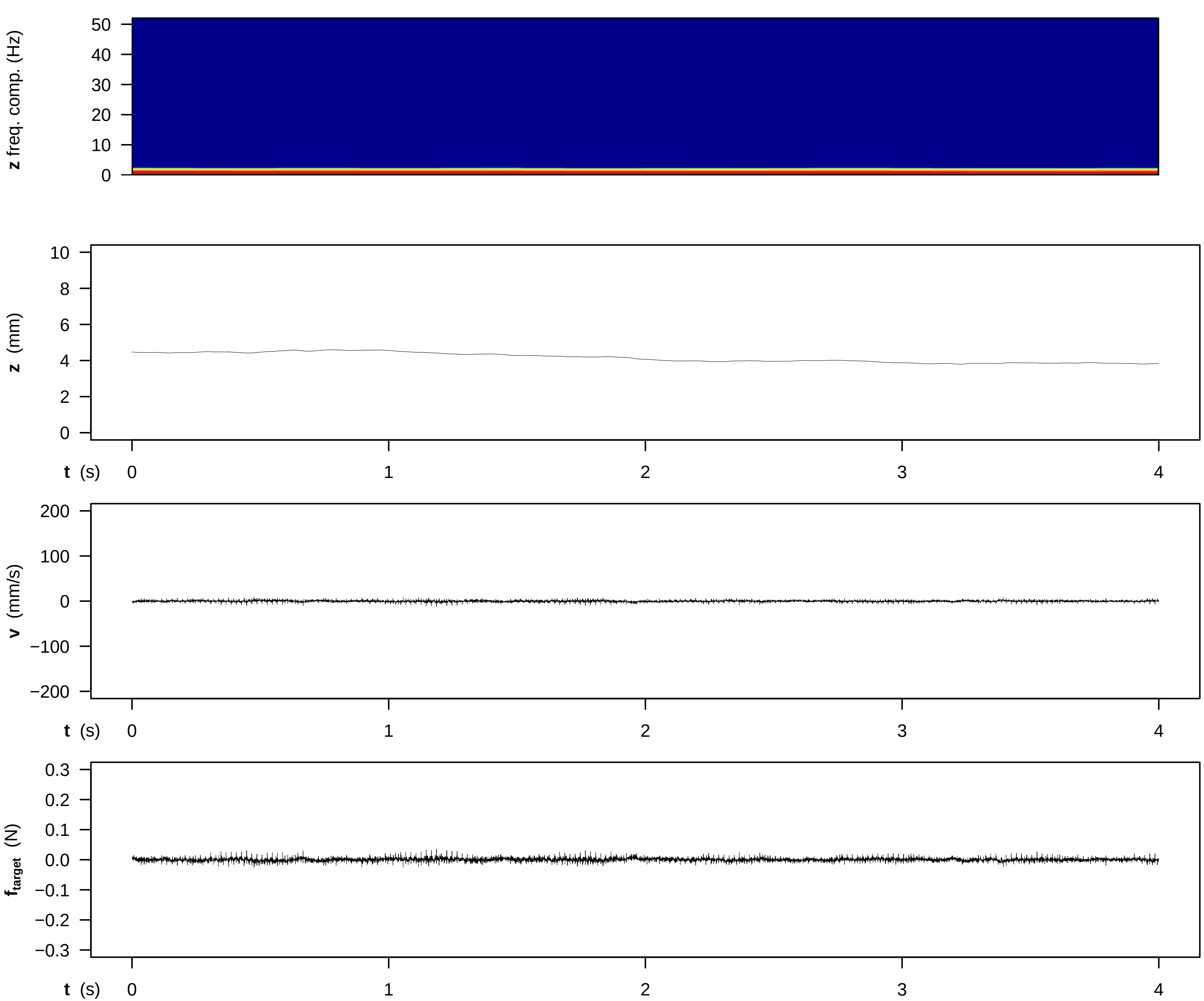

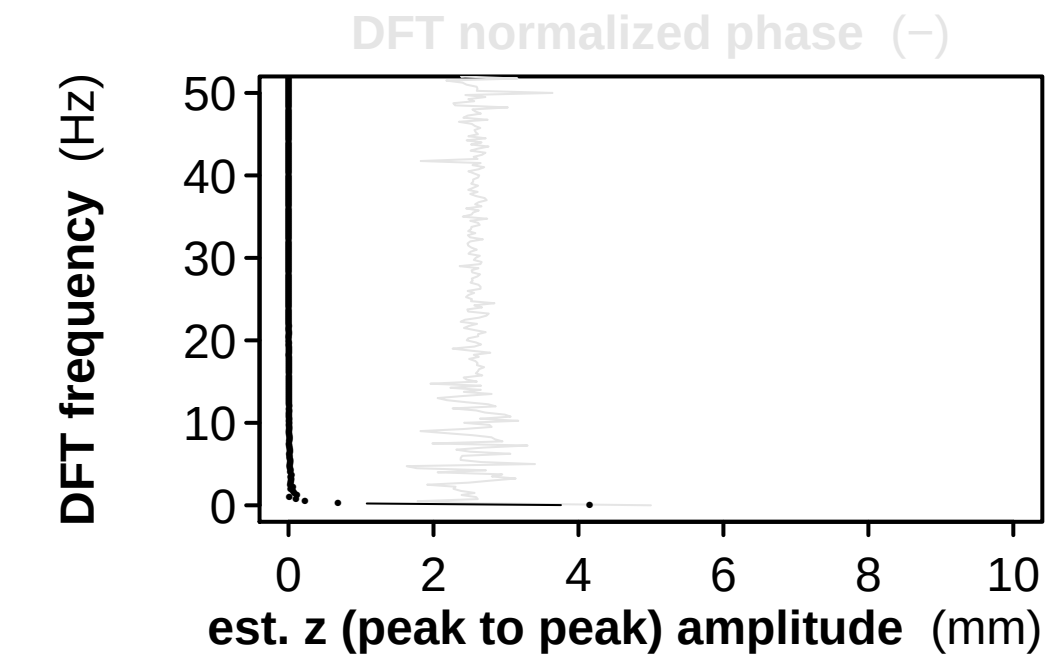

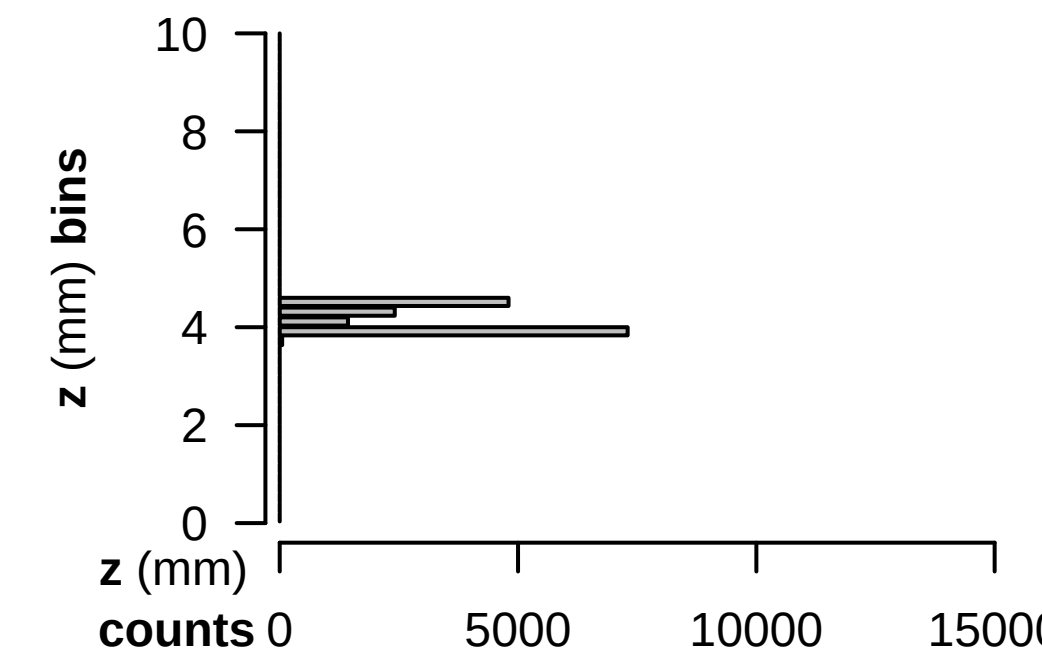

**SUBJECT 1 − RUN 27 − CONDITION 3,0**
**SC_180323_105436_0.AIFF**

```
z_min               : 3.80 mm
z_max               : 4.60 mm
z_travel_amplitude  : 0.80 mm

avg_abs_z_travel    : 3.41 mm/s

z_jarque-bera_jb    : 1711.92
z_jarque-bera_p     : 0.00e+00

z_lin_mod_est_slope: −0.22 mm/s
z_lin_mod_adj_R²    : 89 %

z_poly40_mod_adj_R²: 100 %

z_dft_ampl_thresh   : 0.010 mm
>=threshold_maxfreq: 11.50 Hz
```

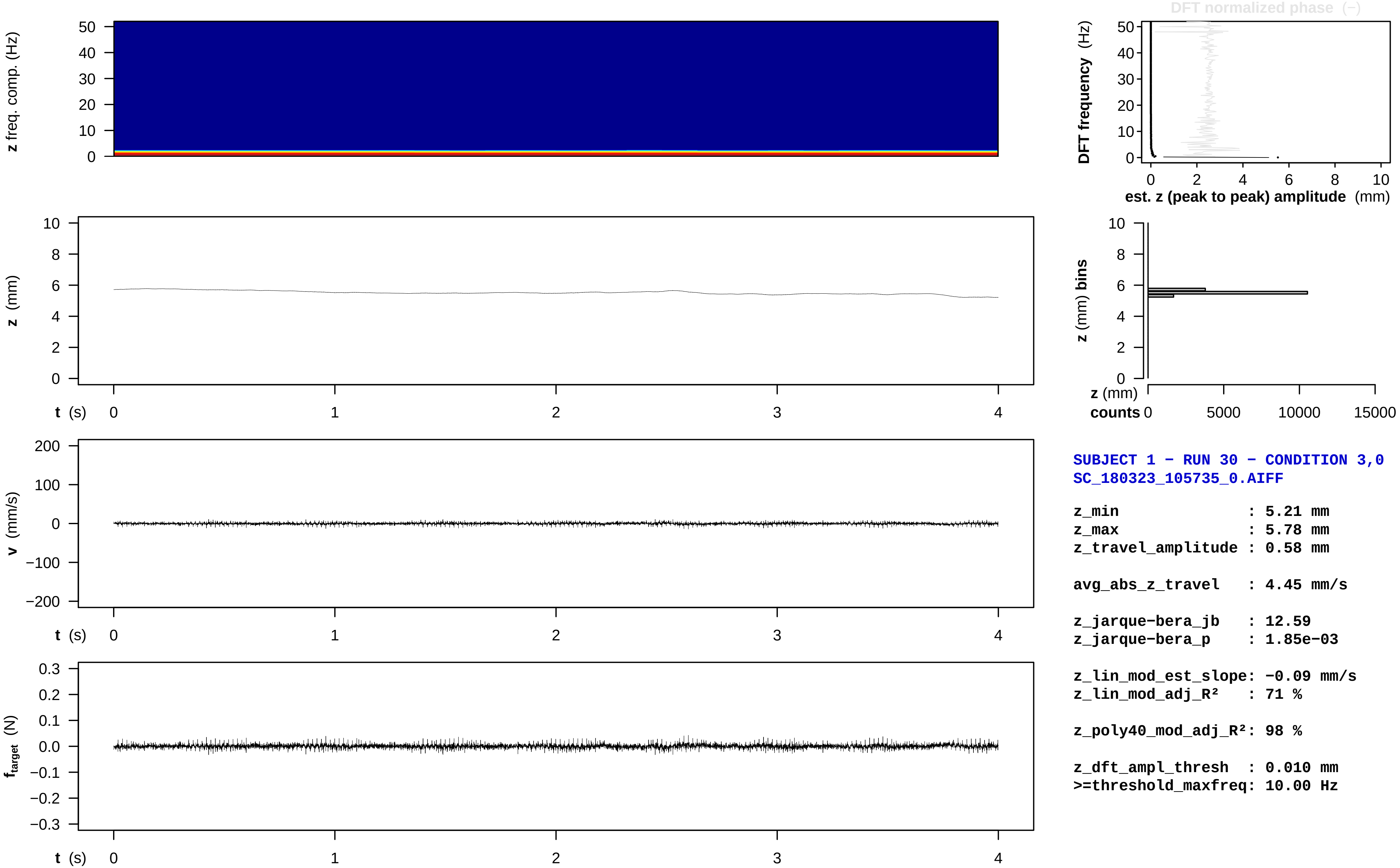

SUBJECT 1 − RUN 30 − CONDITION 3,0
SC_180323_105735_0.AIFF

```
z_min              : 5.21 mm
z_max              : 5.78 mm
z_travel_amplitude : 0.58 mm

avg_abs_z_travel   : 4.45 mm/s

z_jarque-bera_jb   : 12.59
z_jarque-bera_p    : 1.85e-03

z_lin_mod_est_slope: -0.09 mm/s
z_lin_mod_adj_R²   : 71 %

z_poly40_mod_adj_R²: 98 %

z_dft_ampl_thresh  : 0.010 mm
>=threshold_maxfreq: 10.00 Hz
```

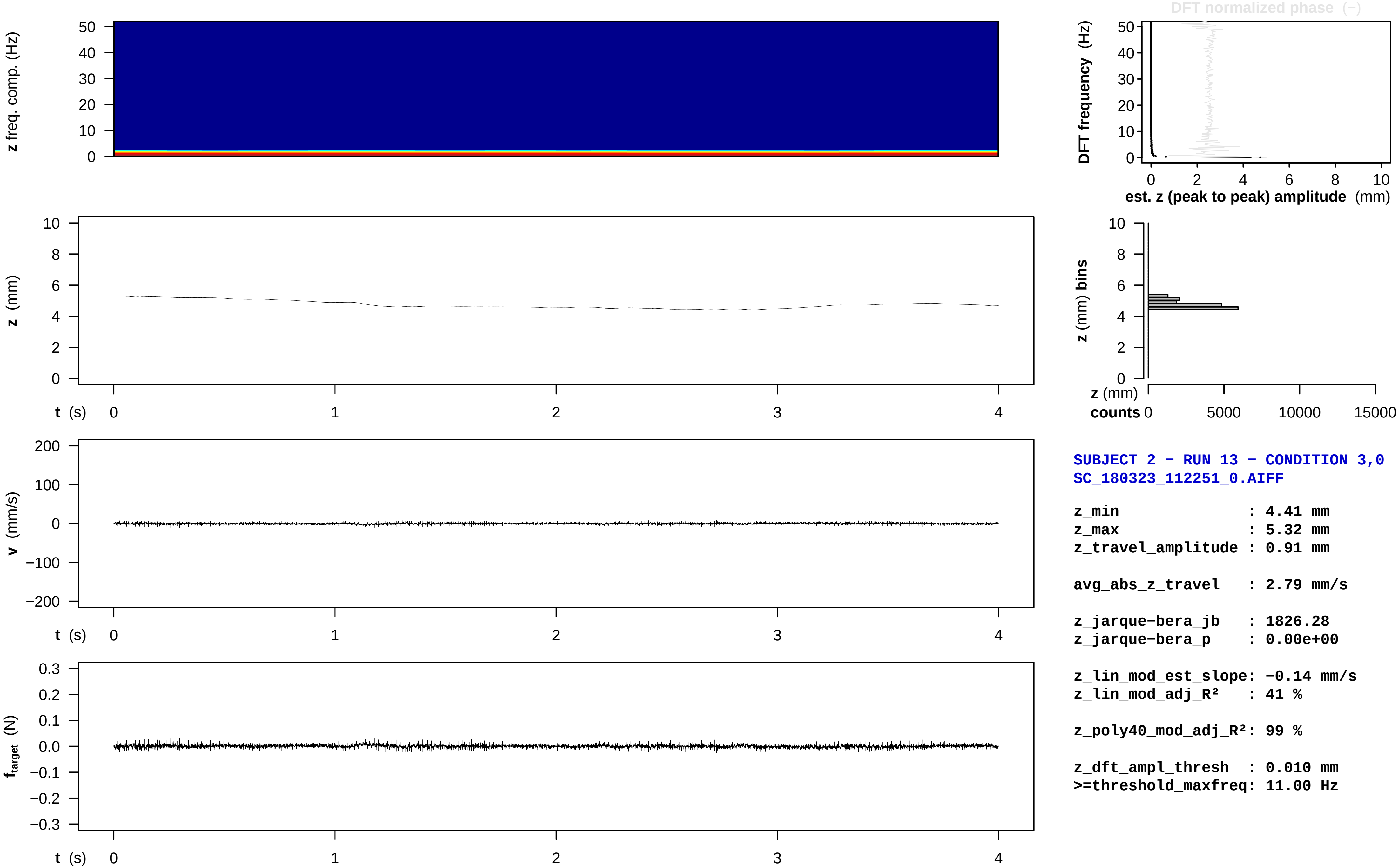

**SUBJECT 2 − RUN 13 − CONDITION 3,0**
**SC_180323_112251_0.AIFF**

```
z_min              : 4.41 mm
z_max              : 5.32 mm
z_travel_amplitude : 0.91 mm

avg_abs_z_travel   : 2.79 mm/s

z_jarque-bera_jb   : 1826.28
z_jarque-bera_p    : 0.00e+00

z_lin_mod_est_slope: −0.14 mm/s
z_lin_mod_adj_R²   : 41 %

z_poly40_mod_adj_R²: 99 %

z_dft_ampl_thresh  : 0.010 mm
>=threshold_maxfreq: 11.00 Hz
```

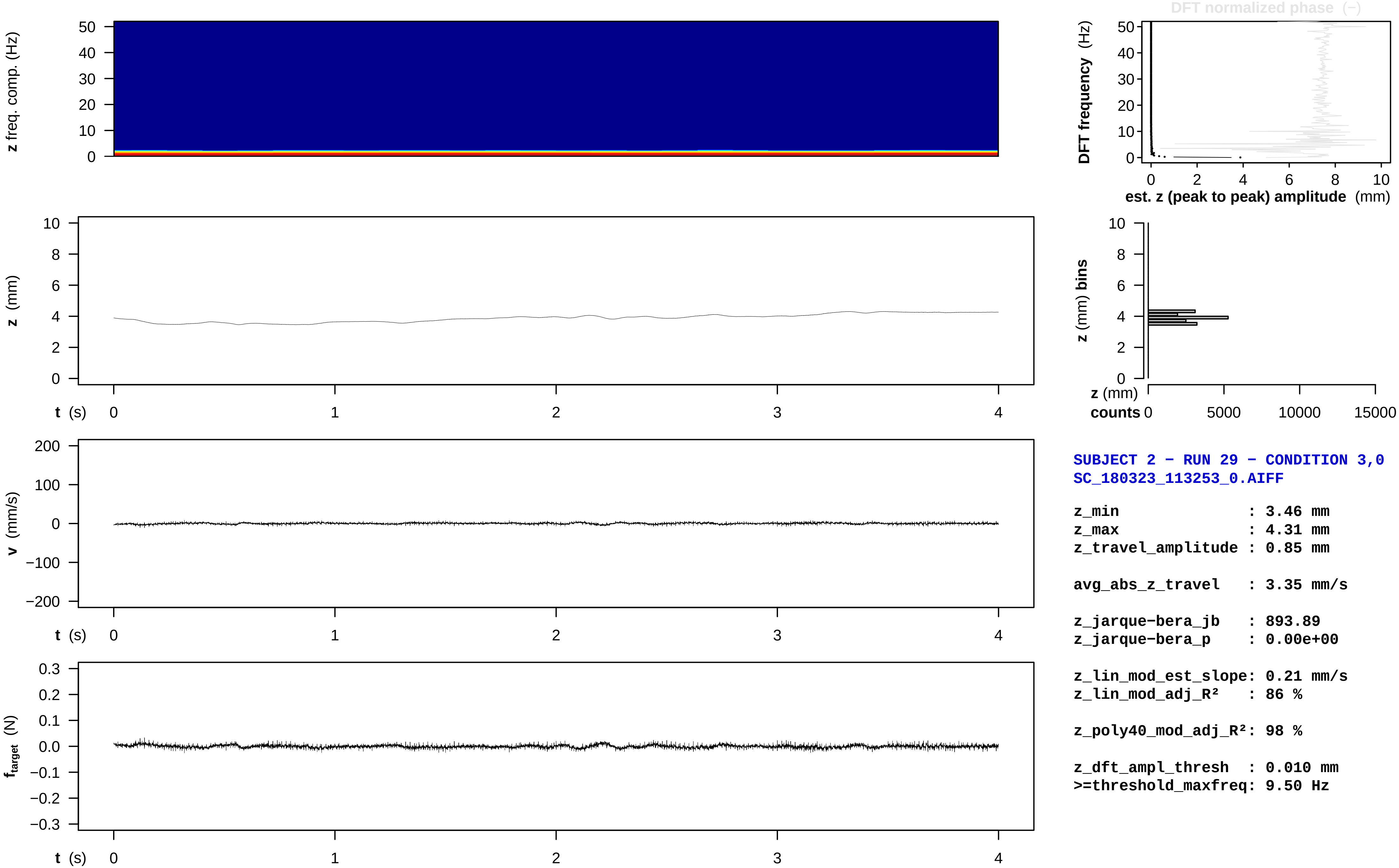

**SUBJECT 2 − RUN 29 − CONDITION 3,0**
**SC_180323_113253_0.AIFF**

```
z_min               : 3.46 mm
z_max               : 4.31 mm
z_travel_amplitude : 0.85 mm

avg_abs_z_travel   : 3.35 mm/s

z_jarque-bera_jb   : 893.89
z_jarque-bera_p    : 0.00e+00

z_lin_mod_est_slope: 0.21 mm/s
z_lin_mod_adj_R²   : 86 %

z_poly40_mod_adj_R²: 98 %

z_dft_ampl_thresh  : 0.010 mm
>=threshold_maxfreq: 9.50 Hz
```

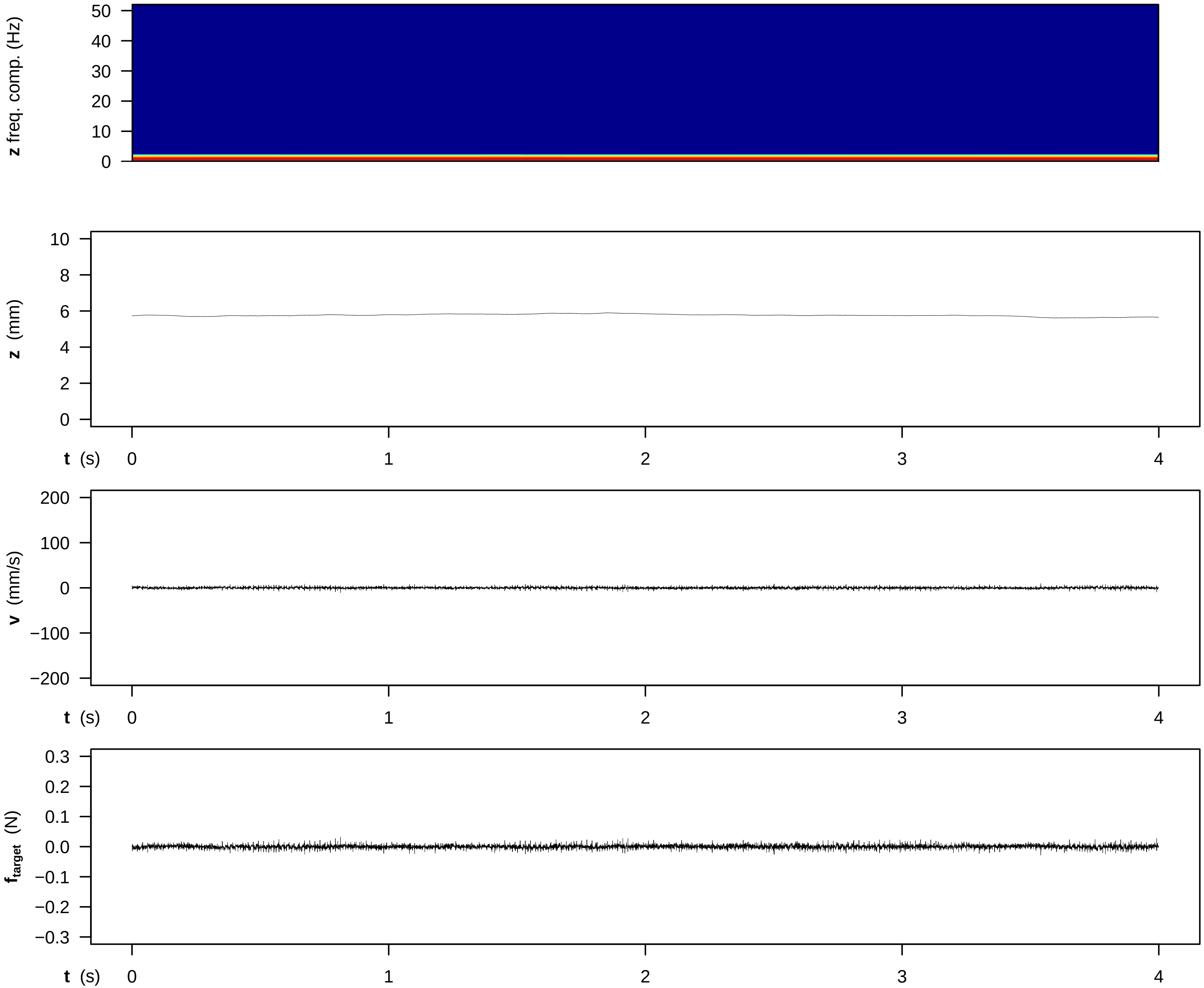

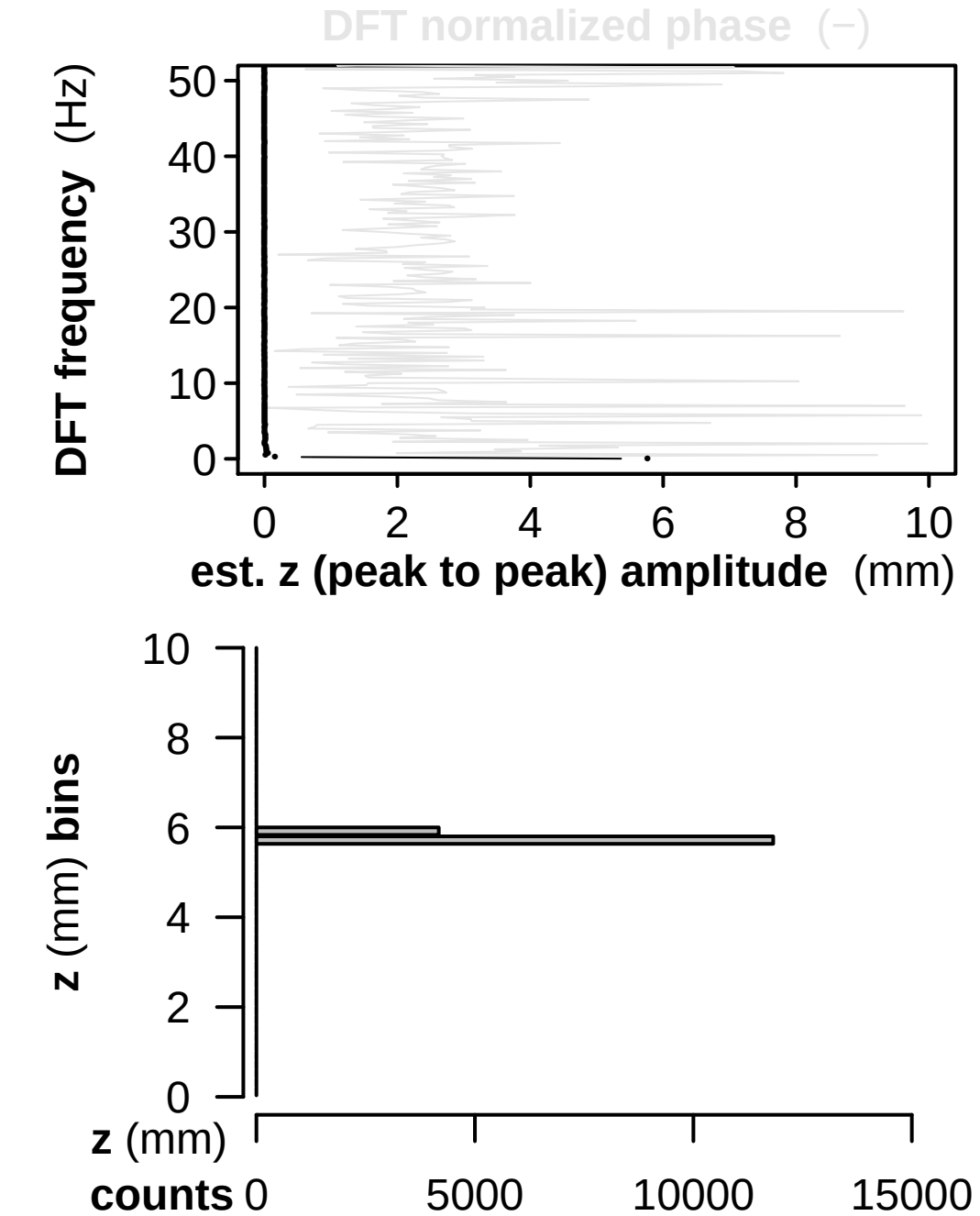

**SUBJECT 2 − RUN 32 − CONDITION 3,0**
**SC_180323_113425_0.AIFF**

```
z_min               : 5.61 mm
z_max               : 5.90 mm
z_travel_amplitude  : 0.29 mm

avg_abs_z_travel    : 3.90 mm/s

z_jarque-bera_jb    : 393.49
z_jarque-bera_p     : 0.00e+00

z_lin_mod_est_slope: −0.02 mm/s
z_lin_mod_adj_R²    : 19 %

z_poly40_mod_adj_R²: 97 %

z_dft_ampl_thresh   : 0.010 mm
>=threshold_maxfreq: 4.50 Hz
```

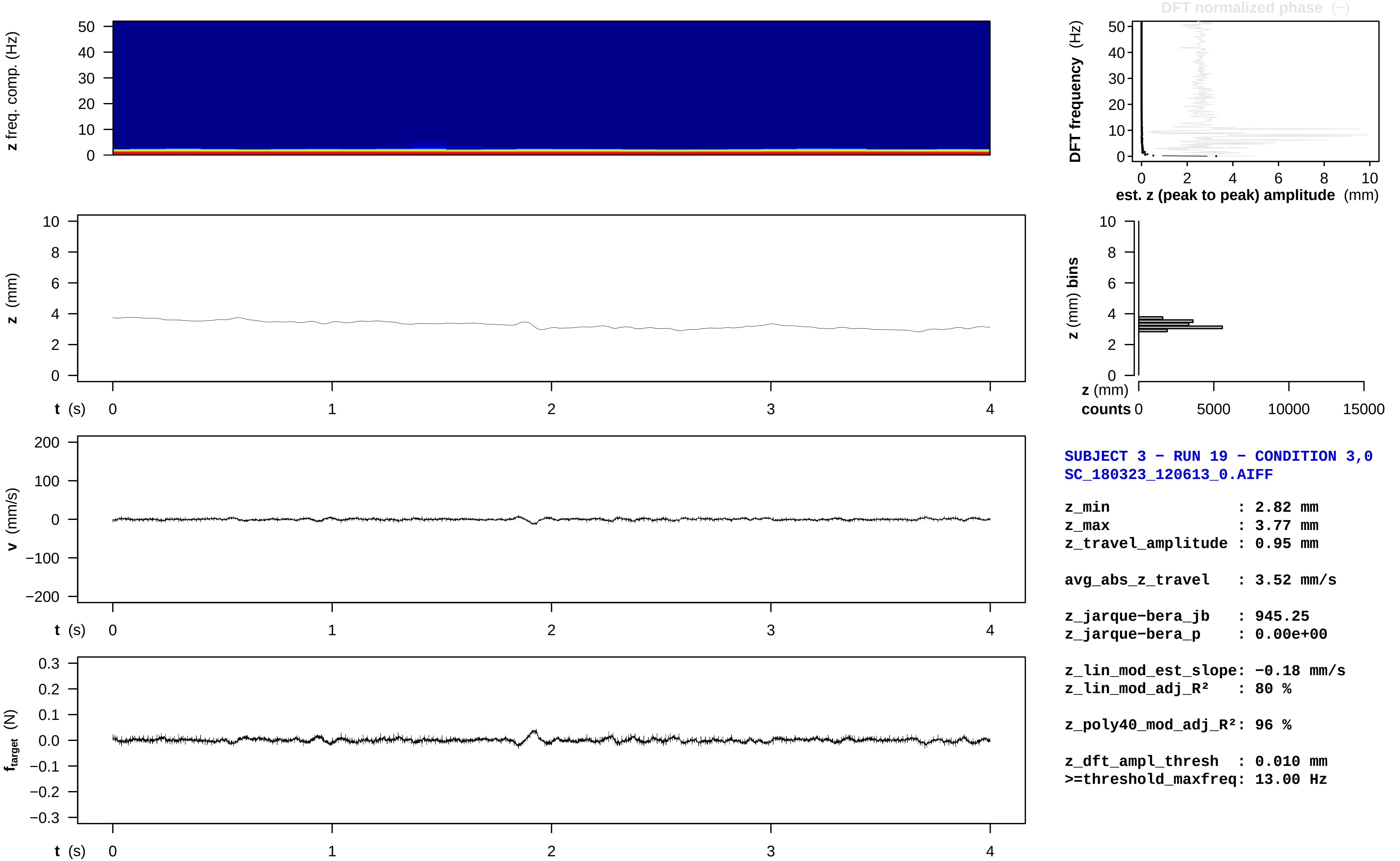

**SUBJECT 3 − RUN 19 − CONDITION 3,0**
**SC_180323_120613_0.AIFF**

```
z_min               : 2.82 mm
z_max               : 3.77 mm
z_travel_amplitude  : 0.95 mm

avg_abs_z_travel    : 3.52 mm/s

z_jarque-bera_jb    : 945.25
z_jarque-bera_p     : 0.00e+00

z_lin_mod_est_slope: -0.18 mm/s
z_lin_mod_adj_R²    : 80 %

z_poly40_mod_adj_R²: 96 %

z_dft_ampl_thresh   : 0.010 mm
>=threshold_maxfreq: 13.00 Hz
```

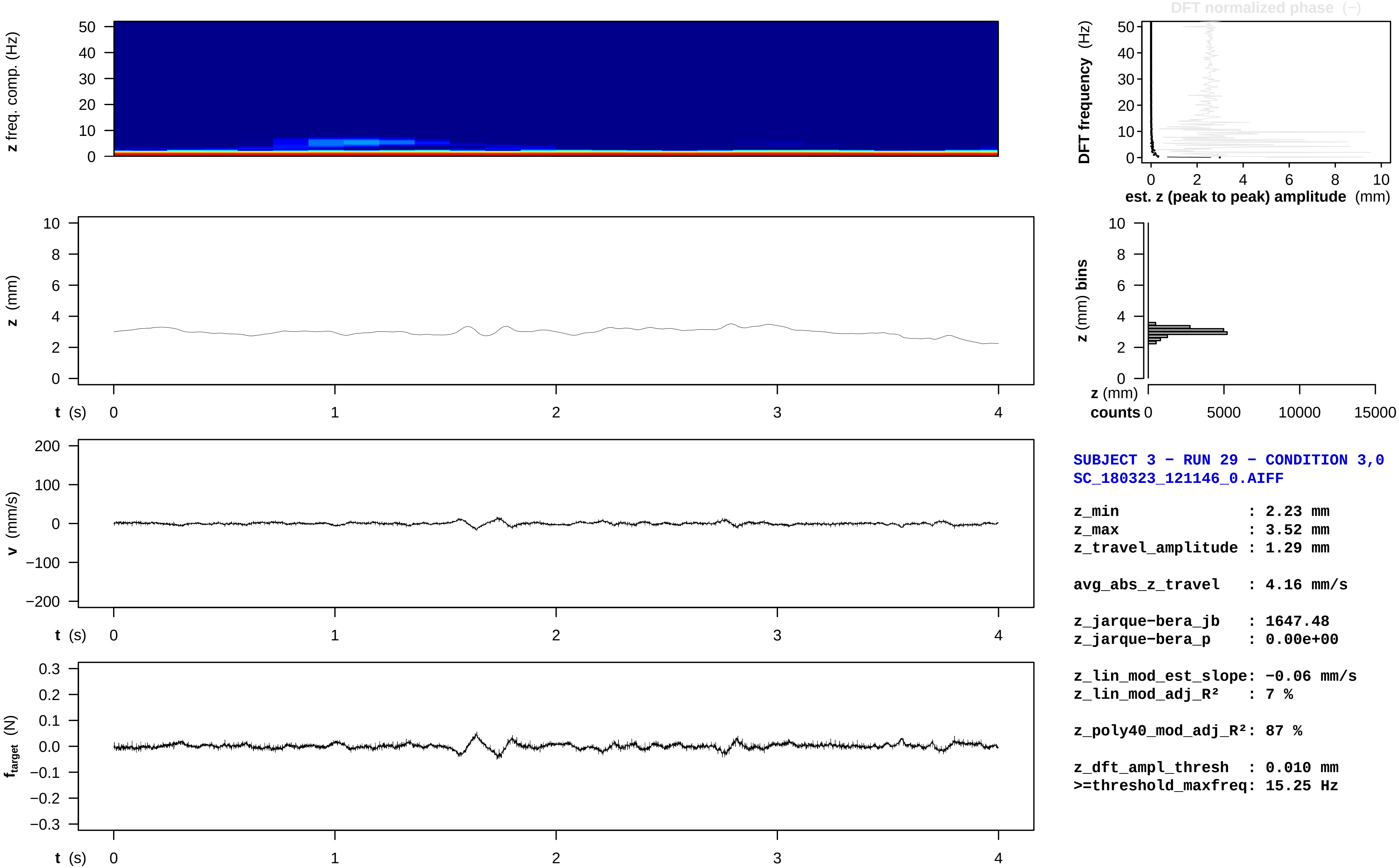

**SUBJECT 3 − RUN 29 − CONDITION 3,0**
**SC_180323_121146_0.AIFF**

```
z_min               : 2.23 mm
z_max               : 3.52 mm
z_travel_amplitude  : 1.29 mm

avg_abs_z_travel    : 4.16 mm/s

z_jarque-bera_jb    : 1647.48
z_jarque-bera_p     : 0.00e+00

z_lin_mod_est_slope: -0.06 mm/s
z_lin_mod_adj_R²    : 7 %

z_poly40_mod_adj_R²: 87 %

z_dft_ampl_thresh   : 0.010 mm
>=threshold_maxfreq: 15.25 Hz
```

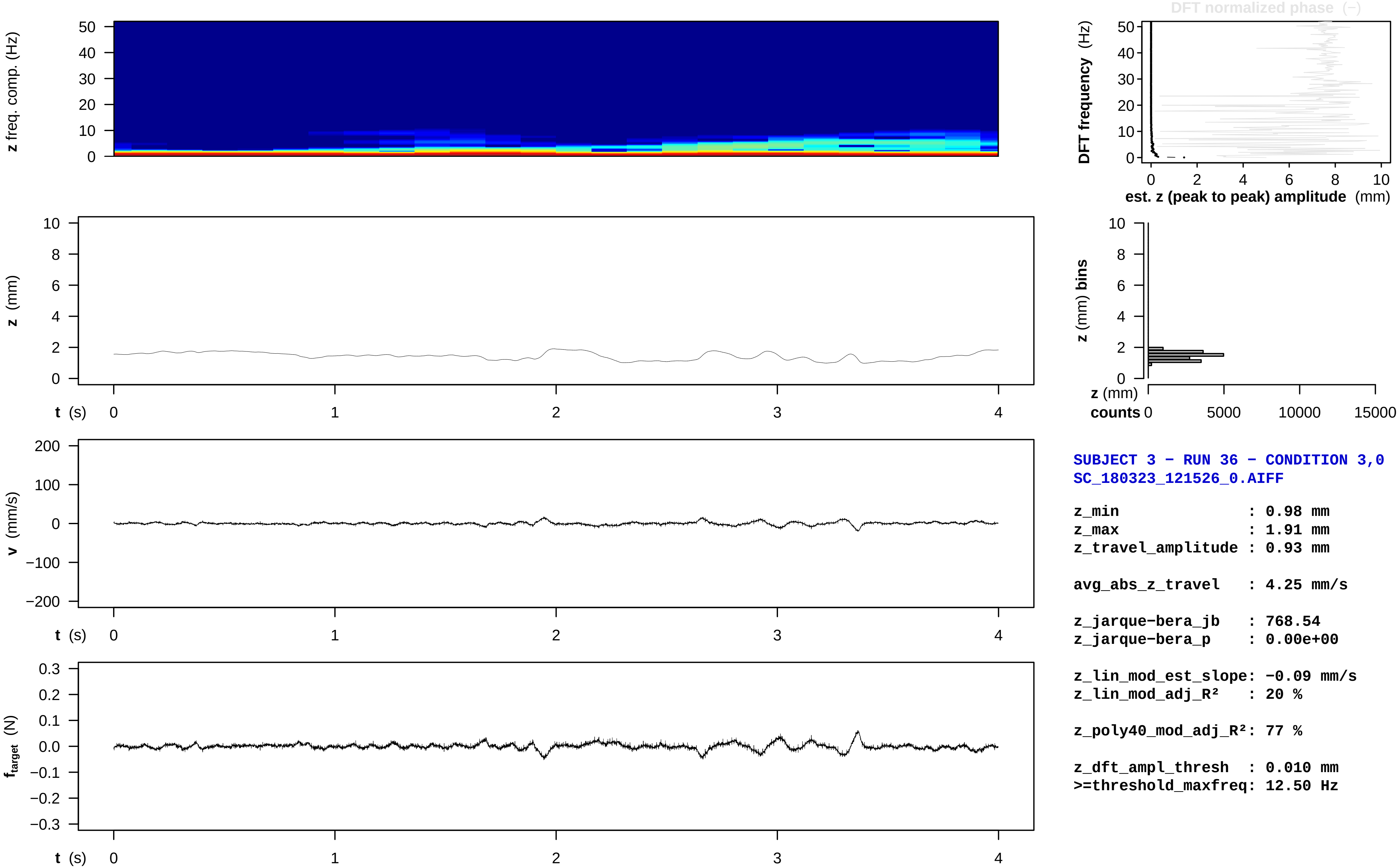

```
SUBJECT 3 − RUN 36 − CONDITION 3,0
SC_180323_121526_0.AIFF

z_min               : 0.98 mm
z_max               : 1.91 mm
z_travel_amplitude  : 0.93 mm

avg_abs_z_travel    : 4.25 mm/s

z_jarque-bera_jb    : 768.54
z_jarque-bera_p     : 0.00e+00

z_lin_mod_est_slope: −0.09 mm/s
z_lin_mod_adj_R²    : 20 %

z_poly40_mod_adj_R²: 77 %

z_dft_ampl_thresh   : 0.010 mm
>=threshold_maxfreq: 12.50 Hz
```

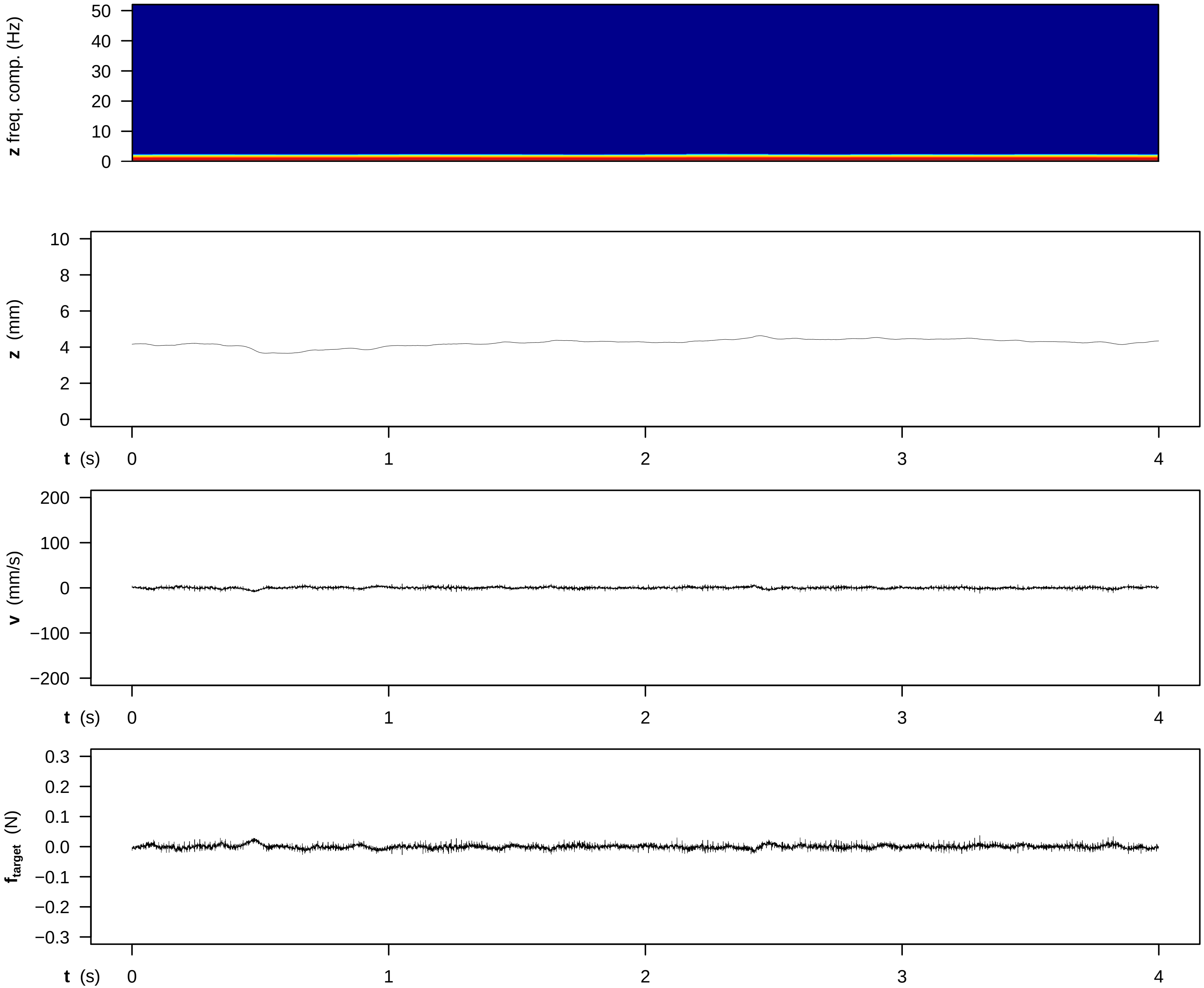

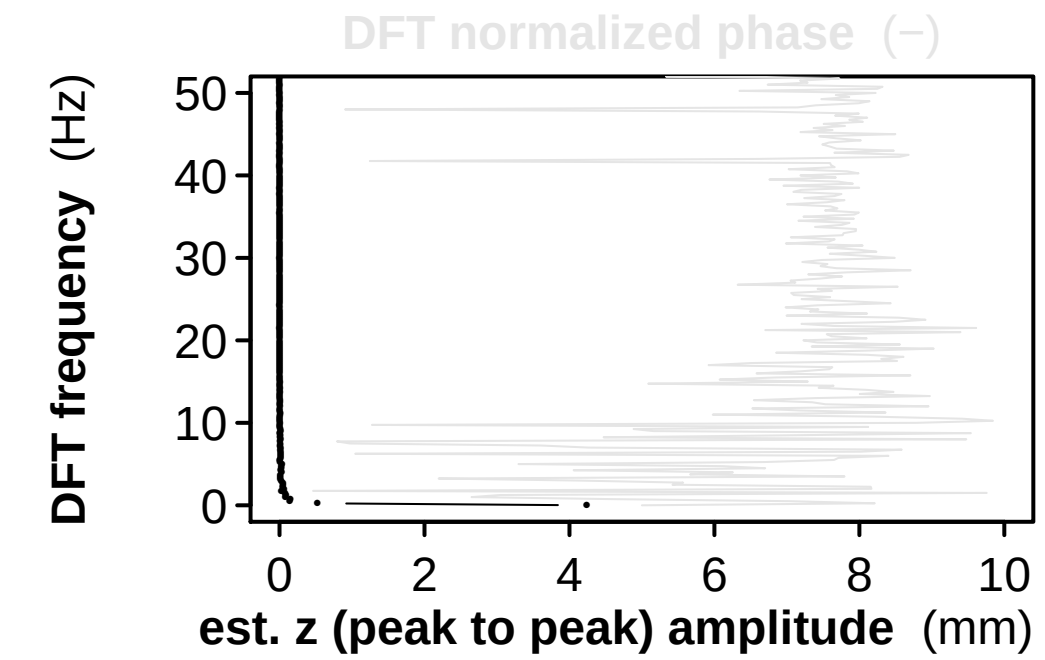

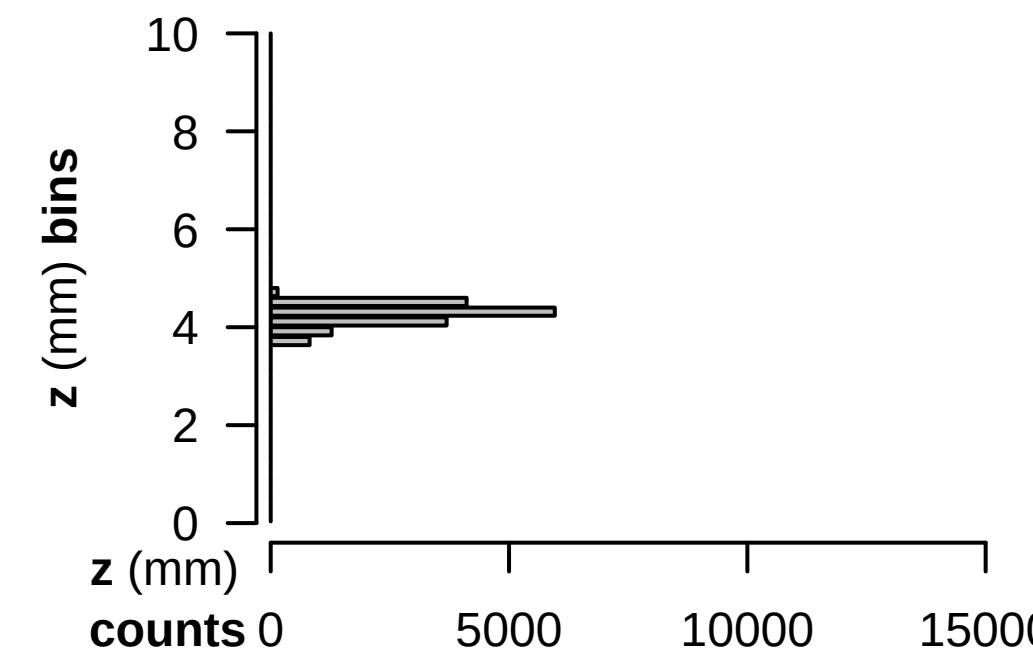

**SUBJECT 4 − RUN 11 − CONDITION 3,0**
**SC_180323_123607_0.AIFF**

```
z_min               : 3.66 mm
z_max               : 4.63 mm
z_travel_amplitude  : 0.98 mm

avg_abs_z_travel    : 2.30 mm/s

z_jarque-bera_jb    : 2592.99
z_jarque-bera_p     : 0.00e+00

z_lin_mod_est_slope: 0.12 mm/s
z_lin_mod_adj_R²    : 44 %

z_poly40_mod_adj_R²: 97 %

z_dft_ampl_thresh   : 0.010 mm
>=threshold_maxfreq: 9.25 Hz
```

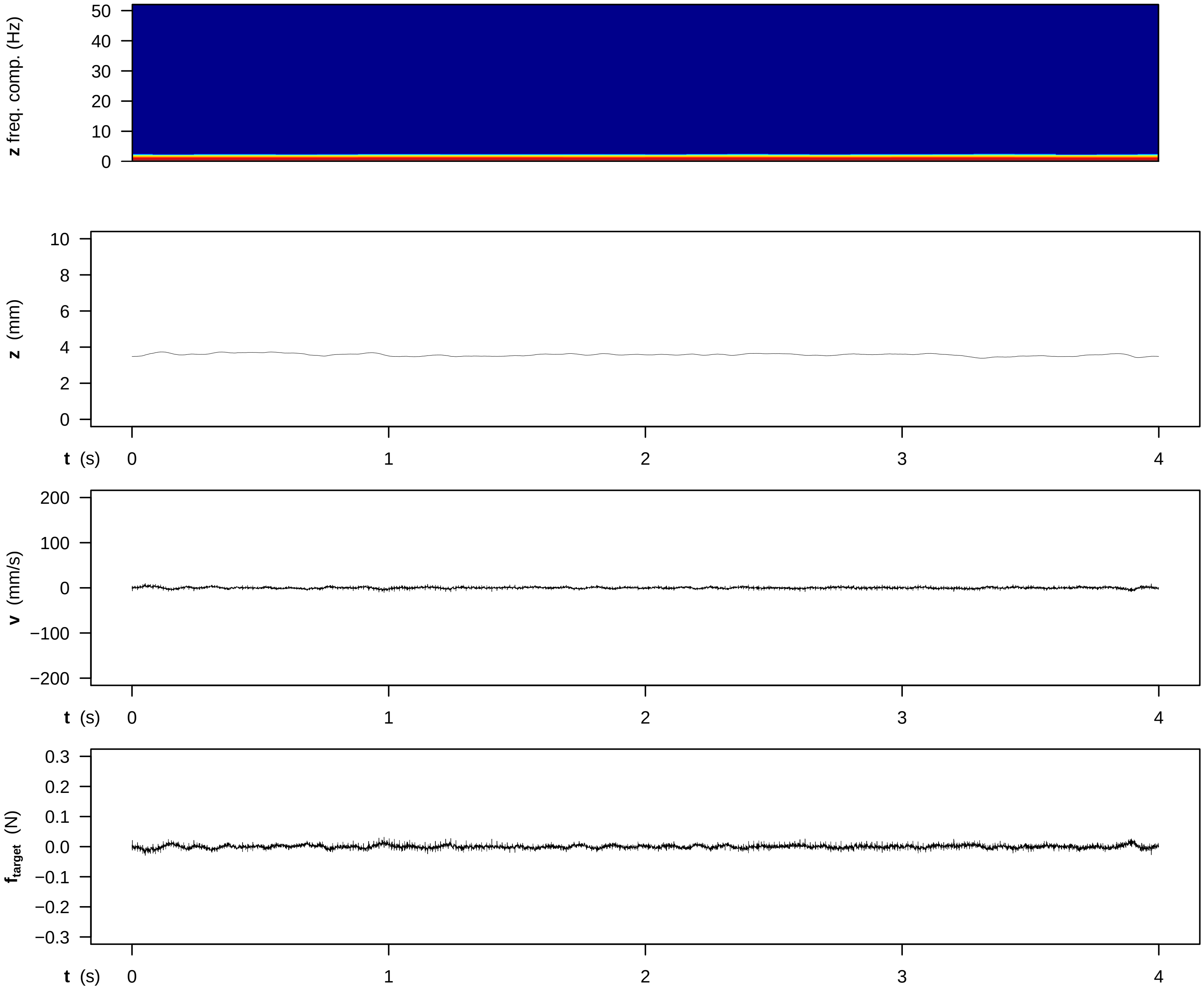

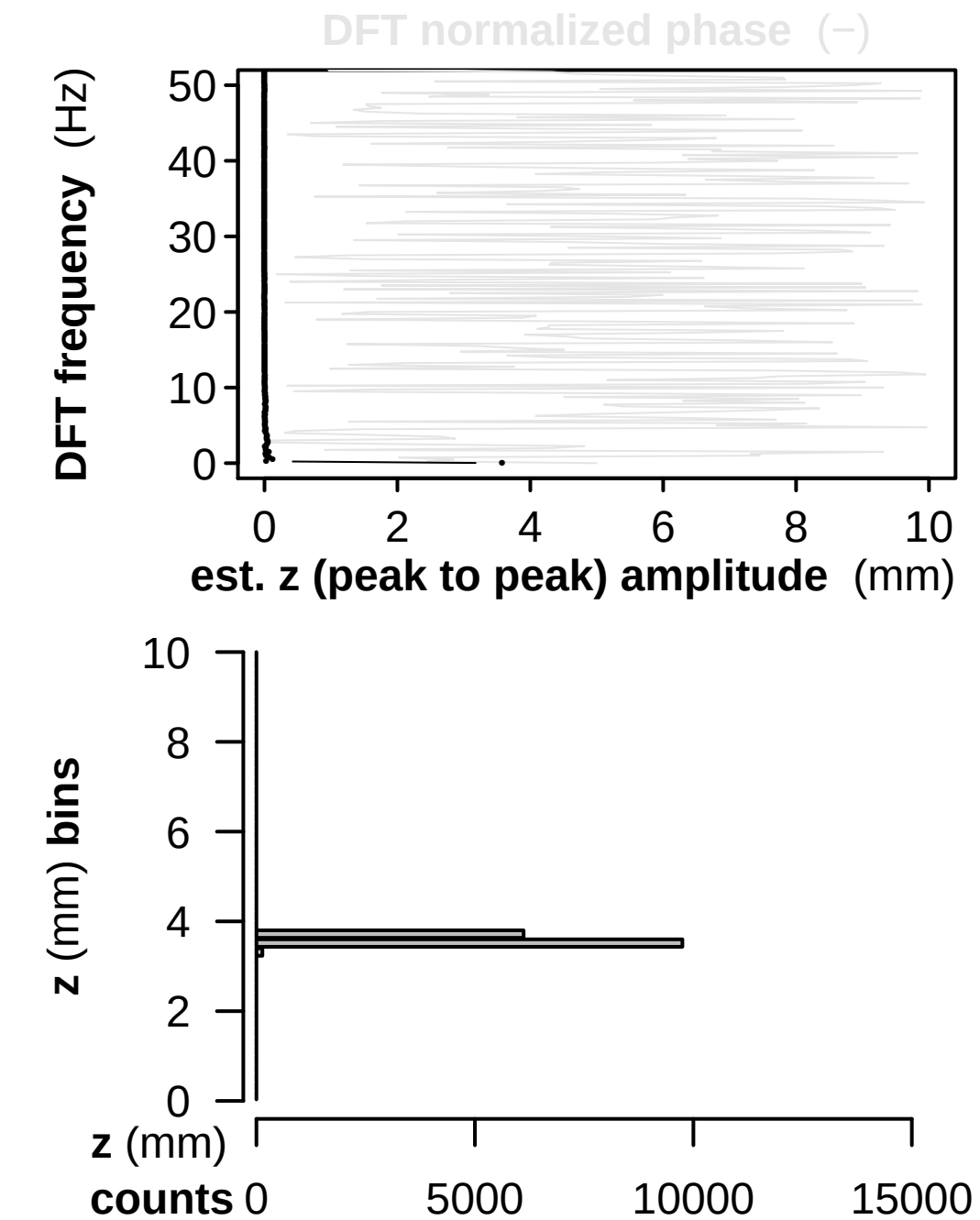

**SUBJECT 4 − RUN 33 − CONDITION 3,0**
**SC_180323_124747_0.AIFF**

```
z_min               : 3.38 mm
z_max               : 3.73 mm
z_travel_amplitude  : 0.35 mm

avg_abs_z_travel    : 2.23 mm/s

z_jarque-bera_jb    : 127.31
z_jarque-bera_p     : 0.00e+00

z_lin_mod_est_slope: −0.03 mm/s
z_lin_mod_adj_R²    : 17 %

z_poly40_mod_adj_R²: 80 %

z_dft_ampl_thresh   : 0.010 mm
>=threshold_maxfreq: 9.25 Hz
```

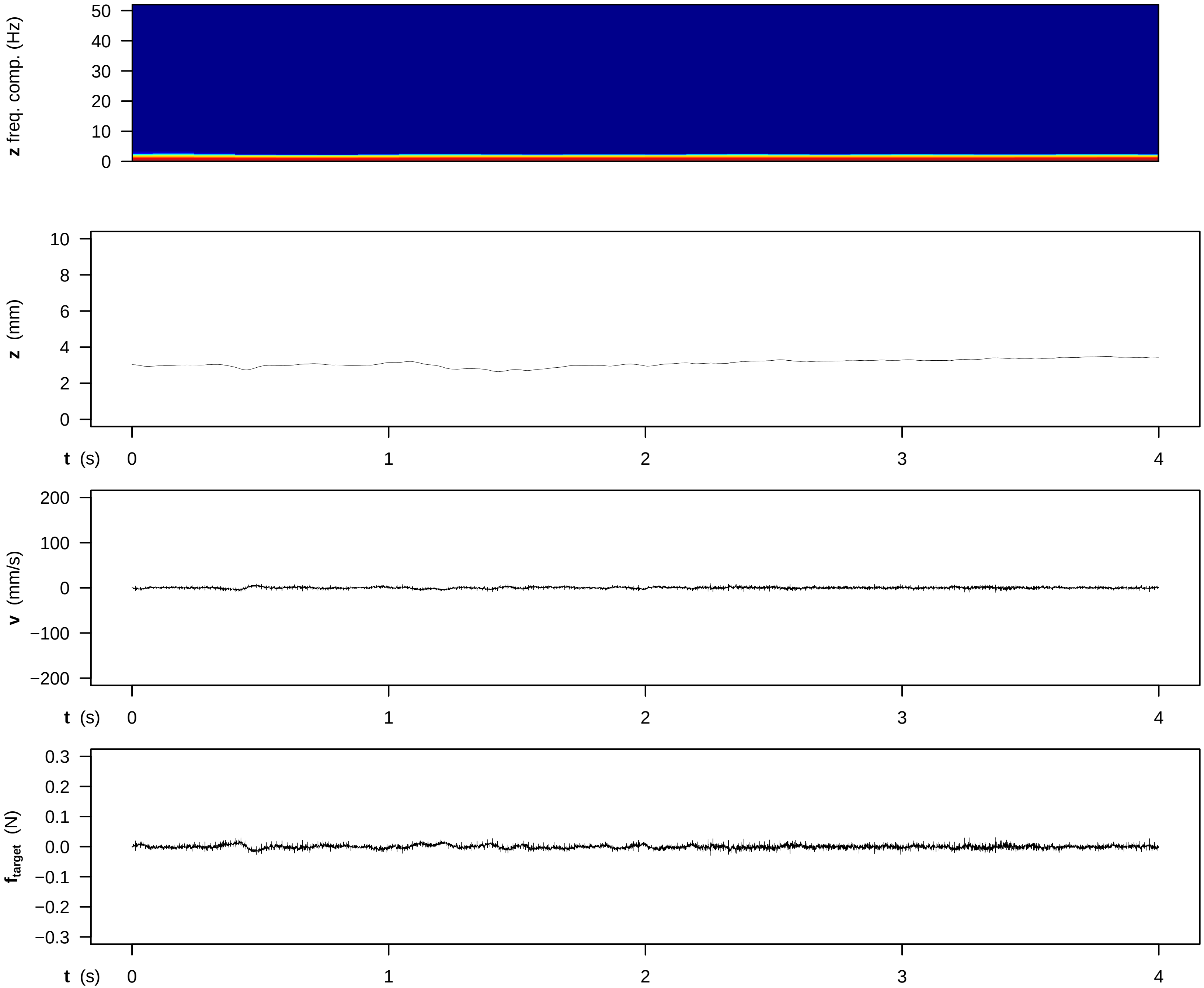

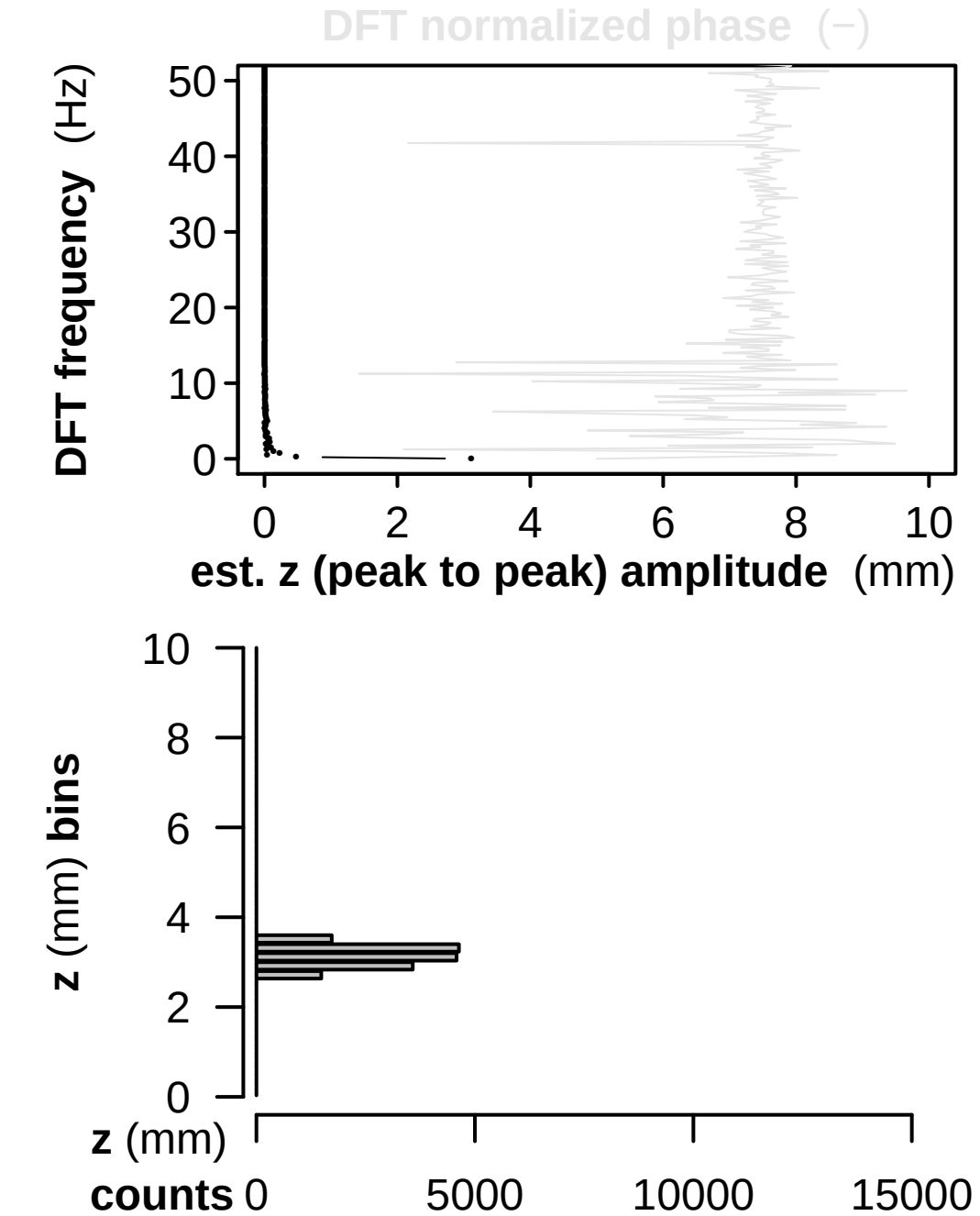

**SUBJECT 4 − RUN 34 − CONDITION 3,0**
**SC_180323_124815_0.AIFF**

```
z_min               : 2.64 mm
z_max               : 3.49 mm
z_travel_amplitude  : 0.85 mm

avg_abs_z_travel    : 2.42 mm/s

z_jarque-bera_jb    : 453.07
z_jarque-bera_p     : 0.00e+00

z_lin_mod_est_slope: 0.14 mm/s
z_lin_mod_adj_R²    : 63 %

z_poly40_mod_adj_R²: 96 %

z_dft_ampl_thresh   : 0.010 mm
>=threshold_maxfreq: 9.75 Hz
```

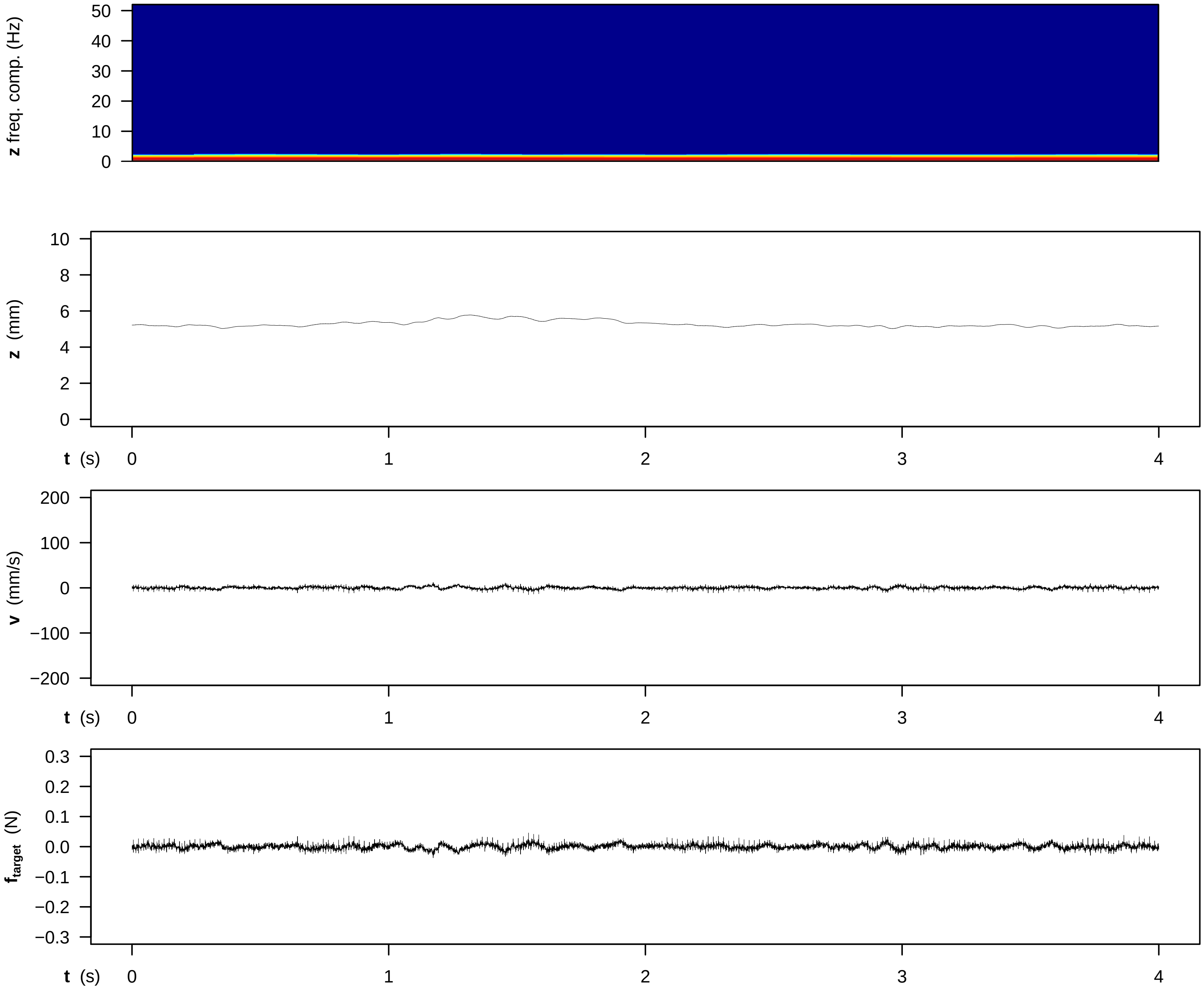

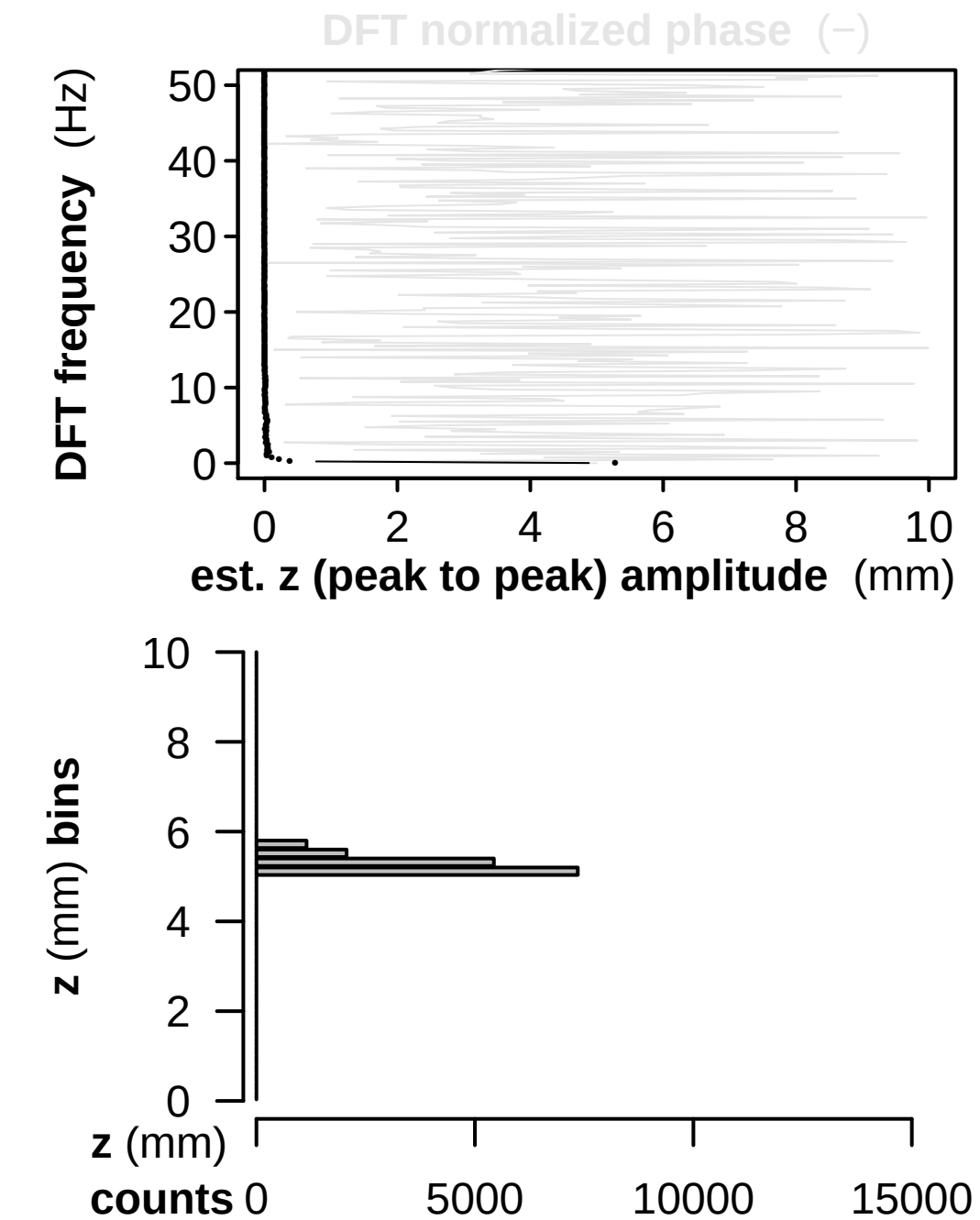

**SUBJECT 5 − RUN 01 − CONDITION 3,0**
**SC_180323_131427_0.AIFF**

```
z_min               : 5.02 mm
z_max               : 5.78 mm
z_travel_amplitude  : 0.76 mm

avg_abs_z_travel    : 4.19 mm/s

z_jarque-bera_jb    : 3713.86
z_jarque-bera_p     : 0.00e+00

z_lin_mod_est_slope: −0.05 mm/s
z_lin_mod_adj_R²    : 10 %

z_poly40_mod_adj_R²: 90 %

z_dft_ampl_thresh   : 0.010 mm
>=threshold_maxfreq: 11.50 Hz
```

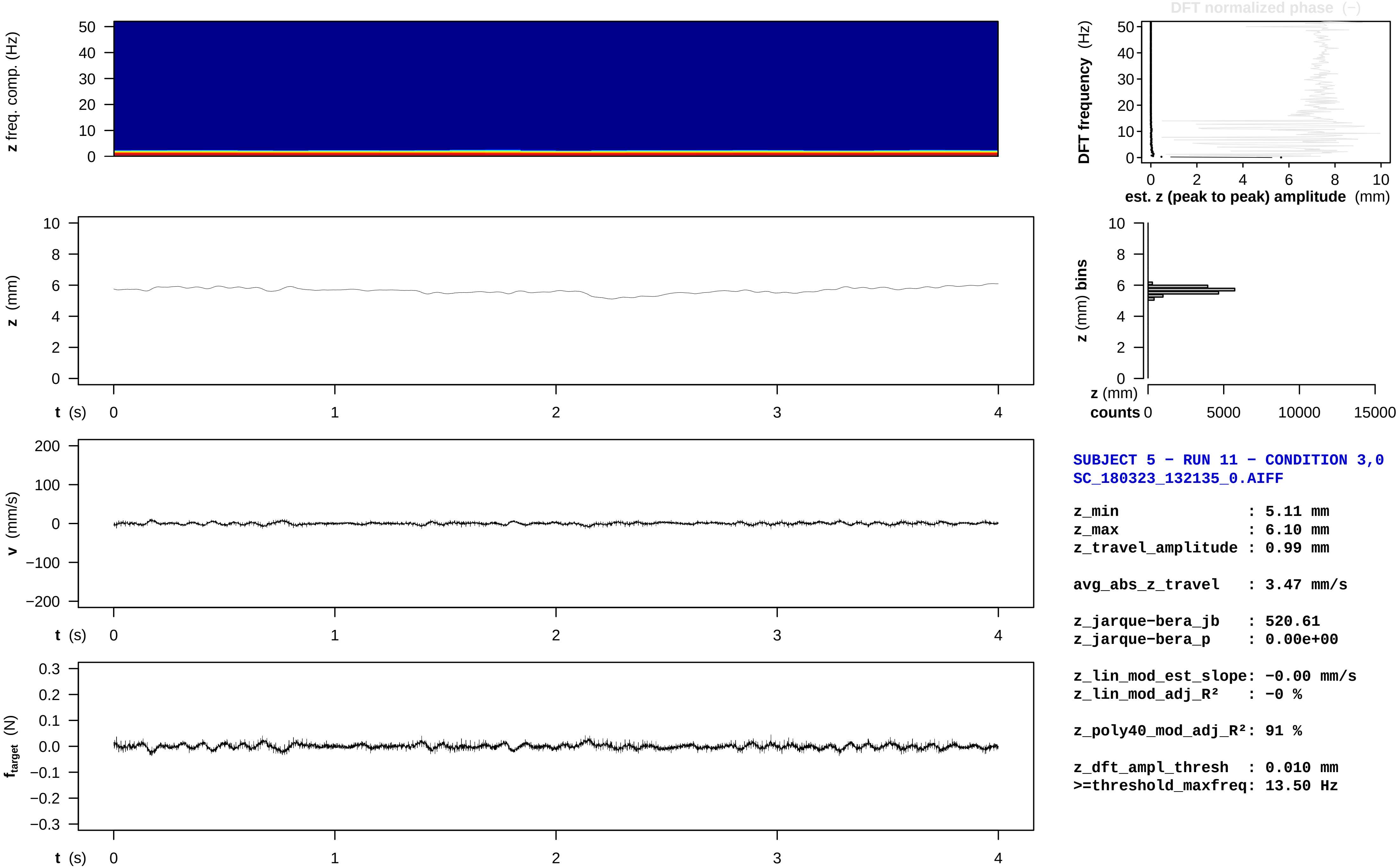

**SUBJECT 5 − RUN 11 − CONDITION 3,0**
**SC_180323_132135_0.AIFF**

```
z_min               : 5.11 mm
z_max               : 6.10 mm
z_travel_amplitude  : 0.99 mm

avg_abs_z_travel    : 3.47 mm/s

z_jarque-bera_jb    : 520.61
z_jarque-bera_p     : 0.00e+00

z_lin_mod_est_slope: −0.00 mm/s
z_lin_mod_adj_R²    : −0 %

z_poly40_mod_adj_R²: 91 %

z_dft_ampl_thresh   : 0.010 mm
>=threshold_maxfreq: 13.50 Hz
```

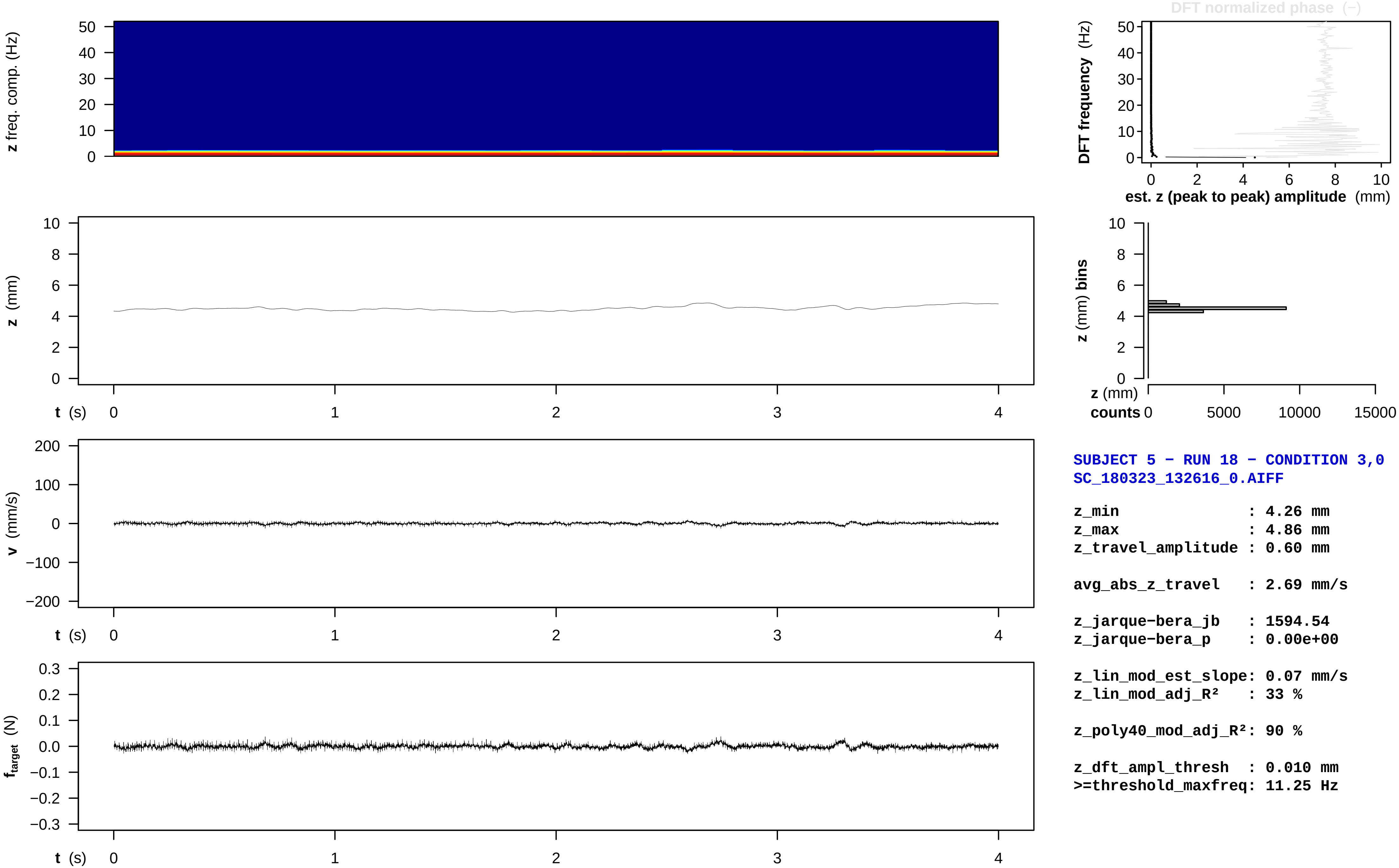

**SUBJECT 5 − RUN 18 − CONDITION 3,0**
**SC_180323_132616_0.AIFF**

```
z_min              : 4.26 mm
z_max              : 4.86 mm
z_travel_amplitude : 0.60 mm

avg_abs_z_travel   : 2.69 mm/s

z_jarque-bera_jb   : 1594.54
z_jarque-bera_p    : 0.00e+00

z_lin_mod_est_slope: 0.07 mm/s
z_lin_mod_adj_R²   : 33 %

z_poly40_mod_adj_R²: 90 %

z_dft_ampl_thresh  : 0.010 mm
>=threshold_maxfreq: 11.25 Hz
```

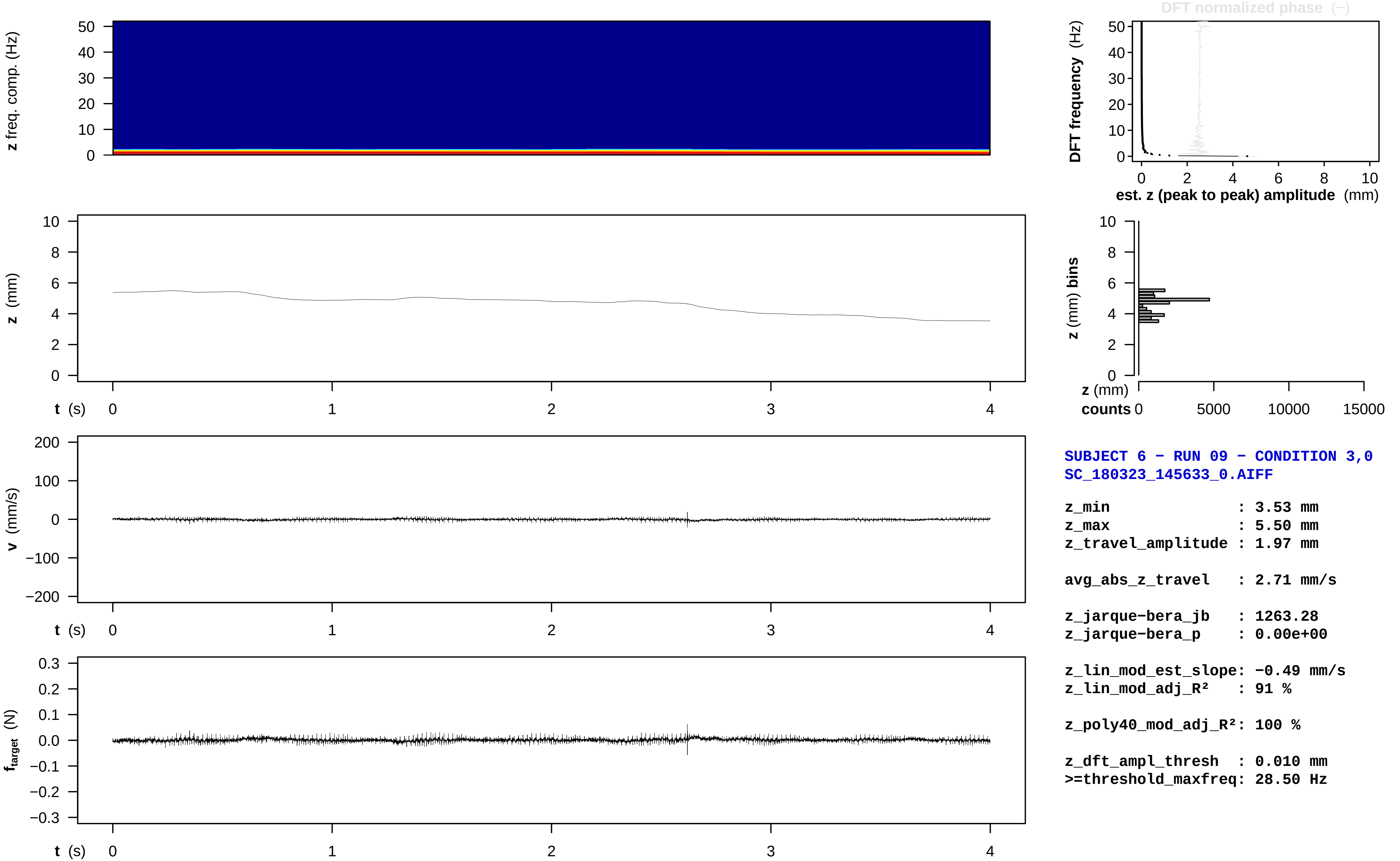

**SUBJECT 6 − RUN 09 − CONDITION 3,0**
**SC_180323_145633_0.AIFF**

```
z_min               : 3.53 mm
z_max               : 5.50 mm
z_travel_amplitude  : 1.97 mm

avg_abs_z_travel    : 2.71 mm/s

z_jarque-bera_jb    : 1263.28
z_jarque-bera_p     : 0.00e+00

z_lin_mod_est_slope: −0.49 mm/s
z_lin_mod_adj_R²    : 91 %

z_poly40_mod_adj_R²: 100 %

z_dft_ampl_thresh   : 0.010 mm
>=threshold_maxfreq: 28.50 Hz
```

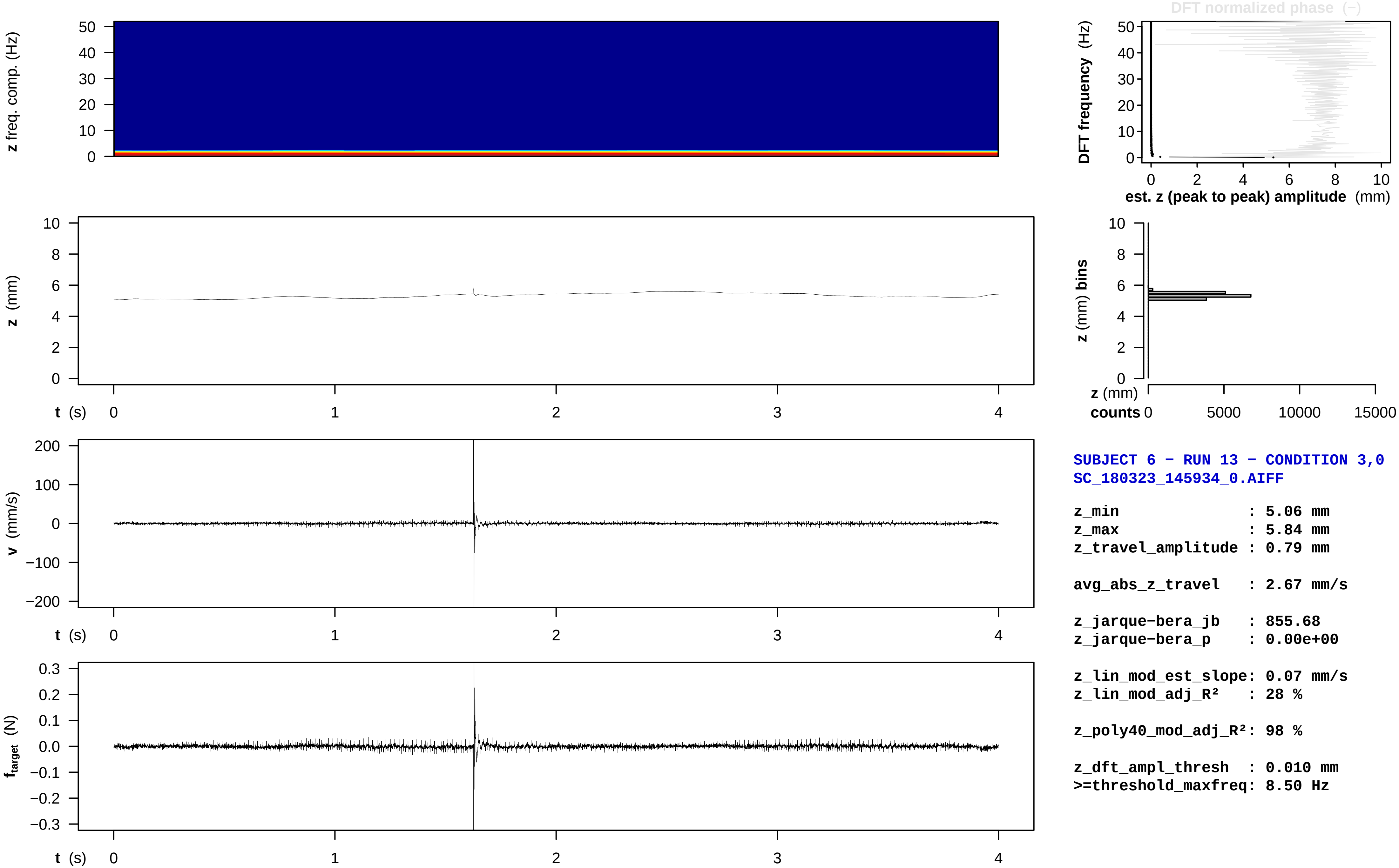

**SUBJECT 6 − RUN 13 − CONDITION 3,0**
**SC_180323_145934_0.AIFF**

```
z_min               : 5.06 mm
z_max               : 5.84 mm
z_travel_amplitude  : 0.79 mm

avg_abs_z_travel    : 2.67 mm/s

z_jarque-bera_jb    : 855.68
z_jarque-bera_p     : 0.00e+00

z_lin_mod_est_slope: 0.07 mm/s
z_lin_mod_adj_R²    : 28 %

z_poly40_mod_adj_R²: 98 %

z_dft_ampl_thresh   : 0.010 mm
>=threshold_maxfreq: 8.50 Hz
```

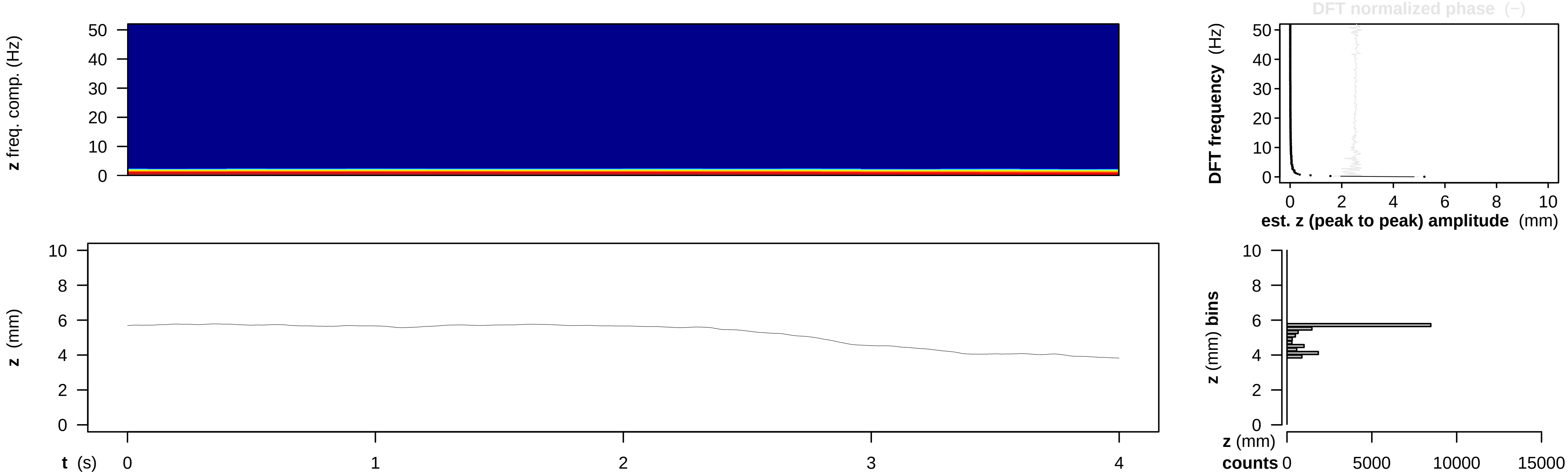

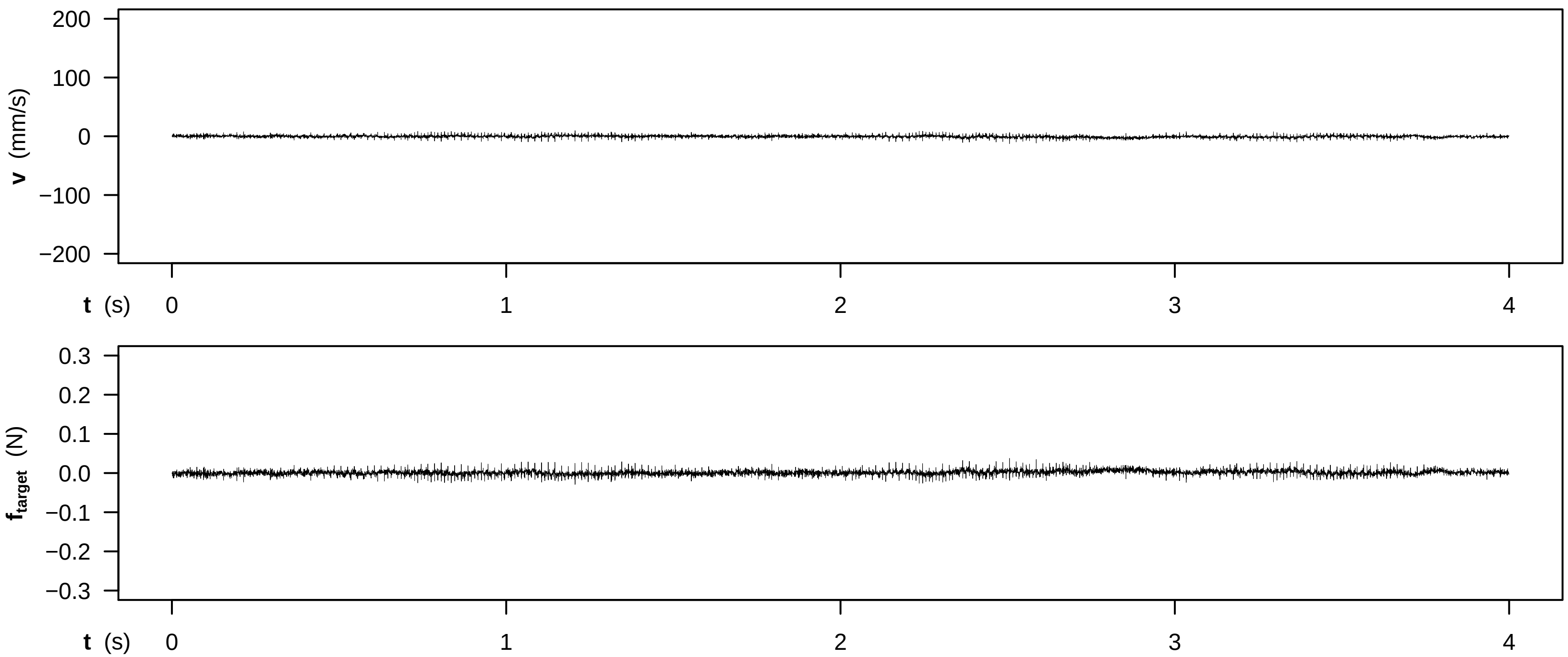

**SUBJECT 6 − RUN 18 − CONDITION 3,0**
**SC_180323_150214_0.AIFF**

```
z_min               : 3.83 mm
z_max               : 5.79 mm
z_travel_amplitude  : 1.96 mm

avg_abs_z_travel    : 2.43 mm/s

z_jarque-bera_jb    : 2693.48
z_jarque-bera_p     : 0.00e+00

z_lin_mod_est_slope: −0.51 mm/s
z_lin_mod_adj_R²    : 77 %

z_poly40_mod_adj_R²: 100 %

z_dft_ampl_thresh   : 0.010 mm
>=threshold_maxfreq: 30.00 Hz
```

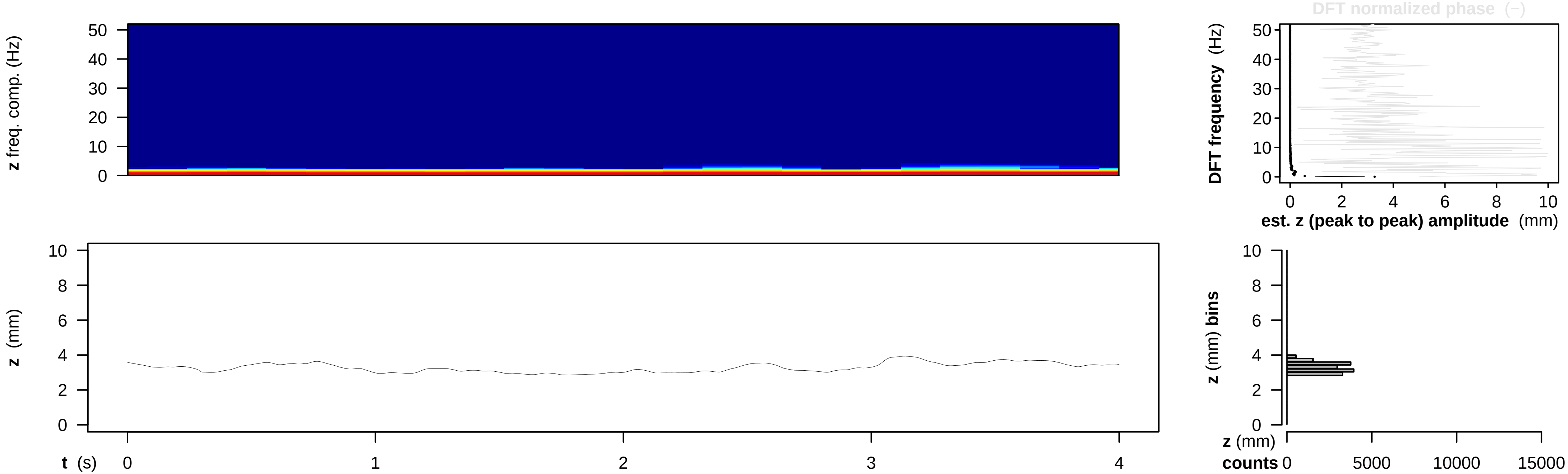

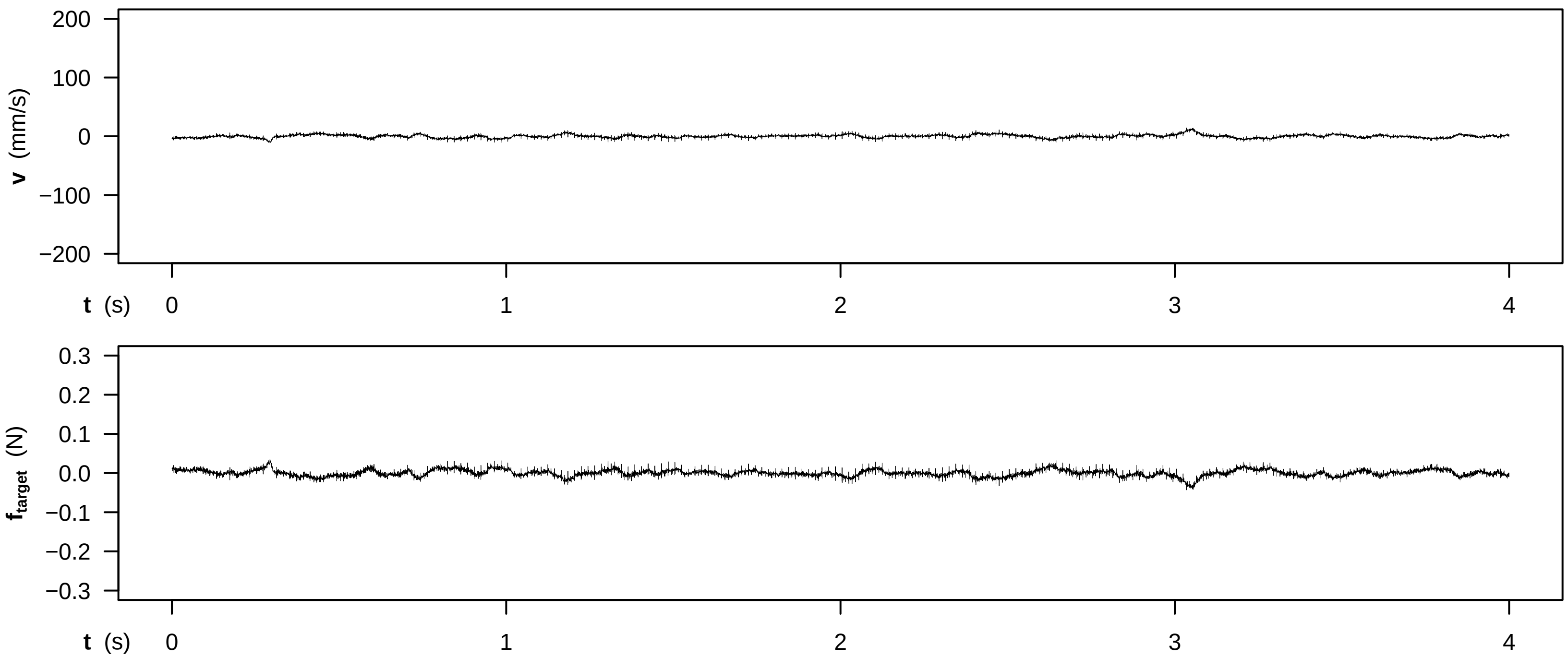

**SUBJECT 7 − RUN 08 − CONDITION 3,0**
**SC_180323_153915_0.AIFF**

```
z_min               : 2.85 mm
z_max               : 3.91 mm
z_travel_amplitude  : 1.06 mm

avg_abs_z_travel    : 2.73 mm/s

z_jarque-bera_jb    : 854.94
z_jarque-bera_p     : 0.00e+00

z_lin_mod_est_slope: 0.08 mm/s
z_lin_mod_adj_R²    : 12 %

z_poly40_mod_adj_R²: 92 %

z_dft_ampl_thresh   : 0.010 mm
>=threshold_maxfreq: 10.75 Hz
```

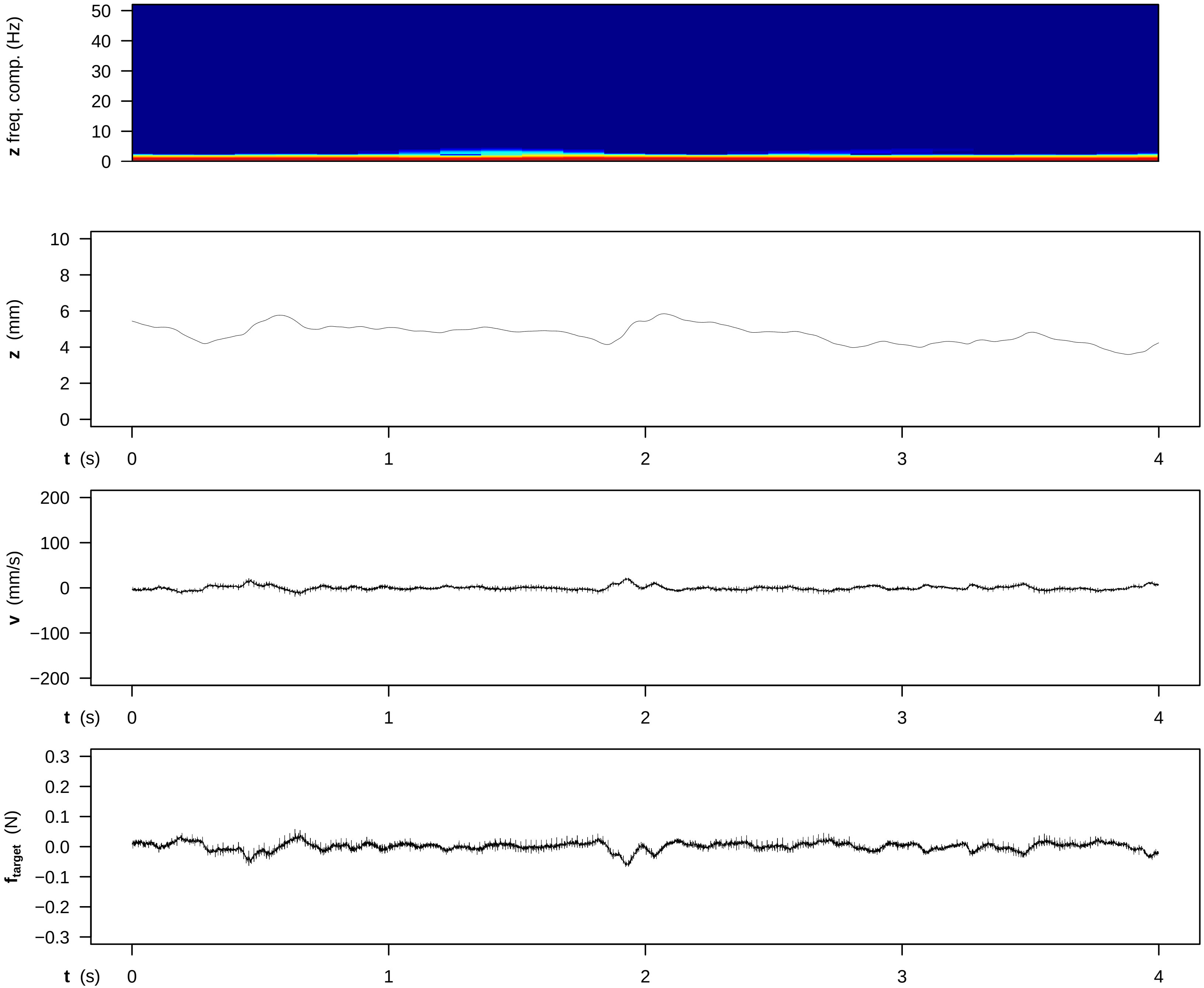

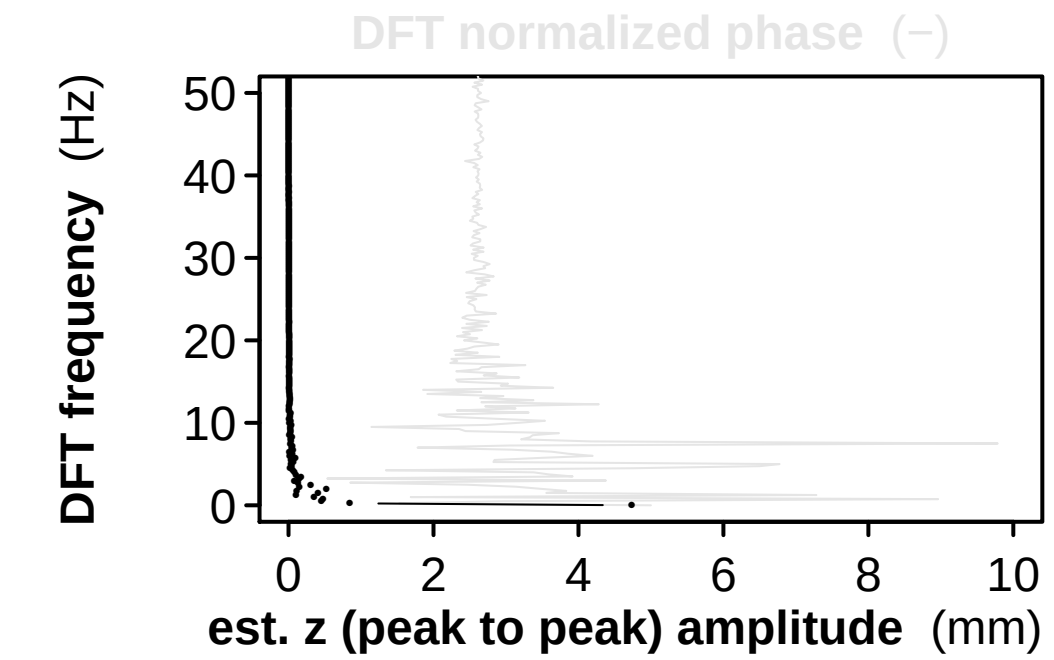

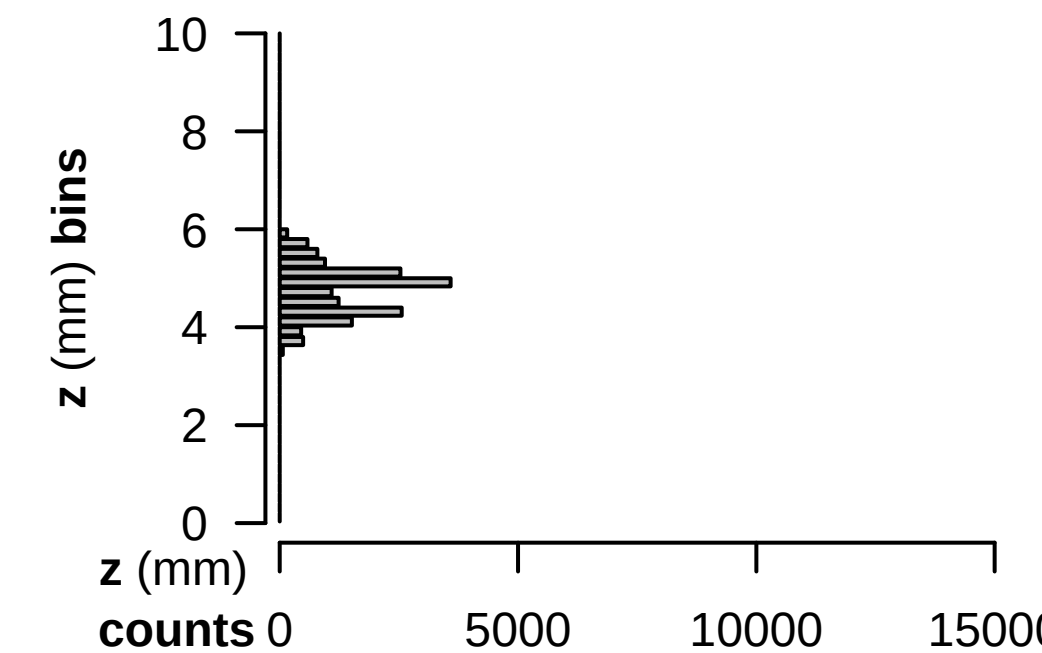

**SUBJECT 7 − RUN 13 − CONDITION 3,0**
**SC_180323_154251_0.AIFF**

```
z_min               : 3.59 mm
z_max               : 5.85 mm
z_travel_amplitude  : 2.26 mm

avg_abs_z_travel    : 4.21 mm/s

z_jarque-bera_jb    : 226.46
z_jarque-bera_p     : 0.00e+00

z_lin_mod_est_slope: −0.26 mm/s
z_lin_mod_adj_R²    : 37 %

z_poly40_mod_adj_R²: 90 %

z_dft_ampl_thresh   : 0.010 mm
>=threshold_maxfreq: 22.25 Hz
```

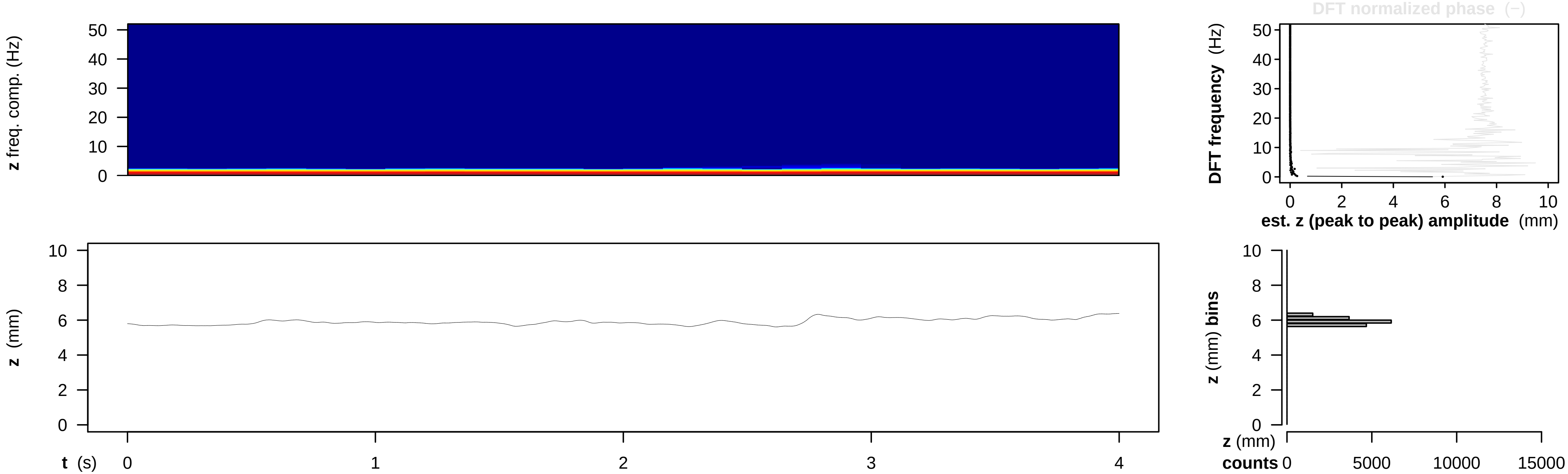

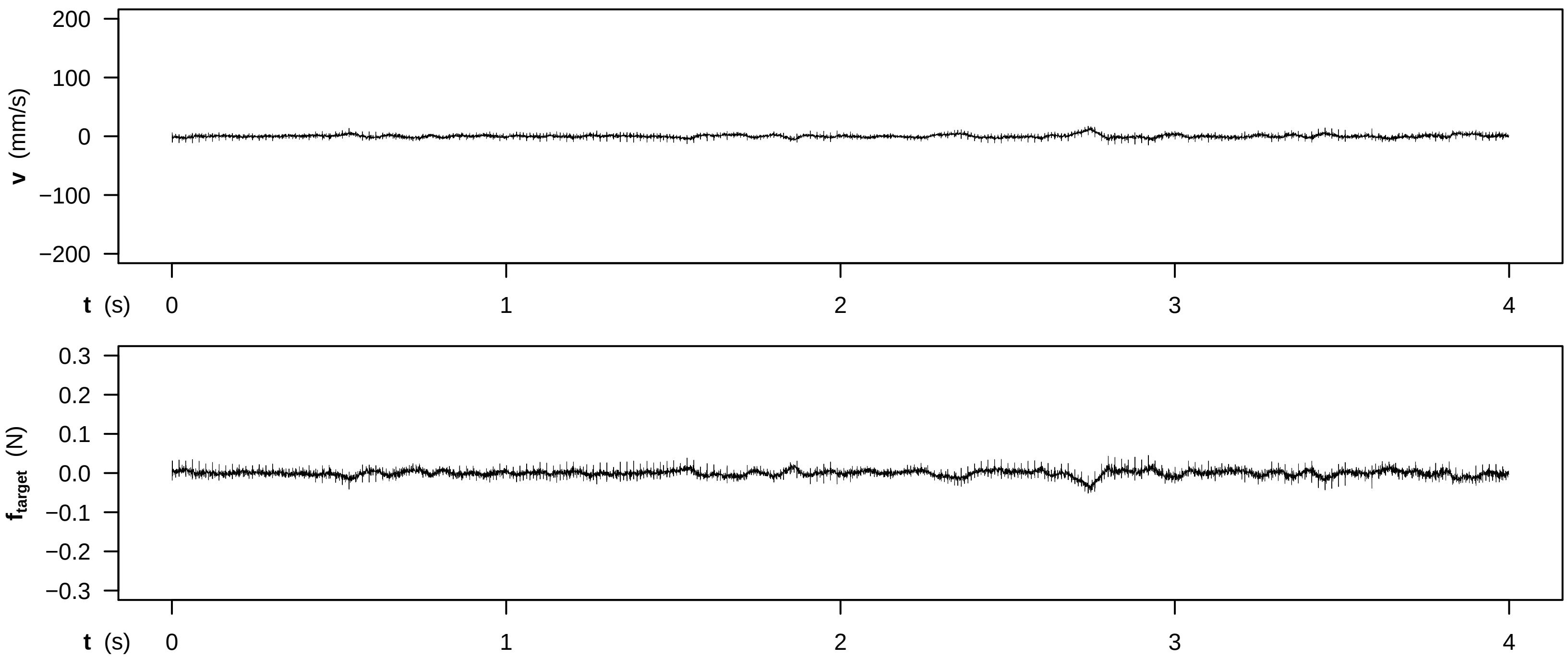

**SUBJECT 7 − RUN 30 − CONDITION 3,0**
**SC_180323_155616_0.AIFF**

```
z_min               : 5.61 mm
z_max               : 6.38 mm
z_travel_amplitude  : 0.78 mm

avg_abs_z_travel    : 2.94 mm/s

z_jarque-bera_jb    : 910.42
z_jarque-bera_p     : 0.00e+00

z_lin_mod_est_slope: 0.11 mm/s
z_lin_mod_adj_R²    : 46 %

z_poly40_mod_adj_R²: 83 %

z_dft_ampl_thresh   : 0.010 mm
>=threshold_maxfreq: 15.00 Hz
```

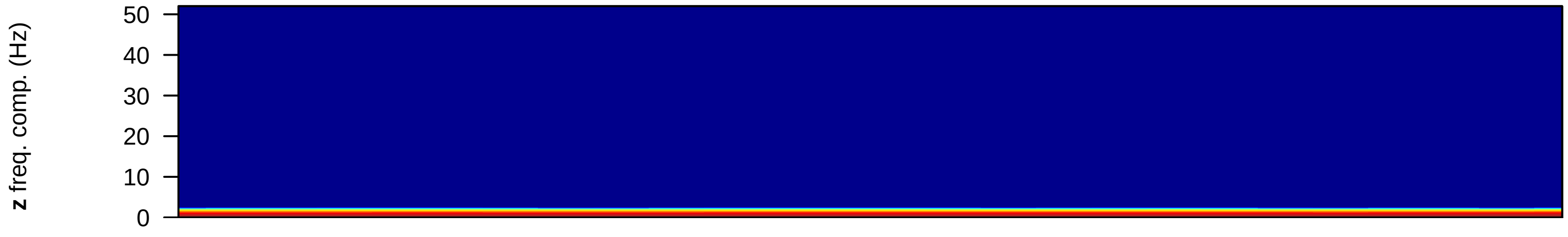

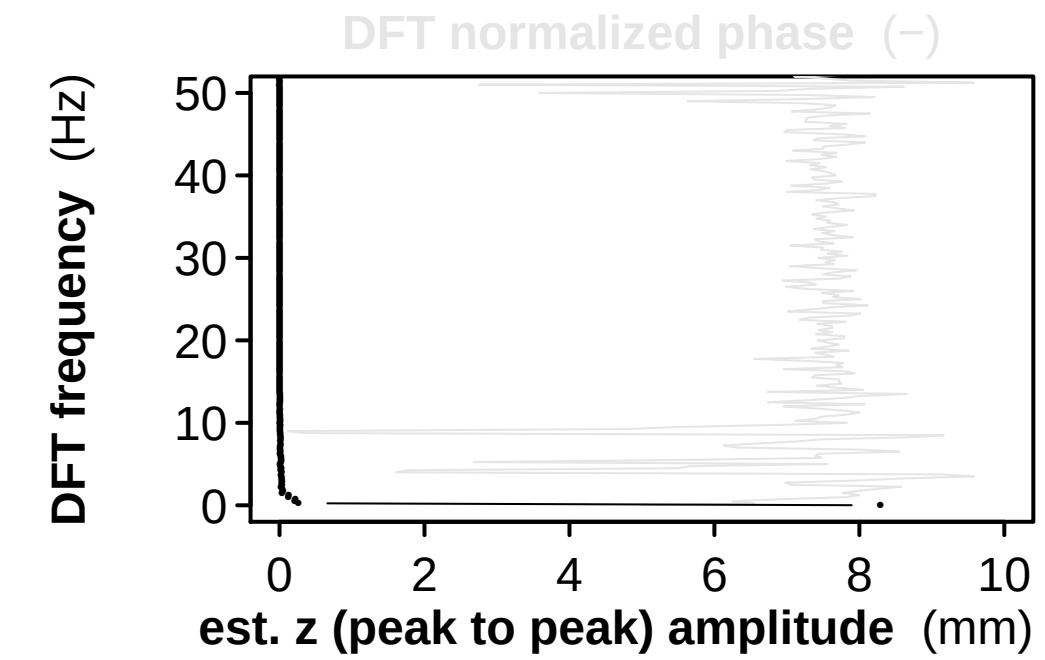

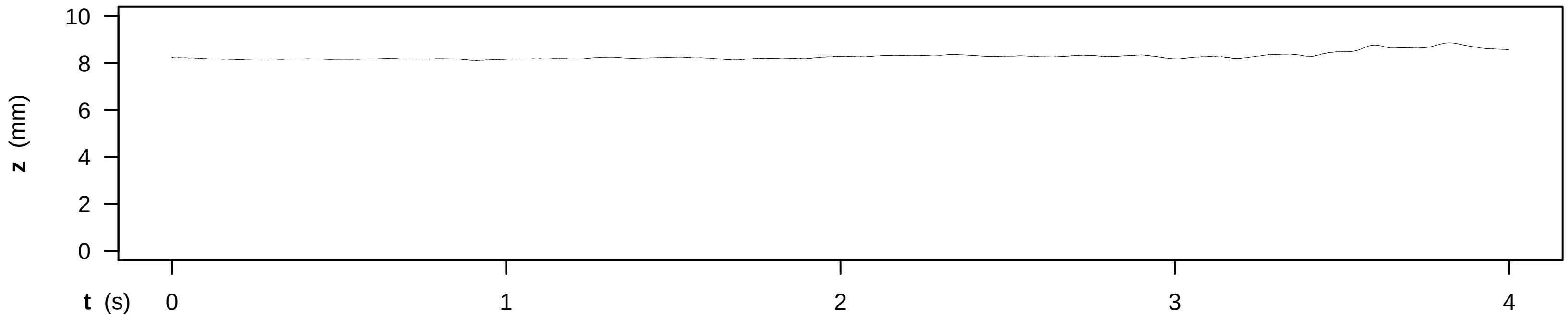

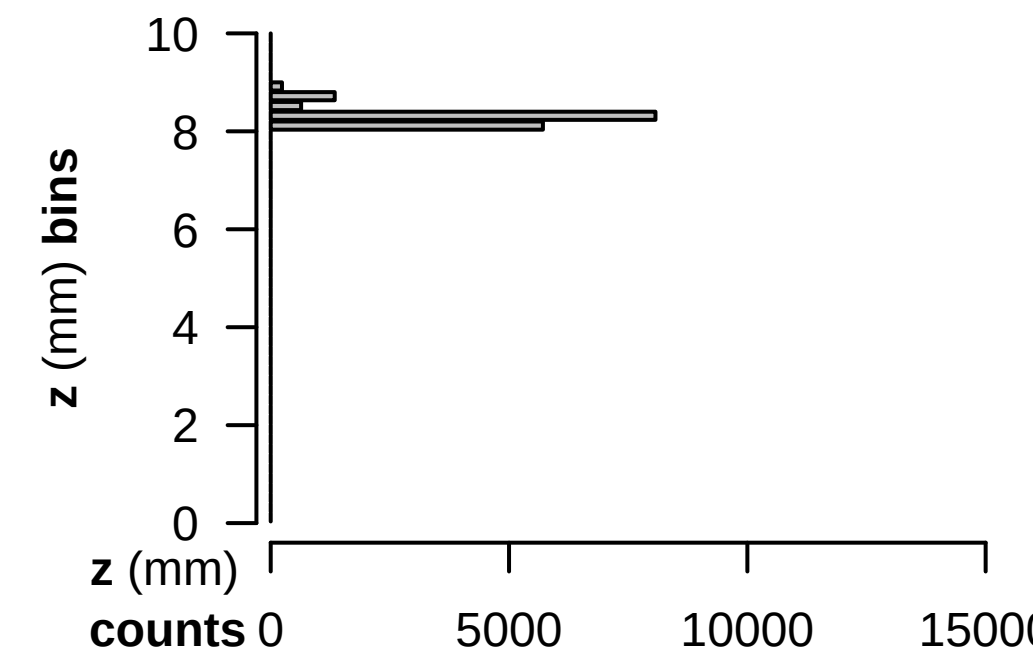

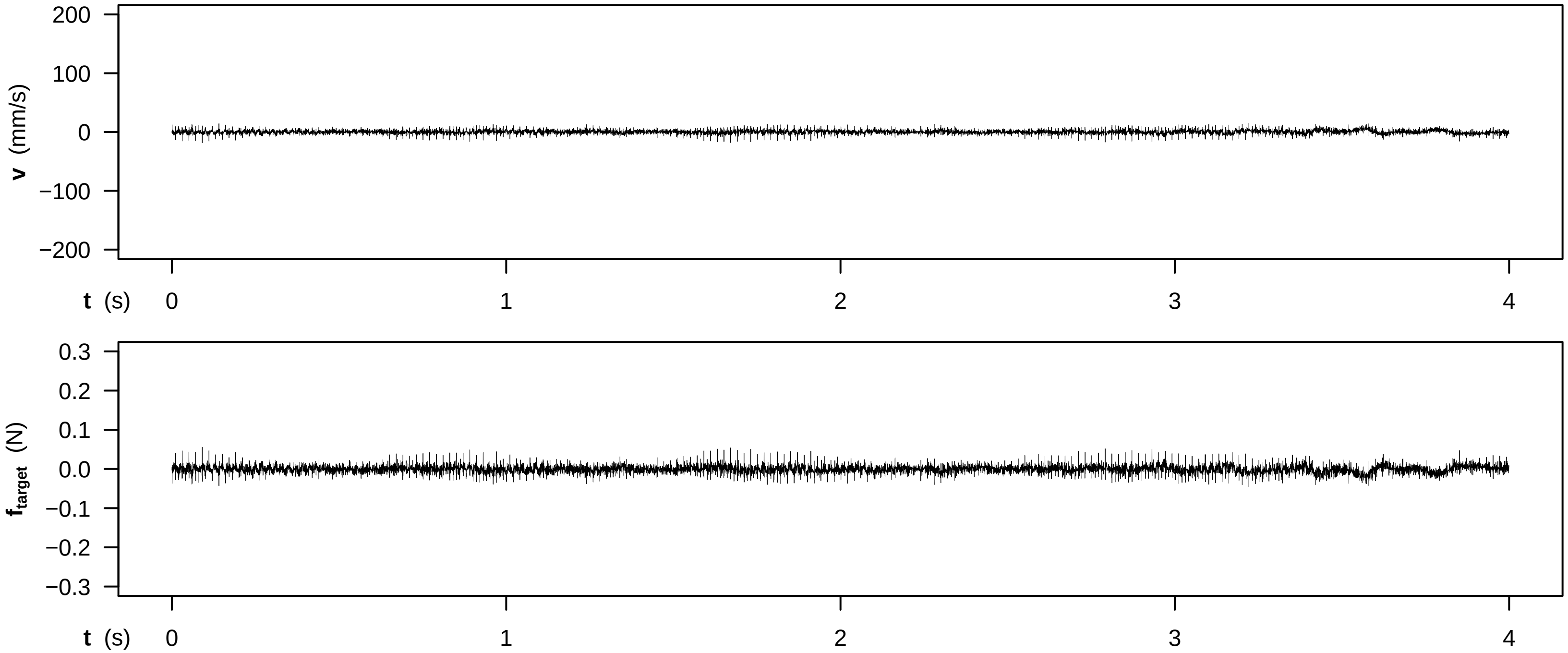

**SUBJECT 8 − RUN 13 − CONDITION 3,0**
**SC_180323_165300_0.AIFF**

```
z_min               : 8.09 mm
z_max               : 8.86 mm
z_travel_amplitude  : 0.77 mm

avg_abs_z_travel    : 6.90 mm/s

z_jarque-bera_jb    : 11993.40
z_jarque-bera_p     : 0.00e+00

z_lin_mod_est_slope: 0.11 mm/s
z_lin_mod_adj_R²    : 60 %

z_poly40_mod_adj_R²: 96 %

z_dft_ampl_thresh   : 0.010 mm
>=threshold_maxfreq: 9.75 Hz
```

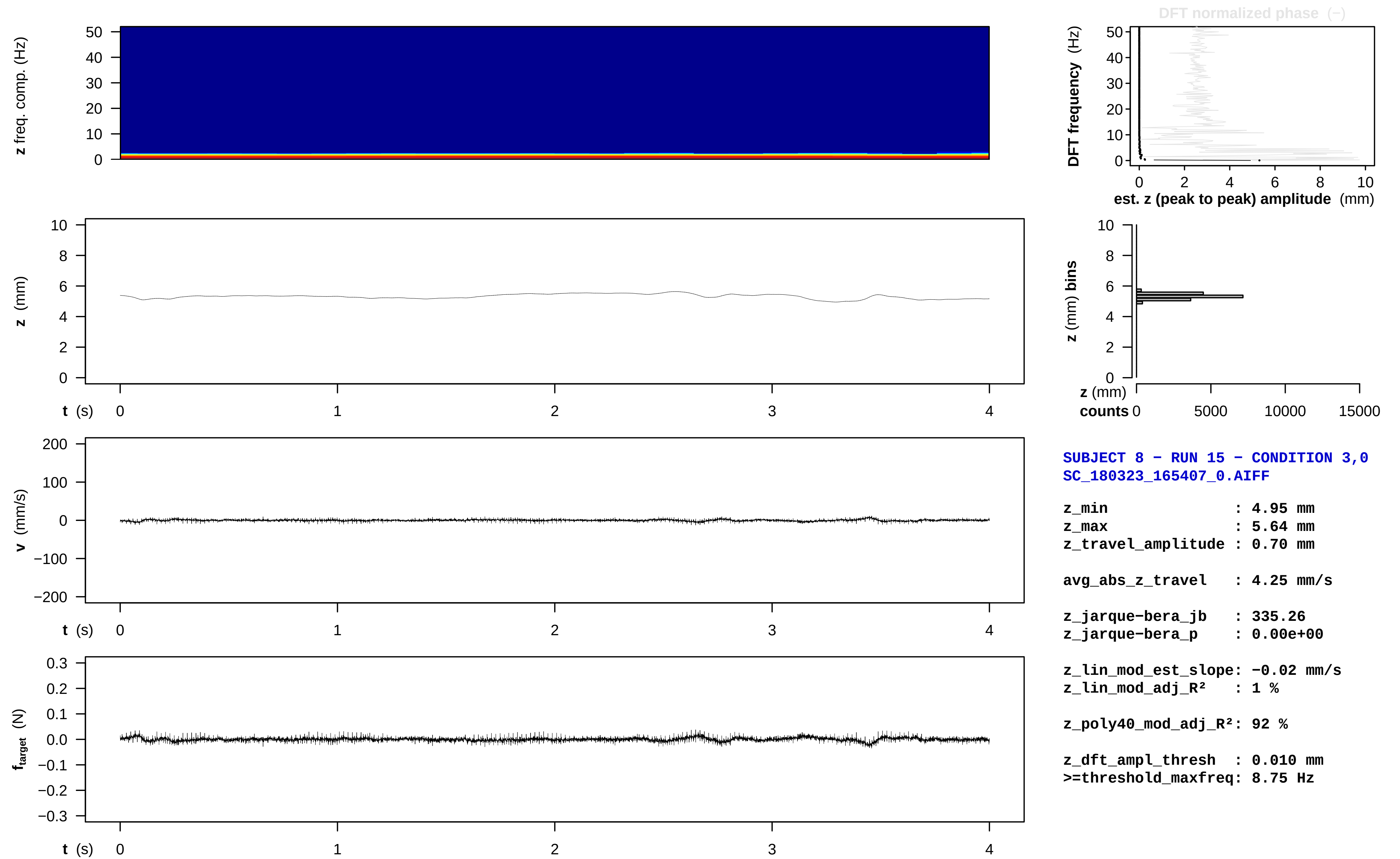

**SUBJECT 8 − RUN 15 − CONDITION 3,0**
**SC_180323_165407_0.AIFF**

```
z_min              : 4.95 mm
z_max              : 5.64 mm
z_travel_amplitude : 0.70 mm

avg_abs_z_travel   : 4.25 mm/s

z_jarque-bera_jb   : 335.26
z_jarque-bera_p    : 0.00e+00

z_lin_mod_est_slope: −0.02 mm/s
z_lin_mod_adj_R²   : 1 %

z_poly40_mod_adj_R²: 92 %

z_dft_ampl_thresh  : 0.010 mm
>=threshold_maxfreq: 8.75 Hz
```

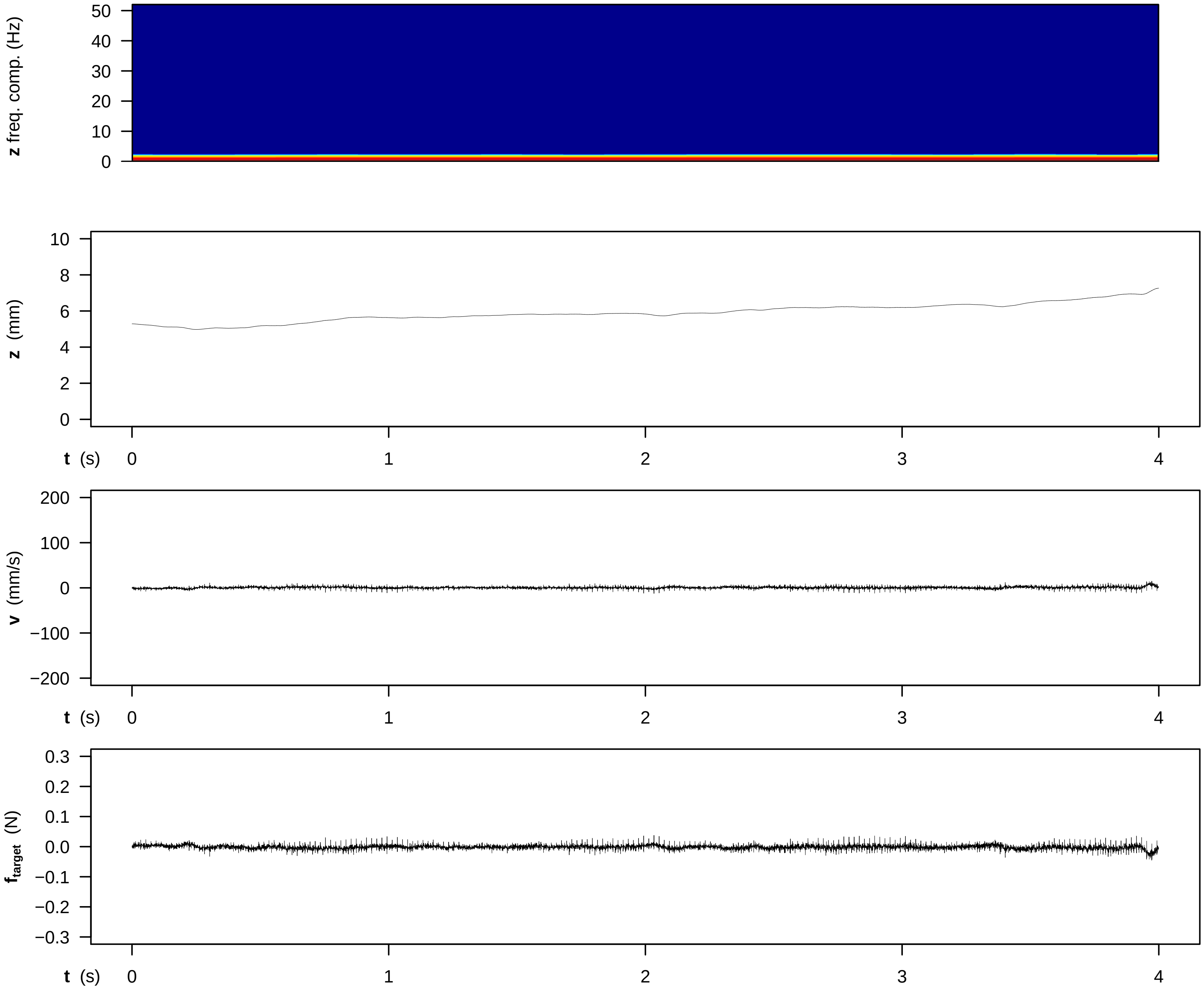

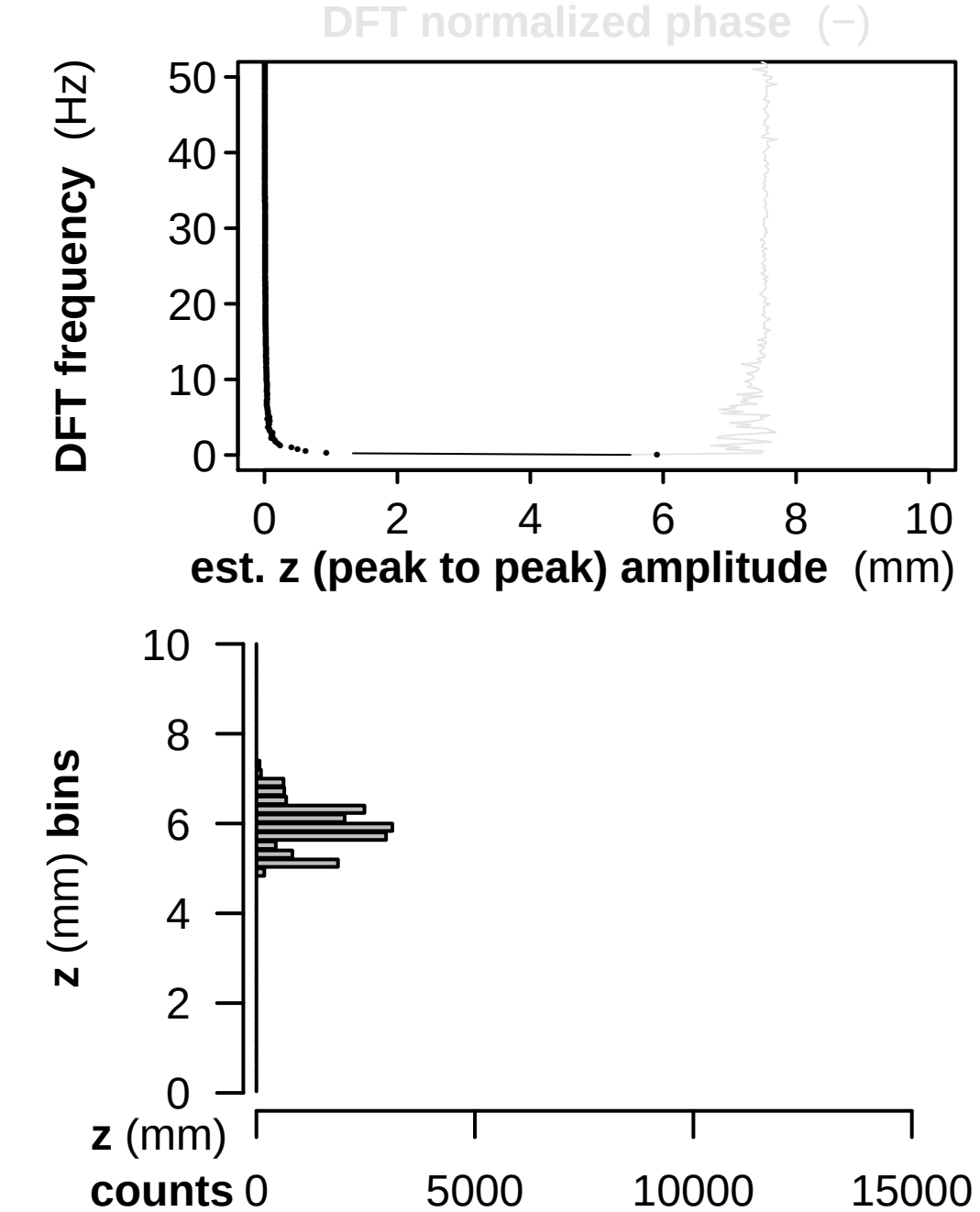

**SUBJECT 8 − RUN 17 − CONDITION 3,0**
**SC_180323_165547_0.AIFF**

```
z_min               : 4.98 mm
z_max               : 7.27 mm
z_travel_amplitude  : 2.29 mm

avg_abs_z_travel    : 2.85 mm/s

z_jarque-bera_jb    : 143.25
z_jarque-bera_p     : 0.00e+00

z_lin_mod_est_slope: 0.42 mm/s
z_lin_mod_adj_R²    : 94 %

z_poly40_mod_adj_R²: 100 %

z_dft_ampl_thresh   : 0.010 mm
>=threshold_maxfreq: 32.00 Hz
```

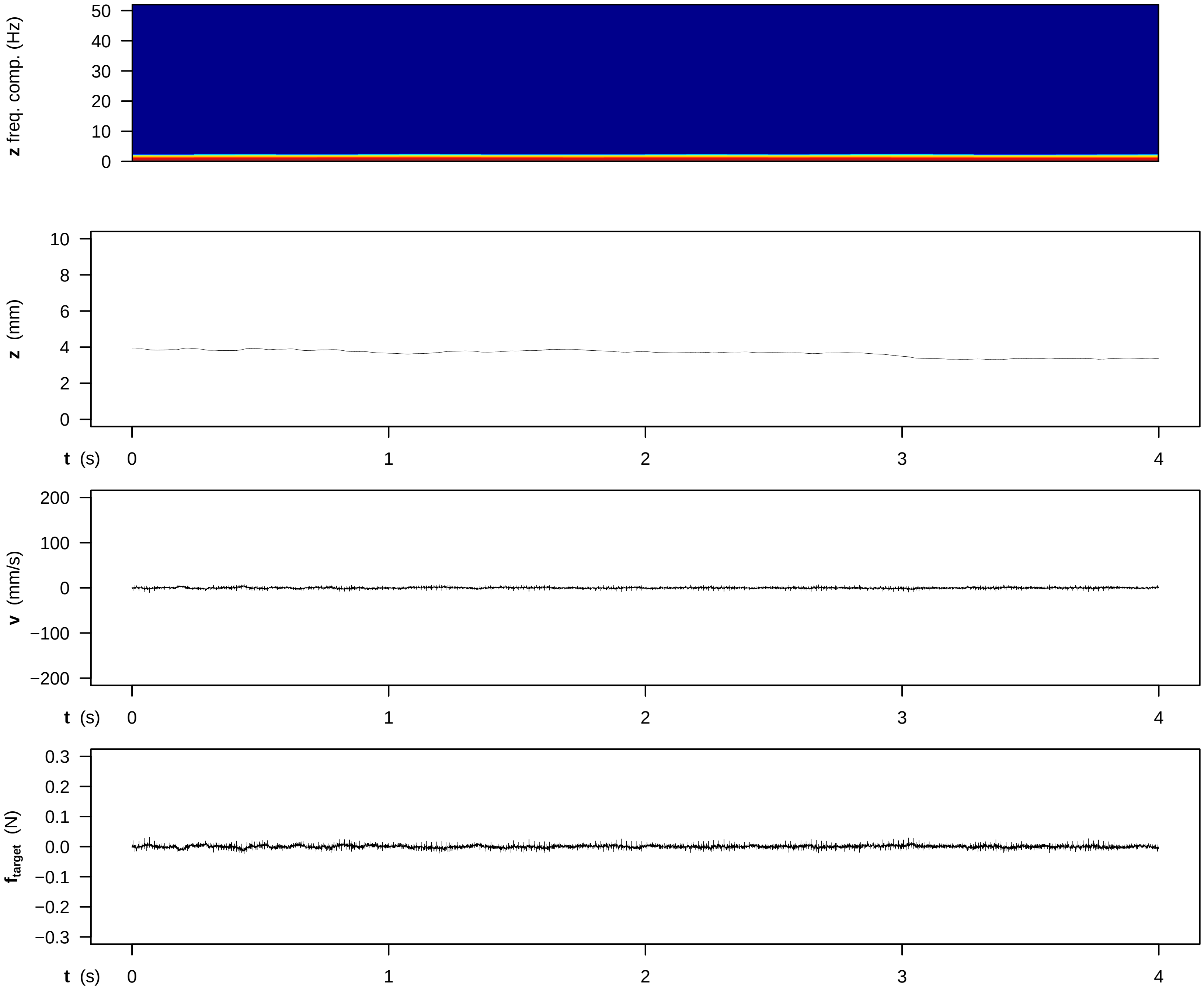

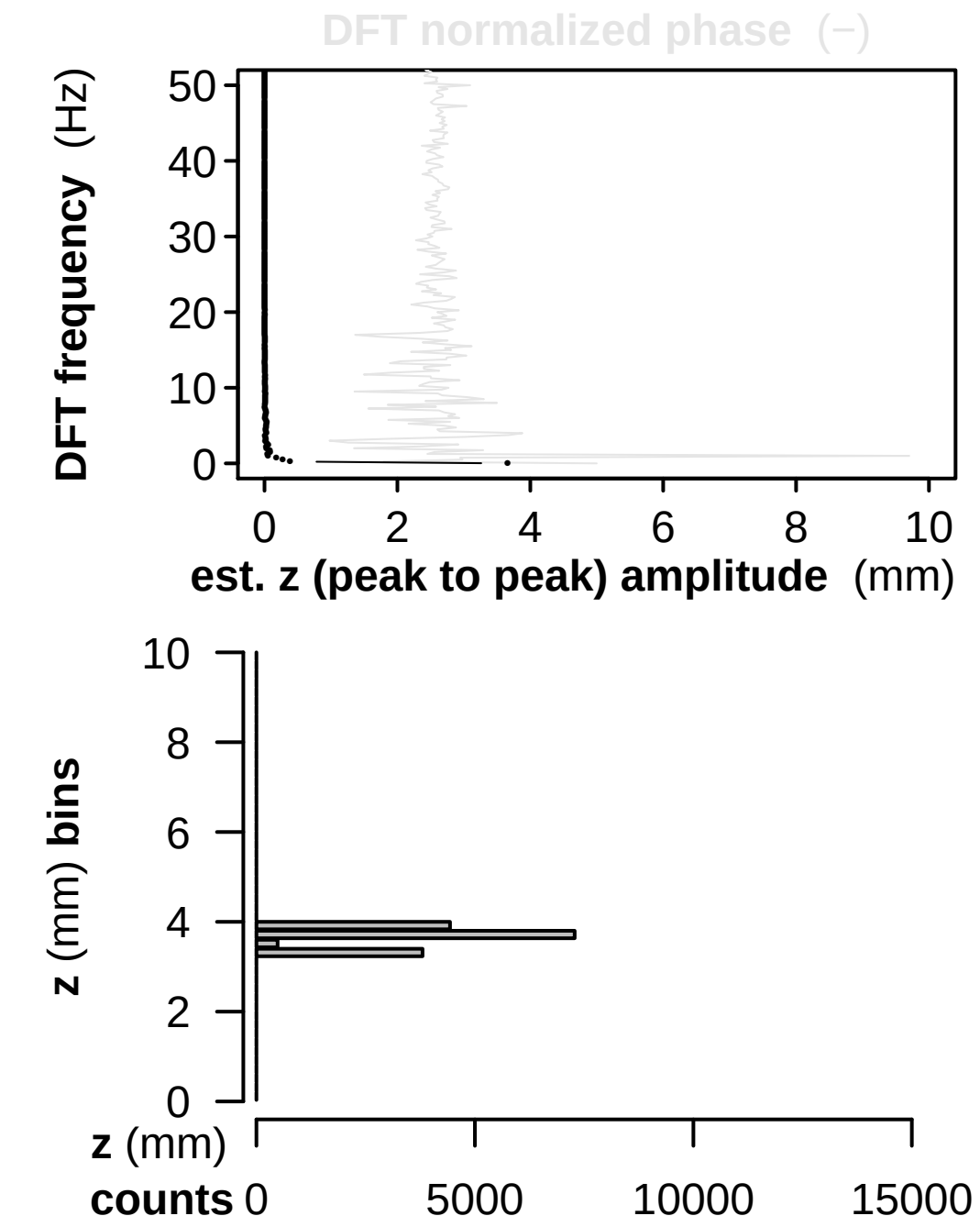

**SUBJECT 1 − RUN 13 − CONDITION 3,1**
**SC_180323_104655_0.AIFF**

```
z_min               : 3.30 mm
z_max               : 3.95 mm
z_travel_amplitude  : 0.65 mm

avg_abs_z_travel    : 3.49 mm/s

z_jarque-bera_jb    : 1685.77
z_jarque-bera_p     : 0.00e+00

z_lin_mod_est_slope: −0.14 mm/s
z_lin_mod_adj_R²    : 77 %

z_poly40_mod_adj_R²: 98 %

z_dft_ampl_thresh   : 0.010 mm
>=threshold_maxfreq: 11.75 Hz
```

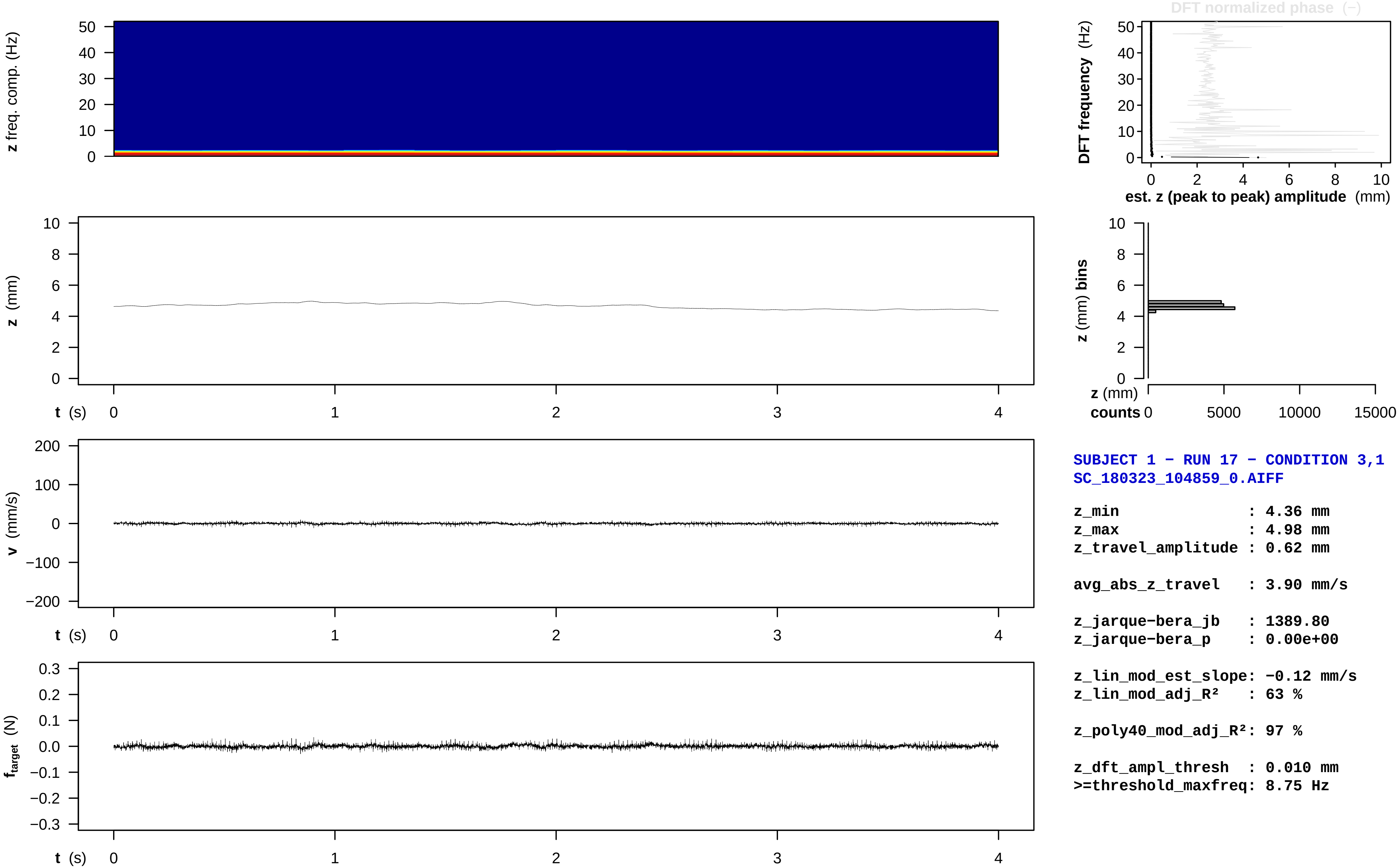

**SUBJECT 1 − RUN 17 − CONDITION 3,1**
**SC_180323_104859_0.AIFF**

```
z_min               : 4.36 mm
z_max               : 4.98 mm
z_travel_amplitude  : 0.62 mm

avg_abs_z_travel    : 3.90 mm/s

z_jarque-bera_jb    : 1389.80
z_jarque-bera_p     : 0.00e+00

z_lin_mod_est_slope: −0.12 mm/s
z_lin_mod_adj_R²    : 63 %

z_poly40_mod_adj_R²: 97 %

z_dft_ampl_thresh   : 0.010 mm
>=threshold_maxfreq: 8.75 Hz
```

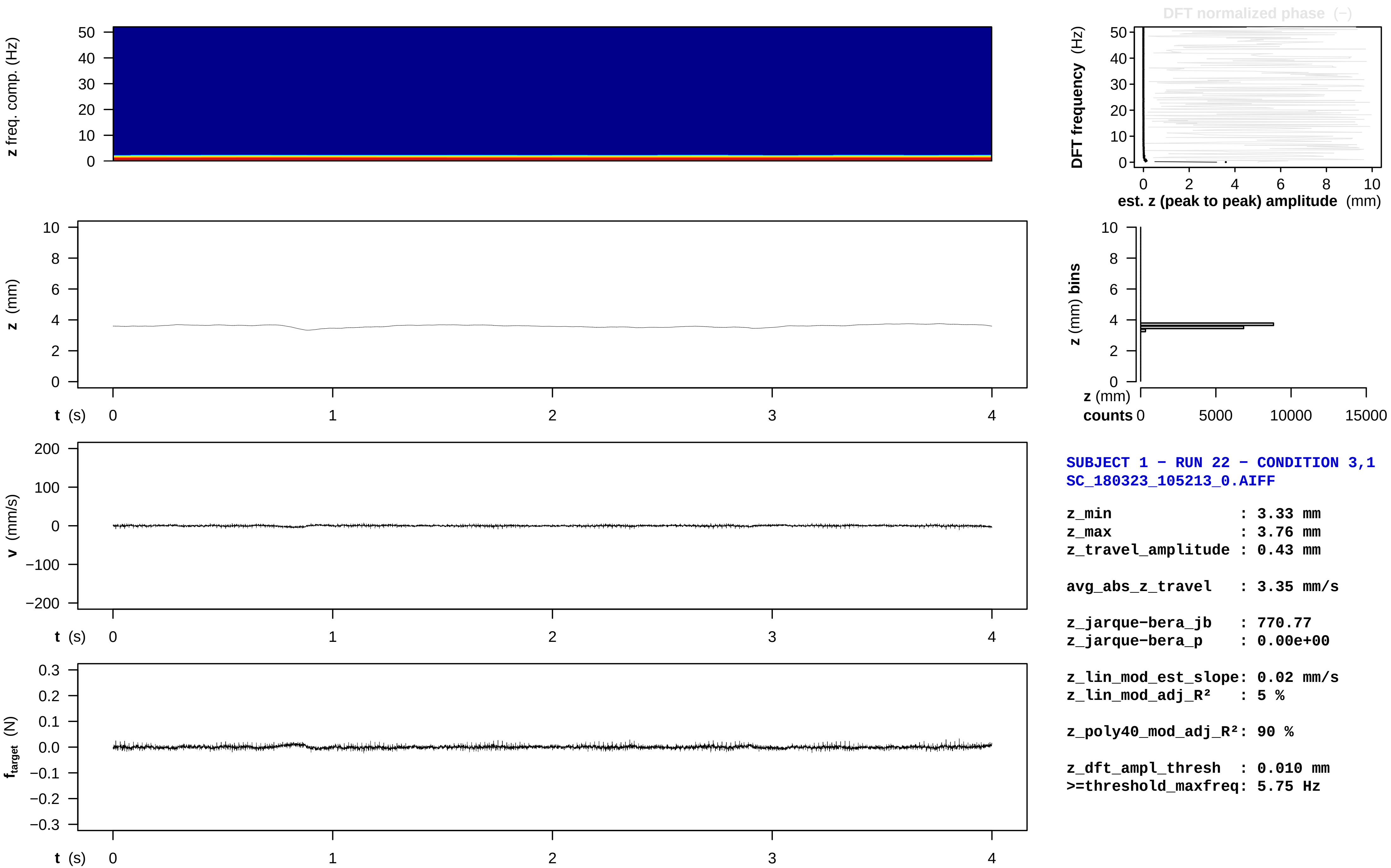

**SUBJECT 1 − RUN 22 − CONDITION 3,1**
**SC_180323_105213_0.AIFF**

```
z_min               : 3.33 mm
z_max               : 3.76 mm
z_travel_amplitude  : 0.43 mm

avg_abs_z_travel    : 3.35 mm/s

z_jarque-bera_jb    : 770.77
z_jarque-bera_p     : 0.00e+00

z_lin_mod_est_slope: 0.02 mm/s
z_lin_mod_adj_R²    : 5 %

z_poly40_mod_adj_R²: 90 %

z_dft_ampl_thresh   : 0.010 mm
>=threshold_maxfreq: 5.75 Hz
```

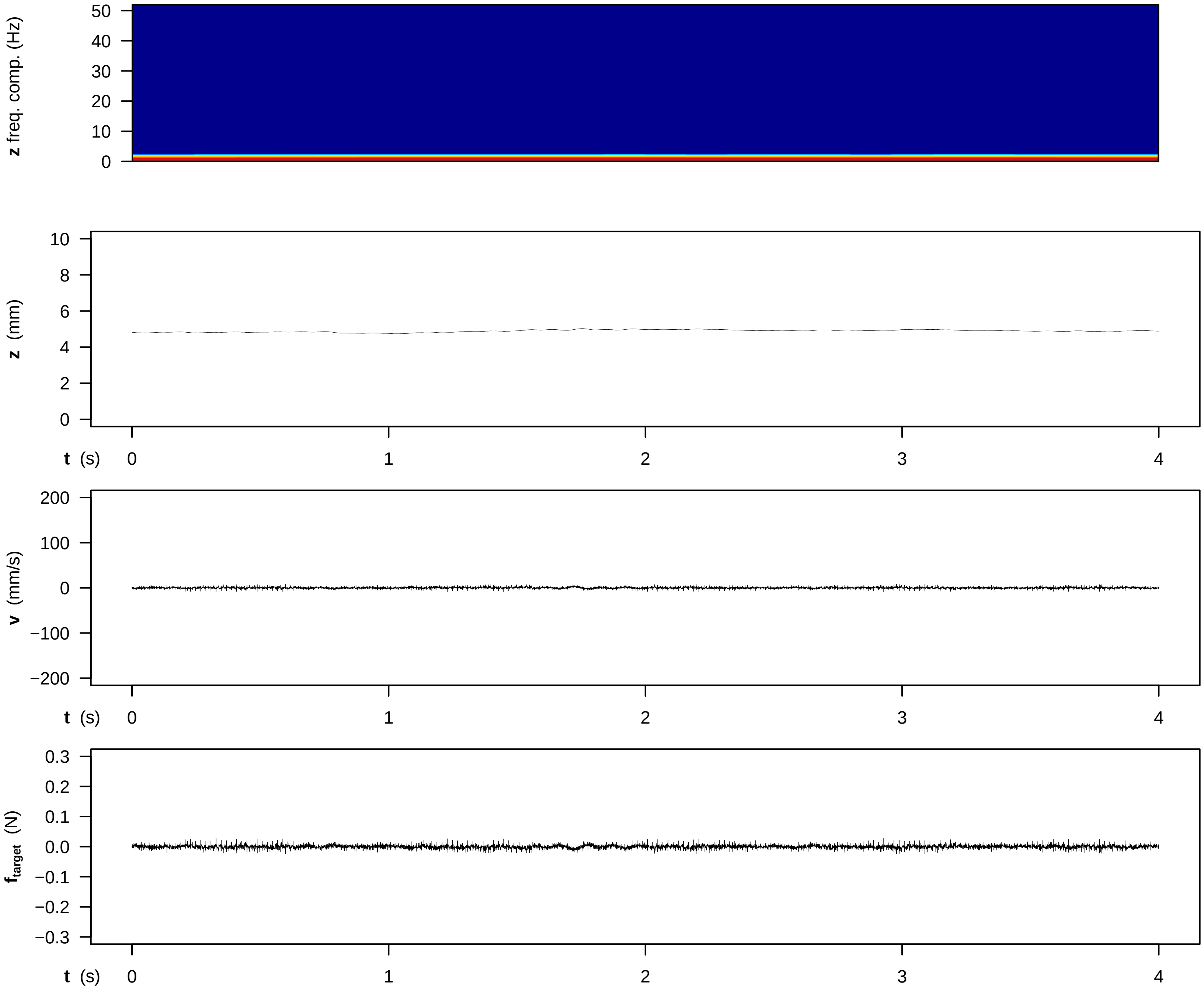

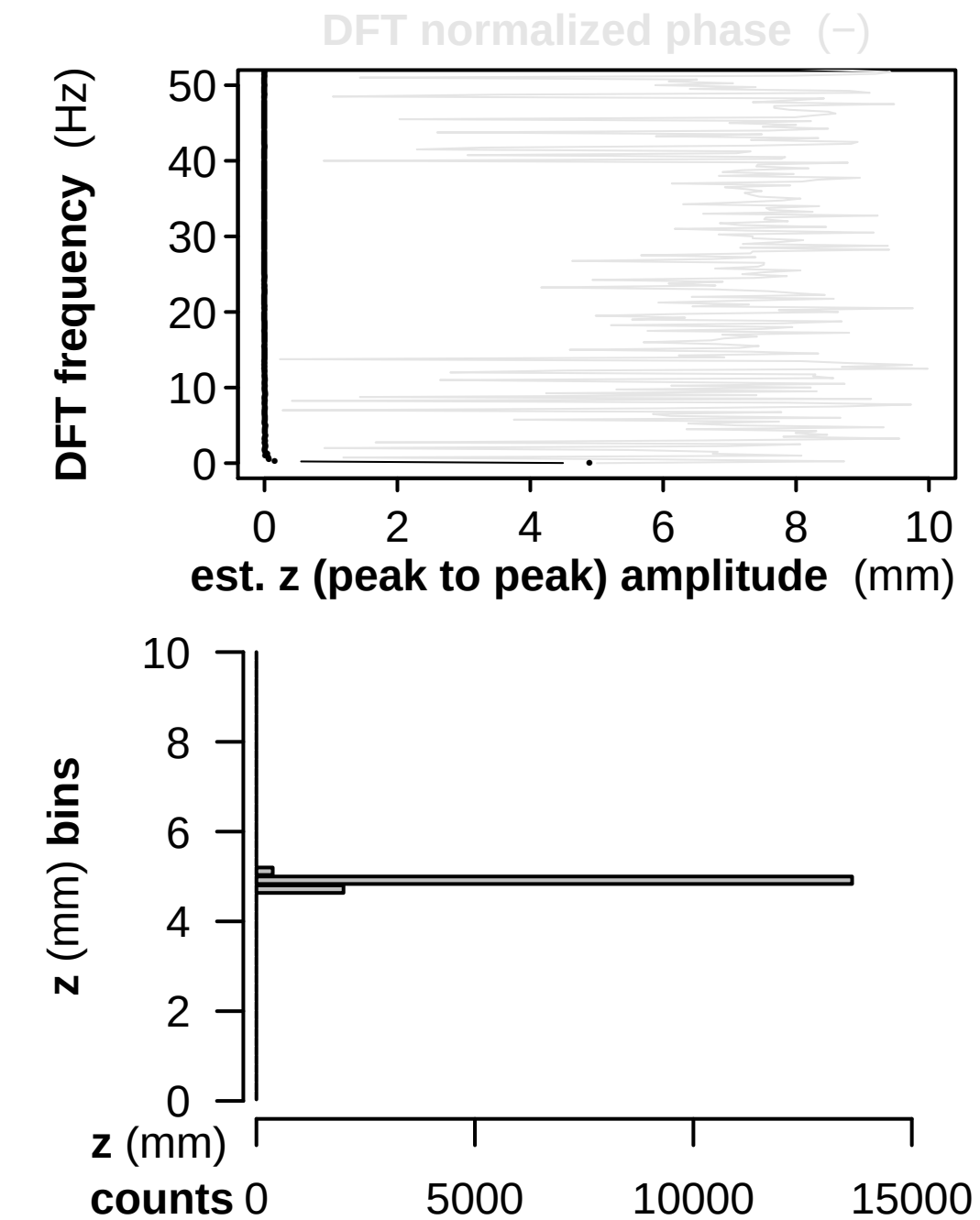

```
SUBJECT 2 − RUN 01 − CONDITION 3,1
SC_180323_111529_0.AIFF

z_min                : 4.74 mm
z_max                : 5.03 mm
z_travel_amplitude : 0.29 mm

avg_abs_z_travel     : 2.01 mm/s

z_jarque-bera_jb     : 686.81
z_jarque-bera_p      : 0.00e+00

z_lin_mod_est_slope: 0.03 mm/s
z_lin_mod_adj_R²     : 32 %

z_poly40_mod_adj_R²: 95 %

z_dft_ampl_thresh    : 0.010 mm
>=threshold_maxfreq: 9.25 Hz
```

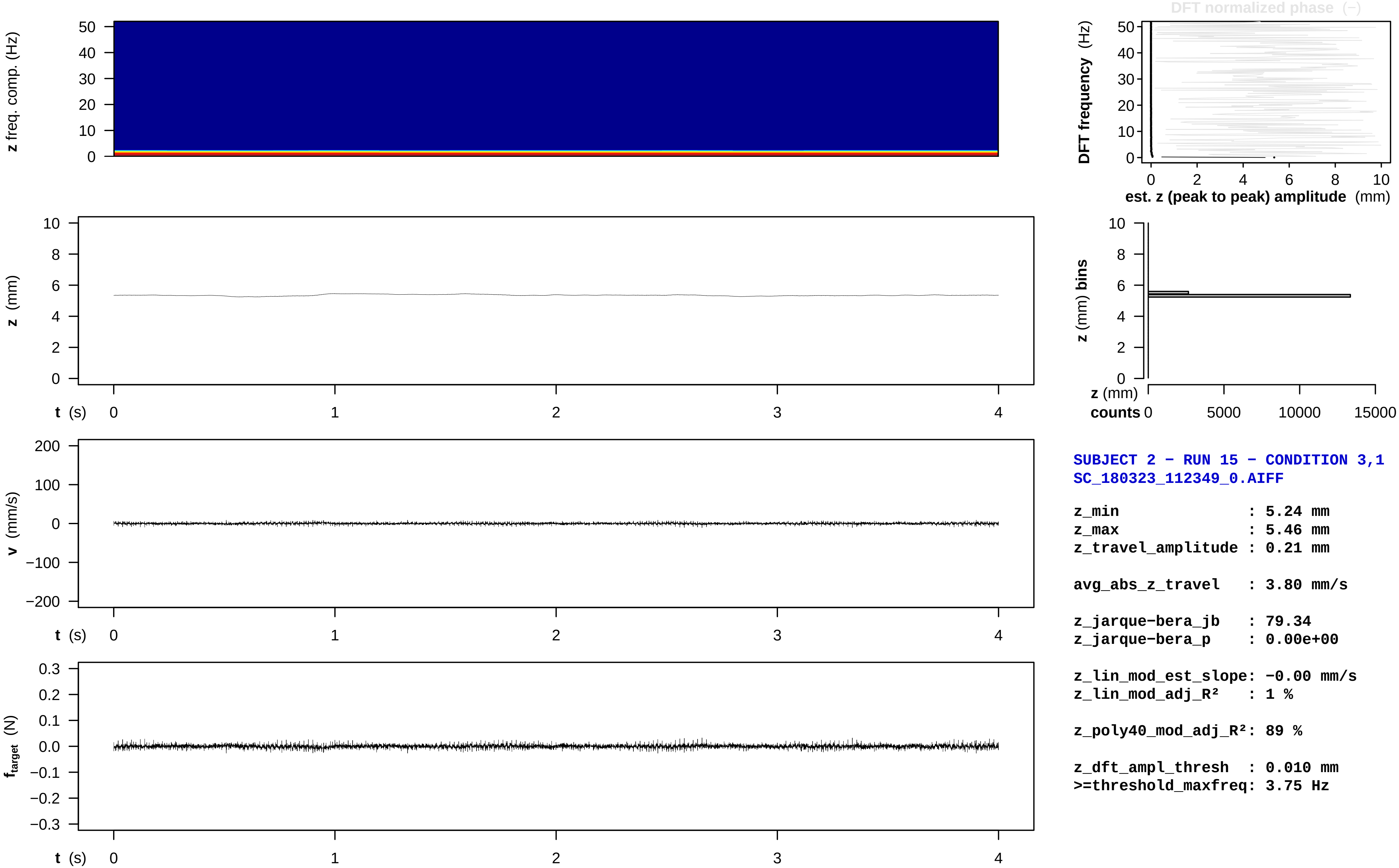

**SUBJECT 2 − RUN 15 − CONDITION 3,1**
**SC_180323_112349_0.AIFF**

```
z_min              : 5.24 mm
z_max              : 5.46 mm
z_travel_amplitude : 0.21 mm

avg_abs_z_travel   : 3.80 mm/s

z_jarque-bera_jb   : 79.34
z_jarque-bera_p    : 0.00e+00

z_lin_mod_est_slope: −0.00 mm/s
z_lin_mod_adj_R²   : 1 %

z_poly40_mod_adj_R²: 89 %

z_dft_ampl_thresh  : 0.010 mm
>=threshold_maxfreq: 3.75 Hz
```

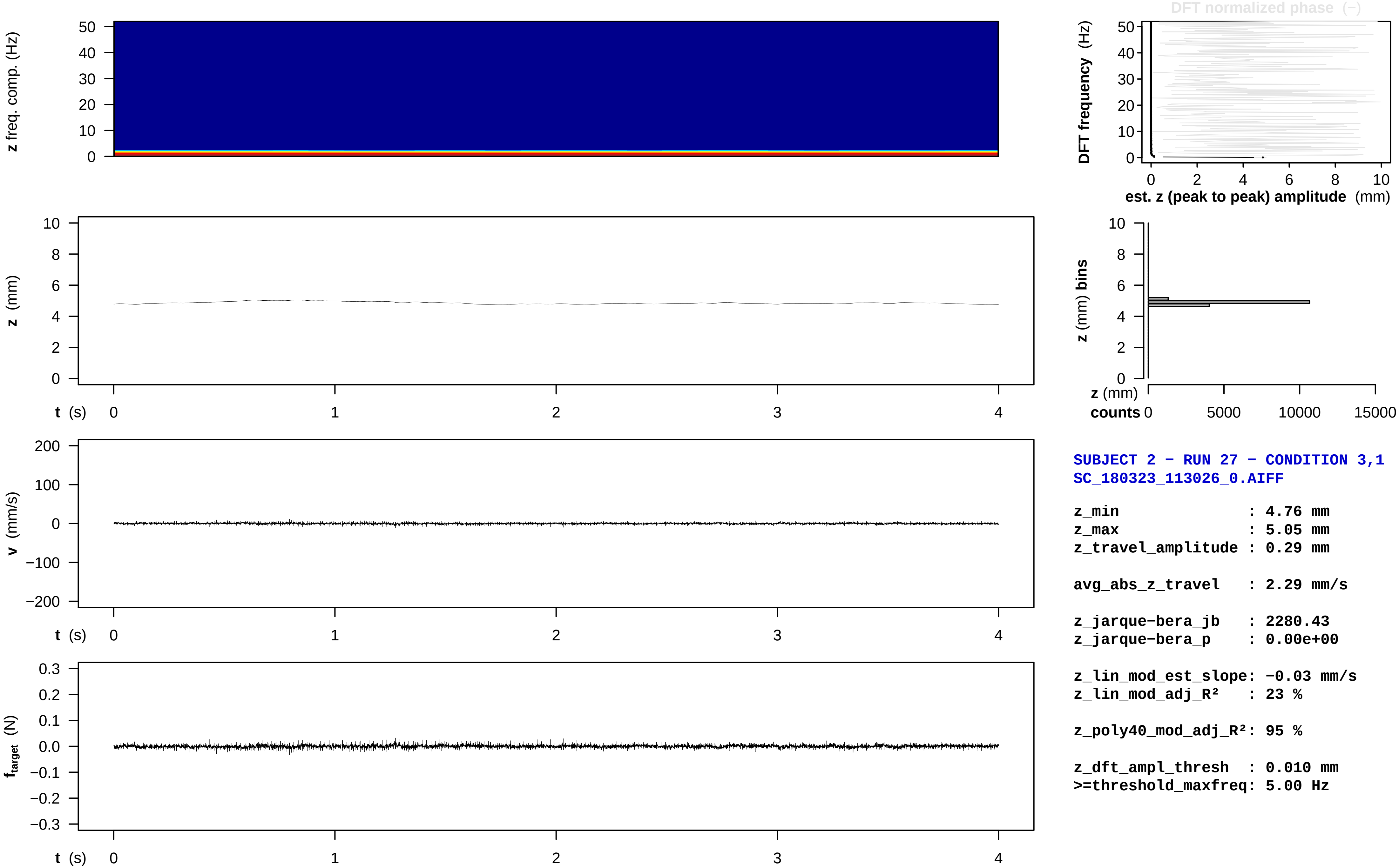

**SUBJECT 2 − RUN 27 − CONDITION 3,1**
**SC_180323_113026_0.AIFF**

```
z_min              : 4.76 mm
z_max              : 5.05 mm
z_travel_amplitude : 0.29 mm

avg_abs_z_travel   : 2.29 mm/s

z_jarque-bera_jb   : 2280.43
z_jarque-bera_p    : 0.00e+00

z_lin_mod_est_slope: −0.03 mm/s
z_lin_mod_adj_R²   : 23 %

z_poly40_mod_adj_R²: 95 %

z_dft_ampl_thresh  : 0.010 mm
>=threshold_maxfreq: 5.00 Hz
```

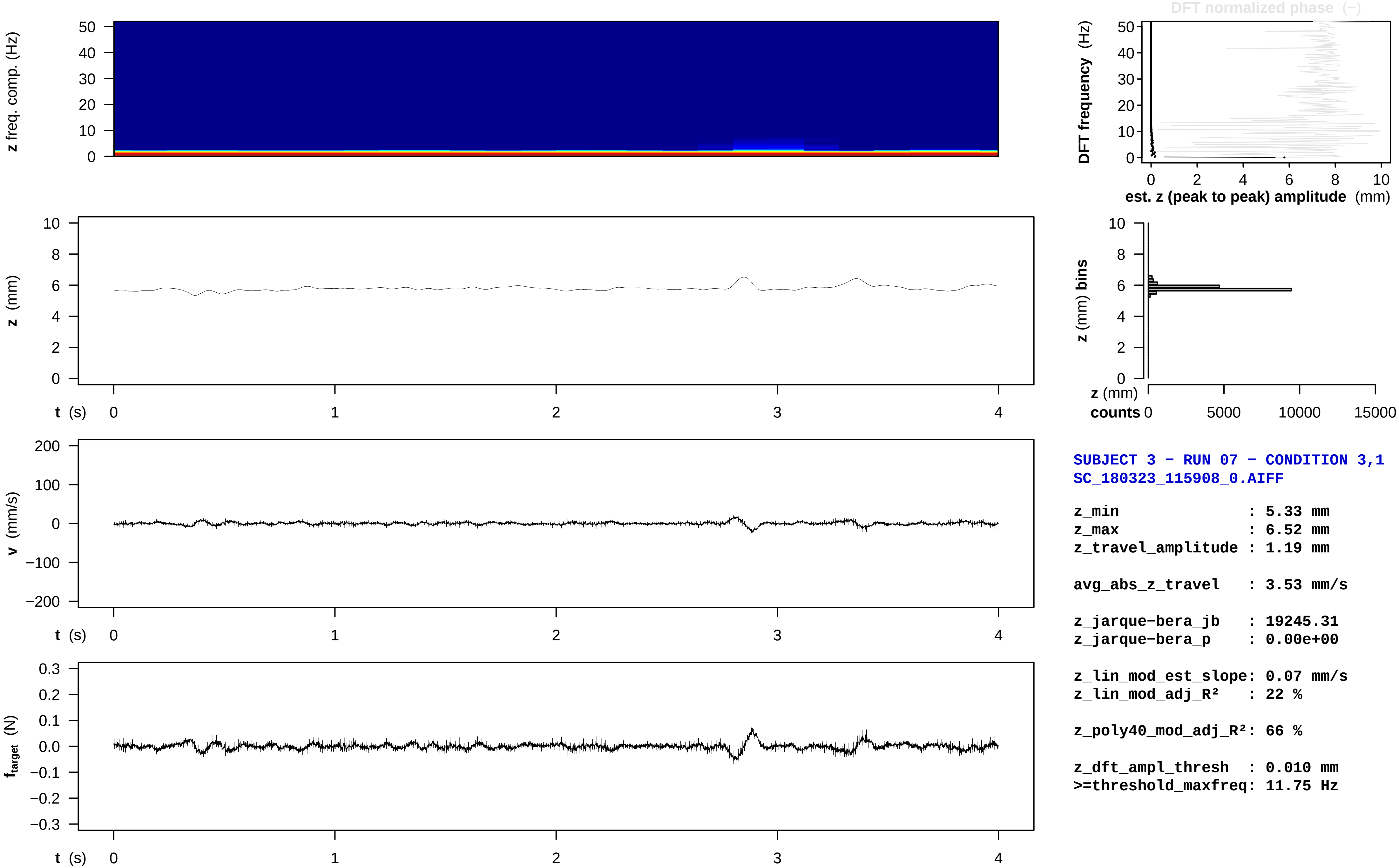

**SUBJECT 3 − RUN 07 − CONDITION 3,1**
**SC_180323_115908_0.AIFF**

```
z_min               : 5.33 mm
z_max               : 6.52 mm
z_travel_amplitude  : 1.19 mm

avg_abs_z_travel    : 3.53 mm/s

z_jarque-bera_jb    : 19245.31
z_jarque-bera_p     : 0.00e+00

z_lin_mod_est_slope: 0.07 mm/s
z_lin_mod_adj_R²    : 22 %

z_poly40_mod_adj_R²: 66 %

z_dft_ampl_thresh   : 0.010 mm
>=threshold_maxfreq: 11.75 Hz
```

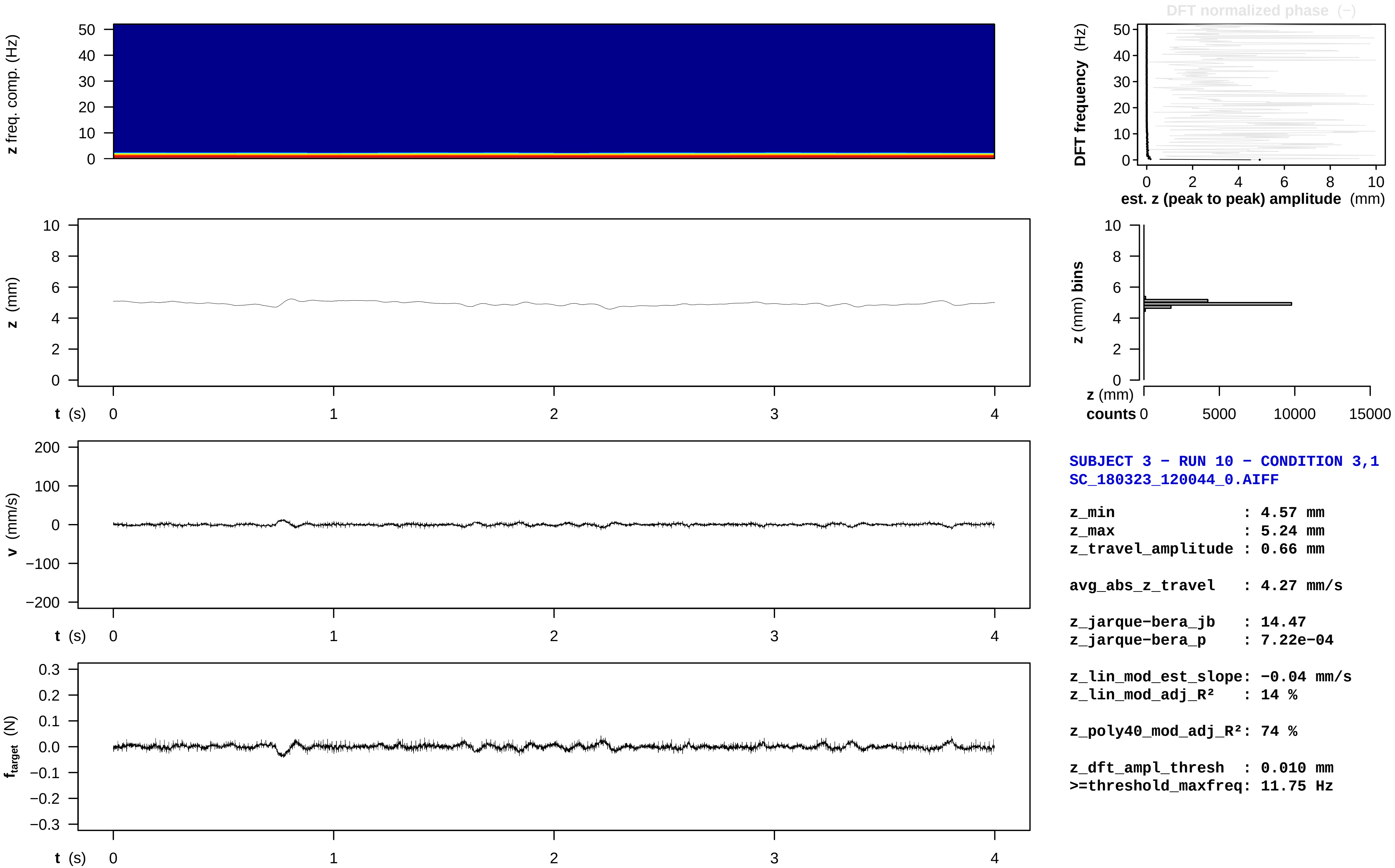

**SUBJECT 3 − RUN 10 − CONDITION 3,1**
**SC_180323_120044_0.AIFF**

```
z_min               : 4.57 mm
z_max               : 5.24 mm
z_travel_amplitude  : 0.66 mm

avg_abs_z_travel    : 4.27 mm/s

z_jarque-bera_jb    : 14.47
z_jarque-bera_p     : 7.22e-04

z_lin_mod_est_slope: -0.04 mm/s
z_lin_mod_adj_R²    : 14 %

z_poly40_mod_adj_R²: 74 %

z_dft_ampl_thresh   : 0.010 mm
>=threshold_maxfreq: 11.75 Hz
```

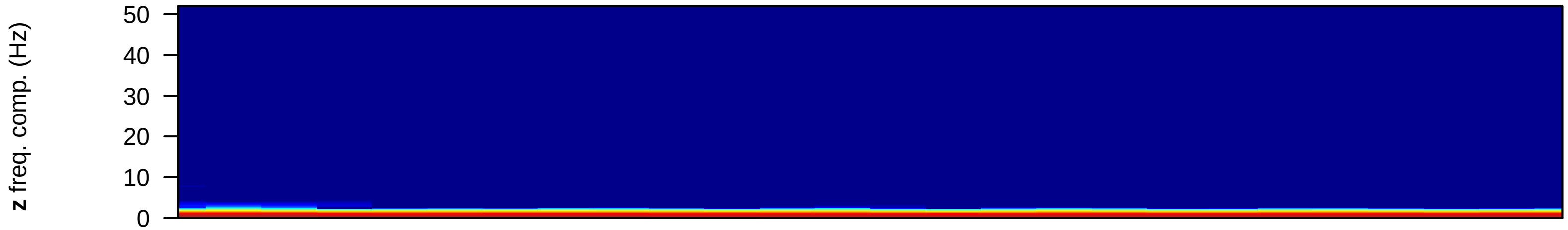

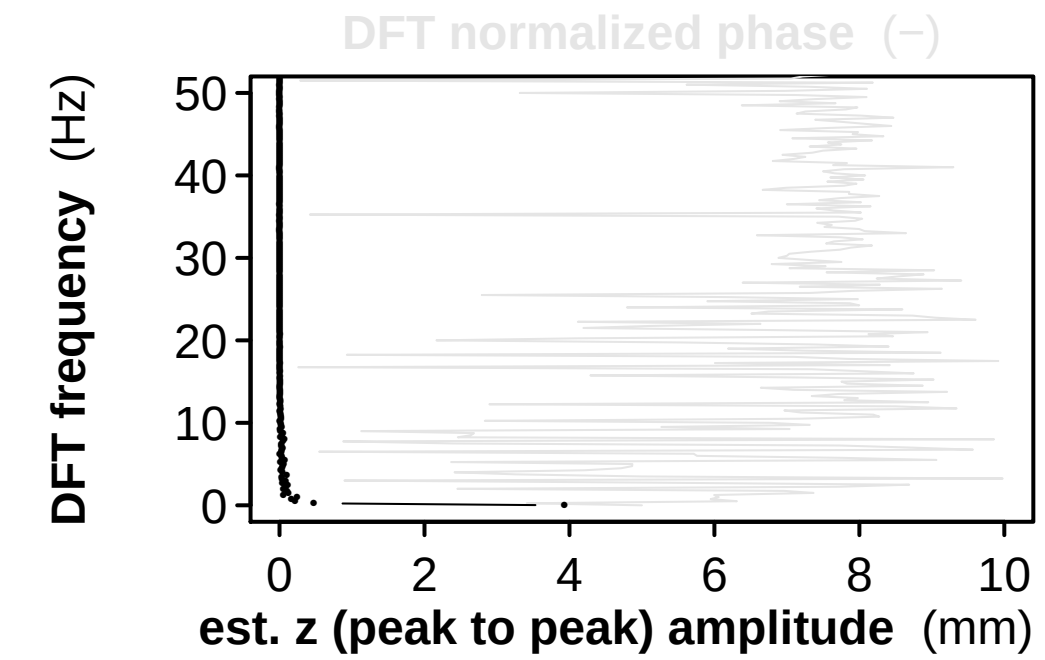

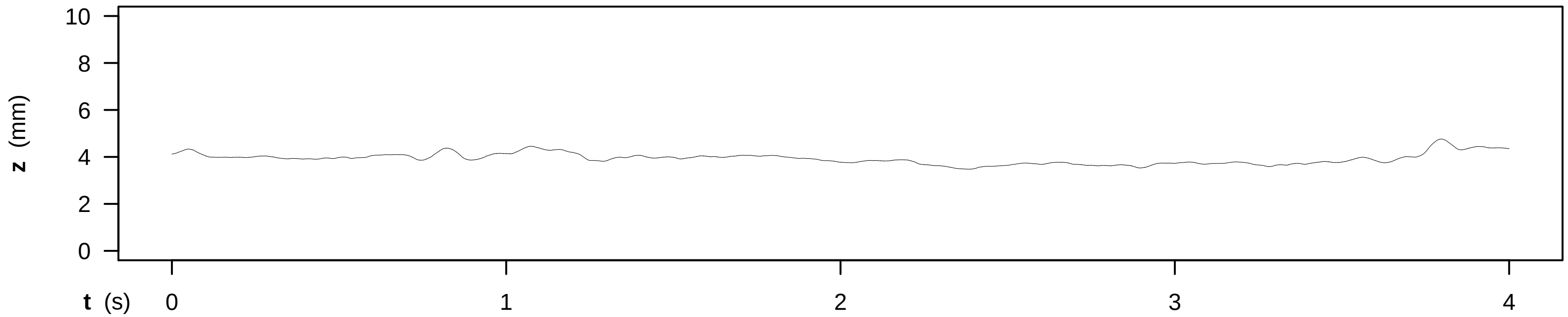

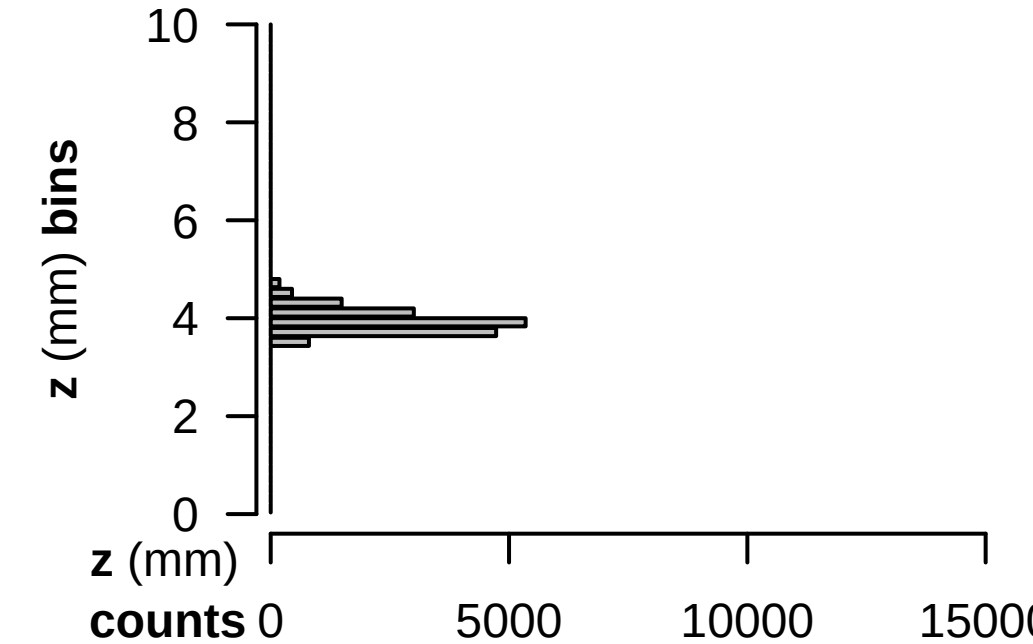

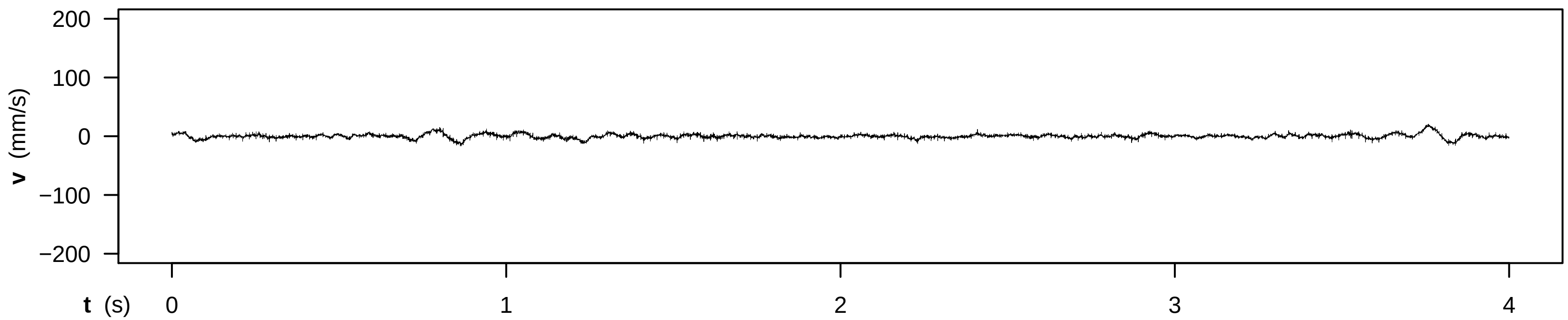

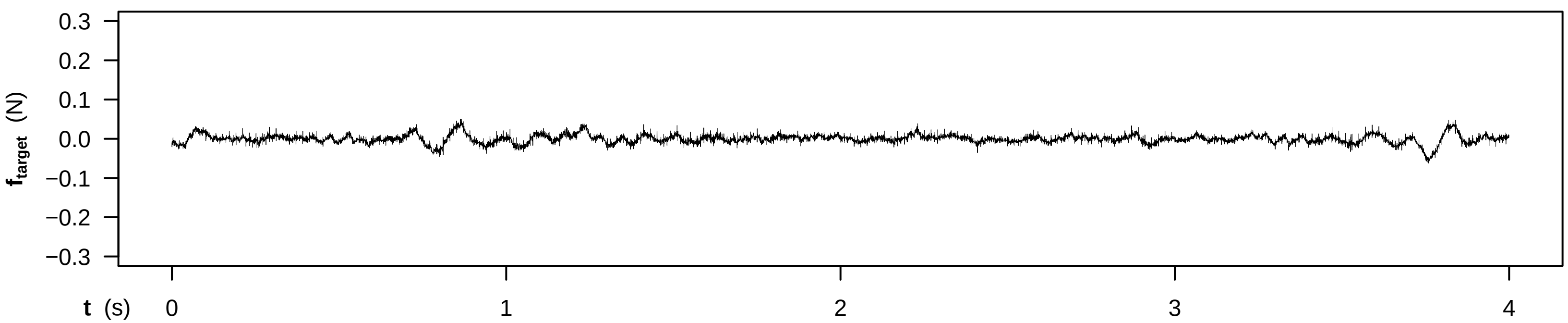

**SUBJECT 3 − RUN 24 − CONDITION 3,1**
**SC_180323_120908_0.AIFF**

```
z_min               : 3.48 mm
z_max               : 4.76 mm
z_travel_amplitude  : 1.29 mm

avg_abs_z_travel    : 4.46 mm/s

z_jarque-bera_jb    : 1360.40
z_jarque-bera_p     : 0.00e+00

z_lin_mod_est_slope: −0.05 mm/s
z_lin_mod_adj_R²    : 6 %

z_poly40_mod_adj_R²: 85 %

z_dft_ampl_thresh   : 0.010 mm
>=threshold_maxfreq: 12.75 Hz
```

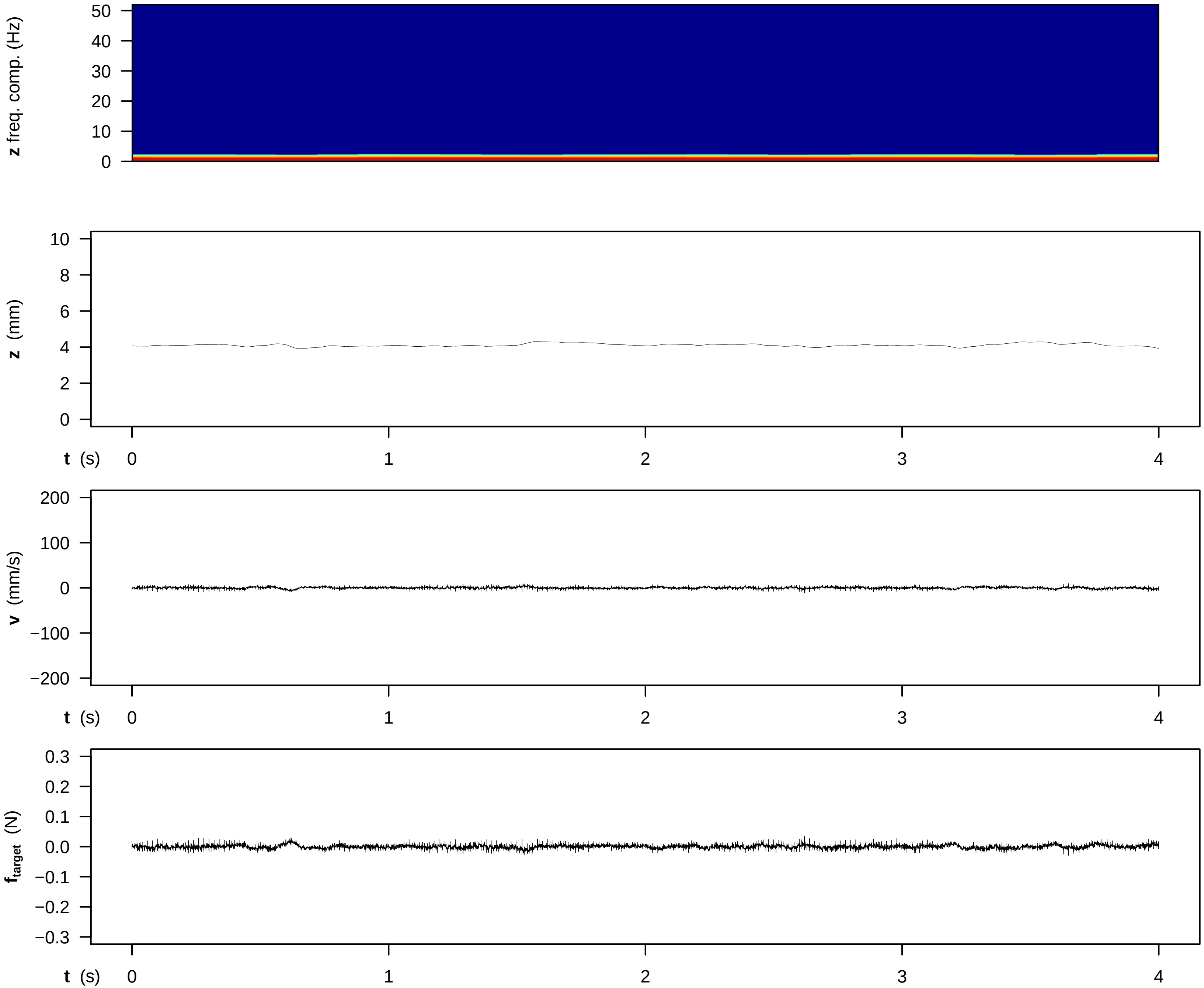

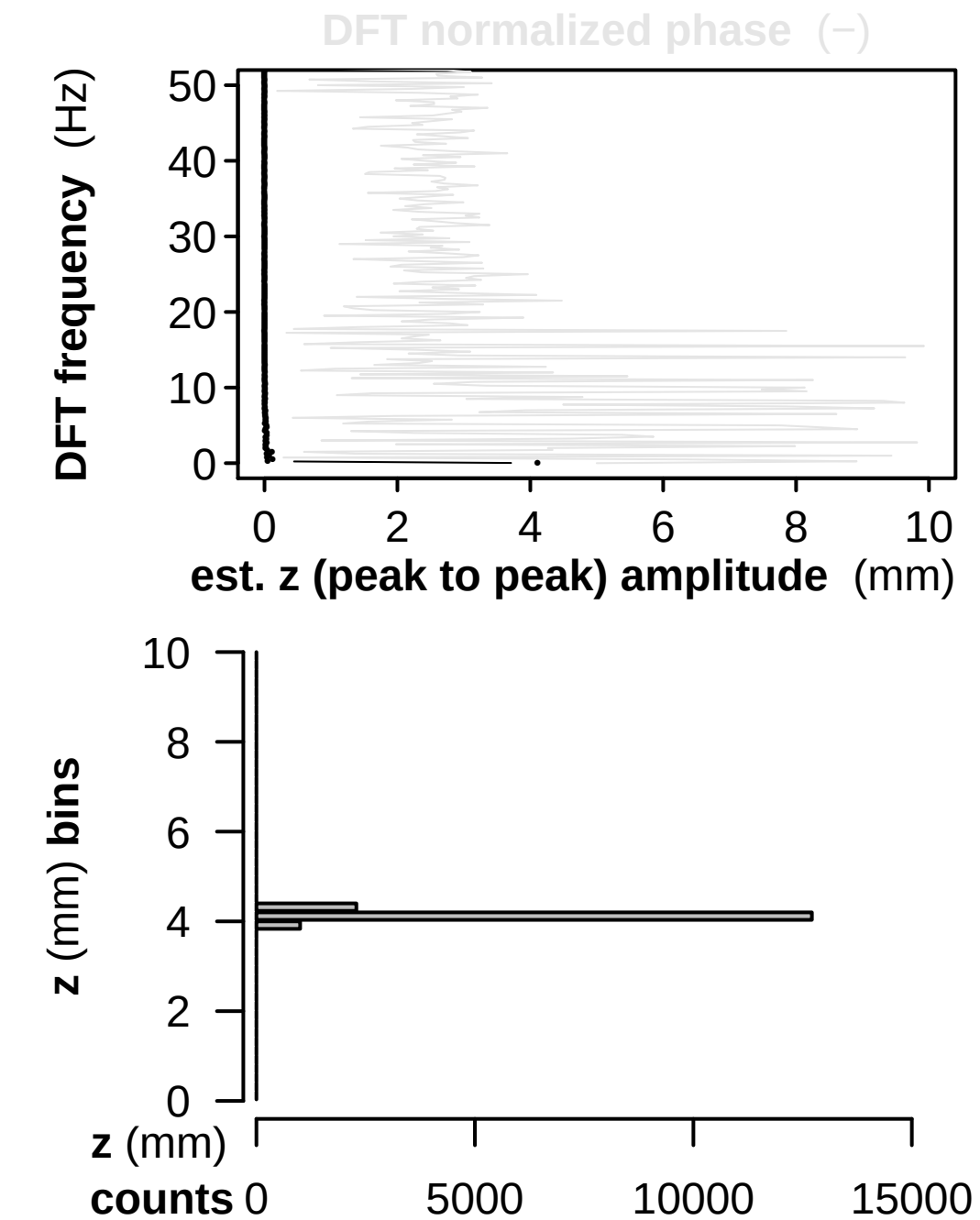

**SUBJECT 4 − RUN 12 − CONDITION 3,1**
**SC_180323_123633_0.AIFF**

```
z_min               : 3.91 mm
z_max               : 4.31 mm
z_travel_amplitude  : 0.40 mm

avg_abs_z_travel    : 2.98 mm/s

z_jarque-bera_jb    : 628.64
z_jarque-bera_p     : 0.00e+00

z_lin_mod_est_slope: 0.01 mm/s
z_lin_mod_adj_R²    : 4 %

z_poly40_mod_adj_R²: 79 %

z_dft_ampl_thresh   : 0.010 mm
>=threshold_maxfreq: 10.50 Hz
```

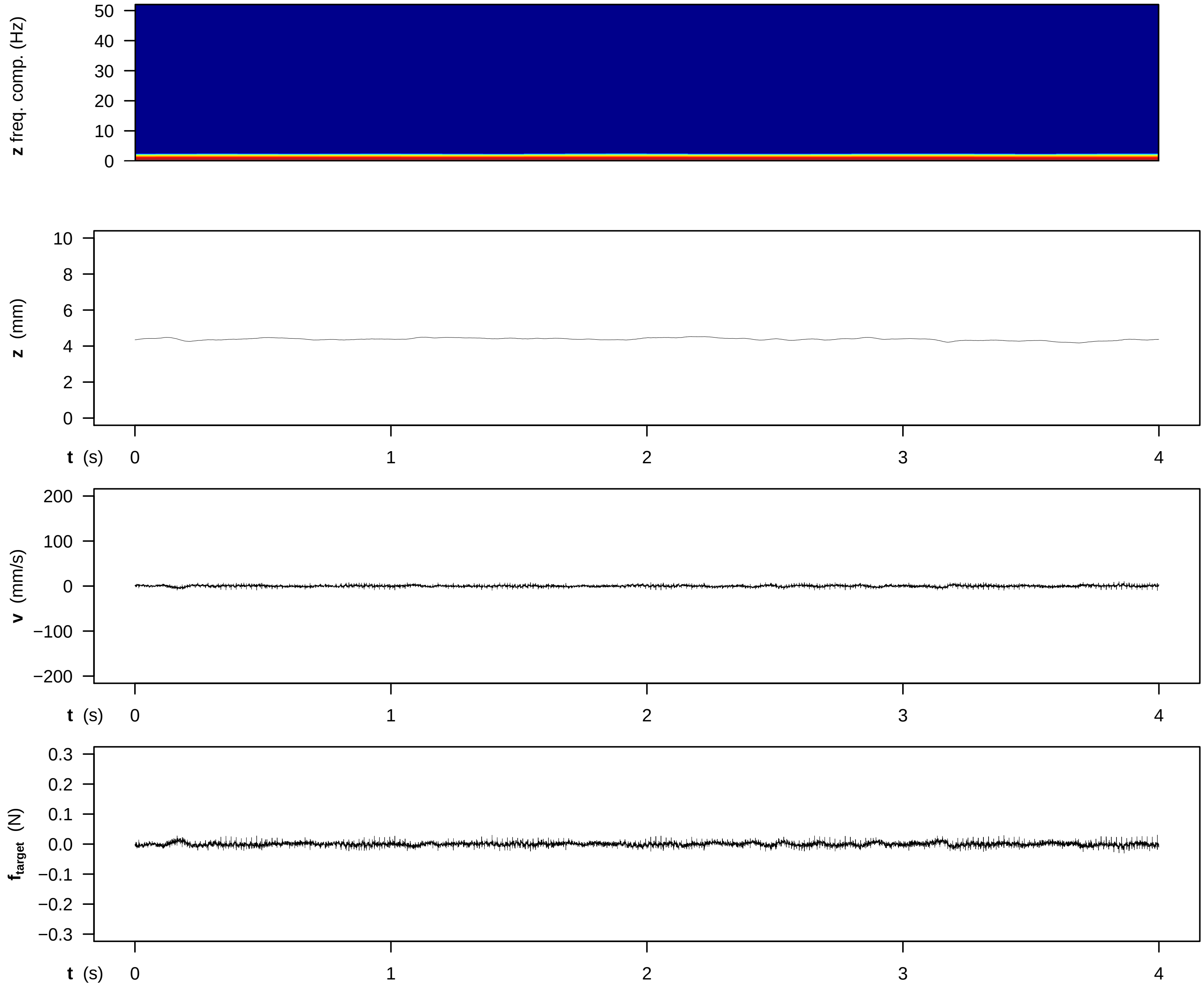

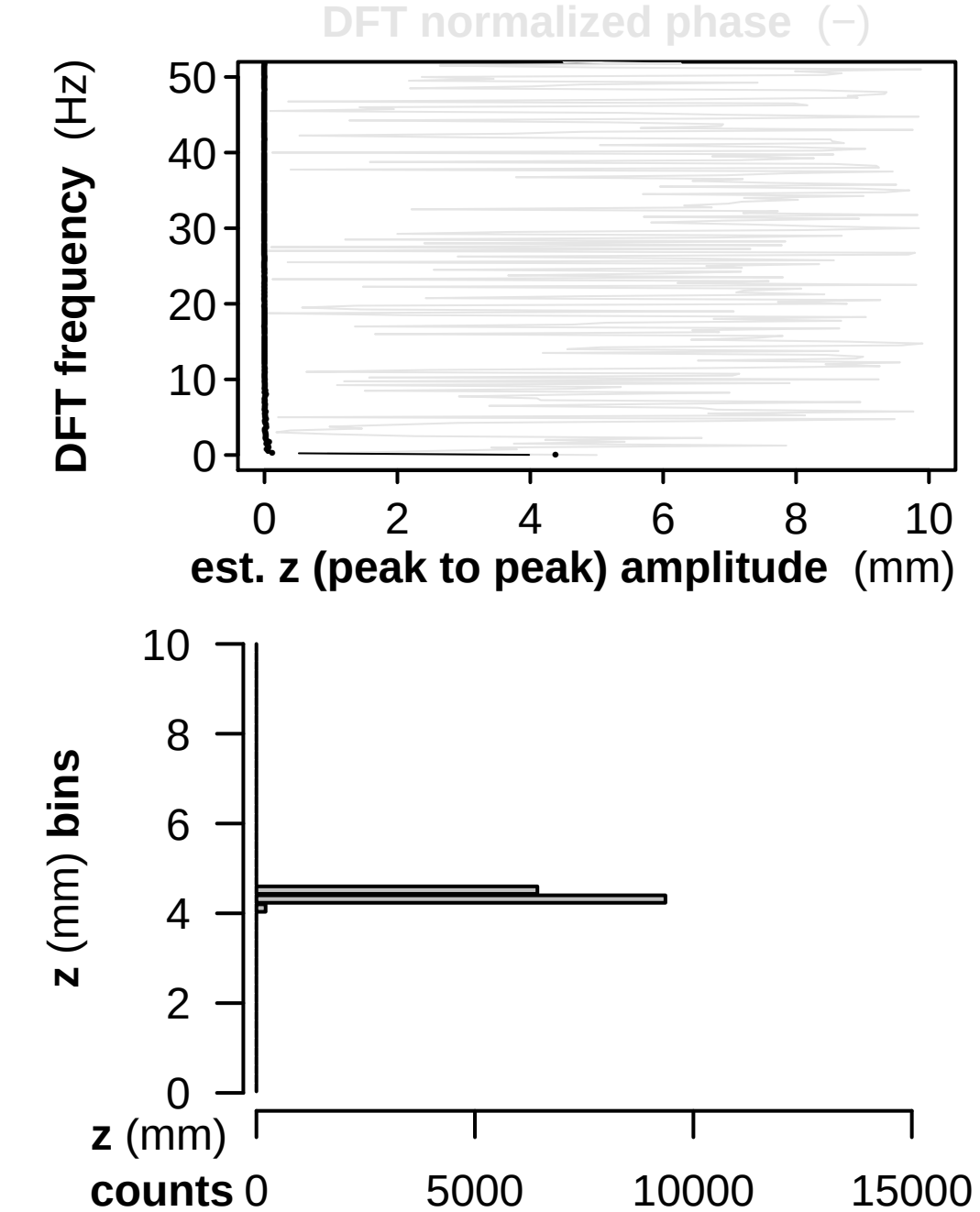

**SUBJECT 4 − RUN 26 − CONDITION 3,1**
**SC_180323_124312_0.AIFF**

```
z_min               : 4.17 mm
z_max               : 4.52 mm
z_travel_amplitude  : 0.35 mm

avg_abs_z_travel    : 2.38 mm/s

z_jarque-bera_jb    : 566.11
z_jarque-bera_p     : 0.00e+00

z_lin_mod_est_slope: −0.03 mm/s
z_lin_mod_adj_R²    : 20 %

z_poly40_mod_adj_R²: 86 %

z_dft_ampl_thresh   : 0.010 mm
>=threshold_maxfreq: 8.50 Hz
```

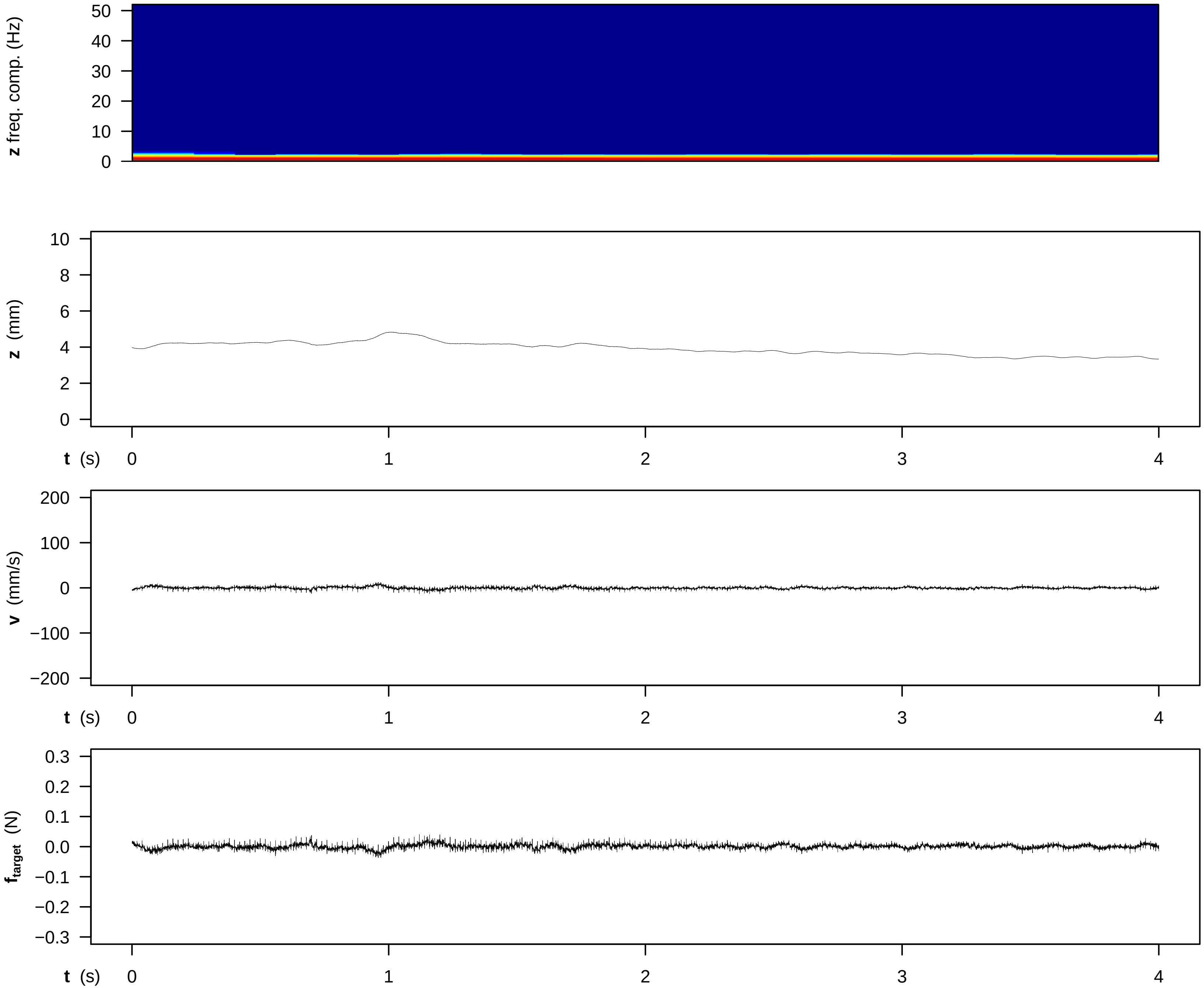

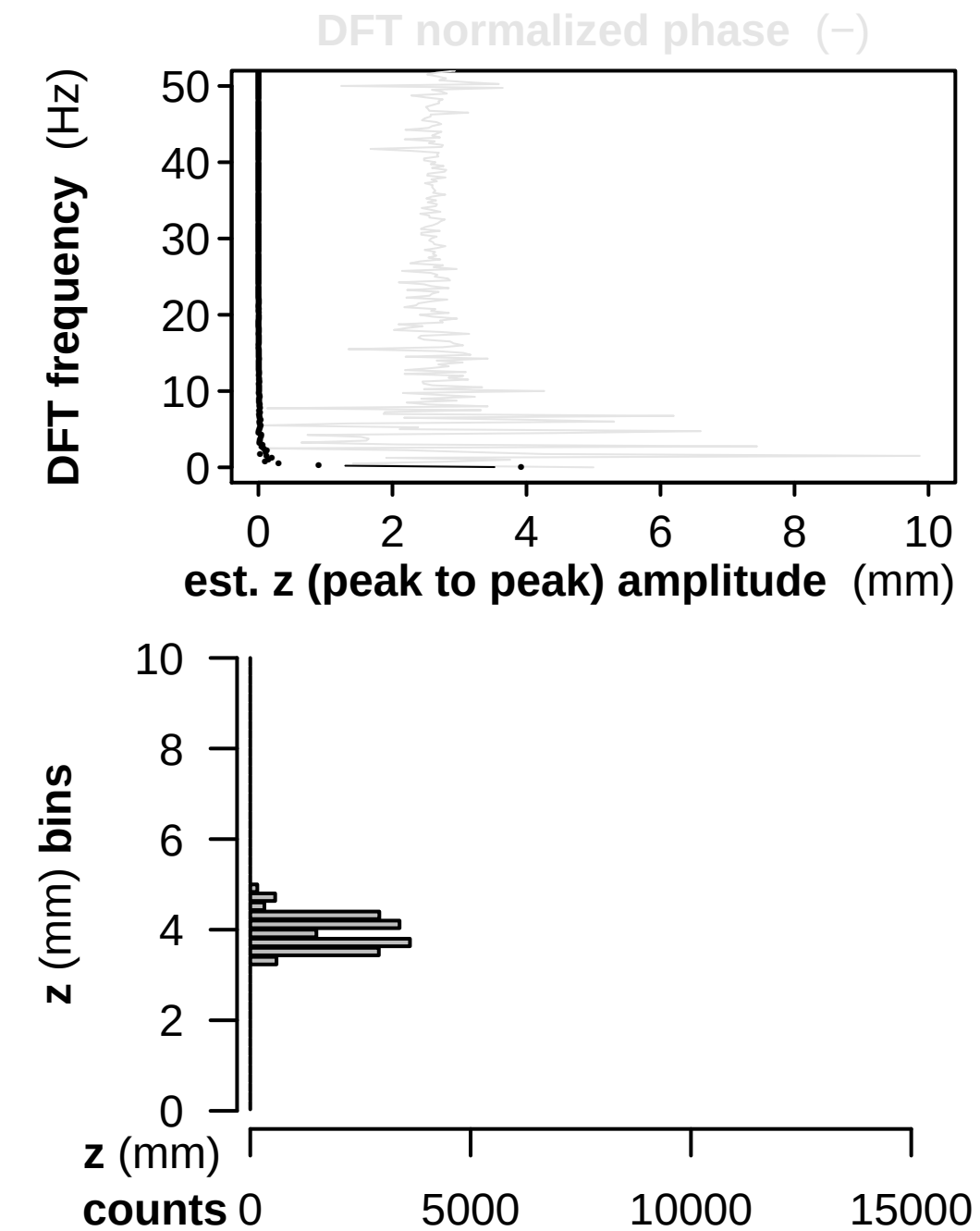

**SUBJECT 4 − RUN 30 − CONDITION 3,1**
**SC_180323_124619_0.AIFF**

```
z_min               : 3.34 mm
z_max               : 4.83 mm
z_travel_amplitude  : 1.49 mm

avg_abs_z_travel    : 3.96 mm/s

z_jarque-bera_jb    : 493.92
z_jarque-bera_p     : 0.00e+00

z_lin_mod_est_slope: −0.27 mm/s
z_lin_mod_adj_R²    : 77 %

z_poly40_mod_adj_R²: 97 %

z_dft_ampl_thresh   : 0.010 mm
>=threshold_maxfreq: 12.25 Hz
```

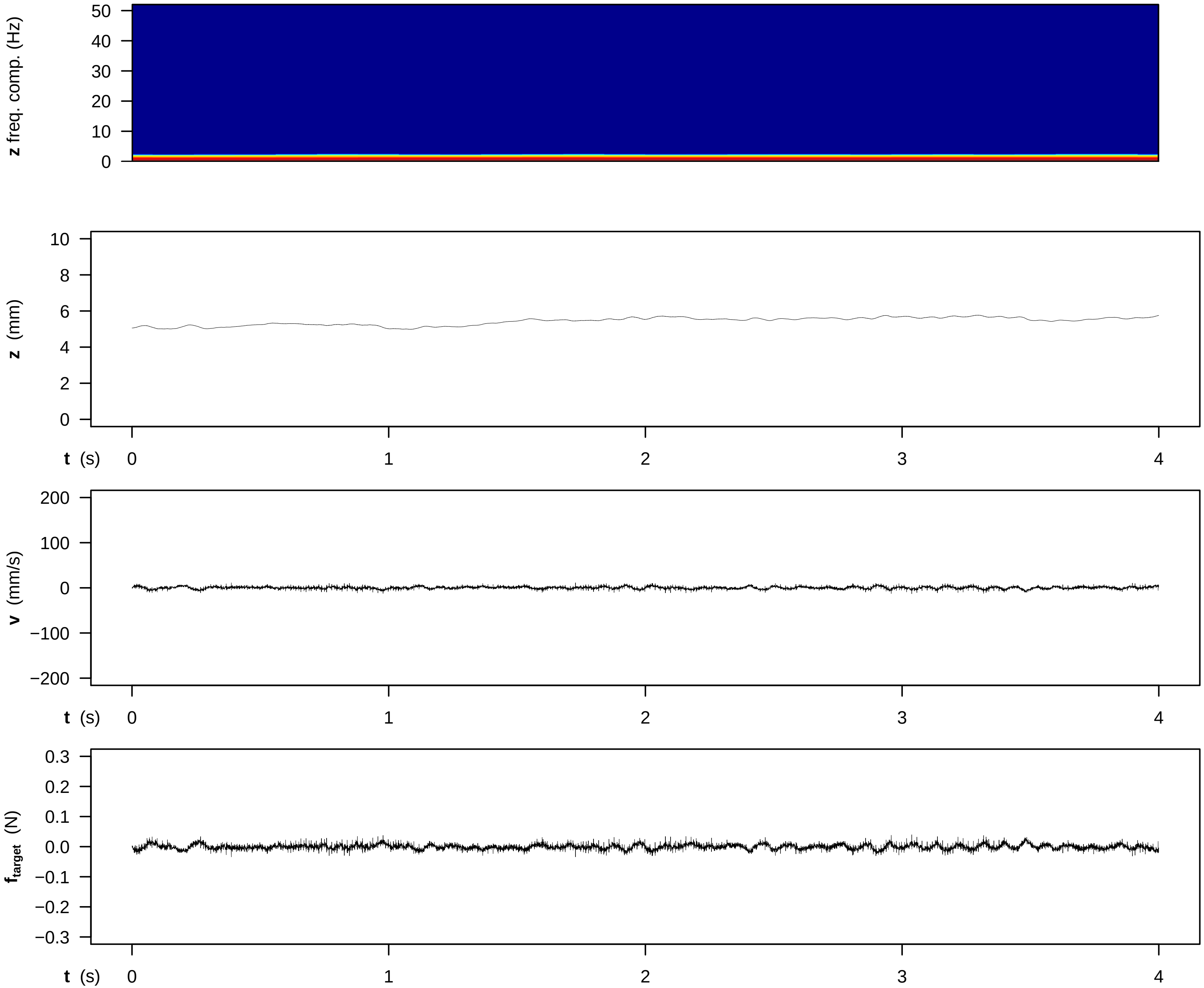

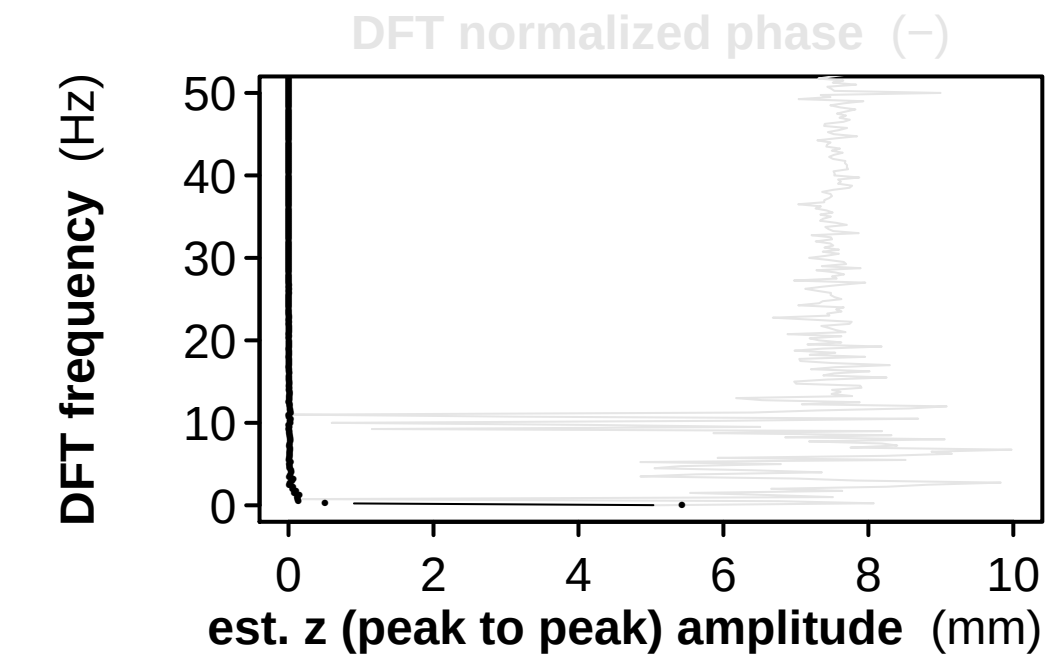

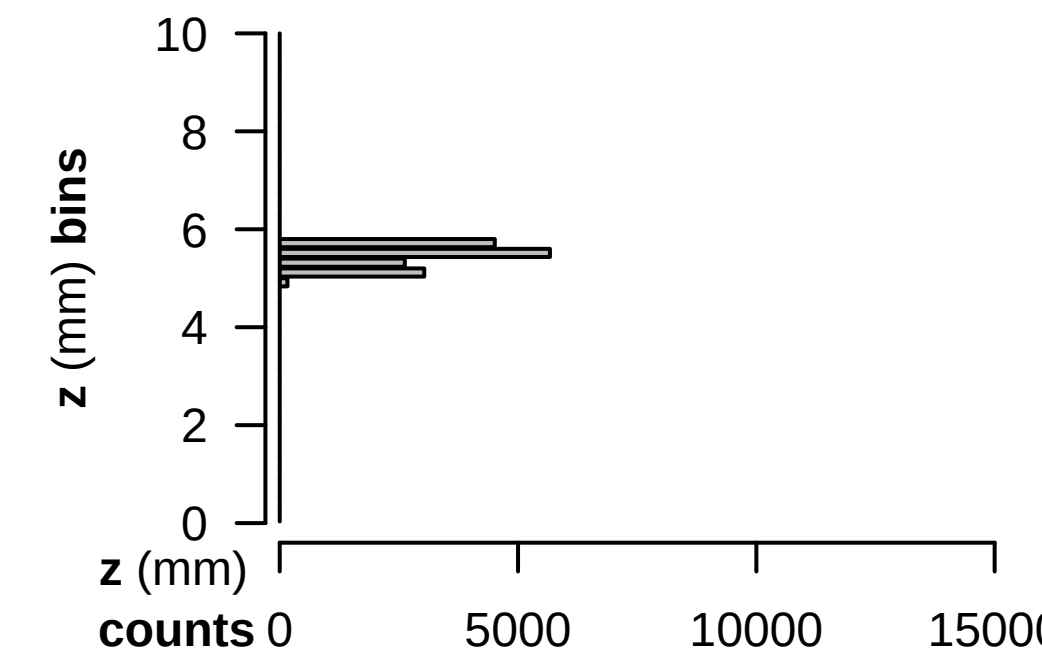

**SUBJECT 5 − RUN 06 − CONDITION 3,1**
**SC_180323_131739_0.AIFF**

```
z_min              : 4.98 mm
z_max              : 5.77 mm
z_travel_amplitude : 0.78 mm

avg_abs_z_travel   : 4.36 mm/s

z_jarque-bera_jb   : 1468.26
z_jarque-bera_p    : 0.00e+00

z_lin_mod_est_slope: 0.16 mm/s
z_lin_mod_adj_R²   : 68 %

z_poly40_mod_adj_R²: 95 %

z_dft_ampl_thresh  : 0.010 mm
>=threshold_maxfreq: 16.00 Hz
```

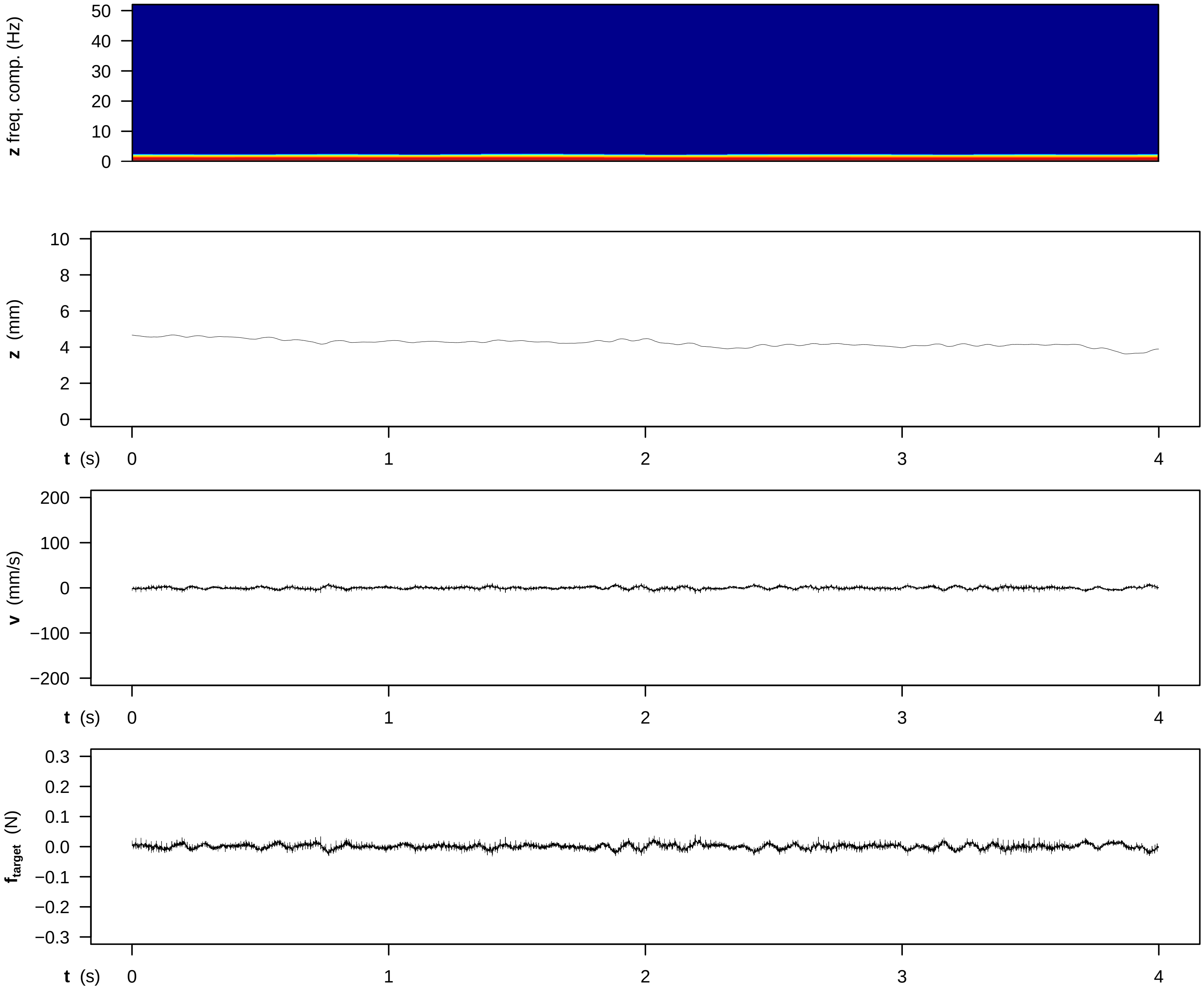

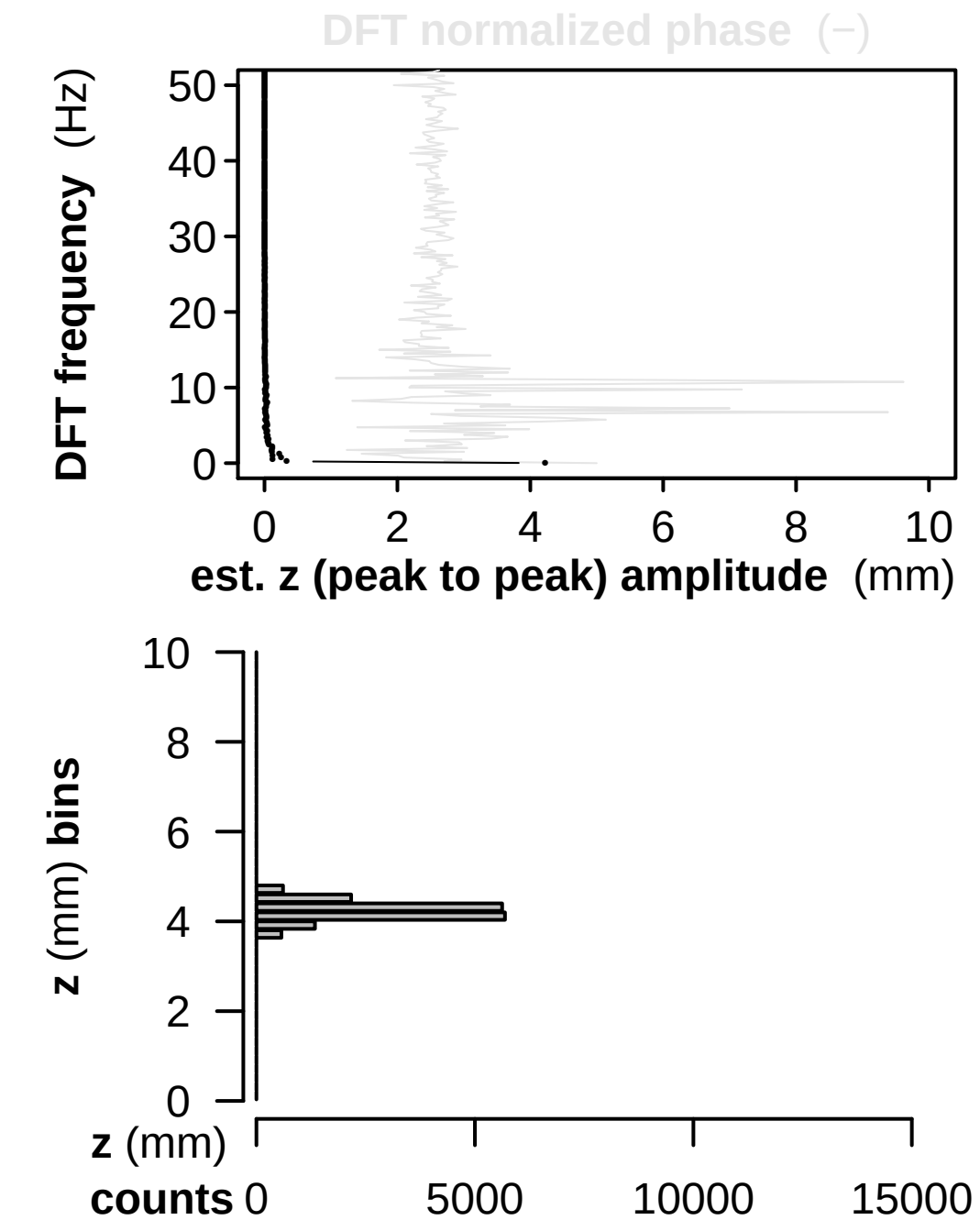

**SUBJECT 5 − RUN 15 − CONDITION 3,1**
**SC_180323_132351_0.AIFF**

```
z_min               : 3.62 mm
z_max               : 4.67 mm
z_travel_amplitude  : 1.05 mm

avg_abs_z_travel    : 2.83 mm/s

z_jarque-bera_jb    : 173.27
z_jarque-bera_p     : 0.00e+00

z_lin_mod_est_slope: −0.15 mm/s
z_lin_mod_adj_R²    : 72 %

z_poly40_mod_adj_R²: 95 %

z_dft_ampl_thresh   : 0.010 mm
>=threshold_maxfreq: 16.25 Hz
```

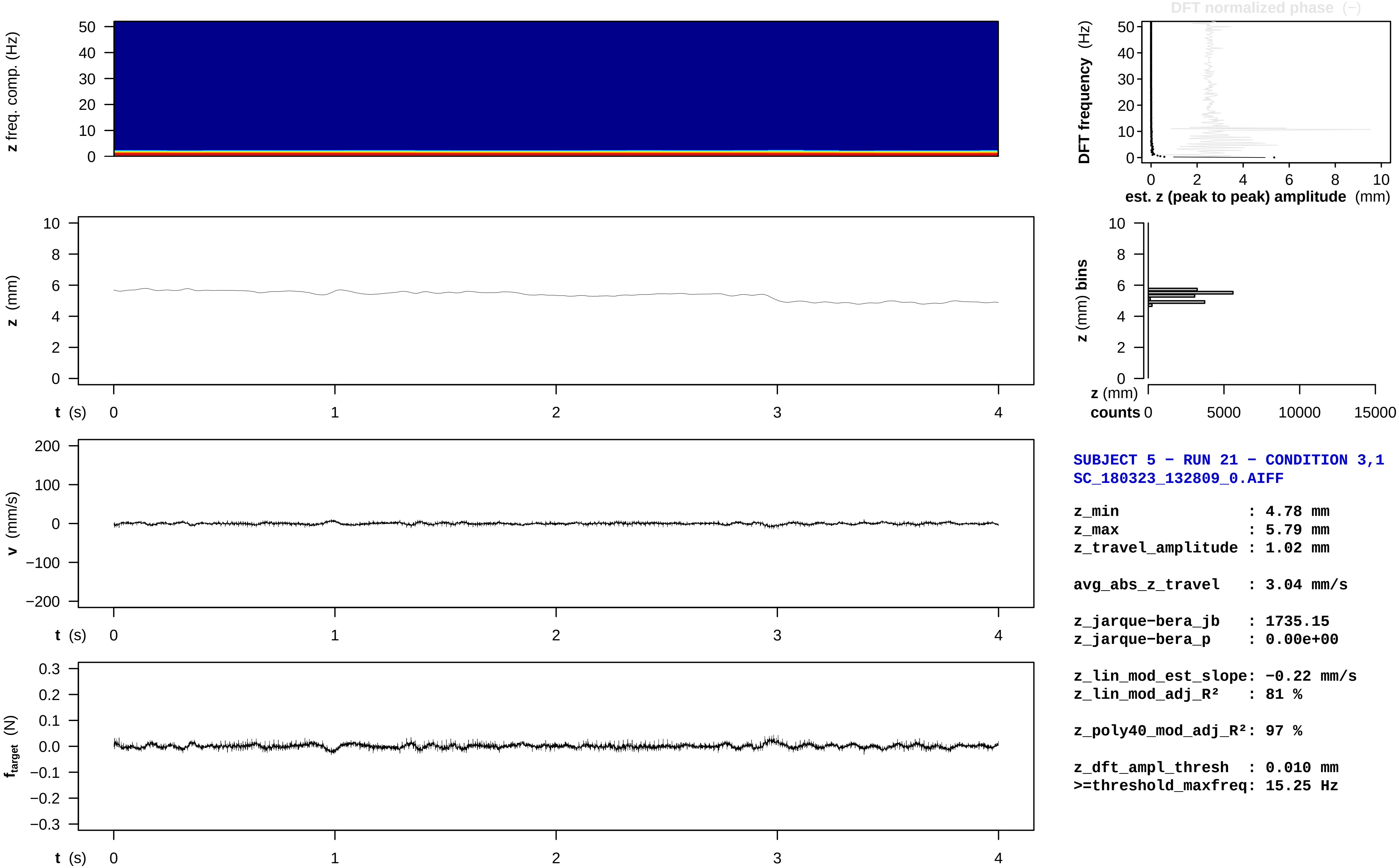

**SUBJECT 5 − RUN 21 − CONDITION 3,1**
**SC_180323_132809_0.AIFF**

```
z_min               : 4.78 mm
z_max               : 5.79 mm
z_travel_amplitude  : 1.02 mm

avg_abs_z_travel    : 3.04 mm/s

z_jarque-bera_jb    : 1735.15
z_jarque-bera_p     : 0.00e+00

z_lin_mod_est_slope: −0.22 mm/s
z_lin_mod_adj_R²    : 81 %

z_poly40_mod_adj_R²: 97 %

z_dft_ampl_thresh   : 0.010 mm
>=threshold_maxfreq: 15.25 Hz
```

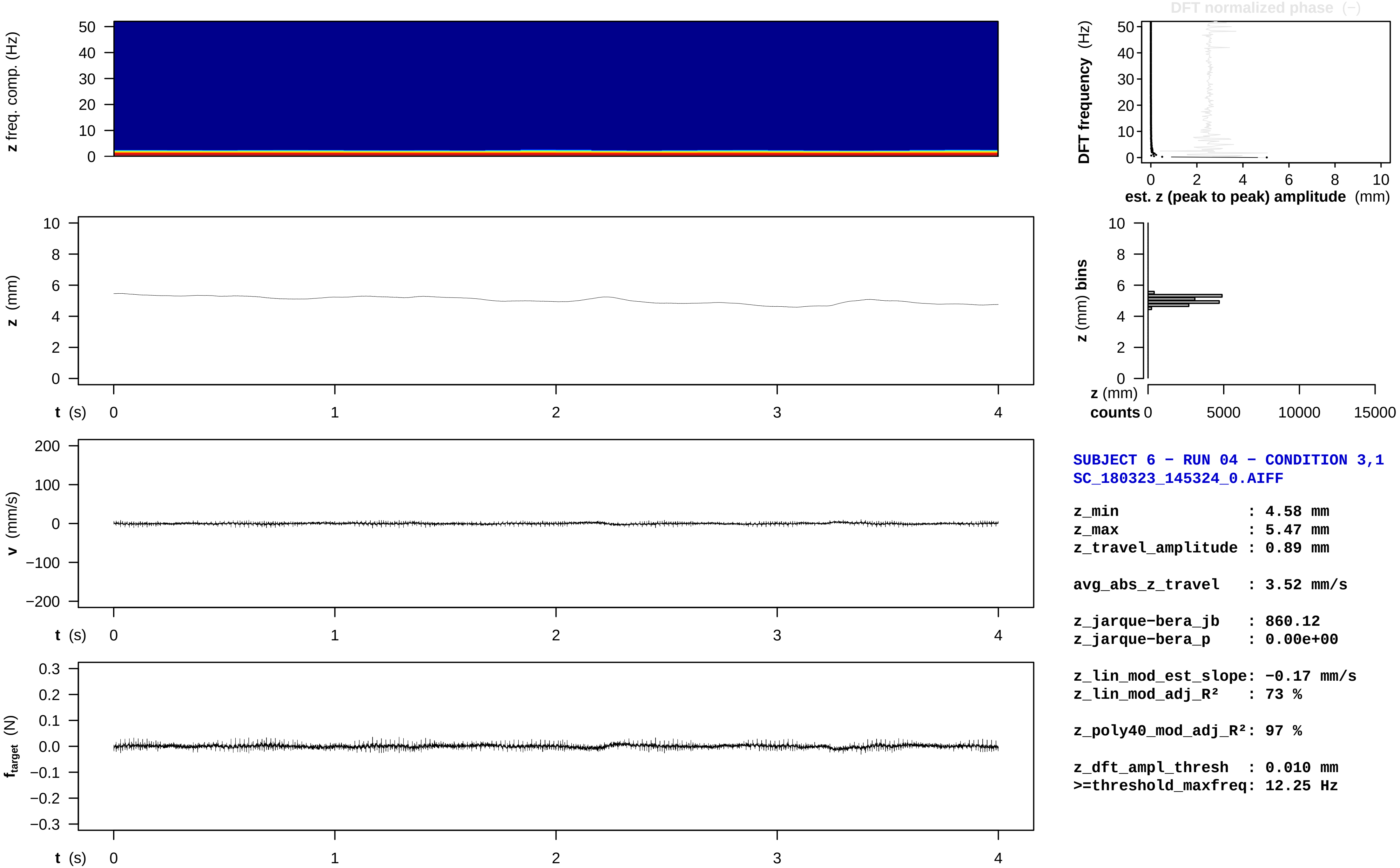

**SUBJECT 6 − RUN 04 − CONDITION 3,1**
**SC_180323_145324_0.AIFF**

```
z_min               : 4.58 mm
z_max               : 5.47 mm
z_travel_amplitude  : 0.89 mm

avg_abs_z_travel    : 3.52 mm/s

z_jarque-bera_jb    : 860.12
z_jarque-bera_p     : 0.00e+00

z_lin_mod_est_slope: −0.17 mm/s
z_lin_mod_adj_R²    : 73 %

z_poly40_mod_adj_R²: 97 %

z_dft_ampl_thresh   : 0.010 mm
>=threshold_maxfreq: 12.25 Hz
```

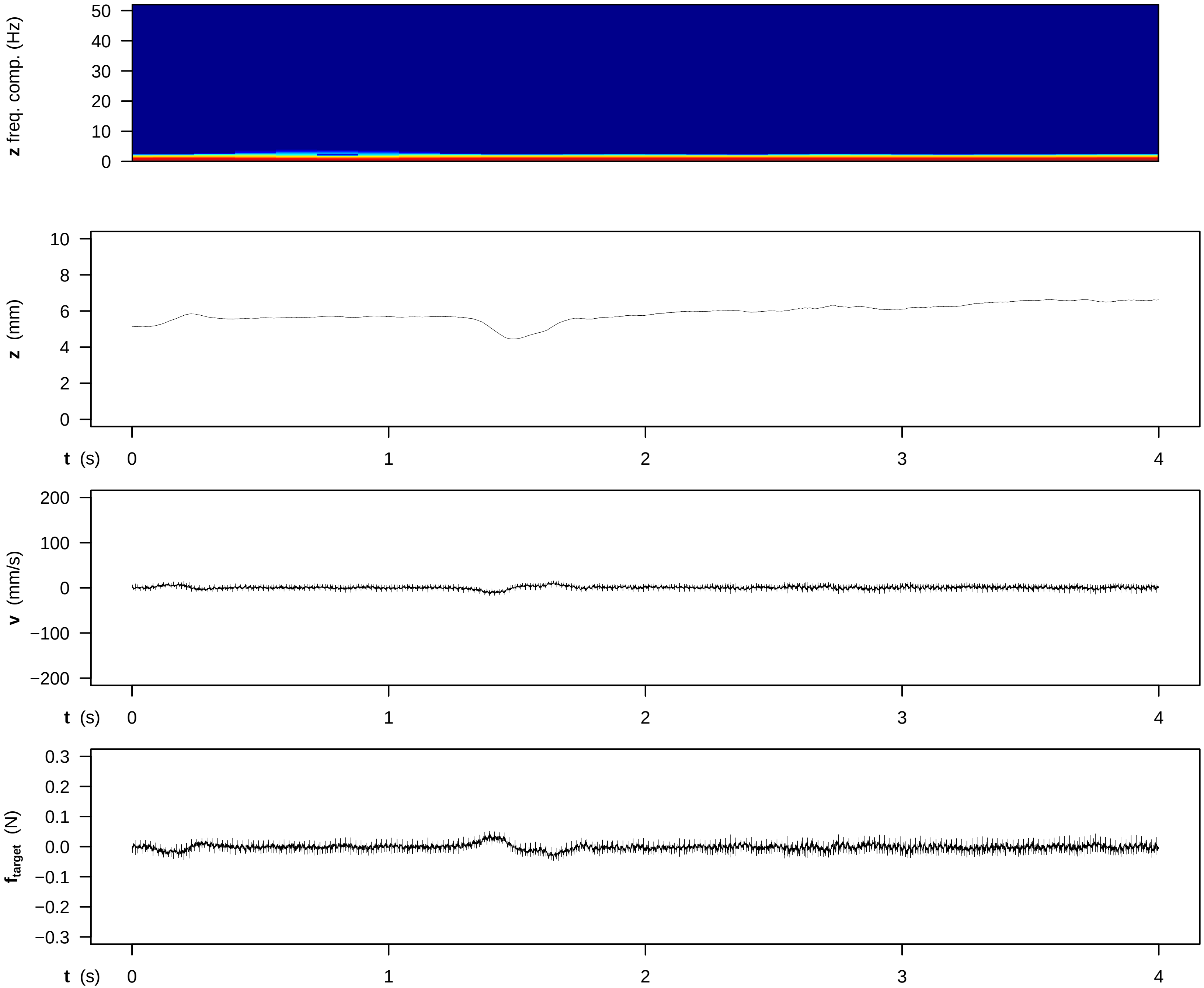

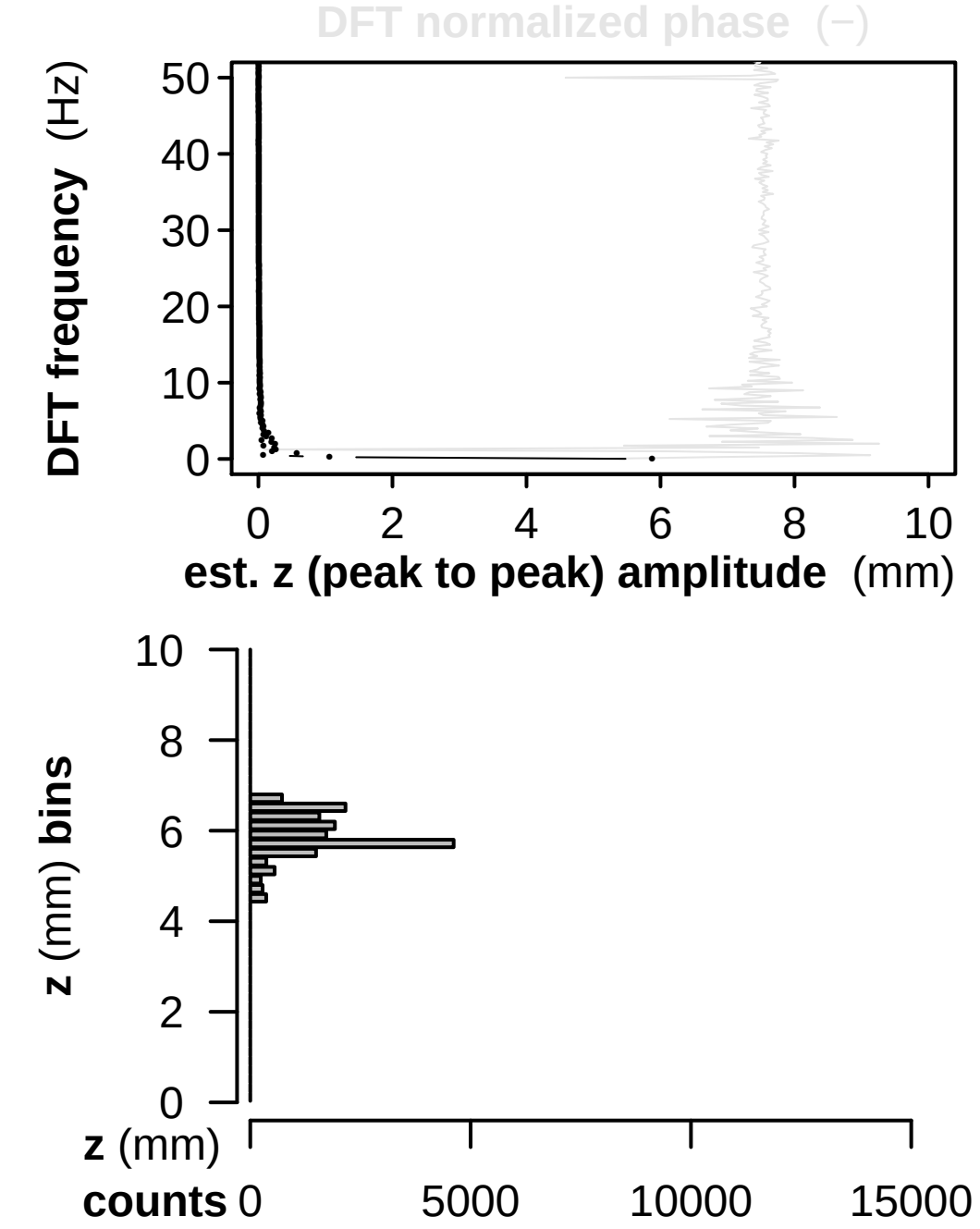

**SUBJECT 6 − RUN 16 − CONDITION 3,1**
**SC_180323_150120_0.AIFF**

```
z_min              : 4.44 mm
z_max              : 6.64 mm
z_travel_amplitude : 2.20 mm

avg_abs_z_travel   : 5.67 mm/s

z_jarque-bera_jb   : 850.83
z_jarque-bera_p    : 0.00e+00

z_lin_mod_est_slope: 0.33 mm/s
z_lin_mod_adj_R²   : 61 %

z_poly40_mod_adj_R²: 95 %

z_dft_ampl_thresh  : 0.010 mm
>=threshold_maxfreq: 24.75 Hz
```

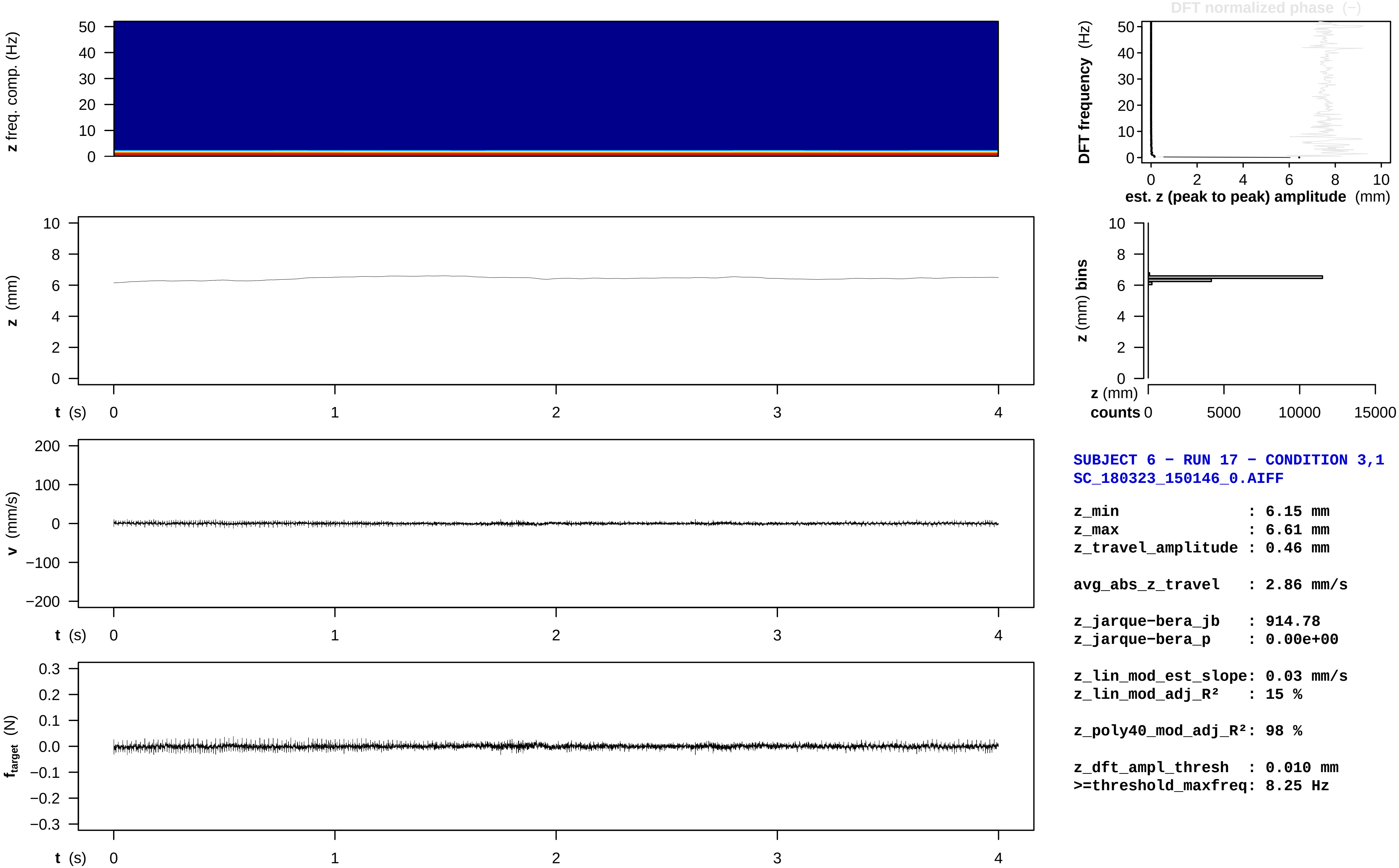

**SUBJECT 6 − RUN 17 − CONDITION 3,1**
**SC_180323_150146_0.AIFF**

```
z_min              : 6.15 mm
z_max              : 6.61 mm
z_travel_amplitude : 0.46 mm

avg_abs_z_travel   : 2.86 mm/s

z_jarque-bera_jb   : 914.78
z_jarque-bera_p    : 0.00e+00

z_lin_mod_est_slope: 0.03 mm/s
z_lin_mod_adj_R²   : 15 %

z_poly40_mod_adj_R²: 98 %

z_dft_ampl_thresh  : 0.010 mm
>=threshold_maxfreq: 8.25 Hz
```

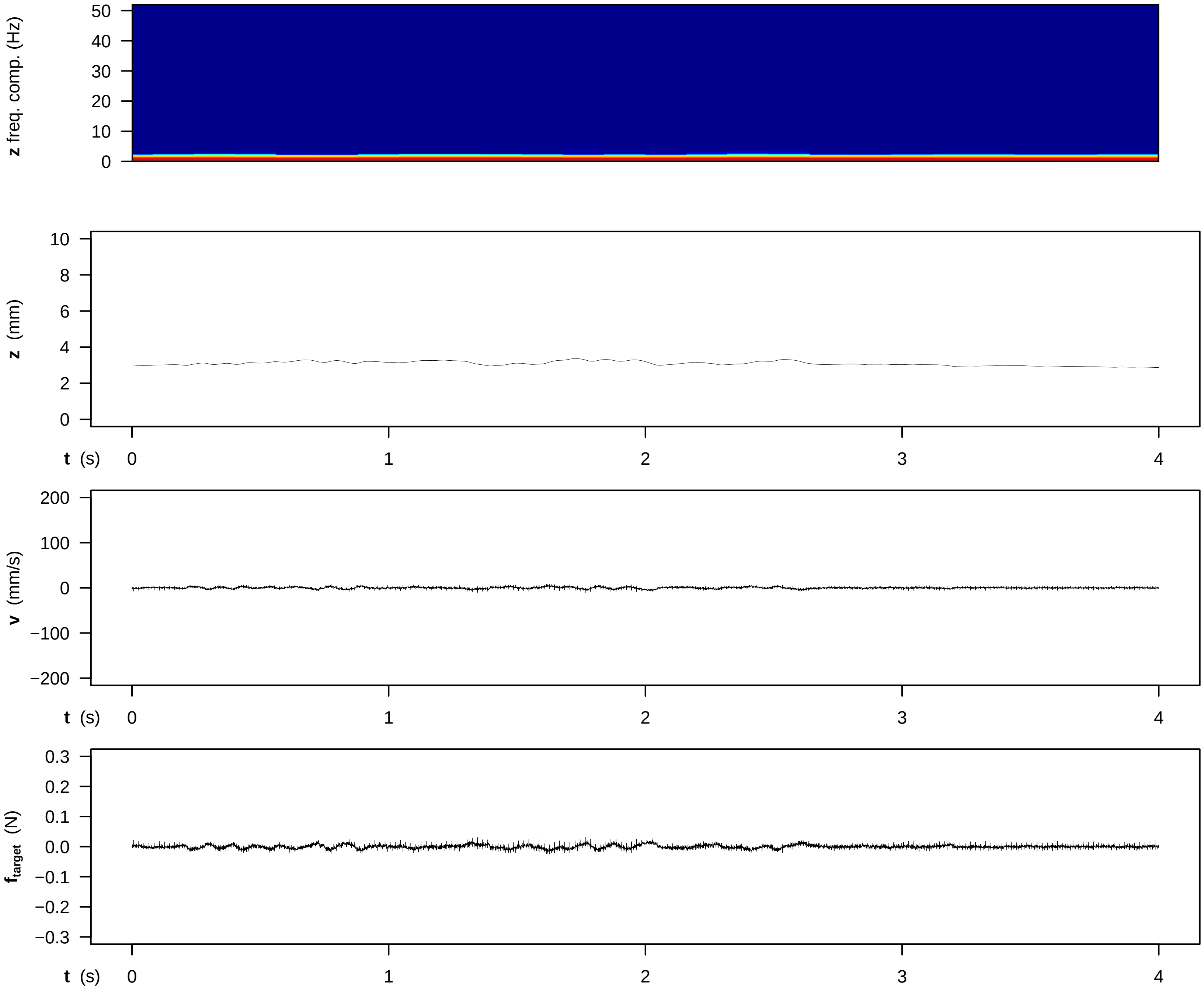

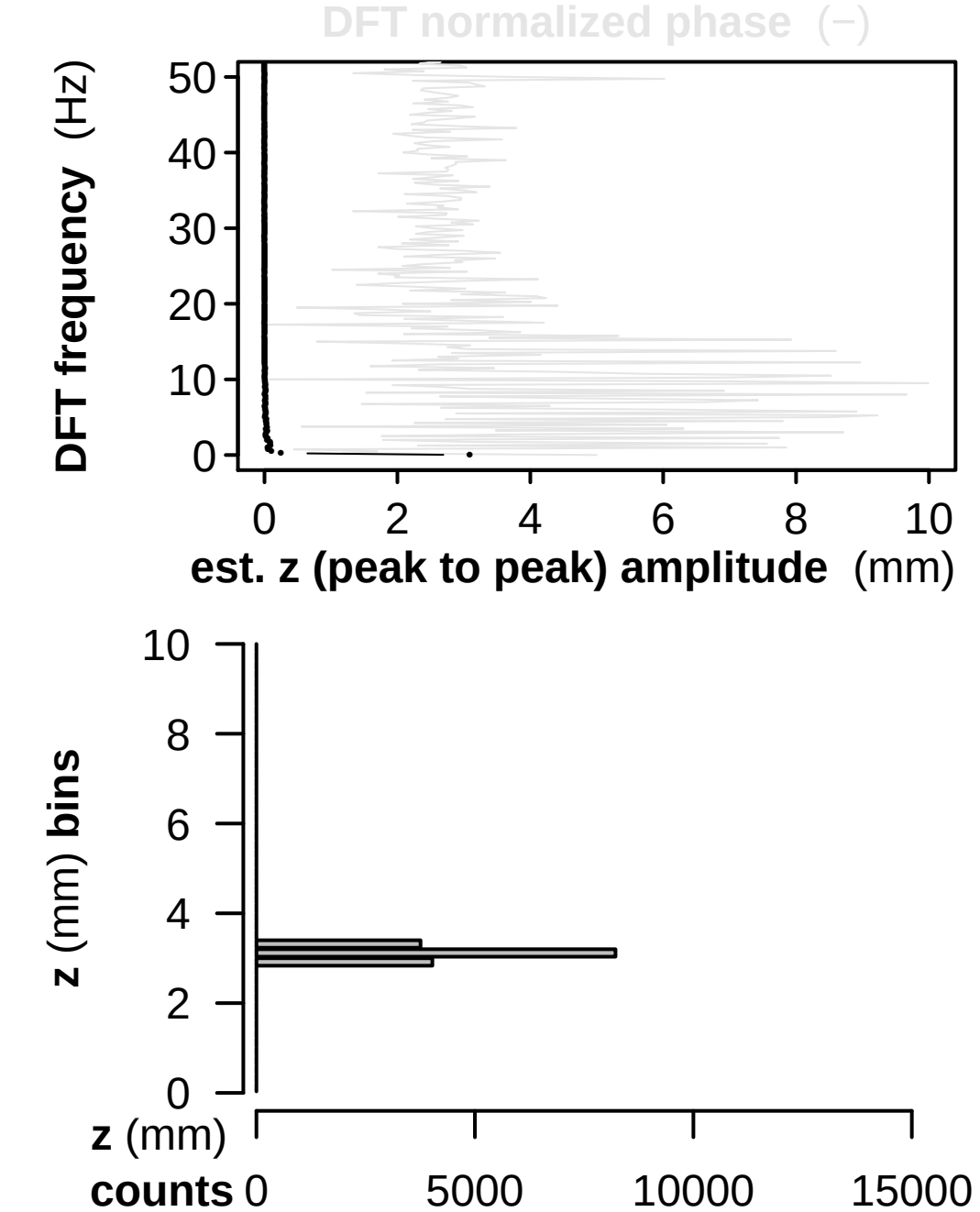

**SUBJECT 7 − RUN 02 − CONDITION 3,1**
**SC_180323_153522_0.AIFF**

```
z_min              : 2.87 mm
z_max              : 3.37 mm
z_travel_amplitude : 0.50 mm

avg_abs_z_travel   : 2.09 mm/s

z_jarque-bera_jb   : 744.15
z_jarque-bera_p    : 0.00e+00

z_lin_mod_est_slope: −0.05 mm/s
z_lin_mod_adj_R²   : 27 %

z_poly40_mod_adj_R²: 86 %

z_dft_ampl_thresh  : 0.010 mm
>=threshold_maxfreq: 11.50 Hz
```

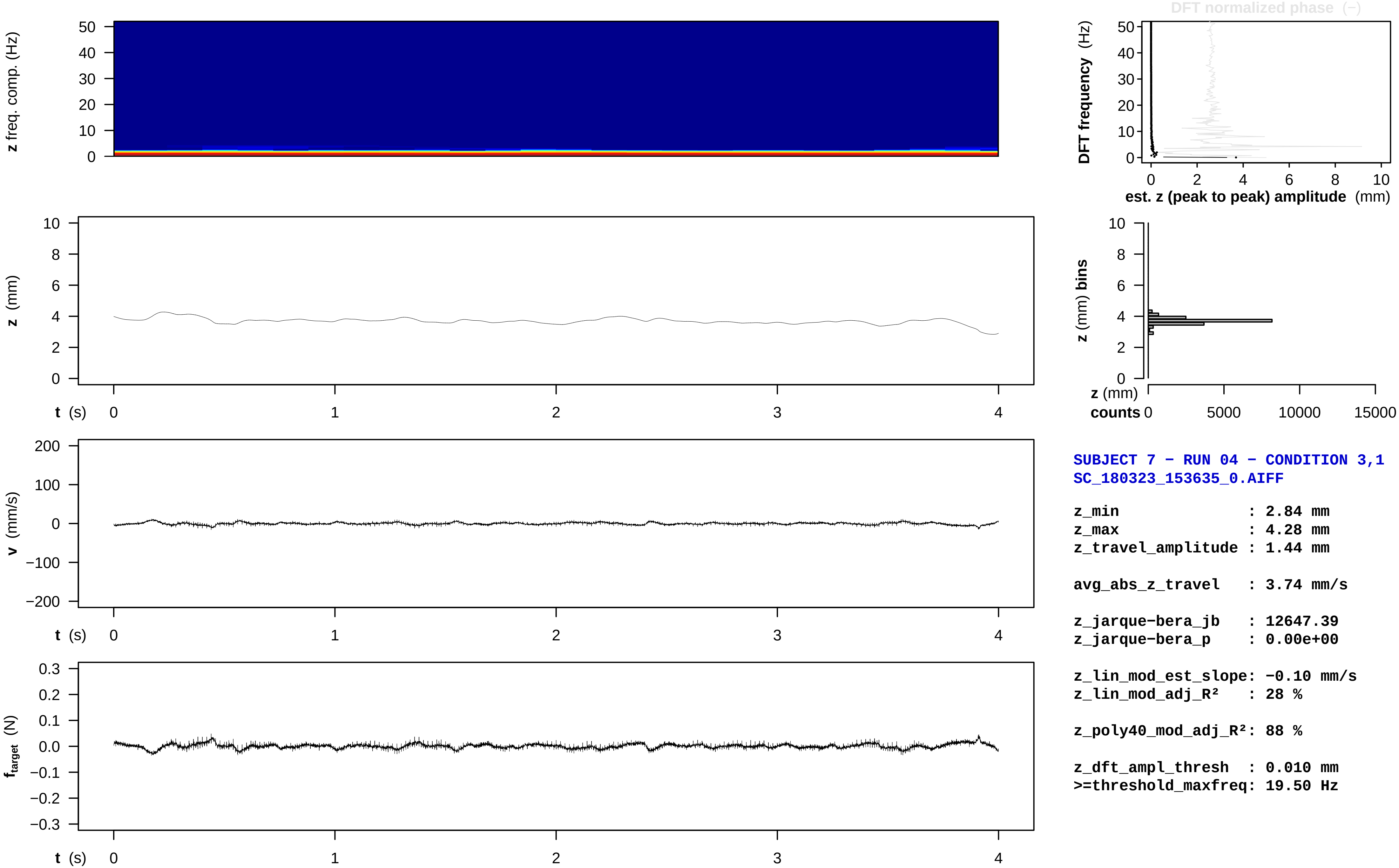

**SUBJECT 7 − RUN 04 − CONDITION 3,1**
**SC_180323_153635_0.AIFF**

```
z_min               : 2.84 mm
z_max               : 4.28 mm
z_travel_amplitude  : 1.44 mm

avg_abs_z_travel    : 3.74 mm/s

z_jarque-bera_jb    : 12647.39
z_jarque-bera_p     : 0.00e+00

z_lin_mod_est_slope: −0.10 mm/s
z_lin_mod_adj_R²    : 28 %

z_poly40_mod_adj_R²: 88 %

z_dft_ampl_thresh   : 0.010 mm
>=threshold_maxfreq: 19.50 Hz
```

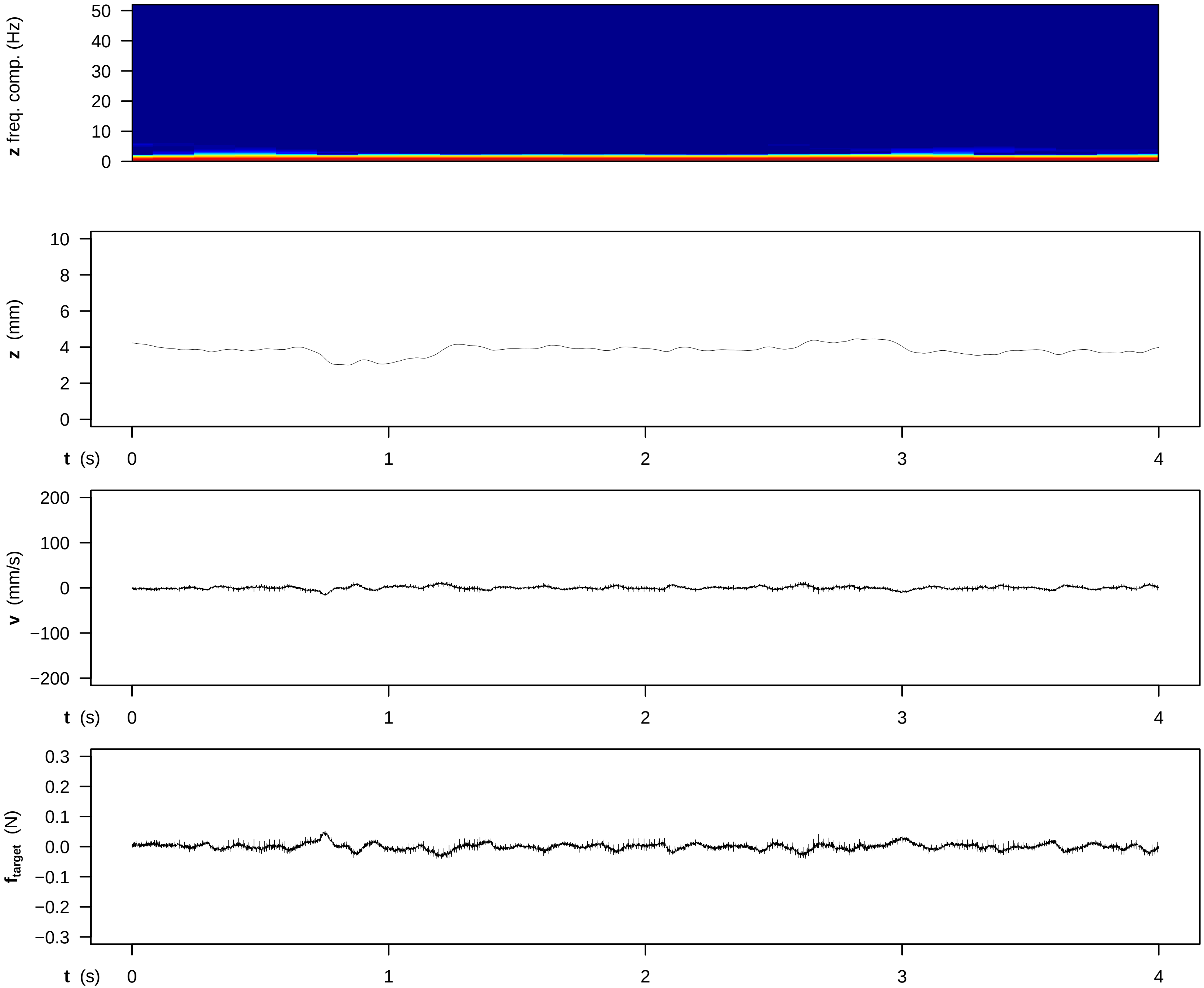

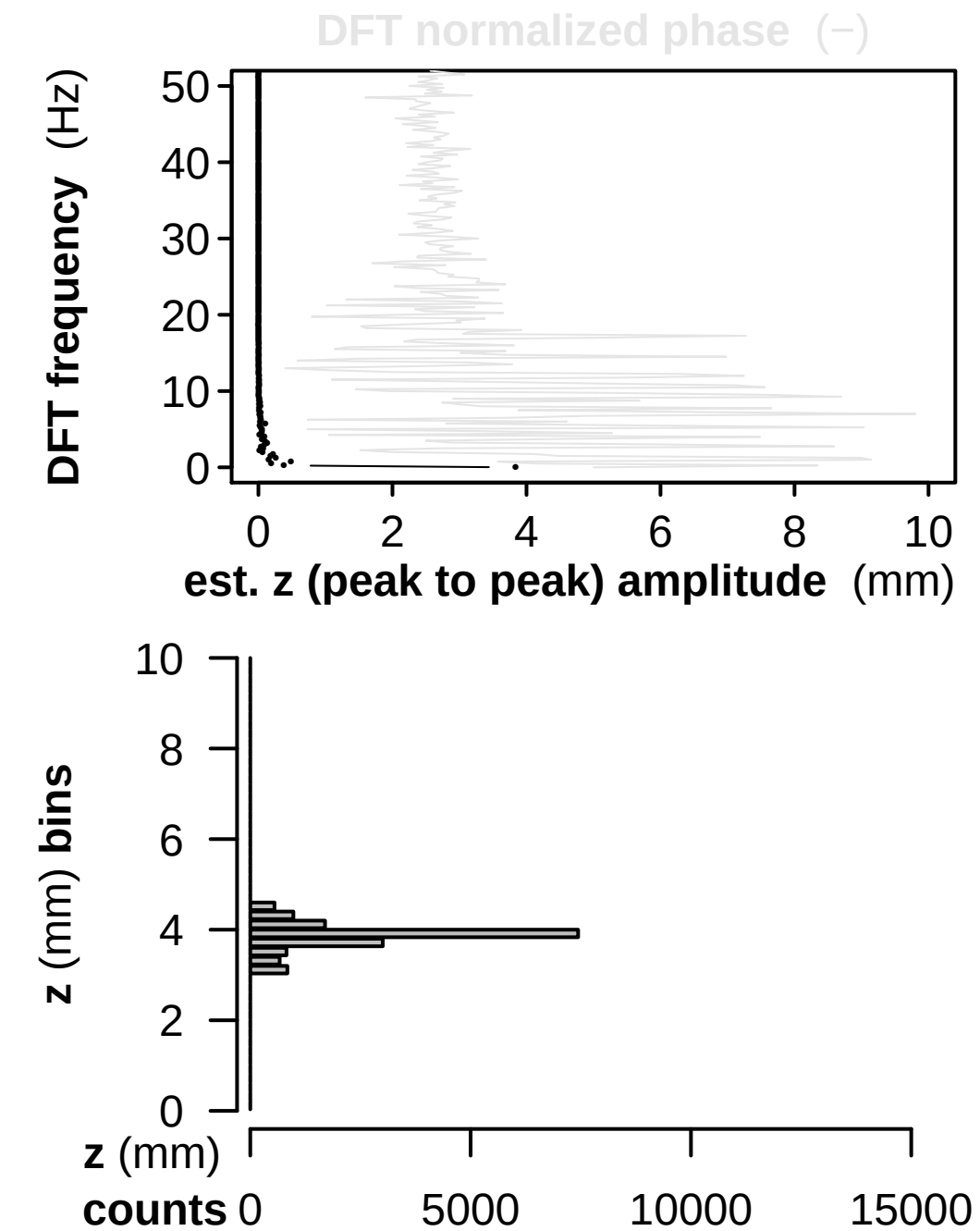

**SUBJECT 7 − RUN 26 − CONDITION 3,1**
**SC_180323_155349_0.AIFF**

```
z_min               : 3.01 mm
z_max               : 4.46 mm
z_travel_amplitude  : 1.45 mm

avg_abs_z_travel    : 3.18 mm/s

z_jarque-bera_jb    : 1886.89
z_jarque-bera_p     : 0.00e+00

z_lin_mod_est_slope: 0.03 mm/s
z_lin_mod_adj_R²    : 2 %

z_poly40_mod_adj_R²: 89 %

z_dft_ampl_thresh   : 0.010 mm
>=threshold_maxfreq: 12.00 Hz
```

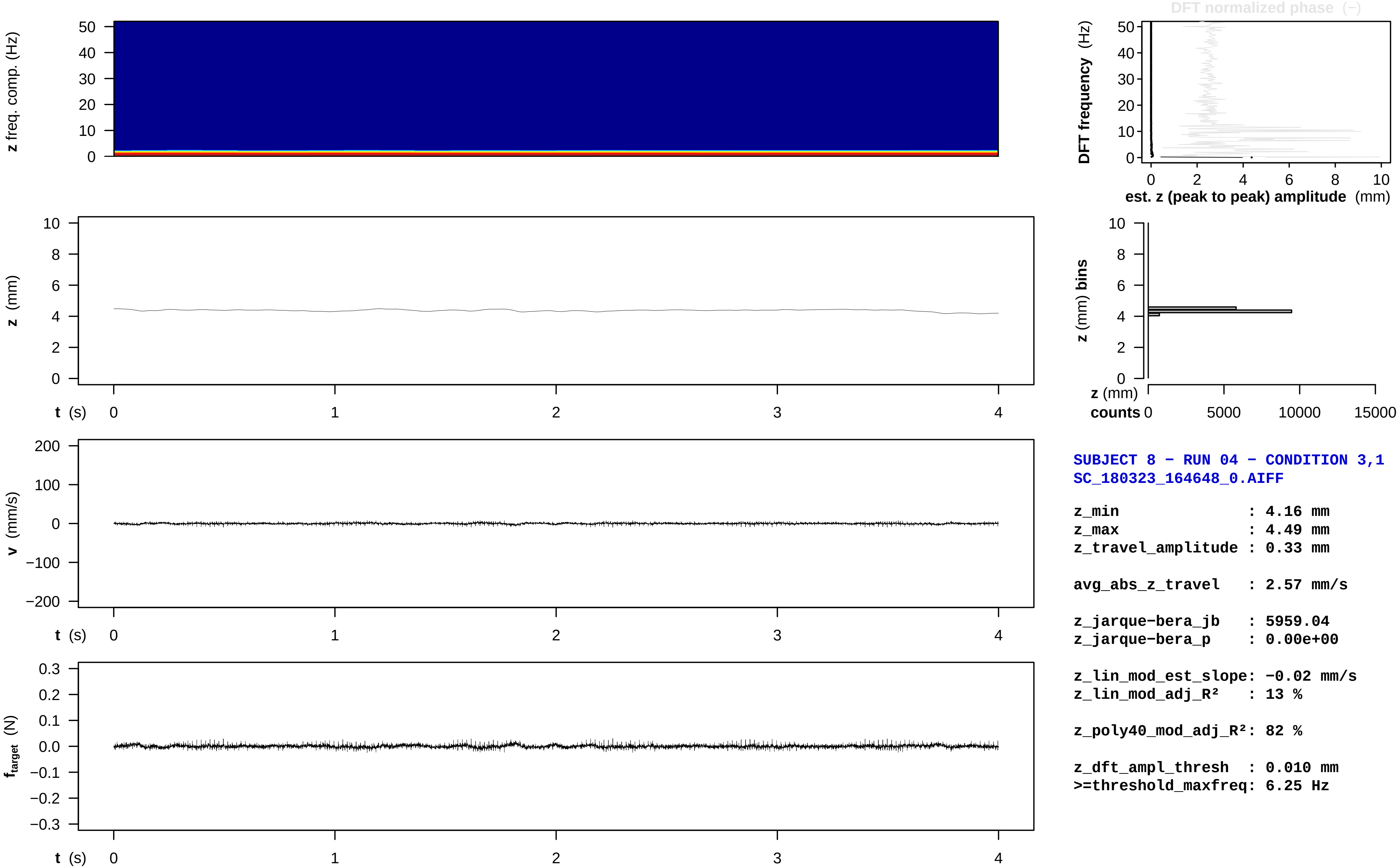

**SUBJECT 8 − RUN 04 − CONDITION 3,1**
**SC_180323_164648_0.AIFF**

```
z_min                : 4.16 mm
z_max                : 4.49 mm
z_travel_amplitude   : 0.33 mm

avg_abs_z_travel     : 2.57 mm/s

z_jarque-bera_jb     : 5959.04
z_jarque-bera_p      : 0.00e+00

z_lin_mod_est_slope: -0.02 mm/s
z_lin_mod_adj_R²   : 13 %

z_poly40_mod_adj_R²: 82 %

z_dft_ampl_thresh  : 0.010 mm
>=threshold_maxfreq: 6.25 Hz
```

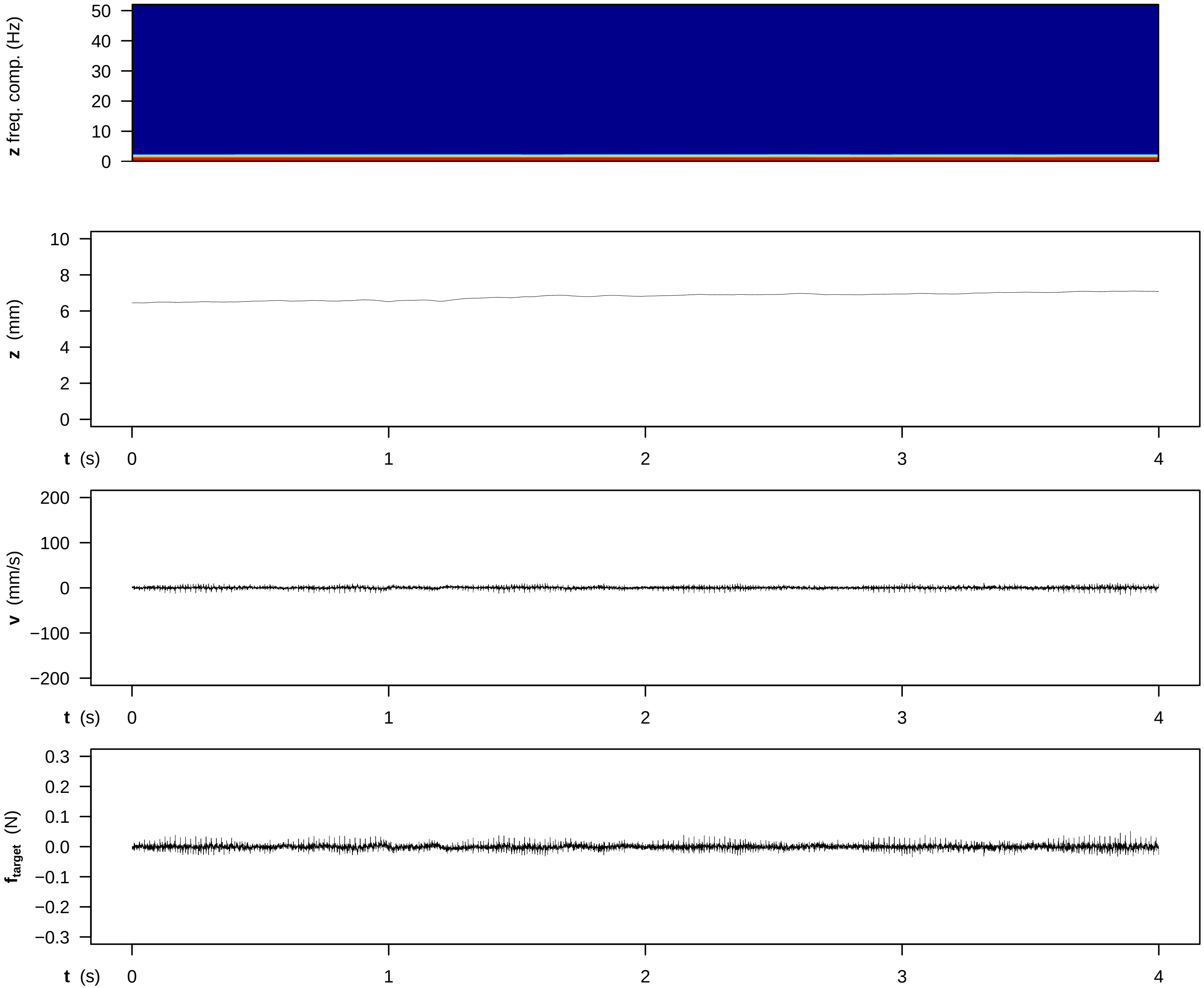

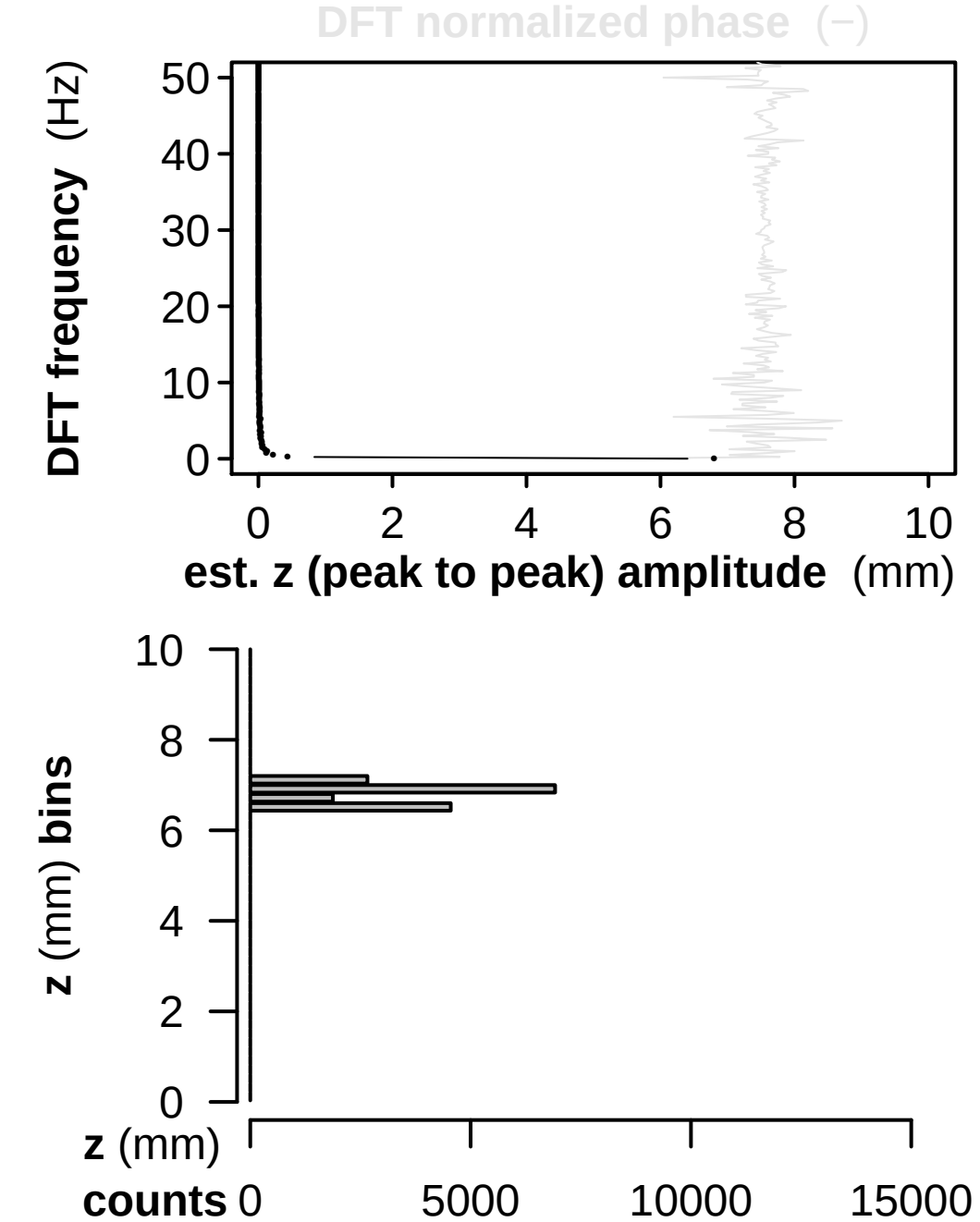

**SUBJECT 8 − RUN 05 − CONDITION 3,1**
**SC_180323_164732_0.AIFF**

```
z_min              : 6.44 mm
z_max              : 7.11 mm
z_travel_amplitude : 0.67 mm

avg_abs_z_travel   : 3.65 mm/s

z_jarque-bera_jb   : 1337.73
z_jarque-bera_p    : 0.00e+00

z_lin_mod_est_slope: 0.17 mm/s
z_lin_mod_adj_R²   : 95 %

z_poly40_mod_adj_R²: 99 %

z_dft_ampl_thresh  : 0.010 mm
>=threshold_maxfreq: 11.75 Hz
```

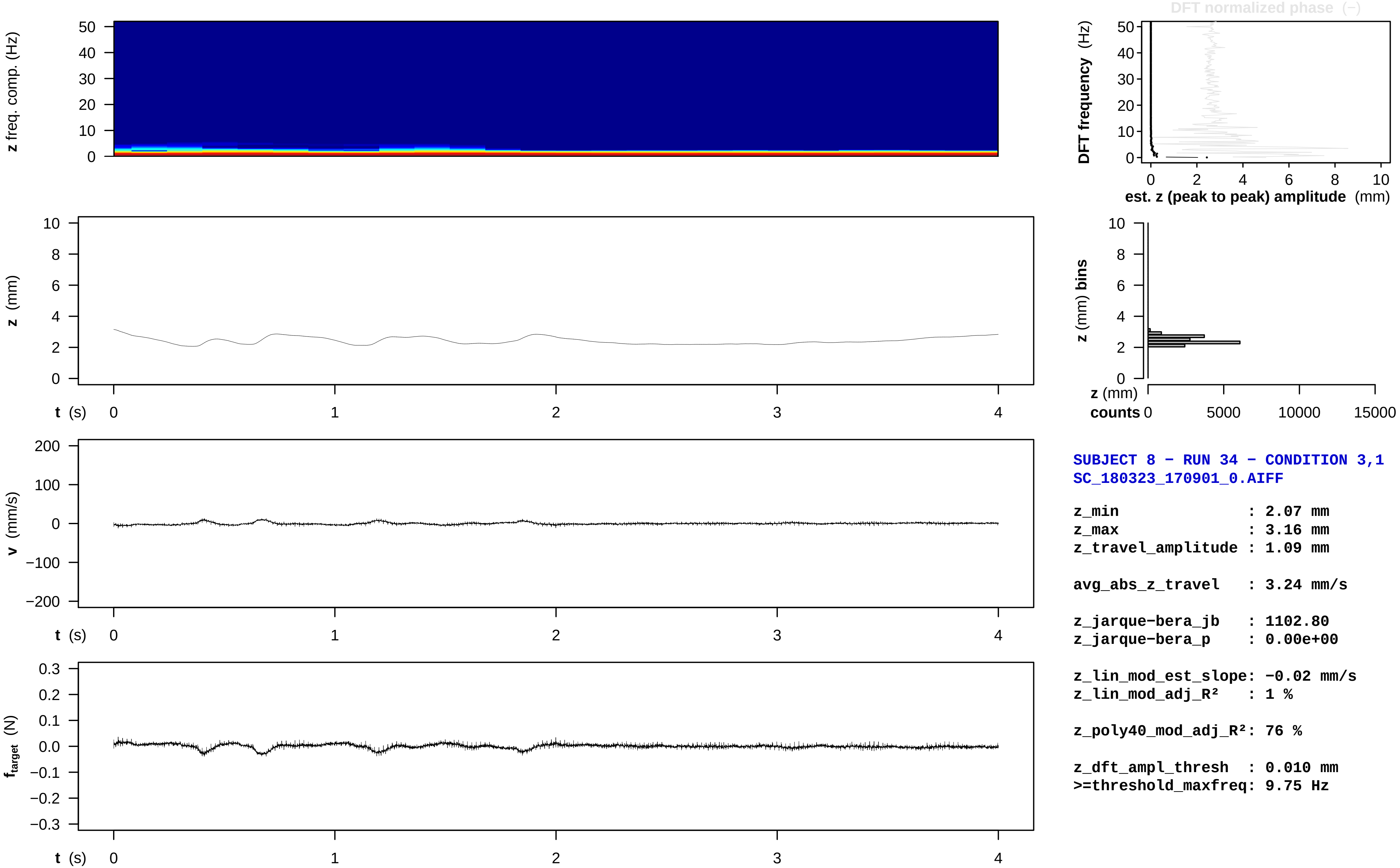

**SUBJECT 8 − RUN 34 − CONDITION 3,1**
**SC_180323_170901_0.AIFF**

```
z_min              : 2.07 mm
z_max              : 3.16 mm
z_travel_amplitude : 1.09 mm

avg_abs_z_travel   : 3.24 mm/s

z_jarque-bera_jb   : 1102.80
z_jarque-bera_p    : 0.00e+00

z_lin_mod_est_slope: -0.02 mm/s
z_lin_mod_adj_R²   : 1 %

z_poly40_mod_adj_R²: 76 %

z_dft_ampl_thresh  : 0.010 mm
>=threshold_maxfreq: 9.75 Hz
```

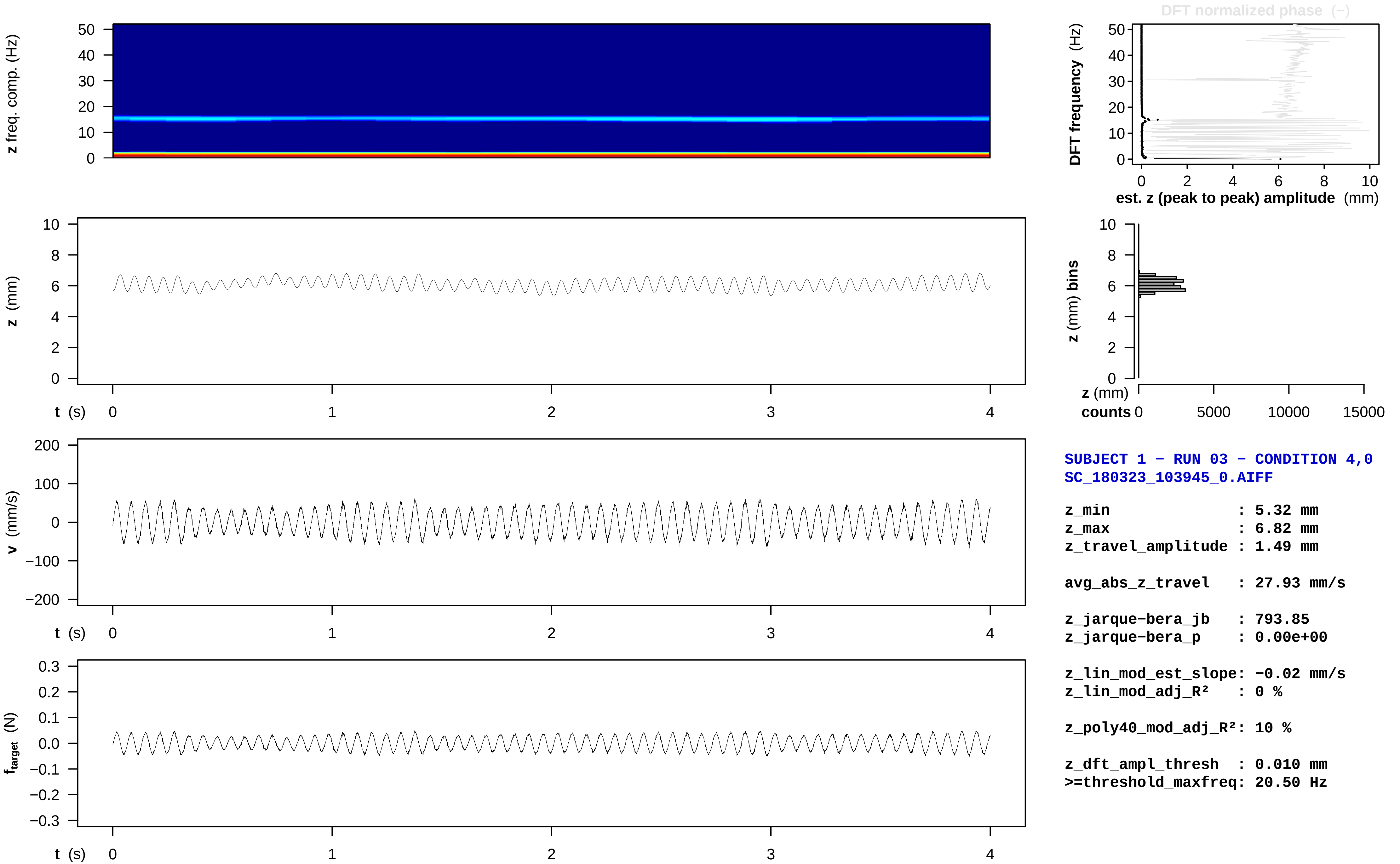

**SUBJECT 1 − RUN 03 − CONDITION 4,0**
**SC_180323_103945_0.AIFF**

```
z_min              : 5.32 mm
z_max              : 6.82 mm
z_travel_amplitude : 1.49 mm

avg_abs_z_travel   : 27.93 mm/s

z_jarque-bera_jb   : 793.85
z_jarque-bera_p    : 0.00e+00

z_lin_mod_est_slope: -0.02 mm/s
z_lin_mod_adj_R²   : 0 %

z_poly40_mod_adj_R²: 10 %

z_dft_ampl_thresh  : 0.010 mm
>=threshold_maxfreq: 20.50 Hz
```

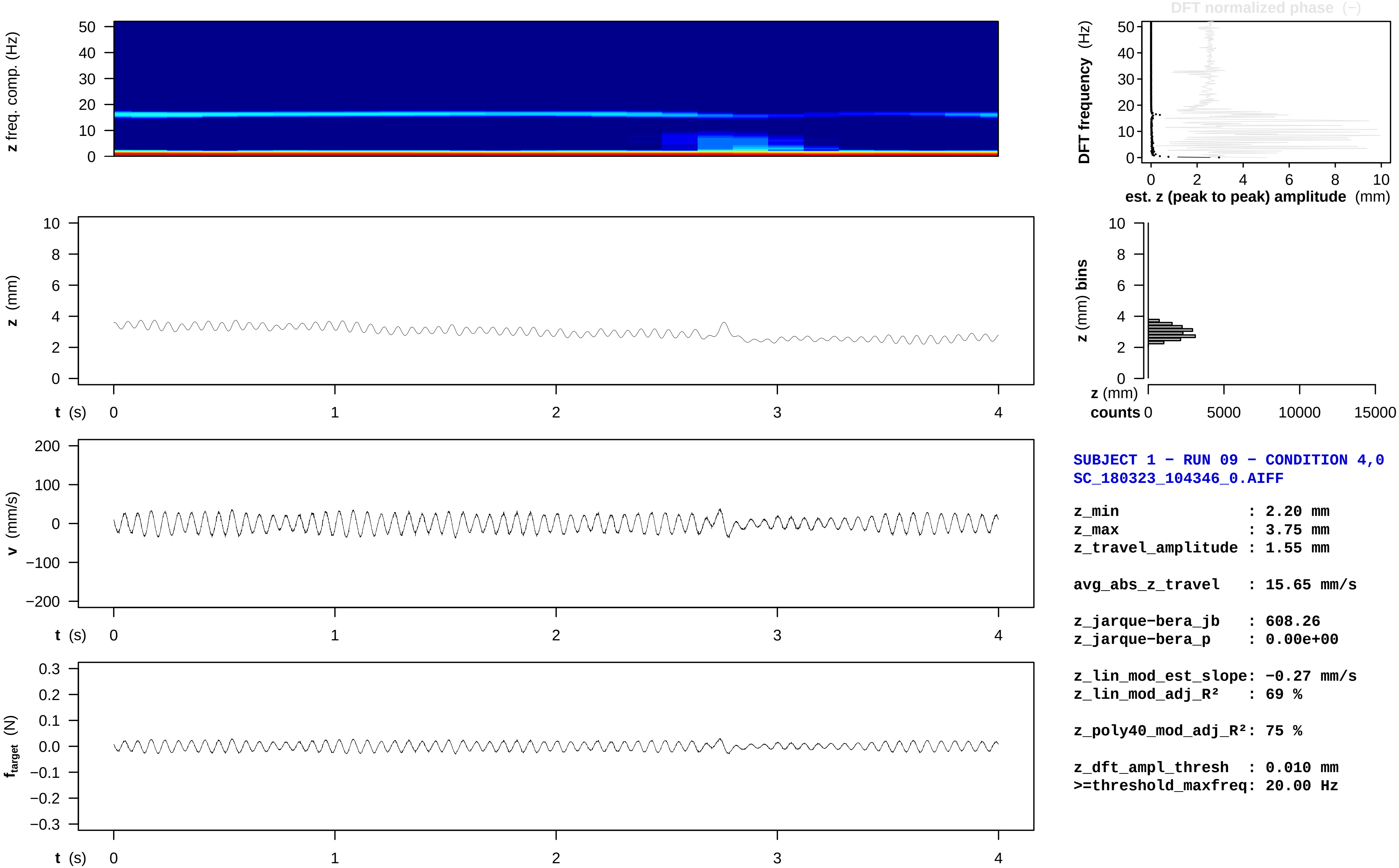

**SUBJECT 1 − RUN 09 − CONDITION 4,0**
**SC_180323_104346_0.AIFF**

```
z_min               : 2.20 mm
z_max               : 3.75 mm
z_travel_amplitude  : 1.55 mm

avg_abs_z_travel    : 15.65 mm/s

z_jarque-bera_jb    : 608.26
z_jarque-bera_p     : 0.00e+00

z_lin_mod_est_slope: -0.27 mm/s
z_lin_mod_adj_R²    : 69 %

z_poly40_mod_adj_R²: 75 %

z_dft_ampl_thresh   : 0.010 mm
>=threshold_maxfreq: 20.00 Hz
```

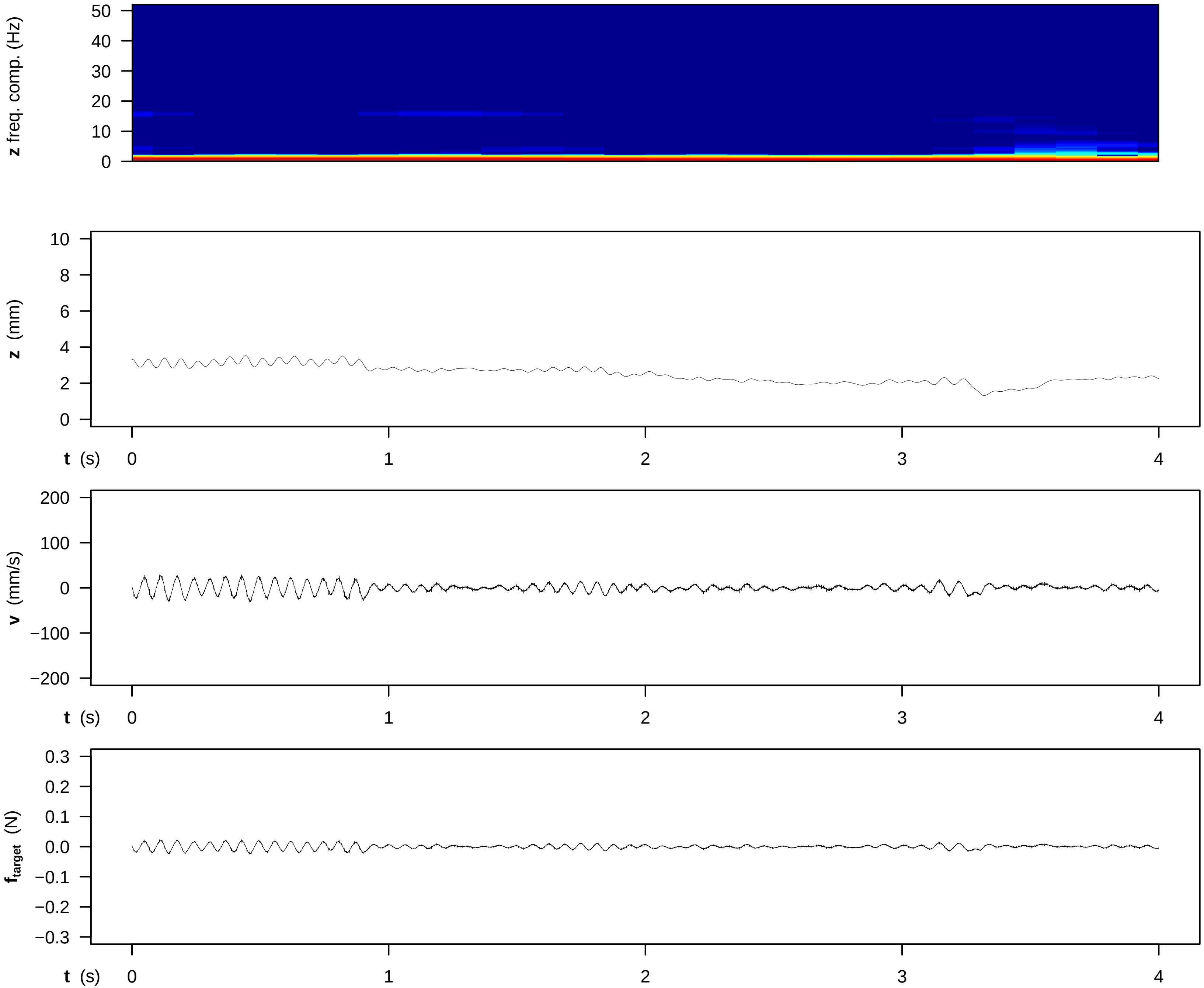

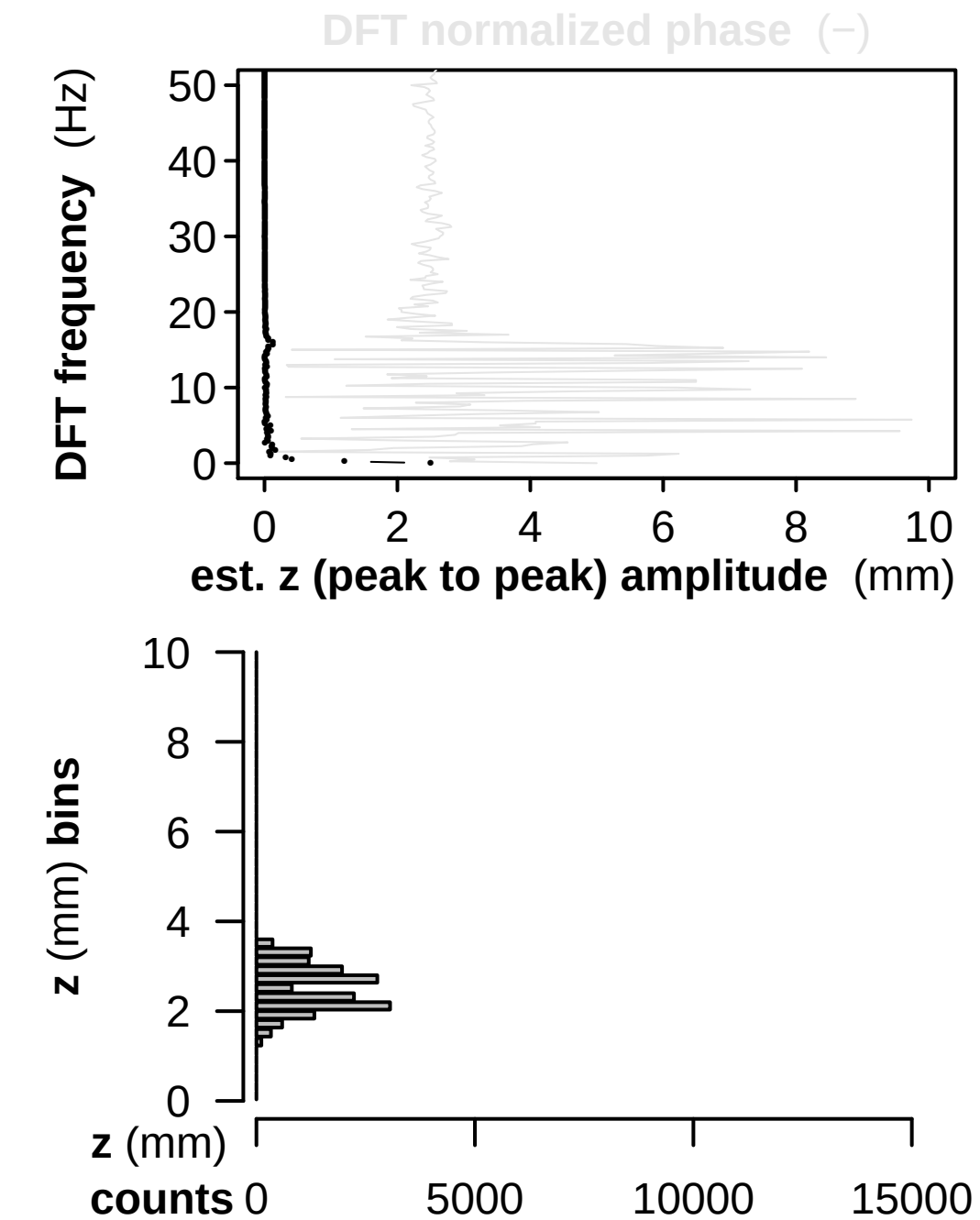

**SUBJECT 1 − RUN 12 − CONDITION 4,0**
**SC_180323_104552_0.AIFF**

```
z_min               : 1.31 mm
z_max               : 3.53 mm
z_travel_amplitude  : 2.22 mm

avg_abs_z_travel    : 6.92 mm/s

z_jarque-bera_jb    : 437.22
z_jarque-bera_p     : 0.00e+00

z_lin_mod_est_slope: -0.37 mm/s
z_lin_mod_adj_R²    : 75 %

z_poly40_mod_adj_R²: 94 %

z_dft_ampl_thresh   : 0.010 mm
>=threshold_maxfreq: 22.25 Hz
```

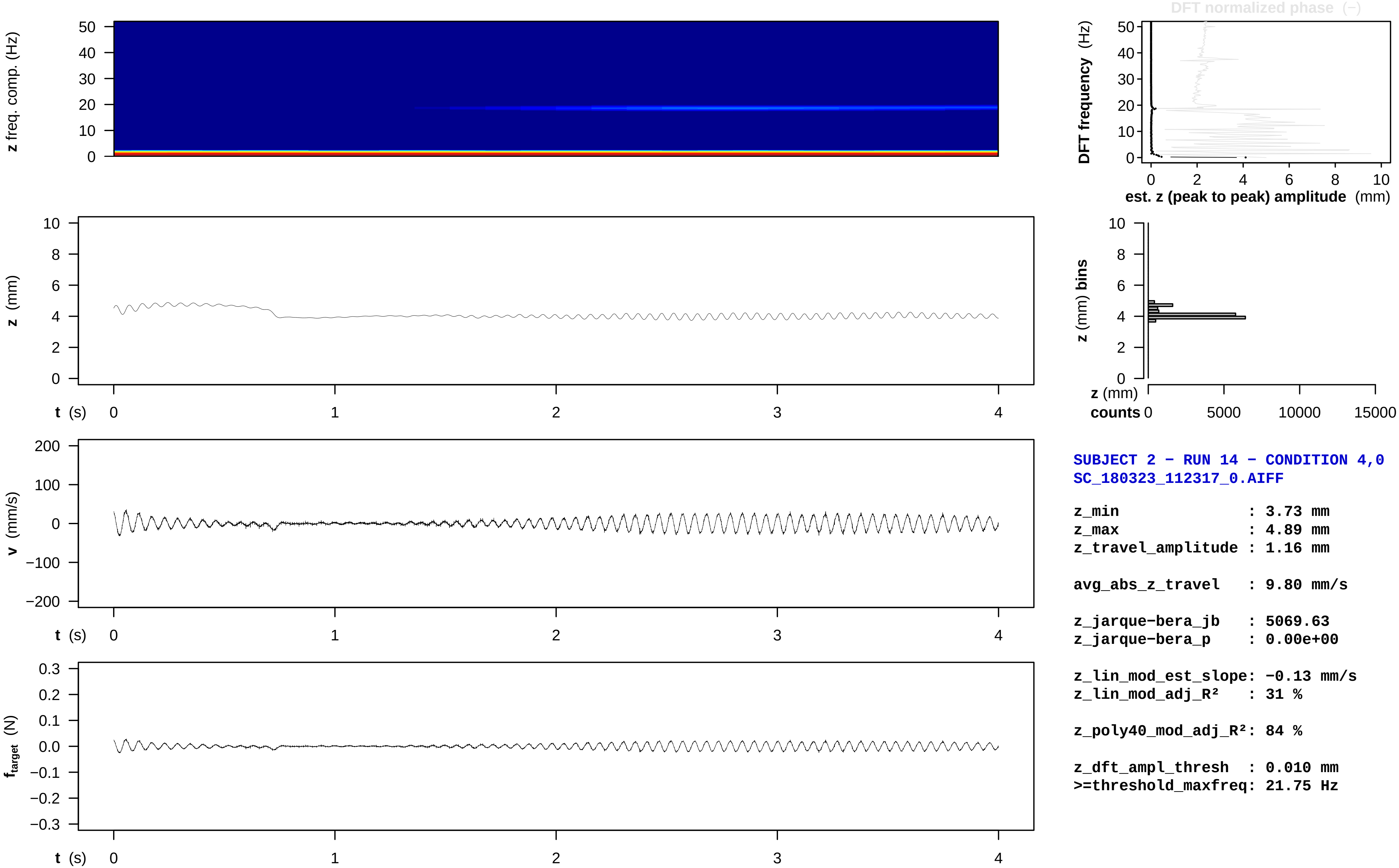

**SUBJECT 2 − RUN 14 − CONDITION 4,0**
**SC_180323_112317_0.AIFF**

```
z_min              : 3.73 mm
z_max              : 4.89 mm
z_travel_amplitude : 1.16 mm

avg_abs_z_travel   : 9.80 mm/s

z_jarque-bera_jb   : 5069.63
z_jarque-bera_p    : 0.00e+00

z_lin_mod_est_slope: −0.13 mm/s
z_lin_mod_adj_R²   : 31 %

z_poly40_mod_adj_R²: 84 %

z_dft_ampl_thresh  : 0.010 mm
>=threshold_maxfreq: 21.75 Hz
```

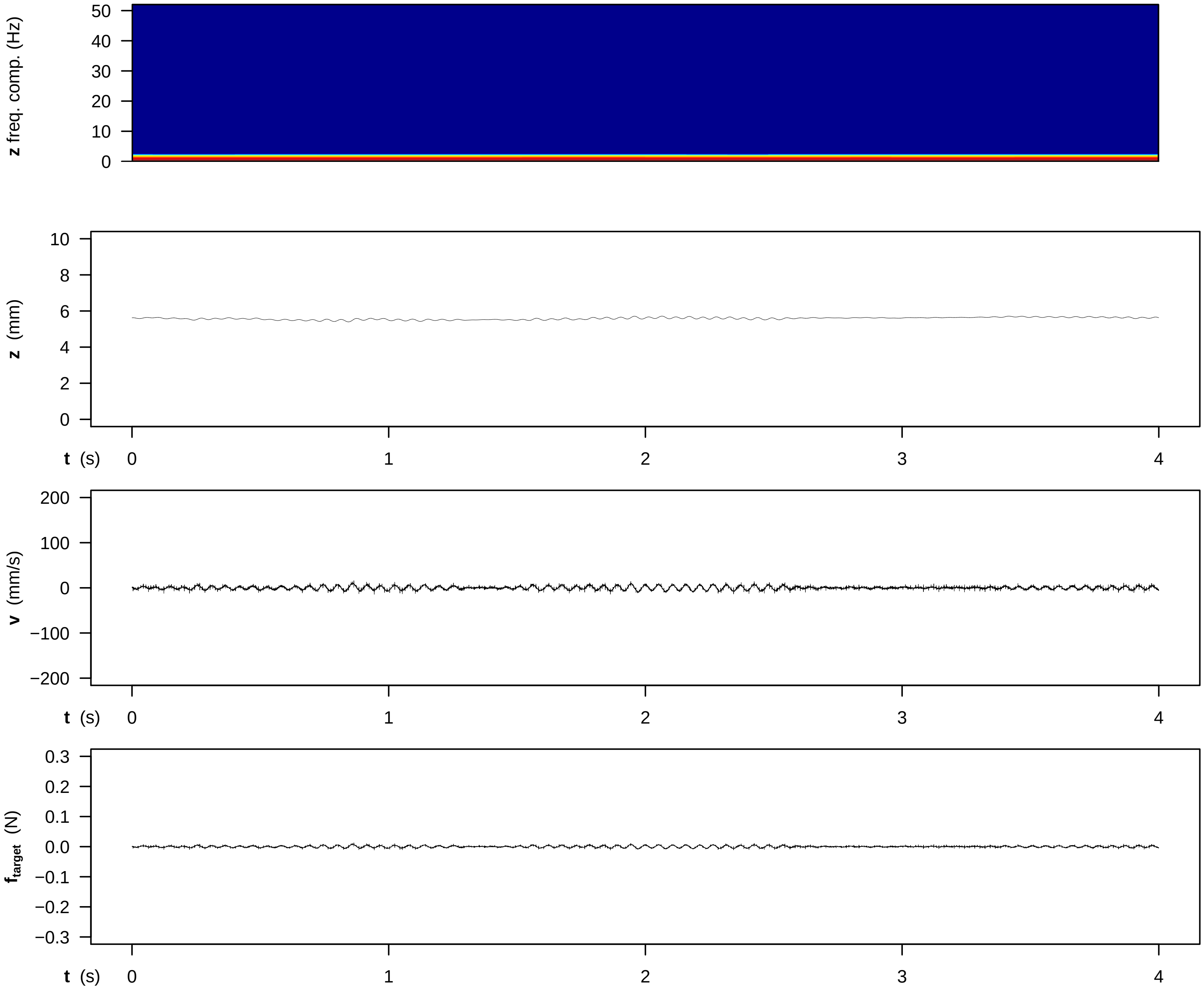

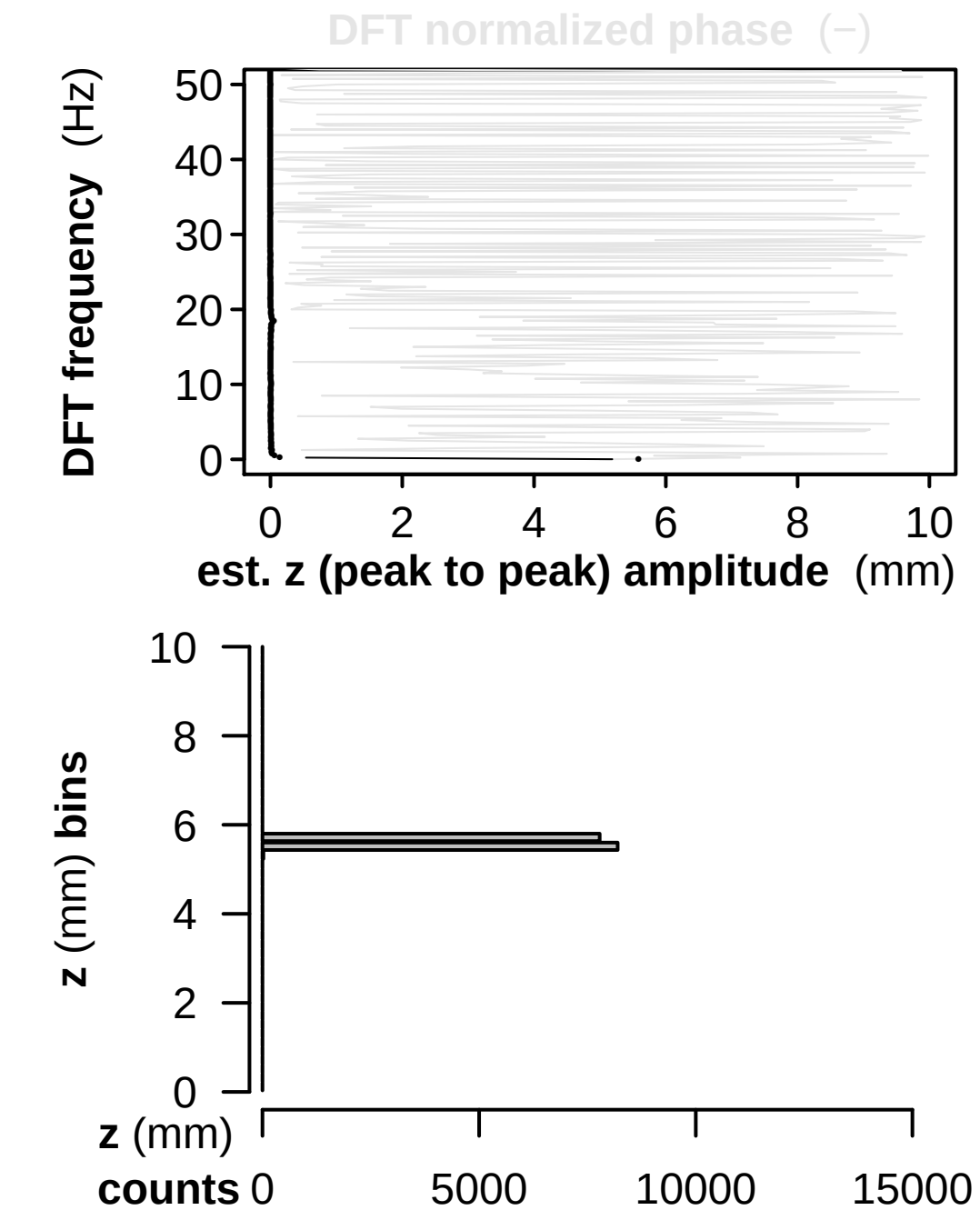

**SUBJECT 2 − RUN 16 − CONDITION 4,0**
**SC_180323_112414_0.AIFF**

```
z_min               : 5.39 mm
z_max               : 5.72 mm
z_travel_amplitude  : 0.32 mm

avg_abs_z_travel    : 3.96 mm/s

z_jarque-bera_jb    : 621.74
z_jarque-bera_p     : 0.00e+00

z_lin_mod_est_slope: 0.04 mm/s
z_lin_mod_adj_R²    : 42 %

z_poly40_mod_adj_R²: 76 %

z_dft_ampl_thresh   : 0.010 mm
>=threshold_maxfreq: 20.00 Hz
```

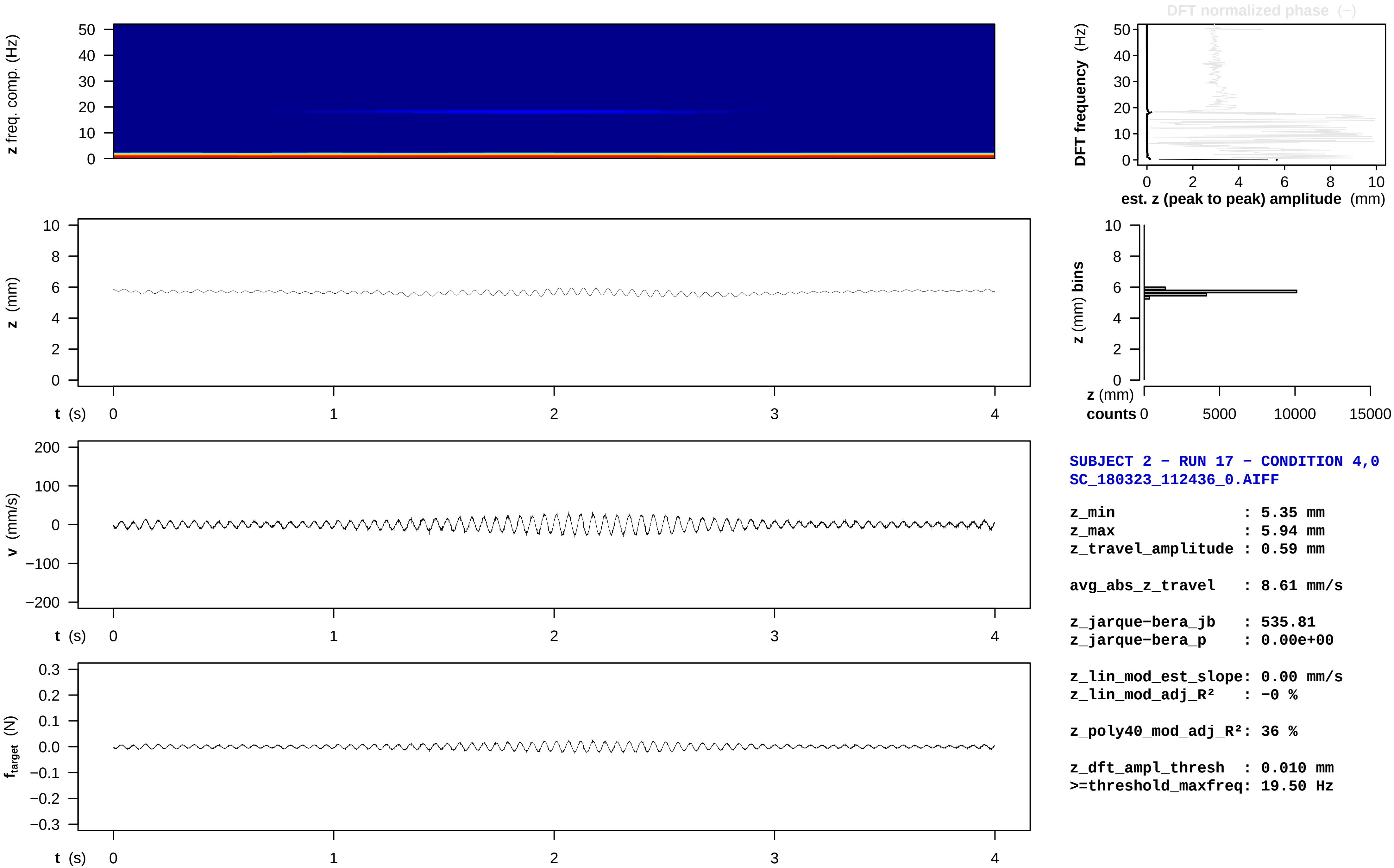

**SUBJECT 2 − RUN 17 − CONDITION 4,0**
**SC_180323_112436_0.AIFF**

```
z_min              : 5.35 mm
z_max              : 5.94 mm
z_travel_amplitude : 0.59 mm

avg_abs_z_travel   : 8.61 mm/s

z_jarque-bera_jb   : 535.81
z_jarque-bera_p    : 0.00e+00

z_lin_mod_est_slope: 0.00 mm/s
z_lin_mod_adj_R²   : −0 %

z_poly40_mod_adj_R²: 36 %

z_dft_ampl_thresh  : 0.010 mm
>=threshold_maxfreq: 19.50 Hz
```

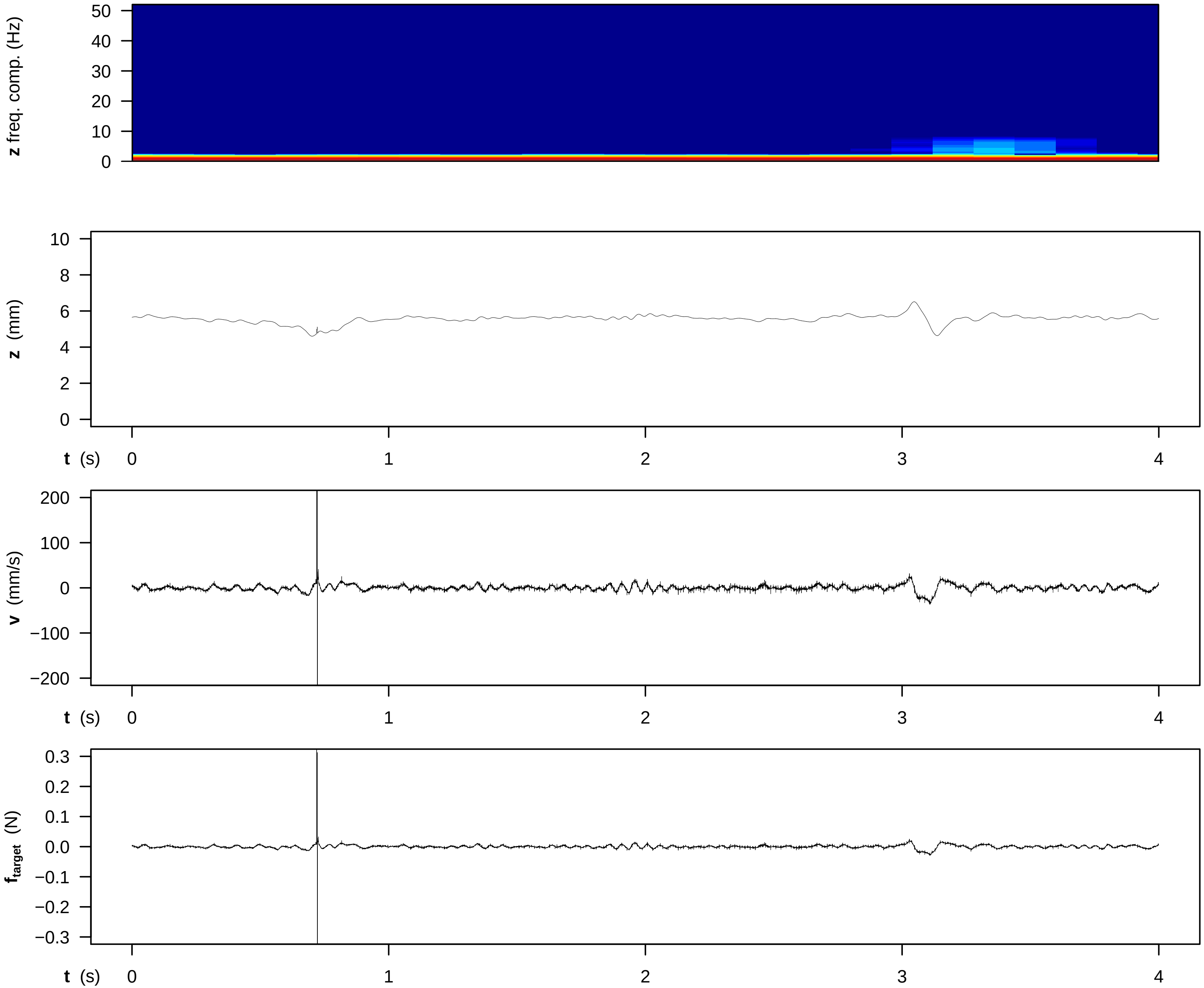

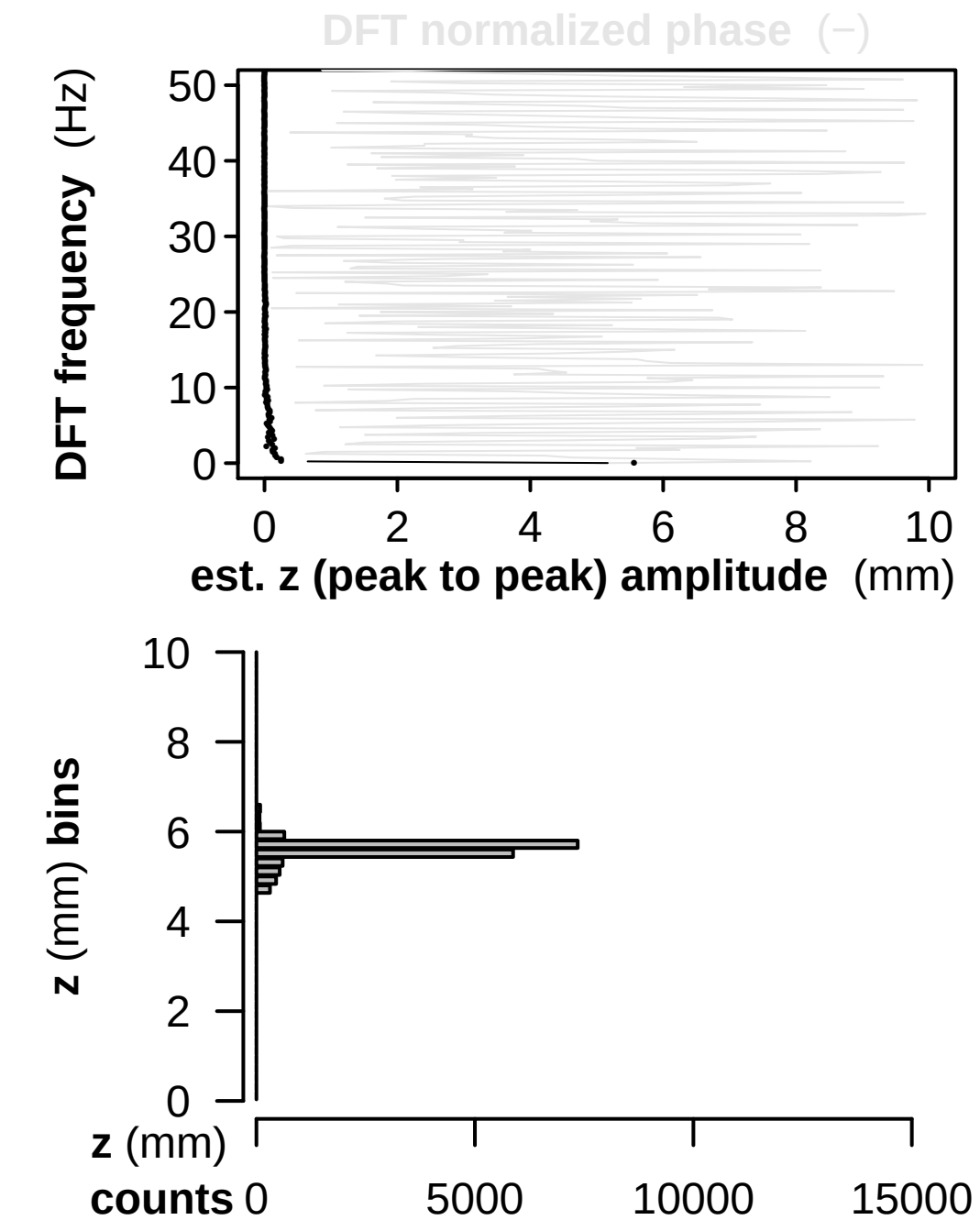

**SUBJECT 3 − RUN 01 − CONDITION 4,0**
**SC_180323_115448_0.AIFF**

```
z_min               : 4.60 mm
z_max               : 6.51 mm
z_travel_amplitude  : 1.91 mm

avg_abs_z_travel    : 5.34 mm/s

z_jarque-bera_jb    : 19283.55
z_jarque-bera_p     : 0.00e+00

z_lin_mod_est_slope: 0.07 mm/s
z_lin_mod_adj_R²    : 11 %

z_poly40_mod_adj_R²: 59 %

z_dft_ampl_thresh   : 0.010 mm
>=threshold_maxfreq: 22.75 Hz
```

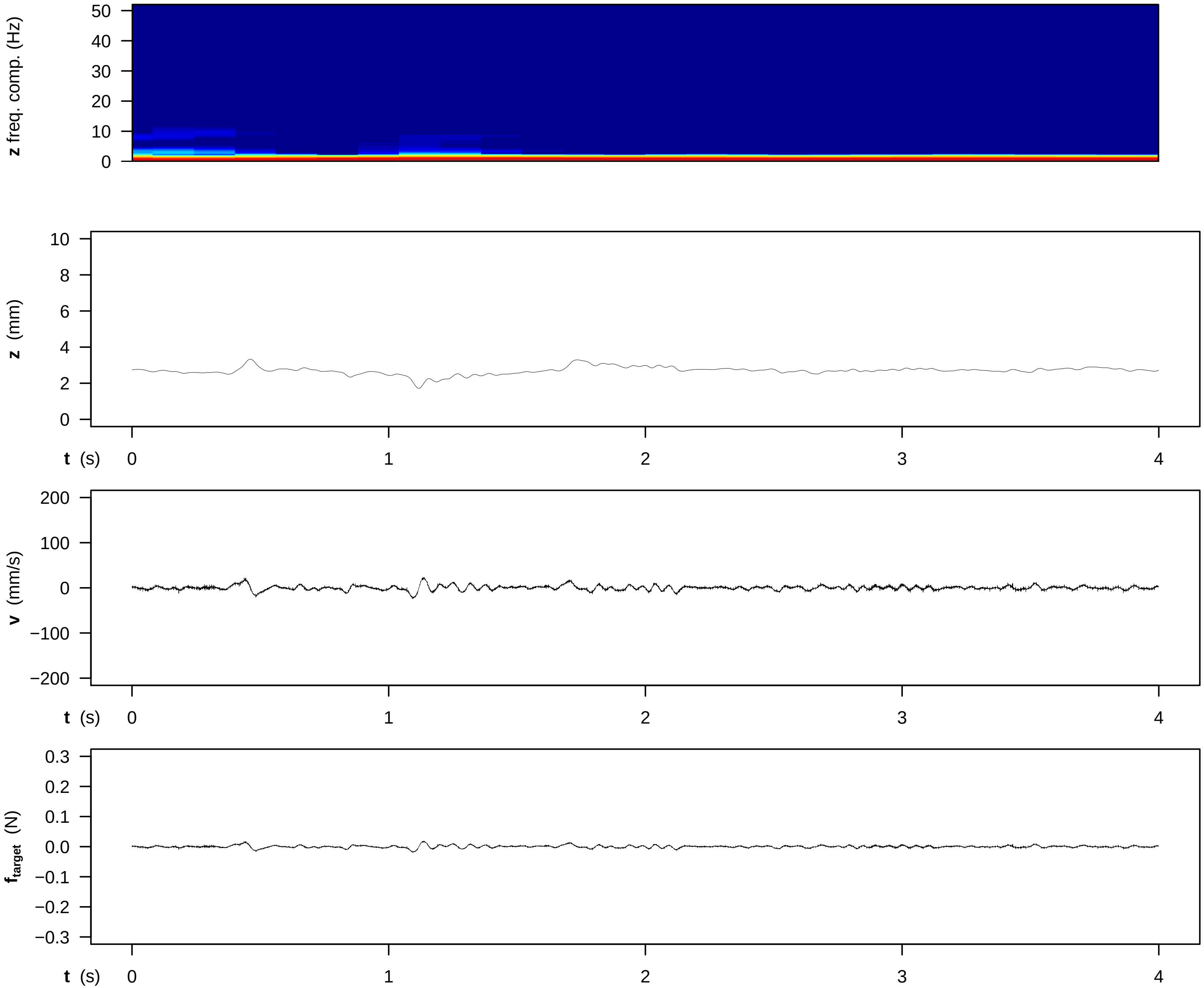

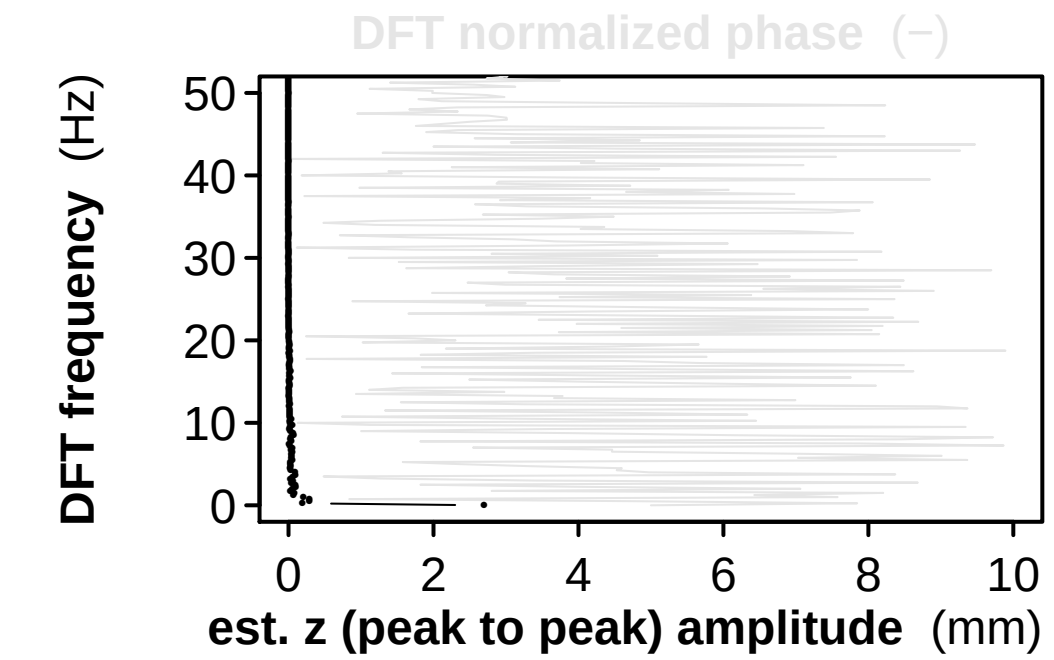

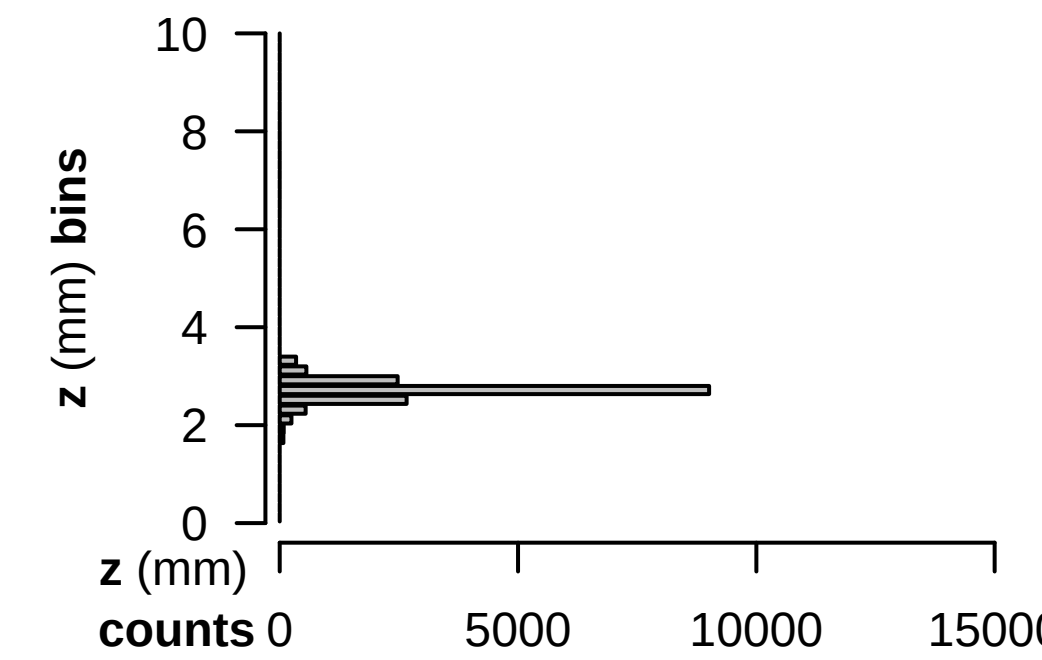

**SUBJECT 3 − RUN 31 − CONDITION 4,0**
**SC_180323_121259_0.AIFF**

```
z_min               : 1.71 mm
z_max               : 3.33 mm
z_travel_amplitude  : 1.62 mm

avg_abs_z_travel    : 4.03 mm/s

z_jarque-bera_jb    : 9987.96
z_jarque-bera_p     : 0.00e+00

z_lin_mod_est_slope: 0.04 mm/s
z_lin_mod_adj_R²    : 5 %

z_poly40_mod_adj_R²: 74 %

z_dft_ampl_thresh   : 0.010 mm
>=threshold_maxfreq: 21.00 Hz
```

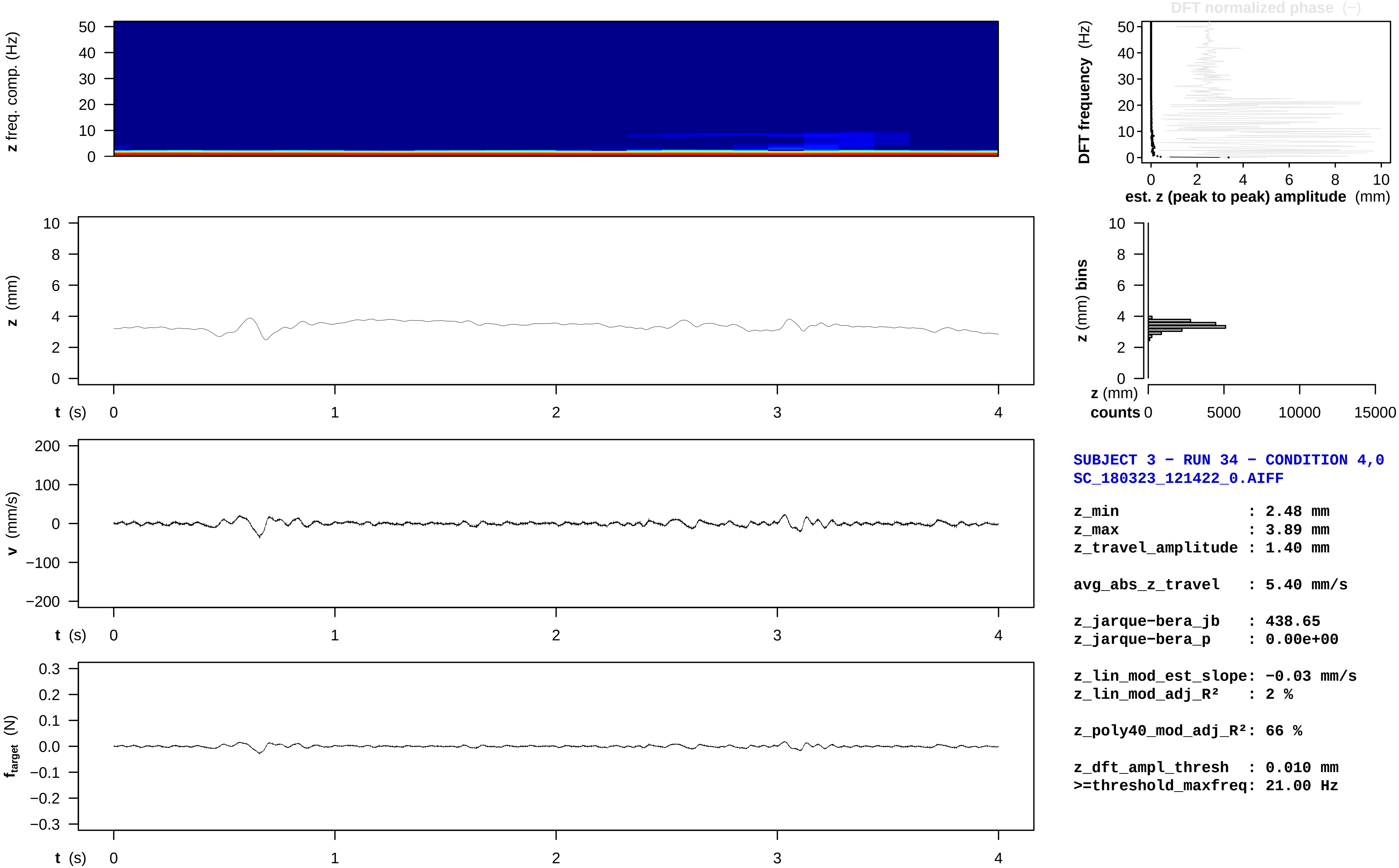

**SUBJECT 3 − RUN 34 − CONDITION 4,0**
**SC_180323_121422_0.AIFF**

```
z_min               : 2.48 mm
z_max               : 3.89 mm
z_travel_amplitude  : 1.40 mm

avg_abs_z_travel    : 5.40 mm/s

z_jarque-bera_jb    : 438.65
z_jarque-bera_p     : 0.00e+00

z_lin_mod_est_slope: −0.03 mm/s
z_lin_mod_adj_R²    : 2 %

z_poly40_mod_adj_R²: 66 %

z_dft_ampl_thresh   : 0.010 mm
>=threshold_maxfreq: 21.00 Hz
```

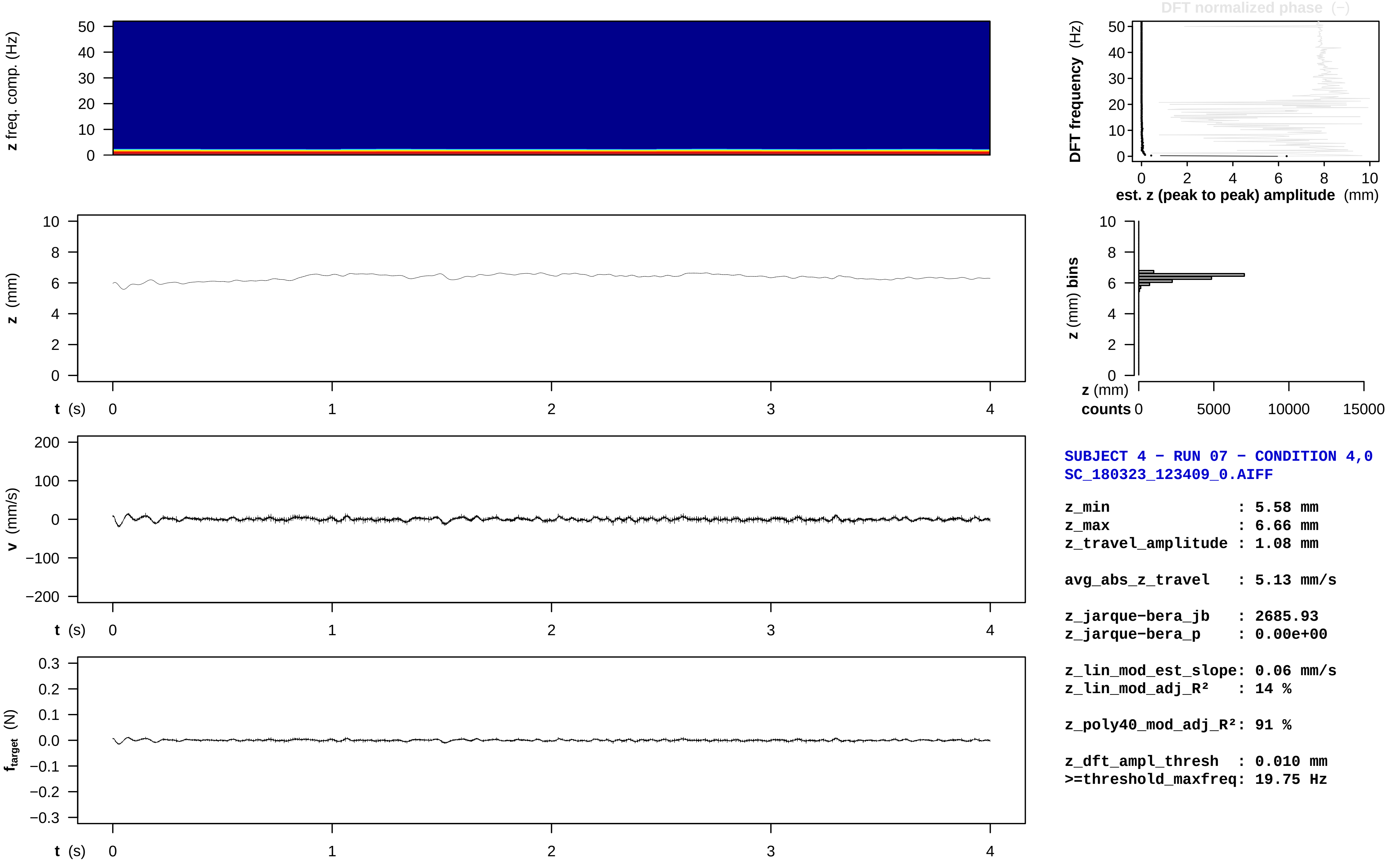

SUBJECT 4 − RUN 07 − CONDITION 4,0
SC_180323_123409_0.AIFF

z_min : 5.58 mm
z_max : 6.66 mm
z_travel_amplitude : 1.08 mm

avg_abs_z_travel : 5.13 mm/s

z_jarque-bera_jb : 2685.93
z_jarque-bera_p : 0.00e+00

z_lin_mod_est_slope: 0.06 mm/s
z_lin_mod_adj_R² : 14 %

z_poly40_mod_adj_R²: 91 %

z_dft_ampl_thresh : 0.010 mm
>=threshold_maxfreq: 19.75 Hz

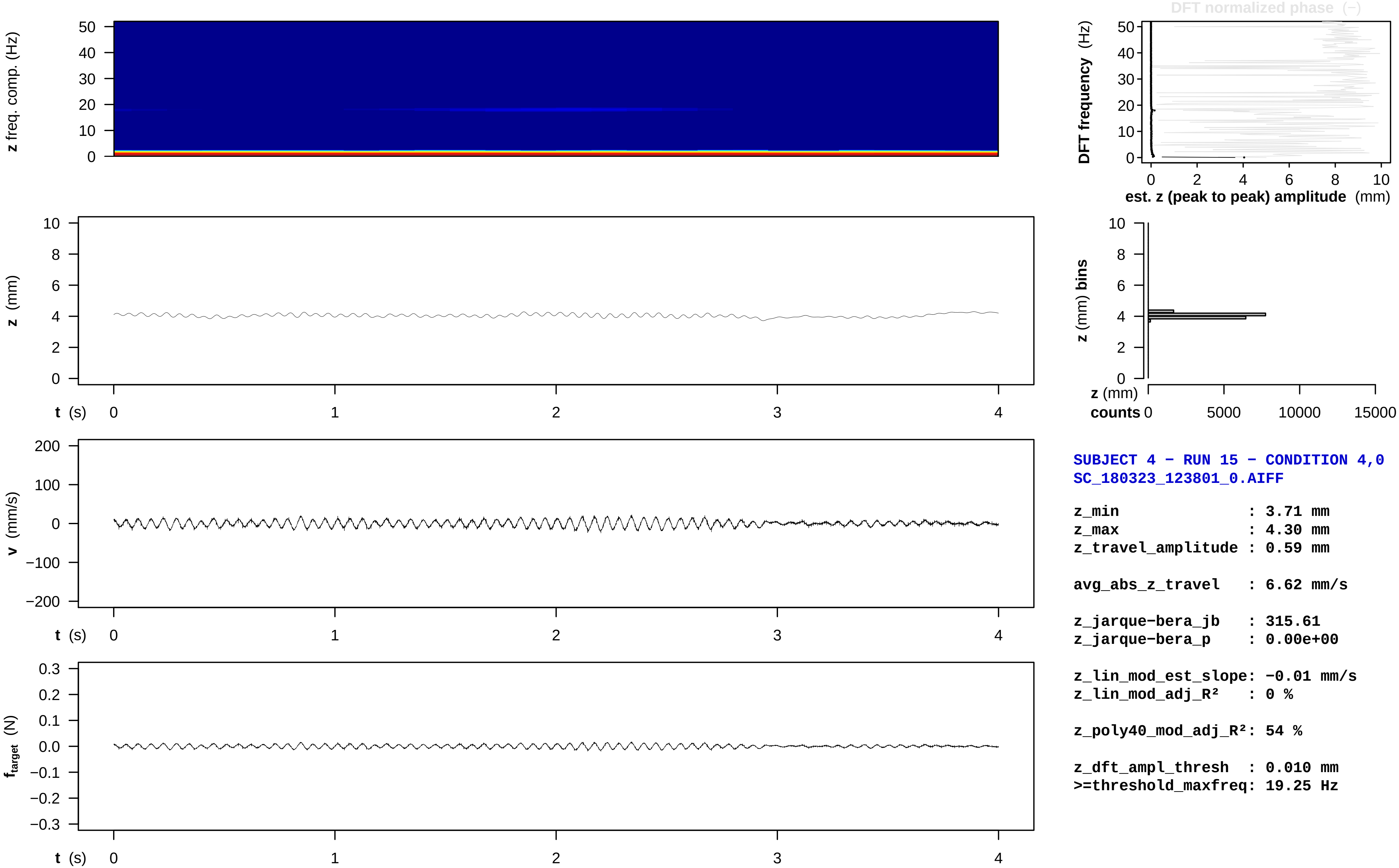

**SUBJECT 4 − RUN 15 − CONDITION 4,0**
**SC_180323_123801_0.AIFF**

```
z_min               : 3.71 mm
z_max               : 4.30 mm
z_travel_amplitude  : 0.59 mm

avg_abs_z_travel    : 6.62 mm/s

z_jarque-bera_jb    : 315.61
z_jarque-bera_p     : 0.00e+00

z_lin_mod_est_slope: −0.01 mm/s
z_lin_mod_adj_R²    : 0 %

z_poly40_mod_adj_R²: 54 %

z_dft_ampl_thresh   : 0.010 mm
>=threshold_maxfreq: 19.25 Hz
```

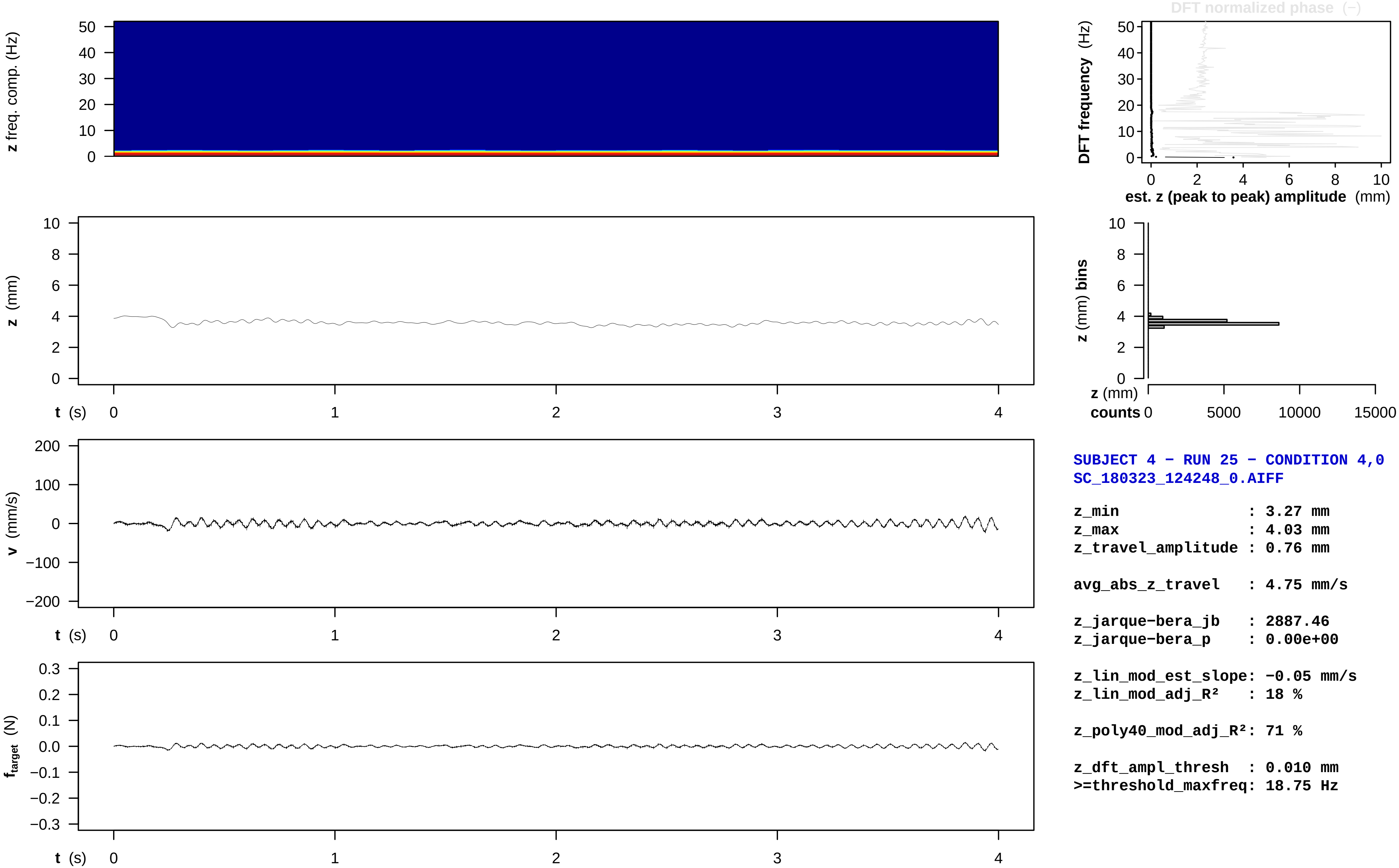

**SUBJECT 4 − RUN 25 − CONDITION 4,0**
**SC_180323_124248_0.AIFF**

```
z_min               : 3.27 mm
z_max               : 4.03 mm
z_travel_amplitude  : 0.76 mm

avg_abs_z_travel    : 4.75 mm/s

z_jarque-bera_jb    : 2887.46
z_jarque-bera_p     : 0.00e+00

z_lin_mod_est_slope: -0.05 mm/s
z_lin_mod_adj_R²    : 18 %

z_poly40_mod_adj_R²: 71 %

z_dft_ampl_thresh   : 0.010 mm
>=threshold_maxfreq: 18.75 Hz
```

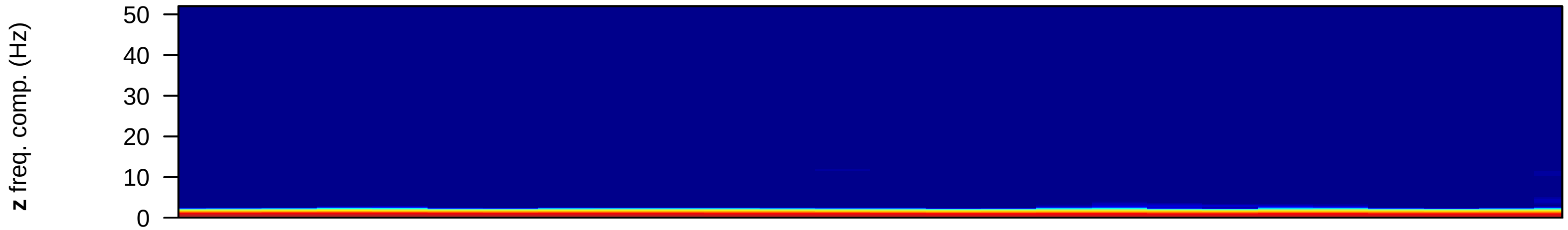

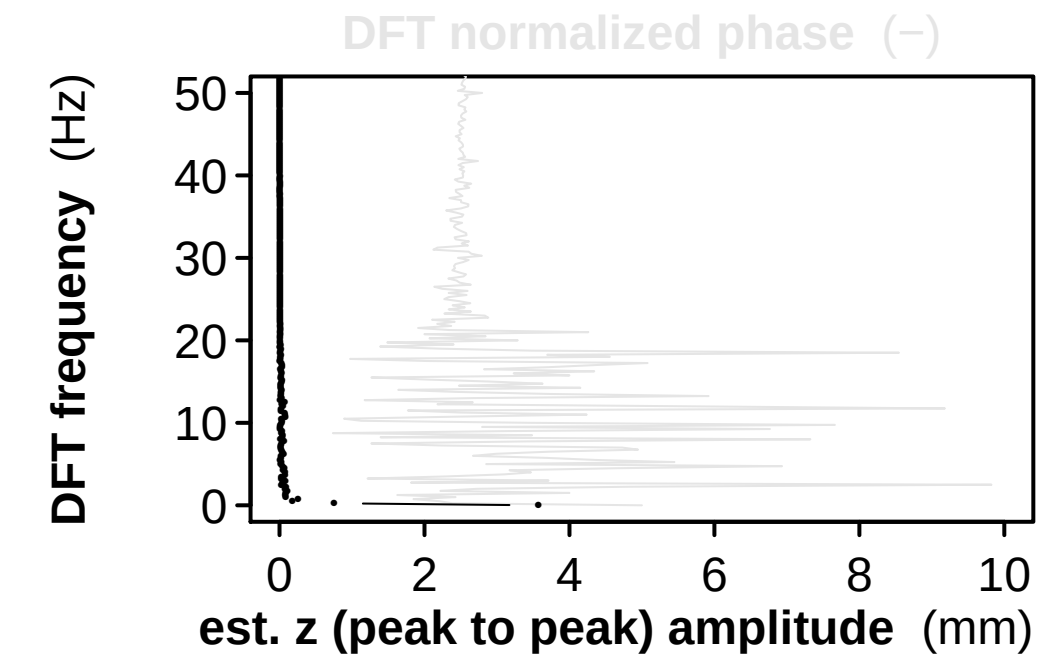

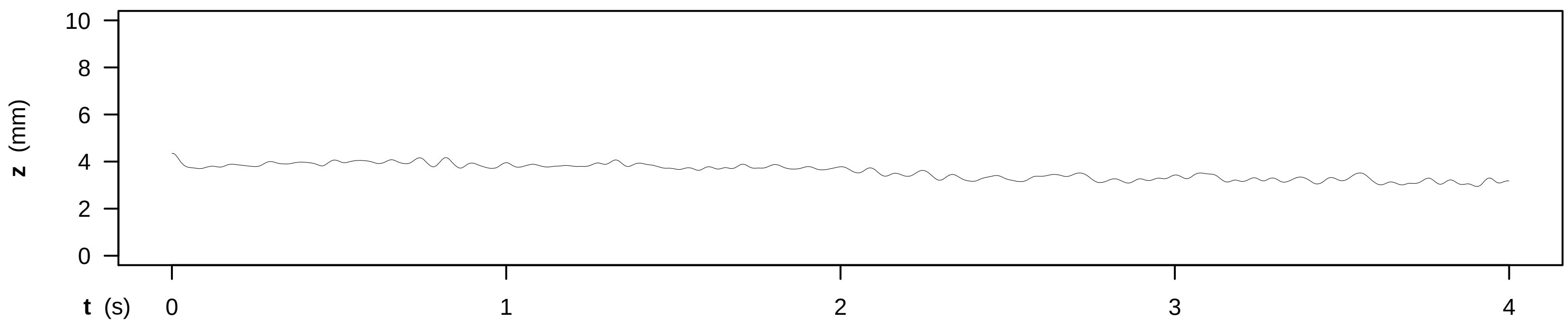

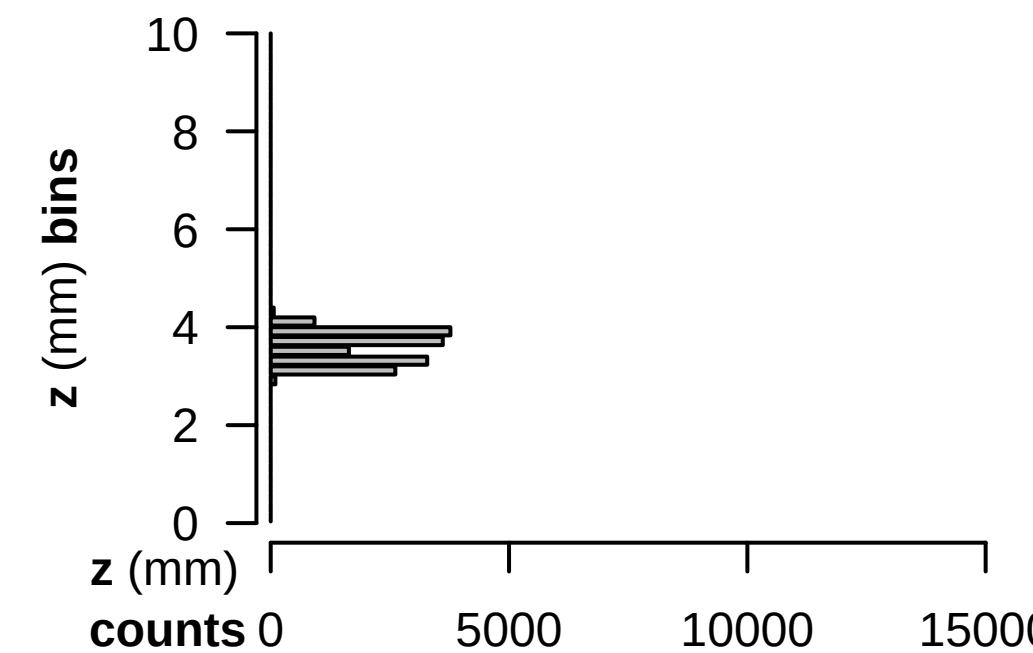

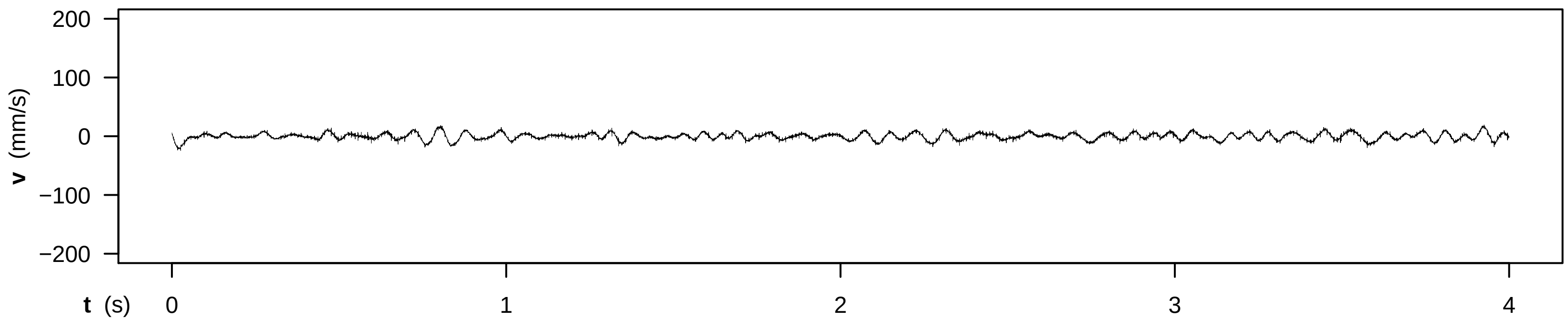

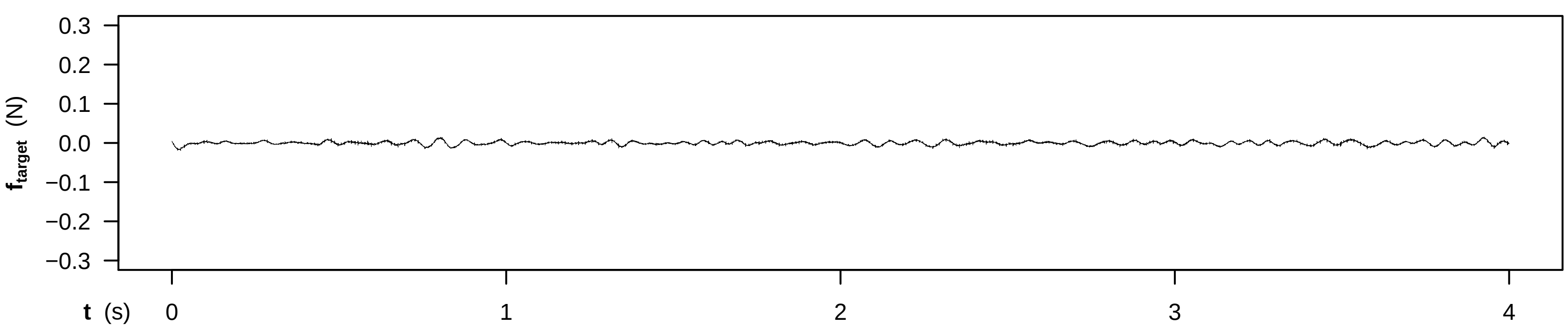

**SUBJECT 5 − RUN 20 − CONDITION 4,0**
**SC_180323_132722_0.AIFF**

```
z_min              : 2.94 mm
z_max              : 4.35 mm
z_travel_amplitude : 1.41 mm

avg_abs_z_travel   : 4.93 mm/s

z_jarque-bera_jb   : 1117.84
z_jarque-bera_p    : 0.00e+00

z_lin_mod_est_slope: −0.25 mm/s
z_lin_mod_adj_R²   : 82 %

z_poly40_mod_adj_R²: 92 %

z_dft_ampl_thresh  : 0.010 mm
>=threshold_maxfreq: 22.00 Hz
```

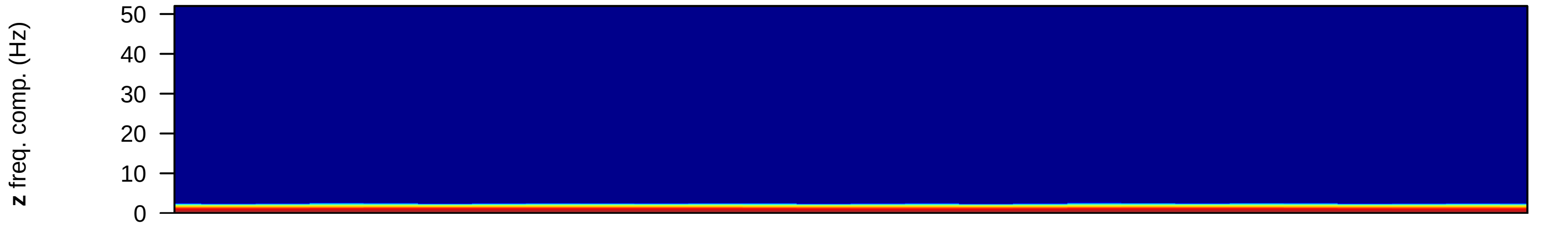

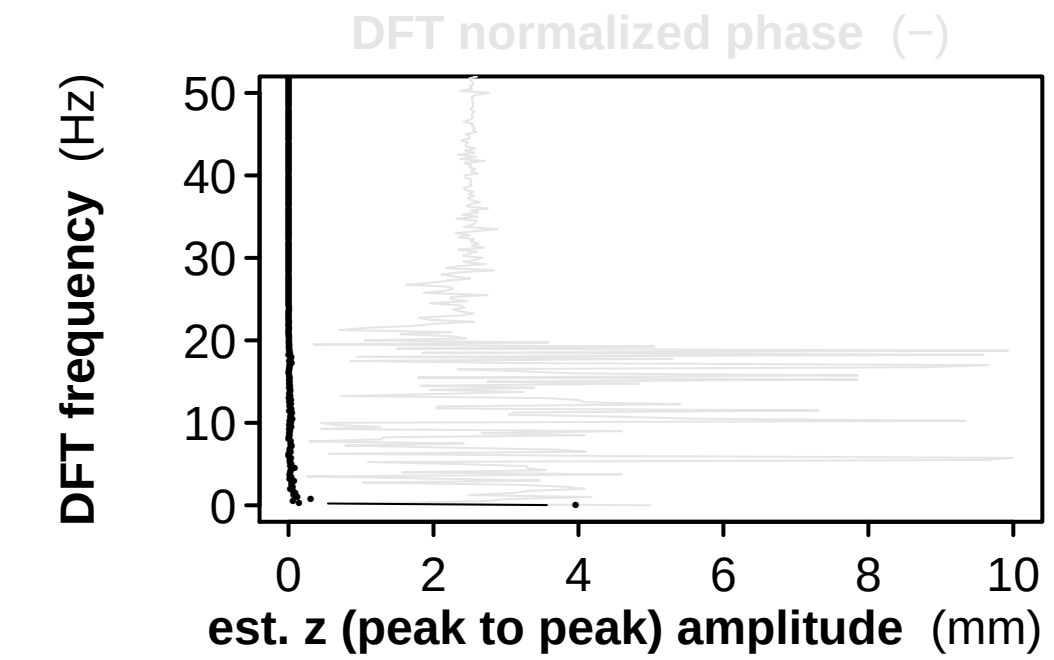

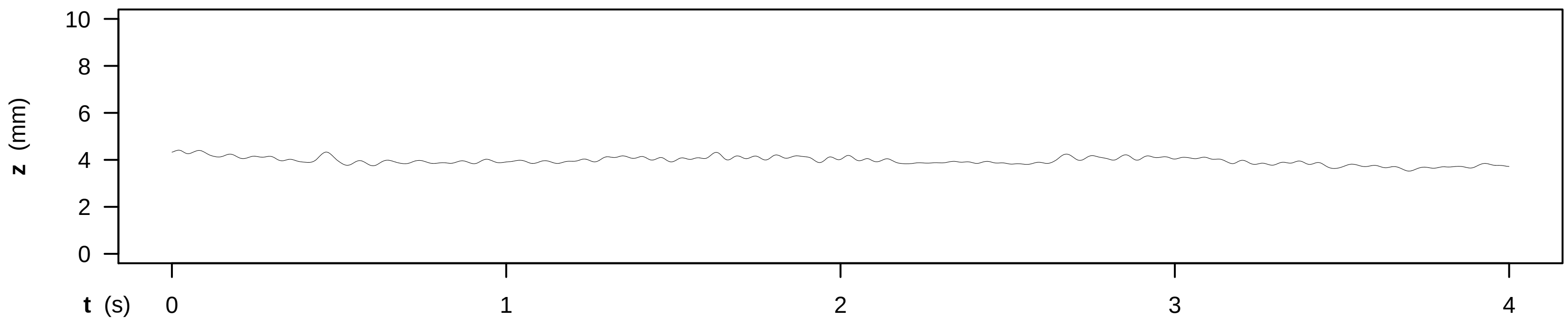

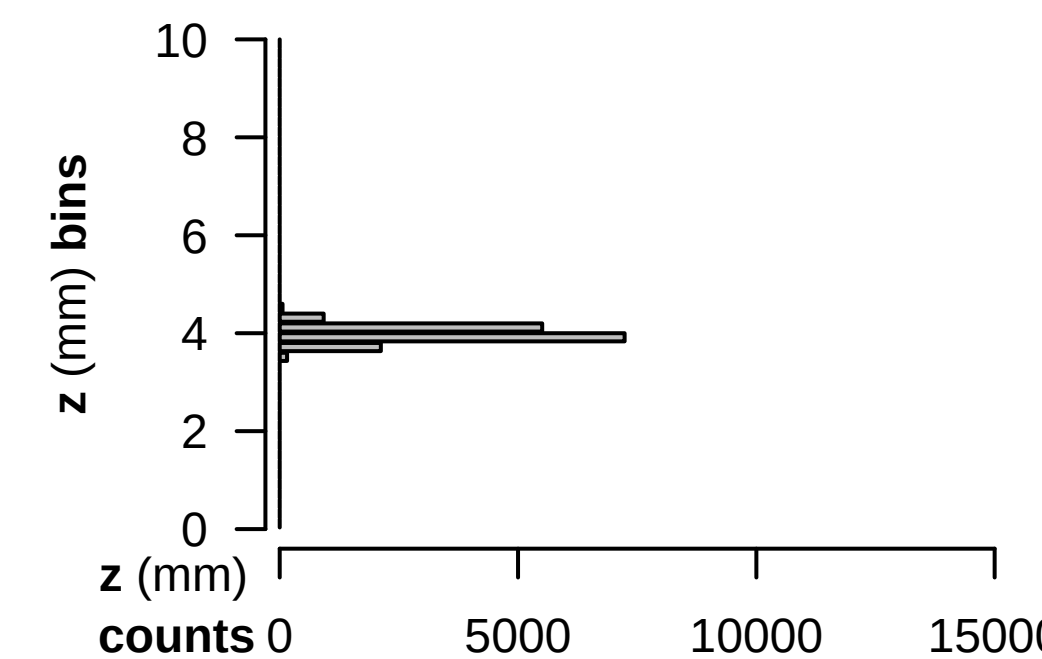

**SUBJECT 5 − RUN 26 − CONDITION 4,0**
**SC_180323_133051_0.AIFF**

```
z_min               : 3.52 mm
z_max               : 4.41 mm
z_travel_amplitude  : 0.89 mm

avg_abs_z_travel    : 4.38 mm/s

z_jarque-bera_jb    : 30.98
z_jarque-bera_p     : 1.87e-07

z_lin_mod_est_slope: −0.07 mm/s
z_lin_mod_adj_R²    : 26 %

z_poly40_mod_adj_R²: 80 %

z_dft_ampl_thresh   : 0.010 mm
>=threshold_maxfreq: 19.50 Hz
```

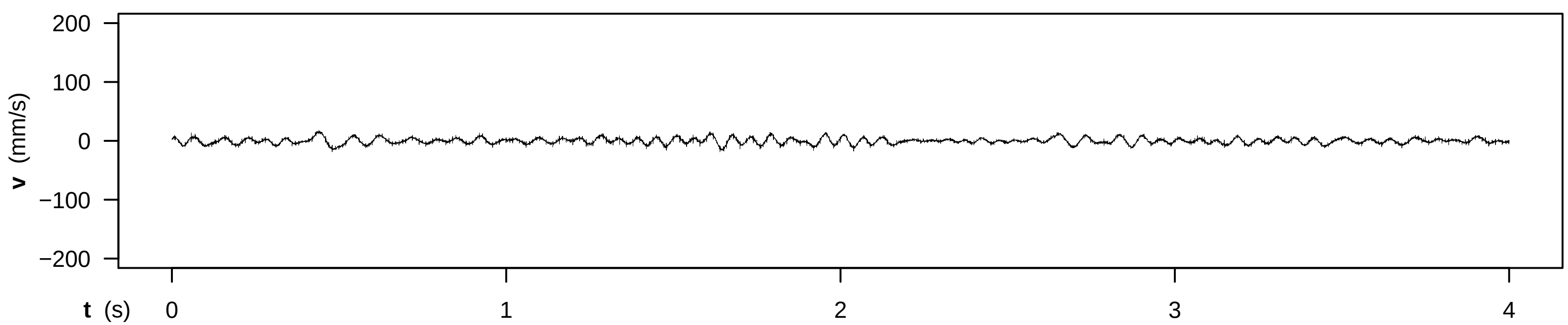

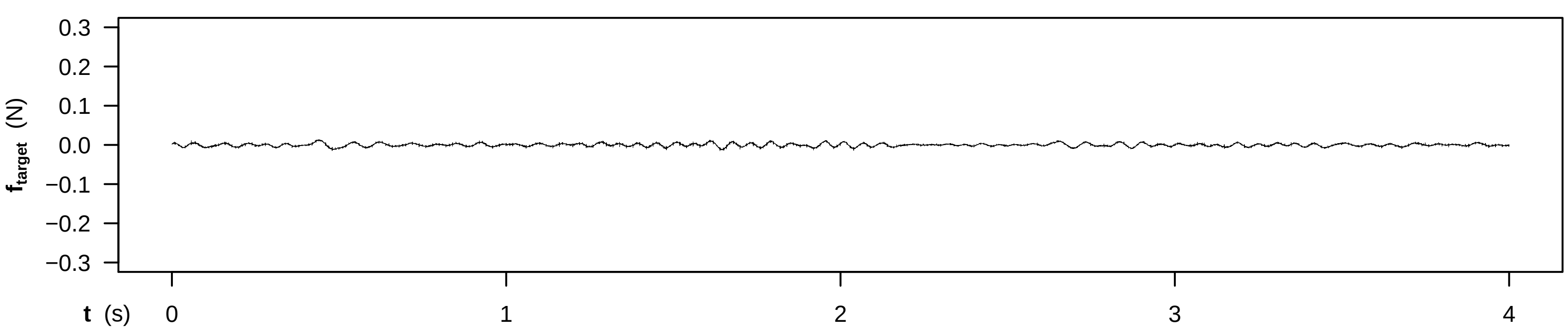

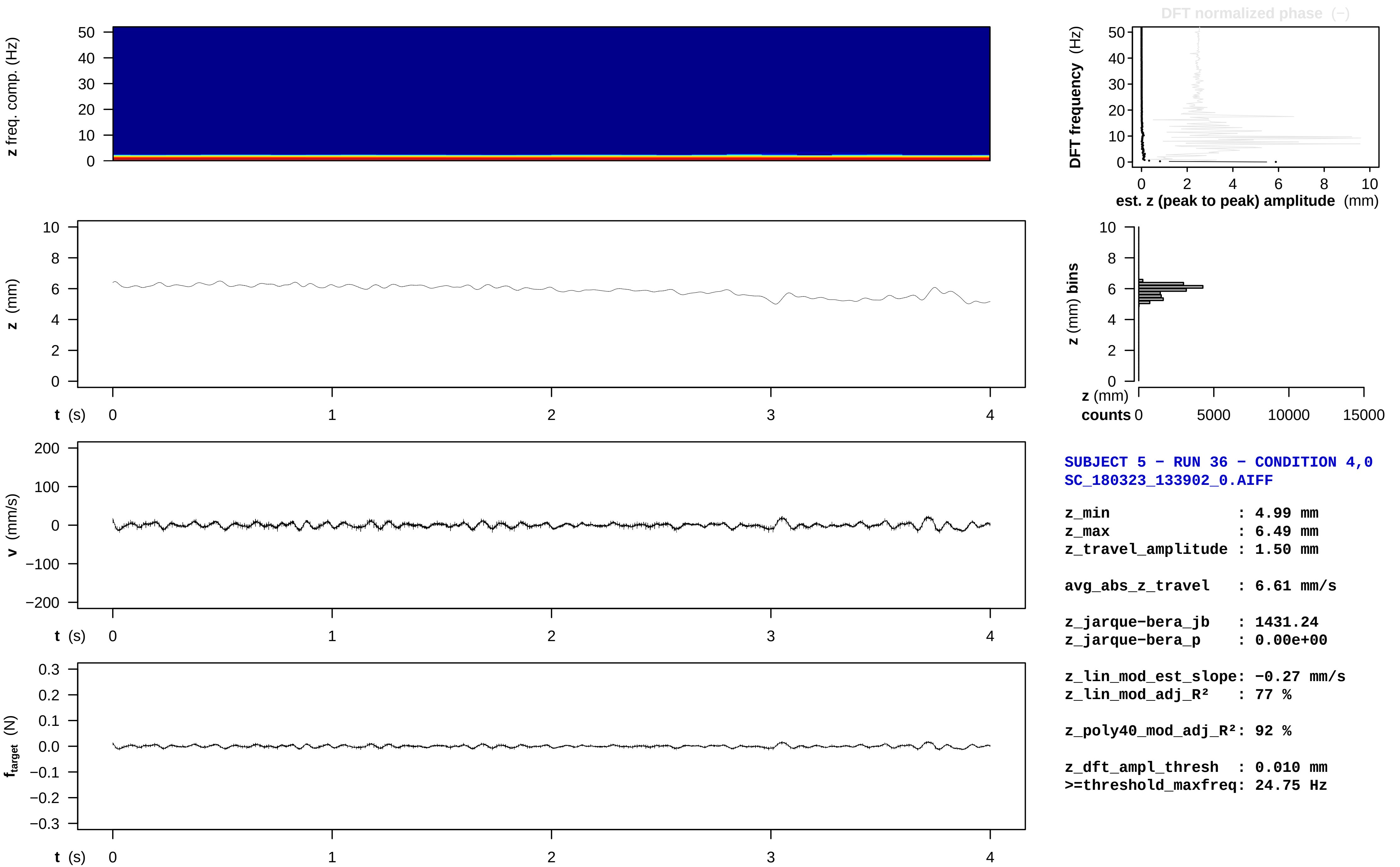

**SUBJECT 5 − RUN 36 − CONDITION 4,0**
**SC_180323_133902_0.AIFF**

```
z_min               : 4.99 mm
z_max               : 6.49 mm
z_travel_amplitude : 1.50 mm

avg_abs_z_travel   : 6.61 mm/s

z_jarque-bera_jb   : 1431.24
z_jarque-bera_p    : 0.00e+00

z_lin_mod_est_slope: −0.27 mm/s
z_lin_mod_adj_R²   : 77 %

z_poly40_mod_adj_R²: 92 %

z_dft_ampl_thresh  : 0.010 mm
>=threshold_maxfreq: 24.75 Hz
```

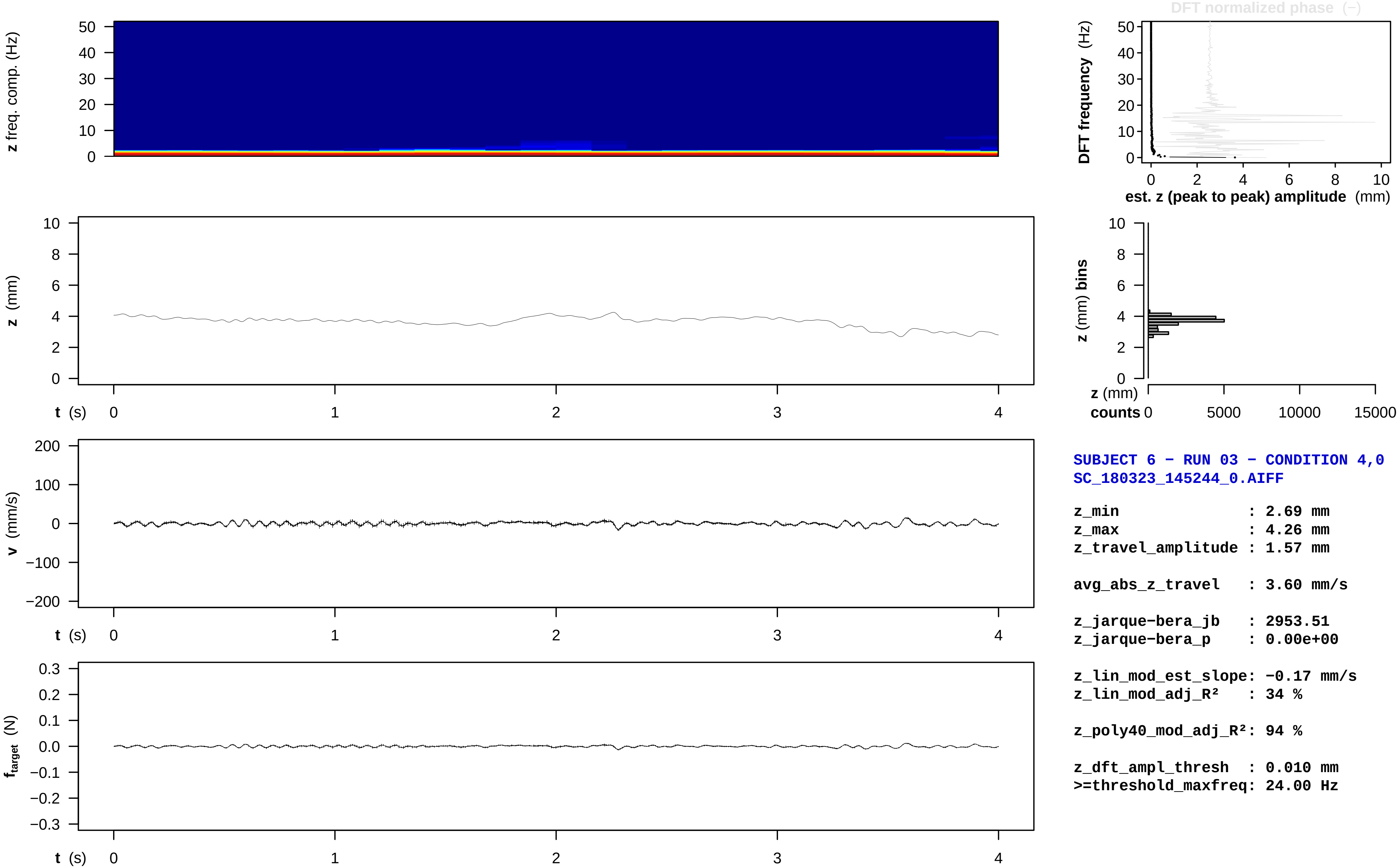

**SUBJECT 6 – RUN 03 – CONDITION 4,0**
**SC_180323_145244_0.AIFF**

```
z_min               : 2.69 mm
z_max               : 4.26 mm
z_travel_amplitude  : 1.57 mm

avg_abs_z_travel    : 3.60 mm/s

z_jarque-bera_jb    : 2953.51
z_jarque-bera_p     : 0.00e+00

z_lin_mod_est_slope: -0.17 mm/s
z_lin_mod_adj_R²    : 34 %

z_poly40_mod_adj_R²: 94 %

z_dft_ampl_thresh   : 0.010 mm
>=threshold_maxfreq: 24.00 Hz
```

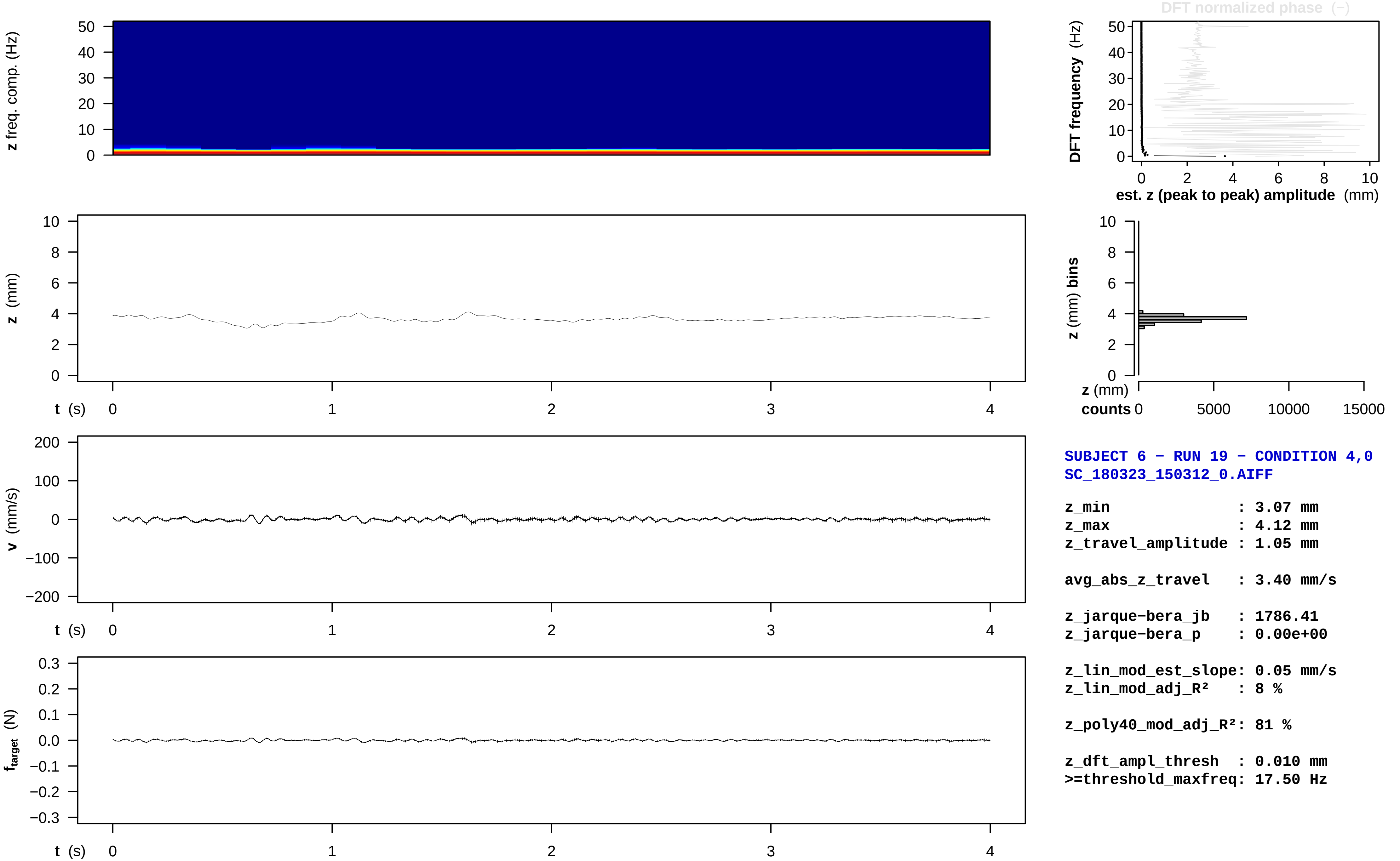

**SUBJECT 6 − RUN 19 − CONDITION 4,0**
**SC_180323_150312_0.AIFF**

```
z_min               : 3.07 mm
z_max               : 4.12 mm
z_travel_amplitude  : 1.05 mm

avg_abs_z_travel    : 3.40 mm/s

z_jarque-bera_jb    : 1786.41
z_jarque-bera_p     : 0.00e+00

z_lin_mod_est_slope: 0.05 mm/s
z_lin_mod_adj_R²    : 8 %

z_poly40_mod_adj_R²: 81 %

z_dft_ampl_thresh   : 0.010 mm
>=threshold_maxfreq: 17.50 Hz
```

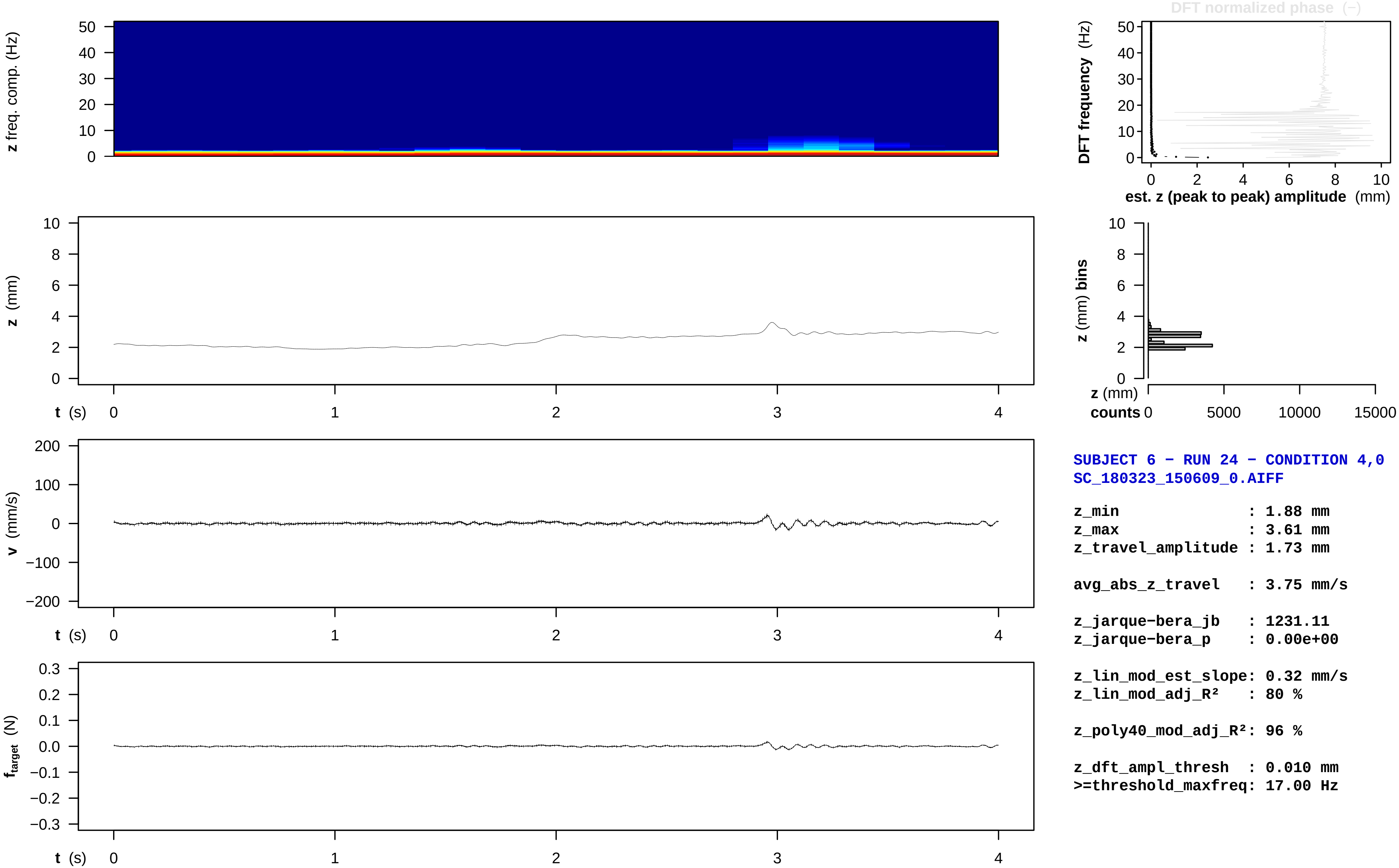

**SUBJECT 6 − RUN 24 − CONDITION 4,0**
**SC_180323_150609_0.AIFF**

```
z_min               : 1.88 mm
z_max               : 3.61 mm
z_travel_amplitude  : 1.73 mm

avg_abs_z_travel    : 3.75 mm/s

z_jarque-bera_jb    : 1231.11
z_jarque-bera_p     : 0.00e+00

z_lin_mod_est_slope: 0.32 mm/s
z_lin_mod_adj_R²    : 80 %

z_poly40_mod_adj_R²: 96 %

z_dft_ampl_thresh   : 0.010 mm
>=threshold_maxfreq: 17.00 Hz
```

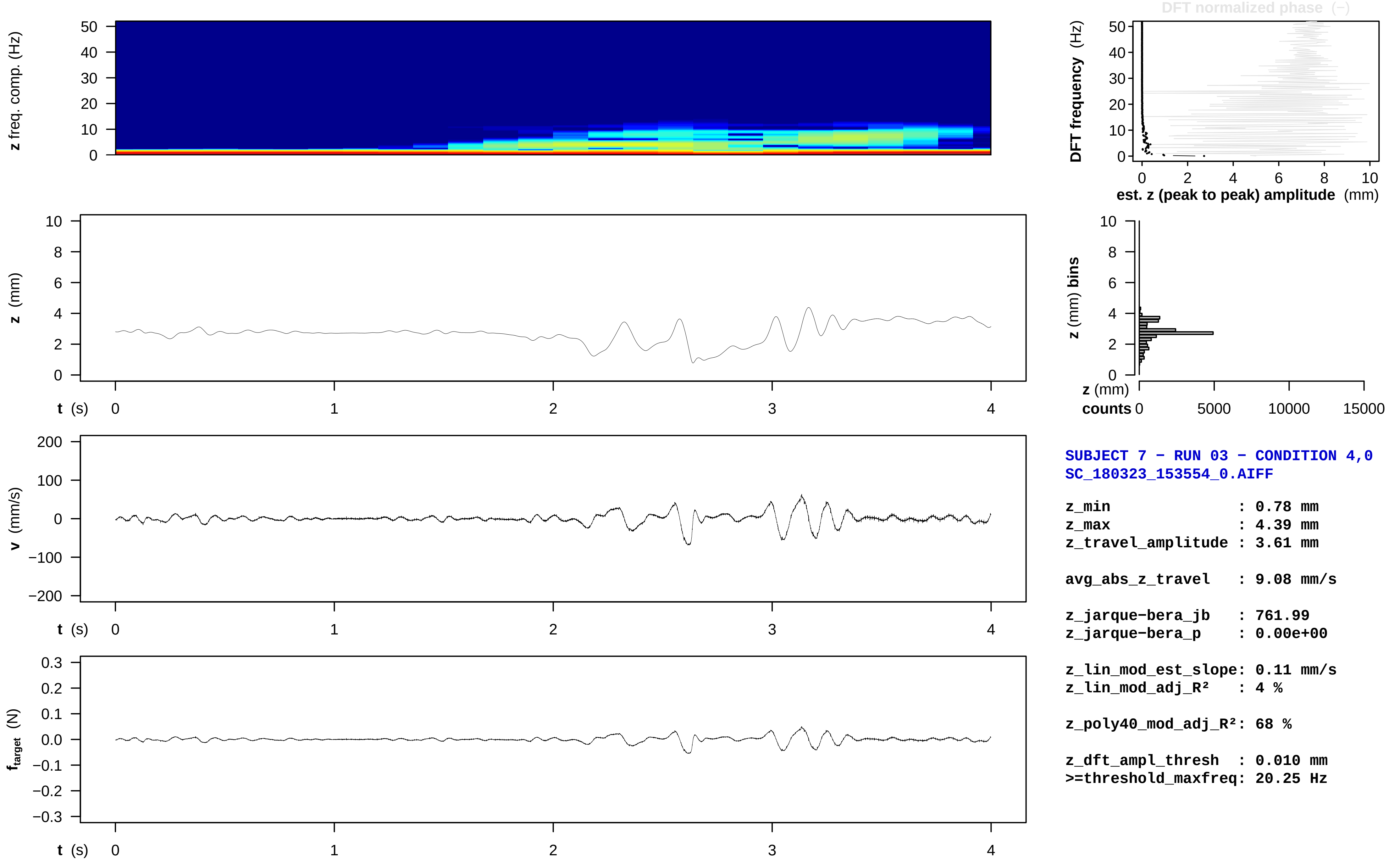

**SUBJECT 7 − RUN 03 − CONDITION 4,0**
**SC_180323_153554_0.AIFF**

```
z_min               : 0.78 mm
z_max               : 4.39 mm
z_travel_amplitude  : 3.61 mm

avg_abs_z_travel    : 9.08 mm/s

z_jarque-bera_jb    : 761.99
z_jarque-bera_p     : 0.00e+00

z_lin_mod_est_slope: 0.11 mm/s
z_lin_mod_adj_R²    : 4 %

z_poly40_mod_adj_R²: 68 %

z_dft_ampl_thresh   : 0.010 mm
>=threshold_maxfreq: 20.25 Hz
```

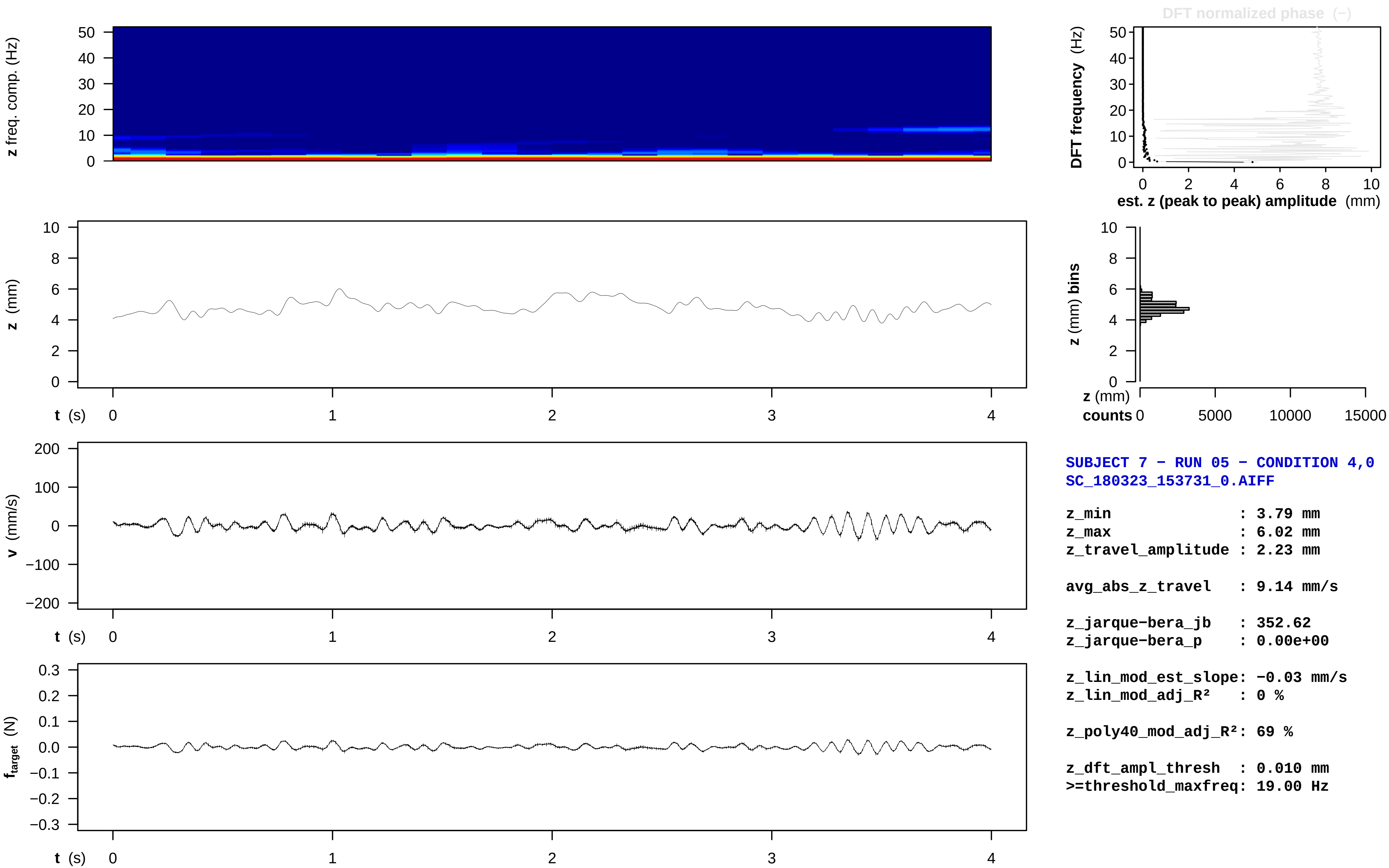

**SUBJECT 7 − RUN 05 − CONDITION 4,0**
**SC_180323_153731_0.AIFF**

```
z_min              : 3.79 mm
z_max              : 6.02 mm
z_travel_amplitude : 2.23 mm

avg_abs_z_travel   : 9.14 mm/s

z_jarque-bera_jb   : 352.62
z_jarque-bera_p    : 0.00e+00

z_lin_mod_est_slope: −0.03 mm/s
z_lin_mod_adj_R²   : 0 %

z_poly40_mod_adj_R²: 69 %

z_dft_ampl_thresh  : 0.010 mm
>=threshold_maxfreq: 19.00 Hz
```

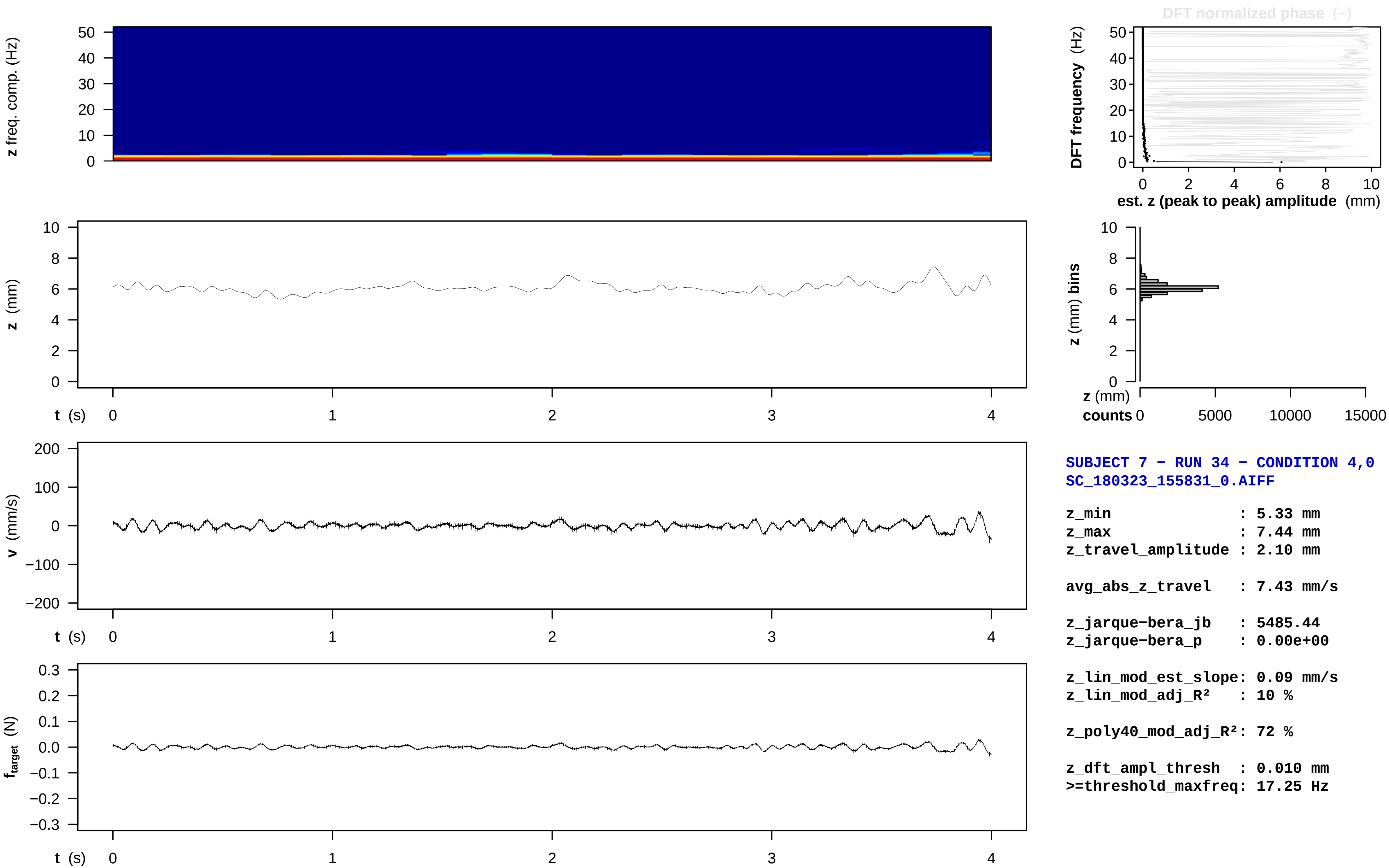

**SUBJECT 7 − RUN 34 − CONDITION 4,0**
**SC_180323_155831_0.AIFF**

```
z_min               : 5.33 mm
z_max               : 7.44 mm
z_travel_amplitude  : 2.10 mm

avg_abs_z_travel    : 7.43 mm/s

z_jarque-bera_jb    : 5485.44
z_jarque-bera_p     : 0.00e+00

z_lin_mod_est_slope: 0.09 mm/s
z_lin_mod_adj_R²    : 10 %

z_poly40_mod_adj_R²: 72 %

z_dft_ampl_thresh   : 0.010 mm
>=threshold_maxfreq: 17.25 Hz
```

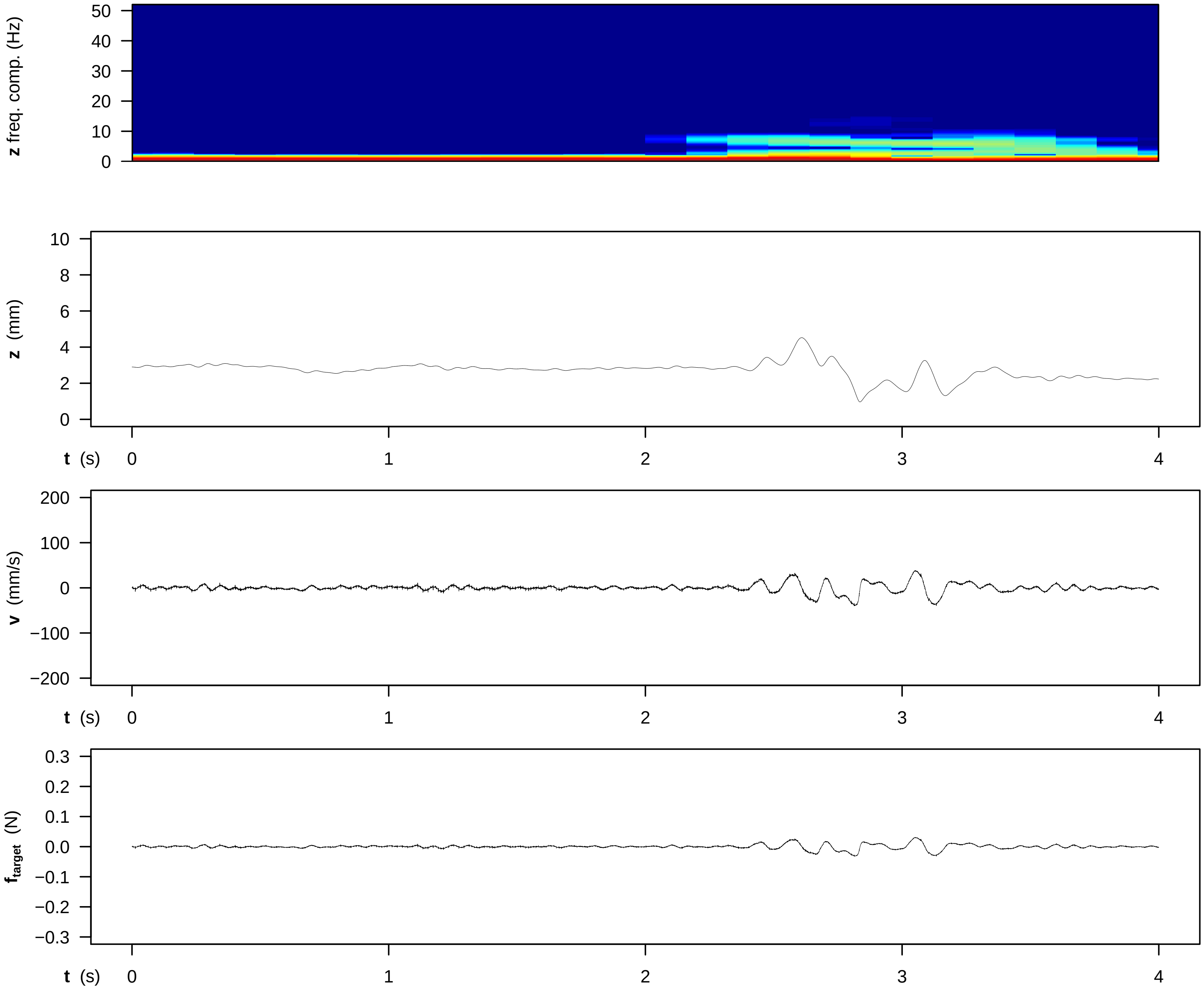

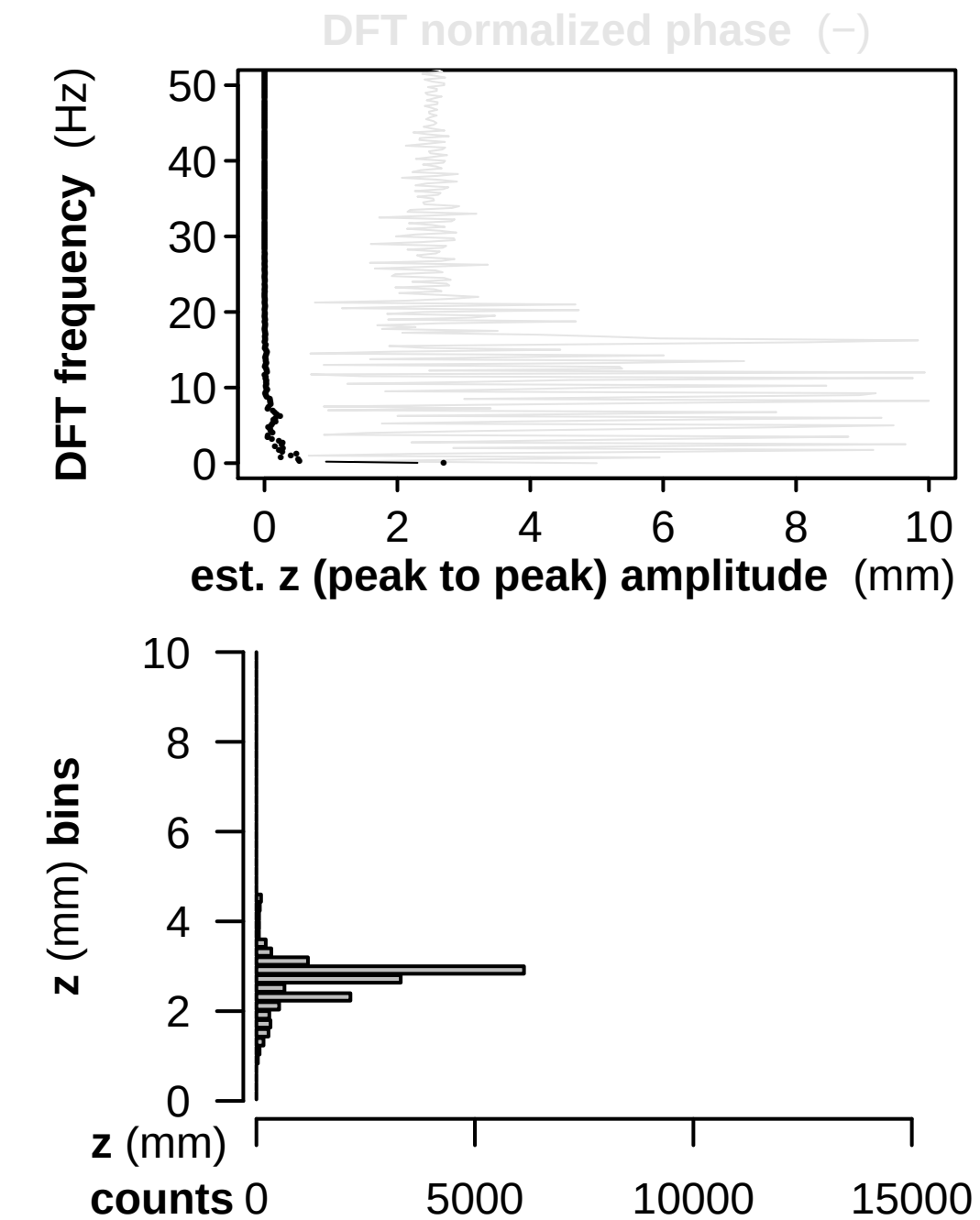

**SUBJECT 8 − RUN 02 − CONDITION 4,0**
**SC_180323_164549_0.AIFF**

**z_min : 0.96 mm**
**z_max : 4.52 mm**
**z_travel_amplitude : 3.56 mm**

**avg_abs_z_travel : 6.31 mm/s**

**z_jarque-bera_jb : 6738.71**
**z_jarque-bera_p : 0.00e+00**

**z_lin_mod_est_slope: −0.18 mm/s**
**z_lin_mod_adj_$R^2$ : 20 %**

**z_poly40_mod_adj_$R^2$: 65 %**

**z_dft_ampl_thresh : 0.010 mm**
**>=threshold_maxfreq: 20.75 Hz**

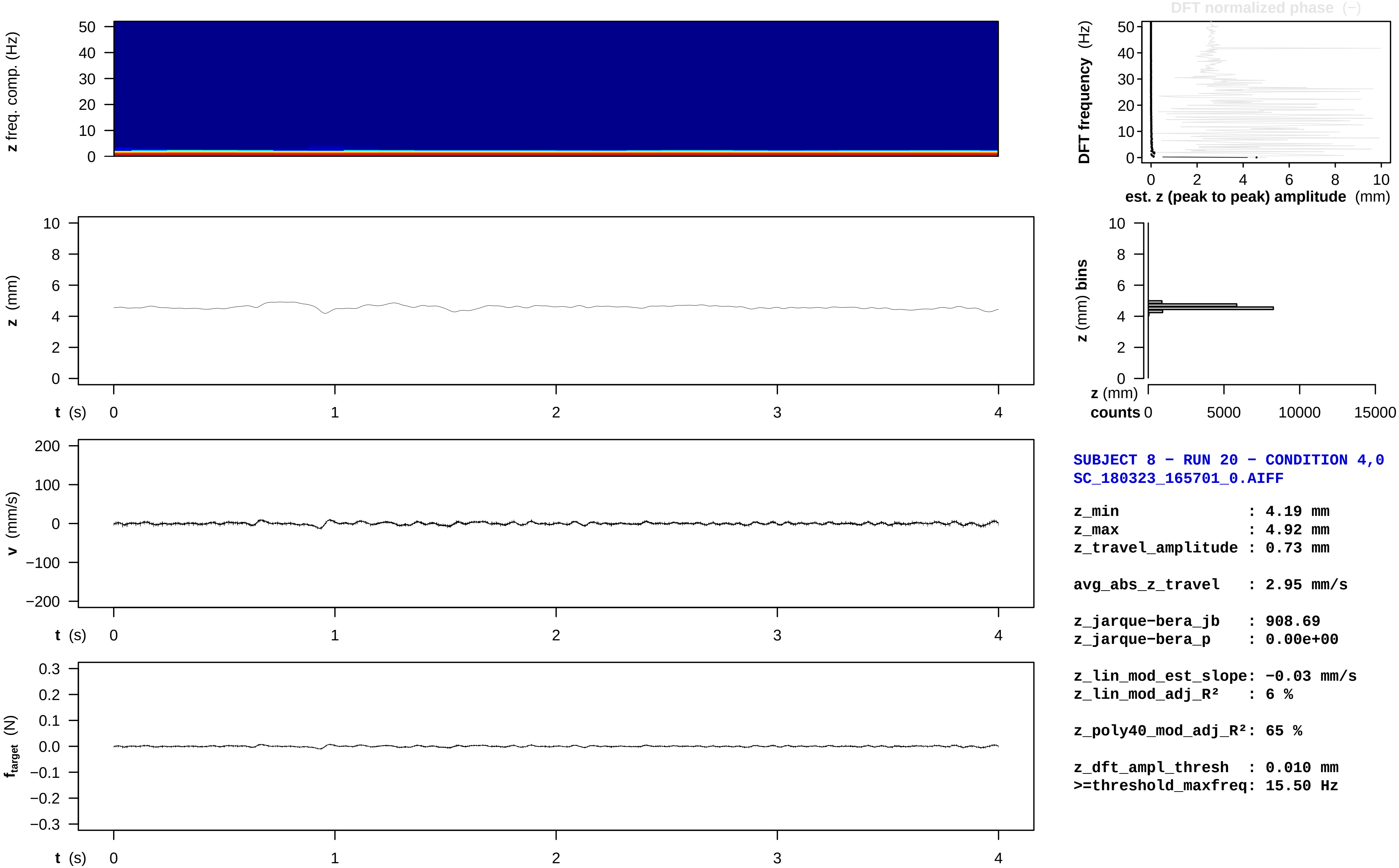

**SUBJECT 8 − RUN 20 − CONDITION 4,0**
**SC_180323_165701_0.AIFF**

```
z_min              : 4.19 mm
z_max              : 4.92 mm
z_travel_amplitude : 0.73 mm

avg_abs_z_travel   : 2.95 mm/s

z_jarque-bera_jb   : 908.69
z_jarque-bera_p    : 0.00e+00

z_lin_mod_est_slope: -0.03 mm/s
z_lin_mod_adj_R²   : 6 %

z_poly40_mod_adj_R²: 65 %

z_dft_ampl_thresh  : 0.010 mm
>=threshold_maxfreq: 15.50 Hz
```

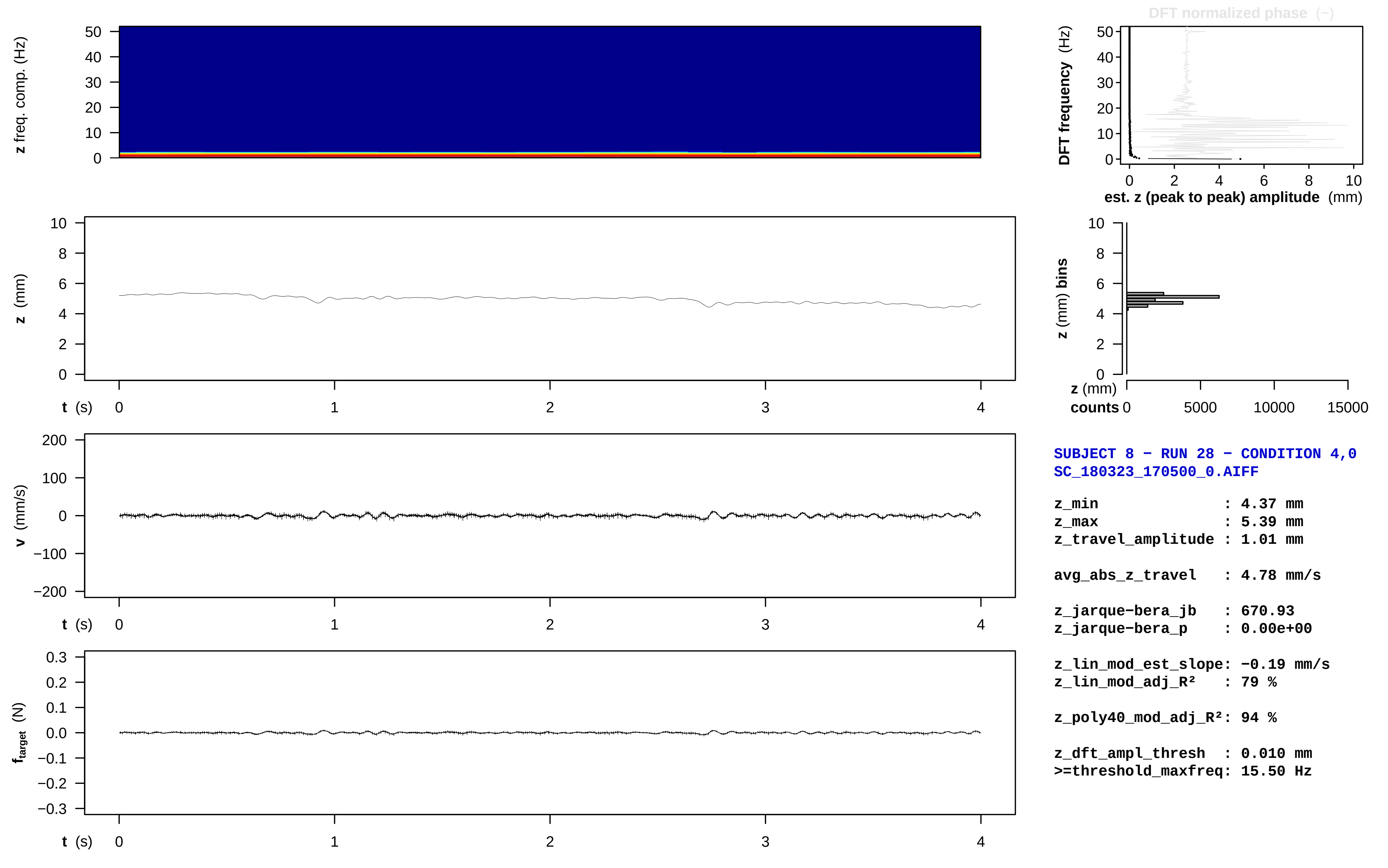
z freq. comp. (Hz)
DFT normalized phase (−)
DFT frequency (Hz)
est. z (peak to peak) amplitude (mm)
z (mm)
t (s)
z (mm) bins
z (mm)
counts
v (mm/s)
f_target (N)
SUBJECT 8 − RUN 28 − CONDITION 4,0
SC_180323_170500_0.AIFF
z_min : 4.37 mm
z_max : 5.39 mm
z_travel_amplitude : 1.01 mm
avg_abs_z_travel : 4.78 mm/s
z_jarque-bera_jb : 670.93
z_jarque-bera_p : 0.00e+00
z_lin_mod_est_slope: −0.19 mm/s
z_lin_mod_adj_R² : 79 %
z_poly40_mod_adj_R²: 94 %
z_dft_ampl_thresh : 0.010 mm
>=threshold_maxfreq: 15.50 Hz

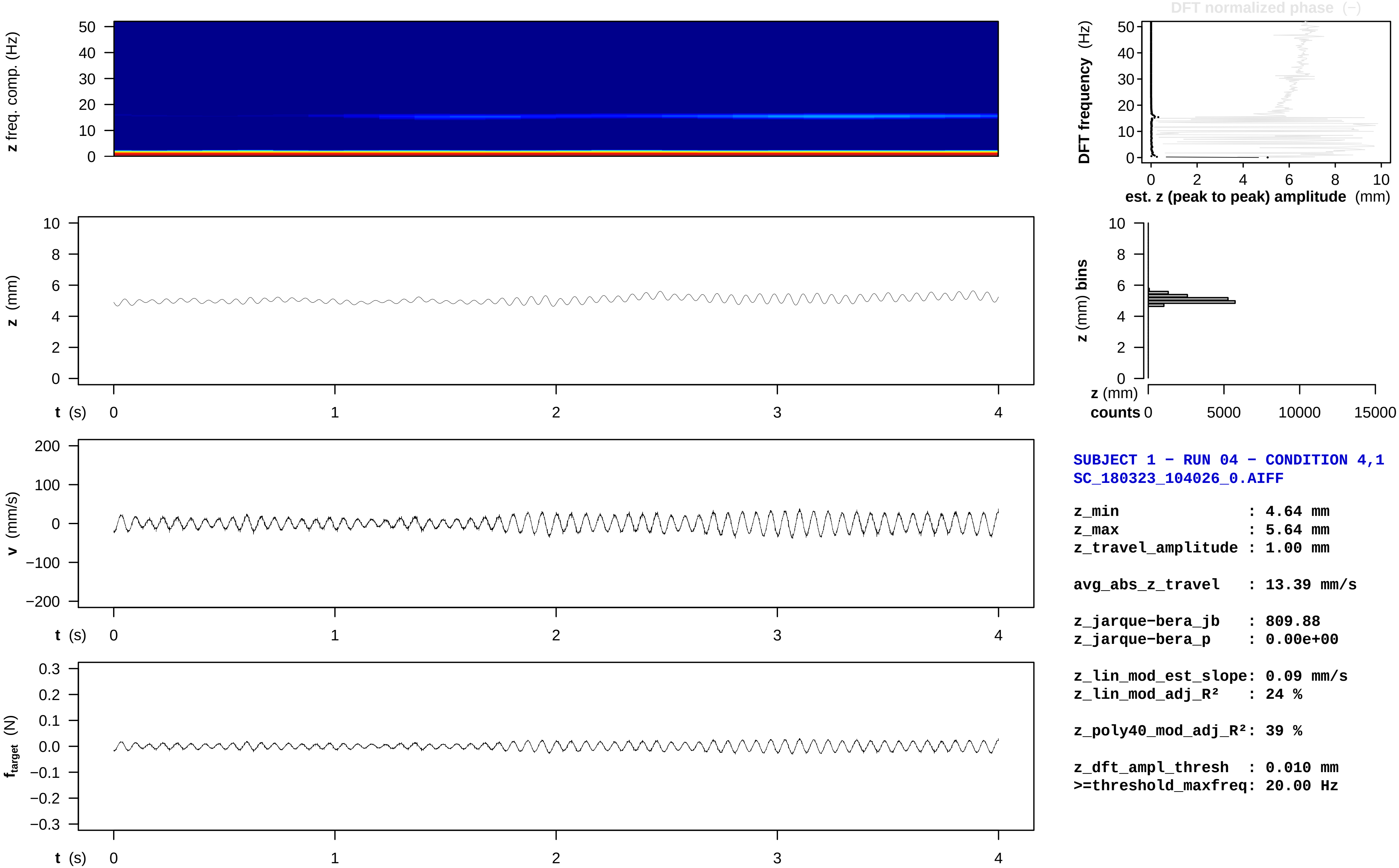

**SUBJECT 1 − RUN 04 − CONDITION 4,1**
**SC_180323_104026_0.AIFF**

```
z_min              : 4.64 mm
z_max              : 5.64 mm
z_travel_amplitude : 1.00 mm

avg_abs_z_travel   : 13.39 mm/s

z_jarque-bera_jb   : 809.88
z_jarque-bera_p    : 0.00e+00

z_lin_mod_est_slope: 0.09 mm/s
z_lin_mod_adj_R²   : 24 %

z_poly40_mod_adj_R²: 39 %

z_dft_ampl_thresh  : 0.010 mm
>=threshold_maxfreq: 20.00 Hz
```

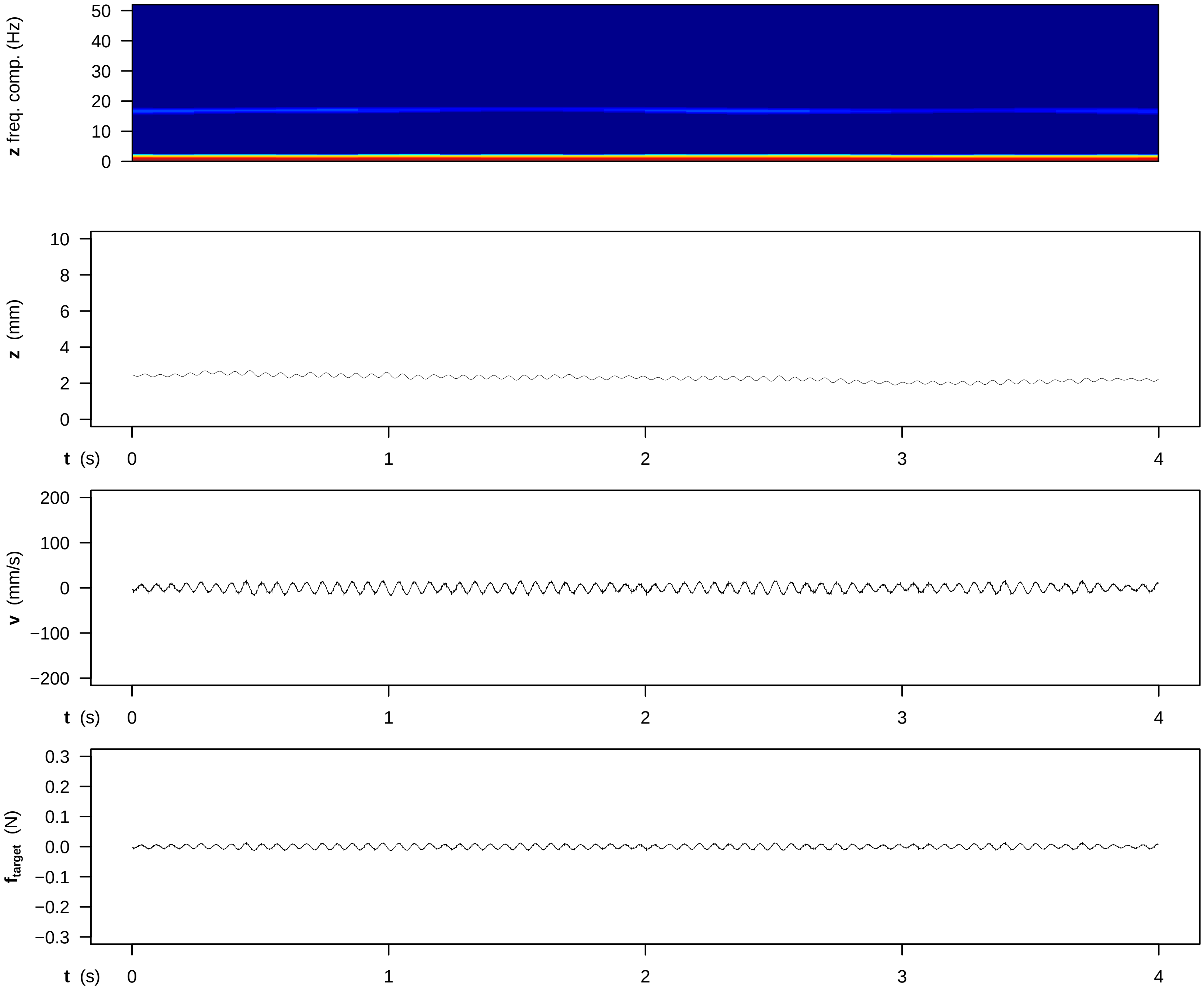

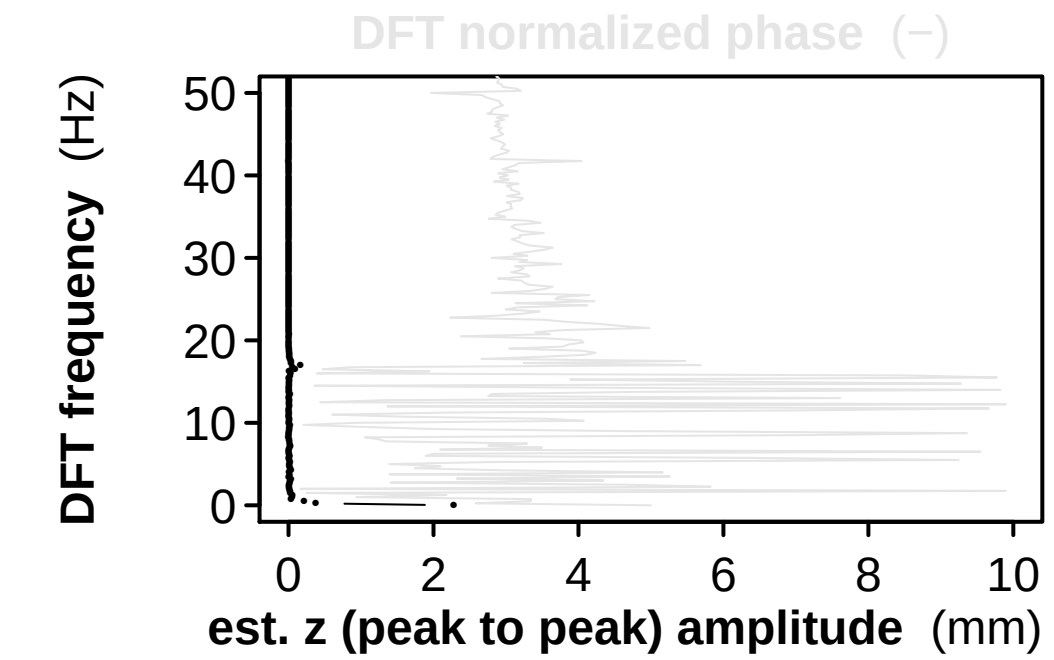

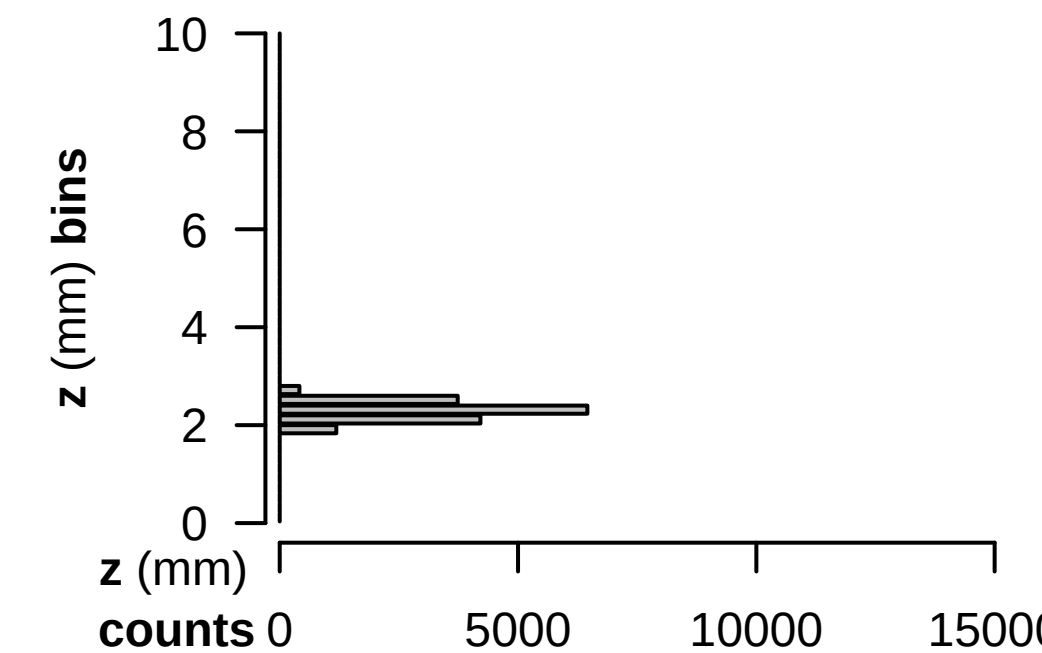

**SUBJECT 1 − RUN 16 − CONDITION 4,1**
**SC_180323_104826_0.AIFF**

```
z_min              : 1.89 mm
z_max              : 2.70 mm
z_travel_amplitude : 0.81 mm

avg_abs_z_travel   : 7.56 mm/s

z_jarque-bera_jb   : 281.11
z_jarque-bera_p    : 0.00e+00

z_lin_mod_est_slope: -0.12 mm/s
z_lin_mod_adj_R²   : 65 %

z_poly40_mod_adj_R²: 80 %

z_dft_ampl_thresh  : 0.010 mm
>=threshold_maxfreq: 18.50 Hz
```

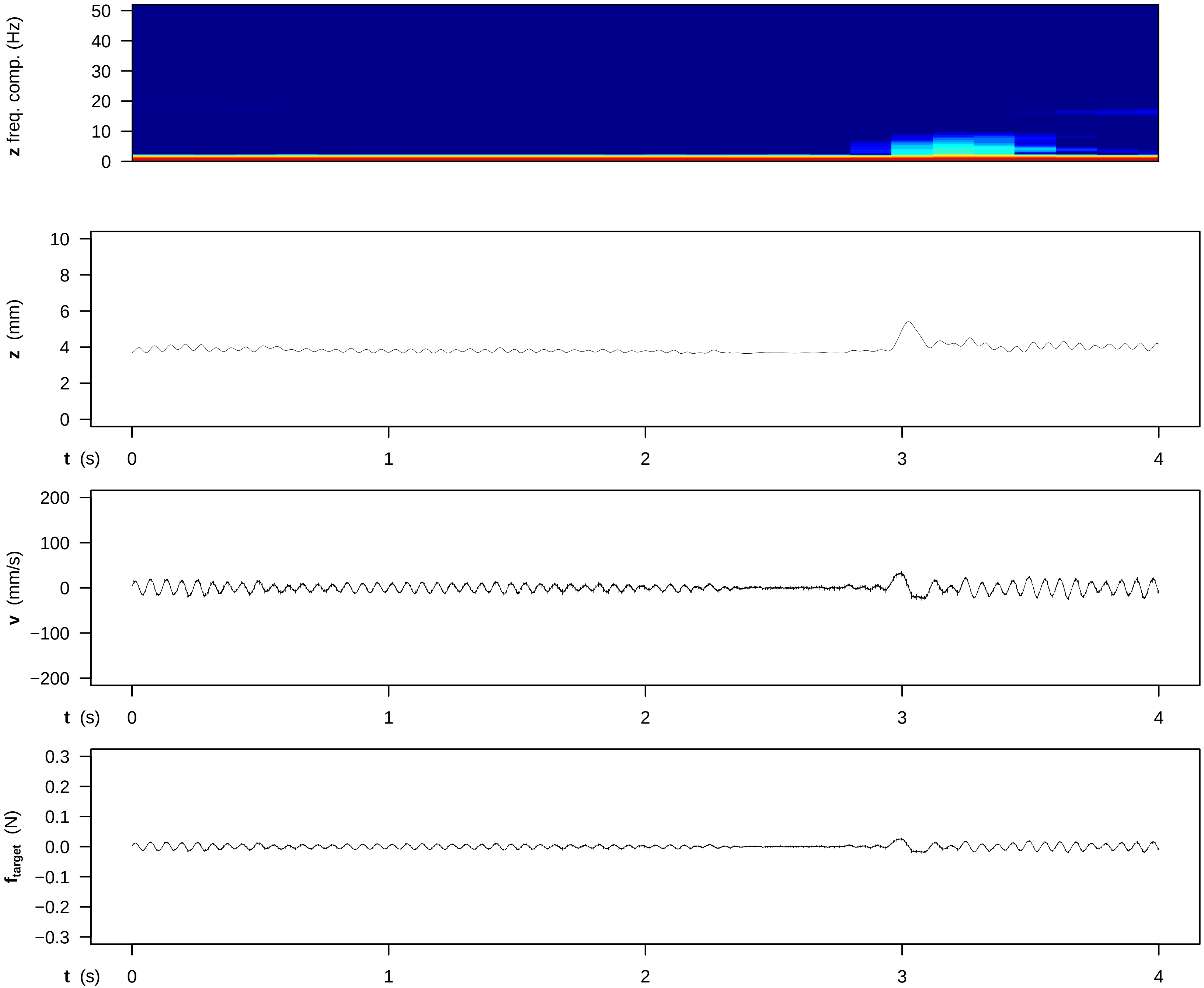

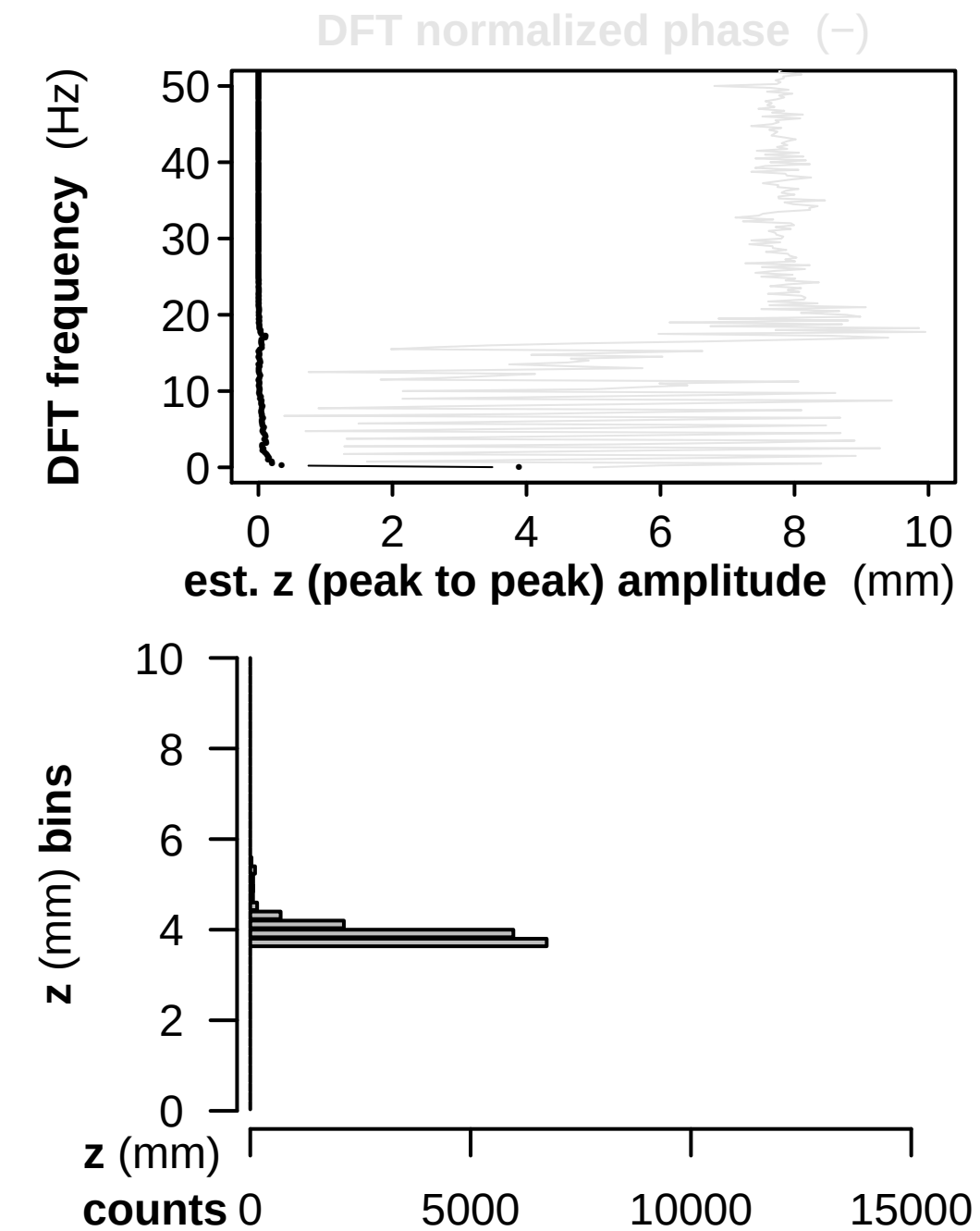

**SUBJECT 1 − RUN 19 − CONDITION 4,1**
**SC_180323_105038_0.AIFF**

```
z_min               : 3.64 mm
z_max               : 5.41 mm
z_travel_amplitude  : 1.77 mm

avg_abs_z_travel    : 7.39 mm/s

z_jarque-bera_jb    : 129157.97
z_jarque-bera_p     : 0.00e+00

z_lin_mod_est_slope: 0.06 mm/s
z_lin_mod_adj_R²    : 9 %

z_poly40_mod_adj_R²: 61 %

z_dft_ampl_thresh   : 0.010 mm
>=threshold_maxfreq: 20.75 Hz
```

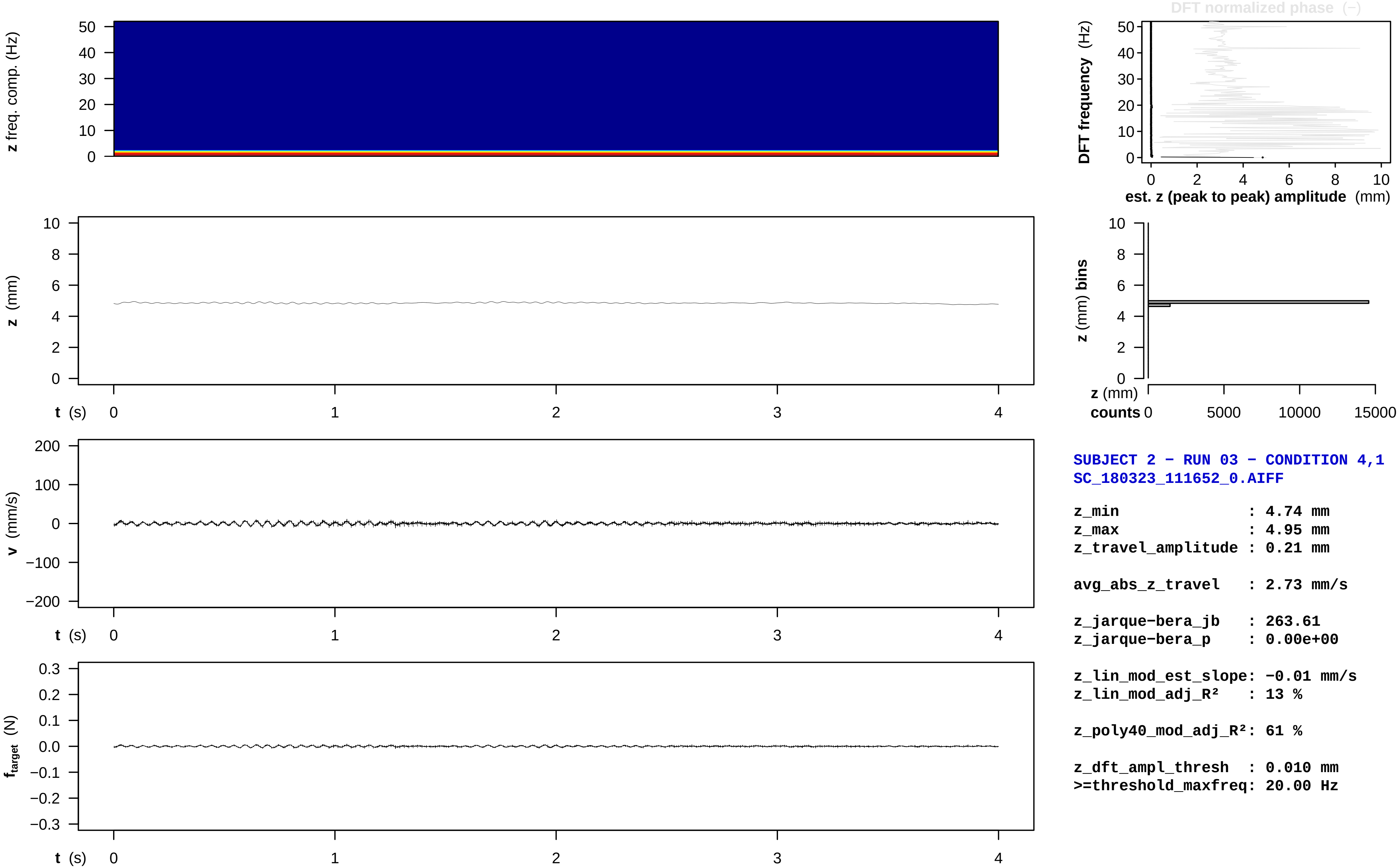

**SUBJECT 2 − RUN 03 − CONDITION 4,1**
**SC_180323_111652_0.AIFF**

```
z_min              : 4.74 mm
z_max              : 4.95 mm
z_travel_amplitude : 0.21 mm

avg_abs_z_travel   : 2.73 mm/s

z_jarque-bera_jb   : 263.61
z_jarque-bera_p    : 0.00e+00

z_lin_mod_est_slope: −0.01 mm/s
z_lin_mod_adj_R²   : 13 %

z_poly40_mod_adj_R²: 61 %

z_dft_ampl_thresh  : 0.010 mm
>=threshold_maxfreq: 20.00 Hz
```

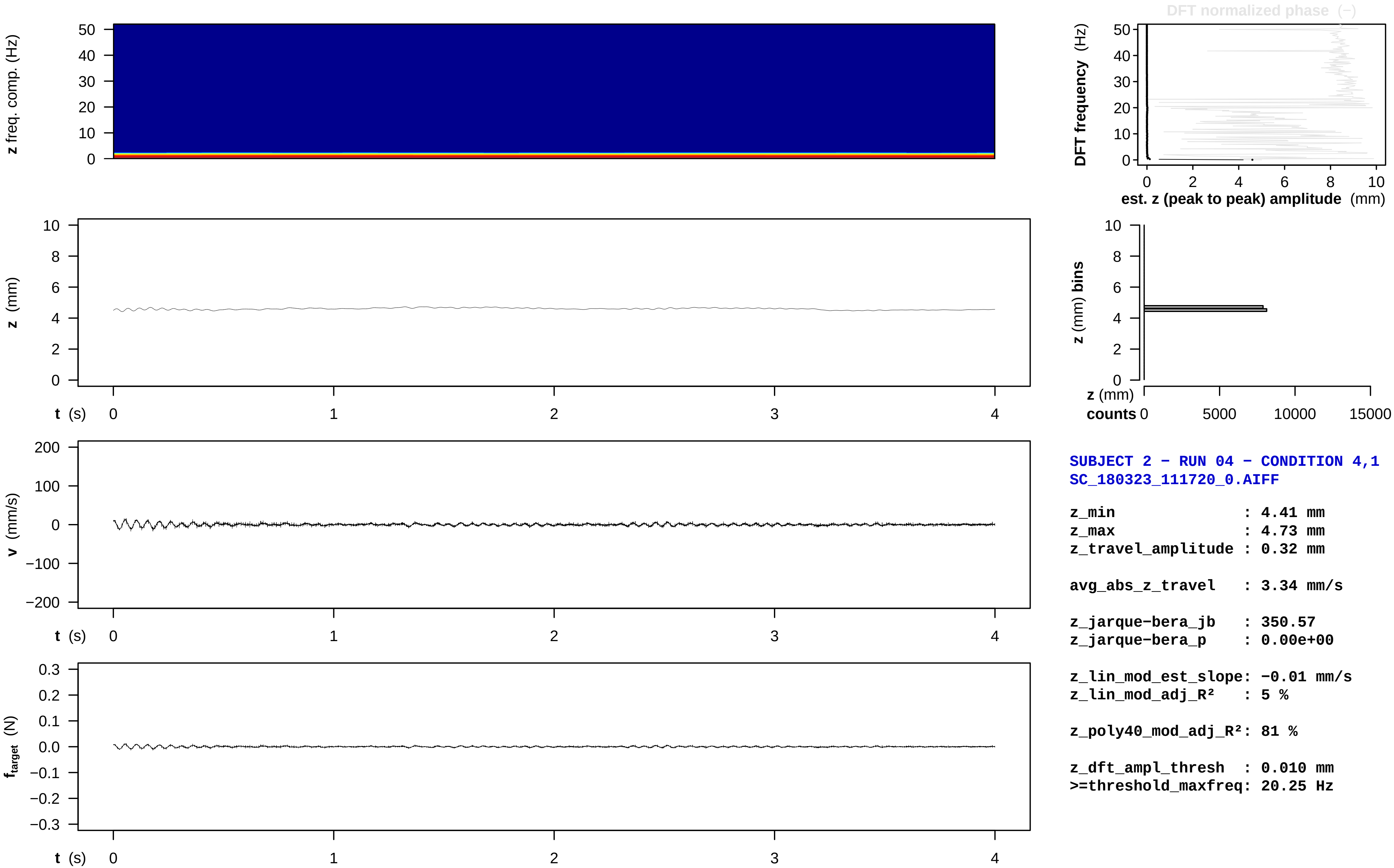

**SUBJECT 2 − RUN 04 − CONDITION 4,1**
**SC_180323_111720_0.AIFF**

```
z_min               : 4.41 mm
z_max               : 4.73 mm
z_travel_amplitude  : 0.32 mm

avg_abs_z_travel    : 3.34 mm/s

z_jarque-bera_jb    : 350.57
z_jarque-bera_p     : 0.00e+00

z_lin_mod_est_slope: -0.01 mm/s
z_lin_mod_adj_R²    : 5 %

z_poly40_mod_adj_R²: 81 %

z_dft_ampl_thresh   : 0.010 mm
>=threshold_maxfreq: 20.25 Hz
```

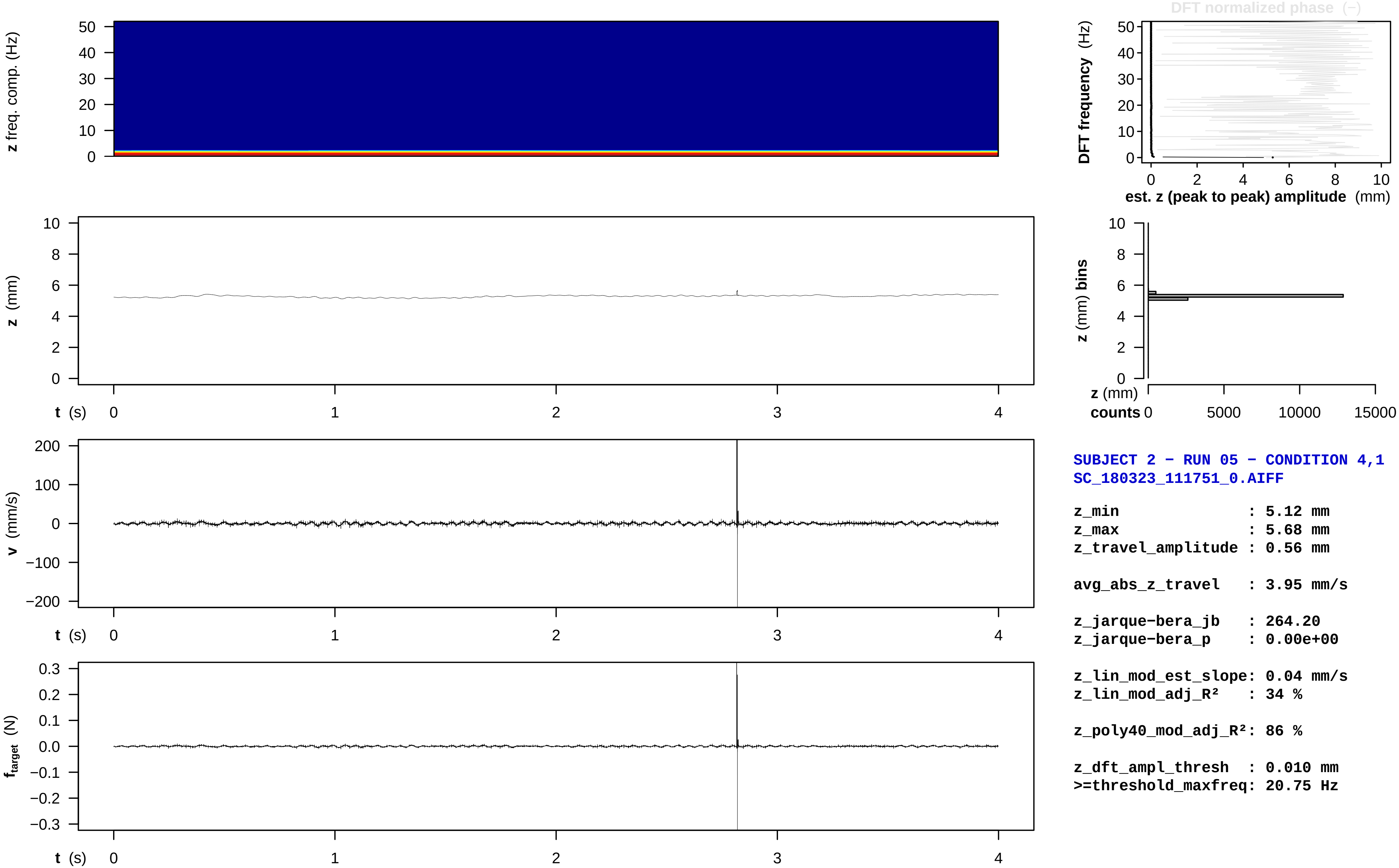

**SUBJECT 2 − RUN 05 − CONDITION 4,1**
**SC_180323_111751_0.AIFF**

```
z_min               : 5.12 mm
z_max               : 5.68 mm
z_travel_amplitude  : 0.56 mm

avg_abs_z_travel    : 3.95 mm/s

z_jarque-bera_jb    : 264.20
z_jarque-bera_p     : 0.00e+00

z_lin_mod_est_slope: 0.04 mm/s
z_lin_mod_adj_R²    : 34 %

z_poly40_mod_adj_R²: 86 %

z_dft_ampl_thresh   : 0.010 mm
>=threshold_maxfreq: 20.75 Hz
```

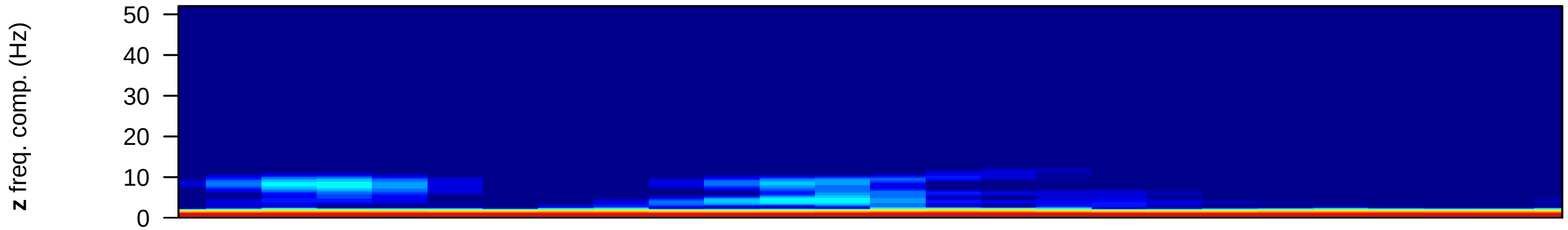

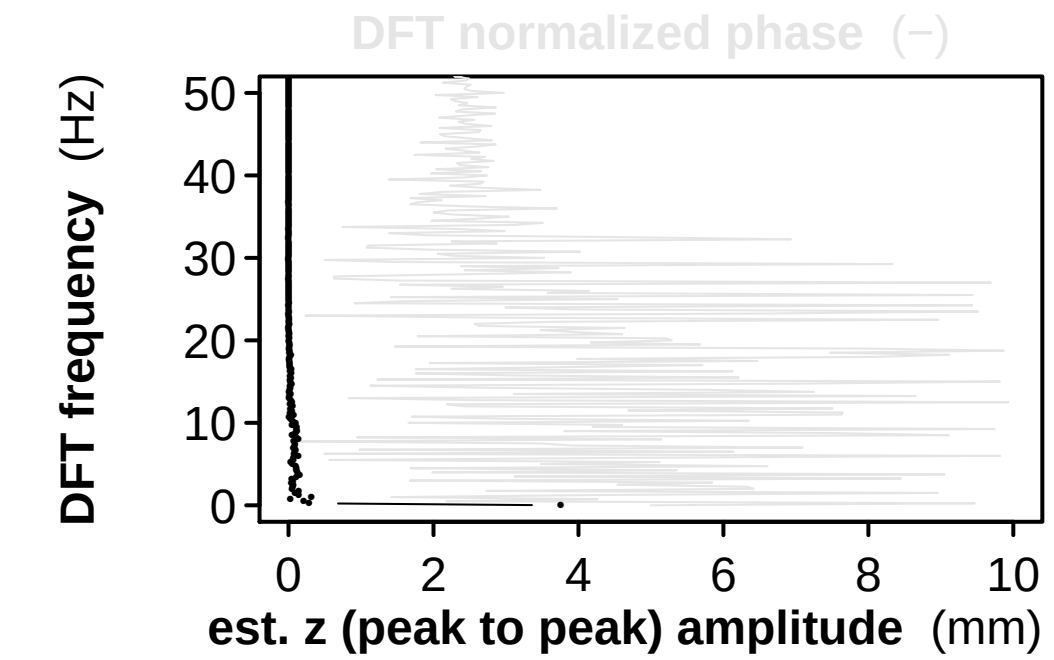

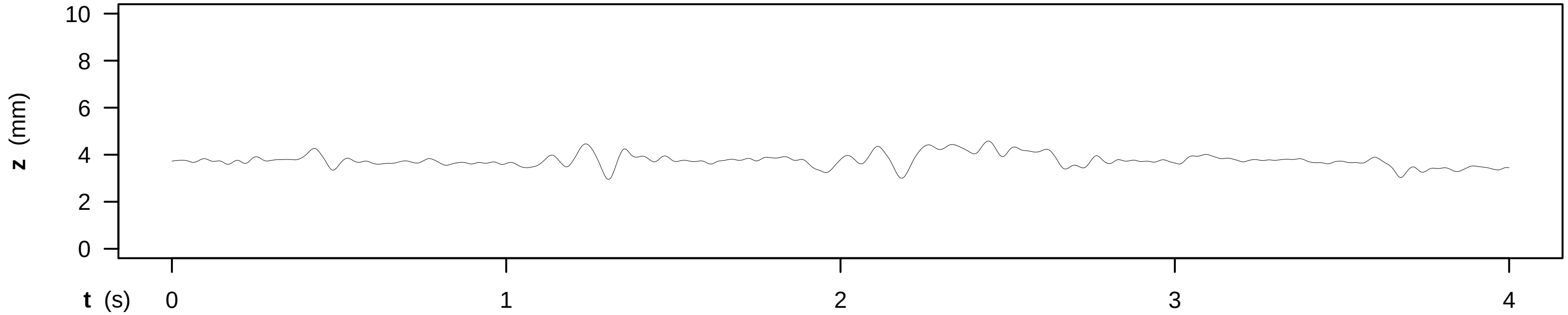

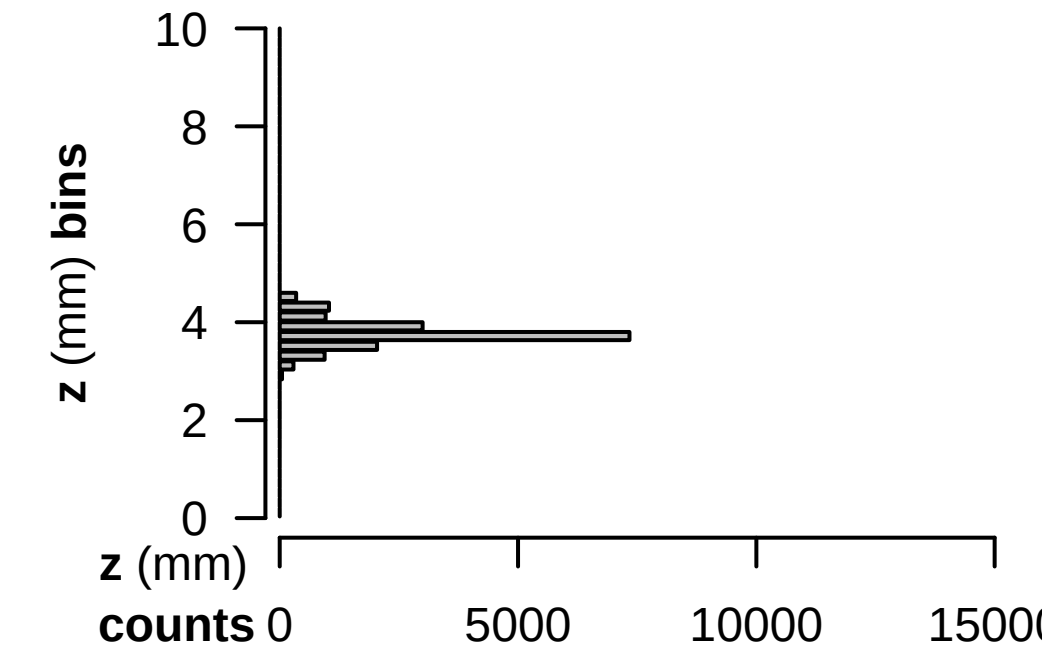

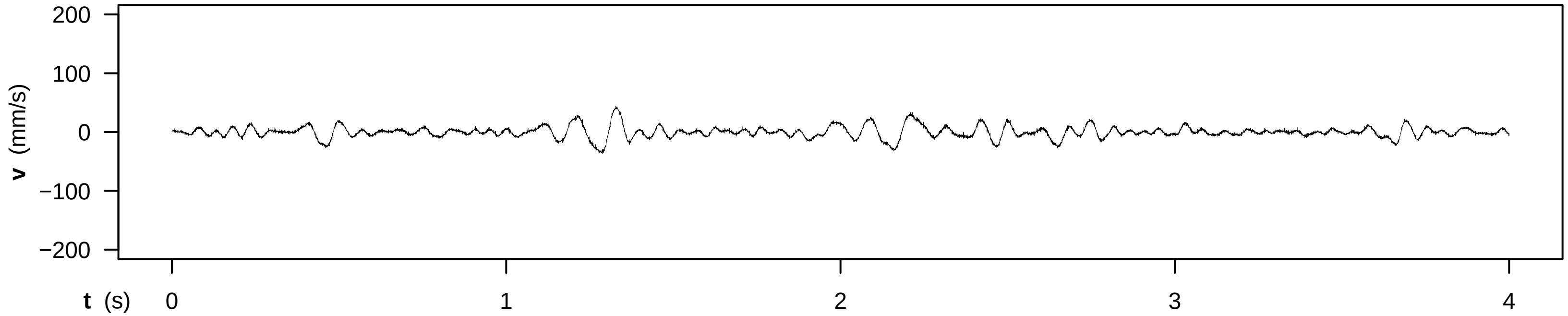

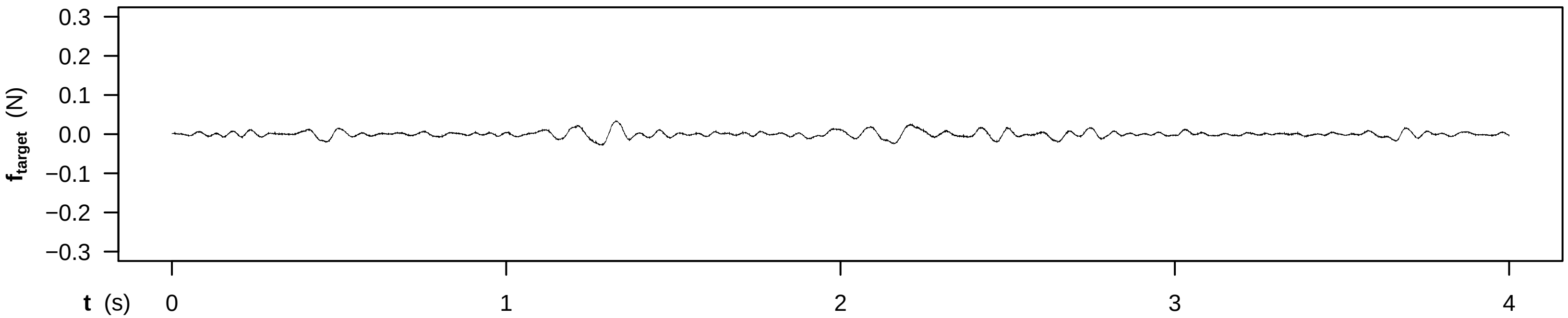

**SUBJECT 3 − RUN 23 − CONDITION 4,1**
**SC_180323_120842_0.AIFF**

z_min : 2.95 mm
z_max : 4.58 mm
z_travel_amplitude : 1.63 mm

avg_abs_z_travel : 7.13 mm/s

z_jarque-bera_jb : 666.61
z_jarque-bera_p : 0.00e+00

z_lin_mod_est_slope: -0.03 mm/s
z_lin_mod_adj_R² : 1 %

z_poly40_mod_adj_R²: 48 %

z_dft_ampl_thresh : 0.010 mm
>=threshold_maxfreq: 20.75 Hz

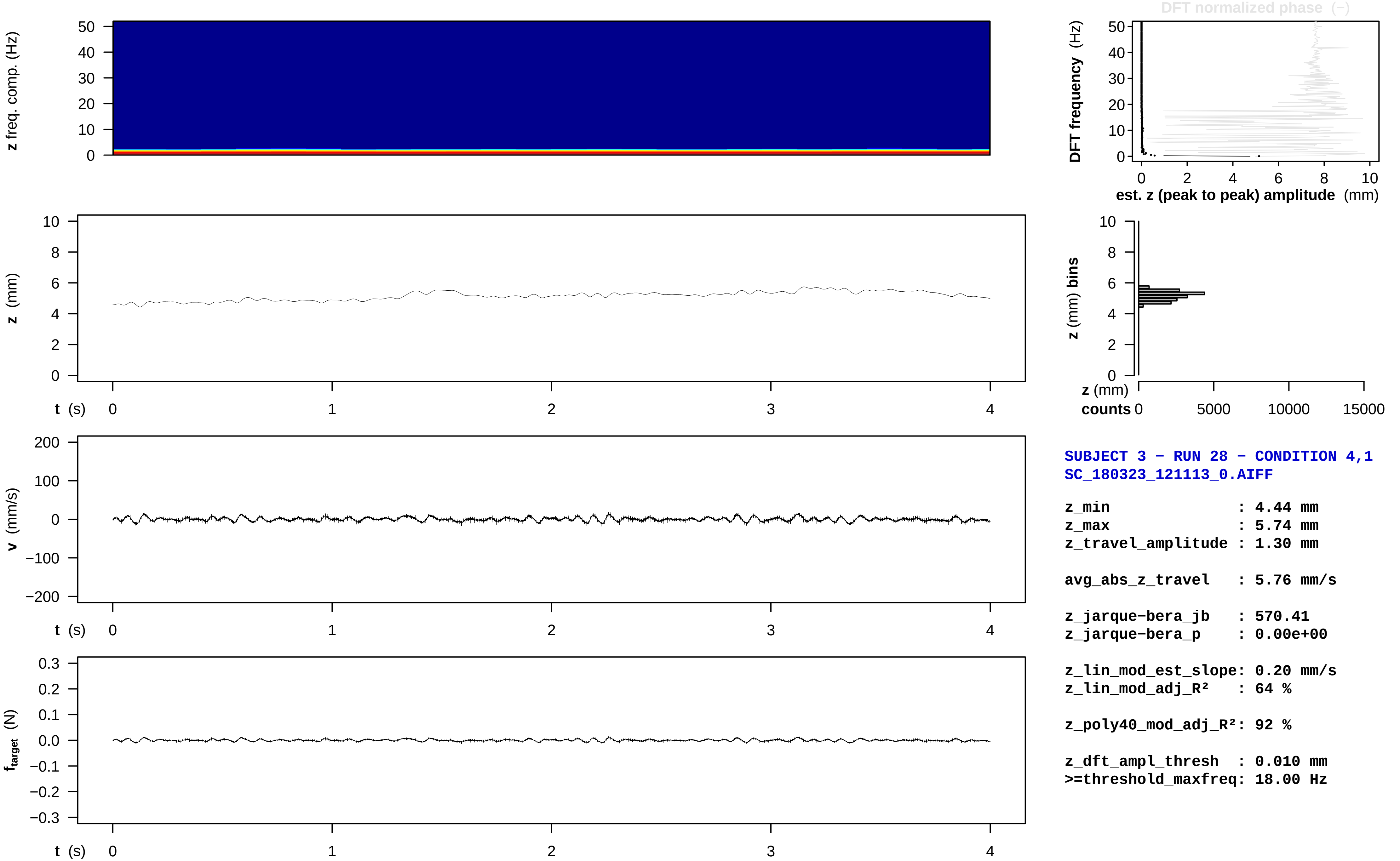

**SUBJECT 3 − RUN 28 − CONDITION 4,1**
**SC_180323_121113_0.AIFF**

```
z_min               : 4.44 mm
z_max               : 5.74 mm
z_travel_amplitude  : 1.30 mm

avg_abs_z_travel    : 5.76 mm/s

z_jarque-bera_jb    : 570.41
z_jarque-bera_p     : 0.00e+00

z_lin_mod_est_slope: 0.20 mm/s
z_lin_mod_adj_R²    : 64 %

z_poly40_mod_adj_R²: 92 %

z_dft_ampl_thresh   : 0.010 mm
>=threshold_maxfreq: 18.00 Hz
```

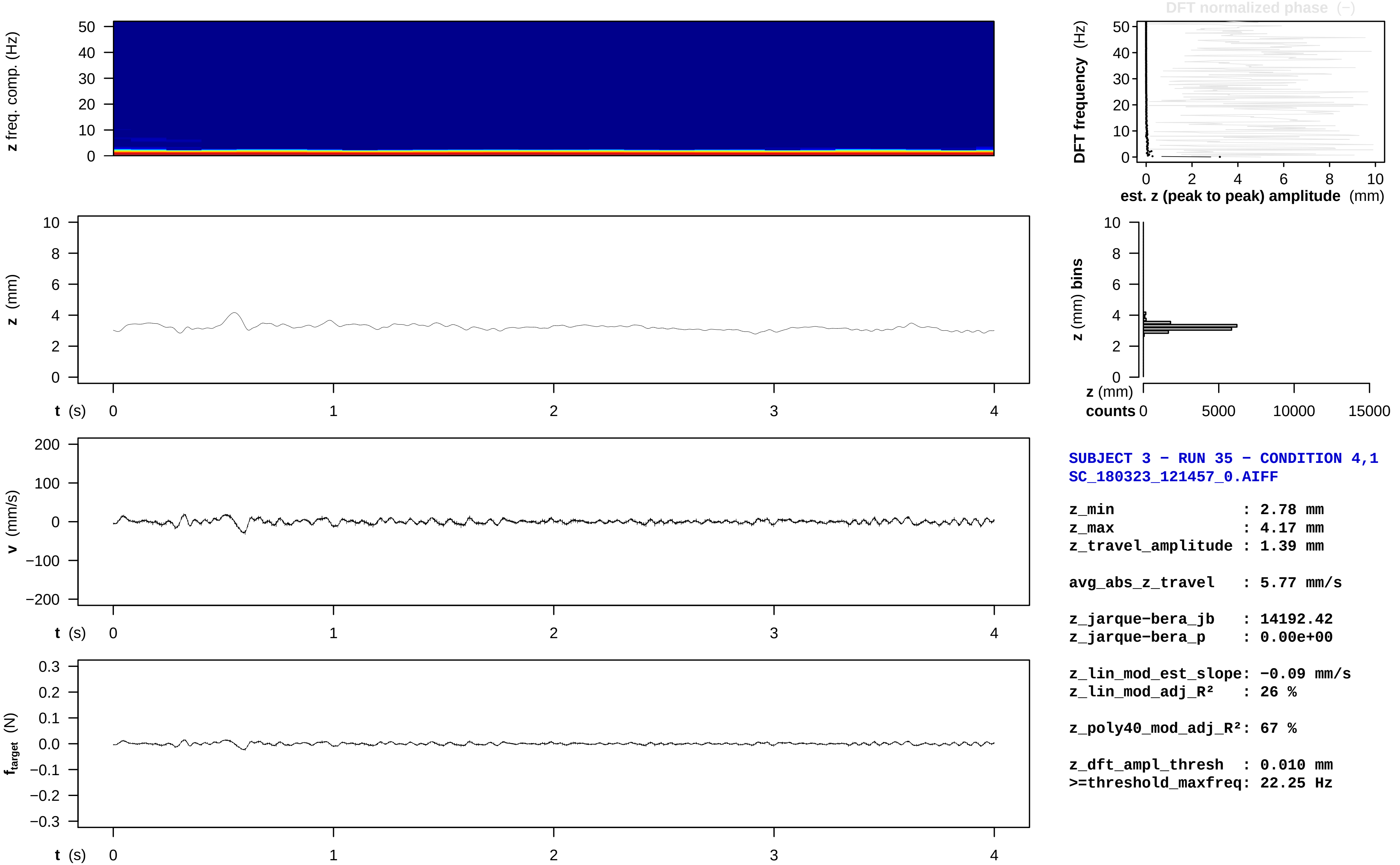

**SUBJECT 3 − RUN 35 − CONDITION 4,1**
**SC_180323_121457_0.AIFF**

```
z_min               : 2.78 mm
z_max               : 4.17 mm
z_travel_amplitude  : 1.39 mm

avg_abs_z_travel    : 5.77 mm/s

z_jarque-bera_jb    : 14192.42
z_jarque-bera_p     : 0.00e+00

z_lin_mod_est_slope: −0.09 mm/s
z_lin_mod_adj_R²    : 26 %

z_poly40_mod_adj_R²: 67 %

z_dft_ampl_thresh   : 0.010 mm
>=threshold_maxfreq: 22.25 Hz
```

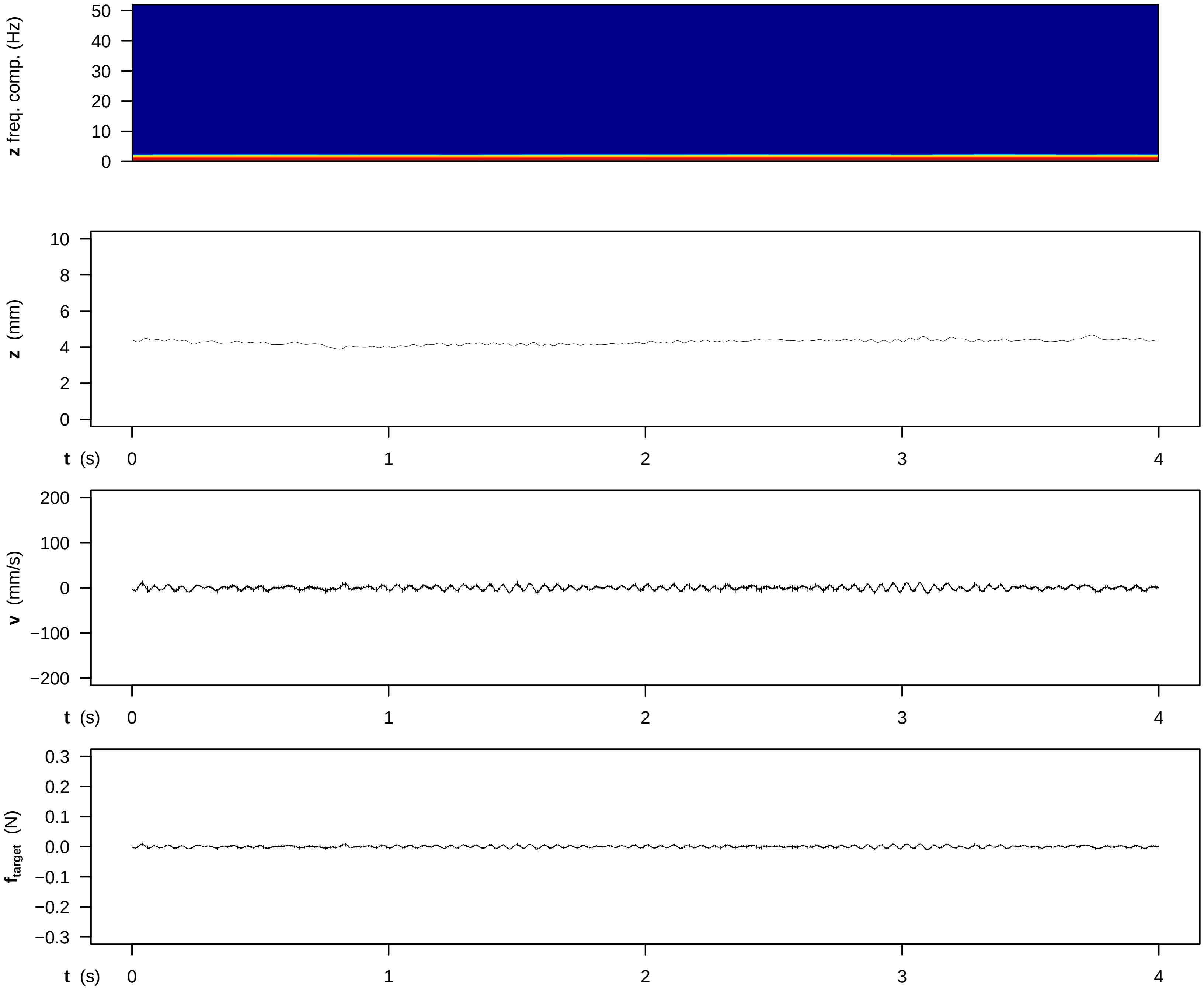

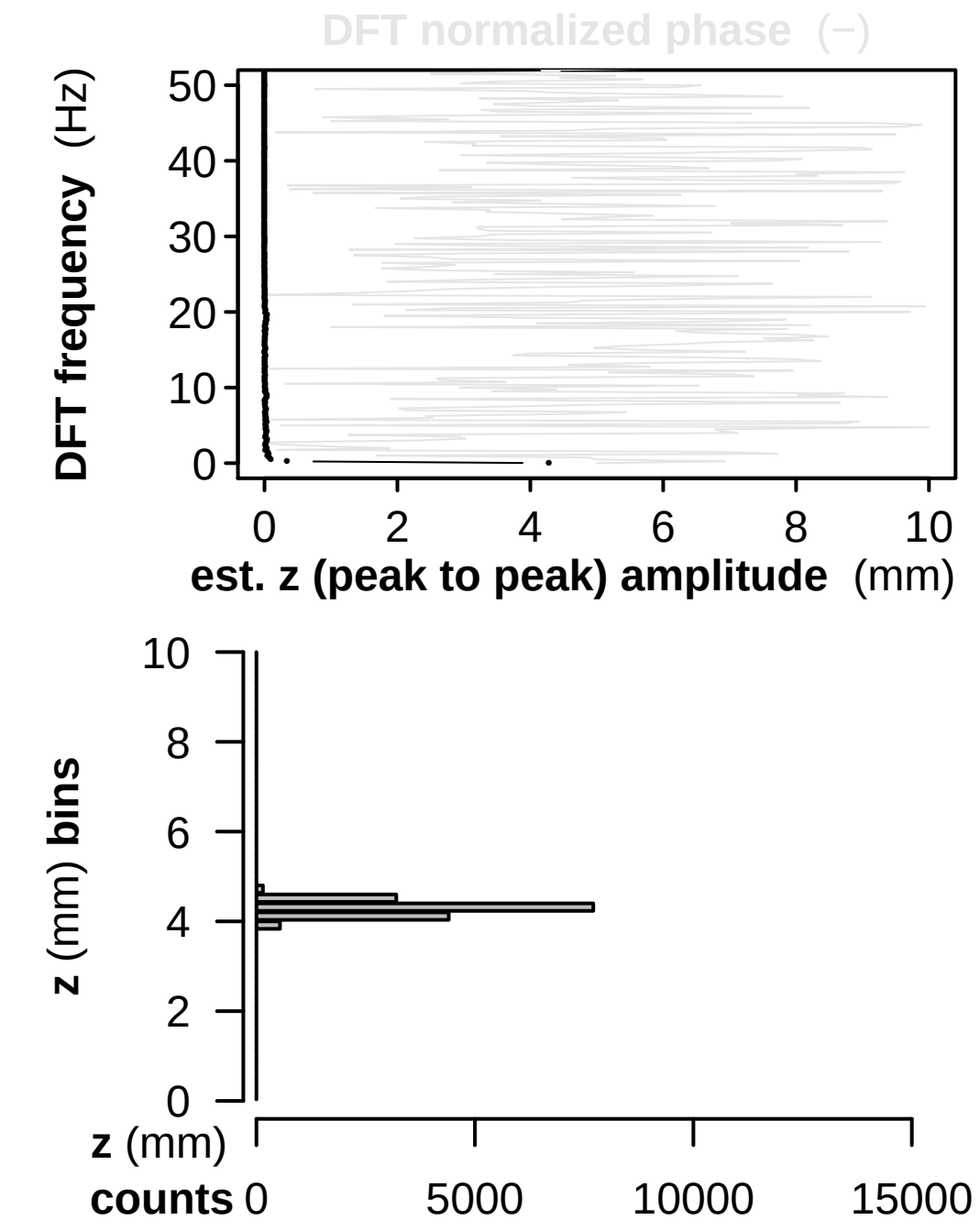

**SUBJECT 4 − RUN 08 − CONDITION 4,1**
**SC_180323_123435_0.AIFF**

```
z_min               : 3.89 mm
z_max               : 4.67 mm
z_travel_amplitude : 0.78 mm

avg_abs_z_travel   : 4.45 mm/s

z_jarque-bera_jb   : 291.10
z_jarque-bera_p    : 0.00e+00

z_lin_mod_est_slope: 0.08 mm/s
z_lin_mod_adj_R²   : 38 %

z_poly40_mod_adj_R²: 86 %

z_dft_ampl_thresh  : 0.010 mm
>=threshold_maxfreq: 20.25 Hz
```

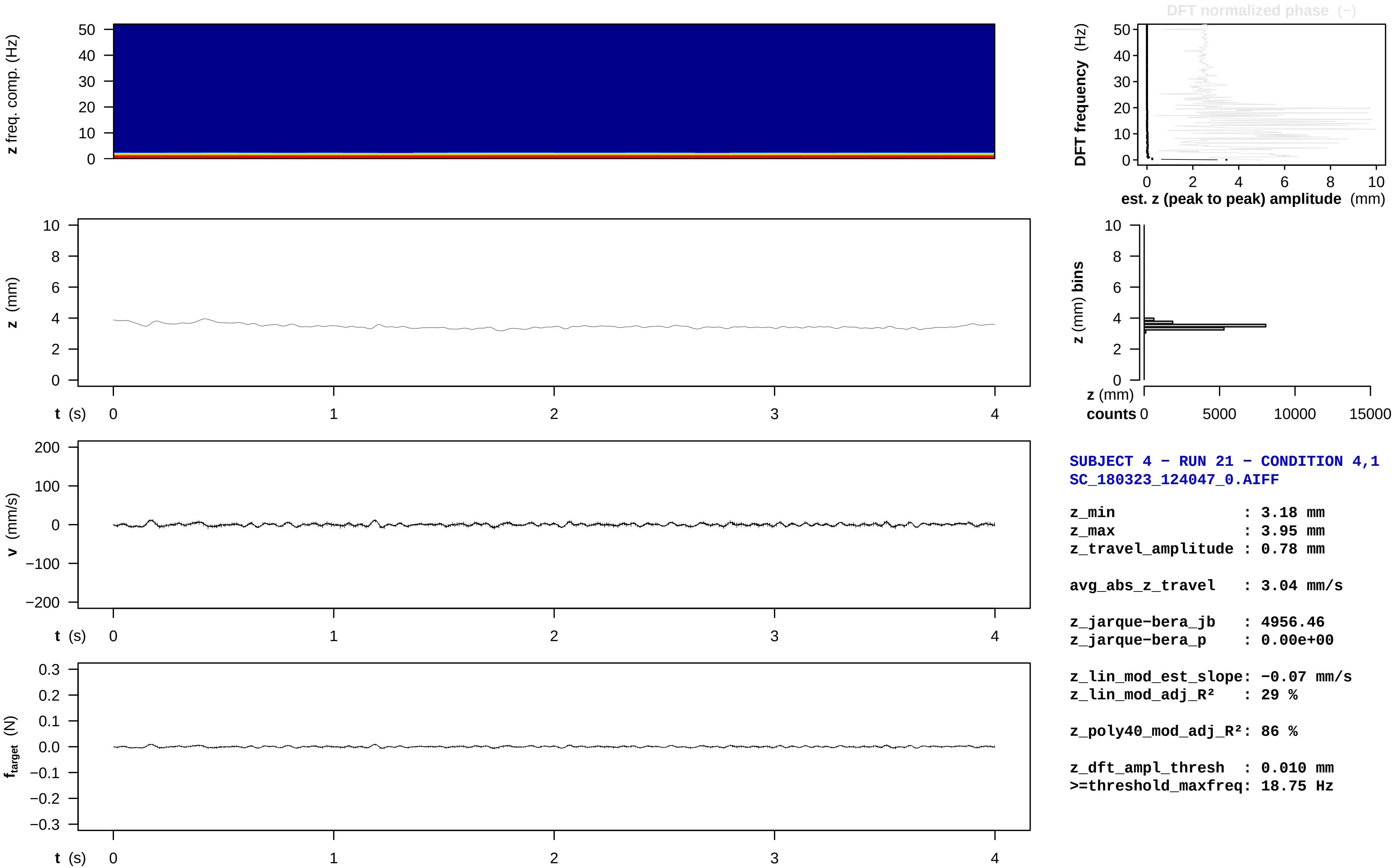

**SUBJECT 4 − RUN 21 − CONDITION 4,1**
**SC_180323_124047_0.AIFF**

```
z_min               : 3.18 mm
z_max               : 3.95 mm
z_travel_amplitude  : 0.78 mm

avg_abs_z_travel    : 3.04 mm/s

z_jarque-bera_jb    : 4956.46
z_jarque-bera_p     : 0.00e+00

z_lin_mod_est_slope: -0.07 mm/s
z_lin_mod_adj_R²    : 29 %

z_poly40_mod_adj_R²: 86 %

z_dft_ampl_thresh   : 0.010 mm
>=threshold_maxfreq: 18.75 Hz
```

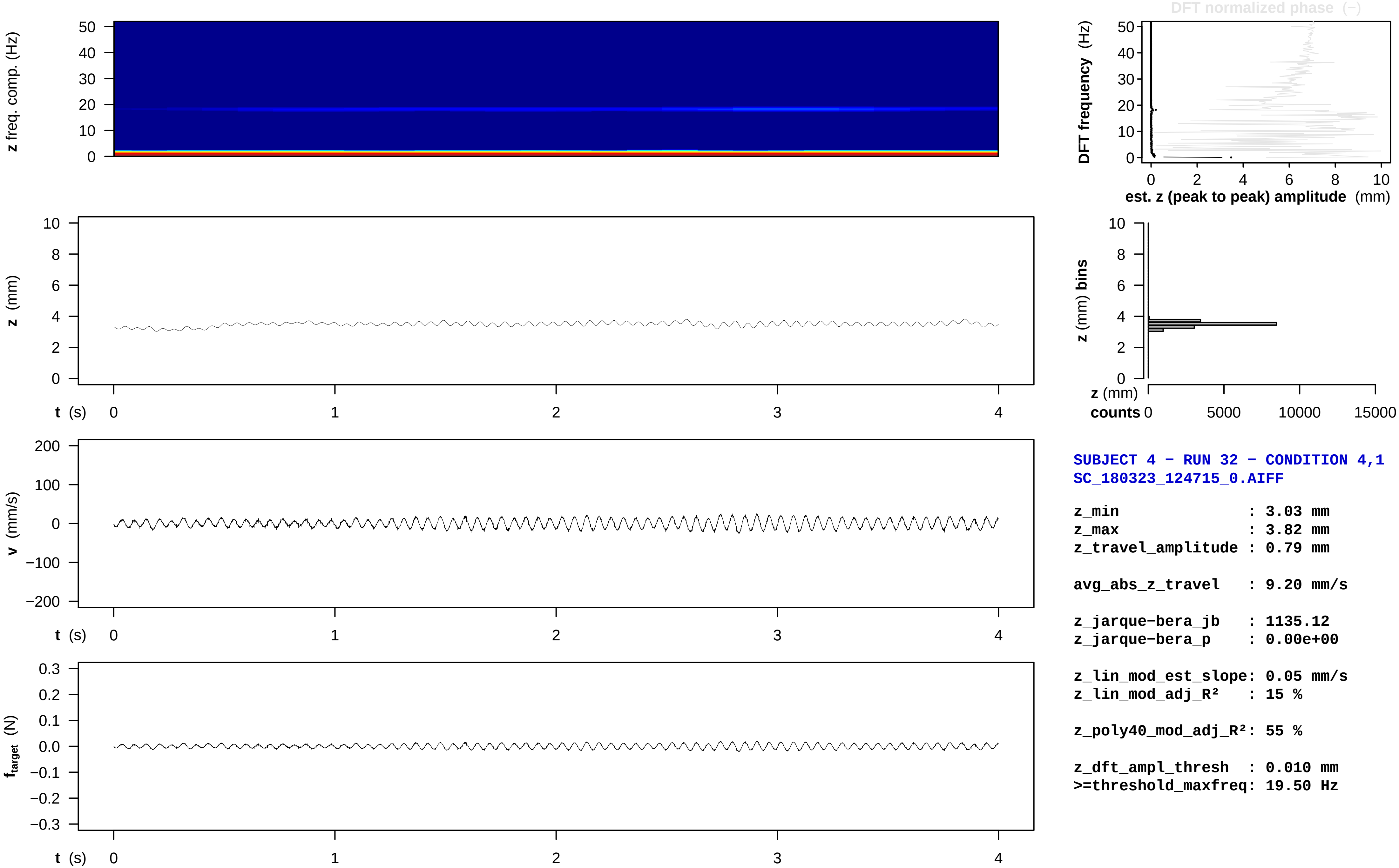

**SUBJECT 4 − RUN 32 − CONDITION 4,1**
**SC_180323_124715_0.AIFF**

```
z_min               : 3.03 mm
z_max               : 3.82 mm
z_travel_amplitude  : 0.79 mm

avg_abs_z_travel    : 9.20 mm/s

z_jarque-bera_jb    : 1135.12
z_jarque-bera_p     : 0.00e+00

z_lin_mod_est_slope: 0.05 mm/s
z_lin_mod_adj_R²    : 15 %

z_poly40_mod_adj_R²: 55 %

z_dft_ampl_thresh   : 0.010 mm
>=threshold_maxfreq: 19.50 Hz
```

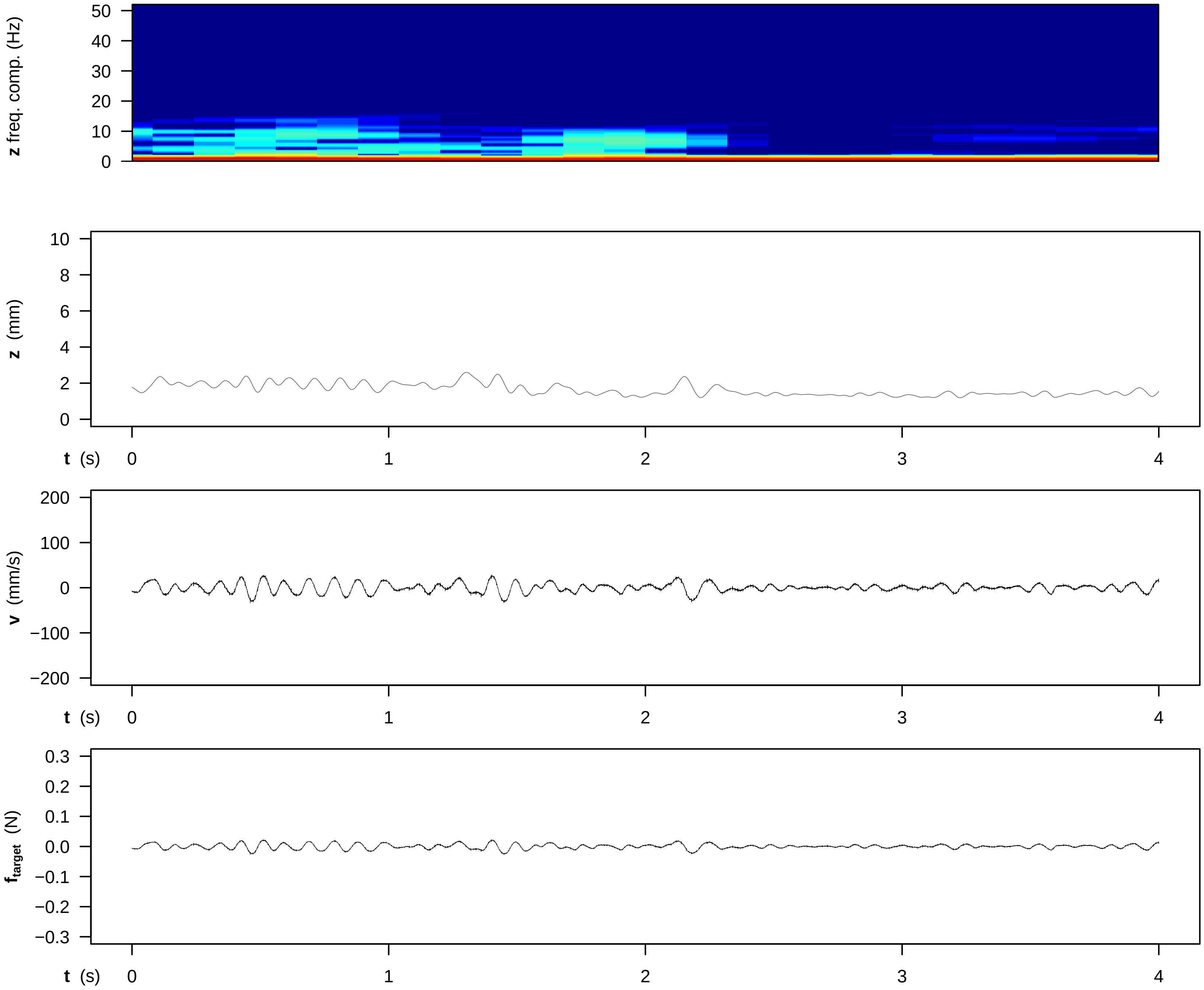

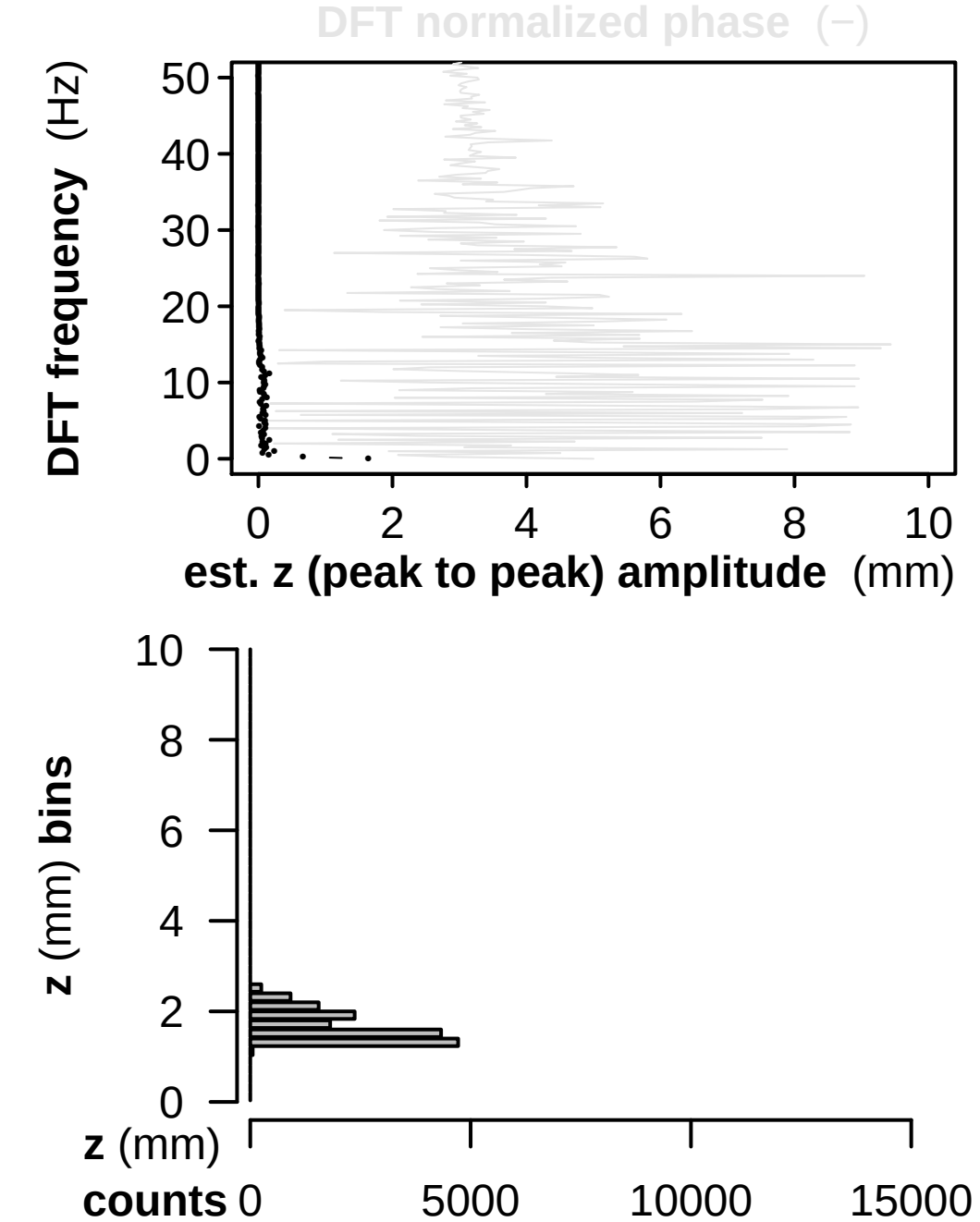

**SUBJECT 5 − RUN 03 − CONDITION 4,1**
**SC_180323_131548_0.AIFF**

```
z_min               : 1.19 mm
z_max               : 2.60 mm
z_travel_amplitude : 1.41 mm

avg_abs_z_travel    : 7.75 mm/s

z_jarque-bera_jb    : 1621.34
z_jarque-bera_p     : 0.00e+00

z_lin_mod_est_slope: −0.19 mm/s
z_lin_mod_adj_R²    : 47 %

z_poly40_mod_adj_R²: 67 %

z_dft_ampl_thresh   : 0.010 mm
>=threshold_maxfreq: 17.75 Hz
```

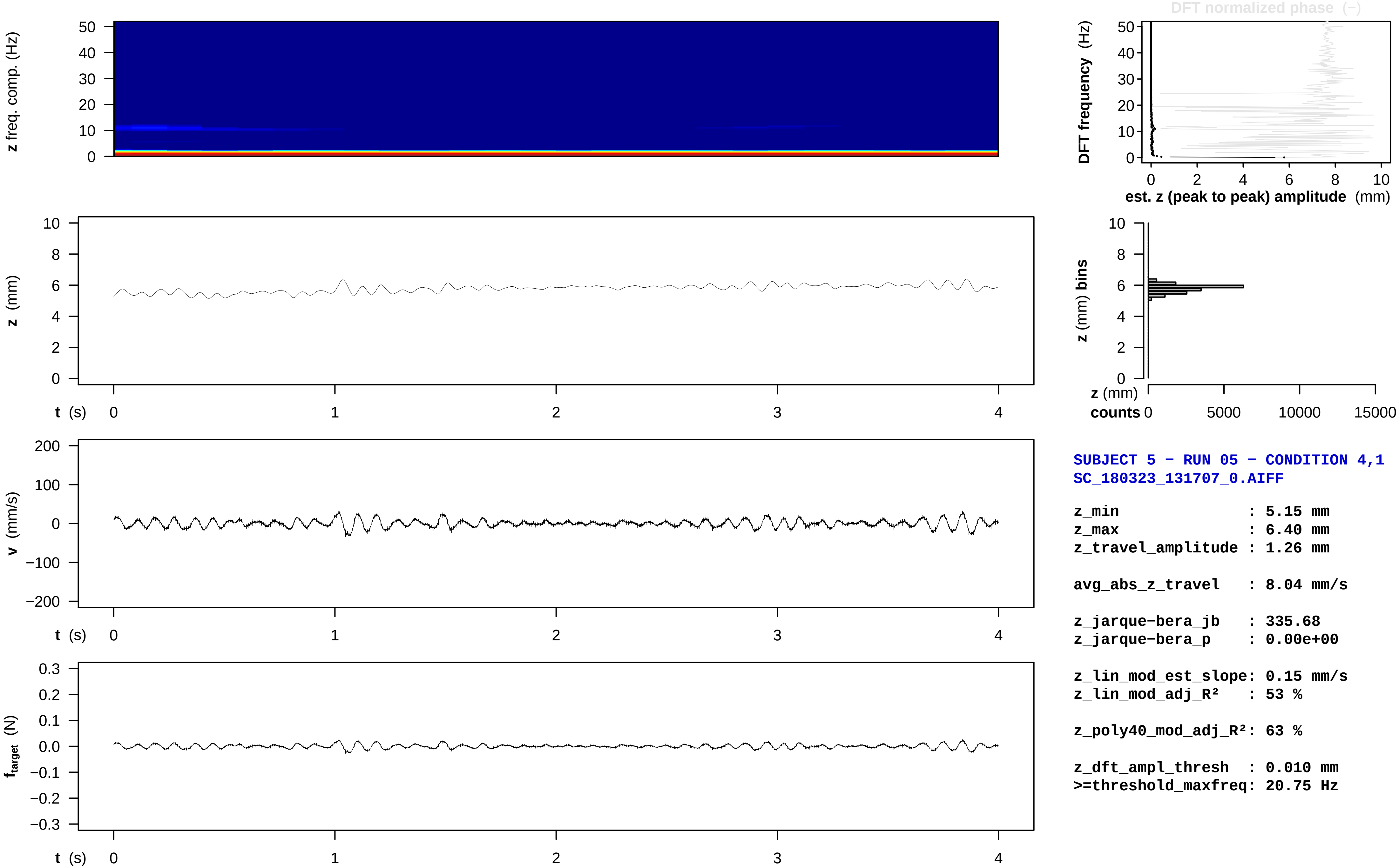

**SUBJECT 5 − RUN 05 − CONDITION 4,1**
**SC_180323_131707_0.AIFF**

```
z_min               : 5.15 mm
z_max               : 6.40 mm
z_travel_amplitude  : 1.26 mm

avg_abs_z_travel    : 8.04 mm/s

z_jarque-bera_jb    : 335.68
z_jarque-bera_p     : 0.00e+00

z_lin_mod_est_slope: 0.15 mm/s
z_lin_mod_adj_R²    : 53 %

z_poly40_mod_adj_R²: 63 %

z_dft_ampl_thresh   : 0.010 mm
>=threshold_maxfreq: 20.75 Hz
```

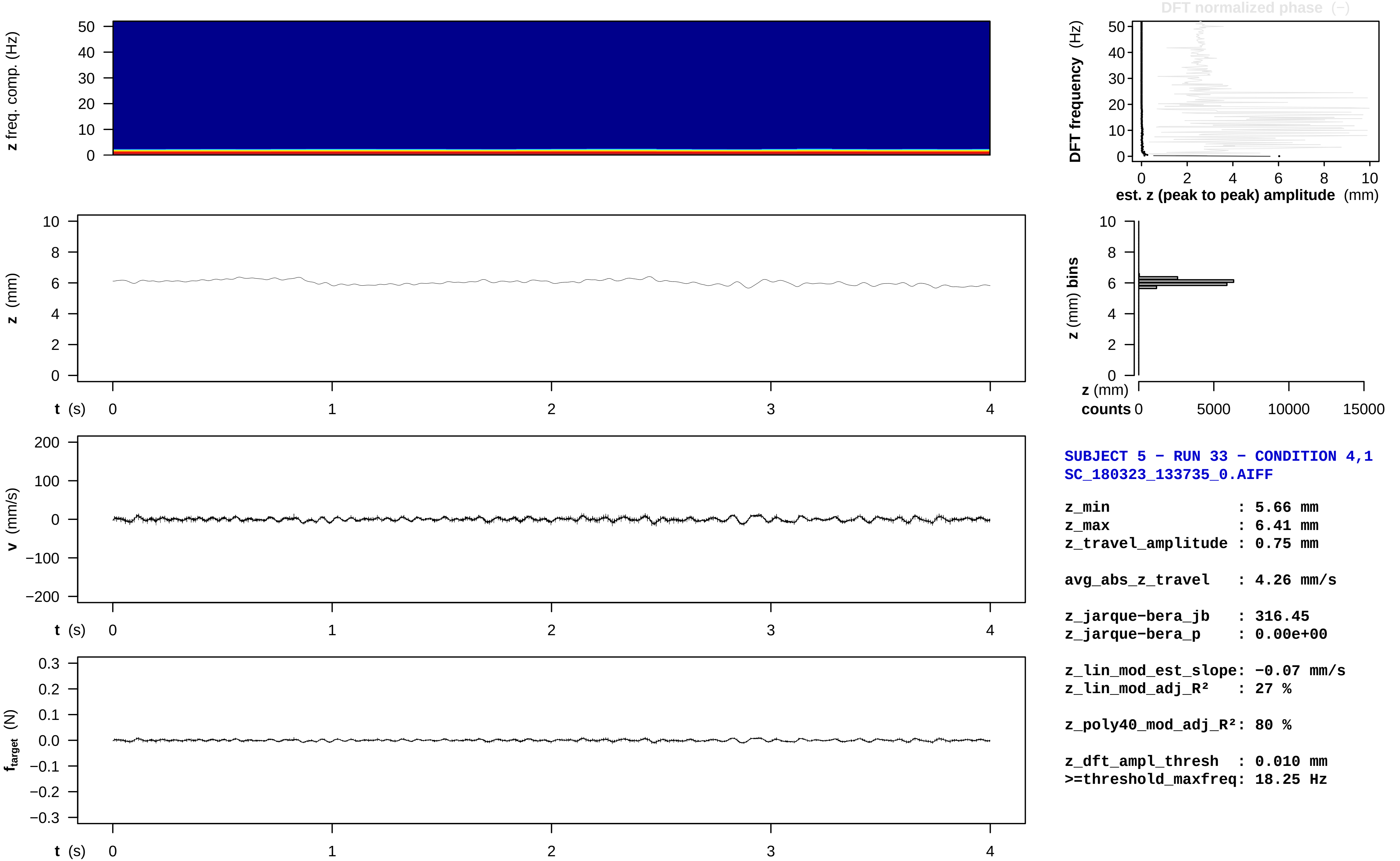

**SUBJECT 5 − RUN 33 − CONDITION 4,1**
**SC_180323_133735_0.AIFF**

```
z_min              : 5.66 mm
z_max              : 6.41 mm
z_travel_amplitude : 0.75 mm

avg_abs_z_travel   : 4.26 mm/s

z_jarque-bera_jb   : 316.45
z_jarque-bera_p    : 0.00e+00

z_lin_mod_est_slope: −0.07 mm/s
z_lin_mod_adj_R²   : 27 %

z_poly40_mod_adj_R²: 80 %

z_dft_ampl_thresh  : 0.010 mm
>=threshold_maxfreq: 18.25 Hz
```

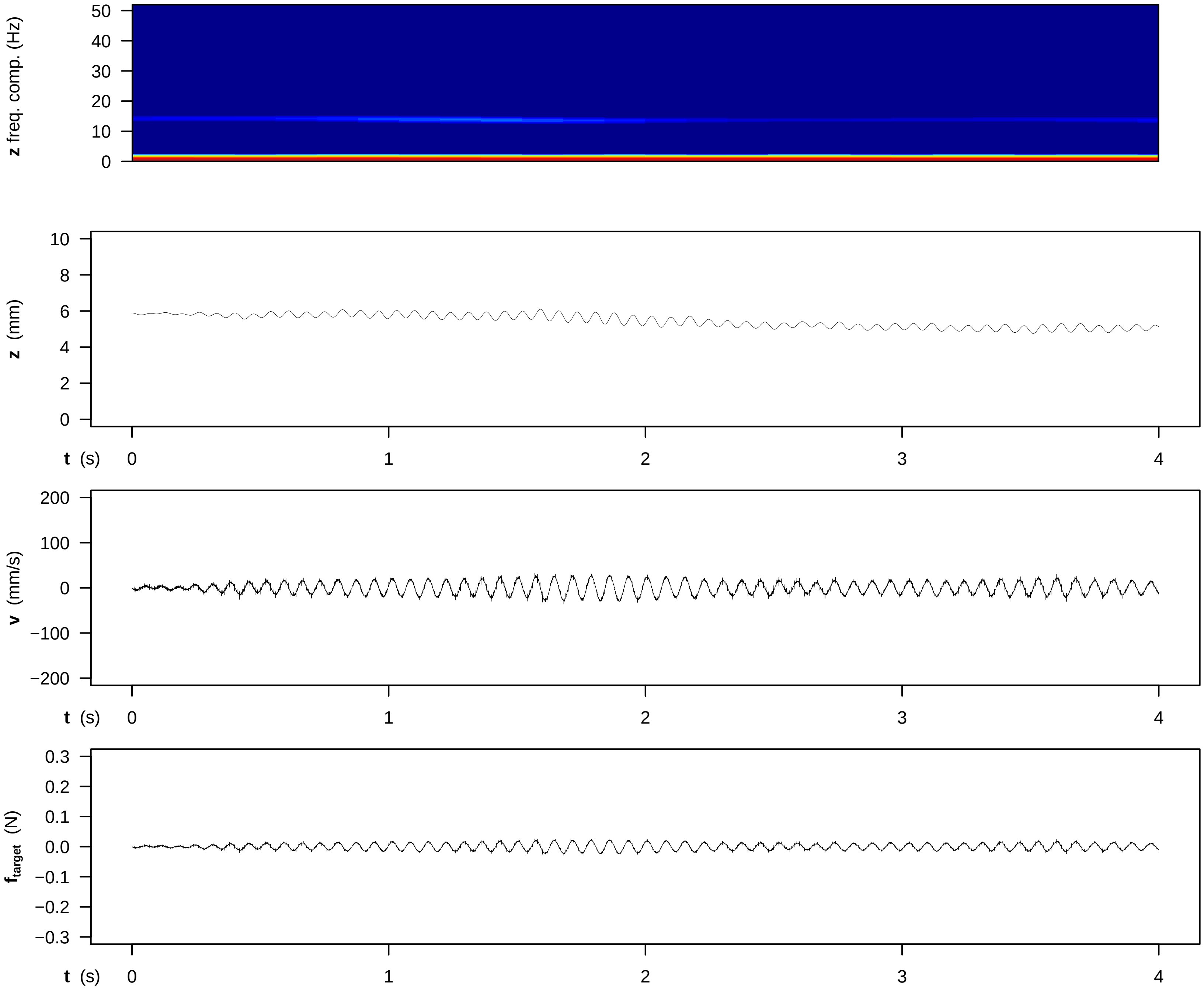

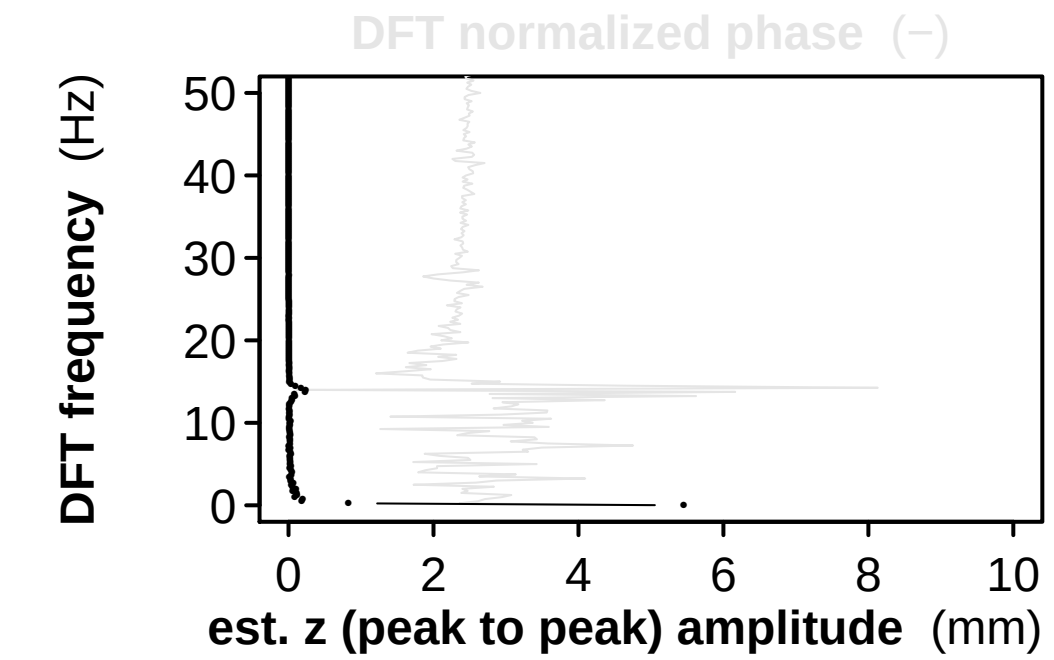

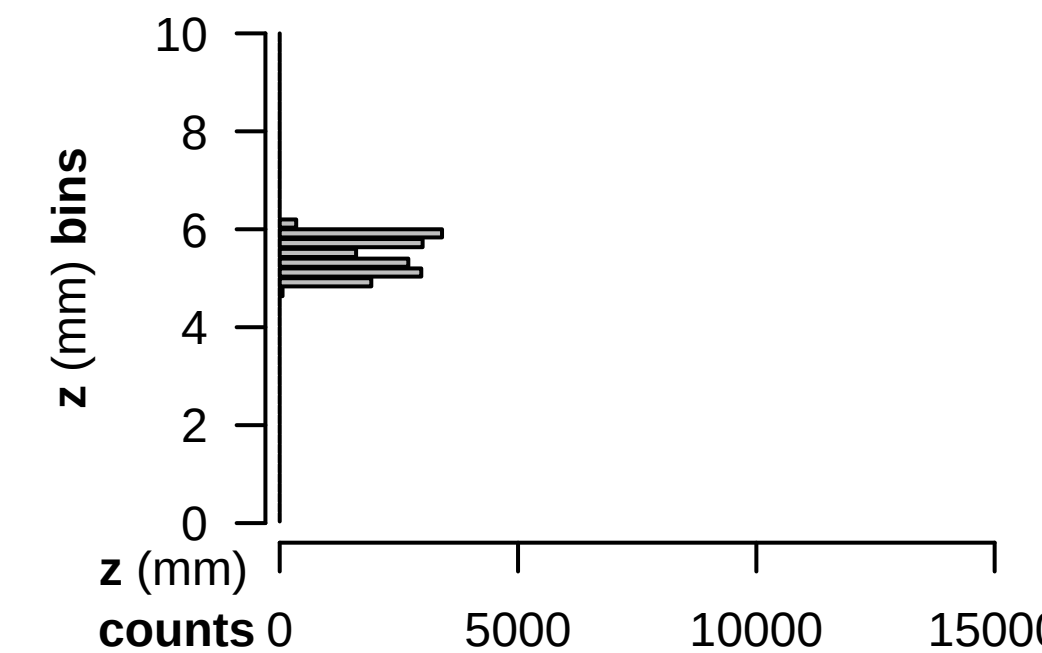

**SUBJECT 6 − RUN 01 − CONDITION 4,1**
**SC_180323_145045_0.AIFF**

```
z_min               : 4.75 mm
z_max               : 6.10 mm
z_travel_amplitude  : 1.35 mm

avg_abs_z_travel    : 11.82 mm/s

z_jarque-bera_jb    : 1170.05
z_jarque-bera_p     : 0.00e+00

z_lin_mod_est_slope: −0.27 mm/s
z_lin_mod_adj_R²    : 75 %

z_poly40_mod_adj_R²: 82 %

z_dft_ampl_thresh   : 0.010 mm
>=threshold_maxfreq: 17.25 Hz
```

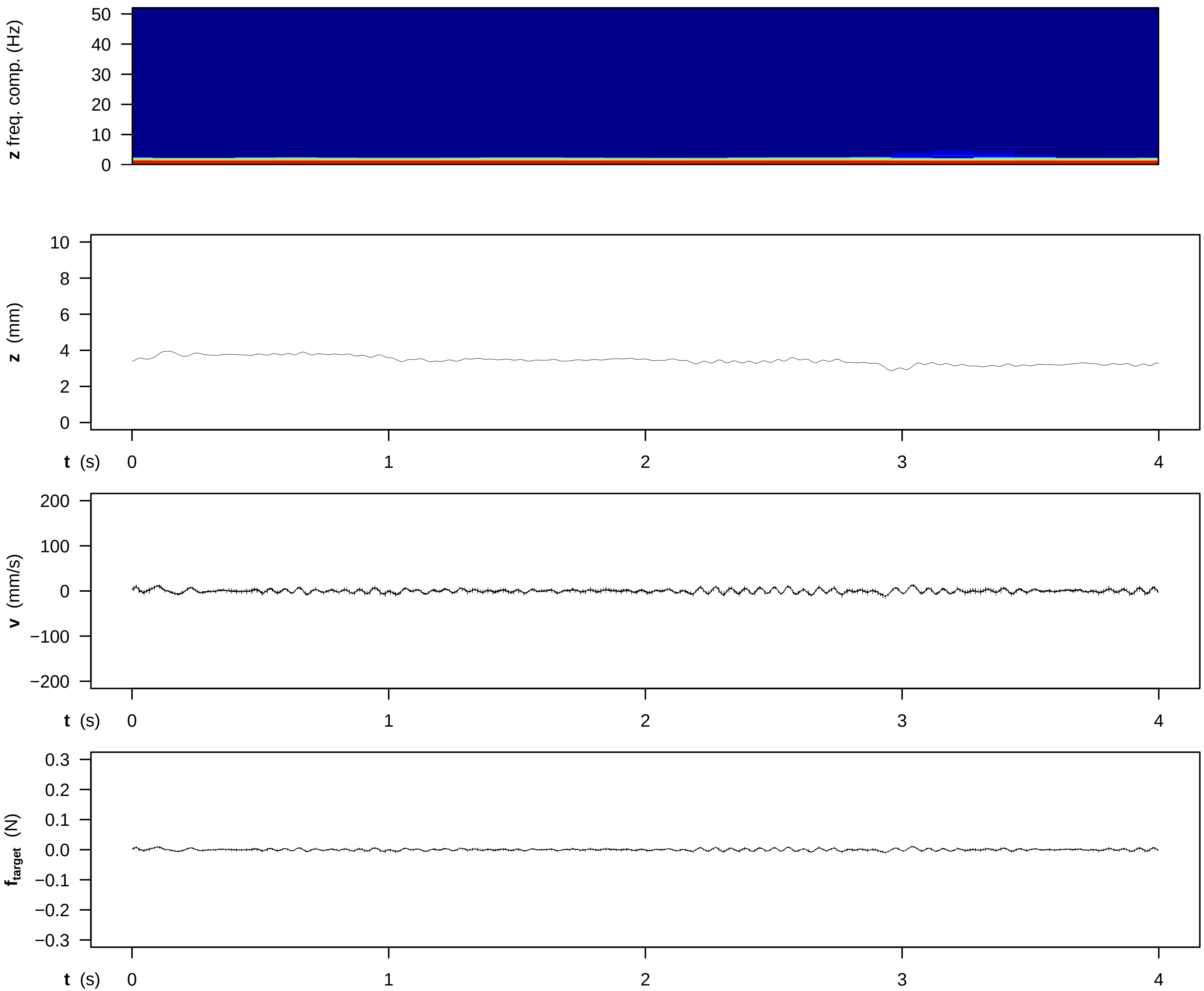

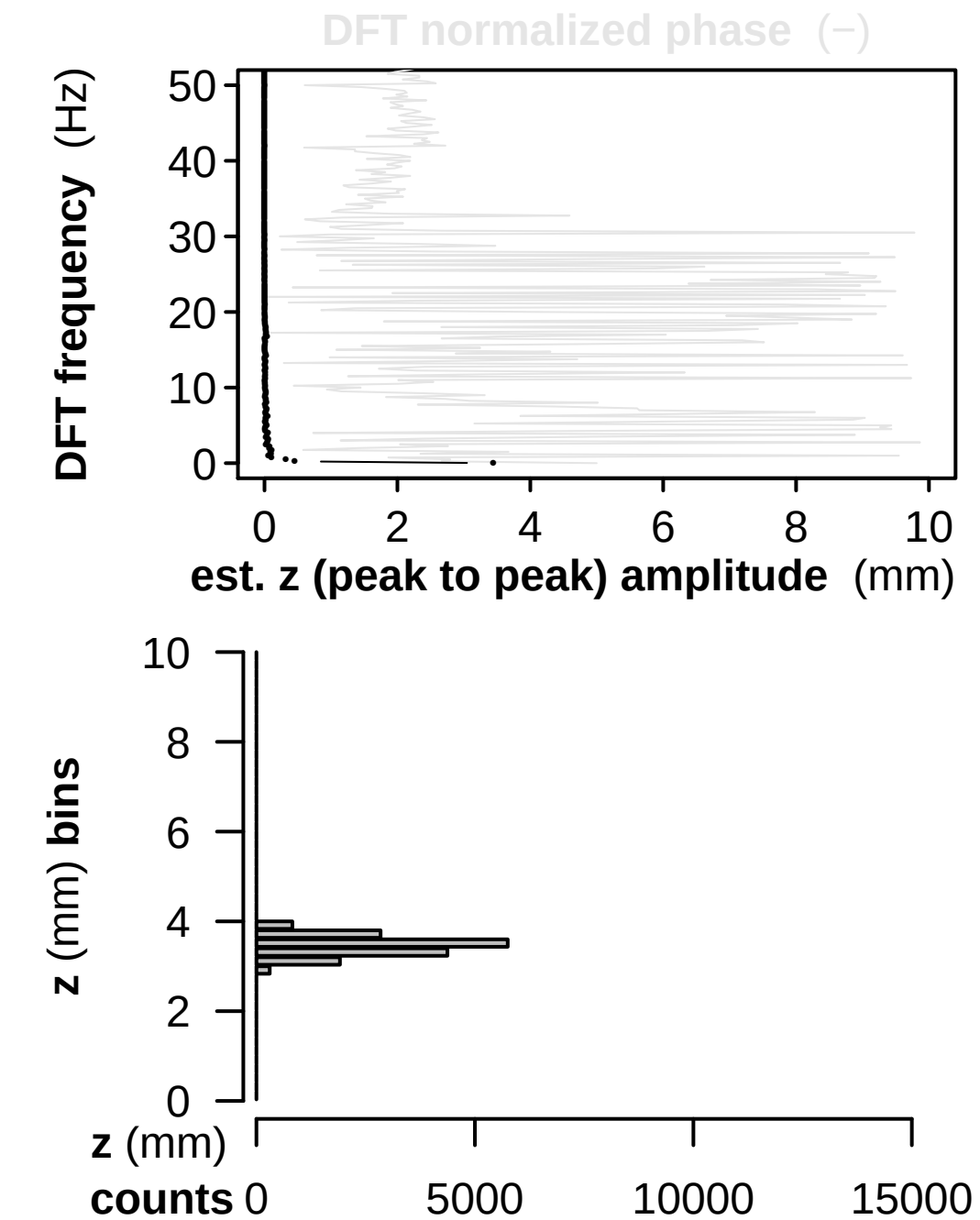

**SUBJECT 6 − RUN 08 − CONDITION 4,1**
**SC_180323_145546_0.AIFF**

```
z_min               : 2.88 mm
z_max               : 3.95 mm
z_travel_amplitude  : 1.07 mm

avg_abs_z_travel    : 4.23 mm/s

z_jarque-bera_jb    : 264.91
z_jarque-bera_p     : 0.00e+00

z_lin_mod_est_slope: −0.17 mm/s
z_lin_mod_adj_R²    : 76 %

z_poly40_mod_adj_R²: 93 %

z_dft_ampl_thresh   : 0.010 mm
>=threshold_maxfreq: 18.25 Hz
```

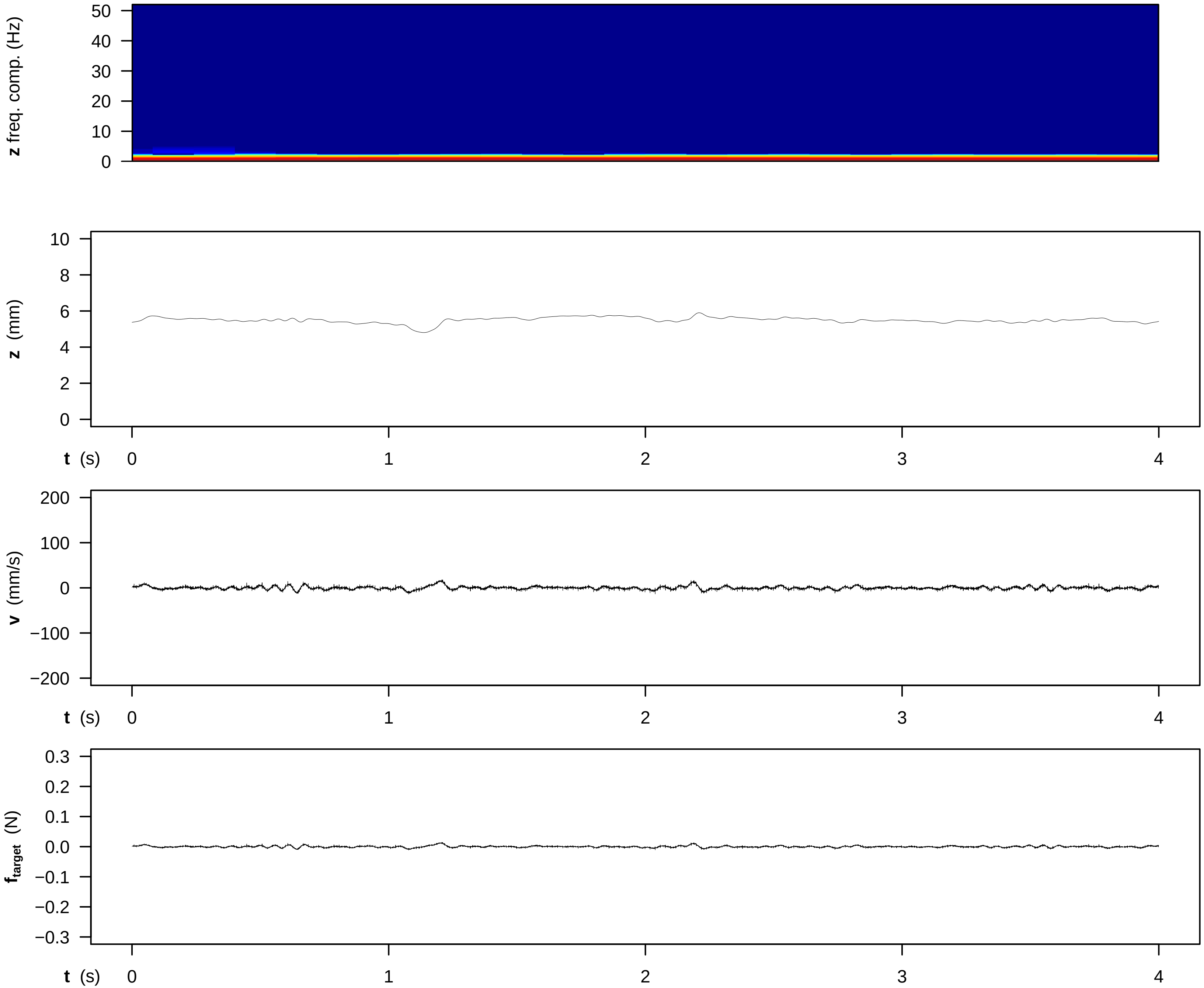

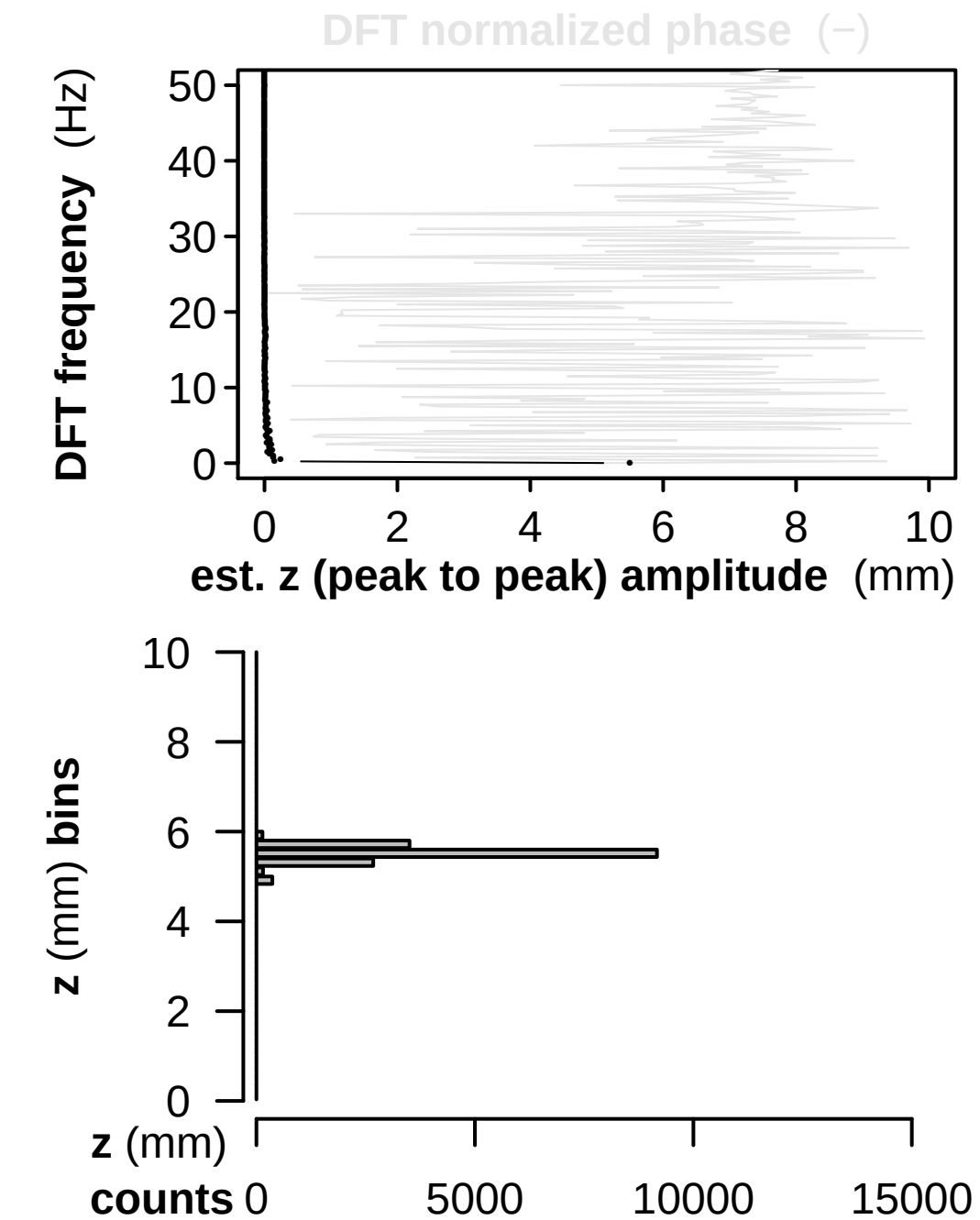

**SUBJECT 6 − RUN 30 − CONDITION 4,1**
**SC_180323_150945_0.AIFF**

```
z_min              : 4.80 mm
z_max              : 5.90 mm
z_travel_amplitude : 1.10 mm

avg_abs_z_travel   : 3.32 mm/s

z_jarque-bera_jb   : 15051.96
z_jarque-bera_p    : 0.00e+00

z_lin_mod_est_slope: -0.00 mm/s
z_lin_mod_adj_R²   : 0 %

z_poly40_mod_adj_R²: 69 %

z_dft_ampl_thresh  : 0.010 mm
>=threshold_maxfreq: 18.00 Hz
```

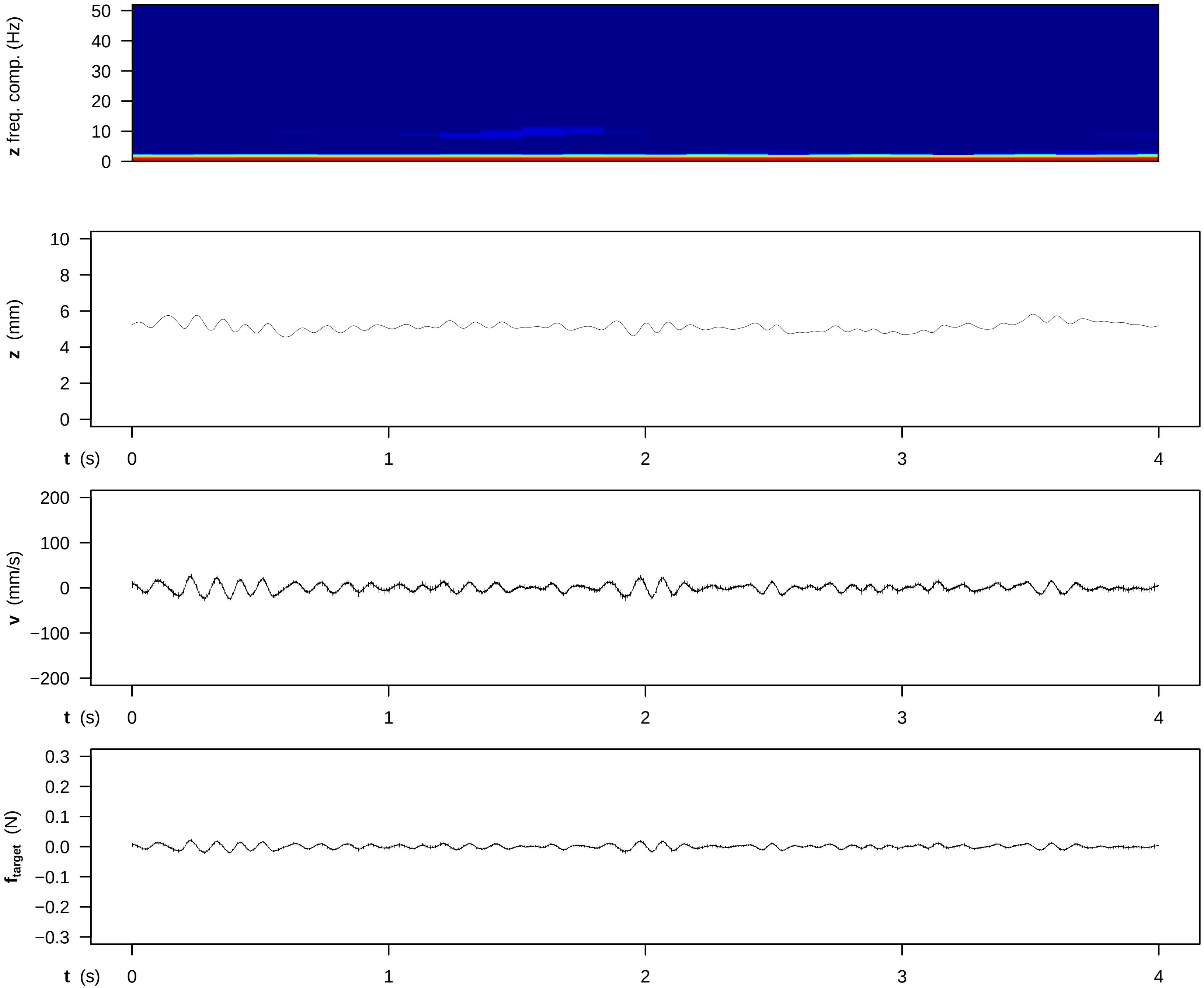

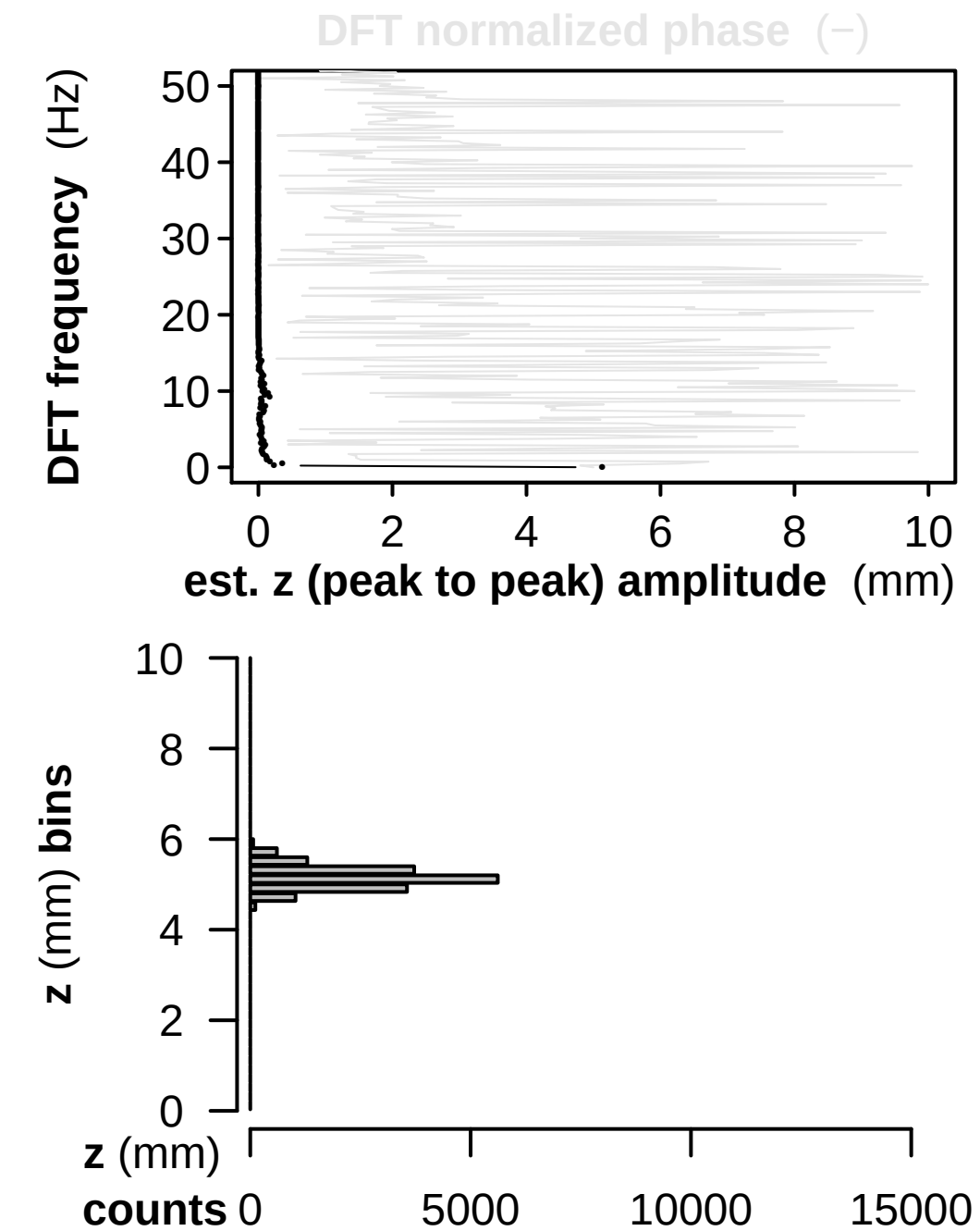

**SUBJECT 7 − RUN 16 − CONDITION 4,1**
**SC_180323_154532_0.AIFF**

```
z_min               : 4.56 mm
z_max               : 5.83 mm
z_travel_amplitude  : 1.27 mm

avg_abs_z_travel    : 7.92 mm/s

z_jarque-bera_jb    : 368.32
z_jarque-bera_p     : 0.00e+00

z_lin_mod_est_slope: 0.02 mm/s
z_lin_mod_adj_R²    : 1 %

z_poly40_mod_adj_R²: 59 %

z_dft_ampl_thresh   : 0.010 mm
>=threshold_maxfreq: 15.75 Hz
```

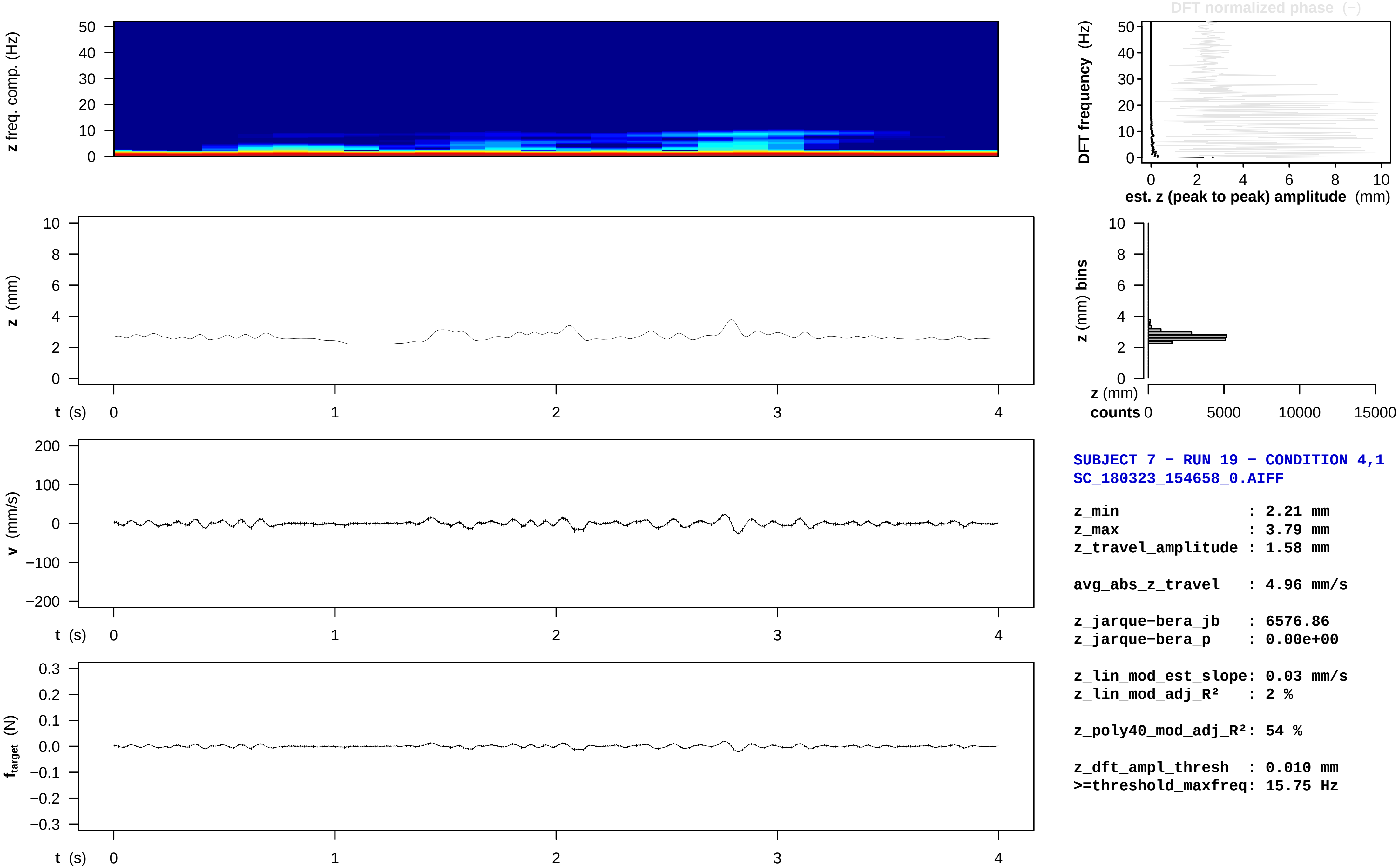

**SUBJECT 7 − RUN 19 − CONDITION 4,1**
**SC_180323_154658_0.AIFF**

```
z_min               : 2.21 mm
z_max               : 3.79 mm
z_travel_amplitude  : 1.58 mm

avg_abs_z_travel    : 4.96 mm/s

z_jarque-bera_jb    : 6576.86
z_jarque-bera_p     : 0.00e+00

z_lin_mod_est_slope: 0.03 mm/s
z_lin_mod_adj_R²    : 2 %

z_poly40_mod_adj_R²: 54 %

z_dft_ampl_thresh   : 0.010 mm
>=threshold_maxfreq: 15.75 Hz
```

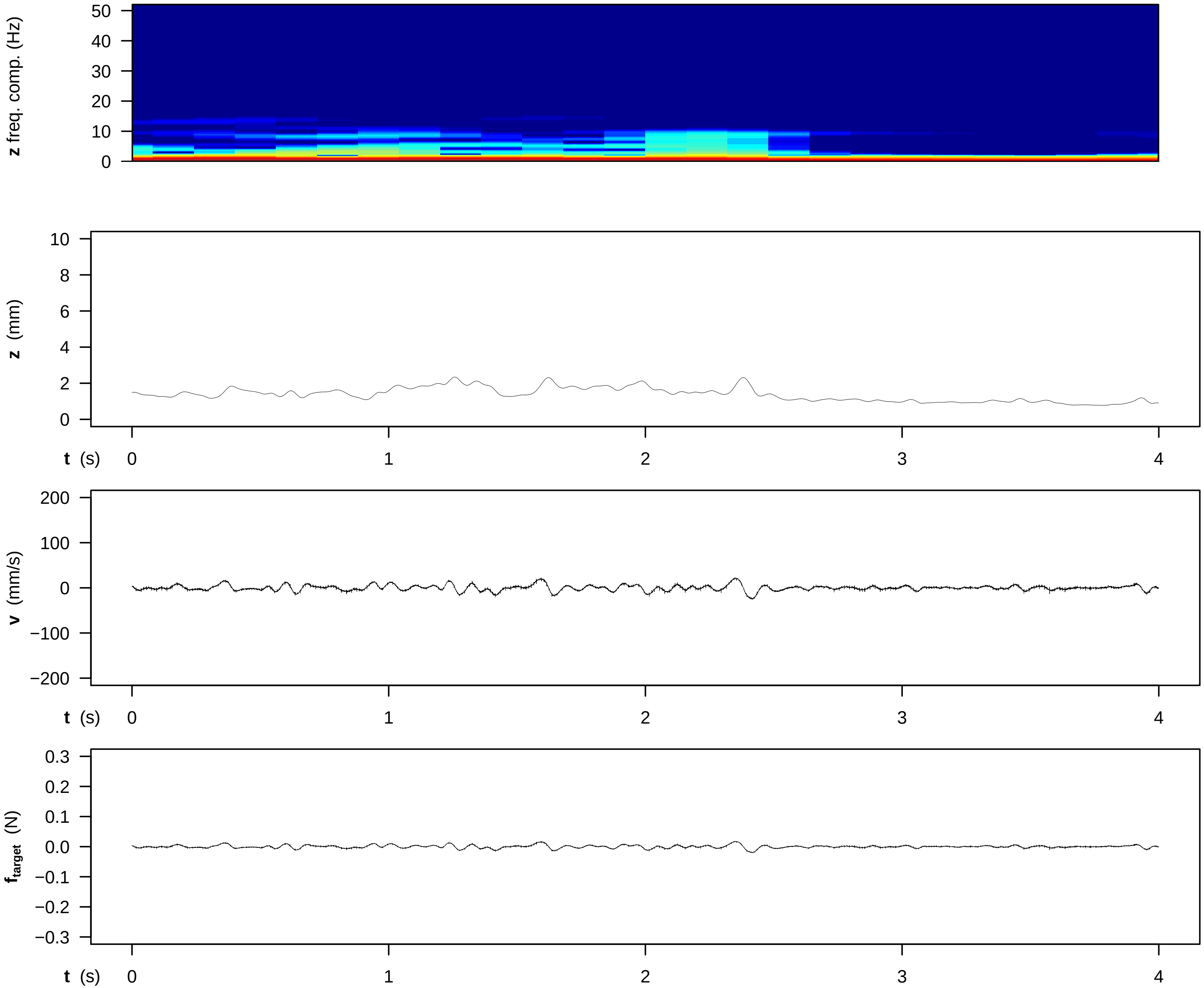

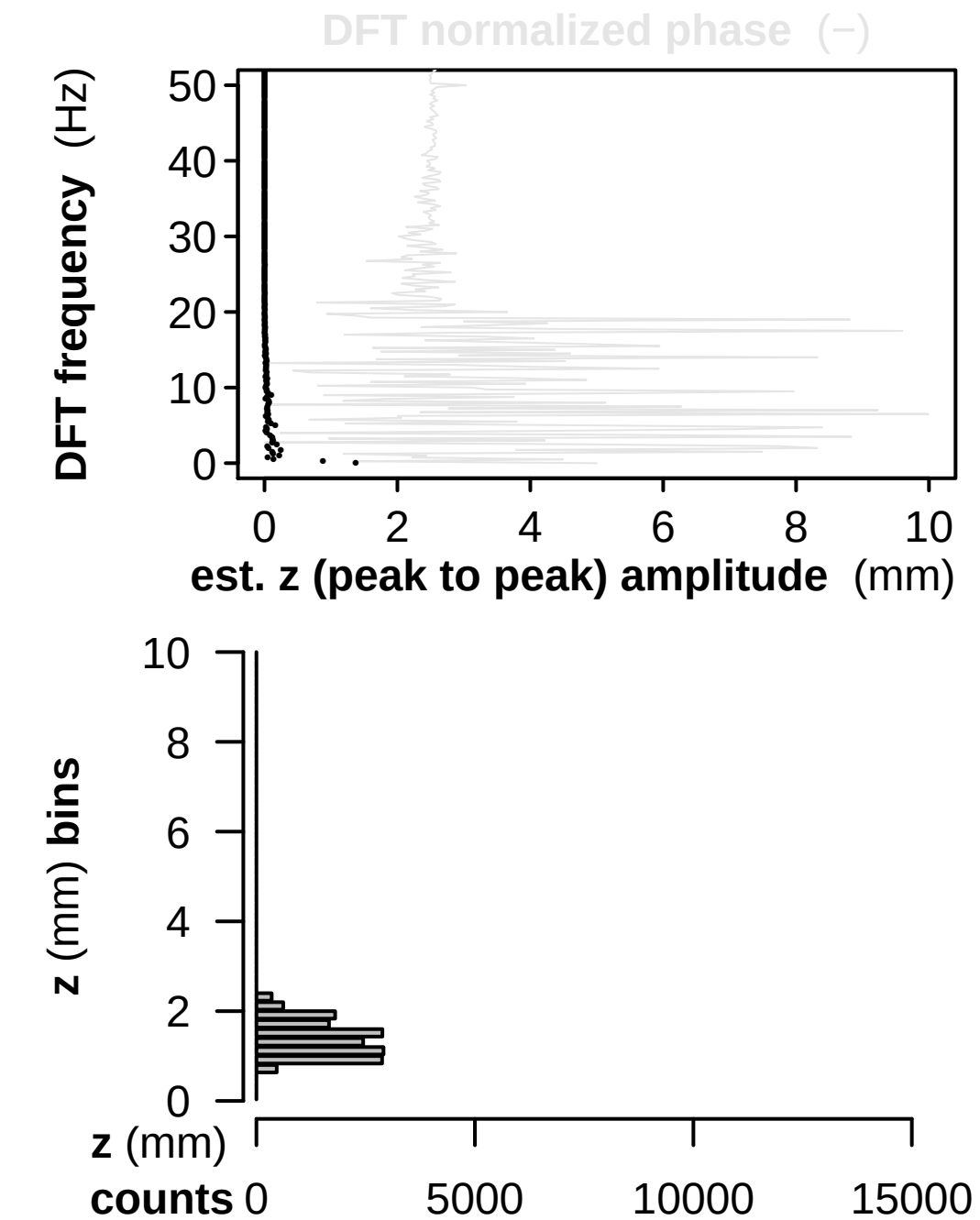

**SUBJECT 7 − RUN 25 − CONDITION 4,1**
**SC_180323_155315_0.AIFF**

```
z_min               : 0.77 mm
z_max               : 2.34 mm
z_travel_amplitude  : 1.57 mm

avg_abs_z_travel    : 5.28 mm/s

z_jarque-bera_jb    : 796.60
z_jarque-bera_p     : 0.00e+00

z_lin_mod_est_slope: −0.19 mm/s
z_lin_mod_adj_R²    : 34 %

z_poly40_mod_adj_R²: 81 %

z_dft_ampl_thresh   : 0.010 mm
>=threshold_maxfreq: 18.00 Hz
```

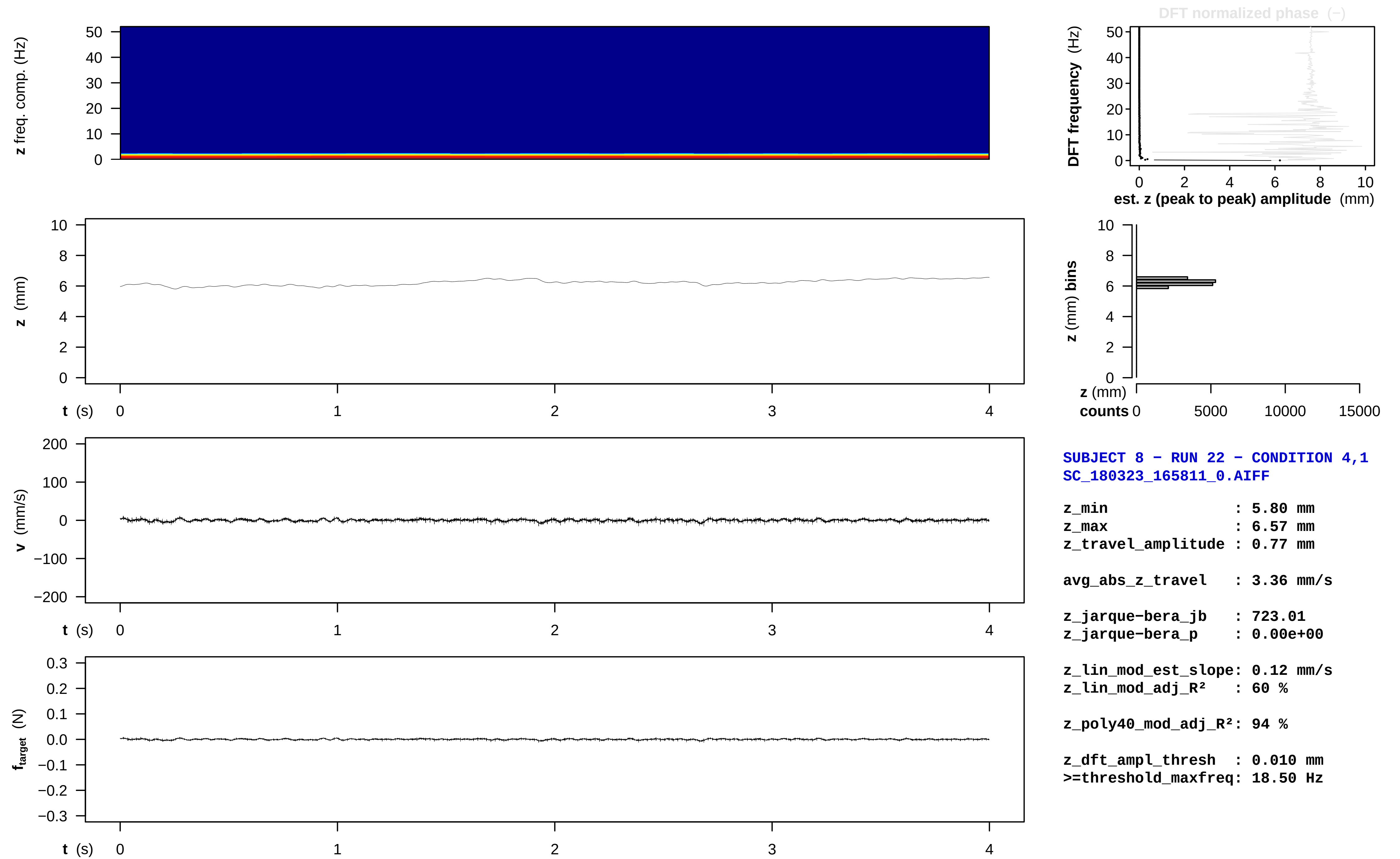

**SUBJECT 8 − RUN 22 − CONDITION 4,1**
**SC_180323_165811_0.AIFF**

```
z_min              : 5.80 mm
z_max              : 6.57 mm
z_travel_amplitude : 0.77 mm

avg_abs_z_travel   : 3.36 mm/s

z_jarque-bera_jb   : 723.01
z_jarque-bera_p    : 0.00e+00

z_lin_mod_est_slope: 0.12 mm/s
z_lin_mod_adj_R²   : 60 %

z_poly40_mod_adj_R²: 94 %

z_dft_ampl_thresh  : 0.010 mm
>=threshold_maxfreq: 18.50 Hz
```

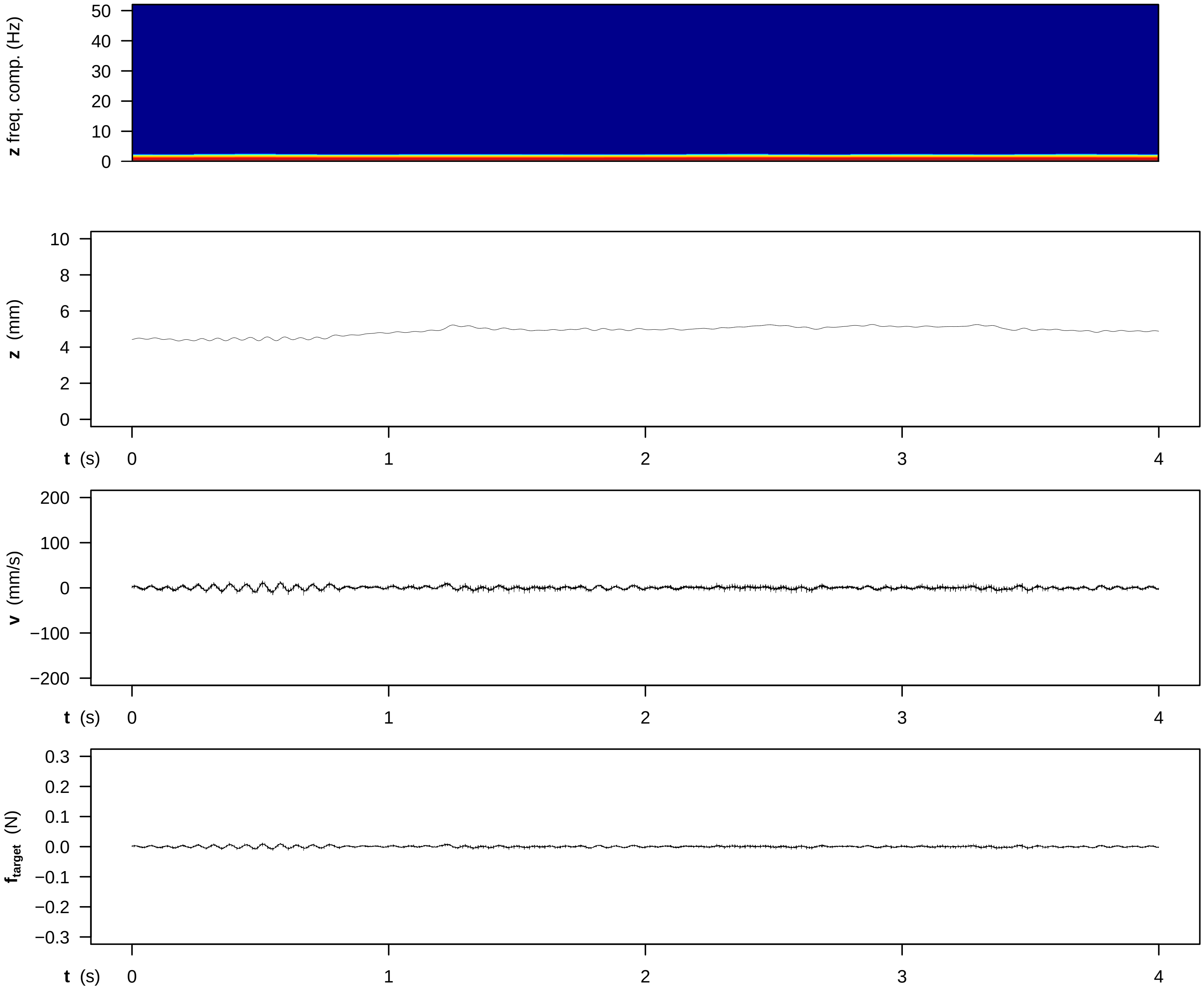

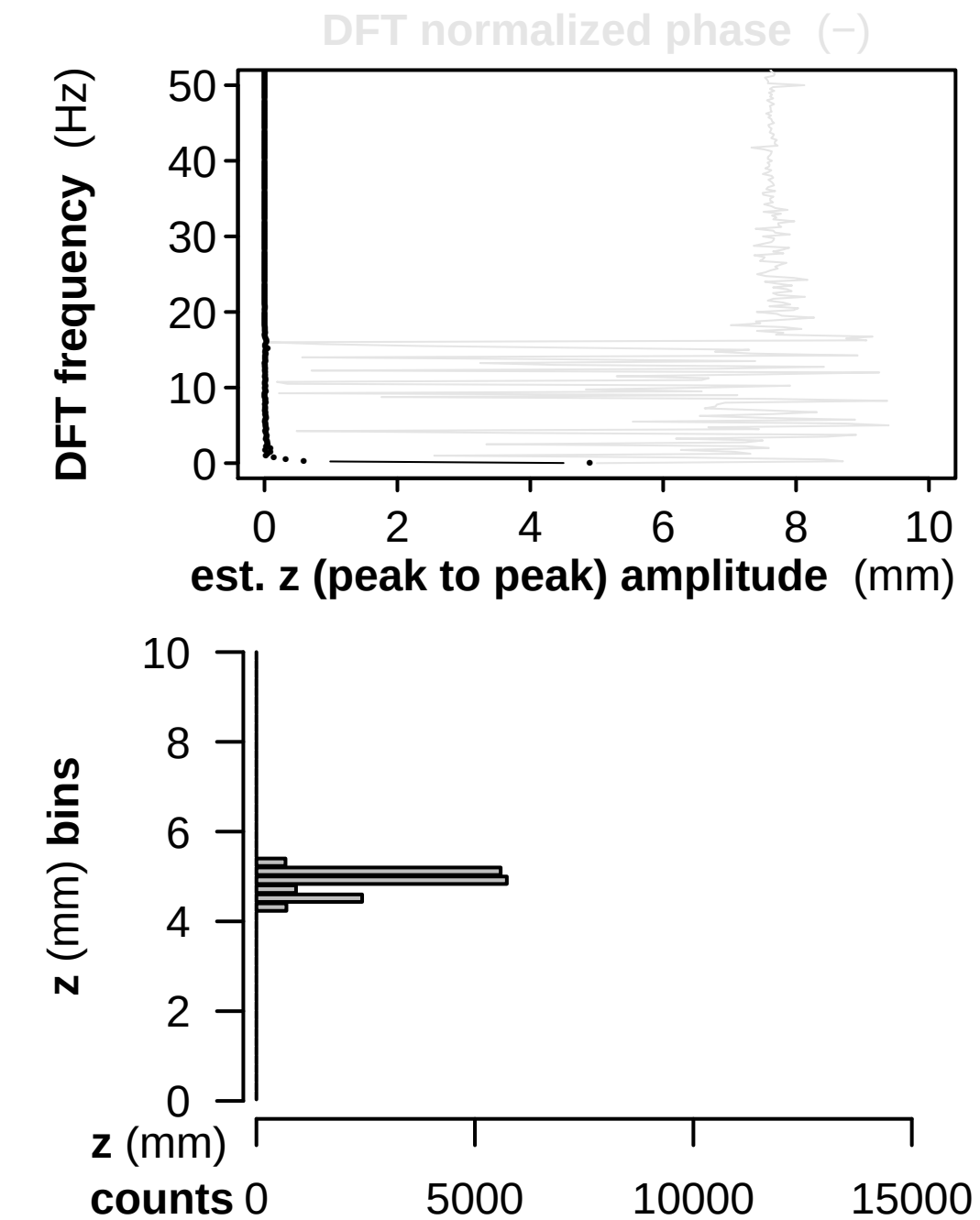

**SUBJECT 8 − RUN 26 − CONDITION 4,1**
**SC_180323_170401_0.AIFF**

```
z_min               : 4.34 mm
z_max               : 5.25 mm
z_travel_amplitude  : 0.91 mm

avg_abs_z_travel    : 3.56 mm/s

z_jarque-bera_jb    : 1988.36
z_jarque-bera_p     : 0.00e+00

z_lin_mod_est_slope: 0.16 mm/s
z_lin_mod_adj_R²    : 51 %

z_poly40_mod_adj_R²: 96 %

z_dft_ampl_thresh   : 0.010 mm
>=threshold_maxfreq: 17.25 Hz
```

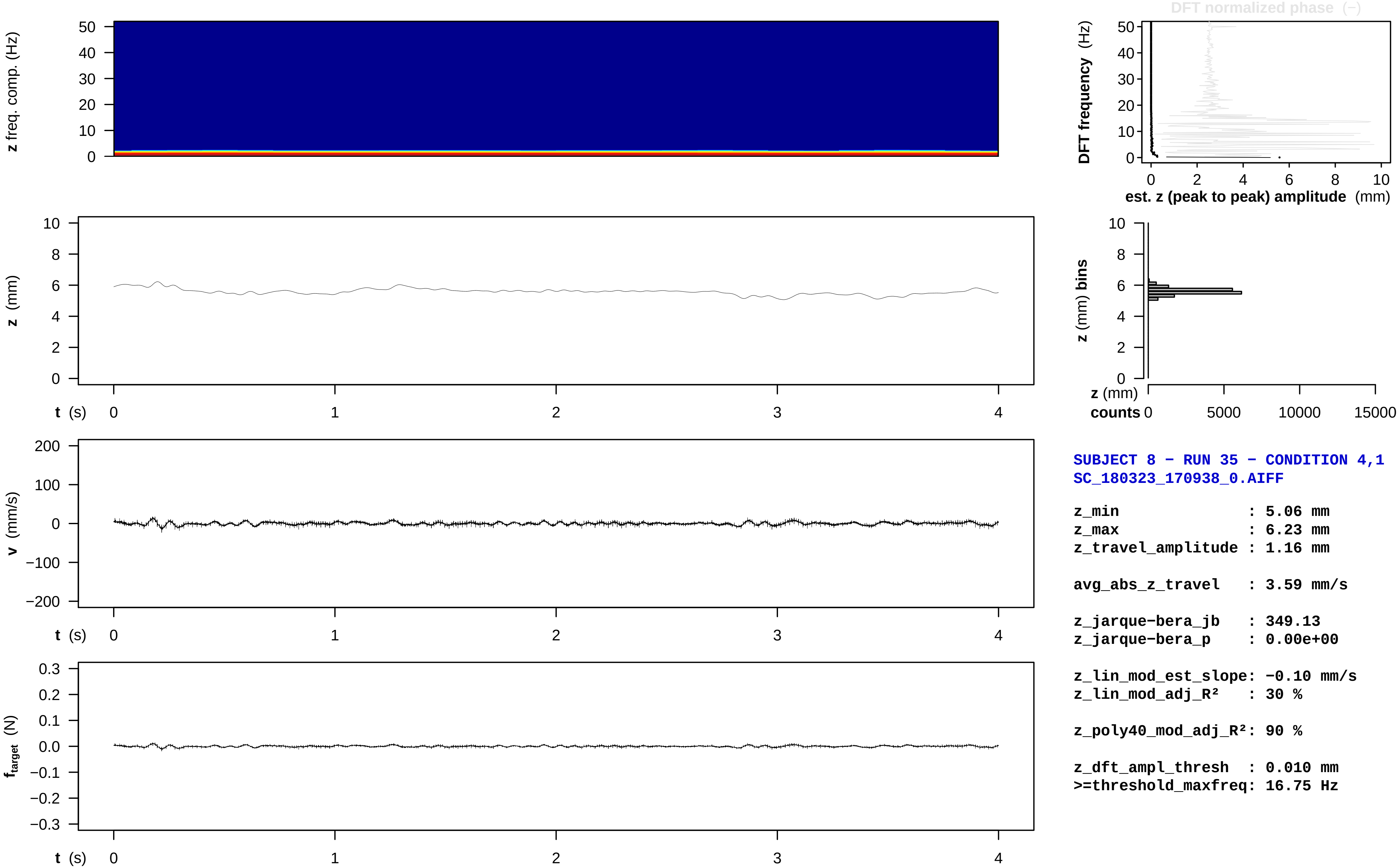

**SUBJECT 8 − RUN 35 − CONDITION 4,1**
**SC_180323_170938_0.AIFF**

```
z_min                : 5.06 mm
z_max                : 6.23 mm
z_travel_amplitude   : 1.16 mm

avg_abs_z_travel     : 3.59 mm/s

z_jarque-bera_jb     : 349.13
z_jarque-bera_p      : 0.00e+00

z_lin_mod_est_slope: −0.10 mm/s
z_lin_mod_adj_R²    : 30 %

z_poly40_mod_adj_R²: 90 %

z_dft_ampl_thresh   : 0.010 mm
>=threshold_maxfreq: 16.75 Hz
```

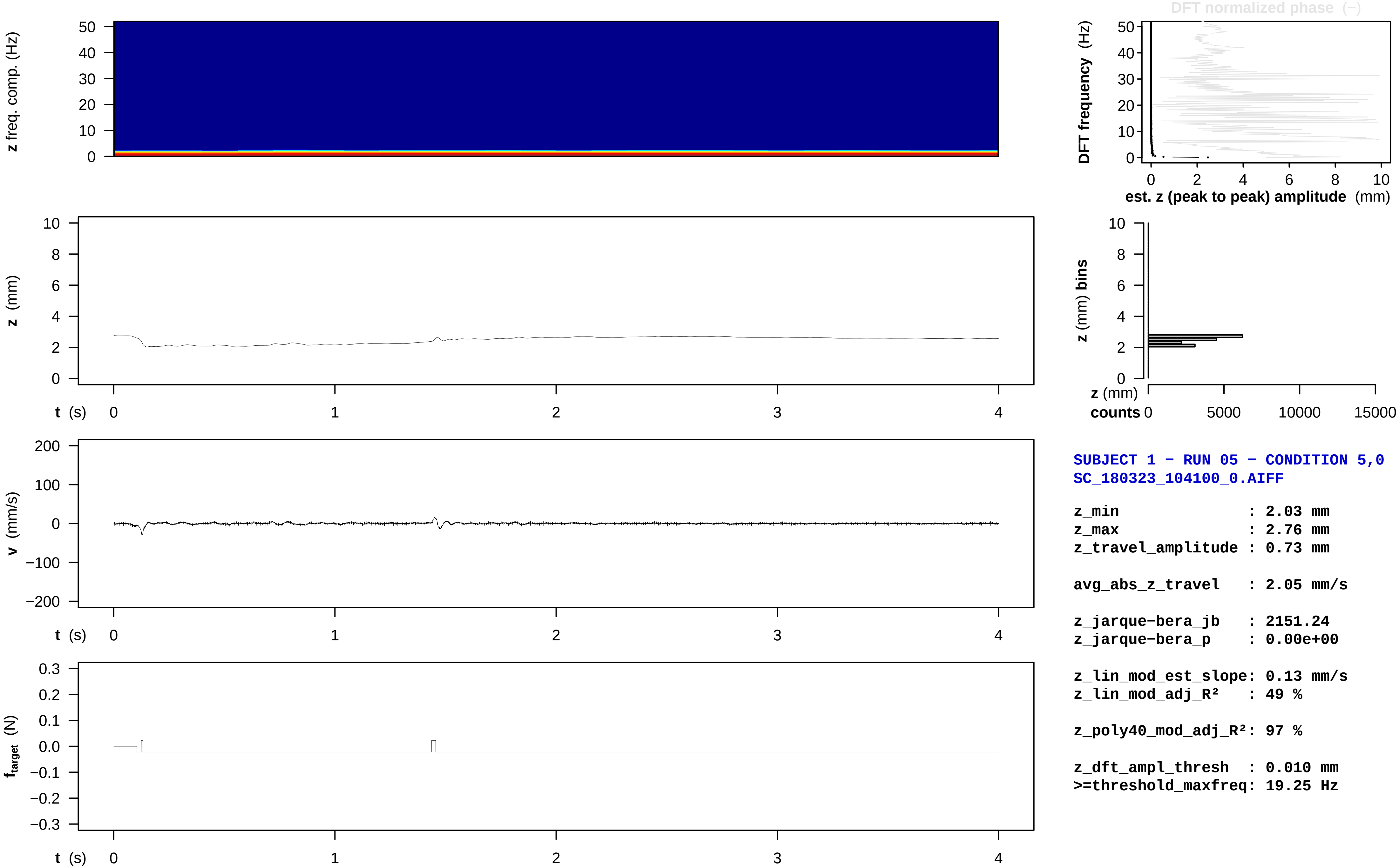

**SUBJECT 1 − RUN 05 − CONDITION 5,0**
**SC_180323_104100_0.AIFF**

```
z_min               : 2.03 mm
z_max               : 2.76 mm
z_travel_amplitude  : 0.73 mm

avg_abs_z_travel    : 2.05 mm/s

z_jarque-bera_jb    : 2151.24
z_jarque-bera_p     : 0.00e+00

z_lin_mod_est_slope: 0.13 mm/s
z_lin_mod_adj_R²    : 49 %

z_poly40_mod_adj_R²: 97 %

z_dft_ampl_thresh   : 0.010 mm
>=threshold_maxfreq: 19.25 Hz
```

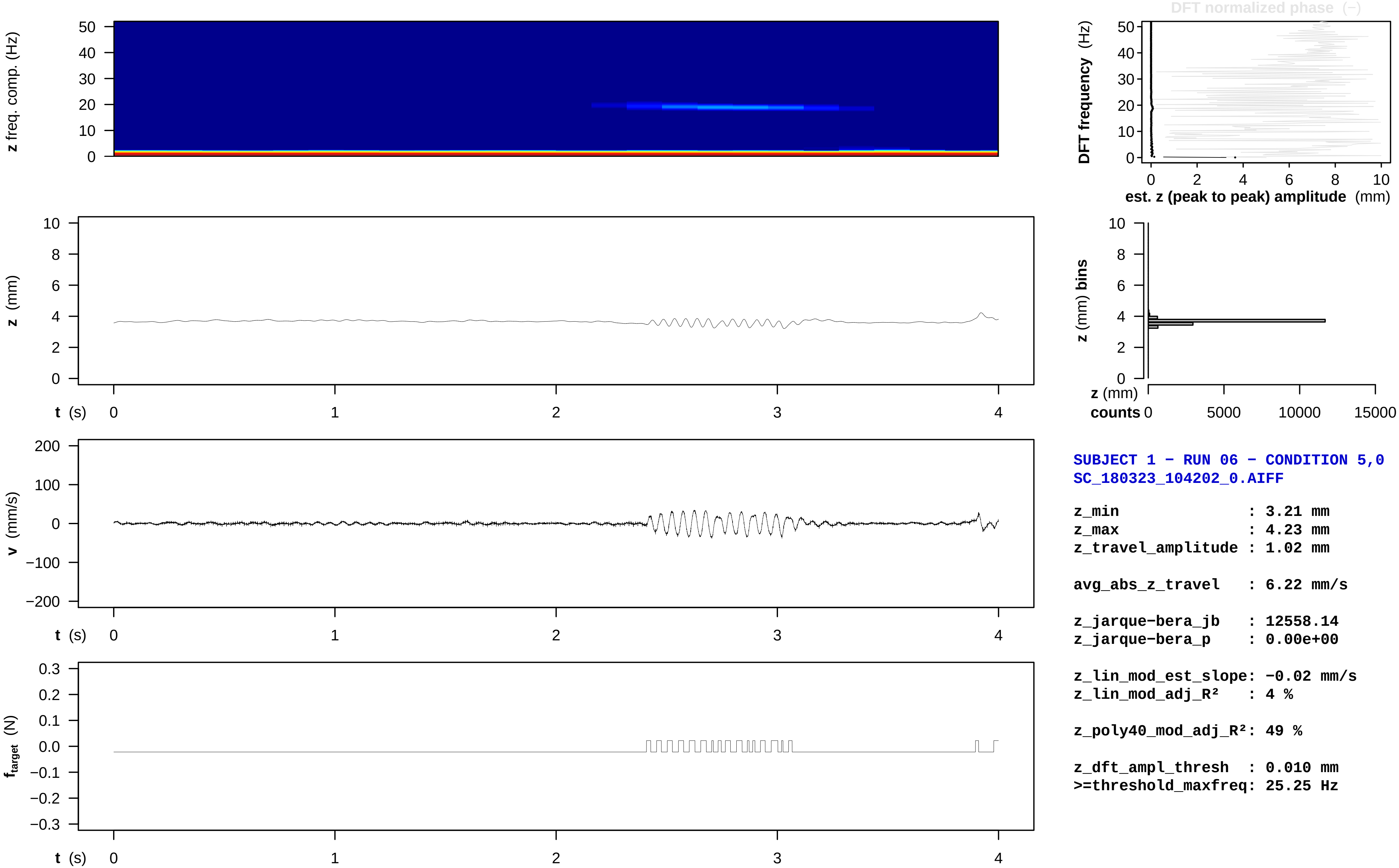

**SUBJECT 1 − RUN 06 − CONDITION 5,0**
**SC_180323_104202_0.AIFF**

```
z_min              : 3.21 mm
z_max              : 4.23 mm
z_travel_amplitude : 1.02 mm

avg_abs_z_travel   : 6.22 mm/s

z_jarque-bera_jb   : 12558.14
z_jarque-bera_p    : 0.00e+00

z_lin_mod_est_slope: −0.02 mm/s
z_lin_mod_adj_R²   : 4 %

z_poly40_mod_adj_R²: 49 %

z_dft_ampl_thresh  : 0.010 mm
>=threshold_maxfreq: 25.25 Hz
```

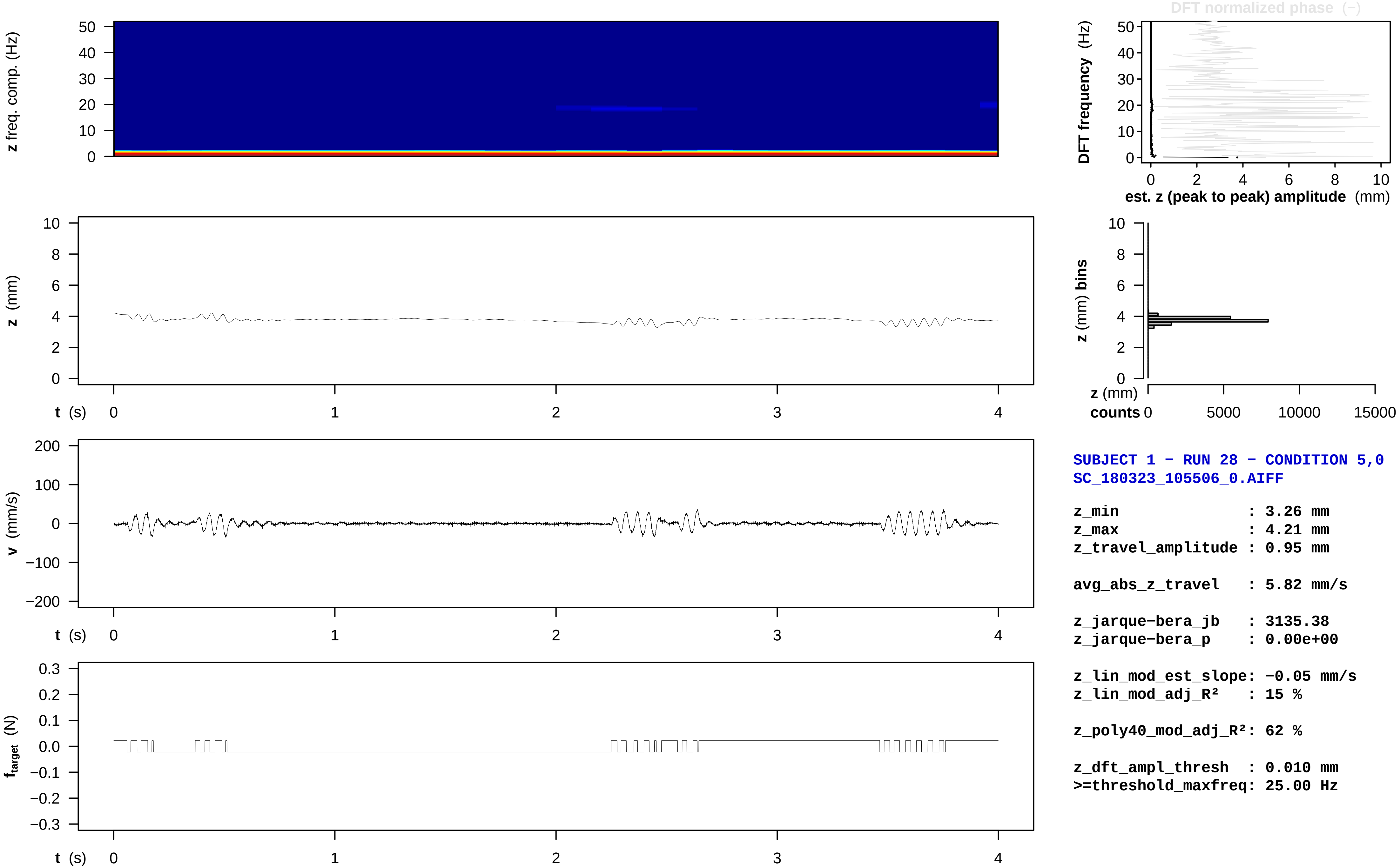

SUBJECT 1 − RUN 28 − CONDITION 5,0
SC_180323_105506_0.AIFF

```
z_min               : 3.26 mm
z_max               : 4.21 mm
z_travel_amplitude : 0.95 mm

avg_abs_z_travel    : 5.82 mm/s

z_jarque-bera_jb    : 3135.38
z_jarque-bera_p     : 0.00e+00

z_lin_mod_est_slope: -0.05 mm/s
z_lin_mod_adj_R²    : 15 %

z_poly40_mod_adj_R²: 62 %

z_dft_ampl_thresh   : 0.010 mm
>=threshold_maxfreq: 25.00 Hz
```

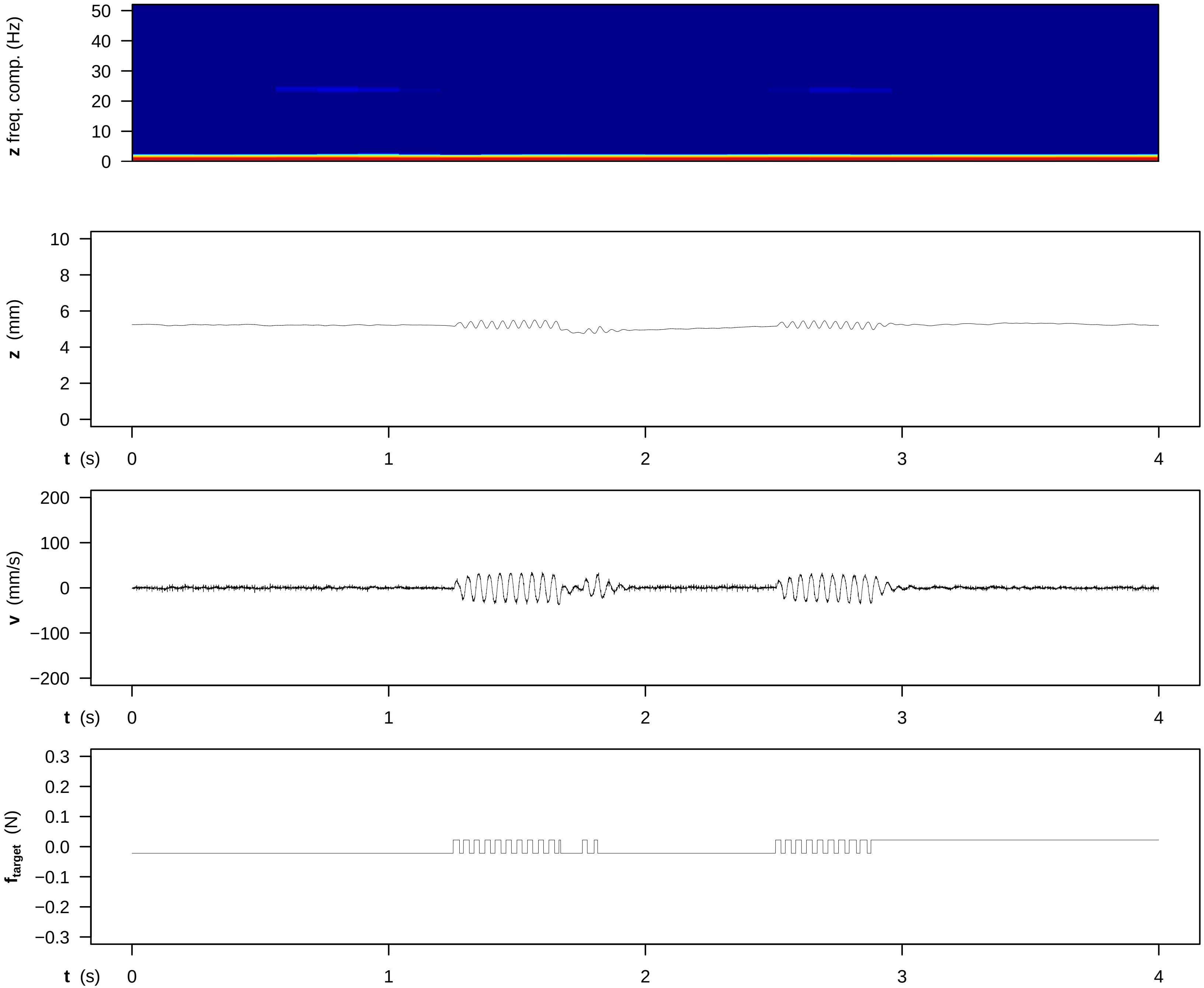

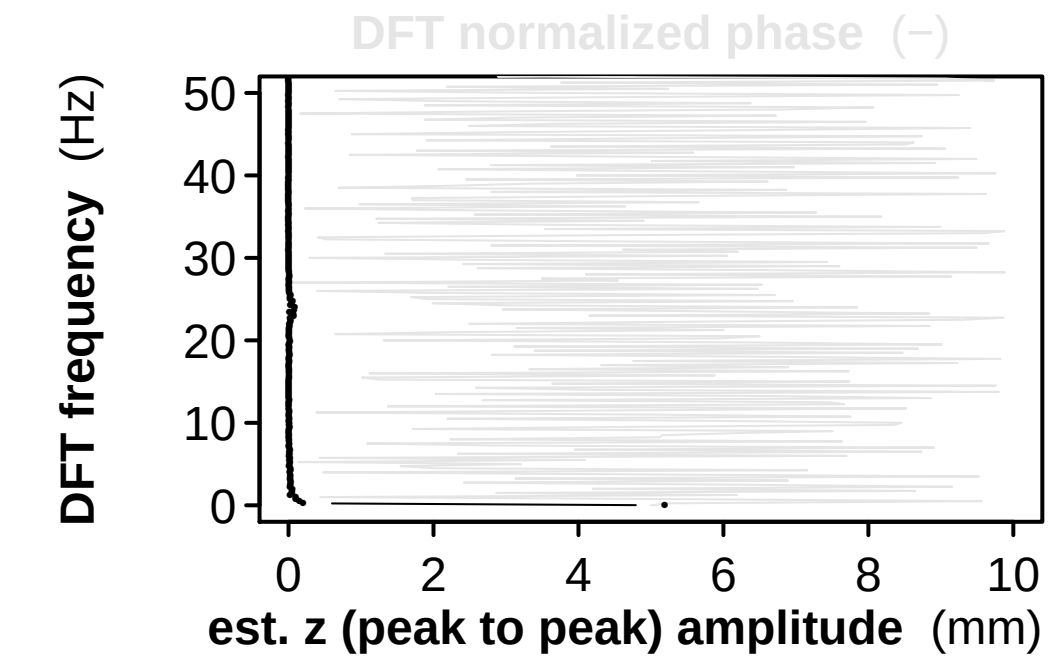

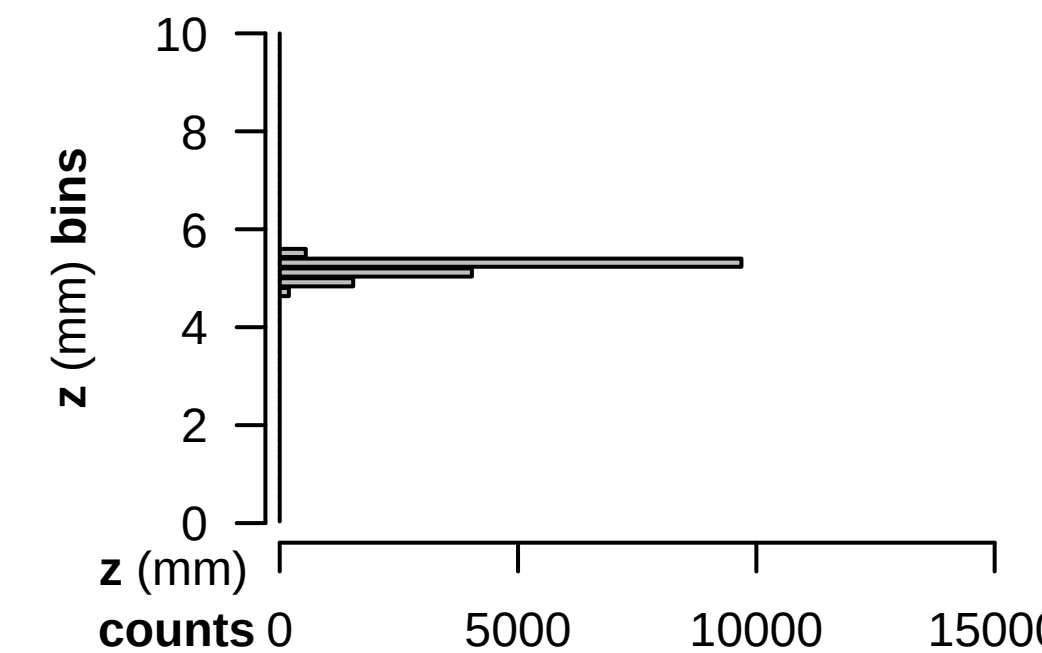

**SUBJECT 2 − RUN 06 − CONDITION 5,0**
**SC_180323_111822_0.AIFF**

```
z_min                : 4.75 mm
z_max                : 5.50 mm
z_travel_amplitude   : 0.75 mm

avg_abs_z_travel     : 7.27 mm/s

z_jarque-bera_jb     : 2526.08
z_jarque-bera_p      : 0.00e+00

z_lin_mod_est_slope: 0.01 mm/s
z_lin_mod_adj_R²     : 1 %

z_poly40_mod_adj_R²: 64 %

z_dft_ampl_thresh    : 0.010 mm
>=threshold_maxfreq: 27.75 Hz
```

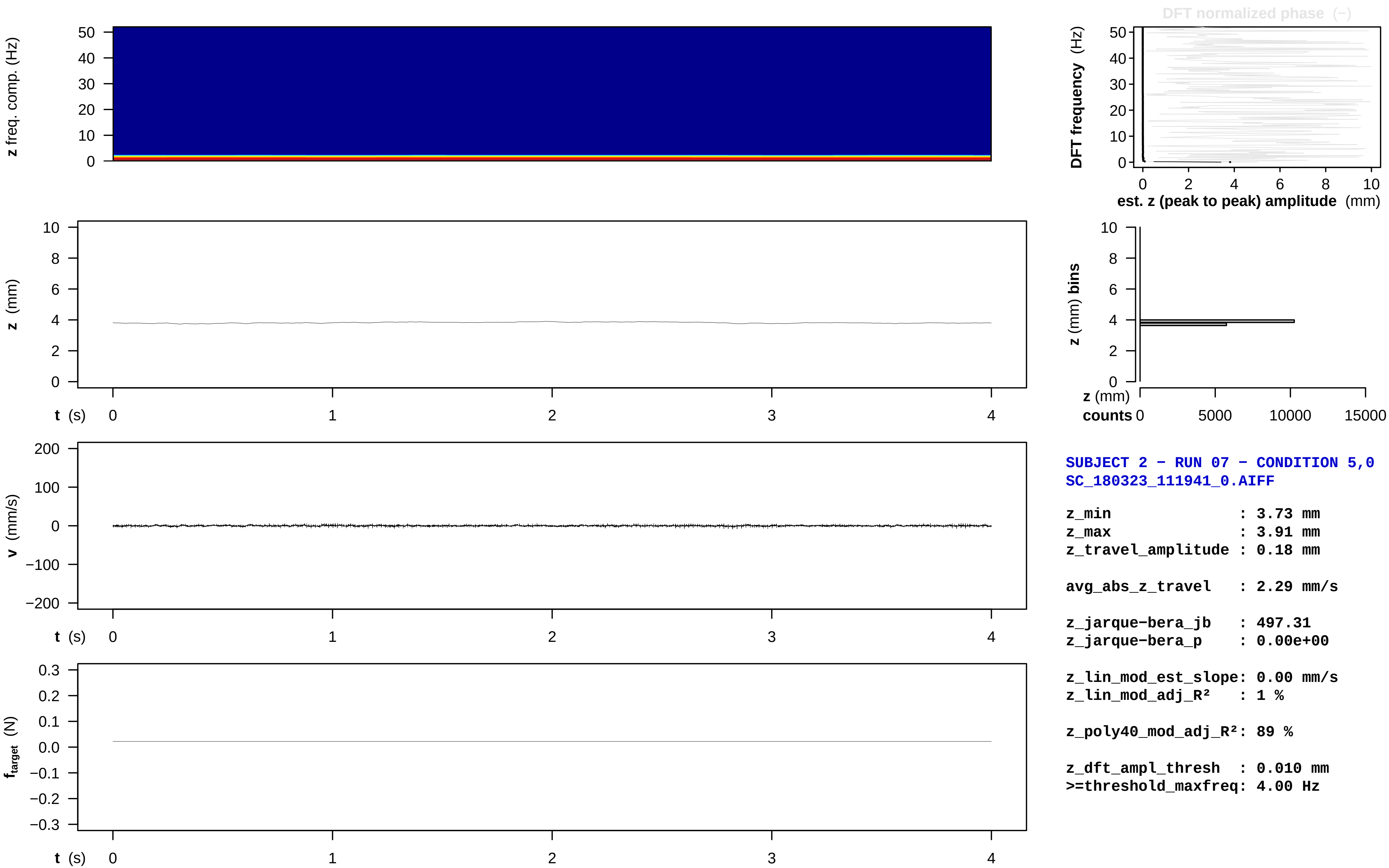

**SUBJECT 2 − RUN 07 − CONDITION 5,0**
**SC_180323_111941_0.AIFF**

```
z_min               : 3.73 mm
z_max               : 3.91 mm
z_travel_amplitude  : 0.18 mm

avg_abs_z_travel    : 2.29 mm/s

z_jarque-bera_jb    : 497.31
z_jarque-bera_p     : 0.00e+00

z_lin_mod_est_slope: 0.00 mm/s
z_lin_mod_adj_R²    : 1 %

z_poly40_mod_adj_R²: 89 %

z_dft_ampl_thresh   : 0.010 mm
>=threshold_maxfreq: 4.00 Hz
```

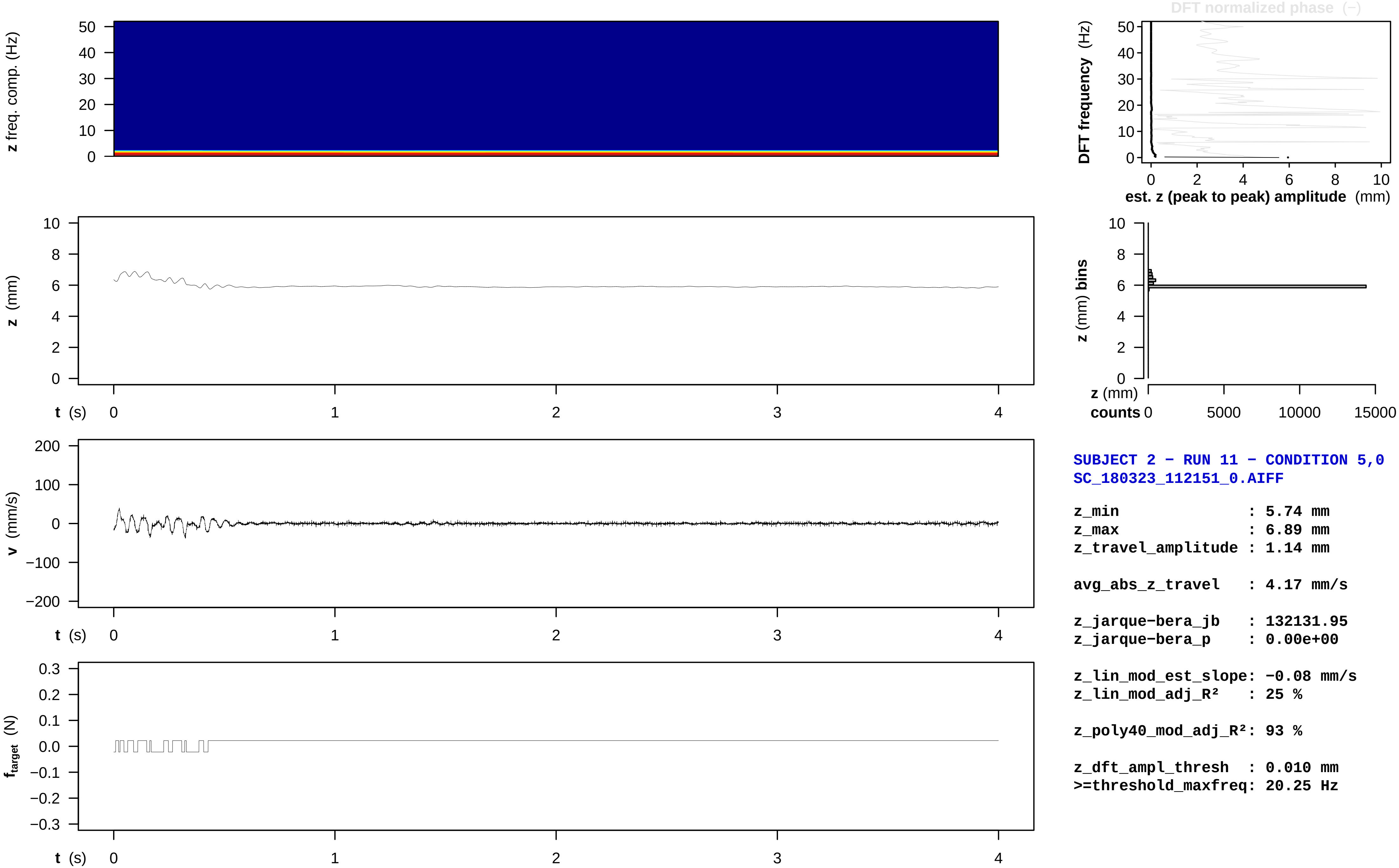

```
SUBJECT 2 − RUN 11 − CONDITION 5,0
SC_180323_112151_0.AIFF

z_min               : 5.74 mm
z_max               : 6.89 mm
z_travel_amplitude : 1.14 mm

avg_abs_z_travel    : 4.17 mm/s

z_jarque-bera_jb    : 132131.95
z_jarque-bera_p     : 0.00e+00

z_lin_mod_est_slope: −0.08 mm/s
z_lin_mod_adj_R²    : 25 %

z_poly40_mod_adj_R²: 93 %

z_dft_ampl_thresh   : 0.010 mm
>=threshold_maxfreq: 20.25 Hz
```

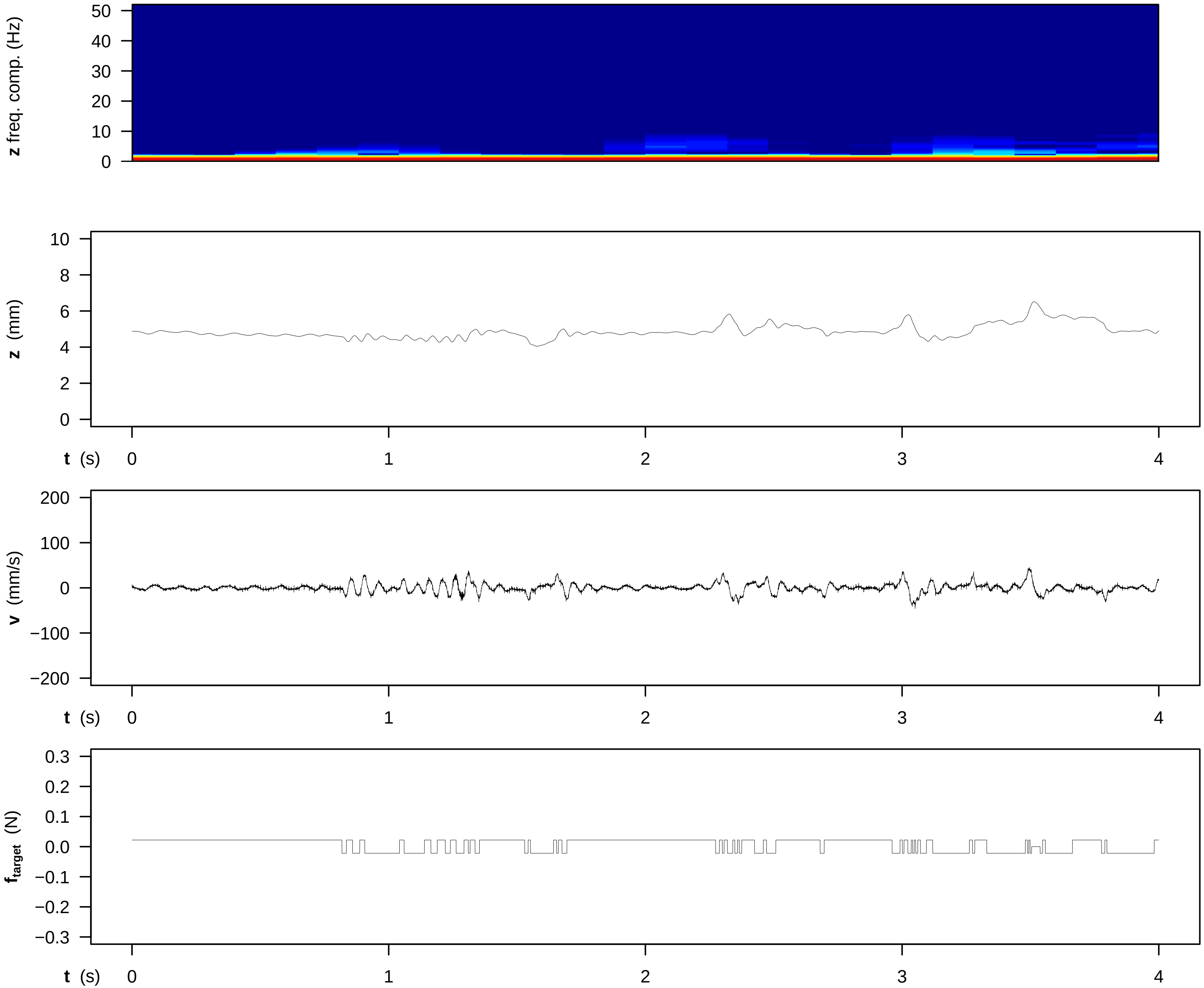

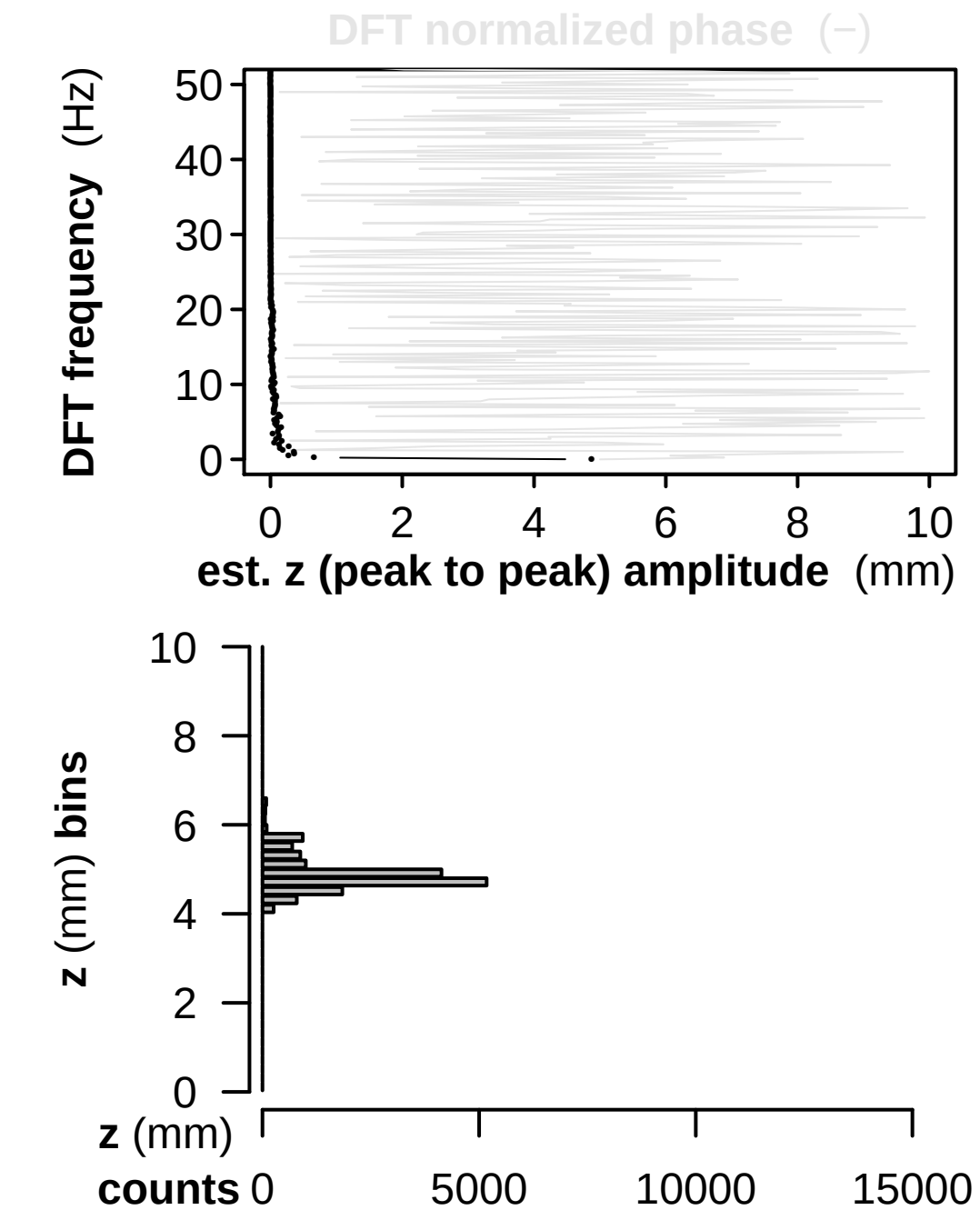

**SUBJECT 3 − RUN 05 − CONDITION 5,0**
**SC_180323_115733_0.AIFF**

```
z_min               : 4.05 mm
z_max               : 6.51 mm
z_travel_amplitude  : 2.46 mm

avg_abs_z_travel    : 7.20 mm/s

z_jarque-bera_jb    : 6481.50
z_jarque-bera_p     : 0.00e+00

z_lin_mod_est_slope: 0.19 mm/s
z_lin_mod_adj_R²    : 30 %

z_poly40_mod_adj_R²: 74 %

z_dft_ampl_thresh   : 0.010 mm
>=threshold_maxfreq: 24.75 Hz
```

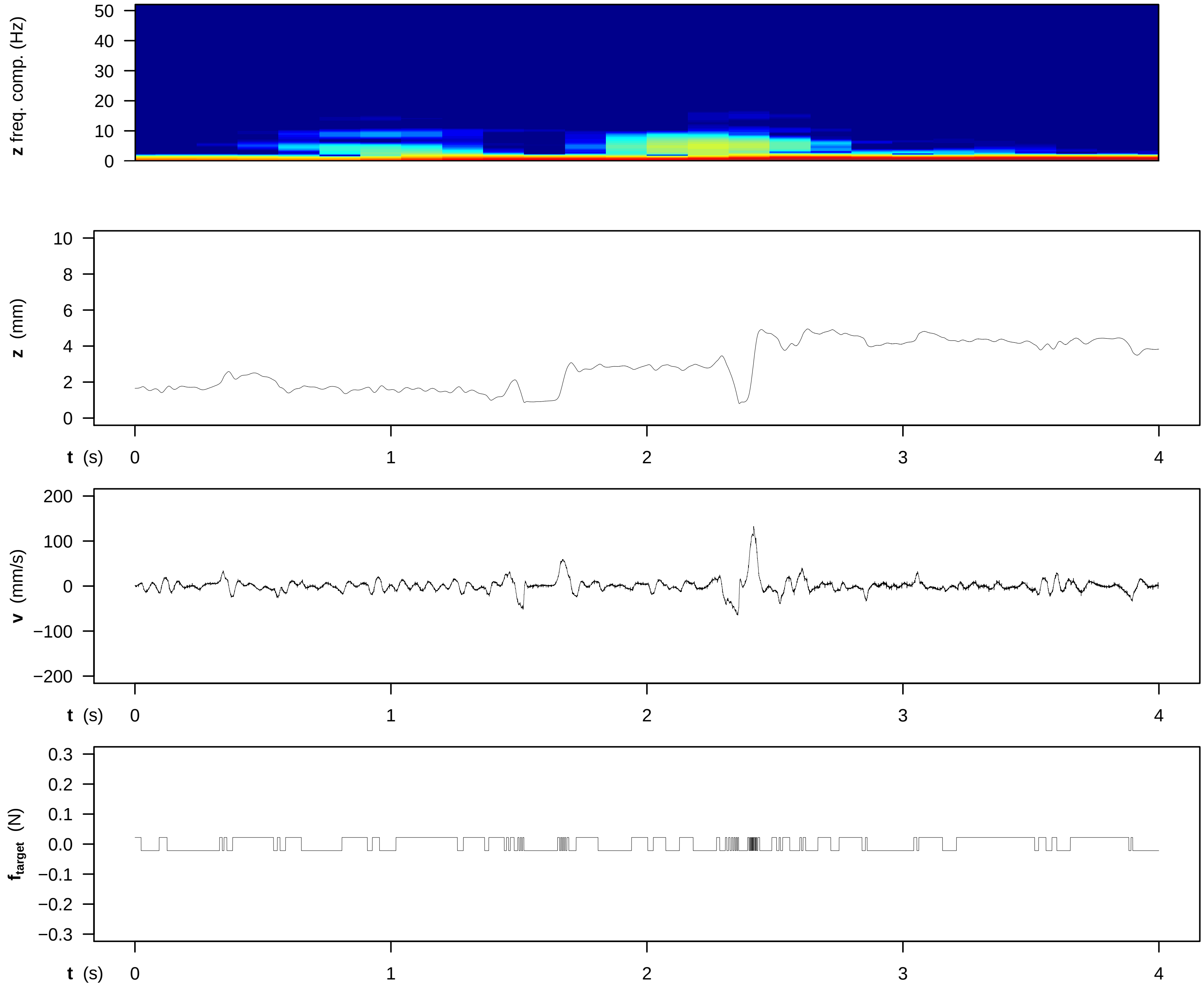

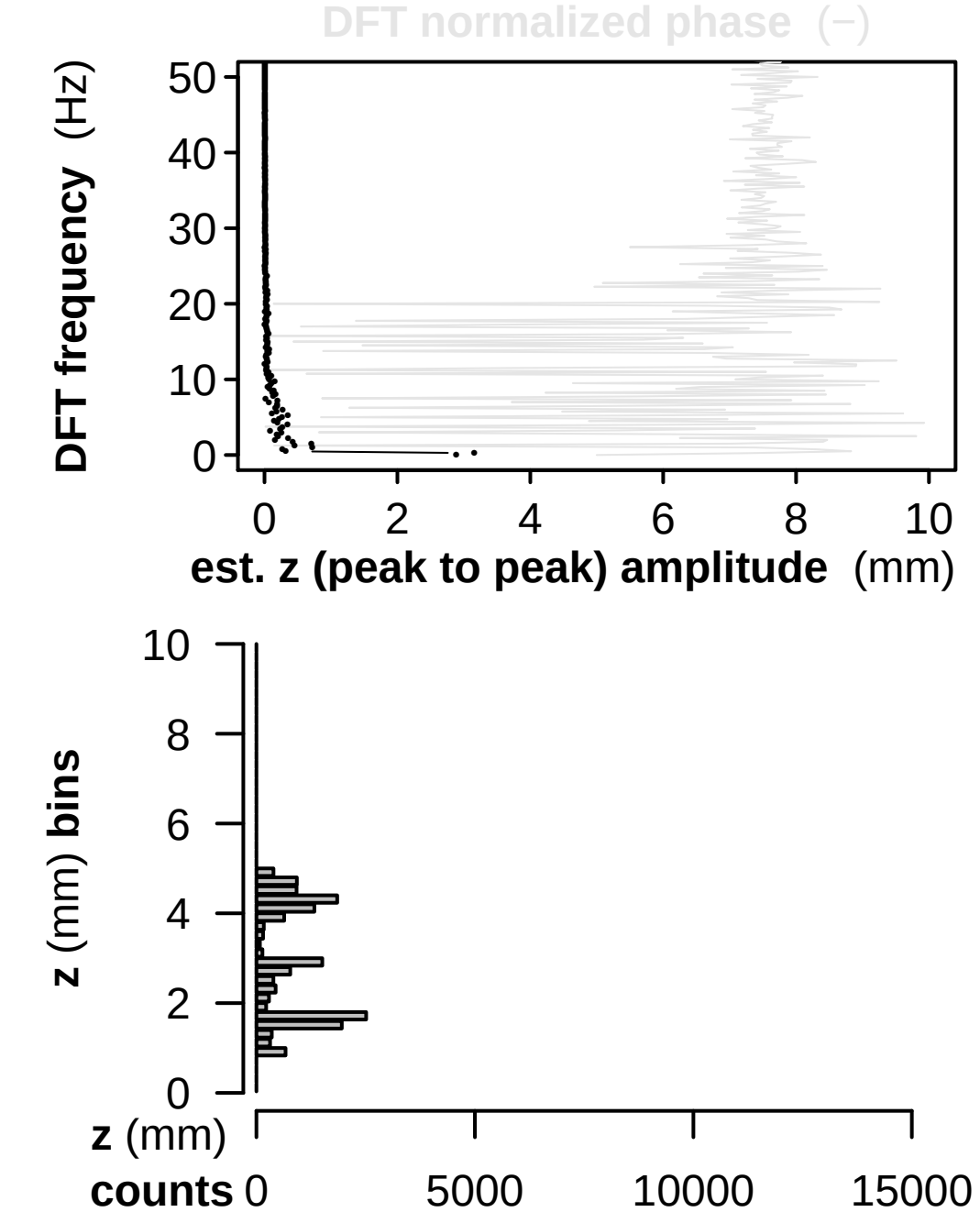

**SUBJECT 3 − RUN 22 − CONDITION 5,0**
**SC_180323_120759_0.AIFF**

```
z_min                : 0.80 mm
z_max                : 4.95 mm
z_travel_amplitude   : 4.15 mm

avg_abs_z_travel     : 9.90 mm/s

z_jarque-bera_jb     : 1577.49
z_jarque-bera_p      : 0.00e+00

z_lin_mod_est_slope: 0.91 mm/s
z_lin_mod_adj_R²     : 68 %

z_poly40_mod_adj_R²: 91 %

z_dft_ampl_thresh    : 0.010 mm
>=threshold_maxfreq: 41.75 Hz
```

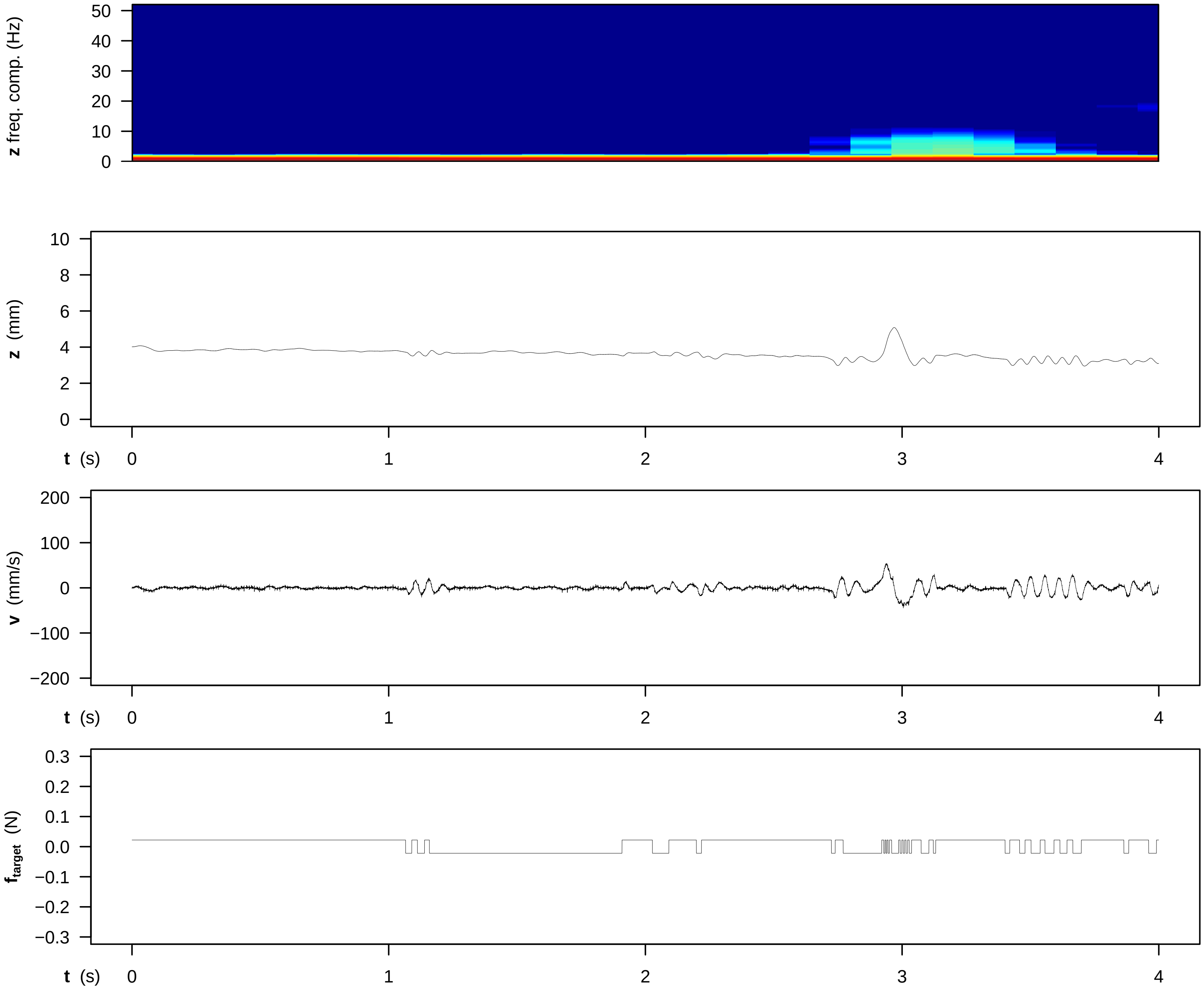

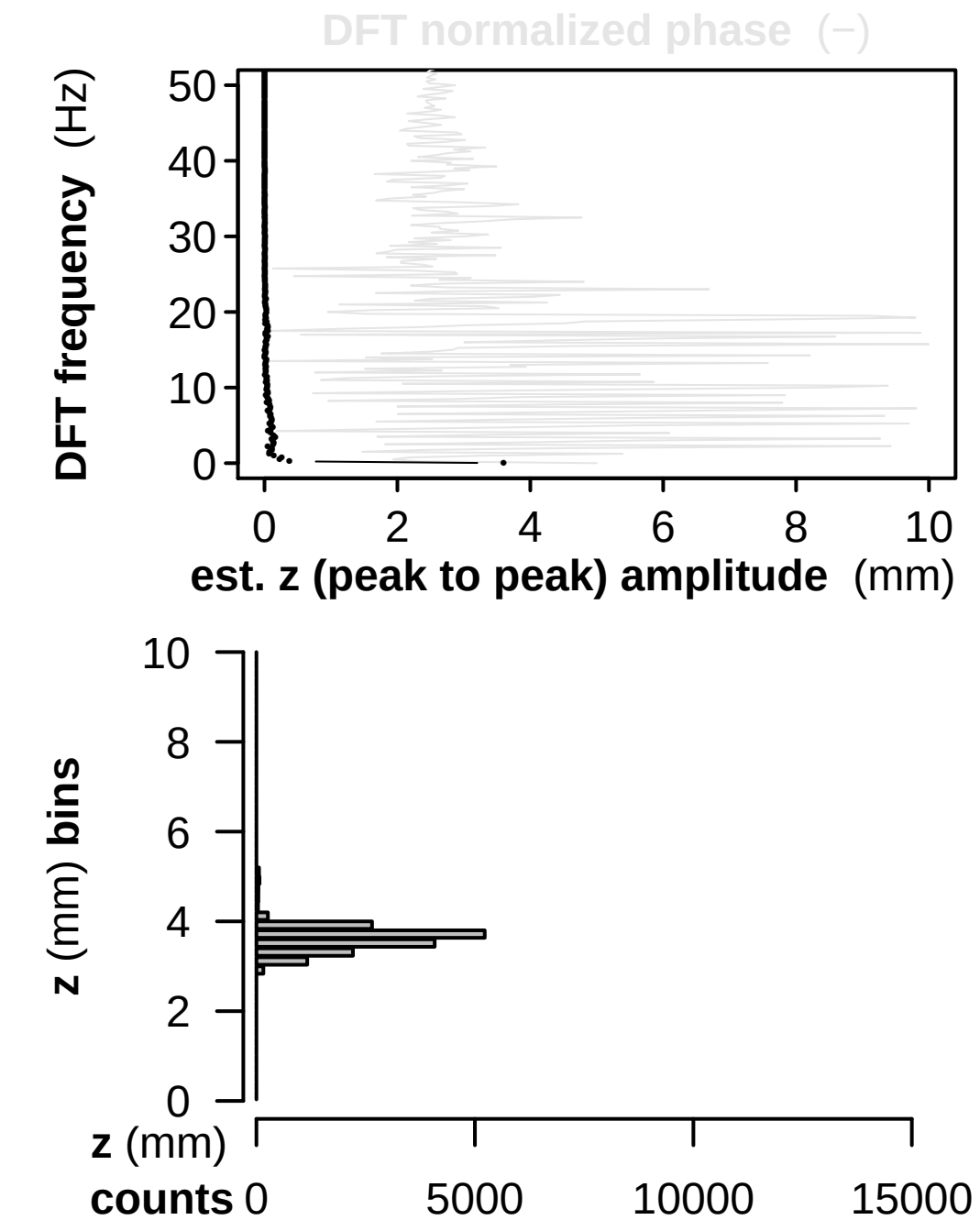

**SUBJECT 3 − RUN 27 − CONDITION 5,0**
**SC_180323_121038_0.AIFF**

```
z_min               : 2.95 mm
z_max               : 5.08 mm
z_travel_amplitude : 2.13 mm

avg_abs_z_travel    : 6.77 mm/s

z_jarque-bera_jb    : 14894.44
z_jarque-bera_p     : 0.00e+00

z_lin_mod_est_slope: −0.16 mm/s
z_lin_mod_adj_R²    : 45 %

z_poly40_mod_adj_R²: 54 %

z_dft_ampl_thresh   : 0.010 mm
>=threshold_maxfreq: 30.00 Hz
```

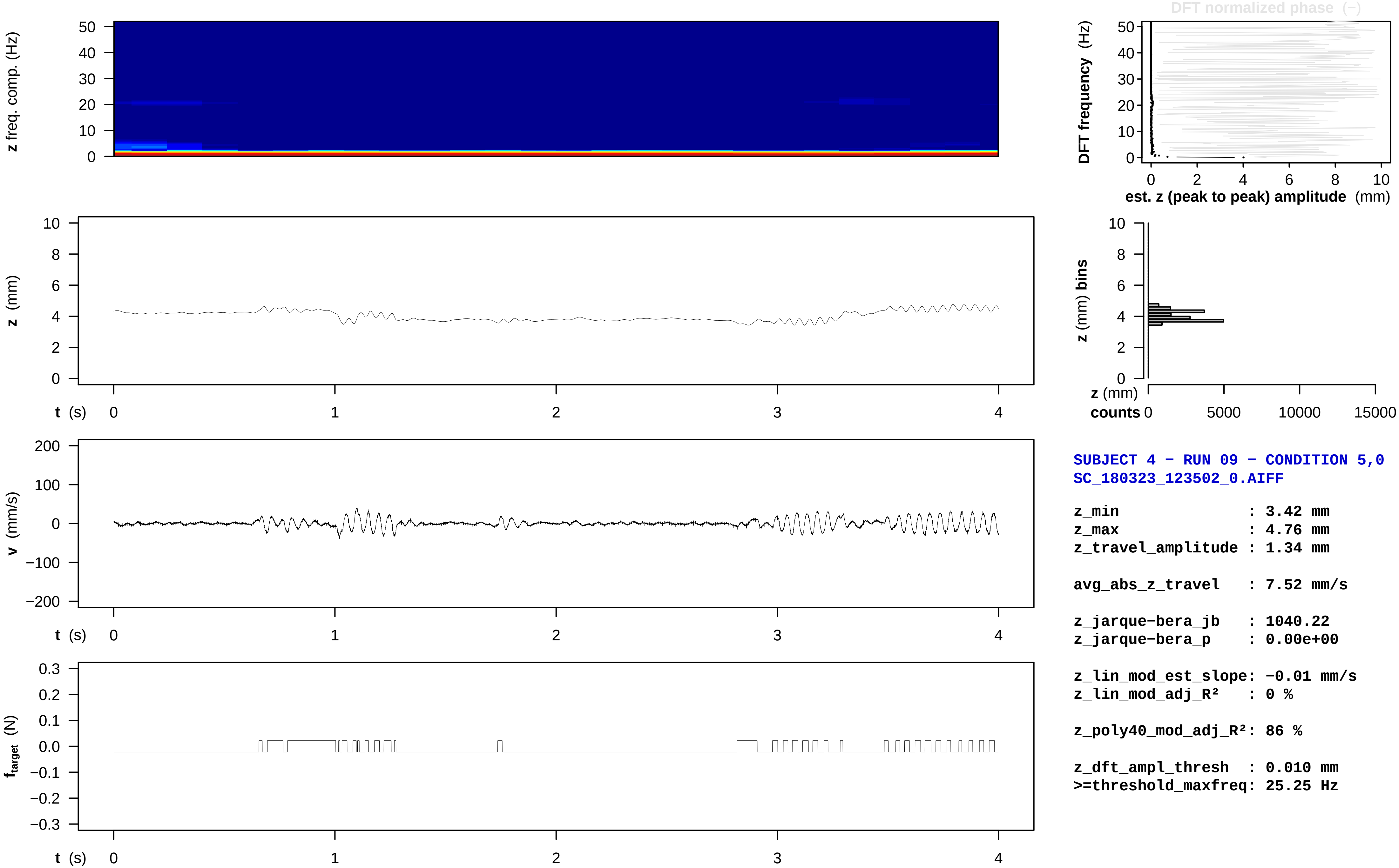

**SUBJECT 4 − RUN 09 − CONDITION 5,0**
**SC_180323_123502_0.AIFF**

```
z_min               : 3.42 mm
z_max               : 4.76 mm
z_travel_amplitude  : 1.34 mm

avg_abs_z_travel    : 7.52 mm/s

z_jarque-bera_jb    : 1040.22
z_jarque-bera_p     : 0.00e+00

z_lin_mod_est_slope: −0.01 mm/s
z_lin_mod_adj_R²    : 0 %

z_poly40_mod_adj_R²: 86 %

z_dft_ampl_thresh   : 0.010 mm
>=threshold_maxfreq: 25.25 Hz
```

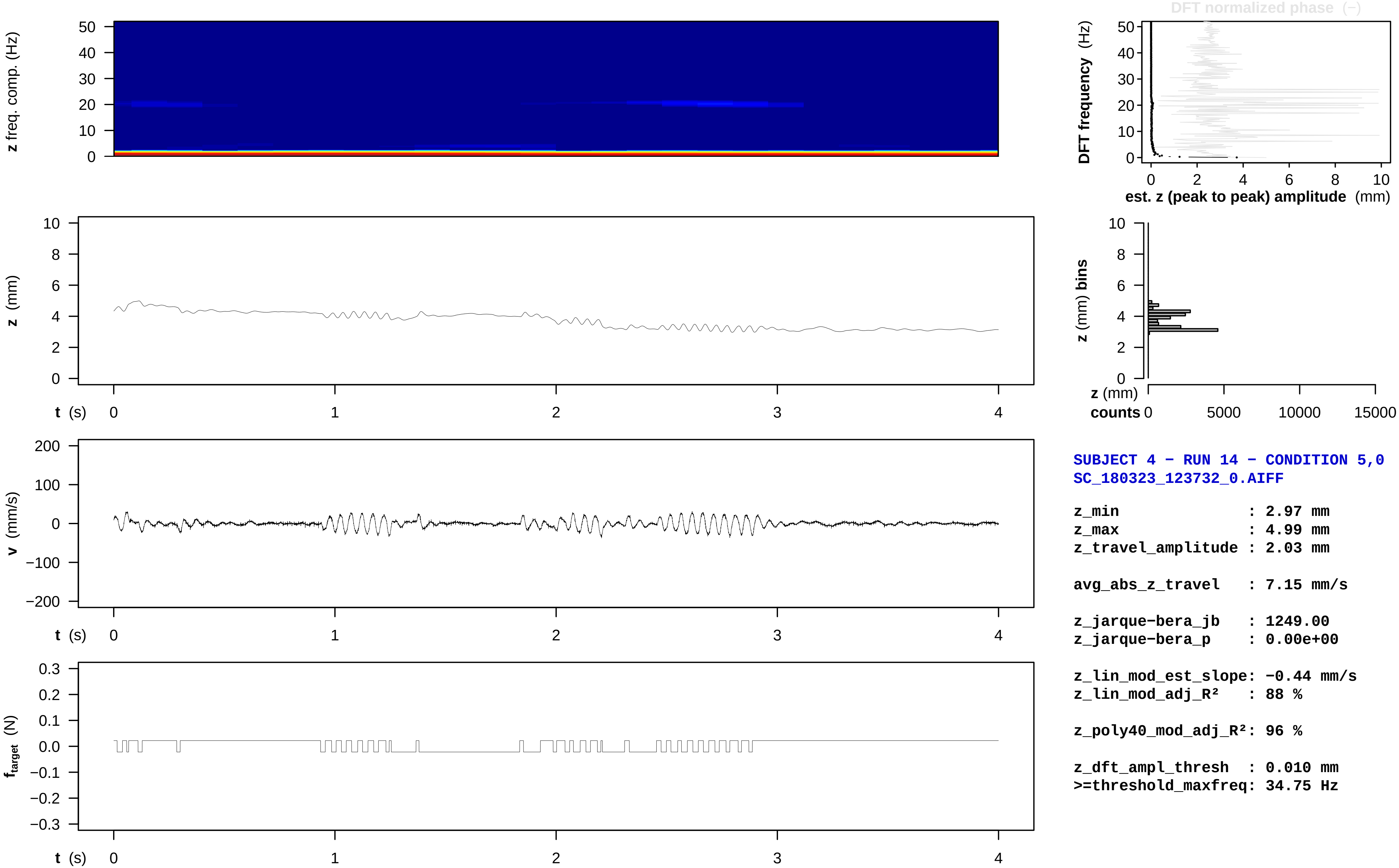

SUBJECT 4 − RUN 14 − CONDITION 5,0
SC_180323_123732_0.AIFF

```
z_min               : 2.97 mm
z_max               : 4.99 mm
z_travel_amplitude : 2.03 mm

avg_abs_z_travel   : 7.15 mm/s

z_jarque-bera_jb   : 1249.00
z_jarque-bera_p    : 0.00e+00

z_lin_mod_est_slope: -0.44 mm/s
z_lin_mod_adj_R²   : 88 %

z_poly40_mod_adj_R²: 96 %

z_dft_ampl_thresh  : 0.010 mm
>=threshold_maxfreq: 34.75 Hz
```

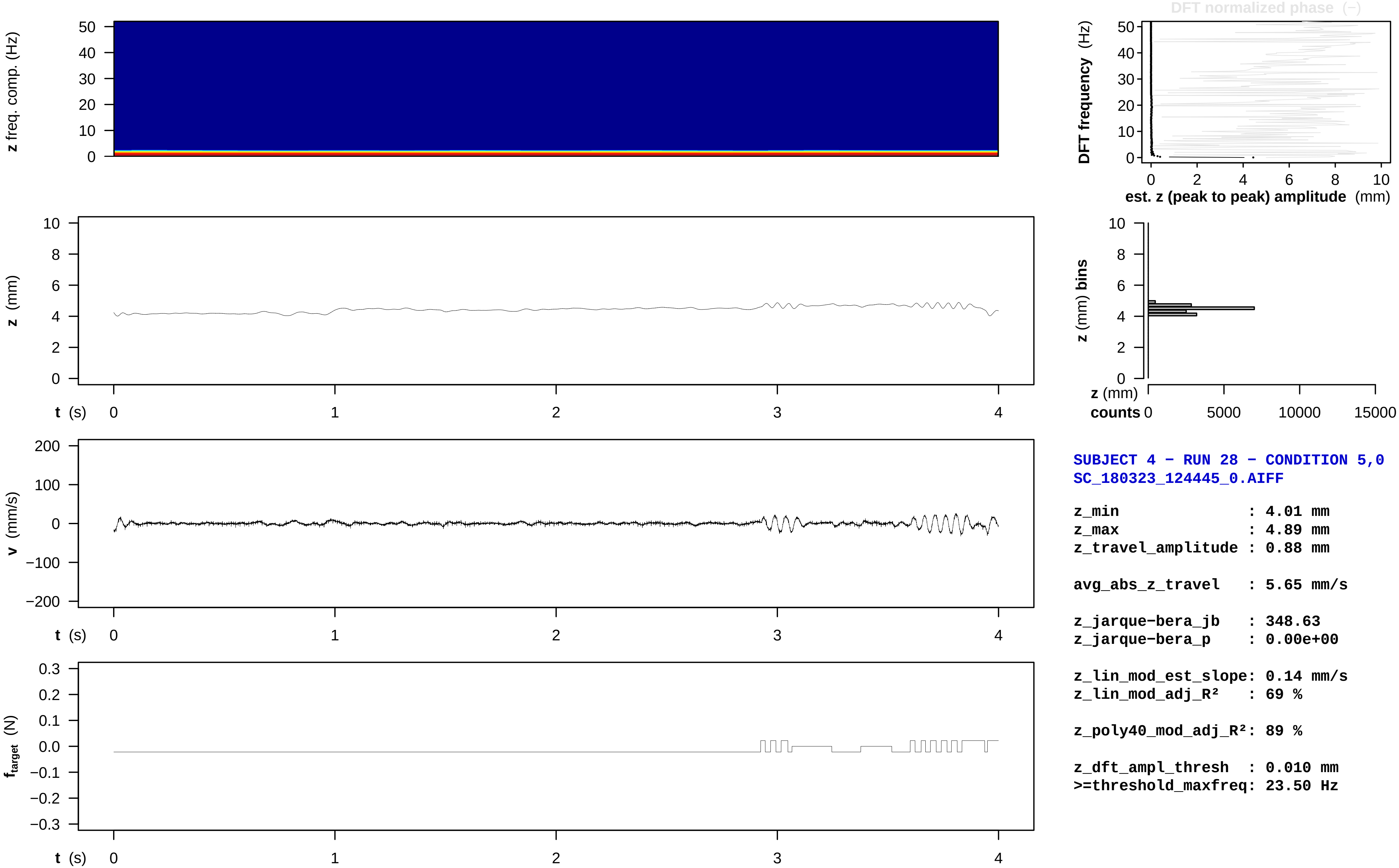

**SUBJECT 4 − RUN 28 − CONDITION 5,0**
**SC_180323_124445_0.AIFF**

```
z_min               : 4.01 mm
z_max               : 4.89 mm
z_travel_amplitude  : 0.88 mm

avg_abs_z_travel    : 5.65 mm/s

z_jarque-bera_jb    : 348.63
z_jarque-bera_p     : 0.00e+00

z_lin_mod_est_slope: 0.14 mm/s
z_lin_mod_adj_R²    : 69 %

z_poly40_mod_adj_R²: 89 %

z_dft_ampl_thresh   : 0.010 mm
>=threshold_maxfreq: 23.50 Hz
```

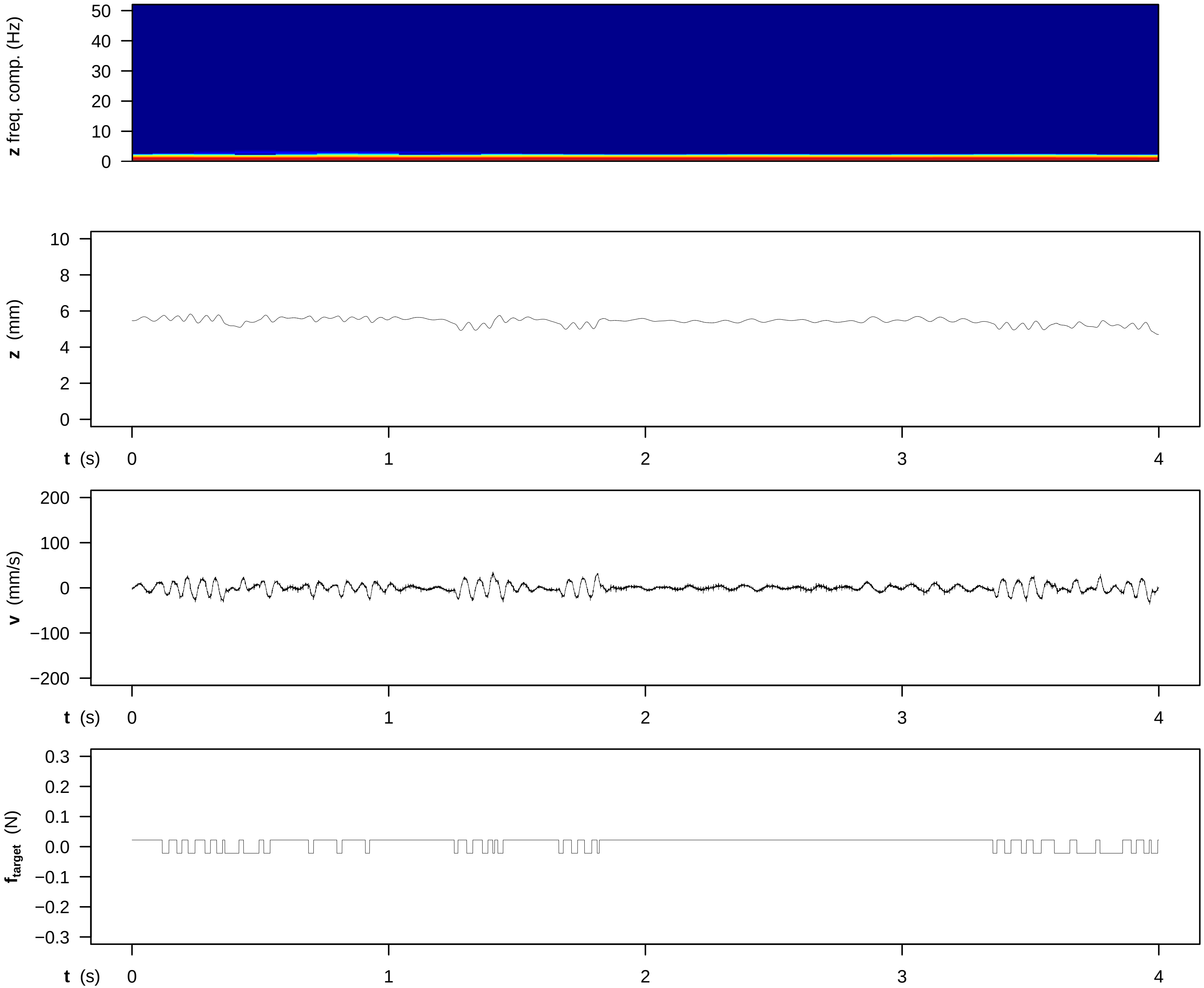

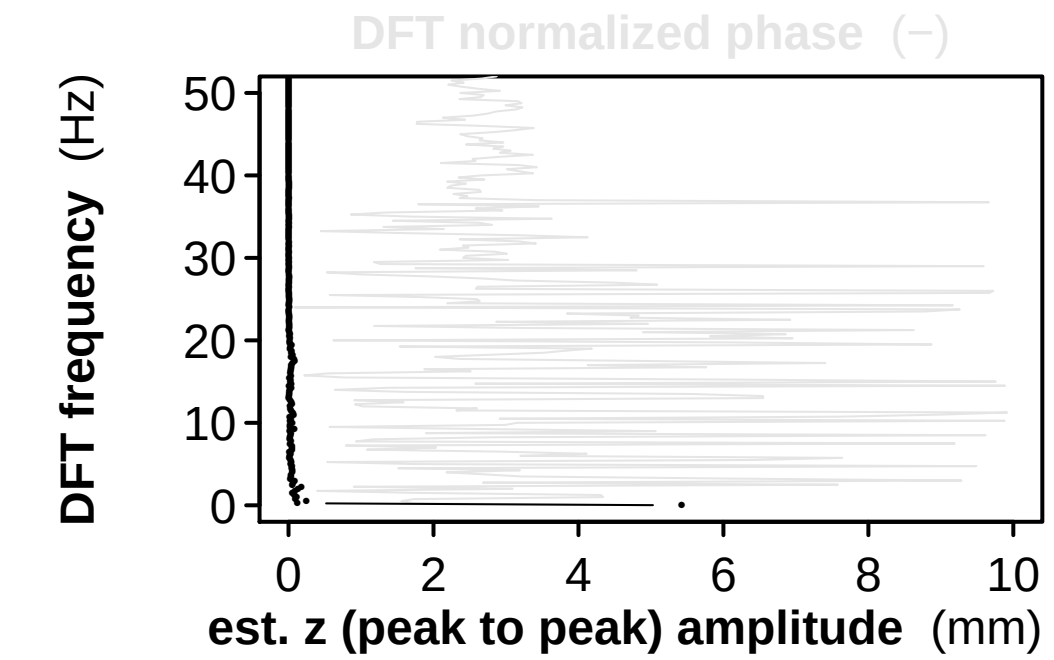

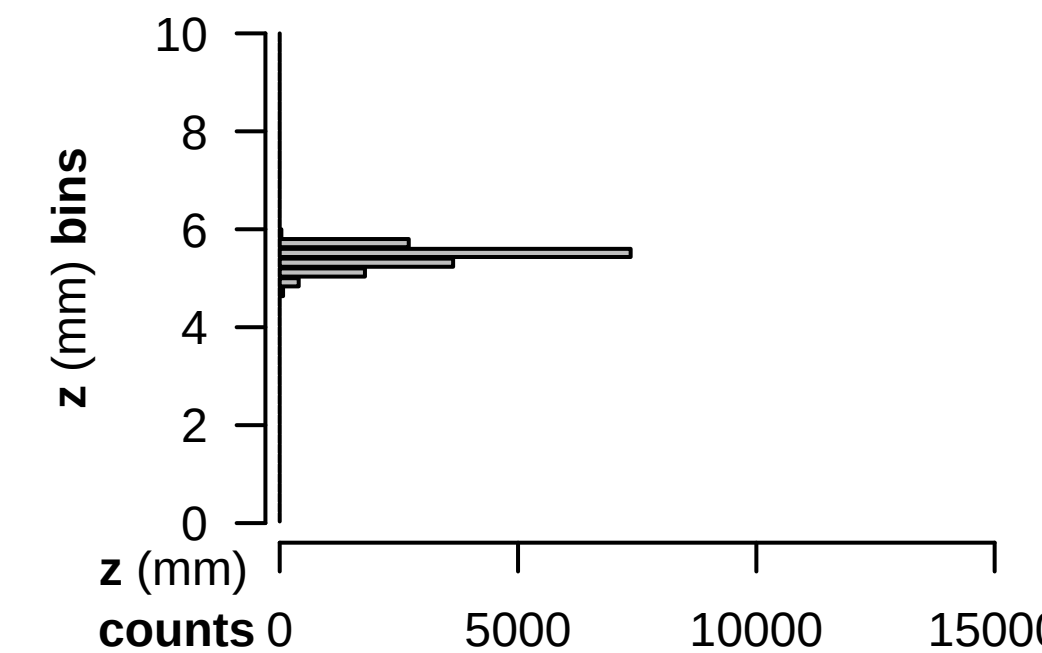

**SUBJECT 5 − RUN 12 − CONDITION 5,0**
**SC_180323_132201_0.AIFF**

```
z_min               : 4.70 mm
z_max               : 5.83 mm
z_travel_amplitude  : 1.13 mm

avg_abs_z_travel    : 8.12 mm/s

z_jarque-bera_jb    : 1941.57
z_jarque-bera_p     : 0.00e+00

z_lin_mod_est_slope: -0.08 mm/s
z_lin_mod_adj_R²    : 23 %

z_poly40_mod_adj_R²: 52 %

z_dft_ampl_thresh   : 0.010 mm
>=threshold_maxfreq: 27.75 Hz
```

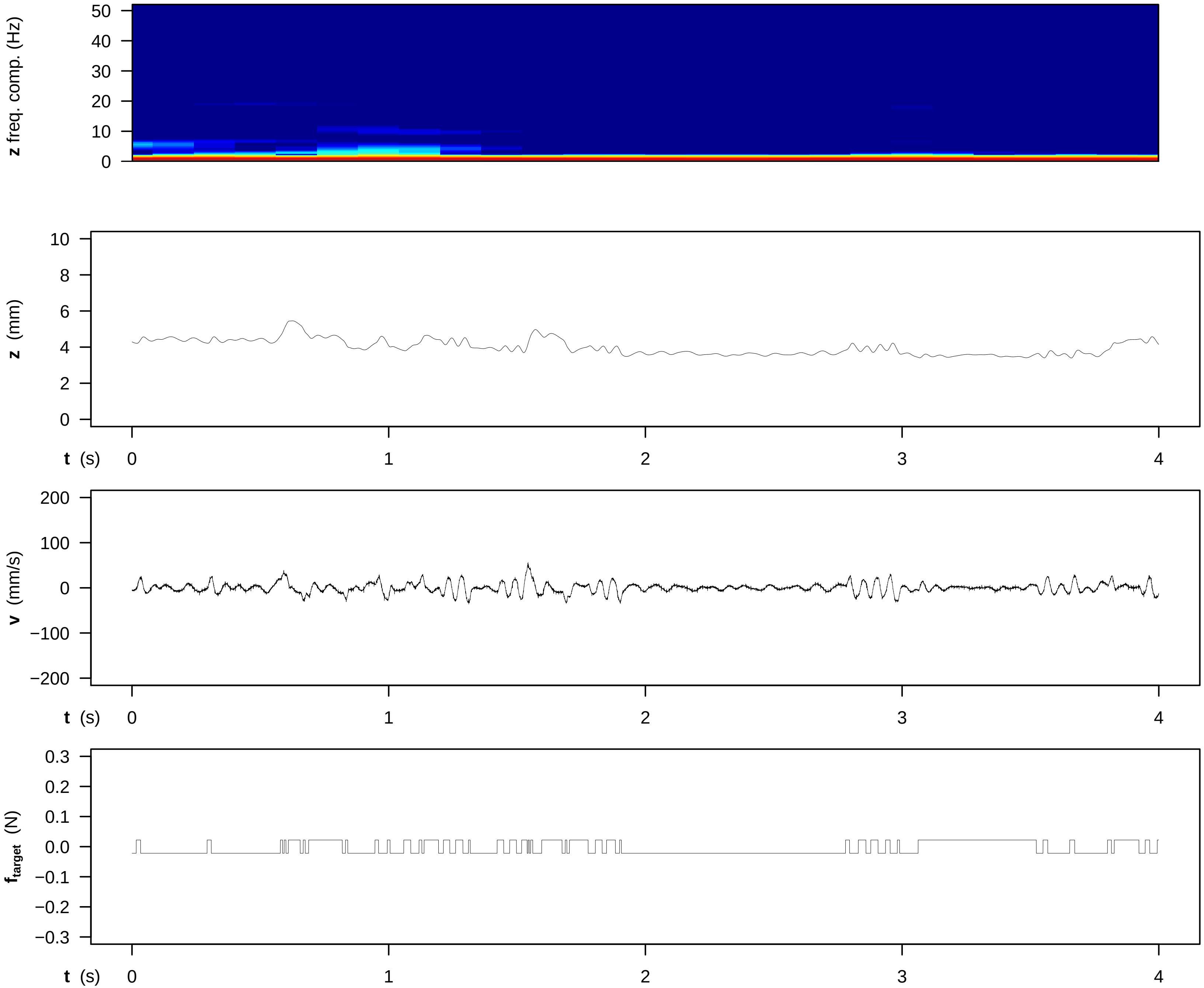

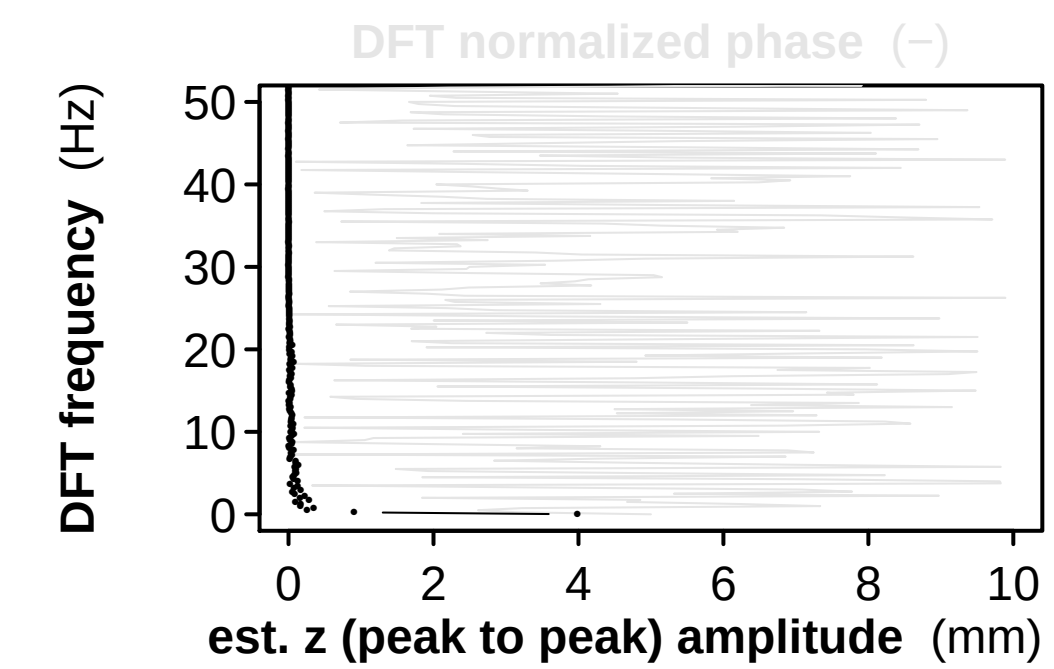

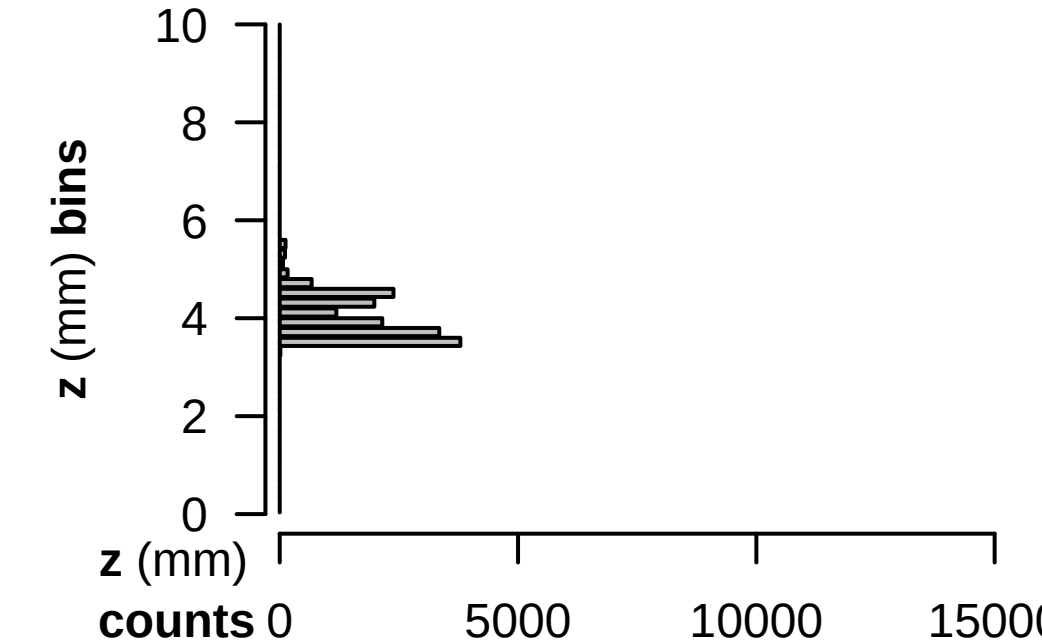

**SUBJECT 5 − RUN 13 − CONDITION 5,0**
**SC_180323_132257_0.AIFF**

```
z_min               : 3.40 mm
z_max               : 5.45 mm
z_travel_amplitude  : 2.06 mm

avg_abs_z_travel    : 8.21 mm/s

z_jarque-bera_jb    : 1529.32
z_jarque-bera_p     : 0.00e+00

z_lin_mod_est_slope: −0.24 mm/s
z_lin_mod_adj_R²    : 40 %

z_poly40_mod_adj_R²: 77 %

z_dft_ampl_thresh   : 0.010 mm
>=threshold_maxfreq: 26.75 Hz
```

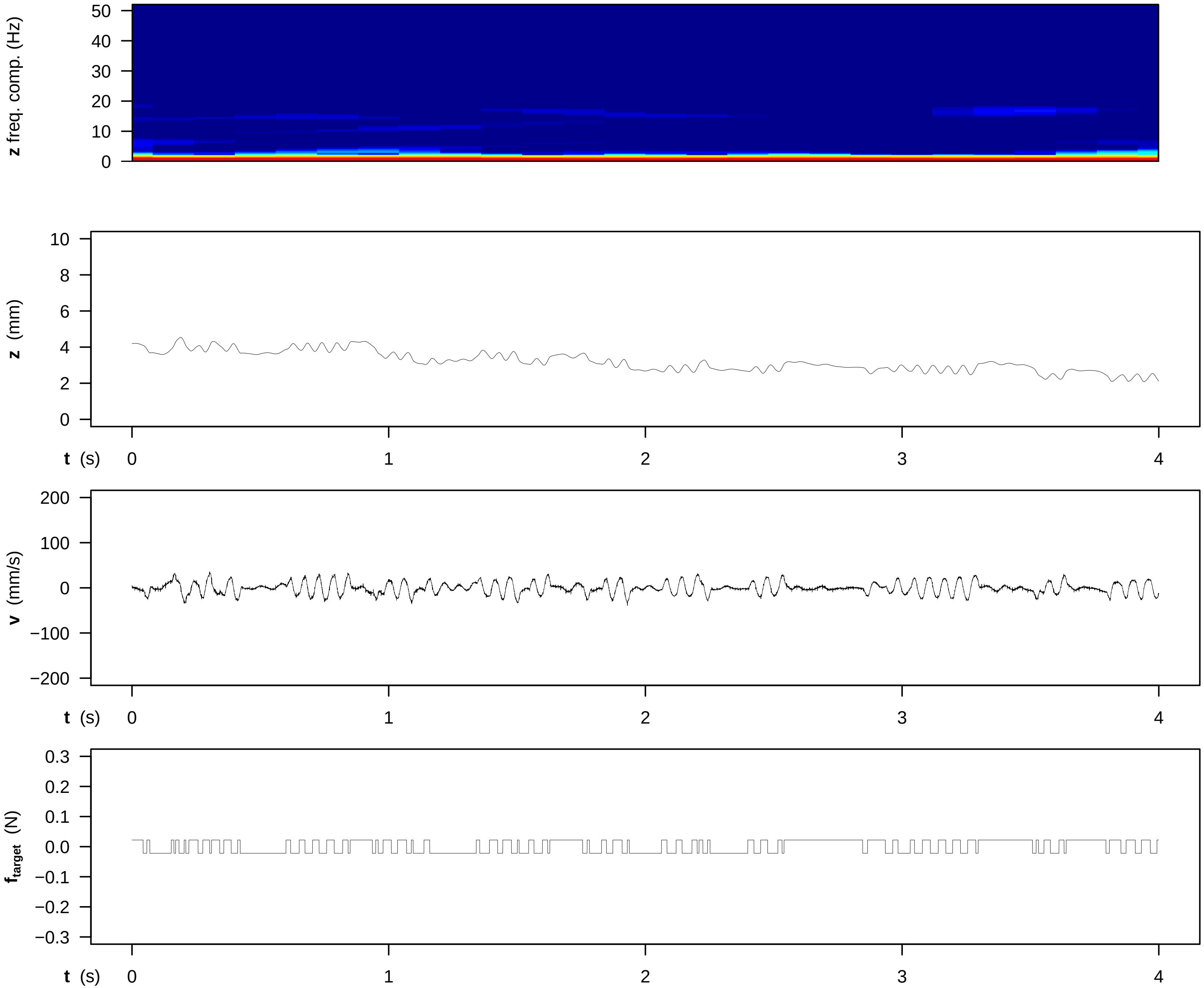

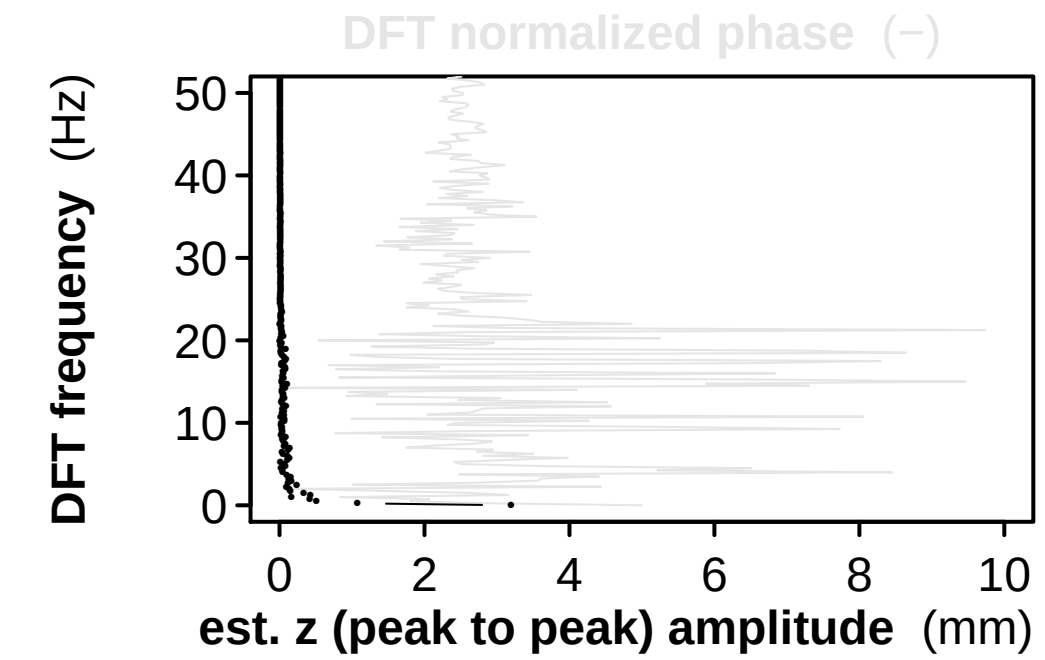

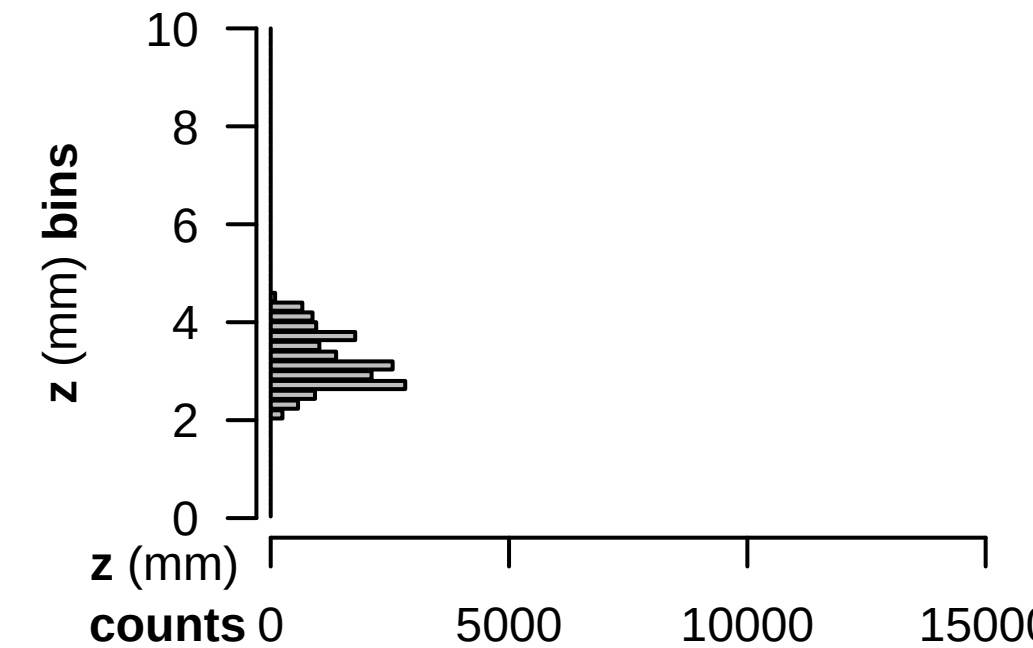

**SUBJECT 5 − RUN 31 − CONDITION 5,0**
**SC_180323_133613_0.AIFF**

```
z_min               : 2.08 mm
z_max               : 4.54 mm
z_travel_amplitude  : 2.46 mm

avg_abs_z_travel    : 9.97 mm/s

z_jarque-bera_jb    : 656.09
z_jarque-bera_p     : 0.00e+00

z_lin_mod_est_slope: −0.41 mm/s
z_lin_mod_adj_R²    : 75 %

z_poly40_mod_adj_R²: 89 %

z_dft_ampl_thresh   : 0.010 mm
>=threshold_maxfreq: 42.75 Hz
```

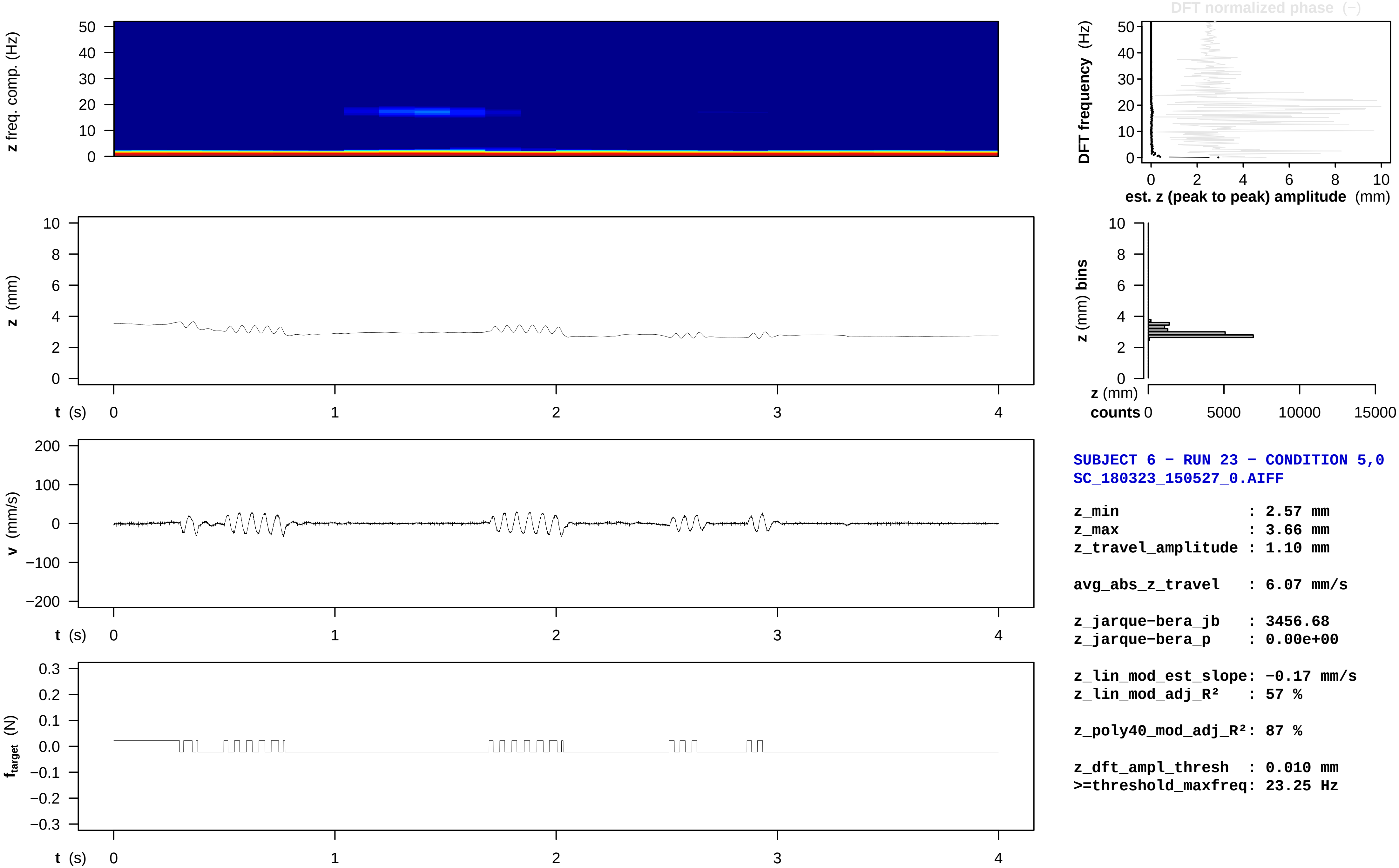

**SUBJECT 6 − RUN 23 − CONDITION 5,0**
**SC_180323_150527_0.AIFF**

```
z_min               : 2.57 mm
z_max               : 3.66 mm
z_travel_amplitude : 1.10 mm

avg_abs_z_travel    : 6.07 mm/s

z_jarque-bera_jb    : 3456.68
z_jarque-bera_p     : 0.00e+00

z_lin_mod_est_slope: −0.17 mm/s
z_lin_mod_adj_R²    : 57 %

z_poly40_mod_adj_R²: 87 %

z_dft_ampl_thresh   : 0.010 mm
>=threshold_maxfreq: 23.25 Hz
```

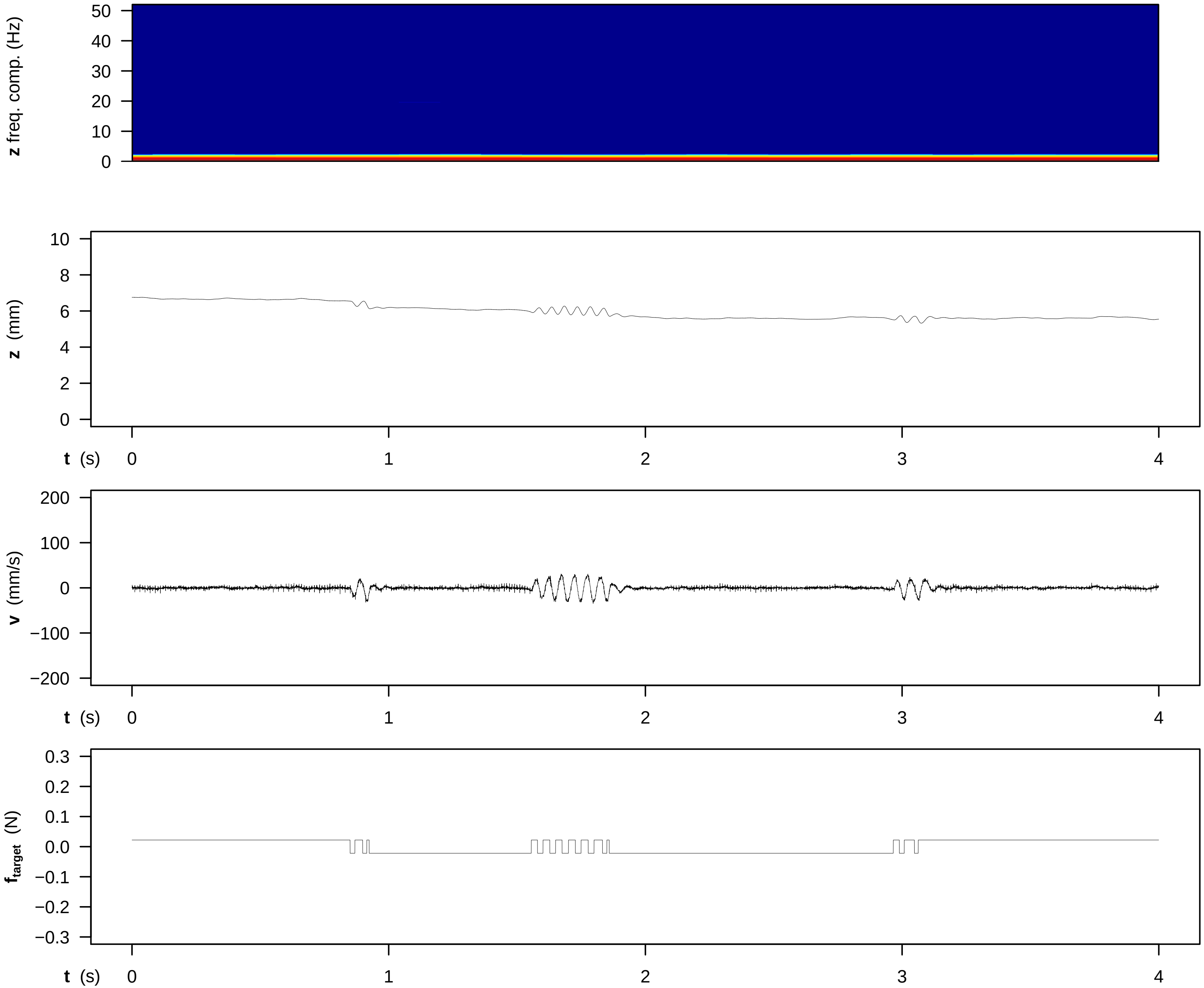

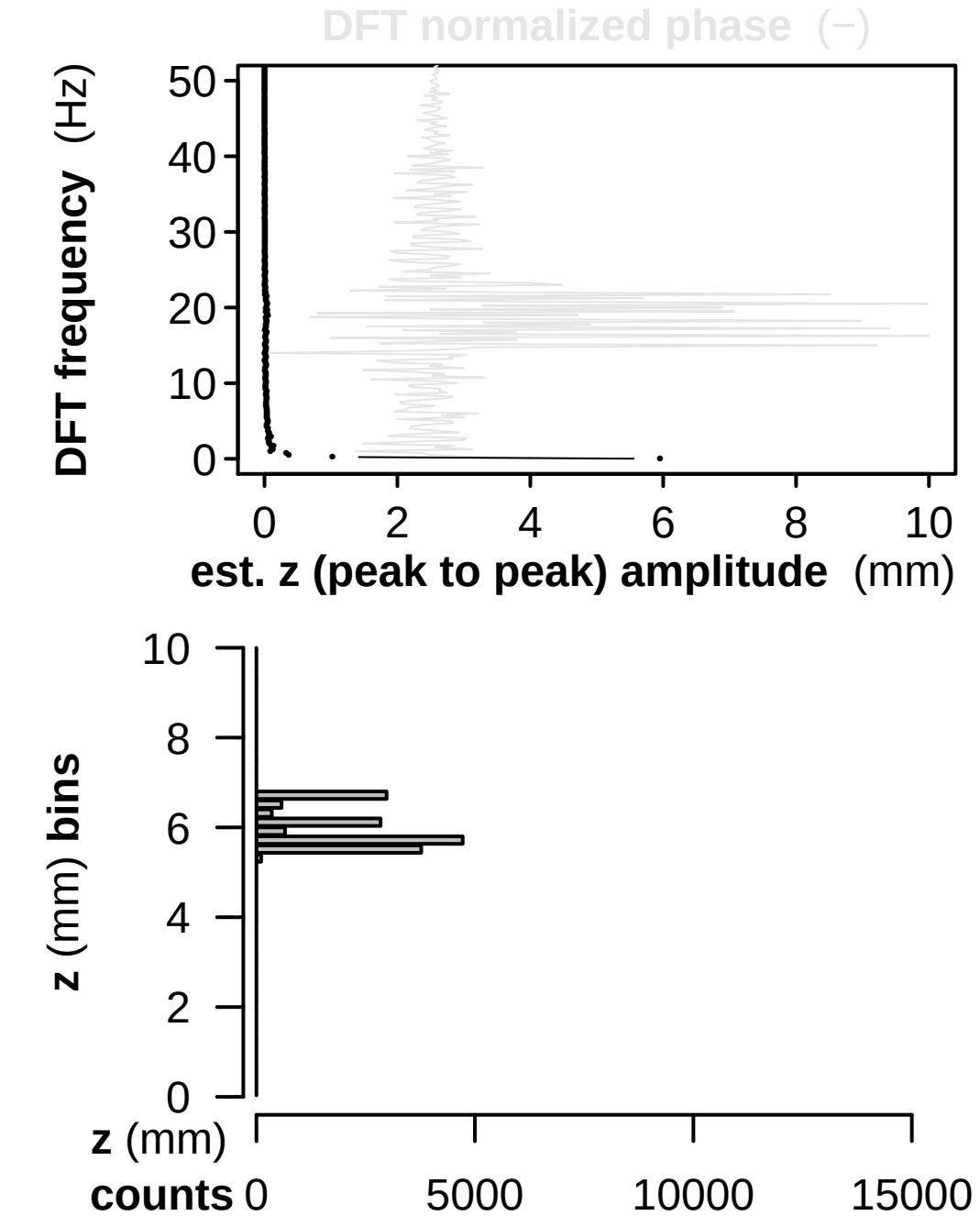

**SUBJECT 6 − RUN 31 − CONDITION 5,0**
**SC_180323_151040_0.AIFF**

**z_min : 5.32 mm**
**z_max : 6.76 mm**
**z_travel_amplitude : 1.44 mm**

**avg_abs_z_travel : 4.60 mm/s**

**z_jarque-bera_jb : 2015.31**
**z_jarque-bera_p : 0.00e+00**

**z_lin_mod_est_slope: −0.33 mm/s**
**z_lin_mod_adj_$R^2$ : 81 %**

**z_poly40_mod_adj_$R^2$: 98 %**

**z_dft_ampl_thresh : 0.010 mm**
**>=threshold_maxfreq: 25.75 Hz**

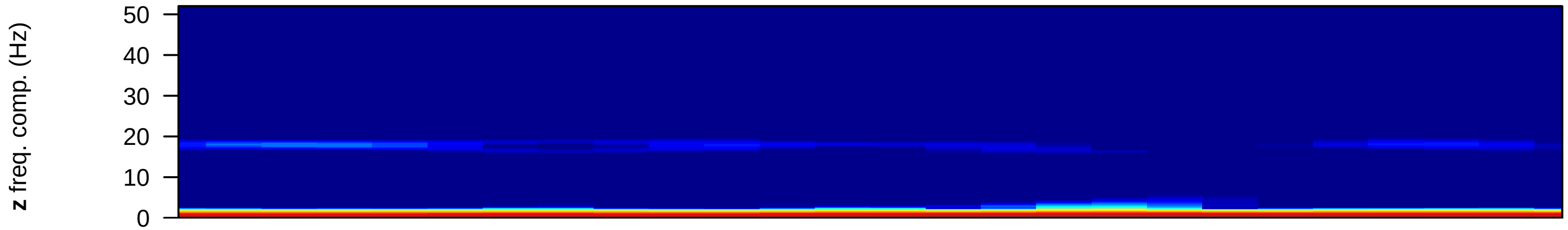

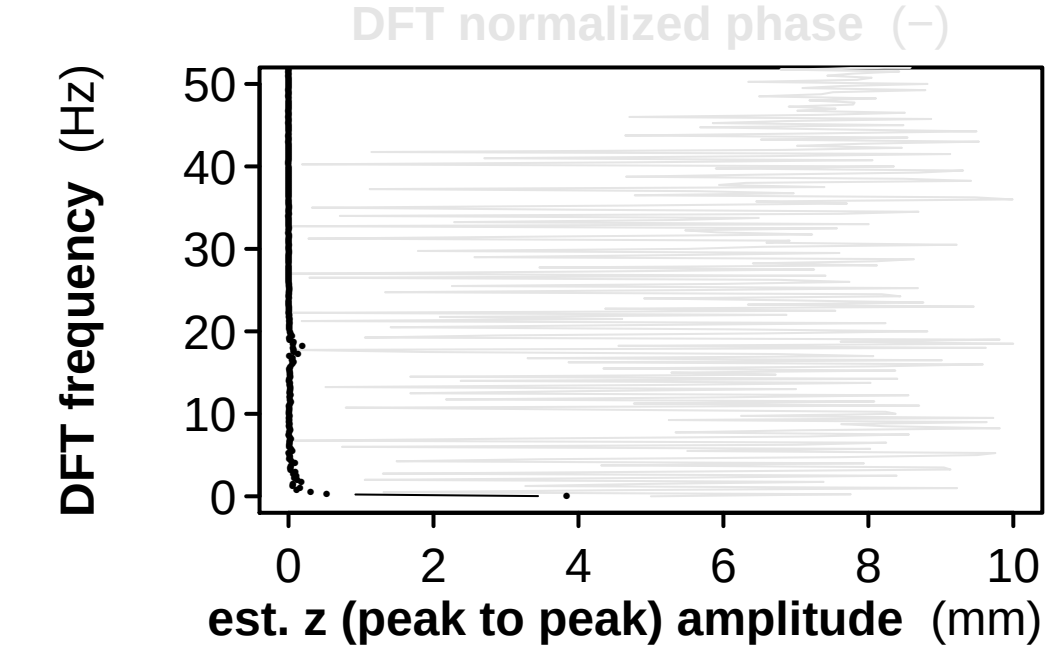

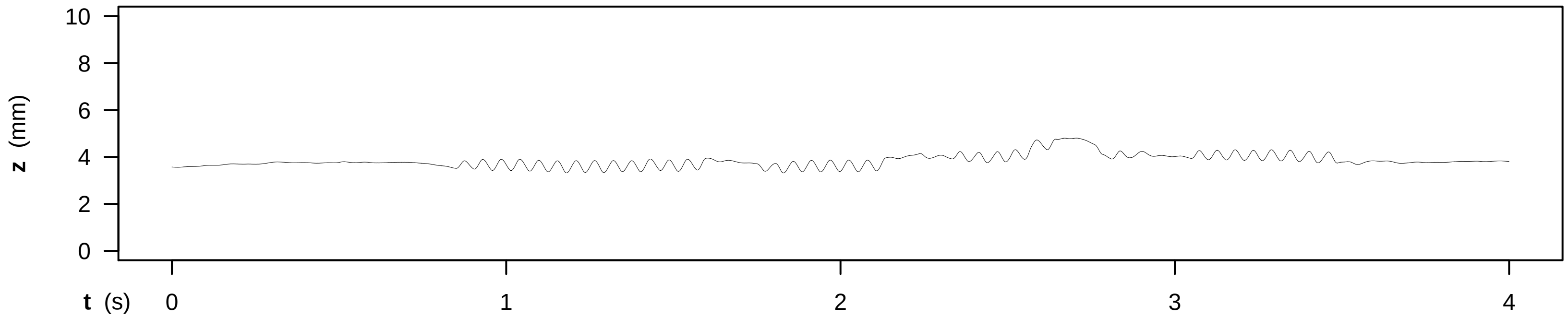

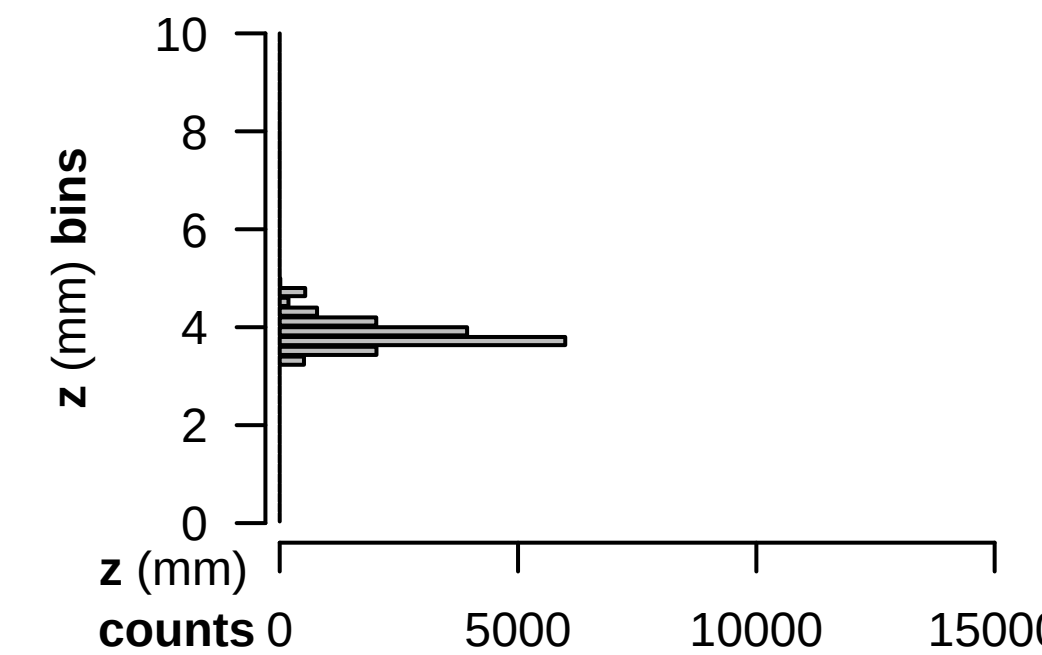

**SUBJECT 6 − RUN 32 − CONDITION 5,0**
**SC_180323_151115_0.AIFF**

```
z_min              : 3.32 mm
z_max              : 4.80 mm
z_travel_amplitude : 1.49 mm

avg_abs_z_travel   : 9.75 mm/s

z_jarque-bera_jb   : 6847.30
z_jarque-bera_p    : 0.00e+00

z_lin_mod_est_slope: 0.11 mm/s
z_lin_mod_adj_R²   : 19 %

z_poly40_mod_adj_R²: 71 %

z_dft_ampl_thresh  : 0.010 mm
>=threshold_maxfreq: 25.25 Hz
```

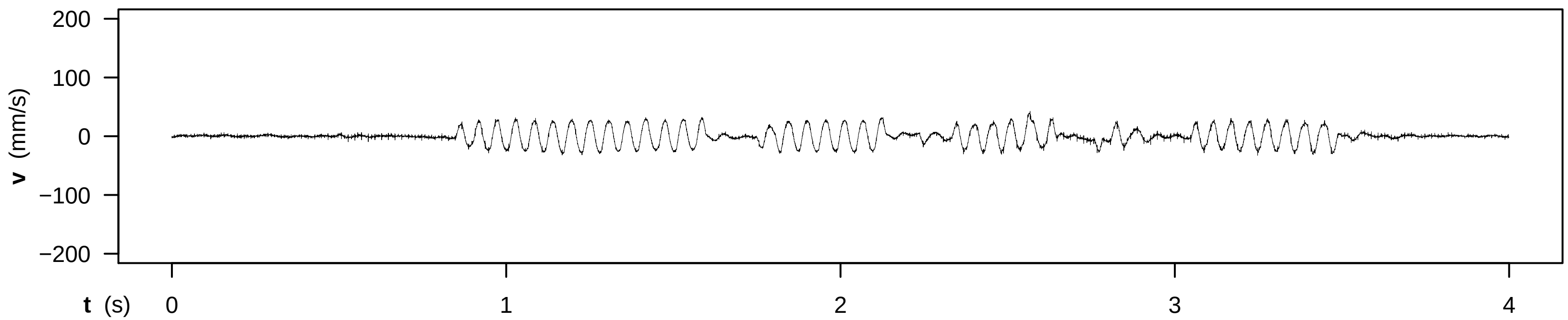

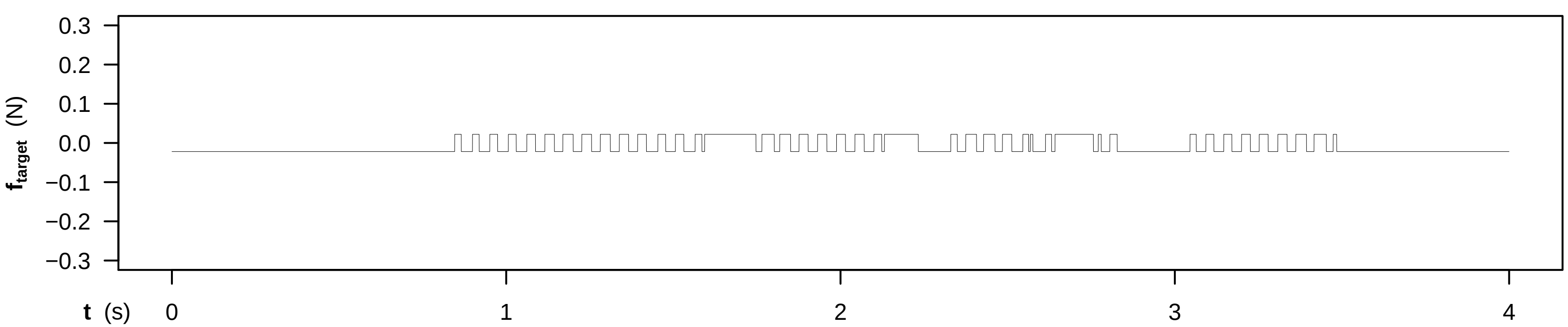

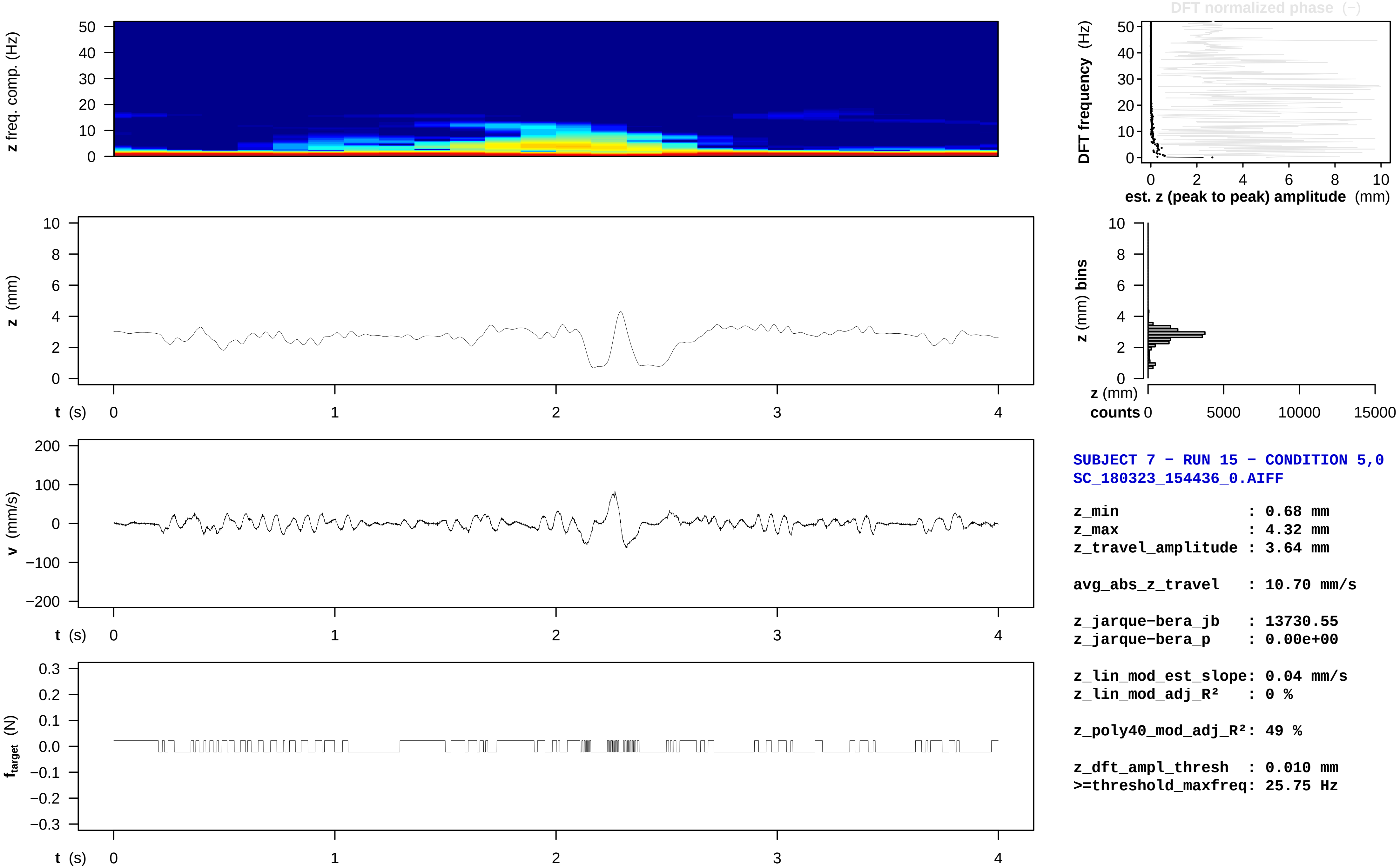

**SUBJECT 7 − RUN 15 − CONDITION 5,0**
**SC_180323_154436_0.AIFF**

```
z_min               : 0.68 mm
z_max               : 4.32 mm
z_travel_amplitude  : 3.64 mm

avg_abs_z_travel    : 10.70 mm/s

z_jarque-bera_jb    : 13730.55
z_jarque-bera_p     : 0.00e+00

z_lin_mod_est_slope: 0.04 mm/s
z_lin_mod_adj_R²    : 0 %

z_poly40_mod_adj_R²: 49 %

z_dft_ampl_thresh   : 0.010 mm
>=threshold_maxfreq: 25.75 Hz
```

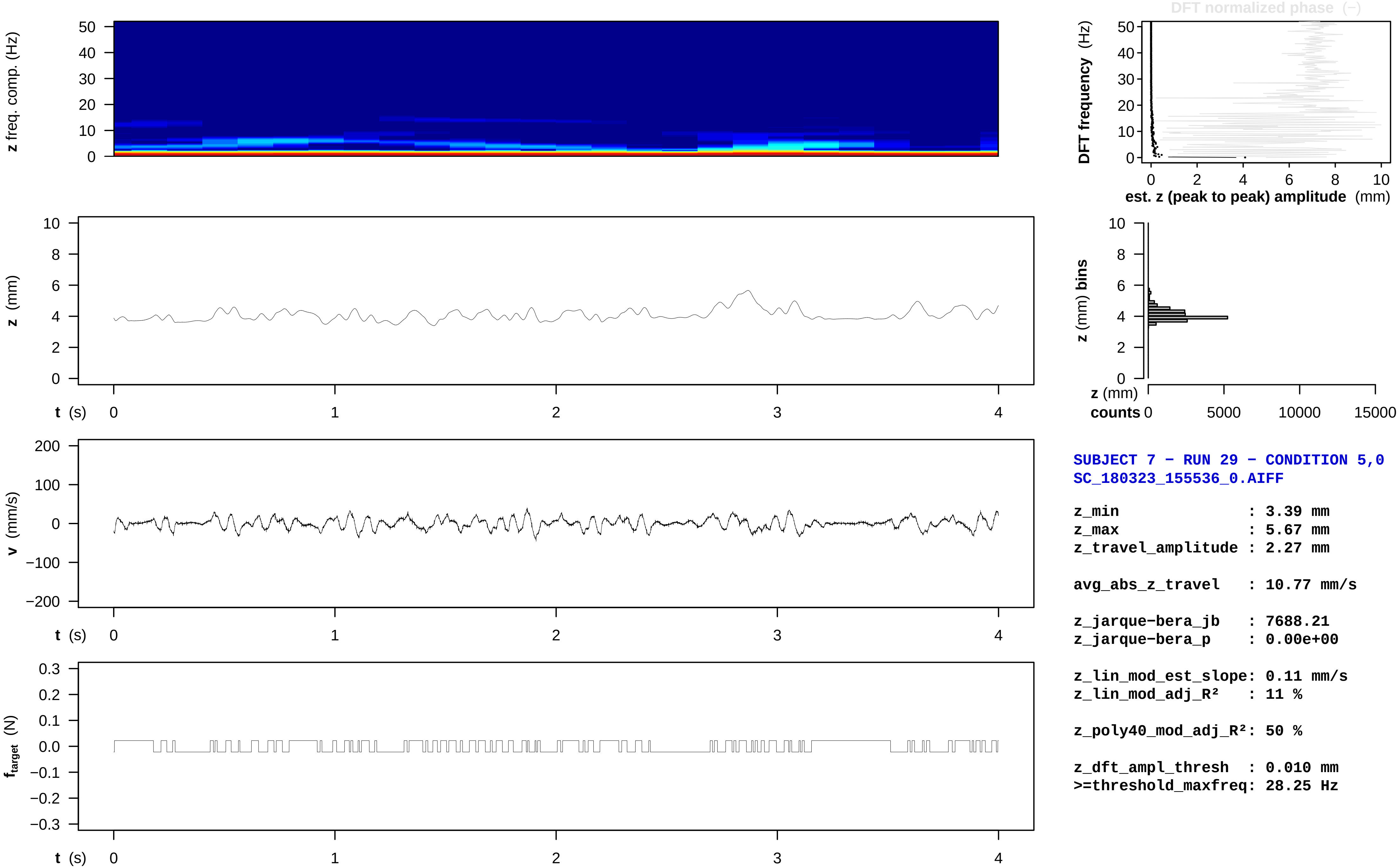

**SUBJECT 7 − RUN 29 − CONDITION 5,0**
**SC_180323_155536_0.AIFF**

```
z_min              : 3.39 mm
z_max              : 5.67 mm
z_travel_amplitude : 2.27 mm

avg_abs_z_travel   : 10.77 mm/s

z_jarque-bera_jb   : 7688.21
z_jarque-bera_p    : 0.00e+00

z_lin_mod_est_slope: 0.11 mm/s
z_lin_mod_adj_R²   : 11 %

z_poly40_mod_adj_R²: 50 %

z_dft_ampl_thresh  : 0.010 mm
>=threshold_maxfreq: 28.25 Hz
```

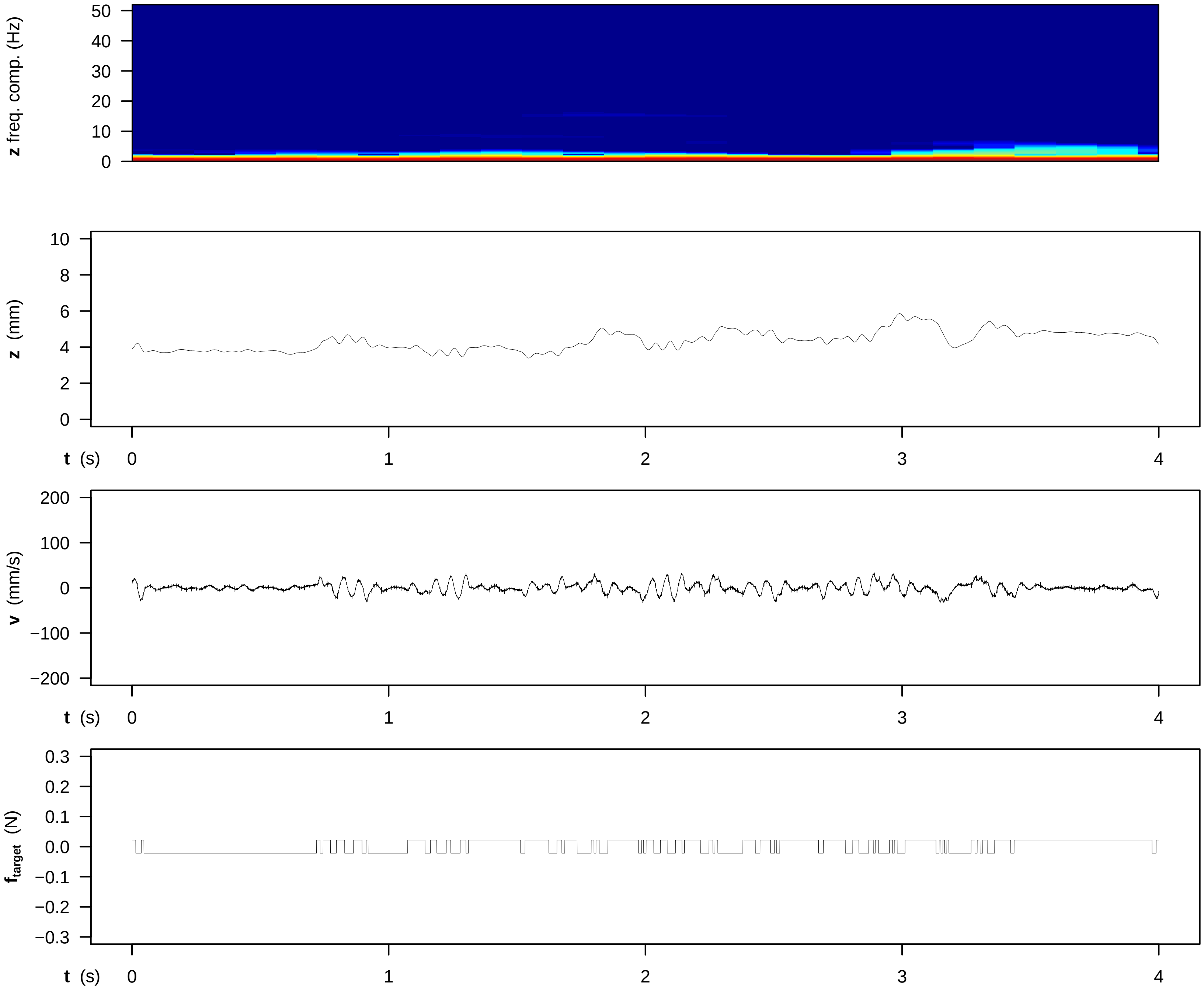

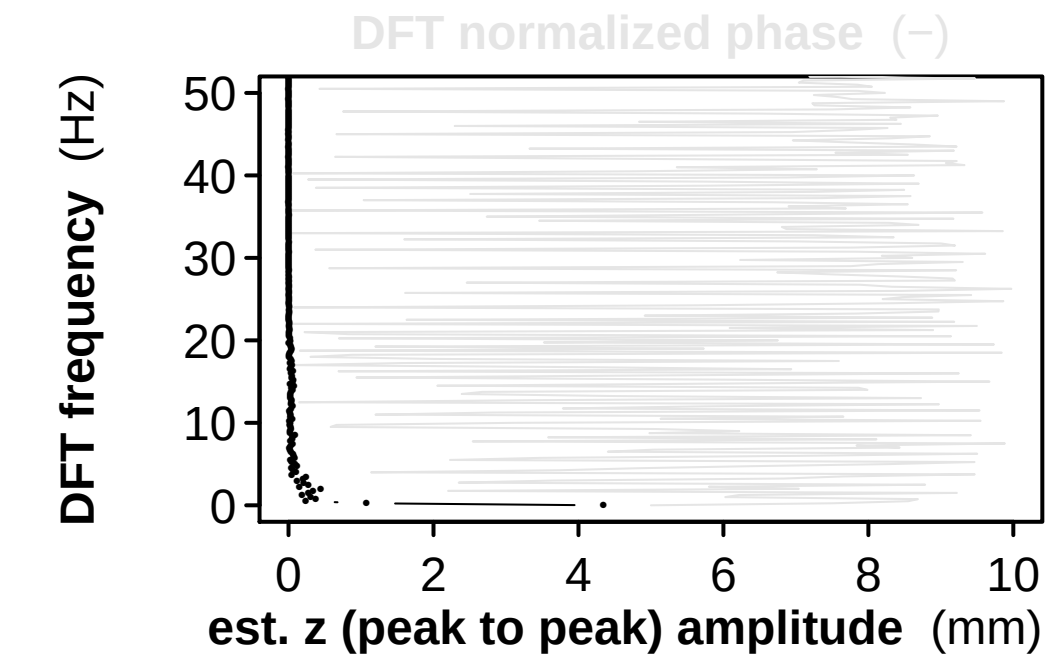

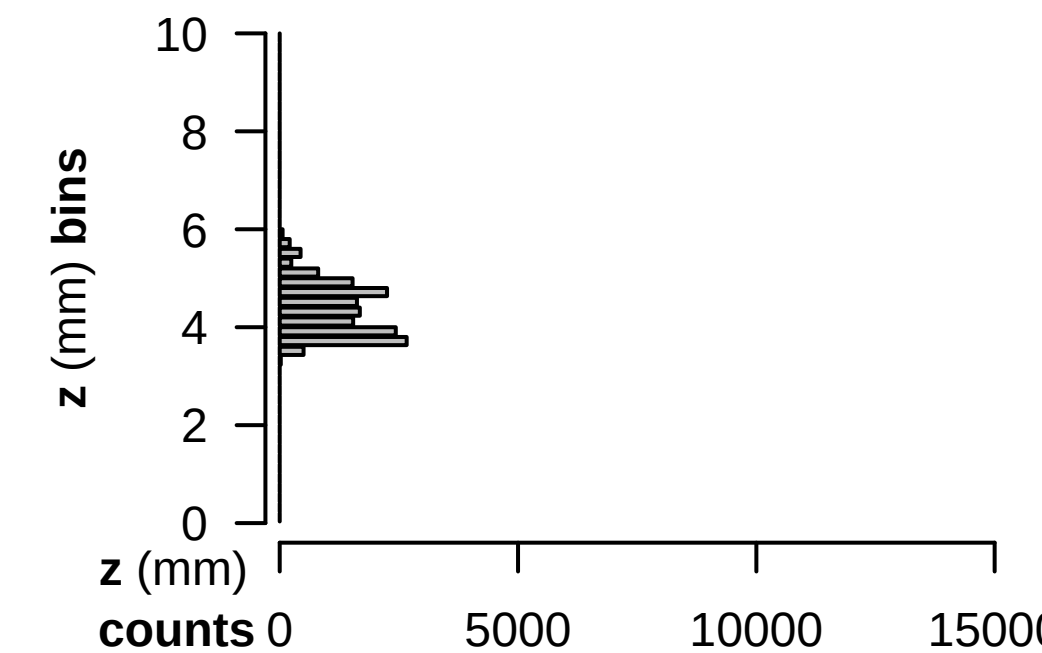

**SUBJECT 7 − RUN 36 − CONDITION 5,0**
**SC_180323_155930_0.AIFF**

```
z_min              : 3.39 mm
z_max              : 5.86 mm
z_travel_amplitude : 2.47 mm

avg_abs_z_travel   : 8.64 mm/s

z_jarque-bera_jb   : 754.37
z_jarque-bera_p    : 0.00e+00

z_lin_mod_est_slope: 0.33 mm/s
z_lin_mod_adj_R²   : 52 %

z_poly40_mod_adj_R²: 77 %

z_dft_ampl_thresh  : 0.010 mm
>=threshold_maxfreq: 24.00 Hz
```

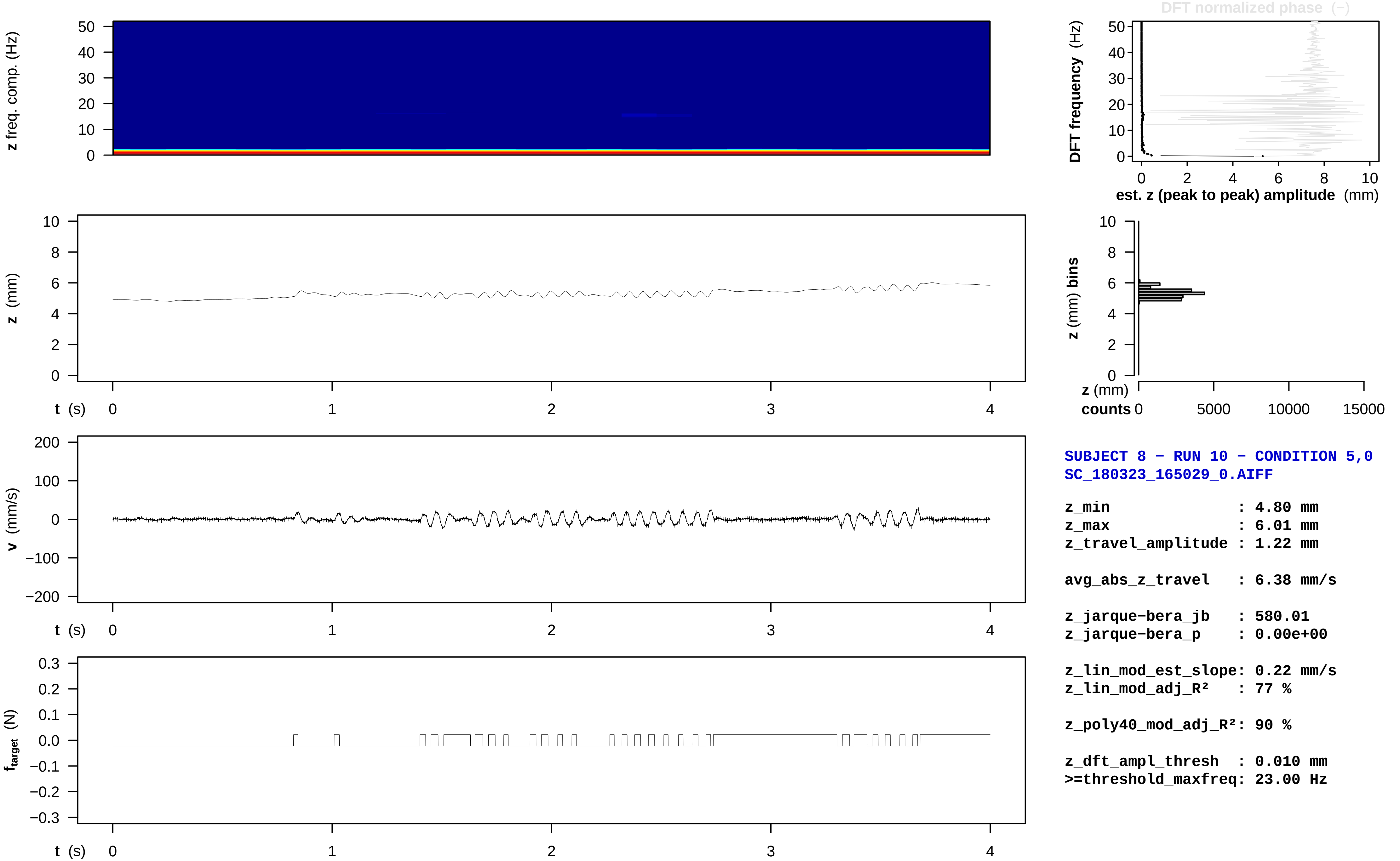

SUBJECT 8 − RUN 10 − CONDITION 5,0
SC_180323_165029_0.AIFF

z_min : 4.80 mm
z_max : 6.01 mm
z_travel_amplitude : 1.22 mm

avg_abs_z_travel : 6.38 mm/s

z_jarque-bera_jb : 580.01
z_jarque-bera_p : 0.00e+00

z_lin_mod_est_slope: 0.22 mm/s
z_lin_mod_adj_R²: 77 %

z_poly40_mod_adj_R²: 90 %

z_dft_ampl_thresh : 0.010 mm
>=threshold_maxfreq: 23.00 Hz

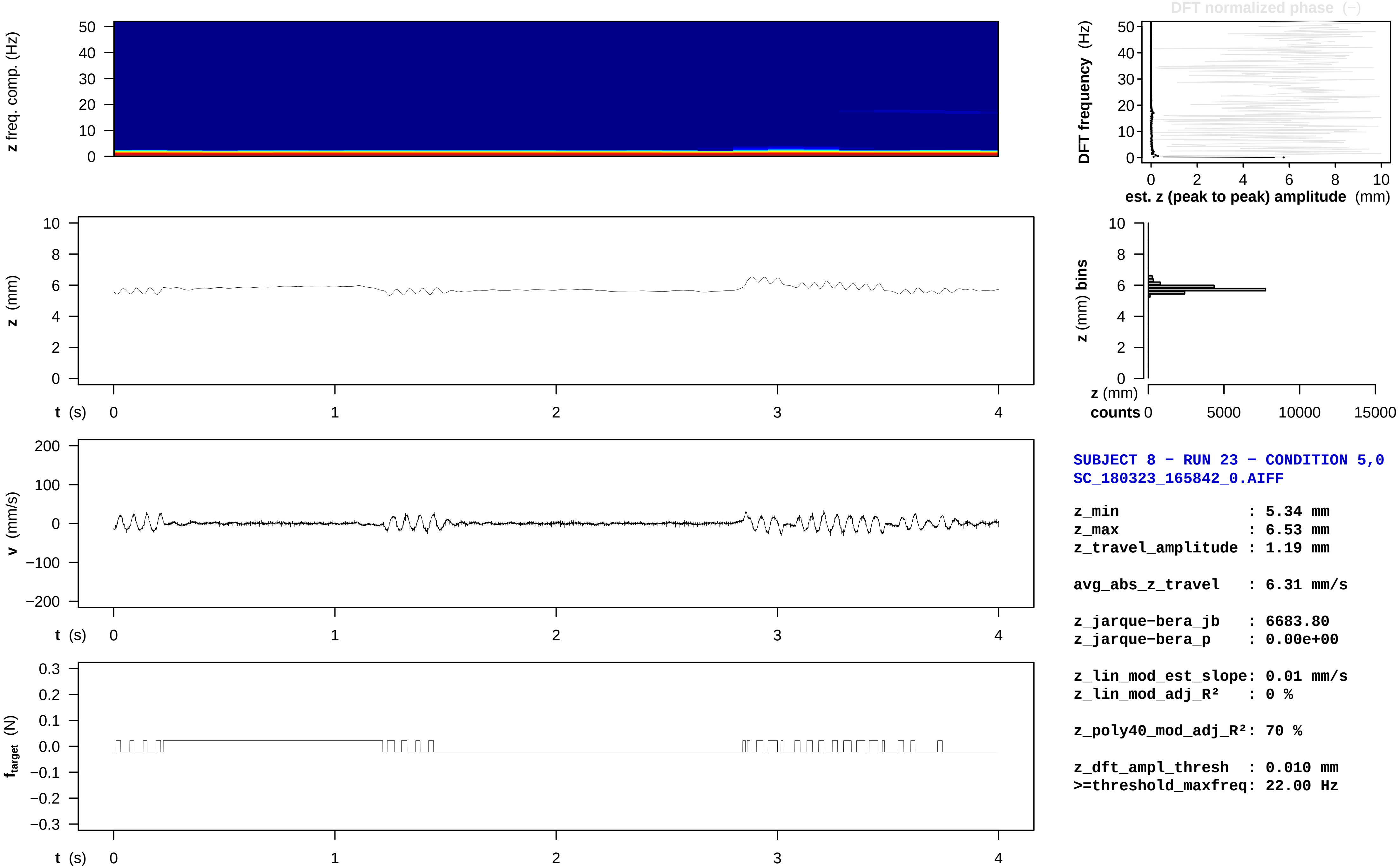

**SUBJECT 8 − RUN 23 − CONDITION 5,0**
**SC_180323_165842_0.AIFF**

```
z_min               : 5.34 mm
z_max               : 6.53 mm
z_travel_amplitude  : 1.19 mm

avg_abs_z_travel    : 6.31 mm/s

z_jarque-bera_jb    : 6683.80
z_jarque-bera_p     : 0.00e+00

z_lin_mod_est_slope: 0.01 mm/s
z_lin_mod_adj_R²    : 0 %

z_poly40_mod_adj_R²: 70 %

z_dft_ampl_thresh   : 0.010 mm
>=threshold_maxfreq: 22.00 Hz
```

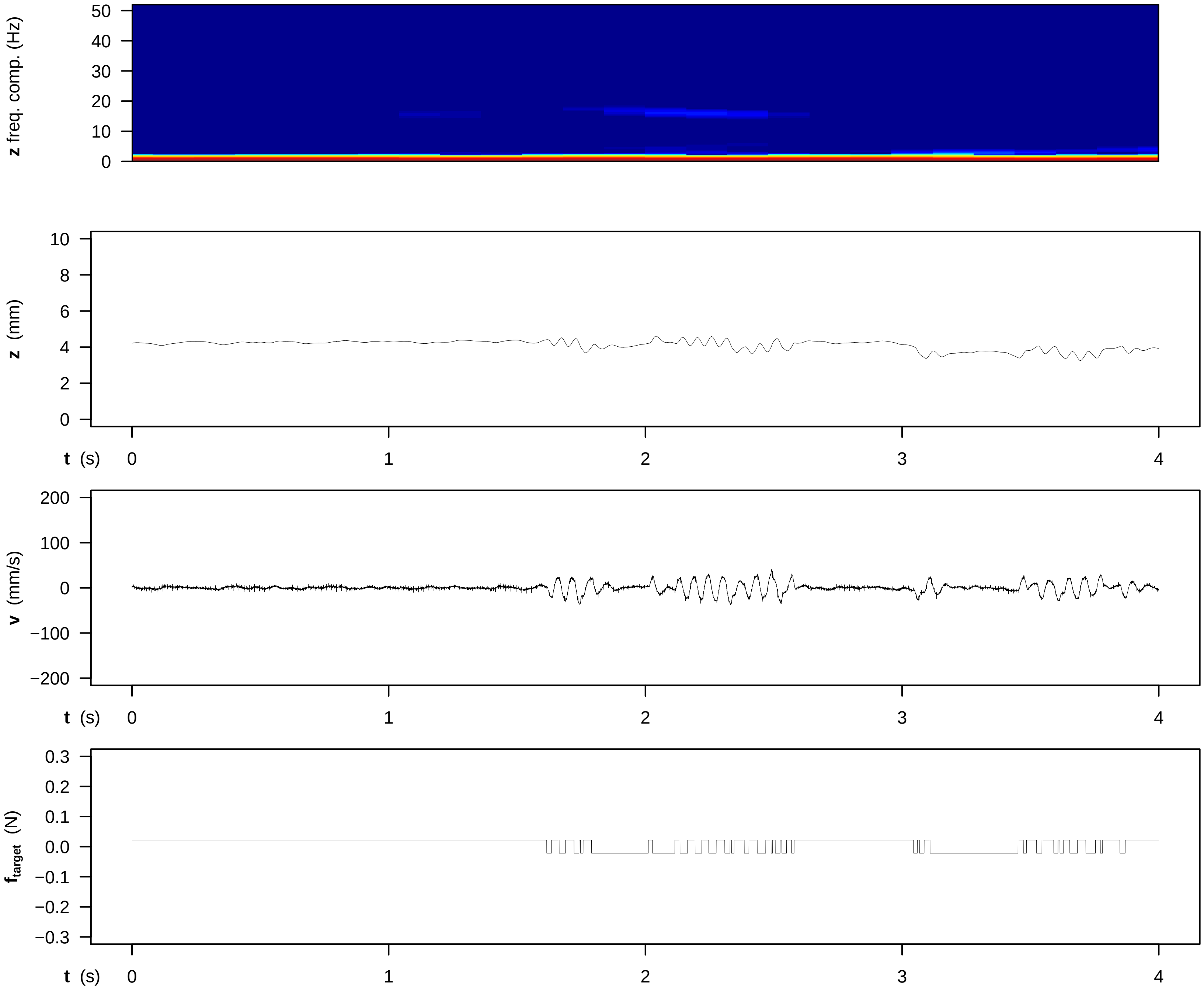

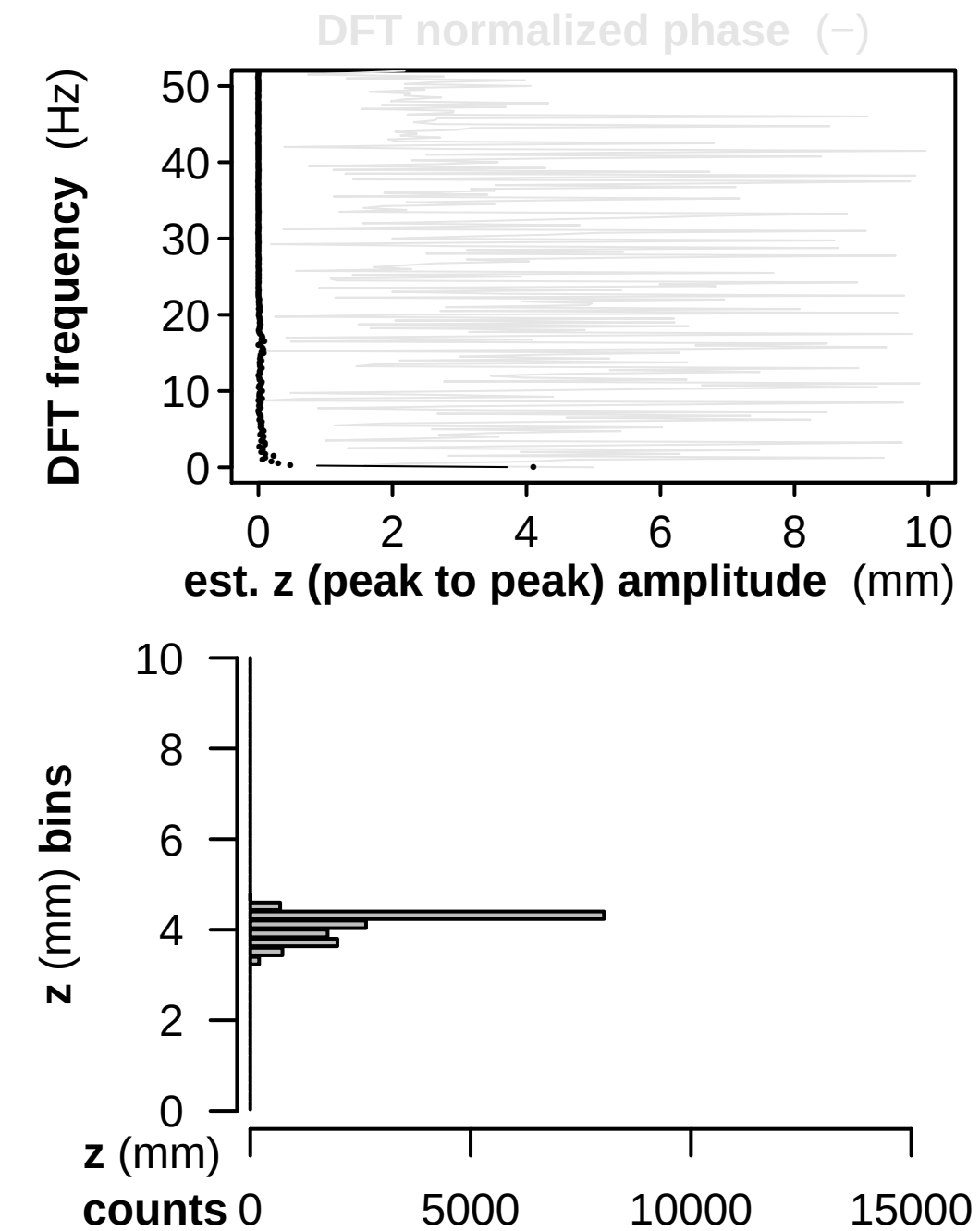

**SUBJECT 8 − RUN 30 − CONDITION 5,0**
**SC_180323_170714_0.AIFF**

```
z_min               : 3.25 mm
z_max               : 4.60 mm
z_travel_amplitude  : 1.35 mm

avg_abs_z_travel    : 7.44 mm/s

z_jarque-bera_jb    : 2536.27
z_jarque-bera_p     : 0.00e+00

z_lin_mod_est_slope: −0.15 mm/s
z_lin_mod_adj_R²    : 41 %

z_poly40_mod_adj_R²: 76 %

z_dft_ampl_thresh   : 0.010 mm
>=threshold_maxfreq: 22.00 Hz
```

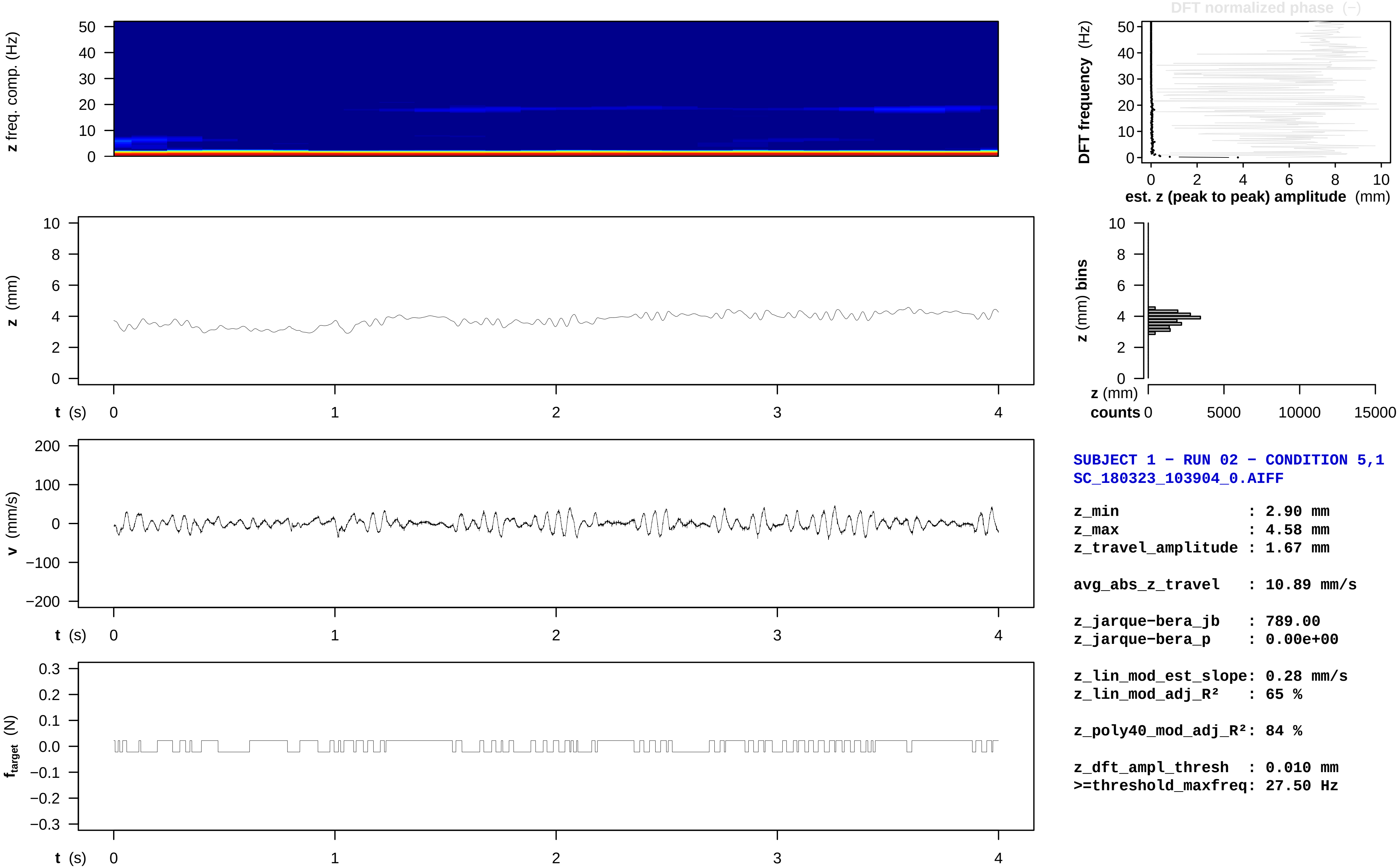

**SUBJECT 1 − RUN 02 − CONDITION 5,1**
**SC_180323_103904_0.AIFF**

```
z_min               : 2.90 mm
z_max               : 4.58 mm
z_travel_amplitude  : 1.67 mm

avg_abs_z_travel    : 10.89 mm/s

z_jarque-bera_jb    : 789.00
z_jarque-bera_p     : 0.00e+00

z_lin_mod_est_slope : 0.28 mm/s
z_lin_mod_adj_R²    : 65 %

z_poly40_mod_adj_R² : 84 %

z_dft_ampl_thresh   : 0.010 mm
>=threshold_maxfreq : 27.50 Hz
```

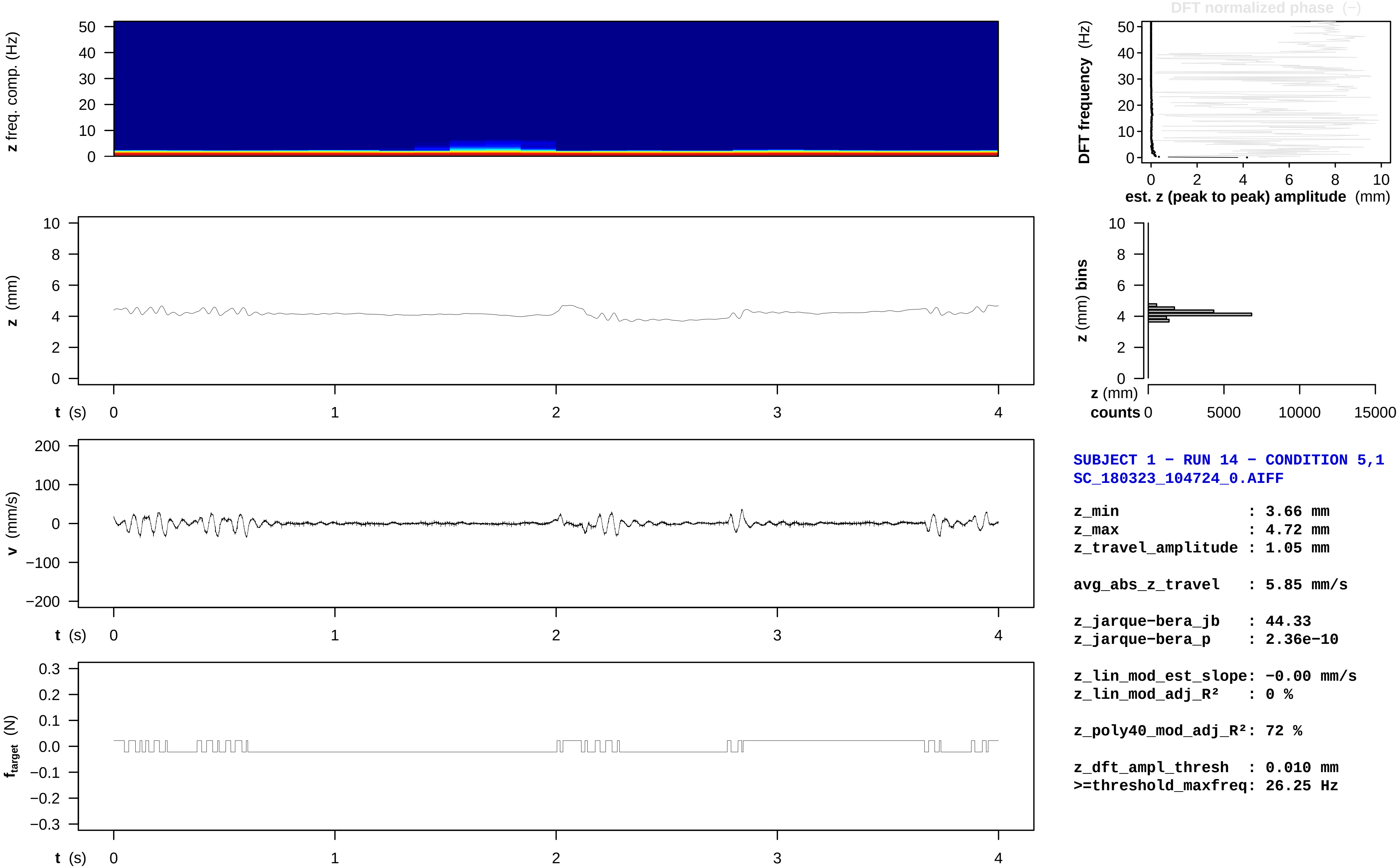

**SUBJECT 1 − RUN 14 − CONDITION 5,1**
**SC_180323_104724_0.AIFF**

```
z_min               : 3.66 mm
z_max               : 4.72 mm
z_travel_amplitude : 1.05 mm

avg_abs_z_travel   : 5.85 mm/s

z_jarque-bera_jb   : 44.33
z_jarque-bera_p    : 2.36e-10

z_lin_mod_est_slope: -0.00 mm/s
z_lin_mod_adj_R²   : 0 %

z_poly40_mod_adj_R²: 72 %

z_dft_ampl_thresh  : 0.010 mm
>=threshold_maxfreq: 26.25 Hz
```

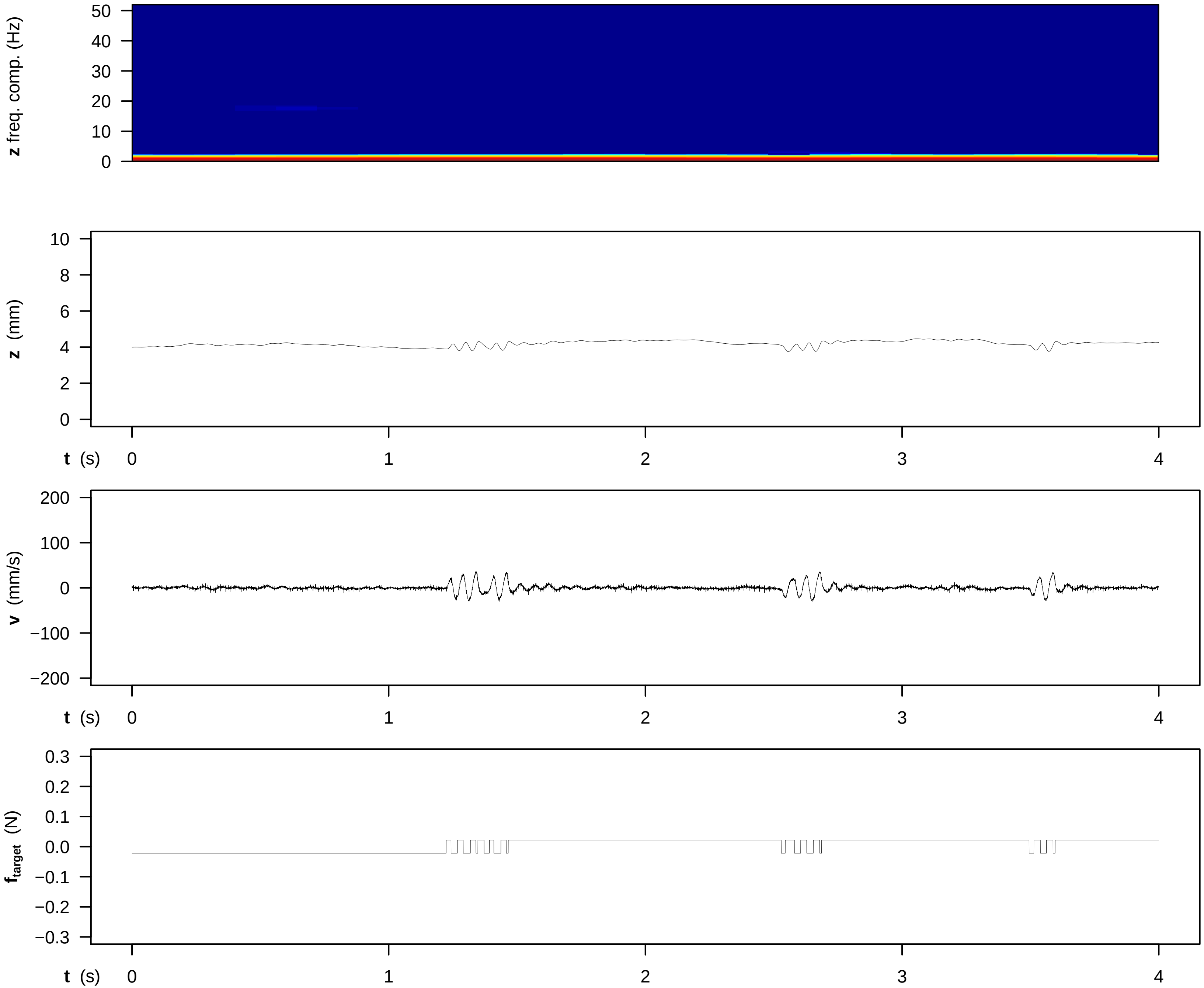

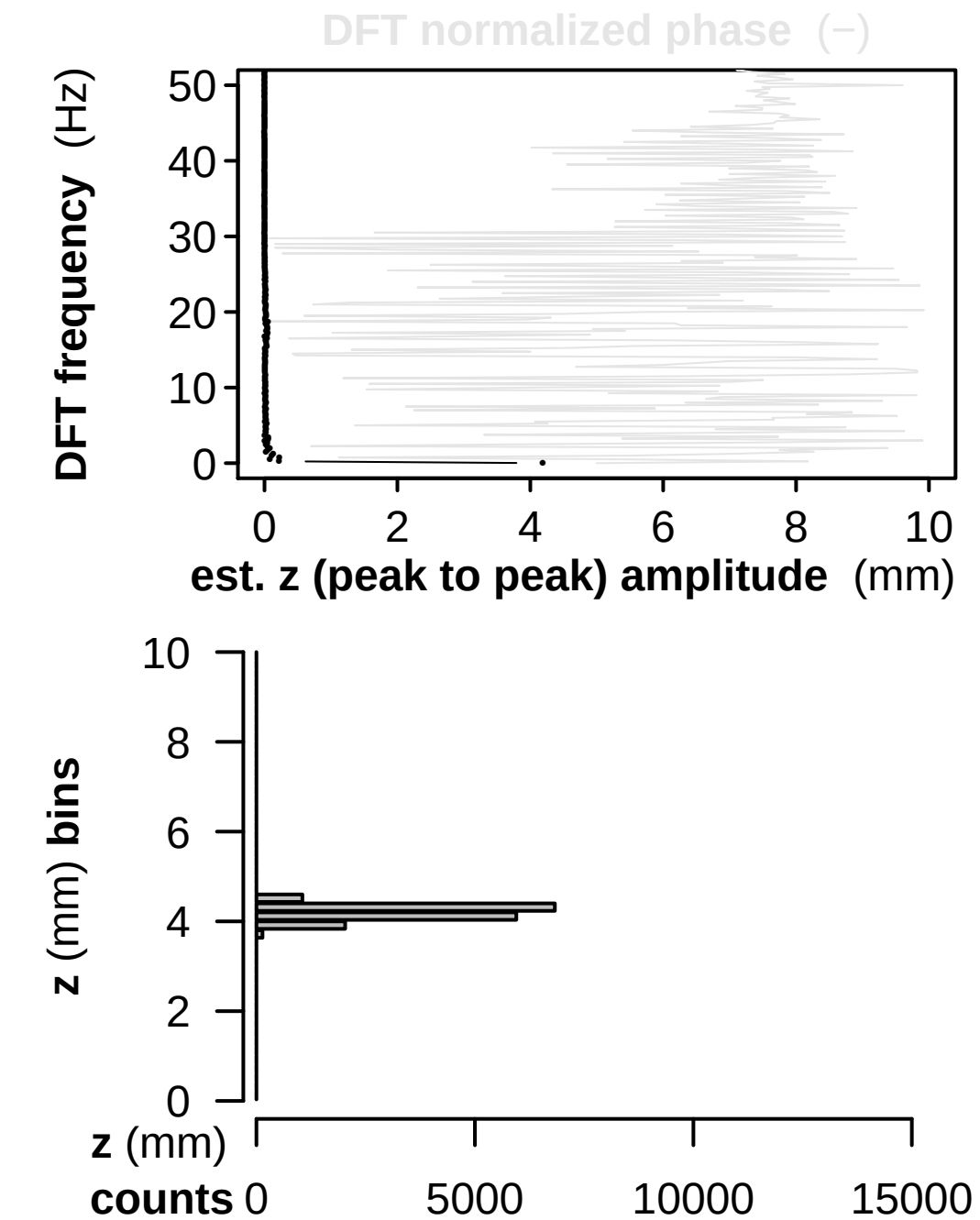

**SUBJECT 1 − RUN 24 − CONDITION 5,1**
**SC_180323_105317_0.AIFF**

```
z_min               : 3.75 mm
z_max               : 4.46 mm
z_travel_amplitude  : 0.71 mm

avg_abs_z_travel    : 4.84 mm/s

z_jarque-bera_jb    : 600.50
z_jarque-bera_p     : 0.00e+00

z_lin_mod_est_slope: 0.05 mm/s
z_lin_mod_adj_R²    : 17 %

z_poly40_mod_adj_R²: 78 %

z_dft_ampl_thresh   : 0.010 mm
>=threshold_maxfreq: 25.00 Hz
```

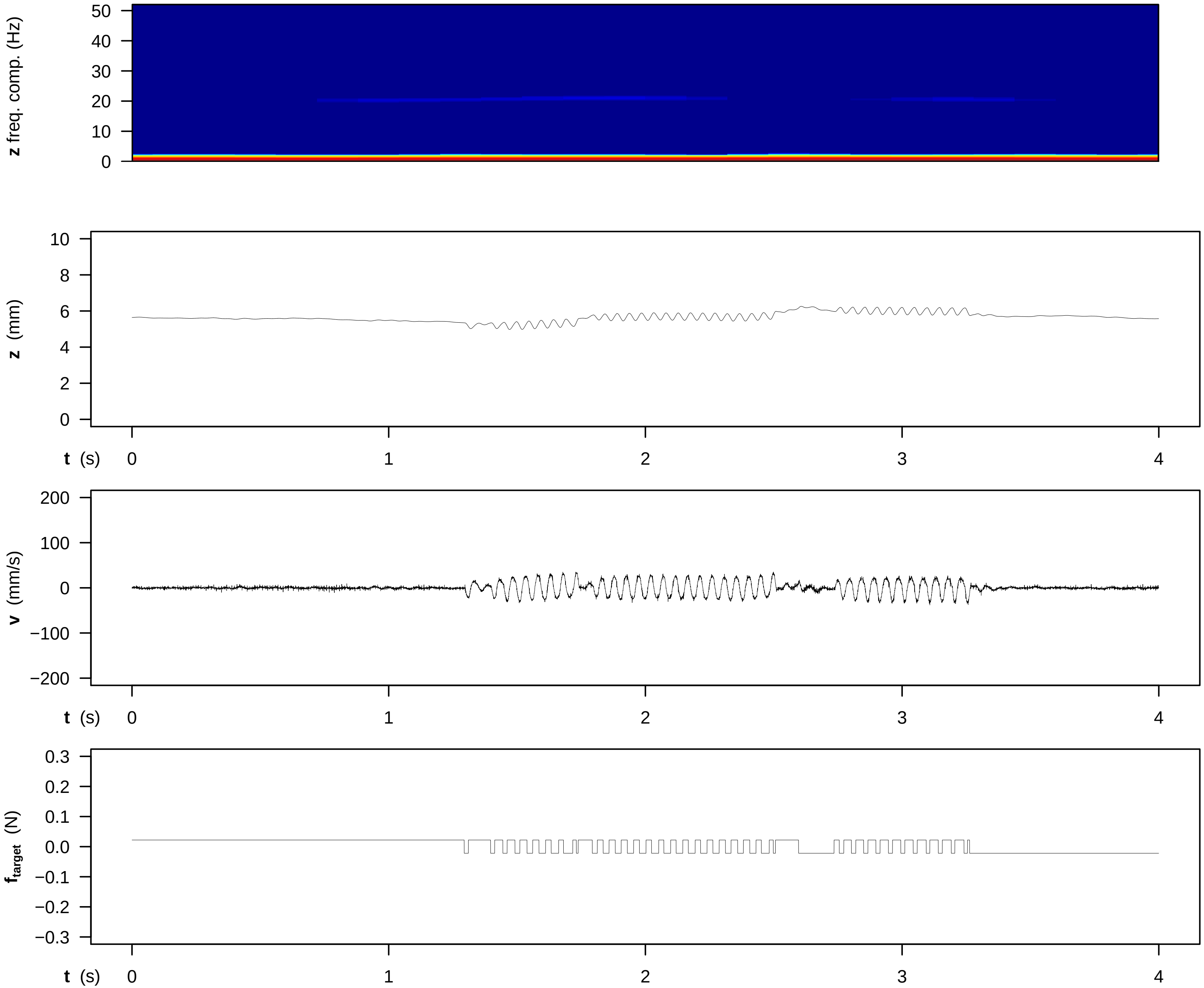

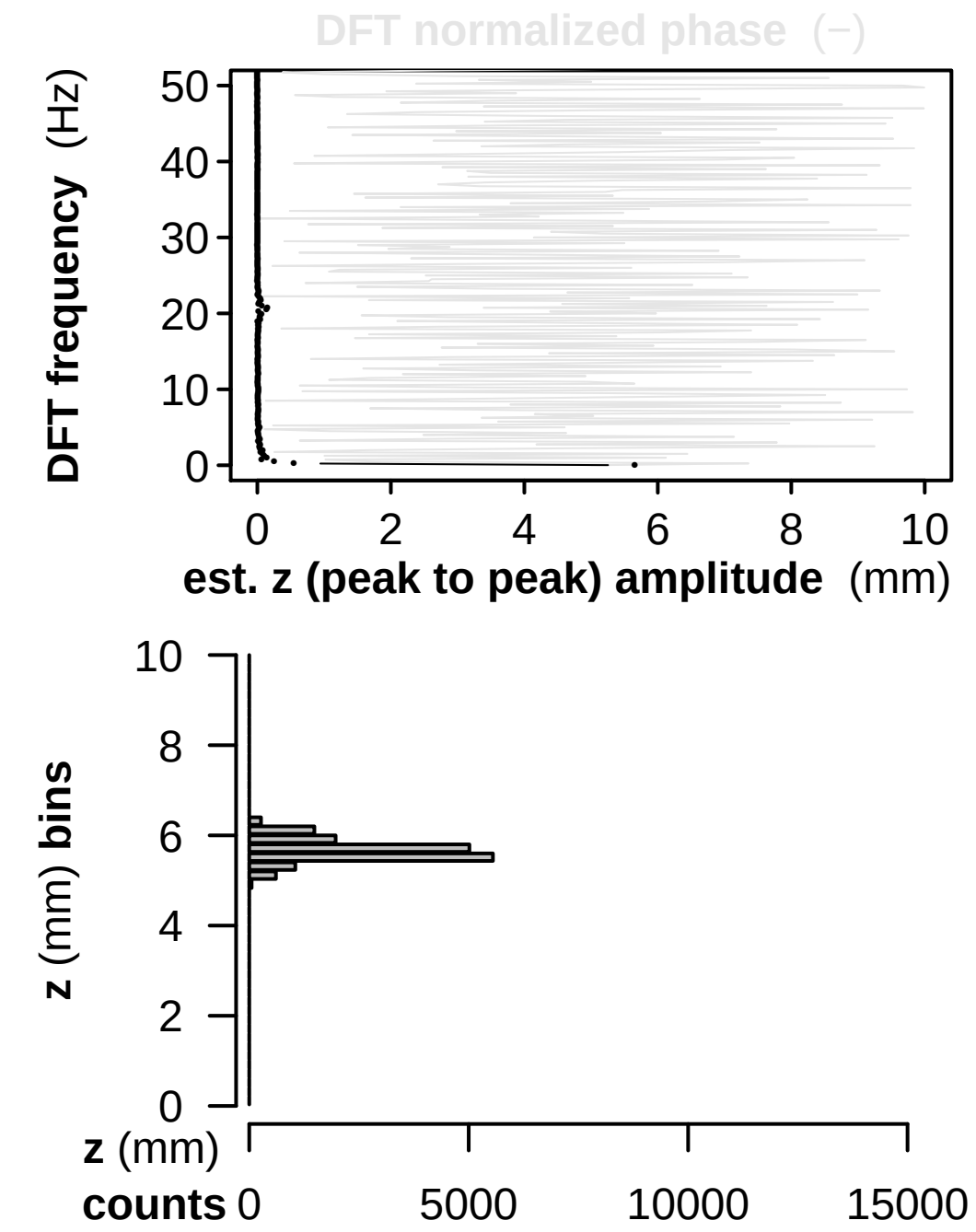

**SUBJECT 2 − RUN 19 − CONDITION 5,1**
**SC_180323_112632_0.AIFF**

```
z_min               : 4.98 mm
z_max               : 6.26 mm
z_travel_amplitude  : 1.28 mm

avg_abs_z_travel    : 8.46 mm/s

z_jarque-bera_jb    : 120.55
z_jarque-bera_p     : 0.00e+00

z_lin_mod_est_slope: 0.10 mm/s
z_lin_mod_adj_R²    : 23 %

z_poly40_mod_adj_R²: 84 %

z_dft_ampl_thresh   : 0.010 mm
>=threshold_maxfreq: 23.25 Hz
```

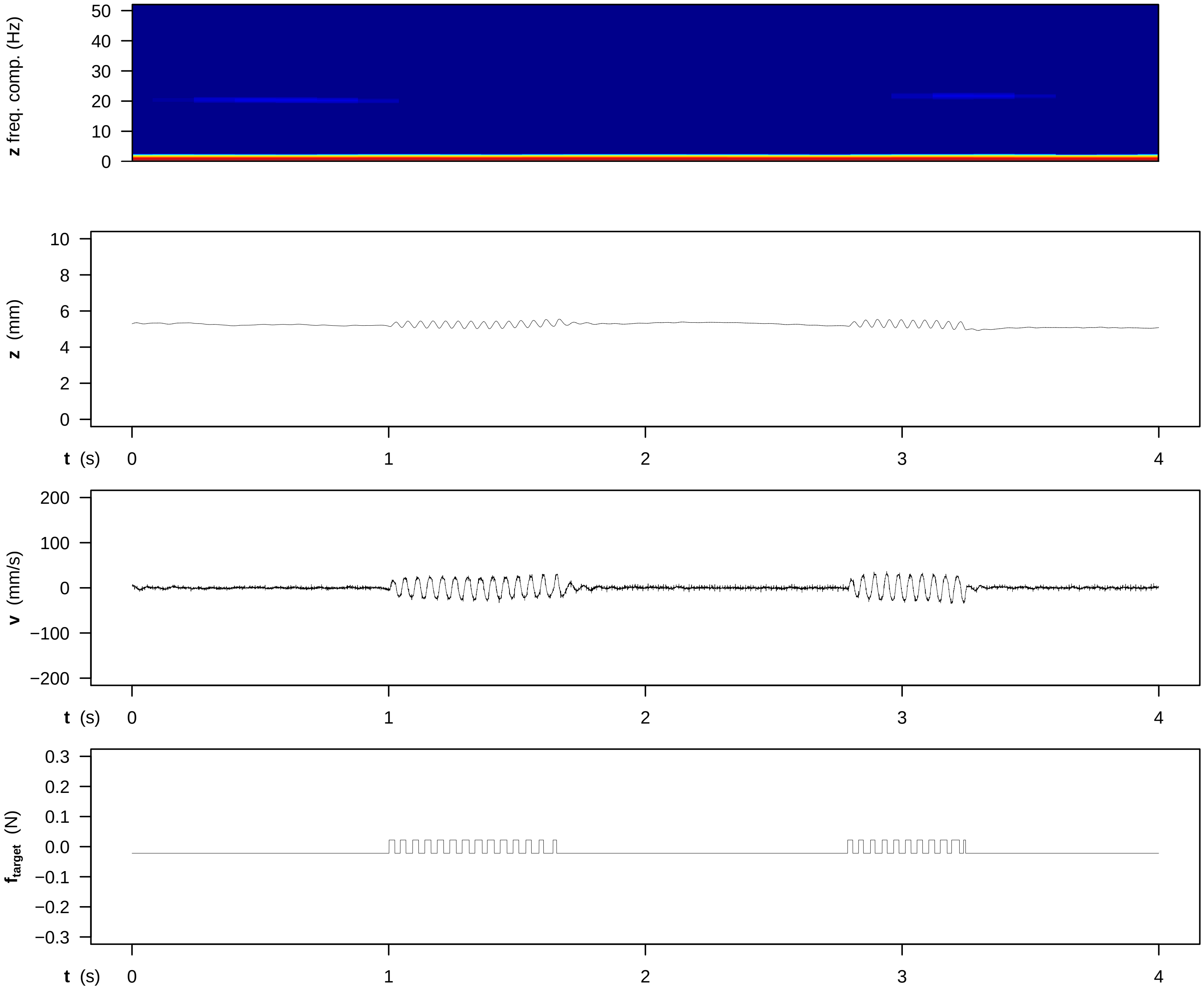

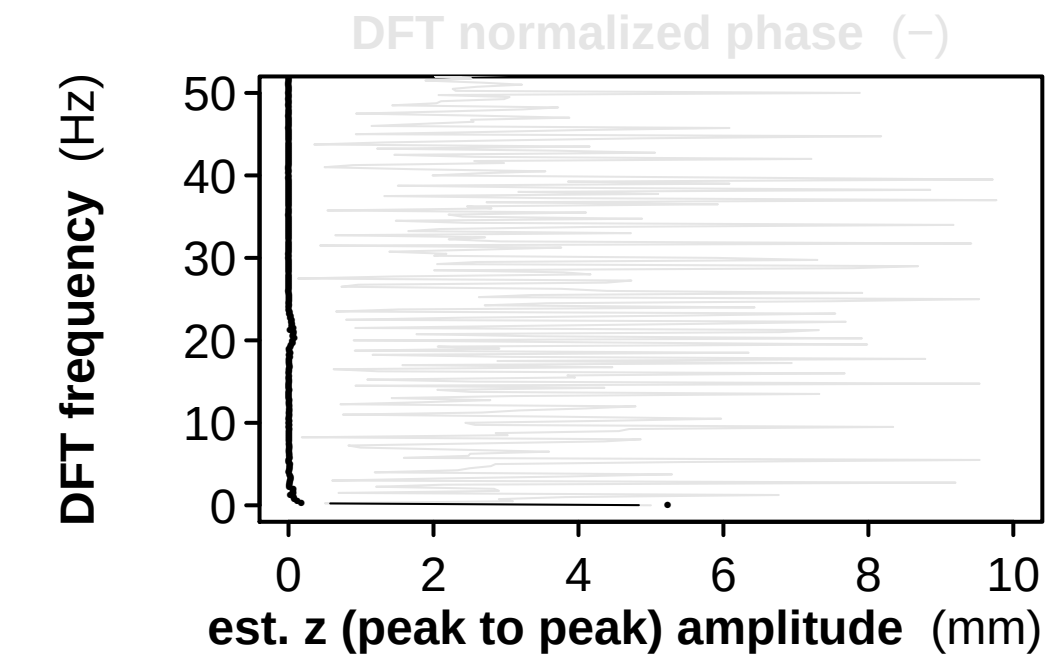

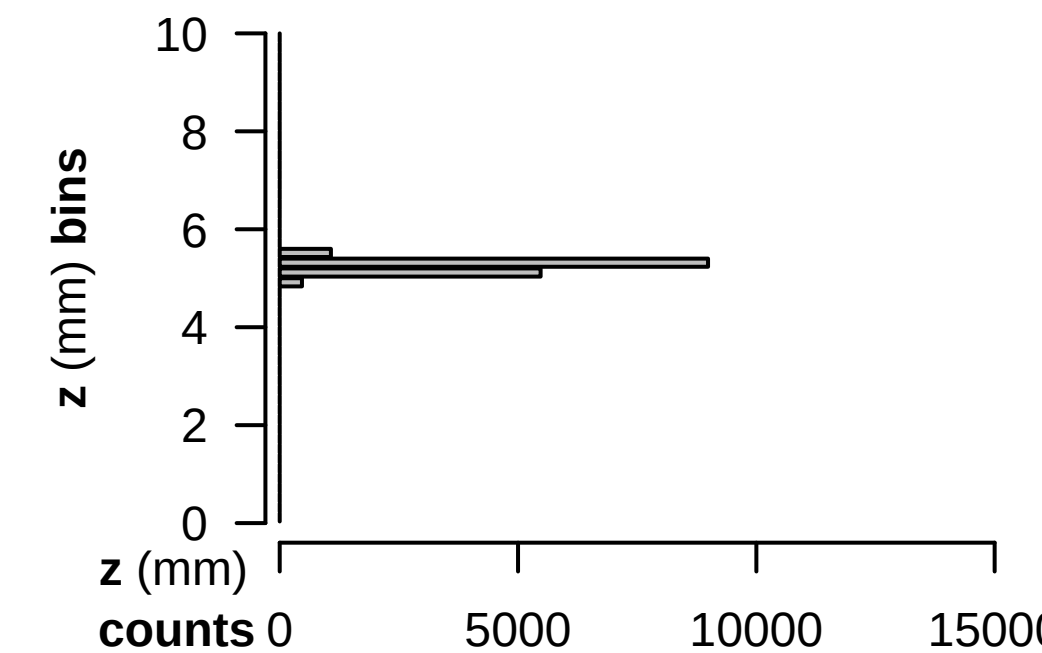

**SUBJECT 2 − RUN 25 − CONDITION 5,1**
**SC_180323_112939_0.AIFF**

```
z_min               : 4.92 mm
z_max               : 5.55 mm
z_travel_amplitude  : 0.63 mm

avg_abs_z_travel    : 7.93 mm/s

z_jarque-bera_jb    : 341.33
z_jarque-bera_p     : 0.00e+00

z_lin_mod_est_slope: −0.04 mm/s
z_lin_mod_adj_R²    : 16 %

z_poly40_mod_adj_R²: 59 %

z_dft_ampl_thresh   : 0.010 mm
>=threshold_maxfreq: 23.50 Hz
```

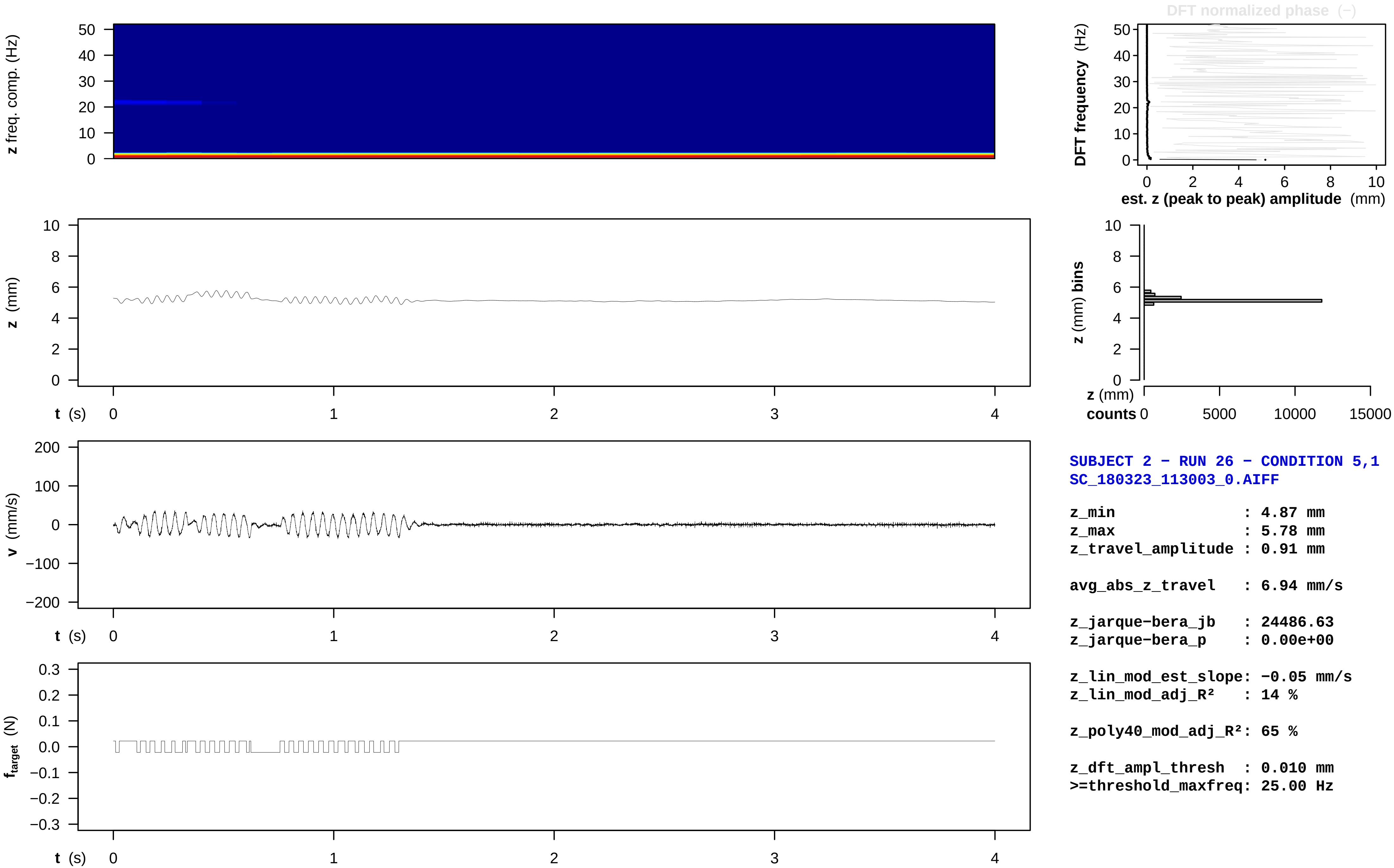

SUBJECT 2 − RUN 26 − CONDITION 5,1
SC_180323_113003_0.AIFF

```
z_min               : 4.87 mm
z_max               : 5.78 mm
z_travel_amplitude : 0.91 mm

avg_abs_z_travel    : 6.94 mm/s

z_jarque-bera_jb    : 24486.63
z_jarque-bera_p     : 0.00e+00

z_lin_mod_est_slope: −0.05 mm/s
z_lin_mod_adj_R²    : 14 %

z_poly40_mod_adj_R²: 65 %

z_dft_ampl_thresh   : 0.010 mm
>=threshold_maxfreq: 25.00 Hz
```

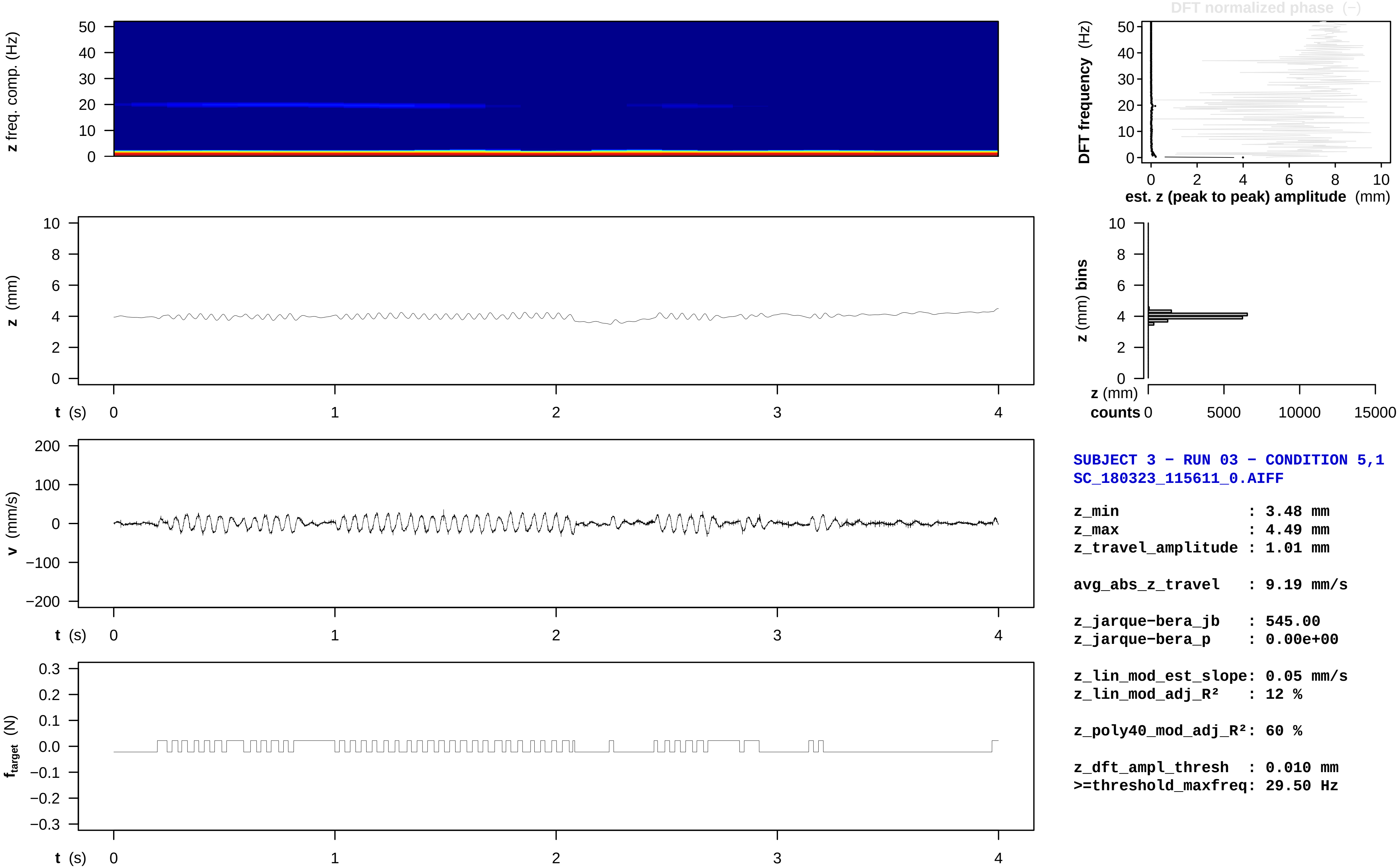

**SUBJECT 3 − RUN 03 − CONDITION 5,1**
**SC_180323_115611_0.AIFF**

```
z_min               : 3.48 mm
z_max               : 4.49 mm
z_travel_amplitude  : 1.01 mm

avg_abs_z_travel    : 9.19 mm/s

z_jarque-bera_jb    : 545.00
z_jarque-bera_p     : 0.00e+00

z_lin_mod_est_slope: 0.05 mm/s
z_lin_mod_adj_R²    : 12 %

z_poly40_mod_adj_R²: 60 %

z_dft_ampl_thresh   : 0.010 mm
>=threshold_maxfreq: 29.50 Hz
```

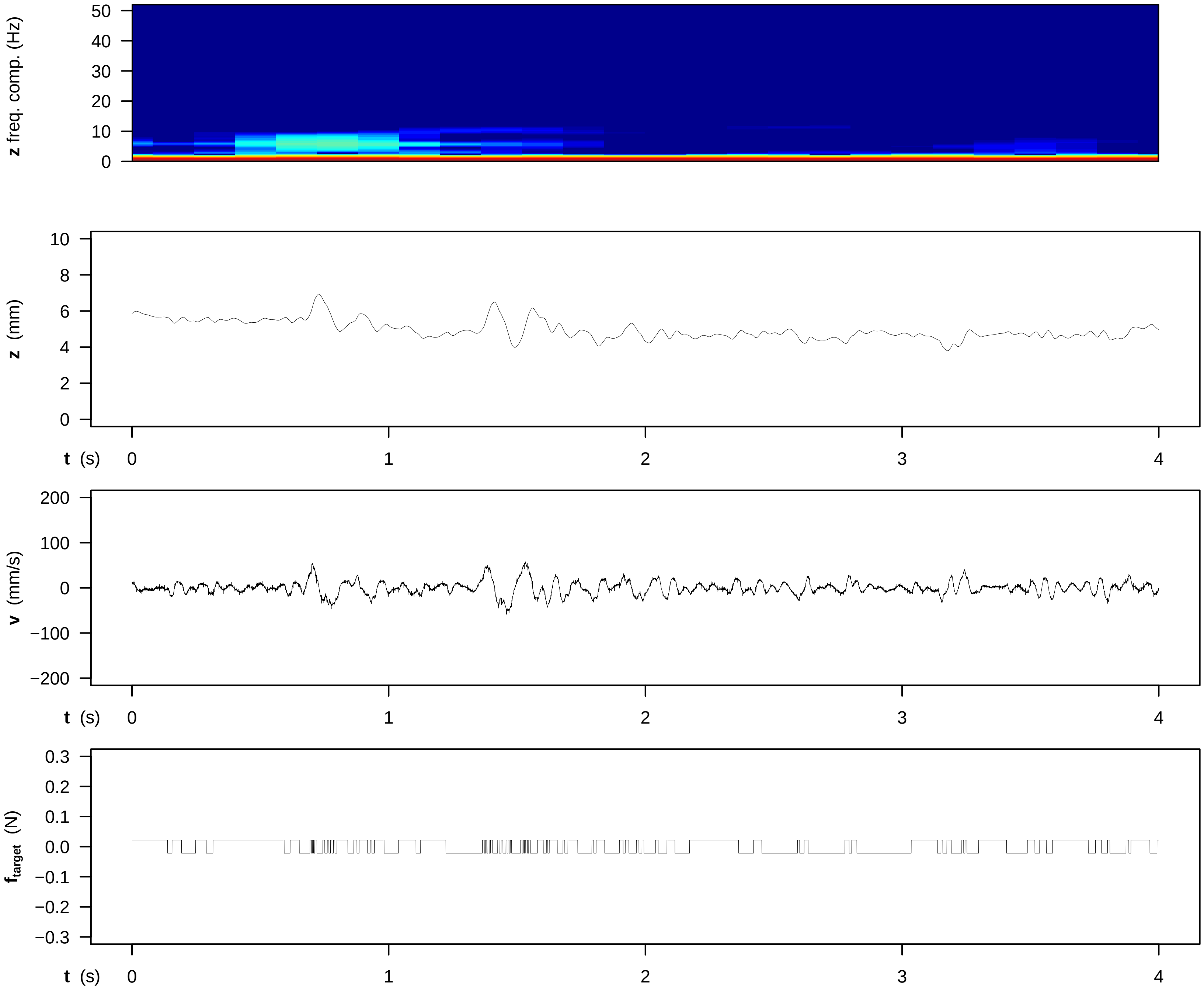

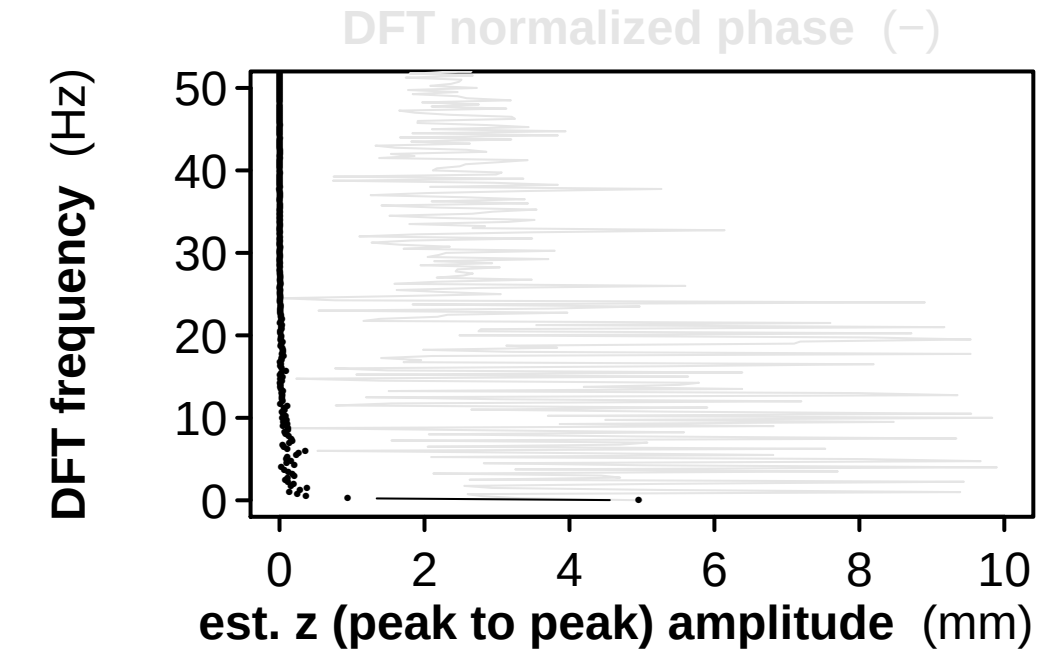

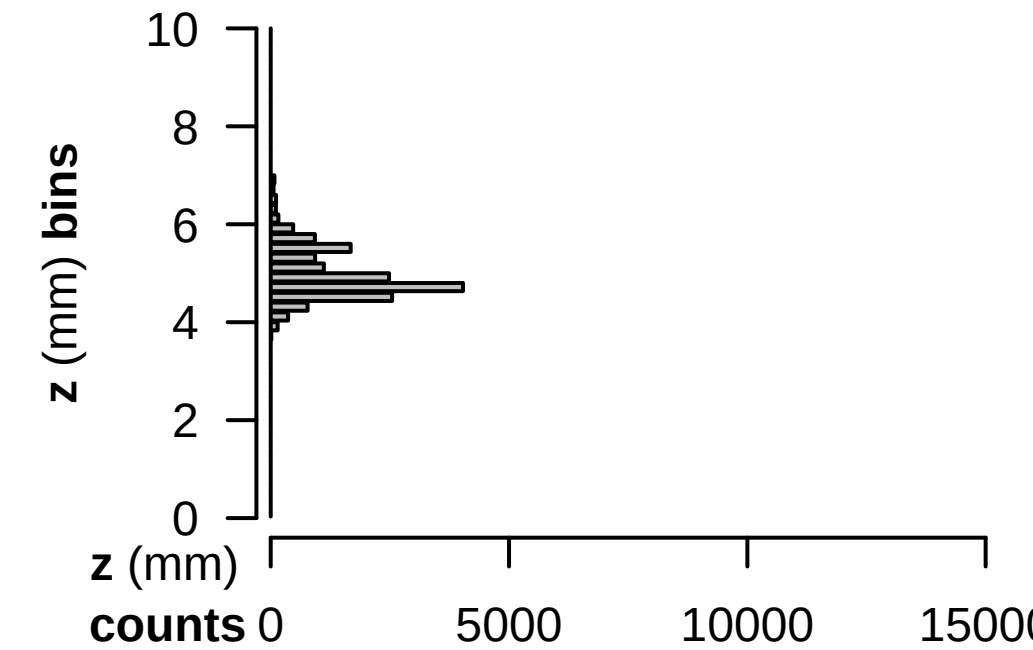

**SUBJECT 3 − RUN 11 − CONDITION 5,1**
**SC_180323_120117_0.AIFF**

```
z_min               : 3.80 mm
z_max               : 6.94 mm
z_travel_amplitude  : 3.14 mm

avg_abs_z_travel    : 10.81 mm/s

z_jarque-bera_jb    : 2597.44
z_jarque-bera_p     : 0.00e+00

z_lin_mod_est_slope: −0.28 mm/s
z_lin_mod_adj_R²    : 40 %

z_poly40_mod_adj_R²: 66 %

z_dft_ampl_thresh   : 0.010 mm
>=threshold_maxfreq: 30.75 Hz
```

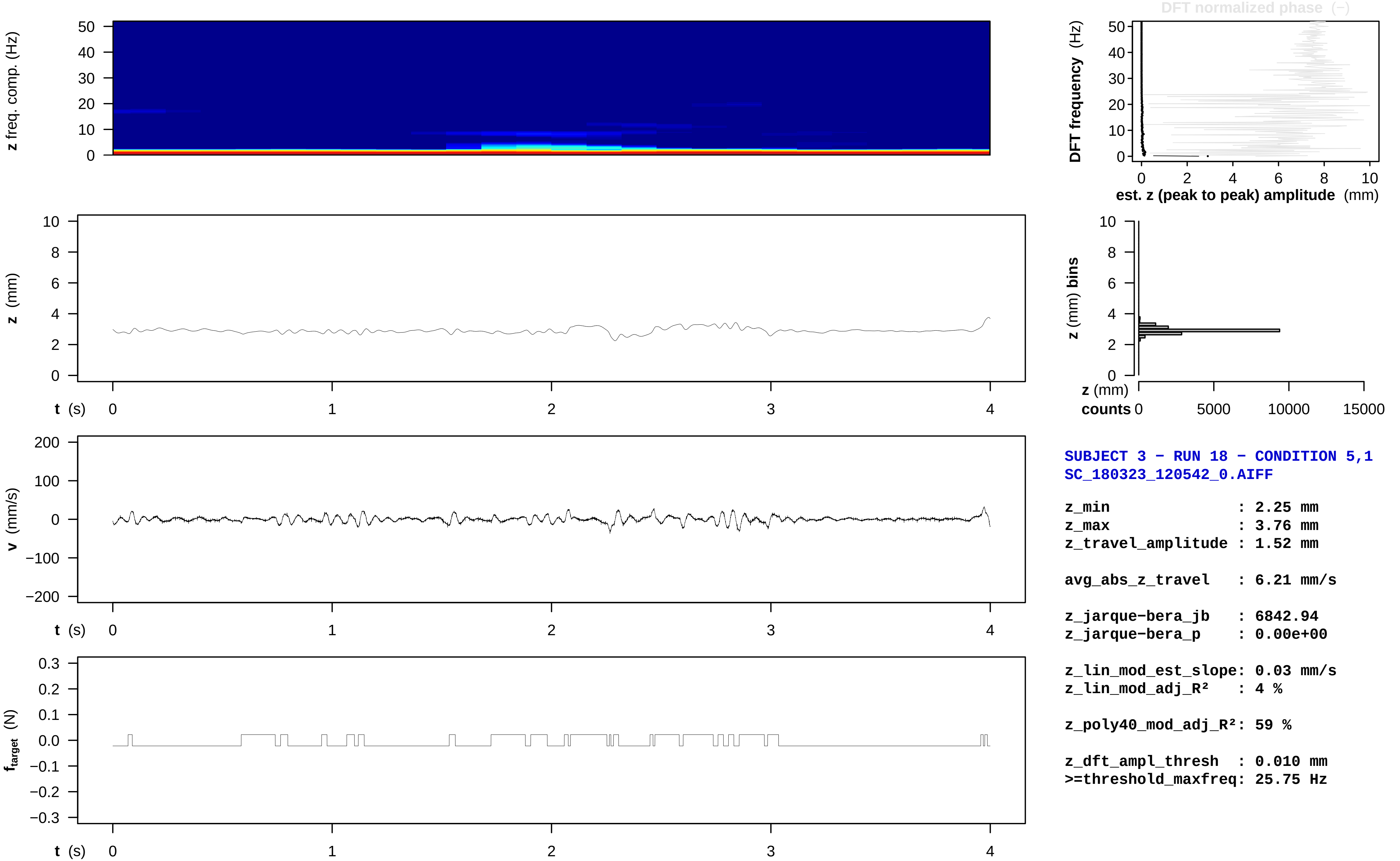

SUBJECT 3 − RUN 18 − CONDITION 5,1
SC_180323_120542_0.AIFF

```
z_min               : 2.25 mm
z_max               : 3.76 mm
z_travel_amplitude  : 1.52 mm

avg_abs_z_travel    : 6.21 mm/s

z_jarque-bera_jb    : 6842.94
z_jarque-bera_p     : 0.00e+00

z_lin_mod_est_slope: 0.03 mm/s
z_lin_mod_adj_R²    : 4 %

z_poly40_mod_adj_R²: 59 %

z_dft_ampl_thresh   : 0.010 mm
>=threshold_maxfreq: 25.75 Hz
```

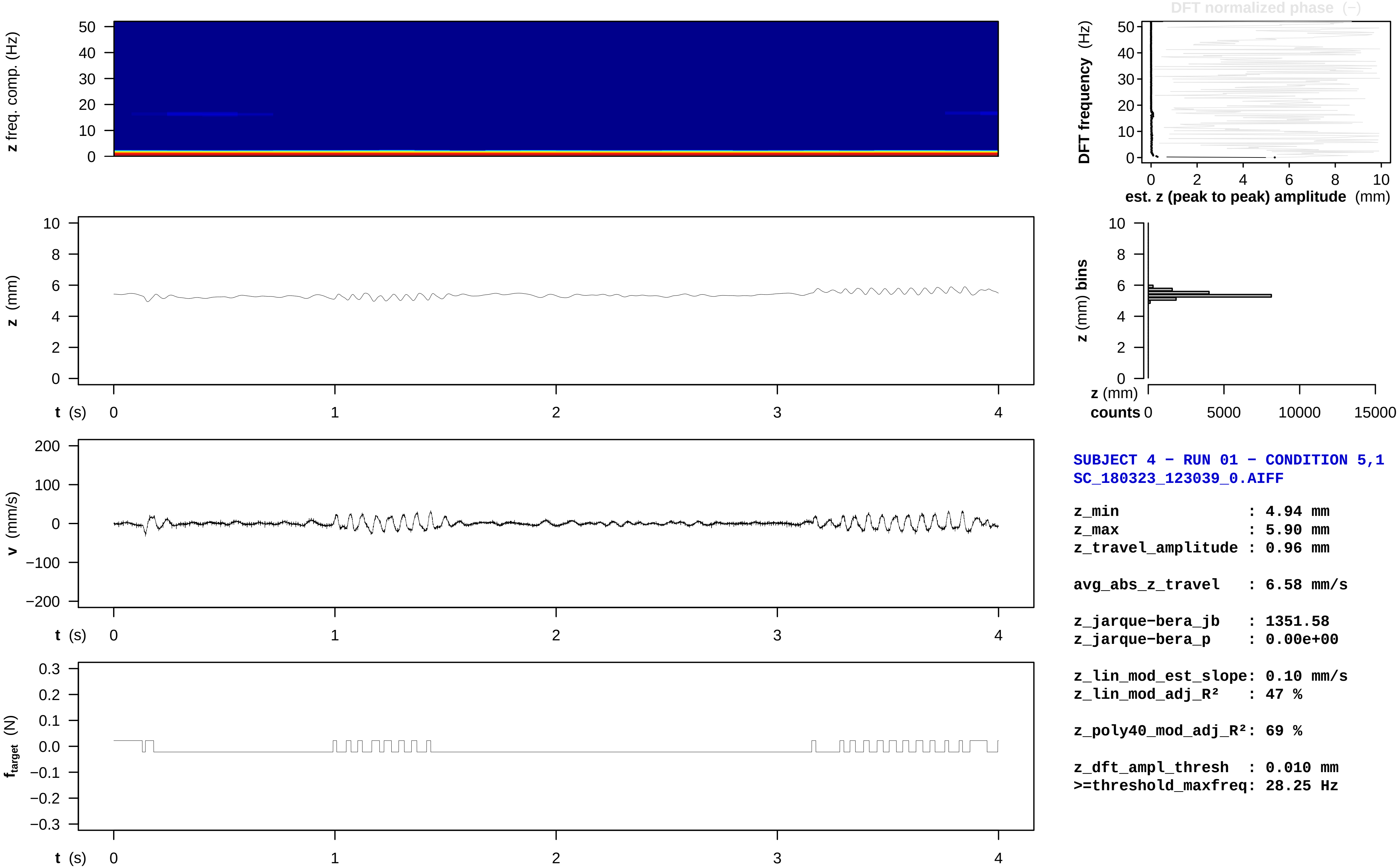

**SUBJECT 4 − RUN 01 − CONDITION 5,1**
**SC_180323_123039_0.AIFF**

```
z_min               : 4.94 mm
z_max               : 5.90 mm
z_travel_amplitude  : 0.96 mm

avg_abs_z_travel    : 6.58 mm/s

z_jarque-bera_jb    : 1351.58
z_jarque-bera_p     : 0.00e+00

z_lin_mod_est_slope: 0.10 mm/s
z_lin_mod_adj_R²    : 47 %

z_poly40_mod_adj_R²: 69 %

z_dft_ampl_thresh   : 0.010 mm
>=threshold_maxfreq: 28.25 Hz
```

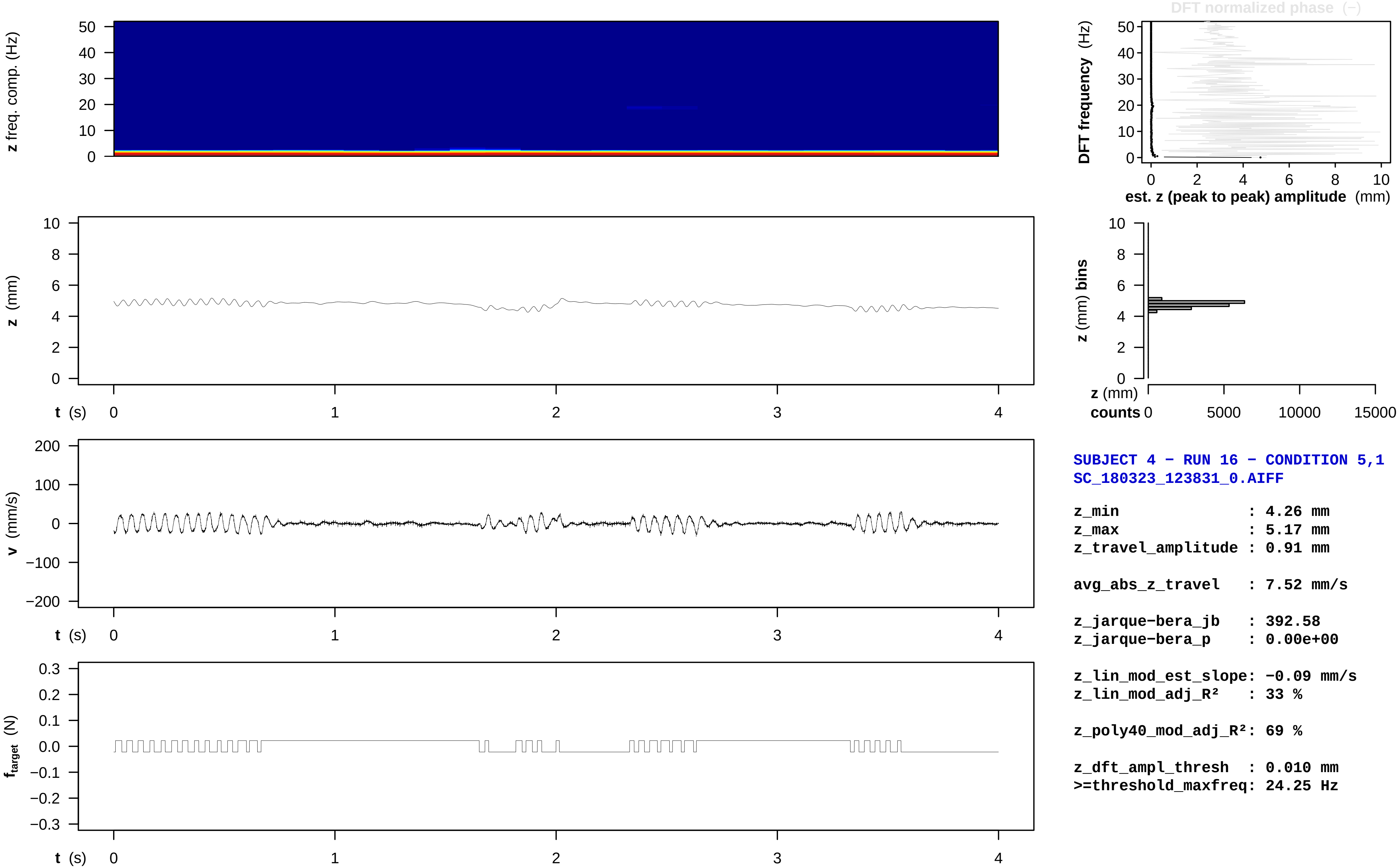

**SUBJECT 4 − RUN 16 − CONDITION 5,1**
**SC_180323_123831_0.AIFF**

```
z_min               : 4.26 mm
z_max               : 5.17 mm
z_travel_amplitude  : 0.91 mm

avg_abs_z_travel    : 7.52 mm/s

z_jarque-bera_jb    : 392.58
z_jarque-bera_p     : 0.00e+00

z_lin_mod_est_slope: -0.09 mm/s
z_lin_mod_adj_R²    : 33 %

z_poly40_mod_adj_R²: 69 %

z_dft_ampl_thresh   : 0.010 mm
>=threshold_maxfreq: 24.25 Hz
```

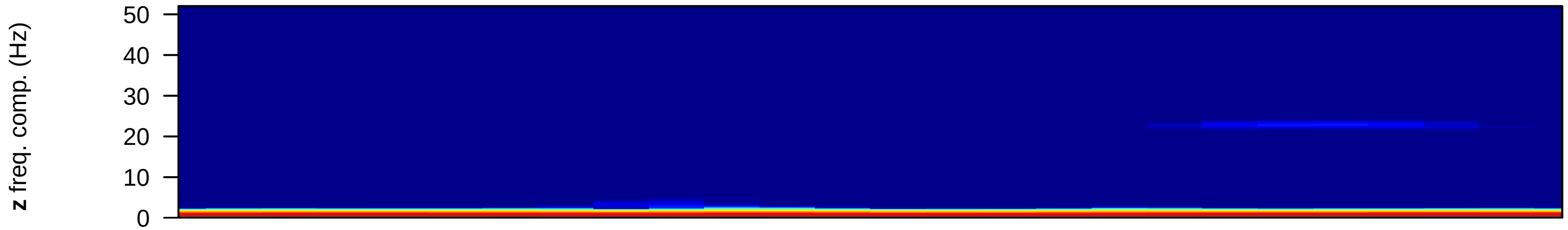

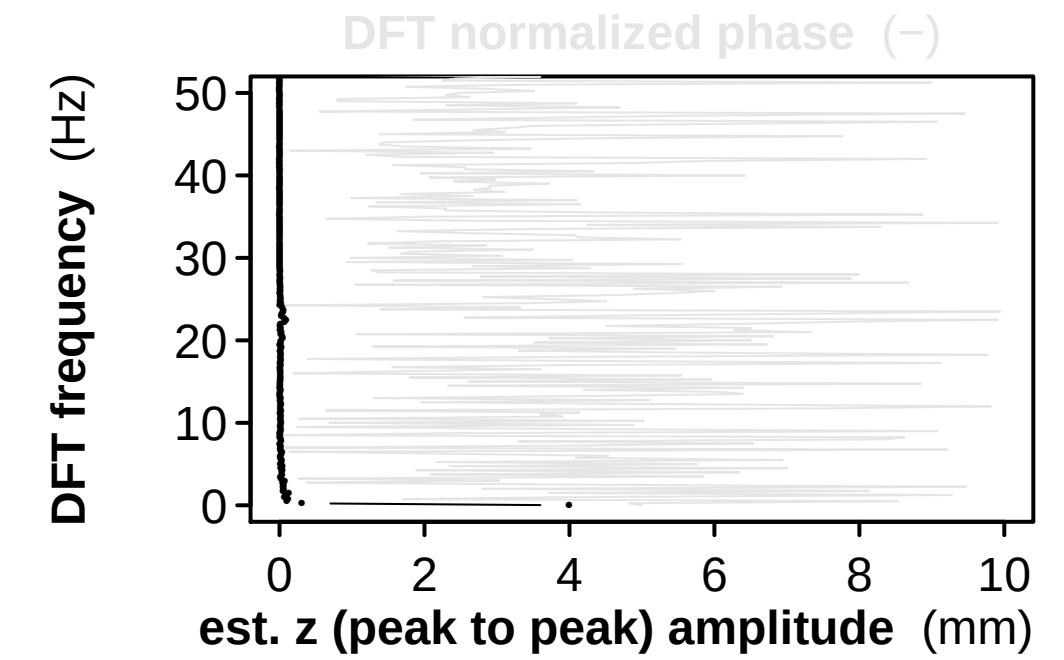

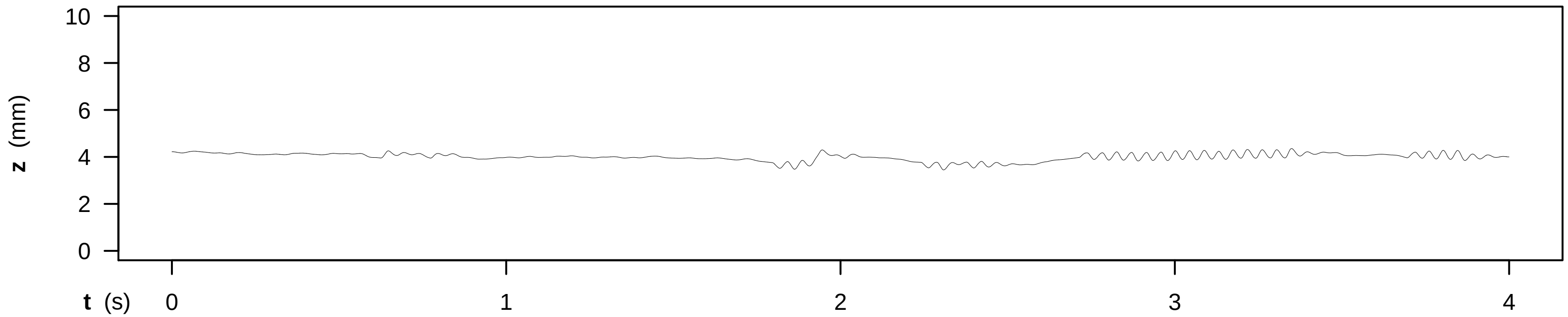

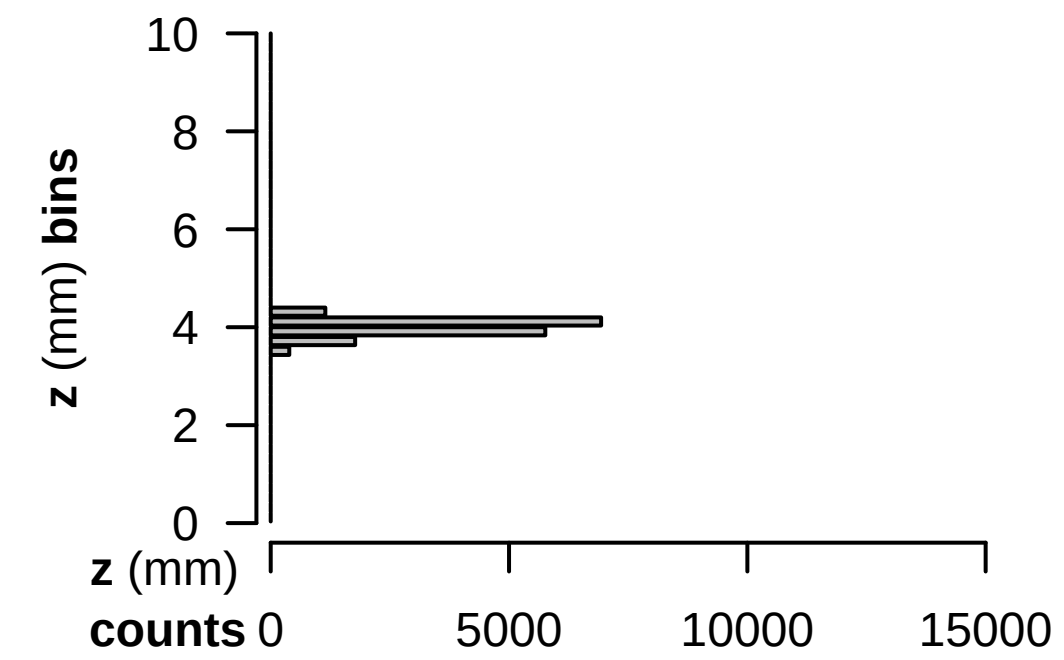

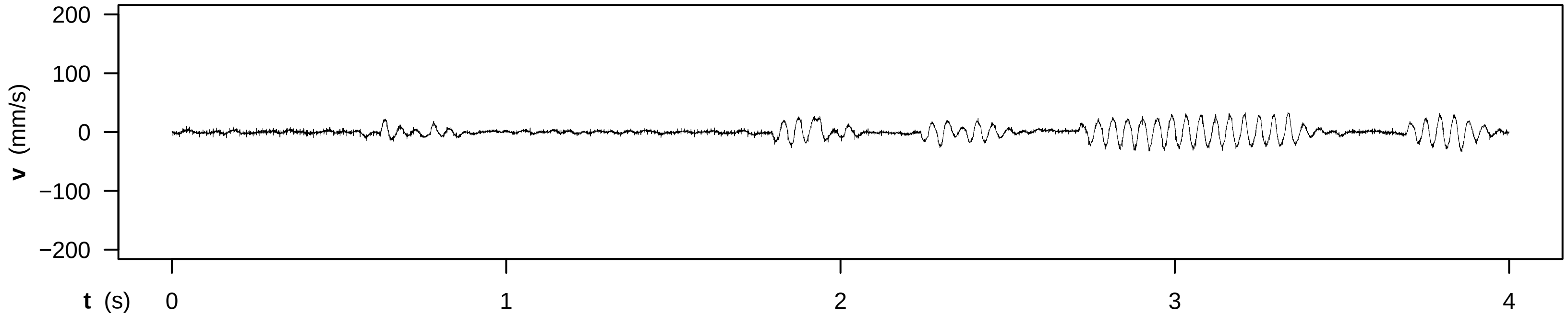

**SUBJECT 4 − RUN 35 − CONDITION 5,1**
**SC_180323_124904_0.AIFF**

```
z_min               : 3.45 mm
z_max               : 4.36 mm
z_travel_amplitude  : 0.91 mm

avg_abs_z_travel    : 6.97 mm/s

z_jarque-bera_jb    : 1704.99
z_jarque-bera_p     : 0.00e+00

z_lin_mod_est_slope: −0.01 mm/s
z_lin_mod_adj_R²    : 1 %

z_poly40_mod_adj_R²: 70 %

z_dft_ampl_thresh   : 0.010 mm
>=threshold_maxfreq: 26.50 Hz
```

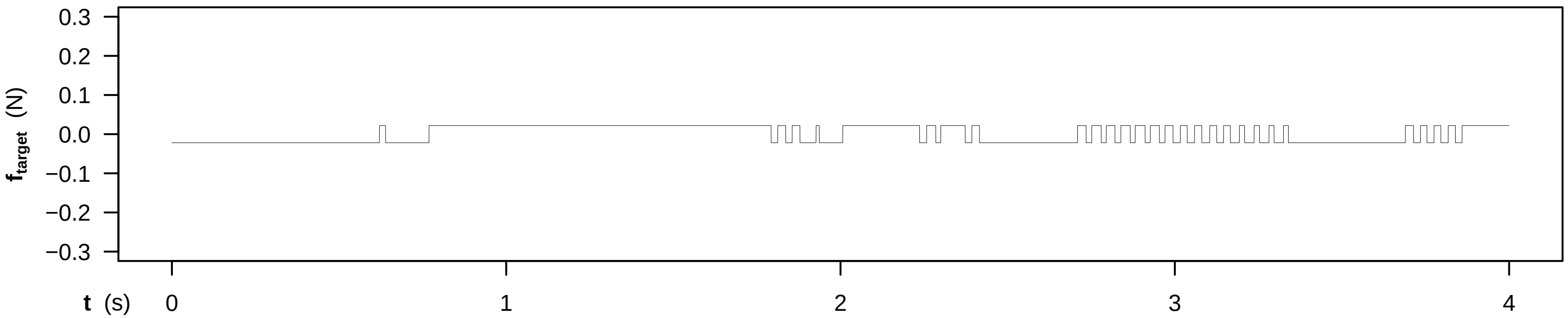

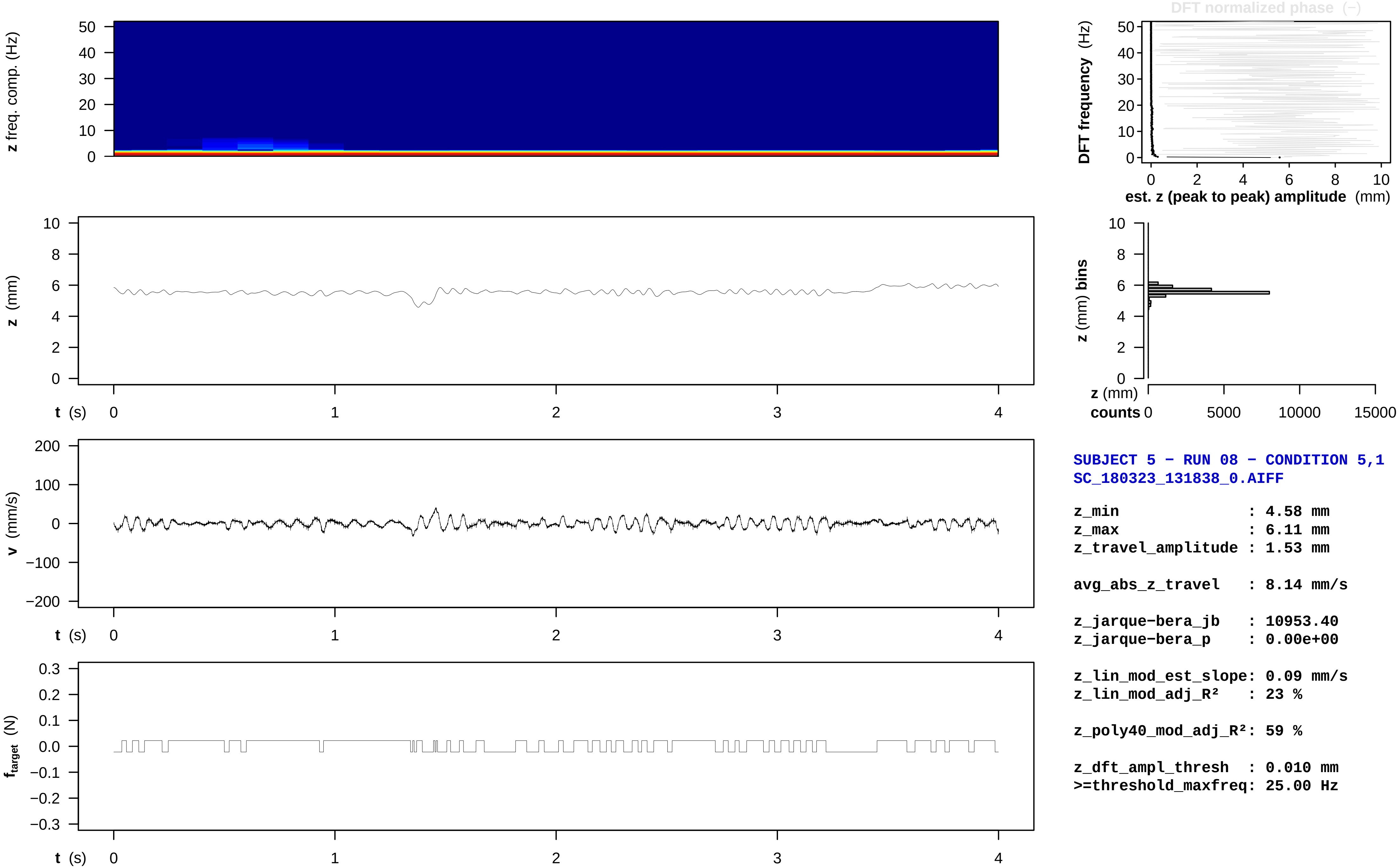

SUBJECT 5 − RUN 08 − CONDITION 5,1
SC_180323_131838_0.AIFF

z_min : 4.58 mm
z_max : 6.11 mm
z_travel_amplitude : 1.53 mm

avg_abs_z_travel : 8.14 mm/s

z_jarque-bera_jb : 10953.40
z_jarque-bera_p : 0.00e+00

z_lin_mod_est_slope: 0.09 mm/s
z_lin_mod_adj_R² : 23 %

z_poly40_mod_adj_R²: 59 %

z_dft_ampl_thresh : 0.010 mm
>=threshold_maxfreq: 25.00 Hz

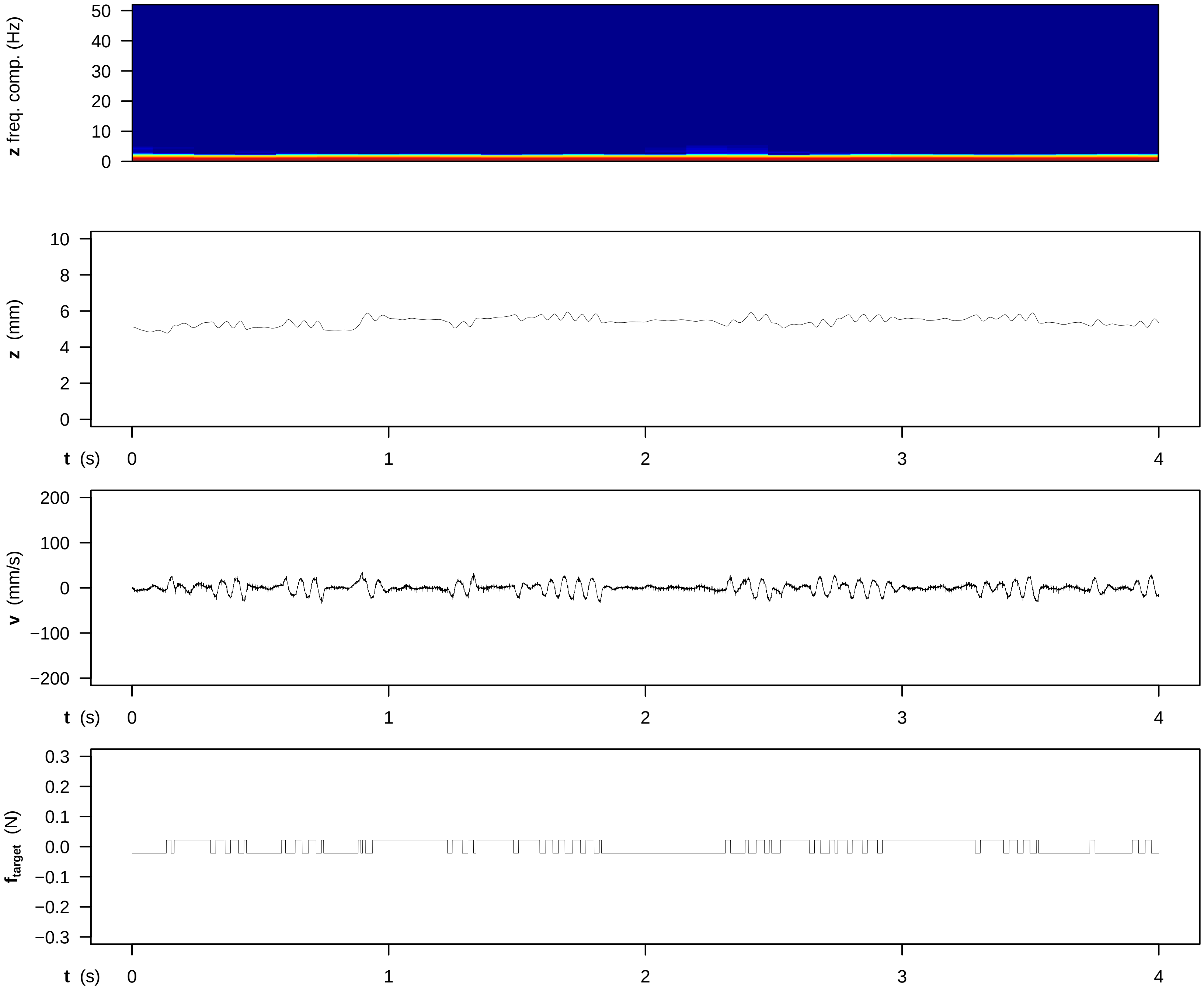

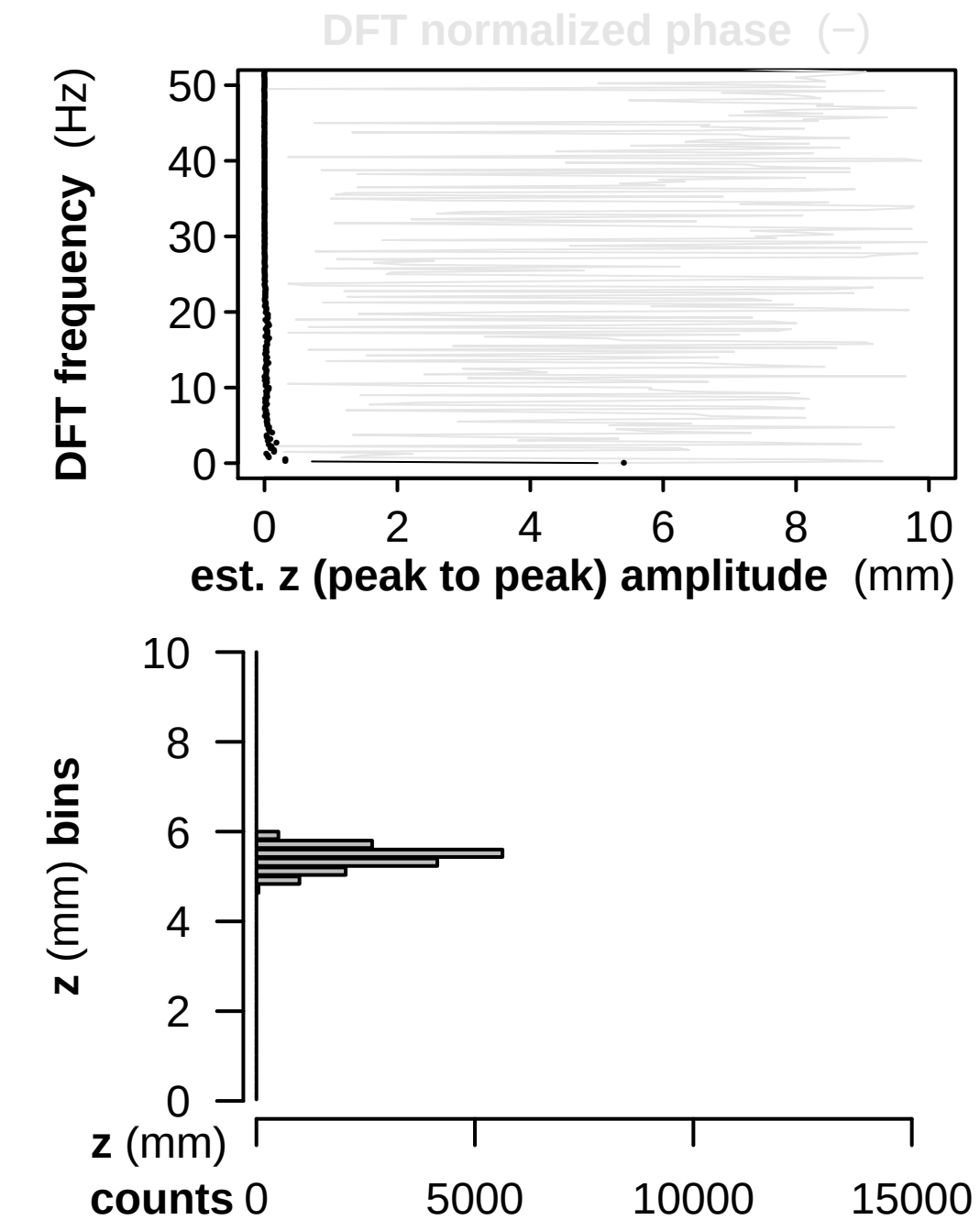

**SUBJECT 5 – RUN 27 – CONDITION 5,1**
**SC_180323_133156_0.AIFF**

z_min : 4.77 mm
z_max : 5.94 mm
z_travel_amplitude : 1.17 mm

avg_abs_z_travel : 8.10 mm/s

z_jarque-bera_jb : 415.29
z_jarque-bera_p : 0.00e+00

z_lin_mod_est_slope: 0.07 mm/s
z_lin_mod_adj_R² : 12 %

z_poly40_mod_adj_R²: 62 %

z_dft_ampl_thresh : 0.010 mm
>=threshold_maxfreq: 24.00 Hz

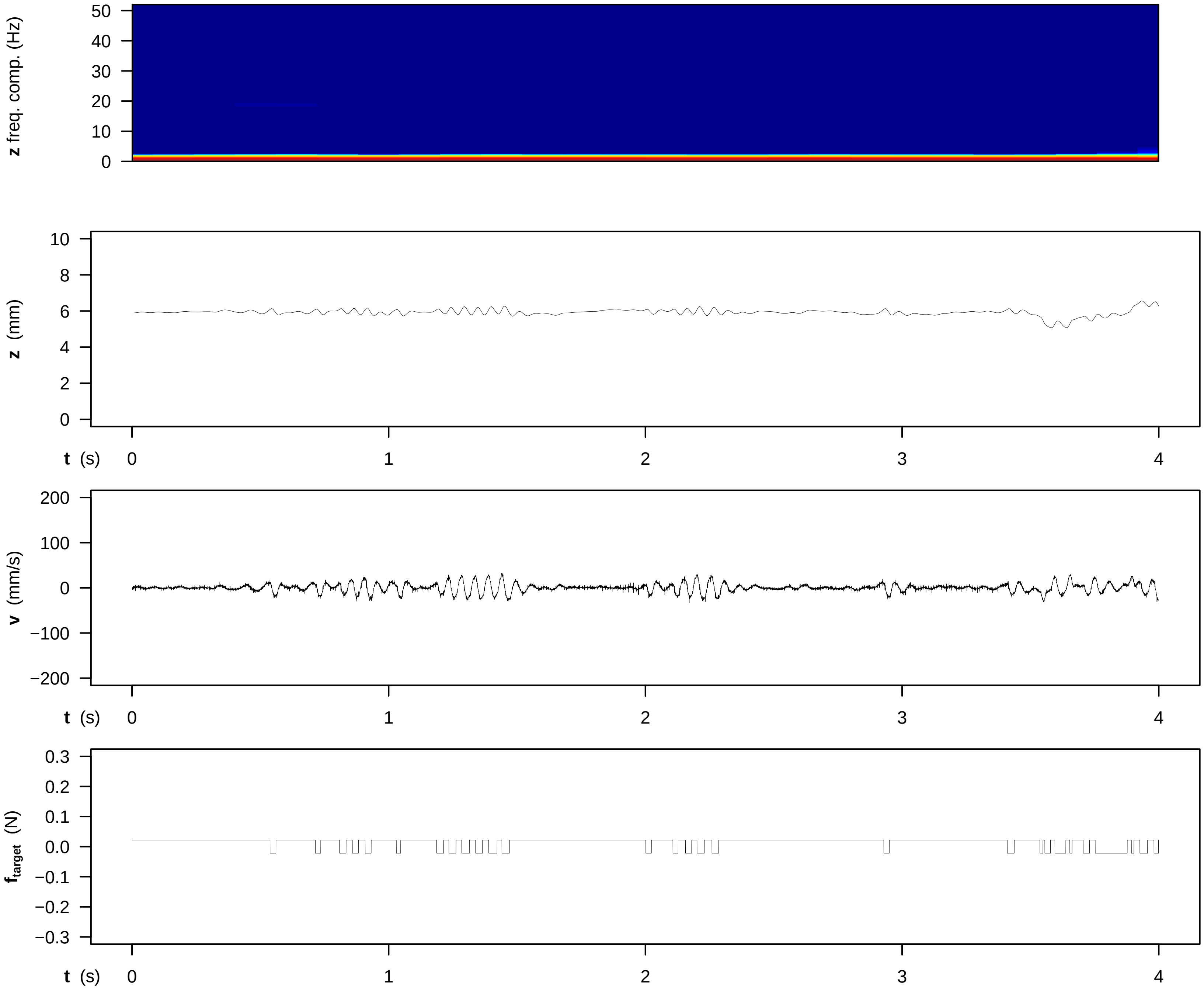

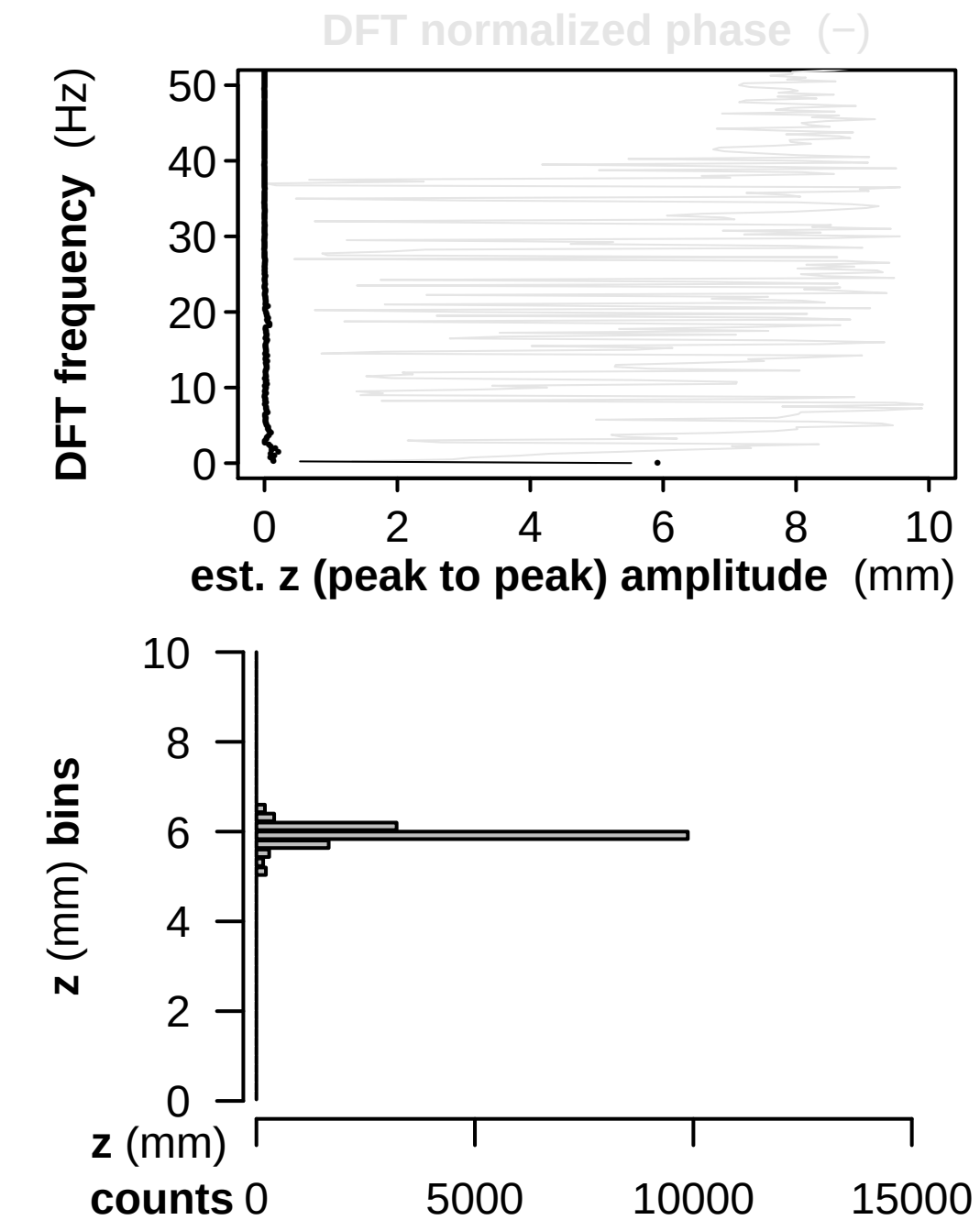

**SUBJECT 5 − RUN 30 − CONDITION 5,1**
**SC_180323_133446_0.AIFF**

```
z_min               : 5.07 mm
z_max               : 6.55 mm
z_travel_amplitude : 1.48 mm

avg_abs_z_travel    : 7.29 mm/s

z_jarque-bera_jb    : 29288.60
z_jarque-bera_p     : 0.00e+00

z_lin_mod_est_slope: −0.03 mm/s
z_lin_mod_adj_R²    : 4 %

z_poly40_mod_adj_R²: 69 %

z_dft_ampl_thresh   : 0.010 mm
>=threshold_maxfreq: 26.75 Hz
```

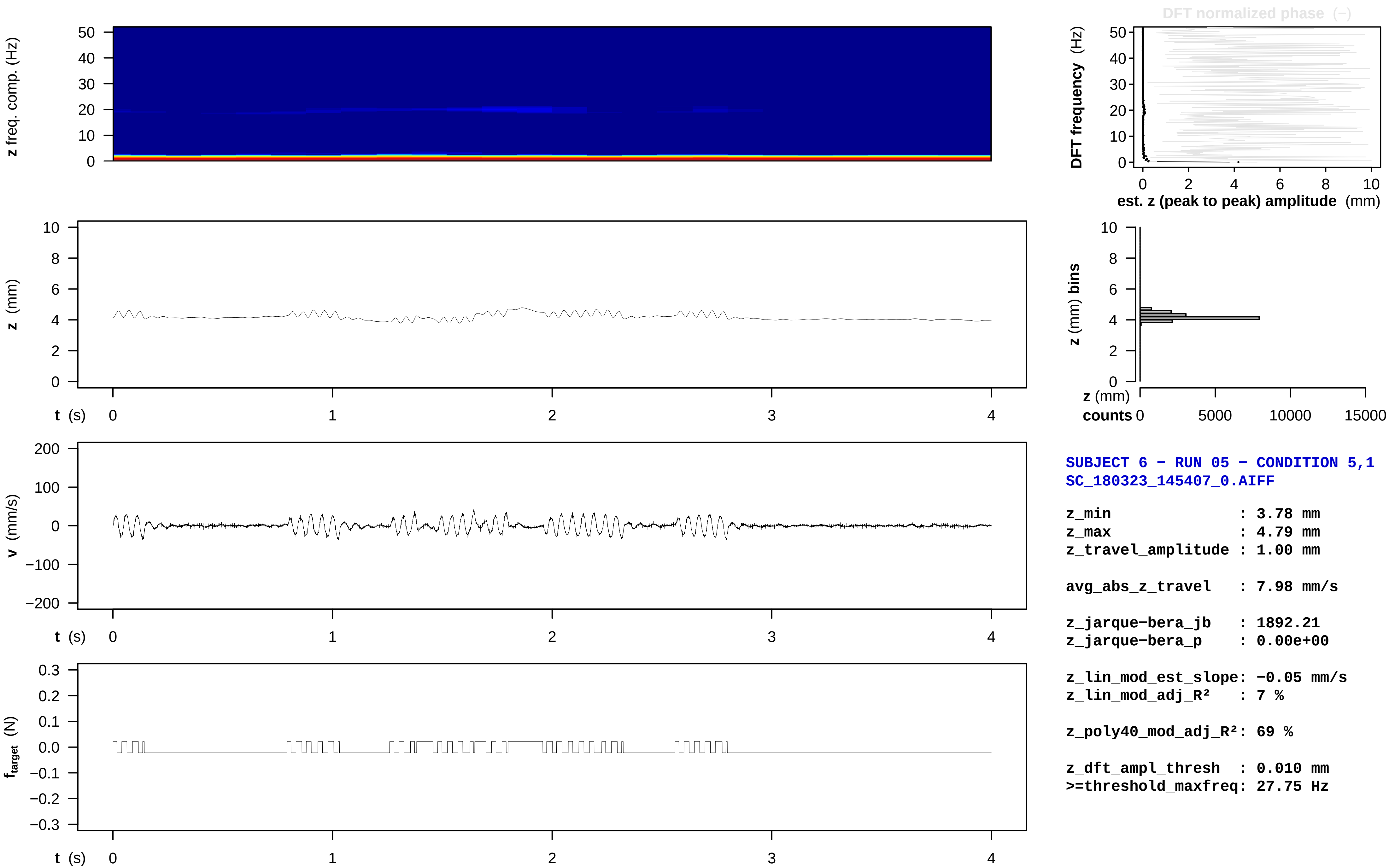

**SUBJECT 6 − RUN 05 − CONDITION 5,1**
**SC_180323_145407_0.AIFF**

```
z_min               : 3.78 mm
z_max               : 4.79 mm
z_travel_amplitude  : 1.00 mm

avg_abs_z_travel    : 7.98 mm/s

z_jarque-bera_jb    : 1892.21
z_jarque-bera_p     : 0.00e+00

z_lin_mod_est_slope: -0.05 mm/s
z_lin_mod_adj_R²    : 7 %

z_poly40_mod_adj_R²: 69 %

z_dft_ampl_thresh   : 0.010 mm
>=threshold_maxfreq: 27.75 Hz
```

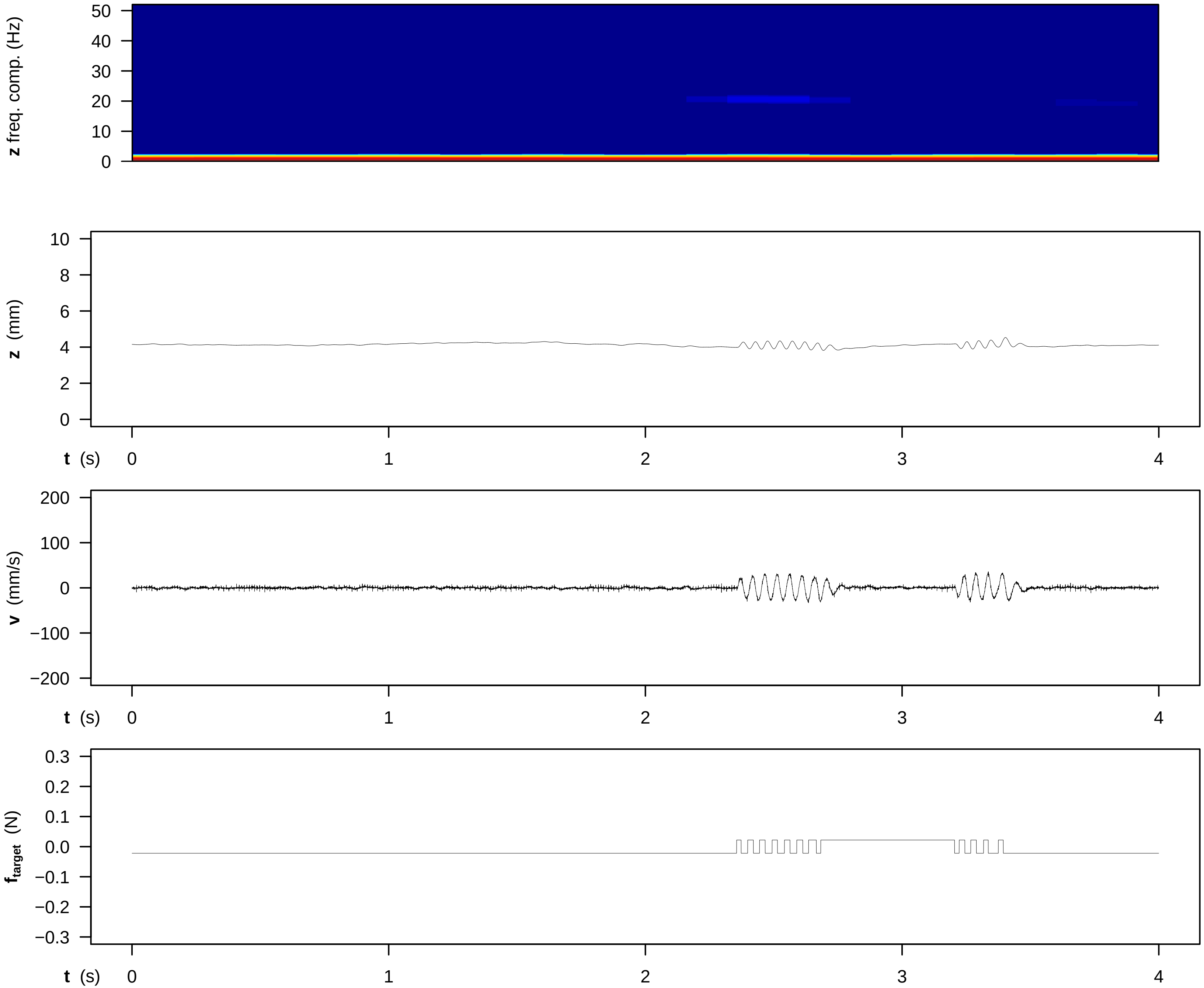

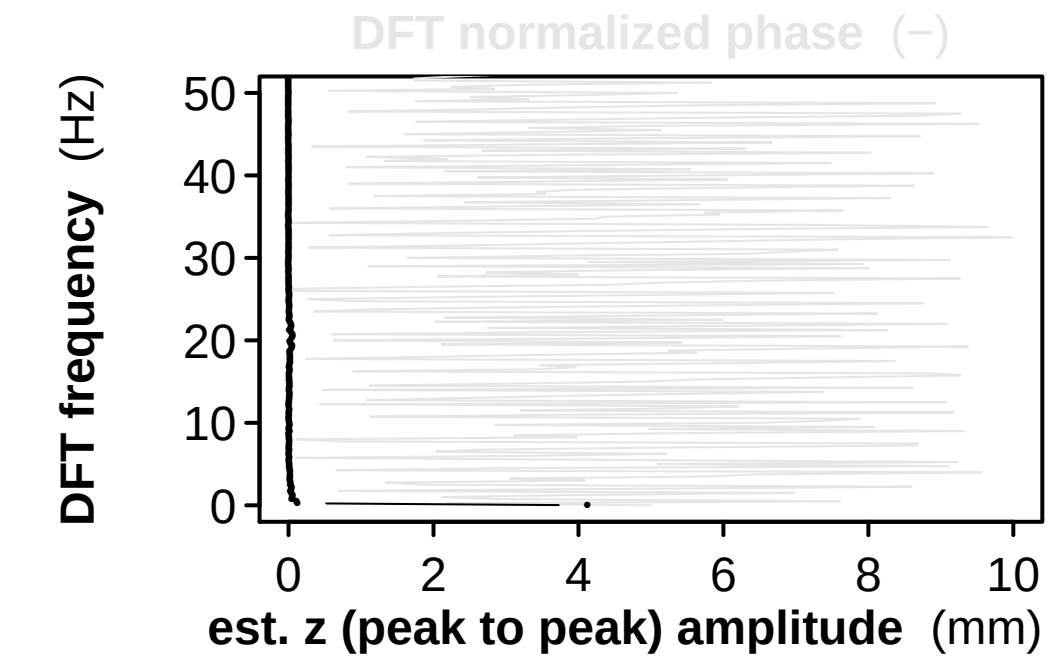

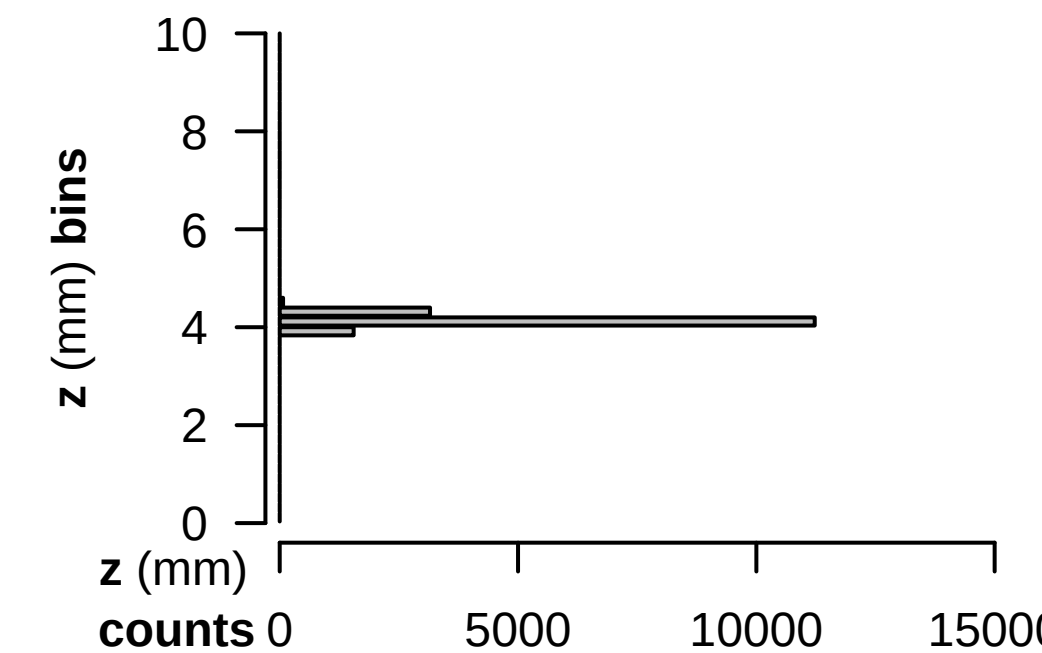

**SUBJECT 6 − RUN 06 − CONDITION 5,1**
**SC_180323_145445_0.AIFF**

```
z_min               : 3.81 mm
z_max               : 4.53 mm
z_travel_amplitude  : 0.72 mm

avg_abs_z_travel    : 4.91 mm/s

z_jarque-bera_jb    : 592.46
z_jarque-bera_p     : 0.00e+00

z_lin_mod_est_slope: −0.02 mm/s
z_lin_mod_adj_R²    : 8 %

z_poly40_mod_adj_R²: 50 %

z_dft_ampl_thresh   : 0.010 mm
>=threshold_maxfreq: 24.25 Hz
```

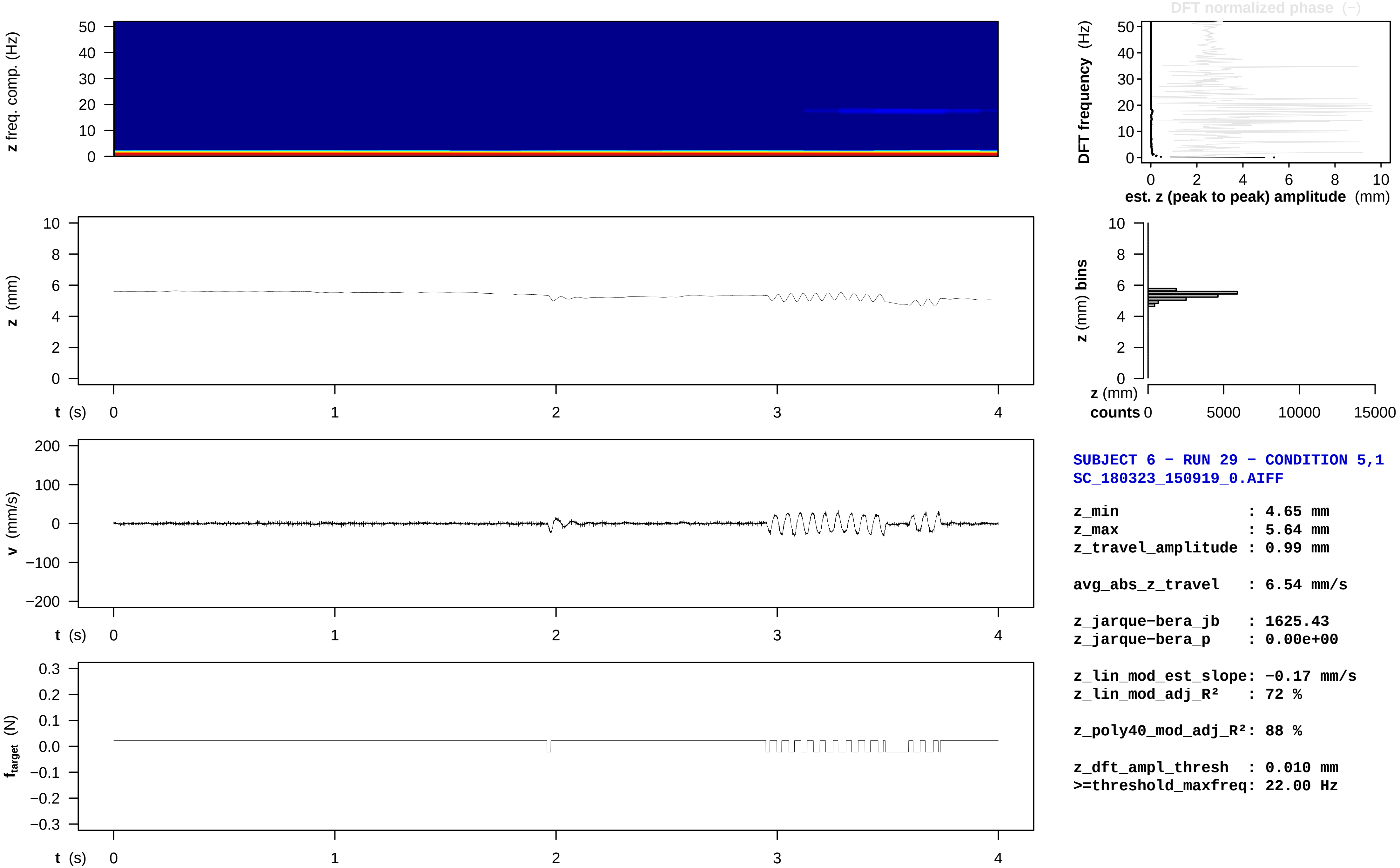

**SUBJECT 6 – RUN 29 – CONDITION 5,1**
**SC_180323_150919_0.AIFF**

```
z_min               : 4.65 mm
z_max               : 5.64 mm
z_travel_amplitude  : 0.99 mm

avg_abs_z_travel    : 6.54 mm/s

z_jarque-bera_jb    : 1625.43
z_jarque-bera_p     : 0.00e+00

z_lin_mod_est_slope: -0.17 mm/s
z_lin_mod_adj_R²    : 72 %

z_poly40_mod_adj_R²: 88 %

z_dft_ampl_thresh   : 0.010 mm
>=threshold_maxfreq: 22.00 Hz
```

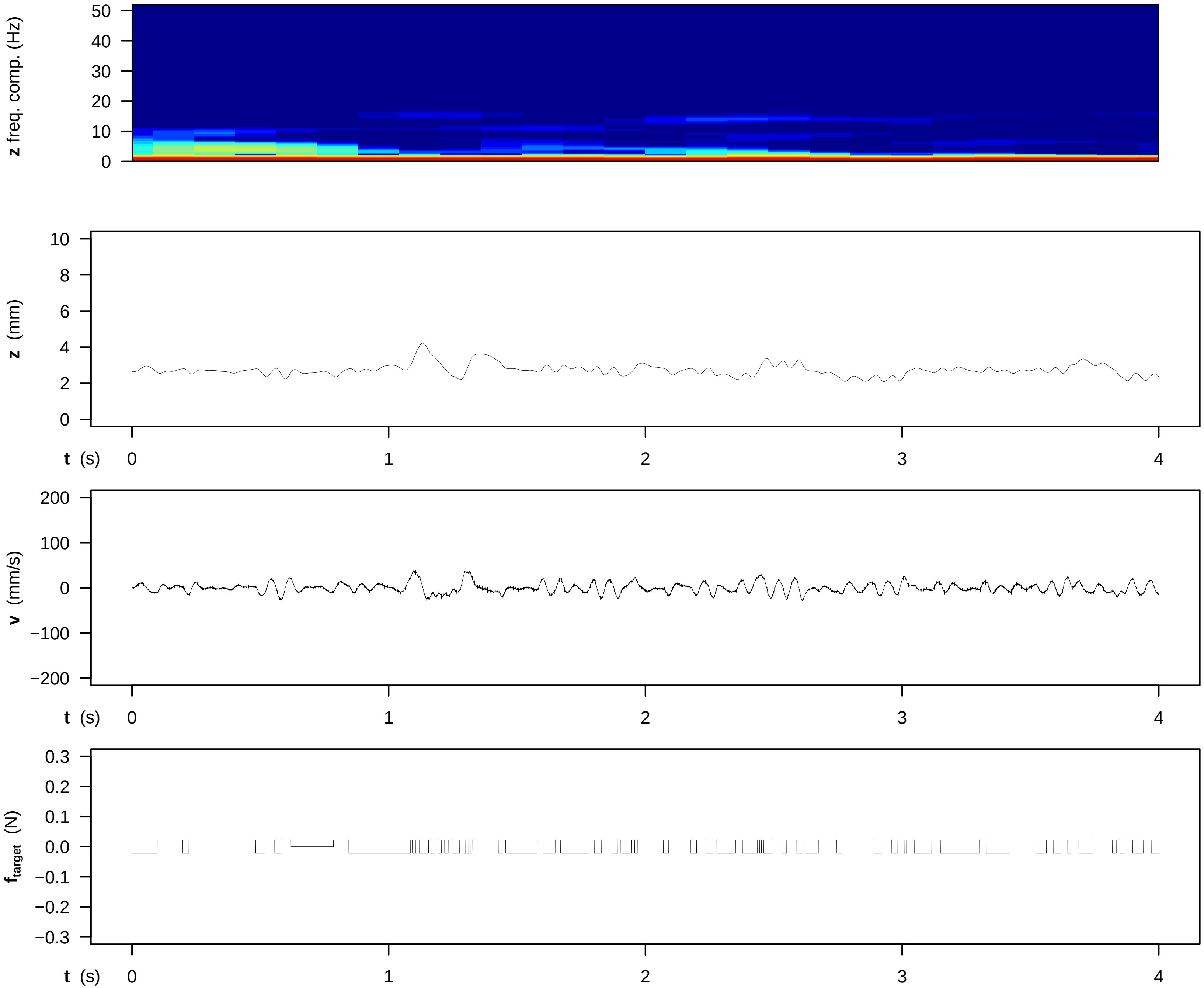

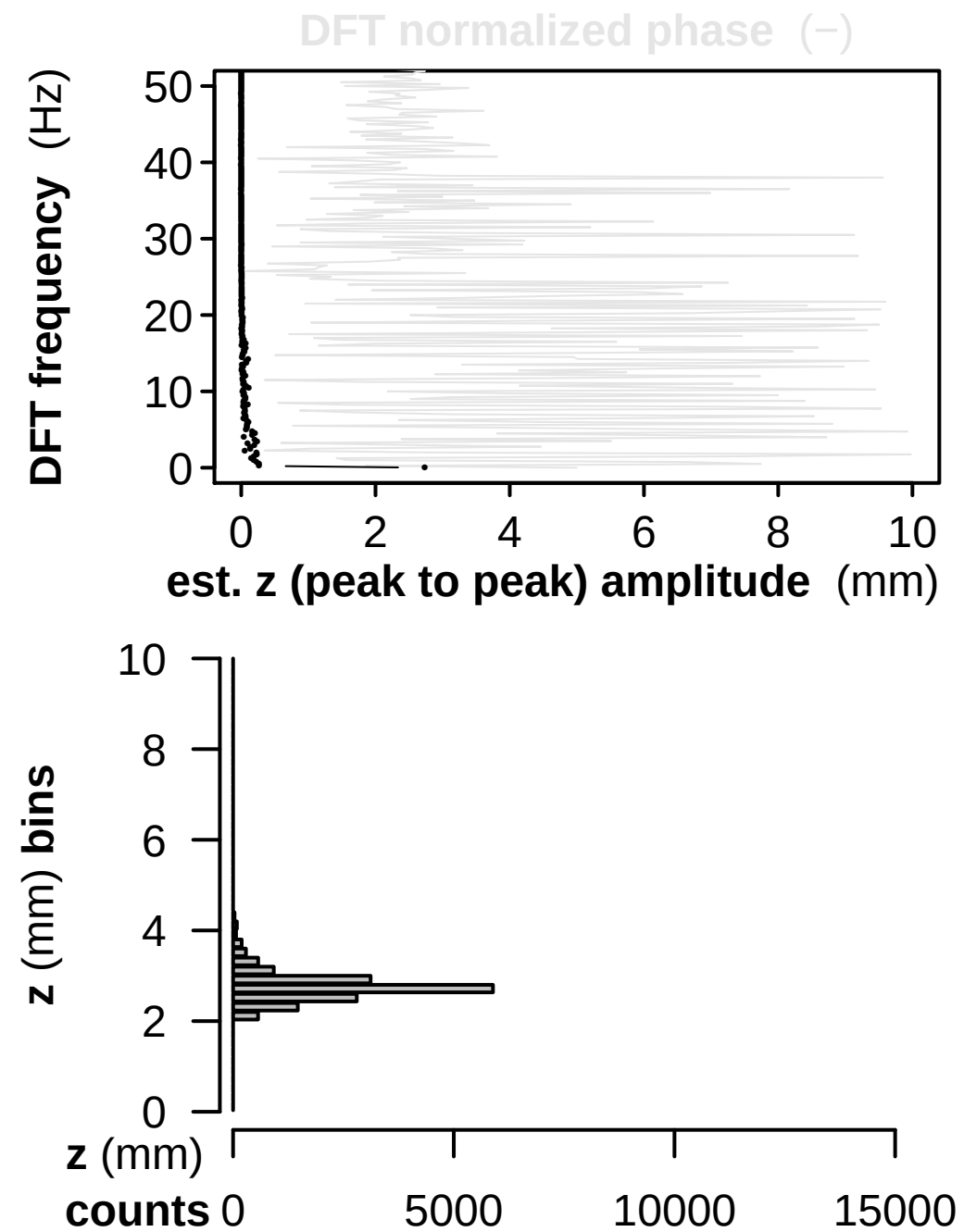

**SUBJECT 7 − RUN 06 − CONDITION 5,1**
**SC_180323_153758_0.AIFF**

```
z_min              : 2.09 mm
z_max              : 4.22 mm
z_travel_amplitude : 2.13 mm

avg_abs_z_travel   : 8.47 mm/s

z_jarque-bera_jb   : 10105.26
z_jarque-bera_p    : 0.00e+00

z_lin_mod_est_slope: −0.04 mm/s
z_lin_mod_adj_R²   : 2 %

z_poly40_mod_adj_R²: 47 %

z_dft_ampl_thresh  : 0.010 mm
>=threshold_maxfreq: 22.25 Hz
```

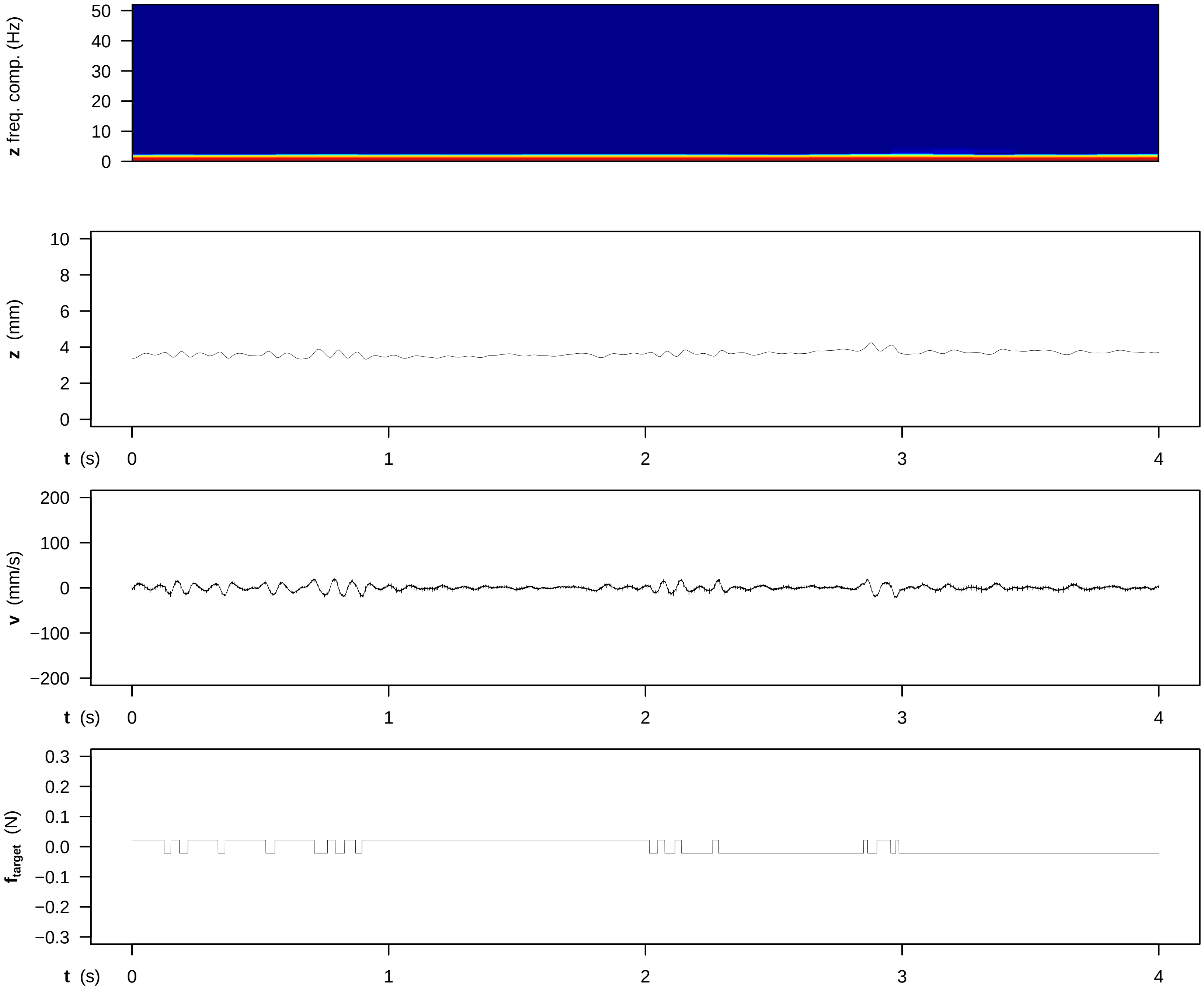

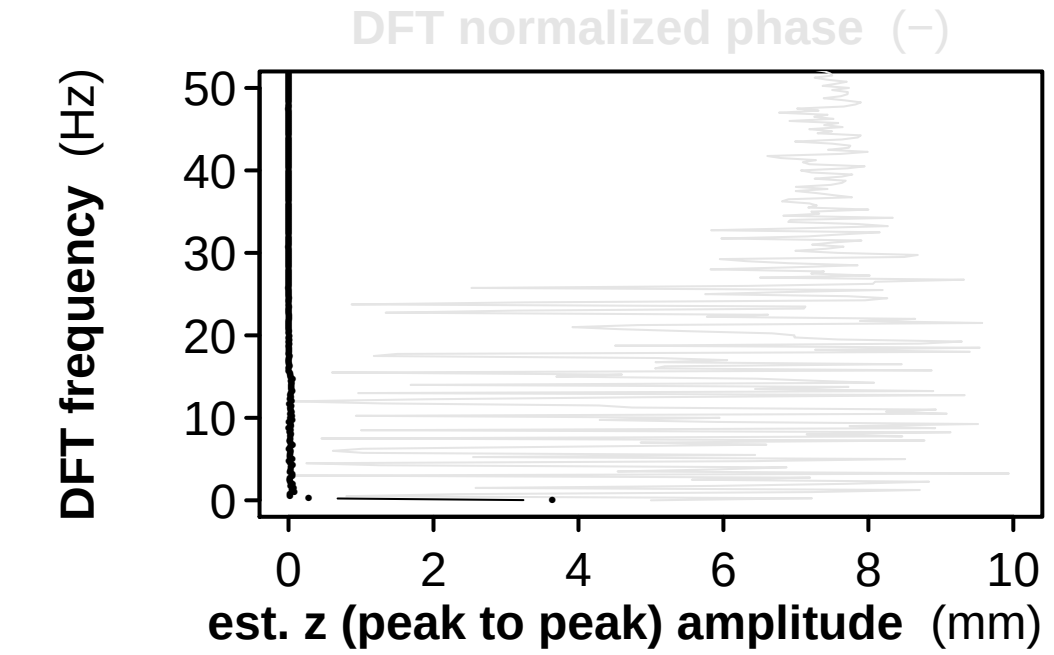

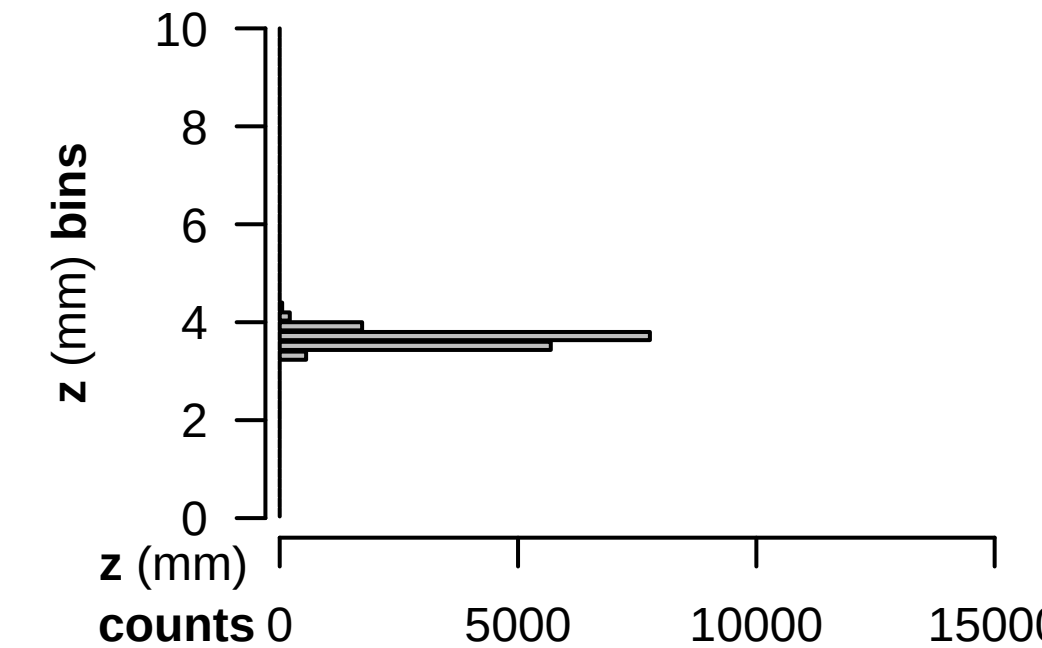

**SUBJECT 7 − RUN 07 − CONDITION 5,1**
**SC_180323_153826_0.AIFF**

```
z_min              : 3.33 mm
z_max              : 4.24 mm
z_travel_amplitude : 0.91 mm

avg_abs_z_travel   : 4.90 mm/s

z_jarque-bera_jb   : 1294.91
z_jarque-bera_p    : 0.00e+00

z_lin_mod_est_slope: 0.07 mm/s
z_lin_mod_adj_R²   : 34 %

z_poly40_mod_adj_R²: 62 %

z_dft_ampl_thresh  : 0.010 mm
>=threshold_maxfreq: 20.00 Hz
```

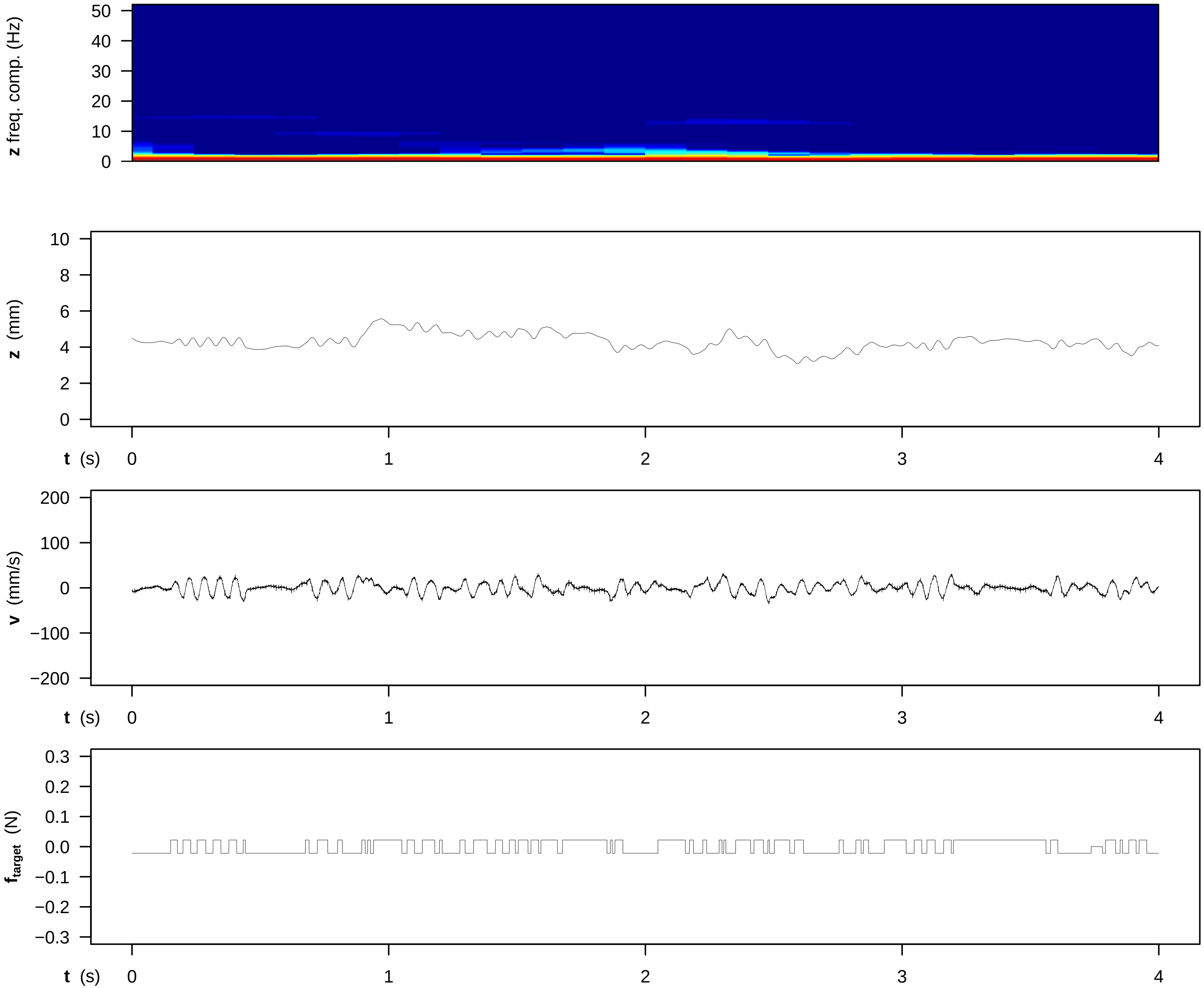

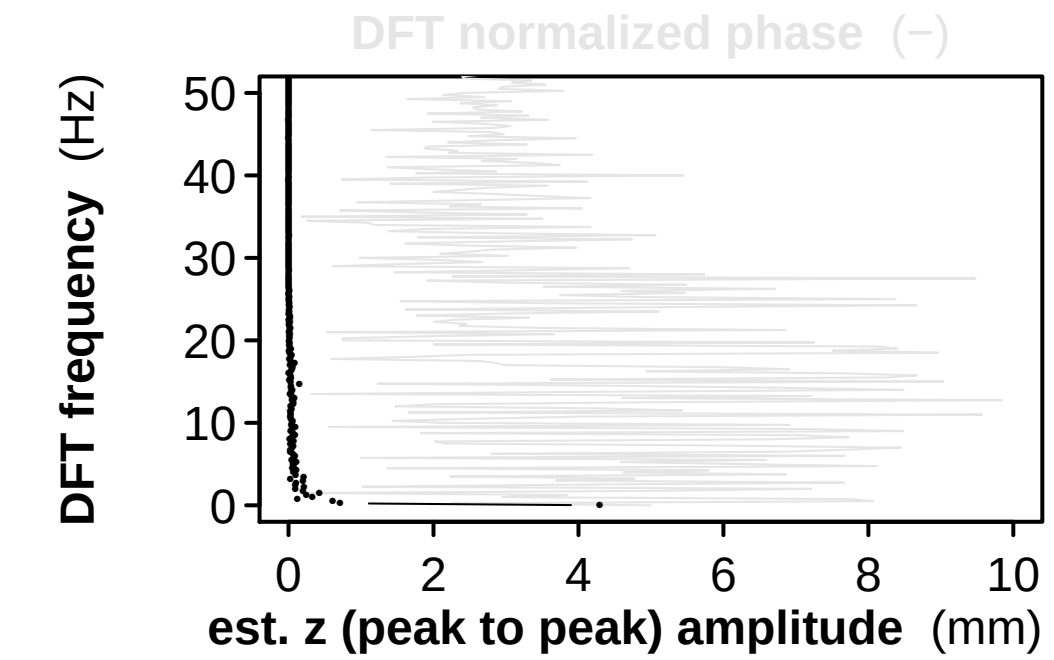

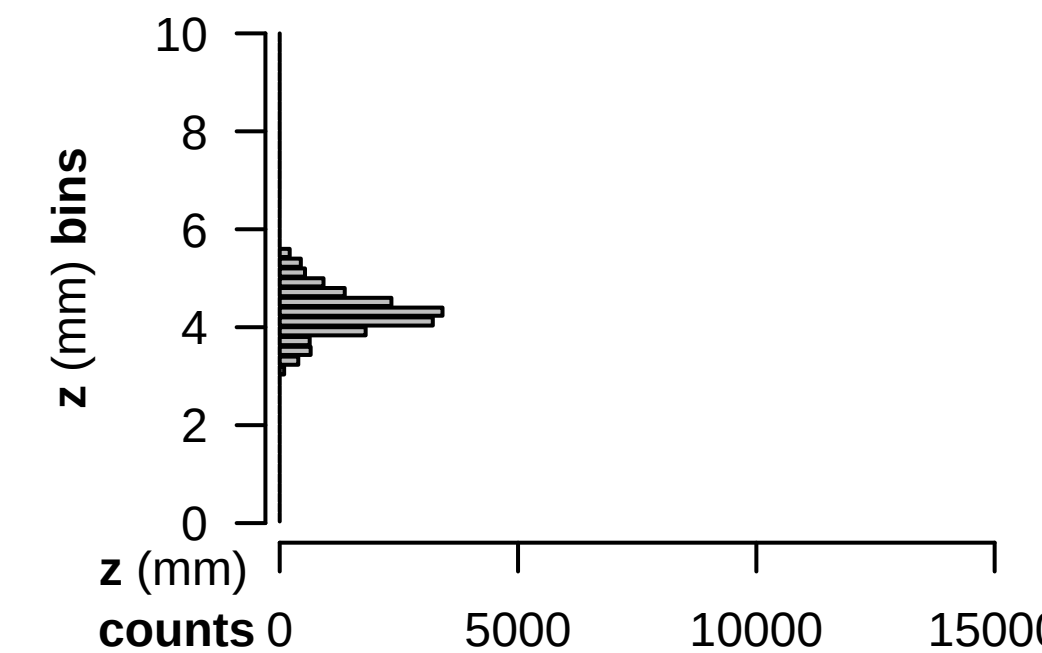

**SUBJECT 7 − RUN 09 − CONDITION 5,1**
**SC_180323_154023_0.AIFF**

z_min : 3.09 mm
z_max : 5.57 mm
z_travel_amplitude : 2.48 mm

avg_abs_z_travel : 9.47 mm/s

z_jarque-bera_jb : 111.73
z_jarque-bera_p : 0.00e+00

z_lin_mod_est_slope: −0.12 mm/s
z_lin_mod_adj_$R^2$ : 10 %

z_poly40_mod_adj_$R^2$: 80 %

z_dft_ampl_thresh : 0.010 mm
>=threshold_maxfreq: 25.25 Hz

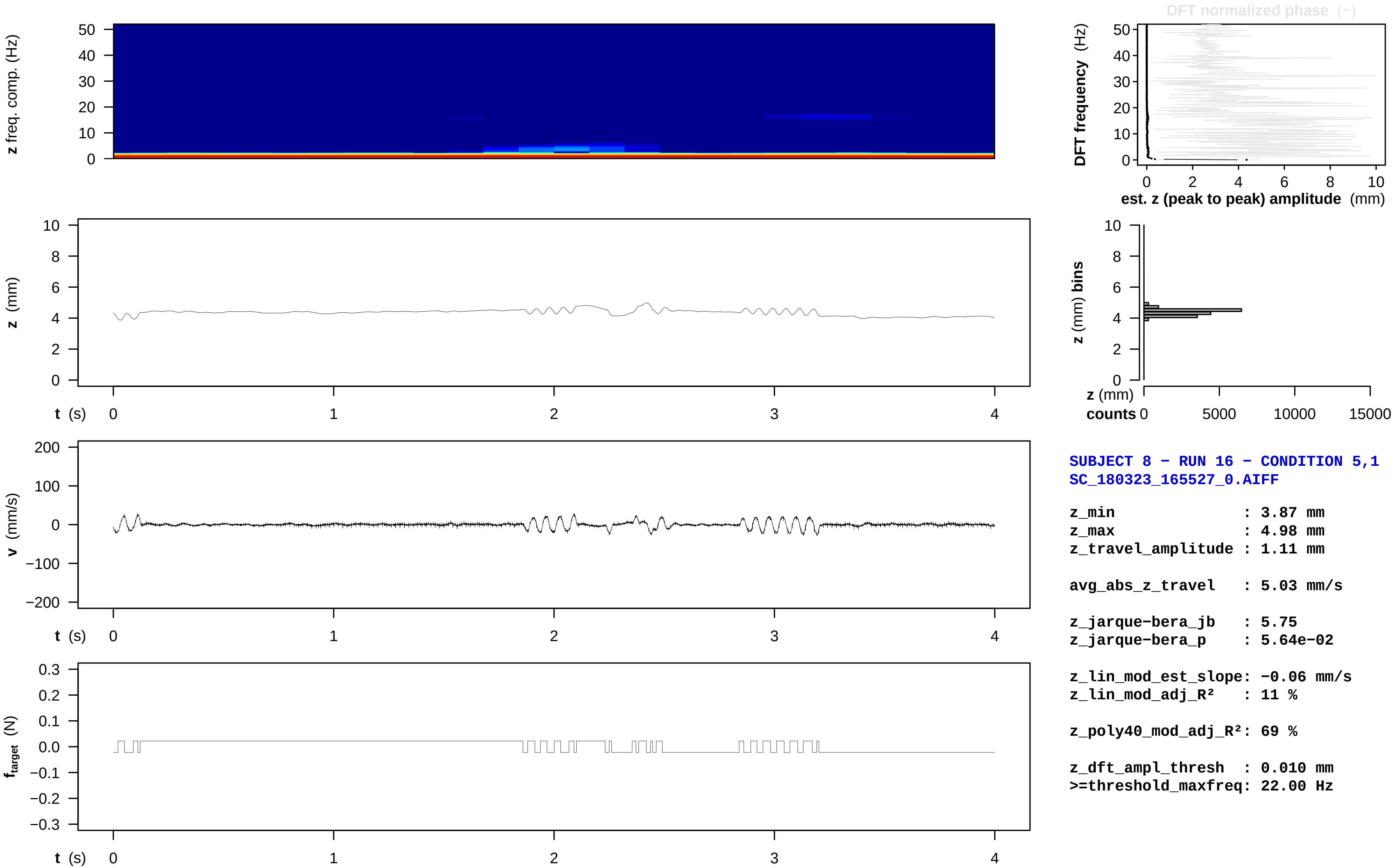

**SUBJECT 8 − RUN 16 − CONDITION 5,1**
**SC_180323_165527_0.AIFF**

```
z_min               : 3.87 mm
z_max               : 4.98 mm
z_travel_amplitude  : 1.11 mm

avg_abs_z_travel    : 5.03 mm/s

z_jarque-bera_jb    : 5.75
z_jarque-bera_p     : 5.64e-02

z_lin_mod_est_slope: -0.06 mm/s
z_lin_mod_adj_R²    : 11 %

z_poly40_mod_adj_R²: 69 %

z_dft_ampl_thresh   : 0.010 mm
>=threshold_maxfreq: 22.00 Hz
```

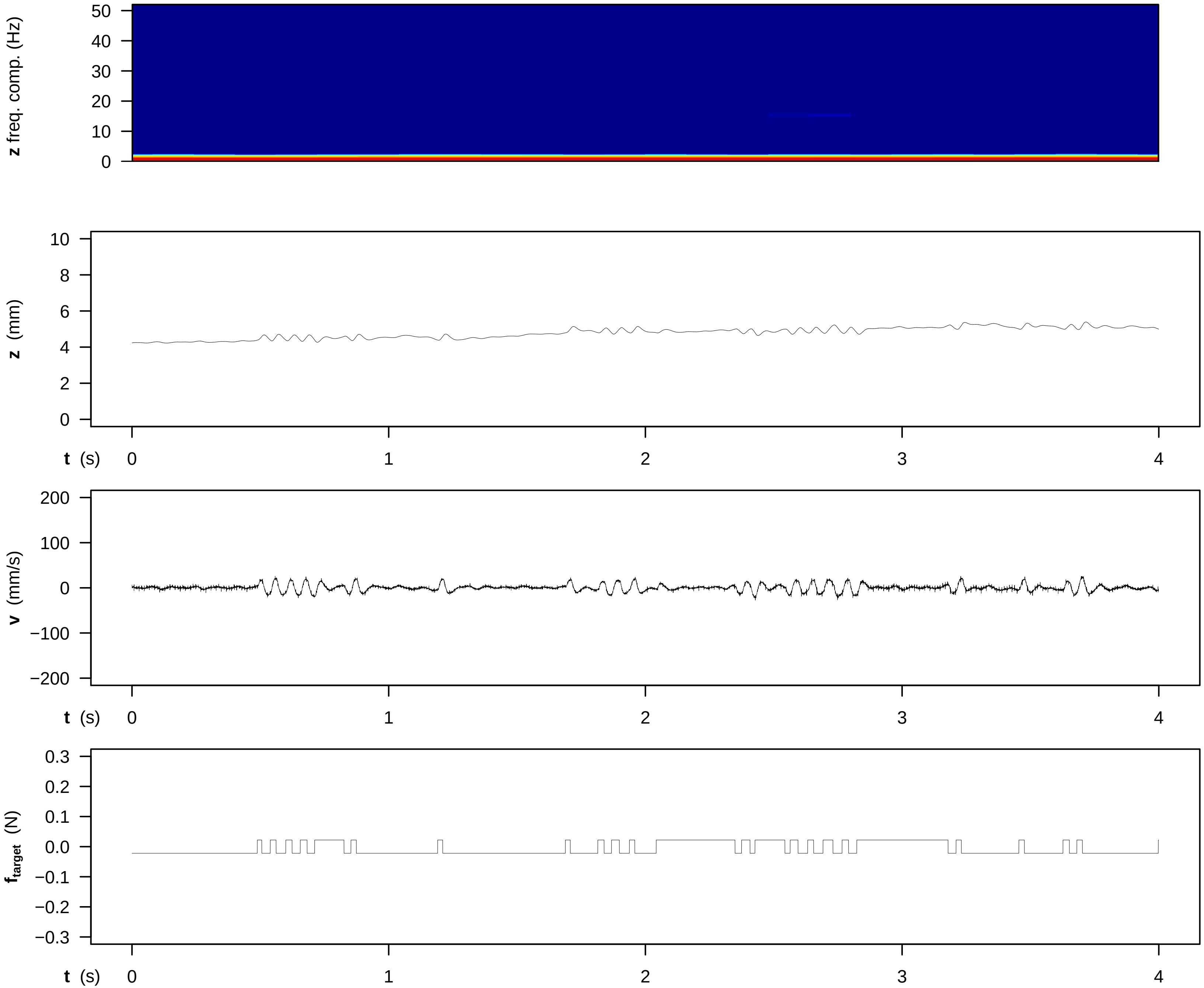

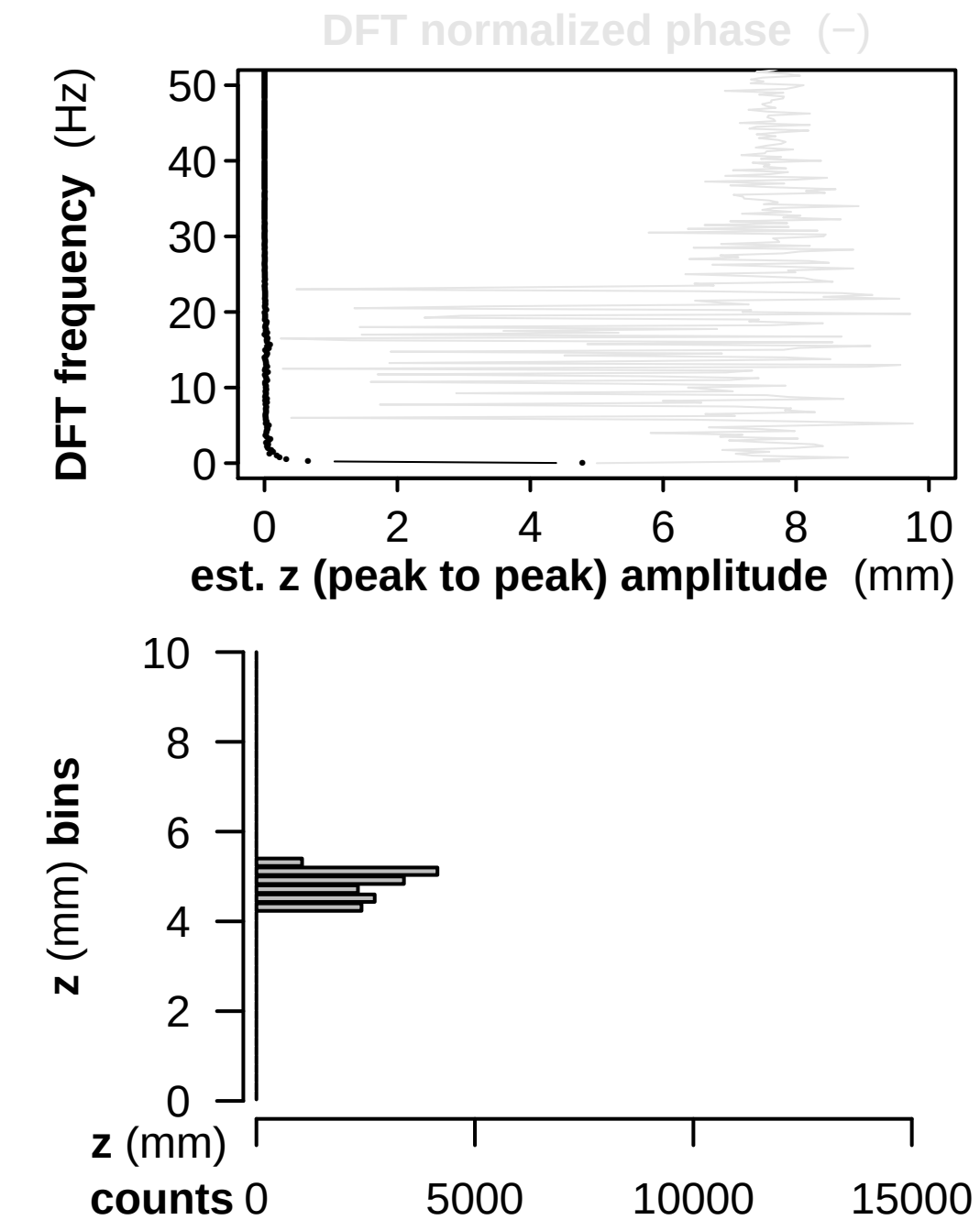

**SUBJECT 8 − RUN 18 − CONDITION 5,1**
**SC_180323_165610_0.AIFF**

```
z_min               : 4.22 mm
z_max               : 5.40 mm
z_travel_amplitude  : 1.17 mm

avg_abs_z_travel    : 5.83 mm/s

z_jarque-bera_jb    : 972.71
z_jarque-bera_p     : 0.00e+00

z_lin_mod_est_slope: 0.25 mm/s
z_lin_mod_adj_R²    : 86 %

z_poly40_mod_adj_R²: 92 %

z_dft_ampl_thresh   : 0.010 mm
>=threshold_maxfreq: 24.25 Hz
```

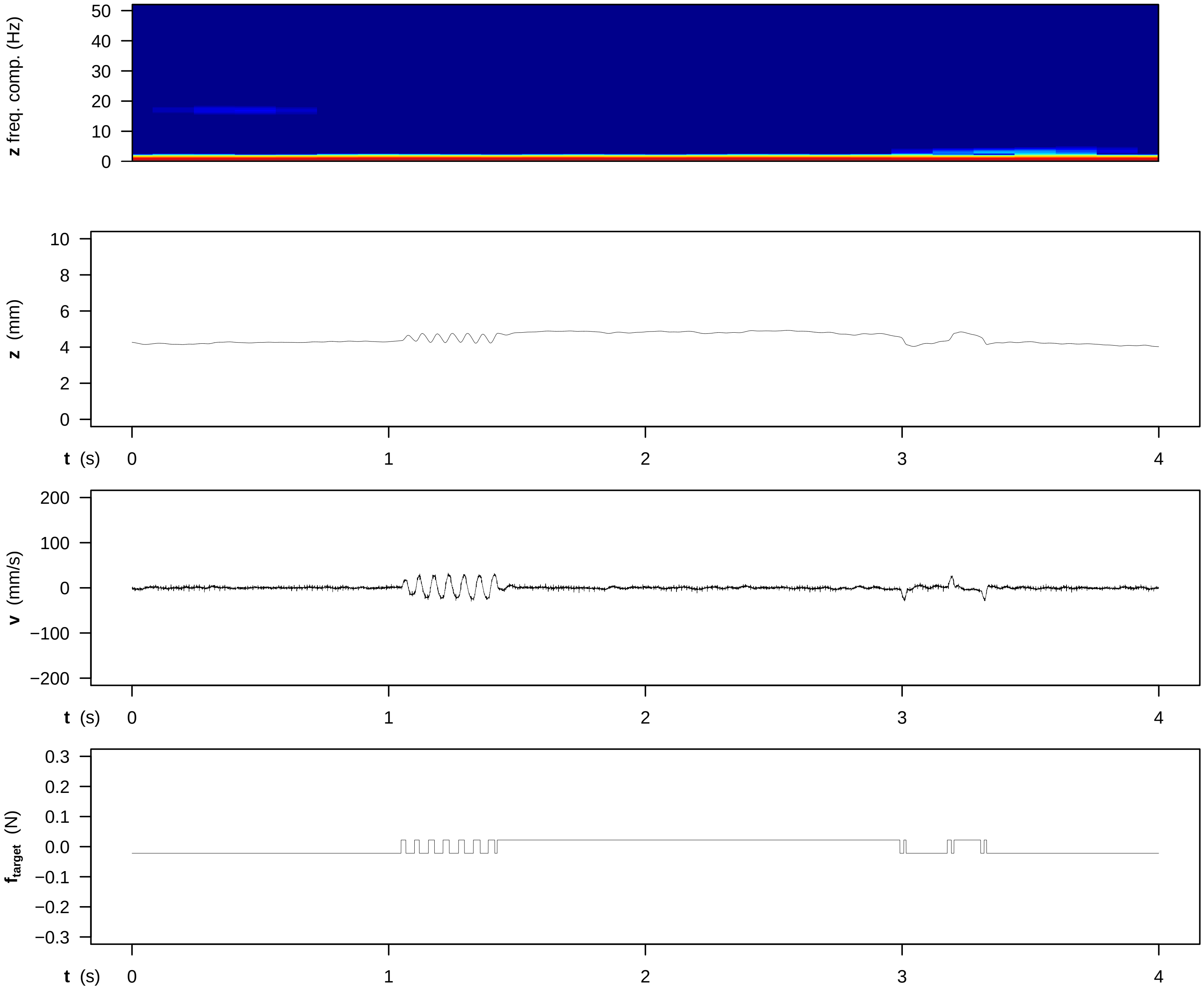

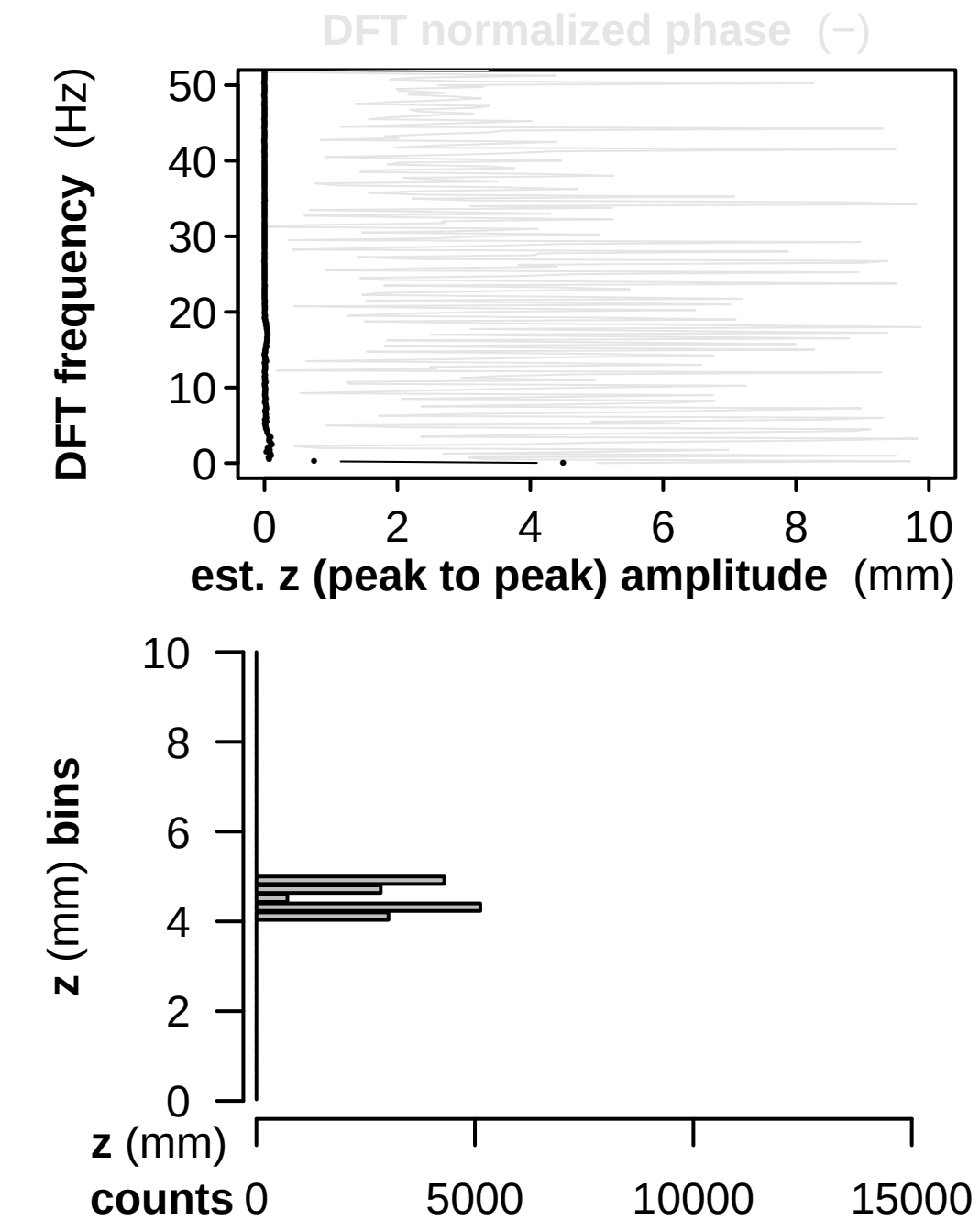

**SUBJECT 8 – RUN 27 – CONDITION 5,1**
**SC_180323_170427_0.AIFF**

```
z_min               : 4.02 mm
z_max               : 4.93 mm
z_travel_amplitude  : 0.91 mm

avg_abs_z_travel    : 4.32 mm/s

z_jarque-bera_jb    : 1909.49
z_jarque-bera_p     : 0.00e+00

z_lin_mod_est_slope: 0.02 mm/s
z_lin_mod_adj_R²    : 0 %

z_poly40_mod_adj_R²: 88 %

z_dft_ampl_thresh   : 0.010 mm
>=threshold_maxfreq: 19.50 Hz
```

```
~gcp = GhostfingerCTIprimitives.run (nightMode: false, signGrid: true, forceOutOfRangeAlerts: true);

// prepare for (de)activating experimental conditions & taking measurements
(

  /* OVERALL PARAMETERS */

  // general experimental parameters
  ~stimBase_mm   =  0.0; // stimulus base   (mm)
  ~stimHeight_mm = 10.0; // stimulus height (mm)

  // experimental parameters specific to the haptic conditions
  ~constPosForce_N =  0.25;
  ~constNegForce_N = -0.25;
  ~viscosity_N_per_mm_per_s     = -0.30 * 0.010;
  ~antiViscosity_N_per_mm_per_s =  0.08 * 0.010;
  ~markersBipolarAmpl_N = 0.022;
  ~markersInterval_mm   = 0.2;

  // experimental parameters specific to the musical control conditions
  ~tone_baseFreq_Hz = 440.0;
  ~tone_capFreq_Hz  = 8000.0;
  ~halftones_per_mm = 4.0;
  ~mm_per_octave    = 12.0 * (1.0 / ~halftones_per_mm);


  /* INDIVIDUAL HAPTIC CONDITIONS: ACTIVATION */

  ~runConstZeroForce_func =
  {
  };

  ~runConstPosForce_func =
  {
    ~gcp.setPrimitive (0, \monoForce);
    ~gcp.setParam     (0, \zBase_mm, ~stimBase_mm);
    ~gcp.setParam     (0, \zSize_mm, ~stimHeight_mm);
    ~gcp.setParam     (0, \zForce_N, ~constPosForce_N);
  };
```

```
~runConstNegForce_func =
{
  ~gcp.setPrimitive (0, \monoForce);
  ~gcp.setParam     (0, \zBase_mm, ~stimBase_mm);
  ~gcp.setParam     (0, \zSize_mm, ~stimHeight_mm);
  ~gcp.setParam     (0, \zForce_N, ~constNegForce_N);
};

~runViscosity_func =
{
  ~gcp.setPrimitive (0, \brickwalledDashpot);
  ~gcp.setParam     (0, \zBase_mm, ~stimBase_mm);
  ~gcp.setParam     (0, \zSize_mm, ~stimHeight_mm);
  ~gcp.setParam     (0, \zViscousDamping_N_per_mm_per_s, ~viscosity_N_per_mm_per_s);
};

~runAntiViscosity_func =
{
  ~gcp.setPrimitive (0, \brickwalledDashpot);
  ~gcp.setParam     (0, \zBase_mm, ~stimBase_mm);
  ~gcp.setParam     (0, \zSize_mm, ~stimHeight_mm);
  ~gcp.setParam     (0, \zViscousDamping_N_per_mm_per_s, ~antiViscosity_N_per_mm_per_s);
};

~runMarkerPrimitives_func =
{
  var n_primitives = ~stimHeight_mm / ~markersInterval_mm,
      markerBase_mm = nil,
      markerHeight_mm = ~markersInterval_mm - 0.001;

  for (0, n_primitives - 1,
  { arg i;

    markerBase_mm = ~stimBase_mm + (i * ~markersInterval_mm);
    ~gcp.setPrimitive (i, \monoForce);
    ~gcp.setParam     (i, \zBase_mm, markerBase_mm);
    ~gcp.setParam     (i, \zSize_mm, markerHeight_mm);
```

```
    if ((i%2) == 1,
    { ~gcp.setParam (i, \zForce_N, ~markersBipolarAmpl_N);
    },
    // else:
    { ~gcp.setParam (i, \zForce_N, ~markersBipolarAmpl_N * -1.0);
    });

  });

};


/* INDIVIDUAL HAPTIC CONDITIONS: DE-ACTIVATION */

~stopNothing_func =
{
};

~stopPrimitiveZero_func =
{
  ~gcp.setPrimitive (0, nil);
};

~stopMarkerPrimitives_func =
{
  var n_primitives = ~stimHeight_mm / ~markersInterval_mm;

  for (0, n_primitives - 1,
  { arg i;

    ~gcp.setPrimitive (i, nil);
  });

};
```

```
/* INDIVIDUAL MUSICAL CONTROL CONDITIONS: ACTIVATION */

~sgprocs = SoundGeneratingProcesses.new (~gcp.ct.postFingertip_group, stereoOutAtt_db: -6);

~noiseWave_synth = nil;
~sineWave_synth  = nil;

~tone_freq_Hz = nil;

~run_noiseWave_func =
{
  if (~noiseWave_synth == nil) // sgproc not running?
  {
    // start sound
    ~noiseWave_synth = ~sgprocs.new_amplModBPnoise_synth
    ( \ampl_nrm,     -15.dbamp,
      \freq_hz,      220,
      \bandwidth_hz, 1000
    );
  };
};

~run_zPosToPitch_sineWave_func =
{
  if (~sineWave_synth == nil) // sgproc not running?
  {
    // start sound
    ~sineWave_synth = ~sgprocs.new_amplModSine_synth
    ( \ampl_nrm, -15.dbamp,
      \freq_hz,  0
    );

    // start modulating fingertip DOF: zPos_rel_mm
    ~gcp.setDOF      (0, \zPos_rel_mm);
    ~gcp.setDOFparam (0, \zBase_mm, ~stimBase_mm);
    ~gcp.setDOFparam (0, \handler,
      { arg value;

        ~tone_freq_Hz = ~tone_baseFreq_Hz * (2 ** (value / ~mm_per_octave));
```

```
        if (~tone_freq_Hz > ~tone_capFreq_Hz)
        { ~tone_freq_Hz = ~tone_capFreq_Hz;
        };

        ~sineWave_synth.set (\freq_hz, ~tone_freq_Hz);
      }
    );

  };
};


/* INDIVIDUAL MUSICAL CONTROL CONDITIONS: DE-ACTIVATION */

~stop_noiseWave_func =
{
  if (~noiseWave_synth != nil) // sgproc running?
  {
    // terminate sound
    ~noiseWave_synth.set (\gate_bit, 0); // completes output, frees synth
    ~noiseWave_synth = nil;
  };
};

~stop_zPosToPitch_sineWave_func =
{
  if (~sineWave_synth != nil) // sgproc running?
  {
    // terminate modulating fingertip DOF
    ~gcp.setDOF (0, nil);

    // terminate sound
    ~sineWave_synth.set (\gate_bit, 0); // completes output, frees synth
    ~sineWave_synth = nil;
  };
};
```

```
/* OVERALL (DE-)ACTIVATION OF EXPERIMENTAL CONDITIONS */

~run_hapticConditions =
[ ~runConstZeroForce_func,
  ~runConstPosForce_func,
  ~runConstNegForce_func,
  ~runViscosity_func,
  ~runAntiViscosity_func,
  ~runMarkerPrimitives_func
];

~stop_hapticConditions =
[ ~stopNothing_func,
  ~stopPrimitiveZero_func,
  ~stopPrimitiveZero_func,
  ~stopPrimitiveZero_func,
  ~stopPrimitiveZero_func,
  ~stopMarkerPrimitives_func
];

~run_musicalControlConditions =
[ ~run_noiseWave_func,
  ~run_zPosToPitch_sineWave_func
];

~stop_musicalControlConditions =
[ ~stop_noiseWave_func,
  ~stop_zPosToPitch_sineWave_func
];

~startCondition_func =
{
  arg i_haptic, i_musicalControl;

  ~run_hapticConditions         [i_haptic]        .value ( );
  ~run_musicalControlConditions [i_musicalControl].value ( );
};
```

```
~stopCondition_func =
{
  arg i_haptic, i_musicalControl;

  ~stop_hapticConditions         [i_haptic]         .value ( );
  ~stop_musicalControlConditions [i_musicalControl].value ( );
};


/* MEASUREMENT TAKING */

~recorder = nil;
~track_zMinMax_synth = nil;

~takeMeasurement_func =
{
  if (~recorder == nil) // no measurement still underway?
  {
    // start recording input  signals:  z (mm),   v_z (mm/s)
    //                & output signals:  f (N),    1 of the (identical) stereo channels
    Server.default = Server.internal;
    ~recorder = MultiChanRecorder ([~gcp.ct.monitorBuses]++[Bus.new (\audio, 12, 1)]);
    ~recorder.prepareForRecord ( );
    ~recorder.record ( ); // NB: recordings store a number of spurious, leading zero-level samples

    // also start tracking the min. and max. z (mm) input values
    ~track_zMinMax_synth = SynthDef
    ( "track_zMinMax",
      {
        arg print_trig = 0;

        var z_mm    = In.ar (~gcp.ct.zPos_mm_continuous_bus),
            zMin_mm = RunningMin.ar (z_mm),
            zMax_mm = RunningMax.ar (z_mm);

        Poll.ar
        ( trig:  K2A.ar (print_trig),
          in:    [ zMin_mm,  zMax_mm],
          label: [\zMin_mm, \zMax_mm]
```

```
          ); // NB: comma-separated output: here impossible to obtain
        }
      ).play (~gcp.ct.postFingertip_group, addAction: \addToTail);

      // then, after a fixed time interval: complete the measurement
      fork
      {
        // wait during the fixed time interval
        (4.0).wait;

        // print the min. & max. z (mm) just seen
        ~track_zMinMax_synth.set (\print_trig, 1);

        (0.7).wait; // empirically yielded recordings of > 4 s duration + actual min./max. display

        // stop tracking the min. & max. z (mm)
        ~track_zMinMax_synth.free ( );

        // stop I/O signals recording
        ~recorder.stop ( );
        ~recorder = nil;
      };
    };

  }; // (end of ~takeMeasurement_func ( ))

)
```

```
// briefing (some key points):
// - "Simple task during measurements: Hold still fingertip +/- 0.5 cm above surface."
// - "Before each measurement: asked to move fingertip up/down."
// - "Then: closing the eyes + concentration -> Stillness measurement."
// - "There will be many measurements. Your precision & effort is crucial."


// dry run, with explanation:
~startCondition_func.value (0,1); // "zero force + position-to-pitch control"
~stopCondition_func.value  (0,1);
~startCondition_func.value (1,1); // "upward force + position-to-pitch control"
~stopCondition_func.value  (1,1);
~startCondition_func.value (2,0); // "downward force + no control (noise)"
~stopCondition_func.value  (2,0);
~startCondition_func.value (3,0); // "viscosity + no control (noise)"
~stopCondition_func.value  (3,0);
~startCondition_func.value (4,0); // "anti-viscosity + no control (noise)"
~stopCondition_func.value  (4,0);
~startCondition_func.value (5,1); // "spatial markers + position-to-pitch control"
~stopCondition_func.value  (5,1);


// randomized repeated measures (EXAMPLE - this code: used for subject #1):

~startCondition_func.value (2,1);
~takeMeasurement_func.value ( );
~stopCondition_func.value  (2,1);

~startCondition_func.value (5,1);
~takeMeasurement_func.value ( );
~stopCondition_func.value  (5,1);

~startCondition_func.value (4,0);
~takeMeasurement_func.value ( );
~stopCondition_func.value  (4,0);

~startCondition_func.value (4,1);
~takeMeasurement_func.value ( );
~stopCondition_func.value  (4,1);
```

```
~startCondition_func.value (5,0);
~takeMeasurement_func.value ( );
~stopCondition_func.value  (5,0);

~startCondition_func.value (5,0);
~takeMeasurement_func.value ( );
~stopCondition_func.value  (5,0);

~startCondition_func.value (1,1);
~takeMeasurement_func.value ( );
~stopCondition_func.value  (1,1);

~startCondition_func.value (1,1);
~takeMeasurement_func.value ( );
~stopCondition_func.value  (1,1);

~startCondition_func.value (4,0);
~takeMeasurement_func.value ( );
~stopCondition_func.value  (4,0);

~startCondition_func.value (0,0);
~takeMeasurement_func.value ( );
~stopCondition_func.value  (0,0);

~startCondition_func.value (1,1);
~takeMeasurement_func.value ( );
~stopCondition_func.value  (1,1);

~startCondition_func.value (4,0);
~takeMeasurement_func.value ( );
~stopCondition_func.value  (4,0);

~startCondition_func.value (3,1);
~takeMeasurement_func.value ( );
~stopCondition_func.value  (3,1);

~startCondition_func.value (5,1);
~takeMeasurement_func.value ( );
```

```
~stopCondition_func.value  (5,1);

~startCondition_func.value (2,1);
~takeMeasurement_func.value ( );
~stopCondition_func.value  (2,1);

~startCondition_func.value (4,1);
~takeMeasurement_func.value ( );
~stopCondition_func.value  (4,1);

~startCondition_func.value (3,1);
~takeMeasurement_func.value ( );
~stopCondition_func.value  (3,1);

~startCondition_func.value (3,0);
~takeMeasurement_func.value ( );
~stopCondition_func.value  (3,0);

~startCondition_func.value (4,1);
~takeMeasurement_func.value ( );
~stopCondition_func.value  (4,1);

~startCondition_func.value (2,1);
~takeMeasurement_func.value ( );
~stopCondition_func.value  (2,1);

~startCondition_func.value (1,0);
~takeMeasurement_func.value ( );
~stopCondition_func.value  (1,0);

~startCondition_func.value (3,1);
~takeMeasurement_func.value ( );
~stopCondition_func.value  (3,1);

~startCondition_func.value (0,1);
~takeMeasurement_func.value ( );
~stopCondition_func.value  (0,1);

~startCondition_func.value (5,1);
```

```
~takeMeasurement_func.value ( );
~stopCondition_func.value  (5,1);

~startCondition_func.value (1,0);
~takeMeasurement_func.value ( );
~stopCondition_func.value  (1,0);

~startCondition_func.value (0,1);
~takeMeasurement_func.value ( );
~stopCondition_func.value  (0,1);

~startCondition_func.value (3,0);
~takeMeasurement_func.value ( );
~stopCondition_func.value  (3,0);

~startCondition_func.value (5,0);
~takeMeasurement_func.value ( );
~stopCondition_func.value  (5,0);

~startCondition_func.value (1,0);
~takeMeasurement_func.value ( );
~stopCondition_func.value  (1,0);

~startCondition_func.value (3,0);
~takeMeasurement_func.value ( );
~stopCondition_func.value  (3,0);

~startCondition_func.value (2,0);
~takeMeasurement_func.value ( );
~stopCondition_func.value  (2,0);

~startCondition_func.value (2,0);
~takeMeasurement_func.value ( );
~stopCondition_func.value  (2,0);

~startCondition_func.value (2,0);
~takeMeasurement_func.value ( );
~stopCondition_func.value  (2,0);
```

```
~startCondition_func.value (0,0);
~takeMeasurement_func.value ( );
~stopCondition_func.value  (0,0);

~startCondition_func.value (0,1);
~takeMeasurement_func.value ( );
~stopCondition_func.value  (0,1);

~startCondition_func.value (0,0);
~takeMeasurement_func.value ( );
~stopCondition_func.value  (0,0);
```

```
~gcp.stop ( );
```

# Input data & outcomes of *z travel amplitude (mm)* normality testing

| condition00 | condition10 | condition20 | condition30 | condition40 | condition50 |
|---|---|---|---|---|---|
| 1.33 | 1.21 | 0.48 | 1.02 | 1.49 | 0.73 |
| 0.48 | 0.70 | 0.72 | 0.80 | 1.55 | 1.02 |
| 2.02 | 0.74 | 1.13 | 0.58 | 2.22 | 0.95 |
| 0.27 | 1.04 | 0.33 | 0.91 | 1.16 | 0.75 |
| 0.67 | 0.78 | 0.51 | 0.85 | 0.32 | 0.18 |
| 0.83 | 0.29 | 0.28 | 0.29 | 0.59 | 1.14 |
| 2.89 | 1.93 | 1.73 | 0.95 | 1.91 | 2.46 |
| 1.13 | 0.67 | 1.67 | 1.29 | 1.62 | 4.15 |
| 1.72 | 1.44 | 1.02 | 0.93 | 1.40 | 2.13 |
| 2.17 | 1.01 | 0.70 | 0.98 | 1.08 | 1.34 |
| 1.51 | 0.99 | 0.91 | 0.35 | 0.59 | 2.03 |
| 2.67 | 1.05 | 1.23 | 0.85 | 0.76 | 0.88 |
| 1.28 | 0.94 | 1.49 | 0.76 | 1.41 | 1.13 |
| 1.04 | 1.50 | 1.86 | 0.99 | 0.89 | 2.06 |
| 1.70 | 0.80 | 1.26 | 0.60 | 1.50 | 2.46 |
| 0.94 | 0.21 | 1.50 | 1.97 | 1.57 | 1.10 |
| 0.48 | 1.55 | 0.83 | 0.79 | 1.05 | 1.44 |
| 0.42 | 1.01 | 2.58 | 1.96 | 1.73 | 1.49 |
| 2.60 | 2.69 | 0.74 | 1.06 | 3.61 | 3.64 |
| 2.13 | 0.76 | 2.77 | 2.26 | 2.23 | 2.27 |
| 0.71 | 0.60 | 1.78 | 0.78 | 2.10 | 2.47 |
| 1.48 | 0.87 | 0.24 | 0.77 | 3.56 | 1.22 |
| 0.96 | 0.75 | 0.98 | 0.70 | 0.73 | 1.19 |
| 1.41 | 0.58 | 2.26 | 2.29 | 1.01 | 1.35 |

| condition01 | condition11 | condition21 | condition31 | condition41 | condition51 |
|---|---|---|---|---|---|
| 1.71 | 0.29 | 0.29 | 0.65 | 1.00 | 1.67 |
| 0.90 | 0.51 | 1.06 | 0.62 | 0.81 | 1.05 |
| 1.33 | 0.47 | 1.29 | 0.43 | 1.77 | 0.71 |
| 0.45 | 0.24 | 0.46 | 0.29 | 0.21 | 1.28 |
| 0.33 | 0.37 | 0.42 | 0.21 | 0.32 | 0.63 |
| 0.69 | 0.13 | 0.53 | 0.29 | 0.56 | 0.91 |
| 1.10 | 1.48 | 2.45 | 1.19 | 1.63 | 1.01 |
| 0.81 | 1.75 | 1.91 | 0.66 | 1.30 | 3.14 |
| 1.07 | 1.43 | 2.40 | 1.29 | 1.39 | 1.52 |
| 0.97 | 2.15 | 1.03 | 0.40 | 0.78 | 0.96 |
| 0.92 | 0.61 | 0.92 | 0.35 | 0.78 | 0.91 |
| 0.75 | 1.47 | 0.82 | 1.49 | 0.79 | 0.91 |
| 1.14 | 0.87 | 1.47 | 0.78 | 1.41 | 1.53 |
| 1.81 | 2.64 | 0.99 | 1.05 | 1.26 | 1.17 |
| 2.06 | 1.27 | 1.01 | 1.02 | 0.75 | 1.48 |
| 1.43 | 0.09 | 3.00 | 0.89 | 1.35 | 1.00 |
| 1.27 | 0.87 | 1.41 | 2.20 | 1.07 | 0.72 |
| 0.60 | 0.21 | 2.80 | 0.46 | 1.10 | 0.99 |
| 2.27 | 0.17 | 1.96 | 0.50 | 1.27 | 2.13 |
| 3.52 | 1.44 | 1.30 | 1.44 | 1.58 | 0.91 |
| 1.50 | 1.19 | 2.25 | 1.45 | 1.57 | 2.48 |
| 0.57 | 0.46 | 0.32 | 0.33 | 0.77 | 1.11 |
| 0.93 | 1.38 | 0.81 | 0.67 | 0.91 | 1.17 |
| 1.36 | 0.65 | 1.10 | 1.09 | 1.16 | 0.91 |

| condition | Shapiro-Wilk_normality_test_W | Shapiro-Wilk_normality_test_p |
|---|---|---|
| condition00 | 0.954 | 0.333 |
| condition10 | 0.879 | 0.008 |
| condition20 | 0.949 | 0.255 |
| condition30 | 0.821 | 0.001 |
| condition40 | 0.895 | 0.017 |
| condition50 | 0.903 | 0.024 |

**← In green: if p >= 0.05.**
**← In red: if p < 0.05.**

| condition | Shapiro-Wilk_normality_test_W | Shapiro-Wilk_normality_test_p |
|---|---|---|
| condition01 | 0.874 | 0.006 |
| condition11 | 0.916 | 0.049 |
| condition21 | 0.926 | 0.080 |
| condition31 | 0.913 | 0.041 |
| condition41 | 0.971 | 0.703 |
| condition51 | 0.796 | 0.000 |